\documentclass[12pt,a4paper,titlepage,twoside,fleqn]{book}
\usepackage{etex,etoolbox}
\usepackage{environ}
\usepackage[inference]{semantic}
\usepackage{setspace}
\usepackage[lmargin=4cm,rmargin=2cm,tmargin=3cm,bmargin=3cm]{geometry}%
\usepackage{lmodern}
\usepackage{amssymb}
\usepackage[fleqn]{amsmath}
\usepackage{dcpic,pictex}
\usepackage{booktabs}
\usepackage{multirow}
\usepackage{enumitem} 
\usepackage{amsthm}
\usepackage{stmaryrd}
\usepackage[color]{circusPP}
\usepackage[british]{babel}
\usepackage[numbers,sort&compress,square,comma]{natbib}
\usepackage{hyperref}
\usepackage{comment}
\usepackage[withpage]{acronym}
\usepackage{graphicx}
\usepackage[labelformat=simple]{subcaption}

\usepackage[capitalise,noabbrev]{cleveref}
\usepackage{tikz}
\usepackage{pdflscape}
\usepackage{booktabs}
\usepackage{xparse}
\usepackage{ifthen}
\usepackage{wasysym}
\usepackage{refcount}
\usepackage{currfile} 
\usepackage{xcolor}
\definecolor{One}{RGB}{112 102 81}
\definecolor{Two}{RGB}{122 39 57}
\definecolor{Three}{RGB}{37 39 99}
\definecolor{Four}{RGB}{39 122 80}
\definecolor{Five}{RGB}{112 106 36}
\usepackage{refcount}
\usepackage{stringstrings}
\usepackage{fmtcount}
 
\def\IINew{\hbox{\texttt{\slshape I \kern -9.5pt I}}}

\makeatletter
\newcommand*{\shifttext}[2]{%
  \settowidth{\@tempdima}{#2}%
  \makebox[\@tempdima]{\hspace*{#1}#2}%
}
\makeatother
\newcommand{\ptext}[1]{\tag*{\normalfont\{\text{#1}\}}} 

\newcommand{\CircusTime}{{\sf\slshape Circus Time}}

\NewDocumentCommand\syncat{m+g}{%
	\IfNoValueTF{#2}
	{\ensuremath{\langle{#1}\rangle}}
	{\ensuremath{\langle{#1}\rangle^{#2}}}%
}

\newcommand*\circled[1]{\tikz[baseline=(char.base)]{
  \node[shape=circle,draw,inner sep=1pt] (char) {#1};}}
\newcommand\circledIn[2]{\circled{$\in$}^{#1}_{#2}}

\newcommand\extchoiceONE\extchoice

\newcommand{\hatop}[1]{\expandafter\hat#1}

\def\seqA{\mathrel{\Semi_{\mathcal{A}}}}

\def\seqRac{\mathrel{\Semi_{\mathcal{R}ac}}}

\newcommand{\IIrea}{\IINew_{rea}}

\newcommand{\IIRac}{\IINew_{\mathbf{RAD}}}
\newcommand{\IIRnew}{\IINew_{\mathbf{AP}}}

\def\seqRac{\mathrel{\mathbf{\circseq_{\mathcal{R}ac}}}}

\newcommand{\IIAPBMH}{\IINew_{\mathbf{A}}}
\newcommand{\IIA}{\IINew_{\mathcal{A}}}
\newcommand{\IID}{\IINew_{\mathcal{D}}}
\newcommand{\IIR}{\IINew_{\mathcal{R}}}
\newcommand{\botD}{\bot_{\mathcal{D}}}
\newcommand{\topD}{\top_{\mathcal{D}}}

\newcommand{\IIDac}{\IINew_{\mathcal{D}ac}}
\newcommand{\botDac}{\bot_{\mathcal{D}ac}}
\newcommand{\topDac}{\top_{\mathcal{D}ac}}
\def\seqAPBMH{\mathrel{\Semi_{\mathbf{A}}}}
\def\seqDac{\mathrel{\Semi_{\mathcal{D}ac}}}

\def\refinedbyDac{\mathrel{\sqsubseteq_{\mathcal{D}}}}
\def\assignDac{\mathrel{:=_{\mathcal{D}ac}}}
\def\sqcupDac{\mathrel{\sqcup_{\mathcal{D}ac}}}
\def\sqcapDac{\mathrel{\sqcap_{\mathcal{D}ac}}}

\def\sqcupBM{\mathrel{\sqcup_{BM}}}

\def\seqBM{\mathrel{\Semi_{BM}}}

\def\refinedbyBM{\mathrel{\sqsubseteq_{BM}}}

\def\sqcupBMbot{\mathrel{\sqcup_{{BM}_\bot}}}
\def\sqcapBMbot{\mathrel{\sqcap_{{BM}_\bot}}}
\def\seqBMbot{\mathrel{\Semi_{BM_\bot}}}
\def\topBMbot{\top_{{BM}_\bot}}
\def\botBMbot{\bot_{{BM}_\bot}}
\def\refinedbyBMbot{\mathrel{\sqsubseteq_{{BM}_\bot}}}

\newcounter{checklabel}

\newcommand{\checkt}[1]{}
\makeatletter
\newcommand{\printcheck}[1]{\@ifundefined{r@check:pfr:#1}{\relax}{$\checkmark_{P}$ }\@ifundefined{r@check:alcc:#1}{\relax}{$\checkmark_{A}$ }}
\makeatother

\newtheoremstyle{mystyle}
{1em}
{1em}
{\itshape}
{0pt}
{\bfseries}
{}
{1em plus 1pt minus 1pt}
{\thmname{#1}\thmnumber{ #2}\thmnote{ (#3) }\refstepcounter{checklabel}\printcheck{\thechecklabel}}

\theoremstyle{mystyle}

\newenvironment{xflalign*}[0]{\csname flalign*\endcsname}{\csname endflalign*\endcsname}

\makeatletter

\newcommand{\gettheoremref}[1]{\csname theoremref#1 \endcsname}%

\def\cs#1{\csname #1\endcsname}%

\long\def\statementdefine#1#2{\expandafter\gdef\csname theoremstmt#1\endcsname{#2}}%
\long\def\proofdefine#1#2{\expandafter\gdef\csname theoremproof#1\endcsname{#2}}%
\long\def\theoremtypedefine#1#2{\expandafter\gdef\csname theoremtypes#1\endcsname{#2}}%
\long\def\theoremnumbermapdefine#1#2{\expandafter\gdef\csname theoremnumbermap#1\endcsname{#2}}%
\long\def\theoremnamedefine#1#2{\expandafter\gdef\csname theoremname#1\endcsname{#2}}%

\def\thecurrentcounter{\relax}%
\newcounter{dummycounterx}%
\newcounter{theoremscounter}
\newif\ifrenamed
\let\oldnewtheorem\newtheorem
\renewcommand{\newtheorem}[3]{%
	\oldnewtheorem{#2in}{#3}[#1]%
	\newenvironment{#2}[1][]{%
 		\@ifundefined{firstseentheorem}{%
 			\renewcommand{\thecurrentcounter}{\the#2in}%
 		}{%
 			\ifx&##1&%
 			\else%
 			\immediate\write\@auxout{\string\theoremnamedefine{\thetheoremscounter}{\unexpanded\expandafter{##1}}}%
 			\fi%
 			\let\oldlabelx\label%
 			\renewcommand{\label}[1]{\oldlabelx{####1}\immediate\write\@auxout{\string\theoremnumbermapdefine{####1}{\thetheoremscounter}}}%
 			\renewcommand{\thecurrentcounter}{\thetheoremscounter}%
 			\@ifundefined{theoremrenaming\thetheoremscounter}{%
				\renamedtrue%
 			}%
  			{%
				\renamedfalse%
  				\setcounter{dummycounterx}{\value{#2in}}%
  				\expandafter\let\expandafter\thetheoreminold\csname the#2in\endcsname
  				\expandafter\renewcommand\csname the#2in\endcsname{\cs{theoremrenaming\thetheoremscounter}}
	  		}%
 		}%
 	\begin{#2in}[##1]%
 	\immediate\write\@auxout{\string\theoremtypedefine{\thecurrentcounter}{#2in}}%
	}%
	{%
 	\end{#2in}%
	\@ifundefined{firstseentheorem}{}{%
		\ifrenamed%
			\expandafter\gdef\csname theoremrenaming\thetheoremscounter\endcsname{2}%
		\else%
			\expandafter\let\csname the#2in\endcsname\thetheoreminold%
			\setcounter{#2in}{\value{dummycounterx}}%
		\fi%
		\stepcounter{theoremscounter}%
		\let\label\oldlabelx%
		}%
	}%
}%

\newcommand*{\gettheoremrefnumber}[1]{%
	\@ifundefined{firstseentheorem}{%
		{\getrefnumber{#1}}%
	}{%
\csname theoremnumbermap#1\endcsname%
	}%
}%

\long\def\gettheoremtype#1{\cs{theoremtypes\gettheoremrefnumber{#1}}}%

\long\def\theoremstatementonlyref#1{\cs{theoremstmt\gettheoremrefnumber{#1}}} 

\long\def\theoremproofonlyref#1{\cs{theoremproof\gettheoremrefnumber{#1}}} 

\long\def\theoremnameref#1{\cs{theoremname\gettheoremrefnumber{#1}}} 

\long\def\theoremref#1{%
	\IfRefUndefinedExpandable{#1}{{\bf ??}}{%
		\@ifundefined{firstseentheorem}{%
			\expandafter\let\expandafter\thecurrenttt\csname the\gettheoremtype{#1}\endcsname%
			\setcounter{dummycounterx}{\value{\gettheoremtype{#1}}}%
			\expandafter\expandafter\expandafter\renewcommand\expandafter{\csname the\gettheoremtype{#1}\endcsname}{\ref{#1}}%
		}{%
			\@ifundefined{theoremrenaming\gettheoremrefnumber{#1}}{%
			\expandafter\let\expandafter\thecurrenttt\csname the\gettheoremtype{#1}\endcsname
			\expandafter\expandafter\expandafter\renewcommand\expandafter{\csname the\gettheoremtype{#1}\endcsname}{\hyperref[#1]{\thecurrenttt}}%
			}{	
 				\expandafter\let\expandafter\thecurrenttt\csname the\gettheoremtype{#1}\endcsname%
 				\setcounter{dummycounterx}{\value{\gettheoremtype{#1}}}%
 				\expandafter\expandafter\expandafter\renewcommand\expandafter{\csname the\gettheoremtype{#1}\endcsname}{\ref{#1}}%
			}
		}
	 \begin{\gettheoremtype{#1}}%
		\theoremstatementonlyref{#1}
		\theoremproofonlyref{#1}
	 \end{\gettheoremtype{#1}}%
	 \@ifundefined{firstseentheorem}{%
		\expandafter\let\csname the\gettheoremtype{#1}\endcsname\thecurrenttt%
		\expandafter\setcounter{\gettheoremtype{#1}}{\value{dummycounterx}}%
		}{
			\@ifundefined{theoremrenaming\gettheoremrefnumber{#1}}{%
				\expandafter\let\csname the\gettheoremtype{#1}\endcsname\thecurrenttt%
				\expandafter\edef\csname theoremrenaming\gettheoremrefnumber{#1}\endcsname{\thecurrenttt}%
			}{
			\expandafter\let\csname the\gettheoremtype{#1}\endcsname\thecurrenttt%
			\expandafter\setcounter{\gettheoremtype{#1}}{\value{dummycounterx}}%
			}
		}
	}%
}%

\long\def\theoremstatementref#1{%
	\IfRefUndefinedExpandable{#1}{{\bf ??}}{%
		\@ifundefined{firstseentheorem}{%
			\expandafter\let\expandafter\thecurrenttt\csname the\gettheoremtype{#1}\endcsname%
			\setcounter{dummycounterx}{\value{\gettheoremtype{#1}}}%
			\expandafter\expandafter\expandafter\renewcommand\expandafter{\csname the\gettheoremtype{#1}\endcsname}{\ref{#1}}%
		}{%
			\@ifundefined{theoremrenaming\gettheoremrefnumber{#1}}{%
			\expandafter\let\expandafter\thecurrenttt\csname the\gettheoremtype{#1}\endcsname
			\expandafter\expandafter\expandafter\renewcommand\expandafter{\csname the\gettheoremtype{#1}\endcsname}{\hyperref[#1]{\thecurrenttt}}%
			}{	
 				\expandafter\let\expandafter\thecurrenttt\csname the\gettheoremtype{#1}\endcsname%
 				\setcounter{dummycounterx}{\value{\gettheoremtype{#1}}}%
 				\expandafter\expandafter\expandafter\renewcommand\expandafter{\csname the\gettheoremtype{#1}\endcsname}{\ref{#1}}%
			}
		}
	 \begin{\gettheoremtype{#1}}%
		\theoremstatementonlyref{#1}
	 \end{\gettheoremtype{#1}}%
	 \@ifundefined{firstseentheorem}{%
		\expandafter\let\csname the\gettheoremtype{#1}\endcsname\thecurrenttt%
		\expandafter\setcounter{\gettheoremtype{#1}}{\value{dummycounterx}}%
		}{
			\@ifundefined{theoremrenaming\gettheoremrefnumber{#1}}{%
				\expandafter\let\csname the\gettheoremtype{#1}\endcsname\thecurrenttt%
				\expandafter\edef\csname theoremrenaming\gettheoremrefnumber{#1}\endcsname{\thecurrenttt}%
			}{
			\expandafter\let\csname the\gettheoremtype{#1}\endcsname\thecurrenttt%
			\expandafter\setcounter{\gettheoremtype{#1}}{\value{dummycounterx}}%
			}
		}
	}%
}%

\newcommand{\theoremstatementwithnameref}[2][]{%
	\IfRefUndefinedExpandable{#2}{{\bf ??}}{%
		\@ifundefined{firstseentheorem}{%
			\expandafter\let\expandafter\thecurrenttt\csname the\gettheoremtype{#2}\endcsname%
			\setcounter{dummycounterx}{\value{\gettheoremtype{#2}}}%
			\expandafter\expandafter\expandafter\renewcommand\expandafter{\csname the\gettheoremtype{#2}\endcsname}{\ref{#2}}%
		}{%
			\@ifundefined{theoremrenaming\gettheoremrefnumber{#2}}{%
			\expandafter\let\expandafter\thecurrenttt\csname the\gettheoremtype{#2}\endcsname
			\expandafter\expandafter\expandafter\renewcommand\expandafter{\csname the\gettheoremtype{#2}\endcsname}{\hyperref[#2]{\thecurrenttt}}%
			}{	
 				\expandafter\let\expandafter\thecurrenttt\csname the\gettheoremtype{#2}\endcsname%
 				\setcounter{dummycounterx}{\value{\gettheoremtype{#2}}}%
 				\expandafter\expandafter\expandafter\renewcommand\expandafter{\csname the\gettheoremtype{#2}\endcsname}{\ref{#2}}%
			}
		}
	\ifx&#1&%
	\begin{\gettheoremtype{#2}}[\theoremnameref{#2}]%
		\theoremstatementonlyref{#2}
	\end{\gettheoremtype{#2}}%
 			\else%
 	\begin{\gettheoremtype{#2}}[#1]%
		\theoremstatementonlyref{#2}
	 \end{\gettheoremtype{#2}}%
 	\fi%
	 \@ifundefined{firstseentheorem}{%
		\expandafter\let\csname the\gettheoremtype{#2}\endcsname\thecurrenttt%
		\expandafter\setcounter{\gettheoremtype{#2}}{\value{dummycounterx}}%
		}{
			\@ifundefined{theoremrenaming\gettheoremrefnumber{#2}}{%
				\expandafter\let\csname the\gettheoremtype{#2}\endcsname\thecurrenttt%
				\expandafter\edef\csname theoremrenaming\gettheoremrefnumber{#2}\endcsname{\thecurrenttt}%
			}{
			\expandafter\let\csname the\gettheoremtype{#2}\endcsname\thecurrenttt%
			\expandafter\setcounter{\gettheoremtype{#2}}{\value{dummycounterx}}%
			}
		}
	}%
}%

\NewEnviron{statement}{\immediate\write\@auxout{\string\statementdefine{\thecurrentcounter}{\unexpanded\expandafter{\BODY}}}\BODY}%
\NewEnviron{proofs}{\immediate\write\@auxout{\string\proofdefine{\thecurrentcounter}{\unexpanded\expandafter{\BODY}}}\@ifundefined{noproofs}{\BODY}{}}%
\NewEnviron{proofsbig}{\@ifundefined{noproofs}{\BODY}{}}%
\NewEnviron{proofs*}{\immediate\write\@auxout{\string\proofdefine{\thecurrentcounter}{\unexpanded\expandafter{\BODY}}}\BODY}%

\makeatother

\newtheorem{section}{lemma}{Lemma}
\newtheorem{}{define}{Definition}
\newtheorem{}{example}{Example}
\newtheorem{}{counter-example}{Counter-example}
\newtheorem{section}{question}{Question}
\newtheorem{section}{law}{Law}
\crefname{lawin}{Law}{Laws}
\crefname{lemmain}{Lemma}{Lemmas}
\crefname{theoremin}{Theorem}{Theorems}
\newtheorem{section}{theorem}{Theorem}
\newtheorem{section}{lemmanew}{Lemma New}
\newtheorem{section}{theoremnew}{Theorem New}

\begin{document}
\bibliographystyle{IEEEtran}
\begin{titlepage}
\begin{center}
	\vskip 15em
	{\huge \bfseries Angelic Processes}
	\vskip 20em
	{\large Pedro Fernando de Oliveira Salazar Ribeiro} \\
	\vskip 20em
	Doctor of Philosophy
	\vskip 0.5em 
	University of York
	\vskip 0.5em   
	Computer Science
	\vskip 2em 
	December 2014
	\vskip 5em
\end{center}
\end{titlepage}
\chapter*{Abstract}
\addcontentsline{toc}{chapter}{Abstract}
In the formal modelling of systems, demonic and angelic nondeterminism play fundamental roles as abstraction mechanisms. The angelic nature of a choice pertains to the property of avoiding failure whenever possible. As a concept, angelic choice first appeared in automata theory and Turing machines, where it can be implemented via backtracking. It has traditionally been studied in the refinement calculus, and has proved to be useful in a variety of applications and refinement techniques. Recently it has been studied within relational, multirelational and higher-order models. It has been employed for modelling user interactions, game-like scenarios, theorem proving tactics, constraint satisfaction problems and control systems. 

When the formal modelling of state-rich reactive systems is considered, it only seems natural that both types of nondeterministic choice should be considered. However, despite several treatments of angelic nondeterminism in the context of process algebras, namely Communicating Sequential Processes, the counterpart to the angelic choice of the refinement calculus has been elusive.

In this thesis, we develop a semantics in the relational setting of Hoare and He's Unifying Theories of Programming that enables the characterisation of angelic nondeterminism in CSP. Since CSP processes are given semantics in the UTP via designs, that is, pre and postcondition pairs, we first introduce a theory of angelic designs, and an isomorphic multirelational model, that is suitable for characterising processes. We then develop a theory of reactive angelic designs by enforcing the healthiness conditions of CSP. Finally, by introducing a notion of divergence that can undo the history of events, we obtain a model where angelic choice avoids divergence. This lays the foundation for a process algebra with both nondeterministic constructs, where existing and novel abstract modelling approaches can be considered. The UTP basis of our work makes it applicable in the wider context of reactive systems.

\tableofcontents

\listoftables
\addcontentsline{toc}{chapter}{List of Tables}
 
\listoffigures
\addcontentsline{toc}{chapter}{List of Figures}

\chapter*{Acknowledgements}
\addcontentsline{toc}{chapter}{Acknowledgements}
\markboth{ACKNOWLEDGEMENTS}{}
Although the work presented in this thesis is mainly the result of three years of hard work, passion and dedication, there is a whole lot more to tell, remember and appreciate. I truly feel that the fact that I was surrounded by the best inspiring minds is the main reason why my seven year-long journey into Computer Science continues to flourish to this day.

I would like to thank my supervisor, Ana Cavalcanti, for her continued support and prompt attention to detail, which in the context of this work, has been of utmost importance. Our insightful discussions have always been extremely productive and have often led to new ideas for future work. I would also like to thank Jim \mbox{Woodcock}, Frank Zeyda and Simon Foster for their insightful discussions and suggestions regarding my work. Frank has not only been a friend and a source of inspiration, but also extremely knowledgeable and helpful in discussing technical aspects, like those of the UTP. I would also like to thank my examiners Professor Steve Schneider and Dr.~\mbox{Andrew} Butterfield for their professional, rigorous and helpful feedback. In addition, I would like to thank my assessor Dr.~Detlef Plump for his prompt feedback and positive contributions over the past three years. I am also grateful for the financial support from EPSRC, UK, which gave me an invaluable sense of comfort in my daily life.

I would not have been able to reach this milestone if it were not for my parents, whose unconditional support for my dreams has been pivotal since an early age. My childhood curiosity has grown and with it so have my science and technology dreams. Despite the geographical distance, their encouragement has always played a key role in being able to study abroad, and for that I will be eternally grateful.

My partner Zhishuang Chen has been an essential source of inspiration and support. It is thanks to her unconditional love and support that I have made through some of the most anxious and tough times during the course of this degree. Being a PhD student herself, I hope to be able to equally and positively contribute towards her achievements.

Finally, I would also like to thank the following friends, and in no particular order, who have in one way or another, contributed positively to my well-being while living and studying in York: Artur Goul\~{a}o Ferreira, Frank Soboczenski, Theodora Lee, Ruofan Jin, Miguel Pr\^{o}a and Luis Carlos Rodrigues.

\chapter*{Author's Declaration}
\addcontentsline{toc}{chapter}{Author's Declaration}
\markboth{AUTHOR'S DECLARATION}{}
I hereby declare that the work presented in this thesis is based on my original contributions, unless otherwise stated. The following material has been previously published.
\\ \\
\citep{Ribeiro2013}
P.~Ribeiro and A.~Cavalcanti, ``{D}esigns with {A}ngelic {N}ondeterminism,'' in
  \emph{{T}heoretical {A}spects of {S}oftware {E}ngineering ({TASE}), 2013
  {I}nternational {S}ymposium on}.\hskip 1em plus 0.5em minus 0.4em\relax IEEE,
  2013, pp. 71--78.
\\ \\
\citep{Ribeiro2014}
------, ``{A}ngelicism in the {T}heory of {R}eactive {P}rocesses,'' in
  \emph{{U}nifying {T}heories of {P}rogramming}, ser. {L}ecture {N}otes in
  {C}omputer {S}cience, D.~Naumann, Ed.\hskip 1em plus 0.5em minus 0.4em\relax
  Springer International Publishing, 2015, vol. 8963, pp. 42--61.
\\ \\
\citep{Ribeiro2014a}
------, ``{UTP} {D}esigns for {B}inary {M}ultirelations,'' in
  \emph{{T}heoretical {A}spects of {C}omputing {ICTAC} 2014}, ser. {L}ecture
  {N}otes in {C}omputer {S}cience, G.~Ciobanu and D.~M\'{e}ry, Eds.\hskip 1em
  plus 0.5em minus 0.4em\relax Springer International Publishing, 2014, vol.
  8687, pp. 388--405.


\chapter{Introduction}\label{chapter:1}
In this chapter we discuss the motivation and objectives underlying our work on a semantic model for~\acs{CSP} processes with angelic nondeterminism. Furthermore, we provide an overview of all semantic models of interest in the context of this thesis and their relationships. Finally, an outline of this document's structure is presented.

\section{Motivation}\label{sec:motivation}
In an increasingly connected world, where software-driven systems are ubiquitous, it is imperative that their behaviour is rigorously studied. Since the software crisis of the seventies~\cite{Dijkstra1972}, significant attention has been devoted to this problem with the development of several theories, techniques and  tools. The earliest contributions can be found in the works of Floyd, Hoare and Dijkstra. In 1967, Floyd~\cite{Floyd1967} proposed techniques for rigorously characterizing and analysing programs specified as flowcharts, by considering propositions associated with the entrance and exit of commands in the flowchart, akin to pre and postconditions. Hoare~\cite{Hoare1969} would later propose a formal system, known as Hoare logic, capable of proving partial correctness of program statements for a sequential programming language. Inspired by Hoare's work, Dijkstra~\cite{Dijkstra1975} introduced weakest precondition semantics with his language of guarded commands, an imperative language that allows for the existence of repetitive and nondeterministic constructs.

As systems present several aspects of interest, ranging from the intended functional behaviour to the actual operating environment, modelling approaches focus on specific properties of interest, at suitable levels of abstraction. For instance, there are several formal notations catering for the specification of functional behaviour, such as Z~\cite{Woodcock1996,Spivey1989}, Object-Z~\cite{Fischer1998}, \ac{VDM}~\cite{Jones1986}, \ac{ASM}~\cite{Borger2003,Borger2005} and B~\cite{Abrial1996,Abrial2007}. Concurrent and reactive systems have also been extensively studied with formalisms such as~\ac{CSP}~\cite{Hoare1985,Roscoe1998,Roscoe2010}, \ac{CCS}~\cite{Milner1989} and \ac{ACP}~\cite{Bergstra1985}. Several works have also focused on combining both state-based and concurrent formalisms as found in the literature~\cite{Fischer1998,Cavalcanti2003,Oliveira2005,Smith2001,Smith2002,Schneider2002,Butler2005,Schneider2010}.

The successful characterisation of a particular system relies on appropriate abstraction mechanisms being available, such that a system can be decomposed into manageable parts with the appropriate level of detail. Formal specifications are, in this sense, at the very top of the hierarchy, and provide the highest-level and most abstract model of a system. Since the foundational works of Back~\cite{Back1978}, Morris~\cite{Morris1987} and Morgan~\cite{Morgan1988}, however, it has been possible to consider both specifications and programs within the same formal model.

An essential abstraction mechanism that is pervasive across modelling approaches is that of nondeterministic choice. It can be used to specify purely nondeterministic behaviour, such that no particular choice is guaranteed, but also to describe concisely a set of choices, such that, if there are options that lead to success, they are guaranteed to be chosen. The former is traditionally referred to as being demonic, while the latter is referred to as angelic. Operationally, both nondeterministic choices embody some notion of failure, and success.

Demonic choice has traditionally been used for the underspecification of behaviour, and plays an essential role in the contractual approach between users and developers. In the context of refinement, the behaviour of a specification can be made more deterministic while adhering to the externally observable behaviour. In other words, given a particular set of choices, the user is unable to force any particular choice and must accept any subset, including failure, if this is a possibility. This corresponds to the semantics of nondeterminism in Dijkstra's~\cite{Dijkstra1975} guarded commands, and internal choice in~\ac{CSP}~\cite{Roscoe1998}, for example.

On the other hand, angelic choice is driven by success. Given a set of choices, as long as there is at least one choice that leads to success, then the angel can achieve a satisfying outcome. Thus, operationally, angelic nondeterminism can be interpreted as a backtracking mechanism. Indeed this is similar to the underlying concept involved in searching for solutions in a given space. Another typical application of this concept can be found in the context of nondeterministic finite state automatons, where acceptance is successful if, and only if, the system reaches an accepting state. 

The concept of angelic nondeterminism has traditionally been studied in the refinement calculus~\cite{Morgan1994,Back1998,Morris1987}, where angelic choice is defined as the least upper bound of the lattice of monotonic predicate transformers. Its dual is demonic choice, which is defined as the greatest lower bound of the lattice. In~\cite{Gardiner1991,Morgan1990} the least upper bound is used to define logical variables, which enable the postcondition of a specification statement to refer to the initial value of a program variable. This is central to the refinement technique of Gardiner and Morgan~\cite{Gardiner1991}, and, in particular, to their calculational data-refinement approach.

In~\cite{Rewitzky2003} Rewitzky introduces binary multirelations for modelling both forms of nondeterminism. Unlike relational models, which relate initial states to final states, multirelations relate initial states to sets of final states. A number of models are explored in~\cite{Martin2004}, of which the model of upward-closed binary multirelations is the most important as it has a lattice-theoretic structure. 
A generalised algebraic structure has also been proposed by Guttmann~\cite{Guttmann2014}, where the monotonic predicate transformers and multirelations are characterised as instances.

Cavalcanti et al.~\cite{Cavalcanti2006} have proposed a predicative encoding of binary multirelations in the context of Hoare and He's~\cite{Hoare1998}~\ac{UTP}, a relational framework suitable for characterising several programming paradigms. This is achieved by encoding program variables as record components. First an isomorphism is established between the new~\ac{UTP} model and a set-based relational model. Afterwards an isomorphism is established between the set-based model and the monotonic predicate transformers. Finally an isomorphism is established between the predicate transformers model and upward-closed binary multirelations. This is then used to establish the correspondence between the semantics of statements in the predicate transformers model and in the proposed~\ac{UTP} model.


Angelic choice has also been considered at the expression, or term, level by Morris~\cite{Morris2007,Morris2004}. In~\cite{Morris2004}, an axiomatic basis is presented for defining operators for both angelic and demonic nondeterminism within a term language. Each type is represented as a partially ordered set, and an ordering is given. This is then lifted into a~\ac{FCD} lattice where the refinement relation corresponds to the ordering relation imposed on the type, demonic choice is the meet, and angelic choice is the join. In~\cite{Morris2007} this model is shown to be isomorphic to higher-order models of predicate transformers, binary multirelations and state transformers. While it is possible to cast typical sequential programming constructs into this theory, its focus is on functional languages. Hesselink~\cite{Hesselink2010} further studies this model and provides a different construction of the FCD.


In~\cite{Tyrrell2006}, Tyrrell et al., inspired on the previous work on the~\ac{FCD} by Morris~\cite{Morris2004}, provide an axiomatization for an algebra, similar to CSP, where external choice is referred to as ``angelic choice''. The definitions are then lifted from a partially ordered set into the FCD lattice. Just as the authors point out, this model is quite different from the traditional~\ac{CSP} model whose complete semantics is based on failures-divergences~\cite{Roscoe1998,Roscoe2010}. In the model proposed, $Stop$ is the bottom of the refinement ordering, rather than divergence. Thus, it is impossible to distinguish divergence from deadlock.

Roscoe~\cite{Roscoe2010} has proposed an angelic choice operator $P \boxcircle Q$ through an operational combinator semantics for CSP. It is an alternative to the external choice operator of~\ac{CSP} that behaves as follows: as long as the environment chooses events offered by both $P$ and $Q$, then the choice between $P$ and $Q$ is unresolved. The possibility of divergence or otherwise has no effect on the choice. 

Despite the various models where angelic nondeterminism is employed in the context of process algebras, and the different semantics considered in the literature~\cite{Tyrrell2006,Roscoe2010}, the counterpart to the angelic choice of the refinement calculus has been elusive. The notion of failure of interest here is that of divergence as required for a characterisation of angelic nondeterminism in the context of state-rich reactive systems for both data and behavioural refinement.

The~\ac{UTP} of Hoare and He~\cite{Hoare1998} provides an ideal framework to study the concept of angelic nondeterminism in a theory of~\ac{CSP}~\cite{Hoare1998,Cavalcanti2006a}. The~\ac{UTP} is a predicative framework of alphabetized relations suitable for characterising different programming aspects, such as functionality, concurrency, logic programming, higher-order programming, object-orientation~\cite{Santos2006,Zeyda2014}, pointers~\cite{Harwood2008}, time~\cite{Sherif2002,Sherif2006,Wei2012} and others. It supports the engineering of theories by enabling results to be related through linking functions, while allowing different concerns to be studied in isolation. The theory of designs~\cite{Hoare1998,Woodcock2004}, which characterises total correctness, is one of the most important. In general, a~\ac{UTP} theory is a complete lattice where we can use joins and meets to model dual choices.

While sequential computations can be characterised by a relation between their initial and final states, the formal characterisation of reactive systems requires a richer model that accounts for the continuous interactions with their environment. In the~\ac{UTP} this is achieved through the theory of reactive processes~\cite{Hoare1998,Cavalcanti2006a}. Together with the theory of designs, these two theories enable the specification of~\ac{CSP} processes in an assertional style, that is, in terms of designs that characterise the pre and postcondition of processes.

The theory of angelic nondeterminism presented in~\cite{Cavalcanti2006} is a starting point for the development of a model of~\ac{CSP} with both nondeterministic constructs. However, this model is focused on correctness of sequential programs and is not directly applicable to reactive processes. It is an encoding that caters for termination, so that designs are not considered as a separate theory.

In summary, a suitable treatment of angelic nondeterminism is yet to be considered in the context of process algebras for state-rich reactive systems. The~\ac{UTP} presents itself as a natural domain for the development of such a model, as existing theories, and their results, can be easily exploited. Our hypothesis is as follows. \\

\noindent\makebox[\textwidth][c]{%
\begin{minipage}[c]{0.82\linewidth}%
\begin{center}\textbf{Research Hypothesis}\end{center}
\textit{\textbf{A model can be defined to give a semantics to~\ac{CSP} that caters for both angelic and demonic nondeterminism, that is applicable in the wider context of any algebra of state-rich reactive systems for refinement, and that preserves the existing semantics of~\ac{CSP} processes, particularly within the subset of nondivergent processes.}}\\
\end{minipage}}\\ \noindent \\
This concludes the discussion of the motivation underlying our work. In what follows we discuss the objectives in more detail.

\section{Objectives}


%

As already mentioned, the overall objective of our work is to define a semantic model suitable for state-rich process algebras, and~\ac{CSP} in particular, where both nondeterministic choices can be expressed. In contrast with some of the existing approaches~\cite{Tyrrell2006}, we do not intend to propose an entirely new semantic model for~\ac{CSP}, rather we aim to extend the current model while conserving the existing semantics. Therefore, our construction must be appropriately justified in the context of the existing model~\cite{Hoare1998,Cavalcanti2006}.

With this in mind, the~\ac{UTP} framework and its~\ac{CSP} model provide a solid basis for studying the concept of angelic nondeterminism in the context of process algebras. We also observe that a~\ac{UTP} theory is a complete lattice where both angelic and demonic choice can be modelled as the meet and join, respectively.

The~\ac{UTP} supports work in the wider context of semantic models that consider behaviour and other aspects, such as data, security, mobility, and so on. Examples of such heterogeneous semantic models built using the~\ac{UTP} include~\Circus~\cite{Oliveira2005}, which combines~\ac{CSP} with the Z specification language. 
Our aim to is to enable such semantic models to benefit from our treatment of angelic nondeterminism. 

We also aim to enable existing modelling approaches and refinement techniques to be reused. This is central to the relevance and applicability of our semantic model. An important factor in~\ac{UTP} theories, for example, is that the refinement order is common across all theories. Our emphasis on maintaining a compatible semantics is essential in order to enable the scenario of reusing existing refinement techniques.

Our goal ultimately consists in developing a conservative extension of the~\ac{CSP} theory~\cite{Hoare1998,Cavalcanti2006a} through a predicative encoding of multirelations that is suitable for characterising~\ac{CSP} processes. Of particular importance is the treatment of divergence where angelic choice can avoid potentially divergent processes. We seek a theory of~\ac{CSP} with both angelic and demonic nondeterminism, which is applicable to any algebra of state-rich reactive processes. In the following section we discuss our theories, by showing their relationship with other semantic models of interest, namely~\ac{CSP}.

\section{Overview of Semantic Models}
In this section we provide an overview of all the semantic models of interest in the context of our work. This includes both existing models as well as those we propose.
\begin{figure}[t]
\begin{center}
\includegraphics[scale=0.58]{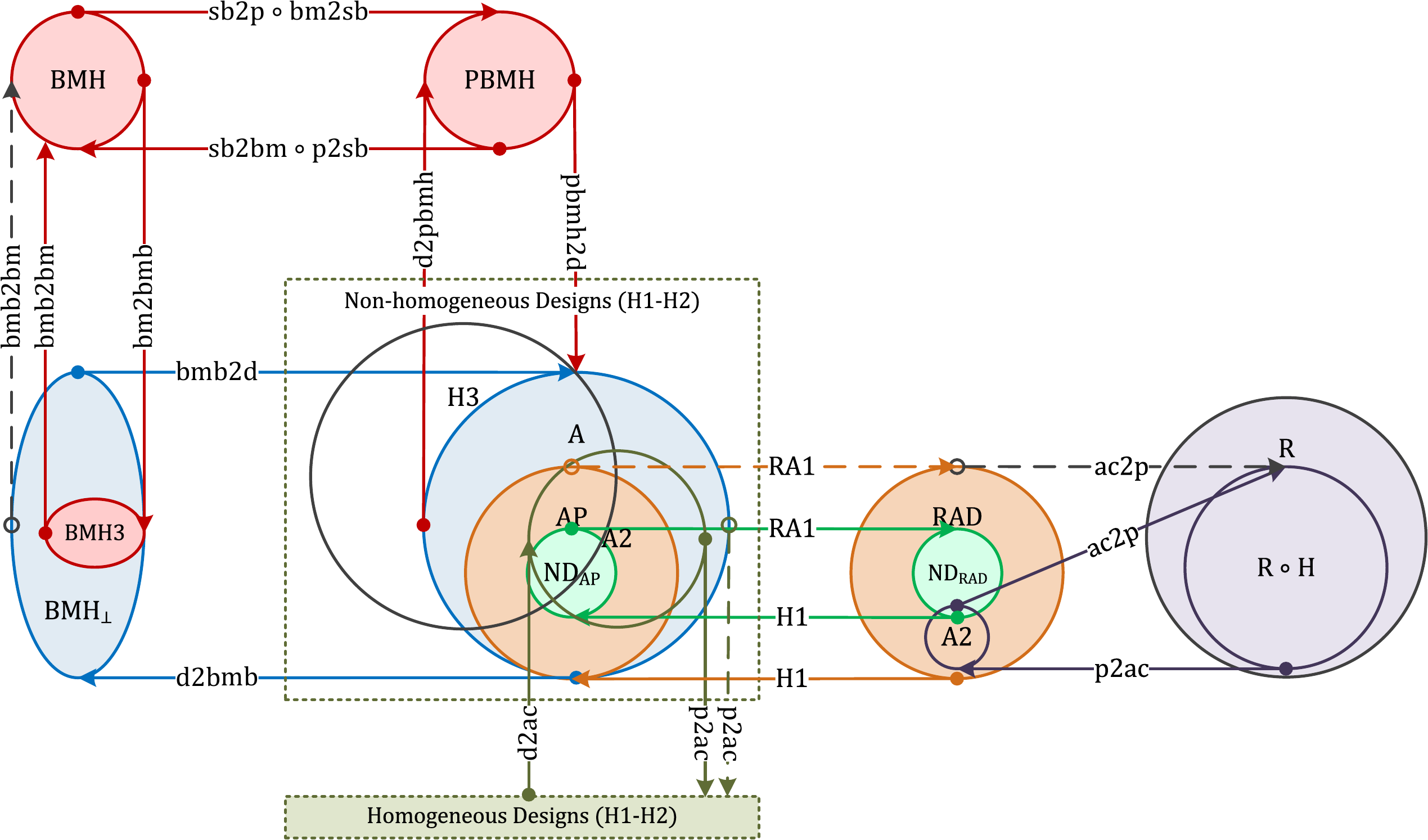}
\caption{\label{fig:theories}Theories and their relationship through linking functions}
\end{center}
\end{figure}

In the~\ac{UTP}~\cite{Hoare1998} theories are characterised by three components: an alphabet, which is a set of variables available for recording the observations of programs in a particular paradigm, including program variables; a set of healthiness conditions, which are idempotent and monotonic functions, usually with a name written in boldface, whose fixed points are the the valid predicates of a theory; and a set of operators. For a relation $P$, the alphabet is split into two disjoint subsets, $in\alpha(P)$ which contains undashed variables corresponding to the initial observations, and $out\alpha(P)$ containing dashed counterparts for after or final observations.

Each theory of interest is depicted in~\cref{fig:theories}, and also individually in the subsequent~\cref{fig:theories:designs,fig:theories:bmh,fig:theories:angelic-designs,fig:theories:reactive-angelic-designs,fig:theories:angelic-processes}, by an ellipse, and labelled according to the name of its characterising healthiness condition. Subset theories correspond to enclosed ellipses. While the formal definition of each healthiness condition is deferred to later chapters, in~\cref{table:H:designs,table:R:processes,table:H:BMbot,table:R:CSP,table:H:A,table:H:RAD} we informally describe the healthiness conditions of each theory. In~\cref{fig:theories} arrows denote linking functions established between theories. Pairs of solid arrows denote isomorphic models, while pairs with a dashed arrow indicate an adjoint (that is part of a Galois connection).

In the next~\cref{sec:ch1:designs} we describe the theory of designs. \cref{sec:ch1:CSP} focuses on the theory of~\ac{CSP} as reactive designs. In~\cref{sec:ch1:BM} we discuss the relationship between the theory of binary multirelations, the predicative encoding of~\cite{Cavalcanti2006}, and the relationship with our theory of extended binary multirelations. In~\cref{sec:ch1:angelic-designs} we discuss our theory of angelic designs, which is the basis for extending the concept of angelic nondeterminism to~\ac{CSP} through the theory of reactive angelic designs, summarized in~\cref{sec:ch1:RAD}. Finally, \cref{sec:ch1:AP} discusses our theory of angelic processes.

\subsection{Designs}\label{sec:ch1:designs}
Since~\ac{CSP} processes are expressed in the~\ac{UTP} through reactive designs, the first theory of interest is that of designs, which models total correctness. Designs are relations whose alphabet contains not only program variables, but also auxiliary Boolean variables to capture termination. Its characterising healthiness conditions are $\mathbf{H1}$ and $\mathbf{H2}$, whose composition is called $\mathbf{H}$, as summarized in~\cref{table:H:designs}.
\begin{figure}
\begin{center}
\includegraphics[scale=0.58]{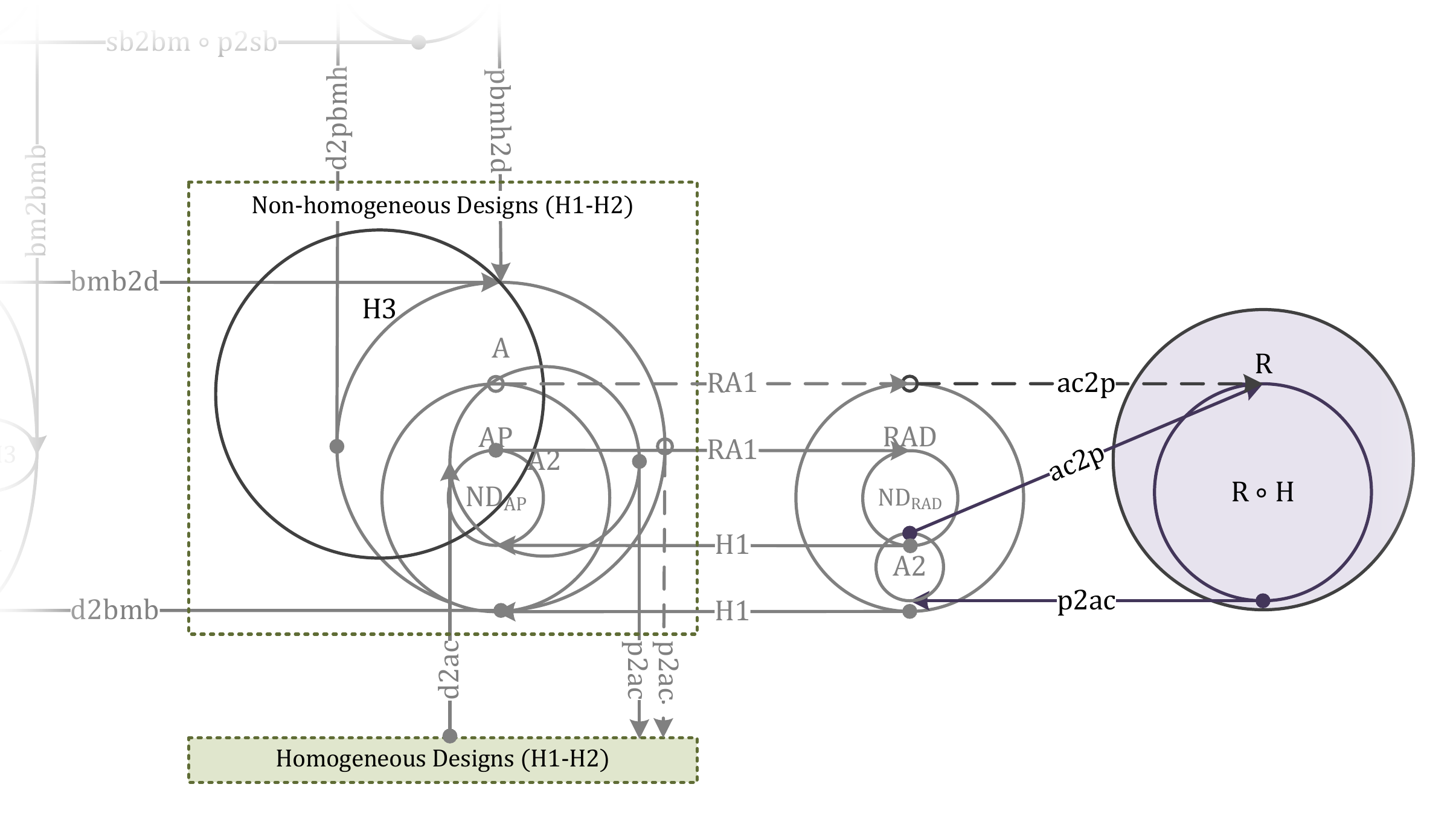}
\caption{\label{fig:theories:designs}Theories of designs and reactive designs}
\end{center}
\end{figure}
\begin{table}[htb]
  \centering
    \begin{tabular}{c|p{13cm}}
	\toprule
    &\multicolumn{1}{c}{\textbf{Description}} \\
    \midrule
    $\mathbf{H1}$ & Meaningful observations can only be made once a design has been started. \\
	\midrule
    $\mathbf{H2}$ & A design may not require non-termination. \\
    \midrule
	$\mathbf{H3}$ & A design must have arbitrary behaviour when it does not terminate. \\
	\bottomrule
    \end{tabular}%
  \caption{\label{table:H:designs} Healthiness Conditions of Designs}%
\end{table}%
In general, this is a theory that encompasses programs whose preconditions can refer to the after or final observations of a computation. As a consequence these observations can be ascertained irrespective of termination. Such designs do not satisfy the healthiness condition $\mathbf{H3}$. This is precisely the case when characterising a~\ac{CSP} process through reactive designs, such as $a \circthen Chaos$, whose precondition requires that no after observation of the trace of events is prefixed by the event $a$ otherwise, it diverges. 

The subset of designs whose preconditions may not refer to the after or final observations of a computation is characterised by $\mathbf{H3}$. These designs correspond to standard pre and postcondition pairs as found in notations like Z~\cite{Woodcock1996} and~\ac{VDM}~\cite{Jones1986}.

In the context of our work, we consider a theory of designs whose relations are not homogeneous, that is, their input and output alphabet differ. This is because of the multirelational nature of our encoding of angelic nondeterminism. In~\cref{fig:theories:designs} we highlight the theories of homogeneous and non-homogeneous designs in the context of other theories previously depicted in~\cref{fig:theories}. 


\subsection{CSP Processes as Reactive Designs}\label{sec:ch1:CSP}
The second theory of interest is that of reactive processes, whose combination with the theory of designs provides the characterisation of~\ac{CSP} processes in the~\ac{UTP}. In the theory of reactive processes the alphabet is extended with observational variables to record the interactions with the environment: a trace of events, a set of events refused, and a Boolean variable that records whether the process is waiting for an interaction. Its healthiness conditions, which we informally describe in~\cref{table:R:processes}, are $\mathbf{R1}$, $\mathbf{R2}$ and $\mathbf{R3}$, whose functional composition is $\mathbf{R}$.

\begin{table}[htbp]
  \centering
    \begin{tabular}{c|p{13cm}}
	\toprule
    &\multicolumn{1}{c}{\textbf{Description}} \\
    \midrule
    $\mathbf{R1}$ & A process can only extend the trace of events. \\
	\midrule
    $\mathbf{R2}$ & A process must be insensitive to the initial trace of events. \\
    \midrule
	$\mathbf{R3}$ & A process must only start executing once any previous interactions with the environment have finished. \\
	\midrule
	$\mathbf{R}$ & Functional composition of $\mathbf{R1}$, $\mathbf{R2}$ and $\mathbf{R3}$ that characterises reactive processes.\\
	\bottomrule
    \end{tabular}%
  \caption{\label{table:R:processes} Healthiness Conditions of Reactive Processes}%
\end{table}%
In order to characterise~\ac{CSP} processes, another two healthiness conditions are necessary. They are $\mathbf{CSP1}$ and $\mathbf{CSP2}$, whose informal description is included in~\cref{table:R:CSP}.
\begin{table}[htbp]
  \centering
    \begin{tabular}{c|p{13cm}}
	\toprule
    &\multicolumn{1}{c}{\textbf{Description}} \\
    \midrule
    $\mathbf{CSP1}$ & A process that is in a divergent state can only extend the trace of events. \\
	\midrule
    $\mathbf{CSP2}$ & A recast of $\mathbf{H2}$ within the model of reactive processes. \\
	\bottomrule
    \end{tabular}%
  \caption{\label{table:R:CSP} Healthiness Conditions of CSP Processes}%
\end{table}%
Together, these healthiness conditions allow the characterisation of~\ac{CSP} processes as the image of designs through the function $\mathbf{R}$~\cite{Hoare1998,Cavalcanti2006a}, that is, in terms of pre and postcondition pairs.

Since it is our goal to keep the semantics unchanged for the subset of nondivergent processes, in each theory of processes that we study, we identify such a subset. This is characterised by the healthiness condition $\mathbf{ND}$, which is tailored to the theory of interest by adding a subscript corresponding to the characterising healthiness condition of the theory it applies to.

\subsection{Binary Multirelations and their UTP Encoding}\label{sec:ch1:BM}
To achieve our goal we have developed a predicative encoding of multirelations suitable for characterising processes. Our starting point was the predicative encoding of Cavalcanti et al.~\cite{Cavalcanti2006}, whose theory is characterised by the healthiness condition $\mathbf{PBMH}$. This is essentially a predicative version of $\mathbf{BMH}$, that characterises a set-based model of upward-closed binary multirelations~\cite{Rewitzky2003}. 

\begin{figure}[t]
\begin{center}
\includegraphics[scale=0.58]{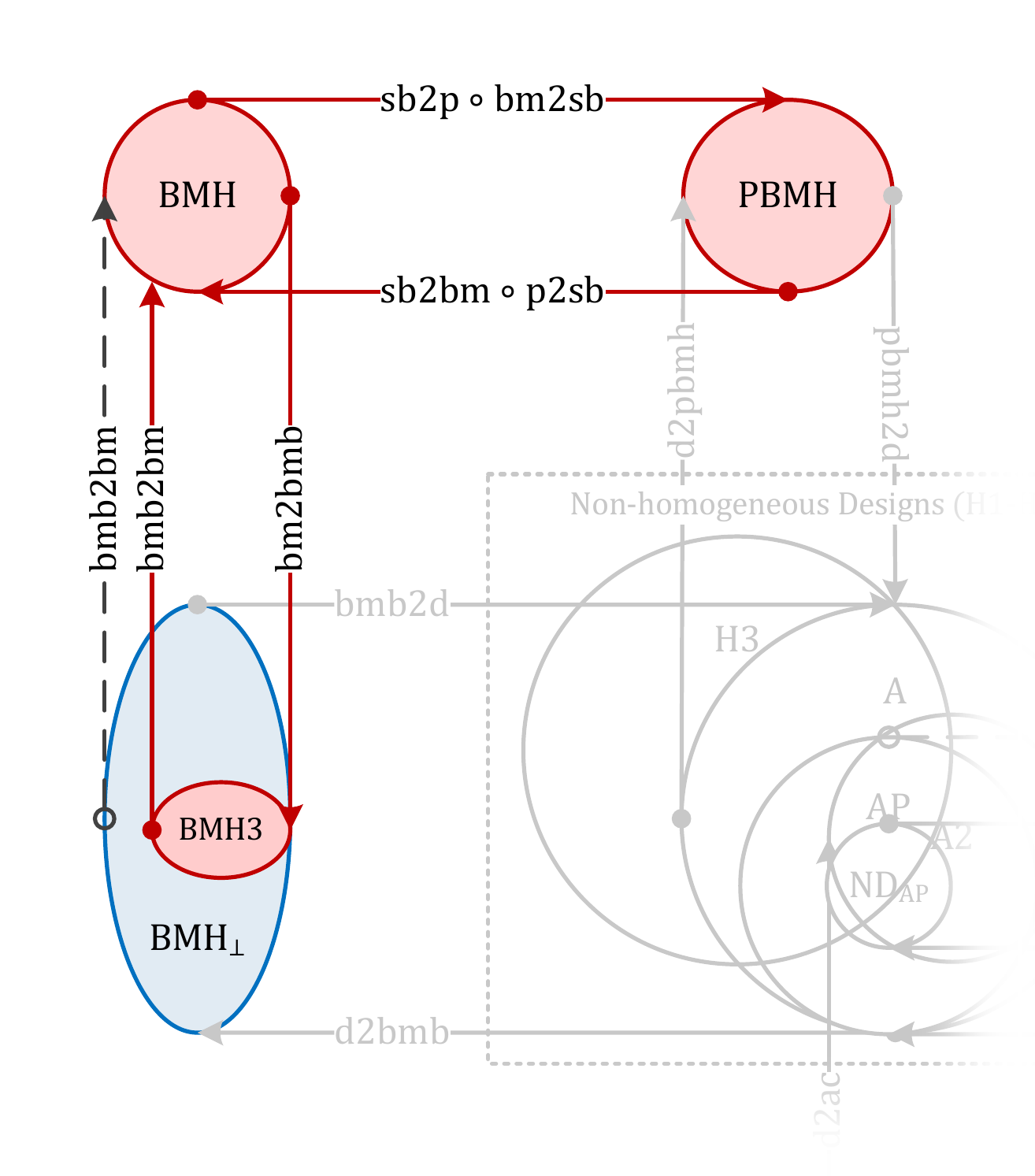}
\caption{\label{fig:theories:bmh}Theories related to binary multirelations}
\end{center}
\end{figure}

In~\cite{Cavalcanti2006} the authors establish that both models are isomorphic through a stepwise construction of models, as previously discussed in~\cref{sec:motivation}. This is achieved through the composition of the linking functions, $sb2p \circ bm2sb$ and $sb2bm \circ p2sb$, which we include in~\cref{fig:theories,fig:theories:bmh} for completeness.
The first contribution of this thesis is a theory of extended binary multirelations that caters for potentially non-terminating computations. This theory is isomorphic to the theory of angelic designs, which we describe in the next section. It is characterised by the healthiness condition $\mathbf{BMH_\bot}$, which corresponds to the conjunction of $\mathbf{BMH0}$, $\mathbf{BMH1}$ and $\mathbf{BMH2}$ as described in~\cref{table:H:BMbot}.
\begin{table}[htbp]
  \centering
    \begin{tabular}{c|p{12.5cm}}
	\toprule
    &\multicolumn{1}{c}{\textbf{Description}} \\
    \midrule
    $\mathbf{BMH0}$ & The set of final states must be upward-closed. \\
	\midrule
    $\mathbf{BMH1}$ & Similarly to $\mathbf{H2}$ forbids the specification of non-termination. \\
	\midrule
	$\mathbf{BMH2}$ & Appropriately characterises two complementary notions of abortion. \\
	\midrule
	$\mathbf{BMH3}$ & Characterises the subset of $\mathbf{BMH_\bot}$ that is isomorphic to the original theory of binary multirelations. \\
	\midrule
	$\mathbf{BMH_\bot}$ & Conjunction of $\mathbf{BMH0}$, $\mathbf{BMH1}$ and $\mathbf{BMH2}$. \\
	\bottomrule
    \end{tabular}%
  \caption{\label{table:H:BMbot} Healthiness Conditions of Extended Binary Multirelations}%
\end{table}%
Finally, we establish that the subset of $\mathbf{BMH3}$ multirelations is isomorphic to the original theory of binary multirelations, via the pair of linking functions $bmb2bm$ and $bm2bmb$. In general, a Galois connection can also be established between $\mathbf{BMH_\bot}$ and $\mathbf{BMH}$. \cref{fig:theories:bmh}, which highlights the theories in the context of~\cref{fig:theories}, illustrates these connections.

\subsection{Angelic Designs}\label{sec:ch1:angelic-designs}
Our approach for developing a model of~\ac{CSP} with angelic nondeterminism closely follows that of the~\ac{UTP} model of~\ac{CSP}. Based on the the encoding proposed in~\cite{Cavalcanti2006}, we have developed a theory of angelic designs where we reintroduce the auxiliary Boolean variables of the original theory of designs. Furthermore, we also generalise that model to cope with non-$\mathbf{H3}$ designs, as required for specifying~\ac{CSP} processes. This theory is characterised by the healthiness conditions $\mathbf{A0}$ and $\mathbf{A1}$, whose functional composition is $\mathbf{A}$ (as described in~\cref{table:H:A}), and $\mathbf{H1}$ and $\mathbf{H2}$ of the original theory of designs.
\begin{table}[htbp]
  \centering
    \begin{tabular}{c|p{13cm}}
	\toprule
    &\multicolumn{1}{c}{\textbf{Description}} \\
    \midrule
    $\mathbf{A0}$ & Whenever the precondition of a design is satisfied, then the set of angelic choices must not be empty. \\
	\midrule
    $\mathbf{A1}$ & The set of angelic choices must be upward-closed. \\
	\midrule
	$\mathbf{A2}$ & Characterises the subset of relations that effectively do not have any angelic choices. \\
	\midrule
	$\mathbf{A}$ & Functional composition of $\mathbf{A0}$ and $\mathbf{A1}$ \\
	\bottomrule
    \end{tabular}%
  \caption{\label{table:H:A} Healthiness Conditions of Angelic Designs}%
\end{table}%

The additional healthiness condition $\mathbf{A2}$ characterises the subset of $\mathbf{A}$-designs that do not exhibit angelic nondeterminism. This is useful to establish that the subset of $\mathbf{A2}$ angelic designs is isomorphic to the original theory of homogeneous designs, via the linking functions $d2ac$ and $p2ac$. In general, these adjoints also enable a Galois connection to be established with the set of $\mathbf{A}$-designs.
\begin{figure}[tbp]
\begin{center}
\includegraphics[scale=0.58]{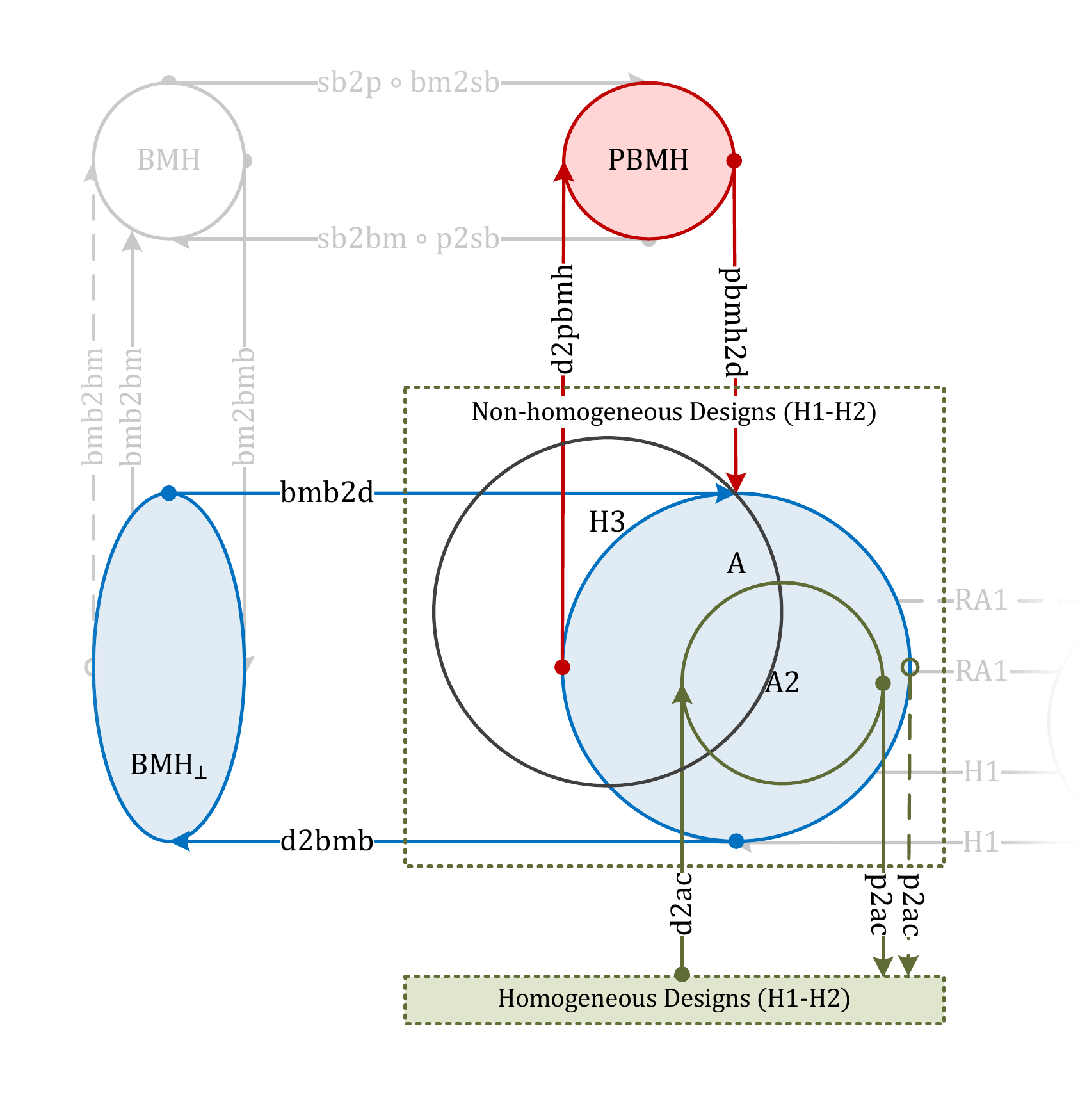}
\caption{\label{fig:theories:angelic-designs}Theory of angelic designs and links}
\end{center}
\end{figure}
As part of validating our approach, we also establish that the subset of angelic designs that is $\mathbf{H3}$-healthy is isomorphic to the theory of $\mathbf{PBMH}$~\cite{Cavalcanti2006}. This is achieved by introducing two linking functions, $d2pbmh$ and $pbmh2d$, that map predicates in that theory to angelic designs, and vice versa. In~\cref{fig:theories:angelic-designs} we highlight the theory of angelic designs in the context of~\cref{fig:theories} and show its relationship with the $\mathbf{PBMH}$ theory, the extended theory of binary multirelations, and the original theory of homogeneous designs.


In addition, and as already discussed, we have developed an extended set-based model of binary multirelations that is isomorphic to $\mathbf{A}$-healthy designs. This complementary model is useful to understand the implications of non-homogeneous relations and also to validate certain aspects of the model of angelic designs, such as the notion of sequential composition, which is not entirely trivial in the context of a predicative encoding of multirelations. We establish that these two models are isomorphic through the pair of linking functions $bmb2d$ and $d2bmb$.

\subsection{Reactive Angelic Designs}\label{sec:ch1:RAD}

Having established a theory of angelic designs, we introduce a conservative extension of~\ac{CSP} with angelic nondeterminism. This is achieved by considering an encoding of the observational variables of reactive processes, based on that used for angelic designs, and expressing every healthiness condition of~\ac{CSP} with this encoding. For each healthiness condition $\mathbf{R1}$, $\mathbf{R2}$, $\mathbf{R3}$, $\mathbf{CSP1}$ and $\mathbf{CSP2}$, we introduce a counterpart in this model, as summarized in~\cref{table:H:RAD}.
\begin{table}[htbp]
  \centering
    \begin{tabular}{c|p{12.5cm}}
	\toprule
    &\multicolumn{1}{c}{\textbf{Description}} \\
    \midrule
    $\mathbf{RA1}$ & There must be some set of angelic choices available to the angel, and in any such set, the trace of events can only be extended. \\
	\midrule
    $\mathbf{RA2}$ & A process must be insensitive to the initial value of the trace of events. \\
	\midrule
	$\mathbf{RA3}$ & A process must not start executing before its predecessor has stopped interacting with its environment. \\
	\midrule
	$\mathbf{RA}$ & Functional composition of $\mathbf{RA1}$, $\mathbf{RA2}$ and $\mathbf{RA3}$. \\
	\midrule
	$\mathbf{CSPA1}$ & When in an unstable state, $\mathbf{RA1}$ must be enforced. \\
	\midrule
	$\mathbf{CSPA2}$ & A recast of $\mathbf{H2}$ within this model. \\
	\midrule
	$\mathbf{RAD}$ & Functional composition of all of the above healthiness conditions and $\mathbf{PBMH}$. \\
	\midrule
	$\mathbf{ND_{RAD}}$ & Characterises the subset of non-divergent reactive angelic designs. \\
	\bottomrule
    \end{tabular}%
  \caption{\label{table:H:RAD} Healthiness Conditions of Reactive Angelic Designs}%
\end{table}%
The theory is characterised by $\mathbf{RAD}$, which is defined by the composition of all healthiness conditions of interest, including $\mathbf{PBMH}$ that guarantees upward-closure for the sets of final states.
\begin{figure}[b]
\begin{center}
\includegraphics[scale=0.58]{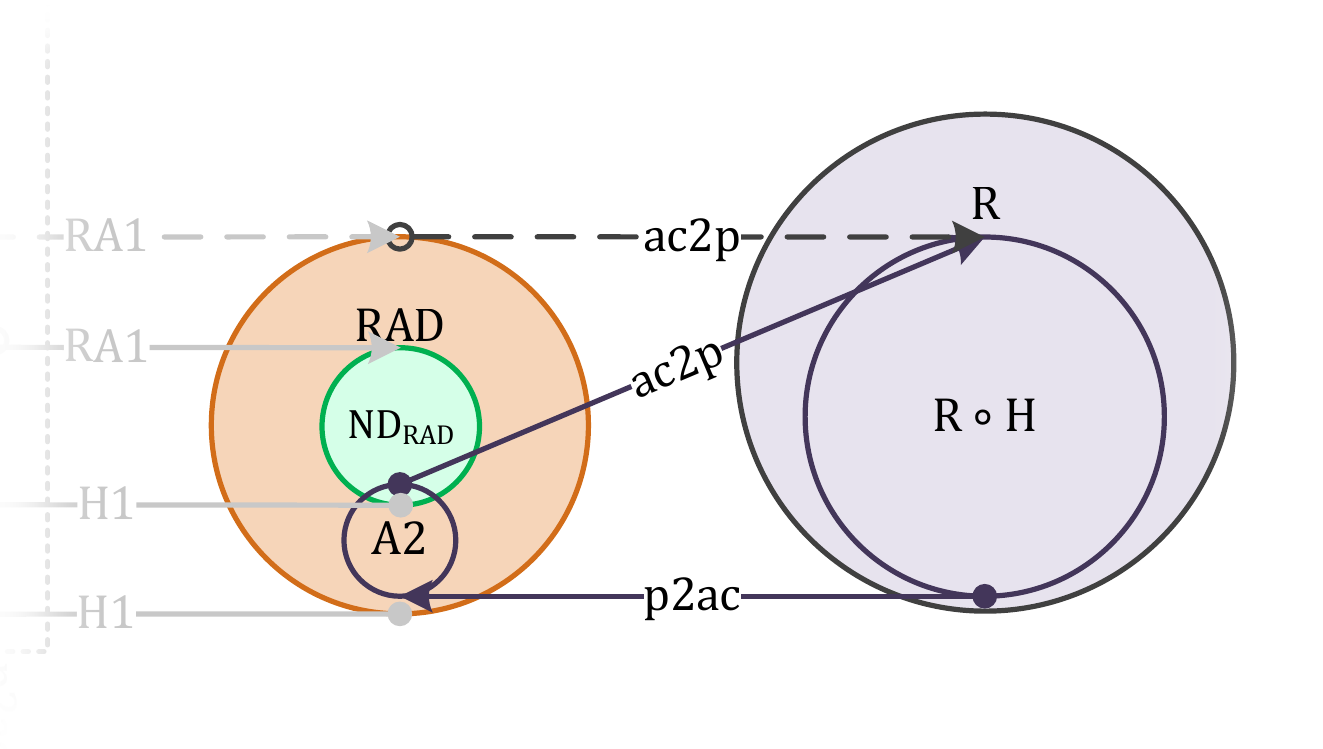}
\caption{\label{fig:theories:reactive-angelic-designs}Theory of reactive angelic designs and links with CSP}
\end{center}
\end{figure}
As part of our validation approach, we establish that the subset of $\mathbf{RAD}$ with no angelic nondeterminism, characterised by $\mathbf{A2}$, is isomorphic to the theory of~\ac{CSP}. This is achieved by introducing the linking functions $ac2p$ and $p2ac$. In general, if we consider the superset $\mathbf{RAD}$, a Galois connection exists between the theories. This relationship is illustrated in~\cref{fig:theories:reactive-angelic-designs}.


The theory of reactive angelic designs corresponds to a natural extension of the~\ac{CSP} theory with both angelic and demonic nondeterminism. In this theory it is possible to establish that angelic choice avoids divergence. For example, the angelic choice $a \circthen Chaos \sqcup b \circthen Skip$ becomes $a \circthen Skip$, provided that $a$ and $b$ are equal. However, since $\mathbf{RA1}$ requires under all circumstances that no trace of events may be undone, if $a$ and $b$ are different events, then the possibility to observe the event $a$ cannot be entirely excluded, and so divergence is still a possibility. In order to lift this restriction we have relaxed $\mathbf{RA1}$ in case of divergence, which is the motivation for the theory of angelic processes that we discuss in the next section.

\subsection{Angelic Processes}\label{sec:ch1:AP}
In order to allow angelic choice to exclude potentially divergent processes, we relax the theory of reactive angelic designs by allowing the history of events to be undone whenever there is the potential to diverge. This is achieved by not enforcing $\mathbf{RA1}$ in all cases. Therefore, we redefine $\mathbf{RA3}$ to cope with this fact as $\mathbf{RA3_{AP}}$, and define the healthiness condition of this theory as $\mathbf{AP}$, as summarized in~\cref{table:H:AP}. 

\begin{table}[htbp]
  \centering
    \begin{tabular}{c|p{12.5cm}}
	\toprule
    &\multicolumn{1}{c}{\textbf{Description}} \\
    \midrule
    $\mathbf{RA3_{AP}}$ & A recast of $\mathbf{RA3}$ in the theory of angelic processes. \\
	\midrule
	$\mathbf{AP}$ & Functional composition of $\mathbf{RA3_{AP}}$, $\mathbf{RA2}$, $\mathbf{A}$ and, $\mathbf{H1}$ and $\mathbf{H2}$ of the theory of designs (with the corresponding alphabet of this theory). \\
	\midrule
	$\mathbf{ND_{AP}}$ & Characterises the subset of non-divergent angelic processes. \\
	\bottomrule
    \end{tabular}%
  \caption{\label{table:H:AP} Healthiness Conditions of Angelic Processes}%
\end{table}%
The consequence of the functional composition underlying $\mathbf{AP}$ is that this model is effectively a theory of angelic designs, where $\mathbf{RA1}$ is only required in the postcondition. This is a direct consequence of the definition of $\mathbf{A}$, as it requires that the set of angelic choices in the postcondition of an $\mathbf{A}$-design is not empty.
\begin{figure}[htbp]
\begin{center}
\includegraphics[scale=0.58]{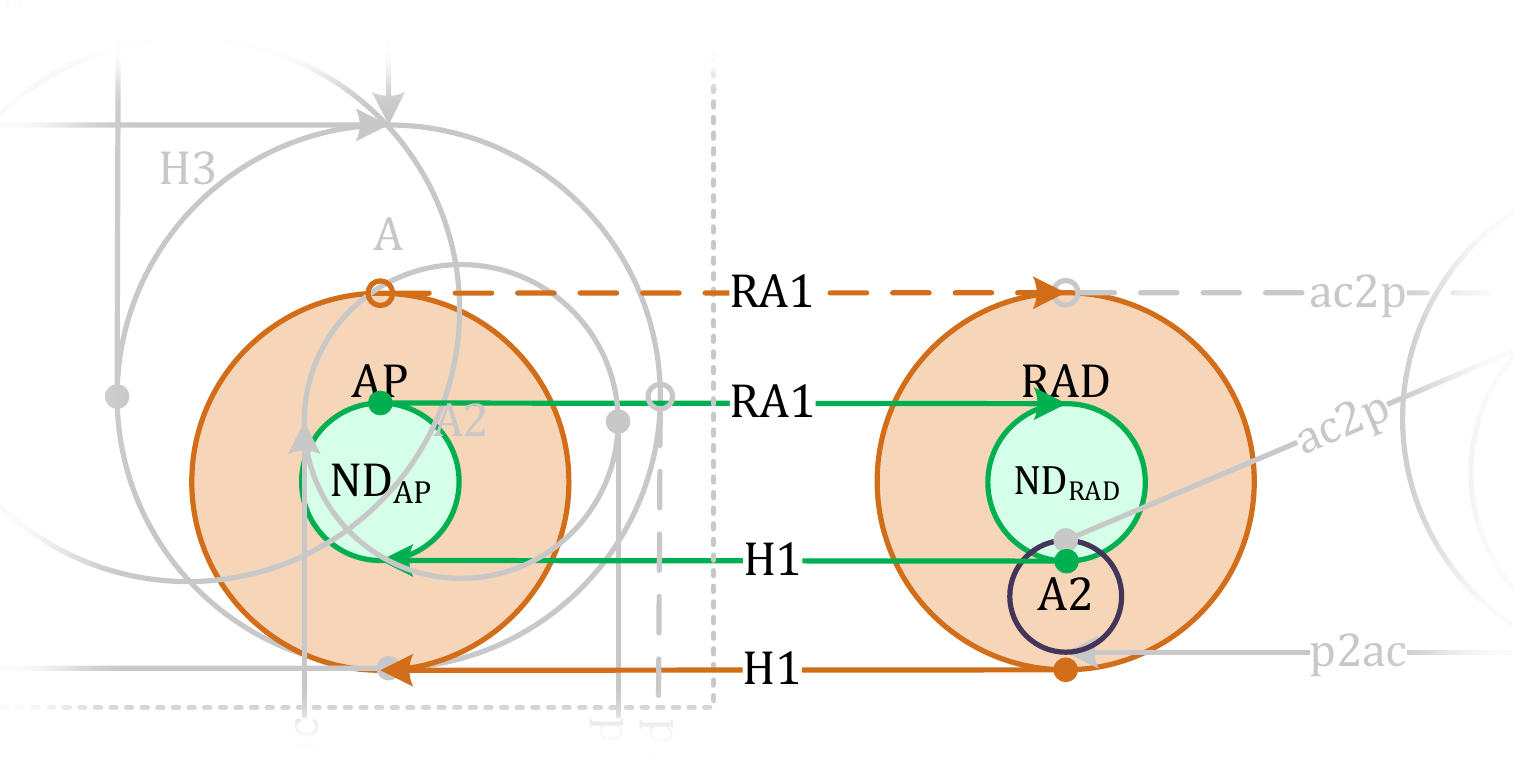}
\caption{\label{fig:theories:angelic-processes}Theory of angelic processes and link with reactive angelic designs}
\end{center}
\end{figure}

The resulting theory is more generic than that of reactive angelic designs, since it does require $\mathbf{RA1}$. As part of our validation approach, we establish a Galois connection with the theory of reactive angelic designs, and also prove that an isomorphism exists with respect to the subsets of non-divergent processes, characterised by $\mathbf{ND_{RAD}}$ and $\mathbf{ND_{AP}}$, respectively. This is achieved by turning reactive angelic designs into designs, through $\mathbf{H1}$, while in the opposite direction we just enforce $\mathbf{RA1}$. These links are depicted in~\cref{fig:theories:angelic-processes} where we highlight both theories in the context of~\cref{fig:theories}.

A detailed account of all the new theories is presented in the sequel as described below.

\section{Outline}

In~\cref{chapter:2}, we provide an overview of the concept of angelic nondeterminism as found in the literature. In addition, we discuss the most important semantic models in the context of our work by introducing: weakest precondition semantics, binary multirelations, the~\ac{UTP}, and the existing models of~\ac{CSP}. 

\Cref{chapter:3} presents the extended model of binary multirelations that handles non-terminating computations. We introduce the healthiness condition $\mathbf{BMH_\bot}$ as well as the most important operators of this theory. Finally, we establish its relationship with the existing model of binary multirelations via linking functions (see \cref{fig:theories:bmh}).

\Cref{chapter:4} introduces the theory of angelic designs, the first new~\ac{UTP} theory developed in this thesis. We introduce the alphabet of the theory, followed by the healthiness conditions $\mathbf{A0}$ to $\mathbf{A2}$. The relationship with the extended model of binary multirelations is studied before introducing the most important operators. We conclude this chapter by studying the relationship of the subset of angelic designs that are $\mathbf{H3}$-healthy and the $\mathbf{PBMH}$ theory of~\cite{Cavalcanti2006}.

In~\cref{chapter:5} the theory of reactive angelic designs is presented. This is a natural extension of the~\ac{UTP} model of~\ac{CSP} in the context of a theory with angelic nondeterminism, where the healthiness conditions of~\ac{CSP} are expressed using this new encoding. The resulting healthiness condition is $\mathbf{RAD}$. Finally, we discuss the operators and study the link with the existing theory of reactive designs.

Our final contribution is found in~\cref{chapter:6}, where we present the theory of angelic processes, whose healthiness condition is $\mathbf{AP}$. This chapter concludes by exploring the relationship with the theory of reactive angelic designs and the main algebraic properties.

Finally, in~\cref{chapter:7} we summarize the main contributions of this thesis and further contextualize our work. We conclude with pointers for future work.


\chapter{Angelic Nondeterminism}\label{chapter:2}

In this chapter we provide an account of angelic nondeterminism as found in the literature, and introduce the foundations upon which our theories are built. \cref{sec:ch2:def-and-applications} discusses the concept of angelic nondeterminism and its applications. In~\cref{sec:ch2:wkp} we introduce Dijkstra's weakest preconditions and the predicate transformers of the refinement calculus. \cref{sec:ch2:binary-multirelations} introduces Rewitzky's theory of binary multirelations. In~\cref{sec:ch2:utp} we provide an introduction to the~\ac{UTP} of Hoare and He. Finally, \cref{sec:ch2:csp-angelic-nondeterminism} contains a short introduction to~\ac{CSP} and a discussion on the different semantic approaches to characterising angelic nondeterminism in~\ac{CSP}.


\section{Definition and Applications}\label{sec:ch2:def-and-applications}

The earliest use of angelic nondeterminism can be found in the theories of computation, more specifically in automata theory~\cite{Rabin1959} and Turing machines~\cite{Cook1971}. For example, in pushdown stack automata, the addition of nondeterminism enables the automaton to accept arbitrary context-free languages~\cite{Schuetzenberger1963}, while for Turing machines it helps characterise the class of NP-problems~\cite{Cook1971} whose solutions can be found efficiently given an angelically nondeterministic machine. 

Angelic nondeterminism has been used as a specification and programming construct in several applications, including parsing~\cite{Hesselink1992}, modelling of game-like scenarios~\cite{Back1998} and user interactions, theorem proving tactics~\cite{Martin1996,Oliveira2003}, constraint programming~\cite{Jagadeesan1991}, logic programming~\cite{Kok1990} and others. These are problems where finding solutions often involves a combination of search and backtracking. For instance, in Angel~\cite{Martin1996,Oliveira2003}, theorem proof tactics can be combined through angelic choice, such that failure leads to backtracking. 

While this is a perfectly reasonable interpretation of angelic choice, backtracking is not the only possibility, nor is it always desired. Irrespective of the actual operation of an angelic choice, its distinguishing feature across the different applications is its capability to provide a high degree of abstraction while still guaranteeing success.

Already in 1967, Floyd~\cite{Floyd1967a} envisioned angelic choice as a mechanism for the abstract specification of algorithms, with actual executable programs being produced mechanically, perhaps by a compiler. In the context of his formal characterisation of programs as flowcharts, Floyd introduced explicit nondeterministic choice points, and appropriate notions of success and failure, in order to avoid implementation details of particular execution strategies. Although angelic nondeterminism is usually interpreted operationally as a backtracking mechanism, it can also be implemented through some form of parallelism~\cite{Ward1991}. 

Almost at the same time, important contributions were being made to the theoretical understanding of programs. In 1969, Hoare proposed his formal system for proving partial correctness in the context of sequential programming languages~\cite{Hoare1969}. While in 1975 Dijkstra~\cite{Dijkstra1975,Dijkstra1976} introduced his language of guarded commands, an imperative language with repetitive and nondeterministic constructs. Unlike Floyd's choice points, Dijkstra's nondeterministic choice was no longer angelic. 

Dijkstra~\cite{Dijkstra1975,Dijkstra1976} fundamentally changed the approach to establishing total correctness by calculation through his weakest precondition semantics. His model restricted itself to feasible programs by excluding the existence of miracles (with the so called ``\emph{Law of the Excluded Miracle}''). Miracle is the theoretical counterpart to abort and corresponds to the infeasible program that can never be executed, while abort represents the worst possible program whose behaviour, in the context of a theory of total correctness, is completely arbitrary.

When Back~\cite{Back1992,Back1978}, Morris~\cite{Morris1987} and Morgan~\cite{Morgan1994} introduced the refinement calculus, miracles were introduced back into their models. This enabled their models to become more generic, and paved the way for the development of models that are complete lattices under the refinement order. The most important was, perhaps, the lattice of monotonic predicate transformers where angelic and demonic choice are modelled as the least upper bound and greatest lower bound of the lattice. Back and von Wright~\cite{Back1998} extensively studied sublattices, where choice can be either angelic or demonic. They have also considered angelic nondeterminism in the context of game-like scenarios and modelling of user interactions.

Angelic choice also plays a significant role amongst data refinement techniques, such as that of Gardiner and Morgan~\cite{Gardiner1991}, where the least upper bound is used to define logical variables. These enable the postcondition of a specification statement to refer to the initial value of a program variable.

Ward and Hayes, in their work~\cite{Ward1991} on applications of angelic nondeterminism, clearly emphasize that unlike Floyd's choice points, the angelic choice of the refinement calculus can ``look ahead" and guide choices to avoid divergence, if at all possible. This is not restricted to explicit choice points, but rather applies to any angelic construct, such as the angelic assignment of values to program variables, which they explore in the refinement of programs from high-level specifications.


In the context of theories of total correctness, computations can also be specified through relations between initial states and final states. This is the notion adopted in formal notations like Z~\cite{Woodcock1996} and~\ac{VDM}~\cite{Jones1986}, where there is an explicit relation between the initial and final value of a computation. However, as Back~\cite{Back1998} and Cavalcanti et al.~\cite{Cavalcanti2004} have noted, relations can only capture one type of nondeterminism, either angelic or demonic, but not both.

When Cavalcanti et al.~\cite{Cavalcanti2004} proposed the introduction of angelic nondeterminism into the relational setting of Hoare and He's~\ac{UTP}~\cite{Hoare1998}, a multirelational encoding had to be considered. They first established that, in general, \ac{UTP} relations are isomorphic to conjunctive predicate transformers. Their solution to the problem consisted in defining a predicative encoding of Rewitzky's~\cite{Rewitzky2003} upward-closed binary multirelations, which is the basis for the work that we describe in this thesis.


As already mentioned, Rewitzky's~\cite{Rewitzky2003} multirelations are relations between initial states and sets of final states. In~\cite{Martin2004} several models of binary multirelations are considered, of which the model of upward-closed multirelations is the most important due to its lattice-theoretic structure. In this model, the refinement order is reverse subset inclusion, and angelic and demonic choice correspond to set union and intersection, respectively. We discuss this model in more detail in~\cref{sec:ch2:binary-multirelations}.

More recently, Guttmann~\cite{Guttmann2014} has proposed a generalised algebraic structure that has both the monotonic predicate transformers and multirelations as instances. Guttmann has also extensively studied the relational properties of multirelations, and proposed an extension catering for non-terminating computations~\cite{Guttmann2014a} in the setting of general correctness. This involves extending the set of final states to record whether a computation does not terminate: a similar idea is used in our extended model of binary multirelations~\cite{Ribeiro2014a} where we record whether a computation may not terminate and still establish some final value. This model is part of the first contribution of this thesis and is discussed in detail in~\cref{chapter:3}.

In~\cref{sec:ch2:csp-angelic-nondeterminism} we come back to the topic of angelic nondeterminism by reviewing the existing approaches to characterising angelic nondeterminism in~\ac{CSP}. Next we introduce Dijkstra's weakest precondition semantics.

\section{Weakest Preconditions}\label{sec:ch2:wkp}

As already discussed, one of the earliest treatments of total correctness is due to Dijkstra~\cite{Dijkstra1975,Dijkstra1976}, through his language of guarded commands and weakest precondition semantics. The underlying idea is that for each program statement $S$ and postcondition $q$, it is possible to establish the weakest precondition $wp(S,q)$, such that, starting $S$ in a state satisfying $wp(S,q)$ achieves postcondition $q$. A weakest precondition characterises all possible initial states that lead to successful termination with the postcondition holding. In Dijkstra's model~\cite{Dijkstra1975,Dijkstra1976}, predicates are characterised by functions on all points of a state space, which in his original presentation~\cite{Dijkstra1976} are defined through Cartesian products.

If we consider the program $Skip$, which does not change the state and always terminates successfully, its weakest precondition semantics is defined as follows.%
\begin{define}
$wp(Skip,q) = q$
\end{define}\noindent%
That is, the weakest precondition corresponds exactly to the intended outcome $q$. A simple assignment statement, where a program variable $x$ is assigned the value of an expression $e$, is given semantics for a postcondition $q$ as follows.%
\begin{define}$wp(x:=e,q) \circdef q[e/x]$
\end{define}\noindent%
In other words, the weakest precondition of the assignment is given as the substitution of expression $e$ for variable $x$ in the corresponding postcondition $q$.

In general, not all possible weakest preconditions are valid, in the sense that the semantic model must obey certain fundamental properties of interest, such as monotonicity. In what follows, we review the original properties of Dijkstra's model~\cite{Dijkstra1976}.

\subsection{Healthiness Conditions}
Dijkstra's semantics~\cite{Dijkstra1976} insist on four healthiness conditions, which we discuss in this section. The first property, reproduced below, corresponds to the ``\emph{Law of the Excluded Miracle}'', which forbids miraculous behaviour from being specified.%
\begin{define}[Non-miraculous] 
$ wp(S,F) = F$
\end{define}\noindent%
If program statement $S$ could achieve $F$, the predicate which is false everywhere, then there must be no such initial state where $wp(S,F)$ that can be satisfied. This is precisely one of the properties that Back~\cite{Back1998}, Morris~\cite{Morris1987} and Morgan~\cite{Morgan1994} relaxed in order to introduce the lattice of monotonic predicate transformers. 

The fundamental property of interest in models for refinement is monotonicity. The definition~\cite{Dijkstra1976} is reproduced below. 
\begin{define}[Monotonicity] 
$(q \implies r) \implies (wp(S,q) \implies wp(S,r))$
\end{define}\noindent%
For every state and program statement $S$, whenever $q$ is a stronger predicate than $r$, then the weakest precondition $wp(S,q)$ is also stronger than $wp(S,r)$. In other words, if $q$ is a postcondition stronger than $r$, then, the set of initial states guaranteed to establish $q$ is a subset of those that establish $r$. 

The next healthiness condition that Dijkstra presents is conjunctivity, whose formal definition is reproduced below~\cite{Dijkstra1976}.%
\begin{define}[Conjunctivity] 
$wp(S,q) \land wp(S,r) \iff wp(S,q \land r)$
\end{define}\noindent%
The right-hand side implication follows directly from monotonicity and properties of the predicate calculus. However, the left-hand side implication is not necessarily satisfied in general. In fact, this property is precisely what prevents angelic nondeterminism from being specified in Dijkstra's model, as noted by Back~\cite{Back1992}. This result follows from the definition of the angelic statement whose semantics, as given, for example, in~\cite{Ward1991,Back1989}, is defined using an existential quantification.

The counterpart to conjunctivity is disjunctivity, whose definition is as follows.%
\begin{define}[Disjunctivity]
$wp(S,q) \lor wp(S,r) \iff wp(S,q \lor r)$
\end{define}\noindent%
Since weakest preconditions observing this property cannot model demonic nondeterminism, Dijkstra~\cite{Dijkstra1976} uses a weaker version where only the left-hand side implication is enforced. Similarly to the angelic statement, the demonic specification statement is defined, for example, in~\cite{Ward1991,Back1989} using a universal quantification. 

In~\cite{Back1992} Back and von Wright extensively study different models of weakest preconditions with different properties, including models with and without miracles, conjunctivity and disjunctivity. They conclude that by considering a model that is neither conjunctive nor disjunctive, both forms of nondeterminism can be modelled together. Furthermore, by considering a model with miracles, a complete lattice exists where angelic and demonic choice correspond to the meet and join, respectively. This is a result explored in all versions of the refinement calculus~\cite{Back1998,Morris1987,Morgan1994}. Our remaining discussion on weakest preconditions is mostly based on Back and von Wright's work~\cite{Back1998}.

\subsection{Predicate Transformers}
The $wp$ function of Dijkstra is a predicate transformer as it maps predicates to predicates. Back and von Wright~\cite{Dijkstra1976}, in their presentation of the refinement calculus introduce the notion of contracts which can be either specifications or programs. The satisfaction of a contract $S$ by establishing postcondition $q$ when started from an initial state $\sigma$ is denoted by $\sigma~\{| S |\}~q$. They characterise $wp : \power \Sigma \fun \power \Sigma$, where the state space is $\Sigma$, for a contract $S$ as follows.%
\begin{define}[Weakest Precondition]
$ wp(S,q) \circdef \{ \sigma | \sigma~\{| S |\}~q \}$
\end{define}\noindent%
That is, the set of all initial states $\sigma$, from which $S$ is guaranteed to establish $q$. Weakest precondition semantics can then be given to their language of contracts~\cite{Back1998}, which we reproduce in the following definition.%
\begin{define}[Basic Weakest Preconditions]
\begin{align*}
wp(\lseq f \rseq,q) 	&= f^{-1} (q) \\
wp(\{ g \},q) 			&= g \cap q \\
wp([g],q) 				&= \lnot g \cup q \\
wp(S_1 \circseq S_2,q)	&= wp(S_1,wp(S_2,q)) \\
wp(S_1 \sqcup S_2,q)	&= wp(S_1,q) \cup wp(S_2,q) \\
wp(S_1 \sqcap S_2,q)	&= wp(S_1,q) \cap wp(S_2,q)
\end{align*}
\end{define}\noindent %
The first construct $\lseq f \rseq$ is a functional update that changes the state according to function $f$. An example is the identity $id$, which does not change the state.

The following construct $\{ g \}$ is an assertion, which has no effect on the state if $g$ holds. Otherwise the program aborts. The assertion $\sigma~\{| \{g\} |\}~q$ holds if, and only if, the state $\sigma$ is in the intersection of $g$ and the postcondition $q$.

Its dual is the assumption $[g]$; it has no effect if $g$ holds and otherwise the contract is satisfied trivially. Hence, the weakest precondition is given by $\sigma \in q$ and otherwise, if $g$ fails to hold then $\sigma \in \lnot g$.

The sequential composition of $S_1$ and $S_2$ is given as the weakest precondition of $S_1$, with respect to the postcondition characterised by the weakest precondition of $S_2$. That is, $wp(S_2, q)$ is an intermediate condition that needs to be satisfied in order to achieve $q$.

Finally, angelic and demonic choice are defined as $\sqcup$ and $\sqcap$, respectively. In an angelic choice, it is sufficient that either the precondition of $S_1$ or $S_2$ is satisfied in order to achieve $q$, whereas in a demonic choice both need to be satisfied.

\subsection{Predicate Transformers Lattice}
In Back and von Wright's model~\cite{Back1998}, the notion of refinement is given for two contracts $S_1$ and $S_2$ as follows.%
\begin{define}
$ S_1 \sqsubseteq S_2 \iff \forall \sigma, q \spot \sigma~\{| S_1 |\}~q \implies \sigma~\{| S_2 |\}~q$
\end{define}\noindent%
A contract $S_1$ is refined by $S_2$ if, and only if, for all initial states $\sigma$ and postconditions $q$, if $\sigma$ is an initial state of contract $S_1$ leading to postcondition $q$, then it is also an initial state of $S_2$ leading to $q$. As this order is reflexive, transitive and antisymmetric~\cite{Back1998,Davey2002}, it is a partial order. The bottom element is the assertion $\{false\}$, which can never be satisfied in any initial state, while the top element is the assumption $[false]$, so that it is trivially satisfied in any initial state for any final condition $q$.

When Back and von Wright~\cite{Back1998} introduce their model of predicate transformers, they actually consider the target state space as being potentially different from the initial state space, as required, for instance, to model states with scoped variables. Thus, the set of predicate transformers from an initial state space $\Sigma$, to a final state space $\Gamma$ is defined by $\power \Gamma \fun \power \Sigma$.

The refinement order for predicate transformers is defined by considering the pointwise extension of the subset ordering; for predicate transformers $T_1$ and $T_2$, we have the following definition.%
\begin{define}
$ T_1 \sqsubseteq T_2 \circdef \forall q \in \power \Gamma \spot T_1(q) \subseteq T_2(q)$
\end{define}\noindent%
That is, $T_1$ is refined by $T_2$, if, and only if, the set of initial states that characterise the weakest precondition for $q$ to be established according to $T_1$ is a subset of that characterised by $T_2$. This order forms a complete Boolean lattice~\cite{Back1998}. Thus the lattice operators on predicate transformers are pointwise extensions of the corresponding operators on predicates~\cite{Back1998}.

Finally, in~\cite{Back1998} Back and von Wright consider the complete sublattice of monotonic predicate transformers. What is particularly important about their result is that every basic statement is monotonic and so are the sequential composition, meet, and join of predicate transformers~\cite{Back1998}.

This concludes our discussion of the lattice of monotonic predicate transformers as the standard model where angelic and demonic nondeterminism have traditionally been studied. In the following~\cref{sec:ch2:binary-multirelations} we discuss the theory of upward-closed binary multirelations, which is effectively a relational characterisation of the predicate transformers model~\cite{Rewitzky2003}.


\section{Binary Multirelations}\label{sec:ch2:binary-multirelations}
As already discussed, it is not possible to model both angelic and demonic nondeterminism in a purely relational model. However, multirelational models can be used to characterise both forms of nondeterminism in a relational setting. 

In~\cite{Rewitzky2003} Rewitzky introduces the theory of binary multirelations, which are relations between initial states and sets of final states. In our presentation we define these relations through the following type $BM$, where $State$ is a type of records with a component for each program variable.%
\begin{define}\label{def:BM}
\begin{statement}
$ BM \circdef State \rel \power State$
\end{statement}
\end{define}\noindent %
An example of a program in this model is the assignment of the value $1$ to the only program variable $x$ when started from any initial state.%
\begin{example}
$x :=_{BM} 1 = \{ s : State, ss : \power State | (x \mapsto 1) \in ss\}$
\end{example}\noindent %
This assignment, which we subscript with $BM$ to distinguish it from assignment statements in other models that we discuss later, is defined by relating every initial state $s$ to a set of final states $ss$ where the component $x$ is set to the value $1$. For conciseness, in the examples and definitions that follow, the types of $s$ and $ss$ may be omitted where it is clear that the composite type is $BM$.

The target set of a binary multirelation can be interpreted as either encoding angelic or demonic choices~\cite{Rewitzky2003,Cavalcanti2004}. Here we present a model where the set of final states encodes angelic choices. This decision is justified in~\cite{Cavalcanti2006} as maintaining the refinement order of the isomorphic~\ac{UTP} model of Cavalcanti et al.~\cite{Cavalcanti2006}, which we discuss in~\cref{sec:ch2:utp:angelic-nondeterminism}. 

Demonic choices are encoded by the different ways in which the set of final states can be chosen. For example, consider the following program which angelically assigns the value $1$ or $2$ to the only program variable $x$; it uses $\sqcup_{BM}$ the angelic choice operator for binary multirelations. %
\begin{example}\label{example:bm:angelic-assignment}
$ x:=_{BM} 1 \sqcupBM x:=_{BM} 2 = \{ s, ss | (x \mapsto 1) \in ss \land (x \mapsto 2) \in ss \} $
\end{example}\noindent %
In this multirelation, every initial state $s$ is associated with all sets $ss$ in which we can find the choice of a final state where $x$ is assigned the value $1$ or $2$. Irrespective of the set of final states chosen by the demon, the angel is always able to enforce this choice. As illustrated, for a particular initial state, the choices available to the angel correspond to those in the distributed intersection over all possible sets of final states.

\subsection{Healthiness Conditions}
\cref{example:bm:angelic-assignment} above illustrates a fundamental property of binary multirelations: upward-closure~\cite{Rewitzky2003}. This property is captured by the following healthiness condition for a multirelation $B$. %
\begin{define}\label{def:bm:bmh}
\begin{statement}
$\mathbf{BMH} \circdef \forall s, ss_0, ss_1 \spot ((s, ss_0) \in B \land ss_0 \subseteq ss_1) \implies (s, ss_1) \in B$
\end{statement}
\end{define}\noindent%
If an initial state $s$ is related to a set of final states $ss_0$, then it is also related to any superset $ss_1$. This reflects the fact that if it is possible to terminate in some final state in $ss_0$, then the addition of any other final states to that set does not change the actual states available for angelic choice. 

Upward-closure ensures that there is a complete lattice under the subset order, with angelic and demonic choice corresponding to the least upper bound and greatest lower bound, respectively. Moreover, in~\cite{Rewitzky2003} Rewitzky establishes that there is a bijection between upward-closed binary multirelations and monotonic unary operators. Since, as explained in~\cref{sec:ch2:wkp} predicate transformer semantics can be given in terms of monotonic unary operators, this establishes that the multirelational model is in fact a relational characterisation for commands with both forms of nondeterminism.

\subsection{Refinement}
In the model of upward-closed binary multirelations, refinement is defined for healthy multirelations $B_0$ and $B_1$ by reverse subset inclusion as follows~\cite{Rewitzky2003}. %
\begin{define}
$B_0 \refinedbyBM B_1 \circdef B_0 \supseteq B_1$
\end{define}\noindent
A multirelation $B_0$ is refined by $B_1$ if, and only if, $B_1$ is a subset of $B_0$. 

This partial order forms a complete lattice. The bottom element $\bot_{BM}$, corresponding to the notion of \textbf{abort}, is defined by the universal relation, which associates every initial state to every possible set of final states. %
\begin{define}
$\bot_{BM} \circdef State \times \power State$
\end{define}\noindent %
The top element $\top_{BM}$ is defined by the empty relation and corresponds to the notion of \textbf{miracle}, the infeasible program. %
\begin{define}
$\top_{BM} \circdef \emptyset$
\end{define}\noindent %
Via refinement, the degree of angelic nondeterminism of a program can be increased, while the degree of demonic nondeterminism can be decreased, that is, a program can be refined into a demonically more deterministic one. In particular, the infeasible program $\top_{BM}$ refines every other program, while every program refines $\bot_{BM}$.

\subsection{Operators}
In this section we present the main operators of the theory of binary multirelations and discuss their most important properties.

\subsubsection{Assignment}
The first operator of interest, which we have briefly discussed in~\cref{example:bm:angelic-assignment}, is assignment. Its complete definition is as follows.

\begin{define}
$ x :=_{BM} e \circdef \{ s, ss | s \oplus (x\mapsto e) \in ss\}$
\end{define}\noindent
Every initial state $s$ is related to every set of final states $ss$ that includes a state where $s$ is overridden to define that $x$ has the value of expression $e$.

\subsubsection{Angelic Choice} The angelic choice operator is defined as set intersection.

\begin{define}
$ B_0 \sqcup_{BM} B_1 = B_0 \cap B_1$
\end{define}\noindent %
This operator corresponds to the least upper bound of the lattice. Intuitively, the final states available for angelic choice are those in the intersection of all choices available for demonic choice. The operator satisfies the following property.
\begin{lemma}
$ B_0 \refinedbyBM B_0 \sqcup_{BM} B_1$
\end{lemma}\noindent
That is, the degree of angelic nondeterminism can be increased.

\subsubsection{Demonic Choice}

Its dual, demonic choice, is the greatest lower bound and is defined as set union.
\begin{define}
$ B_0 \sqcap_{BM} B_1 = B_0 \cup B_1$
\end{define}\noindent %
For a given initial state, the sets of final states available for demonic choice correspond to those in either $B_0$ or $B_1$. Demonic choice observes the following property.

\begin{lemma}
$ B_0 \sqcap_{BM} B_1 \refinedbyBM B_0$
\end{lemma}\noindent %
That is, the degree of demonic nondeterminism can be decreased. Finally, angelic and demonic choice distribute over one another.

\begin{lemma}
$ B_0 \sqcap_{BM} (B_1 \sqcup B_2) = (B_0 \sqcap_{BM} B_1) \sqcup_{BM} (B_0 \sqcap_{BM} B_2)$
\end{lemma}\noindent %
This property follows from the distributive properties of set union and set intersection. It is equally valid in the theory of predicate transformers and the isomorphic~\ac{UTP} model of~\cite{Cavalcanti2006}.

\subsubsection{Sequential Composition}
Although this is a relational model, since states are related to sets of states, the definition of sequential composition is not relational composition. Instead it is defined as follows. %
\begin{define}
\begin{align*}
B_0 \seqBM B_1 \circdef \{ s_0, ss_1 | \exists ss_0 \spot (s_0, ss_0) \in B_0 \land ss_0 \subseteq \{ s_1 | (s_1, ss_1) \in B_1 \} \}
\end{align*}
\end{define}\noindent %
It considers every initial state $s_0$ in $B_0$ and set of final states $ss_1$, such that there is some intermediate set of states $ss_0$ that is related from $s_0$ in $B_0$, and $ss_0$ is a subset of those initial states of $B_1$ that achieve $ss_1$. As noted in~\cite{Cavalcanti2006} for healthy multirelations this definition can be simplified further as shown in the following lemma. %
\begin{lemma}\label{lemma:seqBM} Provided $B_0$ satisfies $\mathbf{BMH}$,
\begin{align*}
B_0 \seqBM B_1 \circdef \{ s_0, ss_1 | (s_0, \{ s_1 | (s_1, ss_1) \in B_1\}) \in B_0 \}
\end{align*}
\begin{proof}
Equation 5 in~\cite{Cavalcanti2006}.
\end{proof} 
\end{lemma}\noindent %
This definition is the basis for the definition of sequential composition in the isomorphic~\ac{UTP} model of~\cite{Cavalcanti2006}, and for the definition of sequential composition in the extended model of binary multirelations that we discuss in~\cref{chapter:3}.


\section{The Unifying Theories of Programming}\label{sec:ch2:utp}
As previously discussed, the~\ac{UTP} of Hoare and He~\cite{Hoare1998} is a framework of alphabetized relations suitable for characterising different programming paradigms. The~\ac{UTP} promotes unification of results while enabling different aspects of programs to be considered in isolation. In~\cite{Hoare1998} a collection of theories is presented that targets multiple aspects of different programming paradigms, such as functionality, concurrency, logic programming and higher-order programming. Several other theories have since been developed which cater for other aspects, such as angelic nondeterminism~\cite{Cavalcanti2006}, object-orientation~\cite{Santos2006,Zeyda2014}, pointers~\cite{Harwood2008} and time~\cite{Sherif2002,Sherif2006,Wei2012}.

The~\ac{UTP} is based on the principle of observation, and so the discourse for recording observations is defined by an alphabet whose variables determine the observable parameters of a system. These can be either program variables, or alternatively, auxiliary variables that capture information like termination and execution time. A~\ac{UTP} theory is characterised by three components: an alphabet, a set of healthiness conditions and a set of operators.

For a given relation $P$, its alphabet is given by $\alpha(P)$. Similar to the conventions of Z, in the~\ac{UTP} an alphabet is split into two disjoint subsets: $in\alpha(P)$, which contains undashed variables for characterising the initial observations, and $out\alpha(P)$, which contains the dashed counterparts of each variable that characterise the final or subsequent observations of a system. For example, a program whose purpose is to increment the initial value of the only program variable $x$ can be specified by the relation: $x' = x + 1$. This relation concisely describes all pairs of values $(x, x')$ that satisfy this predicate. Thus relations characterise the possible observations of a program.

When the input and output alphabets of a relation are exactly the same, except for the fact that variables are undashed and dashed in either set, respectively, a relation is said to be homogeneous. %
\begin{define}[Homogeneous Relation]\label{def:homogeneous} A relation $P$ is homogeneous if, and only if, $(in\alpha(P))' = out\alpha(P)$.\end{define}\noindent
This is captured by~\cref{def:homogeneous}, where $(in\alpha(P))'$ is the set of variables obtained by dashing every variable in the set $in\alpha(P)$.

The remainder of this section is organised as follows. In~\cref{sec:ch2:theories} we discuss the other two components of~\ac{UTP} theories, namely healthiness conditions and operators. In~\cref{sec:ch2:designs} we introduce the theory of designs which captures total correctness. In~\cref{sec:ch2:utp:links} we discuss the approach to linking theories in the~\ac{UTP}. Finally, \cref{sec:ch2:utp:angelic-nondeterminism} discusses the theory of angelic nondeterminism of~\cite{Cavalcanti2006}.

\subsection{Theories}\label{sec:ch2:theories}
The second component of a~\ac{UTP} theory is a set of healthiness conditions that characterise the predicates of a theory. These are normally specified by idempotent and monotonic functions whose fixed points are the valid predicates of the theory.

\subsubsection{Healthiness Conditions}
For instance, in the context of theories concerning time, it is often possible to make observations of a system in discrete-time units recorded using a variable $t$. It is expected that any plausible theory describing such a system must guarantee that time is increasingly monotonic. This property can be described by the following healthiness condition $HC$.
\begin{example}
$ \mathbf{HC} (P) \circdef P \land t \le t'$
\end{example}\noindent
It requires that under all circumstances, it must be the case that the initial value of $t$ is less than or equal to the final or after value $t'$. This healthiness condition is defined in terms of conjunction, so it is called a conjunctive healthiness condition~\cite{Harwood2008}. A general result on conjunctive healthiness conditions~\cite{Harwood2008} enables us to establish that HC is idempotent and monotonic with respect to refinement. An observation in this theory is valid if, and only if, it is a fixed point of HC. 

\subsubsection{Refinement}
The theory of relations forms a complete lattice~\cite{Hoare1998}, with the order given by (reverse) universal implication. The top of the lattice is $false$ and the bottom is $true$. This order corresponds to the notion of refinement. Its definition is presented below, where the square brackets stand for universal quantification over all the variables in the alphabet~\cite{Hoare1998}. %
\begin{define}[Refinement]
$P \sqsubseteq Q \circdef [Q \implies P]$
\end{define}\noindent %
Refinement can be understood as capturing the notion of correctness in the sense that, if a predicate $Q$ refines $P$, then all possible behaviours exhibited by $Q$ are permitted by $P$. This notion is paramount for the~\ac{UTP} framework and it is the same across all theories. The relation $true$ imposes no restriction and permits the observation of any value for all variables in the alphabet, while $false$ permits none.

\subsubsection{Operators}
A~\ac{UTP} theory comprises a number of operators that characterise how the theory may be used algebraically to specify more complex behaviours. In the theory of relations there are a number of core operators that correspond to typical constructs found in programming languages, such as assignment ($:=$), conditional ($A \dres c \rres B$), and sequential composition ($\circseq$). In what follows we present some of the most important operators of the theory of relations.

\subsubsection{Sequential Composition}
In~\ac{UTP} theories whose relations are homogeneous, sequential composition is defined as relational composition. The definition is shown below through substitution.
\begin{define}[Sequential Composition]\label{def:sequential-composition}
$P \circseq Q \circdef \exists v_0 \spot P[v_0/v'] \land Q[v_0/v]$
\end{define}\noindent
The intuition here is that the sequential composition of two relations $P$ and $Q$ involves some intermediate, unobservable state, whose vector of variables is represented by $v_0$. This vector is substituted in place for the final values of $P$, as represented by $v'$, as well as substituted for the initial values of $Q$, as represented by $v$. It is finally hidden by the existential quantifier.

\subsubsection{Skip}
An important construct in the relational theory is the program $\IIR$, otherwise also known as $\mathbf{Skip}$, whose definition is presented below.
\begin{define}[Skip]
$\IIR \circdef (v' = v)$
\end{define}\noindent
This is a program that keeps the value of all variables unchanged. The most interesting property of $\IIR$ is that it is the left-unit for sequential composition~\cite{Hoare1998}.

\subsubsection{Demonic Choice}
Due to the lattice-theoretic approach of the~\ac{UTP}, demonic choice ($\sqcap$) corresponds to the greatest lower bound. This means that its definition is simply disjunction. %
\begin{define}[Demonic choice]
$P \sqcap Q \circdef P \lor Q$
\end{define}\noindent
Unfortunately the least upper bound, which is conjunction, does not correspond to the notion of angelic choice. As mentioned previously, it is not possible to represent both choices directly within the relational model~\cite{Cavalcanti2006}. 

\subsubsection{Recursion}
Recursion is defined in the~\ac{UTP} as the weakest fixed point. Since we have a complete lattice, it is possible to find a complete lattice of fixed points as established by a result due to Tarski~\cite{Hoare1998,Davey2002}. In the following definition, $F$ is a monotonic function and $\bigsqcap$ is the greatest lower bound. %
\begin{define}[Recursion]
$\circmu X \spot F(X) \circdef \bigsqcap \{ X | [F(X) \sqsubseteq X] \}$
\end{define}\noindent
A non-terminating recursion, such as $(\circmu Y \spot Y)$, is equated with the bottom of the lattice, $true$~\cite{Hoare1998}. Intuitively this means that it does not terminate, but if we sequentially compose this recursion with another program, then it becomes possible to recover from the non-terminating recursion as shown in the following example~\cite{Woodcock2004}.
\begin{example}
\begin{align*}
	&(\circmu Y \spot Y) \circseq x' = 0
	&&\ptext{Definition of recursion}\\
	&=\bigsqcap \{ X | [(\circmu Y \spot Y)(X) \sqsubseteq X] \} \circseq x' = 0
	&&\ptext{Function application}\\
	&=\bigsqcap \{ X | [X \sqsubseteq X] \} \circseq x' = 0
	&&\ptext{Reflexivity of $\sqsubseteq$}\\
	&=\bigsqcap \{ X | true \} \circseq x' = 0
	&&\ptext{Property of $\sqcap$}\\
	&=true \circseq x' = 0
	&&\ptext{Definition of sequential composition}\\
	&=\exists v_0 \spot true \land x' = 0
	&&\ptext{Propositional calculus}\\
	&=x' = 0
\end{align*}
\end{example}\noindent
This issue motivated Hoare and He~\cite{Hoare1998} to propose the theory of designs that we present in the following~\cref{sec:ch2:designs}. 
\subsection{Designs}\label{sec:ch2:designs}
As already mentioned, when considering theories of total correctness for reasoning about programs, the theory of relations is not appropriate due to the fact that it allows unrealistic observations of recovery from non-terminating programs~\cite{Hoare1998,Woodcock2004}. In other words, the bottom of the lattice, $true$, is not necessarily a left-zero of sequential composition as would be needed. As a result, Hoare and He~\cite{Hoare1998} have introduced the theory of designs, which addresses this issue. 

\subsubsection{Alphabet}
The theory of designs is defined by considering the addition of two auxiliary Boolean variables to the alphabet: $ok$ and $ok'$. Their purpose is to track whether a program has been started, in which case $ok$ is $true$, and whether a program has successfully terminated, in which case $ok'$ is $true$.

In what follows we present the healthiness conditions that define the theory of designs. Finally we discuss the notion of refinement in the context of designs.

\subsubsection{Healthiness Conditions}
Any valid predicate of this theory has to obey two basic principles: that no guarantees can be made by a program before it has started, and, that no program may require non-termination. These two principles are formally characterised by the healthiness conditions $\mathbf{H1}$, and $\mathbf{H2}$, respectively~\cite{Hoare1998}. We reproduce their definitions below.%
\begin{define}\label{def:H1}
\begin{statement}
$\mathbf{H1} (P) \circdef ok \implies P$
\end{statement}
\end{define}\noindent
The definition of $\mathbf{H1}$ states that any guarantees made by $P$ can only be established once it has started. Otherwise, any observation is permitted and it behaves like the bottom of the lattice, which is the same as the one for relations: $true$. %
\begin{define}\label{def:H2}
\begin{statement}
$\mathbf{H2} (P) \circdef [P[false/ok'] \implies P[true/ok']]$
\end{statement}
\end{define}\noindent
The definition of $\mathbf{H2}$ states that if it is possible for a program $P$ not to terminate, that is for $ok'$ to be $false$, then it must also be possible for it to terminate, that is for $ok'$ to be true $true$. This healthiness condition can alternatively be expressed using the $J$-split of~\cite{Cavalcanti2006a} as $\mathbf{H2} (P) = P \circseq J$, where $J \circdef (ok \implies ok') \land v'=v$. That is, the value of $ok$ can increase monotonically, while every other variable $v$ is unchanged. 

A predicate that is both $\mathbf{H1}$ and $\mathbf{H2}$ satisfies the following property.
\begin{lemma}[Design]
\begin{align*}
	\mathbf{H1} \circ \mathbf{H2} (P) = (ok \land \lnot P[false/ok']) \implies (P[true/ok'] \land ok')
\end{align*}
\begin{proof}
Theorem 3.2.3 in~\cite{Hoare1998}.
\end{proof}
\end{lemma}\noindent
Here the design is split into two parts: a precondition and a postcondition. It is defined using the notation of Hoare and He~\cite{Hoare1998} as shown in the following definition.
\begin{define}[Design]
$(P \vdash Q) \circdef (ok \land P) \implies (ok' \land Q)$
\end{define}\noindent
A design can also be written using the following notation, where we use the shorthand notation $P^a = P[a/ok']$, with $t = true$ and $f = false$, as introduced by Woodcock and Cavalcanti~\cite{Woodcock2004}, which emphasises that we can assume without loss of generality, that $ok'$ is not free in pre and postconditions. Furthermore, it is usually assumed that $ok$ is also not free in either $P$ or $Q$.
\begin{lemma}[Design] A predicate $P$ is a design if, and only if, it can be written in the following form:
$(\lnot P^f \vdash P^t)$.
\begin{proof}
Theorem 3.2.3 in~\cite{Hoare1998} and definition of design.
\end{proof}
\end{lemma}\noindent
We observe that the functions $\mathbf{H1}$ and $\mathbf{H2}$ (and indeed all of the healthiness conditions of designs) are idempotent and monotonic with respect to refinement~\cite{Hoare1998}. Furthermore, none of the proofs establishing these results rely on the property of homogeneity. Therefore it is possible to define a non-homogeneous theory of designs.

Hoare and He~\cite{Hoare1998} identified another two healthiness conditions of interest which we discuss further below. The third healthiness condition $\mathbf{H3}$ requires $\IID$, the $\mathbf{Skip}$ of designs, to be a right-unit for sequential composition~\cite{Hoare1998}.

\begin{define}[Skip]
$\IID \circdef (true \vdash v' = v)$
\end{define}\noindent
$\mathbf{Skip}$ is the program that always terminates successfully and does not change the program variables. It is essentially the counterpart to $\IIR$ in the theory of designs. 
\begin{define}\label{def:H3}
\begin{statement}
$\mathbf{H3} (P) \circdef P \circseq \IID$
\end{statement}
\end{define}\noindent
From this definition it may not be immediately obvious how designs are further restricted by $\mathbf{H3}$. In fact, it requires the precondition not to have any dashed variables (as confirmed by~\cref{theorem:P-sequence-IID}). In order to understand the intuition behind it we consider an example of a design that is not $\mathbf{H3}$-healthy.
\begin{example}\label{example:non-H3:one}
\begin{align*} 
	&(x'\neq2 \vdash true)
	&&\ptext{Definition of designs}\\
	&=(ok \land x'\neq2) \implies ok'
	&&\ptext{Propositional calculus}\\
	&=ok \implies (x'=2 \lor ok')
\end{align*}
\end{example}\noindent
In this case we have a program that upon having started can either terminate and any final values are permitted, or can assign the value $2$ to the variable $x$ and termination is then not required. In the context of a theory of total correctness for sequential programs this is a behaviour that would not normally be expected. However it is worth noting that in the context of~\ac{CSP} non $\mathbf{H3}$-designs are important, since they enable the specification of~\ac{CSP} processes such as $a \circthen Chaos$.

The healthiness condition $\mathbf{H3}$ can also be interpreted as guaranteeing that if a program may not terminate, then it has arbitrary behaviour. Thus a predicate that is $\mathbf{H3}$-healthy is also necessarily $\mathbf{H2}$-healthy~\cite{Cavalcanti2006}.

If we expand the definition of $\mathbf{H3}$ by applying the definition of sequential definition for designs we obtain the following result~\cite{Hoare1998,Woodcock2004}.
\begin{theorem}\label{theorem:P-sequence-IID}
$((\lnot P^f \vdash P^t) = (\lnot P^f \vdash P^t) \circseq \IID) \iff (\lnot P^f = \exists v' \spot \lnot P^f)$
\begin{proof}
Theorem 3.2.4 in~\cite{Hoare1998} and proof in Section 6.3 of~\cite{Woodcock2004}.
\end{proof}
\end{theorem}\noindent
This theorem shows that the value of any dashed variables in $\lnot P^f$ must be irrelevant. Therefore any design that is $\mathbf{H3}$-healthy can only have a condition as its precondition, that is, a predicate that only mentions undashed variables, and thus can only impose restrictions on previous programs.

Finally, the last healthiness condition of interest is $\mathbf{H4}$, which restricts designs to feasible programs. It is defined by the following algebraic equation~\cite{Hoare1998} that requires that $true$ is a right-zero for sequential composition. 
\begin{define}[$\mathbf{H4}$]\label{def:H4}
\begin{statement}
$P \circseq true = true$
\end{statement}
\end{define}\noindent
The intuition here is that this prevents the top of the lattice, $\top_{D}$, itself a trivial refinement of any program, from being healthy. In order to explain the intuition for this, we consider the definition of $\top_{D}$.
\begin{define}[$\mathbf{Miracle}$]
\begin{align*}
	\top_{D} &\circdef (true \vdash false)
	&&\ptext{Property of designs}\\
	&=ok \implies false
	&&\ptext{Propositional calculus}\\
	&=\lnot ok&
\end{align*}
\end{define}\noindent
The top $\top_{D}$ denotes a program that could never be started ($\lnot ok$). Furthermore, if it could, and indeed its precondition makes no restriction, it would establish the impossible: $false$. Any conceivable implementable program must not behave in this way. However, miracle is an important construct in refinement calculi~\cite{Woodcock2004,Cavalcanti2006}.

For completeness we also provide the definition of the bottom of the lattice of designs, which is usually named $\mathbf{Abort}$. 
\begin{define}[$\mathbf{Abort}$]
$\bot_{D} \circdef (false \vdash true)$
\end{define}\noindent
The bottom $\bot_{D}$ provides no guarantees at all: it may fail to terminate, and if it does terminate there are no guarantees on the final values. Indeed it is not required to guarantee anything at all since its precondition is $false$.

\subsubsection{Operators}
In the following theorems we introduce the meet and join of the lattice of designs as presented in~\cite{Woodcock2004}. Like in the lattice of relations, the greatest lower bound corresponds to demonic choice. %
\begin{theorem}[Greatest lower bound]
$\bigsqcap_{i} (P_i \vdash Q_i) = (\bigwedge_{i} P_i) \vdash (\bigvee_{i} Q_i)$
\begin{proof}
Theorem 1 in~\cite{Woodcock2004}.
\end{proof}
\end{theorem}
\begin{theorem}[Least upper bound]
$\bigsqcup_{i} (P_i \vdash Q_i) = (\bigvee_{i} P_i) \vdash (\bigvee_{i} P_i \implies Q_i)$
\begin{proof}
Theorem 1 in~\cite{Woodcock2004}.
\end{proof}
\end{theorem}

\paragraph{Sequential Composition}
The definition of sequential composition for designs can be deduced from Definition~\ref{def:sequential-composition}. Here we present the result as proved in~\cite{Hoare1998,Woodcock2004}. %
\begin{theorem}[Sequential composition of designs]\label{theorem:designs:sequential-composition} Provided $ok$ and $ok'$ are not free in $P_0$, $P_1$, $Q_0$ and $Q_1$,
\begin{align*}
	(P_0 \vdash P_1) \circseq (Q_0 \vdash Q_1) =  
	( \lnot (\lnot P_0 \circseq true ) \land \lnot (P_1 \circseq \lnot Q_0) \vdash P_1 \circseq Q_1)
\end{align*}
\begin{proof}
Law $T3$ in~\cite{Woodcock2004}.
\end{proof}
\end{theorem}\noindent
This definition can be interpreted as establishing $P_1$ followed by $Q_1$ provided that $P_0$ holds and $P_1$ satisfies $Q_0$. As pointed out in~\cite{Woodcock2004}, if $P_0$ is a condition then the definition can be further simplified.
\begin{theorem}[Sequential composition of designs]\label{theorem:designs:sequential-composition-condition} Provided $ok$ and $ok'$ are not free in $P_0$, $P_1$, $Q_0$ and $Q_1$, and $P_0$ is a condition,
\begin{align*}
	(P_0 \vdash P_1) \circseq (Q_0 \vdash Q_1) =  
	(P_0 \land \lnot (P_1 \circseq \lnot Q_0) \vdash P_1 \circseq Q_1)
\end{align*}
\begin{proof}
Law $T3'$ in~\cite{Woodcock2004}.
\end{proof}
\end{theorem}\noindent

\subsubsection{Refinement}
As in all~\ac{UTP} theories, the refinement order in the theory of designs is: universal (reverse) implication. Thus the following result can be established~\cite{Woodcock2004}.
\begin{theorem}[Refinement]\label{theorem:designs:seq-composition}
\[	(P_0 \vdash P_1) \sqsubseteq (Q_0 \vdash Q_1) = [ P_0 \land Q_1 \implies P_1 ] \land [ P_0 \implies Q_0 ] \]
\begin{proof}
Law $5$ in~\cite{Woodcock2004}.
\end{proof}
\end{theorem}\noindent
Theorem~\ref{theorem:designs:seq-composition} confirms the intuition about refinement as found in other calculi: preconditions can be weakened while postconditions can be strengthened. 

This section concludes our overview of the theory of designs. In the following section we focus on how theories can be related and combined.

\subsection{Linking Theories}\label{sec:ch2:utp:links}
The~\ac{UTP} provides a very powerful framework that allows relationships to be established between different theories. This means that results in different theories can be reused. We elaborate on some of principles behind the linking of theories in the following paragraphs. A full account is available in~\cite{Hoare1998}.

Following the convention of Hoare and He~\cite{Hoare1998}, we assume the existence of a pair of functions $L$ and $R$ that map one theory into another: $L$ maps the (potentially) more expressive theory into the (potentially) weaker theory, and $R$, vice-versa.

\subsubsection{Subset Theories}
The simplest form of relationship that can be established is that between subset theories~\cite{Hoare1998}. We consider the case where a theory $T$ is a subset of $S$, it is then possible to find a function $R : T \fun S$: it is simply the identity~\cite{Hoare1998}. Defining $L : S \fun T$ for the reverse direction may be slightly more complicated as the subset theory is normally less expressive.

Hoare and He~\cite{Hoare1998} pinpoint the most important properties of such a function $L : S \fun T$: weakening or strengthening, idempotence and, ideally, monotonicity. As highlighted in~\cite{Hoare1998}, monotonicity is not always necessarily observed. We reproduce the respective definitions below.
\begin{define}[Weakening]
$\forall X \in S \spot L(X) \sqsubseteq X$
\end{define}
\begin{define}[Strengthening]
$\forall X \in S \spot X \sqsubseteq L(X)$
\end{define}\noindent
We follow Hoare and He's convention and refer to a function that is both weakening and idempotent as a link and, if it is also monotonic we refer to it as a retract.

\subsubsection{Bijective Links}
When two theories have equal expressive power, the pair of linking functions between them can be proved to form a bijection. In other words, each function undoes exactly the effect of the application of the other and, thus, as expected, the following identities hold.
\begin{define}[Bijection] A function $L$ is a bijection if, and only if, the inverse function $R = L^{-1}$ exists, and the following hold for all $P$,
\[	L \circ R (P) = P \land R \circ L (P) = P \]
\end{define}\noindent
A bijection constitutes the strongest form of relationship between theories. It can apply even when the alphabets are different or when the theories are presented in different styles~\cite{Hoare1998}. Indeed this is often what is sought: proving that two theories have exactly the same expressive power, yet their shape may suit different applications better.

\subsubsection{Galois Connections}
\begin{figure}
\begin{center}
\includegraphics[scale=0.8]{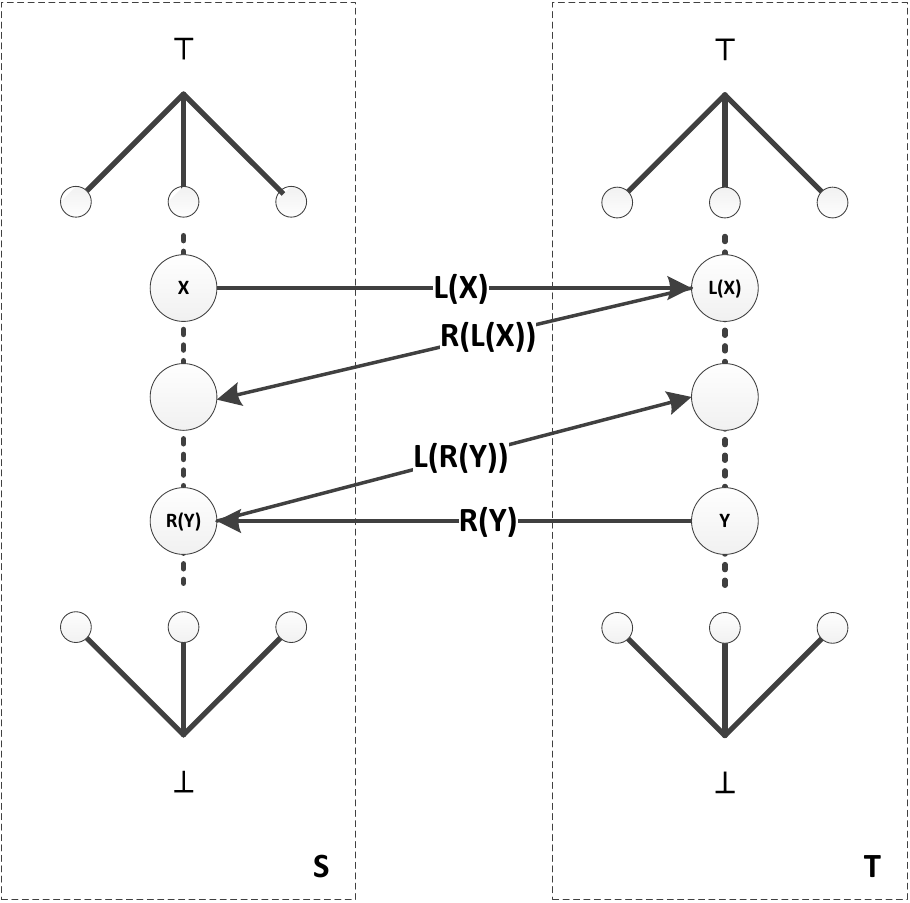}
\caption{\label{fig:galois-connection}Galois connection between two lattices, $S$ and $T$}
\end{center}
\end{figure}

Often, though, and as explained previously in the discussion of subset theories, we want to relate theories with different expressivity. Therefore the linking function is not a bijection, as there has to be some weakening or strengthening in either direction. A pair of functions describing this relationship constitutes what is known as a Galois connection. Here we reproduce the definition of~\cite{Hoare1998} and provide a pictorial illustration in Figure~\ref{fig:galois-connection}.
\begin{define}[Galois Connection] For lattices $S$ and $T$, a pair $(L, R)$ of functions $L : S \fun T$ and $R : T \fun S$ is defined to be a Galois connection if, and only if, for all $X \in S$ and $Y \in T$:
\[	R(Y) \sqsubseteq X \iff Y \sqsubseteq L(X) \]
\end{define}\noindent
\noindent
As pointed out earlier, a bijection presents a stronger relationship than a Galois connection. However, it is not the case that every bijection is a Galois connection~\cite{Hoare1998}. Hoare and He~\cite{Hoare1998} give the example of negation whose inverse is precisely itself, however negation is not monotonic. It is a known property of Galois connections that the functions are monotonic. In addition, the composition of Galois connections is also a Galois connection (Theorem 4.2.5 in~\cite{Hoare1998}).

\subsection{Angelic Nondeterminism}\label{sec:ch2:utp:angelic-nondeterminism}
In order to model both angelic and demonic nondeterminism in the relational setting of the~\ac{UTP}, Cavalcanti et al.~\cite{Cavalcanti2006} have proposed an encoding of upward-closed binary multirelations through non-homogeneous relations. The alphabet of that theory consists of the undashed program variables, whose set is given by $in\alpha$, and of the sole dashed variable $ac'$, which is a set of final states whose components range over $out\alpha$, the output variables of a program. The final states in $ac'$ are those available for angelic choice, while the demonic choices are those over the value of $ac'$. Similarly to our presentation of binary multirelations in~\cref{sec:ch2:binary-multirelations}, a state is a record whose components are program variables.

Despite being a theory which does not include the variables $ok$ and $ok'$, it directly captures termination. The intuition here is that a program may fail to terminate if there are no choices available to the angel. In other words, if $ac'$ may be empty, then non-termination is a possibility. Conversely, if the program terminates, then there must be at least one final state available for angelic choice.

\subsubsection{Healthiness Conditions}
Since the theory is essentially a relational encoding of binary multirelations, in order for it to observe the essential properties of binary multirelations, the set of final choices $ac'$ needs to be upward-closed. So the only healthiness condition of the theory is defined as follows~\cite{Cavalcanti2006}.%
\begin{define}
$ \mathbf{PBMH} (P) \circdef P \circseq ac \subseteq ac'$
\end{define}\noindent
This is a predicative version of $\mathbf{BMH}$, which is defined using the sequential composition operator. If it were possible for $P$ to establish some set of final states $ac'$, then any superset could have also been obtained.

One immediate consequence of $\mathbf{PBMH}$ illustrated is that no well-behaved program can require the set of final states $ac'$ to be empty as illustrated in the following~\cref{lemma:PBMH(ac'-emptyset):true}, which establishes that $ac'\neq\emptyset$ is not a fixed point of $\mathbf{PBMH}$.
\begin{lemma}\label{lemma:PBMH(ac'-emptyset):true}
$\mathbf{PBMH} (ac'=\emptyset) = true$ %
\begin{proof} %
\begin{flalign*}
	&\mathbf{PBMH} (ac'=\emptyset)
	&&\ptext{Definition of $\mathbf{PBMH}$}\\
	&=ac'=\emptyset \circseq ac \subseteq ac'
	&&\ptext{Definition of sequential composition}\\
	&=\exists ac_0 \spot (ac'=\emptyset)[ac_0/ac'] \land (ac\subseteq ac')[ac_0/ac]
	&&\ptext{Substitution}\\
	&=\exists ac_0 \spot ac_0=\emptyset \land ac_0 \subseteq ac' 
	&&\ptext{Property of sets}\\
	&=true
\end{flalign*}
\end{proof}
\end{lemma}\noindent
In other words, this corresponds to the same condition enforced by $\mathbf{H2}$ of the theory of designs. Moreover, because non-termination involves $ac'$ being empty, and since there is a requirement on $ac'$ being upward-closed, this theory also satisfies the condition enforced by $\mathbf{H3}$ of the theory of designs: arbitrary behaviour when there is non-termination. In the following, where we discuss the operators of the theory, we establish this result by proving that the $Skip$ of this theory is a right-unit for sequential composition, essentially a recast of $\mathbf{H3}$.

\subsubsection{Operators}
The operators of the~\ac{UTP} theory presented in~\cite{Cavalcanti2006} are calculated from their corresponding predicate transformer's definition through a composition of linking functions that establish isomorphisms between predicate transformers, binary multirelations and the proposed~\ac{UTP} model. In the following paragraphs we reproduce the most important operators, whose definitions are subscripted with $\mathbf{A}$.

Since this theory is a complete lattice, the angelic choice operator is the least upper bound, conjunction, while demonic choice corresponds to the greatest lower bound, disjunction. Furthermore, the bottom of the lattice is $true$ and corresponds to abort, while $false$ is the top and corresponds to miracle.

\paragraph{Skip}
The program that terminates successfully without changing the state is defined as follows.
\begin{define}
$ \IIAPBMH \circdef (\theta in\alpha)' \in ac'$
\end{define}\noindent
The definition requires that the dashed version of the initial state $\theta in\alpha$ is available for angelic choice in $ac'$. The notation $\theta in\alpha$ is used to denote a state where each name $x$ in $in\alpha$ is a component associated with the corresponding program variable $x$, while the notation $(\theta in\alpha)'$ denotes the state obtained from $\theta in\alpha$ by dashing the name of each state component.

This operator was originally not considered in~\cite{Cavalcanti2006}, but is useful, for example, to show that this theory observes the same property as $\mathbf{H3}$ of the theory of designs. This is presented following the introduction of the sequential composition operator.

\paragraph{Assignment}
The next operator of interest is assignment. An assignment of the value of an expression $e$ to a program variable $x$ is defined as follows.
\begin{define}[Assignment]
$ x :=_{\mathbf{A}} e \circdef (\theta in\alpha)' \oplus (x' \mapsto e) \in ac'$
\end{define}\noindent
The definition requires that there is a final state available for angelic choice in $ac'$, where the dashed version of the initial state $(\theta in\alpha)$ is overridden with a component of name $x'$ with value $e$.

\paragraph{Sequential Composition}
The operator that is perhaps most challenging is sequential composition. Since the theory is non-homogeneous, sequential composition is no longer relational composition as in other~\ac{UTP} theories. Instead, the authors in~\cite{Cavalcanti2006} have calculated the following definition, which uses substitution. %
\begin{define}
$ P \seqAPBMH Q \circdef P[\{ s' | Q[s/in\alpha] \}/ac'] $
\end{define}\noindent
The set of angelic choices resulting from composing $P$ and $Q$ corresponds to the angelic choices of $Q$, such that they can be reached from an initial state $s$ of $Q$ that is available for $P$ as a set $ac'$ of angelic choices. The states in $Q$ are obtained by considering the substitution in $Q$ over all variables $x$ in $in\alpha$ with their corresponding state component $s.x$. Since states in $ac'$ have dashed components, the set construction considers the dashed $s'$ version of $s$. This definition can be interpreted as back propagating the necessary information regarding the final states.

We consider the following example, where there is a choice between angelically assigning the value $1$ or $2$ to the only program variable $x$, followed by a sequential composition with an assumption, where the program terminates successfully only when the initial value of $x$ is $1$ and otherwise aborts. For simplicity, we consider $x$ to be the only program variable.
\begin{example}
\begin{flalign*}
	&(x :=_{\mathbf{A}} 1 \sqcup x :=_{\mathbf{A}} 2) \seqAPBMH (x = 1 \implies \IIAPBMH)
	&&\ptext{Definition of $\sqcup$ and assignment}\\
	&=((x' \mapsto 1) \in ac' \land	(x' \mapsto 2) \in ac') \seqAPBMH (x = 1 \implies \IIAPBMH)
	&&\ptext{Definition of $\seqAPBMH$ and $\IIAPBMH$}\\
	&=\left(\begin{array}{l}
		(x' \mapsto 1) \in ac' 
		\\ \land \\
		(x' \mapsto 2) \in ac'
	\end{array}\right)\left[\{ s' | (x = 1 \implies (x' \mapsto x) \in ac')[s/in\alpha] \}/ac'\right] 
	&&\ptext{Substitution}\\
	&=((x' \mapsto 1) \in ac' \land (x' \mapsto 2) \in ac')[\{ s' | s.x = 1 \implies (x' \mapsto s.x) \in ac' \}/ac']
	&&\ptext{Property of substitution}\\
	&=\left(\begin{array}{l}
		((x' \mapsto 1) \in ac')[\{ s' | s.x = 1 \implies (x' \mapsto s.x) \in ac' \}/ac'] 
		\\ \land \\
		((x' \mapsto 2) \in ac')[\{ s' | s.x = 1 \implies (x' \mapsto s.x) \in ac' \}/ac']
	\end{array}\right)
	&&\ptext{Substitution}\\
	&=\left(\begin{array}{l}
		((x' \mapsto 1) \in \{ s' | s.x = 1 \implies (x' \mapsto s.x) \in ac' \}) 
		\\ \land \\
		((x' \mapsto 2) \in \{ s' | s.x = 1 \implies (x' \mapsto s.x) \in ac' \})
	\end{array}\right)
	&&\ptext{Property of sets}\\
	&=\left(\begin{array}{l}
		(x \mapsto 1).x = 1 \implies (x' \mapsto (x \mapsto 1).x) \in ac' 
		\\ \land \\
		(x \mapsto 2).x = 1 \implies (x' \mapsto (x \mapsto 2).x) \in ac'
	\end{array}\right)
	&&\ptext{Record component $x$}\\
	&=(1 = 1 \implies (x' \mapsto 1) \in ac') \land (2 = 1 \implies (x' \mapsto 2) \in ac')
	&&\ptext{Predicate calculus}\\
	&=(x' \mapsto 1) \in ac'
	&&\ptext{Definition of assignment}\\
	&=x :=_{\mathbf{A}} 1
\end{flalign*}
\end{example}\noindent
The result is that the angel avoids assigning $2$ to $x$, since that would lead to abortion. So effectively, the information regarding the sets available for angelic choice is back propagated from the assumption through the sequential composition.

Finally, we show that this theory observes the property of $\mathbf{H3}$ of the theory of designs by expressing $\mathbf{H3}$ in this model.
\begin{define}
$\mathbf{H3}_\mathbf{A} (P) \circdef P \seqAPBMH \IIAPBMH$
\end{define}\noindent
This requires the identity of the theory $\IIAPBMH$ to be a right-unit, which we prove in the following lemma for healthy predicates.
\begin{lemma}
$P = P \seqAPBMH \IIAPBMH$
\begin{proof}
\begin{flalign*}
	&P \seqAPBMH \IIAPBMH
	&&\ptext{Definition of $\IIAPBMH$ and $\seqAPBMH$}\\
	&=P[\{ s' | ((\theta in\alpha)' \in ac')[s/in\alpha] \}/ac']
	&&\ptext{Expand $\theta in\alpha$ for each $x_i$ in $in\alpha$}\\
	&=P[\{ s' | ((x_0 \mapsto x_0, \ldots, x_i \mapsto x_i)' \in ac')[s/in\alpha] \}/ac']
	&&\ptext{Dash state components}\\
	&=P[\{ s' | ((x_0' \mapsto x_0, \ldots, x_i' \mapsto x_i) \in ac')[s/in\alpha] \}/ac']
	&&\ptext{Substitution}\\
	&=P[\{ s' | (x_0' \mapsto s.x_0, \ldots, x_i' \mapsto s.x_i) \in ac'\}/ac']
	&&\ptext{Dash state components}\\
	&=P[\{ s | (x_0' \mapsto s.x_0', \ldots, x_i' \mapsto s.x_i') \in ac' \}/ac']
	&&\ptext{State components}\\
	&=P[\{ s | s \in ac' \}/ac']
	&&\ptext{Property of sets}\\
	&=P[ac'/ac']
	&&\ptext{Property of substitution}\\
	&=P
\end{flalign*}
\end{proof}
\end{lemma}\noindent
This concludes the discussion of the healthiness conditions of the theory. In what follows we discuss the relationship between this theory, binary multirelations and the predicate transformers.

\subsubsection{Relationship with Binary Multirelations}
As previously discussed, the theory of~\cite{Cavalcanti2006} is isomorphic to the theory of upward-closed binary multirelations. We depict this relationship in~\cref{fig:theories,fig:theories:bmh} where both theories, characterised by their respective healthiness conditions $\mathbf{PBMH}$ and $\mathbf{BMH}$ are related through a pair of composed linking functions~\cite{Cavalcanti2006}. For completeness, we reproduce the result of these linking results in what follows, while the definition of each individual linking function is available in~\cite{Cavalcanti2006}.

The first composition maps from this theory into the model of binary multirelations; this result is reproduced below~\cite{Cavalcanti2006}. 
\begin{theorem}\label{theorem:sb2bm-o-p2sb}
$ sb2bm \circ p2sb (P) \circdef \{ s : State, ss : \power State | P[s,ss/in\alpha,ac'] \}$
\begin{proof}
Part of Theorem 4.8 in~\cite{Cavalcanti2006}, following the definitions of $p2sb$ and $sb2bm$.
\end{proof}
\end{theorem}\noindent
It considers every initial state $s$ and set of final states $ss$, such that $P$ holds when every initial variable $x$ in $in\alpha$ is substituted with its corresponding state component $s.x$, and the set of final states $ss$ is substituted for $ac'$.

The inverse link is established by the composition of the respective inverse linking functions $sb2p$ and $bm2sb$, whose functional composition is shown below~\cite{Cavalcanti2006}.
\begin{theorem}\label{theorem:sb2p-o-bm2sb}
$ sb2p \circ bm2sb (B) \circdef (\theta in\alpha, ac') \in B$
\begin{proof}
Part of Theorem 4.7 in~\cite{Cavalcanti2006}, following the definitions of $bm2sb$ and $sb2p$.
\end{proof}
\end{theorem}\noindent
For a binary multirelation $B$, the corresponding~\ac{UTP} predicate requires that every pair of initial states $\theta in\alpha$ and set of final states $ac'$ is in $B$.

\subsubsection{Relationship with Predicate Transformers}
The last relationship that we discuss in this section pertains to the links between the~\ac{UTP} model of~\cite{Cavalcanti2006} and the monotonic predicate transformers. This is achieved in~\cite{Cavalcanti2006} through a pair of linking functions, $pt2p$, which maps from the predicate transformers model into this one, and a functional composition in the opposite direction, whose combined result we call $p2pt$. The definition of $pt2p$ is the result of Theorem 4.5 in~\cite{Cavalcanti2006}, which we reproduce below.
\begin{theorem}
$ pt2p(PT) = \theta in\alpha \in \lnot PT.(\lnot ac') $
\begin{proof}
Theorem 4.5 in~\cite{Cavalcanti2006}.
\end{proof}
\end{theorem}\noindent
For a predicate transformer $PT$, $pt2p$ defines the predicate that requires that the initial state $\theta in\alpha$ is associated with all postconditions $ac'$ that $PT$ is not guaranteed not to establish from the initial state~\cite{Cavalcanti2006}. In this treatment of predicate transformers, predicates are modelled by their characteristic sets, such that $PT$ is a monotonic function from sets of final states to sets of initial states~\cite{Cavalcanti2006}.

The function mapping in the opposite direction is not presented in~\cite{Cavalcanti2006}, however it can be calculated from the definitions of $p2sb$, $sb2bm$ and $bm2pt$, which leads to the following definition.
\begin{define}
$p2pt(P)(\psi) = \{ s | \lnot P[s, \lnot \psi/in\alpha,ac'] \}$%
\end{define}\noindent
This definition is justified by the following lemma.
\begin{lemma}
$bm2pt(sb2bm \circ p2sb(P),\psi) = \{ s | \lnot P[s, \lnot \psi/in\alpha,ac'] \}$%
\begin{proof}
\begin{flalign*}
	&bm2pt(sb2bm \circ p2sb(P),\psi)
	&&\ptext{\cref{theorem:sb2bm-o-p2sb}}\\
	&=bm2pt(\{ s_1, ss | P[s_1,ss/in\alpha,ac'] \},\psi)
	&&\ptext{Definition of $bm2pt$~\cite{Cavalcanti2006}}\\
	&=\{ s | (s, \lnot \psi) \notin \{ s_1, ss | P[s_1,ss/in\alpha,ac'] \}\}
	&&\ptext{Property of sets}\\
	&=\{ s | \lnot P[s_1,ss/in\alpha,ac'][s,\lnot \psi/s_1,ss]\}
	&&\ptext{Substitution}\\
	&=\{ s | \lnot P[s,\lnot \psi/in\alpha,ac'] \} 
\end{flalign*}
\end{proof}
\end{lemma}\noindent
This result concludes our discussion regarding the theory of angelic nondeterminism in the~\ac{UTP} and its relationship with the standard model of predicate transformers, where angelic and demonic nondeterminism have traditionally been characterised.

\section{Processes: CSP and Angelic Nondeterminism}\label{sec:ch2:csp-angelic-nondeterminism}
Motivated by the advances of concurrency in both hardware and software, and the lack of a clear understanding of the mechanisms involved, in 1978 Hoare~\cite{Hoare1978} proposed the original version of~\acf{CSP}. The idea was to characterise concurrent systems as the result of sequential processes that execute in parallel, and communicate and synchronize through primitive operations of input and output. However, it was not until further contributions by Hoare~\cite{Hoare1980,Hoare1985}, Brookes~\cite{Brookes1984} and Roscoe~\cite{Roscoe1998,Roscoe2010} that the algebra of~\ac{CSP} appeared, together with a complete semantics, presented in all three main flavours: algebraic, denotational and operational. This was followed by the introduction of support for model checking through~\ac{FDR}~\cite{FDR,FDR2014}.

In~\cref{sec:ch2:CSP:notation} we provide an introduction to~\ac{CSP} through a presentation of its most important operators and algebraic laws. In~\cref{sec:ch2:CSP:semantics} we discuss the standard semantics of~\ac{CSP} as found in~\cite{Roscoe2010}. The material presented here is meant as background for understanding both~\ac{CSP} and the existing proposals for handling angelic nondeterminism, which we discuss in~\cref{sec:ch2:CSP:angelic-nondeterminism}. A full account of~\ac{CSP} can be found in~\cite{Roscoe1998,Roscoe2010}. Finally, ~\cref{sec:ch2:CSP:UTP} explores the~\ac{UTP} model of~\ac{CSP}~\cite{Hoare1998,Cavalcanti2006a}. 


\subsection{Notation}\label{sec:ch2:CSP:notation}
As the name processes in~\ac{CSP} suggests, the central notion of~\ac{CSP} is that of processes. These include basic processes, such as $Skip$, the process that terminates successfully without influence from the environment, $Stop$, which behaves as deadlock and hence refuses to do anything, and $Chaos$, which behaves unpredictably.

The other core notion of~\ac{CSP} is that of communication. This is achieved by defining events, which the system can perform only with the cooperation of its environment. That is, once the environment is given the possibility to perform an event, and it agrees to do so, then the event happens instantaneously and atomically. The easiest way to express this behaviour in~\ac{CSP} is through prefixing of events.
\begin{define}[Prefixing]\label{def:CSP:prefixing}
$ a \circthen P$
\end{define}\noindent
This process offers the environment the possibility to perform the event $a$, after which it behaves like $P$, some other~\ac{CSP} process. We consider the process $P_0$.
\begin{example}\label{example:CSP:prefixing}\label{example:CSP:P0}
$ P_0 = up \circthen down \circthen Stop$
\end{example}\noindent
In this case a sequence of $up$ and $down$ events is followed by deadlock. A direct consequence of the definition of processes in this way is that recursion can occur naturally as part of the functional style of~\ac{CSP} as shown in the following example.
\begin{example}[Mutual Recursion]
\begin{align*}
& P_1 = up \circthen P_2 \\
& P_2 = down \circthen P_1
\end{align*}
\end{example}\noindent
These processes are defined by mutual recursion. The set of possible traces of events of $P_1$ is a superset of~\cref{example:CSP:prefixing}. It never terminates nor deadlocks.

\ac{CSP} presents a rich set of operators that allow more complex interactions to be modelled. The first that we consider in the sequel is called external choice.
\begin{define}[External Choice]
$ P \extchoice Q$
\end{define}\noindent
In this case the environment is offered the choice between behaving as either $P$ or $Q$. This operator satisfies a number of laws as reproduced below~\cite{Roscoe1998}. %
\begin{lemma}[Laws of External Choice]
\begin{align*}
	Idempotent: &P \extchoice P = P\\
	Associative: &P \extchoice (Q \extchoice R) = (P \extchoice Q) \extchoice R\\
	Symmetry: &P \extchoice Q = Q \extchoice P \\
	Unit: &P \extchoice Stop = P
\end{align*}
\end{lemma}\noindent
Perhaps the most interesting result here is that $Stop$ is the unit of external choice. When the environment is given the choice between deadlocking or behaving as $P$, it can only choose to behave as $P$.

External choice can be used to generalize the prefixing operator of~\cref{def:CSP:prefixing}. Instead of permitting a single event, prefixing can be of a set of events $E \subseteq \Sigma$ over some alphabet $\Sigma$ as follows.
\begin{define}
$ x : E \circthen P =~\extchoice x : E \spot x \circthen P $
\end{define}\noindent
This is basically a distributed external choice over all possible events in $E$. Moreover, \ac{CSP} permits the definition of channels, which can carry values of a certain type $E$. For a channel name $c$ of type $E$, the set of possible events that represent communications over $c$ is defined by considering events with composed names prefixed by $c$ as follows: $\{ c.x | x \in E \}$. Usually in the~\ac{CSP} syntax, channel communications are prefixed with $?$ to denote input communications while $!$ denotes output communications, as shown in~\cref{example:CSP:buffer}.
\begin{example}[Buffer]\label{example:CSP:buffer}
$ P_3 = in?x \circthen out!x \circthen P_3$
\end{example}\noindent
These annotations are syntactic sugar for the corresponding events $in.x$ and $out.x$. In this example we have an input communication over channel $in$, which is then relayed onto the output channel $out$, effectively behaving as a one place buffer.

In addition to external choice, there is an operator in~\ac{CSP} known as internal choice.
\begin{define}[Internal Choice]
$ P \sqcap Q$
\end{define}\noindent
This choice is also known as demonic choice, since the environment cannot possibly force the system into behaving as either $P$ or $Q$. Indeed the system can choose either at its discretion. For instance, if $Stop$ is offered as a choice, then the system may deadlock. This operator satisfies a number of important laws, of which a summary is included below~\cite{Roscoe1998}.
\begin{lemma}[Laws of Internal Choice]
\begin{align*}
	Idempotent: &P \sqcap P = P\\
	Associative: &P \sqcap (Q \sqcap R) = (P \sqcap Q) \sqcap R\\
	Symmetry: &P \sqcap Q = Q \sqcap P \\
	Distributive: &P \sqcap (Q \extchoice R) = (P \sqcap Q) \extchoice (P \sqcap R)
\end{align*}
\end{lemma}\noindent
Of these, distributivity is perhaps the most important. In fact, most~\ac{CSP} operators distribute through internal choice, except, for example, recursion~\cite{Roscoe1998}.

The next operator of interest is that of sequential composition; it allows the composition of processes sequentially, other than by using prefixing.
\begin{define}[Sequential Composition]
$ P \circseq Q$
\end{define}\noindent
A consequence of~\ac{CSP}'s functional language is that it is not possible to pass local process information through sequential composition. So for instance, the following process $P_4$ does not behave as would intuitively be expected in~\ac{CSP}.
\begin{example}\label{example:CSP:in-out-sec}
$ P_4 = in?x \circthen Skip \circseq out!x \circthen Stop$
\end{example}\noindent
This is because the scope of $x$ is local to both of these processes, and not global. However, this problem can be obviated by the introduction of parallelism in~\ac{CSP}.

\ac{CSP} provides a number of different parallel composition operators~\cite{Roscoe1998}. Here we consider the most generic operator, which is the alphabetised parallel composition. %
\begin{define}[Alphabetised Parallel Composition]
$P \parallel[\alpha P][\alpha Q] Q$
\end{define}\noindent
Alphabetised here means that processes $P$ and $Q$ only need to agree on events in the intersection of the alphabet of events of each process as defined in the operator: $\alpha P$ and $\alpha Q$, respectively. Events not in the intersection do not need the agreement of both processes. For instance, to specify the behaviour that may be expected of the process $P_4$ from~\cref{example:CSP:in-out-sec}, we can consider a third process in parallel that communicates the desired value between the two processes.
\begin{example}[Parallel Composition]\label{example:CSP:in-out-parallel}
\begin{align*}
P_5 = \left(\begin{array}{l}
		((in?x \circthen t!x \circthen Skip) \circseq (t?y \circthen out!y \circthen Stop))
		\\ \parallel[\{|in,out,t|\}][\{|t|\}] \\
		(t?z \circthen t!z \circthen Skip)
	  \end{array}\right)
\end{align*}
\end{example}\noindent
In this example, we add the extra channel $t$ that serves as an internal communication channel. However, in pursuing this style of specification we have added an externally observable set of events $t$, which may not always be desired. \ac{CSP} provides a solution for this kind of modelling problem as well.

Events can effectively be hidden from other processes when they are not needed. This abstraction is achieved in~\ac{CSP} by using the hiding operator. %
\begin{define}[Hiding]
$P \setminus E$
\end{define}\noindent
Here the process $P$ has the events in the set $E$ hidden from other processes, such that events in $E$ become internal events that can happen irrespective of the cooperation from the environment~\cite{Roscoe1998}. In the following example, we give the effect of hiding the communications over $t$ of $P_5$.
\begin{example}[Hiding]
$ P_6 = P_5 \setminus \{| t |\} = in?x \circthen out!x \circthen Stop$
\end{example}\noindent
This new process $P_6$ is equivalent to the process that takes a communication over channel $in$, relays over channel $out$ and then deadlocks. 

This concludes our discussion on the notation of~\ac{CSP} and the most important concepts underlying its operators and algebraic properties. In the following section we focus our attention on the denotational semantics of~\ac{CSP}. 

\subsection{Semantics}\label{sec:ch2:CSP:semantics}
Many interesting properties in~\ac{CSP} are proved using its algebraic laws. For instance, step-laws~\cite{Roscoe1998} provide a mechanism for a stepwise calculation of the behaviour of operators. In addition, \ac{CSP} also has a denotational semantics, which we discuss in this section.

\subsubsection{Traces}
The simplest semantic model proposed for~\ac{CSP} considers the observable sequences of events that a process may produce. For a $CSP$ process, where $\Sigma$ is the set of all possible events, the set of $traces$ is given by the function $traces : CSP \fun \power (\seq \Sigma)$. For instance, the set of traces for process $P_0$ from~\cref{example:CSP:P0} is obtained as follows.
\[ traces(P_0) = \{ \lseq \rseq, \lseq up \rseq, \lseq up, down \rseq \} \]
This includes the empty sequence followed by all possible sequences of events.

Refinement in this model allows reasoning about safety, since a process $P$ is refined by $Q$ if, and only if, the set of trances of $Q$ is a subset of those of $P$
\begin{define}[Traces Refinement]
$ P \sqsubseteq_{T} Q \iff traces(Q) \subseteq traces(P)$
\end{define}\noindent
In other words, every behaviour of $Q$ is a possible behaviour of $P$. In particular, $Stop$, refines every process in the traces model, since it is a possible behaviour of every process. This motivates the definition of the following semantic model. 

\subsubsection{Failures}
The following semantic model of~\ac{CSP} considers the set of events that may be refused by a process after a certain trace of events. This allows reasoning about liveness, in that a process like $Stop$ no longer refines every other process. For a $CSP$ process, the set of failures, is given by the function $failures : CSP \fun \power (\seq \Sigma \times \power \Sigma)$. For example, in the case of process $P_0$, and assuming that the alphabet $\Sigma$ is $\{ up, down\}$ the failures are obtained as follows.
\[ failures(P_0) = \left\{\begin{array}{l}
  (\lseq \rseq, \{ down \}), (\lseq \rseq, \emptyset), (\lseq up \rseq, \{ up \}), (\lseq up \rseq, \emptyset), 
\\(\lseq up, down \rseq, \{ up, down \}), (\lseq up, down \rseq, \{up\}), 
\\(\lseq up, down \rseq, \{down\}), (\lseq up, down \rseq, \emptyset) 
 \end{array}\right\} \]
In other words, once the process deadlocks it refuses every possible event. Failures allow the semantics of external and internal choice to be distinguished~\cite{Roscoe1998}.

Refinement is defined by considering the refusal pairs in addition to the traces.
\begin{define}[Failures Refinement]
\begin{align*}
P \sqsubseteq_{F} Q \iff traces(Q) \subseteq traces(P) \land failures(Q) \subseteq failures(P)
\end{align*}
\end{define}\noindent
A process $P$ is refined by $Q$, if, and only if, in addition to the traces of $Q$ being a subset of those for $P$, the failures of $Q$ are also a subset of $P$. 

This is almost the complete semantics for~\ac{CSP} except, for the treatment of divergence, which requires one final addition to the model~\cite{Roscoe1998}.

\subsubsection{Failures-Divergences}
Divergence can arise in~\ac{CSP} in different ways. For example, the most obvious is through the process $Chaos$, whose arbitrary behaviour includes divergence, while a process such as $P = P$, with an infinite recursion and no visible events, is also a divergence. The $Chaos$ process in~\cite{Roscoe1998} is the most non-deterministic process that does not include divergence. Here we consider the behaviour of $Chaos$ to be completely arbitrary, which corresponds to $\mathbf{div}$ in the standard~\ac{CSP} failures-divergences semantics. The approach followed in~\ac{CSP} is that any two processes that can diverge immediately are equivalent and useless, and that, once a process diverges, it can perform any trace of events and refuse any event~\cite{Roscoe1998}.

The function $divergences : CSP \fun \power (\seq \Sigma)$ gives the set of divergences for a $CSP$ process. We consider the following example, where the process $P_7$ offers the event $a$ followed by divergent behaviour.
\begin{example}[Divergence]
$ P_7 = a \circthen Chaos$
\end{example}\noindent
Its divergences are the set of all traces that lead to divergent behaviour. In the example above this is $\{ s : \seq \Sigma | \lseq a \rseq \le s\}$, that is, every trace that has $a$ as the first event. In addition, because $divergences(P)$ includes every trace on which process $P$ can diverge, the notion of failures needs to be redefined. This is because once a process has diverged it can refuse anything. These failures are obtained by the following function $failures_\bot$.
\begin{define}
$ failures_\bot (P) = failures(P) \cup \{ s : \seq \Sigma, ss : \Sigma | s \in divergences(P) \}$
\end{define}\noindent
A process $P$ can then be characterised through a pair $(failures\bot(P), divergences(P))$.

Finally, the refinement order for processes $P$ and $Q$ in the failures-divergences model is given as follows.
\begin{define}[Failures-Divergences Refinement]
\begin{align*}
	P \sqsubseteq_{FD} Q \iff failures_\bot(Q) \subseteq failures_\bot(P) \land divergences(Q) \subseteq divergences(P)
\end{align*}
\end{define}\noindent
Process $P$ is refined by $Q$ if, and only if, the set of $failures_\bot$ and $divergences$ for $Q$ are a subset of those of $P$. Consequently, $Chaos$ is refined by every other process.

This concludes our discussion on the standard~\ac{CSP} semantic model of failures-divergences~\cite{Roscoe1998}. A full account of the~\ac{CSP} semantics, including the operational semantics, which is the basis for the~\ac{FDR} model checker, is available in~\cite{Roscoe1998}. In~\cref{sec:ch2:CSP:UTP} we present the~\ac{UTP} model of~\ac{CSP}.

\subsection{Angelic Nondeterminism in CSP}\label{sec:ch2:CSP:angelic-nondeterminism}
As we have previously discussed, the concept of angelic nondeterminism has also been considered in the context of~\ac{CSP}. Here we consider in more detail the different approaches proposed and discuss their properties.

\subsubsection{Lattice-Theoretic Model}
In~\cite{Tyrrell2006} Tyrrell et al. present an axiomatized model for an algebra resembling~\ac{CSP}. At the core of their proposal is the notion that external choice, referred to as angelic choice, is a dual of internal choice in a lattice-theoretic model. This is achieved by a stepwise construction that begins with proper processes, that is, processes without choice, parallelism or recursion, which are modelled as finite sequences of events that terminate with either an empty sequence $\lseq \rseq$ or with $\Omega$. This is sufficient to give semantics to the following processes~\cite{Tyrrell2006}, where $[\_] : Proc(\Sigma) \fun \seq \Sigma$ is the semantic denotation for a process, $Proc(\Sigma)$ is the set of all processes constructed from $Skip$, $Stop$ and prefixing of events in $\Sigma$, and $\cat$ is sequence concatenation. %
\begin{define}[Proper Processes]
\begin{align*}
	&[Skip] \circdef \lseq \rseq \\
	&[Stop] \circdef \Omega \\
	&[a \circthen P] \circdef a \cat [P]
\end{align*}
\end{define}\noindent
A partial order $\le_{P}$ is then defined for $[Proc(\Sigma)]$, such that $\Omega$ is the least element, and for any two processes $P$ and $Q$, their order is given recursively in terms of the suffix of the respective sequences of events.
\begin{define}[Refinement of Proper Processes]
\begin{align*}
&	\forall s \in [Proc(\Sigma)] \spot \Omega \le_{P} s \\
&	\forall e \in \Sigma, s, t \in [Proc(\Sigma)] \spot e\cat s \le_{P} e\cat t \iff s \le_{P} t
\end{align*}
\end{define}\noindent
This corresponds to the refinement order for proper processes, where $Stop$ is the least element of the order. The definition for other operators, such as restriction and sequential composition, is further specified in~\cite{Tyrrell2006}.

Having defined the refinement order for proper processes, an order-embedding is defined from the set of sequences into the~\ac{FCD} lattice. A lattice $L$ is a free completely distributive lattice over a partially ordered set $C$, written $FCD(C)$, if, and only if, ``there is a completion $\phi : C \fun L$ such that for every~\ac{FCD} lattice $M$ and function $f : C \fun M$, there is a unique function $\phi^*_M : L \fun M$ which is a complete homomorphism and satisfies $\phi^*_M \circ \phi = f$''~\cite{Tyrrell2006,Morris2004}. We illustrate this functional relationship in~\cref{fig:FCD}. %
\begin{figure}[ht]
\begin{center}
$$
\begindc{\commdiag}
\obj(10,10){$C$}
\obj(30,10){$L$}
\obj(30,30){$M$}
\mor{$C$}{$L$}{$\phi$}[\atright , \solidarrow]
\mor{$L$}{$M$}{$\phi^*_M$}[\atright , \solidarrow]
\mor{$C$}{$M$}{$f$}
\enddc
$$
\caption{\label{fig:FCD}Free Completely Distributive Lattice completion}
\end{center}
\end{figure}\noindent
The~\ac{FCD} provides a number of interesting properties, namely, that each element can be described as the meet of joins of subsets of $\phi C$, or the join of meets of subsets of $\phi C$~\cite{Tyrrell2006}. This is essential in the characterisation of recursive processes, which is achieved through the weakest fixed point of the lattice that excludes the least element~\cite{Tyrrell2006}. Liftings are then defined for unary and binary operators into the~\ac{FCD} lattice, such that internal and angelic choice correspond to the meet and join, respectively. Definitions are also given in~\cite{Tyrrell2006} for the alphabetised parallel operator and recursive processes.

The construction of~\cite{Tyrrell2006} provides for an elegant algebra, whose axiomatic description follows from the construction of the~\ac{FCD} lattice. However, with $Stop$ as the least element of the refinement order, it is not possible to distinguish deadlock from divergence in this model. Thus, the semantics is quite different from the standard model of failures-divergences~\cite{Roscoe1998}.

\subsubsection{Operational CSP Combinators}
In~\cite{Roscoe2010} Roscoe proposes an angelic choice operator through combinator style operational semantics of~\ac{CSP}. Traditionally~\cite{Roscoe1998,Roscoe2010}, the operational semantics of~\ac{CSP} has been defined through a~\ac{LTS}. An~\ac{LTS} is a directed graph, where each edge is labelled with an action that denotes what happens when the system transitions between states. In~\ac{CSP} the set of possible labels includes the events in $\Sigma$ and another two special events: $\tick$ which signals successful termination and does not require the cooperation of the environment (such as in the case of $Skip$), and $\tau$ which is an internal event invisible to the environment. Hence, $\tick$ is always the last event possible and leads to a special end state $\Omega$.

Operational semantics for~\ac{CSP} operators can be given in the style of Plotkin's \ac{SOS} \cite{Plotkin2004}. For example, the process $Stop$ has no actions, while $Skip$ can be given the following rule~\cite{Roscoe2010}.
\[
\inference[]{ \\  \\ }{Skip \xrightarrow{\tick} \Omega }
\]
Since the transition relation always associates $Skip$ to $\Omega$ with action $\tick$, the bar is empty above, while the transition below means that $Skip$ can transition into the final special state $\Omega$ by doing action $\tick$. External choice, on the other hand, requires more rules since an internal event $\tau$ does not decide the choice~\cite{Roscoe2010}.
\[
\inference[]{ P \xrightarrow{\tau} P' \\  \\ }{P \extchoice Q \xrightarrow{\tau} P' \extchoice Q }~ , ~
\inference[]{ Q \xrightarrow{\tau} Q' \\  \\ }{P \extchoice Q \xrightarrow{\tau} P \extchoice Q' }
\]
In these two cases, an internal action can be performed by either $P$ or $Q$, in which case, the $\tau$ event is promoted, while the choice is not resolved. Any other event $a$, including $\tick$, decides the choice between processes $P$ and $Q$.
\[
\inference[]{ P \xrightarrow{a} P' \\ \\ }{P \extchoice Q \xrightarrow{a} P' } (a \neq \tau) , ~
\inference[]{ Q \xrightarrow{a} Q' \\ \\ }{P \extchoice Q \xrightarrow{a} Q' } (a \neq \tau)
\]
Given the number of different rules needed to specify an operator, and the fact that it is actually possible to define operators that are not conformant with the failures-divergences semantics of~\ac{CSP}~\cite{Roscoe2010}, Roscoe proposes an alternative known as combinator style operational rules. The idea is that it is possible to distinguish process arguments whose actions are immediately relevant from those that are not~\cite{Roscoe2010}. The latter are \textbf{off}, while the former are \textbf{on}. Thus the semantics of external choice can be given as
\[ ((a, .), a, \mathbf{1}), ((., a), a, \mathbf{2})~\text{for each}~ a \in \Sigma \]
where each triple is defined by: a tuple that denotes the actions that each \textbf{on} process performs (with . indicating none), ordered according to the indices of the arguments, the overall action performed, and the format of the resulting state given in~\ac{CSP} syntax. In the case of external choice, for each event $a$ in $\Sigma$, either the first process, whose tuple is $(a, .)$, or the second process, whose tuple is $(., a)$ can decide the choice. The resulting event performed by the system is $a$, and the resulting state is either $\mathbf{1}$, which corresponds to the first process or $\mathbf{2}$, which corresponds to the second process.

An assumption of this style of specification is that $\tau$ events are always promoted for arguments that are \textbf{on}, so there is no need to include rules for this~\cite{Roscoe2010}. Finally, the specification of the external choice operator also requires rules for termination:
\[ ((\tick, .), \tick, \Omega), ((., \tick), \tick, \Omega)\]
In this case, the termination of either process leads to termination, in which case the system transitions to the special state $\Omega$, with the visible action being $\tick$.

The interesting result about this style of operational specification, is that every such operator conforms to the failures-divergences semantics of~\ac{CSP}, and Roscoe~\cite{Roscoe2010} envisions this as a mechanism for adding new operators to~\ac{FDR}. Moreover, in~\cite{Roscoe2010} Roscoe also gives a~\ac{CSP} process, which is able to simulate processes specified using combinator style semantics.

Having defined his combinator-style operational rules, Roscoe~\cite{Roscoe2010} proposes an angelic choice operator $P \boxcircle Q$ (Example 9.2 in~\cite{Roscoe2010}), which gives the environment a choice over both actions $P$ and $Q$ as long as the environment picks one that they both offer. In fact, to achieve this definition Roscoe defines a family of operators $P\,_{s}\boxcircle Q$ and $P\,\boxcircle_{s} Q$, where $s$ is a non-empty trace that keeps track of the difference in events performed ``ahead'' by the other operand. The operational semantics of this angelic choice operator is reproduced below~\cite{Roscoe2010}.
\begin{itemize}[itemsep=0pt]
\item For $\boxcircle$: 
	\begin{tabular}[t]{l}
	$\forall a \in \Sigma$: $((a, .), a, \mathbf{1} \boxcircle_{\lseq a \rseq} \mathbf{2})$, $((., a), a, \mathbf{1} _{\lseq a \rseq}\boxcircle \mathbf{2})$ \\
	$((\tick, .), \tick, \Omega)$ and $((., \tick), \tick, \Omega)$
	\end{tabular}
\item For $_{\lseq b \rseq \cat s}\boxcircle$:
	\begin{tabular}[t]{l}
	$\forall a \in \Sigma$: $((b, .), \tau, \mathbf{1} _{s}\boxcircle \mathbf{2})$, $((., a), a, \mathbf{1} _{\lseq a, b \rseq \cat s}\boxcircle \mathbf{2})$ \\
	$((\tick, .), \tau, \mathbf{2})$ and $((., \tick), \tick, \Omega)$
	\end{tabular} 
\item For $\boxcircle_{\lseq b \rseq \cat s}$:
	\begin{tabular}[t]{l}
	$\forall a \in \Sigma$: $((., b), \tau, \mathbf{1} \boxcircle_{s} \mathbf{2})$, $((a, .), a, \mathbf{1} \boxcircle_{\lseq a, b \rseq \cat s} \mathbf{2})$ \\
	$((\tick, .), \tick, \Omega)$ and $((., \tick), \tau, \mathbf{1})$
	\end{tabular} 
\end{itemize}
The first set of rules for $P \boxcircle Q$ considers the case where either $P$ or $Q$ perform the event $a$, in which case the event $a$ is visible. If $P$ performs event $a$, then the resulting process $P \boxcircle_{\lseq a \rseq} Q$ has the sequence $\lseq a \rseq$ corresponding to the events $Q$ could catch up to. Similarly, there is a rule for the case when $Q$ performs the event $a$. If either process terminates, then $\tick$ is observed and the system transitions to $\Omega$.

The second set of rules for $P _{\lseq b \rseq \cat s}\boxcircle Q$ considers the case where process $Q$ is ahead. If $P$ performs the event $b$, then an internal event is observed, and the resulting process $P _{s}\boxcircle Q$ considers the tail $s$ of the sequence. Process $Q$ could perform another $a$ event and step further ahead, in which case $a$ is appended to the initial sequence $\lseq b \rseq \cat s$. If $P$ terminates, then an internal event $\tau$ is observed and the choice is resolved in favour of $Q$. Otherwise if $Q$ terminates, then $\tick$ is observed and the system transitions into $\Omega$. The last set of rules describes the case where $P$ is ahead of $Q$ instead.
 
In summary, a process whose trace is behind the other is allowed to catch up, while if it terminates then the choice resolves in favour of the other process. We consider the following example, with $\Sigma = \{ a, b \}$. %
\begin{example}\label{example:CSP:combinators1}
$ a \circthen Chaos \boxcircle a \circthen Skip $
\end{example}\noindent
Suppose the left-hand side process $a \circthen Chaos$ performs event $a$ first, then we arrive at the configuration $Chaos \boxcircle_{\lseq a \rseq} a \circthen Skip$. Now either $a \circthen Skip$ catches up, in which case the process can then potentially terminate, or we observe events from $Chaos$ with the potential for non-termination. Similar reasoning applies to the case where the right-hand side performs event $a$ first. In other words, an equivalent~\ac{CSP} process describing this behaviour would be $a \circthen (Chaos \sqcap Skip)$, where following the event $a$, it may terminate or diverge. Essentially, this angelic choice operator is a variant of the external choice operator that is able to delay the choice between either branch, as long as the environment can control that choice.

It is clear from~\cref{example:CSP:combinators1} that the angelic choice operator of Roscoe~\cite{Roscoe2010} is not able to avoid divergence. Ideally, a counterpart to the angelic choice of the refinement calculus should avoid divergence and favour successfully terminating processes, just like in most theories of angelic nondeterminism.

%



\subsection{UTP Model}\label{sec:ch2:CSP:UTP}
As we have previously discussed, \ac{CSP} can be characterised in the~\ac{UTP} through the theory of reactive processes~\cite{Hoare1998,Cavalcanti2006a}. In addition to the variables $ok$ and $ok'$ of the theory of designs, this theory includes the variables $wait$, $tr$, $ref$ and their dashed counterparts, that record information about interactions with the environment.

The variable $wait$ records whether the previous process is waiting for an interaction from the environment or, alternatively, has terminated. Similarly, $wait'$ ascertains this for the current process. The variable $ok$ indicates whether the previous process is in a stable state, while $ok'$ records this information for the current process. If a process is not in a stable state, then it is said to have diverged. A process only starts executing in a state where $ok$ and $\lnot wait$ are $true$. Successful termination is characterised by $ok'$ and $\lnot wait'$ being $true$.

Like in standard~\ac{CSP}, the interactions with the environment are represented using sequences of events, recorded by $tr$ and $tr'$. The variable $tr$ records the sequence of events that took place before the current process started, while $tr'$ records all the events that have been observed so far. Finally, $ref$ and $ref'$ record the set of events that may be refused by the process at the start, and currently, as required for the appropriate modelling of deadlock~\cite{Roscoe1998}.

\subsubsection{Healthiness Conditions}

The theory of reactive processes $\mathbf{R}$ is characterised by the functional composition of the following three healthiness conditions, which we reproduce below~\cite{Hoare1998,Cavalcanti2006a}.
\begin{define}[Reactive Process]\label{def:R1-R2-R3}
\begin{statement}
\begin{align*}
	&\mathbf{R1} (P) \circdef P \land tr \le tr' \\
	&\mathbf{R2} (P) \circdef P[\lseq\rseq,tr'-tr/tr,tr'] \\
	&\mathbf{R3} (P) \circdef \IIrea \dres wait \rres P \\
	&\mathbf{R} (P)  \circdef \mathbf{R3} \circ \mathbf{R1} \circ \mathbf{R2} (P)
\end{align*}
\end{statement}
\end{define}\noindent
$\mathbf{R1}$ requires that in all circumstances the only change that can be observed in the final trace of events $tr'$ is an extension of the initial sequence $tr$, while $\mathbf{R2}$ requires that a process must not impose any restriction on the initial value of $tr$. Finally, $\mathbf{R3}$ requires that if the previous process is waiting for an interaction with the environment, that is $wait$ is $true$, then the process behaves as the identity of the theory $\IIrea$~\cite{Hoare1998,Cavalcanti2006a}, otherwise it behaves as $P$. The healthiness condition of the theory of reactive processes is $\mathbf{R}$, the functional composition of $\mathbf{R1}$, $\mathbf{R2}$ and $\mathbf{R3}$.

\subsubsection{CSP Processes as Reactive Designs} 
The theory of~\ac{CSP} can be described by reactive processes that in addition also satisfy two other healthiness conditions, $\mathbf{CSP1}$ and $\mathbf{CSP2}$, whose definitions are reproduced below~\cite{Hoare1998,Cavalcanti2006a}.
\begin{define}[CSP]\label{def:CSP1-CSP2}
\begin{statement}
\begin{align*}
	&\mathbf{CSP1} (P) \circdef P \lor \mathbf{R1} (\lnot ok) \\
	&\mathbf{CSP2} (P) \circdef P \circseq ((ok \implies ok') \land tr'=tr \land ref'=ref \land wait'=wait) 
\end{align*}
\end{statement}
\end{define}\noindent
The first healthiness condition $\mathbf{CSP1}$ requires that if the previous process has diverged, that is, $ok$ is $false$, then extension of the trace is the only guarantee. $\mathbf{CSP2}$ is $\mathbf{H2}$, using the $J$-split of Cavalcanti and Woodcock~\cite{Cavalcanti2006a}, restated with the extended alphabet of reactive processes. 

A process that is $\mathbf{R}$, $\mathbf{CSP1}$ and $\mathbf{CSP2}$-healthy can be described in terms of a design as proved in~\cite{Hoare1998,Cavalcanti2006a}. We reproduce this result below, where we use the notation $P^o_w = P[o,w/ok',wait]$.
\begin{theorem}[Reactive Design]For every $CSP$ process $P$,
\begin{align*}
\mathbf{R} (\lnot P^f_f\vdash P^t_f) = P
\end{align*}
\begin{proof}
Theorem 12 in~\cite{Cavalcanti2006a}, or Theorem 8.2.2 in~\cite{Hoare1998}.
\end{proof}
\end{theorem}\noindent
This result is important as it allows~\ac{CSP} processes to be specified in terms of pre and postconditions, such as is the case for sequential programs, while the healthiness condition $\mathbf{R}$ enforces the required reactive behaviour.

\subsubsection{Operators}\label{sec:ch2:UTP:CSP-Operators}
The operators of~\ac{CSP} can then be defined using reactive designs. In what follows we present the most important~\ac{CSP} operators and discuss their specification, where use the subscript $\mathbf{R}$ to distinguish these definitions from those in other theories.

The first process of interest is $Skip_{\mathbf{R}}$, which terminates successfully. 
\begin{define}[Skip]
$ Skip_{\mathbf{R}} \circdef \mathbf{R} (true \vdash tr'=tr \land \lnot wait')$
\end{define}\noindent
Its precondition is $true$ since it never diverges and its postcondition requires that the trace of events $tr$ is unchanged while it terminates $\lnot wait'$.

On the other hand, the process that never terminates is defined by $Stop_{\mathbf{R}}$.
\begin{define}[Stop]
$ Stop_{\mathbf{R}} \circdef \mathbf{R} (true \vdash tr'=tr \land wait')$
\end{define}\noindent
Its precondition is $true$ while the postcondition requires that not only is the trace of events $tr$ never changed, but the process is always waiting for the environment: $wait'$ is $true$.

Immediate divergence is captured by the process $Chaos_{\mathbf{R}}$. 
\begin{define}[Chaos]
$ Chaos_{\mathbf{R}} \circdef \mathbf{R} (false \vdash true)$
\end{define}\noindent
In this case, the precondition is $false$, since it always diverges, then there is no way to satisfy the precondition of this process, and its postcondition is $true$. In fact, this design becomes just $true$, and the function $\mathbf{R}$ ensures that the only observation that can be made is the extension of the sequence of traces $tr$.

Prefixing can be described in terms of reactive designs as follows.
\begin{define}[Prefixing]
\begin{align*}
a \circthen_{\mathbf{R}} Skip_{\mathbf{R}} 
	\circdef 
	\mathbf{R} (true \vdash (tr'=tr \land a \notin ref') \dres wait' \rres (tr' = tr \cat \lseq a \rseq))
\end{align*}
\end{define}\noindent
The precondition is $true$, while in the postcondition there is a conditional, which defines two possible observations of its behaviour. When the process is still waiting for an interaction from the environment, and $wait'$ is $true$, then the trace of events remains unchanged while the event $a$ is not in the set of refusals $ref'$. When the process is no longer waiting, and $wait'$ is $false$, then the event $a$ is appended to the initial trace of events $tr$.

In the case of internal choice the environment has no control over the choice.
\begin{define}[Internal Choice]
$ P \sqcap_{\mathbf{R}} Q \circdef \mathbf{R} (\lnot P^f_f \land \lnot Q^f_f \vdash P^t_f \lor Q^t_f)$
\end{define}\noindent
In this case the precondition requires that the precondition of both processes $P$ and $Q$, $\lnot P^f_f$ and $\lnot Q^f_f$, holds. Moreover, the postcondition is the disjunction of the postconditions of $P$ and $Q$, $P^t_f$ and $Q^t_f$, respectively, as either postcondition may be established.

External choice, on the other hand, presents a more complex definition as a reactive design.
\begin{define}[External Choice]
\begin{align*}
 P \extchoice_{\mathbf{R}} Q \circdef \mathbf{R} (\lnot P^f_f \land \lnot Q^f_f \vdash (P^t_f \land Q^t_f) \dres tr'=tr \land wait' \rres (P^t_f \lor Q^t_f))
\end{align*}
\end{define}\noindent
Like in the definition for internal choice, both preconditions of $P$ and $Q$ need to be satisfied. The postcondition defines two cases: when the process is waiting and the trace of events has not changed, and the only possible observations of the external choice are those that are admitted by the postconditions of both processes, and, once a choice is made, the observations are either those of $P$ or $Q$, according to the postconditions.

The final, and perhaps most complex, yet fundamental operator that we consider in this discussion is sequential composition.
\begin{define}[Sequential Composition]
\begin{align*}
 P \circseq_{\mathbf{R}} Q \circdef \mathbf{R} \left(\begin{array}{l}
		\left(\begin{array}{l}
			\lnot (\mathbf{R1} (P^f_f) \circseq \mathbf{R1} (true))
			\\ \land \\
			\lnot (\mathbf{R1} (P^t_f) \circseq (\lnot wait \land \mathbf{R1} \circ \mathbf{R2} (Q^f_f)))
		\end{array}\right) 
		\\ \vdash \\
		\mathbf{R1} (P^t_f) \circseq (\II \dres wait \rres \mathbf{R1} \circ \mathbf{R2} (Q^t_f))
	\end{array}\right)
\end{align*}
\end{define}\noindent
The precondition is the conjunction of two terms, the first of which requires that the precondition of $P$ is satisfied. This is similar to the sequential composition of designs (\cref{theorem:designs:sequential-composition}), apart from the fact that $\mathbf{R1}$ is required to hold. The second term requires that the postcondition of $P$ satisfies the precondition of $Q$ when $wait$ is no longer $true$, that is, when it actually starts executing. This is again similar to the result for designs, apart from the fact there is the variable $wait$ and that $\mathbf{R1}$ must hold, and so must $\mathbf{R2}$ for the negation of the precondition of $Q$. Finally, the postcondition is given by the sequential composition of the postcondition of $P$ with a conditional, where: if $P$ is still waiting for the environment, then it behaves as the identity $\II$, otherwise it behaves as the postcondition of $Q$, where both $\mathbf{R1}$ and $\mathbf{R2}$ are required to hold.

This concludes our discussion of the~\ac{UTP} model of~\ac{CSP}. We have covered the definition of the most important operators as reactive designs. In the following section we summarise the main points of this chapter.


\section{Final Considerations}
The concept of angelic nondeterminism has been employed in many different applications as we have discussed. Its original treatment made the abstract specification of algorithms in problems involving backtracking and search possible. In the context of theories of correctness, it has traditionally been studied in the refinement calculus of Back~\cite{Back1998}, Morris~\cite{Morris1987} and Morgan~\cite{Morgan1994} through the universal monotonic predicate transformers, where it can be characterised as the least upper bound of the lattice.

In the context of relational theories, however, capturing both angelic and demonic nondeterminism is not entirely trivial. Rewitzky~\cite{Rewitzky2003} provided the fundamental theory of binary multirelations in which angelic nondeterminism can be characterised in terms of relations between states and sets of states. This has been used by Cavalcanti et al.~\cite{Cavalcanti2006} to encode both angelic and demonic nondeterminism in the relational setting of Hoare and He's~\ac{UTP}~\cite{Hoare1998}, a framework suitable for studying different programming paradigms, including process algebras like~\ac{CSP}.

\ac{CSP} has received some attention regarding the concept of angelic nondeterminism as well. In particular, Tyrrel et al.~\cite{Tyrrell2006} have suggested a lattice-theoretic model for an algebra resembling~\ac{CSP} where angelic choice is the dual of internal choice. However, the semantics is quite different from the standard model of failures-divergences of~\ac{CSP}~\cite{Roscoe1998,Roscoe2010}. Roscoe has also proposed an angelic choice operator, which however, does not avoid divergent behaviour. Ideally, an angelic choice counterpart to the refinement calculus should avoid divergent behaviour. This notion, however, has been elusive. We address this problem in the remainder of this thesis.


\chapter{Extended Binary Multirelations}\label{chapter:3}
In this chapter we introduce an extended model of binary multirelations that caters for sets of final states that are not necessarily terminating. This is achieved by extending Rewitzky's~\cite{Rewitzky2003} model of upward-closed binary multirelations with a special state that denotes the possibility for non-termination.

The following~\cref{sec:ch3:intro} introduces the model. In~\cref{sec:ch3:healthiness-conditions} the healthiness conditions are defined; their characterisation as fixed points is discussed in~\cref{sec:ch3:fixed-points}. In~\cref{sec:ch3:refinement} the refinement order is defined, while the operators are defined in~\cref{sec:ch3:operators}. \cref{sec:ch3:relationship} formalizes the relationship between this model and that of~\cite{Rewitzky2003}. Finally, we summarize our results in~\cref{sec:bmbot:final-considerations}.

\section{Introduction}\label{sec:ch3:intro}
Similarly to the original model of binary multirelations, a relation in this model associates to each initial program state a set of final states. The notion of a final state, however, is different, as formalised by the following type $BM_\bot$.
\begin{define}[Extended Binary Multirelation]\label{def:BMbot}
\begin{statement}
\begin{align*}
&State_\bot ==  State \cup \{\bot\} \\
&BM_\bot == State \rel \power State_\bot
\end{align*}
\end{statement}
\end{define}\noindent
Each initial state is related to a set of final states of type $State_\bot$, a set that may contain the special state $\bot$, which denotes non-termination. If a set of final states does not contain $\bot$, then termination in one of its states is guaranteed.

Similar to the original theory of binary multirelations, the set of final states encodes the choices available to the angel. The demonic choices are encoded by the different ways in which the set of final states can be chosen.

We consider the following example, where the value $1$ is assigned to the program variable $x$, but termination is not guaranteed. This is specified by the following relation, where $:=_{{BM}_\bot}$ is the assignment operator that does not require termination. %
\begin{example}\label{example:bmbot:assignment}
$x :=_{{BM}_\bot} 1 = \{ s : State, ss : \power State_\bot | s \oplus (x \mapsto 1) \in ss \}$
\end{example}\noindent
Every initial state $s$ is related to a set of final states $ss$ where the state obtained from $s$ by overriding the value of the component $x$ with $1$ is included. Since $ss$ is of type $State_\bot$, the sets of final states $ss$ include those with and without $\bot$. The angelic choice, therefore, cannot guarantee termination. In the following examples and definitions we may omit the type of $s$ and $ss$ for conciseness; they always have the same types as in~\cref{example:bmbot:assignment}.

It is also possible to specify a program that must terminate for certain sets of final states but not necessarily for others as shown in the following example, where $\sqcapBMbot$ is the demonic choice operator of the theory. %
\begin{example}
\begin{align*}
	&(x :=_{BM} 1) \sqcapBMbot (x :=_{{BM}_\bot} 2) \\
	&= \\
	&\{ s, ss | (s \oplus (x \mapsto 1) \in ss \land \bot \notin ss) \lor (s \oplus (x \mapsto 2) \in ss) \}
\end{align*}
\end{example}\noindent
Since $BM$ is in fact a subset of $BM_\bot$, it is possible to use some of the existing operators, such as the terminating assignment operator $:=_{BM}$. In this case, there is a demonic choice between the terminating assignment of $1$ to $x$, and the assignment of $2$ to $x$ that does not require termination.

\section{Healthiness Conditions}\label{sec:ch3:healthiness-conditions}
Having defined the type of the extended binary multirelations $BM_{\bot}$, in the following~\cref{ch3:bmbot:bmh0,ch3:bmbot:bmh1,ch3:bmbot:bmh2,ch3:bmbot:bmh3} we introduce the healthiness conditions that characterise the relations in the theory.

\subsection{$\mathbf{BMH0}$}\label{ch3:bmbot:bmh0}
The first healthiness condition of interest is $\mathbf{BMH0}$. It enforces the upward closure of the original theory of binary multirelations~\cite{Rewitzky2003} for sets of final states that are necessarily terminating, and in addition enforces a similar property for sets of final states that are not required to terminate. 
\begin{define}[$\mathbf{BMH0}$]\label{def:BMH0}
\begin{statement}
\begin{align*}
	&\forall s, ss_0, ss_1 \spot ((s, ss_0) \in B \land ss_0 \subseteq ss_1 \land (\bot \in ss_0 \iff \bot \in ss_1)) \implies (s, ss_1) \in B
\end{align*}
\end{statement}
\end{define}\noindent
It states that for every initial state $s$, and for every set of final states $ss_0$ in a relation $B$, any superset $ss_1$ of that final set of states is also associated with $s$ such that $\bot$ is in $ss_0$ if, and only if, it is in $ss_1$. That is, $\mathbf{BMH0}$ requires the upward closure for sets of final states that terminate, and for those that may or may not terminate, but separately.

The definition of $\mathbf{BMH0}$ can be split into two conjunctions as established by the following~\cref{law:bmbot:BMH0-BMH}. $\mathbf{BMH}$ is the healthiness condition of the original theory whose definition was reproduced in~\cref{sec:ch2:binary-multirelations}. Proof of these and other results to follow can be found in~\cref{appendix:bm}.
\theoremstatementref{law:bmbot:BMH0-BMH}\noindent%
This result confirms that for sets of final states that terminate this healthiness condition enforces $\mathbf{BMH}$ exactly as in the original theory of binary multirelations~\cite{Rewitzky2003}.

\subsection{$\mathbf{BMH1}$}\label{ch3:bmbot:bmh1}
The second healthiness condition $\mathbf{BMH1}$ requires that if it is possible to choose a set of final states where termination is not guaranteed, then it must also be possible to choose an equivalent set of states where termination is guaranteed. This healthiness condition is similar in nature to $\mathbf{H2}$ of the theory of designs.
\begin{define}[$\mathbf{BMH1}$]\label{def:BMH1}
\begin{statement}
$\forall s:State, ss : \power State_\bot \spot (s, ss \cup \{\bot\}) \in B \implies (s, ss) \in B$
\end{statement}
\end{define}\noindent
If it is possible to reach a set of final states $(ss \cup \{\bot\})$ from some initial state $s$, then the set of final states $ss$, without $\bot$, so that termination is required, is also associated with $s$.

This healthiness condition excludes relations that only offer sets of final states that may not terminate. We consider the following~\cref{example:bmbot-bot-not-in-ss}.
\begin{example}\label{example:bmbot-bot-not-in-ss}
$\{ s : State, ss : \power State_\bot | s \oplus (x \mapsto 1) \in ss \land \bot \in ss \}$
\end{example}\noindent
This relation describes the assignment of $1$ to the program variable $x$ where termination is not guaranteed. It discards the inclusive situation where termination may indeed occur, and so is not $\mathbf{BMH1}$-healthy. The inclusion of a corresponding final set of states that requires termination does not change the choices available to the angel as it is still impossible to guarantee termination.

\subsection{$\mathbf{BMH2}$}\label{ch3:bmbot:bmh2}
In this model, both the empty set of final states and $\{\bot\}$ characterise abortion. This redundancy, which facilitates the linking between theories, in particular with the original theory of Rewitzky~\cite{Rewitzky2003}, is captured by the following healthiness condition. 
\begin{define}[$\mathbf{BMH2}$]\label{def:BMH2}
\begin{statement}
$\forall s : State \spot (s, \emptyset) \in B \iff (s, \{ \bot \}) \in B$
\end{statement}
\end{define}\noindent
It requires that every initial state $s$ is related to the empty set of final states if, and only if, it is also related to the set of final states $\{ \bot \}$. By allowing $(s, \emptyset)$ to be part of the model, we can easily characterise the original theory of binary multirelations as a subset of ours.

If we consider $\mathbf{BMH1}$ in isolation, it covers the reverse implication of $\mathbf{BMH2}$ because if $(s, \{\bot\})$ is in the relation, so is $(s, \emptyset)$. However, $\mathbf{BMH2}$ is stronger than $\mathbf{BMH1}$ by requiring $(s, \{\bot\})$ to be in the relation if $(s, \emptyset)$ is also in the relation.

This new model of binary multirelations is characterised by the conjunction of the healthiness conditions $\mathbf{BMH0}$, $\mathbf{BMH1}$ and $\mathbf{BMH2}$ to which we refer as $\mathbf{BMH_\bot}$. In~\cref{sec:ch3:fixed-points} we provide alternative definitions of the healthiness conditions in terms of fixed points. This characterisation enables us, for instance, to establish that the healthiness conditions are idempotent and monotonic.

\subsection{$\mathbf{BMH3}$}\label{ch3:bmbot:bmh3}
The fourth healthiness condition characterises a subset of the model that corresponds to the original theory of binary multirelations of Rewitzky~\cite{Rewitzky2003}.
\begin{define}[$\mathbf{BMH3}$]\label{def:BMH3}
\begin{statement}
\begin{align*}
	&\forall s : State \spot 
		(s, \emptyset) \notin B
		\implies
		(\forall ss : \power State_\bot \spot (s, ss) \in B \implies \bot \notin ss)
\end{align*}
\end{statement}
\end{define}\noindent
If an initial state $s$ is not related to the empty set, then it must be the case that for all sets of final states $ss$ related to $s$, $\bot$ is not included in the set $ss$. 

The healthiness condition $\mathbf{BMH3}$ excludes relations that do not guarantee termination for particular initial states, yet establish some set of final states. An example of such a relation is \cref{example:bmbot:assignment}. This is also the case for the original theory of binary multirelations. If it is possible for a program not to terminate when started from some initial state, then execution from that state must lead to arbitrary behaviour. This is the same intuition for $\mathbf{H3}$ of the theory of designs~\cite{Hoare1998}.

\section{Healthiness Conditions as Fixed Points}\label{sec:ch3:fixed-points}
Having defined the healthiness conditions of the theory, in this section we consider their definitions via idempotent functions, whose fixed points are the relations in the theory. This is similar to the approach followed in~\ac{UTP} theories. This dual characterisation is used in~\cref{sec:ch3:relationship} to establish an isomorphism between a subset of this model and the original theory of binary multirelations.

For each healthiness condition of interest, we use the notation $\mathbf{bmh}_\mathbf{x}$ to denote the function whose fixed points correspond exactly to the relations characterised by the healthiness condition $\mathbf{BMHx}$, that is $\mathbf{bmh}_\mathbf{x} (B) = B \iff \mathbf{BMHx}$. Furthermore, the notation $\mathbf{bmh}_\mathbf{x,y}$ denotes the functional composition of the functions $\mathbf{bmh_x}$ and $\mathbf{bmh_y}$, so that $\mathbf{bmh}_\mathbf{x,y} (B) = \mathbf{bmh}_\mathbf{x} \circ \mathbf{bmh}_\mathbf{y} (B)$.

In the next~\cref{sec:bmbot:hc:bmh-0-1-2-3-or}, each healthiness condition is characterised by a corresponding function. A full account of the properties of the functional composition of each function is found in~\cref{appendix:bmbot:healthiness-conditions}. Moreover, in~\cref{sec:bmbot:hc:bmh-0-1-2,sec:bmbot:hc:bmh-0-1-3-2} the two functions that characterise the model as a whole, and its subset of interest, are presented.


\subsection{$\mathbf{bmh}_\mathbf{0}$, $\mathbf{bmh}_\mathbf{1}$, $\mathbf{bmh}_\mathbf{2}$ and $\mathbf{bmh}_\mathbf{3}$}\label{sec:bmbot:hc:bmh-0-1-2-3-or}
The first function of interest is $\mathbf{bmh}_\mathbf{0}$, whose fixed points are the $\mathbf{BMH0}$-healthy binary multirelations. It is defined as follows.
\begin{define}\label{def:bmbot:bmh0}
\begin{align*}
	&\mathbf{bmh}_\mathbf{0} (B) \circdef \{ s, ss | \exists ss_0 \spot (s, ss_0) \in B \land ss_0 \subseteq ss \land (\bot \in ss_0 \iff \bot \in ss)\}
\end{align*}
\end{define}\noindent
For every initial state $s$ in $B$, whenever it is related to a set of final states $ss_0$ it is also related to its superset $ss$, such that $\bot$ is in $ss_0$ if, and only if, $\bot$ is also in $ss$. In other words, $\mathbf{bmh_0}$ enforces the upward closure of a relation $B$ just like $\mathbf{BMH0}$.

The healthiness condition $\mathbf{BMH1}$ is characterised by the fixed points of $\mathbf{bmh}_\mathbf{1}$.
\begin{define}
$\mathbf{bmh}_\mathbf{1} (B) \circdef \{ s, ss | (s, ss \cup \{\bot\}) \in B \lor (s, ss) \in B \}$
\end{define}\noindent
Its definition considers all pairs $(s, ss)$ in $B$, such that if a set of final states includes $\bot$ then there is also a set of final states without $\bot$. 

$\mathbf{BMH2}$-healthy relations are fixed points of the function $\mathbf{bmh}_\mathbf{2}$, whose definition is presented below.
\begin{define}
$\mathbf{bmh}_\mathbf{2} (B) \circdef \{s, ss | (s, ss) \in B \land ((s, \{ \bot \}) \in B \iff (s, \emptyset) \in B)\}$
\end{define}\noindent
The definition considers every pair $(s, ss)$ in $B$ and requires that $(s,\{\bot\})$ is in $B$ if, and only if, $(s, \emptyset)$ is also in $B$. If the equivalence is not satisfied then $\mathbf{bmh}_\mathbf{2}$ yields the empty set.

Finally, the $\mathbf{BMH3}$-healthy relations are characterised by the fixed points of $\mathbf{bmh}_\mathbf{3}$.
\begin{define}
$\mathbf{bmh}_\mathbf{3} (B) \circdef \{ s, ss | ( (s, \emptyset) \in B \lor \bot \notin ss) \land (s, ss) \in B \}$
\end{define}\noindent
The definition considers every pair $(s, ss)$ in $B$ and requires that either $ss$ is a set of final states with guaranteed termination, and so without $\bot$, or $(s, \emptyset)$ is in $B$, and thus the initial state $s$ leads to arbitrary behaviour.

The following~\cref{law:bmbot:bmh0-definition,law:bmh1-definition,law:bmbot:bmh2-definition,law:bmbot:bmh3-definition} establish that the fixed points of each $\mathbf{bmh_x}$ function are exactly those relations that satisfy the corresponding healthiness condition $\mathbf{BMHx}$.
\theoremstatementref{law:bmbot:bmh0-definition}
\theoremstatementref{law:bmh1-definition}
\theoremstatementref{law:bmbot:bmh2-definition}
\theoremstatementref{law:bmbot:bmh3-definition}\noindent%
Furthermore, the following~\cref{law:bmbot:bmh0-idempotent,law:bmh1-idempotent,law:bmbot:bmh-2-idempotent,law:bmbot:bmh3-idempotent} establish that each $\mathbf{bmh_x}$ function is idempotent.
\theoremstatementref{law:bmbot:bmh0-idempotent} 
\theoremstatementref{law:bmh1-idempotent}
\theoremstatementref{law:bmbot:bmh-2-idempotent}
\theoremstatementref{law:bmbot:bmh3-idempotent}\noindent%
This section concludes our discussion regarding the definition of the $\mathbf{bmh}_\mathbf{x}$ functions. Properties of their functional composition are studied in detail in~\cref{appendix:bmbot:healthiness-conditions}. In the following~\cref{sec:bmbot:hc:bmh-0-1-2,sec:bmbot:hc:bmh-0-1-3-2} we focus our attention only on the functional compositions that characterise the theory of $\mathbf{BMH0}$-$\mathbf{BMH2}$ multirelations and the subset, that in addition, satisfies $\mathbf{BMH3}$.

\subsection{$\mathbf{bmh}_\mathbf{0,1,2}$}\label{sec:bmbot:hc:bmh-0-1-2}
The relations in the model of extended binary multirelations are characterised by the conjunction of the healthiness conditions $\mathbf{BMH0}$, $\mathbf{BMH1}$ and $\mathbf{BMH2}$, otherwise also named as $\mathbf{BMH_\bot}$ as depicted in~\cref{fig:theories}. These relations can also be expressed as fixed points of the functional composition of the functions $\mathbf{bmh}_\mathbf{0}$, $\mathbf{bmh}_\mathbf{1}$ and $\mathbf{bmh}_\mathbf{2}$, as shown by the following~\cref{law:bmbot:bmh-0-1-2:definition}.
\theoremstatementref{law:bmbot:bmh-0-1-2:definition}\noindent%
The notation $\mathbf{bmh}_\mathbf{0,1,2}$ denotes the functional composition $\mathbf{bmh_0}\circ\mathbf{bmh_1}\circ\mathbf{bmh_2}$. The order of this functional composition is justified by~\cref{law:bmbot:BMH-0-1-2:iff:bmh-0-1-2}, and results established in~\cref{sec:bmbot:hc:bmh-0-1,sec:bmbot:hc:bmh-1-2}.
\theoremref{law:bmbot:BMH-0-1-2:iff:bmh-0-1-2}\noindent%
That is, a multirelation $B$ is a fixed point of $\mathbf{bmh_{0,1,2}}$, if, and only if, it satisfies the healthiness conditions $\mathbf{BMH0}$, $\mathbf{BMH1}$ and $\mathbf{BMH2}$. The proof of this theorem relies on the results which we discuss in the following paragraphs.

First we establish in~\cref{law:bmbot:bmh-0-1-2-implies-BMH0,law:bmbot:bmh-0-1-2-implies-BMH1,law:bmbot:bmh-0-1-2-implies-BMH2} that a fixed point of $\mathbf{bmh}_\mathbf{0,1,2}$ satisfies each of the healthiness conditions $\mathbf{BMH0}$, $\mathbf{BMH1}$ and $\mathbf{BMH2}$.
\theoremstatementref{law:bmbot:bmh-0-1-2-implies-BMH0}
\theoremstatementref{law:bmbot:bmh-0-1-2-implies-BMH1}
\theoremstatementref{law:bmbot:bmh-0-1-2-implies-BMH2}\noindent%
Moreover, we establish in~\cref{law:bmbot:BMH0-1-2-implies-bmh-0-1-2} that a relation that is $\mathbf{BMH0}$, $\mathbf{BMH1}$ and $\mathbf{BMH2}$-healthy is also a fixed point of $\mathbf{bmh}_\mathbf{0,1,2}$.
\theoremstatementref{law:bmbot:BMH0-1-2-implies-bmh-0-1-2}\noindent%
These lemmas conclude our discussion of the healthiness conditions of the new theory of binary multirelations. In summary, these relations can be characterised either by the predicates $\mathbf{BMH0}$-$\mathbf{BMH2}$ or as fixed points of $\mathbf{bmh}_\mathbf{0,1,2}$. In the following section we focus our attention on the subset of the theory that contains only the multirelations that are in addition $\mathbf{BMH3}$-healthy. 

\subsection{$\mathbf{bmh}_\mathbf{0,1,3,2}$}\label{sec:bmbot:hc:bmh-0-1-3-2}
Relations that are $\mathbf{BMH0}$, $\mathbf{BMH1}$, $\mathbf{BMH2}$ and $\mathbf{BMH3}$-healthy can be characterised as fixed points of the functional composition $\mathbf{bmh}_\mathbf{0,1,3,2}$. The result of this composition is given by the following~\cref{law:bmbot:bmh-0-1-3-2:definition}.
\theoremstatementref{law:bmbot:bmh-0-1-3-2:definition}\noindent%
The set construction considers a disjunction, where, either $s$ is an aborting state, and hence it is related to the empty set and $\{\bot\}$, and otherwise, if it is not aborting, it satisfies the same property of upward-closure as required by $\mathbf{bmh_0}$. The particular order of this functional composition is justified by the following~\cref{theorem:bmbot:BMH0-3:iff:bmh-0-1-3-2}.
\theoremstatementref{theorem:bmbot:BMH0-3:iff:bmh-0-1-3-2}\noindent%
The proof of~\cref{theorem:bmbot:BMH0-3:iff:bmh-0-1-3-2} is split into two implications. First, we establish through~\cref{law:bmbot:BMH0-3:implies:bmh-0-1-3-2} that the conjunction of the predicative healthiness conditions $\mathbf{BMH0}$ to $\mathbf{BMH3}$ implies that $B$ is a fixed point of $\mathbf{bmh_{0,1,3,2}}$.
\theoremstatementref{law:bmbot:BMH0-3:implies:bmh-0-1-3-2}\noindent%
To prove the reserve implication, we first establish through~\cref{law:bmh-0-1-2:fixed-point:bmh-0-1-2-3} that a fixed point of $\mathbf{bmh_{0,1,3,2}}$ is also a fixed point of $\mathbf{bmh_{0,1,2}}$, so that~\cref{law:bmbot:bmh-0-1-2-implies-BMH0,law:bmbot:bmh-0-1-2-implies-BMH1,law:bmbot:bmh-0-1-2-implies-BMH2} are directly applicable. 
\theoremstatementref{law:bmh-0-1-2:fixed-point:bmh-0-1-2-3}\noindent%
Finally, \cref{law:bmbot:bmg-0-1-3-2:implies:BMH3} establishes that every fixed point of $\mathbf{bmh_{0,1,3,2}}$ satisfies the predicative healthiness condition $\mathbf{BMH3}$.
\theoremstatementref{law:bmbot:bmg-0-1-3-2:implies:BMH3}\noindent%
This concludes the proof that the subset of the theory that is in addition $\mathbf{BMH3}$-healthy also has a counterpart characterisation via fixed points of $\mathbf{bmh_{0,1,3,2}}$. This function characterises the subset that corresponds to the original theory of binary multirelations. The relationship with the original theory of binary multirelations is explored in~\cref{sec:ch3:relationship}.
\section{Refinement}\label{sec:ch3:refinement}
The refinement order for the new binary multirelation model is defined exactly as in the original theory of binary multirelations~\cite{Rewitzky2003}.

\begin{define}[Refinement]
$B_1 \refinedbyBMbot B_0 \circdef B_1 \supseteq B_0$
\end{define}\noindent
It is reverse subset inclusion, such that a program characterised by a multirelation $B_0$ refines another characterised by a multirelation $B_1$ if, and only if, $B_0$ is a subset of $B_1$.

The extreme points of the theory as expected of a theory of designs, are the everywhere miraculous program and abort. Their definitions are presented below.


\begin{define}[Miracle]
$\top_{BM_\bot} \circdef \emptyset$
\end{define}\noindent
As in the original theory, miracle is denoted by the absence of any relationship between any input state and any set of final states, that is, the program cannot possibly be executed.

\begin{define}[Abort]
$\bot_{BM_\bot} \circdef State \times \power State_\bot$
\end{define}\noindent
On the other hand, abort is characterised by the universal relation, such that every initial state is related to every possible set of final states.

\section{Operators}\label{sec:ch3:operators}
In this section the most important operators of the theory are introduced. Namely, we define the operators of assignment, angelic and demonic choice, and sequential composition. These enable the discussion of interesting properties observed in this model of extended binary multirelations.

As discussed in~\cref{chapter:1}, the model that we propose here is isomorphic to the theory of angelic designs that we discuss in~\cref{chapter:4}. In that chapter we establish that the operators discussed here are in correspondence with those in the theory of angelic designs, which we prove to be closed. Together with the respective isomorphism that we discuss in~\cref{sec:ch4:bmbot-relationship}, these results are sufficient to establish closure of the operators with respect to the healthiness condition $\mathbf{BMH_\bot}$.

\subsection{Assignment}
The first operator of interest is assignment. As already illustrated, in this new model, there is the possibility to define two distinct assignment operators. The first one behaves exactly as in the original theory of binary multirelations $x :=_{BM} e$. This operator does not need to be redefined, since $BM \subseteq BM_\bot$. The new operator that we define below, however, behaves rather differently, in that it may or may not terminate.
\begin{define}
$x: =_{{BM}_\bot} e \circdef \{ s : State, ss : \power State_\bot | s \oplus (x \mapsto e) \in ss \}$
\end{define}\noindent
This assignment guarantees that for every initial state $s$, there is some set of final states available for angelic choice where $x$ has the value of expression $e$. However, termination is not guaranteed. While the angel can choose the final value of $x$ it cannot possibly guarantee termination in this case.

\subsection{Angelic Choice}
The definition of angelic choice is exactly the same as in the original theory of binary multirelations.
\begin{define} $B_0 \sqcupBMbot B_1 \circdef B_0 \cap B_1$
\end{define}\noindent
For every set of final states available for demonic choice in $B_0$ and $B_1$, only those that can be chosen both in $B_0$ and $B_1$ are available.

An interesting property of angelic choice that is observed in this model is illustrated by the following~\cref{law:bmbot:angelic-two-assignment}. It considers the angelic choice between two assignments of the same expression, yet only one is guaranteed to terminate.
\theoremstatementref{law:bmbot:angelic-two-assignment}\noindent
This result can be interpreted as follows: given an assignment that is guaranteed to terminate, adding a corresponding angelic choice that is potentially non-terminating does not in fact introduce any new choices.

In general, and as expected from the original model of binary multirelations, the angelic choice operator observes the following properties. As the refinement ordering in the new model is exactly the same as in the theory of binary multirelations, the angelic choice operator, being the least upper bound in both theories, has the same properties with respect to the extreme points of the lattice. 
\theoremstatementref{lemma:bmbot:top-sqcup-B}\noindent%
The angelic choice between an everywhere miraculous program and any other program is still miraculous.
\theoremstatementref{lemma:bmbot:bot-sqcup-B}\noindent%
On the other hand, the angelic choice between abort and any other program $B$ is the same as $B$. That is, the angel will avoid choosing an aborting program if possible.
\subsection{Demonic Choice}
The next operator of interest is demonic choice. It is also defined exactly like in the original theory of binary multirelations.
\begin{define}
$B_0 \sqcapBMbot B_1 \circdef B_0 \cup B_1$
\end{define}\noindent
For every initial state, a corresponding set of final states available for demonic choice in either, or both, of $B_0$ and $B_1$, is included in the result.

Similarly to the angelic choice operator, there is a general result regarding the demonic choice over the two assignment operators, terminating and not necessarily terminating. This is established by the following~\cref{law:bmbot:demonic-two-assignment}.
\theoremstatementref{law:bmbot:demonic-two-assignment}\noindent%
If there is an assignment for which termination is not guaranteed, then the demonic choice over this assignment and a corresponding one that is guaranteed to terminate is the same as the assignment that does not require termination. In other words, if it is possible for the demon to choose between two similar sets of final states, one that is possibly non-terminating and one that terminates, then the one for which termination is not guaranteed dominates the choice.

The following two laws show how the demonic choice operator behaves with respect to the extreme points of the lattice.
\theoremstatementref{lemma:bmbot:bot-sqcap-B}\noindent%
\theoremstatementref{lemma:bmbot:top-sqcap-B}\noindent%
As expected, the demonic choice between abort and some other program is abort. In the case of a miracle, the demon will avoid choosing it if possible. 

Since the angelic and demonic choice operators are defined as set intersection and union, respectively, they also distribute through each other. This is exactly the same property as in the original theory of binary multirelations.
\subsection{Sequential Composition}
The definition of sequential composition in this new model is not immediately obvious. We note, however, that one of the reasons for developing this theory is the fact that it allows a more intuitive account of the definition of sequential composition and, as such, an easier route to discover the definition in the theory of angelic designs. To illustrate the issue, we consider the following example from the theory of designs, where a non-$\mathbf{H3}$-design is sequentially composed with $\IID$. %
\begin{example}\label{example:bmbot:non-terminating-seqD-IID}
\begin{flalign*}
	&(x' = 1 \vdash true) \circseq \IID
	&&\ptext{Definition of $\IID$}\\
	&=(x' = 1 \vdash true) \circseq (true \vdash x' = x)
	&&\ptext{Sequential composition for designs}\\
	&=( \lnot (x' \neq 1 \circseq true) \land \lnot (true \circseq false) \vdash true \circseq x' = x)
	&&\raisetag{18pt}\ptext{Sequential composition}\\
	&=(\lnot (\exists x_0 \spot x_0 \neq 1 \land true) \land \lnot (\exists x_0 \spot true \land false) \vdash \exists x_0 \spot true \land x' = x_0)
	&&\ptext{Predicate calculus and one-point rule}\\
	&=(\lnot true \land \lnot false \vdash true)
	&&\ptext{Predicate calculus and property of designs}\\
	&=true
\end{flalign*}
\end{example}\noindent
The result is $true$, the bottom of designs~\cite{Hoare1998}, whose behaviour is arbitrary. This arises because, since the first design can always establish a final value for $x$, namely $1$, where termination is then not guaranteed, the $Skip$ design $\IID$ that follows can never guarantee termination. This result can be generalised for a sequential composition involving any non-$\mathbf{H3}$-design.

This provides the motivation for the definition of sequential composition in the new binary multirelational model.
\begin{define}
\begin{align*}
	&B_0 \seqBMbot B_1
	\circdef
	\left\{
		s_0, ss_0
		\left|\begin{array}{l}
			\exists ss \spot (s_0, ss) \in B_0 
				 \land \\
				(\bot \in ss
					 \lor 
					ss \subseteq \{ s_1 : State | (s_1, ss_0) \in B_1\})
				\end{array}\right.
		\right\}
\end{align*}
\end{define}\noindent%
For sets of final states where termination is guaranteed, that is, $\bot$ is not in the set of intermediate states $ss$, this definition matches that of the original theory. If $\bot$ is in $ss$, and hence termination is not guaranteed, then the result of the sequential composition is arbitrary as it can include any set of final states.
If we assume that $B_0$ is $\mathbf{BMH0}$-healthy, then the definition of sequential composition can be split into the set union of two sets as shown in~\cref{law:bmbot:seqBMbot:BMH0-healthy}.%
\theoremstatementref{law:bmbot:seqBMbot:BMH0-healthy}\noindent%
The first set considers the case when $B_0$ leads to sets of final states where termination is not required and, therefore, to the whole of $State_\bot$, due to upward closure. The second set considers the case where termination is required and matches the result of~\cref{lemma:seqBM}.

For a similar example to~\cref{example:bmbot:non-terminating-seqD-IID} expressed in the new theory, we consider the following example, where a non-terminating assignment is followed by the assignment that requires termination, but does not change the value of $x$.
\begin{example}
\begin{flalign*}
	&(x :=_{{BM}_\bot} e) \seqBMbot (x :=_{BM} x)
	&&\ptext{Definition of $\seqBMbot$ (\cref{law:bmbot:seqBMbot:BMH0-healthy})}\\
	&=\left(\begin{array}{l}
		\left\{
		s_0 : State, ss_0 : \power State_\bot | (s_0, State_\bot) \in (x :=_{{BM}_\bot} e)
		\right\}
		\\ \cup \\
		\left\{\begin{array}{l}
			s_0 : State, ss_0 : \power State_\bot \\
			| (s_0, \{ s_1 : State | (s_1, ss_0) \in (x :=_{BM} x)\}) \in (x :=_{{BM}_\bot} e)
		\end{array}\right\}
	\end{array}\right)
	&&\ptext{Definition of $:=_{BM}$ and $:=_{{BM}_\bot}$}\\
	&=\left(\begin{array}{l}
		\left\{\begin{array}{l}
			s_0 : State, ss_0 : \power State_\bot \\
			| (s_0, State_\bot) \in \{ s : State, ss : \power State_\bot | s \oplus (x \mapsto e) \in ss \}
		\end{array}\right\}
		\\ \cup \\
		\left\{\begin{array}{l}
			s_0 : State, ss_0 : \power State_\bot \\
			\left|\begin{array}{l}
				(s_0, \{ s_1 : State | (s_1, ss_0) \in (x :=_{BM} x)\})
				\\ \in \\ 
				\{ s : State, ss : \power State | s \oplus (x \mapsto e) \in ss \}
			\end{array}\right.
		\end{array}\right\}
	\end{array}\right)
	&&\ptext{Property of sets}\\
	&=\left(\begin{array}{l}
		\{s_0 : State, ss_0 : \power State_\bot | s_0 \oplus (x \mapsto e) \in State_\bot \}
		\\ \cup \\
		\left\{\begin{array}{l}
			s_0 : State, ss_0 : \power State_\bot \\
			\left|\begin{array}{l}
				s_0 \oplus (x \mapsto e) \in \{ s_1 : State | (s_1, ss_0) \in (x :=_{BM} x)\}
			\end{array}\right.
		\end{array}\right\}
	\end{array}\right)
	&&\ptext{Property of sets}\\
	&=\left(\begin{array}{l}
		\{s_0 : State, ss_0 : \power State_\bot | true \}
		\\ \cup \\
		\left\{\begin{array}{l}
			s_0 : State, ss_0 : \power State_\bot \\
			\left|\begin{array}{l}
				s_0 \oplus (x \mapsto e) \in \{ s_1 : State | (s_1, ss_0) \in (x :=_{BM} x)\}
			\end{array}\right.
		\end{array}\right\}
	\end{array}\right)
	&&\ptext{Property of sets and definition of $\botBMbot$}\\
	&=\botBMbot
\end{flalign*}
\end{example}\noindent
The result of this sequential composition is an aborting program. Like in the theory of designs, if it is possible for the first program not to terminate, then the sequential composition cannot provide any guarantees either. The properties observed by the sequential composition operator are explored in what follows.

\subsubsection{Properties}
The first property of interest considers the sequential composition of $\topBMbot$ followed by some program $B$. The result is also a miraculous program as shown in the following~\cref{law:bmbot:top-seqBMbot-B:top}.%
\theoremstatementref{law:bmbot:top-seqBMbot-B:top}\noindent%
The following law expresses that the sequential composition of abort with another program is also abort.
\theoremstatementref{law:bmbot:bot-seqBMbot-B:bot}\noindent%
In the following paragraphs we explore some examples with respect to the extreme points of the lattice. 

The following example describes the general behaviour of some program $B$ that is $\mathbf{BMH0}$-healthy sequentially composed with a miraculous program. 
\begin{example}
\begin{flalign*}
	&B \seqBMbot \topBMbot
	&&\ptext{Definition of $\topBMbot$ and $\seqBMbot$ (\cref{law:bmbot:seqBMbot:BMH0-healthy})}\\
	&=\left(\begin{array}{l}
		\left\{
		s_0 : State, ss_0 : \power State_\bot | (s_0, State_\bot) \in B
		\right\}
		\\ \cup \\
		\left\{
		s_0 : State, ss_0 : \power State_\bot | (s_0, \{ s_1 : State | (s_1, ss_0) \in \emptyset\}) \in B
		\right\}
	\end{array}\right)
	&&\ptext{Property of sets}\\
	&=\left(\begin{array}{l}
		\left\{
		s_0 : State, ss_0 : \power State_\bot | (s_0, State_\bot) \in B
		\right\}
		\\ \cup \\
		\left\{
		s_0 : State, ss_0 : \power State_\bot | (s_0, \emptyset) \in B
		\right\}
	\end{array}\right)
\end{flalign*}
\end{example}\noindent
If $B$ may not terminate for some set of initial states, and it is $\mathbf{BMH0}$-healthy, then the result of the sequential composition is also abort, for those initial states. If $B$ aborts for some particular initial state $s_0$, then that state is related to the empty set in $B$ and the result of the sequential composition is also abort. Otherwise, the result is miraculous as the initial state is not in the domain of either relation in the union above.


The following example describes the behaviour of a program $B$ sequentially composed with abort. 
\begin{example}
\begin{flalign*}
	&B \seqBMbot \botBMbot
	&&\ptext{Definition of $\botBMbot$ and $\seqBMbot$ (\cref{law:bmbot:seqBMbot:BMH0-healthy})}\\	
	&=\left(\begin{array}{l}
		\left\{
		s_0 : State, ss_0 : \power State_\bot | (s_0, State_\bot) \in B
		\right\}
		\\ \cup \\
		\left\{\begin{array}{l}
		s_0 : State, ss_0 : \power State_\bot \\
		| (s_0, \{ s_1 : State | (s_1, ss_0) \in (State \times \power State_\bot)\}) \in B
		\end{array}\right\}
	\end{array}\right)
	&&\ptext{Property of sets}\\
	&=\left(\begin{array}{l}
		\left\{
		s_0 : State, ss_0 : \power State_\bot | (s_0, State_\bot) \in B
		\right\}
		\\ \cup \\
		\left\{
		s_0 : State, ss_0 : \power State_\bot | (s_0, \{ s_1 : State | true\}) \in B
		\right\}
	\end{array}\right)
	&&\raisetag{36pt}\ptext{Property of sets}\\
	&=\{ s_0 : State, ss_0 : \power State_\bot | (s_0, State_\bot) \in B \lor (s_0, State) \in B \}
\end{flalign*}
\end{example}\noindent
Because $B$ is upward closed, if it definitely terminates then $State$ is a superset of all sets of final states and is in $B$. If $B$ may or may not terminate for some particular set of final states, then $State_\bot$ is also in $B$ due to the upward closure guaranteed by $\mathbf{BMH0}$. In either case, the sequential composition behaves as abort. If $B$ is miraculous, then so is the sequential composition.

\section{Relationship with Binary Multirelations}\label{sec:ch3:relationship}
Having presented the most important operators of the theory, in this section we focus our attention on the relationship between the new model and the original theory of binary multirelations. The first step consists in the definition of a pair of linking functions, $bmb2bm$, which maps relations from the new model into the original theory of binary multirelations, and $bm2bm$, a mapping in the opposite direction.

As previously discussed in~\cref{chapter:1}, the relationship is illustrated in~\cref{fig:theories,fig:theories:bmh} where each theory is labelled according to its healthiness conditions. In this case, we have a bijection between the subset of $\mathbf{BMH_\bot}$ characterised by the relations that are $\mathbf{BMH3}$-healthy and the original theory of binary multirelations characterised by $\mathbf{BMH}$. In this section our discussion is focused on this isomorphism, while in~\cref{chapter:4} we discuss the isomorphism with the theory of angelic designs.

\subsection{From $BM_\bot$ to $BM$ ($bmb2bm$)}\label{sec:bmbot:bmb2bm}
The first function of interest is $bmb2bm$ that maps from binary multirelations in the new model, of type $BM_\bot$, to those in the original model of type $BM$.
\begin{define}
\begin{align*}
	&bmb2bm : {BM}_\bot \fun BM \\
	&bmb2bm (B) \circdef \{ s : State, ss : \power State_\bot | (s, ss) \in B \land \bot \notin ss \}
\end{align*}
\end{define}\noindent%
Its definition considers every pair $(s, ss)$ in $B$ such that $\bot$ is not in $ss$. We consider the following example, where $bmb2bm$ is applied to the potentially non-terminating assignment of $e$ to the program variable $x$.
\begin{example}$bmb2bm(x :=_{BM_\bot} e) = (x :=_{BM} e)$\end{example}\noindent
The result corresponds to assignment in the original theory. 

In order to establish that $bmb2bm$ yields a multirelation that is $\mathbf{BMH}$-healthy we use an alternative way to characterise the set of healthy binary multirelations as fixed points of the function $\mathbf{bmh_{up}}$.
\begin{define}
$\mathbf{bmh_{up}} (B) \circdef \{ s, ss | \exists ss_0 : \power State \spot (s, ss_0) \in B \land ss_0 \subseteq ss \}$
\end{define}\noindent
This definition is justified by~\cref{law:bm:bmh-upclosed}.
\theoremstatementref{law:bm:bmh-upclosed}\noindent%
Finally, \cref{theorem:scratchpad:bmh-upclosed-o-bmb2bm-bmh-0-1-3-2} establishes that the application of $bmb2bm$ to a multirelation that is $\mathbf{BMH0}$-$\mathbf{BMH3}$-healthy yields a $\mathbf{BMH}$-healthy relation.
\theoremstatementref{theorem:scratchpad:bmh-upclosed-o-bmb2bm-bmh-0-1-3-2}\noindent
In summary, $bmb2bm$ yields relations that are in the original theory.

\subsection{From $BM$ to $BM\bot$ ($bm2bmb$)}\label{sec:bmbot:bm2bmb}
The mapping in the opposite direction, from $BM$ to $BM_\bot$ is achieved by the function $bmb2bm$, whose definition is presented below.
\begin{define}
\begin{align*}
	&bm2bmb : BM \fun {BM}_\bot \\
	&bm2bmb (B) \circdef \{s : State, ss : \power State_\bot | ((s, ss) \in B \land \bot \notin ss) \lor (s, \emptyset) \in B\}
\end{align*}
\end{define}\noindent
It considers every pair $(s, ss)$ in a relation $B$, where $\bot$ is not in the set of final states $ss$, or if $B$ is aborting for a particular state $s$, that is, $s$ is related to the empty set, then it is related to every possible final state, including $\bot$, so that we have nontermination for $s$.

Similarly to the treatment of $bm2bmb$, \cref{theorem:bmbot:bmh-0-1-3-2-circ-bm2bmb} establishes that the application of $bmb2bm$ to an upward-closed relation, that is $\mathbf{BMH}$-healthy, yields a relation that is $\mathbf{BMH0}$-$\mathbf{BMH3}$-healthy.
\theoremstatementref{theorem:bmbot:bmh-0-1-3-2-circ-bm2bmb}\noindent%
This result completes the proof for healthiness of both linking functions. In the following section we discuss the isomorphism.

\subsection{Isomorphism ($bm2bmb$ and $bmb2bm$)}\label{sec:bmbot:bm2bmb-and-bm2bmb}
Based on the results of the previous~\cref{sec:bmbot:bmb2bm,sec:bmbot:bm2bmb} we can establish that $bm2bmb$ and $bmb2bm$ form a bijection for healthy relations as ascertained by the following~\cref{theorem:bmbot:bm2bmb-circ-bmb2bm,theorem:bmbot:bmb2bm-circ-bm2bmb}.
\theoremstatementref{theorem:bmbot:bm2bmb-circ-bmb2bm}
\theoremstatementref{theorem:bmbot:bmb2bm-circ-bm2bmb}\noindent%
These results show that the subset of the theory that is $\mathbf{BMH0}$-$\mathbf{BMH3}$-healthy is isomorphic to the original theory of binary multirelations~\cite{Rewitzky2003}. This confirms that while our model is more expressive, it is still possible to express every program that could be specified using the original model. 

\section{Final Considerations}\label{sec:bmbot:final-considerations}
In this chapter we have introduced a new model of binary multirelations that allows the specification of sets of final states for which termination is not required. This model extends the theory of Rewiztky~\cite{Rewitzky2003} by considering a special state $\bot$ that denotes the possibility for non-termination. The healthiness conditions have been introduced as predicates and subsequently characterised as fixed points of idempotent functions. This dual characterisation is useful for reasoning about the link between this model and the theory of~\cite{Rewitzky2003}.

The operators of the theory have been introduced and their properties studied. Notable differences with respect to the original theory include the potentially non-terminating assignment and sequential composition. The definition of the latter is perhaps the most unexpected, as the intuition comes from the~\ac{UTP} theory of designs. The full justification for some of the operators and the refinement order is revisited again in~\cref{chapter:4} where we introduce the isomorphic model of angelic designs.

Finally, we have studied the relationship between this new model of binary multirelations and the theory of~\cite{Rewitzky2003}. We have found that the subset of multirelations that are, in addition, $\mathbf{BMH3}$-healthy, is isomorphic to the original theory. While this model is more expressive, we can still reason about the existing model of binary multirelations.

\chapter{Angelic Designs}\label{chapter:4}
In this chapter we introduce a new~\ac{UTP} theory of designs with both angelic and demonic nondeterminism. As already indicated the starting points for this predicative model are the theory of Cavalcanti et al.~\cite{Cavalcanti2006} and the extended model of binary multirelations presented earlier in~\cref{chapter:3}. For this reason, \cref{sec:ch4:alphabet} begins by discussing the choice of alphabet and the relationship with the alphabet of~\cite{Cavalcanti2006}. In~\cref{sec:ch4:healthiness-conditions} the healthiness conditions of the theory are presented. \cref{sec:ch4:bmbot-relationship} discusses the isomorphism with the model of extended binary multirelations. In \cref{sec:ch4:refinement} we explore the notion of refinement and prove that it corresponds exactly to that in the model of~\cref{chapter:3}. In~\cref{sec:ch4:operators} the main operators of the theory are presented, including angelic and demonic choice. In~\cref{sec:ch4:rel-designs} we explore the relationship with the original theory of designs. In~\cref{sec:ch4:pbmh-relationship} we show that the subset of $\mathbf{H3}$-healthy designs is isomorphic to the theory of~\cite{Cavalcanti2006}. Finally, \cref{sec:ch4:final} concludes the chapter with a summary of the main results.

\section{Alphabet}\label{sec:ch4:alphabet}
Our aim is to build a theory of designs. Therefore, the alphabet of our theory includes the observational variables $ok$ and $ok'$, like every theory of designs and two additional variables $s$ and $ac'$, as shown in the following definition, where the notation for a type of $State$ is enriched to carry a parameterised set of variables $S\alpha$ that specifies the names of all the record components considered. The approach followed in our discussion is that a record can be represented as a set of ordered pairs where the first component is the variable name, from a set of all possible variables, and the second component corresponds to the associated value or expression.
\begin{define}[Alphabet]\label{def:angelic-designs:alphabet}
\begin{statement}
\begin{align*}
 &s : State(S\alpha) &\\
 &ac' : \power State(S\alpha) &\\
 &ok, ok' : \{ true, false \}&\\
 &State(S\alpha) = \{ x, e| x \in S\alpha \}
\end{align*}
\end{statement}
\end{define}\noindent%
The variable $s$ encapsulates the initial values of program variables as record components of $s$, just like in the extended model of binary multirelations discussed in~\cref{chapter:3}. The set of final states $ac'$ is similar to that of~\cite{Cavalcanti2006} with the notable difference that we do not dash the variable names in the record components, instead we only consider these as undashed. This deliberate choice bears no consequences, other than simplifying reasoning and proofs. The set of program variables $S\alpha$ recorded in both $s$ and final states of $ac'$ is the same. 

The set of angelic choices $ac'$ of this new model and that of~\cite{Cavalcanti2006} can be related by dashing or undashing the variables of the components of all states in either set. This relationship is formalized by the following pair of functions.
\begin{define}
\begin{align*}
	undashset(ss) 	&\circdef \{z : State(S\alpha) | z \in ss @ undash(z)\}\\
	dashset(ss)		&\circdef \{z : State(S\alpha) | z \in ss @ dash(z)\}
\end{align*}
\end{define}\noindent%
The function $undashset$ maps a set $ss$ of states whose record components are dashed variables into a set where every state has its components undashed. This is achieved by considering every state $z$ in the set $ss$ and applying $undash$, a function which undashes the names of every record component of a state. Similarly, $dashset$ maps in the opposite direction by dashing every state in $ss$. A state $z$ whose components range over the set of variables $S\alpha$ can be dashed and undashed via the functions, $dash$ and $undash$.

The function $dash(z)$ considers every record component $z.x$ of $z$, and dashes the name of $x$ into $x'$. Similarly, the function $undash$ performs the inverse renaming, by undashing every $x'$ to $x$. 
The functions $dash$ and $undash$ are bijective. They are the exact inverse of each other. Useful properties include, for instance, that $undash(z).x = z.x'$ and $dash(z).x' = z.x$. These and other properties of $dash$ and $undash$ are included for completeness in~\cref{sec:appendix-state-substituion:dash-and-undash}.

These functions are important in the development of links between the theories, in particular with the theory of~\cite{Cavalcanti2006}, which we explore in~\cref{sec:ch4:pbmh-relationship}. In the following~\cref{sec:ch4:healthiness-conditions} we introduce the healthiness conditions.

\section{Healthiness Conditions}\label{sec:ch4:healthiness-conditions}
Since the theory we propose is a theory of designs, at the very least predicates need to satisfy $\mathbf{H1}$ and $\mathbf{H2}$. More important for our discussion is the fact that none of the proofs in~\cite{Hoare1998} regarding $\mathbf{H1}$ and $\mathbf{H2}$ require homogeneity, so it is possible to consider a non-homogeneous theory of designs. 

In addition, since we have a theory with $ok$ and $ok'$, the record of termination embedded in the use of $ac'$ must be related to that in $ok$ and $ok'$. This is the concern of the first healthiness condition $\mathbf{A0}$, which we discuss in~\cref{sec:ch4:A0}. Similarly to the theory of~\cite{Cavalcanti2006}, there is a requirement for $ac'$ to be upward-closed. This is the concern of the second healthiness condition $\mathbf{A1}$, which we discuss in~\cref{sec:ch4:A1}. Finally, the composition of both healthiness conditions, named as $\mathbf{A}$, is explored in~\cref{sec:ch4:A}.

\subsection{$\mathbf{A0}$}\label{sec:ch4:A0}
The notion of termination considered in this theory is related to that of~\cite{Cavalcanti2006}. In that model, termination is always guaranteed as long as $ac'$ is not empty. In the theory of designs termination is signalled by $ok'$. In order to reconcile these two notions we introduce the following healthiness condition $\mathbf{A0}$.
\begin{define}\label{def:A0}
\begin{statement}
$\mathbf{A0} (P) \circdef P \land ((ok \land \lnot P^f) \implies (ok' \implies ac'\neq\emptyset))$
\end{statement}
\end{define}\noindent%
It states that when a design is started and its precondition $\lnot P^f$ is satisfied, if it terminates, with $ok'$ being $true$, then it must be the case that $ac'$ is not empty. In other words, there must be at least one state in $ac'$ available for angelic choice. If the precondition $\lnot P^f$ is not satisfied, then the design aborts and there are no guarantees on the outcome, and so $ac'$ may or may not be empty.

The function $\mathbf{A0}$ is idempotent and monotonic as established by the following~\cref{law:A0:idempotent,law:A0:monotonic}. Proof of these and other results to follow can be found in~\cref{appendix:angelic-designs}.
\theoremstatementref{law:A0:idempotent}
\theoremstatementref{law:A0:monotonic}\noindent%
More importantly, the function $\mathbf{A0}$ is closed with respect to designs.
\theoremstatementref{law:A0:design}\noindent%
Therefore a design in this theory can be stated in the usual manner, with a pre and a postcondition which in this case requires $ac'$ not to be empty. In other words, once the precondition of an angelic design is satisfied, it terminates successfully with at least one final state available for angelic choice.

Finally, $\mathbf{A0}$ is closed with respect to conjunction and disjunction as stated in the following~\cref{law:A0:conjunction-closure,law:A0:disjunction-closure}.
\theoremstatementref{law:A0:conjunction-closure}
\theoremstatementref{law:A0:disjunction-closure}\noindent%
The function $\mathbf{A0}$ distributes through conjunction, and provided that the predicate is a design, that is $\mathbf{H1}$ and $\mathbf{H2}$-healthy, it also distributes through disjunction. This extra proviso is not a problem since this is a theory of designs. These properties conclude our discussion regarding $\mathbf{A0}$.

\subsection{$\mathbf{A1}$}\label{sec:ch4:A1}
In addition to requiring a consistent treatment of termination, our theory of designs also requires that both the pre and postcondition observe the upward closure of the set of final states $ac'$. In order to enforce this property in the new theory we extend the original healthiness condition $\mathbf{PBMH}$ of~\cite{Cavalcanti2006} to accommodate the additional variables $ok$ and $ok'$ as follows.
\begin{define}\label{def:PBMH}
\begin{statement}
$ \mathbf{PBMH} (P) \circdef P \circseq ac \subseteq ac' \land ok' = ok$
\end{statement}
\end{define}\noindent%
In addition to requiring that the value of $ac'$ must be upward-closed, the value of $ok'$ is left unchanged. This is the definition of $\mathbf{PBMH}$ adopted throughout our work. Its expanded version given by~\cref{lemma:PBMH:alternative-1} is more often used directly in proofs.
\theoremstatementref{lemma:PBMH:alternative-1}\noindent%
When considering a design, with precondition $P$ and postcondition $Q$, the application of $\mathbf{PBMH}$ yields a design where it is itself applied to the postcondition and the negation of the precondition, as shown in the following~\cref{lemma:PBMH(design):(lnot-PBMH(pre)|-PBMH(post))}.
\theoremstatementref{lemma:PBMH(design):(lnot-PBMH(pre)|-PBMH(post))}\noindent%
The requirement on the postcondition is exactly like in the original theory of~\cite{Cavalcanti2006}. While the requirement on the negation of the precondition follows directly from the definition of designs, where for non-$\mathbf{H3}$ designs it is actually the negation of the precondition that determines what is enforced in the case of non-termination. In~\cref{sec:ch2:utp} we show in~\cref{example:non-H3:one} such a scenario.

The application of $\mathbf{PBMH}$ to a design is precisely the motivation behind the definition of the following healthiness condition $\mathbf{A1}$.
\begin{define}\label{def:A1}
\begin{statement}
$\mathbf{A1} (P \vdash Q) \circdef (\lnot \mathbf{PBMH} (\lnot P) \vdash \mathbf{PBMH} (Q))$
\end{statement}
\end{define}\noindent
Therefore $\mathbf{A1}$ and $\mathbf{PBMH}$ are synonyms and can be used interchangeably. 

The function $\mathbf{A1}$ is idempotent and monotonic as established by the following~\cref{law:A1:idempotent,law:A1:monotonic}.
\theoremstatementref{law:A1:idempotent}
\theoremstatementref{law:A1:monotonic}\noindent%
Furthermore it is closed with respect to both conjunction and disjunction, and distributes through disjunction. In the following section we discuss the functional composition of $\mathbf{A0}$ and $\mathbf{A1}$.

\subsection{$\mathbf{A}$}\label{sec:ch4:A}
The theory of designs we propose is characterised by the functional composition of $\mathbf{A0}$, $\mathbf{A1}$, and $\mathbf{H1}$ and $\mathbf{H2}$ of the original theory of designs. The order in which these functions are composed is important since they to not always necessarily commute. In order to explain the reason behind this we consider the following counter-example.
\begin{counter-example}
\begin{flalign*}
	&\mathbf{A0} \circ \mathbf{A1} (true \vdash ac'=\emptyset)
	&&\ptext{Definition of $\mathbf{A1}$}\\
	&=\mathbf{A0} \left(\begin{array}{l}
		\lnot (false \circseq ac \subseteq ac')
		\\ \vdash \\
		ac'=\emptyset \circseq ac \subseteq ac'
	\end{array}\right)
	&&\ptext{Definition of sequential composition}\\
	&=\mathbf{A0} \left(\begin{array}{l}
		\lnot (false \land \exists ac_0 \spot ac_0 \subseteq ac')
		\\ \vdash \\
		\exists ac_0 \spot ac_0=\emptyset \land ac_0 \subseteq ac'
	\end{array}\right)
	&&\raisetag{36pt}\ptext{One-point rule and predicate calculus}\\
	&=\mathbf{A0} (true \vdash true)
	&&\ptext{Definition of $\mathbf{A0}$ (\cref{law:A0:design})}\\
	&=\mathbf{A0} (true \vdash ac'\neq\emptyset)
\end{flalign*}
\begin{flalign*}
	&\mathbf{A1} \circ \mathbf{A0} (true \vdash ac'=\emptyset)
	&&\ptext{Definition of $\mathbf{A0}$ (\cref{law:A0:design})}\\
	&=\mathbf{A1} (true \vdash ac'=\emptyset \land ac'\neq\emptyset)
	&&\ptext{Predicate calculus}\\
	&=\mathbf{A1} (true \vdash false)
	&&\ptext{Definition of $\mathbf{A1}$}\\
	&=(\lnot (false \circseq ac \subseteq ac') \vdash false \circseq ac \subseteq ac')
	&&\raisetag{18pt}\ptext{Definition of sequential composition}\\
	&=(true \vdash false)
\end{flalign*}
\end{counter-example}\noindent
In this example we apply the healthiness conditions in different orders to an unhealthy design $(true \vdash ac'=\emptyset)$ whose postcondition requires non-termination: $ac'=\emptyset$. In the first case $\mathbf{A1}$ changes the postcondition into $true$, followed by the application of $\mathbf{A0}$. While in the second case, $\mathbf{A0}$ is applied in the first place, making the postcondition $false$, a predicate that satisfies $\mathbf{PBMH}$. The resulting predicate conforms to the definition of $\topD$. Thus the functions do not always commute.

If instead we consider healthy predicates, then we can ensure that $\mathbf{A0}$ and $\mathbf{A1}$ commute. The following~\cref{law:A0:commute-A0-healthy} establishes this result for predicates that are $\mathbf{A1}$-healthy. In fact the only requirement is for the postcondition, $P^t$ to satisfy $\mathbf{PBMH}$.
\theoremstatementref{law:A0:commute-A0-healthy}\noindent%
This indicates that it is appropriate to introduce the definition of $\mathbf{A}$ as the functional composition of $\mathbf{A1}$ followed by $\mathbf{A0}$, since $\mathbf{A0}$ preserves $\mathbf{A1}$-healthiness.
\begin{define}\label{def:A}
\begin{statement}
$\mathbf{A} (P) \circdef \mathbf{A0} \circ \mathbf{A1} (P)$
\end{statement}
\end{define}\noindent
\cref{law:A0:commute-A0-healthy} establishes that once the postcondition of $P$ satisfies $\mathbf{PBMH}$ then the functions commute. Therefore by applying first $\mathbf{A1}$ first we guarantee that this is always the case.

Since the function $\mathbf{A}$ is defined by the functional composition of $\mathbf{A1}$ and $\mathbf{A0}$, and these functions are monotonic, so is $\mathbf{A}$. It is also idempotent as established by the following~\cref{law:A:idempotent}.
\theoremstatementref{law:A:idempotent}\noindent%
More importantly, it commutes with $\mathbf{H1}$ and $\mathbf{H2}$ of the theory of designs as established by the following~\cref{theorem:A-o-H1-o-H2(P):H1-o-H2-o-A(P)}.
\theoremstatementref{theorem:A-o-H1-o-H2(P):H1-o-H2-o-A(P)}\noindent%
The healthiness condition of our theory is $\mathbf{H1} \circ \mathbf{H2} \circ \mathbf{A}$. Since these commute, and they are all idempotents so is their functional composition~\cite{Hoare1998}. Furthermore, monotonicity also follows from the monotonicity of each function.

This concludes the main discussion on the healthiness conditions of the theory of angelic designs. Before exploring the relationship between this theory and the model of extended binary multirelations in~\cref{sec:ch4:bmbot-relationship}, we first discuss how to define the subset of non-angelic designs of this theory in the following~\cref{sec:ch4:A2}.
\subsection{$\mathbf{A2}$}\label{sec:ch4:A2}
In general, in our theory, a relation that does not exhibit angelic nondeterminism always provides at most one angelic choice. In other words, for every initial state, there must be at most one final state available in the distributed intersection over all possible values of $ac'$. That is, without directly considering the upward-closure of $ac'$, there must be at most one state in $ac'$. This leads to the following healthiness condition $\mathbf{A2}$.
\begin{define}\label{def:A2}
\begin{statement}
$\mathbf{A2}	 (P) \circdef \mathbf{PBMH} (P \seqA \{s\} = ac')$
\end{statement}
\end{define}\noindent
This definition is given in terms of the operator $\seqA$, which we previously discussed in~\cref{sec:ch2:utp:angelic-nondeterminism} and whose formal definition in the context of the theory of angelic designs is discussed in~\cref{sec:ch4:operators}. The intuition behind this definition is that $\mathbf{A2}$ requires the set of final states in $P$ to be either empty or a singleton, otherwise it becomes $false$. Since this purposedly breaks the upward-closure, $\mathbf{PBMH}$ must be applied as a result. If we consider the definition of $\mathbf{PBMH}$ and $\seqA$, the definition of $\mathbf{A2}$ can be expanded as established by the following~\cref{lemma:A2:alternative-2:disjunction}.
\theoremstatementref{lemma:A2:alternative-2:disjunction}\noindent%
It confirms our intuition that $ac'$ must be either empty or a singleton.

As expected of a healthiness condition, $\mathbf{A2}$ is idempotent and monotonic as confirmed by~\cref{theorem:A2-o-A2(P):A2(P),theorem:A2:monotonic}.
\theoremstatementref{theorem:A2-o-A2(P):A2(P)}
\theoremstatementref{theorem:A2:monotonic}\noindent%
The function $\mathbf{A2}$ distributes through disjunction as established by~\cref{theorem:A2(P-lor-Q):A2(P)-lor-A2(Q)}.
\theoremstatementref{theorem:A2(P-lor-Q):A2(P)-lor-A2(Q)}\noindent%
Consequently it is also closed under disjunction. However, and as expected, $\mathbf{A2}$ is not necessarily closed under conjunction. As we discuss later in~\cref{sec:ch4:angelic-choice} angelic choice is defined through conjunction, so it is no surprise that the conjunction of two $\mathbf{A2}$-healthy predicates can introduce angelic nondeterminism. Finally, when applied to a design, we obtain the following result of~\cref{lemma:A2-o-H1-o-H2}.
\theoremstatementref{lemma:A2-o-H1-o-H2}\noindent%
That is, $\mathbf{A2}$ can be directly applied to both the negation of the precondition and the postcondition of a design.

This concludes the discussion of the healthiness conditions of the theory, and its subset of non-angelic designs. As highlighted in~\cref{fig:theories}, the function $\mathbf{A2}$ plays a fundamental role in identifying the subset of theories with no angelic nondeterminism, particularly when links are established with other theories.

\section[Relationship with Extended Binary Multirelations]{Relationship with Extended\\ Binary Multirelations}\label{sec:ch4:bmbot-relationship}
As previously discussed, the model of extended binary multirelations developed in~\cref{chapter:3} is a complementary model to that of angelic designs. In this section we show how these two models can be related and prove that they are isomorphic.

In order to do so, we define a pair of linking functions, $d2bmb$ that maps from angelic designs to binary multirelations, and $bmb2d$ mapping in the opposite direction. The latter is defined in~\cref{sec:dac:bmb2d} while the former is defined in~\cref{sec:dac:d2bmb}. Finally, in~\cref{sec:dac:d2bmb-and-bmb2d} the isomorphism is established by proving that these functions form a bijection.

\subsection{From Designs to Binary Multirelations ($d2bmb$)}\label{sec:dac:d2bmb}
The first function of interest is $d2bmb$. It maps from $\mathbf{A}$-healthy designs into relations of type $BM_\bot$ and is defined as follows, where, as before, $s$ is of type $State$ and $ss$ of type $State_\bot$.
\begin{define}[d2bmb]\label{def:d2bmb}
\begin{align*}
	&d2bmb : \mathbf{A} \fun {BM}_\bot \\
	&d2bmb(P) \circdef \left\{\begin{array}{l}
		s , ss \left|\begin{array}{l}
				(\lnot P^f \implies P^t)[true/ok][ss/ac'] \land \bot \notin ss)
				\\ \lor \\
				(P^f[true/ok][(ss \setminus \{ \bot \})/ac'] \land \bot \in ss)
			\end{array}\right.
	\end{array}\right\}
\end{align*}
\end{define}\noindent
For a given design $P$, whose precondition is $\lnot P^f$, and postcondition is $P^t$, the set construction of $d2bmb(P)$ is split into two disjuncts.

The first disjunct considers the case where $P$ is guaranteed to terminate, with $ok$ and $ok'$ both substituted with $true$ in the design $P$ to obtain the implication $\lnot P^f \implies P^t$. The resulting set of final states $ss$, for which termination is required $(\bot \notin ss)$ is obtained by substituting $ss$ for $ac'$ in $P$.

In the second disjunct we consider the case where $ok$ is also $true$, but $ok'$ is false. This corresponds to the situation where $P$ does not terminate. In this case, the set of final states is obtained by substituting $ss \setminus \{\bot\}$ for $ac'$ and requiring $\bot$ to be in the set of final states $ss$.

As a consequence of $P$ satisfying $\mathbf{H2}$, we ensure that if there is some set of final states characterised by the second disjunct, and therefore, containing $\bot$, then there is also an equivalent set of final states without $\bot$ that is characterised by the first disjunct.

In the following~\cref{theorem:bmh-0-1-2:d2bmb(A)} we establish that the application of $d2bmb$ to $\mathbf{A}$-healthy designs yields relations that are $\mathbf{BMH0}$-$\mathbf{BMH2}$-healthy.
\theoremstatementref{theorem:bmh-0-1-2:d2bmb(A)}\noindent%
That is, the application of $d2bmb$ to an $\mathbf{A}$-healthy design is a fixed point of $\mathbf{bmh_{0,1,2}}$.

We consider the following~\cref{example:d2bmb:x-1-or-2} where $d2bmb$ is applied to the program that either assigns the value $1$ to the sole program variable $x$ and terminates, or assigns the value $2$ to $x$, in which the case termination is not required.
\begin{example}\label{example:d2bmb:x-1-or-2}
\begin{flalign*}
	&d2bmb((x \mapsto 2) \notin ac' \vdash (x \mapsto 1) \in ac')
	&&\ptext{Definition of $d2bmb$ (\cref{lemma:d2bmb(P|-Q)})}\\
	&=\left\{\begin{array}{l}
		s, ss \left|\begin{array}{l}
			((x \mapsto 2) \notin ac' \implies (x \mapsto 1) \in ac')[ss/ac'] \land \bot \notin ss)
			\\ \lor \\
			(((x \mapsto 2) \in ac')[ss \setminus \{ \bot \}/ac'] \land \bot \in ss)
		\end{array}\right.
	\end{array}\right\}
	&&\ptext{Predicate calculus and substitution}\\
	&=\left\{\begin{array}{l}
		s, ss \left|\begin{array}{l}
			((x \mapsto 2) \in ss \land \bot \notin ss)
			\\ \lor \\
			((x \mapsto 1) \in ss \land \bot \notin ss)
			\\ \lor \\
			((x \mapsto 2) \in (ss \setminus \{ \bot \}) \land \bot \in ss)
		\end{array}\right.
	\end{array}\right\}
	&&\ptext{Property of sets}\\
	&=\left\{\begin{array}{l}
		s, ss \left|\begin{array}{l}
			((x \mapsto 2) \in ss \land \bot \notin ss)
			\\ \lor \\
			((x \mapsto 1) \in ss \land \bot \notin ss)
			\\ \lor \\
			((x \mapsto 2) \in ss \land (x \mapsto 2) \notin \{ \bot \} \land \bot \in ss)
		\end{array}\right.
	\end{array}\right\}
	&&\ptext{Property of sets}\\
	&=\left\{\begin{array}{l}
		s, ss \left|\begin{array}{l}
			((x \mapsto 2) \in ss \land \bot \notin ss)
			\\ \lor \\
			((x \mapsto 1) \in ss \land \bot \notin ss)
			\\ \lor \\
			((x \mapsto 2) \in ss \land \bot \in ss)
		\end{array}\right.
	\end{array}\right\}
	&&\ptext{Predicate calculus}\\
	&=\{ s, ss | (x \mapsto 2) \in ss \lor ((x \mapsto 1) \in ss \land \bot \notin ss) \}
	&&\ptext{Definition of $\sqcapBMbot$ and $:=_{{BM}_{\bot}}$ and $:=_{BM}$}\\
	&=(x :=_{{BM}_{\bot}} 2) \sqcapBMbot (x :=_{BM} 1)
\end{flalign*}
\end{example}\noindent
As expected, the function $d2bmb$ yields a program with the same behaviour specified using the binary multirelational model. It is the demonic choice over two assignments, one requires termination while the other does not.
\subsection{From Binary Multirelations to Designs ($bmb2d$)}\label{sec:dac:bmb2d}
The second linking function of interest is $bmb2d$, which maps from relations of type $BM_\bot$ to $\mathbf{A}$-healthy predicates. Its definition is presented below.
\begin{define}
\begin{align*}
	&bmb2d : {BM}_\bot \fun \mathbf{A} \\
	&bmb2d(B) \circdef ((s, ac' \cup \{ \bot \}) \notin B \vdash (s, ac') \in B)
\end{align*}
\end{define}\noindent
It is defined as a design, such that for a particular initial state $s$, the precondition requires $(s, ac' \cup \{\bot\})$ not to be in $B$, while the postcondition establishes that $(s, ac')$ is in $B$. This definition can be expanded into a more intuitive representation according to the following~\cref{lemma:bmb2d:alternative-def}.
\theoremstatementref{lemma:bmb2d:alternative-def}\noindent%
The behaviour of $bmb2d$ is split into two disjuncts. The first one considers the case where $B$ requires termination, and hence $\bot$ is not part of the set of final states of the pair in $B$. While the second disjunct considers sets of final states that do not require termination, in which case $ok'$ can be either $true$ or $false$.

\cref{theorem:A(bmb2d(B)):bmb2d(B)} establishes that $bmb2d(B)$ yields $\mathbf{A}$-healthy designs provided that $B$ is $\mathbf{BMH0}$-$\mathbf{BMH2}$-healthy.
\theoremstatementref{theorem:A(bmb2d(B)):bmb2d(B)}\noindent
This result confirms that $bmb2d$ is closed with respect to $\mathbf{A}$ when applied to relations that are $\mathbf{BMH0}$-$\mathbf{BMH2}$-healthy. This concludes our discussion of $bmb2d$. In the following~\cref{sec:dac:d2bmb-and-bmb2d} we focus our attention on the isomorphism.

\subsection{Isomorphism: $d2bmb$ and $bmb2d$}\label{sec:dac:d2bmb-and-bmb2d}
In this section we show that $d2bmb$ and $bmb2d$ form a bijection. The following~\cref{theorem:d2bmb-o-bmb2d} establishes that $d2bmb$ is the inverse function of $bmb2d$ for relations that are $\mathbf{BMH0}$-$\mathbf{BMH2}$-healthy. 
\theoremstatementref{theorem:d2bmb-o-bmb2d}\noindent%
\cref{theorem:bmb2d-o-d2bmb}, on the other hand, establishes that $bmb2d$ is the inverse function of $d2bmb$ for designs that are $\mathbf{A}$-healthy.
\theoremstatementref{theorem:bmb2d-o-d2bmb}\noindent%
Together these results establish that the models are isomorphic.
This result is of fundamental importance since it allows the same programs to be characterised using two different approaches. The binary multirelational model provides a set-theoretic approach, while the predicative theory proposed can be easily linked with other~\ac{UTP} theories of interest, namely the theory of reactive processes.

Furthermore, this dual approach enables us to justify the definition of certain aspects of our theory. This includes the healthiness conditions and the definition of certain operators such as sequential composition. The most intuitive and appropriate model can be used in each case. The results obtained in either model can then be related using the linking functions.

\section{Refinement}\label{sec:ch4:refinement}
The healthiness condition $\mathbf{A}$ can be viewed as a function from the theory of designs into our theory. The theory of designs is a complete lattice~\cite{Hoare1998}. Since $\mathbf{A}$ is monotonic and idempotent, its range is also a complete lattice~\cite{Hoare1998}. Therefore we can assert that the theory we propose is also a complete lattice under the universal reverse implication order.

In the following~\cref{sec:dac:operators:extreme-points} we revisit the least and greatest elements of the of designs lattice and explore their properties within our theory. Next in~\cref{sec:dac:refinement:bmbot} we show that the refinement order of our theory corresponds exactly to subset inclusion in the extended theory of binary multirelations of~\cref{chapter:3}.

\subsection{Extreme Points}\label{sec:dac:operators:extreme-points}
Since we have a theory of designs, the extreme points of the lattice are exactly the same as those of any theory of designs. The bottom is defined by $true$ ($\botD$), whose behaviour is unpredictable and may include non-termination. While the top is the everywhere miraculous program given by $\lnot ok$ ($\topD$). (In the theory of angelic nondeterminism of~\cite{Cavalcanti2006} the top is defined by $false$ and the bottom by $true$.)

The bottom of the lattice $true$ is an angelic design as established by the following~\cref{law:A:extreme-point:true}.
\theoremstatementref{law:A:extreme-point:true}\noindent%
The consequence of $true$ being the bottom of the lattice is that $ac'$ may be empty. This is as expected, since a program for which there is no choice available to the angel corresponds to the possibility of non-termination.

The definition for the top of the lattice is a direct consequence of having the additional variables $ok$ and $ok'$. It is also an angelic design as established by the following~\cref{law:A:extreme-point:not-ok}.
\theoremstatementref{law:A:extreme-point:not-ok}\noindent%
Thus, such a program may never be started and its characterisation as a pre and postcondition pair is just like in the original theory of designs.

This concludes our introduction to the extreme points of the theory. In the following~\cref{sec:dac:refinement:bmbot} we establish the relationship between the refinement order of this theory and that of the binary multirelational model.

\subsection{Relationship with Extended Binary Multirelations}\label{sec:dac:refinement:bmbot}
The model in~\cref{chapter:3} is meant to be as similar as possible to the original model of binary multirelations. In~\cref{sec:ch3:refinement} the refinement order $\refinedbyBMbot$ is defined as subset inclusion, like in the original theory. The following~\cref{theorem:refinement-Dac-BMbot} establishes that in fact the refinement order $\refinedbyBMbot$ corresponds to the refinement order of designs $\refinedbyDac$.
\theoremstatementref{theorem:refinement-Dac-BMbot}\noindent%
It is reassuring to find that the refinement order in our theory of angelic designs corresponds to subset ordering in the binary multirelational model. This is particularly important as it confirms the intuitive definition of the theory of extended binary multirelations.

\section{Operators}\label{sec:ch4:operators}
In this section we define the main operators of the theory of angelic designs. This includes the definition of assignment in the following~\cref{sec:ch4:assignment}, sequential composition in~\cref{sec:ch4:sequential-composition}, demonic choice in~\cref{sec:ch4:angelic-choice}, and finally angelic choice in~\cref{sec:ch4:demonic-choice}. For these operators we show how they relate to their counterpart in the model of extended binary multirelations. In addition we also prove that they are all closed under $\mathbf{A}$.

\subsection{Assignment}\label{sec:ch4:assignment}
The first operator we consider is assignment. The definition, presented below, is similar to that of~\cite{Cavalcanti2006}.
\begin{define}[Assignment]
$(x \assignDac e) \circdef (true \vdash s \oplus (x \mapsto e) \in ac')$
\end{define}\noindent
It is defined by a design whose precondition is $true$, and whose postcondition establishes that every set of final states $ac'$ has a state where the component $x$ is assigned the value of the expression $e$. Every such state is the result of overriding the value of $x$ in the initial state $s$, while leaving every other program variable unchanged.

\subsection{Sequential Composition}\label{sec:ch4:sequential-composition}A challenging aspect of the theory of angelic designs is that it uses non-homogeneous relations. Consequently sequential composition cannot be simply defined as relational composition like in other~\ac{UTP} theories. The definition we propose here is layered upon the sequential composition operator $\seqA$ originally introduced in~\cite{Cavalcanti2006}.

The definition of sequential composition for angelic designs is given by considering the auxiliary variables $ok$ and $ok'$ separately, as follows.
\begin{define}[$\seqDac$-sequence]
$P \seqDac Q \circdef \exists ok_0 \spot P[ok_0/ok'] \seqA Q[ok_0/ok]$
\end{define}\noindent
This definition resembles relational composition with the notable difference that instead of conjunction we use the operator $\seqA$ that handles the non-homogeneous alphabet of the relations. In~\cref{sec:ch2:utp:angelic-nondeterminism} we previously discussed its definition as found in~\cite{Cavalcanti2006}. Since in our theory we have a different alphabet, we redefine the operator $\seqA$ in terms of the input state $s$ as follows.
\begin{define}[$\seqA$-sequence]
$P \seqA Q \circdef P[\{ s : State | Q \}/ac']$
\end{define}\noindent
This is the definition adopted throughout this thesis. Just like before, this sequential composition can be understood as follows: a final state of $P \seqA Q$ is a final state of $Q$ that can be reached from a set of input states $s$ of $Q$ that is available to $P$ as a set $ac'$ of angelic choices.

In~\cref{appendix:seqA} we explore and prove properties observed by the $\seqA$ operator. Based on those results, and the fact that $ok$ and $ok'$ are not free in neither the pre nor postcondition, it is possible to characterise the sequential composition of two angelic designs as follows.
\theoremstatementref{theorem:seqD:sequential-composition}\noindent%
The result obtained is very similar to that of sequential composition for the original theory of designs~\cite{Hoare1998,Woodcock2004}, except for the postcondition and the fact that we use the operator $\seqA$ instead of the sequential composition operator for relations~\cite{Hoare1998}. While the precondition guarantees that it is not the case that $Q$ establishes $\lnot R$, the implication in the postcondition acts as a filter that removes final states available for angelic choice in $Q$ that fail to satisfy $R$. We consider the following~\cref{example:seqDac:1}.
\begin{example}\label{example:seqDac:1}
\begin{xflalign*}
	&(true \vdash \{x\mapsto 1\} \in ac' \land \{x\mapsto 2\} \in ac') \seqDac (s.x\neq 1 \vdash s \in ac')
	&&\ptext{\cref{theorem:seqD:sequential-composition}}\\
	&=\left(
\begin{array}{l}
		\lnot (\lnot true \seqA true) \land \lnot ((\{x\mapsto 1\} \in ac' \land \{x\mapsto 2\} \in ac') \seqA s.x=1)
		\\ \vdash \\
		(\{x\mapsto 1\} \in ac' \land \{x\mapsto 2\} \in ac') \seqA (s.x\neq 1 \implies s \in ac')
	\end{array}\right)
	&&\ptext{Predicate calculus}\\
	&=\left(\begin{array}{l}
		\lnot (false \seqA true) \land \lnot ((\{x\mapsto 1\} \in ac' \land \{x\mapsto 2\} \in ac') \seqA s.x=1)
		\\ \vdash \\
		(\{x\mapsto 1\} \in ac' \land \{x\mapsto 2\} \in ac') \seqA (s.x\neq 1 \implies s \in ac')
	\end{array}\right)
	&&\ptext{Property of $\seqA$}\\
	&=\left(\begin{array}{l}
		\lnot false \land \lnot ((\{x\mapsto 1\} \in ac' \land \{x\mapsto 2\} \in ac') \seqA s.x=1)
		\\ \vdash \\
		(\{x\mapsto 1\} \in ac' \land \{x\mapsto 2\} \in ac') \seqA (s.x\neq 1 \implies s \in ac')
	\end{array}\right)
	&&\ptext{Predicate calculus}\\
	&=\left(\begin{array}{l}
		\lnot ((\{x\mapsto 1\} \in ac' \land \{x\mapsto 2\} \in ac') \seqA s.x=1)
		\\ \vdash \\
		(\{x\mapsto 1\} \in ac' \land \{x\mapsto 2\} \in ac') \seqA (s.x\neq 1 \implies s \in ac')
	\end{array}\right)
	&&\ptext{Definition of $\seqA$ and substitution}\\
	&=\left(\begin{array}{l}
		\lnot (\{x\mapsto 1\} \in \{ s | s.x=1\} \land \{x\mapsto 2\} \in \{ s | s.x=1\})
		\\ \vdash \\
		(\{x\mapsto 1\} \in \{ s | s.x\neq 1 \implies s \in ac'\} \land \{x\mapsto 2\} \in \{ s | s.x\neq 1 \implies s \in ac'\})
	\end{array}\right)
	&&\ptext{Property of sets}\\
	&=\left(\begin{array}{l}
		\lnot (\{x\mapsto 1\}.x=1 \land \{x\mapsto 2\}.x=1)
		\\ \vdash \\
		(\{x\mapsto 1\}.x\neq 1 \implies \{x\mapsto 1\} \in ac') \land (\{x\mapsto 2\}.x\neq 1 \implies \{x\mapsto 2\} \in ac')
	\end{array}\right)
	&&\ptext{Value of component $x$}\\
	&=\left(\begin{array}{l}
		\lnot (1=1 \land 2=1)
		\\ \vdash \\
		(1\neq 1 \implies \{x\mapsto 1\} \in ac') \land (2\neq 1 \implies \{x\mapsto 2\} \in ac')
	\end{array}\right)
	&&\ptext{Predicate calculus}\\
	&=\left(\begin{array}{l}
		true
		\\ \vdash \\
		(false \implies \{x\mapsto 1\} \in ac') \land (true \implies \{x\mapsto 2\} \in ac')
	\end{array}\right)
	&&\ptext{Predicate calculus}\\
	&=(true \vdash \{x\mapsto 2\} \in ac')
\end{xflalign*}
\end{example}\noindent
In this case, there is an angelic choice between the assignment of the value $1$ and $2$ to the program variable $x$, sequentially composed with the program that aborts if $x$ is $1$ and that otherwise behaves as $Skip$. The resulting design is just the assignment of $2$ to $x$ that avoids aborting. In this case, the implication in the postcondition of~\cref{theorem:seqD:sequential-composition} is discarding the angelic choice where $x$ is $1$.

If we consider designs that observe $\mathbf{H3}$, we can simplify the result further as there are no dashed variables in the precondition as established by~\cref{theorem:seqD:sequential-composition:H3}.
\theoremstatementref{theorem:seqD:sequential-composition:H3}\noindent%
This is similar to the definition of sequential composition for designs where the precondition is a condition~\cite{Woodcock2004}, except for the use of the operator $\seqA$. 

\subsubsection{Closure}
It is important that we establish closure of sequential composition ($\seqDac$) with respect to~$\mathbf{A}$. The proof of the following closure theorem relies on results established in~\cref{appendix:pbmh,appendix:seqA}.
\theoremstatementref{law:seqDac:closure}\noindent
This result establishes that $\seqDac$ is closed with respect to $\mathbf{A}$ provided both operands are also $\mathbf{A}$-healthy.


\subsubsection{Sequential Composition in Extended Binary Multirelations}
The following~\cref{theorem:bmb2d:seqBMbot} establishes that for designs that are $\mathbf{A}$-healthy, the definition of sequential composition corresponds to that in the isomorphic model of extended binary multirelations. 
\theoremstatementref{theorem:bmb2d:seqBMbot}\noindent
Together with the closure of $\seqDac$, this result enables us to ascertain the closure of $\seqBMbot$.

In what follows, we concentrate our attention on important properties observed by the sequential composition operator.

\subsubsection{Skip}
Similarly to the original theory of designs, we identify the $\mathbf{Skip}$ of the theory. We denote it by $\IIDac$ and define it as follows.
\begin{define}[Skip]
$\IIDac \circdef (true \vdash s \in ac')$
\end{define}\noindent
This is a design whose precondition is $true$, thus it is always applicable, and upon terminating it establishes that the input state $s$ is in all sets of angelic choices $ac'$. The only results that can be guaranteed by the angel are those that are available in all demonic choices of the value of $ac'$ that can be made. In this case, $s$ is the only guarantee that we have, so the behaviour of $\IIDac$ is to maintain the current state. The following~\cref{theorem:A(IIDac):IIDac,theorem:IIDac-seqDac-P:P} establish that $\IIDac$ is $\mathbf{A}$-healthy and that it is the left-unit for sequential composition ($\seqDac$).
\theoremstatementref{theorem:A(IIDac):IIDac}
\theoremstatementref{theorem:IIDac-seqDac-P:P}\noindent
These results confirm that $\IIDac$ is indeed a suitable definition for the identity. We observe that $\IIDac$ is only a right-identity for angelic designs that are $\mathbf{H3}$-healthy. This is the motivation for the following discussion. 

In what follows we establish that an $\mathbf{H3}$-design in our theory requires the precondition not to mention dashed variables, as expected~\cite{Hoare1998}. We first show the result of sequentially composing an $\mathbf{A}$-healthy design $P$ with $\IIDac$ in~\cref{law:seqD-sequence-Skip}.
\theoremstatementref{law:seqD-sequence-Skip}\noindent%
Finally~\cref{law:seqDac:H3} establishes that $P \seqDac \IIDac = P$ restricts the precondition to a condition.
\theoremstatementref{law:seqDac:H3}\noindent%

\subsubsection{Sequential Composition and the Extreme Points}
We now explore the consequences of sequentially composing a program with the extreme points of the lattice. As expected, we establish the same left-zero laws that hold in the original theory of designs~\cite{Hoare1998}. 

The following~\cref{law:seqD:bot-P} shows that it is impossible to recover from an aborting program. \cref{law:seqD:top-P} establishes that if a design is miraculous then sequentially composing it with another design does not change its behaviour.
\theoremstatementref{law:seqD:bot-P}
\theoremstatementref{law:seqD:top-P}\noindent
Both of these results are expected of a theory of designs~\cite{Hoare1998}. 

This concludes our discussion of sequential composition. In the following~\cref{sec:ch4:angelic-choice,sec:ch4:demonic-choice} we concentrate our attention on nondeterminism.

\subsection{Demonic Choice}\label{sec:ch4:demonic-choice}
The intuition for the demonic choice in our theory is related to the possible ways of choosing a value for $ac'$. In general, this can be described using disjunction like in the original theory of designs~\cite{Hoare1998}.
\begin{define}
$P \sqcapDac Q \circdef P \lor Q$
\end{define}\noindent
This corresponds to the greatest lower bound of the lattice. We consider the following example, where $\oplus$ is the overriding operator~\cite{Spivey1989}.
\begin{example}
\begin{flalign*}
	&(x := 1) \sqcapDac (x := 2)
	&&\ptext{Definition of assignment}\\
	&=(true \vdash s \oplus (x \mapsto 1) \in ac') \sqcapDac (true \vdash s \oplus (x \mapsto 2) \in ac')
	&&\ptext{Definition of $\sqcapDac$ and disjunction of designs}\\
	&=(true \vdash s \oplus (x \mapsto 1) \in ac' \lor s \oplus (x \mapsto 2) \in ac')
\end{flalign*}
\end{example}\noindent
In this example we have at least two choices for the final value of $ac'$: one has a state where $x$ is $1$ and the other has a state where $x$ is $2$. The demon can choose any set $ac'$ satisfying either predicate. In this case, the angel is not guaranteed to be able to choose a particular final value for $x$, since there are no choices in the intersection of all possible choices of $ac'$.

\subsubsection{Closure Properties}
The demonic choice operator is closed with respect to $\mathbf{A}$, provided that both operands are also $\mathbf{A}$-healthy. This result follows from the distributive property of $\mathbf{A}$ with respect to disjunction, as established by the following~\cref{law:A:distribute-disjunction}.
\theoremstatementref{law:A:distribute-disjunction}\noindent
\theoremstatementref{theorem:A:disjunction-closure}\noindent

\subsubsection{Relationship with Extended Binary Multirelations}

The demonic choice operator ($\sqcapDac$) corresponds exactly to the demonic choice operator ($\sqcapBMbot$) of the binary multirelational model. This result is established by the following~\cref{theorem:bmb2p:demonic-choice}.

\theoremstatementref{theorem:bmb2p:demonic-choice}\noindent%
This result confirms the correspondence of demonic choice in both models. In what follows we focus our attention on its properties.

\subsubsection{Properties}

In general, and since demonic choice is the greatest lower bound, if presented with the possibility to abort ($\botD$), we expect the demon to choose the worst possible outcome as shown by the following~\cref{theorem:P-sqcapDac-botDac:botDac}.
\theoremstatementref{theorem:P-sqcapDac-botDac:botDac}\noindent%
As observed in the original theory of designs~\cite{Hoare1998}, the sequential composition operator distributes through demonic choice, but only from the right as established by~\cref{law:seqDac:sqcap-right-distributivity}.
\theoremstatementref{law:seqDac:sqcap-right-distributivity}\noindent
These results conclude our discussion regarding the demonic choice operator and its properties. In the following section we focus our attention on the angelic choice operator and its respective properties.

\subsection{Angelic Choice}\label{sec:ch4:angelic-choice}
Similarly to other models, angelic choice is defined as the least upper bound, which in this case is conjunction.
\begin{define}
$P \sqcupDac Q \circdef P \land Q$
\end{define}\noindent
This definition is justified by the correspondence with the angelic choice operator of the binary multirelational model of~\cref{chapter:3}.

To provide the intuition for this definition we consider the following~\cref{example:sqcupDac}.
\begin{example}\label{example:sqcupDac}
\begin{flalign*}
	&((x \mapsto 1) \notin ac' \vdash (x \mapsto 1) \in ac') \sqcupDac (true \vdash (x \mapsto 2) \in ac')
	&&\ptext{Definition of $\sqcupDac$}\\
	&=\left(\begin{array}{l}
		(x \mapsto 1) \notin ac' \lor true
		\\ \vdash \\
		\left(\begin{array}{l} 
			(x \mapsto 1) \notin ac' \implies (x \mapsto 1) \in ac'
			\\ \land \\
			true \implies (x \mapsto 2) \in ac'
		\end{array}\right)
	\end{array}\right)
	&&\ptext{Predicate calculus}\\
	&=(true \vdash (x \mapsto 1) \in ac' \land (x \mapsto 2) \in ac')
\end{flalign*}
\end{example}\noindent
It considers the angelic choice between a design that assigns $1$ to the only program variable $x$, but does not necessarily terminate, and a design that assigns $2$ to $x$, but terminates. The result is a program that always terminates and, for every set of final states, there is the possibility to choose angelically the assignment of the value $1$ or $2$ to $x$.

\subsubsection{Closure Properties}
Having defined angelic choice as the least upper bound operator, in the following \cref{law:sqcapDac:closure} we prove that it is closed under $\mathbf{A}$, provided that both operands are $\mathbf{A}$-healthy.
\theoremstatementref{law:sqcapDac:closure}\noindent
The proof for this theorem relies on the closure of $\mathbf{PBMH}$ for conjunction.

\subsubsection{Relationship with Extended Binary Multirelations}
\cref{law:bmb2p:sqcup} establishes that the angelic choice operator of the designs and the binary multirelations models are in correspondence. This requires the operands to be $\mathbf{BMH1}$-healthy. This is satisfied by every binary multirelation that is $\mathbf{BMH0}$-$\mathbf{BMH2}$.
\theoremstatementref{law:bmb2p:sqcup}\noindent%
Having established the correspondence of the angelic choice operator in both models, in the following section we focus on its properties.

\subsubsection{Properties}
In general, and since angelic choice is the least upper bound, the angelic choice of a design $P$ and the top of the lattice ($\topD$) is also $\topD$.%
\theoremstatementref{theorem:P-sqcupDac-topD:topD}\noindent
In this model, sequential composition does not necessarily distribute from the right nor from the left. In order to explain the intuition behind this we present the following~\cref{counter-example:angelic-distribute-left} for distribution from the left. %
\begin{counter-example}\label{counter-example:angelic-distribute-left}%
Assuming $\seqDac$ distributes over $\sqcapDac$ from the left,
\begin{flalign*}
	&\left(
\begin{array}{l}
		(true \vdash s \oplus (x \mapsto 1) \in ac') 
		\\ \sqcapDac \\
		(true \vdash s \oplus (x \mapsto -1) \in ac')
	\end{array}\right) \seqDac \left(\begin{array}{l}
		( s.x = 1 \vdash false ) 
		\\ \sqcup \\
		( s.x = -1 \vdash false)
	\end{array}%
\right)
	&&\ptext{Property of $\seqA$ and propositional calculus}\\
	&=\left(%
\right)
	&&\ptext{Property of sets and predicate calculus}\\
	&=\left(%
\right)
	&&\ptext{Propositional calculus}&\\
	&=\left( false \vdash true \right) \sqcupDac \left( false \vdash true \right)
	&&\ptext{Property of $\sqcupDac$}\\
	&=(false \vdash true)
	&&\ptext{Definition of design and propositional calculus}\\
	&=true
	&&\ptext{Definitionf of $\botD$}\\
	&=\botD
\end{flalign*}
\end{counter-example}\noindent
This is a sequential composition. In the first program the precondition always holds and the program presents a choice to the demon. In this case, the demon can choose the set of final states, $ac'$, by guaranteeing that either $x$ is set to $1$ or $-1$ in the final set of states $ac'$. The second program presents an angelic choice, but the precondition makes a restriction on the value of $x$ in the initial state $s$: in either case, if the precondition is satisfied the program is $\topD$, otherwise if no precondition can be satisfied, the program behaves as $\botD$.

It is expected that the angel will avoid $\botD$ if possible. In this case, it is expected, since the angel can avoid aborting irrespective of the choice the demon makes before the angel. However, if we assume that the sequential composition operator $\seqDac$ left-distributes over angelic choice we get a different result as shown above.


In addition, sequential composition does not distribute from the right. We illustrate this in~\cref{counter-example:angelic-distribute-right}. It is the sequential composition of two designs. The first design is the angelic choice between the program that assigns $2$ to $x$, but may not terminate, and the program that always terminates but whose final set of states $ac'$ is unrestricted, except that it cannot be the empty set. The second design is miraculous for $s.x = 2$ and for every other value of $s.x$ it aborts.
\begin{counter-example}\label{counter-example:angelic-distribute-right}
\begin{flalign*}
	&\left(
\begin{array}{l}
		((x \mapsto 2) \notin ac' \vdash (x \mapsto 2) \in ac')
		\\ \sqcupDac \\
		(true \vdash ac'\neq\emptyset)
	\end{array}\right)
	\seqDac 
	\left(\begin{array}{l}
		s.x=2 
		\\ \vdash \\
		s.x\neq2 \land ac'\neq\emptyset
	\end{array}%
\right)
	&&\ptext{Predicate calculus}\\
	&=(true \vdash (x \mapsto 2) \in ac' \land ac'\neq\emptyset)
	  \seqDac
	  (s.x=2 \vdash s.x\neq2 \land ac'\neq\emptyset)
	&&\ptext{Property of sets and predicate calculus}\\
	&=(true \vdash (x \mapsto 2) \in ac')
	  \seqDac
	  (s.x=2 \vdash s.x\neq2 \land ac'\neq\emptyset)
	&&\ptext{\cref{theorem:seqD:sequential-composition}}\\
	&=\left(\begin{array}{l}
		\lnot (false \seqA true) \land \lnot ((x \mapsto 2) \in ac' \seqA s.x\neq2)
		\\ \vdash \\
		   (x \mapsto 2) \in ac' \seqA (s.x=2 \implies (s.x\neq2 \land ac'\neq\emptyset))
	\end{array}\right)
	&&\ptext{Predicate calculus}\\
	&=\left(\begin{array}{l}
		\lnot (false \seqA true) \land \lnot ((x \mapsto 2) \in ac' \seqA s.x\neq2)
		\\ \vdash \\
		   (x \mapsto 2) \in ac' \seqA s.x\neq2)
	\end{array}\right)
	&&\ptext{Definition of $\seqA$ and substitution}\\
	&=\left(\begin{array}{l}
		\lnot false \land \lnot ((x \mapsto 2) \in \{ z | z.x\neq2 \})
		\\ \vdash \\
		(x \mapsto 2) \in \{ z | z.x\neq2 \}
	\end{array}\right)
	&&\ptext{Property of sets}\\
	&=\left(\begin{array}{l}
		\lnot false \land \lnot ((x \mapsto 2).x\neq2)
		\\ \vdash \\
		(x \mapsto 2).x\neq2
	\end{array}\right)
	&&\ptext{Predicate calculus}\\
	&=(\lnot (2\neq2) \vdash 2\neq2)
	&&\ptext{Predicate calculus}\\
	&=(true \vdash false)
	&&\ptext{Predicate calculus and definition of $\topD$}\\
	&=\topD\\
	&\neq\\
	&\left(\begin{array}{l}
		((x \mapsto 2) \notin ac' \vdash (x \mapsto 2) \in ac') \seqDac (s.x=2 \vdash s.x\neq2 \land ac'\neq\emptyset)
		\\ \sqcupDac \\
		(true \vdash ac'\neq\emptyset) \seqDac (s.x=2 \vdash s.x\neq2 \land ac'\neq\emptyset)
	\end{array}%
\right)
	&&\ptext{Predicate calculus and property of sets}\\
	&=\left(%
\right)
	\end{array}\right)
	&&\ptext{Predicate calculus}\\
	&=(false \vdash false) \sqcupDac (false \vdash true)
	&&\ptext{Predicate calculus and definition of $\botD$}\\
	&=\botD \sqcupDac \botD
	&&\ptext{Definition of $\sqcupDac$, $\botD$ and predicate calculus}\\
	&=\botD
\end{flalign*}
\end{counter-example}\noindent
When the angelic choice is resolved first the result is the program that always terminates and whose set of final states $ac'$ has a state where $x$ is assigned the value $2$. Sequentially composing this with the second design results in a miracle ($\topD$) as the only state available for angelic choice is where $x$ has the value $2$. And this is precisely the case in which the design behaves miraculously.

If we distribute the sequential composition through the angelic choice, in the resulting angelic choice there are two sequential compositions. In the first one, the result is $\botD$ as the first design may not terminate. In the second, termination is guaranteed but any final set of states ($ac'\neq\emptyset$) may fail to satisfy the precondition $s.x=2$, in which case the design aborts. In conclusion, angelic choice does not distribute through sequential composition at all.

\section{Relationship with Designs}\label{sec:ch4:rel-designs}
In this section we study the relationship between the model of $\mathbf{A}$-designs and the original theory of homogeneous designs of Hoare and He~\cite{Hoare1998}. As we depict in \cref{fig:theories:angelic-designs,fig:theories}, this is achieved by defining a pair of linking functions: $d2ac$, which maps from designs into angelic designs, and $ac2p$, which maps in the opposite direction.

In the following~\cref{sec:ch4:designs-rel:d2ac} we introduce the definition of $d2ac$. In~\cref{sec:ch4:designs-rel:ac2p} we define $ac2p$ and discuss how the angelic nondeterminism of a theory can be removed. Finally in~\cref{sec:ch4:designs-rel:ac2p-designs} we establish that there is a Galois connection between the theory of $\mathbf{A}$-designs and the original theory of designs, and that there is an isomorphism when we consider the subset of $\mathbf{A2}$-healthy angelic designs.

\subsection{From Designs to Angelic Designs ($d2ac$ and $p2ac$)}\label{sec:ch4:designs-rel:d2ac}
\begin{figure}[t]
\begin{center}
\includegraphics[scale=1.6]{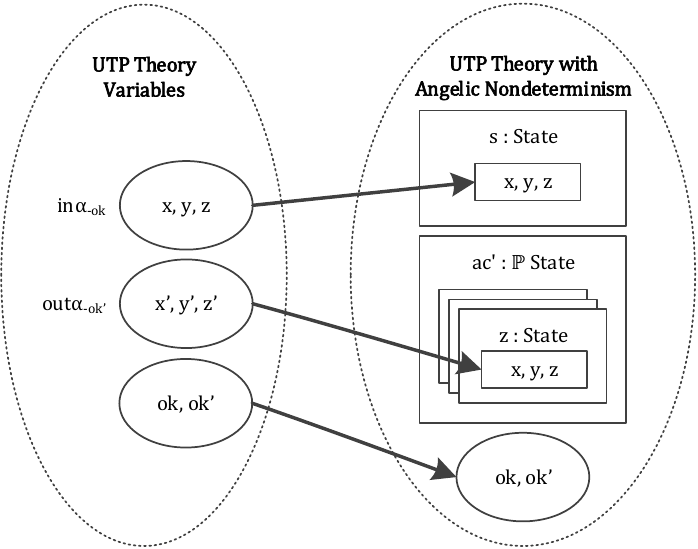}
\caption{\label{fig:p2ac-States}Encoding variables in a theory of angelic designs using $p2ac$}
\end{center}
\end{figure}

The main concern when mapping a design into an angelic design pertains to encoding both the pre and postcondition in terms of a single initial state $s$ and a set of final states $ac'$. Since the model of $\mathbf{A}$-designs is also a theory of designs, $ok$ and $ok'$ retain the same meaning. The function $d2ac$ is defined as follows.
\begin{define}
$d2ac (P) \circdef (\lnot p2ac(P^f) \land (\lnot P^f[\mathbf{s}/in\alpha_{-ok}] \circseq true) \vdash p2ac(P^t))$
\end{define}\noindent
The negation of the precondition $P^f$ and the postcondition are mapped using the auxiliary function $p2ac$, while the second conjunct in the precondition of the angelic design requires that whenever $\lnot P^f$ holds, then there is some final observation of the values of the variables in $out\alpha$. The predicate $\lnot P^f[\mathbf{s}/in\alpha_{-ok}] \circseq true$ can be restated as $\exists out\alpha @ \lnot P^f[\mathbf{s}/in\alpha_{-ok}]$. Essentially this allows the value of $ac'$ to be unspecified when the precondition $\lnot P^f$ is not satisfied. This is defined using the substitution operator $[\mathbf{s}/S\alpha]$, where the boldface indicates that $s$ is a record, and so the substitution is not simply $s$ for $S\alpha$. Instead, for an arbitrary set of variables $S\alpha$, the substitution operator needed is defined as follows. 
\begin{define}\label{def:substitution-state}
\begin{statement}
$P[\mathbf{z}/S\alpha] \circdef P[z.s_0,\ldots,z.s_n/s_0,\ldots,s_n]$
\end{statement}
\end{define}\noindent%
Each variable $s_i$ in $S\alpha$ is replaced with $z.s_i$. As an example, we consider the substitution $(x'=2 \land ok')[\mathbf{s},\mathbf{z}/in\alpha_{-ok},out\alpha_{-ok'}]$, whose result is $z.x'=2 \land ok'$. The substitution $[\mathbf{z}/S\alpha]$ is well-formed whenever $S\alpha$ is a subset of the record components of $z$. In~\cref{appendix:state-substitution} we establish properties satisfied by this operator.

The main purpose of $p2ac$ is to encode predicates in terms of $s$ and $ac'$. For a given predicate $P$ whose input and output alphabets are $in\alpha$ and $out\alpha$, respectively, its encoding in a theory with angelic nondeterminism is given by the following function $p2ac$, which we illustrate in~\cref{fig:p2ac-States}.
\begin{define}\label{def:p2ac}
$p2ac (P) \circdef \exists z \spot P[\mathbf{s},\mathbf{z}/in\alpha_{-ok},out\alpha_{-ok'}] \land undash(z) \in ac'$
\end{define}\noindent%
First, each variable in the set of input and output variables, other than $ok$ and $ok'$, is replaced with the corresponding component of the initial state $s$ and a final state $z$ from the set of final states available for angelic choice. Since in our encoding states have undashed components, we require $undash(z)$ to be in $ac'$.

The result of $p2ac$ is upward-closed, that is, the predicates in the range of $p2ac$ are fixed points of $\mathbf{PBMH}$ as established by the following~\cref{lemma:PBMH-o-p2ac(P):p2ac(P)}.
\theoremstatementref{lemma:PBMH-o-p2ac(P):p2ac(P)}\noindent%
As previously discussed, this property is essential for a theory of angelic nondeterminism. The function $p2ac$ distributes through disjunction as established by the following~\cref{theorem:p2ac(P-lor-Q):p2ac(P)-lor-p2ac(Q)}
\theoremstatementref{theorem:p2ac(P-lor-Q):p2ac(P)-lor-p2ac(Q)}\noindent%
In the case of conjunction there is an implication as established by~\cref{theorem:p2ac(P-land-Q):implies:p2ac(P)-land-p2ac(Q)}, rather than an equality, as $p2ac$ is defined using an existential quantifier.
\theoremstatementref{theorem:p2ac(P-land-Q):implies:p2ac(P)-land-p2ac(Q)}\noindent%
More importantly, the result of $p2ac$ is $\mathbf{A2}$-healthy as established by~\cref{theorem:A2-o-p2ac(P):p2ac(P)}.
\theoremstatementref{theorem:A2-o-p2ac(P):p2ac(P)}\noindent%
This is expected since the original predicates mapped by $p2ac$ do not have angelic nondeterminism.

A consequence of the definition of $p2ac$ is that it requires $ac'$ not to be empty, unless $P$ is itself $false$. In the following~\cref{theorem:p2ac(design)-land-ac'-neq-emptyset}, we consider the application of $p2ac$ to a design $P$ when $ac'$ is not empty.
\theoremstatementref{theorem:p2ac(design)-land-ac'-neq-emptyset}\noindent
In this case $p2ac$ can be applied directly to the negation of the precondition $P^f$ and the postcondition $P^t$ of a design $P$. This result sheds light on the relationship between $p2ac$ and $d2ac$ as established by~\cref{theorem:p2ac(P)-land-ac'-neq-emptyset:d2ac(P)-land-ac'-neq-emptyset}.
\theoremstatementref{theorem:p2ac(P)-land-ac'-neq-emptyset:d2ac(P)-land-ac'-neq-emptyset}\noindent%
When we consider the case of a design whose set of final states $ac'$ is not empty, then $d2ac$ is simply $p2ac$.

Finally, we establish that $d2ac$ yields an $\mathbf{A}$-healthy design, that is, the designs in the range of $d2ac$ are fixed points of the healthiness condition $\mathbf{A}$.
\theoremstatementref{theorem:A-o-d2ac(P):d2ac(P)}\noindent%
This concludes our discussion regarding the definition of $d2ac$ and its most important properties.

\subsection{Removing Angelic Nondeterminism ($ac2p$)}\label{sec:ch4:designs-rel:ac2p}
The mapping from angelic to non-angelic predicates is defined by $ac2p$, whose goal is to collapse the set of final states $ac'$ into a single state, and, introduce the input and output variables as used in other theories. Its definition is presented below.
\begin{define}\label{def:ac2p}
\begin{align*}
	&ac2p(P) \circdef \mathbf{PBMH} (P)[State_{\II}(in\alpha_{-ok})/s] \seqA \bigwedge x : out\alpha_{-ok'} \spot dash(s).x = x
\end{align*}
\end{define}\noindent
First, for a predicate $P$, $ac2p$ takes the result of applying $\mathbf{PBMH}$ to $P$ to achieve upward closure of $ac'$. This is followed by the replacement of $s$ to introduce the corresponding input variables of the set $in\alpha_{-ok}$, which excludes $ok$. As already mentioned, the observational variables $ok$ and $ok'$ retain the same meaning in the theories considered. Finally, the resulting predicate is sequentially composed, using $\seqA$, with a predicate that introduces the corresponding output variables of the resulting final state, except for $ok'$. For a set of variables $S\alpha$, $State_{\II} (S\alpha)$ is an identity record, whose components $s_i$ are mapped to the respective variables $s_i$. 
\begin{define}
$ State_{\II} (S\alpha) \circdef \{ s_0 \mapsto s_0, \ldots, s_n \mapsto s_n \}$
\end{define}\noindent
As an example, we consider the substitution $(s.x = 1 \land ok)[State_{\II_{-ok}} (in\alpha)/s]$, whose result is $x = 1 \land ok$. If we consider the definition of $\mathbf{PBMH}$ and $\seqA$, then $ac2p$ can be rewritten as established by the following~\cref{lemma:ac2p:alternative-3}.
\theoremstatementref{lemma:ac2p:alternative-3}\noindent%
That is, the variable $ac'$ is quantified away, and for each state $z$ in the set $ac'$, the output variables in $out\alpha$, except for $ok'$, are introduced and set to the respective values of the components of $z$. Since in our encoding the components of a state are always undashed, we apply the function $dash(z)$ to $z$. If there is more than one state in $ac'$, then $ac2p$ yields $false$ as no $x$ variable can take on more than one value.

\subsection{Isomorphism and Galois Connection}\label{sec:ch4:designs-rel:ac2p-designs}
Having defined a pair of linking functions between the theory of angelic designs and designs, in this section we show that, in general, there is a Galois connection between the two theories. In addition, when we consider the subset of $\mathbf{A2}$-healthy designs these two theories can be shown to be isomorphic.

\subsubsection{From Designs}
The mapping of a design $P$ through $d2ac$ and then $ac2p$ yields the same design $P$ as established by the following~\cref{theorem:ac2p-o-d2acp(P):P}.
\theoremstatementref{theorem:ac2p-o-d2acp(P):P}\noindent%
That is, in the theory of angelic designs we can model the original designs of Hoare and He~\cite{Hoare1998} without angelic nondeterminism. This is a reassuring result which confirms the suitability of our model.

\subsubsection{From Angelic Designs}
When the linking functions are applied in the reverse order, however, we do not obtain the same design $P$. This result is established by~\cref{theorem:d2ac-o-ac2p:implies:P}.
\theoremstatementref{theorem:d2ac-o-ac2p:implies:P}\noindent%
In general, the result of the application of $ac2p$ followed by $d2ac$ to an $\mathbf{A}$-healthy design $P$ is stronger than $P$. This is because the angelic nondeterminism is removed. For instance, the mapping of an angelic choice over two assignments $x:=1$ and $x:=2$ yields the top of the lattice $\topD$.
\begin{example}
\begin{xflalign*}
	&d2ac\circ ac2p(x := 1 \sqcup x := 2)
	&&\ptext{Definition of assignment and $\sqcup$}\\
	&=d2ac\circ ac2p(true \vdash s \oplus \{x \mapsto 1\} \in ac' \land s \oplus \{x\mapsto 2\} \in ac')
	&&\ptext{\cref{lemma:d2ac-o-ac2p(P)}}\\
	&=\left(\begin{array}{l}
		\lnot p2ac(ac2p(false)) \land (\exists out\alpha \spot \lnot ac2p(false)[\mathbf{s}/in\alpha]) 
		\\ \vdash \\
		p2ac(ac2p(s \oplus \{x \mapsto 1\} \in ac' \land s \oplus \{x\mapsto 2\} \in ac'))
	\end{array}\right)
	&&\ptext{\cref{lemma:ac2p(P)-s-ac'-not-free:P}}\\
	&=\left(\begin{array}{l}
		\lnot p2ac(false) \land (\exists out\alpha \spot \lnot false[\mathbf{s}/in\alpha]) 
		\\ \vdash \\
		p2ac(ac2p(s \oplus \{x \mapsto 1\} \in ac' \land s \oplus \{x\mapsto 2\} \in ac'))
	\end{array}\right)
	&&\ptext{Predicate calculus and~\cref{lemma:p2ac(false)}}\\
	&=\left(\begin{array}{l}
		true
		\\ \vdash \\
		p2ac(ac2p(s \oplus \{x \mapsto 1\} \in ac' \land s \oplus \{x\mapsto 2\} \in ac'))
	\end{array}\right)
	&&\ptext{\cref{lemma:p2ac-o-ac2p(P)}}\\
	&=\left(\begin{array}{l}
		true
		\\ \vdash \\
		\exists ac_0, y @ \left(\begin{array}{l}
			s \oplus \{x \mapsto 1\} \in ac' 
			\\ \land \\
			s \oplus \{x\mapsto 2\} \in ac'
		\end{array}\right)[ac_0/ac'] \land ac_0 \subseteq \{y\} \land y \in ac'
	\end{array}\right)
	&&\ptext{Substitution and property of sets}\\
	&=(true \vdash false)
	&&\ptext{Definition of $\topD$}\\
	&=\topD
\end{xflalign*}
\end{example}\noindent%
The results of~\cref{theorem:d2ac-o-ac2p:implies:P,theorem:ac2p-o-d2acp(P):P} establish that we have a Galois connection between the two theories.
\subsubsection{From $\mathbf{A2}$-healthy Angelic Designs}
If we consider the subset of $\mathbf{A}$-healthy designs that is in addition $\mathbf{A2}$-healthy, then we can prove the reverse implication of~\cref{theorem:d2ac-o-ac2p:implies:P} as established by the following~\cref{theorem:d2ac-o-ac2p:L-implies:P}.
\theoremstatementref{theorem:d2ac-o-ac2p:L-implies:P}\noindent%
Together these results allow us to prove that there is a bijection for the subset of $\mathbf{A2}$-healthy designs.
\theoremref{theorem:d2ac-o-ac2p(P):P}\noindent%
This result confirms that these models are isomorphic as depicted in~\cref{fig:theories}. 

This concludes our discussion on the relationship between the original theory of designs and the model of angelic designs. In the following~\cref{sec:ch4:pbmh-relationship} we focus our attention on the relationship with the $\mathbf{PBMH}$ theory~\cite{Cavalcanti2006}.

\section{Relationship with the $\mathbf{PBMH}$ Theory}\label{sec:ch4:pbmh-relationship}
The final link that we study in this chapter pertains to the relationship between the model of $\mathbf{A}$-designs and the theory of angelic nondeterminism of Cavalcanti et al.~\cite{Cavalcanti2006}. As previously discussed in~\cref{sec:ch2:utp:angelic-nondeterminism}, in that theory the alphabet consists of the input program variables, and a single output variable $ac'$, which is a record whose components range over the dashed output program variables. In addition, termination is captured without considering $ok$ and $ok'$. 

When establishing a link between the theories of interest, the first concern is their alphabets. As we discussed in~\cref{sec:ch4:alphabet}, the $ac'$ of both theories can be related through the functions $undashset$ and $dashset$, which $undash$ and $dash$ the components of every state in a set, respectively. 

In order to relate both theories, we introduce a pair of linking functions, $d2pbmh$, which maps $\mathbf{A}$-healthy designs to $\mathbf{PBMH}$ predicates, and $pbmh2d$, which maps predicates in the opposite direction. We introduce their definitions in the following~\cref{sec:ch4:d2pbmh,sec:ch4:pbmh2d}. Finally in~\cref{sec:ch4:pbmh-rel:galois} we show that there is a Galois connection between the theories, and that in general, the subset of angelic designs that is $\mathbf{H3}$-healthy is isomorphic to the theory of~\cite{Cavalcanti2006}.

\subsection{From Angelic Designs to $\mathbf{PBMH}$ ($d2pbmh$)}\label{sec:ch4:d2pbmh}
In order to map angelic designs into the theory of $\mathbf{PBMH}$, it is necessary to hide the variables $ok$ and $ok'$, introduce the input variables in $in\alpha$, and appropriately $dash$ the set of final states $ac'$. This is captured by the function $d2pbmh$ as follows.
\begin{define}
\begin{align*}
	&d2pbmh : \mathbf{A} \fun \mathbf{PBMH}\\
	&d2pbmh(P) \circdef (\lnot P^f \implies P^t)[true/ok][undashset(ac')/ac'][State_{\II}(in\alpha_{-ok})/s]
\end{align*}
\end{define}\noindent
First we consider the implication between the precondition $\lnot P^f$ and postcondition $P^t$ of a design $P$. We require that $ok$ is $true$ and perform the following substitutions. Since the new variable $ac'$ considers dashed components, the old variable $ac'$ is replaced with an $undashed$ version of $ac'$. Finally, the input variables in $in\alpha_{-ok}$, which excludes $ok$, are introduced via the substitution of $State_{\II}(in\alpha_{-ok})$ for $s$.

We consider~\cref{example:d2pbmh(x-1)}, where $d2pbmh$ is applied to the assignment $x := 1$.%
\begin{example}\label{example:d2pbmh(x-1)}
\begin{xflalign*}
	&d2pbmh(x:=1)
	&&\ptext{Definition of assignment}\\
	&=d2pbmh(true \vdash s\oplus\{ x\mapsto 1\} \in ac')
	&&\ptext{Definition of $d2pbmh$}\\
	&=(true \implies s\oplus\{ x\mapsto 1\} \in ac')[true/ok][undashset(ac')/ac'][State_{\II}(in\alpha_{-ok})/s]
	&&\ptext{Substitution}\\
	&=true \implies State_{\II}(in\alpha_{-ok}) \oplus\{ x\mapsto 1\} \in undashset(ac')
	&&\ptext{Predicate calculus}\\
	&=State_{\II}(in\alpha_{-ok}) \oplus\{ x\mapsto 1\} \in undashset(ac')
	&&\ptext{Definition of $State_{\II}$}\\
	&=\{ x_0 \mapsto x_0, \ldots, x_n \mapsto x_n \} \oplus\{ x\mapsto 1\} \in undashset(ac')
	&&\ptext{Definition of $\theta in\alpha$}\\
	&=\theta in\alpha\oplus \{ x\mapsto 1\} \in undashset(ac')
	&&\ptext{Property of sets, $dash$ and $dashsset$}\\
	&=(\theta in\alpha)'\oplus \{ x'\mapsto 1\} \in ac'
\end{xflalign*}
\end{example}\noindent%
The result is the corresponding assignment in the~$\mathbf{PBMH}$ theory~\cite{Cavalcanti2006}, where the state obtained by dashing every component of the initial state $\theta in\alpha$ is overridden so that the component $x'$ takes the value of $1$.
The following~\cref{theorem:d2pbmh:pbmh} establishes that $d2pbmh$ yields predicates that are $\mathbf{PBMH}$-healthy. 
\theoremstatementref{theorem:d2pbmh:pbmh}\noindent%
That is, when $d2pbmh$ is applied to an angelic design that is $\mathbf{A}$-healthy, then it is also $\mathbf{PBMH}$-healthy. Therefore the application of $d2pbmh$ yields a $\mathbf{PBMH}$-healthy predicate as required.

\subsection{From $\mathbf{PBMH}$ to Angelic Designs ($pbmh2d$)}\label{sec:ch4:pbmh2d}
In order to define a mapping in the opposite direction, we need to consider how to express a precondition in the theory of~\cite{Cavalcanti2006}. In that model, successful termination is guaranteed whenever $ac'$ is not empty. The definition of the mapping from $\mathbf{PBMH}$ into angelic designs, $pbmh2d$, is defined below.
\begin{define}
\begin{align*}
	&pbmh2d : \mathbf{PBMH} \fun \mathbf{A}\\
	&pbmh2d(P) \circdef (\lnot P[\emptyset/ac'] \vdash P[dashset(ac')/ac'])[\mathbf{s}/in\alpha_{-ok}]
\end{align*}
\end{define}\noindent
The precondition of the corresponding $\mathbf{A}$-design requires that $ac'$ is not empty. In the postcondition we substitute the existing set of final states $ac'$ with a dashed version $dashset(ac')$. Finally, we require that the initial variables of $P$ are components of the initial state $s$.
In the following~\cref{theorem:A-o-H3-o-pbmh2d(P):pbmh2d(P)} we prove that $pbmh2d$ yields designs that are $\mathbf{A}$ and $\mathbf{H3}$-healthy.
\theoremstatementref{theorem:A-o-H3-o-pbmh2d(P):pbmh2d(P)}\noindent%
Similarly to the definition of $d2pbmh$, the proviso of~\cref{theorem:A-o-H3-o-pbmh2d(P):pbmh2d(P)} ensures that the function is only applied to predicates that are $\mathbf{PBMH}$-healthy.

\subsection{Galois Connection and Isomorphism}\label{sec:ch4:pbmh-rel:galois}
In general, the model of angelic designs can express every existing program of the theory of~\cite{Cavalcanti2006}. That is, those programs can be specified as angelic designs, where the precondition may not refer to the final set of states $ac'$. This is formally established by the following~\cref{theorem:d2pbm-o-pbmh2d(P):P}.
\theoremstatementref{theorem:d2pbm-o-pbmh2d(P):P}\noindent%
Its only requirement is that the predicate must be $\mathbf{PBMH}$-healthy.

However, when we consider the reverse functional composition of $d2pbmh$ and $pbmh2d$, we obtain a different result as established by~\cref{theorem:pbmh2d-o-d2pbmh(P):sqsubseteq:P}.
\theoremstatementref{theorem:pbmh2d-o-d2pbmh(P):sqsubseteq:P}\noindent%
This is because the theory of~\cite{Cavalcanti2006} cannot model sets of final states where termination is not guaranteed, as is the case for angelic designs which are not $\mathbf{H3}$-healthy. Hence, these two results establish that the two adjoints form a Galois connection.

If we consider the subset of angelic designs that are, in addition, $\mathbf{H3}$-healthy, then we obtain a bijection via the functions $d2pbmh$ and $pbmh2d$, as established by the following~\cref{theorem:pbmh2d-o-d2pbmh(P):P}.
\theoremstatementref{theorem:pbmh2d-o-d2pbmh(P):P}\noindent
While this is an expected result, it is reassuring that the subset of our theory that is $\mathbf{H3}$-healthy is in exact correspondence with the~\ac{UTP} theory of~\cite{Cavalcanti2006}. 

We observe that the subset of the binary multirelational model of \cref{chapter:3} that is $\mathbf{BMH3}$-healthy is isomorphic to the original theory of binary multirelations. Since binary multirelations are also isomorphic to the~\ac{UTP} theory of~\cite{Cavalcanti2006}, the result presented in this section is also in agreement. 

\section{Final Considerations}\label{sec:ch4:final}
In this chapter we have presented a new theory of designs where both angelic and demonic nondeterminism can be modelled. This consists of an extension of the binary multirelational encoding of~\cite{Cavalcanti2006} to include the auxiliary variables $ok$ and $ok'$ of the theory of designs~\cite{Hoare1998}. Our angelic designs are not necessarily $\mathbf{H3}$-healthy as required for a treatment of processes. 

The healthiness conditions of the theory have been presented and their main properties proved. Through the development of the extended theory of binary multirelations of~\cref{chapter:3}, and the subsequent isomorphism, we have been able to justify and explore the definition of the operators and the refinement order. It is reassuring to know that the usual refinement order defined by universal reverse implication corresponds to subset inclusion in the binary multirelational model.

Perhaps the most challenging aspect of the theory is that it relies on non-homogeneous relations. As a consequence, sequential composition cannot be defined as relational composition. While the definition may not be immediately obvious, it is more intuitive when considered in the equivalent binary multirelational model of~\cref{chapter:3}. We have taken advantage of this correspondence to define an operator with the expected properties.

In addition, we have established that every design can be expressed in the theory of angelic designs. Moreover, the subset of $\mathbf{A2}$-healthy designs is isomorphic to the original theory of homogeneous designs of Hoare and He~\cite{Hoare1998}.

Finally, we have also studied the relationship between angelic designs and the \ac{UTP} theory of~\cite{Cavalcanti2006}. This is a complementary result to the link between the model of $BM_\bot$ relations and that of the original theory of binary multirelations. This gives us further assurance as to the capability to express the existing theories as a subset of our own correctly.

\chapter{Reactive Angelic Designs}\label{chapter:5}
Based on the theory of angelic designs and the principles underlying the theory of reactive processes, in this chapter we propose a natural extension to the~\ac{UTP} theory of~\ac{CSP} where both angelic and demonic nondeterminism can be modelled. In~\cref{sec:ch5:introduction} we introduce the principles underlying our approach and justify the encoding proposed for~\ac{CSP}. In~\cref{sec:ch5:healthiness-conditions} the healthiness conditions of the theory are presented. \cref{sec:ch5:rel-CSP} discusses the relationship between the new theory and the existing model of~\ac{CSP}. The operators of the theory are discussed in~\cref{sec:ch5:operators} and, for each operator, we discuss the relationship with their respective counterpart in the original~\ac{CSP} theory. In~\cref{sec:ch5:non-divergent} we characterise the important subset of non-divergent reactive angelic designs. Finally, we summarize our results in~\cref{sec:ch5:final-considerations}.

\section{Introduction}\label{sec:ch5:introduction}
As discussed earlier in~\cref{sec:ch2:CSP:UTP} the observational variables of the~\ac{UTP} theory of~\ac{CSP} are $ok$ and $ok'$ to record stability, and the additional variables $wait$, $tr$ and $ref$, and their respective dashed counterparts. Based on the concept of states originally introduced in~\cref{sec:ch2:binary-multirelations}, we consider a model where the observational variables of the theory of reactive processes are encoded as components of a $State$. We define the alphabet as follows.
\begin{define}[Alphabet]\label{def:RAD:alphabet}
\begin{statement}
\begin{align*}
	&ok, ok' : \{ true, false \}, s : State(\{tr,ref,wait\}), ac' : \power State(\{tr,ref,wait\}) 
\end{align*}
\end{statement}
\end{define}\noindent
In addition to a single initial state $s$, a set of final states $ac'$, and the observational variables $ok$ and $ok'$ that record stability, we require that every $State$ has record components of name $tr$, $wait$ and $ref$. This enables the angelic choice over the final or intermediate observations of $tr$, $ref$ and $wait$.

We next show how we can express every healthiness condition of the original theory of reactive processes, and ultimately~\ac{CSP}, in this new encoding. We then propose linking functions between the theories so that we can reason about the correspondence of the healthiness conditions and operators of both models. These are important aspects for establishing the validity of the model.

\section{Healthiness Conditions}\label{sec:ch5:healthiness-conditions}
Since this is a theory with angelic nondeterminism, the set of final states $ac'$ must be upward-closed, so relations in this theory need to satisfy $\mathbf{PBMH}$. As previously discussed in~\cref{sec:ch2:CSP:UTP}, in the~\ac{UTP}, \ac{CSP} processes are characterised as the image of designs through the function $\mathbf{R}$. In order to preserve the existing semantics, we propose a corresponding construction; in the following~\cref{sec:ch5:RA1,sec:ch5:RA2,sec:ch5:RA3,sec:ch5:RA,sec:ch5:CSP-angelic} we restate all the properties enforced by~$\mathbf{R}$ in this new model. Namely, we define healthiness conditions $\mathbf{RA1}$, $\mathbf{RA2}$ and $\mathbf{RA3}$, whose functional composition is named $\mathbf{RA}$, and, $\mathbf{CSPA1}$ and $\mathbf{CSPA2}$. All the healthiness conditions discussed in this chapter are monotonic and idempotent. In~\cref{sec:ch5:RAD} we show how this construction allows~\ac{CSP} processes with angelic nondeterminism to be expressed as the image of angelic designs through $\mathbf{RA}$, the counterpart to $\mathbf{R}$.

\subsection{$\mathbf{RA1}$}\label{sec:ch5:RA1}
The first property of interest that underpins the theory of reactive processes is the notion that the history of events observed cannot be undone. In general, for any initial state $x$, the set of all final states that satisfy this property is given by $States_{tr\le tr'} (x)$ as defined below.
\begin{define}
$States_{tr\le tr'} (x) \circdef \{ z : State(\{tr,ref,wait\}) | x.tr \le z.tr \} $
\end{define}\noindent%
This definition is used for introducing the first healthiness condition, $\mathbf{RA1}$, that not only enforces this notion for final states in $ac'$, but also requires that there is some final state satisfying this property available for angelic choice.
\begin{define}\label{def:RA1}
\begin{statement}
$\mathbf{RA1} (P) \circdef (P \land ac'\neq\emptyset)[States_{tr\le tr'} (s) \cap ac'/ac']$
\end{statement}
\end{define}\noindent%
A consequence of the definition of $\mathbf{RA1}$ is that it also enforces $\mathbf{A0}$. 
\theoremstatementref{theorem:RA1-o-A0:RA1}\noindent%
Although $\mathbf{A0}$ only requires $ac'$ not to be empty in the postcondition of an angelic design, $\mathbf{RA1}$ requires this under all circumstances. Proof of this and other results not explicitly included in the body of this document can be found in~\cref{appendix:RAD}.

The function $\mathbf{RA1}$ distributes through both conjunction and disjunction as established by the following~\cref{lemma:RA1(P-land-Q):RA1(P)-land-RA1(Q),lemma:RA1(P-lor-Q):RA1(P)-lor-RA1(Q)}.
\theoremstatementref{lemma:RA1(P-land-Q):RA1(P)-land-RA1(Q)}
\theoremstatementref{lemma:RA1(P-lor-Q):RA1(P)-lor-RA1(Q)}\noindent%
Since $\mathbf{RA1}$ is also idempotent, consequently both conjunction and disjunction are also closed under $\mathbf{RA1}$.

Similarly to the theory of angelic designs, in this model, the definition of sequential composition is also based on $\seqA$. In~\cref{theorem:RA1(P-seqA-Q):closure} we establish that this operator is closed under $\mathbf{RA1}$.
\theoremstatementref{theorem:RA1(P-seqA-Q):closure}\noindent%
For every healthiness condition of the theory, the upward-closure enforced by $\mathbf{PBMH}$ must be maintained. \cref{theorem:PBMH-o-RA1(P):RA1(P)} establishes this for $\mathbf{RA1}$.
\theoremstatementref{theorem:PBMH-o-RA1(P):RA1(P)}\noindent%
However, $\mathbf{PBMH}$ and $\mathbf{RA1}$ do not commute in general. We consider the following~\cref{counter-example:RA1-PBMH} where the healthiness conditions are applied to the relation $ac'=\emptyset$, which is not $\mathbf{PBMH}$-healthy.
\begin{counter-example}\label{counter-example:RA1-PBMH}
\begin{xflalign*}
	&\mathbf{RA1} \circ \mathbf{PBMH} (ac'=\emptyset)
	&&\ptext{Definition of $\mathbf{PBMH}$ (\cref{lemma:PBMH:alternative-1})}\\
	&=\mathbf{RA1} (\exists ac_0 \spot ac_0=\emptyset \land ac_0\subseteq ac')
	&&\ptext{One-point rule and property of sets}\\
	&=\mathbf{RA1} (true)
	&&\ptext{\cref{lemma:RA1(true):alternative-1}}\\
	&=States_{tr\le tr'} (s) \cap ac' \neq\emptyset
\end{xflalign*}
\begin{xflalign*}
	&\mathbf{PBMH} \circ \mathbf{RA1} (ac'=\emptyset)
	&&\ptext{Definition of $\mathbf{RA1}$}\\
	&=\mathbf{PBMH} ((ac'=\emptyset \land ac'\neq\emptyset)[States_{tr\le tr'}(s)\cap ac'/ac'])
	&&\raisetag{12pt}\ptext{Predicate calculus}\\
	&=\mathbf{PBMH} (false)
	&&\ptext{Definition of $\mathbf{PBMH}$ (\cref{lemma:PBMH:alternative-1})}\\
	&=false
\end{xflalign*}
\end{counter-example}\noindent%
In the first case, the application of $\mathbf{PBMH}$ yields $true$. The result of the functional composition is then $\mathbf{RA1} (true)$. On the other hand, in the second case, there is a contradiction arising from the application of $\mathbf{RA1}$, which leaves us with the result $false$.

\subsection{$\mathbf{RA2}$}\label{sec:ch5:RA2}
The next healthiness condition of interest is $\mathbf{RA2}$, which requires a process to be insensitive to the initial trace of events $s.tr$. It is the counterpart to $\mathbf{R2}$ of the original theory of reactive processes, and is also defined using substitution. 
\begin{define}\label{def:RA2}
\begin{statement}
\begin{align*}
	&\mathbf{RA2} (P) \circdef
	P\left[s\oplus\{tr \mapsto \lseq\rseq\},\left.\left\{z\left|\begin{array}{l}
		z\in ac'\land s.tr\le z.tr\\
		\spot z\oplus\{tr\mapsto z.tr-s.tr\}
	\end{array}\right.\right\}\right/s,ac'\right]
\end{align*}
\end{statement}
\end{define}\noindent
It defines the component $tr$ in the initial state $s$ to be the empty sequence, and consequently the set of final states $ac'$ is restricted by considering those states $z$ whose traces are a suffix of $s.tr$, and furthermore, defining their trace to be the difference with respect to the initial trace $s.tr$.

Since substitution distributes through conjunction and disjunction, so does the function $\mathbf{RA2}$ as established by the following~\cref{theorem:RA2(P-land-Q):RA2(P)-land-RA2(Q),theorem:RA2(P-lor-Q):RA2(P)-lor-RA2(Q)}.
\theoremstatementref{theorem:RA2(P-land-Q):RA2(P)-land-RA2(Q)}
\theoremstatementref{theorem:RA2(P-lor-Q):RA2(P)-lor-RA2(Q)}\noindent%
As $\mathbf{RA2}$ is idempotent, both conjunction and disjunction are closed under $\mathbf{RA2}$.

Similarly to the case for $\mathbf{RA1}$, the operator $\seqA$ is also closed under $\mathbf{RA2}$.
\theoremstatementref{theorem:RA2(P-seqA-Q)-closure}\noindent%
A consequence of the definition of $\mathbf{RA2}$ is that applying it to the predicate that requires $ac'$ not to be empty is equivalent to applying $\mathbf{RA2}$ to the relation $true$.
\theoremref{lemma:RA2(ac'-neq-emptyset):RA1(true)}\noindent%
This result sheds light on the relationship between $\mathbf{RA2}$ and $\mathbf{RA1}$, as in fact, these functions are commutative as established by~\cref{theorem:RA2-o-RA1:RA1-o-RA2}.
\theoremstatementref{theorem:RA2-o-RA1:RA1-o-RA2}\noindent%
Finally, $\mathbf{RA2}$ preserves the upward closure of $\mathbf{PBMH}$.
\theoremstatementref{theorem:PBMH-o-RA2(P):RA2(P)}\noindent%
These results conclude our discussion of $\mathbf{RA2}$ and its most important properties.

\subsection{$\mathbf{RA3}$}\label{sec:ch5:RA3}

Similarly to the theory of reactive processes, we must ensure that a process cannot be started before the previous process has finished interacting with the environment. The counterpart to $\mathbf{R3}$ in this new theory is $\mathbf{RA3}$. Before exploring its definition, we introduce the identity $\IIRac$ of our theory.
\begin{define}\label{def:IIRac}
$\IIRac \circdef (\mathbf{RA1} (\lnot ok) \lor (ok' \land s \in ac'))$
\end{define}\noindent
Similarly to the reactive identity $\IIrea$, the behaviour for an unstable state $\lnot ok$ is given by $\mathbf{RA1}$, that is, there must be at least one final state in $ac'$ whose trace is a suffix of the initial trace $s.tr$. Otherwise, the process is stable, $ok'$ is $true$, and the initial state $s$ is in the set of final states $ac'$.

Having defined the identity, we introduce the definition of $\mathbf{RA3}$ below.
\begin{define}\label{def:RA3}
\begin{statement}
$\mathbf{RA3} (P) \circdef  \IIRac \dres s.wait \rres P$
\end{statement}
\end{define}\noindent
This definition is similar to that of the original theory, except that we use $\IIRac$ as the identity and use $s.wait$ instead of $wait$ as a condition since in our theory $wait$ is a component of the initial state $s$. Using Leibniz's substitution, it is possible to establish the following~\cref{lemma:RA3:s-oplus-wait-false}, where $P_w = P[s\oplus\{wait\mapsto w\}/s]$.
\theoremstatementref{lemma:RA3:s-oplus-wait-false}\noindent%
This result is in correspondence with a similar property of $\mathbf{R3}$ in the original theory of~\ac{CSP} that is essential in the characterisation of~\ac{CSP} processes via reactive designs.

Similarly to the previous healthiness conditions, $\mathbf{RA3}$ also distributes through both conjunction and disjunction as established by~\cref{theorem:RA3(P-land-Q):RA3(P)-land-RA3(Q),theorem:RA3(P-lor-Q):RA3(P)-lor-RA3(Q)}.
\theoremstatementref{theorem:RA3(P-land-Q):RA3(P)-land-RA3(Q)}
\theoremstatementref{theorem:RA3(P-lor-Q):RA3(P)-lor-RA3(Q)}\noindent%
Consequently, these operators are closed under $\mathbf{RA3}$.

The operator $\seqA$ is also closed under $\mathbf{RA3}$ provided that the second operand is also $\mathbf{RA1}$-healthy as established by~\cref{theorem:RA3(P-seqA-Q)-closure}. 
\theoremstatementref{theorem:RA3(P-seqA-Q)-closure}\noindent%
The proviso is similar to that observed for the closure of $\circseq$ under $\mathbf{R3}$ in the original theory of reactive processes~\cite{Cavalcanti2006a}. The extra restriction on $Q$, which needs to be $\mathbf{RA1}$-healthy, is not a problem since the theory of interest is characterised by the functional composition of all healthiness conditions.

Furthermore, as required, the function $\mathbf{RA3}$ also preserves the upward-closure.
\theoremstatementref{theorem:PBMH-o-RA3(P):RA3(P)}\noindent%
The identity $\IIRac$ is a fixed point of every healthiness condition, including $\mathbf{RA1}$, $\mathbf{RA2}$, $\mathbf{RA3}$ and $\mathbf{PBMH}$ as established by~\cref{theorem:RA1(IIRac):IIRac,theorem:RA2(IIRac):IIRac,theorem:RA3(IIRac):IIRac,theorem:PBMH(IIRac):IIRac}. Finally, $\mathbf{RA3}$ commutes with both $\mathbf{RA1}$ and $\mathbf{RA2}$ as established by~\cref{theorem:RA3-o-RA1:RA1-o-RA3,theorem:RA3-o-RA2:RA2-o-RA3}.
\theoremstatementref{theorem:RA3-o-RA1:RA1-o-RA3}
\theoremstatementref{theorem:RA3-o-RA2:RA2-o-RA3}\noindent%
This concludes our discussion of the most important properties of $\mathbf{RA3}$.

\subsection{$\mathbf{RA}$}\label{sec:ch5:RA}
The healthiness conditions that we have considered so far in this chapter are counterparts to those of the original model of reactive processes. Hence this is a theory that is similarly characterised by the functional composition of the healthiness conditions $\mathbf{RA1}$, $\mathbf{RA2}$, $\mathbf{RA3}$, besides $\mathbf{PBMH}$. In order to provide a parallel with the original theory of reactive processes, we define part of this composition as $\mathbf{RA}$.
\begin{define}\label{def:RA}
\begin{statement}
$\mathbf{RA} (P) \circdef \mathbf{RA1} \circ \mathbf{RA2} \circ \mathbf{RA3} (P)$
\end{statement}
\end{define}\noindent%
The order of the functional composition is not important since these functions commute, except for $\mathbf{PBMH}$ that does not necessarily commute with every function. So when considering the counterpart theory to reactive processes, but with angelic nondeterminism, $\mathbf{PBMH}$ needs to be applied before $\mathbf{RA}$.

As previously stated, every healthiness condition considered in this chapter is idempotent and monotonic. \cref{theorem:RA1-idempotent,theorem:RA2:idempotent,theorem:RA3-idempotent} in~\cref{appendix:RAD} establish that $\mathbf{RA1}$, $\mathbf{RA2}$ and $\mathbf{RA3}$ are idempotent. Similarly monotonicity is established for these functions by~\cref{theorem:RA1-monotonic,theorem:RA2:monotonic,theorem:RA3-monotonic}. As a consequence the functional composition $\mathbf{RA}$ is also idempotent and monotonic.

In addition, since all of the $\mathbf{RA}$ functions distribute through conjunction and disjunction so does the functional composition $\mathbf{RA}$. Finally, $\mathbf{RA}$ maintains the upward-closure enforced by $\mathbf{PBMH}$ since all of the $\mathbf{RA}$ healthiness conditions do so as well. This concludes our discussion of the most important properties of $\mathbf{RA}$.

\subsection{CSP Processes with Angelic Nondeterminism}\label{sec:ch5:CSP-angelic}
In the original theory of~\ac{CSP}, another two healthiness conditions, $\mathbf{CSP1}$ and $\mathbf{CSP2}$, are required, in addition to $\mathbf{R}$, to characterise~\ac{CSP} processes. In order to consider a theory of~\ac{CSP} with angelic nondeterminism we follow a similar approach by defining a counterpart to these functions in what follows.

 
\subsubsection{$\mathbf{CSPA1}$}\label{sec:ch5:CSPA1}
The first healthiness condition of interest is~$\mathbf{CSPA1}$, which is the counterpart to $\mathbf{CSP1}$ in the new theory. Its definition is presented below.
\begin{define}\label{def:CSPA1}
\begin{statement}
$\mathbf{CSPA1} (P)  \circdef P \lor \mathbf{RA1} (\lnot ok)$
\end{statement}
\end{define}\noindent
A~\ac{CSP} process with angelic nondeterminism $P$ is required to observe $\mathbf{RA1}$ when in an unstable state. For a $\mathbf{RA}$-healthy process, this property is already enforced by $\mathbf{RA1}$ under all circumstances. Similarly to the original theory of~\ac{CSP}~\cite{Cavalcanti2006a} the following~\cref{theorem:CSPA1-o-RA1:RA1-o-H1} establishes that this behaviour can also be described as the functional composition of $\mathbf{RA1}$ after $\mathbf{H1}$.
\theoremref{theorem:CSPA1-o-RA1:RA1-o-H1}\noindent%
The function $\mathbf{CSPA1}$ is idempotent and monotonic. Furthermore, it preserves the upward-closure as required by $\mathbf{PBMH}$.
\theoremstatementref{theorem:PBMH-o-CSPA1:CSPA1}\noindent%
This concludes the discussion of the properties of~$\mathbf{CSPA1}$.

\subsubsection{$\mathbf{CSPA2}$}\label{sec:ch5:CSPA2}
The last healthiness condition of interest is the counterpart to $\mathbf{CSP2}$. It is defined as $\mathbf{H2}$ with the extended alphabet that includes $s$ and $ac'$.
\begin{define}\label{def:CSPA2}
\begin{statement}
$\mathbf{CSPA2} (P) \circdef \mathbf{H2} (P)$
\end{statement}
\end{define}\noindent%
This healthiness condition satisfies the same properties as $\mathbf{H2}$, including, for example, those established by~\cref{theorem:A-o-H1-o-H2(P):H1-o-H2-o-A(P),theorem:H2-o-PBMH:PBMH-o-H2}. It can alternatively be defined using the $J$-split of Woodcock and Cavalcanti~\cite{Woodcock2004}. 

\subsection{Reactive Angelic Designs ($\mathbf{RAD}$)}\label{sec:ch5:RAD} 
The theory of~\ac{CSP} processes in the new model is defined by $\mathbf{RAD}$, which is the functional composition of all the healthiness conditions of interest.
\begin{define}\label{def:RAD}
\begin{statement}
$\mathbf{RAD} (P) \circdef \mathbf{RA} \circ \mathbf{CSPA1} \circ \mathbf{CSPA2} \circ \mathbf{PBMH} (P)$
\end{statement}
\end{define}\noindent
Since $\mathbf{PBMH}$ and $\mathbf{RA1}$ do not commute, $\mathbf{PBMH}$ is applied first. The fixed points of $\mathbf{RAD}$ are the reactive angelic designs. Every such process $P$ can be expressed as $\mathbf{RA}\circ\mathbf{A} (\lnot P^f_f \vdash P^t_f)$ as established by the following~\cref{theorem:RA-o-A(design):RA-CSPA-PBMH}, where $P^o_w = P[o,s\oplus\{wait\mapsto w\}/ok',s]$ 
\theoremref{theorem:RA-o-A(design):RA-CSPA-PBMH}\noindent%
That is, such processes can be specified as the image of an $\mathbf{A}$-healthy design through the function $\mathbf{RA}$. This is a result similar to that obtained for~\ac{CSP} processes as the image of designs through $\mathbf{R}$~\cite{Hoare1998,Cavalcanti2006a}. Since both $\mathbf{RA}$ and $\mathbf{A}$ are monotonic and idempotent, and the theory of designs is a complete lattice~\cite{Hoare1998}, so is the theory of reactive angelic designs.

Since $\mathbf{PBMH}$ is just $\mathbf{A1}$, and $\mathbf{RA1}$ enforces $\mathbf{A0}$, a fixed point $P$ of $\mathbf{RAD}$ can alternatively be expressed as shown in the following~\cref{lemma:RAD(P):RA(lnot-PBMH(P)ff|-PBMH(P)tf)}.
\theoremstatementref{lemma:RAD(P):RA(lnot-PBMH(P)ff|-PBMH(P)tf)}\noindent%
That is, an angelic design, with $\mathbf{PBMH}$ applied to the negation of the precondition and postcondition. Furthermore, it is possible to infer that if $P$ is a reactive angelic design, then it is also $\mathbf{PBMH}$-healthy.
\theoremstatementref{theorem:PBMH(P)-RAP:P}\noindent%
This concludes our discussion of the healthiness condition of the theory of reactive angelic designs, $\mathbf{RAD}$, and its respective properties.

\section{Relationship with CSP}\label{sec:ch5:rel-CSP}
The theory of reactive angelic designs can be related to the original~\ac{UTP} theory of~\ac{CSP} through the pair of linking functions $ac2p$ and $p2ac$ previously introduced in~\cref{sec:ch4:rel-designs} and reproduced below.
\begin{align*}
	&ac2p(P) \circdef \mathbf{PBMH} (P)[State_{\II}(in\alpha_{-ok})/s] \seqA \bigwedge x : out\alpha_{-ok'} \spot dash(s).x = x\\
	&p2ac (P) \circdef \exists z \spot P[\mathbf{s},\mathbf{z}/in\alpha_{-ok},out\alpha_{-ok'}] \land undash(z) \in ac'
\end{align*}
We employ $ac2p$ by considering the set of variables $in\alpha$ to be $\{ tr, ref, wait\}$, and a corresponding set of variables $out\alpha$ with dashed counterparts; $State$, therefore, has components ranging over $in\alpha$. Similarly, for the mapping in the opposite direction, from reactive angelic designs to~\ac{CSP} processes we employ $p2ac$ with the same sets of variables $in\alpha$ and $out\alpha$.

\begin{figure}
\centering
\begin{subfigure}[b]{.5\textwidth}
\centering
\includegraphics[width=.94\linewidth]{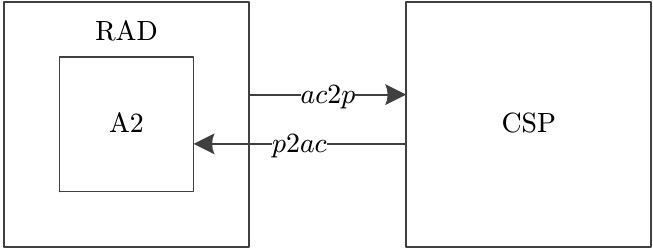}
\caption{\label{fig:ch5:theories}Theories and links}
\end{subfigure}%
\begin{subfigure}[b]{.5\textwidth}
\centering
\includegraphics[width=.94\linewidth]{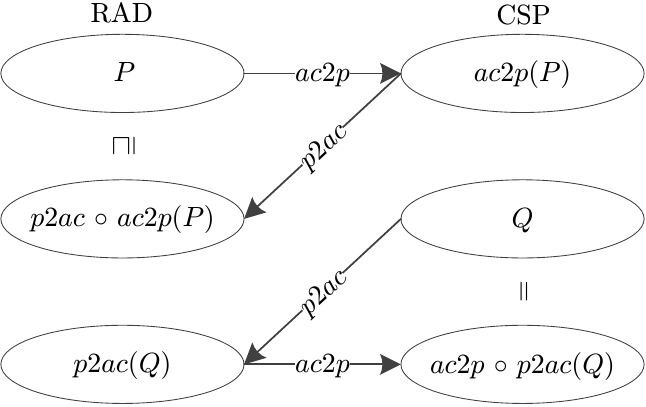}
\caption{\label{fig:ch5:predicates}Predicates and links}
\end{subfigure}
\caption{Relationship between theories}
\end{figure}
The relationship between the models has previously been illustrated in the context of all theories in~\cref{fig:theories}. Here we focus our attention on the relationship with~\ac{CSP}. In~\cref{fig:ch5:theories} each theory is labelled according to its healthiness conditions. The subset of reactive angelic designs that corresponds exactly to~\ac{CSP} processes is characterised by $\mathbf{A2}$, the healthiness condition which we previously discussed in~\cref{sec:ch4:A2} that characterises predicates with no angelic nondeterminism.

In~\cref{fig:ch5:predicates} the relationship between the predicates of each theory is illustrated. For a predicate $P$ of the theory of reactive angelic designs, the functional composition $p2ac \circ ac2p(P)$ yields a stronger predicate since any angelic nondeterminism in $P$ is virtually collapsed into a single final state, while for a predicate $Q$ of the~\ac{CSP} theory, the composition $ac2p \circ p2ac(Q)$ yields exactly the same predicate $Q$. Thus a Galois connection exists between the theories.

\subsection{From Reactive Angelic Designs to CSP ($ac2p$)}
As already stated, the mapping from reactive angelic designs to~\ac{CSP} processes achieved through $ac2p$ defines a Galois connection. Application of this function to a predicate $P$ that is both $\mathbf{RA}$-healthy and $\mathbf{PBMH}$-healthy yields a healthy counterpart in the original theory as established by the following~\cref{theorem:ac2p-o-RA(P):R-o-ac2p(P)}.
\theoremstatementref{theorem:ac2p-o-RA(P):R-o-ac2p(P)}\noindent%
If we consider $P$ to be a reactive angelic design, then we can show that the application of $ac2p$ yields a reactive design as established by~\cref{theorem:ac2p-o-RA-o-A(design):R(lnot-ac2p(pre)|-ac2p(post))}
\theoremref{theorem:ac2p-o-RA-o-A(design):R(lnot-ac2p(pre)|-ac2p(post))}\noindent%
This is a pleasing result that supports the reuse of results across the theories. We consider the following~\cref{example:ac2p}, where $ac2p$ is applied to the angelic choice between a prefixing on the event $a$ followed by deadlock, and on the event $b$ followed by deadlock. The operators of the theory of reactive angelic designs have subscript $_{\mathbf{RAD}}$ in order to distinguish them from those of the original theory of~\ac{CSP} which have subscript $_{\mathbf{R}}$.
\begin{example}\label{example:ac2p}
\begin{align*}
&ac2p(a \circthen_{\mathbf{RAD}} Stop_{\mathbf{RAD}} \sqcup_{\mathbf{RAD}} b \circthen_{\mathbf{RAD}} Stop_{\mathbf{RAD}})\\
	&=\\
	&a \circthen_{\mathbf{R}} Stop_{\mathbf{R}} \sqcup_{\mathbf{R}} b \circthen_{\mathbf{R}} Stop_{\mathbf{R}}
\end{align*}
\begin{proof}
\cref{lemma:ac2p(RAD:a-then-Stop-sqcup-b-then-Stop):R:a-then-Stop-sqcup-b-then-Stop}
\end{proof}
\end{example}\noindent%
The result is the least upper bound of the corresponding~\ac{CSP} process, where $\sqcup_{\mathbf{R}}$ is also defined using conjunction. This is a process that cannot be expressed using the standard operators of~\ac{CSP}. The conjunction of non-divergent~\ac{CSP} processes requires the conjunction of their respective postconditions, and thus an agreement. In this case, both processes can only agree on the trace of events remaining unchanged, and not refusing events $a$ and $b$, while waiting.

\subsection{From CSP to Reactive Angelic Designs ($p2ac$)}
The mapping in the opposite direction, from~\ac{CSP} processes to reactive angelic designs, is  achieved through the function $p2ac$. As discussed in~\cref{sec:ch4:rel-designs} the result of applying $p2ac$ is upward-closed as established by~\cref{lemma:PBMH-o-p2ac(P):p2ac(P)}. The application of $p2ac$ to a process $P$ that is $\mathbf{R}$-healthy, can be described by the functional composition of $\mathbf{RA}$ after $p2ac$ to the original process $P$, as established by the following~\cref{theorem:p2ac-o-R:RA-o-p2ac}.
\theoremstatementref{theorem:p2ac-o-R:RA-o-p2ac}\noindent%
The result of applying $p2ac$ to a reactive design is established in~\cref{theorem:p2ac-o-R(design):RA-o-A(lnot-p2ac(pre)|-p2ac(post))}: $p2ac$ can be directly applied to the pre and postconditions separately, followed by $\mathbf{A}$ and $\mathbf{RA}$.
\theoremref{theorem:p2ac-o-R(design):RA-o-A(lnot-p2ac(pre)|-p2ac(post))}\noindent%
This result enables~\ac{CSP} processes to be easily mapped into the theory of reactive angelic designs by considering the mapping of the pre and postconditions of~\ac{CSP} processes directly.

We consider the following example, where the terminating process $Skip_{\mathbf{R}}$ is mapped through $p2ac$ into the theory of reactive angelic designs.
\begin{example}
\begin{align*}
	&p2ac(Skip_{\mathbf{R}}) 
	=
	\mathbf{RA}\circ\mathbf{A} (true \vdash \exists y @ \lnot y.wait \land y.tr=s.tr \land y \in ac')
\end{align*}
\begin{proof}
\cref{theorem:p2ac(Skip-R):Skip-RAD}
\end{proof}
\end{example}\noindent%
The reactive angelic design also has $true$ as its precondition, while the postcondition asserts that there is a final state $y$ in the set of angelic choices $ac'$ where the trace of events $s.tr$ is kept unchanged and the value of the component $wait$ is $false$, that is, the process has finished interacting with the environment.
 
\subsection{Galois Connection and Isomorphism}
As already mentioned, the pair of linking functions we have considered establish a Galois connection between the theory of~\ac{CSP} and that of reactive angelic designs. When considering the mapping from the original theory of reactive processes, followed by the mapping in the opposite direction, we obtain an exact correspondence as shown in the following~\cref{theorem:ac2p-o-p2ac(P):P}.
\theoremref{theorem:ac2p-o-p2ac(P):P}\noindent%
This results establishes that our theory can accommodate the existing~\ac{CSP} processes appropriately, that is, those without angelic nondeterminism.

When considering the mapping in the opposite direction we obtain the following result in~\cref{lemma:p2ac-o-ac2p(P)}.
\theoremstatementref{lemma:p2ac-o-ac2p(P)}\noindent%
If the set of final states $ac_0$ in $P$ has more than one state, then the result of $p2ac\circ ac2p(P)$ is $false$, otherwise, $ac_0$ is either a singleton, in which case $ac'$ is any set containing its element, or empty, in which case $ac'$ is arbitrary. Most importantly, the functional composition only preserves predicates whose set of angelic choices is either empty or a singleton, otherwise the result is $false$.

We consider the following~\cref{example:p2ac-o-ac2p}, where~\cref{lemma:p2ac-o-ac2p(P)} is applied to the angelic choice between events $a$ or $b$ followed by deadlock.
\begin{example}\label{example:p2ac-o-ac2p}
\begin{align*}
	&p2ac \circ ac2p(a \circthen_{\mathbf{RAD}} Stop_{\mathbf{RAD}} \sqcup b \circthen_{\mathbf{RAD}} Stop_{\mathbf{RAD}}) \\
	&=\\
	&\mathbf{RA} \circ \mathbf{A} \left(true \vdash \exists y @ y \in ac' \land y.wait \land y.tr=s.tr \land a \notin y.ref \land b \notin y.ref\right)	
\end{align*}
\begin{proof}
\cref{lemma:ac2p(RAD:a-then-Stop-sqcup-b-then-Stop):R:a-then-Stop-sqcup-b-then-Stop,lemma:p2ac(R:a-then-Stop-sqcup-b-then-Stop)}
\end{proof}
\end{example}\noindent%
This process corresponds to the application of $p2ac$ to the result obtained in the previous~\cref{example:ac2p}. In this case, the process is always waiting for the environment and keeps the trace of events unchanged, however it requires that neither event $a$ nor $b$ are refused. This is a process whose behaviour cannot be described using the standard operators of~\ac{CSP}.

If we consider the result of~\cref{lemma:p2ac-o-ac2p(P)} in the context of the predicates of our theory, that is, those which are~$\mathbf{PBMH}$-healthy, then we obtain an inequality as shown in the following~\cref{theorem:p2ac-o-ac2p:implies:P}.
\theoremref{theorem:p2ac-o-ac2p:implies:P}\noindent%
This theorem, together with~\cref{theorem:ac2p-o-p2ac(P):P}, establishes the existence of a Galois connection between the theories. In particular, these results also hold between reactive processes, characterised by $\mathbf{R}$, and the reactive angelic designs, characterised by $\mathbf{RAD}$, that is, in general, the Galois connection is not restricted to~\ac{CSP} processes. This is because the proviso of~\cref{theorem:ac2p-o-p2ac(P):P} only requires $P$ to be $\mathbf{PBMH}$-healthy.

The result of~\cref{theorem:p2ac-o-ac2p:implies:P} can be strengthened into an equality by considering the subset of reactive angelic designs that are $\mathbf{A2}$-healthy. These are reactive processes that do not exhibit angelic nondeterminism. If we consider the application of $\mathbf{A2}$ to the process $a \circthen_{\mathbf{RAD}} Stop_{\mathbf{RAD}} \sqcup_{\mathbf{RAD}} b \circthen_{\mathbf{RAD}} Stop_{\mathbf{RAD}}$, we obtain exactly the same result as in~\cref{example:p2ac-o-ac2p}. In other words, for reactive angelic designs, $\mathbf{A2}$ characterises the same fixed points as $p2ac \circ ac2p (P)$. We observe, however, that in general, $\mathbf{A2}$ permits an empty set of final states, whereas in the theory of reactive angelic designs, both $\mathbf{RA1}$ and the mapping $p2ac$ require the set of final states not to be empty. For example, in the theory of angelic designs the bottom $\botD$ of the lattice, which is $true$, is a fixed point of $\mathbf{A2}$ (\cref{lemma:A2(true):true}).

Finally, \cref{theorem:p2ac-o-ac2p-RA-o-A:RA-o-A} establishes that the result $p2ac \circ ac2p(P)$ for a reactive angelic design $P$ that is $\mathbf{A2}$-healthy yields exactly the same reactive angelic design $P$.
\theoremstatementref{theorem:p2ac-o-ac2p-RA-o-A:RA-o-A}\noindent%
In summary, when we consider the theory of reactive angelic designs that are $\mathbf{A2}$-healthy, then we find that there is a bijection with the original theory of reactive designs. Thus this subset is isomorphic to the theory of~\ac{CSP}.

\section{Operators}\label{sec:ch5:operators}
Having discussed the healthiness conditions of our theory, and the relationship with the original model of~\ac{CSP}, in this section we present the definition of some important operators of~\ac{CSP} in the new model. For each of the operators we show how they relate to their original~\ac{CSP} counterparts.

\subsection{Angelic Choice}\label{sec:ch5:operators:angelic-choice}
The first operator of interest is angelic choice. Similarly to the theory of angelic designs, it is also defined as the least upper bound of the lattice, which is conjunction.
\begin{define}
$P \sqcup_{\mathbf{RAD}} Q \circdef P \land Q$
\end{define}\noindent%
For reactive angelic designs $P$ and $Q$, this result can be restated as shown in the following~\cref{theorem:RAP:P-sqcup-Q}.
\theoremstatementref{theorem:RAP:P-sqcup-Q}\noindent%
The precondition of the resulting process is the disjunction of the preconditions of $P$ and $Q$, while the postcondition is the conjunction of two implications. In both cases, if either the precondition of $P$ or $Q$ holds, then the corresponding postcondition is established. This is a result that is similar to that observed for the least upper bound of designs~\cite{Hoare1998,Woodcock2004}.

The least upper bound of this theory can be related with that of~\ac{CSP} as follows. If we consider two~\ac{CSP} processes $P$ and $Q$, apply $p2ac$ followed by the least upper bound $\sqcup_{\mathbf{RAD}}$ and then $ac2p$, then we obtain the same result defined by the original least upper bound operator $\sqcup_{\mathbf{R}}$ of~\ac{CSP} as shown in~\cref{theorem:ac2p(p2ac(P)-sqcup-RAD-p2ac(Q)):P-sqcup-R-Q}.
\theoremref{theorem:ac2p(p2ac(P)-sqcup-RAD-p2ac(Q)):P-sqcup-R-Q}\noindent%
This is expected since we can express every existing~\ac{CSP} process in the new theory. The result in the opposite direction, however, is an inequality as shown in the following~\cref{theorem:p2ac(ac2p(P)-sqcup-R-ac2p(Q)):P-sqcup-RAD-Q}.
\theoremref{theorem:p2ac(ac2p(P)-sqcup-R-ac2p(Q)):P-sqcup-RAD-Q}\noindent%
That is, there is a strengthening of the resulting predicate. This is expected, as in general the application of $ac2p$ collapses the angelic nondeterminism, and $p2ac$ cannot undo such effect completely.

This concludes our discussion of the basic properties of angelic choice. In the following sections, and as we present the definition of the~\ac{CSP} operators, we revisit angelic choice and explore its role when applied together with other operators.

\subsection{Demonic Choice}\label{sec:ch5:operators:demonic-choice}
Similarly to the definition of internal choice in~\ac{CSP}, in our theory, this operator is also defined using the greatest lower bound of the lattice, disjunction. 
\begin{define}
$P \sqcap_{\mathbf{RAD}} Q \circdef P \lor Q$
\end{define}\noindent%
For any two reactive angelic designs $P$ and $Q$, their demonic choice can be described as a reactive angelic design as stated as in~\cref{theorem:RAP:P-sqcap-Q}.
\theoremstatementref{theorem:RAP:P-sqcap-Q}\noindent%
That is, the resulting precondition is the conjunction of the respective preconditions of $P$ and $Q$, while the postcondition is the disjunction of the respective postconditions of $P$ and $Q$. Intuitively, in a demonic choice both preconditions need to be satisfied, while either the postcondition of $P$ or $Q$ may be observed. 

The greatest lower bound of both theories can be related through the pair of linking functions $p2ac$ and $ac2p$. Since $p2ac$ distributes through disjunction we can establish the following general result in~\cref{theorem:p2ac(ac2p(P)-sqcap-ac2p(Q)):p2ac-o-ac2p(P)-sqcap-p2ac-o-ac2p(Q)}.
\theoremref{theorem:p2ac(ac2p(P)-sqcap-ac2p(Q)):p2ac-o-ac2p(P)-sqcap-p2ac-o-ac2p(Q)}\noindent%
If we consider two reactive angelic designs $P$ and $Q$ and apply $ac2p$, followed by the greatest lower bound $\sqcap_{\mathbf{R}}$ and then $p2ac$, then this result can be directly obtained by applying $p2ac \circ ac2p$ followed by the greatest lower bound $\sqcap_{\mathbf{RAD}}$. When $P$ and $Q$ are $\mathbf{A2}$-healthy (\cref{theorem:p2ac-o-ac2p-RA-o-A:RA-o-A}) we obtain the result shown in~\cref{lemma:p2ac(ac2p(P)-sqcap-ac2p(Q)):A2-healthy}.
\theoremstatementref{lemma:p2ac(ac2p(P)-sqcap-ac2p(Q)):A2-healthy}\noindent%
That is, for reactive angelic designs with no angelic nondeterminism, the demonic choice of both theories is in correspondence. Similarly, since $ac2p$ also distributes through disjunction, we can establish the following result in the opposite direction, as shown in~\cref{theorem:ac2p(p2ac(P)-sqcap-RAD-p2ac(Q)):P-sqcap-R-Q}.
\theoremstatementref{theorem:ac2p(p2ac(P)-sqcap-RAD-p2ac(Q)):P-sqcap-R-Q}\noindent%
That is, the greatest lower bound of both theories is in correspondence. Finally, since the least upper bound is conjunction, and the greatest lower bound is disjunction, angelic and demonic choice distribute over each other.

\subsection{Chaos}\label{sec:ch5:operators:chaos}
The following operator of interest is $Chaos_{\mathbf{RAD}}$, which is the bottom of the lattice of reactive angelic designs.
\begin{define}
$Chaos_{\mathbf{RAD}} \circdef \mathbf{RA} \circ \mathbf{A} (false \vdash ac'\neq\emptyset)$
\end{define}\noindent%
Its precondition is $false$ while the postcondition requires that $ac'$ is not empty. The postcondition can alternatively be specified as $true$ since both $\mathbf{A}$ and $\mathbf{RA1}$ ensure that the design is $\mathbf{A0}$-healthy. This process is a zero for demonic choice as established by~\cref{theorem:Chaos-RAD-sqcap-P:Chaor-RAD}.
\theoremstatementref{theorem:Chaos-RAD-sqcap-P:Chaor-RAD}\noindent%
Similarly to the original theory, if a process may diverge immediately in a demonic choice, then this is the only possibility. The dual of this property is the unit law for angelic choice as shown in the following~\cref{theorem:RAP:Chaos-sqcup-P-RAD}.
\theoremref{theorem:RAP:Chaos-sqcup-P-RAD}\noindent%
When the angel is given the choice between diverging immediately or behaving as $P$, then the choice is resolved in favour of $P$. This is one of the fundamental properties underlying an angelic choice, in that, if possible, the angel can avoid divergence.

The bottom of the lattice is also in direct correspondence with that of the original theory of~\ac{CSP} as~\cref{theorem:ac2p(Chaos-RAD):Chaos-R,theorem:p2ac(Chaos-R):Chaos-RAD} establish.
\theoremstatementref{theorem:ac2p(Chaos-RAD):Chaos-R}
\theoremstatementref{theorem:p2ac(Chaos-R):Chaos-RAD}\noindent%
This is a reassuring result in that the bottom of the lattice of~\ac{CSP} also maps into the bottom of the lattice of reactive angelic designs and vice versa.

\subsection{Choice}\label{sec:ch5:operators:choice}
The next operator we introduce in this section corresponds to $Chaos$ in Roscoe's original presentation~\cite{Roscoe1998} of~\ac{CSP}, where it is the most nondeterministic process that does not diverge. In our model, this behaviour is given by $Choice_{\mathbf{RAD}}$.
\begin{define}
$Choice_{\mathbf{RAD}} \circdef \mathbf{RA} \circ \mathbf{A} (true \vdash ac'\neq\emptyset)$
\end{define}\noindent%
The precondition is $true$ while the postcondition allows any non-empty set of final states $ac'$. Similarly to the definition of $Chaos_{\mathbf{RAD}}$, and every other reactive angelic design, we observe that the complete behaviour of a process is constrained by~$\mathbf{RA}$ and thus the final states in $ac'$ must observe the properties enforced by $\mathbf{RA}$, notably that the traces are suffixes of the initial trace $s.tr$.

If we consider the design $Choice = (true \vdash true)$, then we can obtain a similar process in the theory of~\ac{CSP} by applying $\mathbf{R}$ as $Choice_{\mathbf{R}} = \mathbf{R} (true \vdash true)$. The application of $p2ac$ to this process yields $Choice_{\mathbf{RAD}}$ as shown in~\cref{theorem:p2ac(Choice-R):Choice-RAD}.
\theoremstatementref{theorem:p2ac(Choice-R):Choice-RAD}\noindent%
Likewise, \cref{theorem:ac2p(Choice-RAD):Choice-R} shows that applying $ac2p$ to $Choice_{\mathbf{RAD}}$ yields exactly the process $Choice_{\mathbf{R}}$ of the~\ac{CSP} model.
\theoremstatementref{theorem:ac2p(Choice-RAD):Choice-R}\noindent%
As is discussed later in~\cref{sec:ch5:non-divergent} the process $Choice_{\mathbf{RAD}}$ plays an important role in the characterisation of the subset of non-divergent processes. The intuition is that for non-divergent processes, the addition of more choices does not change those that are actually available for angelic choice, which are those in the distributed intersection over all permitted values of $ac'$. Consider the general result of the least upper bound and $Choice_{\mathbf{RAD}}$ in~\cref{theorem:Choice-RAD-sqcup-P}.
\theoremstatementref{theorem:Choice-RAD-sqcup-P}\noindent%
The precondition is $true$, while the postcondition $P^t_f$ is that of $P$. In other words, if $P$ could diverge, this is no longer possible in an angelic choice with $Choice_{\mathbf{RAD}}$.

Finally, when considering the greatest lower bound $\sqcap_{\mathbf{RAD}}$ and $Choice_{\mathbf{RAD}}$ we obtain the following result.
\theoremref{theorem:Choice-RAD-sqcap-P}\noindent%
The precondition of $P$ is maintained, while the postcondition requires a non-empty set of final states $ac'$. In other words, if there was a possibility to diverge in $P$, this is still the case. However, if the precondition $\lnot P^f_f$ is satisfied then the process behaves nondeterministically like $Choice_{\mathbf{RAD}}$.

\subsection{Stop}\label{sec:ch5:operators:stop}
Similarly to~\ac{CSP}, the notion of deadlock is captured by $Stop_{\mathbf{RAD}}$.
\begin{define}
$Stop_{\mathbf{RAD}} \circdef \mathbf{RA} \circ \mathbf{A} (true \vdash \circledIn{y}{ac'} (y.tr=s.tr \land y.wait))$
\end{define}\noindent%
The precondition is $true$ while the postcondition requires the process to always be waiting for the environment and keep the trace of events unchanged. In this definition and others to follow, we introduce the following auxiliary predicate.
\begin{define}\label{def:circledIn}
\begin{statement}
$\circledIn{y}{ac'}(P) \circdef \exists y \spot y \in ac' \land P[\{y\}/ac']$
\end{statement}
\end{define}\noindent%
This definition requires that $P$ admits a state $y$ as a single option for angelic choice. In general, this predicate allows the definition of~\ac{CSP} operators to be lifted into the theory of reactive angelic designs. It can be further extrapolated to other~\ac{CSP} operators, such as external choice.

An angelic choice between a process $P$ and $Stop_{\mathbf{RAD}}$ is, in general, not resolved in favour of either process as shown in~\cref{theorem:Stop-RAD-sqcup-P}.
\theoremref{theorem:Stop-RAD-sqcup-P}\noindent%
However, the possibility for divergence is avoided, since the precondition becomes $true$. If $P$ diverges, then the process behaves as $Stop_{\mathbf{RAD}}$, otherwise there is an angelic choice between $P$ or $Stop_{\mathbf{RAD}}$ which corresponds to the conjunction of their respective postconditions.

Finally, we can establish that the definition of $Stop_{\mathbf{RAD}}$ is in correspondence with $Stop_{\mathbf{R}}$ of~\ac{CSP} as established by~\cref{theorem:p2ac(Stop-R):Stop-RAD,theorem:ac2p(Stop-RAD):Stop-R}.
\theoremstatementref{theorem:p2ac(Stop-R):Stop-RAD}
\theoremstatementref{theorem:ac2p(Stop-RAD):Stop-R}\noindent%
This is a reassuring result that follows our intuition on using the auxiliary predicate $\circledIn{y}{ac'}$ to capture the definition of~\ac{CSP} operators in our new model.

\subsection{Skip}\label{sec:ch5:operators:skip}
The process that always terminates successfully is defined as $Skip_{\mathbf{RAD}}$.
\begin{define}
$Skip_{\mathbf{RAD}} \circdef \mathbf{RA}\circ\mathbf{A} (true \vdash \circledIn{y}{ac'} (\lnot y.wait \land y.tr=s.tr))$
\end{define}\noindent%
Its precondition is $true$ while the postcondition requires that there is a final state in $ac'$ such that the trace of events $s.tr$ is unchanged and that it terminates by requiring the component $wait$ to be $false$.

Similarly to the case with $Stop_{\mathbf{RAD}}$, the angelic choice between a process $P$ and $Skip_{\mathbf{RAD}}$ does not resolve in favour of either as~\cref{theorem:Skip-RAD-sqcup-P} shows.
\theoremstatementref{theorem:Skip-RAD-sqcup-P}\noindent%
However, the possibility for any divergence in $P$ is avoided. If $P$ diverges, then the angelic choice behaves as $Skip_{\mathbf{RAD}}$, otherwise the behaviour is given by the conjunction of the postconditions of $P$ and $Skip_{\mathbf{RAD}}$. We consider in~\cref{example:Stop-RAD-sqcup-Skip-RAD} an angelic choice between terminating and deadlocking.
\begin{example}\label{example:Stop-RAD-sqcup-Skip-RAD}
\begin{xflalign*}
	&Stop_{\mathbf{RAD}} \sqcup_{\mathbf{RAD}} Skip_{\mathbf{RAD}}
	&&\ptext{Definition of $Stop_{\mathbf{RAD}}$ and $Skip_{\mathbf{RAD}}$}\\
	&=\left(\begin{array}{l}
		\mathbf{RA} \circ \mathbf{A} (true \vdash \circledIn{y}{ac'} (y.tr=s.tr \land y.wait))
		\\ \sqcup_{\mathbf{RAD}} \\
		\mathbf{RA}\circ\mathbf{A} (true \vdash \circledIn{y}{ac'} (\lnot y.wait \land y.tr=s.tr))
	\end{array}\right)
	&&\ptext{\cref{theorem:RAP:P-sqcup-Q}}\\
	&=\mathbf{RA} \circ \mathbf{A} \left(\begin{array}{l}
		true \lor true
		\\ \vdash \\
		\left(\begin{array}{l}
			(true \implies \circledIn{y}{ac'} (y.tr=s.tr \land y.wait))
			\\ \land \\
			(true \implies \circledIn{y}{ac'} (\lnot y.wait \land y.tr=s.tr))
		\end{array}\right)
	\end{array}\right)
	&&\ptext{Predicate calculus}\\
	&=\mathbf{RA} \circ \mathbf{A} (true	\vdash
		\circledIn{y}{ac'} (y.tr=s.tr \land y.wait)	
		\land 
		\circledIn{y}{ac'} (\lnot y.wait \land y.tr=s.tr))
\end{xflalign*}
\end{example}\noindent%
In this case, the choice is not resolved by either process. If we map this example into the original theory of~\ac{CSP}, then we obtain the top $\top_{\mathbf{R}}$ of that lattice, defined by $\top_{\mathbf{R}} = \mathbf{R} (true \vdash false)$, as~\cref{lemma:ac2p(Stop-RAD-sqcup-Skip-RAD):top-R} establishes.
\theoremstatementref{lemma:ac2p(Stop-RAD-sqcup-Skip-RAD):top-R}\noindent%
This is because the result of mapping $Stop_{\mathbf{RAD}} \sqcup_{\mathbf{RAD}} Skip_{\mathbf{RAD}}$ through $ac2p$ insists on both waiting for an interaction and terminating. Likewise, if we map $\top_{\mathbf{R}}$ through $p2ac$, the top of the lattice of reactive angelic designs is obtained. Thus, this is an instance of the general strengthening indicated by~\cref{theorem:p2ac(ac2p(P)-sqcup-R-ac2p(Q)):P-sqcup-RAD-Q}. Although the miraculous process $\top_{\mathbf{R}}$ is not part of the standard~\ac{CSP} semantics~\cite{Roscoe1998,Roscoe2010} it plays an important role, for example, in the characterisation of deadline operators in the context of timed versions of process calculi~\cite{Woodcock2010,Wei2010,Wei2011,Wei2013}.

Finally, the definition of $Skip_{\mathbf{RAD}}$ can be be related with the original $Skip_{\mathbf{R}}$ process of~\ac{CSP} by applying $p2ac$ and $p2ac$ as established by~\cref{theorem:p2ac(Skip-R):Skip-RAD,theorem:ac2p(Skip-RAD):Skip-R}.
\theoremstatementref{theorem:p2ac(Skip-R):Skip-RAD}
\theoremstatementref{theorem:ac2p(Skip-RAD):Skip-R}\noindent%
In other words, as expected the two processes are in correspondence.

\subsection{Sequential Composition}\label{sec:ch5:operators:sequential-composition}
The definition of sequential composition is exactly $\seqDac$ from the theory of angelic designs, which is itself layered upon $\seqA$. When considering reactive angelic designs, we obtain the following closure result.
\theoremstatementref{theorem:RAP:seqDac}\noindent%
This is a result that resembles that for~\ac{CSP}, apart from the postcondition of the design. When $s.wait$ is $false$, and hence $P^t_f$ has finished its interaction with the environment, the behaviour is given by $\mathbf{RA2} \circ \mathbf{RA1} (\lnot Q^f_f \implies Q^t_f)$. In contrast with the result in~\ac{CSP} (\cref{sec:ch2:CSP:UTP}), this is an implication between the pre and postcondition of $Q$, instead of its postcondition.

As previously discussed in~\cref{sec:ch4:sequential-composition}, in the theory of angelic designs, the sequential composition operator also has a similar implication in the postcondition that acts as a filter by eliminating final states of $P$ that fail to satisfy the precondition of $Q$. For example, we consider the result established in~\cref{lemma:(Skip-RAD-sqcup-Stop-RAD)-seqDac-Chaos-RAD}.
\theoremstatementref{lemma:(Skip-RAD-sqcup-Stop-RAD)-seqDac-Chaos-RAD}\noindent%
In this case there is an angelic choice between deadlocking and terminating, followed by divergence. The angel avoids the divergence by choosing to deadlock. The precondition of $Chaos_{\mathbf{RAD}}$ is unsatisfiable since it is $false$. Once the preceding process of the sequential composition terminates, that is the component $wait$ is $false$, then the composition diverges. However, because the angel can choose the non-terminating process $Stop_{\mathbf{RAD}}$, the divergence can be avoided.

In general, when considering the result of applying the sequential composition of~\ac{CSP} to two processes $P$ and $Q$ mapped through $ac2p$, followed by $p2ac$, a strengthening is obtained as established by the following~\cref{theorem:p2ac(ac2p(P)-seq-ac2p(Q)):sqsupseteq:P-seqDac-Q}.
\theoremref{theorem:p2ac(ac2p(P)-seq-ac2p(Q)):sqsupseteq:P-seqDac-Q}\noindent%
We consider, for example, the case of the processes of~\cref{lemma:(Skip-RAD-sqcup-Stop-RAD)-seqDac-Chaos-RAD}. As previously discussed in~\cref{sec:ch5:operators:skip}, the result of $ac2p(Skip_{\mathbf{RAD}} \sqcup_{\mathbf{RAD}} Stop_{\mathbf{RAD}})$ is the top $\top_{\mathbf{R}}$ of the lattice of reactive designs (\cref{lemma:ac2p(Stop-RAD-sqcup-Skip-RAD):top-R}). The result of applying $ac2p(Chaos_{\mathbf{RAD}})$ is the bottom $Chaos_{\mathbf{R}}$ as established by~\cref{theorem:ac2p(Chaos-RAD):Chaos-R}. The sequential composition of $\top_{\mathbf{R}}$ followed by $Chaos_{\mathbf{R}}$ is also $\top_{\mathbf{R}}$. Applying $p2ac(\top_{\mathbf{R}})$ yields the top of the lattice of reactive angelic designs $\top_{\mathbf{RAD}} \circdef \mathbf{RA}\circ\mathbf{A} (true \vdash false)$. This is a trivial refinement of any process, including $Stop_{\mathbf{RAD}}$.

If we strengthen the assumption of~\cref{theorem:p2ac(ac2p(P)-seq-ac2p(Q)):sqsupseteq:P-seqDac-Q} by considering the case where both $P$ and $Q$ are, in addition, $\mathbf{A2}$-healthy, then an equality is obtained instead as established by~\cref{theorem:p2ac(ac2p(P)-seq-ac2p(Q))-A2:P-seqDac-Q}.
\theoremstatementref{theorem:p2ac(ac2p(P)-seq-ac2p(Q))-A2:P-seqDac-Q}\noindent%
This is because $\mathbf{A2}$-healthy processes do not have angelic nondeterminism, and so the result obtained in both models is exactly the same.

When considering two~\ac{CSP} processes $P$ and $Q$, we also obtain an equality as shown in the following~\cref{theorem:ac2p(p2ac(P)-seqDac-p2ac(Q)):P-seq-Q}.
\theoremstatementref{theorem:ac2p(p2ac(P)-seqDac-p2ac(Q)):P-seq-Q}\noindent%
This result confirms the correspondence of sequential composition in both models. In particular, the result of sequentially composing two~\ac{CSP} processes with no angelic nondeterminism can be directly calculated in the new model.

Finally, the sequential composition operator is closed under $\mathbf{A2}$ for reactive angelic designs as shown in the following~\cref{theorem:RAD:A2(P-seqDac-Q):P-seqDac-Q}.
\theoremstatementref{theorem:RAD:A2(P-seqDac-Q):P-seqDac-Q}\noindent%
Therefore, given any two reactive angelic designs $P$ and $Q$ with no angelic nondeterminism, their sequential composition does not introduce any angelic choices. This concludes our discussion of the sequential composition operator.

\subsection{Prefixing}\label{sec:ch5:operators:prefixing}
Having discussed the definition of sequential composition, in this section we introduce the definition of event prefixing, which is similar to that of~\ac{CSP}.
\begin{define}
\begin{align*}
	&a \circthen_{\mathbf{RAD}} Skip_{\mathbf{RAD}} \circdef \mathbf{RA} \circ \mathbf{A} 
		\left(true 
				\vdash 
					\circledIn{y}{ac'} \left(\begin{array}{l}(y.tr=s.tr \land a \notin y.ref)
						\\ \dres y.wait \rres \\
					(y.tr = s.tr \cat \lseq a \rseq)
					\end{array}\right)
		\right)
\end{align*}
\end{define}\noindent%
The precondition is $true$, while the postcondition is split into two cases. When the process is waiting for an interaction from the environment, that is, $y.wait$ is $true$, then $a$ is not in the set of refusals and the trace $s.tr$ is kept unchanged. While in the second case, the process has interacted with the environment, and so the only guarantee is that the event $a$ is part of the final trace $y.tr$.

Like for~$Stop_{\mathbf{RAD}}$ and $Skip_{\mathbf{RAD}}$, an angelic choice between a process $P$ and $a \circthen_{\mathbf{RAD}} Skip_{\mathbf{RAD}}$ avoids divergence as established by~\cref{theorem:a-then-Skip-RAD-sqcup-P}.
\theoremstatementref{theorem:a-then-Skip-RAD-sqcup-P}\noindent%
The complete behaviour of this process depends on that of $P$ as well. If $P$ diverges, then the process behaves as $a \circthen_{\mathbf{RAD}} Skip_{\mathbf{RAD}}$, otherwise there is an angelic choice between the behaviour of $a \circthen_{\mathbf{RAD}} Skip_{\mathbf{RAD}}$ and $P$.

Event prefixing in both theories is in exact correspondence as established by the following~\cref{theorem:ac2p(a-then-Skip-RAD):a-then-Skip-R,theorem:p2ac(a-then-Skip-R):a-then-Skip-RAD}.
\theoremstatementref{theorem:ac2p(a-then-Skip-RAD):a-then-Skip-R}
\theoremstatementref{theorem:p2ac(a-then-Skip-R):a-then-Skip-RAD}\noindent%
This is expected since event prefixing, even in the presence of angelic nondeterminism, does not behave differently to prefixing in the original theory of~\ac{CSP}.

In order to illustrate the behaviour of angelic choice we consider the following examples. In~\cref{example:a-stop-angelic-skip} we have a choice between terminating and deadlocking following event $a$, sequentially composed with $Chaos_{\mathbf{RAD}}$. In general, the process $a \circthen_{\mathbf{RAD}} P$ denotes the compound process $a \circthen_{\mathbf{RAD}} Skip_{\mathbf{RAD}} \seqDac P$, whose result as a reactive angelic design is established by~\cref{theorem:RAP:a-circthen-RA-P}.
\theoremstatementref{theorem:RAP:a-circthen-RA-P}\noindent%
The precondition states that it is not the case that once event $a$ occurs the precondition of $P$ fails to be satisfied. While the postcondition considers two cases: when the process is waiting for the environment the trace of events is kept unchanged and event $a$ is not refused; when he process does event $a$, then the result is that of the postcondition of $P$ with initial state $y$, where the trace $y.tr$ includes event $a$.
\begin{example}\label{example:a-stop-angelic-skip}
\begin{align*}
	&((a \circthen_{\mathbf{RAD}} Stop_{\mathbf{RAD}}) \sqcup_{\mathbf{RAD}} Skip_{\mathbf{RAD}}) \seqDac Chaos_{\mathbf{RAD}}\\
	&=\\
	&a \circthen_{\mathbf{RAD}} Stop_{\mathbf{RAD}}
\end{align*}
\begin{proof}
\cref{lemma:RAD:((a-circthen-Stop)-sqcup-Skip)-seqDac-Chaos:a-circthen-Stop}
\end{proof}
\end{example}\noindent%
In the case of~\cref{example:a-stop-angelic-skip}, the angel avoids divergence by choosing non termination by allowing the environment to perform the event $a$ and then deadlocking. In \cref{example:angelic-chaos} there is a choice between terminating or diverging upon performing the event $a$.
\begin{example}\label{example:angelic-chaos}
\begin{flalign*}
	&(a \circthen_{\mathbf{RAD}} Skip_{\mathbf{RAD}}) \sqcup_{\mathbf{RAD}} (a \circthen_{\mathbf{RAD}} Chaos_{\mathbf{RAD}})
	&&\ptext{Definition of prefixing and \cref{theorem:RAP:a-circthenRA-ChaosRA}}\\
	&=\left(\begin{array}{l}
		\mathbf{RA} \circ \mathbf{A} \left(true 
			\vdash 
				\left(\begin{array}{l}
					\circledIn{y}{ac'} (y.wait \land y.tr=s.tr \land a \notin y.ref)
					\\ \lor \\
					\circledIn{y}{ac'} (\lnot y.wait \land y.tr = s.tr \cat \lseq a \rseq)
				\end{array}\right)
			\right)
			\\ \sqcup \\
		\mathbf{RA} \circ \mathbf{A} \left(\begin{array}{l}
			\lnot \circledIn{y}{ac'} (s.tr \cat \lseq a \rseq \le y.tr) 
			\\ \vdash \\
			 \circledIn{y}{ac'} (y.wait \land y.tr=s.tr \land a \notin y.ref)
		\end{array}\right)
	\end{array}\right)
	&&\ptext{\cref{theorem:RAP:P-sqcup-Q} and predicate calculus}\\
	&=\mathbf{RA} \circ \mathbf{A} \left(true 
			\vdash
			\left(\begin{array}{l} 
				\left(\begin{array}{l}
					\circledIn{y}{ac'} (y.wait \land y.tr=s.tr \land a \notin y.ref)
					\\ \lor \\
					\circledIn{y}{ac'} (\lnot y.wait \land y.tr = s.tr \cat \lseq a \rseq)
				\end{array}\right)
				\\ \land \\
				\left(\begin{array}{l}
					\circledIn{y}{ac'} (s.tr \cat \lseq a \rseq \le y.tr)
					\\ \lor \\
					\circledIn{y}{ac'} (y.wait \land y.tr=s.tr \land a \notin y.ref)
				\end{array}\right)
			\end{array}\right)\right)
	&&\ptext{Predicate calculus}\\
	&=\mathbf{RA} \circ \mathbf{A} \left(true 
			\vdash
			\left(\begin{array}{l} 
				\circledIn{y}{ac'} (y.wait \land y.tr=s.tr \land a \notin y.ref)
				\\ \lor \\
				\circledIn{y}{ac'} (\lnot y.wait \land y.tr = s.tr \cat \lseq a \rseq)
				\end{array}\right)
			\right)
	&&\ptext{Definition of prefixing}\\
	&=a \circthen_{\mathbf{RAD}} Skip_{\mathbf{RAD}}
\end{flalign*}
\end{example}\noindent%
The result is a process that following event $a$ can only terminate, and thus avoids divergence. This property illustrates that our angelic choice operator is a counterpart to that of the refinement calculus. It resolves choices to avoid divergence but here we have choices over interactions.

However, if we consider the processes of~\cref{example:angelic-chaos} to be prefixes on different events, the result of the angelic choice is rather different as shown in~\cref{example:a-then-Skip-sqcup-b-then-Chaos}.
\begin{example}\label{example:a-then-Skip-sqcup-b-then-Chaos}
\begin{align*}
	&(a \circthen_{\mathbf{RAD}} Skip_{\mathbf{RAD}}) \sqcup_{\mathbf{RAD}} (b \circthen_{\mathbf{RAD}} Chaos_{\mathbf{RAD}})\\
	&=\\
	&(a \circthen_{\mathbf{RAD}} Skip_{\mathbf{RAD}}) \sqcup_{\mathbf{RAD}} (b \circthen_{\mathbf{RAD}} Choice_{\mathbf{RAD}})
\end{align*}
\begin{proof}
\cref{lemma:RAD:(a-circthen-Skip)-sqcup-(b-circthen-Chaos)}
\end{proof}
\end{example}\noindent%
In this case, the possibility of diverging after the event $a$ is avoided by turning $Chaos_{\mathbf{RAD}}$ into $Choice_{\mathbf{RAD}}$. The possibility for engaging in the event $a$ cannot be avoided by the angel, since $\mathbf{RA1}$ requires that under all circumstances no trace of events may be undone. Ideally for a counterpart to the angelic choice of the refinement calculus, it should be possible to discard any trace of events that lead to divergence. This is the motivation for the theory of angelic processes that we introduce in the following~\cref{chapter:6}.

\subsection{External Choice}\label{sec:ch5:operators:external-choice}
External choice, which offers the environment the choice over the events initially offered by processes $P$ and $Q$, is similarly (\cref{sec:ch2:UTP:CSP-Operators}) defined in our theory as follows.
\begin{define}
\begin{xflalign*}
	&P \extchoice_{\mathbf{RAD}} Q\\%
	&\circdef\\%
	&\mathbf{RA} \circ \mathbf{A} \left(\begin{array}{l}%
		(\lnot P^f_f \land \lnot Q^f_f)%
		\\ \vdash \\%
		\circledIn{y}{ac'}((P^t_f \land Q^t_f) \dres y.tr=s.tr \land y.wait \rres (P^t_f \lor Q^t_f))%
	\end{array}\right)%
\end{xflalign*}
\end{define}\noindent%
The precondition is the conjunction of the preconditions of the processes $P$ and $Q$, while the postcondition is split into two cases. When the process is waiting and the trace of events $s.tr$ is unchanged, then the behaviour is given by the conjunction of both postconditions, otherwise it is given by their disjunction. In other words, before the process performs any event, $P$ and $Q$ must be in agreement. In particular, if there is angelic nondeterminism in either $P$ or $Q$, there must be an agreement on a single common state in $ac'$.

Once the process has finished interacting with the environment or performed an event, there is a choice between $P$ and $Q$. Even if there is angelic nondeterminism in either $P$ or $Q$, then there is also a requirement for there to be an agreement on a final state, as enforced by the lifting $\circledIn{y}{ac'}$. We consider, for example, the following result on the external choice between a reactive angelic design and $Stop_{\mathbf{RAD}}$.
\theoremstatementref{theorem:RAD:P-extchoice-Stop}\noindent%
That is, the angelic nondeterminism of $P$ is collapsed. Unlike in the original theory of~\ac{CSP}, $Stop_{\mathbf{RAD}}$ is not necessarily a unit for external choice. However, when considering the subset of reactive angelic designs corresponding to~\ac{CSP} processes, which are the $\mathbf{A2}$-healthy, then $Stop_{\mathbf{RAD}}$ is a unit as expected.
\theoremstatementref{theorem:RAD:P-A2-extchoice-Stop:P-A2}\noindent%
\cref{theorem:RAD:P-A2-extchoice-Stop:P-A2} follows from the correspondence of the operator in both models, which we discuss below, and the proviso which ensures that there is no angelic nondeterminism in $P$.

As established by the following~\cref{theorem:ac2p(p2ac(P)-extchoiceONE-p2ac(Q)):P-extchoice-R-Q} the result of mapping two~\ac{CSP} processes $P$ and $Q$ through $p2ac$ and composing them with the external choice operator $\extchoice_{\mathbf{RAD}}$ of reactive angelic designs, followed by the mapping $ac2p$ in the opposite direction is exactly the same as applying $\extchoice_{\mathbf{R}}$ to the original processes.
\theoremstatementref{theorem:ac2p(p2ac(P)-extchoiceONE-p2ac(Q)):P-extchoice-R-Q}\noindent%
However, if we consider the application in the opposite direction in the following~\cref{theorem:p2ac(ac2p(P)-extchoice-acp(Q))}, the result obtained is not an equality.
\theoremstatementref{theorem:p2ac(ac2p(P)-extchoice-acp(Q))}\noindent%
This establishes that by considering two reactive angelic designs, applying $ac2p$ to both, composing the result with the external choice operator of~\ac{CSP}, and then mapping back through $p2ac$, the result obtained is stronger than the respective composition using $\extchoice_{\mathbf{RAD}}$. This follows from the fact that, since $P$ and $Q$ can be nondeterministic, and external choice is monotonic with respect to refinement, the application of $ac2p$ may yield stronger processes.
 
We consider the following~\cref{example:RAD:a-then-Chaos-sqcup-b-then-Chaos-extchoice-Stop} in the context of~\cref{theorem:p2ac(ac2p(P)-extchoice-acp(Q))}. Here we have an angelic choice between engaging in an event $a$ or an event $b$ followed by divergence, with $Stop_{\mathbf{RAD}}$ in an external choice.
\begin{example}\label{example:RAD:a-then-Chaos-sqcup-b-then-Chaos-extchoice-Stop}
\begin{align*}
	&(a \circthen_{\mathbf{RAD}} Chaos_{\mathbf{RAD}} \sqcup_{\mathbf{RAD}} b \circthen_{\mathbf{RAD}} Chaos_{\mathbf{RAD}}) \extchoice_{\mathbf{RAD}} Stop_{\mathbf{RAD}}\\
	&=\\
	&\mathbf{RA}\circ\mathbf{A} \left(\begin{array}{l}
		\lnot (
				\circledIn{y}{ac'} (s.tr \cat \lseq a \rseq \le y.tr)
				\land 
				\circledIn{y}{ac'} (s.tr \cat \lseq b \rseq \le y.tr)
			)
		\\ \vdash \\
		\circledIn{y}{ac'} (y.wait \land y.tr=s.tr \land a \notin y.ref \land b \notin y.ref)
	\end{array}\right)
\end{align*}
\begin{proof}
\cref{lemma:RAD:(a-circthen-Chaos-sqcup-b-circthen-Chaos)-extchoice-Stop}
\end{proof}
\end{example}\noindent%
The precondition requires that there is not a final state where the trace includes the event $a$ or the event $b$. The postcondition states that the process is always waiting for the environment, while keeping the trace of events unchanged and not refusing either $a$ or $b$. The mapping through $ac2p$ of the left-hand side of~\cref{example:RAD:a-then-Chaos-sqcup-b-then-Chaos-extchoice-Stop} yields a~\ac{CSP} process whose precondition is $true$ as shown in the following~\cref{example:ac2p(a-circthen-Chaos-sqcuo-b-circthen-Chaos)}.
\begin{example}\label{example:ac2p(a-circthen-Chaos-sqcuo-b-circthen-Chaos)}
\begin{align*}
	&ac2p(a \circthen_{\mathbf{RAD}} Chaos_{\mathbf{RAD}} \sqcup_{\mathbf{RAD}} b \circthen_{\mathbf{RAD}} Chaos_{\mathbf{RAD}})\\
	&=\\
	&\mathbf{R} (true \vdash tr'=tr \land wait' \land a \notin ref' \land b \notin ref')
\end{align*}
\begin{proof}
\cref{lemma:ac2p(a-circthen-Chaos-sqcuo-b-circthen-Chaos)}
\end{proof}
\end{example}\noindent%
The postcondition, expressed in the theory of reactive designs, is similar to that of~\cref{example:RAD:a-then-Chaos-sqcup-b-then-Chaos-extchoice-Stop}. The mapping of~\cref{example:ac2p(a-circthen-Chaos-sqcuo-b-circthen-Chaos)} through $p2ac$ yields a refinement of the reactive angelic design of~\cref{example:RAD:a-then-Chaos-sqcup-b-then-Chaos-extchoice-Stop}. This is an expected result, which follows from the general result of~\cref{theorem:p2ac(ac2p(P)-extchoice-acp(Q))}.

If we consider reactive angelic designs that are in addition $\mathbf{A2}$-healthy, an equality is obtained as established by~\cref{theorem:p2ac(ac2p(P)-exthoice-R-ac2p(Q)):P-extchoiceONE-Q}.
\theoremstatementref{theorem:p2ac(ac2p(P)-exthoice-R-ac2p(Q)):P-extchoiceONE-Q}\noindent%
Furthermore, the external choice operator is also closed under $\mathbf{A2}$ as established by~\cref{theorem:A2(P-extchoiceOne-Q):P-extchoiceOne-Q}.
\theoremstatementref{theorem:A2(P-extchoiceOne-Q):P-extchoiceOne-Q}\noindent%
In other words, the definition of external choice is in correspondence between both models for processes with no angelic nondeterminism.

\section{Non-divergent Reactive Angelic Designs}\label{sec:ch5:non-divergent}
As previously discussed in~\cref{chapter:1}, and as part of our approach to studying the relationship between theories, it is useful to identify the subset of non-divergent reactive angelic designs. These are processes that satisfy the following healthiness condition $\mathbf{ND_{RAD}}$.
\begin{define}\label{def:NDRAD}
\begin{statement}
$\mathbf{ND_{RAD}} (P) = P \sqcup_{\mathbf{RAD}} Choice_{\mathbf{RAD}}$
\end{statement}
\end{define}\noindent%
This function is defined using the least upper bound of the lattice $\sqcup_{\mathbf{RAD}}$ and the most nondeterministic process $Choice_{\mathbf{RAD}}$ that does not diverge. The intuition underlying $\mathbf{ND_{RAD}}$ is that, for a given process $P$, increasing the number of final states available for angelic choice, does not actually add any new choices, unless the process $P$ could itself diverge. We consider the following~\cref{example:NDRAD(Chaos)} where the function $\mathbf{ND_{RAD}}$ is applied to the bottom of the lattice $Chaos_{\mathbf{RAD}}$.
\begin{example}\label{example:NDRAD(Chaos)}
$\mathbf{ND_{RAD}} (Chaos_{\mathbf{RAD}}) = Choice_{\mathbf{RAD}}$
\begin{proof}
\cref{lemma:RAD:NDRAD(Chaos):Choice}
\end{proof}
\end{example}\noindent%
The divergence is avoided and the result is the process $Choice_{\mathbf{RAD}}$. If instead we consider a process that is not divergent, such as $Skip_{\mathbf{RAD}}$, the result is as follows.
\begin{example}\label{example:NDRAD(a-then-Skip)}
$\mathbf{ND_{RAD}} (a \circthen_{\mathbf{RAD}} Skip_{\mathbf{RAD}}) = a \circthen_{\mathbf{RAD}} Skip_{\mathbf{RAD}}$
\begin{proof}
\cref{lemma:RAD:NDRAD(a-then-Skip):a-then-Skip}
\end{proof}
\end{example}\noindent%
The process is a fixed point of $\mathbf{ND_{RAD}}$.

The function $\mathbf{ND_{RAD}}$ is idempotent as shown in the following~\cref{theorem:NDRA-idempotent}.
\theoremref{theorem:NDRA-idempotent}\noindent%
More importantly, when considering a reactive angelic design $P$, \cref{theorem:RAP:P-sqcup-Choice:(true|Pt)} establishes that the application of $\mathbf{ND_{RAD}}$ to a reactive angelic design $P$ requires the precondition of the design to be $true$.
\theoremstatementref{theorem:RAP:P-sqcup-Choice:(true|Pt)}\noindent%
Furthermore, if we consider the fixed points of $\mathbf{ND_{RAD}}$ then we obtain the following result in~\cref{theorem:RAP:P-sqcup-Choice:iff:forall-s-ac'-lnot-Pff}.
\theoremstatementref{theorem:RAP:P-sqcup-Choice:iff:forall-s-ac'-lnot-Pff}\noindent%
That is, it must be the case that the precondition $\lnot P^f_f$ of the reactive angelic design $P$ is satisfied for every possible initial state $s$ and set of final states $ac'$. These complementary results confirm our intuition about the definition of $\mathbf{ND_{RAD}}$.

\section{Final Considerations}\label{sec:ch5:final-considerations}
Based on the underlying principles of the theory of~\ac{CSP}~\cite{Hoare1998,Cavalcanti2006a} and the model of angelic designs presented in~\cref{chapter:4}, in this chapter we have presented a model for~\ac{CSP} where both angelic and demonic nondeterminism can be expressed. The approach we have followed consists of a natural extension to the existing~\ac{CSP} model. First we have encoded the observational variables of the theory of reactive processes and enforced all of the healthiness conditions of the original model in this new theory. Similarly to the original theory of~\ac{CSP} we have shown how~\ac{CSP} processes can be specified through reactive angelic designs. We have then established links with the original theory and studied this relationship.

We have established that there is a Galois connection between the theory of reactive angelic designs and~\ac{CSP}. In addition, when considering the subset of processes that are $\mathbf{A2}$-healthy, this relationship can be strengthened into a bijection. We have studied the most important operators of the theory and shown that they are in correspondence with their~\ac{CSP} counterparts. Furthermore, we have also proposed a natural way for specifying existing~\ac{CSP} operators in this new theory, including, for example, the external choice operator. While the definition of the external choice operator preserves the semantics of~\ac{CSP}, it is not the only one possible. Indeed, we hypothesize that there are other plausible semantic-preserving definitions for external choice with different algebraic properties. For example, when considering an external choice which includes angelic choices it may be desirable to allow the environment to choose any of those choices.

Finally, a number of examples have been presented to illustrate the role of angelic choice in a theory of~\ac{CSP}. In particular, we have shown that whenever possible, angelic choice avoids divergence. This behaviour is closer in spirit to that of the original choice operator of the refinement calculus than that of any other notion of angelic choice for~\ac{CSP} which we are aware. However, this avoidance still preserves any potential sequence of observable events. Ideally, the counterpart to the angelic choice of the refinement calculus should avoid any divergent behaviour altogether. For example, in the case of~\cref{example:a-then-Skip-sqcup-b-then-Chaos} the angelic choice should be resolved in favour of $a \circthen_{\mathbf{RAD}} Skip_{\mathbf{RAD}}$. This is the motivation for the theory of angelic processes which we discuss in the next~\cref{chapter:6}.

\chapter{Angelic Processes}\label{chapter:6}
Following from the impossibility for the angel to completely avoid divergent processes in the theory of reactive angelic designs, and based on its underlying principles, in this chapter we present a different approach to characterising~\ac{CSP} processes with angelic nondeterminism. The result is a theory which better accommodates the angelic choice over divergent processes, in that the resulting algebraic properties are closer in spirit to the angelic choice of the refinement calculus. In~\cref{sec:ch6:introduction} we revisit the motivation for this theory and discuss our approach. \cref{sec:ch6:healthiness-conditions} introduces the healthiness conditions of the theory and discusses their relationship with the theory of reactive angelic designs. In~\cref{sec:ch6:rel-RAD} we study the relationship between the two models and establish that the subsets of non-divergent processes are isomorphic. In~\cref{sec:ch6:operators} we present operators of this model and discuss some of their properties as well as their relationship with counterparts in the theory of reactive angelic designs. Finally, the chapter ends with a summary of the results in~\cref{sec:ch6:final-considerations}.

\section{Introduction}\label{sec:ch6:introduction}
As previously discussed in~\cref{chapter:5}, in the theory of reactive angelic designs, healthy processes, as required by $\mathbf{RA1}$, must never undo the history of events. For example, the definition of $Chaos_{\mathbf{RAD}}$, which diverges immediately, guarantees that there is always a final state in $ac'$ where the trace of events is a suffix of the initial trace $s.tr$. This behaviour is as expected for a theory of processes.

Since angelic choice is defined as the least upper bound, and~$Chaos_\mathbf{RAD}$ is the bottom of the lattice of reactive angelic designs, it follows that immediate divergence is avoided, if possible, by the angel. However, once there is the possibility for interacting with the environment, such as in the case of~\cref{example:a-stop-angelic-skip}, the possibility for performing an event followed by divergence cannot be eliminated completely, as doing so would violate $\mathbf{RA1}$. This is unlike the angelic choice of the refinement calculus and the theory of angelic designs, where angelic choices leading to divergence are pruned altogether.

In this chapter we propose a theory like $\mathbf{RAD}$, but which does not necessarily enforce $\mathbf{RA1}$ when a process diverges. This is a departure from the norm for a theory of~\ac{CSP}. The main consequence of this approach is that divergent processes have a different semantics to standard~\ac{CSP}. However, the subset of non-divergent processes preserves the existing semantics defined by $\mathbf{RAD}$, and by extension, the semantics of non-divergent~\ac{CSP} processes. 


\section{Healthiness Conditions}\label{sec:ch6:healthiness-conditions}
The alphabet of angelic processes is exactly the same as that of reactive angelic designs. Namely, we have variables $ok$, $ok'$, $s$ and $ac'$, where a $State$ is defined with components $tr$, $ref$ and $wait$.

As with every~\ac{UTP} theory, we define the healthiness conditions. Since we aim to define a theory like~$\mathbf{RAD}$, but without necessarily enforcing $\mathbf{RA1}$, we focus our attention on the definition of $\mathbf{RAD}$, which we reproduce below.
\[ \mathbf{RAD} (P) \circdef \mathbf{RA1} \circ \mathbf{RA2} \circ \mathbf{RA3} \circ \mathbf{CSPA1} \circ \mathbf{CSPA2} \circ \mathbf{PBMH} (P) \]
If we simply remove $\mathbf{RA1}$ from the functional composition, then $\mathbf{A0}$ is not necessarily enforced any more, and thus successful termination does not guarantee that $ac'$ is not empty. Furthermore, $\mathbf{CSPA1}$ is also stronger than required, since when in an unstable state, that is $\lnot ok$, $\mathbf{RA1}$ should not be enforced. Equally, the identity $\IIRac$ and, therefore, $\mathbf{RA3}$ also need to be changed, so that divergence no longer requires $\mathbf{RA1}$. This leads us to the following healthiness condition $\mathbf{AP}$.
\begin{define}\label{def:AP}
\begin{statement}
$\mathbf{AP} (P) \circdef \mathbf{RA3_{AP}}\circ\mathbf{RA2}\circ\mathbf{A}\circ\mathbf{H1}\circ\mathbf{CSPA2} (P)$
\end{statement}
\end{define}\noindent%
The healthiness condition $\mathbf{RA3}$ is replaced with $\mathbf{RA3_{AP}}$, which does not require $\mathbf{RA1}$. The function $\mathbf{A}$ is included in the functional composition since it enforces both $\mathbf{A0}$ and $\mathbf{A1}$ (itself $\mathbf{PBMH}$ as previously discussed in~\cref{sec:ch4:A1}) as required. The function $\mathbf{CSPA1}$ is replaced with $\mathbf{H1}$, since in an unstable state, that is when $\lnot ok$ is $true$, $\mathbf{RA1}$ is no longer enforced. Finally $\mathbf{CSPA2}$ is enforced like in $\mathbf{RAD}$.

The definition of $\mathbf{RA3_{AP}}$ is introduced in the following~\cref{sec:ch6:RA3AP}. In~\cref{sec:ch6:AP} the definition of $\mathbf{AP}$ is explored in more detail. Finally in~\cref{sec:ch6:ND-AP} the subset of non-divergent angelic processes is characterised by another healthiness condition $\mathbf{ND_{AP}}$.

\subsection{Redefining $\mathbf{RA3}$ as $\mathbf{RA3_{AP}}$}\label{sec:ch6:RA3AP}
Similarly to the theory of reactive angelic designs, we define a new identity $\IIRnew$ as follows.
\begin{define}
$\IIRnew \circdef \mathbf{H1} (ok' \land s \in ac')$
\end{define}\noindent%
In contrast with the definition for $\IIRac$, there is no longer a requirement for $\mathbf{RA1}$ to be enforced when the process is unstable and $ok$ is $false$. Instead, the only guarantee in this case is that if the process is stable, and $ok$ is $true$, then stability is maintained and the state is kept unchanged, by requiring the initial state $s$ to be in the set of final states $ac'$.

The definition of $\mathbf{RA3_{AP}}$ is similar to $\mathbf{RA3}$ except that we use the identity $\IIRnew$, which does not enforce $\mathbf{RA1}$, instead of $\IIRac$.
\begin{define}\label{def:RA3AP}
\begin{statement}
$\mathbf{RA3_{AP}} (P) \circdef \IIRnew \dres s.wait \rres P$
\end{statement}
\end{define}\noindent%
The function $\mathbf{RA3_{AP}}$ is idempotent and monotonic as established by the following~\cref{theorem:RA3N:idempotent,theorem:RA3N:monotonic}. Proof of these and other theorems to follow, which are not included explicitly in the body of this thesis, can be found in \cref{appendix:AP}.
\theoremstatementref{theorem:RA3N:idempotent}
\theoremstatementref{theorem:RA3N:monotonic}\noindent%
Furthermore, it distributes through both conjunction and disjunction.
\theoremstatementref{theorem:RA3AP(P-land-Q):RA3AP(P)-land-RA3AP(Q)}
\theoremstatementref{theorem:RA3AP(P-lor-Q):RA3AP(P)-lor-RA3AP(Q)}\noindent%
Since $\mathbf{RA3_{AP}}$ is idempotent and distributes through both conjunction and disjunction, conjunction and disjunction are closed under~$\mathbf{RA3_{AP}}$. More importantly, the operator $\seqA$ is closed under $\mathbf{RA3_{AP}}$.
\theoremstatementref{theorem:RA3AP(P-seqA-Q):P-seqA-Q}\noindent%
Finally, $\mathbf{RA3_{AP}}$ commutes with $\mathbf{PBMH}$, and $\mathbf{RA2}$ as established by the following~\cref{theorem:RA3N-o-PBMH(P):PBMH-o-RA3N(P),theorem:RA2-o-RA3N(P):RA3N-o-RA2(P)}
\theoremref{theorem:RA3N-o-PBMH(P):PBMH-o-RA3N(P)}
\theoremstatementref{theorem:RA2-o-RA3N(P):RA3N-o-RA2(P)}\noindent%
\cref{theorem:RA3N-o-PBMH(P):PBMH-o-RA3N(P)} is important in establishing that $\mathbf{RA3_{AP}}$ preserves the upward-closure of $\mathbf{PBMH}$. This is established by~\cref{lemma:PBMH-o-RA3N-o-PBMH(P):RA3N-o-PBMH(P)}.
\theoremstatementref{lemma:PBMH-o-RA3N-o-PBMH(P):RA3N-o-PBMH(P)}\noindent%
This concludes our discussion of the most important properties of $\mathbf{RA3_{AP}}$.

\subsection{Angelic Processes ($\mathbf{AP}$)}\label{sec:ch6:AP}
As already mentioned, the theory of angelic processes is characterised by the functional composition of $\mathbf{RA3_{AP}}$, $\mathbf{RA2}$, $\mathbf{A}$, $\mathbf{H1}$ and $\mathbf{CSPA2}$. A parallel result to that of the theory of reactive angelic designs (\cref{theorem:RA-o-A(design):RA-CSPA-PBMH}) can be obtained as established by the following~\cref{theorem:RAPN(P):RAPN(Pff-vdash-Ptf)}: $\mathbf{AP}$ processes can also be expressed in terms of a design.
\theoremref{theorem:RAPN(P):RAPN(Pff-vdash-Ptf)}\noindent%
This result establishes that an angelic process can also be specified in terms of pre and postconditions, as the image of a design through the functions $\mathbf{RA3_{AP}}$, $\mathbf{RA2}$ and $\mathbf{A}$. Since these functions are all idempotent and monotonic, and the theory of designs is a complete lattice~\cite{Hoare1998}, so is the theory of angelic processes.

The original theory of~\ac{CSP} is not a theory of designs, since when $ok$ is $false$, $\mathbf{R1}$ must hold, unlike in the theory of designs, where $\mathbf{H1}$ requires that no meaningful observations can be made about a design unless it is started, that is, unless $ok$ is $true$. Here, since we have dropped $\mathbf{RA1}$, in fact the theory we propose is a theory of angelic designs as established by the following~\cref{theorem:RAPN(P)}.
\theoremref{theorem:RAPN(P)}\noindent%
The precondition of the design has a conditional on $s.wait$. If the previous process has not terminated interacting with the environment, then this is simply $true$. Otherwise, the original precondition of $P$ must be satisfied, and its negation must be $\mathbf{PBMH}$ and $\mathbf{RA2}$-healthy. We recall that in a non-$\mathbf{H3}$ design it is actually the negation of the precondition that is established irrespective of termination.

The postcondition of an angelic process also has a conditional on $s.wait$. When the previous process has not terminated its interactions with the environment, then the state is kept unchanged by making sure that the initial state $s$ is in the set of final states $ac'$. Otherwise, the original postcondition of $P$ holds and must be $\mathbf{PBMH}$, $\mathbf{RA2}$ and $\mathbf{RA1}$-healthy.

Although we have dropped $\mathbf{RA1}$ because the postcondition requires that the set of final states $ac'$ is not empty, and since we enforce $\mathbf{RA2}$, this means that $\mathbf{RA1}$ is enforced in the postcondition (\cref{lemma:RA2(ac'-neq-emptyset):RA1(true)}). Similarly, if the negation of the precondition imposes any particular set of final states $ac'$, because it must also be $\mathbf{RA2}$-healthy, it will also enforce $\mathbf{RA1}$.

\subsection{Non-divergent Angelic Processes ($\mathbf{ND_{AP}}$)}\label{sec:ch6:ND-AP}
Like in the theory of reactive angelic designs, it is possible to identify the subset of non-divergent angelic processes. These are angelic processes that satisfy the following healthiness condition $\mathbf{ND_{AP}}$. As depicted in~\cref{fig:theories,fig:theories:angelic-processes} we show that the subsets of non-divergent processes of the theory of angelic processes and reactive angelic designs are isomorphic. This is a key result that supports our hypothesis on the preservation of the semantics of a subset of~\ac{CSP}.
\begin{define}\label{def:NDAP}
\begin{statement}
$\mathbf{ND_{AP}} (P) \circdef Choice_{\mathbf{AP}} \sqcup_{\mathbf{AP}} P$
\end{statement}
\end{define}\noindent%
The definition of $\mathbf{ND_{AP}}$ is similar to that of $\mathbf{ND_{RAD}}$, except that here we use the corresponding least upper bound $\sqcup_{\mathbf{AP}}$ and $Choice_{\mathbf{AP}}$ operators of the theory of angelic processes. An angelic process that is non-divergent can be characterised as established by the following~\cref{theorem:RAPN:ChoiceN-sqcup-P}.
\theoremstatementref{theorem:RAPN:ChoiceN-sqcup-P}\noindent%
The precondition is $true$, while the postcondition corresponds to that of $P$. If $P$ could diverge, then by applying $\mathbf{ND_{AP}}$ this is no longer the case. Since in $\mathbf{H3}$-healthy designs the precondition cannot have any free dashed variables, every non-divergent angelic process is also $\mathbf{H3}$-healthy. However, not every $\mathbf{H3}$-healthy angelic process is necessarily non-divergent. For example, the angelic process $(s.wait \vdash s \in ac')$ is $\mathbf{H3}$-healthy, however, it diverges when $s.wait$ is $false$.

\section{Relationship with Reactive Angelic Designs}\label{sec:ch6:rel-RAD}
As part of our approach for validating the theories we propose, in this section we study the relationship between the theory of angelic processes and reactive angelic designs. Through the links previously discussed in~\cref{sec:ch5:rel-CSP} between the theory of reactive angelic designs and~\ac{CSP} these results also link this new theory to that of~\ac{CSP}.

In~\cref{sec:ch6:RAD-to-AP} we discuss how reactive angelic designs can be mapped into the theory of angelic processes. In~\cref{sec:ch6:AP-to-RAD} we present the reverse mapping between angelic processes and reactive angelic designs. Finally in~\cref{sec:ch6:AP-RAD-Iso-Galois} we show that the subsets of non-divergent processes of both theories are isomorphic.
 
\subsection{From Reactive Angelic Designs to Angelic Processes}\label{sec:ch6:RAD-to-AP}
As already mentioned, in defining $\mathbf{AP}$ we have dropped $\mathbf{RA1}$ and thus the theory of angelic processes is a theory of designs that satisfies both $\mathbf{H1}$ and $\mathbf{H2}$. Therefore, a reactive angelic design, can be turned into an angelic process by applying $\mathbf{H1}$. Since $\mathbf{CSPA2}$ is equally enforced in both models, $\mathbf{H2}$ is also satisfied.

The following result characterises the designs obtained when we apply $\mathbf{H1}$ to a reactive angelic design~$\mathbf{RAD}$.
\theoremstatementref{theorem:H1-o-RAP(P)}\noindent%
In words, and considering the general result for angelic processes established by \cref{theorem:RAPN(P)}, the postcondition is exactly the same as that of any other angelic process, while the precondition requires, in addition, that $P^f_f$ is $\mathbf{RA1}$-healthy. This is a property carried over from the theory of reactive angelic designs, where the negation of the precondition must also be $\mathbf{RA1}$-healthy (\cref{lemma:RA1(P|-Q):RA1(lnot-RA1(lnot-P)|-Q)}). 

We consider the following~\cref{example:H1(ChaosRAD)} where $\mathbf{H1}$ is applied to $Chaos_{\mathbf{RAD}}$.
\begin{example}\label{example:H1(ChaosRAD)}
$\mathbf{H1} (Chaos_{\mathbf{RAD}}) = (s.wait \lor \lnot \mathbf{RA1}(true) \vdash s.wait \land s\in ac')$
\begin{proof}
\cref{theorem:H1(ChaosRA)}
\end{proof}
\end{example}\noindent%
In this case, if the previous process is still waiting for the environment, and $s.wait$ is $true$, then the state is kept unchanged by requiring $s$ to be in the set of final states $ac'$. Otherwise, once the process starts, and $s.wait$ is $false$, the design can be restated as $ok \implies \mathbf{RA1} (true)$. 

%

\subsubsection{Non-divergent Processes}
The application of $\mathbf{H1}$ to a reactive angelic design that is non-divergent, that is $\mathbf{ND_{RAD}}$-healthy, is established by~\cref{lemma:H1-o-RA-o-A(true|-Ptf):(true|-Ptf)}.
\theoremstatementref{lemma:H1-o-RA-o-A(true|-Ptf):(true|-Ptf)}\noindent%
The precondition is $true$, similarly to the original reactive angelic design, while the postcondition is that corresponding to the mapping through $\mathbf{H1}$, which follows the general result of~\cref{theorem:H1-o-RAP(P)}. We consider, for example, the mapping of the process $Skip_{\mathbf{RAD}}$ through $\mathbf{H1}$.
\begin{example}
\begin{align*}
	&\mathbf{H1} (Skip_{\mathbf{RAD}})\\
	&=\\
	&(true \vdash s \in ac'\dres s.wait \rres \circledIn{y}{ac'} (\lnot y.wait \land y.tr=s.tr))
\end{align*}
\begin{proof}
\cref{theorem:H1(SkipRA),lemma:RAPN(true|-Ptf)}
\end{proof}
\end{example}\noindent%
The original postcondition of $Skip_{\mathbf{RAD}}$ is kept intact on the right-handside of the conditional on $s.wait$. 


\subsection{From Angelic Processes to Reactive Angelic Designs}\label{sec:ch6:AP-to-RAD}
When considering the mapping in the opposite direction, from angelic processes to reactive angelic designs, we must ensure that $\mathbf{RA1}$ is observed under all circumstances. Therefore, the mapping we need is $\mathbf{RA1}$ itself. The result of applying $\mathbf{RA1}$ to an angelic process is established by~\cref{theorem:RA1-o-RAPN(P):RA-o-A(Pff|-Ptf)}.
\theoremref{theorem:RA1-o-RAPN(P):RA-o-A(Pff|-Ptf)}\noindent%
The reactive angelic design ensures that $\mathbf{RA1}$ applies to the whole angelic design, which by extension also includes the negation of the precondition (\cref{lemma:RA1(P|-Q):RA1(lnot-RA1(lnot-P)|-Q)}). We consider the following~\cref{example:RA1-o-H1(ChaosRAD)}, where we apply $\mathbf{RA1}$ to the design of \cref{example:H1(ChaosRAD)}.
\begin{example}\label{example:RA1-o-H1(ChaosRAD)}
$\mathbf{RA1} (s.wait \lor \lnot \mathbf{RA1}(true) \vdash s.wait \land s\in ac') = Chaos_{\mathbf{RAD}}$
\begin{proof}
\cref{theorem:H1(ChaosRA),theorem:RA1(ChaosCSPAP):ChaosRAD}.
\end{proof}
\end{example}\noindent%
This result shows that it is possible to recover the original $Chaos_{\mathbf{RAD}}$ of reactive angelic designs. In fact, as we discuss in the next~\cref{sec:ch6:AP-RAD-Iso-Galois} this is the case for every reactive angelic design.

\subsection{Galois Connection and Isomorphism}\label{sec:ch6:AP-RAD-Iso-Galois}
The results of the previous section suggest that every reactive angelic design can be expressed as an angelic process. If we consider the application of $\mathbf{H1}$ to a reactive angelic design followed by the application of $\mathbf{RA1}$, then we obtain the same reactive angelic design as established by the following~\cref{theorem:RA1-o-H1-o-RAP(P):RAP(P)}. 
\theoremref{theorem:RA1-o-H1-o-RAP(P):RAP(P)}\noindent%
This is a fundamental result, which together with the links between the theory of reactive angelic designs and~\ac{CSP}, establishes that every~\ac{CSP} process can also be modelled in this theory, following results on the composition of Galois connections (Theorem 4.2.5 in~\cite{Hoare1998}).

When we consider the mapping in the opposite direction, however, an inequality is obtained, as established by~\cref{theorem:H1-o-RA1-o-RAPN(P):sqsupseteq:RAPN(P)}.
\theoremref{theorem:H1-o-RA1-o-RAPN(P):sqsupseteq:RAPN(P)}\noindent%
This is expected, since reactive angelic designs require $\mathbf{RA1}$ to be enforced under all circumstances, whereas angelic processes do not necessarily enforce $\mathbf{RA1}$. Thus there is a Galois connection between the theory of reactive angelic designs and angelic processes. We consider the following example, where $\mathbf{RA1}$ and $\mathbf{H1}$ are applied to the bottom of the lattice $\bot_{\mathbf{AP}} = (s.wait \vdash s \in ac')$ of angelic processes.
\begin{example}
\begin{align*}
	&\mathbf{H1}\circ\mathbf{RA1} (s.wait \vdash s \in ac')\\
	&=\\
	&(s.wait \lor \lnot \mathbf{RA1} (true) \vdash s.wait \land s \in ac')
\end{align*}
\begin{proof}
\cref{theorem:RA1(ChaosCSPAP):ChaosRAD,theorem:H1(ChaosRA)}.
\end{proof}
\end{example}\noindent%
The result is exactly the same as the result of applying $\mathbf{H1}$ to $Chaos_{\mathbf{RAD}}$. This angelic process has a weaker precondition than that of the bottom $\bot_{\mathbf{AP}}$ and is therefore a refinement of $\bot_{\mathbf{AP}}$.

If we restrict our attention to the subset of angelic processes that are non-divergent, then~\cref{theorem:H1-o-RA1-o-RAPN(P):sqsupseteq:RAPN(P)} can be strengthened into an equality as the established by the following~\cref{theorem:H1-o-RA1-o-NDAP-o-AP(P):ND-o-AP(P)}.
\theoremstatementref{theorem:H1-o-RA1-o-NDAP-o-AP(P):ND-o-AP(P)}\noindent%
Therefore, the subsets of non-divergent processes of the theories of angelic processes and of reactive angelic designs are isomorphic. In addition, if we consider the links between~\ac{CSP} and the theory of reactive angelic designs, and in particular, the subset characterised by $\mathbf{A2}$ and $\mathbf{ND_{RAD}}$, then we can also ascertain that there is a subset corresponding exactly to non-divergent~\ac{CSP} processes in our model. 

\section{Operators}\label{sec:ch6:operators}
In this section we present the definition of some important operators of the theory of angelic processes. Similarly to the approach in~\cref{sec:ch5:operators} we study the relationship between these operators and their counterparts as reactive angelic designs.
\subsection{Angelic Choice}
The angelic choice operator of this theory is also defined through the least upper bound of the lattice of angelic processes, which is conjunction.
\begin{define}
$P \sqcup_{\mathbf{AP}} Q \circdef P \land Q$
\end{define}\noindent%
This operator is closed under $\mathbf{AP}$ as established by~\cref{theorem:AP(P-sqcup-Q):P-sqcup-Q}.
\theoremstatementref{theorem:AP(P-sqcup-Q):P-sqcup-Q}\noindent%
It is also closed under the subset of non-divergent angelic processes, characterised by $\mathbf{ND_{AP}}$, as established by~\cref{theorem:NDAP(P-sqcup-Q):P-sqcup-Q}.
\theoremstatementref{theorem:NDAP(P-sqcup-Q):P-sqcup-Q}\noindent%
The angelic choice of two reactive angelic designs can be equally obtained through the least upper bound of the lattice of angelic processes as established by the following~\cref{theorem:RA1(H1(P)-sqcup-H1(Q)):P-sqcup-Q}.
\theoremref{theorem:RA1(H1(P)-sqcup-H1(Q)):P-sqcup-Q}\noindent%
In words, if we consider two reactive angelic designs $P$ and $Q$, and after mapping them through the function $\mathbf{H1}$ we take the least upper bound $\sqcup_{\mathbf{AP}}$, followed by $\mathbf{RA1}$, then we obtain the same result as the least upper bound $\sqcup_{\mathbf{RAD}}$ of $P$ and $Q$. Together with the result of~\cref{theorem:NDAP(P-sqcup-Q):P-sqcup-Q} this establishes that the angelic choice operator for the subset of non-divergent processes is in correspondence with that of the theory of reactive angelic designs.

However, when we consider the result in the opposite direction, that is, by considering two angelic processes $P$ and $Q$ mapped through $\mathbf{RA1}$, followed by the application of $\mathbf{H1}$, then the result is not an equality.
\theoremstatementref{theorem:H1(RA1(P)-sqcup-RA1(Q)):sqsupseteq:P-sqcup-Q}\noindent%
This is expected since the theory of angelic processes is less strict with regards to enforcing $\mathbf{RA1}$.

\subsection{Demonic Choice}
Like in the theory of reactive angelic designs, demonic choice is also defined using the greatest lower bound, which is disjunction.
\begin{define}
$P \sqcap_{\mathbf{AP}} Q \circdef P \lor Q$
\end{define}\noindent
This operator is closed under $\mathbf{AP}$ as established by~\cref{theorem:RAPN(P-sqcap-Q):closure}, and is also closed under the subset of non-divergent processes as established by~\cref{theorem:NDAP(P-sqcap-Q):P-sqcap-Q}.
\theoremstatementref{theorem:RAPN(P-sqcap-Q):closure}
\theoremstatementref{theorem:NDAP(P-sqcap-Q):P-sqcap-Q}\noindent%
The demonic choice of two reactive angelic designs $P$ and $Q$ can be equally obtained through the greatest lower bound of the lattice of angelic processes as the following~\cref{theorem:RA1(H1(P)-sqcap-H1(Q)):P-sqcap-Q} establishes.
\theoremref{theorem:RA1(H1(P)-sqcap-H1(Q)):P-sqcap-Q}\noindent%
If we map $P$ and $Q$ through $\mathbf{H1}$, take the greatest lower bound $\sqcap_{\mathbf{AP}}$, and then apply $\mathbf{RA1}$, then the same result can be obtained by taking the greatest lower bound of reactive angelic designs $\sqcap_{\mathbf{RAD}}$. With this result, together with the closure of $\sqcap_{\mathbf{AP}}$ under $\mathbf{ND_{AP}}$ (\cref{theorem:NDAP(P-sqcap-Q):P-sqcap-Q}) it is possible to ascertain that the demonic choice for non-divergent processes is in correspondence in both models.

In general, the greatest lower bound of the theory of angelic processes cannot be replicated in the theory of reactive angelic designs, as established by the following~\cref{theorem:H1(RA1(P)-sqcap-RA1(Q)):sqsupseteq:P-sqcap-Q}.
\theoremstatementref{theorem:H1(RA1(P)-sqcap-RA1(Q)):sqsupseteq:P-sqcap-Q}\noindent%
This inequality is expected, since the model of angelic processes does not necessarily enforce $\mathbf{RA1}$ under all circumstances, while in the theory of reactive angelic designs this is always the case.

\subsection{Divergence: Chaos and Chaos of CSP}
In our theory of angelic processes, the bottom of the lattice is defined by $Chaos_{\mathbf{AP}}$, whose definition can be given in terms of the bottom of designs as follows.
\begin{define}
$Chaos_{\mathbf{AP}} \circdef \mathbf{AP} (false \vdash true)$
\end{define}\noindent
This result can be expanded into a design as established by~\cref{lemma:ChaosRAPN:design}.
\theoremstatementref{lemma:ChaosRAPN:design}\noindent%
The precondition requires the component $wait$ of the initial state $s$ to be $true$, while the postcondition keeps the state unchanged by requiring $s$ to be in the set of final states $ac'$. In other words, as long as the environment is waiting for an interaction, the state is kept unchanged. However, once the environment is no longer waiting, then $Chaos_{\mathbf{AP}}$ diverges and the behaviour is described by $true$. $Chaos_{\mathbf{AP}}$ is a unit for angelic choice as established by~\cref{theorem:AP:P-sqcup-ChaosAP:P}.
\theoremstatementref{theorem:AP:P-sqcup-ChaosAP:P}\noindent%
In other words, if possible, the angel can avoid divergence.


In this theory, the process that corresponds to $Chaos_{\mathbf{RAD}}$ is $ChaosCSP_{\mathbf{AP}}$, which is defined through a design as follows.
\begin{define}
$ChaosCSP_{\mathbf{AP}} \circdef \mathbf{AP} (\lnot \mathbf{RA1} (true) \vdash true)$
\end{define}\noindent%
Instead of $false$, the precondition requires $\lnot \mathbf{RA1} (true)$. As already discussed, it is the negation of the precondition of a design that gives the behaviour in case of possible non-termination. This design can be expanded as established by the following~\cref{lemma:AP:ChaosCSP}.
\theoremstatementref{lemma:AP:ChaosCSP}\noindent
In words, when the environment is waiting for an interaction, the state is kept unchanged. Otherwise, the design diverges, but still requires that $\mathbf{RA1}$ holds, unlike $Chaos_{\mathbf{AP}}$. This corresponds exactly to the mapping of $Chaos_{\mathbf{RAD}}$ through the linking function $\mathbf{H1}$ as established by~\cref{theorem:H1(ChaosRA)}.
\theoremstatementref{theorem:H1(ChaosRA)}\noindent%
Similarly, if we map $ChaosCSP_{\mathbf{AP}}$ through $\mathbf{RA1}$ we obtain the bottom of the lattice of reactive angelic designs $Chaos_{\mathbf{RAD}}$.
\theoremstatementref{theorem:RA1(ChaosCSPAP):ChaosRAD}\noindent%
This follows from the general result of~\cref{theorem:RA1-o-H1-o-RAP(P):RAP(P)}. 

\subsection{Choice}
The most nondeterministic process that does not diverge is defined as $Choice_{\mathbf{AP}}$ and can be defined through a design as follows.
\begin{define}
$Choice_{\mathbf{AP}} \circdef \mathbf{AP} (true \vdash ac'\neq\emptyset)$
\end{define}\noindent%
The precondition is $true$, while any set of final states $ac'$ is acceptable. The resulting behaviour, constrained by $\mathbf{AP}$, is established through the following~\cref{lemma:AP(true|-ac'neq-emptyset)}.
\theoremstatementref{lemma:AP(true|-ac'neq-emptyset)}\noindent%
The precondition is also $true$, while the postcondition has a conditional on $s.wait$. As is the case for every angelic process, when the process is waiting for the environment, and $s.wait$ is $true$, the state is kept unchanged. Otherwise, the only guarantee is that there is a final state in $ac'$ satisfying $\mathbf{RA1}$.

As previously discussed, the operator $Choice_{\mathbf{AP}}$ is used to characterise algebraically the subset of angelic processes that are non-divergent. Therefore, it is closed under $\mathbf{ND_{AP}}$, and by definition, equally closed under $\mathbf{AP}$. It is the counterpart to $Choice_{\mathbf{RAD}}$ of the theory of reactive angelic designs as established by the following \cref{theorem:H1(ChoiceRAP):ChoiceAP,theorem:RA1(ChoiceAP):ChoiceRAD}.
\theoremstatementref{theorem:H1(ChoiceRAP):ChoiceAP}
\theoremstatementref{theorem:RA1(ChoiceAP):ChoiceRAD}\noindent%
The result of~\cref{theorem:RA1(ChoiceAP):ChoiceRAD} follows directly from~\cref{theorem:H1(ChoiceRAP):ChoiceAP} and the general result of~\cref{theorem:RA1-o-H1-o-RAP(P):RAP(P)}.
\subsection{Stop}
In this theory, deadlock is modelled by $Stop_{\mathbf{AP}}$, whose definition is similar to that of the reactive angelic design $Stop_{\mathbf{RAD}}$.
\begin{define}
$Stop_{\mathbf{AP}} \circdef \mathbf{AP} (true \vdash \circledIn{y}{ac'} (y.tr=s.tr \land y.wait))$
\end{define}\noindent
The precondition is $true$, while the postcondition states that there is a final state $y$ in the set of final states $ac'$ where the trace is kept unchanged and the process is always waiting for the environment. This definition can be directly obtained by applying $\mathbf{H1}$ to $Stop_{\mathbf{RAD}}$ as established by~\cref{theorem:H1(StopRA)}.
\theoremstatementref{theorem:H1(StopRA)}\noindent%
Similarly, $Stop_{\mathbf{RAD}}$ can be obtained by applying $\mathbf{RA1}$ to $Stop_{\mathbf{AP}}$ as established by the following~\cref{theorem:RA1(StopAP)}.
\theoremstatementref{theorem:RA1(StopAP)}\noindent%
This is expected since $Stop_{\mathbf{AP}}$ is a non-divergent angelic process, and so it is in direct correspondence with a reactive angelic design.

\subsection{Skip}
The process that always terminates successfully is characterised by $Skip_{\mathbf{AP}}$. Its definition as a design is presented below.
\begin{define}
$Skip_{\mathbf{AP}} \circdef \mathbf{AP} (true \vdash \circledIn{y}{ac'} (y.tr=s.tr \land \lnot y.wait))$
\end{define}\noindent
The precondition is $true$, while the postcondition states that there is a final state $y$ in $ac'$ where the trace of events is kept unchanged and the component $wait$ is $false$. $Skip_{\mathbf{AP}}$ is in correspondence with $Skip_{\mathbf{RAD}}$ of the theory of reactive angelic designs as established by the following~\cref{theorem:H1(SkipRA),theorem:RA1(SkipAP)}.
\theoremstatementref{theorem:H1(SkipRA)}
\theoremstatementref{theorem:RA1(SkipAP)}\noindent%
These results are expected since $Skip_{\mathbf{AP}}$ and $Skip_{\mathbf{RAD}}$ are both non-divergent processes.

\subsection{Sequential Composition}
In our theory of angelic processes, the definition of sequential composition is also $\seqDac$ from the theory of angelic designs. When we consider two angelic processes $P$ and $Q$, the following closure result is obtained.
\theoremstatementref{theorem:AP(P)-seqDac-AP(Q)}\noindent
This result is similar to that obtained in the theory of reactive angelic designs (\cref{theorem:RAP:seqDac}). The differences are in that $\mathbf{RA1}$ is no longer applied to $P^f_f$ and $Q^f_f$, the negation of the preconditions of $P$ and $Q$, respectively. If $P$ may diverge, then the result is the bottom of the lattice $Chaos_{\mathbf{AP}}$. Similarly, since the precondition of $Q$ does not need to observe $\mathbf{RA1}$, if $Q$ diverges, then the sequential composition also behaves like $Chaos_{\mathbf{AP}}$ once $P$ has finished interacting with the environment. 

Thus, in our theory of angelic processes, $\seqDac$ is a sequential composition operator that behaves differently to that of~\ac{CSP}, in that it can back propagate the divergence of $Q$ through $P$, irrespective of other interactions that happen in $P$, as long as, eventually the environment may terminate its interactions with $P$ and behave as $Q$. We consider the following example~\cref{example:AP:Stop-sqcup-Skip-seqDac-Chaos}.
\begin{example}\label{example:AP:Stop-sqcup-Skip-seqDac-Chaos}
$(Stop_{\mathbf{AP}} \sqcup_{\mathbf{AP}} Skip_{\mathbf{AP}}) \seqDac Chaos_{\mathbf{AP}} = Stop_{\mathbf{AP}}$
\begin{proof}
\cref{lemma:AP:(Skip-AP-sqcup-Stop-AP)-seqDac-Chaos-AP:Stop-AP}.
\end{proof}
\end{example}\noindent%
In this case, the angel avoids the divergence of $Chaos_{\mathbf{AP}}$ by resolving the choice in favour of deadlock. This is similar to the behaviour in the theory of reactive angelic designs, since $Stop_{\mathbf{AP}}$ can prevent $Chaos_{\mathbf{AP}}$ from ever being reached.

In general, the result of applying $\mathbf{RA1}$ to the sequential composition of two reactive angelic designs $P$ and $Q$ mapped through $\mathbf{H1}$ is not equivalent to sequentially composing these two processes in the theory of reactive angelic designs as established by~\cref{theorem:RA1(H1(P)-seqDac-H1(Q)):sqsubseteq:P-seqDac-Q}.
\theoremstatementref{theorem:RA1(H1(P)-seqDac-H1(Q)):sqsubseteq:P-seqDac-Q}\noindent%
This is because the possibility to diverge in $P$, in the theory of angelic processes, can lead to immediate divergence, as already discussed. Thus, when the sequential composition of $\mathbf{H1} (P)$ and $\mathbf{H1} (Q)$ is mapped back through $\mathbf{RA1}$, there is a weakening.

Similarly, the reverse mapping through $\mathbf{H1}$ of the sequential composition of two angelic processes $P$ and $Q$ mapped through $\mathbf{RA1}$ is also an inequality as established by~\cref{theorem:H1(RA1(P)-seqDac-RA1(Q)):sqsupseteq:P-seqDac-Q}.
\theoremstatementref{theorem:H1(RA1(P)-seqDac-RA1(Q)):sqsupseteq:P-seqDac-Q}\noindent%
This is due to the fact that the notion of divergence is different. In a sequential composition of $P$ and the bottom of the lattice $Chaos_{\mathbf{AP}}$, the result is also $Chaos_{\mathbf{AP}}$. If we map $Chaos_{\mathbf{AP}}$ through $\mathbf{RA1}$ the result is $Chaos_{\mathbf{RAD}}$ (\cref{theorem:RA1(ChaosCSPAP):ChaosRAD}), which when sequentially composed after the process $\mathbf{RA1} (P)$, still preserves the history of events in $P$, whereas the corresponding process in the theory of angelic processes does not. Hence, there is a strengthening. 

However, if we consider the subset of non-divergent reactive angelic designs, characterised by $\mathbf{ND_{RAD}}$, then~\cref{theorem:RA1(H1(P)-seqDac-H1(Q)):sqsubseteq:P-seqDac-Q} can be strengthened into an equality as established by~\cref{theorem:RA1(H1(P)-seqDac-H1(Q)):P-seqDac-Q}.
\theoremstatementref{theorem:RA1(H1(P)-seqDac-H1(Q)):P-seqDac-Q}\noindent%
In addition, the operator $\seqDac$ is closed under $\mathbf{ND_{AP}}$ as established by the following~\cref{theorem:NDAP(P-seqDac-Q)}.
\theoremstatementref{theorem:NDAP(P-seqDac-Q)}\noindent%
Thus, as long as $P$ and $Q$ are non-divergent, $\seqDac$ behaves exactly in the same way as in the theory of reactive angelic designs. By extension, this also applies to the subset of $\mathbf{A2}$ processes, which do not exhibit angelic nondeterminism. Therefore, it also applies to the subset of non-divergent~\ac{CSP} processes.

\subsection{Prefixing}
Similarly to the previous non-divergent processes, event prefixing has a definition similar to that of $a \circthen_{\mathbf{RAD}} Skip_{\mathbf{RAD}}$ in the theory of reactive angelic designs.
\begin{define}
\begin{align*}
	a \circthen_{\mathbf{AP}} Skip_{\mathbf{AP}} \circdef \mathbf{AP} \left(true 
				\vdash 
					\circledIn{y}{ac'} \left(\begin{array}{l}(y.tr=s.tr \land a \notin y.ref)
						\\ \dres y.wait \rres \\
					(y.tr = s.tr \cat \lseq a \rseq)
					\end{array}\right)
		\right)
\end{align*}
\end{define}\noindent%
The precondition is $true$, while the postcondition is exactly like that of the corresponding reactive angelic design $a \circthen_{\mathbf{RAD}} Skip_{\mathbf{RAD}}$ (\cref{sec:ch5:operators:prefixing}).

The event prefixing of both theories is in correspondence as established by the following~\cref{lemma:H1(a-then-Skip-RAD):a-then-Skip-AP,lemma:RA1(a-then-Skip-AP):a-then-Skip-RAD}.
\theoremstatementref{lemma:H1(a-then-Skip-RAD):a-then-Skip-AP}
\theoremstatementref{lemma:RA1(a-then-Skip-AP):a-then-Skip-RAD}\noindent
Similarly to the theory of reactive angelic designs, in general, the process $a \circthen_{\mathbf{AP}} P$ denotes the compound process $a \circthen_{\mathbf{AP}} Skip_{\mathbf{AP}} \seqDac P$, whose result as an angelic process is established by~\cref{theorem:AP:a-then-P}.
\theoremstatementref{theorem:AP:a-then-P}\noindent
This result is a counterpart to that of~\cref{theorem:RAP:a-circthen-RA-P}. The difference lies in the precondition of the design: the negation of the precondition of $P$ is not necessarily required to observe $\mathbf{RA1}$. In addition, the application of $\mathbf{PBMH}$ can be simplified by taking into account that every $\mathbf{AP}$-healthy process is also $\mathbf{PBMH}$-healthy.

In order to illustrate the behaviour of prefixing in the presence of divergence, we consider the following~\cref{example:AP:a-then-Chaos}.
\begin{example}\label{example:AP:a-then-Chaos}
$a \circthen_{\mathbf{AP}} Chaos_{\mathbf{AP}} = Chaos_{\mathbf{AP}}$
\begin{proof}
\cref{lemma:AP:a-then-ChaosAP:ChaosAP}.
\end{proof}
\end{example}\noindent%
In this case, the potential for divergence after performing event $a$ leads to immediate divergence. If instead we sequentially compose prefixing on the event $a$ with $ChaosCSP_{\mathbf{AP}}$, the behaviour is different as established by~\cref{lemma:AP:a-then-ChaosCSPAP}.
\theoremstatementref{lemma:AP:a-then-ChaosCSPAP}\noindent%
This result mirrors the behaviour of $a \circthen_{\mathbf{RAD}} Chaos_{\mathbf{RAD}}$ of the theory of reactive angelic designs (\cref{theorem:RAP:a-circthenRA-ChaosRA}). 

We revisit~\cref{example:a-then-Skip-sqcup-b-then-Chaos}, by restating it in the theory of angelic processes as~\cref{example:AP:a-then-Skip-sqcup-b-then-Chaos}.
\begin{example}\label{example:AP:a-then-Skip-sqcup-b-then-Chaos}
$a \circthen_{\mathbf{AP}} Chaos_{\mathbf{AP}} \sqcup_{\mathbf{AP}} b \circthen_{\mathbf{AP}} Skip_{\mathbf{AP}} = b \circthen_{\mathbf{AP}} Skip_{\mathbf{AP}}$
\begin{proof}
\cref{lemma:AP:a-then-ChaosAP:ChaosAP,theorem:AP:P-sqcup-ChaosAP:P}.
\end{proof}
\end{example}\noindent%
Now, in the context of the theory of angelic processes, the possibility for divergence is avoided altogether, and the result is the prefixing on the event $b$. As required, the angel can avoid processes that may lead to divergence altogether, a property that is not observed in the theory of reactive angelic designs.

\section{Final Considerations}\label{sec:ch6:final-considerations}
The motivation for the theory of angelic processes stems from the limitations of the angelic choice of reactive angelic designs, which is unable to avoid divergence completely, as in the case of~\cref{example:a-then-Skip-sqcup-b-then-Chaos}. The possibility to avoid divergence is a desirable property that is much closer in spirit to the refinement calculus. In order to tackle this aspect, we have pursued a theory that drops $\mathbf{RA1}$, and thus, is able to undo the history of events if necessary. The result is a theory of angelic designs, whose pre and postconditions observe a subset of the healthiness conditions of the theory of reactive angelic designs, such as $\mathbf{RA2}$ and $\mathbf{PBMH}$.

We have studied the relationship between the theories and established that there is a Galois connection between them. As illustrated in~\cref{fig:theories,fig:theories:angelic-processes}, reactive angelic designs can be mapped into this theory by turning them into designs, through $\mathbf{H1}$, while angelic processes can be mapped in the opposite direction by applying $\mathbf{RA1}$. We have found that the subset of non-divergent angelic processes, characterised by $\mathbf{ND_{AP}}$, is isomorphic to the subset of non-divergent reactive angelic designs characterised by $\mathbf{ND_{RAD}}$. Together with the linking results from~\cref{chapter:5} between $\mathbf{RAD}$ and~\ac{CSP}, this implies that the subset of non-divergent~\ac{CSP} processes has exactly the same semantics in this model.

Since every reactive angelic design can be mapped into the model of angelic processes and back, we can ascertain that there is a subset in~$\mathbf{AP}$ that characterises all reactive angelic designs. This is essentially a subset whose negated preconditions satisfy $\mathbf{RA1}$. If we consider the subset of $\mathbf{RAD}$ that is isomorphic to~\ac{CSP} (characterised by~$\mathbf{A2}$), it is possible to postulate that there is also a subset in $\mathbf{AP}$ characterising every~\ac{CSP} process.

However, since we allow the history of events to be undone when $ok$ is $false$, not all operators are necessarily in correspondence, as is the case, for example, with sequential composition. A parallel can be drawn in the theory of~\ac{CSP}, where this problem corresponds to the possibility of characterising~\ac{CSP} processes as designs, rather than reactive designs. The difference between these two can clearly be seen from the fact that $\mathbf{H1}$ and $\mathbf{R1}$ are not commutative. While such a theory of designs could possibly characterise~\ac{CSP} processes, this would mean that the definition of the operators would need to change in order to accommodate such a model, thus negating the benefits of unification in the~\ac{UTP}. 




%

\chapter{Conclusions}\label{chapter:7}
In this chapter we conclude this thesis by summarizing our contributions. In addition, we discuss lines for future work.

\section{Contributions}
As previously discussed, angelic nondeterminism has been used in a variety of different contexts, such as in problems whose solutions may involve a combination of search and backtracking. This is the case, for example, when modelling game-like scenarios, theorem-proving tactics, or constraint satisfaction problems. In general, angelic nondeterminism enables a great degree of abstraction in the context of formal models and specifications. Its characterisation in the context of process algebras, such as~\ac{CSP}, however, has to the best of our knowledge, been elusive. The existing approaches have either considered notions of angelic nondeterminism~\cite{Roscoe2010} different from that of refinement calculi, or different~\ac{CSP} semantics~\cite{Tyrrell2006}.

Angelic nondeterminism has traditionally been studied in the context of theories of correctness for sequential computations, such as in the refinement calculus~\cite{Back1998,Morris1987,Morgan1994}, where it is characterised as the least upper bound of the lattice of monotonic predicate transformers. Isomorphic models include Rewitzky's theory of binary multirelations~\cite{Rewitzky2003}, which is the foundation of our approach. 

Our first contribution in~\cref{chapter:3} is an extended model of binary multirelations that caters for possibly non-terminating computations. This model provides a complementary view of our theory of angelic designs, which allows for preconditions that refer to the later or final values of a computation, as required for characterising~\ac{CSP} processes. Unlike purely sequential computations, in a reactive system, there is a rich sequence of interactions, whose history cannot be undone even in the case of divergence, such as in the case of the process $a \circthen Chaos$.

Our work is based on the~\ac{UTP} of Hoare and He~\cite{Hoare1998}, a relational framework suitable for characterising different programming paradigms. As such, our results are applicable not only to~\ac{CSP}, but also to any other algebra of (state-rich) reactive systems whose semantics is or can be described in the~\ac{UTP}. Our theories are complete lattices and angelic and demonic choice are modelled as the meet and join, respectively. Each and every one of them is appropriately justified by studying its relationship with the established theories, which is central to the unification of theories in the~\ac{UTP}.

Our theory of angelic designs generalises the theory of Cavalcanti et al.~\cite{Cavalcanti2006} to include the variables $ok$ and $ok'$ for capturing termination. It caters for non-$\mathbf{H3}$ designs, as required for specifying~\ac{CSP} processes like $Chaos$, whose precondition, as a reactive design, refers to the after value of the trace of events. Its relationship with the theories of~\cite{Cavalcanti2006} and of extended binary multirelations sheds light on the definition of less trivial operators. Sequential composition, for instance, due to the use of non-homogeneous relations, is not relational composition like in other~\ac{UTP} theories. Apart from the relational characterisation of $ok$ and $ok'$, this suggests itself as a form of a Kleisli composition through the results established between the theory of angelic designs and binary multirelations, and its respective characterisation as the category of multirelations or multifunctions~\cite{Martin2013}. The result obtained for the sequential composition of angelic designs is pleasing, in that, using the operators $\seqDac$ and $\seqA$, we have a definition similar to that in the original theory of designs.

The theory of reactive angelic designs considers the encoding of the observational variables $ref$, $tr$ and $wait$ of~\ac{CSP} as state components. This enables angelic choice over the value of these components in final or after states. Rather pleasingly, like the processes in the theory of~\ac{CSP}~\cite{Hoare1998,Cavalcanti2006a}, every $\mathbf{RAD}$ process can be specified in terms of designs, that is, pre and postcondition pairs, but now we use angelic designs. Unlike other attempts~\cite{Roscoe2010,Tyrrell2006}, our approach consists of a natural extension of the concept of angelic nondeterminism from a theory of sequential correctness to a model of processes. This approach is strongly justified by the relationship between the theories, their isomorphic subsets, and by the correspondence of operators in both theories. We have a theory of~\ac{CSP} that preserves its existing semantics and that can be used to describe both angelic and demonic nondeterminism.

An important result obtained in the theory of reactive angelic designs pertains to the capability of the angel to avoid divergence. However, unlike in a theory of correctness for sequential computations, the history of interactions, as recorded by traces, cannot simply be undone, even in the presence of divergence. The healthiness condition $\mathbf{RA1}$, the counterpart to $\mathbf{R1}$ of~\ac{CSP} in the model of reactive angelic designs, ensures that this is the case under all circumstances.

Our final theory does not adopt $\mathbf{RA1}$ as a healthiness condition and as such allows the angel to discard traces of events leading to divergence. It is a theory of angelic designs: a complete lattice whose bottom $Chaos_{\mathbf{AP}}$ is not the $Chaos$ of~\ac{CSP}. It is a process that once executed behaves arbitrarily, and may even undo the history of interactions. More importantly, in an angelic choice involving other interactions, it becomes possible for the angel to undo the history of events, if necessary, and avoid divergence. This is a property much closer in spirit to the angelic choice of the refinement calculus.

As a consequence not every operator preserves the original semantics of~\ac{CSP}. That is the case of the sequential composition operator, for instance. However, the subset of non-divergent angelic processes is isomorphic to the subset of non-divergent reactive angelic designs. Moreover, each of the operators studied is closed within this subset.

In summary, we have two closely related theories for characterising angelic nondeterminism in~\ac{CSP} whose algebraic properties are clearly distinct. The theory of reactive angelic designs is a natural extension of~\ac{CSP}, where the angelic choice cannot undo the history of events, but which preserves the semantics of~\ac{CSP}. On the other hand, the theory of angelic processes possesses algebraic properties closer to those of the refinement calculus, but does not necessarily preserve the semantics of all~\ac{CSP} processes. Nevertheless, the semantics of the subset of non-divergent processes is maintained, and so our initial hypothesis is satisfied.











\section{Future Work}
The work presented in this thesis lays the foundation for the complete development of process algebras with angelic nondeterminism in the wider context of state-rich reactive systems. Our approach has focused mainly on~\ac{CSP}, however due to the~\ac{UTP} basis of our work, our results are equally applicable to other process calculi, including, for example, \Circus, which is a combination of~\ac{CSP} and Z, and whose semantics~\cite{Oliveira2005} is also given using the~\ac{UTP}. Depending on the desired properties of the algebra, a future approach to incorporating our results in~\Circus~needs to consider the implications of the treatment of divergence, which in the case of our model of angelic processes, is rather different from the~\ac{CSP} theory.

A practical application of angelic nondeterminism in~\Circus~can be found, for instance, in the modelling strategy of~\cite{Cavalcanti2013}, which uses~\CircusTime, a timed version of~\Circus. Therefore, an interesting avenue for future work includes studying the role of angelic nondeterminism in timed versions of process calculi, such as Timed CSP~\cite{Schneider2000} and \CircusTime~\cite{Wei2010,Wei2011,Wei2012,Wei2013}. A concern that is likely to surface is whether the angel should be allowed to change time in order to avoid divergence, an issue similar to the problem posed by $\mathbf{RA1}$. Such a construction would enable angelic nondeterminism to be employed as a specification abstraction in a theory that also includes time. 

While we have studied a number of~\ac{CSP} operators, a complete theory of angelic nondeterminism for~\ac{CSP} requires other important operators to be considered, such as hiding and parallel composition. Recursion can be treated in a similar way to other~\ac{UTP} theories as the weakest fixed point. For many of these, the use of our lifting operator $\circledIn{y}{ac'}$ is likely to be useful and give rise to definitions similar to those in the original theory of~\ac{CSP}, however, some operators, such as parallel composition, require further work. For instance, in the~\ac{CSP} theory, parallel composition is defined using the parallel by merge technique~\cite{Hoare1998} which, in the context of our theory, requires further support for renaming and changing the fields of records.

Furthermore, the algebraic properties of many of the operators have yet to be fully explored. For example, in the case of the external choice operator, there are other alternative and plausible definitions that preserve the~\ac{CSP} semantics, whose algebraic properties, in the context of processes with angelic nondeterminism, are different. In the case of hiding, and similarly to the case of sequential composition, we hypothesize that angelic choice is likely not to be distributive, however future work is necessary in order to propose and establish further laws. A related, and interesting, path for future work is the study of the encoding of additional healthiness conditions~\cite{Hoare1998,Cavalcanti2006a} of~\ac{CSP} and whether the addition of angelic choice may be needed to enable or simplify the algebraic specification of these.

Even in the context of the theory of angelic designs there is a wide scope for further work. While we have established links between that theory, the extended model of binary multirelations and the $\mathbf{PBMH}$ theory, it would also be beneficial to have a direct link with the weakest precondition model. The model of extended binary multirelations is also ameanable to further study. For instance, recently Guttmann~\cite{Guttmann2014a} has proposed a model of binary multirelations in the context of general correctness. A link could be established with this theory, and perhaps, with other models of binary multirelations~\cite{Martin2004}. The links with the $\mathbf{BMH_\bot}$ theory open the door for our theories to be studied in the context of multirelations.

From a practitioner's point of view a theory becomes significantly more useful once there is a toolkit. There may be different approaches for tackling this aspect. For instance, one approach could involve the mechanisation of our theories using a theorem prover, which would not only help practitioners, but also help further validate our theories, proofs and examples. 
Approaches for mechanising~\ac{UTP} theories include those of Foster et al.~\cite{Foster2014} and Feliachi et al.~\cite{Feliachi2010} using Isabelle/HOL, Zeyda et al.~\cite{Zeyda2010} and Oliveira et al.~\cite{Oliveira2005} using ProofPower/Z, and others~\cite{Butterfield2010,Butterfield2013}. Particular issues that would need to be considered include reasoning about families of theories and encoding record types, with the capability to change and rename fields as well as type check them, as required to appropriately model sets of final states. 

Finally, since the concept of angelic nondeterminism has been used in a variety of different contexts, it would be useful to conduct case studies. For example, in~\cite{Cavalcanti2013} angelic nondeterminism is employed to facilitate the faithful characterisation of idealised time models of control systems using~\CircusTime. In that context, the specification models are constructed from Simulink counterparts which, embody a notion of infinitely fast computations, while the respective implementation models capture the constraints of actual real-time computers. The link between these two is established through an assertion that requires the values output by the implementation to be in agreement with the values of the simulation model. Angelic nondeterminism is employed as an abstract specification mechanism, which, through back propagation enforces the correct choices in the model. A necessary prerequisite for such a case study is the treatment of parallel composition which features prominently. 


We envision that many problems that have traditionally been tackled using angelic nondeterminism could be just as easily modelled using our theories, with the added benefit that they can be modelled in the context of process algebras. It remains to be seen how the inclusion of angelic nondeterminism can be fully exploited in the development of refinement strategies for the formal specification and verification of complex state-rich reactive systems. An example to be considered is the refinement of a specification with angelic nondeterminism to an algorithm which uses explicit backtracking. Related to this construction is the relationship between our theories and that of concurrent logic programming~\cite{Hoare1998}, which has yet to be explored.

In summary, we have now presented the first extension of~\ac{CSP} that includes a notion of angelic nondeterminism compatible with that of refinement calculi. It is a solid foundation for the extension of state-rich process algebra for refinement. As such, it provides a basis for further work on theory, so as to explore the algebra, techniques, and applications.













\appendix
\chapter{UTP: Relations, Designs and CSP}\label{appendix:UTP}\label{appendix-designs}\label{appendix-relations}

\section{Theory of Relations}

\subsection{Conditional}

\begin{lemma}\label{lemma:conditional:(Q-implies-R)}
$P \dres c \rres (Q \implies R) = (true \dres c \rres Q) \implies (P \dres c \rres R)$
\begin{proofs}
\begin{proof}\checkt{alcc}\checkt{pfr}
\begin{xflalign*}
	&P \dres c \rres (Q \implies R)
	&&\ptext{Predicate calculus}\\
	&=(false \lor P) \dres c \rres (\lnot Q \lor R)
	&&\ptext{Property of conditional}\\
	&=(false \dres c \rres \lnot Q) \lor (P \dres c \rres R)
	&&\ptext{Predicate calculus and property of conditional}\\
	&=(true \dres c \rres Q) \implies (P \dres c \rres R)
\end{xflalign*}
\end{proof}
\end{proofs}
\end{lemma}

\begin{lemma}\label{lemma:seqA:conditional-right} Provided $ac'$ is not free in $c$,
\begin{align*}
	&(P \dres c \rres Q) \seqA R = (P \seqA R) \dres c \rres (Q \seqA R)
\end{align*}
\begin{proofs}
\begin{proof}\checkt{alcc}\checkt{pfr}
\begin{xflalign*}
	&(P \dres c \rres Q) \seqA R
	&&\ptext{Definition of conditional}\\
	&=((c \land P) \lor (\lnot c \land Q)) \seqA R
	&&\ptext{Distributivity of $\seqA$ (\cref{law:seqA-right-distributivity})}\\
	&=((c \land P) \seqA R) \lor ((\lnot c \land Q) \seqA R)
	&&\ptext{Distributivity of $\seqA$ (\cref{law:seqA-right-distributivity-conjunction})}\\
	&=((c \seqA R) \land (P\seqA R)) \lor ((\lnot c \seqA R) \land (Q\seqA R))
	&&\ptext{$ac'$ not free in $c$ (\cref{law:seqA-ac'-not-free})}\\
	&=(c \land (P\seqA R)) \lor (\lnot c \land (Q\seqA R))
	&&\ptext{Definition of conditional}\\
	&=(P\seqA R) \dres c \rres (Q\seqA R)
\end{xflalign*}
\end{proof}
\end{proofs}
\end{lemma}

\begin{lemma}\label{lemma:lnot(conditional)}
$\lnot (P \dres c \rres Q) = (\lnot P \dres c \rres \lnot Q)$
\begin{proofs}
\begin{proof}\checkt{pfr}\checkt{alcc}
\begin{xflalign*}
	&\lnot (P \dres c \rres Q)
	&&\ptext{Definition of conditional}\\
	&=\lnot ((c \land P) \lor (\lnot c \land Q))
	&&\ptext{Predicate calculus}\\
	&=(\lnot c \lor \lnot P) \land (c \lor \lnot Q)
	&&\ptext{Predicate calculus}\\
	&=(\lnot c \land c) \lor (\lnot c \land \lnot Q) \lor (\lnot P \land c) \lor (\lnot P \land \lnot Q)
	&&\ptext{Predicate calculus}\\
	&=(\lnot c \land \lnot Q) \lor (\lnot P \land c) \lor (\lnot P \land \lnot Q)
	&&\ptext{Predicate calculus}\\
	&=(\lnot c \land \lnot Q) \lor (\lnot P \land (c \lor \lnot Q))
	&&\ptext{Predicate calculus}\\
	&=(\lnot c \lor (\lnot P \land (c \lor \lnot Q))) \land (\lnot Q \lor (\lnot P \land (c \lor \lnot Q)))
	&&\ptext{Predicate calculus}\\
	&=(\lnot c \lor \lnot P) \land (\lnot c \lor c \lor \lnot Q) \land (\lnot Q \lor \lnot P) \land (\lnot Q \lor c)
	&&\ptext{Predicate calculus}\\
	&=(\lnot c \lor \lnot P) \land (\lnot c \lor c) \land (\lnot Q \lor \lnot P) \land (\lnot Q \lor c)
	&&\ptext{Predicate calculus}\\
	&=(c \land \lnot P) \lor (\lnot c \land \lnot Q)
	&&\ptext{Definition of conditional}\\
	&=(\lnot P) \dres c \rres (\lnot Q)
\end{xflalign*}
\end{proof}
\end{proofs}
\end{lemma}

\begin{lemma}\label{lemma:conditional-distribute-disjunctive-right}
$P \dres c \rres (Q \lor R) = (P \dres c \rres Q) \lor (P \dres c \rres R)$
\begin{proofs}
\begin{proof}
\begin{flalign*}
	&P \dres c \rres (Q \lor R)
	&&\ptext{Definition of conditional}\\
	&=(c \land P) \lor (\lnot c \land (Q \lor R))
	&&\ptext{Predicate calculus}\\
	&=(c \land P) \lor (\lnot c \land Q) \lor (\lnot c \land R)
	&&\ptext{Predicate calculus}\\
	&=(c \land P) \lor (\lnot c \land Q) \lor (c \land P) \lor (\lnot c \land R)
	&&\ptext{Definition of conditional}\\
	&=(P \dres c \rres Q) \lor (P \dres c \rres R)
\end{flalign*}
\end{proof}
\end{proofs}
\end{lemma}

\begin{lemma}\label{lemma:lnot-(false-cond-Q):(true-cond-lnot-Q)}
$\lnot (false \dres c \rres Q) = true \dres c \rres \lnot Q$
\begin{proofs}
\begin{proof}\checkt{pfr}\checkt{alcc}
\begin{flalign*}
	&\lnot (false \dres c \rres Q)
	&&\ptext{\cref{lemma:lnot(conditional)}}\\
	&=(true \dres c \rres \lnot Q)
\end{flalign*}
\end{proof}
\end{proofs}
\end{lemma}

\begin{lemma}\label{lemma:lnot-(true-cond-Q):(false-cond-lnot-Q)}
$\lnot (true \dres c \rres Q) = false \dres c \rres \lnot Q$
\begin{proofs}
\begin{proof}
\begin{flalign*}
	&\lnot (true \dres c \rres Q)
	&&\ptext{\cref{lemma:lnot-(false-cond-Q):(true-cond-lnot-Q)}}\\
	&=\lnot \lnot (false \dres c \rres \lnot Q)
	&&\ptext{Predicate calculus}\\
	&=false \dres c \rres \lnot Q
\end{flalign*}
\end{proof}
\end{proofs}
\end{lemma}

\subsection{Predicate Calculus}

\begin{lemma}\label{lemma:(P-land-Q)-iff-P:P-implies-Q}
$(P \land Q) \iff P = P \implies Q$
\begin{proofs}
\begin{proof}\checkt{alcc}\checkt{pfr}
\begin{xflalign*}
	&(P \land Q) \iff P
	&&\ptext{Predicate calculus}\\
	&=((P \land Q) \implies P) \land (P \implies (P \land Q))
	&&\ptext{Predicate calculus}\\
	&=(P \implies (P \land Q))
	&&\ptext{Predicate calculus}\\
	&=P \implies Q
\end{xflalign*}
\end{proof}
\end{proofs}
\end{lemma}

\begin{lemma}\label{lemma:(P-lor-Q)-iff-(P-lor-R):P-lor-(Q-iff-R)}
$(P \lor Q) \iff (P \lor R) = P \lor (Q \iff R)$
\begin{proofs}
\begin{proof}\checkt{alcc}\checkt{pfr}
\begin{xflalign*}
	&(P \lor Q) \iff (P \lor R)
	&&\ptext{Predicate calculus}\\
	&=((P \lor Q) \implies (P \lor R)) \land ((P \lor R) \implies (P \lor Q))
	&&\ptext{Predicate calculus}\\
	&=(P \implies (P \lor R)) \land (Q \implies (P \lor R)) \land (P \implies (P \lor Q)) \land (R \implies (P \lor Q))
	&&\ptext{Predicate calculus}\\
	&=(Q \implies (P \lor R)) \land (R \implies (P \lor Q))
	&&\ptext{Predicate calculus}\\
	&=(\lnot Q \lor P \lor R) \land (\lnot R \lor P \lor Q)
	&&\ptext{Predicate calculus}\\
	&=P \lor ((\lnot Q \lor R) \land (\lnot R \lor Q))
	&&\ptext{Predicate calculus}\\
	&=P \lor (Q \iff R)
\end{xflalign*}
\end{proof}
\end{proofs}
\end{lemma}

\section{Theory of Designs}
\label{appendix:designs:healthiness-conditions}

\subsection{Healthiness Conditions}

\subsubsection{$\mathbf{H1}$}

\begin{lemma}\label{lemma:H1(P-cond-Q):H1(P)-cond-H1(Q)}
$\mathbf{H1} (P \dres c \rres Q) = \mathbf{H1} (P) \dres c \rres \mathbf{H1} (Q)$
\begin{proofs}
\begin{proof}\checkt{alcc}
\begin{xflalign*}
	&\mathbf{H1} (P) \dres c \rres \mathbf{H1} (Q)
	&&\ptext{Definition of $\mathbf{H1}$}\\
	&=(\lnot ok \lor P) \dres c \rres (\lnot ok \lor Q)
	&&\ptext{Property of conditional}\\
	&=(\lnot ok \dres c \rres \lnot ok) \lor (P \dres s.wait \rres Q)
	&&\ptext{Property of conditional}\\
	&=\lnot ok \lor (P \dres s.wait \rres Q)
	&&\ptext{Definition of $\mathbf{H1}$}\\
	&=\mathbf{H1} (P \dres s.wait \rres Q)
\end{xflalign*}
\end{proof}
\end{proofs}
\end{lemma}

\begin{lemma}\label{lemma:H1(P-land-Q):H1(P)-land-H1(Q)}
$\mathbf{H1} (P \land Q) = \mathbf{H1} (P) \land \mathbf{H1} (Q)$
\begin{proofs}
\begin{proof}\checkt{alcc}
\begin{xflalign*}
	&\mathbf{H1} (P \land Q)
	&&\ptext{Definition of $\mathbf{H1}$}\\
	&=ok \implies (P \land Q)
	&&\ptext{Predicate calculus}\\
	&=(ok \implies P) \land (ok \implies Q)
	&&\ptext{Definition of $\mathbf{H1}$}\\
	&=\mathbf{H1} (P) \land \mathbf{H1} (Q)
\end{xflalign*}
\end{proof}
\end{proofs}
\end{lemma}

\begin{lemma}\label{lemma:H1(P-lor-Q):H1(P)-lor-H1(Q)}
$\mathbf{H1} (P \lor Q) = \mathbf{H1} (P) \lor \mathbf{H1} (Q)$
\begin{proofs}
\begin{proof}\checkt{alcc}
\begin{xflalign*}
	&\mathbf{H1} (P \lor Q)
	&&\ptext{Definition of $\mathbf{H1}$}\\
	&=ok \implies (P \lor Q)
	&&\ptext{Predicate calculus}\\
	&=(ok \implies P) \lor (ok \implies Q)
	&&\ptext{Definition of $\mathbf{H1}$}\\
	&=\mathbf{H1} (P) \lor \mathbf{H1} (Q)
\end{xflalign*}
\end{proof}
\end{proofs}
\end{lemma}

\subsubsection{$\mathbf{H2}$}

\begin{define}
$\mathbf{H2A} (P) \circdef \lnot P^f \implies (P^t \land ok')$
\end{define}

\begin{lemma}[$\mathbf{H2A} \iff \mathbf{H2}$]
The definition of $\mathbf{H2A}$ implies that the fixed points are the same as those of $\mathbf{H2}$,
\begin{proofs}
\begin{proof}[Proof for implication] The following proof is based on~\cite{Woodcock1996}.
\begin{flalign*}
	&P
	&&\ptext{Introduce fresh variable and substitution}\\
	&= \exists ok_0 \spot P \land ok'=ok_0
	&&\ptext{Case-split on $ok_0$}\\
	&= (\lnot ok' \land P^f) \lor (ok' \land P^t)
	&&\ptext{Assumption: P is $\mathbf{H2}$-healthy}\\
	&= (\lnot ok' \land P^f \land P^t) \lor (ok' \land P^t)
	&&\ptext{Propositional calculus}\\
	&= (((\lnot ok' \land P^f) \lor ok') \land P^t)
	&&\ptext{Propositional calculus}\\
	&= ((P^f \lor ok') \land P^t)
	&&\ptext{Propositional calculus}\\
	&= (P^f \land P^t) \lor (ok' \land P^t)
	&&\ptext{Assumption: P is $\mathbf{H2}$-healthy}&\\
	&= P^f \lor (ok' \land P^t)
	&&\ptext{Propositional calculus}\\
	&= \lnot P^f \implies (P^t \land ok')
\end{flalign*}
\end{proof}
\begin{proof}[Proof for reverse implication]
\begin{flalign*}
	&[(\mathbf{H2A} (P))^f \implies (\mathbf{H2A} (P))^t]
	&&\ptext{Definition of $\mathbf{H2A}$}\\
	&=[(\lnot P^f \implies (P^t \land ok'))^f \implies (\lnot P^f \implies (P^t \land ok'))^t]
	&&\ptext{Substitution}\\
	&=[(P^f \implies (\lnot P^f \implies P^t)]
	&&\ptext{Propositional calculus}\\
	&=[\lnot P^f \lor P^f \lor P^t]
	&&\ptext{Propositional calculus}\\
	&=true
\end{flalign*}
\end{proof}
\end{proofs}
\end{lemma}

\subsection{Lemmas}

\begin{lemma}\label{law:design:true-ok'}
Provided $ok \land P$ and $ok'$ is not free in $P$,
$(P \vdash Q)^t = Q$.
\begin{proofs}\begin{proof}\checkt{alcc}As stated and proved in~\cite{Cavalcanti2011a} (Lemma 4.2).\end{proof}\end{proofs}
\end{lemma}

\begin{lemma}\label{law:design:false-ok'}
Provided $ok'$ is not free in $P$,
$ok \land \lnot(P \vdash Q)^f = ok \land P$.
\begin{proofs}\begin{proof}\checkt{alcc}As stated and proved in~\cite{Cavalcanti2011a} (Lemma 4.3).\end{proof}\end{proofs}
\end{lemma}

\begin{lemma}\label{law:design:exists-ok'}
$\exists ok' \spot (P \vdash Q) = (ok \land P) \implies Q$
\begin{proofs}
\begin{proof}
\begin{flalign*}
	&\exists ok' \spot (P \vdash Q)
	&&\ptext{Definition of design}\\
	&=\exists ok' \spot (ok \land P) \implies (Q \land ok')
	&&\ptext{Case-split on $ok'$}\\
	&=((ok \land P) \implies Q) \lor \lnot (ok \land P)
	&&\ptext{Propositional calculus}\\
	&=(ok \land P) \implies Q
\end{flalign*}
\end{proof}
\end{proofs}
\end{lemma}

\begin{lemma}\label{law:design:sqcup}
\begin{align*}
	&(\lnot P^f \vdash P^t) \sqcup (\lnot Q^f \vdash Q^t) \\
	&=\\
	&(\lnot P^f \lor \lnot Q^f \vdash (\lnot P^f \implies P^t) \land (\lnot Q^f \implies Q^t))
\end{align*}
\begin{proofs}
\begin{proof}
\begin{flalign*}
	&(\lnot P^f \vdash P^t) \sqcup (\lnot Q^f \vdash Q^t)
	&&\ptext{Definition of design}\\
	&=((ok \land \lnot P^f) \implies (P^t \land ok')) \sqcup ((ok \land \lnot Q^f) \implies (Q^t \land ok'))
	&&\ptext{Definition of $\sqcup$}\\
	&=((ok \land \lnot P^f) \implies (P^t \land ok')) \land ((ok \land \lnot Q^f) \implies (Q^t \land ok'))
	&&\ptext{Propositional calculus}\\
	&=ok \implies ((P^t \land ok') \lor P^f) \land ((Q^t \land ok') \lor Q^f)
	&&\ptext{Propositional calculus}\\
	&=ok \implies (P^t \lor P^f) \land (ok' \lor P^f) \land (Q^t \lor Q^f) \land (ok' \lor Q^f)
	&&\ptext{Propositional calculus}\\
	&=ok \implies (P^t \lor P^f) \land (Q^t \lor Q^f) \land (ok' \lor (P^f \land Q^f))
	&&\ptext{Propositional calculus: absorption law}\\
	&=ok \implies ((P^f \land Q^f) \lor P^t \lor P^f) \land ((P^f \land Q^f) \lor Q^t \lor Q^f) \land (ok' \lor (P^f \land Q^f))
	&&\ptext{Propositional calculus}\\
	&=ok \implies (P^f \land Q^f) \lor ((P^t \lor P^f) \land (Q^t \lor Q^f) \land ok')
	&&\ptext{Propositional calculus}\\
	&=(ok \land \lnot (P^f \land Q^f)) \implies ((\lnot P^f \implies P^t) \land (\lnot Q^f \implies Q^t) \land ok')
	&&\ptext{Definition of design}\\
	&=(\lnot P^f \lor \lnot Q^f \vdash (\lnot P^f \implies P^t) \land (\lnot Q^f \implies Q^t))
\end{flalign*}
\end{proof}
\end{proofs}
\end{lemma}

\begin{lemma}\label{law:design:exists-ok':sqcup}
Provided $P$ and $Q$ are designs,
\begin{align*}
	&\exists ok' \spot (P \land Q) = (\exists ok' \spot P) \land (\exists ok' \spot Q)
\end{align*}
\begin{proofs}
\begin{proof}
\begin{flalign*}
	&(\exists ok' \spot P) \land (\exists ok' \spot Q)
	&&\ptext{Assumption: $P$ and $Q$ are designs}\\
	&=(\exists ok' \spot (\lnot P^f \vdash P^t)) \land (\exists ok' \spot (\lnot Q^f \vdash Q^t))
	&&\ptext{\cref{law:design:exists-ok'}}\\
	&=((ok \land \lnot P^f ) \implies P^t) \land ((ok \land \lnot Q^f) \implies Q^t)
	&&\ptext{Propositional calculus}\\
	&=(ok \implies (P^t \lor P^f)) \land (ok \implies (Q^t \lor Q^f))
	&&\ptext{Propositional calculus}\\
	&=ok \implies ((P^t \lor P^f) \land (Q^t \lor Q^f))
	&&\ptext{Propositional calculus: absorption law}\\
	&=ok \implies (((P^f \land Q^f) \lor P^t \lor P^f) \land ((P^f \land Q^f) \lor Q^t \lor Q^f))
	&&\ptext{Propositional calculus}\\
	&=ok \implies ((P^f \land Q^f) \lor ((P^t \lor P^f) \land (Q^t \lor Q^f)))
	&&\ptext{Propositional calculus}\\
	&=(ok \land \lnot (P^f \land Q^f)) \implies ((\lnot P^f \implies P^t) \land (\lnot Q^f \implies Q^t))
	&&\ptext{\cref{law:design:exists-ok'}}\\
	&=\exists ok' \spot (\lnot (P^f \land Q^f) \vdash (\lnot P^f \implies P^t) \land (\lnot Q^f \implies Q^t))
	&&\ptext{Conjunction of designs}\\
	&=\exists ok' \spot (\lnot P^f \vdash P^t) \land (\lnot Q^f \vdash Q^t)
	&&\ptext{Assumption: $P$ and $Q$ are designs}\\
	&=\exists ok' \spot (P \land Q)
\end{flalign*}
\end{proof}
\end{proofs}
\end{lemma}

\begin{lemma}\label{law:designs:conjunction-alternative}
\begin{align*}
	&(\lnot P^f \vdash P^t) \sqcup (\lnot Q^f \vdash Q^t)\\
	&=\\
	&(\lnot P^f \lor \lnot Q^f \vdash (P^f \land Q^t) \lor (P^t \land Q^f) \lor (P^t \land Q^t))
\end{align*}
\begin{proofs}
\begin{proof}\checkt{alcc}
\begin{flalign*}
	&(\lnot P^f \vdash P^t) \sqcup (\lnot Q^f \vdash Q^t)
	&&\ptext{Conjunction of designs}\\
	&=(\lnot P^f \lor \lnot Q^f \vdash (\lnot P^f \implies P^t) \land (\lnot Q^f \implies Q^t))
	&&\ptext{Propositional calculus}\\
	&=(\lnot P^f \lor \lnot Q^f \vdash (P^f \lor P^t) \land (Q^f \lor Q^t))
	&&\ptext{Predicate calculus}\\
	&=(\lnot (P^f \land Q^f) \vdash (P^f \land Q^f) \lor (P^f \land Q^t) \lor (P^t \land Q^f) \lor (P^t \land Q^t))
	&&\ptext{Definition of design}\\
	&=\left(\begin{array}{l}
		(ok \land \lnot (P^f \land Q^f)) 
		\\ \implies \\
		(((P^f \land Q^f) \lor (P^f \land Q^t) \lor (P^t \land Q^f) \lor (P^t \land Q^t)) \land ok')
	\end{array}\right)
	&&\ptext{Predicate calculus}\\
	&=\left(\begin{array}{l}
		(ok \land \lnot (P^f \land Q^f) \land (\lnot (P^f \land Q^f) \lor \lnot ok')) 
		\\ \implies \\
		(((P^f \land Q^t) \lor (P^t \land Q^f) \lor (P^t \land Q^t)) \land ok')
	\end{array}\right)
	&&\ptext{Predicate calculus: absorption law}\\
	&=(ok \land \lnot (P^f \land Q^f)) \implies (((P^f \land Q^t) \lor (P^t \land Q^f) \lor (P^t \land Q^t)) \land ok')
	&&\ptext{Definition of design}\\
	&=(\lnot (P^f \land Q^f) \vdash (P^f \land Q^t) \lor (P^t \land Q^f) \lor (P^t \land Q^t))
	&&\ptext{Predicate calculus}\\
	&=(\lnot P^f \lor \lnot Q^f \vdash (P^f \land Q^t) \lor (P^t \land Q^f) \lor (P^t \land Q^t))
\end{flalign*}
\end{proof}
\end{proofs}
\end{lemma}

\begin{lemma}\label{lemma:design:(P|-Q)f:ok-implies-lnot-Pf}
$(P \vdash Q)^f = ok \implies \lnot P^f$
\begin{proofs}
\begin{proof}
\begin{xflalign*}
	&(P \vdash Q)^f
	&&\ptext{Definition of design}\\
	&=((ok \land P) \implies (Q \land ok'))^f
	&&\ptext{Substitution}\\
	&=((ok \land P^f) \implies (Q^f \land false))
	&&\ptext{Predicate calculus}\\
	&=\lnot (ok \land P^f)
	&&\ptext{Predicate calculus}\\
	&=ok \implies \lnot P^f
\end{xflalign*}
\end{proof}
\end{proofs}
\end{lemma}

\begin{lemma}\label{lemma:design(P|-Q)t:ok-land-Pt-implies-Qt}
$(P \vdash Q)^t = (ok \land P^t) \implies Q^t$
\begin{proofs}
\begin{proof}
\begin{xflalign*}
	&(P \vdash Q)^t
	&&\ptext{Definition of design}\\
	&=((ok \land P) \implies (Q \land ok'))^t
	&&\ptext{Substitution}\\
	&=((ok \land P^t) \implies (Q^t \land true))
	&&\ptext{Predicate calculus}\\
	&=(ok \land P^t) \implies Q^t
\end{xflalign*}
\end{proof}
\end{proofs}
\end{lemma}

\begin{lemma}\label{lemma:design:ok-land-lnot-exists(P|-Q)f:ok-land-lnot-exists-Pf}
$ok \land \lnot \exists ac' \spot (P \vdash Q)^f = ok \land \lnot \exists ac' \spot \lnot P^f$
\begin{proofs}
\begin{proof}
\begin{xflalign*}
	&ok \land \lnot \exists ac' \spot (P \vdash Q)^f
	&&\ptext{\cref{lemma:design:(P|-Q)f:ok-implies-lnot-Pf}}\\
	&=ok \land \lnot \exists ac' \spot (ok \implies \lnot P^f)
	&&\ptext{Predicate calculus}\\
	&=\lnot (\lnot ok \lor \exists ac' \spot (ok \implies \lnot P^f))
	&&\ptext{Predicate calculus}\\
	&=\lnot \exists ac' \spot (\lnot ok \lor (ok \implies \lnot P^f))
	&&\ptext{Predicate calculus}\\
	&=\lnot \exists ac' \spot (\lnot ok \lor \lnot P^f)
	&&\ptext{Predicate calculus}\\
	&=\lnot (\lnot ok \lor \exists ac' \spot \lnot P^f)
	&&\ptext{Predicate calculus}\\
	&=(ok \lor \lnot \exists ac' \spot \lnot P^f)
	&&\ptext{Predicate calculus}\\
	&=ok \land \lnot \exists ac' \spot \lnot P^f
\end{xflalign*}
\end{proof}
\end{proofs}
\end{lemma}

\begin{lemma}\label{lemma:design:(P|-Q)f-|-(P|-Q)t} Provided $ok$ is not free in $P$ and $Q$,
\begin{align*}
	&((P \vdash Q)^f \vdash (P \vdash Q)^t) = (P \vdash Q)
\end{align*}
\begin{proofs}
\begin{proof}\checkt{alcc}
\begin{xflalign*}
	&((P \vdash Q)^f \vdash (P \vdash Q)^t)
	&&\ptext{\cref{law:design:false-ok'}}\\
	&=(P \vdash (P \vdash Q)^t)
	&&\ptext{\cref{law:design:true-ok'}}\\
	&=(P \vdash Q)
\end{xflalign*}
\end{proof}
\end{proofs}
\end{lemma}

\begin{lemma}\label{lemma:design:(lnot-exists-ac'-(P|-Q)f-|-(P|-Q)t)} Provided $ok'$ is not free in $P$ and $Q$,
\begin{align*}
	&(\lnot \exists ac' \spot (P \vdash Q)^f \vdash (P \vdash Q)^t) = (\lnot \exists ac' \spot \lnot P \vdash Q)
\end{align*}
\begin{proofs}
\begin{proof}
\begin{xflalign*}
	&(\lnot \exists ac' \spot (P \vdash Q)^f \vdash (P \vdash Q)^t)
	&&\ptext{Definition of design and~\cref{lemma:design:ok-land-lnot-exists(P|-Q)f:ok-land-lnot-exists-Pf}}\\
	&=(\lnot \exists ac' \spot \lnot P^f \vdash (P \vdash Q)^t)
	&&\ptext{Assumption: $ok'$ is not free in $P$}\\
	&=(\lnot \exists ac' \spot \lnot P \vdash (P \vdash Q)^t)
	&&\ptext{\cref{lemma:design(P|-Q)t:ok-land-Pt-implies-Qt}}\\
	&=(\lnot \exists ac' \spot \lnot P \vdash (ok \land P^t) \implies Q^t)
	&&\ptext{Assumption: $ok'$ not free in $P$ and $Q$}\\
	&=(\lnot \exists ac' \spot \lnot P \vdash (ok \land P) \implies Q)
	&&\ptext{Definition of design and predicate calculus}\\
	&=(\lnot \exists ac' \spot \lnot P \vdash P \implies Q)
	&&\ptext{Predicate calculus}\\
	&=(\lnot \exists ac' \spot \lnot P \vdash (\lnot P \land (\exists ac' \spot \lnot P)) \lor Q)
	&&\ptext{Definition of design and predicate calculus}\\
	&=(\lnot \exists ac' \spot \lnot P \vdash Q)
\end{xflalign*}
\end{proof}
\end{proofs}
\end{lemma}

\begin{lemma}\label{lemma:design:(lnot-(P|-Q)ff|-(P|-Q)tf)}
\begin{statement}
Provided $ok'$ is not free in $P$ and $Q$,
\begin{align*}
	&(\lnot (P \vdash Q)^f_f \vdash (P \vdash Q)^t_f) = (P_f \vdash Q_f)
\end{align*}
\end{statement}
\begin{proofs}
\begin{proof}
\begin{xflalign*}
	&(\lnot (P \vdash Q)^f_f \vdash (P \vdash Q)^t_f)
	&&\ptext{Definition of design}\\
	&=(\lnot ((ok \land P) \implies (Q \land ok'))^f_f \vdash ((ok \land P) \implies (Q \land ok'))^t_f)
	&&\ptext{Substitution}\\
	&=(\lnot ((ok \land P_f) \implies (Q_f \land false)) \vdash ((ok \land P_f) \implies (Q_f \land true)))
	&&\ptext{Predicate calculus}\\
	&=(ok \land P_f \vdash ((ok \land P_f) \implies Q_f))
	&&\ptext{Definition of design}\\
	&=(P_f \vdash ((ok \land P_f) \implies Q_f))
	&&\ptext{Definition of design and predicate calculus}\\
	&=(P_f \vdash Q_f)
\end{xflalign*}
\end{proof}
\end{proofs}
\end{lemma}

\section{Theory of CSP}

\subsection{Operators}

\begin{lemma}\label{lemma:Top-R-extchoice-Skip-R:Skip-R}
\begin{statement}
$\top_{\mathbf{R}} \extchoice_{\mathbf{R}} Skip_{\mathbf{R}} = Skip_{\mathbf{R}}$
\end{statement}
\begin{proofs}
\begin{proof}
\begin{xflalign*}
	&\top_{\mathbf{R}} \extchoice_{\mathbf{R}} Skip_{\mathbf{R}}
	&&\ptext{Definition of $\top_{\mathbf{R}}$ and $Skip_{\mathbf{R}}$}\\
	&=\mathbf{R} (true \vdash false) \extchoice_{\mathbf{R}} \mathbf{R} (true \vdash \lnot wait' \land tr'=tr)
	&&\ptext{Definition of $\extchoice_{\mathbf{R}}$}\\
	&=\mathbf{R} \left(\begin{array}{l} 
		true \land true
		\\ \vdash \\
		(false \land \lnot wait' \land tr'=tr) \dres tr'=tr \land wait' \rres (false \lor (\lnot wait' \land tr'=tr))
	\end{array}\right)
	&&\ptext{Predicate calculus}\\
	&=\mathbf{R} \left(\begin{array}{l} 
		true
		\\ \vdash \\
		false \dres tr'=tr \land wait' \rres (\lnot wait' \land tr'=tr)
	\end{array}\right)
	&&\ptext{Definition of conditional and predicate calculus}\\
	&=\mathbf{R} \left(\begin{array}{l} 
		true
		\\ \vdash \\
		\lnot (tr'=tr \land wait') \land (\lnot wait' \land tr'=tr)
	\end{array}\right)
	&&\ptext{Predicate calculus}\\
	&=\mathbf{R} \left(\begin{array}{l} 
		true
		\\ \vdash \\
		(tr'\neq tr \lor \lnot wait') \land (\lnot wait' \land tr'=tr)
	\end{array}\right)
	&&\ptext{Predicate calculus}\\
	&=\mathbf{R} (true \vdash \lnot wait' \land tr'=tr)
	&&\ptext{Definition of $Skip_{\mathbf{R}}$}
	&=Skip_{\mathbf{R}}
\end{xflalign*}
\end{proof}
\end{proofs}
\end{lemma}

\begin{lemma}\label{lemma:P-extchoice-Stop-R:P}
\begin{statement}
Provided $P$ is a CSP process,
\begin{align*}
	&P \extchoice_{\mathbf{R}} Stop_{\mathbf{R}} = P
\end{align*}
\end{statement}
\begin{proofs}
\begin{proof}
\begin{xflalign*}
	&P \extchoice_{\mathbf{R}} Stop_{\mathbf{R}}
	&&\ptext{Assumption: $P$ is a~\ac{CSP} process and definition of $Stop_{\mathbf{R}}$}\\
	&=\left(\begin{array}{l} 
		\mathbf{R} (\lnot P^f_f \vdash P^t_f)
		\\ \extchoice_{\mathbf{R}} \\
		\mathbf{R} (true \vdash wait' \land tr'=tr)
	\end{array}\right)
	&&\ptext{Definition of $\extchoice_{\mathbf{R}}$}\\
	&=\mathbf{R} \left(\begin{array}{l} 
		true \land \lnot P^f_f
		\\ \vdash \\
		(P^t_f \land wait' \land tr'=tr) \dres tr'=tr \land wait' \rres (P^t_f \lor (wait' \land tr'=tr))
	\end{array}\right)
	&&\ptext{Definition of conditional and predicate calculus}\\
	&=\mathbf{R} \left(\begin{array}{l} 
		\lnot P^f_f
		\\ \vdash \\
		\left(\begin{array}{l} 
			(P^t_f \land tr'=tr \land wait')
			\\ \lor \\
			(tr'\neq tr \land (P^t_f \lor (wait' \land tr'=tr)))
			\\ \lor \\
			(\lnot wait' \land (P^t_f \lor (wait' \land tr'=tr)))
		\end{array}\right)
	\end{array}\right)
	&&\ptext{Predicate calculus}\\
	&=\mathbf{R} \left(\begin{array}{l} 
		\lnot P^f_f
		\\ \vdash \\
		\left(\begin{array}{l} 
			(P^t_f \land tr'=tr \land wait')
			\\ \lor \\
			(tr'\neq tr \land P^t_f)
			\\ \lor \\
			(\lnot wait' \land P^t_f)
		\end{array}\right)
	\end{array}\right)
	&&\ptext{Predicate calculus}\\
	&=\mathbf{R} \left(\begin{array}{l} 
		\lnot P^f_f
		\\ \vdash \\
		(P^t_f \land ((tr'=tr \land wait') \lor tr'\neq tr \lor \lnot wait'))
	\end{array}\right)
	&&\ptext{Predicate calculus}\\
	&=\mathbf{R} \left(\begin{array}{l} 
		\lnot P^f_f
		\\ \vdash \\
		(P^t_f \land ((tr'=tr \land wait') \lor \lnot (tr'=tr \land wait')))
	\end{array}\right)
	&&\ptext{Predicate calculus}\\
	&=\mathbf{R} (\lnot P^f_f \vdash P^t_f)
	&&\ptext{Assumption: $P$ is a~\ac{CSP} process}\\
	&=P
\end{xflalign*}
\end{proof}
\end{proofs}
\end{lemma}

\begin{lemma}\label{lemma:exp:a-circthen-Stop}
\begin{align*}
	&a \circthen_{\mathbf{R}} Stop_{\mathbf{R}} = \mathbf{R} (true \vdash wait' \land ((a\notin ref' \land tr'=tr) \lor (tr'=tr\cat \lseq a \rseq)))
\end{align*}
\begin{proofs}\begin{proof}
\begin{flalign*}
	&a \circthen_{\mathbf{R}} Stop_{\mathbf{R}}
	&&\ptext{Definition of event prefixing}\\
	&=a \circthen Skip \circseq Stop
	&&\ptext{Definition of event prefixing and $Stop$}\\
	&=\left(
\begin{array}{l}
		\mathbf{R} (true \vdash (tr'=tr\land a\notin ref')\dres wait' \rres (tr'=tr\cat \lseq a \rseq))
		\circseq
		\mathbf{R} (true \vdash tr'=tr \land wait')
	\end{array}\right)
	&&\ptext{Definition of sequential composition}\\
	&=\mathbf{R} \left(%
\right)
	&&\ptext{Predicate calculus and definition of $\mathbf{R1}$}\\
	&=\mathbf{R} \left(%
\right)
	&&\ptext{Definition of sequential composition}\\
	&=\mathbf{R} \left(\begin{array}{l}
		(\lnot false \land \lnot false)
		\\ \vdash \\
		\mathbf{R1} ((tr'=tr\land a\notin ref')\dres wait' \rres (tr'=tr\cat \lseq a \rseq)) \circseq (\II \dres wait \rres \mathbf{R1} (tr'=tr \land wait'))
	\end{array}\right)
	&&\ptext{Predicate calculus}\\
	&=\mathbf{R} \left(\begin{array}{l}
		true
		\\ \vdash \\
		\mathbf{R1} ((tr'=tr\land a\notin ref')\dres wait' \rres (tr'=tr\cat \lseq a \rseq)) \circseq (\II \dres wait \rres \mathbf{R1} (tr'=tr \land wait'))
	\end{array}%
\right)
	&&\ptext{Definition of conditional}\\
	&=\mathbf{R} \left(%
\right)
	&&\ptext{Distributivity of sequential composition}\\
	&=\mathbf{R} \left(%
\right)
	&&\ptext{Property of sequential composition}\\
	&=\mathbf{R} \left(%
\right)
	&&\ptext{Definition of sequential composition}\\
	&=\mathbf{R} \left(%
\right)
\end{align*}
\begin{proofs}\begin{proof}
\begin{flalign*}
	&a \circthen_{\mathbf{R}} Choice_{\mathbf{R}}
	&&\ptext{Definition of prefixing}\\
	&=(a \circthen_{\mathbf{R}} Skip_{\mathbf{R}}) \circseq_{\mathbf{R}} Choice_{\mathbf{R}}
	&&\ptext{Definition of prefixing and $Choice$}\\
	&=\left(\begin{array}{l}
		\mathbf{R} (true \vdash (tr'=tr \land a\notin ref')\dres wait' \rres (tr'=tr\cat \lseq a \rseq))
		\\ \circseq \\
		\mathbf{R} (true \vdash true)
	\end{array}%
\right)
	&&\ptext{Property of sequential composition}\\
	&=\mathbf{R} \left(\begin{array}{l}
		\lnot false \land \lnot false	
		\\ \vdash \\
		\mathbf{R1} ((tr'=tr \land a\notin ref')\dres wait' \rres (tr'=tr\cat \lseq a \rseq)) \circseq (\II \dres wait \rres \mathbf{R1} (true))
	\end{array}\right)
	&&\ptext{Predicate calculus}\\
	&=\mathbf{R} \left(\begin{array}{l}
		true	
		\\ \vdash \\
		\mathbf{R1} ((tr'=tr \land a\notin ref')\dres wait' \rres (tr'=tr\cat \lseq a \rseq)) \circseq (\II \dres wait \rres \mathbf{R1} (true))
	\end{array}\right)
	&&\ptext{Definition of $\mathbf{R1}$ and predicate is $\mathbf{R1}$-healthy}\\
	&=\mathbf{R} \left(\begin{array}{l}
		true	
		\\ \vdash \\
		((tr'=tr \land a\notin ref')\dres wait' \rres (tr'=tr\cat \lseq a \rseq)) \circseq (\II \dres wait \rres tr\le tr')
	\end{array}\right)
	&&\ptext{Definition of conditional and predicate calculus}\\
	&=\mathbf{R} \left(%
\right)
	&&\ptext{Property of sequential composition}\\
	&=\mathbf{R} \left(%
\right)
	&&\ptext{One-point rule and definition of $\II$}\\
	&=\mathbf{R} \left(\begin{array}{l}
		true	
		\\ \vdash \\
		(tr'=tr \land a\notin ref' \land wait')\lor (tr\cat \lseq a \rseq\le tr')	
	\end{array}\right)
\end{flalign*}
\end{proof}\end{proofs}
\end{lemma}

\begin{lemma}\label{lemma:exp:a-circthen-Chaos}
\begin{align*}
	&a \circthen_{\mathbf{R}} Chaos_{\mathbf{R}} = \mathbf{R} (\lnot (tr \cat \lseq a \rseq \le tr') \vdash wait' \land tr'=tr \land a \notin ref')
\end{align*}
\begin{proofs}\begin{proof}
\begin{flalign*}
	&a \circthen_{\mathbf{R}} Chaos_{\mathbf{R}}
	&&\ptext{Definition of prefixing}\\
	&=a \circthen_{\mathbf{R}} Skip_{\mathbf{R}} \circseq \mathbf{R} (false \vdash true)
	&&\ptext{Definition of prefixing}\\
	&=\mathbf{R} (true \vdash (tr'=tr \land a\notin ref')\dres wait' \rres (tr'=tr\cat\lseq a \rseq)) \circseq \mathbf{R} (false \vdash true)
	&&\ptext{Definition of sequence}\\
	&=\mathbf{R} \left(
\right)
	&&\ptext{Property of $\mathbf{R1}$ and sequential composition}\\
	&=\mathbf{R} \left(%
\right)
	&&\ptext{Property of conditional}\\
	&=\mathbf{R} \left(\begin{array}{l}
		\lnot ((tr'=tr\cat\lseq a \rseq \land \lnot wait') \circseq \mathbf{R1} (true))
		\\ \vdash \\
		((tr'=tr \land a\notin ref')\dres wait' \rres (tr'=tr\cat\lseq a \rseq)) \circseq (\II \dres wait \rres \mathbf{R1} (true))
	\end{array}\right)
	&&\ptext{Definition of $\mathbf{R1} (true)$}\\
	&=\mathbf{R} \left(\begin{array}{l}
		\lnot ((tr'=tr\cat\lseq a \rseq \land \lnot wait') \circseq tr\le tr')
		\\ \vdash \\
		((tr'=tr \land a\notin ref')\dres wait' \rres (tr'=tr\cat\lseq a \rseq)) \circseq (\II \dres wait \rres tr \le tr')
	\end{array}\right)
	&&\ptext{Definition of conditional}\\
	&=\mathbf{R} \left(%
\right)
	&&\ptext{Property of sequential composition}\\
	&=\mathbf{R} \left(%
\right)
	&&\ptext{Definition of sequential composition}\\
	&=\mathbf{R} \left(%
\right)
	&&\ptext{Predicate calculus}\\
	&=\mathbf{R} (\lnot (tr\cat\lseq a \rseq\le tr') \vdash (a\notin ref' \land wait' \land tr=tr'))
\end{flalign*}
\end{proof}\end{proofs}
\end{lemma}

\chapter{Extended Binary Multirelations}\label{appendix:bm}

\section{Healthiness Conditions}

\subsection{$\mathbf{BMH0}$}

\theoremref{def:bm:bmh}

\begin{lemma}\label{law:bmbot:BMH0-BMH}
\begin{statement}
\begin{align*}
	&\mathbf{BMH0}\\
	&\iff \\
	&\left(%
\right)
	&&\ptext{Predicate calculus: type restriction}\\
	&\iff \left(%
\right)
	&&\ptext{Propositional calculus}\\
	&\implies \exists s_0 : State, ss_1 : \power State_\bot \spot (\bot \in ss_1 \land (s_0, ss_1) \in B)
\end{flalign*}
\end{proof}
\begin{proof}(Reverse implication)
\begin{flalign*}
	&\exists s_0 : State, ss_1 : \power State_\bot \spot (\bot \in ss_1 \land (s_0, ss_1) \in B)\\
	&&\ptext{Propositional calculus: introduce fresh variable}\\
	&=\left(\begin{array}{l}
		\exists s_0 : State, ss_0, ss_1 : \power State_\bot \spot \\
		(\bot \in ss_1 \land (s_0, ss_1) \in B \land ss_0 = ss_1 \land (s_0, ss_0) \in B \land \bot \in ss_0)
	\end{array}\right)\\
	&&\ptext{Propositional calculus: weaken predicate}\\
	&\implies \left(\begin{array}{l}
		\exists s_0 : State, ss_0, ss_1 : \power State_\bot \spot \\
		(\bot \in ss_1 \land (s_0, ss_1) \in B \land ss_0 \subseteq ss_1 \land (s_0, ss_0) \in B \land \bot \in ss_0)
	\end{array}\right)\\
\end{flalign*}
\end{proof}\end{proofs}
\end{lemma}

\subsection{$\mathbf{BMH1}$}

\begin{lemma}\label{law:bmh1:bot-notin-ss}
\begin{align*}
	&\mathbf{BMH1} \\
	&\iff \\
	&\forall s : State, ss : \power State_\bot \spot (s, ss \cup \{\bot\}) \in B \land \bot \notin ss \implies (s, ss) \in B
\end{align*}
\begin{proofs}\begin{proof}
\begin{flalign*}
	&\mathbf{BMH1}
	&&\ptext{Definition of $\mathbf{BMH1}$}\\
	&\iff \forall s : State, ss : \power State_\bot \spot (s, ss \cup \{\bot\}) \in B \implies (s, ss) \in B
	&&\ptext{Predicate calculus}\\
	&\iff \forall s : State, ss : \power State_\bot \spot 
		\left(\begin{array}{l}
			(s, ss \cup \{\bot\}) \in B \land (\bot \in ss \lor \bot \notin ss))
			\\ \implies \\
			(s, ss) \in B
		\end{array}\right)
	&&\ptext{Predicate calculus}\\
	&\iff \forall s : State, ss : \power State_\bot \spot 
		\left(\begin{array}{l}
		((s, ss \cup \{\bot\}) \in B \land \bot \in ss) \implies (s, ss) \in B
		\\ \land \\
		((s, ss \cup \{\bot\}) \in B \land \bot \notin ss) \implies (s, ss) \in B
	\end{array}\right)
	&&\ptext{Property of sets (\cref{law:set-theory:A-B:x-in-B})}\\
	&\iff \forall s : State, ss : \power State_\bot \spot 
		\left(\begin{array}{l}
		((s, ss) \in B \land \bot \in ss) \implies (s, ss) \in B
		\\ \land \\
		((s, ss \cup \{\bot\}) \in B \land \bot \notin ss) \implies (s, ss) \in B
	\end{array}\right)
	&&\ptext{Predicate calculus}\\
	&\iff \forall s : State, ss : \power State_\bot \spot ((s, ss \cup \{\bot\}) \in B \land \bot \notin ss) \implies (s, ss) \in B
\end{flalign*}
\end{proof}\end{proofs}
\end{lemma}

\section{Healthiness Conditions as Fixed Points}\label{appendix:bmbot:healthiness-conditions}
\subsection{$\mathbf{bmh}_\mathbf{0}$}

\begin{lemma}\label{law:bmbot:bmh0-definition}
\begin{statement}
$\mathbf{BMH0} \iff \mathbf{bmh}_\mathbf{0} (B) = B$
\end{statement}
\begin{proofs}
\begin{proof}
\begin{flalign*}
	&\mathbf{BMH0}
	&&\ptext{Definition of $\mathbf{BMH0}$}\\
	&\iff \left(
\begin{array}{l}
		\forall s_0 : State, ss_0 , ss_1 : \power State_\bot \spot \\
		((s_0, ss_0) \in B \land ss_0 \subseteq ss_1 
		\land (\bot \in ss_0 \iff \bot \in ss_1)) \implies (s_0, ss_1) \in B
	\end{array}\right)
	&&\ptext{Predicate calculus: quantifier scope}\\
	&\iff \left(%
\right)
	&&\ptext{Property of sets: subset inclusion}\\
	&\iff \left\{%
\right\} \subseteq B
	&&\ptext{Property of sets}\\
	&\iff \left(\left\{%
\right\} \cup B \right) =  B
	&&\ptext{Property of sets}\\
	&\iff \left(\left\{%
\right\}\right) = B
	&&\ptext{Instantiation of existential quantifier for $ss_0 = ss_1$}\\
	&\iff \left(\left\{%
\right\}
	&&\ptext{Property of sets}\\
	&=\left\{%
\right\}
	&&\ptext{Predicate calculus and transitivity of subset inclusion}\\
	&=\left\{%
\right.
	\end{array}\right\}
	&&\ptext{Definition of $\mathbf{bmh}_\mathbf{0}$}\\
	&=\mathbf{bmh}_\mathbf{0} (B)
\end{flalign*}
\end{proof}
\end{proofs}
\end{lemma}

\subsection{$\mathbf{bmh}_\mathbf{1}$}

\begin{lemma}\label{law:bmh1-definition}
\begin{statement}
$\mathbf{BMH1} \iff \mathbf{bmh}_\mathbf{1} (B) = B$
\end{statement}
\begin{proofs}
\begin{proof}
\begin{flalign*}
	&\mathbf{BMH1}
	&&\ptext{Definition of $\mathbf{BMH1}$}\\
	&\iff \forall s : State; ss : \power State_\bot \spot (s, ss \cup \{\bot\}) \in B \implies (s, ss) \in B
	&&\ptext{Property of sets and definition of subset inclusion}\\
	&\iff \{ s : State; ss : \power State_\bot | (s, ss \cup \{\bot\}) \in B \} \subseteq B
	&&\ptext{Property of sets}\\
	&\iff (\{ s : State; ss : \power State_\bot | (s, ss \cup \{\bot\}) \in B \} \cup B) = B
	&&\ptext{Property of sets}\\
	&\iff (\{ s : State; ss : \power State_\bot | (s, ss \cup \{\bot\}) \in B \lor (s, ss) \in B \}) = B
	&&\ptext{Definition of $\mathbf{bmh}_\mathbf{1}$}\\
	&\iff \mathbf{bmh}_\mathbf{1} (B) = B
\end{flalign*}
\end{proof}
\end{proofs}
\end{lemma}

\begin{lemma}\label{law:bmh1-idempotent}
\begin{statement}
	$\mathbf{bmh}_\mathbf{1} \circ \mathbf{bmh}_\mathbf{1} (B) = \mathbf{bmh}_\mathbf{1} (B)$
\end{statement}
\begin{proofs}
\begin{proof}
\begin{flalign*}
	&\mathbf{bmh}_\mathbf{1} \circ \mathbf{bmh}_\mathbf{1} (B)
	&&\ptext{Definition of $\mathbf{bmh}_\mathbf{1}$}\\
	&=\{ s : State, ss : \power State_\bot | (s, ss \cup \{\bot\}) \in \mathbf{bmh}_\mathbf{1} (B) \lor (s, ss) \in \mathbf{bmh}_\mathbf{1} (B) \}
	&&\ptext{Definition of $\mathbf{bmh}_\mathbf{1}$}\\
	&=\left\{\begin{array}{l}
		s : State, ss : \power State_\bot \\
		\left|\begin{array}{l}
			(s, ss \cup \{\bot\}) \in \left\{\begin{array}{l}
						  s : State, ss : \power State_\bot \\
						  | (s, ss \cup \{\bot\}) \in B \lor (s, ss) \in B 
					  	\end{array}\right\}
			\\ \lor \\
			(s, ss) \in \{ s : State, ss : \power State_\bot | (s, ss \cup \{\bot\}) \in B \lor (s, ss) \in B \}
		\end{array}\right.
	\end{array}\right\}
	&&\ptext{Property of sets}\\
	&=\left\{\begin{array}{l}
		s : State, ss : \power State_\bot \\
		\left|\begin{array}{l}
			(s, ss \cup \{\bot\} \cup \{\bot\}) \in B \lor (s, ss \cup \{\bot\}) \in B
			\\ \lor \\
			(s, ss \cup \{\bot\}) \in B \lor (s, ss) \in B
		\end{array}\right.
	\end{array}\right\}
	&&\ptext{Property of sets and predicate calculus}\\
	&=\{ s : State, ss : \power State_\bot | (s, ss \cup \{\bot\}) \in B \lor (s, ss) \in B \}
	&&\ptext{Definition of $\mathbf{bmh}_\mathbf{1}$}\\
	&=\mathbf{bmh}_\mathbf{1} (B)
\end{flalign*}
\end{proof}
\end{proofs}
\end{lemma}

\subsection{$\mathbf{bmh}_\mathbf{2}$}

\begin{lemma}\label{law:bmbot:bmh2-definition}
\begin{statement}
	$\mathbf{BMH2} \iff \mathbf{bmh}_\mathbf{2} (B) = B$
\end{statement}
\begin{proofs}
\begin{proof}
\begin{flalign*}
	&\mathbf{BMH2}
	&&\ptext{Definition of $\mathbf{BMH2}$}\\
	&\iff \forall s : State \spot (s, \emptyset) \in B \iff (s, \{ \bot \}) \in B
	&&\ptext{Predicate calculus}\\
	&\iff \forall s : State \spot \left(\begin{array}{l}
		(s, \emptyset) \in B \implies (s, \{ \bot \}) \in B
		\\ \land \\
		(s, \{ \bot \}) \in B \implies (s, \emptyset) \in B
		\end{array}\right)
	&&\ptext{Predicate calculus}\\
	&\iff \forall s : State \spot \left(\begin{array}{l}
		(\exists ss_0 : \power State_\bot \spot (s, \emptyset) \in B \land (s, ss_0) \in B) \implies (s, \{ \bot \}) \in B
		\\ \land \\
		(\exists ss_0 : \power State_\bot \spot (s, \{ \bot \}) \in B \land (s, ss_0) \in B) \implies (s, \emptyset) \in B
		\end{array}\right)
	&&\ptext{Predicate calculus}\\
	&\iff \forall s : State, ss_0 : \power State_\bot \spot \left(\begin{array}{l}
		((s, \emptyset) \in B \land (s, ss_0) \in B) \implies (s, \{ \bot \}) \in B
		\\ \land \\
		((s, \{ \bot \}) \in B \land (s, ss_0) \in B) \implies (s, \emptyset) \in B
		\end{array}\right)
	&&\ptext{Predicate calculus}\\
	&\iff \forall s : State, ss_0 : \power State_\bot \spot \left(\begin{array}{l}
		(s, ss_0) \in B \implies ((s, \{ \bot \}) \in B \lor (s, \emptyset) \notin B)
		\\ \land \\
		(s, ss_0) \in B \implies ((s, \emptyset) \in B \lor (s, \{ \bot \}) \notin B)
		\end{array}\right)
	&&\ptext{Predicate calculus}\\
	&\iff \forall s : State, ss_0 : \power State_\bot \spot 
		(s, ss_0) \in B \implies \left(\begin{array}{l}
				(s, \{ \bot \}) \in B \lor (s, \emptyset) \notin B) 
				\\ \land \\ 
				((s, \emptyset) \in B \lor (s, \{ \bot \}) \notin B)
			\end{array}\right)
	&&\ptext{Predicate calculus}\\
	&\iff \forall s : State, ss_0 : \power State_\bot \spot (s, ss_0) \in B \implies ((s, \{ \bot \}) \in B \iff (s, \emptyset) \in B)
	&&\ptext{Property of sets}\\
	&\iff B \subseteq \{ s : State, ss : \power State_\bot | (s, \{ \bot \}) \in B \iff (s, \emptyset) \in B \}
	&&\ptext{Property of sets}\\
	&\iff B = (B \cap \{ s : State, ss : \power State_\bot | (s, \{ \bot \}) \in B \iff (s, \emptyset) \in B \})
	&&\ptext{Property of sets}\\
	&\iff B = \{ s : State, ss : \power State_\bot | (s, ss) \in B \land ((s, \{ \bot \}) \in B \iff (s, \emptyset) \in B) \}
	&&\ptext{Definition of $\mathbf{bmh}_\mathbf{2}$}\\
	&\iff B = \mathbf{bmh}_\mathbf{2} (B)
\end{flalign*}
\end{proof}
\end{proofs}
\end{lemma}

\begin{lemma}\label{law:bmbot:bmh-2-idempotent}
\begin{statement}
	$\mathbf{bmh}_\mathbf{2} \circ \mathbf{bmh}_\mathbf{2} (B) = \mathbf{bmh}_\mathbf{2} (B)$
\end{statement}
\begin{proofs}
\begin{proof}
\begin{flalign*}
	&\mathbf{bmh}_\mathbf{2} \circ \mathbf{bmh}_\mathbf{2} (B)
	&&\ptext{Definition of $\mathbf{bmh}_\mathbf{2}$}\\
	&=\left\{
\right\}
	&&\ptext{Property of sets}\\
	&=\left\{%
\right\}
	&&\ptext{Definition of $\mathbf{bmh}_\mathbf{2}$}\\
	&=\mathbf{bmh}_\mathbf{2} (B)
\end{flalign*}
\end{proof}
\end{proofs}
\end{lemma}

\subsection{$\mathbf{bmh}_\mathbf{3}$}

\begin{lemma}\label{law:bmbot:bmh3-definition}
\begin{statement}
$\mathbf{BMH3} \iff \mathbf{bmh}_\mathbf{3} (B) = B$
\end{statement}
\begin{proofs}
\begin{proof}
\begin{flalign*}
	&\mathbf{BMH3}
	&&\ptext{Definition of $\mathbf{BMH3}$}\\
	&\iff \forall s : State \spot ((s, \emptyset) \notin B) \implies (\forall ss : \power State_\bot \spot (s, ss) \in B \implies \bot \notin ss)
	&&\ptext{Predicate calculus}\\
	&\iff \forall s : State, ss : \power State_\bot \spot ((s, \emptyset) \notin B) \implies ((s, ss) \in B \implies \bot \notin ss)
	&&\ptext{Predicate calculus}\\
	&\iff \forall s : State, ss : \power State_\bot \spot ((s, ss) \in B \land \bot \in ss) \implies (s, \emptyset) \in B
	&&\ptext{Predicate calculus}\\
	&\iff \forall s : State, ss : \power State_\bot \spot (s, ss) \in B \implies ((s, \emptyset) \in B \lor \bot \notin ss)
	&&\ptext{Property of sets and subset inclusion}\\
	&\iff B \subseteq \{ s : State, ss : \power State_\bot | ((s, \emptyset) \in B \lor \bot \notin ss) \}
	&&\ptext{Property of sets}\\
	&\iff B = (B \cap \{ s : State, ss : \power State_\bot | ((s, \emptyset) \in B \lor \bot \notin ss) \})
	&&\ptext{Property of sets}\\
	&\iff B = \{ s : State, ss : \power State_\bot | ((s, \emptyset) \in B \lor \bot \notin ss) \land (s, ss) \in B\}
	&&\ptext{Definition of $\mathbf{bmh}_\mathbf{3}$}\\
	&\iff B = \mathbf{bmh}_\mathbf{3} (B)
\end{flalign*}
\end{proof}
\end{proofs}
\end{lemma}

\begin{lemma}\label{law:bmbot:bmh3-idempotent}
\begin{statement}
	$\mathbf{bmh}_\mathbf{3} \circ \mathbf{bmh}_\mathbf{3} (B) = \mathbf{bmh}_\mathbf{3} (B)$
\end{statement}
\begin{proofs}
\begin{proof}
\begin{flalign*}
	&\mathbf{bmh}_\mathbf{3} \circ \mathbf{bmh}_\mathbf{3} (B)
	&&\ptext{Definition of $\mathbf{bmh}_\mathbf{3}$}\\
	&=\left\{
\right\}
	&&\ptext{Property of sets}\\
	&=\left\{%
\right\}
	&&\ptext{Predicate calculus: absorption law}\\
	&=\left\{%
\right\}
	&&\ptext{Predicate calculus and	definition of $\mathbf{bmh}_\mathbf{3}$}\\
	&=\mathbf{bmh}_\mathbf{3} (B)
\end{flalign*}
\end{proof}
\end{proofs}
\end{lemma}

\subsection{$\mathbf{bmh}_\mathbf{0}$ and $\mathbf{bmh}_\mathbf{1}$}\label{sec:bmbot:hc:bmh-0-1}

\begin{lemma}\label{law:bmbot:bmh0-circ-bmh1}
\begin{align*}
	&\mathbf{bmh}_\mathbf{0} \circ \mathbf{bmh}_\mathbf{1} (B)\\
	&=\\
	&\left\{%
\right\}
	&&\ptext{Property of sets}\\
	&=\left\{%
\right\}
\end{flalign*}
\end{proof}\end{proofs}
\end{lemma}

\subsubsection{Properties}

\begin{lemma}
$\mathbf{bmh}_\mathbf{0} \circ \mathbf{bmh}_\mathbf{1} (B) = \mathbf{bmh}_\mathbf{1} \circ \mathbf{bmh}_\mathbf{0} (B)$
\begin{proofs}\begin{proof}
\begin{flalign*}
	&\mathbf{bmh}_\mathbf{1} \circ \mathbf{bmh}_\mathbf{0} (B)
	&&\ptext{Definition of $\mathbf{bmh}_\mathbf{1}$}\\
	&=\left\{\begin{array}{l}
		s : State, ss : \power State_\bot | (s, ss \cup \{\bot\}) \in \mathbf{bmh}_\mathbf{0} (B) \lor (s, ss) \in \mathbf{bmh}_{0} (B)
	\end{array}%
\right\}
	&&\ptext{Property of sets}\\
	&=\left\{%
\right\}
	&&\ptext{Property of sets and predicate calculus}\\
	&=\left\{%
\right.
	\end{array}\right\}
	&&\ptext{\cref{law:bmbot:bmh0-circ-bmh1}}\\
	&=\mathbf{bmh}_\mathbf{0} \circ \mathbf{bmh}_\mathbf{1} (B)
\end{flalign*}
\end{proof}\end{proofs}
\end{lemma}

\subsection{$\mathbf{bmh}_\mathbf{1}$ and $\mathbf{bmh}_\mathbf{2}$}\label{sec:bmbot:hc:bmh-1-2}


\begin{lemma}\label{law:bmbot:bmh1-o-bmh2}
\begin{align*}
	&\mathbf{bmh}_\mathbf{1} \circ \mathbf{bmh}_\mathbf{2} (B)\\
	&=\left\{
\right\}
	&&\ptext{Property of sets}\\
	&=\left\{%
\right\}
	&&\ptext{Property of sets}\\
	&=\left\{%
\right\}
	&&\ptext{Property of sets and predicate calculus}\\
	&=\left\{%
\right.
	\end{array}\right\}	
\end{flalign*}
\end{proof}\end{proofs}
\end{lemma}


It can be conclued from~\cref{law:bmbot:bmh2-o-bmh1} and~\cref{law:bmbot:bmh1-o-bmh2} that the functional application of $\mathbf{bmh}_\mathbf{1} \circ \mathbf{bmh}_\mathbf{2}$ is stronger than that of $\mathbf{bmh}_\mathbf{2} \circ \mathbf{bmh}_\mathbf{1}$. The order in which these two healthiness conditions are functionally composed is important, since they are not necessarily commutative. The following counter-example illustrates the issue for a relation that is not $\mathbf{BMH2}$-healthy.

\begin{counter-example}
\begin{flalign*}
	&\mathbf{bmh}_\mathbf{2} \circ \mathbf{bmh}_\mathbf{1} (\{ s : State, ss : \power State_\bot | ss = \{\bot\} \})
	&&\ptext{\cref{law:bmbot:bmh2-o-bmh1}}\\
	&=\{ s : State, ss : \power State_\bot | ss = \{ \bot \} \lor ss = \emptyset \}\\
&\\
	&\mathbf{bmh}_\mathbf{1} \circ \mathbf{bmh}_\mathbf{2} (\{ s : State, ss : \power State_\bot | ss = \{\bot\} \})
	&&\ptext{\cref{law:bmbot:bmh1-o-bmh2}}\\
	&=\emptyset
\end{flalign*}
\end{counter-example}\noindent

\subsection{$\mathbf{bmh}_\mathbf{2}$ and $\mathbf{bmh}_\mathbf{3}$}\label{sec:bmbot:hc:bmh-2-3}

\begin{lemma}\label{law:bmbot:bmh2-o-bmh3}
\begin{align*}
	&\mathbf{bmh}_\mathbf{2} \circ \mathbf{bmh}_\mathbf{3} (B)\\
	&=\\
	&\left\{
\right\}
	&&\ptext{Property of sets}\\
	&=\left\{%
\right\}
	&&\ptext{Property of sets and predicate calculus}\\
	&=\left\{%
\right\}
	&&\ptext{Property of sets}\\
	&=\left\{%
\right\}
	&&\ptext{Predicate calculus: absorption law}\\
	&=\left\{%
\right.
	\end{array}\right\}
\end{flalign*}
\end{proof}\end{proofs}
\end{lemma}

The functions $\mathbf{bmh}_\mathbf{2}$ and $\mathbf{bmh}_\mathbf{3}$ are not in general commutative. The following counter-example illustrates the issue for a relation that is not $\mathbf{BMH2}$-healthy. 

\begin{counter-example}
\begin{flalign*}
	&\mathbf{bmh}_\mathbf{2} \circ \mathbf{bmh}_\mathbf{3} (\{ s : State, ss : \power State_\bot | ss = \{\bot\} \lor ss = \{ s \} \})
	&&\raisetag{18pt}\ptext{\cref{law:bmbot:bmh2-o-bmh3}}\\
	&=\{ s : State, ss : \power State_\bot | ss = \{ s \} \}
\end{flalign*}
\begin{flalign*}
	&\mathbf{bmh}_\mathbf{3} \circ \mathbf{bmh}_\mathbf{2} (\{ s : State, ss : \power State_\bot | ss = \{\bot\} \lor ss = \{ s \} \}) 
	&&\raisetag{18pt}\ptext{\cref{law:bmbot:bmh3-o-bmh2}}\\
	&=\emptyset
\end{flalign*}
\end{counter-example}

\subsection{$\mathbf{bmh}_\mathbf{1}$ and $\mathbf{bmh}_\mathbf{3}$}\label{sec:bmbot:hc:bmh-1-3}

\begin{lemma}\label{law:bmbot:bmh3-o-bmh1}
\begin{align*}
	&\mathbf{bmh}_\mathbf{3} \circ \mathbf{bmh}_\mathbf{1} (B)\\
	&=\\
	&\left\{
\right\}
	&&\ptext{Property of sets}\\
	&=\left\{%
\right\}
	&&\ptext{Property of sets}\\
	&=\left\{%
\right\}
	&&\ptext{Property of sets and predicate calculus}\\
	&=\left\{%
\right.
	\end{array}\right\}	
\end{flalign*}
\end{proof}\end{proofs}
\end{lemma}

The functions $\mathbf{bmh}_\mathbf{3}$ and $\mathbf{bmh}_\mathbf{1}$ do not necessarily commute. The following counter-example shows this for a relation that is not $\mathbf{BMH3}$-healthy. In fact, the functional application $\mathbf{bmh}_\mathbf{3} \circ \mathbf{bmh}_\mathbf{1}$ is not suitable as the counter-example shows that we have a fixed point.

\begin{counter-example}
\begin{flalign*}
	&\mathbf{bmh}_\mathbf{3} \circ \mathbf{bmh}_\mathbf{1} (\{ s : State, ss : \power State_\bot | ss = \{\bot, s\} \lor ss = \{\bot\} \}) 
	&&\raisetag{18pt}\ptext{\cref{law:bmbot:bmh3-o-bmh1}}\\
	&=\{ s : State, ss : \power State_\bot | ss = \{\bot, s\} \lor ss = \{\bot\} \}
\end{flalign*}
\begin{flalign*}
	&\mathbf{bmh}_\mathbf{1} \circ \mathbf{bmh}_\mathbf{3} (\{ s : State, ss : \power State_\bot | ss = \{\bot, s\} \lor ss = \{\bot\} \})
	&&\raisetag{18pt}\ptext{\cref{law:bmbot:bmh1-o-bmh3}}\\
	&=\emptyset
\end{flalign*}
\end{counter-example}

\subsection{$\mathbf{bmh}_\mathbf{0,1,2}$}

\begin{lemma}\label{law:bmbot:bmh-0-1-2:definition}
\begin{statement}
\begin{align*}
	&\mathbf{bmh}_\mathbf{0,1,2} (B)
	=
	\left\{
\right\}
	&&\ptext{Property of sets}\\
	&=\left\{%
\right.
	\end{array}\right\}
\end{flalign*}
\end{proof}
\end{proofs}
\end{lemma}

\begin{theorem}\label{law:bmbot:BMH-0-1-2:iff:bmh-0-1-2}
\begin{statement}
$\mathbf{BMH0} \land \mathbf{BMH1} \land \mathbf{BMH2} \iff \mathbf{bmh_{0,1,2}} (B) = B$
\end{statement}
\begin{proofs}
\begin{proof}
Follows from~\cref{law:bmbot:bmh-0-1-2-implies-BMH0,law:bmbot:bmh-0-1-2-implies-BMH1,law:bmbot:bmh-0-1-2-implies-BMH2,law:bmbot:BMH0-1-2-implies-bmh-0-1-2} below.
\end{proof}
\end{proofs}
\end{theorem}

\begin{lemma}\label{law:bmbot:bmh-0-1-2-implies-BMH0}
\begin{statement}
$(\mathbf{bmh}_\mathbf{0,1,2} (B) = B) \implies \mathbf{BMH0}$
\end{statement}
\begin{proofs}
\begin{proof}
\begin{flalign*}
	&\mathbf{BMH0}
	&&\ptext{Definition of $\mathbf{BMH0}$}\\
	&=\left(
\begin{array}{l}
		\forall s_0 : State, ss_0 , ss_1 : \power State_\bot \spot \\
		  ((s_0, ss_0) \in B \land ss_0 \subseteq ss_1 \land (\bot \in ss_0 \iff \bot \in ss_1)) \implies (s_0, ss_1) \in B
	\end{array}\right)
	&&\ptext{Predicate calculus: quantifier scope}\\
	&=\left(\begin{array}{l}
		\forall s_0 : State, ss_1 : \power State_\bot \spot \\
		  (\exists ss_0 : \power State_\bot \spot (s_0, ss_0) \in B \land ss_0 \subseteq ss_1 \land (\bot \in ss_0 \iff \bot \in ss_1)) 
		  \\ \implies \\
		  (s_0, ss_1) \in B
	\end{array}%
\right)
	&&\ptext{Property of sets}\\
	&=\left(%
\right)
	&&\ptext{Predicate calculus and property of sets}\\
	&=\left(%
\right)
	&&\ptext{Introduce fresh variable}\\
	&=\left(%
\right)
	&&\ptext{Property of sets and predicate calculus}\\
	&=\left(%
\right)
		\end{array}\right)
	\end{array}\right)
	&&\ptext{Variable renaming and predicate calculus}\\
	&=true
\end{flalign*}
\end{proof}
\end{proofs}
\end{lemma}

\begin{lemma}\label{law:bmbot:bmh-0-1-2-implies-BMH2}
\begin{statement}
$(\mathbf{bmh}_\mathbf{0,1,2} (B) = B) \implies \mathbf{BMH2}$
\end{statement}
\begin{proofs}
\begin{proof}
\begin{flalign*}
	&\mathbf{BMH2}
	&&\ptext{Definition of $\mathbf{BMH2}$}\\
	&=\forall s : State \spot (s, \emptyset) \in B \iff (s, \{ \bot \}) \in B
	&&\ptext{Assumption: $\mathbf{bmh}_\mathbf{0,1,2} (B) = B$}\\
	&=\forall s : State \spot (s, \emptyset) \in \mathbf{bmh}_\mathbf{0,1,2} (B) \iff (s, \{ \bot \}) \in \mathbf{bmh}_\mathbf{0,1,2} (B)
	&&\ptext{\cref{law:bmbot:aux:emptyset-in-bmh-0-1-2} and \cref{law:bmbot:aux:bot-in-bmh-0-1-2}}\\
	&=\forall s : State \spot ((s,\emptyset) \in B \land (s,\{\bot\}) \in B) \iff ((s,\emptyset) \in B \land (s,\{\bot\}) \in B)
	&&\ptext{Predicate calculus}\\
	&=true
\end{flalign*}
\end{proof}
\end{proofs}
\end{lemma}

\begin{lemma}\label{law:bmbot:BMH0-1-2-implies-bmh-0-1-2}
\begin{statement}
Provided $B$ is $\mathbf{BMH0}-\mathbf{BMH2}$-healthy, $\mathbf{bmh}_\mathbf{0,1,2} (B) = B$.
\end{statement}
\begin{proofs}
\begin{proof}
\begin{flalign*}
	&\mathbf{bmh}_\mathbf{0,1,2} (B) = B
	&&\ptext{Definition of $\mathbf{bmh}_\mathbf{0,1,2}$}\\
	&\iff \left\{
\right\} = B	
	&&\ptext{Assumption: $B$ is $\mathbf{BMH1}$-healthy and predicate calculus}\\
	&\iff \left\{%
\right\} = B
	&&\ptext{Predicate calculus: absorption law}\\
	&\iff \left\{%
\right\} = B	
	&&\ptext{Instantiation of existential quantifier for $ss_0 = ss$}\\
	&\iff \left\{%
\right) \\ \land (s, ss) \in B
		\end{array}\right.
	\end{array}\right\} = B	
	&&\ptext{Predicate calculus: absorption law}\\
	&\iff \{ s : State, ss : \power State_\bot | (s, ss) \in B \} = B
	&&\ptext{Property of sets}\\
	&\iff true
\end{flalign*}
\end{proof}
\end{proofs}
\end{lemma}

\begin{lemma}\label{law:bmh-0-1-2:idempotent}
\begin{statement}
$\mathbf{bmh}_\mathbf{0,1,2} \circ \mathbf{bmh}_\mathbf{0,1,2} (B) = \mathbf{bmh}_\mathbf{0,1,2} (B)$
\end{statement}
\begin{proofs}
\begin{proof}
\begin{flalign*}
	&\mathbf{bmh}_\mathbf{0,1,2} \circ \mathbf{bmh}_\mathbf{0,1,2} (B)
	&&\ptext{Definition of $\mathbf{bmh}_\mathbf{0,1,2}$}\\
	&=\left\{
\right\}
	&&\ptext{\cref{law:bmbot:aux:bot-in-bmh-0-1-2}, \cref{law:bmbot:aux:emptyset-in-bmh-0-1-2} and predicate calculus}\\
	&=\left\{%
\right\}
	&&\ptext{Property of sets and predicate calculus}\\
	&=\left\{%
\right\}
	&&\ptext{Property of sets}\\
	&=\left\{%
\right\}
	&&\ptext{Predicate calculus: absorption law}\\
	&=\left\{%
\right\}
	&&\ptext{Property of sets}\\
	&=\left\{%
\right\}
	&&\ptext{Introduce fresh variable}\\
	&=\left\{%
\right\}
	&&\ptext{Variable renaming and substitution}\\
	&=\left\{%
\right\}
	&&\ptext{Instatiation: consider case where $t = \emptyset$}\\
	&=\left\{%
\right\}
	&&\ptext{Predicate calculus: absorption law and distribution}\\
	&=\left\{%
\right\}
	&&\ptext{Instatiation: consider case where $t = \emptyset$}\\
	&=\left\{%
\right\}
	&&\ptext{Variable renaming and definition of sequential composition}\\
	&=\left\{%
\right\}
	&&\ptext{Property of sets}\\
	&=\left(%
\right)
	&&\ptext{Property of sets}\\
	&=\exists ss_1 \spot \left(%
\right)
	&&\ptext{Predicate calculus: quantifier scope}\\
	&=\left(%
\right\}
	&&\ptext{Property of sets}\\
	&=\left(\begin{array}{l}
		\exists ss_0 \spot 
			((s, ss_0) \in B \lor (s, ss_0 \cup \{\bot\}) \in B) \\
			\land ((s, \{ \bot \}) \in B \iff (s, \emptyset) \in B) \\
			\land ss_0 \subseteq \emptyset \land (\bot \in ss_0 \iff \bot \in \emptyset)
	\end{array}\right)
	&&\ptext{Predicate calculus}\\
	&=\left(\begin{array}{l}
		\exists ss_0 \spot 
			((s, ss_0) \in B \lor (s, ss_0 \cup \{\bot\}) \in B) \\
			\land ((s, \{ \bot \}) \in B \iff (s, \emptyset) \in B) \\
			\land ss_0 \subseteq \emptyset \land \bot \notin ss_0
	\end{array}\right)
	&&\ptext{Case analysis on $ss_0$ and one-point rule}\\
	&=((s, \emptyset) \in B \lor (s, \emptyset \cup \{\bot\}) \in B) \land ((s, \{ \bot \}) \in B \iff (s, \emptyset) \in B)
	&&\ptext{Property of sets and predicate calculus}\\
	&=(s, \{\bot\}) \in B \land (s, \emptyset) \in B
\end{flalign*}
\end{proof}\end{proofs}
\end{lemma}

\begin{lemma}\label{law:bmbot:aux:bot-in-bmh-0-1-2}
$(s, \{\bot\}) \in \mathbf{bmh}_\mathbf{0,1,2} (B) = (s,\emptyset) \in B \land (s,\{\bot\}) \in B$
\begin{proofs}\begin{proof}
\begin{flalign*}
	&(s, \{\bot\}) \in \mathbf{bmh}_\mathbf{0,1,2} (B)
	&&\ptext{Definition of $\mathbf{bmh}_\mathbf{0,1,2}$}\\
	&=(s, \{\bot\}) \in \left\{
\right\}
	&&\ptext{Property of sets}\\
	&=\left(\begin{array}{l}
		\exists ss_0 \spot 
			((s, ss_0) \in B \lor (s, ss_0 \cup \{\bot\}) \in B) \\
			\land ((s, \{ \bot \}) \in B \iff (s, \emptyset) \in B) \\
			\land ss_0 \subseteq \{\bot\} \land (\bot \in ss_0 \iff \bot \in \{\bot\})
	\end{array}\right)
	&&\ptext{Predicate calculus}\\
	&=\left(\begin{array}{l}
		\exists ss_0 \spot 
			((s, ss_0) \in B \lor (s, ss_0 \cup \{\bot\}) \in B) \\
			\land ((s, \{ \bot \}) \in B \iff (s, \emptyset) \in B) \\
			\land ss_0 \subseteq \{\bot\} \land \bot \in ss_0
	\end{array}\right)
	&&\ptext{Case analysis on $ss_0$ and one-point rule}\\
	&=((s, \{\bot\}) \in B \lor (s, \{\bot\} \cup \{\bot\}) \in B)	\land ((s, \{ \bot \}) \in B \iff (s, \emptyset) \in B)
	&&\ptext{Property of sets and predicate calculus}\\
	&=(s, \{\bot\}) \in B \land (s, \emptyset) \in B
\end{flalign*}
\end{proof}\end{proofs}
\end{lemma}

\begin{lemma}
\begin{align*}
	&B_1 \subseteq B_0 \\
	&\iff \\
	&\forall s : State, ss : \power State \spot \left(\begin{array}{l}
		(s, ss) \in B_1 \implies (s, ss) \in B_0
		\\ \land \\
		(s, ss \cup \{\bot\}) \in B_1 \implies (s, ss \cup \{\bot\}) \in B_0
	\end{array}\right)
\end{align*}
\begin{proofs}\begin{proof}
\begin{flalign*}
	&B_1 \subseteq B_0
	&&\ptext{Definition of subset inclusion}\\
	&\iff \forall s : State, ss : \power State_\bot \spot (s, ss) \in B_1 \implies (s, ss) \in B_0
	&&\ptext{Predicate calculus}\\
	&\iff \forall s : State, ss : \power State_\bot \spot ((s, ss) \in B_1 \implies (s, ss) \in B_0) \land (\bot \in ss \lor \bot \notin ss)
	&&\ptext{Predicate calculus}\\
	&\iff \forall s : State, ss : \power State_\bot \spot 
	\left(\begin{array}{l}
		(\bot \in ss\implies ((s, ss) \in B_1 \implies (s, ss) \in B_0))
		\\ \land \\
		(\bot \notin ss\implies ((s, ss) \in B_1 \implies (s, ss) \in B_0))
	\end{array}\right)
	&&\ptext{Introduce fresh variable}\\
	&\iff 
	\left(%
\right)
	&&\ptext{Predicate calculus: quantifier scope}\\
	&\iff 
	\left(%
\right)
	&&\ptext{Predicate calculus: quantifier scope}\\
	&\iff 
	\left(%
\right)
\end{flalign*}
\end{proof}\end{proofs}
\end{lemma}

\subsection{$\mathbf{bmh}_\mathbf{0,1,3}$}\label{sec:bmbot:hc:bmh-0-1-3}

\begin{lemma}\label{law:bmbot:bmh0-1-3}
\begin{align*}
	&\mathbf{bmh}_\mathbf{0} \circ \mathbf{bmh}_\mathbf{1} \circ \mathbf{bmh}_\mathbf{3} (B)\\
	&=\\
	&\left\{%
\right\}
	&&\ptext{Property of sets}\\
	&=\left\{%
\right\}
	&&\ptext{Property of sets and predicate calculus}\\
	&=\left\{%
\right)
	&&\ptext{Instantiation of existential quantification for $ss_0 = \{ \bot \}$ and $ss_0 = \emptyset$}\\
	&=\left(\begin{array}{l}
		\exists ss_0 \spot (s, ss_0) \in B \land ss_0 \subseteq ss \land (\bot \in ss_0 \iff \bot \in ss)
		\\ \lor \\
		\exists ss_0 \spot (s, ss_0 \cup \{\bot\}) \in B \land ss_0 \subseteq ss \land (\bot \in ss_0 \iff \bot \in ss)
		\\ \lor \\
		((s, \{ \bot \} \cup \{\bot\}) \in B \land \{ \bot \} \subseteq ss \land (\bot \in \{ \bot \} \iff \bot \in ss))
		\\ \lor \\
		((s, \emptyset \cup \{\bot\}) \in B \land \emptyset \subseteq ss \land (\bot \in \emptyset \iff \bot \in ss))
	\end{array}\right)
	&&\ptext{Property of sets}\\
	&=\left(\begin{array}{l}
		\exists ss_0 \spot (s, ss_0) \in B \land ss_0 \subseteq ss \land (\bot \in ss_0 \iff \bot \in ss)
		\\ \lor \\
		\exists ss_0 \spot (s, ss_0 \cup \{\bot\}) \in B \land ss_0 \subseteq ss \land (\bot \in ss_0 \iff \bot \in ss)
		\\ \lor \\
		((s, \{\bot\}) \in B \land \{ \bot \} \subseteq ss \land \bot \in ss)
		\\ \lor \\
		((s, \{\bot\}) \in B \land \bot \notin ss)
	\end{array}\right)
	&&\ptext{\cref{law:set-theory:set-membership-subset-1} and predicate calculus}\\
	&=\left(\begin{array}{l}
		\exists ss_0 \spot (s, ss_0) \in B \land ss_0 \subseteq ss \land (\bot \in ss_0 \iff \bot \in ss)
		\\ \lor \\
		\exists ss_0 \spot (s, ss_0 \cup \{\bot\}) \in B \land ss_0 \subseteq ss \land (\bot \in ss_0 \iff \bot \in ss)
		\\ \lor \\
		((s, \{\bot\}) \in B \land \bot \in ss)
		\\ \lor \\
		((s, \{\bot\}) \in B \land \bot \notin ss)
	\end{array}\right)
	&&\ptext{Predicate calculus}\\
	&=\exists ss_0 \spot
		\left(\begin{array}{l}
			((s, ss_0) \in B \lor (s, ss_0 \cup \{\bot\}) \in B)
			\\ \land \\
			ss_0 \subseteq ss \land (\bot \in ss_0 \iff \bot \in ss)
		\end{array}\right) \lor (s, \{\bot\}) \in B
\end{flalign*}
\end{proof}\end{proofs}
\end{lemma}
\subsection{$\mathbf{bmh}_\mathbf{0,1,3,2}$}

\begin{lemma}\label{law:bmbot:bmh-0-1-3-2:definition}
\begin{statement}
\begin{align*}
	&\mathbf{bmh}_\mathbf{0} \circ \mathbf{bmh}_\mathbf{1} \circ \mathbf{bmh}_\mathbf{3} \circ \mathbf{bmh}_\mathbf{2} (B)\\
	&=\\
	&\left\{%
\right\}
	&&\ptext{Property of sets}\\
	&=\left\{%
\right\}
	&&\ptext{Predicate calculus: absorption law}\\
	&=\left\{%
\right\}
	&&\ptext{Predicate calculus: absorption law}\\
	&=\left\{%
\right)
			\end{array}\right)
		\end{array}\right.
	\end{array}\right\}	
\end{align*}	
\end{proof}
\end{proofs}
\end{lemma}

\begin{theorem}\label{theorem:bmbot:BMH0-3:iff:bmh-0-1-3-2}
\begin{statement}
$\mathbf{BMH0} \land \mathbf{BMH1} \land \mathbf{BMH2} \land \mathbf{BMH3} \iff \mathbf{bmh}_\mathbf{0,1,3,2} (B) = B$
\end{statement}
\begin{proofs}
\begin{proof}
The implication follows from~\cref{law:bmbot:BMH0-3:implies:bmh-0-1-3-2}.
While the reverse implication follows from the fact that $\mathbf{bmh}_\mathbf{0,1,3,2}$ is a fixed point of $\mathbf{bmh}_{0,1,2}$ (\cref{law:bmh-0-1-2:fixed-point:bmh-0-1-2-3}) and~\cref{law:bmbot:bmh-0-1-2-implies-BMH0,law:bmbot:bmh-0-1-2-implies-BMH1,law:bmbot:bmh-0-1-2-implies-BMH2,law:bmbot:bmg-0-1-3-2:implies:BMH3}.
\end{proof}
\end{proofs}
\end{theorem}

\begin{lemma}\label{law:bmbot:BMH0-3:implies:bmh-0-1-3-2}
\begin{statement}
$\mathbf{BMH0} \land \mathbf{BMH1} \land \mathbf{BMH2} \land \mathbf{BMH3} \implies \mathbf{bmh}_\mathbf{0,1,3,2} (B) = B$
\end{statement}
\begin{proofs}
\begin{proof}
\begin{flalign*}
	&\mathbf{bmh}_\mathbf{0,1,3,2} (B)
	&&\ptext{Definition of $\mathbf{bmh}_\mathbf{0,1,3,2}$}\\
	&=\left\{
\right\}
	&&\ptext{Predicate calculus: instatiation of existential quantifier for $ss_0 = ss$}\\
	&=\left\{%
\right\}
	&&\ptext{Predicate calculus: absorption law}\\
	&=\left\{%
\right.
	\end{array}\right\}
	&&\ptext{Assumption: $B$ is $\mathbf{BMH2}$-healthy}\\
	&=\{ s : State, ss : \power State_\bot | (s, \emptyset) \in B \lor ((s, ss) \in B \land \bot \notin ss) \}
	&&\ptext{Assumption: $B$ is $\mathbf{BMH0}$, $\mathbf{BMH2}$ and $\mathbf{BMH3}$-healthy and~\cref{law:bmbot:aux:BMH3-provided-BMH0-2-fixed-point}}\\
	&=B
\end{flalign*}
\end{proof}
\end{proofs}
\end{lemma}

\begin{lemma}\label{law:bmh-0-1-2:fixed-point:bmh-0-1-2-3}
\begin{statement}
	$\mathbf{bmh}_\mathbf{0,1,2} \circ \mathbf{bmh}_\mathbf{0,1,3,2} (B) = \mathbf{bmh}_\mathbf{0,1,3,2} (B)$
\end{statement}
\begin{proofs}
\begin{proof}
\begin{flalign*}
	&\mathbf{bmh}_\mathbf{0,1,2} \circ \mathbf{bmh}_\mathbf{0,1,3,2} (B)
	&&\ptext{Definition of $\mathbf{bmh}_\mathbf{0,1,2}$}\\
	&=\left\{
\right\}
	&&\ptext{\cref{law:bmbot:emptyset-in-bmh0-1-3-2} and \cref{law:bmbot:bot-in-bmh0-1-3-2} and predicate calculus}\\
	&=\left\{%
\right)
	&&\ptext{Case analysis on $ss_0$}\\
	&=\forall s_0 : State \spot \left(\begin{array}{l}
		((s, \emptyset) \in B \land (s, \{ \bot \}) \in B)
		\\ \implies \\
		((s,\emptyset) \in B \land (s,\{\bot\}) \in B)
	\end{array}%
\right\}
	&&\ptext{Predicate calculus and definition of $\mathbf{bmh}_\mathbf{0,1,3,2}$}\\
	&=\left\{%
\right\}
	&&\ptext{Variable renaming and property of sets}\\
	&=\left\{%
\right\}
	&&\ptext{Predicate calculus: quantifier scope}\\
	&=\left\{%
\right\}
	&&\ptext{Predicate calculus: absorption law}\\
	&=\left\{%
\right\}
	&&\ptext{Property of sets}\\
	&=\left(%
\right)
	\end{array}\right)
\end{flalign*}
\end{proof}\end{proofs}
\end{lemma}

\begin{lemma}\label{law:bmbot:aux:exists-ss-cup-bot-in-bmh-0-1-3-2}
\begin{align*}
	&\exists ss_1 : \power State_\bot \spot (s, ss_1 \cup \{\bot\}) \in \mathbf{bmh}_\mathbf{0,1,3,2} (B) \land ss_1 \subseteq ss \land (\bot \in ss_1 \iff \bot \in ss)\\
	&\iff\\
	&((s, \emptyset) \in B \land (s, \{ \bot \}) \in B)
\end{align*}
\begin{proofs}\begin{proof}
\begin{flalign*}
	&\exists ss_1 : \power State_\bot \spot (s, ss_1 \cup \{\bot\}) \in \mathbf{bmh}_\mathbf{0,1,3,2} (B) \land ss_1 \subseteq ss \land (\bot \in ss_1 \iff \bot \in ss)\\
	&&\ptext{Definition of $\mathbf{bmh}_\mathbf{0,1,3,2}$}\\
	&\iff \left(
\right)
	&&\ptext{Property of sets}\\
	&\iff \left(%
\right)
	&&\ptext{Property of sets and predicate calculus}\\
	&\iff \left(\begin{array}{l}
		\exists ss_1 : \power State_\bot \spot \\
		((s, \emptyset) \in B \land (s, \{ \bot \}) \in B) \land
		ss_1 \subseteq ss \land (\bot \in ss_1 \iff \bot \in ss)
	\end{array}\right)
	&&\ptext{Predicate calculus: instatiation of existential quantifier for $ss_1 = ss$}\\
	&\iff ((s, \emptyset) \in B \land (s, \{ \bot \}) \in B)
\end{flalign*}
\end{proof}\end{proofs}
\end{lemma}

\begin{lemma}\label{law:bmbot:aux:exists-ss-in-bmh-0-1-3-2}
\begin{align*}
	&\exists ss_1 : \power State_\bot \spot (s, ss_1) \in \mathbf{bmh}_\mathbf{0,1,3,2} (B) \land ss_1 \subseteq ss \land (\bot \in ss_1 \iff \bot \in ss)\\
	&\iff\\
	&\left(%
\right)
	&&\ptext{Property of sets}\\
	&\iff \left(%
\right)
	&&\ptext{Predicate calculus: instatiation of existential quantifier for $ss_1 = ss$}\\
	&\iff \left(%
\right\}
	&&\ptext{Property of sets}\\
	&=\left(%
\right)
	&&\ptext{Property of sets and one-point rule}\\
	&=\left(\begin{array}{l}
		((s, \emptyset) \in B \land (s, \{ \bot \}) \in B)
		\\ \lor \\
		((s, \{\bot\}) \notin B \land (s, \emptyset) \notin B \land (s, \emptyset) \in B)
	\end{array}%
\right\}
	&&\ptext{Property of sets}\\
	&=\left(%
\right)
	&&\ptext{Property of sets}\\
	&=\left(\begin{array}{l}
		((s, \emptyset) \in B \land (s, \{ \bot \}) \in B)
		\\ \lor \\
		((s, \{\bot\}) \notin B \land (s, \emptyset) \notin B \land false)
	\end{array}\right)
	&&\ptext{Predicate calculus}\\
	&=(s, \emptyset) \in B \land (s, \{ \bot \}) \in B
\end{flalign*}
\end{proof}\end{proofs}
\end{lemma}

\begin{lemma}\label{law:bmbot:aux:BMH3-provided-BMH0-2-fixed-point}
Provided $B$ is $\mathbf{BMH0}$ and $\mathbf{BMH2}$-healthy,
\begin{align*}
	&B = (B \nrres \{ ss : \power State_\bot | \bot \in ss \}) \cup \{ s_0 : State, ss : \power State_\bot | (s_0, \emptyset) \in B\}\\
	&\iff \\
	&\mathbf{BMH3}
\end{align*}
\begin{proofs}\begin{proof}
\begin{flalign*}
	&B = (B \nrres \{ ss : \power State_\bot | \bot \in ss \}) \cup \{ s_0 : State, ss : \power State_\bot | (s_0, \emptyset) \in B\}
	&&\ptext{Property of sets}\\
	&\iff (B = \{ s : State, ss : State_\bot | ((s, ss) \in B \land \bot \notin ss) \lor (s, \emptyset) \in B \})
	&&\ptext{Property of sets}\\
	&\iff \forall s, ss \spot \left(
\begin{array}{l}
		(s, ss) \in B \implies (((s, ss) \in B \land \bot \notin ss) \lor (s, \emptyset) \in B)
		\\ \land \\
		((((s, ss) \in B \land \bot \notin ss) \lor (s, \emptyset) \in B) \implies (s, ss) \in B)
	\end{array}\right)
	&&\ptext{Propositional calculus}\\
	&\iff \forall s, ss \spot \left(\begin{array}{l}
		((s, \emptyset) \notin B \implies ((s, ss) \notin B \lor ((s, ss) \in B \land \bot \notin ss)))
		\\ \land \\
		((((s, ss) \notin B \lor \bot \in ss) \land (s, \emptyset) \notin B) \lor (s, ss) \in B)
	\end{array}\right)
	&&\ptext{Propositional calculus: absorption law}\\
	&\iff \forall s, ss \spot \left(\begin{array}{l}
		((s, \emptyset) \notin B \implies ((s, ss) \notin B \lor \bot \notin ss))
		\\ \land \\
		((s, \emptyset) \notin B \lor (s, ss) \in B)
	\end{array}\right)
	&&\ptext{Propositional calculus}\\
	&\iff \left(\begin{array}{l}
		\forall s, ss \spot (s, \emptyset) \notin B \implies ((s, ss) \in B \implies \bot \notin ss)
		\\ \land \\
		\forall s, ss \spot (s, \emptyset) \in B \implies (s, ss) \in B
	\end{array}\right)
	&&\ptext{Propositional calculus: introduce term}\\
	&\iff \left(\begin{array}{l}
		\forall s, ss \spot (s, \emptyset) \notin B \implies ((s, ss) \in B \implies \bot \notin ss)
		\\ \land \\
		\forall s, ss \spot (s, \emptyset) \in B \implies ((s, ss) \in B \lor (\bot \in ss \land \bot \notin ss))
	\end{array}\right)
	&&\ptext{Propositional calculus}\\
	&\iff \left(\begin{array}{l}
		\forall s, ss \spot (s, \emptyset) \notin B \implies ((s, ss) \in B \implies \bot \notin ss)
		\\ \land \\
		\forall s, ss \spot (s, \emptyset) \in B \implies ((s, ss) \in B \lor \bot \in ss)
		\\ \land \\
		\forall s, ss \spot (s, \emptyset) \in B \implies ((s, ss) \in B \lor \bot \notin ss)
	\end{array}\right)
	&&\ptext{Propositional calculus}\\
	&\iff \left(\begin{array}{l}
		\forall s, ss \spot (s, \emptyset) \notin B \implies ((s, ss) \in B \implies \bot \notin ss)
		\\ \land \\
		\forall s, ss \spot ((s, \emptyset) \in B \land \bot \notin ss) \implies (s, ss) \in B
		\\ \land \\
		\forall s, ss \spot ((s, \emptyset) \in B \land \bot \in ss) \implies (s, ss) \in B
	\end{array}\right)
	&&\ptext{Property of sets}\\
	&\iff \left(\begin{array}{l}
		\forall s, ss \spot (s, \emptyset) \notin B \implies ((s, ss) \in B \implies \bot \notin ss)
		\\ \land \\
		\forall s, ss \spot ((s, \emptyset) \in B \land \emptyset \subseteq ss \land \bot \notin \emptyset \land \bot \notin ss) \implies (s, ss) \in B
		\\ \land \\
		\forall s, ss \spot ((s, \emptyset) \in B \land \bot \in ss) \implies (s, ss) \in B
	\end{array}\right)
	&&\ptext{Assumption: $B$ is $\mathbf{BMH2}$-healthy and \cref{law:set-theory:set-membership-subset-1}}\\\\
	&\iff \left(\begin{array}{l}
		\forall s, ss \spot (s, \emptyset) \notin B \implies ((s, ss) \in B \implies \bot \notin ss)
		\\ \land \\
		\forall s, ss \spot ((s, \emptyset) \in B \land \emptyset \subseteq ss \land \bot \notin \emptyset \land \bot \notin ss) \implies (s, ss) \in B
		\\ \land \\
		\forall s, ss \spot ((s, \{\bot\}) \in B \land \{\bot\} \subseteq ss \land \bot \in \{\bot\} \land \bot \in ss) \implies (s, ss) \in B
	\end{array}\right)
	&&\ptext{Assumption: $B$ is $\mathbf{BMH0}$-healthy}\\
	&\iff \forall s, ss \spot (s, \emptyset) \notin B \implies ((s, ss) \in B \implies \bot \notin ss)
	&&\ptext{Propositional calculus: move quantifier}\\
	&\iff \forall s \spot (s, \emptyset) \notin B \implies \forall ss \spot ((s, ss) \in B \implies \bot \notin ss)
	&&\ptext{Definition of $\mathbf{BMH3}$}\\
	&\iff \mathbf{BMH3}
\end{flalign*}
\end{proof}\end{proofs}
\end{lemma}
 
\section{Operators}

\subsection{Angelic Choice}

\begin{lemma}\label{law:bmbot:angelic-two-assignment}
\begin{statement}
$(x :=_{{BM}_\bot} e) \sqcupBMbot (x :=_{BM} e) = (x :=_{BM} e)$
\end{statement}
\begin{proofs}
\begin{proof}
\begin{flalign*}
	&(x :=_{{BM}_\bot} e) \sqcupBMbot (x :=_{BM} e)
	&&\ptext{Definition of $:=_{{BM}_\bot}$, $:=_{BM}$ and $\sqcupBMbot$}\\
	&=\left(\begin{array}{l}
		\{ s : State, ss : \power State_\bot | s \oplus (x \mapsto e) \in ss\} 
		\\ \cap \\
		\{ s : State, ss : \power State | s \oplus (x \mapsto e) \in ss\}
	\end{array}\right)
	&&\ptext{Type: $\bot \notin \power State$}\\
	&=\left(\begin{array}{l}
		\{ s : State, ss : \power State_\bot | s \oplus (x \mapsto e) \in ss\} 
		\\ \cap \\
		\{ s : State, ss : \power State_\bot | s \oplus (x \mapsto e) \in ss \land \bot \notin ss \}
	\end{array}\right)
	&&\ptext{Property of sets and predicate calculus}\\
	&=\{ s : State, ss : \power State_\bot | s \oplus (x \mapsto e) \in ss \land \bot \notin ss\} 
	&&\ptext{Type: $\bot \notin \power State$}\\
	&=\{ s : State, ss : \power State | s \oplus (x \mapsto e) \in ss \}
	&&\ptext{Definition of $:=_{BM}$}\\
	&=(x :=_{BM} e)
\end{flalign*}
\end{proof}
\end{proofs}
\end{lemma}

\begin{lemma}\label{lemma:bmbot:top-sqcup-B}
\begin{statement}
$\topBMbot \sqcupBMbot B = \topBMbot$
\end{statement}
\begin{proofs}
\begin{proof}
\begin{flalign*}
	&\topBMbot \sqcupBMbot B
	&&\ptext{Definition of $\topBMbot$ and $\sqcupBMbot$}\\
	&=\emptyset \cap B
	&&\ptext{Property of sets}\\
	&=\emptyset
	&&\ptext{Definition of $\topBMbot$}\\
	&=\topBMbot
\end{flalign*}
\end{proof}
\end{proofs}
\end{lemma}

\begin{lemma}\label{lemma:bmbot:bot-sqcup-B}
\begin{statement}
$\botBMbot \sqcupBMbot B = B$
\end{statement}
\begin{proofs}
\begin{proof}
\begin{flalign*}
	&\botBMbot \sqcupBMbot B
	&&\ptext{Definition of $\botBMbot$ and $\sqcupBMbot$}\\
	&=(State \times \power State_\bot) \cap B
	&&\ptext{Property of sets}\\
	&=B
\end{flalign*}
\end{proof}
\end{proofs}
\end{lemma}

\subsection{Demonic Choice}

\begin{lemma}\label{law:bmbot:demonic-two-assignment}
\begin{statement}
$(x :=_{BM} e) \sqcapBMbot (x :=_{{BM}_\bot} e) = (x :=_{{BM}_\bot} e)$
\end{statement}
\begin{proofs}
\begin{proof}
\begin{flalign*}
	&(x :=_{BM} e) \sqcapBMbot (x :=_{{BM}_\bot} e)
	&&\ptext{Definition of $:=_{BM}$, $:=_{{BM}_\bot}$ and $\sqcapBMbot$}\\
	&=\left(\begin{array}{l}
		\{ s : State, ss : \power State | s \oplus (x \mapsto e) \in ss \}
		\\ \cup \\
		\{ s : State, ss : \power State_\bot | s \oplus (x \mapsto e) \in ss \}
	\end{array}\right)
	&&\ptext{Type: $\bot \notin \power State$}\\
	&=\left(\begin{array}{l}
		\{ s : State, ss : \power State | s \oplus (x \mapsto e) \in ss \land \bot \notin ss\}
		\\ \cup \\
		\{ s : State, ss : \power State_\bot | s \oplus (x \mapsto e) \in ss \}
	\end{array}\right)
	&&\ptext{Property of sets}\\
	&=\{ s : State, ss : \power State | (s \oplus (x \mapsto e) \in ss \land \bot \notin ss) \lor s \oplus (x \mapsto e) \in ss \}
	&&\ptext{Predicate calculus: absorption law}\\
	&=\{ s : State, ss : \power State | s \oplus (x \mapsto e) \in ss \}
	&&\ptext{Definition of $:=_{{BM}_\bot}$}\\
	&=(x :=_{{BM}_\bot} e)
\end{flalign*}
\end{proof}
\end{proofs}
\end{lemma}

\begin{lemma}\label{lemma:bmbot:bot-sqcap-B}
\begin{statement}
$\botBMbot \sqcapBMbot B = \botBMbot$
\end{statement}
\begin{proofs}
\begin{proof}
\begin{flalign*}
	&\botBMbot \sqcapBMbot B
	&&\ptext{Definition of $\botBMbot$ and $\sqcapBMbot$}\\
	&=(State \times \power State_\bot) \cup B
	&&\ptext{Property of sets}\\
	&=(State \times \power State_\bot)
	&&\ptext{Definition of $\botBMbot$}\\
	&=\botBMbot
\end{flalign*}
\end{proof}
\end{proofs}
\end{lemma}

\begin{lemma}\label{lemma:bmbot:top-sqcap-B}
\begin{statement}
$\topBMbot \sqcapBMbot B = B$
\end{statement}
\begin{proofs}
\begin{proof}
\begin{flalign*}
	&\topBMbot \sqcapBMbot B
	&&\ptext{Definition of $\topBMbot$ and $\sqcapBMbot$}\\
	&=\emptyset \cup B
	&&\ptext{Property of sets}\\
	&=B
\end{flalign*}
\end{proof}
\end{proofs}
\end{lemma}\noindent

\subsection{Sequential Composition}

\begin{theorem}\label{law:bmbot:seqBMbot:BMH0-healthy}
\begin{statement}
Provided $B_0$ is $\mathbf{BMH0}$-healthy,
\begin{align*}
	&B_0 \seqBMbot B_1
	=
	\left(
\right\}
	&&\ptext{Predicate calculus and property of sets}\\
	&=\left(%
\right) \\
	&&\ptext{Propositional calculus and property of sets}\\
	&=\left(%
\right)
	&&\ptext{Property of sets}\\
	&=\{ s_0 : State, ss_0 : \power State_\bot | false \} \cup \{ s_0 : State, ss_0 : \power State_\bot | false \}
	&&\ptext{Property of sets}\\
	&=\emptyset \cup \emptyset
	&&\ptext{Property of sets and definition of $\topBMbot$}\\
	&=\topBMbot
\end{flalign*}
\end{proof}
\end{proofs}
\end{lemma}\noindent

\begin{lemma}\label{law:bmbot:bot-seqBMbot-B:bot}
\begin{statement}
$\botBMbot \seqBMbot B = \botBMbot$
\end{statement}
\begin{proofs}
\begin{proof}
\begin{flalign*}
	&\botBMbot \seqBMbot B
	&&\ptext{Definition of $\botBMbot$}\\
	&=(State \times \power State_\bot) \seqBMbot B
	&&\ptext{Definition of $\seqBMbot$ (\cref{law:bmbot:seqBMbot:BMH0-healthy} as $\botBMbot$ is $\mathbf{BMH0}$-healthy)}\\
	&=\left(\begin{array}{l}
		\left\{\begin{array}{l}
		s_0 : State, ss_0 : \power State_\bot | (s_0, State_\bot) \in (State \times \power State_\bot)
		\end{array}\right\}
		\\ \cup \\
		\left\{\begin{array}{l}
		s_0 : State, ss_0 : \power State_\bot \\
		| (s_0, \{ s_1 : State | (s_1, ss_0) \in B\}) \in (State \times \power State_\bot)
		\end{array}\right\}
	\end{array}\right)
	&&\ptext{Property of sets}\\
	&=\{ s_0 : State, ss_0 : \power State_\bot | true \} \cup \{ s_0 : State, ss_0 : \power State_\bot | true \}
	&&\ptext{Property of sets}\\
	&=(State \times \power State_\bot)
	&&\ptext{Definition of $\botBMbot$}\\
	&=\botBMbot
\end{flalign*}
\end{proof}
\end{proofs}
\end{lemma}\noindent

\section{Relationship with Binary Multirelations}

\subsection{$bmb2bm$}

\begin{theorem}[$\mathbf{bmb2bm}$-is-$\mathbf{bmh}_\mathbf{up}$]
\label{theorem:scratchpad:bmh-upclosed-o-bmb2bm-bmh-0-1-3-2}
\begin{statement}
\begin{align*}
	&\mathbf{bmh}_\mathbf{up} \circ bmb2bm(\mathbf{bmh}_\mathbf{0,1,3,2} (B)) = bmb2bm(\mathbf{bmh}_\mathbf{0,1,3,2} (B))
\end{align*}
\end{statement}
\begin{proofs}
\begin{proof}
\begin{flalign*}
	&\mathbf{bmh}_\mathbf{up} \circ bmb2bm(\mathbf{bmh}_\mathbf{0,1,3,2} (B))
	&&\ptext{Definition of $\mathbf{bmh}_\mathbf{up}$}\\
	&=\left\{
\right\}
	&&\ptext{Variable renaming and property of sets}\\
	&=\left\{%
\right\}
	&&\ptext{Predicate calculus: distributivity and quantifier scope}\\
	&=\left\{%
\right\}
	&&\ptext{Predicate calculus: case-analysis on $ss_0$}\\
	&=\left\{%
\right)
			\end{array}\right)
		\end{array}\right.
	\end{array}\right\}
	&&\ptext{\cref{law:scratchpad:bmb2bm:bmh-0-1-3-2}}\\
	&=bmb2bm(\mathbf{bmh}_\mathbf{0,1,3,2} (B))
\end{flalign*}
\end{proof}
\end{proofs}
\end{theorem}

\begin{lemma}\label{law:bm:bmh-upclosed}
\begin{statement}
$\mathbf{BMH} \iff \mathbf{bmh_{up}} (B) = B$
\end{statement}
\begin{proofs}
\begin{proof}
\begin{flalign*}
	&\mathbf{BMH}
	&&\ptext{Definition of $\mathbf{BMH}$}\\
	&\iff \forall s : State ; ss_0, ss_1 : \power State \spot 
		((s,ss_0) \in B \land ss_0 \subseteq ss_1) \implies (s, ss_1) \in B
	&&\ptext{Predicate calculus: quantifier scope}\\
	&\iff \left(
\begin{array}{l}
		\forall s : State ; ss_1 : \power State \spot \\
		(\exists ss_0 : \power State \spot (s,ss_0) \in B \land ss_0 \subseteq ss_1) \implies (s, ss_1) \in B
		\end{array}\right)
	&&\ptext{Property of sets: subset inclusion}\\
	&\iff   \{ s : State, ss : \power State 
		 | \exists ss_0 : \power State \spot (s, ss_0) \in B \land ss_0 \subseteq ss 
		\} \subseteq B
	&&\ptext{Property of sets: subset inclusion}\\
	&\iff   \{ s : State, ss : \power State 
		 | \exists ss_0 : \power State \spot (s, ss_0) \in B \land ss_0 \subseteq ss 
		\} \cup B = B
	&&\ptext{Property of sets: set union}\\
	&\iff   \left\{%
\right\} = B
	&&\ptext{Predicate calculus: instantiation of existential quantifier for $ss_0 = ss$}\\
	&\iff   \{ s : State, ss : \power State 
		 | \exists ss_0 : \power State \spot (s, ss_0) \in B \land ss_0 \subseteq ss
		\} = B
	&&\ptext{Definition of $\mathbf{bmh_{up}}$}\\
	&\iff \mathbf{bmh_{up}} (B)
\end{flalign*}	
\end{proof}
\end{proofs}
\end{lemma}\noindent

\begin{lemma}\label{law:bmb2bm:bmh-0-1-3-2}
\begin{align*}
	&bmb2bm(\mathbf{bmh}_\mathbf{0,1,3,2} (B))\\
	&= \\
	&\left\{%
\right\}
	&&\ptext{Property of sets}\\
	&=\left\{%
\right\}
	&&\ptext{Property of sets}\\
	&=\left(\begin{array}{l}
		\exists ss_0 \spot (s, ss_0) \in B \land \bot \notin ss_0 \land ss_0 \subseteq \emptyset \land \bot \notin \emptyset
		\\ \lor \\
		(s, \emptyset) \in B
	\end{array}\right)
	&&\ptext{Predicate calculus and one-point rule}\\
	&=(s, \emptyset) \in B
\end{flalign*}
\end{proof}
\end{proofs}
\end{lemma}

\begin{lemma}\label{law:bot-in-bmb2bm-bmh-upclosed}
$(s,\{\bot\}) \in bmb2bm(\mathbf{bmh}_\mathbf{up}) = (s, \emptyset) \in B$
\begin{proofs}
\begin{proof} 
\begin{flalign*}
	&(s,\{\bot\}) \in bmb2bm(\mathbf{bmh}_\mathbf{up})
	&&\ptext{Definition of $bmb2bm(\mathbf{bmh}_\mathbf{up})$ (\cref{law:bm2bmb-bmhupclosed})}\\
	&=(s,\{\bot\}) \in \left\{%
\right\}
	&&\ptext{Property of sets}\\
	&=\left(\begin{array}{l}
		\exists ss_0 \spot (s, ss_0) \in B \land \bot \notin ss_0 \land ss_0 \subseteq \{\bot\} \land \bot \notin \{\bot\}
		\\ \lor \\
		(s, \emptyset) \in B
		\end{array}\right)
	&&\ptext{Property of sets and predicate calculus}\\
	&=(s, \emptyset) \in B
\end{flalign*}
\end{proof}
\end{proofs}
\end{lemma}

\begin{theorem}\label{theorem:bmh-upclosed-o-bmb2bm-bmh-0-1-3-2}
\begin{statement}
Provided $B$ is $\mathbf{BMH_{0,1,2,3}}$-healthy,
\begin{align*}\mathbf{bmh}_\mathbf{up} \circ bmb2bm(B) = bmb2bm(B)\end{align*}
\end{statement}
\begin{proofs}
\begin{proof}
\begin{flalign*}
	&\mathbf{bmh}_\mathbf{up} \circ bmb2bm(B)
	&&\ptext{Assumption: $B$ is $\mathbf{BMH_{0,1,2,3}}$-healthy}\\
	&=\mathbf{bmh}_\mathbf{up} \circ bmb2bm(\mathbf{bmh}_\mathbf{0,1,3,2} (B))
	&&\ptext{\cref{theorem:scratchpad:bmh-upclosed-o-bmb2bm-bmh-0-1-3-2}}\\
	&=bmb2bm(\mathbf{bmh}_\mathbf{0,1,3,2} (B))
	&&\ptext{Assumption: $B$ is $\mathbf{BMH_{0,1,2,3}}$-healthy}\\
	&=bmb2bm(B)
\end{flalign*}
\end{proof}
\end{proofs}
\end{theorem}

\begin{lemma}\label{law:scratchpad:bmb2bm:bmh-0-1-3-2}
\begin{statement}
\begin{align*}
	&bmb2bm(\mathbf{bmh}_\mathbf{0,1,3,2} (B))\\
	&= \\
	&\left\{
\right\}
	&&\ptext{Property of sets}\\
	&=\left\{%
\right)
		\end{array}\right)
		\end{array}\right.
	\end{array}\right\}
\end{flalign*}	
\end{proof}
\end{proofs}
\end{lemma}

\subsection{$bm2bmb$}

\begin{theorem}\label{theorem:bmbot:bmh-0-1-3-2-circ-bm2bmb}
\begin{statement}
\begin{align*}
	&\mathbf{bmh}_\mathbf{0,1,3,2} \circ bm2bmb(\mathbf{bmh}_\mathbf{up} (B)) = bm2bmb(\mathbf{bmh}_\mathbf{up} (B))
\end{align*}
\end{statement}
\begin{proofs}
\begin{proof}
\begin{flalign*}
	&\mathbf{bmh}_\mathbf{0,1,3,2} \circ bm2bmb(\mathbf{bmh}_\mathbf{up} (B))
	&&\ptext{Definition of $\mathbf{bmh}_\mathbf{0,1,3,2}$}\\
	&=\left\{
\right\}
	&&\ptext{Predicate calculus and definition of $bm2bmb(\mathbf{bmh}_\mathbf{up} (B))$ (\cref{law:bm2bmb-bmhupclosed})}\\
	&=\left\{%
\right\}
	&&\ptext{Variable renaming and property of sets}\\
	&=\left\{%
\right\}
	&&\ptext{Predicate calculus: absorption law}\\
	&=\left\{%
\right.
	\end{array}\right\}
	&&\ptext{\cref{law:bm2bmb-bmhupclosed}}\\
	&=bm2bmb(\mathbf{bmh}_\mathbf{up} (B))
\end{flalign*}
\end{proof}
\end{proofs}
\end{theorem}\noindent

\begin{theorem}\label{theorem:bmbot:bm2bmb-circ-bmb2bm}
\begin{statement}
Provided $B$ is $\mathbf{BMH_{0,1,2,3}}$-healthy, $bm2bmb \circ bmb2bm(B) = B$,
\end{statement}
\begin{proofs}
\begin{proof}
\begin{flalign*}
	&bm2bmb \circ bmb2bm(B)
	&&\ptext{Assumption: $B$ is $\mathbf{BMH0}$-$\mathbf{BMH3}$-healthy}\\
	&=bm2bmb \circ bmb2bm(\mathbf{bmh}_\mathbf{0,1,3,2} (B))
	&&\ptext{Definition of $bm2bmb$}\\
	&=\left\{
\right\}
	&&\ptext{Property of sets}\\
	&=\left\{%
\right\}
	&&\ptext{Property of sets, predicate calculus and one-point rule}\\
	&=\left\{%
\right\}
	&&\ptext{Predicate calculus: absorption law}\\
	&=\left\{%
\right)
			\end{array}\right)
		\end{array}\right.
	\end{array}\right\}
	&&\ptext{Definition of $\mathbf{bmh}_\mathbf{0,1,3,2}$}\\
	&=\mathbf{bmh}_\mathbf{0,1,3,2} (B)
	&&\ptext{Assumption: $B$ is $\mathbf{BMH0}$-$\mathbf{BMH3}$-healthy}\\	
	&=B
\end{flalign*}
\end{proof}
\end{proofs}
\end{theorem}

\begin{theorem}\label{theorem:bmbot:bmb2bm-circ-bm2bmb}
\begin{statement}
Provided $B$ is $\mathbf{BMH}$-healthy, $bmb2bm \circ bm2bmb(B) = B$,
\end{statement}
\begin{proofs}
\begin{proof}
\begin{flalign*}
	&bmb2bm \circ bm2bmb(B)
	&&\ptext{Assumption: $B$ is $\mathbf{BMH}$-healthy}\\
	&=bmb2bm \circ bm2bmb(\mathbf{bmh}_\mathbf{upclosed} (B))
	&&\ptext{Definition of $bmb2bm$}\\
	&=\{ s : State, ss : \power State_\bot | ((s, ss) \in bm2bmb(\mathbf{bmh}_\mathbf{upclosed} (B)) \land \bot \notin ss)\}
	&&\ptext{\cref{law:bm2bmb-bmhupclosed}}\\
	&=\left\{
\right\}
	&&\ptext{Property of sets}\\
	&=\left\{%
\right\}
	&&\ptext{Instantiation: consider case where $ss_0 = \emptyset$}\\
	&=\left\{%
\right\}
	&&\ptext{Property of sets and predicate calculus}\\
	&=\left\{%
\right\}
	&&\ptext{Property of sets and predicate calculus}\\
	&=\left\{%
\right.
		\end{array}\right\}
\end{flalign*}
\end{proof}
\end{proofs}
\end{lemma}

\section{Set Theory}

\begin{lemma}\label{law:bmbot:ss0}
\begin{align*}
	&\exists ss_0 \spot (s, ss_0 \cup \{\bot\}) \in B \land ss_0 \subseteq ss \land (\bot \in ss_0 \iff \bot \in ss)\\
	&\iff\\
	&\exists ss_0 \spot (s, ss_0) \in B \land ss_0 \subseteq (ss \cup \{ \bot \}) \land \bot \in ss_0
\end{align*}
\begin{proofs}
\begin{proof}
\begin{flalign*}
	&\exists ss_0 \spot (s, ss_0 \cup \{\bot\}) \in B \land ss_0 \subseteq ss \land (\bot \in ss_0 \iff \bot \in ss)
	&&\ptext{Predicate calculus}\\
	&=\exists ss_0 \spot \left(
\begin{array}{l}
		(s, ss_0 \cup \{\bot\}) \in B \land ss_0 \subseteq ss \land \bot \in ss_0 \land \bot \in ss
		\\ \lor \\
		(s, ss_0 \cup \{\bot\}) \in B \land ss_0 \subseteq ss \land \bot \notin ss_0 \land \bot \notin ss
	\end{array}\right)
	&&\ptext{Predicate calculus and property of sets}\\
	&=\left(\begin{array}{l}
		\exists ss_0 \spot (s, ss_0) \in B \land ss_0 \subseteq (ss \cup \{ \bot \}) \land \bot \in ss_0 \land \bot \in ss
		\\ \lor \\
		\exists ss_0 \spot (s, ss_0 \cup \{\bot\}) \in B \land ss_0 \subseteq ss \land \bot \notin ss_0 \land \bot \notin ss
	\end{array}\right)
	&&\ptext{Introduce fresh variable $t$ and substitution}\\
	&=\left(\begin{array}{l}
		\exists ss_0 \spot (s, ss_0) \in B \land ss_0 \subseteq (ss \cup \{ \bot \}) \land \bot \in ss_0 \land \bot \in ss
		\\ \lor \\
		\exists t, ss_0 \spot (s,t) \in B \land t = ss_0 \cup \{\bot\} \land ss_0 \subseteq ss \land \bot \notin ss_0 \land \bot \notin ss
	\end{array}\right)
	&&\ptext{Property of sets (\cref{law:set-theory:A-setminus-x})}\\
	&=\left(\begin{array}{l}
		\exists ss_0 \spot (s, ss_0) \in B \land ss_0 \subseteq (ss \cup \{ \bot \}) \land \bot \in ss_0 \land \bot \in ss
		\\ \lor \\
		\exists t, ss_0 \spot (s,t) \in B \land t \setminus \{ \bot \} = ss_0 \land ss_0 \subseteq ss \land \bot \in t \land \bot \notin ss
	\end{array}\right)
	&&\ptext{One-point rule and subsitutiton}\\
	&=\left(\begin{array}{l}
		\exists ss_0 \spot (s, ss_0) \in B \land ss_0 \subseteq (ss \cup \{ \bot \}) \land \bot \in ss_0 \land \bot \in ss
		\\ \lor \\
		\exists t \spot (s,t) \in B \land (t \setminus \{ \bot \}) \subseteq ss \land \bot \in t \land \bot \notin ss
	\end{array}\right)
	&&\ptext{Property of sets}\\
	&=\left(\begin{array}{l}
		\exists ss_0 \spot (s, ss_0) \in B \land ss_0 \subseteq (ss \cup \{ \bot \}) \land \bot \in ss_0 \land \bot \in ss
		\\ \lor \\
		\exists t \spot (s,t) \in B \land t \subseteq (ss \cup \{ \bot \}) \land \bot \in t \land \bot \notin ss
	\end{array}\right)
	&&\ptext{Rename variables}\\
	&=\left(\begin{array}{l}
		\exists ss_0 \spot (s, ss_0) \in B \land ss_0 \subseteq (ss \cup \{ \bot \}) \land \bot \in ss_0 \land \bot \in ss
		\\ \lor \\
		\exists ss_0 \spot (s, ss_0) \in B \land ss_0 \subseteq (ss \cup \{ \bot \}) \land \bot \in ss_0 \land \bot \notin ss
	\end{array}\right)
	&&\ptext{Predicate calculus}\\
	&=\exists ss_0 \spot ((s, ss_0) \in B \land ss_0 \subseteq (ss \cup \{ \bot \}) \land \bot \in ss_0) \land (\bot \in ss \lor \bot \notin ss)
	&&\ptext{Propositional calculus}\\
	&=\exists ss_0 \spot ((s, ss_0) \in B \land ss_0 \subseteq (ss \cup \{ \bot \}) \land \bot \in ss_0)
\end{flalign*}
\end{proof}
\end{proofs}
\end{lemma}

\begin{lemma}
\label{law:set-theory:A-setminus-x}
$(A = B \cup \{ x \} \land x \notin B) \iff (A \setminus \{ x \} = B \land x \in A)$
\begin{proofs}
\begin{proof}
\begin{flalign*}
	&A = B \cup \{ x \} \land x \notin B
	&&\ptext{Set equality}\\
	&=(\forall y \spot y \in A \iff y \in (B \cup \{x\})) \land x \notin B
	&&\ptext{Propositional calculus}\\
	&=(\forall y \spot (y \in A \implies y \in (B \cup \{x\})) \land (y \in (B \cup \{x\})) \implies y \in A) \land x \notin B
	&&\ptext{Property of sets}\\
	&=\left(\begin{array}{l}
		(\forall y \spot (y \in A \implies (y \in B \lor y \in \{x\})) \land ((y \in B \lor y \in \{x\}) \implies y \in A))
		\\ \land x \notin B
	\end{array}\right)
	&&\ptext{Propositional calculus}\\
	&=\left(\begin{array}{l}
		\forall y \spot \left(\begin{array}{l}
			 ((y \in A \land y \notin \{x\}) \implies y \in B) \\
			\land (y \in B \implies y \in A) \land (y \in \{x\} \implies y \in A)
		\end{array}\right)
		\\ \land x \notin B
	\end{array}\right)
	&&\ptext{\cref{law:set-theory:set-membership-subset-2} and propositional calculus}\\
	&=\left(\begin{array}{l}
		\forall y \spot \left(\begin{array}{l}
			 ((y \in A \land y \notin \{x\}) \implies y \in B) \\
			 \land (y \in B \implies y \in A) \land (y \in \{x\} \implies y \in A) \land (y \in B \implies y \notin \{x\})
		\end{array}\right)
	\end{array}\right)
	&&\ptext{Propositional calculus}\\
	&=\left(\begin{array}{l}
		\forall y \spot ((y \in A \land y \notin \{x\}) \iff (y \in B)) \land (y \in \{x\} \implies y \in A)
	\end{array}\right)
	&&\ptext{Property of sets}\\
	&=(A \setminus \{ x \} = B \land \{x\} \subseteq A)
	&&\ptext{\cref{law:set-theory:set-membership-subset-1} and propositional calculus}\\
	&=(A \setminus \{ x \} = B \land x \in A)
\end{flalign*}
\end{proof}
\end{proofs}
\end{lemma}

\begin{lemma}
\label{law:set-theory:set-membership-subset-1}
$\{ x \} \subseteq A \iff x \in A$
\begin{proofs}
\begin{proof}
\begin{flalign*}
	&\{ x \} \subseteq A
	&&\ptext{Definition of subset inclusion}\\
	&=\forall y \spot y \in \{ x \} \implies y \in A
	&&\ptext{Propositional calculus}\\
	&=\forall y \spot \lnot (y \in \{ x \} \land y \notin A)
	&&\ptext{Propositional calculus}\\
	&=\lnot \exists y \spot y = x \land y \notin A
	&&\ptext{One-point rule}\\
	&=\lnot (x \notin A)
	&&\ptext{Propositional calculus}\\
	&=x \in A
\end{flalign*}
\end{proof}
\end{proofs}
\end{lemma}

\begin{lemma}
\label{law:set-theory:set-membership-subset-2}
$x \notin A \iff (\forall y \spot y \in A \implies y \notin \{ x \})$
\begin{proofs}
\begin{proof}
\begin{flalign*}
	&x \notin A
	&&\ptext{Propositional calculus}\\
	&=\lnot (x \in A)
	&&\ptext{Introduce fresh variable}\\
	&=\lnot (\exists y \spot y = x \land y \in A)
	&&\ptext{Property of sets}\\
	&=\lnot (\exists y \spot y \in \{x\} \land y \in A)
	&&\ptext{Propositional cauclus}\\
	&=\forall y \spot y \in A \implies y \notin \{x\}
\end{flalign*}
\end{proof}
\end{proofs}
\end{lemma}

\begin{lemma}\label{law:set-theory:A-B:x-in-B}
$(A = (B \cup \{ x \}) \land x \in B) \iff (A = B \land x \in B)$
\begin{proofs}
\begin{proof}(Implication)
\begin{flalign*}
	&A = B \cup \{ x \} \land x \in B
	&&\ptext{Property of sets}\\
	&=(A \subseteq (B \cup \{ x \}) \land (B \cup \{x\}) \subseteq A \land x \in B)
	&&\ptext{\cref{law:set-theory:set-membership-subset-1}}\\
	&=(A \subseteq (B \cup \{ x \}) \land (B \cup \{x\}) \subseteq A \land \{x\} \subseteq B)
	&&\ptext{Property of sets}\\
	&=(A \subseteq (B \cup \{ x \}) \land B \subseteq A \land \{x\} \subseteq A \land \{x\} \subseteq B)
	&&\ptext{Property of sets}\\
	&=(A \subseteq (B \cup \{ x \}) \land B \subseteq A \land \{x\} \subseteq A \land (\{x\} \cup B = B))
	&&\ptext{Propositional calculus}\\
	&= (A \subseteq (B \cup \{ x \}) \land B \subseteq A \land \{x\} \subseteq A \land (\{x\} \cup B) \subseteq B) \land B \subseteq (\{x\} \cup B)
	&&\ptext{Transitivity of subset inclusion and propositional calculus}\\
	&=\left(\begin{array}{l}
		(A \subseteq (B \cup \{ x \}) \land B \subseteq A \land \{x\} \subseteq A 
		\\ \land \\
		(\{x\} \cup B) \subseteq B \land A \subseteq B \land B \subseteq (\{x\} \cup B))
	\end{array}\right)
	&&\ptext{Propositional calculus}\\
	&\implies B \subseteq A \land A \subseteq B \land (\{x\} \cup B) \subseteq B \land B \subseteq (\{x\} \cup B)
	&&\ptext{Property of sets}\\
	&=(B = A \land \{x\} \cup B)
	&&\ptext{\cref{law:set-theory:set-membership-subset-1}}\\
	&=(B = A \land x \in B)
\end{flalign*}
\end{proof}
\begin{proof}(Reverse implication)
\begin{flalign*}
	&(B = A \land x \in B)
	&&\ptext{\cref{law:set-theory:set-membership-subset-1}}\\
	&(B = A \land \{x\} \subseteq B)
	&&\ptext{Property of sets}\\
	&=(A \subseteq B \land B \subseteq A \land \{x\} \subseteq B)
	&&\ptext{Transitivity of subset inclusion and propositional calculus}\\
	&=(A \subseteq B \land B \subseteq A \land \{x\} \subseteq B \land \{x\} \subseteq A)
	&&\ptext{Property of sets}\\
	&=(A \subseteq B \land B \subseteq A \land \{x\} \subseteq B \land \{x\} \subseteq A \land (B \cup \{x\}) \subseteq A \land (A \cup \{x\}) \subseteq B
	&&\ptext{Property of sets}\\
	&=\left(\begin{array}{l}
		(A \subseteq B \land B \subseteq A \land \{x\} \subseteq B \land (\{x\} \cup B = B)
		\\ \land \\
		\{x\} \subseteq A \land (B \cup \{x\}) \subseteq A \land (A \cup \{x\}) \subseteq B
	\end{array}\right)
	&&\ptext{Property of sets and weaken predicate}\\
	&\implies \left(\begin{array}{l}
		(A \subseteq B \land B \subseteq A \land \{x\} \subseteq B \land B \subseteq (\{x\} \cup B) 
		\\ \land \\
		\{x\} \subseteq A \land (B \cup \{x\}) \subseteq A \land (A \cup \{x\}) \subseteq B
	\end{array}\right)
	&&\ptext{Transitivity of subset inclusion and propositional calculus}\\
	&\implies \left(\begin{array}{l}
		(A \subseteq B \land B \subseteq A \land \{x\} \subseteq B \land B \subseteq (\{x\} \cup B) 
		\\ \land \\
		A \subseteq (\{x\} \cup B) \land \{x\} \subseteq A \land (B \cup \{x\}) \subseteq A
	\end{array}\right)
	&&\ptext{Property of sets}\\
	&=(A = B \land B \subseteq (\{x\} \cup B) \land \{x\} \subseteq B \land \{x\} \subseteq A \land (B \cup \{x\}) = A
	&&\ptext{Propositional calculus}\\
	&\implies (\{x\} \subseteq B \land (B \cup \{x\}) = A)
	&&\ptext{\cref{law:set-theory:set-membership-subset-1}}\\
	&=((B \cup \{x\}) = A \land x \in B)
\end{flalign*}
\end{proof}
\end{proofs}
\end{lemma}

\begin{lemma}\label{law:set-theory:A-cup-x:subseteq:B-cup-x}
\begin{align*}
	&((A \cup \{x\}) \subseteq (B \cup \{x\}) \land x \notin A \land x \notin B) \iff (A \subseteq B \land x \notin A \land x \notin B)
\end{align*}
\begin{proofs}
\begin{proof}
\begin{flalign*}
	&(A \cup \{x\}) \subseteq (B \cup \{x\}) \land x \notin A \land x \notin B
	&&\ptext{Definition of subset inclusion}\\
	&=\forall y \spot y \in (A \cup \{x \}) \implies y \in (B \cup \{x\}) \land x \notin A \land x \notin B
	&&\ptext{Property of sets}\\
	&=\forall y \spot (y \in A \lor y \in \{x \}) \implies (y \in B \lor y \in \{x\}) \land x \notin A \land x \notin B
	&&\ptext{Propositional calculus}\\
	&=\forall y \spot y \in A \implies (y \in B \lor y \in \{x\}) \land x \notin A \land x \notin B
	&&\ptext{\cref{law:set-theory:set-membership-subset-2}}\\
	&=\left(\begin{array}{l}
		\forall y \spot y \in A \implies (y \in B \lor y \in \{x\})
		\\ \land \\
		\forall y \spot y \in A \implies y \notin \{x\}
		\\ \land \\
		\forall y \spot y \in B \implies y \notin \{x\}
	\end{array}\right)
	&&\ptext{Propositional calculus}\\
	&=\forall y \spot
	 	\left(\begin{array}{l}
			y \in A \implies (y \in B \land y \notin \{x\})
			\\ \land \\
			y \in B \implies y \notin \{x\}
		\end{array}\right)
	&&\ptext{Propositional calculus}\\
	&=\forall y \spot (y \in A \implies y \in B) \land ((y \in A \lor y \in B) \implies (y \notin \{x\}))
	&&\ptext{Propositional calculus and definition of subset inclusion}\\
	&=A \subseteq B \land \forall y \spot ((y \in A \lor y \in B) \implies (y \notin \{x\}))
	&&\ptext{Property of sets and \cref{law:set-theory:set-membership-subset-2}}\\
	&=A \subseteq B \land x \notin (A \cup B)
	&&\ptext{Propositional calculus and property of sets}\\
	&=A \subseteq B \land x \notin A \land x \notin B
\end{flalign*}
\end{proof}
\end{proofs}
\end{lemma}
\chapter{Angelic Designs ($\mathbf{A}$)}\label{appendix:angelic-designs}

\section{Healthiness Conditions}

\subsection{$\mathbf{A0}$}

\theoremstatementref{def:A0}

\begin{theorem}\label{law:A0:idempotent}
\begin{statement}$\mathbf{A0} \circ \mathbf{A0} (P) = \mathbf{A0} (P)$\end{statement}
\begin{proofs}
\begin{proof}
\begin{flalign*}
	&\mathbf{A0} \circ \mathbf{A0} (P)
	&&\ptext{\cref{law:A0:design}}\\
	&=\mathbf{A0} (\lnot P^f \vdash P^t \land ac'\neq\emptyset)
	&&\ptext{\cref{law:A0:design}}\\
	&=(\lnot P^f \vdash P^t \land ac'\neq\emptyset \land ac'\neq\emptyset)
	&&\ptext{Propositional calculus}\\
	&=(\lnot P^f \vdash P^t \land ac'\neq\emptyset)
	&&\ptext{Definition of $\mathbf{A0}$}\\
	&=\mathbf{A0} (P)
\end{flalign*}
\end{proof}
\end{proofs}
\end{theorem}
\begin{theorem}\label{law:A0:monotonic}
\begin{statement}$(P \sqsubseteq Q) \implies (\mathbf{A0} (P) \sqsubseteq \mathbf{A0} (Q))$\end{statement}
\begin{proofs}
\begin{proof}
\begin{flalign*}
	&\mathbf{A0} (Q)
	&&\ptext{Definition of $\mathbf{A0}$}\\
	&=Q \land ((ok \land \lnot Q^f) \implies (ok' \implies ac'\neq\emptyset))
	&&\ptext{Assumption: $[Q \implies P] \iff [\lnot P \implies \lnot Q] $}\\
	&\implies P \land ((ok \land \lnot P^f) \implies (ok' \implies ac'\neq\emptyset))
	&&\ptext{Definition of $\mathbf{A0}$}\\
	&=\mathbf{A0} (P)
\end{flalign*}
\end{proof}
\end{proofs}
\end{theorem}\noindent

\begin{theorem}\label{law:A0:design}
\begin{statement}If $P$ is a design so is $\mathbf{A0} (P)$.
\begin{align*}
	\mathbf{A0} (P) = (\lnot P^f \vdash P^t \land ac'\neq\emptyset)	
\end{align*}
\end{statement}
\begin{proofs}
\begin{proof}
\begin{flalign*}
	&\mathbf{A0} (P)
	&&\ptext{Definition of design and $\mathbf{A0}$}\\
	&=(\lnot P^f \vdash P^t) \land ((ok \land \lnot P^f) \implies (ok' \implies ac'\neq\emptyset))
	&&\ptext{Definition of design and propositional calculus}\\
	&=(ok \land \lnot P^f) \implies (P^t \land ok' \land (ok' \implies ac'\neq\emptyset))
	&&\ptext{Propositional calculus}\\
	&=(ok \land \lnot P^f) \implies (P^t \land ok' \land ac'\neq\emptyset)
	&&\ptext{Definition of design}\\
	&=(\lnot P^f \vdash P^t \land ac'\neq\emptyset)
\end{flalign*}
\end{proof}
\end{proofs}
\end{theorem}\noindent

\begin{theorem}\label{law:A0:conjunction-closure}
\begin{statement}
Provided $P$ and $Q$ are $\mathbf{A0}$-healthy,
\begin{align*}
	&\mathbf{A0} (P \land Q) = P \land Q
\end{align*}
\end{statement}
\begin{proofs}
\begin{proof}
\begin{flalign*}
	&P \land Q
	&&\ptext{Assumption: $P$ and $Q$ are $\mathbf{A0}$-healthy}\\
	&=\mathbf{A0} (P) \land \mathbf{A0} (Q)
	&&\ptext{Definition of $\mathbf{A0}$}&\\
	&=\left(\begin{array}{l}
		(P \land ((ok \land \lnot P^f) \implies (ok' \implies ac'\neq\emptyset))) \\
		\land \\
		(Q \land ((ok \land \lnot Q^f) \implies (ok' \implies ac'\neq\emptyset)))
	\end{array}\right)
	&&\ptext{Propositional calculus}\\
	&=(P \land Q) \land (((ok \land \lnot P^f) \lor (ok \land \lnot Q^f)) \implies (ok' \implies ac'\neq\emptyset))
	&&\ptext{Propositional calculus}\\
	&=(P \land Q) \land ((ok \land \lnot (P^f \land Q^f)) \implies (ok' \implies ac'\neq\emptyset))
	&&\ptext{Definition of $\mathbf{A0}$}\\
	&=\mathbf{A0} (P \land Q)
\end{flalign*}
\end{proof}
\end{proofs}
\end{theorem}
\begin{theorem}\label{law:A0:disjunction-closure}
\begin{statement}
Provided $P$ and $Q$ are $\mathbf{A0}$-healthy designs,
\begin{align*} 
	&\mathbf{A0} (P \lor Q) = P \lor Q
\end{align*}
\end{statement}
\begin{proofs}
\begin{proof}
\begin{flalign*}
	&P \lor Q
	&&\ptext{Assumption: $P$ and $Q$ are $\mathbf{A0}$-healthy}\\
	&=\mathbf{A0} (P) \lor \mathbf{A0} (Q)
	&&\ptext{Definition of $\mathbf{A0}$}\\
	&=(\lnot P^f \vdash P^t \land ac'\neq\emptyset) \lor (\lnot Q^f \vdash Q^t \land ac'\neq\emptyset)
	&&\ptext{Disjunction of designs}\\
	&=(\lnot P^f \land \lnot Q^f \vdash (P^t \land ac'\neq\emptyset) \lor (Q^t \land ac'\neq\emptyset))
	&&\ptext{Propositional calculus}\\
	&=(\lnot (P^f \lor Q^f) \vdash (P^t \lor Q^t) \land ac'\neq\emptyset)
	&&\ptext{Property of substitution}\\
	&=(\lnot (P \lor Q)^f \vdash (P \lor Q)^t \land ac'\neq\emptyset)
	&&\ptext{Definition of $\mathbf{A0}$}\\
	&=\mathbf{A0} (P \lor Q)
\end{flalign*}
\end{proof}
\end{proofs}
\end{theorem}\noindent
\begin{theorem}\label{theorem:A0(P-land-Q):A0(P)-land-A0(Q)}
\begin{statement}
$ \mathbf{A0} (P \land Q) = \mathbf{A0} (P) \land \mathbf{A0} (Q)$
\end{statement}
\begin{proofs}
\begin{proof}
\begin{flalign*}
	&\mathbf{A0} (P \land Q)
	&&\ptext{Definition of $\mathbf{A0}$}\\
	&=(P \land Q) \land ((ok \land \lnot (P \land Q)^f) \implies (ok' \implies ac'\neq\emptyset))
	&&\ptext{Property of substitution}\\
	&=(P \land Q) \land ((ok \land \lnot (P^f \land Q^f)) \implies (ok' \implies ac'\neq\emptyset))
	&&\ptext{Predicate calculus}\\
	&=(P \land Q) \land ((ok \land (\lnot P^f \lor \lnot Q^f)) \implies (ok' \implies ac'\neq\emptyset))
	&&\ptext{Predicate calculus}\\
	&=(P \land Q) \land (((ok \land \lnot P^f) \lor (ok \land \lnot Q^f)) \implies (ok' \implies ac'\neq\emptyset))
	&&\ptext{Predicate calculus}\\
	&=\left(\begin{array}{l} 
		P \land ((ok \land \lnot P^f) \implies (ok' \implies ac'\neq\emptyset)) 
		\\ \land \\
		Q \land ((ok \land \lnot Q^f) \implies (ok' \implies ac'\neq\emptyset))
	\end{array}\right)
	&&\ptext{Definition of $\mathbf{A0}$}\\
	&=\mathbf{A0} (P) \land \mathbf{A0} (Q)
\end{flalign*}
\end{proof}
\end{proofs}
\end{theorem}


\begin{theorem}\label{theorem:A0-o-H1-o-H2(P):(lnot-Pf|-Pt-land-ac'-neq-emptyset)}
\begin{statement}$\mathbf{A0} \circ \mathbf{H1} \circ \mathbf{H2} (P) = (\lnot P^f \vdash P^t \land ac'\neq\emptyset)$\end{statement}
\begin{proofs}
\begin{proof}
\begin{xflalign*}
	&\mathbf{A0} \circ \mathbf{H1} \circ \mathbf{H2} (P)
	&&\ptext{Definition of design}\\
	&=\mathbf{A0} (\lnot P^f \vdash P^t)
	&&\ptext{Definition of $\mathbf{A0}$}\\
	&=(\lnot P^f \vdash P^t) \land ((ok \land \lnot (\lnot P^f \vdash P^t)^f) \implies (ok' \implies ac'\neq\emptyset))
	&&\ptext{Definition of design}\\
	&=(\lnot P^f \vdash P^t) \land \left(\begin{array}{l}
			(ok \land \lnot ((ok \land \lnot P^f) \implies (P^t \land ok'))^f)
			\\ \implies \\ 
			(ok' \implies ac'\neq\emptyset)
	\end{array}\right)
	&&\ptext{Substitution}\\
	&=(\lnot P^f \vdash P^t) \land \left(\begin{array}{l}
			(ok \land \lnot ((ok \land \lnot P^f) \implies (P^t \land false)))
			\\ \implies \\ 
			(ok' \implies ac'\neq\emptyset)
	\end{array}\right)	
	&&\ptext{Predicate calculus}\\
	&=(\lnot P^f \vdash P^t) \land ((ok \land \lnot P^f) \implies (ok' \implies ac'\neq\emptyset))
	&&\ptext{Definition of design}\\
	&=((ok \land \lnot P^f) \implies (P^t \land ok')) \land ((ok \land \lnot P^f) \implies (ok' \implies ac'\neq\emptyset))
	&&\ptext{Predicate calculus}\\
	&=((ok \land \lnot P^f) \implies (P^t \land ok' \land (ok' \implies ac'\neq\emptyset))
	&&\ptext{Predicate calculus}\\
	&=((ok \land \lnot P^f) \implies (P^t \land ok' \land ac'\neq\emptyset)
	&&\ptext{Definition of design}\\
	&=(\lnot P^f \vdash P^t \land ac'\neq\emptyset)
\end{xflalign*}
\end{proof}
\end{proofs}
\end{theorem}

\begin{theorem}\label{theorem:A0-o-H1-o-H2(P):H1-o-H2-o-A0(P)}
\begin{statement}$\mathbf{H1} \circ \mathbf{H2} \circ \mathbf{A0} (P) = \mathbf{A0} \circ \mathbf{H1} \circ \mathbf{H2} (P)$\end{statement}
\begin{proofs}
\begin{proof}
\begin{xflalign*}
	&\mathbf{H1} \circ \mathbf{H2} \circ \mathbf{A0} (P)
	&&\ptext{Definition of design}\\
	&=(\lnot \mathbf{A0} (P)^f \vdash \mathbf{A0} (P)^t)
	&&\ptext{\cref{lemma:A0(P)-subs-f:Pf,lemma:A0(P)-subs-t:Pt-land}}\\
	&=(\lnot P^f \vdash P^t \land ((ok \land \lnot P^f) \implies ac'\neq\emptyset))
	&&\ptext{Definition of design and predicate calculus}\\
	&=(\lnot P^f \vdash P^t \land ac'\neq\emptyset)
	&&\ptext{\cref{theorem:A0-o-H1-o-H2(P):(lnot-Pf|-Pt-land-ac'-neq-emptyset)}}\\
	&=\mathbf{A0} \circ \mathbf{H1} \circ \mathbf{H2} (P)
\end{xflalign*}
\end{proof}
\end{proofs}
\end{theorem}

\begin{lemma}\label{lemma:A0-substitution-s} 
\begin{statement}
Provided $ok'$ not free in $e$, $\mathbf{A0} (P)[e/s] = \mathbf{A0} (P[e/s])$.
\end{statement}
\begin{proofs}
\begin{proof}\checkt{alcc}
\begin{flalign*}
	&\mathbf{A0} (P)[e/s]
	&&\ptext{Definition of $\mathbf{A0}$}\\
	&=(P \land (ok \land \lnot P^f) \implies (ok' \implies ac'\neq\emptyset))[e/s]
	&&\ptext{Property of substitution}\\
	&=(P[e/s] \land (ok \land \lnot P^f[e/s]) \implies (ok' \implies ac'\neq\emptyset))
	&&\ptext{Property of substitution: $ok'$ not free in $e$}\\
	&=(P[e/s] \land (ok \land \lnot P[e/s]^f) \implies (ok' \implies ac'\neq\emptyset))
	&&\ptext{Definition of $\mathbf{A0}$}\\
	&=\mathbf{A0} (P[e/s])
\end{flalign*}
\end{proof}
\end{proofs}
\end{lemma}

\begin{lemma}\label{lemma:A0(P)-subs-o}
\begin{statement}$\mathbf{A0} (P)^o = P^o \land ((ok \land \lnot P^f) \implies (o \implies ac'\neq\emptyset))$\end{statement}
\begin{proofs}
\begin{proof}
\begin{xflalign*}
	&\mathbf{A0} (P)^o
	&&\ptext{Definition of $\mathbf{A0}$}\\
	&=(P \land ((ok \land \lnot P^f) \implies (ok' \implies ac'\neq\emptyset)))^o
	&&\ptext{Substitution}\\
	&=P^o \land ((ok \land \lnot P^f) \implies (o \implies ac'\neq\emptyset))
\end{xflalign*}
\end{proof}
\end{proofs}
\end{lemma}

\begin{lemma}\label{lemma:A0(P)-subs-f:Pf}
\begin{statement}
$\mathbf{A0} (P)^f = P^f$
\end{statement}
\begin{proofs}
\begin{proof}
\begin{xflalign*}
	&\mathbf{A0} (P)^f
	&&\ptext{\cref{lemma:A0(P)-subs-o}}\\
	&=P^f \land ((ok \land \lnot P^f) \implies (false \implies ac'\neq\emptyset))
	&&\ptext{Predicate calculus}\\
	&=P^f
\end{xflalign*}
\end{proof}
\end{proofs}
\end{lemma}

\begin{lemma}\label{lemma:A0(P)-subs-t:Pt-land}
\begin{statement}
$\mathbf{A0} (P)^t = P^t \land ((ok \land \lnot P^f) \implies ac'\neq\emptyset)$
\end{statement}
\begin{proofs}
\begin{proof}
\begin{xflalign*}
	&\mathbf{A0} (P)^t
	&&\ptext{\cref{lemma:A0(P)-subs-o}}\\
	&=P^t \land ((ok \land \lnot P^f) \implies (true \implies ac'\neq\emptyset))
	&&\ptext{Predicate calculus}\\
	&=P^t \land ((ok \land \lnot P^f) \implies ac'\neq\emptyset)
\end{xflalign*}
\end{proof}
\end{proofs}
\end{lemma}

\subsection{$\mathbf{A1}$}

\begin{theorem}\label{law:A1:idempotent}
\begin{statement}$\mathbf{A1} \circ \mathbf{A1} (P_0 \vdash P_1) = \mathbf{A1} (P_0 \vdash P_1)$\end{statement}
\begin{proofs}
\begin{proof}
\begin{flalign*}
	&\mathbf{A1} \circ \mathbf{A1} (P_0 \vdash P_1)
	&&\ptext{Definition of $\mathbf{A1}$}\\
	&=\mathbf{A1} \circ (\lnot \mathbf{PBMH} (\lnot P_0) \vdash \mathbf{PBMH} (P_1))
	&&\ptext{Definition of $\mathbf{A1}$}\\
	&=(\lnot (\mathbf{PBMH} (\lnot \lnot \mathbf{PBMH} (\lnot P_0))) \vdash \mathbf{PBMH} \circ \mathbf{PBMH} (P_1))
	&&\ptext{Propositional calculus}\\
	&=(\lnot (\mathbf{PBMH} \circ \mathbf{PBMH} (\lnot P_0)) \vdash \mathbf{PBMH} \circ \mathbf{PBMH} (P_1))
	&&\ptext{\cref{law:pbmh:idempotent}}\\
	&=(\lnot (\mathbf{PBMH} (\lnot P_0)) \vdash \mathbf{PBMH} (P_1))
	&&\ptext{Definition of $\mathbf{A1}$}\\
	&=\mathbf{A1} (P_0 \vdash P_1)
\end{flalign*}
\end{proof}
\end{proofs}
\end{theorem}

\begin{theorem}\label{law:A1:monotonic}
\begin{statement}$(P \sqsubseteq Q) \implies \mathbf{A1} (P) \sqsubseteq \mathbf{A1} (Q)$\end{statement}
\begin{proofs}
\begin{proof}
\begin{flalign*}
	&\mathbf{A1} (Q)
	&&\ptext{Definition of design}\\
	&=\mathbf{A1} (\lnot Q^f \vdash Q^t)
	&&\ptext{Definition of design and propositional calculus}\\
	&=\mathbf{A1} ( (\lnot ok \lor Q^f) \lor (Q^t \land ok') )
	&&\ptext{Assumption: $[Q \implies P]$ holds}\\
	&=\mathbf{A1} ( (\lnot ok \lor (Q^f \land (Q^f \implies P^f))) \lor (Q^t \land (Q^t \implies P^t) \land ok'))
	&&\ptext{Predicate calculus and definition of design}\\
	&=\mathbf{A1} (\lnot (Q^f \land P^f) \vdash Q^t \land P^t)
	&&\ptext{Definition of $\mathbf{A1}$}\\
	&=(\lnot \mathbf{PBMH} (Q^f \land P^f) \vdash \mathbf{PBMH} (Q^t \land P^t))
	&&\ptext{Definition of $\mathbf{PBMH}$}\\
	&=(\lnot \mathbf{PBMH} (Q^f \land P^f) \vdash \mathbf{PBMH} (Q^t \land P^t))
	&&\ptext{Definition of sequential composition}\\
	&=(\lnot \exists ac_0 \spot Q^f[ac_0/ac'] \land P^f[ac_0/ac'] \land ac_0 \subseteq ac' \vdash (Q^t \land P^t) \circseq ac \subseteq ac')
	&&\ptext{Predicate calculus}\\
	&=\left(\begin{array}{l}
		\forall ac_0 \spot \lnot Q^f[ac_0/ac'] \lor \lnot P^f[ac_0/ac'] \lor \lnot (ac_0 \subseteq ac') 
		\\ \vdash \\
		(Q^t \land P^t) \circseq ac \subseteq ac'
	\end{array}\right)
	&&\ptext{Predicate calculus}\\
	&=\left(\begin{array}{l}
		\forall ac_0 \spot \left(\begin{array}{l}
			(\lnot Q^f[ac_0/ac'] \lor \lnot (ac_0 \subseteq ac'))
			\\ \lor \\
			(\lnot P^f[ac_0/ac'] \lor \lnot (ac_0 \subseteq ac'))
			\end{array}\right)
		\\ \vdash \\
		(Q^t \land P^t) \circseq ac \subseteq ac'
	\end{array}\right)
	&&\ptext{Weaken precondition}\\
	&\sqsupseteq \left(\begin{array}{l}
		\forall ac_0 \spot (\lnot Q^f[ac_0/ac'] \lor \lnot (ac_0 \subseteq ac')) 
		\\ \lor \\
		\forall ac_0 \spot (\lnot P^f[ac_0/ac'] \lor \lnot (ac_0 \subseteq ac'))
		\\ \vdash \\
		(Q^t \land P^t) \circseq ac \subseteq ac'
	\end{array}\right)
	&&\ptext{Weaken precondition}\\
	&\sqsupseteq (\forall ac_0 \spot (\lnot P^f[ac_0/ac'] \lor \lnot (ac_0 \subseteq ac')) \vdash (Q^t \land P^t) \circseq ac \subseteq ac')
	&&\ptext{Predicate calculus}\\
	&= (\lnot \exists ac_0 \spot P^f[ac_0/ac'] \land ac_0 \subseteq ac' \vdash (Q^t \land P^t) \circseq ac \subseteq ac')
	&&\ptext{Definition of sequential composition}\\
	&= (\lnot (P^f \circseq ac \subseteq ac') \vdash (Q^t \land P^t) \circseq ac \subseteq ac')
	&&\ptext{Strengthen postcondition}\\
	&\sqsupseteq (\lnot (P^f \circseq ac \subseteq ac') \vdash P^t \circseq ac \subseteq ac')
	&&\ptext{Definition of $\mathbf{PBMH}$}\\
	&= (\lnot \mathbf{PBMH} (P^f) \vdash \mathbf{PBMH} (P^t))
	&&\ptext{Definition of $\mathbf{A1}$}\\
	&=\mathbf{A1} (\lnot P^f \vdash P^t)
	&&\ptext{Definition of designs}\\
	&=\mathbf{A1} (P)
\end{flalign*}
\end{proof}
\end{proofs}
\end{theorem}\noindent

\subsection{$\mathbf{A}$}

\begin{theorem}\label{law:A0:commute-A0-healthy}
\begin{statement}
Provided $P^t$ satisfies $\mathbf{PBMH}$,
$\mathbf{A0} \circ \mathbf{A1} (P) = \mathbf{A1} \circ \mathbf{A0} (P)$
\end{statement}
\begin{proofs}
\begin{proof}
\begin{flalign*}
	&\mathbf{A0} \circ \mathbf{A1} (P)
	&&\ptext{Definition of design}\\
	&=\mathbf{A0} \circ \mathbf{A1} (\lnot P^f \vdash P^t)
	&&\ptext{Definition of $\mathbf{A1}$}\\
	&=\mathbf{A0} (\lnot \mathbf{PBMH}(P^f) \vdash \mathbf{PBMH} (P^t))
	&&\ptext{\cref{law:A0:design}}\\
	&=(\lnot \mathbf{PBMH}(P^f) \vdash \mathbf{PBMH} (P^t) \land ac'\neq\emptyset)
	&&\ptext{$ac'\neq\emptyset$ satisfies $\mathbf{PBMH}$ (\cref{law:pbmh:ac'-neq-emptyset})}\\
	&=(\lnot \mathbf{PBMH}(P^f) \vdash \mathbf{PBMH} (P^t) \land \mathbf{PBMH} (ac'\neq\emptyset))
	&&\ptext{Closure of $\mathbf{PBMH}$ w.r.t. conjunction (\cref{law:pbmh:conjunction-closure})}\\
	&=(\lnot \mathbf{PBMH}(P^f) \vdash \mathbf{PBMH} (\mathbf{PBMH} (P^t) \land \mathbf{PBMH} (ac'\neq\emptyset)))
	&&\ptext{$ac'\neq\emptyset$ satisfies $\mathbf{PBMH}$ (\cref{law:pbmh:ac'-neq-emptyset})}\\
	&=(\lnot \mathbf{PBMH}(P^f) \vdash \mathbf{PBMH} (\mathbf{PBMH} (P^t) \land ac'\neq\emptyset))
	&&\ptext{Assumption: $P^t$ satisfies $\mathbf{PBMH}$}\\
	&=(\lnot \mathbf{PBMH}(P^f) \vdash \mathbf{PBMH} (P^t \land ac'\neq\emptyset))
	&&\ptext{Definition of $\mathbf{A1}$}\\
	&=\mathbf{A1} (\lnot P^f \vdash P^t \land ac'\neq\emptyset)
	&&\ptext{Definition of $\mathbf{A0}$}\\
	&=\mathbf{A1} \circ \mathbf{A0} (\lnot P^f \vdash P^t)
	&&\ptext{Definition of design}\\
	&=\mathbf{A1} \circ \mathbf{A0} (P)
\end{flalign*}
\end{proof}
\end{proofs}
\end{theorem}\noindent

\begin{theorem}\label{law:A:idempotent}
\begin{statement}$\mathbf{A} \circ \mathbf{A} (P) = \mathbf{A} (P)$\end{statement}
\begin{proofs}
\begin{proof}
\begin{flalign*}
	&\mathbf{A} \circ \mathbf{A} (P)
	&&\ptext{Definition of $\mathbf{A}$ twice}\\
	&=\mathbf{A0} \circ \mathbf{A1} \circ \mathbf{A0} \circ \mathbf{A1} (P)
	&&\ptext{\cref{law:A0:commute-A0-healthy} and $\mathbf{A1} (P)$ ensures $P^t$ satisfies $\mathbf{PBMH}$}\\
	&=\mathbf{A0} \circ \mathbf{A0} \circ \mathbf{A1} \circ \mathbf{A1} (P)
	&&\ptext{$\mathbf{A0}$-idempotent (\cref{law:A0:idempotent}) and $\mathbf{A1}$-idempotent (\cref{law:A1:idempotent})}\\
	&=\mathbf{A0} \circ \mathbf{A1} (P)
	&&\ptext{Definition of $\mathbf{A}$}\\
	&=\mathbf{A} (P)
\end{flalign*}
\end{proof}
\end{proofs}
\end{theorem}\noindent

\begin{theorem}\label{theorem:A-o-H1-o-H2(P):H1-o-H2-o-A(P)}
\begin{statement}
$\mathbf{H1} \circ \mathbf{H2} \circ \mathbf{A} (P) = \mathbf{A} \circ \mathbf{H1} \circ \mathbf{H2} (P)$
\end{statement}
\begin{proofs}
\begin{proof}
\begin{xflalign*}
	&\mathbf{H1} \circ \mathbf{H2} \circ \mathbf{A} (P)
	&&\ptext{Definition of $\mathbf{A}$}\\
	&=\mathbf{H1} \circ \mathbf{H2} \circ \mathbf{A0} \circ \mathbf{A1} (P)
	&&\ptext{\cref{theorem:A0-o-H1-o-H2(P):H1-o-H2-o-A0(P)}}\\
	&=\mathbf{A0} \circ \mathbf{H1} \circ \mathbf{H2} \circ \mathbf{A1} (P)
	&&\ptext{$\mathbf{A1}$ is $\mathbf{PBMH}$}\\
	&=\mathbf{A0} \circ \mathbf{H1} \circ \mathbf{H2} \circ \mathbf{PBMH} (P)
	&&\ptext{\cref{theorem:H2-o-PBMH:PBMH-o-H2}}\\
	&=\mathbf{A0} \circ \mathbf{H1} \circ \mathbf{PBMH} \circ \mathbf{H2} (P)
	&&\ptext{\cref{theorem:H1-o-PBMH:PBMH-o-H1}}\\
	&=\mathbf{A0} \circ \mathbf{PBMH} \circ \mathbf{H1} \circ \mathbf{H2} (P)
	&&\ptext{$\mathbf{A1}$ is $\mathbf{PBMH}$}\\
	&=\mathbf{A0} \circ \mathbf{A1} \circ \mathbf{H1} \circ \mathbf{H2} (P)
\end{xflalign*}
\end{proof}
\end{proofs}
\end{theorem}

\begin{theorem}\label{law:A:monotonic}
\begin{statement}$P \sqsubseteq Q \implies \mathbf{A} (P) \sqsubseteq \mathbf{A} (Q)$\end{statement}
\begin{proofs}
\begin{proof}
Follows from $\mathbf{A0}$-monotonic (\cref{law:A0:monotonic}) and $\mathbf{A1}$-monotonic (\cref{law:A1:monotonic}).
\end{proof}
\end{proofs}
\end{theorem}\noindent

\begin{lemma}\label{lemma:A-substitution-s} Provided $ok'$ is not free in $e$,
$\mathbf{A} (P)[e/s] = \mathbf{A} (P[e/s])$
\begin{proofs}\begin{proof}\checkt{alcc}
\begin{xflalign*}
	&\mathbf{A} (P)[e/s]
	&&\ptext{Definition of $\mathbf{A}$}\\
	&=(\mathbf{A0} \circ \mathbf{PBMH} (P))[e/s]
	&&\ptext{\cref{lemma:A0-substitution-s}}\\
	&=\mathbf{A0} \circ (\mathbf{PBMH} (P))[e/s]
	&&\ptext{\cref{lemma:PBMH-substitution-s}}\\
	&=\mathbf{A0} \circ \mathbf{PBMH} (P[e/s])
	&&\ptext{Definition of $\mathbf{A}$}\\
	&=\mathbf{A} (P[e/s])
\end{xflalign*}
\end{proof}\end{proofs}
\end{lemma}

\begin{lemma}\label{lemma:state-s-Leibniz-P}
$s.x = v \land P \iff s.x = v \land P[s\oplus\{x \mapsto v\}/s]$
\begin{proofs}\begin{proof}
\begin{flalign*}
	&s.x = v \land P
	&&\ptext{Predicate calculus for fresh variable $z$}\\
	&\iff \exists z \spot s.x = v \land z = s \land P[z/s]
	&&\ptext{Relational calculus}\\
	&\iff \exists z \spot s.x = v \land z = s \oplus \{ x \mapsto v \} \land P[z/s]
	&&\ptext{One-point rule}\\
	&\iff s.x = v \land P[z/s][s \oplus \{ x \mapsto v \}/z]
	&&\ptext{Substitution}\\
	&\iff s.x = v \land P[s \oplus \{ x \mapsto v \}/s]
\end{flalign*}
\end{proof}\end{proofs}
\end{lemma}

\begin{lemma}\label{lemma:A:P-f} Provided $P$ is an $\mathbf{A}$-healthy design,
$P^f = ok \implies P^f$.
\begin{proofs}\begin{proof}\checkt{alcc}\checkt{pfr}
\begin{xflalign*}
	&P^f
	&&\ptext{Assumption: $P$ is an $\mathbf{A}$-healthy design}\\
	&=(\mathbf{A} (\lnot P^f \vdash P^t))^f
	&&\ptext{Definition of $\mathbf{A}$}\\
	&=(\lnot \mathbf{PBMH} (P^f) \vdash \mathbf{PBMH} (P^t) \land ac'\neq\emptyset)^f
	&&\ptext{Definition of design}\\
	&=((ok \land \lnot \mathbf{PBMH} (P^f)) \implies (\mathbf{PBMH} (P^t) \land ac'\neq\emptyset \land ok'))^f
	&&\ptext{Substitution}\\
	&=((ok \land \lnot \mathbf{PBMH} (P^f)) \implies (\mathbf{PBMH} (P^t) \land ac'\neq\emptyset \land false))
	&&\ptext{Predicate calculus}\\
	&=\lnot ok \lor \mathbf{PBMH} (P^f)
	&&\ptext{\cref{lemma:PBMH(P)-ow:PBMH(P-ow)}}\\
	&=\lnot ok \lor \mathbf{PBMH} (P)^f
	&&\ptext{Assumption: $P$ is $\mathbf{A}$-healthy}\\
	&=\lnot ok \lor P^f
	&&\ptext{Predicate calculus}\\
	&=ok \implies P^f
\end{xflalign*}
\end{proof}\end{proofs}
\end{lemma}

\begin{lemma}\label{lemma:A:P-t} Provided $P$ is an $\mathbf{A}$-healthy design,
\begin{align*}
	&P^t = ((ok \land \lnot P^f) \implies (P^t \land ac'\neq\emptyset))
\end{align*}
\begin{proofs}\begin{proof}\checkt{alcc}\checkt{pfr}
\begin{xflalign*}
	&P^t
	&&\ptext{Assumption: $P$ is an $\mathbf{A}$-healthy design}\\
	&=(\mathbf{A} (\lnot P^f \vdash P^t))^t
	&&\ptext{Definition of $\mathbf{A}$}\\
	&=(\lnot \mathbf{PBMH} (P^f) \vdash \mathbf{PBMH} (P^t) \land ac'\neq\emptyset)^t
	&&\ptext{Definition of design}\\
	&=((ok \land \lnot \mathbf{PBMH} (P^f)) \implies (\mathbf{PBMH} (P^t) \land ac'\neq\emptyset \land ok'))^t
	&&\ptext{Substitution}\\
	&=((ok \land \lnot \mathbf{PBMH} (P^f)) \implies (\mathbf{PBMH} (P^t) \land ac'\neq\emptyset \land true))
	&&\ptext{Predicate calculus}\\
	&=((ok \land \lnot \mathbf{PBMH} (P^f)) \implies (\mathbf{PBMH} (P^t) \land ac'\neq\emptyset))
	&&\ptext{\cref{lemma:PBMH(P)-ow:PBMH(P-ow)}}\\
	&=((ok \land \lnot \mathbf{PBMH} (P)^f) \implies (\mathbf{PBMH} (P)^t \land ac'\neq\emptyset))
	&&\ptext{Assumption: $P$ is $\mathbf{A}$-healthy}\\
	&=((ok \land \lnot P^f) \implies (P^t \land ac'\neq\emptyset))
\end{xflalign*}
\end{proof}\end{proofs}
\end{lemma}

\begin{lemma}\label{lemma:A(exists-ac'-Pf|-Pt):(exists-ac'-Pf|-Pt)} Provided $P$ is an $\mathbf{A}$-healthy design,
\begin{align*}
	&(\lnot \exists ac' \spot \mathbf{PBMH} (P^f) \vdash \mathbf{PBMH} (P^t) \land ac'\neq\emptyset) \\
	&=\\
	&(\lnot \exists ac' \spot P^f \vdash P^t)
\end{align*}
\begin{proofs}\begin{proof}\checkt{alcc}\checkt{pfr}
\begin{xflalign*}
	&(\lnot \exists ac' \spot \mathbf{PBMH} (P^f) \vdash \mathbf{PBMH} (P^t) \land ac'\neq\emptyset)
	&&\ptext{\cref{lemma:PBMH(P)-ow:PBMH(P-ow)}}\\
	&=(\lnot \exists ac' \spot \mathbf{PBMH} (P)^f \vdash \mathbf{PBMH} (P)^t \land ac'\neq\emptyset)
	&&\ptext{Assumption: $P$ is $\mathbf{A}$-healthy (and hence $\mathbf{PBMH}$-healthy)}\\
	&=(\lnot \exists ac' \spot P^f \vdash P^t \land ac'\neq\emptyset)
	&&\ptext{Assumption: $P$ is $\mathbf{A}$-healthy and \cref{lemma:A:P-t,lemma:A:P-f}}\\
	&=\left(\begin{array}{l} 
		\lnot \exists ac' \spot ok \implies P^f
		\\ \vdash \\
		((ok \land \lnot P^f) \implies (P^t \land ac'\neq\emptyset)) \land ac'\neq\emptyset
	\end{array}\right)
	&&\ptext{Predicate calculus}\\
	&=\left(\begin{array}{l} 
		\lnot (ok \implies \exists ac' \spot P^f)
		\\ \vdash \\
		(\lnot ok \lor P^f \lor (P^t \land ac'\neq\emptyset)) \land ac'\neq\emptyset
	\end{array}\right)
	&&\ptext{Predicate calculus}\\
	&=\left(\begin{array}{l} 
		ok \land \lnot \exists ac' \spot P^f
		\\ \vdash \\
		(\lnot ok \land ac'\neq\emptyset) \lor (P^f \land ac'\neq\emptyset) \lor (P^t \land ac'\neq\emptyset) 
	\end{array}\right)
	&&\ptext{Definition of design and predicate calculus}\\
	&=\left(\begin{array}{l} 
		\lnot \exists ac' \spot P^f
		\\ \vdash \\
		(\lnot ok \land ac'\neq\emptyset) \lor (P^f \land (\exists ac' \spot P^f) \land ac'\neq\emptyset) \lor (P^t \land ac'\neq\emptyset) 
	\end{array}\right)
	&&\ptext{Definition of design and predicate calculus}\\
	&=\left(\begin{array}{l} 
		\lnot \exists ac' \spot P^f
		\\ \vdash \\
		\left(\begin{array}{l}
			 (ok \land \lnot \exists ac' \spot P^f)
			 \\ \land \\
			((\lnot ok \land ac'\neq\emptyset) \lor (P^f \land (\exists ac' \spot P^f) \land ac'\neq\emptyset) \lor (P^t \land ac'\neq\emptyset))
		\end{array}\right) 
	\end{array}\right)
	&&\ptext{Predicate calculus}\\
	&=\left(\begin{array}{l} 
		\lnot \exists ac' \spot P^f
		\\ \vdash \\
			 (ok \land \lnot \exists ac' \spot P^f)
			  \land 
			((\lnot ok) \lor (P^f) \lor (P^t \land ac'\neq\emptyset)) 
	\end{array}\right)
	&&\ptext{Definition of design and predicate calculus}\\
	&=\left(\begin{array}{l} 
		\lnot \exists ac' \spot P^f
		 \vdash 
		(ok \land \lnot P^f) \implies (P^t \land ac'\neq\emptyset) 
	\end{array}\right)
	&&\ptext{Assumption: $P$ is $\mathbf{A}$-healthy and~\cref{lemma:A:P-t}}\\
	&=(\lnot \exists ac' \spot P^f \vdash P^t)
\end{xflalign*}
\end{proof}\end{proofs}
\end{lemma}

\begin{theorem}\label{theorem:P-seqDac-II} Provided $P$ is an $\mathbf{A}$-healthy design,
\begin{align*}
	&\mathbf{H3_{\mathcal{D}ac}} (P) = (\lnot \exists ac' \spot P^f \vdash P^t)
\end{align*}
\begin{proofs}\begin{proof}\checkt{alcc}\checkt{pfr}
\begin{xflalign*}
	&\mathbf{H3_{\mathcal{D}ac}} (P)
	&&\ptext{Definition of $\mathbf{H3}$ for angelic designs}\\
	&=P \seqDac (true \vdash s \in ac')
	&&\ptext{Assumption: $P$ is an $\mathbf{A}$-healthy design}\\
	&=\mathbf{A} (\lnot P^f \vdash P^t) \seqDac (true \vdash s \in ac')
	&&\ptext{Definition of $\mathbf{A}$}\\
	&=(\lnot \mathbf{PBMH} (P^f) \vdash \mathbf{PBMH} (P^t) \land ac'\neq\emptyset) \seqDac (true \vdash s \in ac')
	&&\ptext{Sequential composition for $\mathbf{A}$-designs}\\
	&=\left(\begin{array}{l}
		\lnot (\mathbf{PBMH} (P^f) \seqA true) \land \lnot ((\mathbf{PBMH} (P^t) \land ac'\neq\emptyset) \seqA \lnot true)
		\\ \vdash \\
		(\mathbf{PBMH} (P^t) \land ac'\neq\emptyset) \seqA (true \implies s \in ac')
	\end{array}\right)
	&&\ptext{Predicate calculus}\\
	&=\left(\begin{array}{l}
		\lnot (\mathbf{PBMH} (P^f) \seqA true) \land \lnot ((\mathbf{PBMH} (P^t) \land ac'\neq\emptyset) \seqA false)
		\\ \vdash \\
		(\mathbf{PBMH} (P^t) \land ac'\neq\emptyset) \seqA s \in ac'
	\end{array}\right)
	&&\ptext{\cref{law:seqA-right-distributivity-conjunction}}\\
	&=\left(\begin{array}{l}
		\lnot (\mathbf{PBMH} (P^f) \seqA true) \land \lnot ((\mathbf{PBMH} (P^t)\seqA false) \land (ac'\neq\emptyset\seqA false))
		\\ \vdash \\
		(\mathbf{PBMH} (P^t) \land ac'\neq\emptyset) \seqA s \in ac'
	\end{array}\right)
	&&\ptext{Definition of $\seqA$ and substitution}\\
	&=\left(\begin{array}{l}
		\lnot (\mathbf{PBMH} (P^f) \seqA true) \land \lnot ((\mathbf{PBMH} (P^t)\seqA false) \land (\emptyset\neq\emptyset))
		\\ \vdash \\
		(\mathbf{PBMH} (P^t) \land ac'\neq\emptyset) \seqA s \in ac'
	\end{array}\right)
	&&\ptext{Predicate calculus}\\
	&=\left(\begin{array}{l}
		\lnot (\mathbf{PBMH} (P^f) \seqA true)
		\\ \vdash \\
		(\mathbf{PBMH} (P^t) \land ac'\neq\emptyset) \seqA s \in ac'
	\end{array}\right)
	&&\ptext{\cref{law:seqA:IIA:right-unit}}\\
	&=(\lnot (\mathbf{PBMH} (P^f) \seqA true) \vdash \mathbf{PBMH} (P^t) \land ac'\neq\emptyset)
	&&\ptext{\cref{lemma:PBMH(P)-seqA-true:exists-ac'-P}}\\
	&=(\lnot \exists ac' \spot \mathbf{PBMH} (P^f) \vdash \mathbf{PBMH} (P^t) \land ac'\neq\emptyset)
	&&\ptext{\cref{lemma:A(exists-ac'-Pf|-Pt):(exists-ac'-Pf|-Pt)}}\\
	&=(\lnot \exists ac' \spot P^f \vdash P^t)
\end{xflalign*}
\end{proof}\end{proofs}
\end{theorem}

\subsection{$\mathbf{A2}$}

\begin{theorem}\label{lemma:A2:alternative-2:disjunction}
\begin{statement}$\mathbf{A2} (P) = P[\emptyset/ac'] \lor (\exists y \spot P[\{y\}/ac'] \land y \in ac')$\end{statement}
\begin{proofs}
\begin{proof}\checkt{alcc}
\begin{flalign*}
	&\mathbf{A2} (P)
	&&\ptext{Predicate calculus}\\
	&=\mathbf{A2} (P \land (ac'=\emptyset \lor ac'\neq\emptyset))
	&&\ptext{Predicate calculus}\\
	&=\mathbf{A2} ((P \land ac'=\emptyset) \lor (P \land ac'\neq\emptyset))
	&&\ptext{\cref{theorem:A2(P-lor-Q):A2(P)-lor-A2(Q)}}\\
	&=\mathbf{A2} (P \land ac'=\emptyset) \lor \mathbf{A2} (P \land ac'\neq\emptyset)
	&&\ptext{\cref{lemma:A2(P-land-ac'-neq-emptyset),lemma:A2(P-land-ac'-emptyset)}}\\
	&=P[\emptyset/ac'] \lor (\exists z \spot P[\{z\}/ac'] \land z \in ac')
\end{flalign*}
\end{proof}
\end{proofs}
\end{theorem}\noindent

\begin{theorem}\label{theorem:A2-o-A2(P):A2(P)}
\begin{statement}$\mathbf{A2} \circ \mathbf{A2} (P) = \mathbf{A2} (P)$\end{statement}
\begin{proofs}
\begin{proof}
\begin{xflalign*}
	&\mathbf{A2} \circ \mathbf{A2} (P)
	&&\ptext{Definition of $\mathbf{A2}$ (\cref{lemma:A2:alternative-2:disjunction})}\\
	&=\mathbf{A2} (P)[\emptyset/ac'] \lor (\exists y \spot \mathbf{A2} (P)[\{y\}/ac'] \land y \in ac')
	&&\ptext{Definition of $\mathbf{A2}$ (\cref{lemma:A2:alternative-2:disjunction})}\\
	&=\left(\begin{array}{l}
		(P[\emptyset/ac'] \lor (\exists y \spot P[\{y\}/ac'] \land y \in ac'))[\emptyset/ac']
		\\ \lor \\
		(\exists y \spot (P[\emptyset/ac'] \lor (\exists y \spot P[\{y\}/ac'] \land y \in ac'))[\{y\}/ac'] \land y \in ac')
	\end{array}\right)
	&&\ptext{Variable renaming}\\
	&=\left(\begin{array}{l}
		(P[\emptyset/ac'] \lor (\exists y \spot P[\{y\}/ac'] \land y \in ac'))[\emptyset/ac']
		\\ \lor \\
		(\exists y \spot (P[\emptyset/ac'] \lor (\exists z \spot P[\{z\}/ac'] \land z \in ac'))[\{y\}/ac'] \land y \in ac')
	\end{array}\right)
	&&\ptext{Substitution}\\
	&=\left(\begin{array}{l}
		(P[\emptyset/ac'] \lor (\exists y \spot P[\{y\}/ac'] \land y \in \emptyset))
		\\ \lor \\
		(\exists y \spot (P[\emptyset/ac'] \lor (\exists z \spot P[\{z\}/ac'] \land z \in \{y\})) \land y \in ac')
	\end{array}\right)
	&&\ptext{Property of sets and predicate calculus}\\
	&=P[\emptyset/ac'] \lor (\exists y \spot (P[\emptyset/ac'] \lor (\exists z \spot P[\{z\}/ac'] \land z = y)) \land y \in ac')
	&&\ptext{One-point rule}\\
	&=P[\emptyset/ac'] \lor (\exists y \spot (P[\emptyset/ac'] \lor P[\{y\}/ac']) \land y \in ac')
	&&\ptext{Predicate calculus}\\
	&=P[\emptyset/ac'] \lor (\exists y \spot P[\emptyset/ac'] \land y \in ac') \lor	(\exists y \spot P[\{y\}/ac'] \land y \in ac')
	&&\ptext{Predicate calculus: $y$ not free in $P$}\\
	&=P[\emptyset/ac'] \lor (P[\emptyset/ac'] \land \exists y \spot y \in ac') \lor	(\exists y \spot P[\{y\}/ac'] \land y \in ac')
	&&\ptext{Predicate calculus: absorption law}\\
	&=P[\emptyset/ac'] \lor (\exists y \spot P[\{y\}/ac'] \land y \in ac')
	&&\ptext{Definition of $\mathbf{A2}$ (\cref{lemma:A2:alternative-2:disjunction})}\\
	&=\mathbf{A2} (P)
\end{xflalign*}
\end{proof}
\end{proofs}
\end{theorem}

\begin{theorem}\label{theorem:A2:monotonic}
\begin{statement}$P \sqsubseteq Q \implies \mathbf{A2} (P) \sqsubseteq \mathbf{A2} (Q)$\end{statement}
\begin{proofs}
\begin{proof}
\begin{xflalign*}
	&\mathbf{A2} (Q)
	&&\ptext{Definition of $\mathbf{A2}$}\\
	&=\mathbf{PBMH} (Q \seqA \{s\} = ac')
	&&\ptext{Assumption: $P \sqsubseteq Q = [Q \implies P]$}\\
	&=\mathbf{PBMH} ((P \land Q) \seqA \{s\} = ac')
	&&\ptext{Distributivity of $\seqA$}\\
	&=\mathbf{PBMH} ((P \seqA \{s\} = ac') \land (Q \seqA \{s\} = ac'))
	&&\ptext{Definition of $\mathbf{PBMH}$ (\cref{lemma:PBMH:alternative-1})}\\
	&=\exists ac_0 \spot ((P \seqA \{s\} = ac') \land (Q \seqA \{s\} = ac'))[ac_0/ac'] \land ac_0 \subseteq ac'
	&&\ptext{Substitution}\\
	&=\exists ac_0 \spot ((P \seqA \{s\} = ac')[ac_0/ac'] \land (Q \seqA \{s\} = ac')[ac_0/ac']) \land ac_0 \subseteq ac'
	&&\ptext{Predicate calculus}\\
	&\sqsupseteq \exists ac_0 \spot (P \seqA \{s\} = ac')[ac_0/ac'] \land ac_0 \subseteq ac'
	&&\ptext{Definition of $\mathbf{PBMH}$ (\cref{lemma:PBMH:alternative-1})}\\
	&=\mathbf{PBMH} (P \seqA \{s\} = ac')
	&&\ptext{Definition of $\mathbf{A2}$}\\
	&=\mathbf{A2} (P)
\end{xflalign*}
\end{proof}
\end{proofs}
\end{theorem}

\begin{theorem}\label{theorem:A2(P-lor-Q):A2(P)-lor-A2(Q)}
\begin{statement}$\mathbf{A2} (P \lor Q) = \mathbf{A2} (P) \lor \mathbf{A2} (Q)$\end{statement}
\begin{proofs}
\begin{proof}\checkt{alcc}\checkt{pfr}
\begin{xflalign*}
	&\mathbf{A2} (P \lor Q)
	&&\ptext{Definition of $\mathbf{A2}$}\\
	&=\mathbf{PBMH} ((P \lor Q) \seqA \{s\} = ac'\})
	&&\ptext{Distributivity of $\seqA$ (\cref{law:seqA-right-distributivity})}\\
	&=\mathbf{PBMH} ((P \seqA \{s\} = ac'\}) \lor (Q \seqA \{s\} = ac'\}))
	&&\ptext{Distributivity of $\mathbf{PBMH}$}\\
	&=\mathbf{PBMH} (P \seqA \{s\} = ac'\}) \lor  \mathbf{PBMH} (Q \seqA \{s\} = ac'\})
	&&\ptext{Definition of $\mathbf{A2}$}\\
	&=\mathbf{A2} (P) \lor \mathbf{A2} (Q)
\end{xflalign*}
\end{proof}
\end{proofs}
\end{theorem}

\begin{theorem}[$\mathbf{A2}$-idempotent] Provided $P$ is $\mathbf{PBMH}$-healthy,
\begin{align*}
	&\mathbf{A2} \circ \mathbf{A2} (P) = \mathbf{A2} (P) 
\end{align*}
\begin{proofs}
\begin{proof}
\begin{flalign*}
	&\mathbf{A2} \circ \mathbf{A2} (P)
	&&\ptext{Definition of $\mathbf{A2}$ twice}\\
	&=\mathbf{PBMH} (\mathbf{PBMH} (P \seqA \{ s | \{s\} = ac'\}) \seqA \{ s | \{s\} = ac'\})
	&&\ptext{$P$ is $\mathbf{PBMH}$-healthy and~\cref{lemma:PBMH(P-seqA-s-ac')-seqA-s-ac':P-seqA-s-ac'}}\\
	&=\mathbf{PBMH} (P \seqA \{ s | \{s\} = ac' \})
	&&\ptext{Definition of $\mathbf{A2}$}\\
	&=\mathbf{A2} (P)
\end{flalign*}
\end{proof}
\end{proofs}
\end{theorem}

\subsubsection{Lemmas}

\begin{lemma}\label{lemma:A2-o-H1-o-H2}
\begin{statement}
$\mathbf{A2} (P \vdash Q) = (\lnot \mathbf{A2} (\lnot P) \vdash \mathbf{A2} (Q))$
\end{statement}
\begin{proofs}
\begin{proof}\checkt{alcc}
\begin{xflalign*}
	&\mathbf{A2} \circ \mathbf{A} (P \vdash Q)
	&&\ptext{Definition of design}\\
	&=\mathbf{A2} ((ok \land P) \implies (Q \land ok'))
	&&\ptext{Predicate calculus}\\
	&=\mathbf{A2} (\lnot ok \lor \lnot P \lor (Q \land ok'))
	&&\ptext{Distributivity of $\mathbf{A2}$ (\cref{theorem:A2(P-lor-Q):A2(P)-lor-A2(Q)})}\\
	&=\mathbf{A2} (\lnot ok) \lor \mathbf{A2} (\lnot P) \lor \mathbf{A2} (Q \land ok')
	&&\ptext{\cref{lemma:A2(P-land-Q)-ac'-not-free:P-land-A2(Q),lemma:A2(P)-ac'-not-free:P}}\\
	&=\lnot ok \lor \mathbf{A2} (\lnot P) \lor (\mathbf{A2} (Q) \land ok')
	&&\ptext{Predicate calculus}\\
	&=(ok \land \lnot \mathbf{A2} (\lnot P)) \implies (\mathbf{A2} (Q) \land ok')
	&&\ptext{Definition of design}\\
	&=(\lnot \mathbf{A2} (\lnot P) \vdash \mathbf{A2} (Q))
\end{xflalign*}
\end{proof}
\end{proofs}
\end{lemma}

\begin{lemma}\label{lemma:A2:alternative-1}
$\mathbf{A2} (P) = \exists ac_0 \spot P[\{s | \{s\}=ac_0\}/ac'] \land ac_0\subseteq ac'$
\begin{proofs}
\begin{proof}\checkt{alcc}
\begin{xflalign*}
	&\mathbf{A2} (P)
	&&\ptext{Definition of $\mathbf{A2}$}\\
	&=\mathbf{PBMH} (P \seqA \{s\} = ac')
	&&\ptext{Definition of $\seqA$ and substitution}\\
	&=\mathbf{PBMH} (P[\{ s | \{s\} = ac'\}/ac'])
	&&\ptext{Definition of $\mathbf{PBMH}$ (\cref{lemma:PBMH:alternative-1})}\\
	&=\exists ac_0 \spot P[\{ s | \{s\} = ac'\}/ac'][ac_0/ac'] \land ac_0 \subseteq ac'
	&&\ptext{Property of substitution}\\
	&=\exists ac_0 \spot P[\{ s | \{s\} = ac_0\}/ac'] \land ac_0 \subseteq ac'
\end{xflalign*}
\end{proof}
\end{proofs}
\end{lemma}

\begin{lemma}\label{lemma:A2-o-A}
\begin{statement}
\begin{align*}
	&\mathbf{A2} \circ \mathbf{A} (\lnot P^f \vdash P^t) \\
	&=\\
	&(\lnot \mathbf{A2} \circ \mathbf{PBMH} (P^f) \vdash \mathbf{A2} (\mathbf{PBMH} (P^t) \land ac'\neq\emptyset))
\end{align*}
\end{statement}
\begin{proofs}
\begin{proof}
\begin{xflalign*}
	&\mathbf{A2} \circ \mathbf{A} (\lnot P^f \vdash P^t)
	&&\ptext{Definition of $\mathbf{A}$}\\
	&=\mathbf{A2} (\lnot \mathbf{PBMH} (P^f) \vdash \mathbf{PBMH} (P^t) \land ac'\neq\emptyset)
	&&\ptext{Definition of design}\\
	&=\mathbf{A2} ((ok \land \lnot \mathbf{PBMH} (P^f)) \implies (\mathbf{PBMH} (P^t) \land ac'\neq\emptyset \land ok'))
	&&\ptext{Predicate calculus}\\
	&=\mathbf{A2} (\lnot ok \lor \mathbf{PBMH} (P^f) \lor (\mathbf{PBMH} (P^t) \land ac'\neq\emptyset \land ok'))
	&&\ptext{Distributivity of $\mathbf{A2}$ (\cref{theorem:A2(P-lor-Q):A2(P)-lor-A2(Q)})}\\
	&=\mathbf{A2} (\lnot ok) \lor \mathbf{A2} \circ \mathbf{PBMH} (P^f) \lor \mathbf{A2} (\mathbf{PBMH} (P^t) \land ac'\neq\emptyset \land ok')
	&&\ptext{\cref{lemma:A2(P-land-Q)-ac'-not-free:P-land-A2(Q),lemma:A2(P)-ac'-not-free:P}}\\
	&=\lnot ok \lor \mathbf{A2} \circ \mathbf{PBMH} (P^f) \lor (\mathbf{A2} (\mathbf{PBMH} (P^t) \land ac'\neq\emptyset) \land ok')
	&&\ptext{Predicate calculus}\\
	&=(ok \land \lnot \mathbf{A2} \circ \mathbf{PBMH} (P^f)) \implies (\mathbf{A2} (\mathbf{PBMH} (P^t) \land ac'\neq\emptyset) \land ok')
	&&\ptext{Definition of design}\\
	&=(\lnot \mathbf{A2} \circ \mathbf{PBMH} (P^f) \vdash \mathbf{A2} (\mathbf{PBMH} (P^t) \land ac'\neq\emptyset))
\end{xflalign*}
\end{proof}
\end{proofs}
\end{lemma}

\begin{lemma}\label{lemma:A2(false):false}
$\mathbf{A2} (false) = false$
\begin{proofs}
\begin{proof}
\begin{flalign*}
	&\mathbf{A2} (false)
	&&\ptext{Definition of $\mathbf{A2}$}\\
	&=\mathbf{PBMH} (false \seqA \{s\} = ac')
	&&\ptext{Property of $\seqA$}\\
	&=\mathbf{PBMH} (false)
	&&\ptext{Property of $\mathbf{PBMH}$}\\
	&=false
\end{flalign*}
\end{proof}
\end{proofs}
\end{lemma}

\begin{lemma}\label{lemma:A2(true):true}
$\mathbf{A2} (true) = true$
\begin{proofs}
\begin{proof}
\begin{flalign*}
	&\mathbf{A2} (true)
	&&\ptext{Definition of $\mathbf{A2}$}\\
	&=\mathbf{PBMH} (true \seqA \{s\} = ac')
	&&\ptext{Property of $\seqA$}\\
	&=\mathbf{PBMH} (true)
	&&\ptext{Property of $\mathbf{PBMH}$}\\
	&=true
\end{flalign*}
\end{proof}
\end{proofs}
\end{lemma}

\begin{lemma}\label{lemma:A2(exists-y-in-ac'-land-P)} Provided $ac'$ is not free in $P$,
\begin{align*}
	&\mathbf{A2} (\exists y \spot y \in ac' \land P) = \exists y \spot y \in ac' \land P
\end{align*}
\begin{proofs}
\begin{proof}
\begin{xflalign*}
	&\mathbf{A2} (\exists y \spot y \in ac' \land P)
	&&\ptext{Definition of $\mathbf{A2}$}\\
	&=\mathbf{PBMH} ((\exists y \spot y \in ac' \land P) \seqA \{s\} = ac')
	&&\ptext{Definition of $\seqA$ and substitution: $ac'$ not free in $P$}\\
	&=\mathbf{PBMH} (\exists y \spot y \in \{ s | \{s\} = ac'\} \land P)
	&&\ptext{Property of sets}\\
	&=\mathbf{PBMH} (\exists y \spot \{y\}=ac' \land P)
	&&\ptext{Definition of $\mathbf{PBMH}$ (\cref{lemma:PBMH:alternative-1})}\\
	&=\exists ac_0 \spot (\exists y \spot \{y\}=ac' \land P)[ac_0/ac'] \land ac_0 \subseteq ac'
	&&\ptext{Substitution: $ac'$ not free in $P$}\\
	&=\exists ac_0, y \spot \{y\}=ac_0 \land P \land ac_0 \subseteq ac'
	&&\ptext{Predicate calculus}\\
	&=\exists y \spot \{y\} \subseteq ac' \land P 
	&&\ptext{Property of sets}\\
	&=\exists y \spot y \in ac' \land P
\end{xflalign*}
\end{proof}
\end{proofs}
\end{lemma}

\subsubsection{Properties}

\begin{lemma}\label{lemma:A2(P-land-Q)-ac'-not-free:P-land-A2(Q)} Provided $ac'$ is not free in $P$,
\begin{statement}
$\mathbf{A2} (P \land Q) = P \land \mathbf{A2} (Q)$.
\end{statement}
\begin{proofs}
\begin{proof}\checkt{alcc}
\begin{xflalign*}
	&\mathbf{A2} (P \land Q)
	&&\ptext{Definition of $\mathbf{A2}$}\\
	&=\mathbf{PBMH} ((P \land Q) \seqA \{s\}=ac')
	&&\ptext{Distributivity of $\seqA$ (\cref{law:seqA-right-distributivity-conjunction})}\\
	&=\mathbf{PBMH} ((P\seqA \{s\}=ac') \land (Q\seqA \{s\}=ac'))
	&&\ptext{Assumption: $ac'$ not free in $P$}\\
	&=\mathbf{PBMH} (P \land (Q\seqA \{s\}=ac'))
	&&\ptext{Assumption: $ac'$ not free in $P$ and \cref{lemma:PBMH(c-land-P):c-land-PBMH(P)}}\\
	&=P \land \mathbf{PBMH} (Q\seqA \{s\}=ac')
	&&\ptext{Definition of $\mathbf{A2}$}\\
	&=P \land \mathbf{A2} (Q)
\end{xflalign*}
\end{proof}
\end{proofs}
\end{lemma}

\begin{lemma}\label{lemma:A2(P)-ac'-not-free:P} Provided $ac'$ not free in $P$,
$\mathbf{A2} (P) = P$.
\begin{proofs}
\begin{proof}\checkt{alcc}
\begin{xflalign*}
	&\mathbf{A2} (P)
	&&\ptext{Definition of $\mathbf{A2}$}\\
	&=\mathbf{PBMH} (P \seqA \{s\}=ac')
	&&\ptext{Assumption: $ac'$ not free in $P$}\\
	&=\mathbf{PBMH} (P)	
	&&\ptext{Assumption: $ac'$ not free in $P$ and~\cref{law:pbmh:P:ac'-not-free}}\\
	&=P	
\end{xflalign*}
\end{proof}
\end{proofs}
\end{lemma}

\begin{lemma}\label{lemma:A2(P-land-ac'-neq-emptyset)}
$\mathbf{A2} (P \land ac'\neq\emptyset) = \exists z \spot P[\{z\}/ac'] \land z \in ac'$
\begin{proofs}
\begin{proof}\checkt{alcc}
\begin{xflalign*}
	&\mathbf{A2} (P \land ac'\neq\emptyset)
	&&\ptext{Definition of $\mathbf{A2}$ (\cref{lemma:A2:alternative-1})}\\
	&=\exists ac_0 \spot ((P \land ac'\neq\emptyset)[\{s | \{s\} = ac_0\}/ac'] \land ac_0 \subseteq ac')
	&&\ptext{Substitution}\\
	&=\exists ac_0 \spot P[\{s | \{s\} = ac_0\}/ac'] \land \{s | \{s\} = ac_0\}\neq\emptyset \land ac_0 \subseteq ac')
	&&\ptext{Property of sets}\\
	&=\exists ac_0 \spot P[\{s | \{s\} = ac_0\}/ac'] \land \exists z \spot \{z\} = ac_0 \land ac_0 \subseteq ac'
	&&\ptext{Predicate calculus}\\
	&=\exists ac_0, z \spot P[\{s | \{s\} = ac_0\}/ac'] \land \{z\} = ac_0 \land ac_0 \subseteq ac'
	&&\ptext{One-point rule}\\
	&=\exists z \spot P[\{s | \{s\} = \{z\}\}/ac'] \land \{z\} \subseteq ac'
	&&\ptext{Property of sets}\\
	&=\exists z \spot P[\{z\}/ac'] \land z \in ac'
\end{xflalign*}
\end{proof}
\end{proofs}
\end{lemma}

\begin{lemma}\label{lemma:A2(P-land-ac'-emptyset)}
$\mathbf{A2} (P \land ac'=\emptyset) = P[\emptyset/ac']$
\begin{proofs}
\begin{proof}\checkt{alcc}
\begin{xflalign*}
	&\mathbf{A2} (P \land ac'=\emptyset)
	&&\ptext{Definition of $\mathbf{A2}$ (\cref{lemma:A2:alternative-1})}\\
	&=\exists ac_0 \spot ((P \land ac'=\emptyset)[\{s | \{s\} = ac_0\}/ac'] \land ac_0 \subseteq ac')
	&&\ptext{Substitution}\\
	&=\exists ac_0 \spot P[\{s | \{s\} = ac_0\}/ac'] \land \{s | \{s\} = ac_0\}=\emptyset \land ac_0 \subseteq ac')
	&&\ptext{Property of sets}\\
	&=\exists ac_0 \spot P[\{s | \{s\} = ac_0\}/ac'] \land \lnot (\exists z \spot \{z\} = ac_0) \land ac_0 \subseteq ac'
	&&\ptext{Predicate calculus}\\
	&=\exists ac_0 \spot P[\{s | \{s\} = ac_0\}/ac'] \land (\forall z \spot \{z\} \neq ac_0) \land ac_0 \subseteq ac'
	&&\ptext{Predicate calculus}\\
	&=\exists ac_0 \spot P[\{s | false\}/ac'] \land (\forall z \spot \{z\} \neq ac_0) \land ac_0 \subseteq ac'
	&&\ptext{Property of sets}\\
	&=\exists ac_0 \spot P[\emptyset/ac'] \land (\forall z \spot \{z\} \neq ac_0) \land ac_0 \subseteq ac'
	&&\ptext{Predicate calculus}\\
	&=P[\emptyset/ac'] \land \exists ac_0 \spot (\forall z \spot \{z\} \neq ac_0) \land ac_0 \subseteq ac'
	&&\ptext{Predicate calculus}\\
	&=P[\emptyset/ac']
\end{xflalign*}
\end{proof}
\end{proofs}
\end{lemma}

\begin{lemma}\label{lemma:A2(P)-emptyset-ac':P-emptyset-ac'}
$\mathbf{A2} (P)[\emptyset/ac'] = P[\emptyset/ac']$
\begin{proofs}
\begin{proof}\checkt{alcc}
\begin{xflalign*}
	&\mathbf{A2} (P)[\emptyset/ac']
	&&\ptext{Definition of $\mathbf{A2}$ (\cref{lemma:A2:alternative-1})}\\
	&=(\exists ac_0 \spot P[\{s | \{s\}=ac_0\}/ac'] \land ac_0 \subseteq ac')[\emptyset/ac']
	&&\ptext{Substitution}\\
	&=\exists ac_0 \spot P[\{s | \{s\}=ac_0\}/ac'] \land ac_0 \subseteq \emptyset
	&&\ptext{Property of sets}\\
	&=\exists ac_0 \spot P[\{s | \{s\}=ac_0\}/ac'] \land ac_0 = \emptyset
	&&\ptext{One-point rule}\\
	&=P[\{s | \{s\}=\emptyset\}/ac']
	&&\ptext{Property of sets}\\
	&=P[\emptyset/ac']
\end{xflalign*}
\end{proof}
\end{proofs}
\end{lemma}

\begin{lemma}\label{lemma:A2(P-cond-Q):A2(P)-cond-A2(Q)}
\begin{statement}
Provided $ac'$ is not free in $c$,
\begin{align*}
	&\mathbf{A2} (P \dres c \rres Q) = \mathbf{A2} (P) \dres c \rres \mathbf{A2} (Q)
\end{align*}
\end{statement}
\begin{proofs}
\begin{proof}\checkt{alcc}
\begin{xflalign*}
	&\mathbf{A2} (P \dres c \rres Q)
	&&\ptext{Definition of conditional}\\
	&=\mathbf{A2} ((c \land P) \lor (\lnot c \land Q))
	&&\ptext{\cref{theorem:A2(P-lor-Q):A2(P)-lor-A2(Q)}}\\
	&=\mathbf{A2} (c \land P) \lor \mathbf{A2} (\lnot c \land Q)
	&&\ptext{Assumption: $ac'$ is not free in $c$ and~\cref{lemma:A2(P-land-Q)-ac'-not-free:P-land-A2(Q)}}\\
	&=(c \land \mathbf{A2} (P)) \lor (\lnot c \land \mathbf{A2} (Q))
	&&\ptext{Definition of conditional}\\
	&=\mathbf{A2} (P) \dres c \rres \mathbf{A2} (Q)
\end{xflalign*}
\end{proof}
\end{proofs}
\end{lemma}

\begin{lemma}\label{lemma:A2(x-in-ac'):x-in-ac'}
\begin{statement}
$\mathbf{A2} (x \in ac') = x \in ac'$
\end{statement}
\begin{proofs}
\begin{proof}\checkt{alcc}
\begin{xflalign*}
	&\mathbf{A2} (x \in ac')
	&&\ptext{Definition of $\mathbf{A2}$ (\cref{lemma:A2:alternative-2:disjunction})}\\
	&=(x \in ac')[\emptyset/ac'] \lor (\exists y @ (x \in ac')[\{y\}/ac'] \land y \in ac')
	&&\ptext{Substitution}\\
	&=(x \in \emptyset) \lor (\exists y @ x \in \{y\} \land y \in ac')
	&&\ptext{Property of sets and predicate calculus}\\
	&=\exists y @ x = y \land y \in ac'
	&&\ptext{One-point rule}\\
	&=x \in ac'
\end{xflalign*}
\end{proof}
\end{proofs}
\end{lemma}

\begin{lemma}\label{lemma:A2(P)-o-w:A2(P-o-w)}
\begin{statement}
$ \mathbf{A2} (P)^o_w = \mathbf{A2} (P^o_w)$
\end{statement}
\begin{proofs}
\begin{proof}\checkt{alcc}
\begin{xflalign*}
	&\mathbf{A2} (P)^o_w
	&&\ptext{Definition of $\mathbf{A2}$}\\
	&=(\mathbf{PBMH} (P \seqA \{s\} = ac'))^o_w
	&&\ptext{\cref{lemma:PBMH(P)-ow:PBMH(P-ow)}}\\
	&=\mathbf{PBMH} ((P \seqA \{s\} = ac')^o_w)
	&&\ptext{Definition of $\seqA$}\\
	&=\mathbf{PBMH} ((P[\{s | \{s\} = ac'\}/ac'])^o_w)
	&&\ptext{Definition of $^o_w$}\\
	&=\mathbf{PBMH} ((P[\{s | \{s\} = ac'\}/ac'])[s\oplus\{wait \mapsto w\},ok/o,s])
	&&\ptext{Substitution}\\
	&=\mathbf{PBMH} (P[s\oplus\{wait \mapsto w\},ok/o,s][\{s | \{s\} = ac'\}/ac'])
	&&\ptext{Definition of $^o_w$}\\
	&=\mathbf{PBMH} (P^o_w[\{s | \{s\} = ac'\}/ac'])
	&&\ptext{Definition of $\seqA$}\\
	&=\mathbf{PBMH} (P^o_w \seqA \{s\} = ac'\})
	&&\ptext{Definition of $\mathbf{A2}$}\\
	&=\mathbf{A2} (P^o_w)
\end{xflalign*}
\end{proof}
\end{proofs}
\end{lemma}

\begin{lemma}\label{lemma:A2(P)-o-ok:A2(P-o-ok)}
\begin{statement} Provided $ac'$ is not free in $o$,
$\mathbf{A2} (P)[o/ok] = \mathbf{A2} (P[o/ok])$.
\end{statement}
\begin{proofs}
\begin{proof}\checkt{alcc}
\begin{xflalign*}
	&\mathbf{A2} (P)[o/ok]
	&&\ptext{Definition of $\mathbf{A2}$ (\cref{lemma:A2:alternative-2:disjunction})}\\
	&=(P[\emptyset/ac'] \lor (\exists y \spot P[\{y\}/ac'] \land y \in ac'))[o/ok]
	&&\ptext{Substitution}\\
	&=(P[o/ok][\emptyset/ac'] \lor (\exists y \spot P[o/ok][\{y\}/ac'] \land y \in ac'))
	&&\ptext{Definition of $\mathbf{A2}$ (\cref{lemma:A2:alternative-2:disjunction})}\\
	&=\mathbf{A2} (P[o/ok])
\end{xflalign*}
\end{proof}
\end{proofs}
\end{lemma}

\begin{lemma}\label{lemma:A2(exists-x-P):exists-x-A2(P)}
\begin{statement}
Provided that $x$ is not $ac'$, $\mathbf{A2} (\exists x @ P) = \exists x @ \mathbf{A2} (P)$
\end{statement}
\begin{proofs}
\begin{proof}\checkt{alcc}
\begin{xflalign*}
	&\mathbf{A2} (\exists x @ P)
	&&\ptext{Definition of $\mathbf{A2}$ (\cref{lemma:A2:alternative-2:disjunction})}\\
	&=(\exists x @ P)[\emptyset/ac'] \lor (\exists y \spot (\exists x @ P)[\{y\}/ac'] \land y \in ac')
	&&\ptext{Assumption: $x$ is not $ac'$ and substitution}\\
	&=(\exists x @ P[\emptyset/ac']) \lor (\exists y \spot (\exists x @ P[\{y\}/ac']) \land y \in ac')
	&&\ptext{Predicate calculus}\\
	&=(\exists x @ P[\emptyset/ac']) \lor (\exists x @ \exists y \spot P[\{y\}/ac'] \land y \in ac')
	&&\ptext{Predicate calculus}\\
	&=\exists x @ (P[\emptyset/ac'] \lor (\exists y \spot P[\{y\}/ac'] \land y \in ac'))
	&&\ptext{Definition of $\mathbf{A2}$ (\cref{lemma:A2:alternative-2:disjunction})}\\
	&=\exists x @ \mathbf{A2} (P)
\end{xflalign*}
\end{proof}
\end{proofs}
\end{lemma}

\subsubsection{Properties with respect to $\mathbf{PBMH}$}

\begin{theorem}\label{theorem:A2-o-PBMH(P):A2(P)}
\begin{statement}
$\mathbf{A2}\circ\mathbf{PBMH} (P) = \mathbf{A2} (P)$
\end{statement}
\begin{proofs}
\begin{proof}
\begin{xflalign*}
	&\mathbf{A2}\circ\mathbf{PBMH} (P)
	&&\ptext{Definition of $\mathbf{A2}$ (\cref{lemma:A2:alternative-2:disjunction})}\\
	&=\left(\begin{array}{l}
		\mathbf{PBMH} (P)[\emptyset/ac']
		\\ \lor \\
		(\exists z @ \mathbf{PBMH} (P)[\{z\}/ac'] \land z \in ac')
	\end{array}\right)
	&&\ptext{Definition of $\mathbf{PBMH}$ (\cref{lemma:PBMH:alternative-1})}\\
	&=\left(\begin{array}{l}
		(\exists ac_0 @ P[ac_0/ac'] \land ac_0\subseteq ac')[\emptyset/ac']
		\\ \lor \\
		(\exists z @ (\exists ac_0 @ P[ac_0/ac'] \land ac_0\subseteq ac')[\{z\}/ac'] \land z \in ac')
	\end{array}\right)
	&&\ptext{Substitution}\\
	&=\left(\begin{array}{l}
		(\exists ac_0 @ P[ac_0/ac'] \land ac_0\subseteq \emptyset)
		\\ \lor \\
		(\exists z @ (\exists ac_0 @ P[ac_0/ac'] \land ac_0\subseteq \{z\}) \land z \in ac')
	\end{array}\right)
	&&\ptext{Property of sets and one-point rule}\\
	&=\left(\begin{array}{l}
		(P[\emptyset/ac'])
		\\ \lor \\
		(\exists z @ (P[\emptyset/ac'] \lor P[\{z\}/ac']) \land z \in ac')
	\end{array}\right)
	&&\ptext{Predicate calculus}\\
	&=\left(\begin{array}{l}
		(P[\emptyset/ac'])
		\\ \lor \\
		(\exists z @ P[\emptyset/ac'] \land z \in ac')
		\\ \lor \\
		(\exists z @ P[\{z\}/ac'] \land z \in ac')
	\end{array}\right)
	&&\ptext{Predicate calculus and property of sets}\\
	&=\left(\begin{array}{l}
		(P[\emptyset/ac'])
		\\ \lor \\
		(P[\emptyset/ac'] \land ac'\neq\emptyset)
		\\ \lor \\
		(\exists z @ P[\{z\}/ac'] \land z \in ac')
	\end{array}\right)
	&&\ptext{Predicate calculus: absorption law}\\
	&=P[\emptyset/ac']\lor (\exists z @ P[\{z\}/ac'] \land z \in ac')
	&&\ptext{Definition of $\mathbf{A2}$ (\cref{lemma:A2:alternative-2:disjunction})}\\
	&=\mathbf{A2} (P)
\end{xflalign*}
\end{proof}
\end{proofs}
\end{theorem}

\begin{lemma}\label{lemma:PBMH(P-seqA-s-ac')-seqA-s-ac':P-seqA-s-ac'} Provided $P$ is $\mathbf{PBMH}$-healthy,
\begin{align*}
	&\mathbf{PBMH} (P \seqA \{ s | \{s\} = ac'\}) \seqA \{ s | \{s\} = ac' \} \\
	&=\\
	&P \seqA \{ s | \{s\} = ac' \}
\end{align*}
\begin{proofs}
\begin{proof}
\begin{xflalign*}
	&\mathbf{PBMH} (P \seqA \{ s | \{s\} = ac'\}) \seqA \{ s | \{s\} = ac' \}
	&&\ptext{Assumption: $P$ is $\mathbf{PBMH}$-healthy and~\cref{lemma:A2-o-PBMH(P)}}\\
	&=\left(
\begin{array}{l}
		(\exists ac_1, ac_0 \spot P[ac_0/ac'] \land ac_0 \subseteq \{ s | \{s\} = ac_1 \} \land ac_1 \subseteq ac')
		\\ \seqA \\
		\{ s | \{s\} = ac' \}
	\end{array}\right)
	&&\ptext{Definition of $\seqA$ and substitution}\\
	&=\left(\begin{array}{l}
		\exists ac_1, ac_0 \spot P[ac_0/ac'] 
		\land ac_0 \subseteq \{ s | \{s\} = ac_1 \} 
		\land ac_1 \subseteq \{ s | \{s\} = ac' \}
	\end{array}\right)
	&&\ptext{\cref{lemma:set-theory:ac0-subseteq-A2:ac0-subseteq-ac1-land-ac0-subseteq-A2}}\\
	&=\left(\begin{array}{l}
		\exists ac_1, ac_0 \spot P[ac_0/ac'] 
		\\ \land ac_0 \subseteq ac_1 \land ac_0 \subseteq \{ s | ac_1 \subseteq \{s\} \} 
		\\ \land ac_1 \subseteq \{ s | \{s\} = ac' \}
	\end{array}\right)
	&&\ptext{\cref{lemma:set-theory:ac0-subseteq-A2:ac1-subseteq-A2}}\\
	&=\left(\begin{array}{l}
		\exists ac_1, ac_0 \spot P[ac_0/ac'] 
		\\ \land ac_0 \subseteq ac_1 \land ac_1 \subseteq \{ s | ac_0 \subseteq \{s\} \} 
		\\ \land ac_1 \subseteq \{ s | \{s\} = ac' \}
	\end{array}\right)
	&&\ptext{Property of sets}\\
	&=\left(\begin{array}{l}
		\exists ac_1, ac_0 \spot P[ac_0/ac'] 
		\land ac_0 \subseteq ac_1 \land ac_1 \subseteq \{ s | ac_0 \subseteq \{s\} \land \{s\} = ac' \} 
	\end{array}\right)
	&&\ptext{Transitivity of subset inclusion}\\
	&=\left(\begin{array}{l}
		\exists ac_0 \spot P[ac_0/ac'] 
		\land ac_0 \subseteq \{ s | ac_0 \subseteq \{s\} \land \{s\} = ac' \} 
	\end{array}\right)
	&&\ptext{Predicate calculus}\\
	&=\left(\begin{array}{l}
		\exists ac_0 \spot P[ac_0/ac'] 
		\land ac_0 \subseteq \{ s | ac_0 \subseteq ac' \land \{s\} = ac' \} 
	\end{array}\right)
	&&\ptext{Property of sets}\\
	&=\left(\begin{array}{l}
		\exists ac_0 \spot P[ac_0/ac'] 
		\land ac_0 \subseteq \{ s | ac_0 \subseteq ac' \}
		\land ac_0 \subseteq \{ s | \{s\} = ac' \} 
	\end{array}\right)
	&&\ptext{\cref{lemma:set-theory:ac0-subseteq-A2:ac0-emptyset-lor-ac0-subseteq-ac'}}\\
	&=\left(\begin{array}{l}
		\exists ac_0 \spot P[ac_0/ac'] 
		\land (ac_0 = \emptyset \lor ac_0 \subseteq ac') 
		\land ac_0 \subseteq \{ s | \{s\} = ac' \}  
	\end{array}\right)
	&&\ptext{Predicate calculus}\\
	&=\left(\begin{array}{l}
		(\exists ac_0 \spot P[ac_0/ac'] \land ac_0 = \emptyset \land ac_0 \subseteq \{ s | \{s\} = ac' \} )
		\\ \lor \\
		(\exists ac_0 \spot P[ac_0/ac'] \land ac_0 \subseteq ac' \land ac_0 \subseteq \{ s | \{s\} = ac' \} ) 
	\end{array}\right)
	&&\ptext{One-point rule}\\
	&=\left(\begin{array}{l}
		P[\emptyset/ac']
		\\ \lor \\
		(\exists ac_0 \spot P[ac_0/ac'] \land ac_0 \subseteq ac' \land ac_0 \subseteq \{ s | \{s\} = ac' \}) 
	\end{array}\right)
	&&\ptext{$P[\emptyset/ac']$ is an instance of existential quantification}\\
	&=\exists ac_0 \spot P[ac_0/ac'] \land ac_0 \subseteq ac' \land ac_0 \subseteq \{ s | \{s\} = ac' \}
	&&\ptext{Predicate calculus and~\cref{lemma:set-theory:ac0-subseteq-A2:ac0-subseteq-ac1-land-ac0-subseteq-A2}}\\
	&=\exists ac_0 \spot P[ac_0/ac'] \land ac_0 \subseteq \{ s | \{s\} = ac' \}
	&&\ptext{Definition of $\seqA$ and substitution}\\
	&=(\exists ac_0 \spot P[ac_0/ac'] \land ac_0 \subseteq ac') \seqA \{ s | \{s\} = ac' \}
	&&\ptext{Definition of $\mathbf{PBMH}$ (\cref{lemma:PBMH:alternative-1})}\\
	&=\mathbf{PBMH} (P) \seqA \{ s | \{s\} = ac' \}
	&&\ptext{Assumption: $P$ is $\mathbf{PBMH}$-healthy}\\
	&=P \seqA \{ s | \{s\} = ac' \}
\end{xflalign*}
\end{proof}
\end{proofs}
\end{lemma}

\begin{lemma}\label{lemma:PBMH-o-A2(P):A2(P)}
$\mathbf{PBMH} \circ \mathbf{A2} (P) = \mathbf{A2} (P)$
\begin{proofs}\begin{proof}\checkt{alcc}
\begin{xflalign*}
	&\mathbf{PBMH} \circ \mathbf{A2} (P)
	&&\ptext{Definition of $\mathbf{A2}$}\\
	&=\mathbf{PBMH} \circ \mathbf{PBMH} (P \seqA \{s\} = ac')
	&&\ptext{$\mathbf{PBMH}$-idempotent (\cref{law:pbmh:idempotent})}\\
	&=\mathbf{PBMH} (P \seqA \{s\} = ac')
	&&\ptext{Definition of $\mathbf{A2}$}\\
	&=\mathbf{A2} (P)
\end{xflalign*}
\end{proof}\end{proofs}
\end{lemma}

\subsubsection{Properties with respect to $\seqA$}

\begin{theorem}\label{theorem:A2(P-seqA-Q):P-seqA-Q}
\begin{statement}
Provided $P$ and $Q$ are $\mathbf{A2}$-healthy,
$\mathbf{A2} (P \seqA Q) = P \seqA Q$
\end{statement}
\begin{proofs}
\begin{proof}\checkt{alcc}
\begin{xflalign*}
	&\mathbf{A2} (P \seqA Q)
	&&\ptext{Assumption: $P$ and $Q$ are $\mathbf{A2}$-healthy}\\
	&=\mathbf{A2} (\mathbf{A2} (P) \seqA \mathbf{A2} (Q))
	&&\ptext{\cref{lemma:A2(A2(P)-seqA-A2(Q)):A2(P)-seqA-A2(Q)}}\\
	&=\mathbf{A2} (P) \seqA \mathbf{A2} (Q)
	&&\ptext{Assumption: $P$ and $Q$ are $\mathbf{A2}$-healthy}\\
	&=P \seqA Q
\end{xflalign*}
\end{proof}
\end{proofs}
\end{theorem}

\begin{lemma}\label{lemma:A2(P)-seqA-A2(Q)}
\begin{statement}
\begin{align*}
	&\mathbf{A2} (P) \seqA \mathbf{A2} (Q)\\
	&=\\
	&\left(
\begin{array}{l}
		P[\emptyset/ac'] \lor (\exists y \spot P[\{y\}/ac'] \land Q[\emptyset/ac'][y/s]) 
		\\ \lor \\
		(\exists y \spot P[\{y\}/ac'] \land (\exists y \spot Q[\{y\}/ac'][y/s] \land y \in ac'))
	\end{array}\right)
\end{align*}
\end{statement}
\begin{proofs}
\begin{proof}\checkt{alcc}
\begin{xflalign*}
	&\mathbf{A2} (P) \seqA \mathbf{A2} (Q)
	&&\ptext{Definition of $\mathbf{A2}$ (\cref{lemma:A2:alternative-2:disjunction})}\\
	&=\left(\begin{array}{l}
		(P[\emptyset/ac'] \lor (\exists y \spot P[\{y\}/ac'] \land y \in ac'))
		\\ \seqA \\
		(Q[\emptyset/ac'] \lor (\exists y \spot Q[\{y\}/ac'] \land y \in ac'))
	\end{array}%
\right)
	&&\ptext{Property of sets and substitution}\\
	&=\left(%
\right)
	&&\ptext{Property of sets and predicate calculus}\\
	&=\left(%
\right)
	&&\ptext{Predicate calculus: absorption law}\\
	&=\left(\begin{array}{l}
		P[\emptyset/ac'] \lor (\exists y \spot P[\{y\}/ac'] \land Q[\emptyset/ac'][y/s]) 
		\\ \lor \\
		(\exists y \spot P[\{y\}/ac'] \land (\exists z @ Q[\{z\}/ac'][z/s] \land z \in ac'))
	\end{array}\right)
	&&\ptext{\cref{lemma:A2(P)-seqA-A2(Q)}}\\
	&=\mathbf{A2} (P) \seqA \mathbf{A2} (Q)
\end{xflalign*}
\end{proof}
\end{proofs}
\end{lemma}

\subsubsection{Properties with respect to links ($p2ac$ and $ac2p$)}

\begin{lemma}\label{lemma:p2ac-o-ac2p-o-A2(P):A2(P)-land-ac'-neq-emptyset}
$p2ac \circ ac2p \circ \mathbf{A2} (P) = \mathbf{A2} (P) \land ac'\neq\emptyset$
\begin{proofs}\begin{proof}\checkt{alcc}
\begin{flalign*}
	&p2ac \circ ac2p \circ \mathbf{A2} (P)
	&&\ptext{\cref{lemma:p2ac-o-ac2p-o-A2(P):disjunction}}\\
	&=(P[\emptyset/ac'] \land ac'\neq\emptyset) \lor (\exists y \spot P[\{y\}/ac'] \land y \in ac')
	&&\ptext{Predicate calculus}\\
	&=(P[\emptyset/ac'] \land ac'\neq\emptyset) \lor (\exists y \spot P[\{y\}/ac'] \land y \in ac' \land ac'\neq\emptyset)
	&&\ptext{Predicate calculus}\\
	&=(P[\emptyset/ac'] \lor (\exists y \spot P[\{y\}/ac'] \land y \in ac')) \land ac'\neq\emptyset
	&&\ptext{\cref{lemma:A2:alternative-2:disjunction}}\\
	&=\mathbf{A2} (P) \land ac'\neq\emptyset
\end{flalign*}
\end{proof}\end{proofs}
\end{lemma}

\begin{lemma}\label{lemma:p2ac-o-ac2p-o-PBMH:p2ac-o-ac2p}
$p2ac \circ ac2p \circ \mathbf{PBMH} (P) = p2ac \circ ac2p (P)$
\begin{proofs}
\begin{proof}\checkt{alcc}
\begin{xflalign*}
	&p2ac \circ ac2p \circ \mathbf{PBMH} (P)
	&&\ptext{\cref{lemma:p2ac-o-ac2p(P)}}\\
	&=\exists ac_0, y \spot \mathbf{PBMH} (P)[ac_0/ac'] \land ac_0 \subseteq \{y\} \land y \in ac'
	&&\ptext{Definition of $\mathbf{PBMH}$ (\cref{lemma:PBMH:alternative-1})}\\
	&=\exists ac_0, y \spot (\exists ac_1 \spot P[ac_1/ac'] \land ac_1 \subseteq ac')[ac_0/ac'] \land ac_0 \subseteq \{y\} \land y \in ac'
	&&\ptext{Substitution}\\
	&=\exists ac_0, y, ac_1 \spot P[ac_1/ac'] \land ac_1 \subseteq ac_0 \land ac_0 \subseteq \{y\} \land y \in ac'
	&&\ptext{Predicate calculus}\\
	&=\exists y, ac_1 \spot P[ac_1/ac'] \land ac_1 \subseteq \{y\} \land y \in ac'
	&&\ptext{\cref{lemma:p2ac-o-ac2p(P)}}\\
	&=p2ac \circ ac2p (P)
\end{xflalign*}
\end{proof}\end{proofs}
\end{lemma}

\begin{lemma}\label{lemma:p2ac-o-ac2p-o-A2:p2ac-o-ac2p-(P-seqA-s-ac')}
$p2ac \circ ac2p \circ \mathbf{A2} (P) = p2ac \circ ac2p (P \seqA \{s\} = ac')$
\begin{proofs}\begin{proof}\checkt{alcc}
\begin{xflalign*}
	&p2ac \circ ac2p \circ \mathbf{A2} (P)
	&&\ptext{Definition of $\mathbf{A2}$}\\
	&=p2ac \circ ac2p \circ \mathbf{PBMH} (P \seqA \{s\} = ac')
	&&\ptext{\cref{lemma:p2ac-o-ac2p-o-PBMH:p2ac-o-ac2p}}\\
	&=p2ac \circ ac2p \circ (P \seqA \{s\} = ac')
\end{xflalign*}
\end{proof}\end{proofs}
\end{lemma}

\begin{lemma}\label{lemma:p2ac-o-ac2p-o-A2(P):disjunction}
\begin{align*}
	&p2ac \circ ac2p \circ \mathbf{A2} (P) \\
	&=\\
	&(P[\emptyset/ac'] \land ac'\neq\emptyset) \lor (\exists y \spot P[\{y\}/ac'] \land y \in ac') 
\end{align*}
\begin{proofs}
\begin{proof}\checkt{alcc}
\begin{xflalign*}
	&p2ac \circ ac2p \circ \mathbf{A2} (P)
	&&\ptext{\cref{lemma:p2ac-o-ac2p-o-A2:p2ac-o-ac2p-(P-seqA-s-ac')}}\\
	&=p2ac \circ ac2p (P \seqA \{s\} = ac')
	&&\ptext{Definition of $\seqA$ and substitution}\\
	&=p2ac \circ ac2p (P[\{s | \{s\} = ac'\}/ac'])
	&&\ptext{\cref{lemma:p2ac-o-ac2p(P)}}\\
	&=\exists ac_0, y \spot (P[\{s | \{s\} = ac'\}/ac'])[ac_0/ac'] \land ac_0 \subseteq \{y\} \land y \in ac'
	&&\ptext{Substitution}\\
	&=\exists ac_0, y \spot P[\{s | \{s\} = ac_0\}/ac'] \land ac_0 \subseteq \{y\} \land y \in ac'
	&&\ptext{Property of sets}\\
	&=\exists ac_0, y \spot P[\{s | \{s\} = ac_0\}/ac'] \land (ac_0 = \emptyset \lor ac_0 = \{y\}) \land y \in ac'
	&&\ptext{Predicate calculus}\\
	&=\left(\begin{array}{l}
		(\exists ac_0, y \spot P[\{s | \{s\} = ac_0\}/ac'] \land ac_0 = \emptyset \land y \in ac')
		\\ \lor \\
		(\exists ac_0, y \spot P[\{s | \{s\} = ac_0\}/ac'] \land ac_0 = \{y\} \land y \in ac')
	\end{array}\right)
	&&\ptext{One-point rule}\\
	&=\left(\begin{array}{l}
		(\exists y \spot P[\{s | \{s\} = \emptyset\}/ac'] \land y \in ac')
		\\ \lor \\
		(\exists y \spot P[\{s | \{s\} = \{y\}\}/ac'] \land y \in ac')
	\end{array}\right)
	&&\ptext{Property of sets}\\
	&=(\exists y \spot P[\emptyset/ac'] \land y \in ac') \lor (\exists y \spot P[\{y\}/ac'] \land y \in ac')
	&&\ptext{Predicate calculus}\\
	&=(P[\emptyset/ac'] \land \exists y \spot y \in ac') \lor (\exists y \spot P[\{y\}/ac'] \land y \in ac')
	&&\ptext{Property of sets}\\
	&=(P[\emptyset/ac'] \land ac'\neq\emptyset) \lor (\exists y \spot P[\{y\}/ac'] \land y \in ac')
\end{xflalign*}
\end{proof}\end{proofs}
\end{lemma}

\section[Relationship with Extended Binary Multirelations]{Relationship with Extended\\ Binary Multirelations}

\subsection{$d2bmb$}

\begin{theorem}\label{theorem:bmh-0-1-2:d2bmb(A)}
\begin{statement}
Provided $P$ is a design,
\begin{align*}
	&\mathbf{bmh}_\mathbf{0,1,2} \circ d2bmb(\mathbf{A} (P)) = d2bmb(\mathbf{A} (P))
\end{align*}
\end{statement}
\begin{proofs}
\begin{proof}
\begin{flalign*}
	&\mathbf{bmh}_\mathbf{0,1,2} \circ d2bmb(\mathbf{A} (P))
	&&\ptext{Definition of $\mathbf{bmh}_\mathbf{0,1,2}$}\\
	&=\left\{
\right\}	
	&&\ptext{Predicate calculus and~\cref{law:d2bmb:A-healthy}}\\
	&=d2bmb(\mathbf{A} (P))
\end{flalign*}
\end{proof}
\end{proofs}
\end{theorem}\noindent

\begin{lemma}[$d2bmb$-$\mathbf{A}$-healthy]\label{law:d2bmb:A-healthy}
\begin{statement}
Provided $P$ is a design,
\begin{align*}
	&d2bmb(\mathbf{A} (P))\\
	&=&\\
	&\left\{%
\right\}
	&&\ptext{Definition of sequential composition}\\
	&=\left\{%
\right\}
	&&\ptext{Propositional calculus and property of sets}\\
	&=\left\{%
\right\}
	&&\ptext{Property of sets}\\
	&=\left\{%
\right\}
	&&\ptext{Propositional calculus, property of sets and~\cref{law:set-theory:set-membership-subset-2}}\\
	&=\left\{%
\right)
	&&\ptext{Property of sets}\\
	&=\left(\begin{array}{l}
		\exists ss_0 : \power State_\bot, ac_0 : \power State \spot 
			\\ (P^f[ac_0/ac'] \lor (P^t[ac_0/ac'] \land (ss_0 \cup \{\bot\})\neq\emptyset \land \bot \notin (ss_0 \cup \{\bot\})))
			\\ \land ac_0 \subseteq (ss_0 \cup \{\bot\}) \land ss_0 \subseteq ss \land (\bot \in ss_0 \iff \bot \in ss)
	\end{array}\right)
	&&\ptext{Property of sets and predicate calculus}\\
	&=\left(\begin{array}{l}
		\exists ss_0 : \power State_\bot, ac_0 : \power State \spot 
			\\ P^f[ac_0/ac'] \land ac_0 \subseteq (ss_0 \cup \{\bot\}) \land ss_0 \subseteq ss \land (\bot \in ss_0 \iff \bot \in ss)
	\end{array}\right)
	&&\ptext{Property of sets}\\
	&=\left(\begin{array}{l}
		\exists ss_0 : \power State_\bot, ac_0 : \power State \spot 
		\\ P^f[ac_0/ac'] \land (ac_0 \setminus \{\bot\}) \subseteq ss_0 \land ss_0 \subseteq ss \land (\bot \in ss_0 \iff \bot \in ss)
	\end{array}\right)
	&&\ptext{Type of $ac'$ : $\bot \notin ac'$, and property of sets}\\
	&=\left(\begin{array}{l}
		\exists ss_0 : \power State_\bot, ac_0 : \power State \spot 
		\\ P^f[ac_0/ac'] \land ac_0 \subseteq ss_0 \land ss_0 \subseteq ss \land (\bot \in ss_0 \iff \bot \in ss)
	\end{array}%
\right)
	&&\ptext{Predicate calculus}\\
	&=\exists ac_0 : \power State \spot P^f[ac_0/ac'] \land ac_0 \subseteq ss 
\end{flalign*}
\end{proof}\end{proofs}
\end{lemma}

\begin{lemma}\label{law:bmbot:ss-in-d2bmb(A(P))-subset}
Provided $P$ is a design,
\begin{align*}
	&\exists ss_0 : \power State_\bot \spot (s, ss_0) \in d2bmb(\mathbf{A} (P)) \land ss_0 \subseteq ss \land (\bot \in ss_0 \iff \bot \in ss)\\
	&=\\
	&\exists ac_0 : \power State \spot (P^f[ac_0/ac'] \lor (P^t[ac_0/ac'] \land ss\neq\emptyset \land \bot \notin ss)) \land ac_0 \subseteq ss
\end{align*}
\begin{proofs}\begin{proof}
\begin{flalign*}
	&\exists ss_0 : \power State_\bot \spot (s, ss_0) \in d2bmb(\mathbf{A} (P)) \land ss_0 \subseteq ss \land (\bot \in ss_0 \iff \bot \in ss)
	&&\ptext{Definition of $d2bmb(\mathbf{A} (P))$}\\
	&=\left(
\right)
	&&\ptext{Property of sets}\\
	&=\left(\begin{array}{l}
		\exists ss_0 : \power State_\bot, ac_0 : \power State \spot
		\\ (P^f[ac_0/ac'] \lor (P^t[ac_0/ac'] \land ss_0\neq\emptyset \land \bot \notin ss_0))
		\\ \land ac_0 \subseteq ss_0 \land ss_0 \subseteq ss \land (\bot \in ss_0 \iff \bot \in ss)
	\end{array}%
\right)
\end{flalign*}
\end{proof}\end{proofs}
\end{lemma}

\begin{lemma}\label{law:bmbot:bot-in-d2bmb(A):iff:emptyset-in-d2bmb(A)}
Provided $P$ is a design,
\begin{align*}
	&(s, \{\bot\}) \in d2bmb(\mathbf{A} (P)) \iff (s, \emptyset) \in d2bmb(\mathbf{A} (P))
\end{align*}
\begin{proofs}\begin{proof}
\begin{flalign*}
	&(s, \{\bot\}) \in d2bmb(\mathbf{A} (P)) \iff (s, \emptyset) \in d2bmb(\mathbf{A} (P))
	&&\ptext{\cref{law:bmbot:bot-in-d2bmb(A)} and \cref{law:bmbot:emptyset-in-d2bmb(A)}}\\
	&=true
\end{flalign*}
\end{proof}\end{proofs}
\end{lemma}

\begin{lemma}\label{law:bmbot:bot-in-d2bmb(A)}
Provided $P$ is a design,
\begin{align*}
	&(s, \{\bot\}) \in d2bmb(\mathbf{A} (P)) = P^f[\emptyset/ac']
\end{align*}
\begin{proofs}\begin{proof}
\begin{flalign*}
	&(s, \{\bot\}) \in d2bmb(\mathbf{A} (P))
	&&\ptext{\cref{law:d2bmb:A-healthy}}\\
	&=(s, \{\bot\}) \in \left\{\begin{array}{l}
		s : State, ss : \power State_\bot \\
		\left|\begin{array}{l}
			\exists ac_0 : \power State \spot 
			\\ (P^f[ac_0/ac'] \lor (P^t[ac_0/ac'] \land ss\neq\emptyset \land \bot \notin ss))
			\\ \land ac_0 \subseteq ss
		\end{array}\right.
	\end{array}\right\}
	&&\ptext{Property of sets}\\
	&=\exists ac_0 : \power State \spot (P^f[ac_0/ac'] \lor (P^t[ac_0/ac'] \land \{\bot\}\neq\emptyset \land \bot \notin \{\bot\})) \land ac_0 \subseteq \{\bot\}
	&&\ptext{Property of sets and predicate calculus}\\
	&=\exists ac_0 : \power State \spot P^f[ac_0/ac'] \land ac_0 \subseteq \{\bot\}
	&&\ptext{Case-analysis on $ac_0$ and one-point rule}\\
	&=P^f[\emptyset/ac']
\end{flalign*}
\end{proof}\end{proofs}
\end{lemma}

\begin{lemma}\label{law:bmbot:emptyset-in-d2bmb(A)}
Provided $P$ is a design,
\begin{align*}
	&(s, \emptyset) \in d2bmb(\mathbf{A} (P)) = P^f[\emptyset/ac']
\end{align*}
\begin{proofs}\begin{proof}
\begin{flalign*}
	&(s, \emptyset) \in d2bmb(\mathbf{A} (P))
	&&\ptext{Definition of $d2bmb$ for $P$ that is $\mathbf{A}$-healthy}\\
	&=(s, \emptyset) \in \left\{\begin{array}{l}
		s : State, ss : \power State_\bot \\
		\left|\begin{array}{l}
			\exists ac_0 : \power State \spot 
			\\ (P^f[ac_0/ac'] \lor (P^t[ac_0/ac'] \land ss\neq\emptyset \land \bot \notin ss))
			\\ \land ac_0 \subseteq ss
		\end{array}\right.
	\end{array}\right\}
	&&\ptext{Property of sets}\\
	&=\exists ac_0 : \power State \spot (P^f[ac_0/ac'] \lor (P^t[ac_0/ac'] \land \emptyset\neq\emptyset \land \bot \notin \emptyset)) \land ac_0 \subseteq \emptyset
	&&\ptext{Property of sets and predicate calculus}\\
	&=\exists ac_0 : \power State \spot P^f[ac_0/ac'] \land ac_0 \subseteq \emptyset
	&&\ptext{Property of sets and one-point rule}\\
	&=P^f[\emptyset/ac']
\end{flalign*}
\end{proof}\end{proofs}
\end{lemma}

\begin{lemma}\label{law:bmbot:emptyset-iff-bot-in-d2bmb(A)}
Provided $P$ is a design,
\begin{align*}
	&(s, \emptyset) \in d2bmb(\mathbf{A} (P)) \iff (s, \{\bot\}) \in d2bmb(\mathbf{A} (P)) = true
\end{align*}
\begin{proofs}\begin{proof}
\begin{flalign*}
	&(s, \emptyset) \in d2bmb(\mathbf{A} (P)) \iff (s, \{\bot\}) \in d2bmb(\mathbf{A} (P))
	&&\ptext{\cref{law:bmbot:emptyset-in-d2bmb(A)} and \cref{law:bmbot:bot-in-d2bmb(A)}}\\
	&=true
\end{flalign*}
\end{proof}\end{proofs}
\end{lemma}

\begin{lemma}\label{lemma:d2bmb(P|-Q)}
\begin{statement}
Provided $ok$ and $ok'$ are not free in $P$ and $Q$,
\begin{align*}
	&d2bmb(P \vdash Q) =\left\{
\right\}
	&&\ptext{Assumption: $ok'$ is not free in $P$ and $Q$}\\
	&=\left\{%
\right\}
	&&\ptext{Substitution and assumption $ok$ not free in $P$ and $Q$}\\
	&=\left\{%
\right\}
\end{xflalign*}
\end{proof}
\end{proofs}
\end{lemma}

\subsection{$bmb2d$}

\begin{lemma}\label{lemma:bmb2d:alternative-def}
\begin{statement}
$bmb2d(B) = ok \implies \left(\begin{array}{l}
			((s, ac') \in B \land \bot\notin ac' \land ok')
			\\ \lor \\
			(s, ac' \cup \{ \bot \}) \in B
		\end{array}\right)$
\end{statement}
\begin{proofs}
\begin{proof}
Follows from the definition of design and type restriction on $ac'$.
\end{proof}
\end{proofs}
\end{lemma}\noindent

\begin{theorem}\label{theorem:A(bmb2d(B)):bmb2d(B)}
\begin{statement}
Provided $B$ satisfies $\mathbf{bmh}_\mathbf{0,1,2}$,
$\mathbf{A} \circ bmb2d(B) = bmb2d(B)$.
\end{statement}
\begin{proofs}
\begin{proof}
\begin{flalign*}
	&\mathbf{A} \circ bmb2d(B)
	&&\ptext{Assumption: $B = \mathbf{bmh}_\mathbf{0,1,2} (B)$ and \cref{lemma:bmb2d-B-BMH0-2}}\\
	&=\mathbf{A} \left(\begin{array}{l}
		\lnot ((s, ac' \cup \{\bot\}) \in B \circseq ac \subseteq ac')
	 	\\ \vdash \\
		((s, ac') \in B \circseq ac \subseteq ac') \land (s, \emptyset) \notin B
	\end{array}\right)
	&&\ptext{\cref{lemma:a-model:pbmh:s-emptyset-in-B}}\\
	&=\mathbf{A} \left(\begin{array}{l}
		\lnot ((s, ac' \cup \{\bot\}) \in B \circseq ac \subseteq ac')
	 	\\ \vdash \\
		((s, ac') \in B \circseq ac \subseteq ac') \land ac'\neq\emptyset \land (s, \emptyset) \notin B
	\end{array}\right)
	&&\ptext{Definition of $\mathbf{PBMH}$}\\
	&=\mathbf{A} \left(\begin{array}{l}
		\lnot \mathbf{PBMH} ((s, ac' \cup \{\bot\}) \in B)
	 	\\ \vdash \\
		\mathbf{PBMH} ((s, ac') \in B) \land ac'\neq\emptyset \land (s, \emptyset) \notin B
	\end{array}\right)
	&&\ptext{Definition of $\mathbf{A}$}\\
	&=\left(\begin{array}{l}
		\lnot (\mathbf{PBMH} \circ \mathbf{PBMH} ((s, ac' \cup \{\bot\}) \in B))
	 	\\ \vdash \\
		\mathbf{PBMH} (\mathbf{PBMH} ((s, ac') \in B) \land ac'\neq\emptyset \land (s, \emptyset) \notin B) \land ac'\neq\emptyset
	\end{array}\right)
	&&\ptext{($\mathbf{PBMH}$-idempotent) \cref{law:pbmh:idempotent}}\\
	&=\left(\begin{array}{l}
		\lnot \mathbf{PBMH} ((s, ac' \cup \{\bot\}) \in B)
	 	\\ \vdash \\
		\mathbf{PBMH} (\mathbf{PBMH} ((s, ac') \in B) \land ac'\neq\emptyset \land (s, \emptyset) \notin B) \land ac'\neq\emptyset
	\end{array}%
\right)
	&&\ptext{\cref{law:pbmh:ac'-neq-emptyset} and \cref{law:pbmh:P:ac'-not-free} and predicate calculus}\\
	&=\left(\begin{array}{l}
		\lnot \mathbf{PBMH} ((s, ac' \cup \{\bot\}) \in B)
	 	\\ \vdash \\
		\mathbf{PBMH} ((s, ac') \in B) \land ac'\neq\emptyset \land (s, \emptyset) \notin B
	\end{array}\right)
	&&\ptext{Definition of $\mathbf{PBMH}$ and~\cref{lemma:a-model:pbmh:s-emptyset-in-B}}\\
	&=\left(\begin{array}{l}
		\lnot ((s, ac' \cup \{\bot\}) \in B \circseq ac \subseteq ac')
	 	\\ \vdash \\
		((s, ac') \in B \circseq ac \subseteq ac') \land (s, \emptyset) \notin B
	\end{array}\right)
	&&\ptext{Assumption: $B = \mathbf{bmh}_\mathbf{0,1,2} (B)$ and \cref{lemma:bmb2d-B-BMH0-2}}\\
	&=bmb2d(B)
\end{flalign*}
\end{proof}
\end{proofs}
\end{theorem}

\begin{lemma}\label{lemma:a-model:pbmh:s-emptyset-in-B}
\begin{align*}
	&((s, ac') \in B \circseq ac \subseteq ac') \land (s, \emptyset) \notin B \\
	&\iff\\
	&((s, ac') \in B \circseq ac \subseteq ac') \land ac'\neq\emptyset \land (s, \emptyset) \notin B
\end{align*}
\begin{proofs}\begin{proof}
\begin{flalign*}
	&((s, ac') \in B \circseq ac \subseteq ac') \land (s, \emptyset) \notin B
	&&\ptext{Definition of sequential composition}\\
	&\iff (\exists ac_0 : \power State \spot (s, ac_0) \in B \land ac_0 \subseteq ac') \land (s, \emptyset) \notin B
	&&\ptext{Predicate calculus}\\
	&\iff \left(
\begin{array}{l}
		(\exists ac_0 : \power State \spot (s, ac_0) \in B \land ac_0 \subseteq ac' \land (ac' = \emptyset \lor ac'\neq\emptyset)) 
		\\ \land \\
		(s, \emptyset) \notin B
	\end{array}\right)
	&&\ptext{Predicate calculus}\\
	&\iff \left(\begin{array}{l}
		(\exists ac_0 : \power State \spot (s, ac_0) \in B \land ac_0 \subseteq ac' \land ac'=\emptyset)
		\\ \lor \\
		(\exists ac_0 : \power State \spot (s, ac_0) \in B \land ac_0 \subseteq ac' \land ac'\neq\emptyset)
	\end{array}\right) \land (s, \emptyset) \notin B
	&&\ptext{Property of sets and case analysis on $ac_0$}\\
	&\iff \left(\begin{array}{l}
		((s, \emptyset) \in B \land ac'=\emptyset)
		\\ \lor \\
		(\exists ac_0 : \power State \spot (s, ac_0) \in B \land ac_0 \subseteq ac' \land ac'\neq\emptyset)
	\end{array}\right) \land (s, \emptyset) \notin B
	&&\ptext{Predicate calculus}\\
	&\iff (\exists ac_0 : \power State \spot (s, ac_0) \in B \land ac_0 \subseteq ac' \land ac'\neq\emptyset) \land (s, \emptyset) \notin B
	&&\ptext{Definition of sequential composition}\\
	&\iff ((s, ac') \in B \circseq ac \subseteq ac' \land ac'\neq\emptyset) \land (s, \emptyset) \notin B
\end{flalign*}
\end{proof}\end{proofs}
\end{lemma}

\begin{lemma}\label{lemma:bmb2d-B-BMH0-2}
Provided $B$ satisfies $\mathbf{bmh}_\mathbf{0,1,2}$,
\begin{align*}
	&bmb2d(B) =
	\left(%
\right)
	&&\ptext{Definition of sequential composition}\\
	&=\left(%
\right)
	&&\ptext{Instantiation of existential quantifier for $ac_0 = \emptyset$}\\
	&=\left(\begin{array}{l}
		\lnot (\exists ac_0 : \power State \spot (s, ac_0 \cup \{\bot\}) \in B \land ac_0 \subseteq ac')
	 	\\ \vdash \\
		((s, ac') \in B \circseq ac \subseteq ac') \land (s, \emptyset) \notin B
	\end{array}\right)
	&&\ptext{Definition of sequential composition}\\
	&=\left(\begin{array}{l}
		\lnot ((s, ac' \cup \{\bot\}) \in B \circseq ac \subseteq ac')
	 	\\ \vdash \\
		((s, ac') \in B \circseq ac \subseteq ac') \land (s, \emptyset) \notin B
	\end{array}%
\right)
	&&\ptext{Property of sets and predicate calculus}\\
	&=ok \implies \left(%
\right)
	&&\ptext{Property of sets}\\
	&=ok \implies \left(%
\right)
	&&\ptext{Predicate calculus: introduce fresh variable}\\
	&=ok \implies \left(%
\right)
	&&\ptext{One-point rule and substitution}\\
	&=ok \implies \left(%
\right)
	&&\ptext{Property of sets and variable renaming}\\
	&=ok \implies \left(%
\right)
	&&\ptext{Predicate calculus: absorption law}\\
	&=ok \implies \left(%
\right)
	&&\ptext{Instantiation: consider case where $ss_0 = \emptyset$}\\
	&=ok \implies \left(%
\right)
	&&\ptext{Predicate calculus: distribution}\\
	&=ok \implies \left(%
\right)
	&&\ptext{Predicate calculus: absorption law}\\
	&=ok \implies \left(%
\right)
	&&\ptext{Instantiation: consider case where $ss_0 = \emptyset$}\\
	&=ok \implies \left(%
\right)
	&&\ptext{Predicate calculus: absorption law}\\
	&=ok \implies \left(%
\right)
	&&\ptext{Variable renaming and substitution}\\
	&=\left(%
\right)
	&&\ptext{Definition of sequential composition and type of $ac'$ : $\bot \notin ac'$}\\
	&=\left(%
\right)
	&&\ptext{Definition of design}\\
	&=\left(%
\right)
	&&\ptext{Property of sets}\\
	&=\left(%
\right)
	\end{array}\right)
	&&\ptext{Property of sets and propositional calculus}\\
	&=P^f[ss \setminus \{ \bot \}/ac'][\{ s_1 : State_\bot | true \}/ss]
	&&\ptext{Substitution}\\
	&=P^f[\{ s_1 : State_\bot | true \} \setminus \{ \bot \}/ac']
	&&\ptext{Property of sets}\\
	&=P^f[\{ s_1 : State | true \}/ac']
\end{flalign*}
\end{proof}\end{proofs}
\end{lemma}

\begin{lemma}\label{law:s-ac'-cup-bot-in-d2bmb(P)}
Provided $\bot \notin ac'$ and $P$ is a design,
\begin{align*}
	&\{ s : State | (s, ac' \cup \{\bot\}) \in d2bmb(P)\} = \{ s : State | P^f \}
\end{align*}
\begin{proofs}\begin{proof}
\begin{flalign*}
	&\{ s : State | (s, ac' \cup \{\bot\}) \in d2bmb(P)\}
	&&\ptext{Definition of $d2bmb$}\\
	&=\left\{
\right\}
	&&\ptext{Property of sets}\\
	&=\left\{%
\right\}
	&&\ptext{Property of sets}\\
	&=\left\{%
\right\}
	&&\ptext{Property of sets, and assumption that $\bot \notin ac'$}\\
	&=\{ s : State | P^f \}
\end{flalign*}
\end{proof}\end{proofs}
\end{lemma}

\begin{lemma}\label{law:s-ac'-in-d2bmb(P)}
Provided $\bot \notin ac'$ and $P$ is a design,
\begin{align*}
	&\{ s : State | (s, ac') \in d2bmb(P)\} = \{ s : State | (\lnot P^f \implies P^t) \}
\end{align*}
\begin{proofs}\begin{proof}
\begin{flalign*}
	&\{ s : State | (s, ac') \in d2bmb(P)\}
	&&\ptext{Definition of $d2bmb$}\\
	&=\left\{%
\right\}
	&&\ptext{Property of sets}\\
	&=\left\{%
\right.
	\end{array}\right\}
	&&\ptext{Assumption: $\bot \notin ac'$}\\
	&=\{ s : State | (\lnot P^f \implies P^t) \}
\end{flalign*}
\end{proof}\end{proofs}
\end{lemma}

\begin{lemma}\label{law:s-ac'-cup-in-d2bmb(P)-in-d2bmb(Q)}
Provided $P$ and $Q$ are designs,
\begin{align*}
	&(s, \{ s : State | (s, ac' \cup \{\bot\}) \in d2bmb(P)\}) \in d2bmb(Q)\\
	&=\\
	&(\lnot Q^f \implies Q^t)[\{ s : State | P^f \}/ac']
\end{align*}
\begin{proofs}\begin{proof}
\begin{flalign*}
	&(s, \{ s : State | (s, ac' \cup \{\bot\}) \in d2bmb(P)\}) \in d2bmb(Q)
	&&\ptext{\cref{law:s-ac'-cup-bot-in-d2bmb(P)}}\\
	&=(s, \{ s : State | P^f \}) \in d2bmb(Q)
	&&\ptext{Definition of $d2bmb$}\\
	&=(s,\{ s : State | P^f \}) \in \left\{\begin{array}{l}
		s : State, ss : \power State_\bot \\
		\left|\begin{array}{l}
			(\lnot Q^f \implies Q^t)[ss/ac'] \land \bot \notin ss)
			\\ \lor \\
			(Q^f[ss \setminus \{ \bot \}/ac'] \land \bot \in ss)
		\end{array}\right.
		\end{array}\right\}
	&&\ptext{Property of sets}\\
	&=\left(\begin{array}{l}
		((\lnot Q^f \implies Q^t)[ss/ac'][\{ s : State | P^f \}/ss] \land \bot \notin \{ s : State | P^f \})
		\\ \lor \\
		(Q^f[ss \setminus \{ \bot \}/ac'][\{ s : State | P^f \}/ss] \land \bot \in \{ s : State | P^f \})
	\end{array}\right)
	&&\ptext{Property of sets}\\
	&=(\lnot Q^f \implies Q^t)[ss/ac'][\{ s : State | P^f \}/ss]
	&&\ptext{Substitution}\\
	&=(\lnot Q^f \implies Q^t)[\{ s : State | P^f \}/ac']
\end{flalign*}
\end{proof}\end{proofs}
\end{lemma}

\begin{lemma}\label{law:s-ac'-in-d2bmb(P)-in-d2bmb(Q)}
Provided $P$ and $Q$ are designs,
\begin{align*}
	&(s, \{ s : State | (s, ac') \in d2bmb(P)\}) \in d2bmb(Q)\\
	&=\\
	&(\lnot Q^f \implies Q^t)[\{ s : State | (\lnot P^f \implies P^t) \}/ac']
\end{align*}
\begin{proofs}\begin{proof}
\begin{flalign*}
	&(s, \{ s : State | (s, ac') \in d2bmb(P)\}) \in d2bmb(Q)
	&&\ptext{\cref{law:s-ac'-in-d2bmb(P)}}\\
	&=(s, \{ s : State | (\lnot P^f \implies P^t) \}) \in d2bmb(Q)
	&&\ptext{Definition of $d2bmb$}\\
	&=(s, \{ s : State | (\lnot P^f \implies P^t) \}) \in
		\left\{
\right\}
	&&\ptext{Property of sets}\\
	&=\left(%
\right)
	&&\ptext{Property of sets and propositional calculus}\\
	&=(\lnot Q^f \implies Q^t)[ss/ac'][\{ s : State | (\lnot P^f \implies P^t) \}/ss]
	&&\ptext{Substitution}\\
	&=(\lnot Q^f \implies Q^t)[\{ s : State | (\lnot P^f \implies P^t) \}/ac']
\end{flalign*}
\end{proof}\end{proofs}
\end{lemma}

\begin{lemma}\label{law:bmb2d-B0-seq-B1}
\begin{align*}
	&bmb2d(B_0 \circseq B_1)\\
	&=\\
	&ok \implies \left(\begin{array}{l}
		((s, \{ s_1 : State | (s_1, ac') \in B_1\}) \in B_0 \land \bot \notin ac' \land ok')
		\\ \lor \\
		((s, \{ s_1 : State_\bot | true \}) \in B_0 \land \bot \notin ac')
		\\ \lor \\
		((s, \{ s_1 : State | (s_1, ac' \cup \{ \bot \}) \in B_1\}) \in B_0 \land \bot \notin ac')
	\end{array}\right)
\end{align*}
\begin{proofs}\begin{proof}
\begin{flalign*}
	&bmb2d(B_0 \circseq B_1)
	&&\ptext{Definition of $bmb2d$}\\
	&=ok \implies \left(\begin{array}{l}
		((s, ac') \in (B_0 \circseq B_1) \land \bot \notin ac' \land ok')
		\\ \lor \\
		((s, ac' \cup \{ \bot \}) \in (B_0 \circseq B_1) \land \bot \notin ac')
	\end{array}\right)
	&&\ptext{Definition of sequential composition}\\
	&=ok \implies \left(%
\right)
	&&\ptext{Property of sets and propositional calculus}\\
	&=ok \implies \left(\begin{array}{l}
		((s, \{ s_1 : State_\bot | true \}) \in B_0 \land \bot \notin ac' \land ok')
		\\ \lor \\
		((s, \{ s_1 : State | (s_1, ac') \in B_1\}) \in B_0 \land \bot \notin ac' \land ok')
		\\ \lor \\
		((s, \{ s_1 : State_\bot | true \}) \in B_0 \land \bot \notin ac')
		\\ \lor \\
		((s, \{ s_1 : State | (s_1, ac' \cup \{ \bot \}) \in B_1\}) \in B_0 \land \bot \notin ac')
	\end{array}\right)
	&&\ptext{Propositional calculus: absorption law}\\
	&=ok \implies \left(\begin{array}{l}
		((s, \{ s_1 : State | (s_1, ac') \in B_1\}) \in B_0 \land \bot \notin ac' \land ok')
		\\ \lor \\
		((s, \{ s_1 : State_\bot | true \}) \in B_0 \land \bot \notin ac')
		\\ \lor \\
		((s, \{ s_1 : State | (s_1, ac' \cup \{ \bot \}) \in B_1\}) \in B_0 \land \bot \notin ac')
	\end{array}\right)	
\end{flalign*}
\end{proof}\end{proofs}
\end{lemma}

\subsection{Isomorphism: $d2bmb$ and $bmb2d$}

\begin{theorem}\label{theorem:d2bmb-o-bmb2d}
\begin{statement}
Provided $B$ is $\mathbf{BMH0}$-$\mathbf{BMH2}$-healthy,
\begin{align*}
	&d2bmb \circ bmb2d(B) = B
\end{align*}
\end{statement}
\begin{proofs}
\begin{proof}
\begin{flalign*}
	&d2bmb \circ bmb2d(B)
	&&\ptext{Assumption: $B$ is $\mathbf{BMH0}$-$\mathbf{BMH2}$-healthy}\\
	&=d2bmb \circ bmb2d(\mathbf{bmh}_\mathbf{0,1,2} (B))
	&&\ptext{\cref{lemma:bmb2d-B-BMH0-2}}\\
	&=d2bmb\left(
\right)
			\end{array}\right)
		\end{array}\right.
	\end{array}\right\}
	&&\ptext{\cref{law:aux:bmh-0-1-2:circseq}}\\
	&=\mathbf{bmh}_\mathbf{0,1,2} (B)
	&&\ptext{Assumption: $B$ is $\mathbf{BMH0}$-$\mathbf{BMH2}$-healthy}\\
	&=B
\end{flalign*}
\end{proof}
\end{proofs}
\end{theorem}

\begin{theorem}\label{theorem:bmb2d-o-d2bmb}
\begin{statement}Provided $P$ is an $\mathbf{A}$-healthy design,
\begin{align*}
	&bmb2d \circ d2bmb(P) = P
\end{align*}
\end{statement}
\begin{proofs}
\begin{proof}
\begin{flalign*}
	&bmb2d \circ d2bmb(P)
	&&\ptext{Assumption: $P$ is $\mathbf{A}$-healthy}\\
	&=bmb2d \circ d2bmb(\mathbf{A} (P))
	&&\ptext{Definition of $bmb2d$}\\
	&=ok \implies \left(
\begin{array}{l}
		((s, ac') \in d2bmb(\mathbf{A} (P)) \land \bot \notin ac' \land ok')
		\\ \lor \\
		((s, ac' \cup \{ \bot \}) \in d2bmb(\mathbf{A} (P)) \land \bot \notin ac')
	\end{array}%
\right)
	&&\ptext{Property of sets}\\
	&=ok \implies \left(%
\right)
	&&\ptext{Property of sets and predicate calculus}\\
	&=ok \implies \left(%
\right)
	&&\ptext{Type restriction: $\bot \notin ac_0$ and property of sets}\\
	&=ok \implies \left(%
\right)
	&&\ptext{Definition of sequential composition}\\
	&=ok \implies \left(\begin{array}{l}
		((P^t \circseq ac \subseteq ac') \land ac'\neq\emptyset \land ok')
		\\ \lor \\
		(P^f \circseq ac \subseteq ac')
	\end{array}\right)
	&&\ptext{Predicate calculus}\\
	&=(ok \land \lnot (P^f \circseq ac \subseteq ac')) \implies ((P^t \circseq ac \subseteq ac') \land ac'\neq\emptyset \land ok')
	&&\ptext{Definition of design}\\
	&=(\lnot (P^f \circseq ac \subseteq ac') \vdash (P^t \circseq ac \subseteq ac') \land ac'\neq\emptyset)
	&&\ptext{Definition of $\mathbf{PBMH}$}\\
	&=(\lnot \mathbf{PBMH} (P^f) \vdash \mathbf{PBMH} (P^t) \land ac'\neq\emptyset)
	&&\ptext{Definition of $\mathbf{A}$}\\
	&=\mathbf{A} (P)
	&&\ptext{Assumption: $P$ is $\mathbf{A}$-healthy}\\
	&=P
\end{flalign*}
\end{proof}
\end{proofs}
\end{theorem}

\section{Refinement and Extreme Points}

\begin{theorem}\label{law:A:extreme-point:true}
\begin{statement}$\mathbf{A} (\botD) = \botD$\end{statement}
\begin{proofs}
\begin{proof}
\begin{flalign*}
	&\mathbf{A} (\botD)
	&&\ptext{Definition of $\botD$}\\
	&=\mathbf{A} (true)
	&&\ptext{Property of designs}\\
	&=\mathbf{A} (false \vdash true)
	&&\ptext{Definition of $\mathbf{A}$}\\
	&=(\lnot \mathbf{PBMH} (true) \vdash \mathbf{PBMH} (true) \land ac'\neq\emptyset)
	&&\ptext{Definition of $\mathbf{PBMH}$ and sequential composition}\\
	&=\left(\begin{array}{l}
		\exists ac_0, ok_0 \spot true[ac_0,ok_0/ac',ok'] \land ac_0 \subseteq ac' \\
		\vdash \\
		\exists ac_0, ok_0 \spot true[ac_0,ok_0/ac',ok'] \land ac_0 \subseteq ac' \land ac'\neq\emptyset
	\end{array}\right)\\
	&&\ptext{Property of substitution and propositional calculus}\\
	&=(false \vdash ac'\neq\emptyset)
	&&\ptext{Definition of design and propositional calculus}\\
	&=\botD
\end{flalign*}
\end{proof}
\end{proofs}
\end{theorem}\noindent

\begin{theorem}\label{law:A:extreme-point:not-ok}
\begin{statement}$\mathbf{A} (\topD) = \topD$\end{statement}
\begin{proofs}
\begin{proof}
\begin{flalign*}
	&\mathbf{A} (\topD)
	&&\ptext{Definition of $\topD$}\\
	&=\mathbf{A} (\lnot ok)
	&&\ptext{Property of designs}\\
	&=\mathbf{A} (true \vdash false)
	&&\ptext{Definition of $\mathbf{A}$}\\
	&=(\lnot \mathbf{PBMH} (false) \vdash \mathbf{PBMH} (false) \land ac'\neq\emptyset)
	&&\ptext{Definition of $\mathbf{PBMH}$}\\
	&=\left(\begin{array}{l}
		\exists ac_0, ok_0 \spot false[ac_0,ok_0/ac',ok'] \land ac_0 \subseteq ac' \\
		\vdash \\
		(\exists ac_0, ok_0 \spot false[ac_0,ok_0/ac',ok'] \land ac_0 \subseteq ac') \land ac'\neq\emptyset
	\end{array}\right)
	&&\ptext{Property of substitution and propositional calculus}\\
	&=(true \vdash false)
	&&\ptext{Property of designs and propositional calculus}\\
	&=\topD
\end{flalign*}
\end{proof}
\end{proofs}
\end{theorem}\noindent

\begin{theorem}\label{theorem:refinement-Dac-BMbot}
\begin{statement}
Provided $B_0$ and $B_1$ are $\mathbf{BMH0}$-$\mathbf{BMH2}$-healthy,
\begin{align*}
	&bmb2d(B_0) \refinedbyDac bmb2d(B_1) \iff B_0 \refinedbyBMbot B_1
\end{align*}
\end{statement}
\begin{proofs}
\begin{proof}
\begin{flalign*}
	&bmb2d(B_0) \refinedbyDac bmb2d(B_1)
	&&\ptext{Definition of $bmb2d$}\\
	&=\left(
\begin{array}{l}
		((s, ac' \cup \{\bot\}) \notin B_0 \vdash (s, ac') \in B_0)
		\\ \refinedbyDac \\
		((s, ac' \cup \{\bot\}) \notin B_1 \vdash (s, ac') \in B_1)
	\end{array}\right)
	&&\ptext{Refinement of designs}\\
	&=\left[\begin{array}{l}
		((s, ac' \cup \{\bot\}) \notin B_0 \land (s, ac') \in B_1) \implies (s, ac') \in B_0
		\\ \land \\
		(s, ac' \cup \{\bot\}) \notin B_0 \implies (s, ac' \cup \{\bot\}) \notin B_1
	\end{array}%
\right]
	&&\ptext{\cref{lemma:B1-subseteq-B0}}\\
	&=B_1 \subseteq B_0
	&&\ptext{Definition of $\refinedbyBMbot$}\\
	&=B_0 \refinedbyBMbot B_1
\end{flalign*}
\end{proof}
\end{proofs}
\end{theorem}

\begin{theorem}\label{theorem:A:refinement-order} 
\begin{statement}Provided that $P$ is an angelic design,
$\botDac \refinedbyDac P \refinedbyDac \topDac$
\end{statement}
\begin{proofs}
\begin{proof}
Follows from $\mathbf{A}$ monotonic, the definition of $\topDac$, $\botDac$ and the implication ordering.
\end{proof}
\end{proofs}
\end{theorem}\noindent

\begin{lemma}\label{lemma:H3-exists-ac':not-P}
$[(\exists ac' \spot P^f) = P^f] \iff [(\exists ac' \spot \lnot P^f) = \lnot P^f]$
\begin{proofs}\begin{proof}
\begin{flalign*}
	&[(\exists ac' \spot \lnot P^f) = \lnot P^f]
	&&\ptext{Universal quantification}\\
	&\iff \left(\begin{array}{l}
		(\forall ok, ok', ac', s \spot (\exists ac' \spot \lnot P^f) \implies \lnot P^f)
		\\ \land \\
		(\forall ok, ok', ac', s \spot \lnot P^f \implies (\exists ac' \spot \lnot P^f))
	\end{array}\right)
	&&\ptext{Predicate calculus}\\
	&\iff \forall ok, ok', ac', s \spot (\exists ac' \spot \lnot P^f) \implies \lnot P^f
	&&\ptext{Predicate calculus}\\
	&\iff \forall ok, s \spot (\exists ac' \spot \lnot P^f) \implies (\forall ac' \spot \lnot P^f)
	&&\ptext{Predicate calculus}\\
	&\iff \forall ok, s \spot \lnot (\forall ac' \spot \lnot P^f) \implies \lnot (\exists ac' \spot \lnot P^f)
	&&\ptext{Predicate calculus}\\
	&\iff \forall ok, s \spot (\exists ac' \spot P^f) \implies (\forall ac' \spot P^f)
	&&\ptext{Predicate calculus}\\
	&\iff \forall ok, s, ac', ok' \spot (\exists ac' \spot P^f) \implies P^f
	&&\ptext{Predicate calculus}\\
	&\iff \left(
\begin{array}{l}
		\iff \forall ok, s, ac', ok' \spot (\exists ac' \spot P^f) \implies P^f
		\\ \land \\
		\iff \forall ok, s, ac', ok' \spot P^f \implies (\exists ac' \spot P^f)
	\end{array}\right)
	&&\ptext{Universal quantification}\\
	&=[(\exists ac' \spot P^f) = P^f]
\end{flalign*}
\end{proof}\end{proofs}
\end{lemma}

\begin{lemma}\label{lemma:B1-subseteq-B0}
Provided $B_0$ and $B_1$ are of type ${BM}_\bot$,
\begin{align*}
	&\left[\begin{array}{l}
		(s, ac') \in B_1 \implies (s, ac') \in B_0
		\\ \land \\
		(s, ac' \cup \{\bot\}) \in B_1 \implies (s, ac' \cup \{\bot\}) \in B_0
	\end{array}\right] 
	\iff
	B_1 \subseteq B_0
\end{align*}
\begin{proofs}\begin{proof}
\begin{flalign*}
	&B_1 \subseteq B_0
	&&\ptext{Definition of subset inclusion}\\
	&\iff \forall s : State, ss : \power State_\bot \spot (s, ss) \in B_1 \implies (s, ss) \in B_0
	&&\ptext{Predicate calculus}\\
	&\iff \left(\begin{array}{l}
		\forall s : State, ss : \power State_\bot \spot \\
		((s, ss) \in B_1 \implies (s, ss) \in B_0) \land (\bot \in ss \lor \bot \notin ss)
		\end{array}%
\right)	
	&&\ptext{Predicate calculus: one-point rule}\\
	&\iff \left(%
\right)
	&&\ptext{Variable renaming and predicate calculus}\\
	&\iff \forall s : State, ac' : \power State \spot
		\left(\begin{array}{l}
		((s, ac' \cup \{\bot\}) \in B_1 \implies (s, ac' \cup \{\bot\}) \in B_0)
		\\ \land \\
		((s, ac') \in B_1 \implies (s, ac') \in B_0)
		\end{array}\right)
	&&\ptext{Universal quantification}\\
	&\iff 	\left[\begin{array}{l}
		((s, ac' \cup \{\bot\}) \in B_1 \implies (s, ac' \cup \{\bot\}) \in B_0)
		\\ \land \\
		((s, ac') \in B_1 \implies (s, ac') \in B_0)
	\end{array}\right]	
\end{flalign*}
\end{proof}\end{proofs}
\end{lemma}

\section{Operators}

\subsection{Sequential Composition}

\begin{theorem}\label{theorem:seqD:sequential-composition} 
\begin{statement}Provided $ok$ and $ok'$ are not free in $P$, $Q$, $R$ and $S$, and that $\lnot P$ and $Q$ are $\mathbf{PBMH}$-healthy,
\begin{align*}
	&(P \vdash Q) \seqDac (R \vdash S) = (\lnot (\lnot P \seqA true) \land \lnot (Q \seqA \lnot R) \vdash Q \seqA (R \implies S))
\end{align*}
\end{statement}
\begin{proofs}
\begin{proof}
\begin{xflalign*}
	&(P \vdash Q) \seqDac (R \vdash S)
	&&\ptext{Definition of $\seqDac$}\\
	&=\exists ok_0 \spot (P \vdash Q)[ok_0/ok'] \seqA (R \vdash S)[ok_0/ok]
	&&\ptext{Definition of design}\\
	&=\exists ok_0 \spot ((ok \land P) \implies (Q \land ok'))[ok_0/ok'] \seqA ((ok \land R) \implies (S \land ok'))[ok_0/ok]
	&&\ptext{Substitution and assumption}\\
	&=\exists ok_0 \spot ((ok \land P) \implies (Q \land ok_0)) \seqA ((ok_0 \land R) \implies (S \land ok'))
	&&\ptext{Case-analysis on $ok_0$ and predicate calculus}\\
	&=\left(\begin{array}{l}
		(((ok \land P) \implies Q) \seqA (R \implies (S \land ok')))
		\\ \lor \\
		(\lnot (ok \land P) \seqA true)
	\end{array}\right)
	&&\ptext{Predicate calculus}\\
	&=\left(\begin{array}{l}
		((\lnot ok \lor \lnot P \lor Q) \seqA (R \implies (S \land ok')))
		\\ \lor \\
		((\lnot ok \lor \lnot P) \seqA true)
	\end{array}\right)
	&&\ptext{Right-distributivity of $\seqA$ (\cref{law:seqA-right-distributivity})}\\
	&=\left(\begin{array}{l}
		(\lnot ok \seqA (R \implies (S \land ok')))
		\\ \lor \\
		(\lnot P \seqA (R \implies (S \land ok')))
		\\ \lor \\
		(Q \seqA (R \implies (S \land ok')))
		\\ \lor \\
		(\lnot ok \seqA true) \lor (\lnot P \seqA true)
	\end{array}\right)
	&&\ptext{\cref{law:seqA-ac'-not-free} and predicate calculus}\\
	&=\left(\begin{array}{l}
		\lnot ok \lor (\lnot P \seqA (R \implies (S \land ok')))
		\\ \lor \\
		(Q \seqA (R \implies (S \land ok')))
		\\ \lor \\
		(\lnot P \seqA true)
	\end{array}\right)
	&&\ptext{Assumption: $\lnot P$ is $\mathbf{PBMH}$-healthy and~\cref{theorem:(P-seqA-Q)-lor-(P-seqA-true):P-seqA-true}}\\
	&=\left(\begin{array}{l}
		\lnot ok \lor (Q \seqA (R \implies (S \land ok')))
		\\ \lor \\
		(\lnot P \seqA true)
	\end{array}\right)
	&&\ptext{Assumption: $Q$ is $\mathbf{PBMH}$-healthy and~\cref{theorem:P-seqA-(Q-implies-(R-land-ok'))}}\\
	&=\left(\begin{array}{l}
		\lnot ok \lor (Q \seqA \lnot R) \lor ((Q \seqA (R \implies S)) \land ok')
		\\ \lor \\
		(\lnot P \seqA true)
	\end{array}\right)
	&&\ptext{Predicate calculus}\\
	&=\left(\begin{array}{l}
		(ok \land \lnot (\lnot P \seqA true) \land \lnot (Q \seqA \lnot R))
		\\ \implies \\
		((Q \seqA (R \implies S)) \land ok')
	\end{array}\right)
	&&\ptext{Definition of design}\\
	&=\left(\begin{array}{l}
		\lnot (\lnot P \seqA true) \land \lnot (Q \seqA \lnot R)
		\\ \vdash \\
		Q \seqA (R \implies S)
	\end{array}\right)
\end{xflalign*}
\end{proof}
\end{proofs}
\end{theorem}

\begin{theorem}\label{theorem:seqD:sequential-composition:H3}
\begin{statement}
Provided $ok$ and $ok'$ are not free in $P$, $Q$, $R$ and $S$, and that $\lnot P$ and $Q$ are $\mathbf{PBMH}$-healthy, and that $ac'$ is not free in $P$,
\begin{align*}
	&(P \vdash Q) \seqDac (R \vdash S) = (P \land \lnot (Q \seqA \lnot R) \vdash Q \seqA (R \implies S))
\end{align*}
\end{statement}
\begin{proofs}
\begin{proof}
\begin{xflalign*}
	&(P \vdash Q) \seqDac (R \vdash S)
	&&\ptext{theorem:seqD:sequential-composition}\\
	&=(\lnot (\lnot P \seqA true) \land \lnot (Q \seqA \lnot R) \vdash Q \seqA (R \implies S))
	&&\ptext{Assumption: $ac'$ is not free in $P$ and~\cref{law:seqA-ac'-not-free}}\\
	&=(\lnot (\lnot P) \land \lnot (Q \seqA \lnot R) \vdash Q \seqA (R \implies S))
	&&\ptext{Predicate calculus}\\
	&=(P \land \lnot (Q \seqA \lnot R) \vdash Q \seqA (R \implies S))
\end{xflalign*}
\end{proof}
\end{proofs}
\end{theorem}

\begin{theorem}[$\seqDac$-$\mathbf{A}$-closure]\label{law:seqDac:closure}
\begin{statement}
Provided $P$ and $Q$ are $\mathbf{A}$-healthy and $ok, ok'$ are not free in $P$ and $Q$,
\begin{align*}
	&\mathbf{A} (P \seqDac Q) = P \seqDac Q	
\end{align*}
\end{statement}
\begin{proofs}
\begin{proof}
\begin{xflalign*}
	&P \seqDac Q
	&&\ptext{Assumption: $P$ and $Q$ are $\mathbf{A}$-healthy}\\
	&=\mathbf{A} (\lnot P^f \vdash P^t) \seqDac \mathbf{A} (\lnot Q^f \vdash Q^t)
	&&\ptext{Definition of $\mathbf{A}$}\\
	&=\left(
\begin{array}{l}
		(\lnot \mathbf{PBMH} (P^f) \vdash \mathbf{PBMH} (P^t) \land ac'\neq\emptyset) 
		\\ \seqDac \\
		(\lnot \mathbf{PBMH} (Q^f) \vdash \mathbf{PBMH} (Q^t) \land ac'\neq\emptyset)
	\end{array}\right)
	&&\ptext{Definition of $\seqDac$}\\
	&=\exists ok_0 \spot \left(\begin{array}{l}
		(\lnot \mathbf{PBMH} (P^f) \vdash \mathbf{PBMH} (P^t) \land ac'\neq\emptyset)[ok_0/ok'] 
		\\ \seqA \\
		(\lnot \mathbf{PBMH} (Q^f) \vdash \mathbf{PBMH} (Q^t) \land ac'\neq\emptyset)[ok_0/ok]
	\end{array}\right)
	&&\ptext{Definition of design}\\
	&=\exists ok_0 \spot \left(\begin{array}{l}
		((ok \land \lnot \mathbf{PBMH} (P^f)) \implies (\mathbf{PBMH} (P^t) \land ac'\neq\emptyset \land ok')[ok_0/ok'] 
		\\ \seqA \\
		((ok \land \lnot \mathbf{PBMH} (Q^f)) \implies (\mathbf{PBMH} (Q^t) \land ac'\neq\emptyset \land ok')[ok_0/ok]
	\end{array}\right)
	&&\ptext{Substitution and assumption}\\
	&=\exists ok_0 \spot \left(\begin{array}{l}
		((ok \land \lnot \mathbf{PBMH} (P^f)) \implies (\mathbf{PBMH} (P^t) \land ac'\neq\emptyset \land ok_0) 
		\\ \seqA \\
		((ok_0 \land \lnot \mathbf{PBMH} (Q^f)) \implies (\mathbf{PBMH} (Q^t) \land ac'\neq\emptyset \land ok')
	\end{array}\right)
	&&\ptext{Case-analysis on $ok_0$ and predicate calculus}\\
	&=\left(%
\right)
	&&\ptext{\cref{law:seqA-ac'-not-free} and predicate calculus}\\
	&=\left(%
\right)
	&&\ptext{Definition of design}\\
	&=\left(%
\right)
	&&\ptext{Predicate calculus and~\cref{law:pbmh:disjunction-closure}}\\
	&=\mathbf{A0} \left(%
\right)
	&&\ptext{Definition of $\mathbf{A1}$ and predicate calculus}\\
	&=\mathbf{A0}\circ\mathbf{A1} \left(%
\right)
	&&\ptext{Definition of $\mathbf{A}$}\\
	&=\mathbf{A} (\mathbf{A} (\lnot P^f \vdash P^t) \seqDac \mathbf{A} (\lnot Q^f \vdash Q^t))
	&&\ptext{Assumption: $P$ and $Q$ are $\mathbf{A}$-healthy}\\
	&=\mathbf{A} (P \seqDac Q)
\end{xflalign*}
\end{proof}
\end{proofs}
\end{theorem}

\subsubsection{Relationship with Extended Binary Multirelations}

\begin{theorem}\label{theorem:bmb2d:seqBMbot}
\begin{statement}
Provided $P$ and $Q$ are $\mathbf{A}$-healthy designs,
\begin{align*}
	&bmb2d(d2bmb(P) \seqBMbot d2bmb(Q)) = P \seqDac Q
\end{align*}
\end{statement}
\begin{proofs}
\begin{proof}
\begin{xflalign*}
	&bmb2d(d2bmb(P) \seqBMbot d2bmb(Q))
	&&\ptext{\cref{law:bmb2d-B0-seq-B1}}\\
	&=ok \implies \left(
\begin{array}{l}
		((s, \{ s_1 : State | (s_1, ac') \in d2bmb(Q)\}) \in d2bmb(P) \land \bot \notin ac' \land ok')
		\\ \lor \\
		((s, \{ s_1 : State_\bot | true \}) \in d2bmb(P) \land \bot \notin ac')
		\\ \lor \\
		((s, \{ s_1 : State | (s_1, ac' \cup \{ \bot \}) \in d2bmb(Q)\}) \in d2bmb(P) \land \bot \notin ac')
	\end{array}\right)
	&&\ptext{\cref{law:s-ac'-in-d2bmb(P)-in-d2bmb(Q)}}\\
	&=ok \implies \left(\begin{array}{l}
		((\lnot P^f \implies P^t)[\{ s : State | (\lnot Q^f \implies Q^t) \}/ac'] \land \bot \notin ac' \land ok')
		\\ \lor \\
		((s, \{ s_1 : State_\bot | true \}) \in d2bmb(P) \land \bot \notin ac')
		\\ \lor \\
		((s, \{ s_1 : State | (s_1, ac' \cup \{ \bot \}) \in d2bmb(Q)\}) \in d2bmb(P) \land \bot \notin ac')
	\end{array}\right)
	&&\ptext{\cref{law:s-ac'-cup-in-d2bmb(P)-in-d2bmb(Q)}}\\
	&=ok \implies \left(\begin{array}{l}
		((\lnot P^f \implies P^t)[\{ s : State | (\lnot Q^f \implies Q^t) \}/ac'] \land \bot \notin ac' \land ok')
		\\ \lor \\
		((s, \{ s_1 : State_\bot | true \}) \in d2bmb(P) \land \bot \notin ac')
		\\ \lor \\
		((\lnot P^f \implies P^t)[\{ s : State | Q^f \}/ac'] \land \bot \notin ac')
	\end{array}\right)
	&&\ptext{\cref{law:s:bot:true:in-d2bmb(P)}}\\
	&=ok \implies \left(\begin{array}{l}
		((\lnot P^f \implies P^t)[\{ s : State | (\lnot Q^f \implies Q^t) \}/ac'] \land \bot \notin ac' \land ok')
		\\ \lor \\
		(P^f[\{ s_1 : State | true \}/ac'] \land \bot \notin ac')
		\\ \lor \\
		((\lnot P^f \implies P^t)[\{ s : State | Q^f \}/ac'] \land \bot \notin ac')
	\end{array}\right)
	&&\ptext{Assumption: $\bot \notin ac'$}\\
	&=ok \implies \left(\begin{array}{l}
		((\lnot P^f \implies P^t)[\{ s : State | (\lnot Q^f \implies Q^t) \}/ac'] \land ok')
		\\ \lor \\
		(P^f[\{ s_1 : State | true \}/ac'])
		\\ \lor \\
		((\lnot P^f \implies P^t)[\{ s : State | Q^f \}/ac'])
	\end{array}\right)
	&&\ptext{Definition of $\seqA$}\\
	&=ok \implies \left(\begin{array}{l}
		(((\lnot P^f \implies P^t) \seqA (\lnot Q^f \implies Q^t)) \land ok')
		\\ \lor \\
		(P^f \seqA true)
		\\ \lor \\
		((\lnot P^f \implies P^t) \seqA Q^f)
	\end{array}\right)
	&&\ptext{Predicate calculus and~\cref{law:seqA-right-distributivity}}\\
	&=ok \implies \left(\begin{array}{l}
		(((P^f \seqA (\lnot Q^f \implies Q^t)) \lor (P^t \seqA (\lnot Q^f \implies Q^t))) \land ok')
		\\ \lor \\
		(P^f \seqA true)
		\\ \lor \\
		(P^f \seqA Q^f) \lor (P^t \seqA Q^f)
	\end{array}\right)
	&&\ptext{Predicate calculus}\\
	&=ok \implies \left(\begin{array}{l}
		((P^f \seqA (\lnot Q^f \implies Q^t)) \land ok')
		\\ \lor \\
		((P^t \seqA (\lnot Q^f \implies Q^t)) \land ok')
		\\ \lor \\
		(P^f \seqA true) \lor (P^f \seqA Q^f) \lor (P^t \seqA Q^f)
	\end{array}\right)
	&&\ptext{Predicate calculus}\\
	&=ok \implies \left(\begin{array}{l}
		(((P^f \seqA (\lnot Q^f \implies Q^t)) \lor (P^f \seqA true)) \land ok')
		\\ \lor \\
		((P^t \seqA (\lnot Q^f \implies Q^t)) \land ok')
		\\ \lor \\
		(P^f \seqA true) \lor (P^f \seqA Q^f) \lor (P^t \seqA Q^f)
	\end{array}\right)
	&&\ptext{Assumption: $P$ is $\mathbf{PBMH}$-healthy and \cref{lemma:PBMH(P)-ow:PBMH(P-ow),theorem:(P-seqA-Q)-lor-(P-seqA-true):P-seqA-true}}\\
	&=ok \implies \left(\begin{array}{l}
		((P^f \seqA true) \land ok')
		\\ \lor \\
		((P^t \seqA (\lnot Q^f \implies Q^t)) \land ok')
		\\ \lor \\
		(P^f \seqA true) \lor (P^t \seqA Q^f)
	\end{array}\right)
	&&\ptext{Predicate calculus: absorption law}\\
	&=ok \implies \left(\begin{array}{l}
		((P^t \seqA (\lnot Q^f \implies Q^t)) \land ok')
		\\ \lor \\
		(P^f \seqA true) \lor (P^t \seqA Q^f)
	\end{array}\right)
	&&\ptext{Predicate calculus}\\
	&=\left(\begin{array}{l}
		(ok \land \lnot (P^f \seqA true) \land \lnot (P^t \seqA Q^f))
		\\ \implies \\
		((P^t \seqA (\lnot Q^f \implies Q^t)) \land ok')
	\end{array}\right)
	&&\ptext{Definition of design}\\
	&=\left(\begin{array}{l}
		\lnot (P^t \seqA Q^f) \land \lnot (P^f \seqA true)
		\\ \vdash \\
		P^t \seqA (\lnot Q^f \implies Q^t)
	\end{array}\right)
	&&\ptext{\cref{theorem:seqD:sequential-composition}}\\
	&=(\lnot P^f \vdash P^t) \seqDac (\lnot Q^f \vdash Q^t)
	&&\ptext{Assumption: $P$ and $Q$ are $\mathbf{A}$-healthy designs}\\
	&=P \seqDac Q	
\end{xflalign*}
\end{proof}
\end{proofs}
\end{theorem}

\subsubsection{Skip}

\begin{theorem}\label{theorem:A(IIDac):IIDac}
\begin{statement}
$\mathbf{A} (\IIDac) = \IIDac$
\end{statement}
\begin{proofs}
\begin{proof}
\begin{flalign*}
	&\mathbf{A} (\IIDac)
	&&\ptext{Definition of $\IIDac$}\\
	&=\mathbf{A} (true \vdash s \in ac')
	&&\ptext{Definition of $\mathbf{A}$}\\
	&=(\lnot \mathbf{PBMH} (\lnot true) \vdash \mathbf{PBMH} (s \in ac') \land ac'\neq\emptyset)
	&&\ptext{\cref{law:pbmh:false}}\\
	&=(\lnot false \vdash \mathbf{PBMH} (s \in ac') \land ac'\neq\emptyset)
	&&\ptext{\cref{law:pbmh:s-in-ac'}}\\
	&=(\lnot false \vdash s \in ac' \land ac'\neq\emptyset)
	&&\ptext{Property of sets and predicate calculus}\\
	&=(true \vdash s \in ac')
	&&\ptext{Definition of $\IIDac$}\\
	&=\IIDac
\end{flalign*}
\end{proof}
\end{proofs}
\end{theorem}

\begin{theorem}\label{theorem:IIDac-seqDac-P:P}
\begin{statement}
Provided $P$ is a design,
$\IIDac \seqDac P= P$
\end{statement}
\begin{proofs}
\begin{proof}
\begin{flalign*}
	&\IIDac \seqDac P
	&&\ptext{Definition of $\IIDac$ and design}\\
	&=(true \vdash s \in ac') \seqDac (\lnot P^f \vdash P^t)
	&&\ptext{\cref{theorem:seqD:sequential-composition}}\\
	&=(\lnot (\lnot true \seqA true) \land \lnot (s \in ac' \seqA P^f) \vdash s \in ac' \seqA (\lnot P^f \implies P^t))
	&&\ptext{Predicate calculus}\\
	&=(\lnot (false \seqA true) \land \lnot (s \in ac' \seqA P^f) \vdash s \in ac' \seqA (\lnot P^f \implies P^t))
	&&\ptext{Definition of $\seqA$ and substitution}\\
	&=(\lnot false \land \lnot (s \in ac' \seqA P^f) \vdash s \in ac' \seqA (\lnot P^f \implies P^t))
	&&\ptext{Predicate calculus}\\
	&=(\lnot (s \in ac' \seqA P^f) \vdash s \in ac' \seqA (\lnot P^f \implies P^t))
	&&\ptext{\cref{law:seqA:IIA:left-unit}}\\
	&=(\lnot P^f \vdash (\lnot P^f \implies P^t))
	&&\ptext{Predicate calculus}\\
	&=(\lnot P^f \vdash P^t)
	&&\ptext{Definition of design}\\
	&=P
\end{flalign*}
\end{proof}
\end{proofs}
\end{theorem}

\begin{theorem}\label{law:seqD-sequence-Skip}
\begin{statement}
Provided $P$ is an $\mathbf{A}$-healthy design,
\begin{align*}
	&P \seqDac \IIDac = ((\lnot \exists ac' \spot P^f) \vdash P^t)
\end{align*}
\end{statement}
\begin{proofs}
\begin{proof}
\begin{flalign*}
	&P \seqDac \IIDac
	&&\ptext{Definition of design and $\IIDac$}\\
	&=(\lnot P^f \vdash P^t) \seqDac (true \vdash s \in ac')
	&&\ptext{\cref{theorem:seqD:sequential-composition}}\\
	&=(\lnot (P^f \seqA true) \land \lnot (P^t \seqA false) \vdash P^t \seqA (true \implies s \in ac'))
	&&\ptext{Predicate calculus}\\
	&=(\lnot (P^f \seqA true) \land \lnot (P^t \seqA false) \vdash P^t \seqA s \in ac')
	&&\ptext{Assumption: $P$ is $\mathbf{A}$-healthy}\\
	&=\left(\begin{array}{l}
		\lnot (P^f \seqA true) \land \lnot ((P^t \land ac'\neq\emptyset) \seqA false)
		\\ \vdash \\
		(P^t \land ac'\neq\emptyset) \seqA (true \implies s \in ac')
	\end{array}\right)
	&&\ptext{Right-distributivity of $\seqA$ (\cref{law:seqA-right-distributivity-conjunction})}\\
	&=\left(\begin{array}{l}
		\lnot (P^f \seqA true) \land \lnot ((P^t \seqA false) \land (ac'\neq\emptyset \seqA false))
		\\ \vdash \\
		(P^t \seqA s \in ac') \land (ac'\neq\emptyset \seqA s \in ac')
	\end{array}\right)
	&&\ptext{Definition of $\seqA$ and substitution}\\
	&=\left(\begin{array}{l}
		\lnot (P^f \seqA true) \land \lnot ((P^t \seqA false) \land \emptyset\neq\emptyset)
		\\ \vdash \\
		(P^t \seqA s \in ac') \land (ac'\neq\emptyset \seqA s \in ac')
	\end{array}\right)
	&&\ptext{Property of sets and predicate calculus}\\
	&=(\lnot (P^f \seqA true) \vdash (P^t \seqA s \in ac') \land (ac'\neq\emptyset \seqA s \in ac'))
	&&\ptext{$s \in ac'$ is right-unit of $\seqA$ (\cref{law:seqA:IIA:right-unit})}\\
	&=(\lnot (P^f \seqA true) \vdash P^t \land ac'\neq\emptyset)
	&&\ptext{\cref{law:seqA-P-sequence-true}}\\
	&=(\lnot \exists ac' \spot P^f \vdash P^t \land ac'\neq\emptyset)
	&&\ptext{Assumption: $P$ is $\mathbf{A}$-healthy}\\
	&=(\lnot \exists ac' \spot P^f \vdash P^t)
\end{flalign*}
\end{proof}
\end{proofs}
\end{theorem}

\begin{theorem}\label{law:seqDac:H3}
\begin{statement}
Provided $P$ is an $\mathbf{A}$-healthy design, it is $\mathbf{H3}$-healthy if, and only if, its precondition does not mention $ac'$,
\begin{align*}
	&(P \seqDac \IIDac) = P \iff ((\exists ac' \spot \lnot P^f) = \lnot P^f)
\end{align*}
\end{statement}
\begin{proofs}
\begin{proof}
\begin{flalign*}
	&(P \seqDac \IIDac) = P
	&&\ptext{Assumption: $P$ is $\mathbf{A}$-healthy}\\
	&\iff (P \seqDac \IID) = (\lnot P^f \vdash P^t \land ac'\neq\emptyset)
	&&\ptext{\cref{law:seqD-sequence-Skip}}\\
	&\iff (\lnot \exists ac' \spot P^f \vdash P^t \land ac'\neq\emptyset) = (\lnot P^f \vdash P^t \land ac'\neq\emptyset)
	&&\ptext{Equality of designs}\\
	&\iff [(\lnot \exists ac' \spot P^f) = \lnot P^f]
	&&\ptext{Predicate calculus}\\
	&\iff [(\exists ac' \spot P^f) = P^f]
	&&\ptext{Predicate calculus (\cref{lemma:H3-exists-ac':not-P})}\\
	&\iff [(\exists ac' \spot \lnot P^f) = \lnot P^f]
\end{flalign*}
\end{proof}
\end{proofs}
\end{theorem}

\subsubsection{Properties with respect to the Extreme Points}

\begin{theorem}\label{law:seqD:bot-P}
\begin{statement}$\botD \seqDac P = \botD$\end{statement}
\begin{proofs}
\begin{proof}
\begin{flalign*}
	&\botD \seqDac P
	&&\ptext{Definition of $\botD$}\\
	&=true \seqDac P
	&&\ptext{Definition of $\seqDac$}\\
	&=\exists ok_0 \spot true[ok_0/ok'] \seqA P[ok_0/ok]
	&&\ptext{Case-split on $ok_0$ and property of substitution}\\
	&=(true \seqA P[true/ok]) \lor (true \seqA P[false/ok])
	&&\ptext{Definition of $\seqA$}\\
	&=true \lor true
	&&\ptext{Propositional calculus and definition of $\botD$}\\
	&=\botD
\end{flalign*}
\end{proof}
\end{proofs}
\end{theorem}

\begin{theorem}\label{law:seqD:top-P}
\begin{statement}$\topD \seqDac P = \topD$\end{statement}
\begin{proofs}
\begin{proof}
\begin{flalign*}
	&\topD \seqDac P
	&&\ptext{Definition of $\topD$}\\
	&=(\lnot ok) \seqDac P
	&&\ptext{Definition of $\seqDac$}\\
	&=\exists ok_0 \spot (\lnot ok)[ok_0/ok'] \seqA P[ok_0/ok]
	&&\ptext{Substitution and case-split on $ok_0$}\\
	&=(\lnot ok \seqA P[true/ok]) \lor (\lnot ok \seqA P[false/ok])
	&&\ptext{Definition of $\seqA$ and substitution}\\
	&=\lnot ok
	&&\ptext{Definition of $\topD$}\\
	&=\topD
\end{flalign*}
\end{proof}
\end{proofs}
\end{theorem}

\subsubsection{Properties with respect to $\mathbf{A2}$}

\begin{theorem}\label{theorem:A2(P-seqDac-Q):P-seqDac-Q}
\begin{statement}
Provided $P$ and $Q$ are $\mathbf{A2}$-healthy,
$\mathbf{A2} (P \seqDac Q) = P \seqDac Q$
\end{statement}
\begin{proofs}
\begin{proof}\checkt{alcc}
\begin{xflalign*}
	&P \seqDac Q
	&&\ptext{Assumption: $P$ and $Q$ are $\mathbf{A2}$-healthy}\\
	&=\mathbf{A2} (P) \seqDac \mathbf{A2} (Q)
	&&\ptext{Definition of $\seqDac$}\\
	&=\exists ok_0 @ \mathbf{A2} (P)[ok_0/ok'] \seqA \mathbf{A2} (Q)[ok_0/ok]
	&&\ptext{\cref{lemma:A2(P)-o-w:A2(P-o-w),lemma:A2(P)-o-ok:A2(P-o-ok)}}\\
	&=\exists ok_0 @ \mathbf{A2} (P[ok_0/ok']) \seqA \mathbf{A2} (Q[ok_0/ok])
	&&\ptext{\cref{lemma:A2(A2(P)-seqA-A2(Q)):A2(P)-seqA-A2(Q)}}\\
	&=\exists ok_0 @ \mathbf{A2} (\mathbf{A2} (P[ok_0/ok']) \seqA \mathbf{A2} (Q[ok_0/ok]))
	&&\ptext{\cref{lemma:A2(exists-x-P):exists-x-A2(P)}}\\
	&=\mathbf{A2} (\exists ok_0 @ \mathbf{A2} (P[ok_0/ok']) \seqA \mathbf{A2} (Q[ok_0/ok]))
	&&\ptext{\cref{lemma:A2(P)-o-w:A2(P-o-w),lemma:A2(P)-o-ok:A2(P-o-ok)}}\\
	&=\mathbf{A2} (\exists ok_0 @ \mathbf{A2} (P)[ok_0/ok'] \seqA \mathbf{A2} (Q)[ok_0/ok])
	&&\ptext{Definition of $\seqDac$}\\
	&=\mathbf{A2} (\mathbf{A2} (P) \seqDac \mathbf{A2} (Q))
	&&\ptext{Assumption: $P$ and $Q$ are $\mathbf{A2}$-healthy}\\
	&=\mathbf{A2} (P \seqDac Q)
\end{xflalign*}
\end{proof}
\end{proofs}
\end{theorem}

\subsubsection{Other Properties}

\begin{lemma}\label{lemma:P-seqDac-Q:implies:P-seqA-(exists-ok-Q)}
\begin{statement}
Provided $P$ is $\mathbf{PBMH}$-healthy and $ok'$ is not free in $P$.
\begin{align*}
	&P \seqDac Q \implies P \seqA (\exists ok @ Q)
\end{align*}
\end{statement}
\begin{proofs}
\begin{proof}
\begin{xflalign*}
	&P \seqDac Q
	&&\ptext{Definition of $\seqDac$}\\
	&=\exists ok_0 @ P[ok_0/ok'] \seqA Q[ok_0/ok]
	&&\ptext{Assumption: $ok'$ is not free in $P$}\\
	&=\exists ok_0 @ P \seqA Q[ok_0/ok]
	&&\ptext{Assumption: $P$ is $\mathbf{PBMH}$-healthy}\\
	&=\exists ok_0 @ \mathbf{PBMH} (P) \seqA Q[ok_0/ok]
	&&\ptext{Definition of $\mathbf{PBMH}$ (\cref{lemma:PBMH:alternative-1})}\\
	&=\exists ok_0 @ (\exists ac_0 @ P[ac_0/ac'] \land ac_0 \subseteq ac') \seqA Q[ok_0/ok]
	&&\ptext{Definition of $\seqA$ and substitution}\\
	&=\exists ok_0 @ \exists ac_0 @ P[ac_0/ac'] \land ac_0 \subseteq \{ s | Q[ok_0/ok] \}
	&&\ptext{Predicate calculus}\\
	&=\exists ac_0 @ P[ac_0/ac'] \land \exists ok_0 @ ac_0 \subseteq \{ s | Q[ok_0/ok] \}
	&&\ptext{Property of sets}\\
	&=\exists ac_0 @ P[ac_0/ac'] \land \exists ok_0 @ (\forall z @ z \in ac_0 \implies z \in \{ s | Q[ok_0/ok] \})
	&&\ptext{Predicate calculus}\\
	&\implies \exists ac_0 @ P[ac_0/ac'] \land \forall z @ \exists ok_0 @ (z \in ac_0 \implies z \in \{ s | Q[ok_0/ok] \})
	&&\ptext{Predicate calculus}\\
	&=\exists ac_0 @ P[ac_0/ac'] \land \forall z @ z \in ac_0 \implies (\exists ok_0 @ z \in \{ s | Q[ok_0/ok] \})
	&&\ptext{Property of sets}\\
	&=\exists ac_0 @ P[ac_0/ac'] \land \forall z @ z \in ac_0 \implies (\exists ok_0 @ Q[ok_0/ok][z/s])
	&&\ptext{Substitution}\\
	&=\exists ac_0 @ P[ac_0/ac'] \land \forall z @ z \in ac_0 \implies ((\exists ok_0 @ Q[ok_0/ok])[z/s])
	&&\ptext{Property of sets}\\
	&=\exists ac_0 @ P[ac_0/ac'] \land \forall z @ z \in ac_0 \implies z \in \{ s | \exists ok_0 @ Q[ok_0/ok]\}
	&&\ptext{Property of sets}\\
	&=\exists ac_0 @ P[ac_0/ac'] \land ac_0 \subseteq \{ s | \exists ok_0 @ Q[ok_0/ok]\}
	&&\ptext{Definition of $\seqA$ and substitution}\\
	&=(\exists ac_0 @ P[ac_0/ac'] \land ac_0 \subseteq ac') \seqA (\exists ok_0 @ Q[ok_0/ok])
	&&\ptext{Definition of $\mathbf{PBMH}$ (\cref{lemma:PBMH:alternative-1})}\\
	&=\mathbf{PBMH} (P) \seqA (\exists ok_0 @ Q[ok_0/ok])
	&&\ptext{Assumption: $P$ is $\mathbf{PBMH}$-healthy}\\
	&=P \seqA (\exists ok_0 @ Q[ok_0/ok])
	&&\ptext{Predicate calculus}\\
	&=P \seqA (\exists ok @ Q)
\end{xflalign*}
\end{proof}
\end{proofs}
\end{lemma}

\subsection{Demonic Choice}

\subsubsection{Properties}

\begin{theorem}\label{law:A:distribute-disjunction}
\begin{statement}
Provided $P$ and $Q$ are designs,
\begin{align*}
	&\mathbf{A} (P \lor Q) = \mathbf{A} (P) \lor \mathbf{A} (Q)
\end{align*}
\end{statement}
\begin{proofs}
\begin{proof}
\begin{flalign*}
	&\mathbf{A} (P \lor Q)
	&&\ptext{Definition of design}\\
	&=\mathbf{A} ((\lnot P^f \vdash P^t) \lor (\lnot Q^f \vdash Q^t))
	&&\ptext{Disjunction of designs}\\
	&=\mathbf{A} (\lnot P^f \land \lnot Q^f \vdash P^t \lor Q^t)
	&&\ptext{Predicate calculus}\\
	&=\mathbf{A} (\lnot (P^f \lor Q^f) \vdash P^t \lor Q^t)
	&&\ptext{Definition of $\mathbf{A}$}\\
	&=(\lnot \mathbf{PBMH} (P^f \lor Q^f) \vdash \mathbf{PBMH} (P^t \lor Q^t) \land ac'\neq\emptyset)
	&&\ptext{Distributivity of $\mathbf{PBMH}$ w.r.t. disjunction~\cref{law:pbmh:distribute-disjunction}}\\
	&=\left(\begin{array}{l}
		\lnot (\mathbf{PBMH} (P^f) \lor \mathbf{PBMH} (Q^f))
		\\ \vdash \\
		(\mathbf{PBMH} (P^t) \lor \mathbf{PBMH} (Q^t)) \land ac'\neq\emptyset
	\end{array}\right)
	&&\ptext{Predicate calculus}\\
	&=\left(\begin{array}{l}
		\lnot \mathbf{PBMH} (P^f) \land \lnot \mathbf{PBMH} (Q^f) 
		\\ \vdash \\
		(\mathbf{PBMH} (P^t) \land ac'\neq\emptyset) \lor (\mathbf{PBMH} (Q^t) \land ac'\neq\emptyset)
	\end{array}\right)
	&&\ptext{Disjunction of designs}\\
	&=\left(\begin{array}{l}
		(\lnot \mathbf{PBMH} (P^f) \vdash \mathbf{PBMH} (P^t) \land ac'\neq\emptyset)
		\\ \lor \\
		(\lnot \mathbf{PBMH} (Q^f) \vdash \mathbf{PBMH} (Q^t) \land ac'\neq\emptyset)
	\end{array}\right)
	&&\ptext{Definition of $\mathbf{A}$}\\
	&=\mathbf{A} (\lnot P^f \vdash P^t) \lor \mathbf{A} (\lnot Q^f \vdash Q^t)
\end{flalign*}
\end{proof}
\end{proofs}
\end{theorem}

\begin{theorem}\label{theorem:A:disjunction-closure}
\begin{statement}
Provided $P$ and $Q$ are $\mathbf{A}$-healthy designs,
\begin{align*}
	&\mathbf{A} (P \sqcapDac Q) = P \sqcapDac Q
\end{align*}
\end{statement}
\begin{proofs}\begin{proof}
\begin{flalign*}
	&\mathbf{A} (P \sqcapDac Q)
	&&\ptext{Definition of $\sqcapDac$ and~\cref{law:A:distribute-disjunction}}\\
	&=\mathbf{A} (P) \lor \mathbf{A} (Q)
	&&\ptext{Assumption: $P$ and $Q$ are $\mathbf{A}$-healthy}\\
	&=P \sqcapDac Q
\end{flalign*}
\end{proof}\end{proofs}
\end{theorem}

\subsubsection{Relationship with Extended Binary Multirelations}

\begin{theorem}\label{theorem:bmb2p:demonic-choice}
\begin{statement}
$bmb2p(B_0 \sqcapBMbot B_1) = bmb2p(B_0) \sqcapDac bmb2p(B_1)$
\end{statement}
\begin{proofs}
\begin{proof}
\begin{flalign*}
	&bmb2p(B_0 \sqcapBMbot B_1)
	&&\ptext{Definition of $\sqcapBMbot$}\\
	&=bmb2p(B_0 \cup B_1)
	&&\ptext{Definition of $bmb2p$}\\
	&=ok \implies \left(
\begin{array}{l}
		((s, ac') \in (B_0 \cup B_1) \land \bot \notin ac' \land ok')
		\\ \lor \\
		((s, ac' \cup \{ \bot \}) \in (B_0 \cup B_1) \land \bot \notin ac')
	\end{array}\right)
	&&\ptext{Property of sets}\\
	&=ok \implies \left(\begin{array}{l}
		(((s, ac') \in B_0 \lor (s, ac') \in B_1) \land \bot \notin ac' \land ok')
		\\ \lor \\
		((((s, ac' \cup \{ \bot \}) \in B_0) \lor (s, ac' \cup \{ \bot \}) \in B_1) \land \bot \notin ac')
	\end{array}%
\right)
	&&\ptext{Property of designs}\\
	&=\left(%
\right)
	&&\ptext{Disjunction of designs and definition of $\sqcapDac$}\\
	&=\left(\begin{array}{l}
		(\lnot ((s, ac' \cup \{ \bot \}) \in B_0 \land \bot \notin ac') \vdash (s, ac') \in B_0 \land \bot \notin ac')
		\\ \sqcapDac \\
		(\lnot ((s, ac' \cup \{ \bot \}) \in B_1 \land \bot \notin ac') \vdash (s, ac') \in B_1 \land \bot \notin ac')
	\end{array}\right)
	&&\ptext{Definition of $bmb2p$}\\
	&=bmb2p(B_0) \sqcapDac bmb2p(B_1)
\end{flalign*}
\end{proof}
\end{proofs}
\end{theorem}

\subsubsection{Other Properties}

\begin{theorem}\label{theorem:P-sqcapDac-botDac:botDac}
\begin{statement}$P \sqcapDac \botD = \botD$\end{statement}
\begin{proofs}
\begin{proof}
\begin{flalign*}
	&P \sqcapDac \botD
	&&\ptext{Definition of $\sqcapDac$ and $\botD$}&\\
	&=P \lor true
	&&\ptext{Propositional calculus and definition of $\botD$}\\
	&=\botD
\end{flalign*}
\end{proof}
\end{proofs}
\end{theorem}\noindent

\begin{theorem}\label{law:seqDac:sqcap-right-distributivity}
\begin{statement}$(P \sqcapDac Q) \seqDac R = (P \seqDac R) \sqcapDac (Q \seqDac R)$
\end{statement}
\begin{proofs}
\begin{proof}
\begin{flalign*}
	&(P \seqDac R) \sqcapDac (Q \seqDac R)
	&&\ptext{Definition of $\seqDac$ and $\sqcapDac$}\\
	&=(\exists ok_0 \spot P[ok_0/ok'] \seqA R[ok_0/ok']) \lor (\exists ok_0 \spot Q[ok_0/ok'] \seqA R[ok_0/ok])
	&&\ptext{Propositional calculus}\\
	&=\exists ok_0 \spot (P[ok_0/ok'] \seqA R[ok_0/ok']) \lor (Q[ok_0/ok'] \seqA R[ok_0/ok])
	&&\ptext{Right-distributivity of $\seqA$ (\cref{law:seqA-right-distributivity})}\\
	&=\exists ok_0 \spot ((P[ok_0/ok'] \lor Q[ok_0/ok']) \seqA R[ok_0/ok])
	&&\ptext{Definition of $\seqA$ and $\sqcapDac$}\\
	&=(P \sqcapDac Q) \seqDac R
\end{flalign*}
\end{proof}
\end{proofs}
\end{theorem}

\subsubsection{Other Properties}

\begin{lemma}\label{lemma:P-seqDac-Q:implies:R-seqDac-Q}
\begin{statement}
Provided $P \implies R$,
$P \seqDac Q \implies R \seqDac Q$.
\end{statement}
\begin{proofs}\begin{proof}\checkt{alcc}
\begin{xflalign*}
	&P \seqDac Q
	&&\ptext{Definition of $\seqDac$}\\
	&=\exists ok_0 @ P[ok_0/ok'] \seqA Q[ok_0/ok]
	&&\ptext{Assumption: $P \implies R$}\\
	&=\exists ok_0 @ (P \land R)[ok_0/ok'] \seqA Q[ok_0/ok]
	&&\ptext{Substitution}\\
	&=\exists ok_0 @ (P[ok_0/ok'] \land R[ok_0/ok']) \seqA Q[ok_0/ok]
	&&\ptext{\cref{law:seqA-right-distributivity-conjunction}}\\
	&=\exists ok_0 @ (P[ok_0/ok'] \seqA Q[ok_0/ok]) \land (R[ok_0/ok'] \seqA Q[ok_0/ok])
	&&\ptext{Predicate calculus}\\
	&\implies \exists ok_0 @ (R[ok_0/ok'] \seqA Q[ok_0/ok])
	&&\ptext{Definition of $\seqDac$}\\
	&=R \seqDac Q
\end{xflalign*}
\end{proof}\end{proofs}
\end{lemma}

\begin{lemma}\label{lemma:P-seqDac-Q:implies:P-seqDac-R}
\begin{statement}
Provided $Q \implies R$,
$P \seqDac Q \implies P \seqDac R$.
\end{statement}
\begin{proofs}\begin{proof}\checkt{alcc}
\begin{xflalign*}
	&P \seqDac Q
	&&\ptext{Definition of $\seqDac$}\\
	&=\exists ok_0 @ P[ok_0/ok'] \seqA Q[ok_0/ok]
	&&\ptext{Assumption: $Q \implies R$}\\
	&=\exists ok_0 @ P[ok_0/ok'] \seqA (Q \land R)[ok_0/ok]
	&&\ptext{Substitution}\\
	&=\exists ok_0 @ P[ok_0/ok'] \seqA (Q[ok_0/ok] \land R[ok_0/ok])
	&&\ptext{Predicate calculus and~\cref{lemma:seqA:P-seqA(Q-land-R):implies:(P-seqA-Q)-land-(P-seqA-R)}}\\
	&\implies (\exists ok_0 @ P[ok_0/ok'] \seqA Q[ok_0/ok]) \land (\exists ok_0 @ P[ok_0/ok'] \seqA R[ok_0/ok])
	&&\ptext{Predicate calculus}\\
	&\implies (\exists ok_0 @ P[ok_0/ok'] \seqA R[ok_0/ok])
	&&\ptext{Definition of $\seqDac$}\\
	&=P \seqDac R
\end{xflalign*}
\end{proof}\end{proofs}
\end{lemma}

\begin{lemma}\label{lemma:P-seqDac-Q:ok-ok'-not-free:P-seqA-Q}
Provided $ok'$ is not free in $P$ and $ok$ is not free in $Q$,
\begin{statement}
\begin{align*}
	&P \seqDac Q = P \seqA Q
\end{align*}
\end{statement}
\begin{proofs}
\begin{proof}
\begin{xflalign*}
	&P \seqDac Q
	&&\ptext{Definition of $\seqDac$}\\
	&=\exists ok_0 @ P[ok_0/ok'] \seqA Q[ok_0/ok]
	&&\ptext{Assumption: $ok'$ is not free in $P$}\\
	&=\exists ok_0 @ P \seqA Q[ok_0/ok]
	&&\ptext{Assumption: $ok$ is not free in $Q$}\\
	&=\exists ok_0 @ P \seqA Q
	&&\ptext{Definition of $\seqA$}\\
	&=\exists ok_0 @ P[\{s | Q\}/ac']
	&&\ptext{Predicate calculus}\\
	&=P[\{s | Q\}/ac']
	&&\ptext{Definition of $\seqA$}\\
	&=P \seqA Q
\end{xflalign*}
\end{proof}
\end{proofs}
\end{lemma}

\subsection{Angelic Choice}

\subsubsection{Closure}

\begin{theorem}\label{law:sqcapDac:closure}\label{law:sqcapDac:closure-design}
\begin{statement}
Provided $P$ and $Q$ are $\mathbf{A}$-healthy,
\begin{align*}
	&\mathbf{A} (P \sqcupDac Q) = P \sqcupDac Q
\end{align*}
\end{statement}
\begin{proofs}
\begin{proof}
\begin{xflalign*}
	&P \sqcupDac Q
	&&\ptext{Assumption: $P$ and $Q$ are $\mathbf{A}$-healthy}\\
	&=\mathbf{A} (P) \sqcupDac \mathbf{A} (Q)
	&&\ptext{Definition of $\sqcupDac$ and $\mathbf{A}$}\\
	&=\mathbf{A0}\circ\mathbf{A1} (P) \land \mathbf{A0}\circ\mathbf{A1} (Q)
	&&\ptext{\cref{theorem:A0(P-land-Q):A0(P)-land-A0(Q),law:A0:idempotent}}\\
	&=\mathbf{A0} (\mathbf{A0}\circ\mathbf{A1} (P) \land \mathbf{A0}\circ\mathbf{A1} (Q))
	&&\ptext{\cref{law:A1:idempotent,law:A0:commute-A0-healthy}}\\
	&=\mathbf{A0} (\mathbf{A1}\circ\mathbf{A0}\circ\mathbf{A1} (P) \land \mathbf{A1}\circ\mathbf{A0}\circ\mathbf{A1} (Q))
	&&\ptext{$\mathbf{A1}$ is $\mathbf{PBMH}$ and~\cref{law:pbmh:conjunction-closure}}\\
	&=\mathbf{A0} \circ \mathbf{A1} (\mathbf{A1}\circ\mathbf{A0}\circ\mathbf{A1} (P) \land \mathbf{A1}\circ\mathbf{A0}\circ\mathbf{A1} (Q))
	&&\ptext{\cref{law:A1:idempotent,law:A0:commute-A0-healthy}}\\
	&=\mathbf{A0} \circ \mathbf{A1} (\mathbf{A0}\circ\mathbf{A1} (P) \land \mathbf{A0}\circ\mathbf{A1} (Q))
	&&\ptext{Definition of $\sqcupDac$ $\mathbf{A}$}\\	
	&=\mathbf{A} (\mathbf{A} (P) \sqcupDac \mathbf{A} (Q))
	&&\ptext{Assumption: $P$ and $Q$ are $\mathbf{A}$-healthy}\\
	&=\mathbf{A} (P \sqcupDac Q)
\end{xflalign*}
\end{proof}
\end{proofs}
\end{theorem}

\subsubsection{Relationship with Extended Binary Multirelations}

\begin{theorem}\label{law:bmb2p:sqcup}
\begin{statement}
Provided $B_0$ and $B_1$ are $\mathbf{BMH1}$-healthy,
\begin{align*}
	&bmb2p(B_0 \sqcupBMbot B_1) = bmb2p(B_0) \sqcupDac bmb2p(B_1)
\end{align*}
\end{statement}
\begin{proofs}
\begin{proof}
\begin{flalign*}
	&bmb2p(B_0) \sqcupDac bmb2p(B_1)
	&&\ptext{Definition of $bmb2p$ and $\sqcupDac$}\\
	&=\left(
\begin{array}{l}
		((s, ac' \cup \{\bot\}) \notin B_0 \lor \bot \in ac' \vdash (s, ac') \in B_0 \land \bot \notin ac')
		\\ \vdash \\
		((s, ac' \cup \{\bot\}) \notin B_1 \lor \bot \in ac' \vdash (s, ac') \in B_1 \land \bot \notin ac')
	\end{array}%
\right)
	&&\ptext{Propositional calculus: absorption law}\\
	&=\left(\begin{array}{l}
		\left((s, ac' \cup \{\bot\}) \notin B_0 \lor \bot \in ac' \lor (s, ac' \cup \{\bot\}) \notin B_1 \right)
		\\ \vdash \\
		(s, ac') \in B_0 \land (s, ac') \in B_1 \land \bot \notin ac'
	\end{array}\right)
	&&\ptext{Propositional calculus}\\
	&=\left(\begin{array}{l}
		\lnot ((s, ac' \cup \{\bot\}) \in B_0 \land (s, ac' \cup \{\bot\}) \in B_1) \lor \bot \in ac'
		\\ \vdash \\
		(s, ac') \in B_0 \land (s, ac') \in B_1 \land \bot \notin ac'
	\end{array}\right)
	&&\ptext{Property of sets}\\
	&=\left(\begin{array}{l}
		(s, ac' \cup \{\bot\}) \notin (B_0 \cap B_1) \lor \bot \in ac'
		\\ \vdash \\
		(s, ac') \in (B_0 \cap B_1) \land \bot \notin ac'
	\end{array}\right)
	&&\ptext{Definition of $bmb2p$}\\
	&=bmb2p(B_0 \cap B_1)
	&&\ptext{Definition of $\sqcupBMbot$}\\
	&=bmb2p(B_0 \sqcupBMbot B_1)
\end{flalign*}
\end{proof}
\end{proofs}
\end{theorem}

\subsubsection{Properties with respect to the Extreme Points}

\begin{theorem}\label{theorem:P-sqcupDac-topD:topD}
\begin{statement}
Provided $P$ is a design,
$P \sqcupDac \topD = \topD$.
\end{statement}
\begin{proofs}
\begin{proof}
\begin{flalign*}
	&P \sqcupDac \topD
	&&\ptext{Definition of $\sqcupDac$ and $\topD$}\\
	&=P \land \lnot ok
	&&\ptext{Definition of design}\\
	&=(\lnot P^f \vdash P^t) \land \lnot ok
	&&\ptext{Definition of design}\\
	&=((ok \land \lnot P^f) \implies (P^t \land ok')) \land \lnot ok
	&&\ptext{Predicate calculus}\\
	&=(\lnot ok \lor P^f \lor (P^t \land ok')) \land \lnot ok
	&&\ptext{Predicate calculus: absorption law}\\
	&=\lnot ok
	&&\ptext{Definition of $\topD$}\\
	&=\topD
\end{flalign*}
\end{proof}
\end{proofs}
\end{theorem}

\section{Relationship with Angelic Designs}

\subsection{$d2ac$}

\begin{theorem}\label{theorem:A-o-d2ac(P):d2ac(P)}
\begin{statement}
$\mathbf{A} \circ d2ac(P) = d2ac(P)$
\end{statement}
\begin{proofs}
\begin{proof}
\begin{xflalign*}
	&\mathbf{A} \circ d2ac(P)
	&&\ptext{Definition of $d2ac$}\\
	&=\mathbf{A} (\lnot p2ac(P^f) \land (\lnot P^f[\mathbf{s}/in\alpha_{-ok}] \circseq true) \vdash p2ac(P^t))
	&&\ptext{Definition of $\mathbf{A}$}\\
	&=\mathbf{A0} \circ \mathbf{A1} (\lnot p2ac(P^f) \land (\lnot P^f[\mathbf{s}/in\alpha_{-ok}] \circseq true) \vdash p2ac(P^t))
	&&\ptext{Definition of $\mathbf{A1}$}\\
	&=\mathbf{A0} \left(\begin{array}{l}
		\lnot \mathbf{PBMH} (\lnot (\lnot p2ac(P^f) \land (\lnot P^f[\mathbf{s}/in\alpha_{-ok}] \circseq true))) 
		\\ \vdash \\
		\mathbf{PBMH} \circ p2ac(P^t)
	\end{array}\right)
	&&\ptext{Predicate calculus}\\
	&=\mathbf{A0} \left(\begin{array}{l}
		\lnot \mathbf{PBMH} (p2ac(P^f) \lor \lnot (\lnot P^f[\mathbf{s}/in\alpha_{-ok}] \circseq true)) 
		\\ \vdash \\
		\mathbf{PBMH} \circ p2ac(P^t)
	\end{array}\right)
	&&\ptext{\cref{law:pbmh:distribute-disjunction}}\\
	&=\mathbf{A0} \left(\begin{array}{l}
		\lnot (\mathbf{PBMH} \circ p2ac(P^f) \lor \mathbf{PBMH} (\lnot (\lnot P^f[\mathbf{s}/in\alpha_{-ok}] \circseq true))) 
		\\ \vdash \\
		\mathbf{PBMH} \circ p2ac(P^t)
	\end{array}\right)
	&&\ptext{$ac'$ not free in $P^f$ and~\cref{law:pbmh:P:ac'-not-free}}\\
	&=\mathbf{A0} \left(\begin{array}{l}
		\lnot (\mathbf{PBMH} \circ p2ac(P^f) \lor \lnot (\lnot P^f[\mathbf{s}/in\alpha_{-ok}] \circseq true)) 
		\\ \vdash \\
		\mathbf{PBMH} \circ p2ac(P^t)
	\end{array}\right)
	&&\ptext{\cref{lemma:PBMH-o-p2ac(P):p2ac(P)}}\\
	&=\mathbf{A0} \left(\begin{array}{l}
		\lnot (p2ac(P^f) \lor \lnot (\lnot P^f[\mathbf{s}/in\alpha_{-ok}] \circseq true)) 
		\\ \vdash \\
		p2ac(P^t)
	\end{array}\right)
	&&\ptext{Definition of $\mathbf{A0}$ and~\cref{law:A0:design}}\\
	&=\left(\begin{array}{l}
		\lnot (p2ac(P^f) \lor \lnot (\lnot P^f[\mathbf{s}/in\alpha_{-ok}] \circseq true)) 
		\\ \vdash \\
		p2ac(P^t) \land ac'\neq\emptyset
	\end{array}\right)
	&&\ptext{\cref{lemma:p2ac(P):implies:ac'-neq-emptyset} and predicate calculus}\\
	&=\left(\begin{array}{l}
		\lnot p2ac(P^f) \land (\lnot P^f[\mathbf{s}/in\alpha_{-ok}] \circseq true)) 
		\\ \vdash \\
		p2ac(P^t)
	\end{array}\right)
	&&\ptext{Definition of $d2ac$}\\
	&=d2ac(P)
\end{xflalign*}
\end{proof}
\end{proofs}
\end{theorem}
\subsection{$p2ac$}

\subsubsection{Properties}

\begin{lemma}\label{lemma:PBMH-o-p2ac(P):p2ac(P)}
\begin{statement}$\mathbf{PBMH} \circ p2ac(P) = p2ac(P)$\end{statement}
\begin{proofs}
\begin{proof}\checkt{alcc}
\begin{xflalign*}
	&\mathbf{PBMH} \circ p2ac(P)
	&&\ptext{Definition of $\mathbf{PBMH}$ (\cref{lemma:PBMH:alternative-1})}\\
	&=\exists ac_0 \spot p2ac(P)[ac_0/ac'] \land ac_0 \subseteq ac'
	&&\ptext{Definition of $p2ac$}\\
	&=\exists ac_0 \spot (\exists z \spot P[\mathbf{s},\mathbf{z'}/in\alpha_{-ok},out\alpha_{-ok'}] \land z \in ac')[ac_0/ac'] \land ac_0 \subseteq ac'
	&&\ptext{Substitution}\\
	&=\exists ac_0 \spot (\exists z \spot P[\mathbf{s},\mathbf{z'}/in\alpha_{-ok},out\alpha_{-ok'}] \land z \in ac_0) \land ac_0 \subseteq ac'
	&&\ptext{Property of sets}\\
	&=\exists z \spot P[\mathbf{s},\mathbf{z'}/in\alpha_{-ok},out\alpha_{-ok'}] \land z \in ac'
	&&\ptext{Definition of $p2ac$}\\
	&=p2ac(P)
\end{xflalign*}
\end{proof}
\end{proofs}
\end{lemma}

\begin{theorem}\label{theorem:p2ac(P-lor-Q):p2ac(P)-lor-p2ac(Q)}
\begin{statement}$p2ac(P \lor Q) = p2ac(P) \lor p2ac(Q)$\end{statement}
\begin{proofs}\begin{proof}\checkt{pfr}\checkt{alcc}
\begin{flalign*}
	&p2ac(P \lor Q)
	&&\ptext{Definition of $p2ac$}\\
	&=\exists z \spot (P \lor Q)[\mathbf{s},\mathbf{z}/in\alpha_{-ok},out\alpha_{-ok'}] \land undash(z) \in ac'
	&&\ptext{Property of substitution}\\
	&=\exists z \spot (P[\mathbf{s},\mathbf{z}/in\alpha_{-ok},out\alpha_{-ok'}] \lor Q[\mathbf{s},\mathbf{z}/in\alpha_{-ok},out\alpha_{-ok'}]) \land undash(z) \in ac'
	&&\ptext{Predicate calculus}\\
	&=\exists z \spot \left(\begin{array}{l}
		(P[\mathbf{s},\mathbf{z}/in\alpha_{-ok},out\alpha_{-ok'}] \land undash(z) \in ac')
		\\ \lor \\
		(Q[\mathbf{s},\mathbf{z}/in\alpha_{-ok},out\alpha_{-ok'}] \land undash(z) \in ac')
	\end{array}\right)
	&&\ptext{Predicate calculus}\\
	&=\left(\begin{array}{l}
		(\exists z \spot P[\mathbf{s},\mathbf{z}/in\alpha_{-ok},out\alpha_{-ok'}] \land undash(z) \in ac')
		\\ \lor \\
		(\exists z \spot Q[\mathbf{s},\mathbf{z}/in\alpha_{-ok},out\alpha_{-ok'}] \land undash(z) \in ac')
	\end{array}\right)
	&&\ptext{Definition of $p2ac$}\\
	&=p2ac(P) \lor p2ac(Q)
\end{flalign*}
\end{proof}\end{proofs}
\end{theorem}

\begin{theorem}\label{theorem:p2ac(P-land-Q):implies:p2ac(P)-land-p2ac(Q)}
\begin{statement}$p2ac(P \land Q) \implies p2ac(P) \land p2ac(Q)$\end{statement}
\begin{proofs}\begin{proof}\checkt{pfr}\checkt{alcc}
\begin{flalign*}
	&p2ac(P \land Q)
	&&\ptext{Definition of $p2ac$}\\
	&=\exists z \spot (P \land Q)[\mathbf{s},\mathbf{z}/in\alpha_{-ok},out\alpha_{-ok'}] \land undash(z) \in ac' 
	&&\ptext{Property of substitution}\\
	&=\exists z \spot (P[\mathbf{s},\mathbf{z}/in\alpha_{-ok},out\alpha_{-ok'}] \land Q[\mathbf{s},\mathbf{z}/in\alpha_{-ok},out\alpha_{-ok'}]) \land undash(z) \in ac'
	&&\ptext{Predicate calculus}\\
	&=\exists z \spot \left(\begin{array}{l}
		(P[\mathbf{s},\mathbf{z}/in\alpha_{-ok},out\alpha_{-ok'}] \land undash(z) \in ac')
		\\ \land \\
		(Q[\mathbf{s},\mathbf{z}/in\alpha_{-ok},out\alpha_{-ok'}] \land undash(z) \in ac')
	\end{array}\right)
	&&\ptext{Predicate calculus}\\
	&\implies \left(\begin{array}{l}
		(\exists z \spot P[\mathbf{s},\mathbf{z}/in\alpha_{-ok},out\alpha_{-ok'}] \land undash(z) \in ac')
		\\ \land \\
		(\exists z \spot Q[\mathbf{s},\mathbf{z}/in\alpha_{-ok},out\alpha_{-ok'}] \land undash(z) \in ac')
	\end{array}\right)
	&&\ptext{Definition of $p2ac$}\\
	&=p2ac(P) \land p2ac(Q)
\end{flalign*}
\end{proof}\end{proofs}
\end{theorem}

\begin{theorem}\label{theorem:A2-o-p2ac(P):p2ac(P)}
\begin{statement}
$\mathbf{A2} \circ p2ac(P) = p2ac(P)$
\end{statement}
\begin{proofs}
\begin{proof}
\begin{xflalign*}
	&\mathbf{A2} \circ p2ac(P)
	&&\ptext{Definition of $\mathbf{A2}$}\\
	&=\mathbf{PBMH} (p2ac(P) \seqA \{ s \} = ac')
	&&\ptext{Definition of $p2ac$}\\
	&=\mathbf{PBMH} ((\exists z \spot P[\mathbf{s},\mathbf{z}/in\alpha_{-ok},out\alpha_{-ok'}] \land undash(z) \in ac') \seqA \{ s \} = ac')
	&&\ptext{Definition of $\seqA$ and substitution}\\
	&=\mathbf{PBMH} (\exists z \spot P[\mathbf{s},\mathbf{z}/in\alpha_{-ok},out\alpha_{-ok'}] \land undash(z) \in \{ s | \{ s \} = ac' \})
	&&\ptext{Property of sets}\\
	&=\mathbf{PBMH} (\exists z \spot P[\mathbf{s},\mathbf{z}/in\alpha_{-ok},out\alpha_{-ok'}] \land \{ undash(z) \} = ac')
	&&\ptext{Definition of $\mathbf{PBMH}$ and substitution}\\
	&=\exists ac_0 @ \exists z \spot P[\mathbf{s},\mathbf{z}/in\alpha_{-ok},out\alpha_{-ok'}] \land \{ undash(z) \} = ac_0 \land ac_0 \subseteq ac'
	&&\ptext{One-point rule}\\
	&=\exists z \spot P[\mathbf{s},\mathbf{z}/in\alpha_{-ok},out\alpha_{-ok'}] \land \{ undash(z) \} \subseteq ac'
	&&\ptext{Property of sets}\\
	&=\exists z \spot P[\mathbf{s},\mathbf{z}/in\alpha_{-ok},out\alpha_{-ok'}] \land undash(z) \in ac'
	&&\ptext{Definition of $p2ac$}\\
	&=p2ac(P)
\end{xflalign*}
\end{proof}
\end{proofs}
\end{theorem}

\begin{theorem}\label{theorem:p2ac(design)-land-ac'-neq-emptyset}
\begin{statement}
\begin{align*}ac'\neq\emptyset \land p2ac(\lnot P^f \vdash P^t) = ac'\neq\emptyset \land (\lnot p2ac(P^f) \vdash p2ac(P^t))
\end{align*}
\end{statement}
\begin{proofs}
\begin{proof}\checkt{alcc}
\begin{xflalign*}
	&ac'\neq\emptyset \land p2ac(\lnot P^f \vdash P^t)
	&&\ptext{Definition of design}\\
	&=ac'\neq\emptyset \land p2ac((ok \land \lnot P^f) \implies (P^t \land ok'))
	&&\ptext{Predicate calculus}\\
	&=ac'\neq\emptyset \land p2ac(\lnot ok \lor P^f \lor (P^t \land ok'))
	&&\ptext{Distributivity of $p2ac$ (\cref{theorem:p2ac(P-lor-Q):p2ac(P)-lor-p2ac(Q)})}\\
	&=ac'\neq\emptyset \land (p2ac(\lnot ok) \lor p2ac(P^f) \lor p2ac(P^t \land ok'))
	&&\ptext{\cref{lemma:p2ac(P)-inoutalpha-not-free:P,lemma:p2ac(P-land-Q)-inoutalpha-not-free:P-land-p2ac(Q)}}\\
	&=ac'\neq\emptyset \land ((\lnot ok \land ac'\neq\emptyset) \lor p2ac(P^f) \lor (p2ac(P^t) \land ok'))
	&&\ptext{Predicate calculus}\\
	&=ac'\neq\emptyset \land (\lnot ok \lor p2ac(P^f) \lor (p2ac(P^t) \land ok'))
	&&\ptext{Predicate calculus}\\
	&=ac'\neq\emptyset \land ((ok \land \lnot p2ac(P^f)) \implies (p2ac(P^t) \land ok'))
	&&\ptext{Definition of design}\\
	&=ac'\neq\emptyset \land (\lnot p2ac(P^f) \vdash p2ac(P^t))
\end{xflalign*}
\end{proof}
\end{proofs}
\end{theorem}

\begin{theorem}\label{theorem:p2ac(P)-land-ac'-neq-emptyset:d2ac(P)-land-ac'-neq-emptyset}
\begin{statement}
Provided $P$ is a design,
\begin{align*}
	&ac'\neq\emptyset \land p2ac(P) = ac'\neq\emptyset \land d2ac(P) 
\end{align*}
\end{statement}
\begin{proofs}
\begin{proof}\checkt{alcc}
\begin{xflalign*}
	&ac'\neq\emptyset \land p2ac(P)
	&&\ptext{Assumption: $P$ is a design}\\
	&=ac'\neq\emptyset \land p2ac((ok \land \lnot P^f) \implies (P^t \land ok'))
	&&\ptext{Predicate calculus}\\
	&=ac'\neq\emptyset \land p2ac((ok \land \lnot P^f \land \exists out\alpha \spot \lnot P^f) \implies (P^t \land ok'))
	&&\ptext{Predicate calculus}\\
	&=ac'\neq\emptyset \land p2ac(\lnot ok \lor P^f \lor \lnot (\exists out\alpha \spot \lnot P^f) \lor (P^t \land ok'))
	&&\ptext{Distributivity of $p2ac$ (\cref{theorem:p2ac(P-lor-Q):p2ac(P)-lor-p2ac(Q)})}\\
	&=ac'\neq\emptyset \land \left(\begin{array}{l}
		p2ac(\lnot ok) \lor p2ac(P^f) 
		\\ \lor \\
		p2ac(\lnot (\exists out\alpha \spot \lnot P^f)) \lor p2ac(P^t \land ok')
	\end{array}\right)
	&&\ptext{\cref{lemma:p2ac(P)-inoutalpha-not-free:P,lemma:p2ac(P-land-Q)-inoutalpha-not-free:P-land-p2ac(Q)}}\\
	&=ac'\neq\emptyset \land \left(\begin{array}{l}
		(\lnot ok \land ac'\neq\emptyset) \lor p2ac(P^f) 
		\\ \lor \\
		\lor p2ac(\lnot (\exists out\alpha \spot \lnot P^f)) \lor (p2ac(P^t) \land ok')
	\end{array}\right)
	&&\ptext{\cref{lemma:p2ac(P)-outalpha-not-free}}\\
	&=ac'\neq\emptyset \land \left(\begin{array}{l}
		(\lnot ok \land ac'\neq\emptyset) \lor p2ac(P^f) 
		\\ \lor \\
		((\lnot (\exists out\alpha \spot \lnot P^f))[\mathbf{s}/in\alpha] \land ac'\neq\emptyset) \lor (p2ac(P^t) \land ok')
	\end{array}\right)
	&&\ptext{Predicate calculus}\\
	&=ac'\neq\emptyset \land \left(\begin{array}{l}
		\lnot ok \lor p2ac(P^f) 
		\\ \lor \\
		(\lnot (\exists out\alpha \spot \lnot P^f))[\mathbf{s}/in\alpha]) \lor (p2ac(P^t) \land ok')
	\end{array}\right)
	&&\ptext{Property of substitution}\\
	&=ac'\neq\emptyset \land \left(\begin{array}{l}
		\lnot ok \lor p2ac(P^f) 
		\\ \lor \\
		\lnot (\exists out\alpha \spot \lnot P^f[\mathbf{s}/in\alpha]) \lor (p2ac(P^t) \land ok')
	\end{array}\right)
	&&\ptext{Predicate calculus}\\
	&=ac'\neq\emptyset \land ((ok \land \lnot p2ac(P^f) \land \exists out\alpha \spot \lnot P^f[\mathbf{s}/in\alpha]) \implies
		(p2ac(P^t) \land ok'))
	&&\ptext{Definition of design}\\
	&=ac'\neq\emptyset \land (\lnot p2ac(P^f) \land \exists out\alpha \spot \lnot P^f[\mathbf{s}/in\alpha] \vdash p2ac(P^t))
	&&\ptext{Predicate calculus and definition of sequential composition}\\
	&=ac'\neq\emptyset \land (\lnot p2ac(P^f) \land (\lnot P^f[\mathbf{s}/in\alpha] \circseq true) \vdash p2ac(P^t))
	&&\ptext{Definition of $d2ac$}\\
	&=ac'\neq\emptyset \land d2ac(P)
\end{xflalign*}
\end{proof}
\end{proofs}
\end{theorem}

\subsubsection{Lemmas}

\begin{lemma}\label{lemma:p2ac(conditional)}Provided $c$ is a condition.
\begin{align*}
	&p2ac(P \dres c \rres Q) = p2ac(P) \dres s.c \rres p2ac(Q)
\end{align*}
\begin{proofs}\begin{proof}\checkt{pfr}\checkt{alcc}
\begin{flalign*}
	&p2ac(P \dres c \rres Q)
	&&\ptext{Definition of $p2ac$}\\
	&=\exists z \spot (P \dres c \rres Q)[\mathbf{s},\mathbf{z}/in\alpha_{ok},out\alpha_{ok'}] \land undash(z) \in ac'
	&&\ptext{Substitution: $c$ is a condition}\\
	&=\exists z \spot (P[\mathbf{s},\mathbf{z}/in\alpha_{ok},out\alpha_{ok'}] \dres s.c \rres Q[\mathbf{s},\mathbf{z}/in\alpha_{ok},out\alpha_{ok'}] )\land undash(z) \in ac'
	&&\ptext{Predicate calculus}\\
	&=\left(\begin{array}{l}
		(\exists z \spot P[\mathbf{s},\mathbf{z}/in\alpha_{ok},out\alpha_{ok'}] \land undash(z) \in ac') 
		\\ \dres s.c \rres \\
		(\exists z \spot Q[\mathbf{s},\mathbf{z}/in\alpha_{ok},out\alpha_{ok'}] \land undash(z) \in ac')
	\end{array}\right)
	&&\ptext{Definition of $p2ac$}\\
	&=p2ac(P) \dres s.c \rres p2ac(Q)
\end{flalign*}
\end{proof}\end{proofs}
\end{lemma}

\begin{lemma}\label{lemma:p2ac(true)}
$p2ac(true) = ac'\neq\emptyset$
\begin{proofs}\begin{proof}\checkt{alcc}
\begin{flalign*}
	&p2ac(true)
	&&\ptext{Definition of $p2ac$}\\
	&=\exists z \spot true[\mathbf{s},\mathbf{z}/in\alpha_{-ok},out\alpha_{-ok'}] \land undash(z) \in ac'
	&&\ptext{Substitution}\\
	&=\exists z \spot true \land undash(z) \in ac'
	&&\ptext{Predicate calculus}\\
	&=\exists z \spot undash(z) \in ac'
	&&\ptext{Property of sets}\\
	&=ac'\neq\emptyset
\end{flalign*}
\end{proof}\end{proofs}
\end{lemma}

\begin{lemma}\label{lemma:p2ac(false)}
$p2ac(false) = false$
\begin{proofs}\begin{proof}\checkt{alcc}
\begin{flalign*}
	&p2ac(false)
	&&\ptext{Definition of $p2ac$}\\
	&=\exists z \spot false[\mathbf{s},\mathbf{z}/in\alpha_{-ok},out\alpha_{-ok'}] \land undash(z) \in ac'
	&&\ptext{Predicate calculus}\\
	&=false
\end{flalign*}
\end{proof}\end{proofs}
\end{lemma}

\begin{lemma}\label{lemma:exists-outalpha-P:exists-z-P[z/outalpha]}
$\exists out\alpha_{-ok'} \spot P = \exists z \spot P[\mathbf{z}/out\alpha_{-ok'}]$
\begin{proofs}\begin{proof}
\begin{xflalign*}
	&\exists out\alpha_{-ok'} \spot P
	&&\ptext{Introduce fresh state variable $z$}\\
	&=\exists z, out\alpha \spot P \land z.x_0 = x_0 \land \ldots \land z.x_n = x.n
	&&\ptext{One-point rule for each $x_i$ in $out\alpha_{-ok'}$}\\
	&=\exists z \spot P[z.x_0,\ldots,z.x_n/x_0,\ldots,x_n]
	&&\ptext{Definition of state substitution}\\
	&=\exists z \spot P[\mathbf{z}/out\alpha_{-ok'}]
\end{xflalign*}
\end{proof}\end{proofs}
\end{lemma}

\begin{lemma}\label{lemma:p2ac(P-land-Q)-inoutalpha-not-free:P-land-p2ac(Q)} Provided that no variable in $in\alpha_{-ok} \cup out\alpha_{-ok'}$ is free in $P$,
\begin{align*}
	&p2ac(P \land Q) = P \land p2ac(Q)
\end{align*}
\begin{proofs}\begin{proof}\checkt{alcc}\checkt{pfr}
\begin{flalign*}
	&p2ac(P \land Q)
	&&\ptext{Definition of $ac2p$}\\
	&=\exists z \spot (P \land Q)[\mathbf{s},\mathbf{z}/in\alpha_{-ok},out\alpha_{-ok'}] \land undash(z) \in ac'
	&&\ptext{Substitution: assumption}\\
	&=\exists z \spot (P \land Q[\mathbf{s},\mathbf{z}/in\alpha_{-ok},out\alpha_{-ok'}]) \land undash(z) \in ac'
	&&\ptext{Predicate calculus}\\
	&=P \land (\exists z \spot Q[\mathbf{s},\mathbf{z}/in\alpha_{-ok},out\alpha_{-ok'}] \land undash(z) \in ac')
	&&\ptext{Definition of $ac2p$}\\
	&=P \land ac2p(Q)
\end{flalign*}
\end{proof}\end{proofs}
\end{lemma}

\begin{lemma}\label{lemma:p2ac(P)-inoutalpha-not-free:P} Provided that no variable in $in\alpha_{-ok} \cup out\alpha_{-ok'}$ is free in $P$,
\begin{align*}
	&p2ac(P) = P \land ac'\neq\emptyset
\end{align*}
\begin{proofs}\begin{proof}\checkt{alcc}
\begin{flalign*}
	&p2ac(P)
	&&\ptext{Definition of $ac2p$}\\
	&=\exists z \spot P[\mathbf{s},\mathbf{z}/in\alpha_{-ok},out\alpha_{-ok'}] \land undash(z) \in ac'
	&&\ptext{Substitution: variables of $out\alpha_{-ok'} \cup in\alpha_{-ok}$ not free in $P$}\\
	&=\exists z \spot P \land undash(z) \in ac'
	&&\ptext{Predicate calculus: $z$ not free in $P$}\\
	&=P \land \exists z \spot undash(z) \in ac'
	&&\ptext{Property of sets}\\
	&=P \land ac'\neq\emptyset
\end{flalign*}
\end{proof}\end{proofs}
\end{lemma}

\begin{lemma}\label{lemma:p2ac(P)-outalpha-not-free} Provided that no dashed variable in $out\alpha_{-ok}$ is free in $P$,
\begin{align*}
	&p2ac(P) = P[\mathbf{s}/in\alpha] \land ac'\neq\emptyset
\end{align*}
\begin{proofs}\begin{proof}\checkt{alcc}
\begin{flalign*}
	&p2ac(P)
	&&\ptext{Definition of $p2ac$}\\
	&=\exists z \spot P[\mathbf{s},\mathbf{z}/in\alpha_{-ok},out\alpha_{-ok'}] \land undash(z) \in ac'
	&&\ptext{Substitution: variables of $out\alpha$ not free in $P$}\\
	&=\exists z \spot P[\mathbf{s}/in\alpha_{-ok}] \land undash(z) \in ac'
	&&\ptext{Predicate calculus: $z$ not free in $P$}\\
	&=P[\mathbf{s}/in\alpha_{-ok}] \land \exists z \spot undash(z) \in ac'
	&&\ptext{Property of sets}\\
	&=P[\mathbf{s}/in\alpha_{-ok}] \land ac'\neq\emptyset
\end{flalign*}
\end{proof}\end{proofs}
\end{lemma}

\begin{lemma}\label{lemma:p2ac-o-ac2p(P)}
\begin{statement}
$p2ac \circ ac2p (P) = \exists ac_0, y \spot P[ac_0/ac'] \land ac_0 \subseteq \{ y \} \land y \in ac'$
\end{statement}
\begin{proofs}
\begin{proof}\checkt{pfr}\checkt{alcc}
\begin{xflalign*}
	&p2ac \circ ac2p (P)
	&&\ptext{Definition of $p2ac$}\\
	&=\exists z \spot ac2p (P)[\mathbf{s},\mathbf{z}/in\alpha_{-ok},out\alpha_{-ok'}] \land undash(z) \in ac'
	&&\ptext{Definition of $ac2p$ (\cref{lemma:ac2p:alternative-2})}\\
	&=\left(\begin{array}{l}
		\exists z \spot \left(\begin{array}{l}
			\exists ac' \spot P[State_{\II}(in\alpha_{-ok})/s] 
			\\ \land \\
			ac' \subseteq \{ z | \bigwedge x : out\alpha_{-ok'} \spot dash(z).x = x \}
			\end{array}\right)[\mathbf{s},\mathbf{z}/in\alpha_{-ok},out\alpha_{-ok'}] 
		\\ \land undash(z) \in ac'
	\end{array}\right)
	&&\ptext{Variable renaming}\\
	&=\left(\begin{array}{l}
		\exists z \spot \left(\begin{array}{l}
			\exists ac' \spot P[State_{\II}(in\alpha)/s] 
			\\ \land \\
			ac' \subseteq \{ y | \bigwedge x : out\alpha_{-ok'} \spot dash(y).x = x \}
			\end{array}\right)[\mathbf{s},\mathbf{z}/in\alpha_{-ok},out\alpha_{-ok'}] 
		\\ \land undash(z) \in ac'
	\end{array}\right)
	&&\ptext{Substitution}\\
	&=\exists z \spot \left(\begin{array}{l}
			\exists ac' \spot P[State_{\II}(in\alpha_{-ok})/s][\mathbf{s}/in\alpha_{-ok}] 
			\\ \land \\
			ac' \subseteq \{ y | \bigwedge x : out\alpha_{-ok'} \spot dash(y).x = z.x \}
			\end{array}\right) \land undash(z) \in ac'
	&&\ptext{\cref{lemma:state-sub:P-S-z:z-S}}\\
	&=\exists z \spot \left(\begin{array}{l}
			\exists ac' \spot P
			\\ \land \\
			ac' \subseteq \{ y | \bigwedge x : out\alpha_{-ok'} \spot dash(y).x = z.x \}
			\end{array}\right) \land undash(z) \in ac'	
	&&\ptext{Equality of records}\\
	&=\exists z \spot (\exists ac' \spot P \land ac' \subseteq \{ y | dash(y) = z \}) \land undash(z) \in ac'
	&&\ptext{Property of $dash$ and $undash$}\\
	&=\exists z \spot (\exists ac' \spot P \land ac' \subseteq \{ y | y = undash(z) \}) \land undash(z) \in ac'
	&&\ptext{Property of sets}\\
	&=\exists z \spot (\exists ac' \spot P \land ac' \subseteq \{ undash(z) \}) \land undash(z) \in ac'
	&&\ptext{Introduce fresh variable $y$}\\
	&=\exists y, z \spot (\exists ac' \spot P \land ac' \subseteq \{ undash(z) \}) \land undash(z) \in ac' \land undash(z) = y
	&&\ptext{One-point rule: $z$ not free in $P$}\\
	&=\exists y \spot (\exists ac' \spot P\land ac' \subseteq \{ y \}) \land y \in ac'
	&&\ptext{Variable renaming}\\
	&=\exists y \spot (\exists ac_0 \spot P[ac_0/ac'] \land ac_0 \subseteq \{ y \}) \land y \in ac'
	&&\ptext{Predicate calculus}\\
	&=\exists ac_0, y \spot P[ac_0/ac'] \land ac_0 \subseteq \{ y \} \land y \in ac'
\end{xflalign*}
\end{proof}
\end{proofs}
\end{lemma}

\begin{lemma}\label{lemma:p2ac(P)-sub-ok'-wait:p2ac(sub-ok'-wait)}
$p2ac(P)^o_w = p2ac(P^o_w)$
\begin{proofs}\begin{proof}\checkt{alcc}
\begin{xflalign*}
	&p2ac(P)^o_w
	&&\ptext{Substitution abbreviation}\\
	&=p2ac(P)[o,s\oplus\{wait\mapsto w\}/ok',s]
	&&\ptext{Definition of $p2ac$}\\
	&=(\exists z \spot P[\mathbf{s},\mathbf{z'}/in\alpha,out\alpha] \land z \in ac')[o,s\oplus\{wait\mapsto w\}/ok',s]
	&&\ptext{Substitution: $ok'$ not in $out\alpha$}\\
	&=(\exists z \spot P[o/ok'][\mathbf{s},\mathbf{z'}/in\alpha,out\alpha] \land z \in ac')[s\oplus\{wait\mapsto w\}/s]
	&&\ptext{$wait$ is not $w$ and \cref{lemma:state-substitution:S-si-mapsto-e:si-not-free-in-e}}\\
	&=\exists z \spot P[o,w/ok',wait][\mathbf{s},\mathbf{z'}/in\alpha,out\alpha] \land z \in ac'
	&&\ptext{Substitution abbreviation}\\
	&=\exists z \spot P^o_w[\mathbf{s},\mathbf{z'}/in\alpha,out\alpha] \land z \in ac'
	&&\ptext{Definition of $p2ac$}\\
	&=p2ac(P^o_w)
\end{xflalign*}
\end{proof}\end{proofs}
\end{lemma}

\begin{lemma}\label{lemma:p2ac(P):implies:ac'-neq-emptyset}
$p2ac(P) \implies ac'\neq\emptyset$
\begin{proofs}\begin{proof}
\begin{xflalign*}
	&p2ac(P)
	&&\ptext{Definition of $p2ac$}\\
	&=\exists z \spot P[\mathbf{s},\mathbf{z}/in\alpha_{-ok},out\alpha_{-ok'}] \land undash(z) \in ac'
	&&\ptext{Predicate calculus}\\
	&\implies \exists z \spot undash(z) \in ac'
	&&\ptext{Predicate calculus}\\
	&=ac'\neq\emptyset
\end{xflalign*}
\end{proof}\end{proofs}
\end{lemma}

\begin{lemma}\label{lemma:p2ac(P-land-Q)-y-cap-ac':(p2ac(P)-land-p2ac(Q))-y-cap-ac'}
\begin{statement}
Provided $ac'$ is not free in $P$ nor $Q$,
\begin{align*}
	&p2ac(P \land Q)[\{y\}\cap ac'/ac'] = (p2ac(P) \land p2ac(Q))[\{y\}\cap ac'/ac']
\end{align*}
\end{statement}
\begin{proofs}
\begin{proof}\checkt{alcc}
\begin{xflalign*}
	&(p2ac(P) \land p2ac(Q))[\{y\}\cap ac'/ac']
	&&\ptext{Definition of $p2ac$}\\
	&=\left(
\begin{array}{l}
		(\exists z @ P[\mathbf{s},\mathbf{z}/in\alpha_{-ok},out\alpha_{-ok'}] \land undash(z) \in ac')
		\\ \land \\
		(\exists z @ Q[\mathbf{s},\mathbf{z}/in\alpha_{-ok},out\alpha_{-ok'}] \land undash(z) \in ac')
	\end{array}\right)[\{y\}\cap ac'/ac']
	&&\ptext{Assumption: $ac'$ is not free in $P$ nor $Q$ and substitution}\\
	&=\left(\begin{array}{l}
		(\exists z @ P[\mathbf{s},\mathbf{z}/in\alpha_{-ok},out\alpha_{-ok'}] \land undash(z) \in \{y\}\cap ac')
		\\ \land \\
		(\exists z @ Q[\mathbf{s},\mathbf{z}/in\alpha_{-ok},out\alpha_{-ok'}] \land undash(z) \in \{y\}\cap ac')
	\end{array}\right)
	&&\ptext{Property of sets}\\
	&=\left(\begin{array}{l}
		(\exists z @ P[\mathbf{s},\mathbf{z}/in\alpha_{-ok},out\alpha_{-ok'}] \land undash(z) \in \{y\} \land undash(z) \in ac')
		\\ \land \\
		(\exists z @ Q[\mathbf{s},\mathbf{z}/in\alpha_{-ok},out\alpha_{-ok'}] \land undash(z) \in \{y\} \land undash(z) \in ac')
	\end{array}\right)
	&&\ptext{Property of sets}\\
	&=\left(\begin{array}{l}
		(\exists z @ P[\mathbf{s},\mathbf{z}/in\alpha_{-ok},out\alpha_{-ok'}] \land undash(z)=y \land undash(z) \in ac')
		\\ \land \\
		(\exists z @ Q[\mathbf{s},\mathbf{z}/in\alpha_{-ok},out\alpha_{-ok'}] \land undash(z)=y \land undash(z) \in ac')
	\end{array}\right)
	&&\ptext{Property of $dash$ and $undash$}\\
	&=\left(\begin{array}{l}
		(\exists z @ P[\mathbf{s},\mathbf{z}/in\alpha_{-ok},out\alpha_{-ok'}] \land z=dash(y) \land undash(z) \in ac')
		\\ \land \\
		(\exists z @ Q[\mathbf{s},\mathbf{z}/in\alpha_{-ok},out\alpha_{-ok'}] \land z=dash(y) \land undash(z) \in ac')
	\end{array}\right)
	&&\ptext{One-point rule}\\
	&=\left(\begin{array}{l}
		P[\mathbf{s},\mathbf{z}/in\alpha_{-ok},out\alpha_{-ok'}][dash(y)/z]
		\\ \land \\
		Q[\mathbf{s},\mathbf{z}/in\alpha_{-ok},out\alpha_{-ok'}][dash(y)/z]
		\\ \land \\
		undash(dash(y)) \in ac'
	\end{array}\right)
	&&\ptext{Introduce fresh variable $t$}\\
	&=\exists t @ \left(\begin{array}{l} 
		P[\mathbf{s},\mathbf{z}/in\alpha_{-ok},out\alpha_{-ok'}][t/z]
		\\ \land \\
		Q[\mathbf{s},\mathbf{z}/in\alpha_{-ok},out\alpha_{-ok'}][t/z]
		\\ \land \\
		undash(t) \in ac' \land dash(y) = t
	\end{array}\right)
	&&\ptext{Property of $dash$ and $undash$}\\
	&=\exists t @ \left(\begin{array}{l}
		P[\mathbf{s},\mathbf{z}/in\alpha_{-ok},out\alpha_{-ok'}][t/z]
		\\ \land \\
		Q[\mathbf{s},\mathbf{z}/in\alpha_{-ok},out\alpha_{-ok'}][t/z]
		\\ \land \\
		undash(t) \in ac' \land y = undash(t)
	\end{array}\right)
	&&\ptext{Substitution}\\
	&=\exists t @ \left(\begin{array}{l}
		P[\mathbf{s},\mathbf{t}/in\alpha_{-ok},out\alpha_{-ok'}]
		\\ \land \\
		Q[\mathbf{s},\mathbf{t}/in\alpha_{-ok},out\alpha_{-ok'}]
		\\ \land \\
		undash(t) \in ac' \land y = undash(t)
	\end{array}\right)
	&&\ptext{Property of sets}\\
	&=\exists t @ \left(\begin{array}{l}
		P[\mathbf{s},\mathbf{t}/in\alpha_{-ok},out\alpha_{-ok'}]
		\\ \land \\
		Q[\mathbf{s},\mathbf{t}/in\alpha_{-ok},out\alpha_{-ok'}]
		\\ \land \\
		undash(t) \in ac' \land undash(t) \in \{y\}
	\end{array}\right)
	&&\ptext{Property of sets}\\
	&=\exists t @ \left(\begin{array}{l}
		P[\mathbf{s},\mathbf{t}/in\alpha_{-ok},out\alpha_{-ok'}]
		\\ \land \\
		Q[\mathbf{s},\mathbf{t}/in\alpha_{-ok},out\alpha_{-ok'}]
		\\ \land \\
		undash(t) \in (\{y\}\cap ac')
	\end{array}\right)
	&&\ptext{Substitution}\\
	&=\exists t @ \left(\begin{array}{l}
		P[\mathbf{s},\mathbf{t}/in\alpha_{-ok},out\alpha_{-ok'}]
		\\ \land \\
		Q[\mathbf{s},\mathbf{t}/in\alpha_{-ok},out\alpha_{-ok'}]
		\\ \land \\
		undash(t) \in ac'
	\end{array}\right)[\{y\}\cap ac'/ac']
	&&\ptext{Definition of $p2ac$}\\
	&=p2ac(P \land Q)[\{y\}\cap ac'/ac']
\end{xflalign*}
\end{proof}
\end{proofs}
\end{lemma}

\begin{lemma}\label{lemma:p2ac(P)-undash-out-alpha-cap-ac'}
\begin{statement}
\begin{align*}
	&p2ac(P)[\{undash(State_{\II}(out\alpha_{-ok'}))\}\cap ac'/ac']\\
	&=\\
	&P[\mathbf{s}/in\alpha_{-ok}] \land undash(State_{\II}({out\alpha_{-ok'}})) \in ac'
\end{align*}
\end{statement}
\begin{proofs}
\begin{proof}
\begin{xflalign*}
	&p2ac(P)[\{undash(State_{\II}(out\alpha_{-ok'}))\}\cap ac'/ac']
	&&\ptext{Definition of $p2ac$}\\
	&=(\exists z @ P[\mathbf{s},\mathbf{z}/in\alpha_{-ok},out\alpha_{-ok'}] \land undash(z) \in ac')[\{undash(State_{\II}(out\alpha_{-ok'}))\}\cap ac'/ac']
	&&\ptext{Substitution}\\
	&=\exists z @ P[\mathbf{s},\mathbf{z}/in\alpha_{-ok},out\alpha_{-ok'}] \land undash(z) \in \{undash(State_{\II}(out\alpha_{-ok'}))\}\cap ac'
	&&\ptext{Property of sets}\\
	&=\exists z @ P[\mathbf{s},\mathbf{z}/in\alpha_{-ok},out\alpha_{-ok'}] \land undash(z) \in \{undash(State_{\II}(out\alpha_{-ok'}))\} \land undash(z) \in ac'
	&&\ptext{Property of sets}\\
	&=\exists z @ P[\mathbf{s},\mathbf{z}/in\alpha_{-ok},out\alpha_{-ok'}] \land undash(z) = undash(State_{\II}(out\alpha_{-ok'})) \land undash(z) \in ac'
	&&\ptext{Property of $undash$}\\
	&=\exists z @ P[\mathbf{s},\mathbf{z}/in\alpha_{-ok},out\alpha_{-ok'}] \land z = State_{\II}(out\alpha_{-ok'}) \land undash(z) \in ac'
	&&\ptext{One-point rule}\\
	&=P[\mathbf{s},\mathbf{z}/in\alpha_{-ok},out\alpha_{-ok'}][State_{\II}(out\alpha_{-ok'})/z] \land undash(State_{\II}(out\alpha_{-ok'})) \in ac'
	&&\ptext{Assumption: $z$ is fresh and~\cref{lemma:state-sub:P-z-S:S-z}}\\
	&=P[\mathbf{s}/in\alpha_{-ok}] \land undash(State_{\II}(out\alpha_{-ok'})) \in ac'
\end{xflalign*}
\end{proof}
\end{proofs}
\end{lemma}

\begin{lemma}\label{lemma:p2ac(P)-e-y-cap-ac':p2ac(P-land-e-z-y)}
\begin{statement}
Provided $ac'$ is not free in $P$,
\begin{align*}
	&p2ac(P)[\{y | e\}\cap ac'/ac'] = p2ac(P \land e[z/y])
\end{align*}
\end{statement}
\begin{proofs}
\begin{proof}
\begin{xflalign*}
	&p2ac(P)[\{z | e\}\cap ac'/ac']
	&&\ptext{Definition of $p2ac$}\\
	&=(\exists z @ P[\mathbf{s},\mathbf{z}/in\alpha_{-ok},out\alpha_{-ok'}] \land z \in ac')[\{z | e\}\cap ac'/ac']
	&&\ptext{Assumption: $ac'$ is not free in $P$ and substitution}\\
	&=\exists z @ P[\mathbf{s},\mathbf{z}/in\alpha_{-ok},out\alpha_{-ok'}] \land z \in \{y | e\}\cap ac'
	&&\ptext{Property of sets}\\
	&=\exists z @ P[\mathbf{s},\mathbf{z}/in\alpha_{-ok},out\alpha_{-ok'}] \land z \in \{y | e\} \land z \in ac'
	&&\ptext{Property of sets}\\
	&=\exists z @ P[\mathbf{s},\mathbf{z}/in\alpha_{-ok},out\alpha_{-ok'}] \land e[z/y] \land z \in ac'
	&&\ptext{Assumption: $(in\alpha_{-ok}\cup out\alpha_{-ok'}) \cap fv(e)=\emptyset$}\\
	&=\exists z @ (P \land e[z/y])[\mathbf{s},\mathbf{z}/in\alpha_{-ok},out\alpha_{-ok'}] \land z \in ac'
	&&\ptext{Definition of $p2ac$}\\
	&=p2ac(P \land e[z/y])
\end{xflalign*}
\end{proof}
\end{proofs}
\end{lemma}

\begin{lemma}\label{lemma:p2ac(P-land-Q):exists-x-p2ac(P)-x-ac'-land-p2ac(Q)-x-ac'}
\begin{statement}
Provided $ac'$ is not free in $P$ nor in $Q$,
\begin{align*}
	&p2ac(P \land Q) = \exists x @ p2ac(P)[\{x\}/ac'] \land p2ac(Q)[\{x\}/ac'] \land x \in ac'
\end{align*}
\end{statement}
\begin{proofs}
\begin{proof}
\begin{xflalign*}
	&\exists x @ p2ac(P)[\{x\}/ac'] \land p2ac(Q)[\{x\}/ac'] \land x \in ac'
	&&\ptext{Definition of $p2ac$}\\
	&=\exists x @ \left(
\begin{array}{l}
		(\exists z \spot P[\mathbf{s},\mathbf{z}/in\alpha_{-ok},out\alpha_{-ok'}] \land undash(z) \in ac')[\{x\}/ac']
		\\ \land \\
		(\exists y \spot Q[\mathbf{s},\mathbf{y}/in\alpha_{-ok},out\alpha_{-ok'}] \land undash(y) \in ac')[\{x\}/ac']
		\\ \land x \in ac'
	\end{array}\right)
	&&\ptext{Assumption: $ac'$ not free in $P$ nor $Q$ and substitution}\\
	&=\exists x @ \left(\begin{array}{l}
		(\exists z \spot P[\mathbf{s},\mathbf{z}/in\alpha_{-ok},out\alpha_{-ok'}] \land undash(z) \in \{x\})
		\\ \land \\
		(\exists y \spot Q[\mathbf{s},\mathbf{y}/in\alpha_{-ok},out\alpha_{-ok'}] \land undash(y) \in \{x\})
		\\ \land x \in ac'
	\end{array}\right)
	&&\ptext{Property of sets}\\
	&=\exists x @ \left(\begin{array}{l}
		(\exists z \spot P[\mathbf{s},\mathbf{z}/in\alpha_{-ok},out\alpha_{-ok'}] \land undash(z) = x)
		\\ \land \\
		(\exists y \spot Q[\mathbf{s},\mathbf{y}/in\alpha_{-ok},out\alpha_{-ok'}] \land undash(y) = x)
		\\ \land x \in ac'
	\end{array}\right)
	&&\ptext{Property of $dash$ and $undash$}\\
	&=\exists x @ \left(\begin{array}{l}
		(\exists z \spot P[\mathbf{s},\mathbf{z}/in\alpha_{-ok},out\alpha_{-ok'}] \land z = dash(x))
		\\ \land \\
		(\exists y \spot Q[\mathbf{s},\mathbf{y}/in\alpha_{-ok},out\alpha_{-ok'}] \land y = dash(x))
		\\ \land x \in ac'
	\end{array}\right)
	&&\ptext{Introduce fresh variable $t$}\\
	&=\exists t, x @ \left(\begin{array}{l}
		(\exists z \spot P[\mathbf{s},\mathbf{z}/in\alpha_{-ok},out\alpha_{-ok'}] \land z = t)
		\\ \land \\
		(\exists y \spot Q[\mathbf{s},\mathbf{y}/in\alpha_{-ok},out\alpha_{-ok'}] \land y = t)
		\\ \land t = dash(x) \land x \in ac'
	\end{array}\right)
	&&\ptext{One-point rule}\\
	&=\exists t, x @ \left(\begin{array}{l}
		P[\mathbf{s},\mathbf{z}/in\alpha_{-ok},out\alpha_{-ok'}][t/z]
		\\ \land \\
		Q[\mathbf{s},\mathbf{y}/in\alpha_{-ok},out\alpha_{-ok'}][t/y]
		\\ \land t = dash(x) \land x \in ac'
	\end{array}\right)
	&&\ptext{Property of $dash$ and $undash$}\\
	&=\exists t, x @ \left(\begin{array}{l}
		P[\mathbf{s},\mathbf{z}/in\alpha_{-ok},out\alpha_{-ok'}][t/z]
		\\ \land \\
		Q[\mathbf{s},\mathbf{y}/in\alpha_{-ok},out\alpha_{-ok'}][t/y]
		\\ \land undash(t) = x \land x \in ac'
	\end{array}\right)
	&&\ptext{One-point rule}\\
	&=\exists t @ \left(\begin{array}{l}
		P[\mathbf{s},\mathbf{z}/in\alpha_{-ok},out\alpha_{-ok'}][t/z]
		\\ \land \\
		Q[\mathbf{s},\mathbf{y}/in\alpha_{-ok},out\alpha_{-ok'}][t/y]
		\\ \land undash(t) \in ac'
	\end{array}\right)
	&&\ptext{Substitution}\\
	&=\exists t @ \left(\begin{array}{l}
		P[\mathbf{s},\mathbf{t}/in\alpha_{-ok},out\alpha_{-ok'}]
		\\ \land \\
		Q[\mathbf{s},\mathbf{t}/in\alpha_{-ok},out\alpha_{-ok'}]
		\\ \land undash(t) \in ac'
	\end{array}\right)
	&&\ptext{Substitution}\\
	&=\exists t @ (P \land Q)[\mathbf{s},\mathbf{t}/in\alpha_{-ok},out\alpha_{-ok'}] \land undash(t) \in ac'
	&&\ptext{Definition of $p2ac$}\\
	&=p2ac(P \land Q)
\end{xflalign*}
\end{proof}
\end{proofs}
\end{lemma}

\begin{lemma}\label{lemma:p2ac(P-land-Q):exists-z-P-land-p2ac(Q)-undash-z}
\begin{statement}
Provided that $ac'$ is not free in $P$,
\begin{align*}
	&p2ac(P \land Q) = \exists z @ \left(%
\right)
	&&\ptext{Introduce fresh variable $y$}\\
	&=\exists z @ \left(%
\right)
	&&\ptext{Property of sets}\\
	&=\exists z @ \left(%
\right)
	&&\ptext{Assumption: $ac'$ not free in $Q$ and substitution}\\
	&=\exists z @ \left(%
\right)[\{undash(z)\}/ac']
		\end{array}\right)
	\land undash(z) \in ac' \end{array}\right)
	&&\ptext{Definition of $p2ac$}\\
	&=\exists z @ P[\mathbf{s},\mathbf{z}/in\alpha_{-ok},out\alpha_{-ok'}] \land p2ac(Q)[\{undash(z)\}/ac'] \land undash(z) \in ac'
\end{xflalign*}
\end{proof}
\end{proofs}
\end{lemma}

\begin{lemma}\label{lemma:exists-x-p2ac(P):p2ac(exists-x-P)}
\begin{statement}
Provided $z$ is not $ac'$,
$\exists x @ p2ac(P) = p2ac(\exists x @ p2ac(P))$.
\end{statement}
\begin{proofs}
\begin{proof}\checkt{alcc}
\begin{xflalign*}
	&\exists x @ p2ac(P)
	&&\ptext{Definition of $p2ac$}\\
	&=\exists x @ (\exists z @ P[\mathbf{s},\mathbf{z}/in\alpha_{-ok},out\alpha_{-ok'}] \land undash(z) \in ac')
	&&\ptext{Assumption: $x \notin (in\alpha_{-ok} \cup out\alpha_{-ok'})$ and predicate calculus}\\
	&=(\exists z @ (\exists x @ P)[\mathbf{s},\mathbf{z}/in\alpha_{-ok},out\alpha_{-ok'}] \land undash(z) \in ac')
	&&\ptext{Definition of $p2ac$}\\
	&=p2ac(\exists x @ p2ac(P))
\end{xflalign*}
\end{proof}
\end{proofs}
\end{lemma}

\begin{lemma}\label{lemma:p2ac(P)-o-ok:p2ac(P-o-ok)}
\begin{statement}
$p2ac(P)[o/ok] = p2ac([o/ok])$
\end{statement}
\begin{proofs}\begin{proof}\checkt{alcc}
\begin{xflalign*}
	&p2ac(P)[o/ok]
	&&\ptext{Definition of $p2ac$}\\
	&=(\exists z @ P[\mathbf{s},\mathbf{z}/in\alpha_{-ok},out\alpha_{-ok'}] \land undash(z) \in ac')[o/ok]
	&&\ptext{Substitution}\\
	&=(\exists z @ P[o/ok][\mathbf{s},\mathbf{z}/in\alpha_{-ok},out\alpha_{-ok'}] \land undash(z) \in ac')
	&&\ptext{Definition of $p2ac$}\\
	&=p2ac([o/ok])
\end{xflalign*}
\end{proof}\end{proofs}
\end{lemma}

\begin{lemma}\label{theorem:p2ac(P-seq-Q)}
\begin{statement}
\begin{align*}
	&p2ac(P \circseq Q) = \exists z @ (P[\mathbf{s}/in\alpha_{-ok}] \circseq Q[\mathbf{z}/out\alpha_{-ok'}]) \land undash(z) \in ac'
\end{align*}
\end{statement}
\begin{proofs}
\begin{proof}\checkt{alcc}
\begin{xflalign*}
	&p2ac(P \circseq Q)
	&&\ptext{Definition of sequential composition}\\
	&=p2ac(\exists v_0 @ P[v_0/v'] \land Q[v_0/v])
	&&\ptext{Definition of $p2ac$}\\
	&=\exists z @ (\exists v_0 @ P[v_0/v'] \land Q[v_0/v])[\mathbf{s},\mathbf{z}/in\alpha_{-ok},out\alpha_{-ok'}] \land undash(z) \in ac'
	&&\ptext{Substitution}\\
	&=\exists z @ (\exists v_0 @ P[\mathbf{s}/in\alpha_{-ok}][v_0/v'] \land Q[\mathbf{z}/out\alpha_{-ok'}][v_0/v]) \land undash(z) \in ac'
	&&\ptext{Definition of sequential composition}\\
	&=\exists z @ (P[\mathbf{s}/in\alpha_{-ok}] \circseq Q[\mathbf{z}/out\alpha_{-ok'}]) \land undash(z) \in ac' 
\end{xflalign*}
\end{proof}
\end{proofs}
\end{lemma}

\begin{lemma}\label{lemma:p2ac(P-seq-Q):alternative-1}
\begin{statement}
Provided $ac'$ is not free in $P$,
\begin{align*}
	&p2ac(P \circseq Q) = P[\mathbf{s}/in\alpha_{-ok}] \circseq (\exists z @ Q[\mathbf{z}/out\alpha_{-ok'}] \land undash(z) \in ac') 
\end{align*}
\end{statement}
\begin{proofs}
\begin{proof}
\begin{xflalign*}
	&p2ac(P \circseq Q)
	&&\ptext{\cref{theorem:p2ac(P-seq-Q)}}\\
	&=\exists z @ (P[\mathbf{s}/in\alpha_{-ok}] \circseq Q[\mathbf{z}/out\alpha_{-ok'}]) \land undash(z) \in ac'
	&&\ptext{Definition of sequential composition}\\
	&=\exists z, v_0, ok_0 @ \left(\begin{array}{l}
		P[\mathbf{s}/in\alpha_{-ok}][ok_0,v_0/ok',v'] 
		\\ \land \\
		Q[\mathbf{z}/out\alpha_{-ok'}][ok_0,v_0/ok',v'] \land undash(z) \in ac'
	\end{array}\right)
	&&\ptext{Predicate calculus}\\
	&=\exists v_0, ok_0 @ \left(\begin{array}{l}
		P[\mathbf{s}/in\alpha_{-ok}][ok_0,v_0/ok',v'] 
		\\ \land \\
		(\exists z @ Q[\mathbf{z}/out\alpha_{-ok'}][ok_0,v_0/ok',v'] \land undash(z) \in ac')
	\end{array}\right)
	&&\ptext{Property of substitution}\\
	&=\exists v_0, ok_0 @ \left(\begin{array}{l}
		P[\mathbf{s}/in\alpha_{-ok}][ok_0,v_0/ok',v'] 
		\\ \land \\
		(\exists z @ Q[\mathbf{z}/out\alpha_{-ok'}] \land undash(z) \in ac')[ok_0,v_0/ok',v']
	\end{array}\right)
	&&\ptext{Definition of sequential composition}\\
	&=P[\mathbf{s}/in\alpha_{-ok}] \circseq (\exists z @ Q[\mathbf{z}/out\alpha_{-ok'}] \land undash(z) \in ac')
\end{xflalign*}
\end{proof}
\end{proofs}
\end{lemma}

\subsection{$ac2p$}

\subsubsection{Properties}

\begin{theorem}\label{theorem:ac2p(P-lor-Q):ac2p(P)-lor-ac2p(Q)}
$ac2p(P \lor Q) = ac2p(P) \lor ac2p(Q)$
\begin{proofs}\begin{proof}\checkt{pfr}\checkt{alcc}
\begin{xflalign*}
	&ac2p(P \lor Q)
	&&\ptext{Definition of $ac2p$}\\
	&=\mathbf{PBMH} (P \lor Q)[State_{\II}(in\alpha_{-ok})/s] \seqA \bigwedge x : out\alpha_{-ok'} \spot dash(s).x = x
	&&\ptext{Distributivity of $\mathbf{PBMH}$ (\cref{law:pbmh:distribute-disjunction})}\\
	&=\left(\begin{array}{l}
		\mathbf{PBMH} (P) 
		\\ \lor \\
		\mathbf{PBMH} (Q)
	\end{array}\right)[State_{\II}(in\alpha_{-ok})/s] \seqA \bigwedge x : out\alpha_{-ok'} \spot dash(s).x = x
	&&\ptext{Property of substitution}\\
	&=\left(\begin{array}{l}
		(\mathbf{PBMH} (P)[State_{\II}(in\alpha_{-ok})/s]  \lor \mathbf{PBMH} (Q)[State_{\II}(in\alpha_{-ok})/s]) 
		\\ \seqA \\
		\bigwedge x : out\alpha_{-ok'} \spot dash(s).x = x
	\end{array}\right)
	&&\ptext{Distributivity of $\seqA$ (\cref{law:seqA-right-distributivity})}\\
	&=\left(\begin{array}{l}
		(\mathbf{PBMH} (P)[State_{\II}(in\alpha_{-ok})/s] \seqA \bigwedge x : out\alpha_{-ok'} \spot dash(s).x = x) 
		\\ \lor \\
		(\mathbf{PBMH} (Q)[State_{\II}(in\alpha_{-ok})/s] \seqA \bigwedge x : out\alpha_{-ok'} \spot dash(s).x = x)
	\end{array}\right)
	&&\ptext{Definition of $ac2p$}\\
	&=ac2p(P) \lor ac2p(Q)
\end{xflalign*}
\end{proof}\end{proofs}
\end{theorem}

\begin{theorem}\label{theorem:ac2p(P-land-Q):ac2p(P)-land-ac2p(Q)} Provided $P$ and $Q$ are $\mathbf{PBMH}$-healthy,
\begin{align*}
	&ac2p(P \land Q) = ac2p(P) \land ac2p(Q)
\end{align*}
\begin{proofs}\begin{proof}\checkt{pfr}\checkt{alcc}
\begin{xflalign*}
	&ac2p(P \land Q)
	&&\ptext{Definition of $ac2p$}\\
	&=\mathbf{PBMH} (P \land Q)[State_{\II}(in\alpha_{-ok})/s] \seqA \bigwedge x : out\alpha_{-ok'} \spot dash(s).x = x
	&&\ptext{Assumption: $P$ and $Q$ are $\mathbf{PBMH}$-healthy and \cref{law:pbmh:distribute-conjunction}}\\
	&=\left(
\begin{array}{l}
		\mathbf{PBMH} (P) 
		\\ \land \\
		\mathbf{PBMH} (Q)
	\end{array}\right)[State_{\II}(in\alpha_{-ok})/s] \seqA \bigwedge x : out\alpha_{-ok'} \spot dash(s).x = x
	&&\ptext{Property of substitution}\\
	&=\left(\begin{array}{l}
		\mathbf{PBMH} (P)[State_{\II}(in\alpha_{-ok})/s] 
		\\ \land \\
		\mathbf{PBMH} (Q)[State_{\II}(in\alpha_{-ok})/s]
	\end{array}\right) \seqA \bigwedge x : out\alpha_{-ok'} \spot dash(s).x = x
	&&\ptext{Distributivity of $\seqA$ (\cref{law:seqA-right-distributivity-conjunction})}\\
	&=\left(\begin{array}{l}
		(\mathbf{PBMH} (P)[State_{\II}(in\alpha_{-ok})/s] \seqA \bigwedge x : out\alpha_{-ok'} \spot dash(s).x = x)
		\\ \land \\ 
		(\mathbf{PBMH} (Q)[State_{\II}(in\alpha_{-ok})/s] \seqA \bigwedge x : out\alpha_{-ok'} \spot dash(s).x = x)
	\end{array}\right)
	&&\ptext{Definition of $ac2p$}\\
	&=ac2p(P) \land ac2p(Q)
\end{xflalign*}
\end{proof}\end{proofs}
\end{theorem}

\subsubsection{Lemmas}

\begin{lemma}[$ac2p$-alternative-1]\label{lemma:ac2p:alternative-3}
\begin{statement}
\begin{align*}
	&ac2p(P) = \exists ac' \spot \left(\begin{array}{l}
		P[State_{\II}(in\alpha)/s]
		\\ \land \\
		\forall z \spot z \in ac' \implies (\bigwedge x : out\alpha \spot dash(z).x = x)
	\end{array}\right)
\end{align*}
\end{statement}
\begin{proofs}
\begin{proof}
\begin{xflalign*}
	&ac2p(P)
	&&\ptext{\cref{lemma:ac2p:alternative-2}}\\
	&=\exists ac' @ P[State_{\II}(in\alpha_{-ok})/s] \land ac'\subseteq \{ s | \bigwedge x : out\alpha_{-ok'} \spot dash(s).x = x\}
	&&\ptext{Property of sets}\\
	&=\exists ac' @ \left(\begin{array}{l}
		 P[State_{\II}(in\alpha_{-ok})/s] 
		 \\ \land \\
		 \forall z @ z \in ac' \implies (\bigwedge x : out\alpha_{-ok'} \spot dash(s).x = x)[z/s]
	\end{array}\right)
	&&\ptext{Substitution}\\
	&=\exists ac' @ \left(\begin{array}{l}
		P[State_{\II}(in\alpha_{-ok})/s]
		\\  \land \\
		\forall z @ z \in ac' \implies (\bigwedge x : out\alpha_{-ok'} \spot dash(z).x = x)
	\end{array}\right)
\end{xflalign*}
\end{proof}
\end{proofs}
\end{lemma}

\begin{lemma}[$ac2p$-alternative-2]\label{lemma:ac2p:alternative-1}
\begin{statement}
\begin{align*}
	&ac2p(P) 
	\circdef 
	\left(\begin{array}{l}
		\exists ac',s \spot P \land 
			(\forall z \spot z \in ac' \implies \bigwedge x : out\alpha_{-ok'} \spot dash(z).x = x)
			\\ \land \\
			(\bigwedge x : in\alpha_{-ok} \spot s.x = x)
	\end{array}\right)
\end{align*}
\end{statement}
\begin{proofs}
\begin{proof}\checkt{alcc}
\begin{xflalign*}
	&\left(\begin{array}{l}
		\exists ac',s \spot P \land 
			(\forall z \spot z \in ac' \implies \bigwedge x : out\alpha \spot dash(z).x = x)
			\\ \land \\
			(\bigwedge x : in\alpha \spot s.x = x)
	\end{array}\right)
	&&\ptext{\cref{lemma:state-sub:exists-z-State-P}}\\
	&=\exists ac' \spot P[State_{\II}(in\alpha)/s] \land	(\forall z \spot z \in ac' \implies \bigwedge x : out\alpha \spot dash(z).x = x)
	&&\ptext{Property of sets}\\
	&=\exists ac' \spot P[State_{\II}(in\alpha)/s] \land (\forall z \spot z \in ac' \implies z \in \{ s | \bigwedge x : out\alpha \spot dash(s).x = x \})
	&&\ptext{Property of subset inclusion}\\
	&=\exists ac' \spot P[State_{\II}(in\alpha)/s] \land ac' \subseteq \{ s | \bigwedge x : out\alpha \spot dash(s).x = x \}
	&&\ptext{Introduce fresh variable}\\
	&=\exists ac_0 \spot P[State_{\II}(in\alpha)/s][ac_0/ac'] \land ac_0 \subseteq \{ s | \bigwedge x : out\alpha \spot dash(s).x = x \}
	&&\ptext{Substitution}\\
	&=\exists ac_0 \spot P[ac_0/ac'][State_{\II}(in\alpha)/s] \land ac_0 \subseteq \{ s | \bigwedge x : out\alpha \spot dash(s).x = x \}
	&&\ptext{Introduce $ac'$ and definition of $\seqA$}\\
	&=(\exists ac_0 \spot P[ac_0/ac'][State_{\II}(in\alpha)/s] \land ac_0 \subseteq ac') \seqA \bigwedge x : out\alpha \spot dash(s).x = x
	&&\ptext{Substitution}\\
	&=(\exists ac_0 \spot P[ac_0/ac'] \land ac_0 \subseteq ac')[State_{\II}(in\alpha)/s] \seqA \bigwedge x : out\alpha \spot dash(s).x = x
	&&\ptext{Definition of $\mathbf{PBMH}$ (\cref{lemma:PBMH:alternative-1})}\\
	&=\mathbf{PBMH} (P)[State_{\II}(in\alpha)/s] \seqA \bigwedge x : out\alpha \spot dash(s).x = x
\end{xflalign*}
\end{proof}
\end{proofs}
\end{lemma}

\begin{lemma}[$ac2p$-alternative-3]\label{lemma:ac2p:alternative-2}
\begin{align*}
	&ac2p(P) \\
	&=\\
	&\exists ac' @ P[State_{\II}(in\alpha_{-ok})/s] \land ac'\subseteq \{ s | \bigwedge x : out\alpha_{-ok'} \spot dash(s).x = x\}
\end{align*}
\begin{proofs}
\begin{proof}
\begin{xflalign*}
	&ac2p(P)
	&&\ptext{Definition of $ac2p$}\\
	&=\mathbf{PBMH} (P)[State_{\II}(in\alpha_{-ok})/s] \seqA \bigwedge x : out\alpha_{-ok'} \spot dash(s).x = x
	&&\ptext{Definition of $\mathbf{PBMH}$ (\cref{lemma:PBMH:alternative-1})}\\
	&=\left(\begin{array}{l}
		(\exists ac_0 @ P[ac_0/ac'] \land ac_0\subseteq ac')[State_{\II}(in\alpha_{-ok})/s]
		\\ \seqA \\
		\bigwedge x : out\alpha_{-ok'} \spot dash(s).x = x
	\end{array}\right)
	&&\ptext{Definition of $\seqA$ and substitution}\\
	&=\left(\begin{array}{l}
		\exists ac_0 @ P[ac_0/ac'][State_{\II}(in\alpha_{-ok})/s] 
		\\ \land \\
		ac_0\subseteq \{ s | \bigwedge x : out\alpha_{-ok'} \spot dash(s).x = x\}
	\end{array}\right)
	&&\ptext{Substitution}\\
	&=\left(\begin{array}{l}
		\exists ac_0 @ P[State_{\II}(in\alpha_{-ok})/s][ac_0/ac']
		\\ \land \\
		ac_0\subseteq \{ s | \bigwedge x : out\alpha_{-ok'} \spot dash(s).x = x\}
	\end{array}\right)
	&&\ptext{Predicate calculus}\\
	&=\exists ac' @ P[State_{\II}(in\alpha_{-ok})/s] \land ac'\subseteq \{ s | \bigwedge x : out\alpha_{-ok'} \spot dash(s).x = x\}
\end{xflalign*}
\end{proof}
\end{proofs}
\end{lemma}

\begin{lemma}\label{lemma:ac2p(exists-y-in-ac'-e)} Provided $ac'$ is not free in $e$,
\begin{align*}
	&ac2p(\exists y \spot y \in ac' \land e) = e[State_{\II}(in\alpha_{-ok}),undash(State_{\II}(out\alpha_{-ok'}))/s,y]
\end{align*}
\begin{proofs}\begin{proof}\checkt{alcc}
\begin{xflalign*}
	&ac2p(\exists y \spot y \in ac' \land e)
	&&\ptext{Definition of $ac2p$}\\
	&=\mathbf{PBMH} (\exists y \spot y \in ac' \land e)[State_{\II}(in\alpha_{-ok})/s] \seqA \bigwedge x:out\alpha_{-ok'} \spot dash(s).x = x
	&&\ptext{Assumption: $ac'$ not free in $e$ and \cref{lemma:PBMH(exists-y-in-ac'-land-e)}}\\
	&=(\exists y \spot y \in ac' \land e)[State_{\II}(in\alpha_{-ok})/s] \seqA \bigwedge x:out\alpha_{-ok'} \spot dash(s).x = x
	&&\ptext{Substitution}\\
	&=(\exists y \spot y \in ac' \land e[State_{\II}(in\alpha_{-ok})/s]) \seqA \bigwedge x:out\alpha_{-ok'} \spot dash(s).x = x
	&&\ptext{Definition of $\seqA$ and substitution, $ac'$ not free in $e$}\\
	&=\exists y \spot y \in \left\{ s | \bigwedge x:out\alpha_{-ok'} \spot dash(s).x = x \right\} \land e[State_{\II}(in\alpha_{-ok})/s]
	&&\ptext{Property of sets}\\
	&=\exists y \spot \left(\bigwedge x:out\alpha_{-ok'} \spot dash(y).x = x\right) \land e[State_{\II}(in\alpha_{-ok})/s]
	&&\ptext{Introduce fresh variable}\\
	&=\exists z, y \spot \left(\bigwedge x:out\alpha_{-ok'} \spot z.x = x\right) \land z = dash(y) \land e[State_{\II}(in\alpha_{-ok})/s]
	&&\ptext{Property of $dash$}\\
	&=\exists z, y \spot \left(\bigwedge x:out\alpha_{-ok'} \spot z.x = x\right) \land undash(z) = y \land e[State_{\II}(in\alpha_{-ok})/s]
	&&\ptext{\cref{lemma:state-sub:exists-z-State-P} and substitution}\\
	&=\exists y \spot undash(State_{\II}(out\alpha_{-ok'})) = y \land e[State_{\II}(in\alpha_{-ok})/s]
	&&\ptext{One-point rule}\\
	&=e[State_{\II}(in\alpha_{-ok})/s][undash(State_{\II}(out\alpha_{-ok'}))/y]
	&&\ptext{Substitution}\\
	&=e[State_{\II}(in\alpha_{-ok}),undash(State_{\II}(out\alpha_{-ok'}))/s,y]
\end{xflalign*}
\end{proof}\end{proofs}
\end{lemma}

\begin{lemma} Provided $P$ is $\mathbf{A2}$-healthy,
\begin{align*}
	&ac2p(P) = \left(
\begin{array}{l}
		\exists ac_0 \spot P[\{s | \{s\} = ac_0\}/ac'][State_{\II}(in\alpha_{-ok})/s] 
		\\ \land \\
		ac_0 \subseteq \{ s | \bigwedge x : out\alpha_{-ok'} \spot dash(s).x = x \} 		
	\end{array}\right)
\end{align*}
\begin{proofs}\begin{proof}
\begin{xflalign*}
	&ac2p(P)
	&&\ptext{Definition of $ac2p$}\\
	&=\mathbf{PBMH} (P)[State_{\II}(in\alpha_{-ok})/s] \seqA \bigwedge x : out\alpha_{-ok'} \spot dash(s).x = x
	&&\ptext{Assumption: $P$ is $\mathbf{A2}$-healthy}\\
	&=\left(\begin{array}{l}
		\mathbf{PBMH} (\mathbf{PBMH} (P \seqA \{s\} = ac'))[State_{\II}(in\alpha_{-ok})/s]
		\\ \seqA \\
		\bigwedge x : out\alpha_{-ok'} \spot dash(s).x = x
	\end{array}\right)
	&&\ptext{$\mathbf{PBMH}$-idempotent}\\
	&=\left(\begin{array}{l}
		\mathbf{PBMH} (P \seqA \{s\} = ac')[State_{\II}(in\alpha_{-ok})/s]
		\\ \seqA \\
		\bigwedge x : out\alpha_{-ok'} \spot dash(s).x = x
	\end{array}\right)
	&&\ptext{Definition of $\seqA$ and substitution}\\
	&=\left(\begin{array}{l}
		\mathbf{PBMH} (P[\{s | \{s\} = ac'\}/ac'])[State_{\II}(in\alpha_{-ok})/s]
		\\ \seqA \\
		\bigwedge x : out\alpha_{-ok'} \spot dash(s).x = x
	\end{array}\right)
	&&\ptext{Definition of $\mathbf{PBMH}$ (\cref{lemma:PBMH:alternative-1})}\\
	&=\left(\begin{array}{l}
		(\exists ac_0 \spot (P[\{s | \{s\} = ac'\}/ac'])[ac_0/ac'] \land ac_0 \subseteq ac')[State_{\II}(in\alpha_{-ok})/s]
		\\ \seqA \\
		\bigwedge x : out\alpha_{-ok'} \spot dash(s).x = x
	\end{array}\right)
	&&\ptext{Substitution}\\
	&=\left(\begin{array}{l}
		(\exists ac_0 \spot P[\{s | \{s\} = ac_0\}/ac'][State_{\II}(in\alpha_{-ok})/s] \land ac_0 \subseteq ac')
		\\ \seqA \\
		\bigwedge x : out\alpha_{-ok'} \spot dash(s).x = x
	\end{array}\right)
	&&\ptext{Definition of $\seqA$ and substitution}\\
	&=\left(\begin{array}{l}
		\exists ac_0 \spot P[\{s | \{s\} = ac_0\}/ac'][State_{\II}(in\alpha_{-ok})/s] 
		\\ \land \\
		ac_0 \subseteq \{ s | \bigwedge x : out\alpha_{-ok'} \spot dash(s).x = x \} 		
	\end{array}\right)
\end{xflalign*}
\end{proof}\end{proofs}
\end{lemma}

\begin{lemma}\label{lemma:exists-out-lnot-ac2p:implies:lnot-P(emptyset-ac')}
\begin{align*}
	&\exists out\alpha \spot \lnot ac2p (P)[\mathbf{s}/in\alpha] \implies \lnot P[\emptyset/ac']
\end{align*}
\begin{proofs}\begin{proof}
\begin{xflalign*}
	&\exists out\alpha \spot \lnot ac2p (P)[\mathbf{s}/in\alpha]
	&&\ptext{Definition of $ac2p$}\\
	&=\exists out\alpha \spot \lnot \left(\begin{array}{l}
		\mathbf{PBMH} (P)[State_{\II}(in\alpha)/s]
		\\ \seqA \\
		\bigwedge x : out\alpha \spot dash(s).x = x
	\end{array}\right)[\mathbf{s}/in\alpha]
	&&\ptext{Definition of $\mathbf{PBMH}$ (\cref{lemma:PBMH:alternative-1})}\\
	&=\exists out\alpha \spot \lnot \left(\begin{array}{l}
		(\exists ac_0 \spot P[ac_0/ac'] \land ac_0\subseteq ac')[State_{\II}(in\alpha)/s]
		\\ \seqA \\
		\bigwedge x : out\alpha \spot dash(s).x = x
	\end{array}\right)[\mathbf{s}/in\alpha]
	&&\ptext{Substitution}\\
	&=\exists out\alpha \spot \lnot \left(\begin{array}{l}
		(\exists ac_0 \spot P[ac_0/ac'][State_{\II}(in\alpha)/s] \land ac_0\subseteq ac')
		\\ \seqA \\
		\bigwedge x : out\alpha \spot dash(s).x = x
	\end{array}\right)[\mathbf{s}/in\alpha]
	&&\ptext{Definition of $\seqA$ and substitution}\\
	&=\exists out\alpha \spot \lnot \left(\begin{array}{l}
		\exists ac_0 \spot P[ac_0/ac'][State_{\II}(in\alpha)/s] 
		\\ \land \\
		ac_0\subseteq \{ s | \bigwedge x : out\alpha \spot dash(s).x = x \} 		
	\end{array}\right)[\mathbf{s}/in\alpha]
	&&\ptext{Substitution}\\
	&=\exists out\alpha \spot \lnot \left(\begin{array}{l}
		\exists ac_0 \spot P[ac_0/ac'][State_{\II}(in\alpha)/s][\mathbf{s}/in\alpha] 
		\\ \land \\
		ac_0\subseteq \{ s | \bigwedge x : out\alpha \spot dash(s).x = x \} 		
	\end{array}\right)
	&&\ptext{\cref{lemma:state-sub:P-S-z:z-S}}\\
	&=\exists out\alpha \spot \lnot (
		\exists ac_0 \spot P[ac_0/ac'] 
		\land 
		ac_0\subseteq \{ s | \bigwedge x : out\alpha \spot dash(s).x = x \})
	&&\ptext{Predicate calculus}\\
	&=\exists out\alpha \spot (
		\forall ac_0 \spot \lnot P[ac_0/ac'] 
		\lor 
		\lnot (ac_0\subseteq \{ s | \bigwedge x : out\alpha \spot dash(s).x = x \}))
	&&\ptext{Predicate calculus}\\
	&\implies 
		\forall ac_0 \spot (
		\exists out\alpha \spot \lnot P[ac_0/ac'] 
		\lor 
		\lnot (ac_0\subseteq \{ s | \bigwedge x : out\alpha \spot dash(s).x = x \}))
	&&\ptext{Predicate calculus: $out\alpha$ not free in P}\\
	&=	\forall ac_0 \spot (
		\lnot P[ac_0/ac'] 
		\lor 
		\exists out\alpha \spot \lnot (ac_0\subseteq \{ s | \bigwedge x : out\alpha \spot dash(s).x = x \}))
	&&\ptext{Definition of subset inclusion}\\
	&=	\forall ac_0 \spot \left(\begin{array}{l}
		\lnot P[ac_0/ac'] 
		\\ \lor \\ 
		\exists out\alpha \spot \lnot (\forall y \spot y \in ac_0 \implies (\bigwedge x : out\alpha \spot dash(y).x = x))
	\end{array}\right)
	&&\ptext{Predicate calculus}\\
	&=	\forall ac_0 \spot \left(\begin{array}{l}
		\lnot P[ac_0/ac'] 
		\\ \lor \\ 
		\exists out\alpha \spot (\exists y \spot y \in ac_0 \land \lnot (\bigwedge x : out\alpha \spot dash(y).x = x))
	\end{array}\right)
	&&\ptext{Predicate calculus}\\
	&\implies \forall ac_0 \spot (\lnot P[ac_0/ac'] \lor (\exists out\alpha \spot \exists y \spot y \in ac_0))
	&&\ptext{Predicate calculus}\\
	&=\lnot \exists ac_0 \spot P[ac_0/ac'] \land ac_0 = \emptyset
	&&\ptext{One-point rule}\\
	&=\lnot P[\emptyset/ac']
\end{xflalign*}
\end{proof}\end{proofs}
\end{lemma}

The following lemma can be restated in a few different ways. Namely it can also imply:
\[
	\exists out\alpha \spot (\lnot P[State_{\II}(in\alpha)/s] \seqA \bigwedge x : out\alpha \spot dash(s).x = x)
\]

\begin{lemma}\label{lemma:lnot-ac2p(P)-exists-outalpha:implies:ac2p(lnot-P)} Provided $P$ is $\mathbf{PBMH}$-healthy,
\begin{align*}
	&\exists out\alpha \spot \lnot ac2p(P) \implies \exists out\alpha \spot ac2p(\lnot P)
\end{align*}
\begin{proofs}\begin{proof}
\begin{xflalign*}
	&\exists out\alpha \spot \lnot ac2p(P)
	&&\ptext{Definition of $ac2p$}\\
	&=\exists out\alpha \spot \lnot (\mathbf{PBMH} (P)[State_{\II}(in\alpha)/s] \seqA \bigwedge x : out\alpha \spot dash(s).x = x)
	&&\ptext{Assumption: $P$ is $\mathbf{PBMH}$-healthy}\\
	&=\exists out\alpha \spot \lnot (P[State_{\II}(in\alpha)/s] \seqA \bigwedge x : out\alpha \spot dash(s).x = x)
	&&\ptext{Property of $\seqA$}\\
	&=\exists out\alpha \spot ((\lnot P[State_{\II}(in\alpha)/s]) \seqA \bigwedge x : out\alpha \spot dash(s).x = x)
	&&\ptext{Predicate calculus (\cref{lemma:P-implies-PBMH})}\\
	&=\exists out\alpha \spot \left(\begin{array}{l}
		(\lnot P \land \mathbf{PBMH} (\lnot P))[State_{\II}(in\alpha)/s] 
		\\ \seqA \\
		\bigwedge x : out\alpha \spot dash(s).x = x
	\end{array}\right)
	&&\ptext{Property of substitution}\\
	&=\exists out\alpha \spot \left(\begin{array}{l}
		(\lnot P[State_{\II}(in\alpha)/s] \land \mathbf{PBMH} (\lnot P)[State_{\II}(in\alpha)/s] )
		\\ \seqA \\
		\bigwedge x : out\alpha \spot dash(s).x = x
	\end{array}\right)
	&&\ptext{Distributivity of $\seqA$ (\cref{law:seqA-right-distributivity-conjunction}) and substitution}\\
	&=\exists out\alpha \spot \left(\begin{array}{l}
		(\lnot P[State_{\II}(in\alpha)/s] \seqA \bigwedge x : out\alpha \spot dash(s).x = x)
		\\ \land \\
		(\mathbf{PBMH} (\lnot P)[State_{\II}(in\alpha)/s] \seqA \bigwedge x : out\alpha \spot dash(s).x = x)
	\end{array}\right)
	&&\ptext{Predicate calculus}\\
	&\implies \exists out\alpha \spot \mathbf{PBMH} (\lnot P)[State_{\II}(in\alpha)/s] \seqA \bigwedge x : out\alpha \spot dash(s).x = x
	&&\ptext{Definition of $ac2p$}\\
	&=\exists out\alpha \spot ac2p(\lnot P)
\end{xflalign*}
\end{proof}\end{proofs}
\end{lemma}

\begin{lemma}\label{lemma:ac2p(P)-exists-outalpha:exists-ac'-P-StateII} Provided none of the variables in $out\alpha$ are free in $P$,
\begin{align*}
	&\exists out\alpha \spot ac2p(P) \implies \exists ac' \spot P[State_{\II}(in\alpha)/s]
\end{align*}
\begin{proofs}\begin{proof}
\begin{xflalign*}
	&\exists out\alpha \spot ac2p(P)
	&&\ptext{Definition of $ac2p$}\\
	&=\exists out\alpha \spot \left(
\begin{array}{l}
		\mathbf{PBMH} (P)[State_{\II}(in\alpha)/s] 
		\\ \seqA \\
		\bigwedge x : out\alpha \spot dash(s).x = x
	\end{array}\right)
	&&\ptext{Definition of $\mathbf{PBMH}$ (\cref{lemma:PBMH:alternative-1})}\\
	&=\exists out\alpha \spot \left(\begin{array}{l}
		(\exists ac_0 \spot P[ac_0/ac'] \land ac_0 \subseteq ac')[State_{\II}(in\alpha)/s] 
		\\ \seqA \\
		\bigwedge x : out\alpha \spot dash(s).x = x
	\end{array}\right)
	&&\ptext{Substitution}\\
	&=\exists out\alpha \spot \left(\begin{array}{l}
		(\exists ac_0 \spot P[ac_0/ac'][State_{\II}(in\alpha)/s]  \land ac_0 \subseteq ac')
		\\ \seqA \\
		\bigwedge x : out\alpha \spot dash(s).x = x
	\end{array}\right)
	&&\ptext{Definition of $\seqA$ and substitution}\\
	&=\exists out\alpha \spot \left(\begin{array}{l}
		\exists ac_0 \spot P[ac_0/ac'][State_{\II}(in\alpha)/s] \land 
		\\ ac_0 \subseteq \{ s | \bigwedge x : out\alpha \spot dash(s).x = x \}
	\end{array}\right)
	&&\ptext{Property of sets}\\
	&=\exists out\alpha \spot \left(\begin{array}{l}
		\exists ac_0 \spot P[ac_0/ac'][State_{\II}(in\alpha)/s] \land 
		\\ \forall z \spot z \in ac_0 \implies (\bigwedge x : out\alpha \spot dash(z).x = x)
	\end{array}\right)
	&&\ptext{Predicate calculus: $out\alpha$ not free in $P$}\\
	&=\left(\begin{array}{l}
		\exists ac_0 \spot P[ac_0/ac'][State_{\II}(in\alpha)/s] 
		\\ \land \\
		\exists out\alpha \spot \forall z \spot z \in ac_0 \implies (\bigwedge x : out\alpha \spot dash(z).x = x)
	\end{array}\right)
	&&\ptext{Predicate calculus}\\
	&\implies \left(\begin{array}{l}
		\exists ac_0 \spot P[ac_0/ac'][State_{\II}(in\alpha)/s] 
		\\ \land \\
		\forall z \spot \exists out\alpha \spot (z \in ac_0 \implies (\bigwedge x : out\alpha \spot dash(z).x = x))
	\end{array}\right)
	&&\ptext{Predicate calculus}\\
	&=\left(\begin{array}{l}
		\exists ac_0 \spot P[ac_0/ac'][State_{\II}(in\alpha)/s] 
		\\ \land \\
		\forall z \spot z \in ac_0 \implies (\exists out\alpha \spot (\bigwedge x : out\alpha \spot dash(z).x = x))
	\end{array}\right)
	&&\ptext{One-point rule}\\
	&=\left(\begin{array}{l}
		\exists ac_0 \spot P[ac_0/ac'][State_{\II}(in\alpha)/s] 
		\\ \land \\
		\forall z \spot z \in ac_0 \implies true
	\end{array}\right)
	&&\ptext{Predicate calculus}\\
	&=\exists ac_0 \spot P[ac_0/ac'][State_{\II}(in\alpha)/s]
	&&\ptext{Predicate calculus}\\
	&=\exists ac' \spot P[State_{\II}(in\alpha)/s]
\end{xflalign*}
\end{proof}\end{proofs}
\end{lemma}

\begin{lemma}\label{lemma:ac2p(P-land-Q)-s-ac'-not-free:P-land-ac2p(Q)} Provided that $s$ and $ac'$ are not free in $P$,
\begin{align*}
	&ac2p(P \land Q) = P \land ac2p(Q)
\end{align*}
\begin{proofs}\begin{proof}\checkt{pfr}\checkt{alcc}
\begin{xflalign*}
	&ac2p(P \land Q)
	&&\ptext{Definition of $ac2p$}\\
	&=\exists ac' \spot (P \land Q)[State_{\II}(in\alpha)/s] \land ac' \subseteq \{ z | \bigwedge x : out\alpha \spot dash(z).x = x \}
	&&\ptext{Subtitution: $s$ not free in $P$}\\
	&=\exists ac' \spot P \land Q[State_{\II}(in\alpha)/s] \land ac' \subseteq \{ z | \bigwedge x : out\alpha \spot dash(z).x = x \}
	&&\ptext{Predicate calculus: $ac'$ not free in $P$}\\
	&=P \land \exists ac' \spot Q[State_{\II}(in\alpha)/s] \land ac' \subseteq \{ z | \bigwedge x : out\alpha \spot dash(z).x = x \}
	&&\ptext{Definition of $ac2p$}\\
	&=P \land ac2p(Q)
\end{xflalign*}
\end{proof}\end{proofs}
\end{lemma}

\begin{lemma}\label{lemma:ac2p(P)-s-ac'-not-free:P} Provided that $s$ and $ac'$ are not free in $P$,
\begin{align*}
	&ac2p(P) = P
\end{align*}
\begin{proofs}\begin{proof}\checkt{pfr}\checkt{alcc}
\begin{xflalign*}
	&ac2p(P)
	&&\ptext{Definition of $ac2p$}\\
	&=\exists ac' \spot P[State_{\II}(in\alpha)/s] \land ac' \subseteq \{ z | \bigwedge x : out\alpha \spot dash(z).x = x \}
	&&\ptext{Subtitution: $s$ not free in $P$}\\
	&=\exists ac' \spot P \land ac' \subseteq \{ z | \bigwedge x : out\alpha \spot dash(z).x = x \}
	&&\ptext{Predicate calculus: $ac'$ not free in $P$}\\
	&=P \land \exists ac' \spot ac' \subseteq \{ z | \bigwedge x : out\alpha \spot dash(z).x = x \}
	&&\ptext{Property of subset inclusion}\\
	&=P
\end{xflalign*}
\end{proof}\end{proofs}
\end{lemma}

\begin{lemma}\label{lemma:ac2p(P-design):(ac2p(lnot-Pf)--ac2p(Pt))} Provided $P$ is a design,
\begin{align*}
	&ac2p(P) = (\lnot ac2p(P^f) \vdash ac2p(P^t))
\end{align*}
\begin{proofs}\begin{proof}\checkt{pfr}\checkt{alcc}
\begin{xflalign*}
	&ac2p(P)
	&&\ptext{Assumption: $P$ is a design}\\
	&=ac2p(\lnot P^f \vdash P^t)
	&&\ptext{Definition of design}\\
	&=ac2p((ok \land \lnot P^f) \implies (P^t \land ok'))
	&&\ptext{Predicate calculus and distributivity of $ac2p$ (\cref{theorem:ac2p(P-lor-Q):ac2p(P)-lor-ac2p(Q)})}\\
	&=ac2p(\lnot ok) \lor ac2p(P^f) \lor ac2p(P^t \land ok')
	&&\ptext{\cref{lemma:ac2p(P)-s-ac'-not-free:P,lemma:ac2p(P-land-Q)-s-ac'-not-free:P-land-ac2p(Q)}}\\
	&=\lnot ok \lor ac2p(P^f) \lor (ac2p(P^t) \land ok')
	&&\ptext{Predicate calculus}\\
	&=(ok \land \lnot ac2p(P^f)) \implies (ac2p(P^t) \land ok')
	&&\ptext{Definition of design}\\
	&=(\lnot ac2p(P^f) \vdash ac2p(P^t))
\end{xflalign*}
\end{proof}\end{proofs}
\end{lemma}

\begin{lemma}\label{lemma:ac2p(P):implies:exists-ac'-P[StateII]}
$ac2p(P) \implies \exists ac' \spot P[State_{\II}(in\alpha)/s]$
\begin{proofs}\begin{proof}\checkt{pfr}\checkt{alcc}
\begin{flalign*}
	&ac2p(P)
	&&\ptext{Definition of $ac2p$}\\
	&=\exists ac' \spot P[State_{\II}(in\alpha)/s] \land ac' \subseteq \{ z | \bigwedge x : out\alpha \spot dash(z).x = x \}
	&&\ptext{Predicate calculus}\\
	&\implies (\exists ac' \spot P[State_{\II}(in\alpha)/s]) \land (\exists ac' \spot ac' \subseteq \{ z | \bigwedge x : out\alpha \spot dash(z).x = x \})
	&&\ptext{Property of sets}\\
	&=\exists ac' \spot P[State_{\II}(in\alpha)/s]
\end{flalign*}
\end{proof}\end{proofs}
\end{lemma}

\begin{lemma}\label{lemma:ac2p(P):ac'-not-free} Provided $ac'$ is not free in $P$,
\begin{align*}
	&ac2p(P) = P[State_{II}(in\alpha)/s]
\end{align*}
\begin{proofs}\begin{proof}
\begin{xflalign*}
	&ac2p(P)
	&&\ptext{Definition of $ac2p$}\\
	&=\mathbf{PBMH} (P)[State_{II}(in\alpha)/s] \seqA \bigwedge x : out\alpha \spot dash(s).x = x
	&&\ptext{Assumption: $ac'$ not free in $P$ and property of $\mathbf{PBMH}$}\\
	&=P[State_{II}(in\alpha)/s] \seqA \bigwedge x : out\alpha \spot dash(s).x = x
	&&\ptext{Definition of $\seqA$ and substitution}\\
	&=P[State_{II}(in\alpha)/s][\{ s | \bigwedge x : out\alpha \spot dash(s).x = x \}/ac']
	&&\ptext{Assumption: $ac'$ not free in $P$}\\
	&=P[State_{II}(in\alpha)/s]
\end{xflalign*}
\end{proof}\end{proofs}
\end{lemma}

\begin{lemma}\label{lemma:ac2p(P)-sub-ok'-wait:ac2p(sub-ok'-wait)}
$ac2p(P)^o_w = ac2p(P^o_w)$
\begin{proofs}\begin{proof}\checkt{alcc}
\begin{xflalign*}
	&ac2p(P)^o_w
	&&\ptext{Substitution abbreviation}\\
	&=ac2p(P)[o,w/ok',wait]
	&&\ptext{Definition of $ac2p$ (\cref{lemma:ac2p:alternative-2})}\\
	&=(\exists ac' \spot P[State_{\II}(in\alpha)/s] \land ac' \subseteq \{ s | \bigwedge x' : out\alpha \spot s.x=x'\})[o,w/ok',wait]
	&&\ptext{Substitution: $ok'$ and $wait$ not in $out\alpha$}\\
	&=\exists ac' \spot P[State_{\II}(in\alpha)/s][o,w/ok',wait] \land ac' \subseteq \{ s | \bigwedge x' : out\alpha \spot s.x=x'\}
	&&\ptext{Substitution: $ok'$ not in $in\alpha$}\\
	&=\exists ac' \spot P[o/ok'][State_{\II}(in\alpha)/s][w/wait] \land ac' \subseteq \{ s | \bigwedge x' : out\alpha \spot s.x=x'\}
	&&\ptext{\cref{lemma:state-sub:S-z-e-xi:z-oplus-xi-e-S-z}}\\
	&=\exists ac' \spot P[o/ok'][s\oplus\{wait\mapsto w\}/s][State_{\II}(in\alpha)/s] \land ac' \subseteq \{ s | \bigwedge x' : out\alpha \spot s.x=x'\}
	&&\ptext{Substitution abbreviation}\\
	&=\exists ac' \spot P^o_w[State_{\II}(in\alpha)/s] \land ac' \subseteq \{ s | \bigwedge x' : out\alpha \spot s.x=x'\}
	&&\ptext{Definition of $ac2p$ (\cref{lemma:ac2p:alternative-2})}\\
	&=ac2p(P^o_w)
\end{xflalign*}
\end{proof}\end{proofs}
\end{lemma}

\begin{lemma}\label{lemma:ac2p(conditional)} Provided $ac'$ is not free in $c$,
\begin{align*}
	&ac2p(P \dres c \rres Q) = ac2p(P) \dres c[State_{\II}(in\alpha_{-ok})/s] \rres ac2p(Q)
\end{align*}
\begin{proofs}\begin{proof}
\begin{xflalign*}
	&ac2p(P \dres c \rres Q)
	&&\ptext{Definition of conditional}\\
	&=ac2p((c \land P) \lor (\lnot c \land Q))
	&&\ptext{Distributivity of $ac2p$ (\cref{theorem:ac2p(P-land-Q):ac2p(P)-land-ac2p(Q)})}\\
	&=ac2p(c \land P) \lor ac2p(\lnot c \land Q)
	&&\ptext{Assumption: $ac'$ not free in $c$ and~\cref{lemma:ac2p(P-land-Q)-ac'-not-free:P-land-ac2p(Q)}}\\
	&=(c[State_{\II}(in\alpha_{-ok})/s] \land ac2p(P)) \lor (\lnot c[State_{\II}(in\alpha_{-ok})/s] \land ac2p(Q))
	&&\ptext{Property of substitution}\\
	&=(c[State_{\II}(in\alpha_{-ok})/s] \land ac2p(P)) \lor (\lnot (c[State_{\II}(in\alpha_{-ok})/s]) \land ac2p(Q))
	&&\ptext{Definition of conditional}\\
	&=ac2p(P) \dres c[State_{\II}(in\alpha_{-ok})/s] \rres ac2p(Q)
\end{xflalign*}
\end{proof}\end{proofs}
\end{lemma}

\begin{lemma}\label{lemma:ac2p(P-land-Q)-ac'-not-free:P-land-ac2p(Q)} Provided $ac'$ is not free in $P$,
\begin{align*}
	&ac2p(P \land Q) = P[State_{\II}(in\alpha_{-ok})/s] \land ac2p(Q)
\end{align*}
\begin{proofs}\begin{proof}\checkt{alcc}
\begin{xflalign*}
	&ac2p(P \land Q)
	&&\ptext{Definition of $ac2p$}\\
	&=\mathbf{PBMH} (P \land Q)[State_{\II}(in\alpha_{-ok})/s] \seqA \bigwedge x:out\alpha_{-ok'} \spot dash(s).x = x
	&&\ptext{Assumption: $ac'$ not free in $P$ and~\cref{lemma:PBMH(c-land-P):c-land-PBMH(P)}}\\
	&=(P \land \mathbf{PBMH}(Q))[State_{\II}(in\alpha_{-ok})/s] \seqA \bigwedge x:out\alpha_{-ok'} \spot dash(s).x = x
	&&\ptext{Property of substitution}\\
	&=\left(\begin{array}{l}
		(P[State_{\II}(in\alpha_{-ok})/s] \land \mathbf{PBMH}(Q)[State_{\II}(in\alpha_{-ok})/s]) 
		\\ \seqA \\
		\bigwedge x:out\alpha_{-ok'} \spot dash(s).x = x
	\end{array}\right)
	&&\ptext{Distributivity of $\seqA$ (\cref{law:seqA-right-distributivity-conjunction})}\\
	&=\left(\begin{array}{l}
		(P[State_{\II}(in\alpha_{-ok})/s] \seqA \bigwedge x:out\alpha_{-ok'} \spot dash(s).x = x)
		\\ \land \\
		(\mathbf{PBMH}(Q)[State_{\II}(in\alpha_{-ok})/s] \seqA \bigwedge x:out\alpha_{-ok'} \spot dash(s).x = x)
	\end{array}\right)
	&&\ptext{Assumption: $ac'$ not free in $P$ and~\cref{law:seqA-ac'-not-free}}\\
	&=\left(\begin{array}{l}
		P[State_{\II}(in\alpha_{-ok})/s]
		\\ \land \\
		(\mathbf{PBMH}(Q)[State_{\II}(in\alpha_{-ok})/s] \seqA \bigwedge x:out\alpha_{-ok'} \spot dash(s).x = x)
	\end{array}\right)
	&&\ptext{Definition of $ac2p$}\\
	&=P[State_{\II}(in\alpha_{-ok})/s] \land ac2p(Q)
\end{xflalign*}
\end{proof}\end{proofs}
\end{lemma}

\begin{lemma}\label{lemma:ac2p(s-in-ac'):x0-xi} Provided $in\alpha_{-ok} = \{x_0,\ldots,x_i\}$ and $in\alpha_{-ok}' = out\alpha_{-ok'}$,
\begin{align*}
	&ac2p(s \in ac') = x_0 = x_0' \land \ldots \land x_i = x_i'
\end{align*}
\begin{proofs}\begin{proof}
\begin{xflalign*}
	&ac2p(s \in ac')
	&&\ptext{Definition of $ac2p$}\\
	&=\mathbf{PBMH} (s \in ac')[State_{\II}(in\alpha_{-ok})/s] \seqA \bigwedge x:out\alpha_{-ok'} \spot dash(s).x = x
	&&\ptext{\cref{law:pbmh:s-in-ac'}}\\
	&=(s \in ac')[State_{\II}(in\alpha_{-ok})/s] \seqA \bigwedge x:out\alpha_{-ok'} \spot dash(s).x = x
	&&\ptext{Substitution}\\
	&=State_{\II}(in\alpha_{-ok}) \in ac' \seqA \bigwedge x:out\alpha_{-ok'} \spot dash(s).x = x
	&&\ptext{Definition of $\seqA$ and sustitution}\\
	&=State_{\II}(in\alpha_{-ok}) \in \{ s | \bigwedge x:out\alpha_{-ok'} \spot dash(s).x = x \}
	&&\ptext{Property of sets}\\
	&=\bigwedge x:out\alpha_{-ok'} \spot dash(State_{\II}(in\alpha_{-ok})).x = x
	&&\ptext{Definition of $State_{\II}$}\\
	&=\bigwedge x:out\alpha_{-ok'} \spot dash(\{x_0\mapsto x_0,\ldots,x_i\mapsto x_i\}).x = x
	&&\ptext{Application of $dash$}\\
	&=\bigwedge x:out\alpha_{-ok'} \spot \{x_0'\mapsto x_0,\ldots,x_i'\mapsto x_i\}.x = x
	&&\ptext{Expansion of conjunction}\\
	&=\{x_0'\mapsto x_0,\ldots,x_i'\mapsto x_i\}.x_0' = x_0' \land \ldots \land \{x_0'\mapsto x_0,\ldots,x_i'\mapsto x_i\}.x_i' = x_i'
	&&\ptext{Value of record component}\\ 
	&=x_0 = x_0' \land \ldots \land x_i = x_i'
\end{xflalign*}
\end{proof}\end{proofs}
\end{lemma}

\begin{lemma}\label{lemma:ac2p(P-land-ac'-neq-emptyset):ac2p(P)} Provided $P$ is $\mathbf{PBMH}$-healthy,
\begin{align*}
	&ac2p(P \land ac'\neq\emptyset) = ac2p(P)
\end{align*}
\begin{proofs}\begin{proof}\checkt{alcc}
\begin{xflalign*}
	&ac2p(P \land ac'\neq\emptyset)
	&&\ptext{Definition of $ac2p$}\\
	&=\mathbf{PBMH} (P \land ac'\neq\emptyset)[State_{\II}(in\alpha)/s] \seqA \bigwedge x':out\alpha \spot s.x=x'
	&&\ptext{Assumption: $P$ is $\mathbf{PBMH}$-healthy}\\
	&=\mathbf{PBMH} (\mathbf{PBMH} (P) \land ac'\neq\emptyset)[State_{\II}(in\alpha)/s] \seqA \bigwedge x':out\alpha \spot s.x=x'
	&&\ptext{$ac'\neq\emptyset$ is $\mathbf{PBMH}$-healthy}\\
	&=\mathbf{PBMH} \left(\begin{array}{l}
		\mathbf{PBMH} (P) 
		\\ \land \\ 
		\mathbf{PBMH} (ac'\neq\emptyset)
	\end{array}\right)[State_{\II}(in\alpha)/s] \seqA \bigwedge x':out\alpha \spot s.x=x'	
	&&\ptext{Closure of conjunction under $\mathbf{PBMH}$ (\cref{law:pbmh:conjunction-closure})}\\
	&=\left(\begin{array}{l}
		\mathbf{PBMH} (P) 
		\\ \land \\ 
		\mathbf{PBMH} (ac'\neq\emptyset)
	\end{array}\right)[State_{\II}(in\alpha)/s] \seqA \bigwedge x':out\alpha \spot s.x=x'	
	&&\ptext{$ac'\neq\emptyset$ is $\mathbf{PBMH}$-healthy}\\
	&=(\mathbf{PBMH} (P) \land ac'\neq\emptyset)[State_{\II}(in\alpha)/s] \seqA \bigwedge x':out\alpha \spot s.x=x'
	&&\ptext{Property of substitution}\\
	&=(\mathbf{PBMH} (P)[State_{\II}(in\alpha)/s] \land ac'\neq\emptyset) \seqA \bigwedge x':out\alpha \spot s.x=x'
	&&\ptext{Right-distributivity of $\seqA$ (\cref{law:seqA-right-distributivity-conjunction})}\\
	&=\left(\begin{array}{l}
		(\mathbf{PBMH} (P)[State_{\II}(in\alpha)/s] \seqA \bigwedge x':out\alpha \spot s.x=x') 
		\\ \land \\
		(ac'\neq\emptyset \seqA \bigwedge x':out\alpha \spot s.x=x')
	\end{array}\right)
	&&\ptext{Definition of $ac2p$}\\
	&=ac2p(P) \land (ac'\neq\emptyset \seqA \bigwedge x':out\alpha \spot s.x=x')
	&&\ptext{Property of sets}\\
	&=ac2p(P) \land	((\exists z \spot z \in ac') \seqA \bigwedge x':out\alpha \spot s.x=x')
	&&\ptext{Definition of $\seqA$ and substitution}\\
	&=ac2p(P) \land (\exists z \spot z \in \{ s | \bigwedge x':out\alpha \spot s.x=x' \})
	&&\ptext{Property of sets}\\
	&=ac2p(P) \land (\exists z \spot \bigwedge x':out\alpha \spot z.x=x')
	&&\ptext{One-point rule}\\
	&=ac2p(P)
\end{xflalign*}
\end{proof}\end{proofs}
\end{lemma}

\begin{lemma}\label{lemma:ac2p-o-PBMH(P):ac2p(P)}
$ac2p \circ \mathbf{PBMH} (P) = ac2p(P)$
\begin{proofs}\begin{proof}\checkt{alcc}
\begin{xflalign*}
	&ac2p \circ \mathbf{PBMH} (P)
	&&\ptext{Definition of $ac2p$}\\
	&=\mathbf{PBMH} (\mathbf{PBMH} (P))[State_{\II}(in\alpha)/s] \seqA \bigwedge x' : out\alpha \spot s.x=x'
	&&\ptext{\cref{law:pbmh:idempotent}}\\
	&=\mathbf{PBMH} (P)[State_{\II}(in\alpha)/s] \seqA \bigwedge x' : out\alpha \spot s.x=x'
	&&\ptext{Definition of $ac2p$}\\
	&=ac2p(P)
\end{xflalign*}
\end{proof}\end{proofs}
\end{lemma}

\begin{lemma}\label{lemma:ac2p(exists-x):exists-x(ac2p(P))}
\begin{statement}
Provided that $x$ is not $s$ nor $ac'$,
$ac2p(\exists x @ P) = \exists x @ ac2p(P)$
\end{statement}
\begin{proofs}
\begin{proof}\checkt{alcc}
\begin{xflalign*}
	&ac2p(\exists x @ P)
	&&\ptext{Definition of $ac2p$}\\
	&=\mathbf{PBMH} (\exists x @ P)[State_{\II}(in\alpha_{-ok})/s] \seqA \bigwedge x : out\alpha_{-ok'} \spot dash(s).x = x
	&&\ptext{Assumption: $x$ is not $ac'$ and~\cref{lemma:PBMH(exsits-x):exists-x(PBMH(x))}}\\
	&=(\exists x @ \mathbf{PBMH} (P))[State_{\II}(in\alpha_{-ok})/s] \seqA \bigwedge x : out\alpha_{-ok'} \spot dash(s).x = x
	&&\ptext{Assumption: $x$ is not $s$ and substitution}\\
	&=(\exists x @ \mathbf{PBMH} (P)[State_{\II}(in\alpha_{-ok})/s]) \seqA \bigwedge x : out\alpha_{-ok'} \spot dash(s).x = x
	&&\ptext{Definition of $\seqA$}\\
	&=(\exists x @ \mathbf{PBMH} (P)[State_{\II}(in\alpha_{-ok})/s])[\{ s | \bigwedge x : out\alpha_{-ok'} \spot dash(s).x = x\}/ac']
	&&\ptext{Assumption: $x$ is not $ac'$ and substitution}\\
	&=\exists x @ \mathbf{PBMH} (P)[State_{\II}(in\alpha_{-ok})/s][\{ s | \bigwedge x : out\alpha_{-ok'} \spot dash(s).x = x\}/ac']
	&&\ptext{Definition of $\seqA$}\\
	&=\exists x @ \mathbf{PBMH} (P)[State_{\II}(in\alpha_{-ok})/s] \seqA \bigwedge x : out\alpha_{-ok'} \spot dash(s).x = x\}
	&&\ptext{Definition of $ac2p$}\\
	&=\exists x @ ac2p(P)
\end{xflalign*}
\end{proof}
\end{proofs}
\end{lemma}

\begin{lemma}\label{lemma:ac2p(y-in-ac')}
\begin{statement}
$ac2p(y \in ac') = \bigwedge x : out\alpha_{-ok'} \spot dash(y[State_{\II}(in\alpha_{-ok})/s]).x = x$
\end{statement}
\begin{proofs}
\begin{proof}\checkt{alcc}
\begin{xflalign*}
	&ac2p(y \in ac')
	&&\ptext{Definition of $ac2p$}\\
	&=\mathbf{PBMH} (y \in ac')[State_{\II}(in\alpha_{-ok})/s] \seqA \bigwedge x : out\alpha_{-ok'} \spot dash(s).x = x
	&&\ptext{\cref{lemma:PBMH(x-in-ac'):x-in-ac'}}\\
	&=(y \in ac')[State_{\II}(in\alpha_{-ok})/s] \seqA \bigwedge x : out\alpha_{-ok'} \spot dash(s).x = x
	&&\ptext{Substitution}\\
	&=(y[State_{\II}(in\alpha_{-ok})/s] \in ac') \seqA \bigwedge x : out\alpha_{-ok'} \spot dash(s).x = x
	&&\ptext{Definition of $\seqA$ and substitution}\\
	&=y[State_{\II}(in\alpha_{-ok})/s] \in \{ z | \bigwedge x : out\alpha_{-ok'} \spot dash(z).x = x\}
	&&\ptext{Property of sets}\\
	&=\bigwedge x : out\alpha_{-ok'} \spot dash(y[State_{\II}(in\alpha_{-ok})/s]).x = x
\end{xflalign*}
\end{proof}
\end{proofs}
\end{lemma}

\begin{lemma}\label{lemma:ac2p(y-in-ac):y-not-s}
\begin{statement}
Provided $y$ is not $s$,
$ac2p(y \in ac') = \bigwedge x : out\alpha_{-ok'} \spot dash(y).x = x$
\end{statement}
\begin{proofs}
\begin{proof}\checkt{alcc}
\begin{xflalign*}
	&ac2p(y \in ac')
	&&\ptext{\cref{lemma:ac2p(y-in-ac')}}\\
	&=\bigwedge x : out\alpha_{-ok'} \spot dash(y[State_{\II}(in\alpha_{-ok})/s]).x = x
	&&\ptext{Assumption: $y$ is not $s$}\\
	&=\bigwedge x : out\alpha_{-ok'} \spot dash(y).x = x
\end{xflalign*}
\end{proof}
\end{proofs}
\end{lemma}

\begin{lemma}\label{lemma:exists-y-ac2p(P-land-y-in-ac'):ac2p(P)-y-subs}
\begin{statement}
Provided $P$ is $\mathbf{PBMH}$-healthy and $y$ is not $s$,
\begin{align*}
&\exists y @ ac2p(P \land y \in ac') = ac2p(P)[undash(State_{\II}(out\alpha_{-ok'})/y]
\end{align*}
\end{statement}
\begin{proofs}
\begin{proof}\checkt{alcc}
\begin{xflalign*}
	&\exists y @ ac2p(P \land y \in ac')
	&&\ptext{Assumption: $P$ is $\mathbf{PBMH}$-healthy and~\cref{theorem:ac2p(P-land-Q):ac2p(P)-land-ac2p(Q)}}\\
	&=\exists y @ ac2p(P) \land ac2p(y \in ac')
	&&\ptext{\cref{lemma:ac2p(y-in-ac):y-not-s}}\\
	&=\exists y @ ac2p(P) \land \bigwedge x : out\alpha_{-ok'} \spot dash(y).x = x
	&&\ptext{Predicate calculus, introduce fresh variable $z$}\\
	&=\exists y,z @ ac2p(P) \land \bigwedge x : out\alpha_{-ok'} \spot z.x = x \land dash(y) = z
	&&\ptext{Property of $dash$}\\
	&=\exists y,z @ ac2p(P) \land \bigwedge x : out\alpha_{-ok'} \spot z.x = x \land y = undash(z)
	&&\ptext{\cref{lemma:state-sub:exists-z-State-P}}\\
	&=\exists y @ (ac2p(P) \land y = undash(z))[State_{\II}(out\alpha_{-ok'})/z]
	&&\ptext{Substitution}\\
	&=\exists y @ ac2p(P) \land y = undash(State_{\II}(out\alpha_{-ok'}))
	&&\ptext{One-point rule}\\
	&=ac2p(P)[undash(State_{\II}(out\alpha_{-ok'})/y]
\end{xflalign*}
\end{proof}
\end{proofs}
\end{lemma}

\begin{lemma}\label{lemma:ac2p(circledIn(P)):ac2p(P)-y-subs:new}
\begin{statement}
Provided $P$ is $\mathbf{PBMH}$-healthy,
\begin{align*}
	&ac2p(\circledIn{y}{ac'} (P)) = ac2p(P[\{y\}\cap ac'/ac'])[undash(State_{\II}(out\alpha_{-ok'})/y]
\end{align*}
\end{statement}
\begin{proofs}
\begin{proof}\checkt{alcc}
\begin{xflalign*}
	&ac2p(\circledIn{y}{ac'} (P))
	&&\ptext{Definition of $\circledIn{y}{ac'}$}\\
	&=ac2p(\exists y @ P[\{y\}\cap ac'/ac'] \land y \in ac')
	&&\ptext{\cref{lemma:ac2p(exists-x):exists-x(ac2p(P))}}\\
	&=\exists y @ ac2p(P[\{y\}\cap ac'/ac'] \land y \in ac')
	&&\ptext{Assumption: $P$ is $\mathbf{PBMH}$-healthy and~\cref{lemma:PBMH(P-y-cap-ac'):PBMH(P)-y-cap-ac',lemma:exists-y-ac2p(P-land-y-in-ac'):ac2p(P)-y-subs}}\\
	&=ac2p(P[\{y\}\cap ac'/ac'])[undash(State_{\II}(out\alpha_{-ok'})/y]
\end{xflalign*}
\end{proof}
\end{proofs}
\end{lemma}

\begin{lemma}\label{lemma:ac2p(circledIn(P)):ac2p(P-undash-II-out-alpha-cap-ac')}
\begin{statement}
Provided $P$ is $\mathbf{PBMH}$-healthy,
\begin{align*}
	&ac2p(\circledIn{y}{ac'} (P))\\
	&=\\
	&ac2p(P[undash(State_{\II}(out\alpha_{-ok'}))/y][\{undash(State_{\II}(out\alpha_{-ok'}))\}\cap ac'/ac'])
\end{align*}
\end{statement}
\begin{proofs}
\begin{proof}\checkt{alcc}
\begin{xflalign*}
	&ac2p(\circledIn{y}{ac'} (P))
	&&\ptext{Assumption: $P$ is $\mathbf{PBMH}$-healthy and~\cref{lemma:ac2p(circledIn(P)):ac2p(P)-y-subs:new}}\\
	&=ac2p(P[\{y\}\cap ac'/ac'])[undash(State_{\II}(out\alpha_{-ok'}))/y]
	&&\ptext{\cref{lemma:ac2p(P)-e-y-subs:ac2p(P-e-y-subs)}}\\
	&=ac2p(P[\{y\}\cap ac'/ac'][undash(State_{\II}(out\alpha_{-ok'}))/y])
	&&\ptext{Substitution}\\
	&=ac2p(P[undash(State_{\II}(out\alpha_{-ok'}))/y][\{undash(State_{\II}(out\alpha_{-ok'}))\}\cap ac'/ac'])
\end{xflalign*}
\end{proof}
\end{proofs}
\end{lemma}

\begin{lemma}\label{lemma:ac2p(circledIn(P-land-Q)):ac'-not-free-in-Q-y-not-free-in-P}
\begin{statement}
Provided $P$ and $Q$ are $\mathbf{PBMH}$-healthy, $y$ is not free in $P$ and $ac'$ is not free in $Q$,
\begin{align*}
	&ac2p(\circledIn{y}{ac'} (P \land Q))\\
	&=\\
	&\left(\begin{array}{l}
		ac2p(P[\{undash(State_{\II}(out\alpha_{-ok'}))\}\cap ac'/ac']) 
		\\ \land \\
		Q[undash(State_{\II}(out\alpha_{-ok'}))/y][State_{\II}(in\alpha_{-ok})/s]
	\end{array}\right)
\end{align*}
\end{statement}
\begin{proofs}
\begin{proof}\checkt{alcc}
\begin{xflalign*}
	&ac2p(\circledIn{y}{ac'} (P \land Q))
	&&\ptext{Assumption: $P$ and $Q$ are $\mathbf{PBMH}$-healthy~\cref{law:pbmh:conjunction-closure,lemma:ac2p(circledIn(P)):ac2p(P-undash-II-out-alpha-cap-ac')}}\\
	&=ac2p((P \land Q)[undash(State_{\II}(out\alpha_{-ok'}))/y][\{undash(State_{\II}(out\alpha_{-ok'}))\}\cap ac'/ac'])
	&&\ptext{Assumption: $y$ is not free in $P$}\\
	&=ac2p\left(\begin{array}{l}
		\left(\begin{array}{l}
			P 
			\\ \land \\
			Q[undash(State_{\II}(out\alpha_{-ok'}))/y]
		\end{array}\right)[\{undash(State_{\II}(out\alpha_{-ok'}))\}\cap ac'/ac']
	\end{array}\right)
	&&\ptext{Assumption: $ac'$ is not free in $Q$}\\
	&=ac2p\left(\begin{array}{l}
		P[\{undash(State_{\II}(out\alpha_{-ok'}))\}\cap ac'/ac']
		\\ \land \\
		Q[undash(State_{\II}(out\alpha_{-ok'}))/y]
	\end{array}\right)
	&&\ptext{Assumption: $ac'$ is not free in $Q$ and~\cref{lemma:ac2p(P-land-Q)-ac'-not-free:P-land-ac2p(Q)}}\\
	&=\left(\begin{array}{l}
		ac2p(P[\{undash(State_{\II}(out\alpha_{-ok'}))\}\cap ac'/ac'])
		\\ \land \\
		Q[undash(State_{\II}(out\alpha_{-ok'}))/y][State_{\II}(in\alpha_{-ok})/s]
	\end{array}\right)
\end{xflalign*}
\end{proof}
\end{proofs}
\end{lemma}

\begin{lemma}\label{lemma:ac2p(P)-e-y-subs:ac2p(P-e-y-subs)}
\begin{statement}
Provided that $ac'$ is not free in $P$, and $s$ and $ac'$ are not free in $e$, and that $y$ is not $ac'$ nor $s$,
\begin{align*}
	&ac2p(P)[e/y] = ac2p(P[e/y])
\end{align*}
\end{statement}
\begin{proofs}
\begin{proof}\checkt{alcc}
\begin{xflalign*}
	&ac2p(P)[e/y]
	&&\ptext{Definition of $ac2p$}\\
	&=(\mathbf{PBMH} (P)[State_{\II}(in\alpha_{-ok})/s] \seqA \bigwedge x : out\alpha_{-ok'} \spot dash(s).x = x)[e/y]
	&&\ptext{Definition of $\seqA$}\\
	&=(\mathbf{PBMH} (P)[State_{\II}(in\alpha_{-ok})/s][\{s | \bigwedge x : out\alpha_{-ok'} \spot dash(s).x = x\}/ac'])[e/y]
	&&\ptext{Assumption: $y$ is not $ac'$ and $ac'$ is not free in $e$}\\
	&=\mathbf{PBMH} (P)[State_{\II}(in\alpha_{-ok})/s][e/y][\{s | \bigwedge x : out\alpha_{-ok'} \spot dash(s).x = x\}/ac']
	&&\ptext{Definition of $\seqA$}\\
	&=\mathbf{PBMH} (P)[State_{\II}(in\alpha_{-ok})/s][e/y] \seqA \bigwedge x : out\alpha_{-ok'} \spot dash(s).x = x
	&&\ptext{Assumption: $y$ is not $s$ and $s$ is not free in $e$}\\
	&=\mathbf{PBMH} (P)[e/y][State_{\II}(in\alpha_{-ok})/s] \seqA \bigwedge x : out\alpha_{-ok'} \spot dash(s).x = x
	&&\ptext{Definition of $\mathbf{PBMH}$ (\cref{lemma:PBMH:alternative-1})}\\
	&=\left(\begin{array}{l}
		(\exists ac_0 @ P[ac_0/ac'] \land ac_0\subseteq ac')[e/y][State_{\II}(in\alpha_{-ok})/s]
		\\ \seqA \\
		\bigwedge x : out\alpha_{-ok'} \spot dash(s).x = x
	\end{array}\right)
	&&\ptext{Assumption: $y$ is not $ac'$ and $ac'$ is not free in $e$}\\
	&=\left(\begin{array}{l}
		(\exists ac_0 @ P[e/y][ac_0/ac'] \land ac_0\subseteq ac')[State_{\II}(in\alpha_{-ok})/s]
		\\ \seqA \\
		\bigwedge x : out\alpha_{-ok'} \spot dash(s).x = x
	\end{array}\right)
	&&\ptext{Definition of $\mathbf{PBMH}$ (\cref{lemma:PBMH:alternative-1})}\\
	&=\mathbf{PBMH} (P[e/y])[State_{\II}(in\alpha_{-ok})/s]\seqA \bigwedge x : out\alpha_{-ok'} \spot dash(s).x = x
	&&\ptext{Definition of $ac2p$}\\
	&=ac2p(P[e/y])
\end{xflalign*}
\end{proof}
\end{proofs}
\end{lemma}

\begin{lemma}\label{lemma:ac2p(P-in-alpha-s-land-undash-state-II-out-alpha-in-ac'):P}
\begin{statement}
Provided $ac'$ is not free in $P$,
\begin{align*}
	&ac2p(P[\mathbf{s}/in\alpha_{-ok}] \land undash(State_{\II}({out\alpha_{-ok'}})) \in ac') = P
\end{align*}
\end{statement}
\begin{proofs}
\begin{proof}\checkt{alcc}
\begin{xflalign*}
	&ac2p(P[\mathbf{s}/in\alpha_{-ok}] \land undash(State_{\II}({out\alpha_{-ok'}})) \in ac')
	&&\ptext{\cref{lemma:ac2p(P-land-Q)-ac'-not-free:P-land-ac2p(Q)}}\\
	&=P[\mathbf{s}/in\alpha_{-ok}][State_{\II}(in\alpha_{-ok}/s] \land ac2p(undash(State_{\II}({out\alpha_{-ok'}})) \in ac'))
	&&\ptext{\cref{lemma:state-sub:P-z-S:S-z}}\\
	&=P \land ac2p(undash(State_{\II}({out\alpha_{-ok'}})) \in ac'))
	&&\ptext{\cref{lemma:ac2p(undash-State-II-out-alpha):true}}\\
	&=P \land true
	&&\ptext{Predicate calculus}\\
	&=P
\end{xflalign*}
\end{proof}
\end{proofs}
\end{lemma}

\begin{lemma}\label{lemma:ac2p(undash-State-II-out-alpha):true}
\begin{statement}
$ac2p(undash(State_{\II}({out\alpha_{-ok'}})) \in ac') = true$
\end{statement}
\begin{proofs}
\begin{proof}\checkt{alcc}
\begin{xflalign*}
	&ac2p(undash(State_{\II}({out\alpha_{-ok'}})) \in ac')
	&&\ptext{\cref{lemma:ac2p(y-in-ac):y-not-s}}\\
	&=\bigwedge x : out\alpha_{-ok'} \spot dash(undash(State_{\II}({out\alpha_{-ok'}}))).x = x
	&&\ptext{Property of $dash$ and $undash$}\\
	&=\bigwedge x : out\alpha_{-ok'} \spot State_{\II}({out\alpha_{-ok'}}).x = x
	&&\ptext{Definition of $Sate_{\II}$ and $x$ ranges over $out\alpha_{-ok'}$}\\
	&=\bigwedge x : out\alpha_{-ok'} \spot (\{ x_0' \mapsto x_0', \ldots, x_n' \mapsto x_n' \}).x = x
	&&\ptext{$x$ ranges over $out\alpha_{-ok'}$}\\
	&=\left(\begin{array}{l}
		(\{ x_0' \mapsto x_0', \ldots, x_n' \mapsto x_n' \}).x_0' = x_0'
		\\ \land \ldots \land \\
		(\{ x_0' \mapsto x_0', \ldots, x_n' \mapsto x_n' \}).x_n' = x_n'
	\end{array}\right)
	&&\ptext{Value of record components}\\
	&=x_0' = x_0' \land \ldots \land x_n' = x_n'
	&&\ptext{Predicate calculus}\\
	&=true
\end{xflalign*}
\end{proof}
\end{proofs}
\end{lemma}

\subsubsection{Properties with respect to Angelic Designs}

\begin{theorem}\label{theorem:ac2p-o-A(P):ac2p(design)} Provided that $P$ is a design,
\begin{align*}
	&ac2p \circ \mathbf{A} (P) = (\lnot ac2p(P^f) \vdash ac2p(P^t))
\end{align*}
\begin{proofs}\begin{proof}
\begin{xflalign*}
	&ac2p \circ \mathbf{A} (P)
	&&\ptext{Assumption: $P$ is a design}\\
	&=ac2p \circ \mathbf{A} (\lnot P^f \vdash P^t)
	&&\ptext{Definition of $\mathbf{A}$}\\
	&=ac2p(\lnot \mathbf{PBMH} (P^f) \vdash \mathbf{PBMH} (P^t) \land ac'\neq\emptyset)
	&&\ptext{Definition of design}\\
	&=ac2p((ok \land \lnot \mathbf{PBMH} (P^f)) \implies (\mathbf{PBMH} (P^t) \land ac'\neq\emptyset \land ok'))
	&&\ptext{Predicate calculus}\\
	&=ac2p(\lnot ok \lor \mathbf{PBMH} (P^f) \lor (\mathbf{PBMH} (P^t) \land ac'\neq\emptyset \land ok'))
	&&\ptext{Distributivity of $ac2p$ (\cref{theorem:ac2p(P-lor-Q):ac2p(P)-lor-ac2p(Q)})}\\
	&=ac2p(\lnot ok) \lor ac2p \circ \mathbf{PBMH} (P^f) \lor ac2p(\mathbf{PBMH} (P^t) \land ac'\neq\emptyset \land ok')
	&&\ptext{\cref{lemma:ac2p(P)-s-ac'-not-free:P}}\\
	&=\lnot ok \lor ac2p \circ \mathbf{PBMH} (P^f) \lor ac2p(\mathbf{PBMH} (P^t) \land ac'\neq\emptyset \land ok')
	&&\ptext{\cref{lemma:ac2p(P-land-Q)-s-ac'-not-free:P-land-ac2p(Q)}}\\
	&=\lnot ok \lor ac2p \circ \mathbf{PBMH} (P^f) \lor (ac2p(\mathbf{PBMH} (P^t) \land ac'\neq\emptyset) \land ok')
	&&\ptext{\cref{lemma:ac2p(P-land-ac'-neq-emptyset):ac2p(P)}}\\
	&=\lnot ok \lor ac2p \circ \mathbf{PBMH} (P^f) \lor (ac2p \circ \mathbf{PBMH} (P^t) \land ok')
	&&\ptext{\cref{lemma:ac2p-o-PBMH(P):ac2p(P)}}\\
	&=\lnot ok \lor ac2p(P^f) \lor (ac2p(P^t) \land ok')
	&&\ptext{Predicate calculus}\\
	&=(ok \land \lnot ac2p(P^f)) \implies (ac2p(P^t) \land ok')
	&&\ptext{Definition of design}\\
	&=(\lnot ac2p(P^f) \vdash ac2p(P^t))
\end{xflalign*}
\end{proof}\end{proofs}
\end{theorem}

\subsection{Isomorphism and Galois Connection ($d2ac$ and $ac2p$)}

\begin{theorem}\label{theorem:ac2p-o-d2acp(P):P}
\begin{statement}Provided that $P$ is a design,
$ac2p \circ d2ac(P) = P$.
\end{statement}
\begin{proofs}
\begin{proof}
\begin{xflalign*}
	&ac2p \circ d2ac(P)
	&&\ptext{Assumption: $P$ is a design}\\
	&=ac2p \circ d2ac(\lnot P^f \vdash P^t)
	&&\ptext{Definition of $d2ac$}\\
	&=ac2p (\lnot p2ac(P^f) \land (\lnot P^f[\mathbf{s}/in\alpha] \circseq true) \vdash p2ac(P^t))
	&&\ptext{Definition of design}\\
	&=ac2p ((ok \land \lnot p2ac(P^f) \land (\lnot P^f[\mathbf{s}/in\alpha] \circseq true)) \implies (p2ac(P^t) \land ok'))
	&&\ptext{Predicate calculus}\\
	&=ac2p (\lnot ok \lor p2ac(P^f) \lor \lnot (\lnot P^f[\mathbf{s}/in\alpha] \circseq true) \lor (p2ac(P^t) \land ok'))
	&&\ptext{Distributivity of $ac2p$ (\cref{theorem:ac2p(P-lor-Q):ac2p(P)-lor-ac2p(Q)})}\\
	&=\left(\begin{array}{l}
		ac2p (\lnot ok) \lor ac2p \circ p2ac(P^f) \lor ac2p(\lnot (\lnot P^f[\mathbf{s}/in\alpha] \circseq true)) 
		\\ \lor \\
		ac2p(p2ac(P^t) \land ok')
	\end{array}\right)
	&&\ptext{\cref{lemma:ac2p(P-land-Q)-s-ac'-not-free:P-land-ac2p(Q),lemma:ac2p(P)-s-ac'-not-free:P}}\\
	&=\left(\begin{array}{l}
		\lnot ok \lor ac2p \circ p2ac(P^f) \lor ac2p(\lnot (\lnot P^f[\mathbf{s}/in\alpha] \circseq true)) 
		\\ \lor \\
		(ac2p \circ p2ac(P^t) \land ok')
	\end{array}\right)
	&&\ptext{\cref{theorem:ac2p-o-p2ac(P):P}}\\
	&=(	\lnot ok \lor P^f \lor ac2p(\lnot (\lnot P^f[\mathbf{s}/in\alpha] \circseq true)) 
		\lor 
		(P^t \land ok'))
	&&\ptext{$ac'$ not free in $P^f$ and~\cref{lemma:ac2p(P):ac'-not-free}}\\
	&=(	\lnot ok \lor P^f \lor \lnot (\lnot P^f[\mathbf{s}/in\alpha] \circseq true))[State_{\II}(in\alpha)/s] 
		\lor 
		(P^t \land ok'))
	&&\ptext{Property of substitution}\\
	&=( \lnot ok \lor P^f \lor \lnot (\lnot P^f[\mathbf{s}/in\alpha][State_{\II}(in\alpha)/s] \circseq true)) 
		\lor
		(P^t \land ok'))
	&&\ptext{\cref{lemma:state-sub:P-z-S:S-z}}\\
	&=(	\lnot ok \lor P^f \lor \lnot (\lnot P^f \circseq true)) 
		\lor 
		(P^t \land ok'))
	&&\ptext{Predicate calculus and definition of design}\\
	&=(\lnot P^f \land (\lnot P^f \circseq true) \vdash P^t)
	&&\ptext{Definition of sequential composition}\\
	&=(\lnot P^f \land (\exists out\alpha \spot \lnot P^f) \vdash P^t)
	&&\ptext{Predicate calculus}\\
	&=(\lnot P^f \vdash P^t)
	&&\ptext{Assumption: $P$ is a design}\\
	&=P
\end{xflalign*}
\end{proof}
\end{proofs}
\end{theorem}


\begin{theorem}\label{theorem:d2ac-o-ac2p:implies:P}
\begin{statement}
Provided $P$ is an $\mathbf{A}$-healthy design, $d2ac \circ ac2p(P) \sqsupseteq P$.
\end{statement}
\begin{proofs}
\begin{proof}
\begin{xflalign*}
	&d2ac \circ ac2p(P) 
	&&\ptext{\cref{lemma:d2ac-o-ac2p(P)}}\\
	&=(\lnot p2ac(ac2p(P^f)) \land (\exists out\alpha \spot \lnot ac2p(P^f)[\mathbf{s}/in\alpha]) \vdash p2ac(ac2p(P^t)))
	&&\ptext{Assumption: $P^f$ and $P^t$ are $\mathbf{PBMH}$-healthy and~\cref{theorem:p2ac-o-ac2p:implies:P}}\\
	&=\left(
\begin{array}{l}
		\lnot (p2ac(ac2p(P^f)) \land P^f) \land (\exists out\alpha \spot \lnot ac2p(P^f)[\mathbf{s}/in\alpha]) 
		\\ \vdash \\
		p2ac(ac2p(P^t)) \land P^t
	\end{array}\right)
	&&\ptext{\cref{lemma:ac2p(P):implies:exists-ac'-P[StateII]} and predicate calculus}\\
	&=\left(%
\right)
	&&\ptext{Property of substitution}\\
	&=\left(%
\right)
	&&\ptext{Predicate calculus: $out\alpha$ not free in $P$}\\
	&=\left(%
\right)
		\\ \vdash \\
		p2ac(ac2p(P^t)) \land P^t
	\end{array}\right)
	&&\ptext{Refinement of designs}\\
	&\sqsupseteq (\lnot P^f \land (\lnot \exists ac' \spot P^f) \vdash P^t)
	&&\ptext{Predicate calculus}\\
	&=(\lnot P^f \vdash P^t)
	&&\ptext{Definition of design}\\
	&=P
\end{xflalign*}
\end{proof}
\end{proofs}
\end{theorem}

\begin{theorem}\label{theorem:d2ac-o-ac2p:L-implies:P}
\begin{statement}
Provided $P$ is an $\mathbf{A0}$-$\mathbf{A2}$-healthy design,
$d2ac \circ ac2p(P) \sqsubseteq P$.
\end{statement}
\begin{proofs}
\begin{proof}
\begin{xflalign*}
	&d2ac \circ ac2p(P)
	&&\ptext{Assumption: $P$ is an $\mathbf{A0}$-$\mathbf{A2}$-healthy design}\\
	&=d2ac \circ ac2p(\lnot \mathbf{A2} \circ \mathbf{PBMH} (P^f) \vdash \mathbf{A2} (\mathbf{PBMH} (P^t) \land ac'\neq\emptyset))
	&&\ptext{\cref{lemma:d2ac-o-ac2p(P)}}\\
	&=\left(
\right)
	&&\ptext{\cref{lemma:exists-out-lnot-ac2p:implies:lnot-P(emptyset-ac')} and refinement of designs}\\
	&\sqsubseteq 
	 \left(\begin{array}{l}
		\lnot p2ac \circ ac2p (\mathbf{A2} \circ \mathbf{PBMH} (P^f)) \land \lnot \mathbf{A2} \circ \mathbf{PBMH} (P^f)[\emptyset/ac']
		\\ \vdash \\
		p2ac \circ ac2p (\mathbf{A2} (\mathbf{PBMH} (P^t) \land ac'\neq\emptyset))
	\end{array}\right)
	&&\ptext{\cref{lemma:A2(P)-emptyset-ac':P-emptyset-ac'}}\\
	&=\left(\begin{array}{l}
		\lnot p2ac \circ ac2p (\mathbf{A2} \circ \mathbf{PBMH} (P^f)) \land \lnot \mathbf{PBMH} (P^f)[\emptyset/ac']
		\\ \vdash \\
		p2ac \circ ac2p (\mathbf{A2} (\mathbf{PBMH} (P^t) \land ac'\neq\emptyset))
	\end{array}\right)
	&&\ptext{Predicate calculus}\\
	&=\left(\begin{array}{l}
		\lnot (p2ac \circ ac2p (\mathbf{A2} \circ \mathbf{PBMH} (P^f)) \lor \mathbf{PBMH} (P^f)[\emptyset/ac'])
		\\ \vdash \\
		p2ac \circ ac2p (\mathbf{A2} (\mathbf{PBMH} (P^t) \land ac'\neq\emptyset))
	\end{array}%
\right)
	&&\ptext{Predicate calculus: absorption law}\\
	&=\left(%
\right)
	&&\ptext{Property of sets}\\
	&=\left(\begin{array}{l}
		\lnot \mathbf{A2} \circ \mathbf{PBMH} (P^f)
		\\ \vdash \\
		(\exists y \spot \mathbf{PBMH} (P^t)[\{y\}/ac'] \land y \in ac')
	\end{array}\right)
	&&\ptext{\cref{lemma:A2(P-land-ac'-neq-emptyset)}}\\
	&=( \lnot \mathbf{A2} \circ \mathbf{PBMH} (P^f)
		\vdash 
		\mathbf{A2} (\mathbf{PBMH} (P^t) \land ac'\neq\emptyset) )
	&&\ptext{Assumption: $P$ is an $\mathbf{A0}$-$\mathbf{A2}$-healthy design}\\
	&=P
\end{xflalign*}
\end{proof}
\end{proofs}
\end{theorem}

\begin{theorem}\label{theorem:d2ac-o-ac2p(P):P}
\begin{statement}
Provided $P$ is a design that is $\mathbf{A0}$-$\mathbf{A2}$-healthy,
\begin{align*}
	&d2ac \circ ac2p(P) = P
\end{align*}
\end{statement}
\begin{proofs} 
\begin{proof}\checkt{alcc}
Follows from~\cref{theorem:d2ac-o-ac2p:implies:P,theorem:d2ac-o-ac2p:L-implies:P}.
\end{proof}
\end{proofs}
\end{theorem}

\begin{lemma}\label{lemma:d2ac-o-ac2p(P)}
\begin{align*}
	&d2ac \circ ac2p(P)\\
	&=\\
	&(\lnot p2ac(ac2p(P^f)) \land (\exists out\alpha \spot \lnot ac2p(P^f)[\mathbf{s}/in\alpha]) \vdash p2ac(ac2p(P^t)))
\end{align*}
\begin{proofs}\begin{proof}\checkt{alcc}
\begin{xflalign*}
	&d2ac \circ ac2p(P)
	&&\ptext{\cref{lemma:ac2p(P-design):(ac2p(lnot-Pf)--ac2p(Pt))}}\\
	&=d2ac(\lnot ac2p(P^f) \vdash ac2p(P^t))
	&&\ptext{Definition of $d2ac$}\\
	&=(\lnot p2ac(ac2p(P^f)) \land (\lnot ac2p(P^f)[\mathbf{s}/in\alpha] \circseq true) \vdash p2ac(ac2p(P^t)))
	&&\ptext{Definition of sequential composition}\\
	&=(\lnot p2ac(ac2p(P^f)) \land (\exists out\alpha \spot \lnot ac2p(P^f)[\mathbf{s}/in\alpha]) \vdash p2ac(ac2p(P^t)))
\end{xflalign*}
\end{proof}\end{proofs}
\end{lemma}

\section{Relationship with the $\mathbf{PBMH}$ Theory}

\subsection{$d2pbmh$}

\begin{theorem}\label{theorem:d2pbmh:pbmh}
\begin{statement}Provided $P$ is $\mathbf{PBMH}$-healthy,
\begin{align*}
	&\mathbf{PBMH} \circ d2pbmh(P) = d2pbmh(P)
\end{align*}
\end{statement}
\begin{proofs}
\begin{proof}
\begin{xflalign*}
	&\mathbf{PBMH} \circ d2pbmh(P)
	&&\ptext{Assumption: $P$ is $\mathbf{PBMH}$-healthy and~\cref{lemma:d2pbmh-o-PBMH(P)}}\\
	&=\exists ac_0 @ \left(\begin{array}{l}
			(\lnot P^f \implies P^t)[true/ok][State_{\II}(in\alpha_{-ok})/s][ac_0/ac']
			\\ \land \\
			ac_0\subseteq undashset(ac')
	\end{array}\right)
	&&\ptext{Introduce fresh variable $ac_1$}\\
	&=\exists ac_0, ac_1 @ \left(\begin{array}{l}
			(\lnot P^f \implies P^t)[true/ok][State_{\II}(in\alpha_{-ok})/s][ac_0/ac']
			\\ \land \\
			ac_0\subseteq undashset(ac_1) \land	undashset(ac_1)\subseteq undashset(ac')
	\end{array}\right)
	&&\ptext{Property of $undashset$}\\
	&=\exists ac_0, ac_1 @ \left(\begin{array}{l}
			(\lnot P^f \implies P^t)[true/ok][State_{\II}(in\alpha_{-ok})/s][ac_0/ac']
			\\ \land \\
			ac_0\subseteq undashset(ac_1) \land	ac_1\subseteq ac'
	\end{array}\right)
	&&\ptext{Substitution and predicate calculus}\\
	&=\exists ac_1 @ \left(\begin{array}{l}\exists ac_0 @ \left(\begin{array}{l}
			(\lnot P^f \implies P^t)[true/ok][State_{\II}(in\alpha_{-ok})/s][ac_0/ac']
			\\ \land \\
			ac_0\subseteq undashset(ac')\end{array}\right)[ac_1/ac'] 
			\\ \land \\	ac_1\subseteq ac'
	\end{array}\right)
	&&\ptext{Definition of $\mathbf{PBMH}$}\\
	&=\left(\exists ac_0 @ \left(\begin{array}{l}
			(\lnot P^f \implies P^t)[true/ok][State_{\II}(in\alpha_{-ok})/s][ac_0/ac']
			\\ \land \\
			ac_0\subseteq undashset(ac')
	\end{array}\right)\right)
	&&\ptext{Assumption: $P$ is $\mathbf{PBMH}$-healthy and~\cref{lemma:d2pbmh-o-PBMH(P)}}\\
	&=d2pbmh(P)
\end{xflalign*}
\end{proof}
\end{proofs}
\end{theorem}\noindent

\begin{lemma}\label{lemma:d2pbmh-o-PBMH(P)}
\begin{statement}
\begin{align*}
	&d2pbmh \circ \mathbf{PBMH} (P)\\
	&=\\
	&\exists ac_0 @ \left(\begin{array}{l}
			(\lnot P^f \implies P^t)[true/ok][State_{\II}(in\alpha_{-ok})/s][ac_0/ac']
			\\ \land \\
			ac_0\subseteq undashset(ac')
	\end{array}\right)
\end{align*}
\end{statement}
\begin{proofs}
\begin{proof}
\begin{xflalign*}
	&d2pbmh \circ \mathbf{PBMH} (P)
	&&\ptext{Definition of $d2pbmh$}\\
	&=(\lnot \mathbf{PBMH} (P)^f \implies \mathbf{PBMH} (P)^t)[true/ok][undashset(ac')/ac'][State_{\II}(in\alpha_{-ok})/s]
	&&\ptext{\cref{lemma:PBMH(P)-ow:PBMH(P-ow)}}\\
	&=(\lnot \mathbf{PBMH} (P^f) \implies \mathbf{PBMH} (P^t))[true/ok][undashset(ac')/ac'][State_{\II}(in\alpha_{-ok})/s]
	&&\ptext{Predicate calculus and~\cref{law:pbmh:distribute-disjunction}}\\
	&=\mathbf{PBMH} (\lnot P^f \implies P^t)[true/ok][undashset(ac')/ac'][State_{\II}(in\alpha_{-ok})/s]
	&&\ptext{Definition of $\mathbf{PBMH}$}\\
	&=\left(\begin{array}{l}
		\exists ac_0 @ (\lnot P^f \implies P^t)[ac_0/ac']
		\\ \land \\
		ac_0\subseteq ac'
	\end{array}\right)[true/ok][undashset(ac')/ac'][State_{\II}(in\alpha_{-ok})/s]
	&&\ptext{Substitution}\\
	&=\exists ac_0 @ \left(\begin{array}{l}
			(\lnot P^f \implies P^t)[true/ok][State_{\II}(in\alpha_{-ok})/s][ac_0/ac']
			\\ \land \\
			ac_0\subseteq undashset(ac')
	\end{array}\right)
\end{xflalign*}
\end{proof}
\end{proofs}
\end{lemma}

\subsection{$pbmh2d$}

\begin{theorem}\label{theorem:A-o-H3-o-pbmh2d(P):pbmh2d(P)}
\begin{statement}
Provided $P$ is $\mathbf{PBMH}$-healthy,
\begin{align*}
	&\mathbf{A} \circ \mathbf{H3} \circ pbmh2d(P) = pbmh2d(P)
\end{align*}
\end{statement}
\begin{proofs}
\begin{proof}
\begin{xflalign*}
	&\mathbf{A} \circ \mathbf{H3} \circ pbmh2d(P)
	&&\ptext{Definition of $pbmh2d$}\\
	&=\mathbf{A} \circ \mathbf{H3} ((\lnot P[\emptyset/ac'] \vdash P[dashset(ac')/ac'])[\mathbf{s}/in\alpha_{-ok}])
	&&\ptext{Substitution}\\
	&=\mathbf{A} \circ \mathbf{H3} ((\lnot P[\emptyset/ac'][\mathbf{s}/in\alpha_{-ok}] \vdash P[dashset(ac')/ac'][\mathbf{s}/in\alpha_{-ok}]))
	&&\ptext{Definition of $\mathbf{A}$ and $\mathbf{H3}$}\\
	&=\left(
\begin{array}{l}
		\exists ac' @ \lnot P[\emptyset/ac'][\mathbf{s}/in\alpha_{-ok}]
		\\ \vdash \\
		\mathbf{PBMH} (P[dashset(ac')/ac'][\mathbf{s}/in\alpha_{-ok}]) \land ac'\neq\emptyset
	\end{array}\right)
	&&\ptext{$ac'$ not free}\\
	&=\left(\begin{array}{l}
		\lnot P[\emptyset/ac'][\mathbf{s}/in\alpha_{-ok}]
		\\ \vdash \\
		\mathbf{PBMH} (P[dashset(ac')/ac'][\mathbf{s}/in\alpha_{-ok}]) \land ac'\neq\emptyset
	\end{array}\right)
	&&\ptext{Substitution and definition of $\mathbf{PBMH}$}\\
	&=\left(\begin{array}{l}
		\lnot P[\emptyset/ac'][\mathbf{s}/in\alpha_{-ok}]
		\\ \vdash \\
		\mathbf{PBMH} (P[dashset(ac')/ac'])[\mathbf{s}/in\alpha_{-ok}] \land ac'\neq\emptyset
	\end{array}\right)
	&&\ptext{\cref{lemma:PBMH(P)-subs-f-ac'}}\\
	&=\left(\begin{array}{l}
		\lnot P[\emptyset/ac'][\mathbf{s}/in\alpha_{-ok}]
		\\ \vdash \\
		\mathbf{PBMH} (P)[dashset(ac')/ac'][\mathbf{s}/in\alpha_{-ok}] \land ac'\neq\emptyset
	\end{array}\right)
	&&\ptext{Assumption: $P$ is $\mathbf{PBMH}$-healthy}\\
	&=\left(\begin{array}{l}
		\lnot P[\emptyset/ac'][\mathbf{s}/in\alpha_{-ok}]
		\\ \vdash \\
		P[dashset(ac')/ac'][\mathbf{s}/in\alpha_{-ok}] \land ac'\neq\emptyset
	\end{array}\right)
	&&\ptext{Property of designs}\\
	&=\left(\begin{array}{l}
		\lnot P[\emptyset/ac'][\mathbf{s}/in\alpha_{-ok}]
		\\ \vdash \\
		(P[\emptyset/ac'][\mathbf{s}/in\alpha_{-ok}]) \lor (P[dashset(ac')/ac'][\mathbf{s}/in\alpha_{-ok}] \land ac'\neq\emptyset)
	\end{array}\right)
	&&\ptext{\cref{lemma:P-land-ac'-neq-emptyset:P-ac'-subs-emptyset}}\\
	&=\left(\begin{array}{l}
		\lnot P[\emptyset/ac'][\mathbf{s}/in\alpha_{-ok}]
		\\ \vdash \\
		(P[\mathbf{s}/in\alpha_{-ok}] \land ac'=\emptyset) \lor (P[dashset(ac')/ac'][\mathbf{s}/in\alpha_{-ok}] \land ac'\neq\emptyset)
	\end{array}%
\right)
	&&\ptext{Definition of $pbmh2d$}\\
	&=pbmh2d(P)
\end{xflalign*}
\end{proof}
\end{proofs}
\end{theorem}

\subsection[Galois Connection and Isomorphism ($d2pbmh$ and $pbmh2d$)]{Galois Connection and Isomorphism \\($d2pbmh$ and $pbmh2d$)}

\begin{theorem}\label{theorem:d2pbm-o-pbmh2d(P):P}
\begin{statement}
Provided $P$ is $\mathbf{PBMH}$-healthy, $d2pbmh \circ pbmh2d(P) = P$.
\end{statement}
\begin{proofs}
\begin{proof}
\begin{xflalign*}
	&d2pbmh \circ pbmh2d(P)
	&&\ptext{Definition of $d2pbmh$}\\
	&=(\lnot (pbmh2d(P))^f \implies (pbmh2d(P))^t)[true/ok][undashset(ac')/ac'][State_{\II}(in\alpha_{-ok})/s]
	&&\ptext{Definition of $pbmh2d$ and~\cref{lemma:design:(P|-Q)f:ok-implies-lnot-Pf,lemma:design(P|-Q)t:ok-land-Pt-implies-Qt}}\\
	&=\left(\begin{array}{l}
		\lnot (ok \implies P[\emptyset/ac'][\mathbf{s}/in\alpha_{-ok}]^f)
		\\ \implies \\
		\left(\begin{array}{l}
			(ok \land P[\emptyset/ac'][\mathbf{s}/in\alpha_{-ok}]^f)
			\\ \implies \\
			P[dashset(ac')/ac'][\mathbf{s}/in\alpha_{-ok}]^t)
		\end{array}\right)
	\end{array}\right)[true/ok][undashset(ac')/ac'][State_{\II}(in\alpha_{-ok})/s]
	&&\ptext{Substitution}\\
	&=\left(\begin{array}{l}
		\lnot P[\emptyset/ac'][\mathbf{s}/in\alpha_{-ok}]^f
		\\ \implies \\
		\left(\begin{array}{l}
			(P[\emptyset/ac'][\mathbf{s}/in\alpha_{-ok}]^f)
			\\ \implies \\
			P[dashset(ac')/ac'][\mathbf{s}/in\alpha_{-ok}]^t)
		\end{array}\right)
	\end{array}\right)[undashset(ac')/ac'][State_{\II}(in\alpha_{-ok})/s]
	&&\ptext{$ok'$ not free in $P$}\\
	&=\left(\begin{array}{l}
		\lnot P[\emptyset/ac'][\mathbf{s}/in\alpha_{-ok}]
		\\ \implies \\
		\left(\begin{array}{l}
			(P[\emptyset/ac'][\mathbf{s}/in\alpha_{-ok}])
			\\ \implies \\
			(P[dashset(ac')/ac'][\mathbf{s}/in\alpha_{-ok}])
		\end{array}\right)
	\end{array}\right)[undashset(ac')/ac'][State_{\II}(in\alpha_{-ok})/s]
	&&\ptext{Predicate calculus}\\
	&=\left(\begin{array}{l}
		\lnot P[\emptyset/ac'][\mathbf{s}/in\alpha_{-ok}]
		\\ \implies \\
		P[dashset(ac')/ac'][\mathbf{s}/in\alpha_{-ok}]
	\end{array}\right)[undashset(ac')/ac'][State_{\II}(in\alpha_{-ok})/s]
	&&\ptext{Substitution}\\
	&=\left(\begin{array}{l}
		\lnot P[\emptyset/ac'][\mathbf{s}/in\alpha_{-ok}][State_{\II}(in\alpha_{-ok})/s]
		\\ \implies \\
		P[dashset \circ undashset(ac')/ac'][\mathbf{s}/in\alpha_{-ok}][State_{\II}(in\alpha_{-ok})/s]
	\end{array}\right)
	&&\ptext{Property of $dashset$ and $undashset$}\\
	&=\left(\begin{array}{l}
		\lnot P[\emptyset/ac'][\mathbf{s}/in\alpha_{-ok}][State_{\II}(in\alpha_{-ok})/s]
		\\ \implies \\
		P[\mathbf{s}/in\alpha_{-ok}][State_{\II}(in\alpha_{-ok})/s]
	\end{array}\right)
	&&\ptext{\cref{lemma:state-sub:P-z-S:S-z}}\\
	&=(\lnot P[\emptyset/ac'] \implies P)
	&&\ptext{Predicate calculus}\\
	&=P[\emptyset/ac'] \lor P
	&&\ptext{Assumption: $P$ is $\mathbf{PBMH}$-healthy and~\cref{law:pmbh:P-empty-ac'-or-P}}\\
	&=P
\end{xflalign*}
\end{proof}
\end{proofs}
\end{theorem}

\begin{theorem}\label{theorem:pbmh2d-o-d2pbmh(P):sqsubseteq:P}
\begin{statement}
Provided $P$ is an $\mathbf{A}$-healthy design,
\begin{align*}
	&pbmh2d \circ d2pbmh(P) \sqsubseteq P
\end{align*}
\end{statement}
\begin{proofs}
\begin{proof}
\begin{xflalign*}
	&pbmh2d \circ d2pbmh(P)
	&&\ptext{\cref{lemma:pbmh2d-o-d2pbmh(P)}}\\
	&=(\lnot P^f[\emptyset/ac'] \land \lnot P^t[\emptyset/ac'] \vdash (\lnot P^f \implies P^t))
	&&\ptext{Assumption: $P$ is an $\mathbf{A}$-healthy design}\\
	&=\left(
\right)
	&&\ptext{Property of sets and predicate calculus}\\
	&=\left(%
\right)
	&&\ptext{Property of designs and predicate calculus}\\
	&=\left(%
\right)
	&&\ptext{Property of designs and predicate calculus}\\
	&=\left(\begin{array}{l}
			\lnot \mathbf{PBMH} (P^f)[\emptyset/ac'] \land \lnot \mathbf{PBMH} (P^f)
			\\ \vdash \\
			\mathbf{PBMH} (P^t) \land ac'\neq\emptyset
	\end{array}\right)
	&&\ptext{Weaken precondition}\\
	&\sqsubseteq (\lnot \mathbf{PBMH} (P^f) \vdash \mathbf{PBMH} (P^t) \land ac'\neq\emptyset)
	&&\ptext{Definition of $\mathbf{A}$}\\
	&=\mathbf{A} (\lnot P^f \vdash P^t)
	&&\ptext{Assumption: $P$ is an~$\mathbf{A}$-healthy design}\\
	&=P
\end{xflalign*}
\end{proof}
\end{proofs}
\end{theorem}

\begin{theorem}\label{theorem:pbmh2d-o-d2pbmh(P):P}
\begin{statement}
Provided $P$ is design that is $\mathbf{A}$ and $\mathbf{H3}$-healthy,
\begin{align*}
	&pbmh2d \circ d2pbmh(P) = P
\end{align*}
\end{statement}
\begin{proofs}
\begin{proof}
\begin{xflalign*}
	&pbmh2d \circ d2pbmh(P)
	&&\ptext{\cref{lemma:pbmh2d-o-d2pbmh(P)}}\\
	&=(\lnot P^f[\emptyset/ac'] \land \lnot P^t[\emptyset/ac'] \vdash (\lnot P^f \implies P^t))
	&&\ptext{Assumption: $P$ is an design that is $\mathbf{A}$ and $\mathbf{H3}$-healthy}\\
	&=\left(
\right)
	&&\ptext{Property of sets and predicate calculus}\\
	&=\left(%
\right)
	&&\ptext{Property of designs and predicate calculus}\\
	&=\left(\begin{array}{l}
			ok \land \exists ac' @ \lnot P^f 
			\\ \vdash \\
			\mathbf{PBMH} (P^t) \land ac'\neq\emptyset
	\end{array}\right)
	&&\ptext{Property of designs and predicate calculus}\\
	&=(\exists ac' @ \lnot P^f \vdash \mathbf{PBMH} (P^t) \land ac'\neq\emptyset)
	&&\ptext{Definition of $\mathbf{A}$ and $\mathbf{H3}$}\\
	&=\mathbf{A} \circ \mathbf{H3} (\lnot P^f \vdash P^t)
	&&\ptext{Assumption: $P$ is an design that is $\mathbf{A}$ and $\mathbf{H3}$-healthy}\\
	&=P
\end{xflalign*}
\end{proof}
\end{proofs}
\end{theorem}

\begin{lemma}\label{lemma:PBMH(P)-subs-f-ac'}
\begin{statement}
Provided $f$ is bijective,
\begin{align*}
	&\mathbf{PBMH} (P)[f(ac')/ac'] = \mathbf{PBMH} (P[f(ac')/ac'])
\end{align*}
\end{statement}
\begin{proofs}
\begin{proof}
\begin{xflalign*}
	&\mathbf{PBMH} (P[f(ac')/ac'])
	&&\ptext{Definition of $\mathbf{PBMH}$}\\
	&=\exists ac_0 @ P[f(ac')/ac'][ac_0/ac'] \land ac_0\subseteq ac'
	&&\ptext{Substitution}\\
	&=\exists ac_0 @ P[f(ac_0)/ac'] \land ac_0\subseteq ac'
	&&\ptext{Predicate calculus}\\
	&=\exists ac_0 @ (\exists ac_1 @ P[ac_1/ac'] \land ac_1 = f(ac_0)) \land ac_0\subseteq ac'
	&&\ptext{Assumption: $f$ is bijective}\\
	&=\exists ac_0 @ (\exists ac_1 @ P[ac_1/ac'] \land f^{-1} (ac_1) = ac_0) \land ac_0\subseteq ac'
	&&\ptext{One-point rule}\\
	&=\exists ac_1 @ P[ac_1/ac'] \land f^{-1} (ac_1)\subseteq ac'
	&&\ptext{Assumption: $f$ is bijective}\\
	&=\exists ac_1 @ P[ac_1/ac'] \land ac_1\subseteq f(ac')
	&&\ptext{Substitution}\\
	&=(\exists ac_1 @ P[ac_1/ac'] \land ac_1\subseteq ac')[f(ac')/ac']
	&&\ptext{Definition of $\mathbf{PBMH}$}\\
	&=\mathbf{PBMH} (P)[f(ac')/ac']
\end{xflalign*}
\end{proof}
\end{proofs}
\end{lemma}

\begin{lemma}\label{lemma:P-land-ac'-neq-emptyset:P-ac'-subs-emptyset}
\begin{statement}
$P \land ac'=\emptyset = P[\emptyset/ac'] \land ac'=\emptyset$
\end{statement}
\begin{proofs}
\begin{proof}
\begin{xflalign*}
	&P \land ac'=\emptyset
	&&\ptext{Predicate calculus and fresh variable}\\
	&=\exists ac_0 @ P[ac_0/ac'] \land ac'=ac_0 \land ac'=\emptyset
	&&\ptext{Transitivity of equality}\\
	&=\exists ac_0 @ P[ac_0/ac'] \land ac'=\emptyset \land ac'=\emptyset
	&&\ptext{One-point rule}\\
	&=P[\emptyset/ac'] \land ac'=\emptyset
\end{xflalign*}
\end{proof}
\end{proofs}
\end{lemma}

\begin{lemma}\label{lemma:pbmh2d-o-d2pbmh(P)}
\begin{statement}
\begin{align*}
	&pbmh2d \circ d2pbmh(P) = (\lnot P^f[\emptyset/ac'] \land \lnot P^t[\emptyset/ac'] \vdash (\lnot P^f \implies P^t))
\end{align*}
\end{statement}
\begin{proofs}
\begin{proof}
\begin{xflalign*}
	&pbmh2d \circ d2pbmh(P)
	&&\ptext{Definition of $pbmh2d$}\\
	&=(\lnot d2pbmh(P)[\emptyset/ac'] \vdash d2pbmh(P)[dashset(ac')/ac'])[\mathbf{s}/in\alpha_{-ok}]
	&&\ptext{Definition of $d2pbmh$}\\
	&=\left(\begin{array}{l}
		\lnot (\lnot P^f \implies P^t)[true/ok][undashset(ac')/ac'][State_{\II}(in\alpha_{-ok})/s][\emptyset/ac']
		\\ \vdash \\
		(\lnot P^f \implies P^t)[true/ok][undashset(ac')/ac'][State_{\II}(in\alpha_{-ok})/s][dashset(ac')/ac']
	\end{array}\right)[\mathbf{s}/in\alpha_{-ok}]
	&&\ptext{Substitution}\\
	&=\left(\begin{array}{l}
		\lnot (\lnot P^f \implies P^t)[true/ok][undashset(\emptyset)/ac'][State_{\II}(in\alpha_{-ok})/s]
		\\ \vdash \\
		(\lnot P^f \implies P^t)[true/ok][undashset \circ dashset(ac')/ac'][State_{\II}(in\alpha_{-ok})/s]
	\end{array}\right)[\mathbf{s}/in\alpha_{-ok}]
	&&\ptext{\cref{lemma:state-sub:P-S-z:z-S}}\\
	&=\left(\begin{array}{l}
		\lnot (\lnot P^f \implies P^t)[true/ok][undashset(\emptyset)/ac']
		\\ \vdash \\
		(\lnot P^f \implies P^t)[true/ok][undashset \circ dashset(ac')/ac']
	\end{array}\right)
	&&\ptext{$ok$ not free}\\
	&=\left(\begin{array}{l}
		\lnot (\lnot P^f \implies P^t)[undashset(\emptyset)/ac']
		\\ \vdash \\
		(\lnot P^f \implies P^t)[undashset \circ dashset(ac')/ac']
	\end{array}\right)
	&&\ptext{Property of $undashset$ and $dashset$}\\
	&=\left(\begin{array}{l}
		\lnot (\lnot P^f \implies P^t)[\emptyset/ac']
		\\ \vdash \\
		(\lnot P^f \implies P^t)
	\end{array}\right)
	&&\ptext{Substitution}\\
	&=(\lnot P^f[\emptyset/ac'] \land \lnot P^t[\emptyset/ac'] \vdash (\lnot P^f \implies P^t))
\end{xflalign*}
\end{proof}
\end{proofs}
\end{lemma}

\chapter{State Substitution Rules}\label{appendix:state-substitution}

\section{State Substitution} 

The substitution operator $[\mathbf{s}/S\alpha]$, where the boldface indicates that $s$ is a record, is defined for an arbitrary set of variables $S\alpha$ as follows. 
\theoremstatementref{def:substitution-state}\noindent
Each variable $s_i$ in $S\alpha$ is replaced with $z.s_i$. As an example, we consider the substitution $(x'=2 \land ok')[\mathbf{s},\mathbf{z}/in\alpha_{-ok},out\alpha_{-ok'}]$, whose result is $z.x'=2 \land ok'$. The substitution $[\mathbf{z}/S\alpha]$ is well-formed whenever $S\alpha$ is a subset of the record components of $z$. 

\begin{lemma}\label{lemma:substitution-state:A-B-1} Provided that $A\alpha \cap B\alpha = \emptyset$, $A\alpha \subseteq S\alpha$ and $B\alpha \subseteq S\alpha$, 
\begin{align*}
	&P[\mathbf{z}/S\alpha] = P[\mathbf{z}/A\alpha][\mathbf{z}/B\alpha]
\end{align*}
\begin{proofs}\begin{proof} 
Suppose:
\begin{itemize}
	\item $S\alpha = \{ s_0,\ldots,s_n,\ldots,s_m \}$, $A\alpha = \{ s_0,\ldots,s_n \}$, $B\alpha = \{ s_{n+1},\ldots,s_m \}$ 
\end{itemize}
Then:
\begin{flalign*}
	&P[\mathbf{z}/S\alpha]
	&&\ptext{\cref{def:substitution-state}}\\
	&=P[z.s_0,\ldots,z.s_n,z.s_{n+1},\ldots,s_m/s_0,\ldots,s_n,s_{n+1},\ldots,s_m]
	&&\ptext{Property of substitution}\\
	&=P[z.s_0,\ldots,z.s_n/s_0,\ldots,s_n][z.s_{n+1},\ldots,s_m/s_{n+1},\ldots,s_m]
	&&\ptext{\cref{def:substitution-state}}\\
	&=P[\mathbf{z}/A\alpha][\mathbf{z}/B\alpha]
\end{flalign*}
\end{proof}\end{proofs}
\end{lemma}

\begin{lemma} Provided that $A\alpha \cap B\alpha = \emptyset$, $A\alpha \subseteq S\alpha$ and $B\alpha \subseteq S\alpha$,
\begin{align*}
	&P[\mathbf{z}/S\alpha] = \left(
\right)
	&&\ptext{Equality of records}\\
	&=\left(%
\right)
	&&\ptext{One-point rule and substitution}\\
	&=P\left[%
\right\}.s_m\end{array}\right/s_{n+1},\ldots,s_m
		\end{array}\right]
	&&\ptext{Record component}\\
	&=P[z.s_0,\ldots,z.s_n/s_0,\ldots,s_n][z.s_{n+1},\ldots,z.s_m/s_{n+1},\ldots,s_m]
	&&\ptext{\cref{def:substitution-state}}\\
	&=P[\mathbf{z}/A\alpha][\mathbf{z}/B\alpha]
	&&\ptext{\cref{lemma:substitution-state:A-B-1}}\\
	&=P[\mathbf{z}/S\alpha]
\end{flalign*}
\end{proof}\end{proofs}
\end{lemma}

\begin{lemma}\label{lemma:state-substitution:S-si-mapsto-e} Provided $z,y : State(S\alpha)$,
\begin{align*}
	&P[\mathbf{z}/S\alpha][y\oplus\{s_i \mapsto e\}/z] = P[\mathbf{y}/(S\alpha\setminus\{s_i\})][e/s_i]
\end{align*}
\begin{proofs}\begin{proof}
\begin{flalign*}
	&P[\mathbf{z}/S\alpha][y\oplus\{s_i \mapsto e\}/z]
	&&\ptext{\cref{def:substitution-state}}\\
	&=P[z.s_0,\ldots,z.s_i,\ldots,z.s_n/s_0,\ldots,s_i,\ldots,s_n][y\oplus\{s_i \mapsto e\}/z]
	&&\ptext{Substitution}\\
	&=P[(y\oplus\{s_i \mapsto e\}).s_0,\ldots,(y\oplus\{s_i \mapsto e\}).s_i,\ldots,(y\oplus\{s_i \mapsto e\}).s_n/s_0,\ldots,s_i,\ldots,s_n]
	&&\ptext{Property of record components}\\
	&=P[y.s_0,\ldots,e,\ldots,y.s_n/s_0,\ldots,s_i,\ldots,s_n]
	&&\ptext{Property of substitution}\\
	&=P[y.s_0,\ldots,y.s_n/s_0,\ldots,s_n][e/s_i]
	&&\ptext{\cref{def:substitution-state}}\\
	&=P[\mathbf{y}/(S\alpha\setminus\{s_i\})][e/s_i]
\end{flalign*}
\end{proof}\end{proofs}
\end{lemma}

\begin{lemma}\label{lemma:state-substitution:S-si-mapsto-e:si-not-free-in-e} Provided $z,y : State(S\alpha)$ and $s_i$ not free in $e$,
\begin{align*}
	&P[\mathbf{z}/S\alpha][y\oplus\{s_i \mapsto e\}/z] = P[e/s_i][\mathbf{y}/(S\alpha)]
\end{align*}
\begin{proofs}\begin{proof}\checkt{alcc}
\begin{flalign*}
	&P[\mathbf{z}/S\alpha][y\oplus\{s_i \mapsto e\}/z]
	&&\ptext{\cref{def:substitution-state}}\\
	&=P[z.s_0,\ldots,z.s_i,\ldots,z.s_n/s_0,\ldots,s_i,\ldots,s_n][y\oplus\{s_i \mapsto e\}/z]
	&&\ptext{Substitution}\\
	&=P[(y\oplus\{s_i \mapsto e\}).s_0,\ldots,(y\oplus\{s_i \mapsto e\}).s_i,\ldots,(y\oplus\{s_i \mapsto e\}).s_n/s_0,\ldots,s_i,\ldots,s_n]
	&&\ptext{Property of record components}\\
	&=P[y.s_0,\ldots,e,\ldots,y.s_n/s_0,\ldots,s_i,\ldots,s_n]
	&&\ptext{Property of substitution}\\
	&=P[e/s_i][y.s_0,\ldots,y.s_n/s_0,\ldots,s_n]
	&&\ptext{Substitution: $s_i$ not free in $e$}\\
	&=P[e/s_i][z.s_0,\ldots,z.s_i,\ldots,z.s_n/s_0,\ldots,s_i,\ldots,s_n]
	&&\ptext{\cref{def:substitution-state}}\\
	&=P[e/s_i][\mathbf{y}/S\alpha]
\end{flalign*}
\end{proof}\end{proofs}
\end{lemma}

\begin{lemma}\label{lemma:state-substitution:e-si:z-Salpha} Provided $s_i \in S\alpha$,
\begin{align*}
	&P[e/s_i][\mathbf{z}/S\alpha] = P[\mathbf{z}/S\alpha\setminus\{s_i\}][e[\mathbf{z}/S\alpha]/s_i]
\end{align*}
\begin{proofs}\begin{proof}
\begin{flalign*}
	&P[e/s_i][\mathbf{z}/S\alpha]
	&&\ptext{\cref{def:substitution-state}}\\
	&=P[e/s_i][z.s_0,\ldots,z.s_i,\ldots,z.s_n/s_0,\ldots,s_i,\ldots,s_n]
	&&\ptext{Substitution: $s_i$ not free in $P$}\\
	&=P[z.s_0,\ldots,z.s_n/s_0,\ldots,s_n][e[z.s_0,\ldots,z.s_i,\ldots,z.s_n/s_0,\ldots,s_i,\ldots,s_n]/s_i]
	&&\ptext{\cref{def:substitution-state}}\\
	&=P[\mathbf{z}/S\alpha\setminus\{s_i\}][e[\mathbf{z}/S\alpha]/s_i]
\end{flalign*}
\end{proof}\end{proofs}
\end{lemma}

%
%
%
%

\begin{lemma}\label{lemma:substitution:sub(S-cup-T):sub(S)sub(T)}
$P[z/(S\alpha \cup T\alpha)] = P[z/S\alpha][z/T\alpha]$
\begin{proofs}\begin{proof}
\begin{flalign*}
	&P[z/(S\alpha \cup T\alpha)]
	&&\ptext{\cref{def:substitution-state}}\\
	&=P[z.s_0,z.t_0,\ldots,z.s_n,z.t_n/s_0',t_0',\ldots,s_n',t_n']
	&&\ptext{Substitution}\\
	&=P[z.s_0,\ldots,z.s_n/s_0',\ldots,s_n'][z.t_0,\ldots,z.t_n/t_0',\ldots,t_n']
	&&\ptext{\cref{def:substitution-state}}\\
	&=P[z/S\alpha][z/T\alpha]
\end{flalign*}
\end{proof}\end{proofs}
\end{lemma}

\begin{lemma} 
\begin{align*}
	&P[e_0,\ldots,e_n/x_0,\ldots,x_n][z/S\alpha] \\
	&=\\
	&P[z/(S\alpha \setminus T\alpha)][e_0[z/T\alpha],\ldots,e_n[z/T\alpha]/x_0,\ldots,x_n]
\end{align*}
Provided that:
\begin{enumerate}
	\item $T\alpha \subseteq S\alpha$
	\item $T\alpha = \{ x_0, \ldots, x_n \}$ 
	\item $\forall y \spot y \in (S\alpha\setminus T\alpha) \implies y \notin fv(e_0,\ldots,e_n)$
\end{enumerate}
\begin{proofs}\begin{proof}
\begin{flalign*}
	&P[e_0,\ldots,e_n/x_0,\ldots,x_n][z/S\alpha]
	&&\ptext{Property of sets}\\
	&=P[e_0,\ldots,e_n/x_0,\ldots,x_n][z/(S\alpha \setminus T\alpha) \cup T\alpha]
	&&\ptext{\cref{lemma:substitution:sub(S-cup-T):sub(S)sub(T)}}\\
	&=P[e_0,\ldots,e_n/x_0,\ldots,x_n][z/(S\alpha \setminus T\alpha)][z/T\alpha]
	&&\ptext{Substitution: Assumption 1}\\
	&=P[z/(S\alpha \setminus T\alpha)][e_0,\ldots,e_n/x_0,\ldots,x_n][z/T\alpha]
	&&\ptext{Substitution}\\
	&=P[z/(S\alpha \setminus T\alpha)][e_0[z/T\alpha],\ldots,e_n[z/T\alpha]/x_0,\ldots,x_n]
	&&\ptext{Property of substitution}\\
	&=P[z/(S\alpha \setminus T\alpha)][z/T\alpha][T\alpha/z][e_0[z/T\alpha],\ldots,e_n[z/T\alpha]/x_0,\ldots,x_n]
	&&\ptext{\cref{lemma:substitution:sub(S-cup-T):sub(S)sub(T)}}\\
	&=P[z/(S\alpha \setminus T\alpha) \cup T\alpha][T\alpha/z][e_0[z/T\alpha],\ldots,e_n[z/T\alpha]/x_0,\ldots,x_n]
	&&\ptext{Property of sets}\\
	&=P[z/S\alpha][T\alpha/z][e_0[z/T\alpha],\ldots,e_n[z/T\alpha]/x_0,\ldots,x_n]
	&&\ptext{Definition}\\
	&=P[z/S\alpha][x_0,\ldots,x_n/z.x,\ldots,z.n][e_0[z/T\alpha],\ldots,e_n[z/T\alpha]/x_0,\ldots,x_n]
	&&\ptext{Property of substitution}\\
	&=P[z/S\alpha][e_0[z/T\alpha],\ldots,e_n[z/T\alpha]/z.x,\ldots,z.n]
	&&\ptext{Definition}\\
	&=P[z/S\alpha][e_0[z/T\alpha],\ldots,e_n[z/T\alpha]/z.x,\ldots,z.n]
\end{flalign*}
\end{proof}\end{proofs}
\end{lemma}

\begin{define}\checkt{alcc} For $S\alpha = \{ x_0, \ldots, x_n \}$,
\begin{align*}
	&State_{\II} (S\alpha) \circdef \{ x_0 \mapsto x_0, \ldots, x_n \mapsto x_n \}
\end{align*}
\end{define}

\begin{lemma}
$State_{\II} (S\alpha)' = \{ x_0' \mapsto x_0, \ldots, x_n' \mapsto x_n \}$
\begin{proofs}\begin{proof}\checkt{alcc}
\begin{flalign*}
	&State_{\II} (S\alpha)'
	&&\ptext{Definition of $State_{\II} (S\alpha)$}\\
	&=(\{ x_0' \mapsto x_0, \ldots, x_n' \mapsto x_n \})'
	&&\ptext{Definition of $'$ on $State$}\\
	&=\{ x_0' \mapsto x_0, \ldots, x_n' \mapsto x_n \}
\end{flalign*}
\end{proof}\end{proofs}
\end{lemma}

\begin{lemma}\label{lemma:state-sub:exists-z-State-P}
\begin{align*}
	&\exists z : State(S\alpha) \spot P \land (\bigwedge x : S\alpha \spot z.x = x) = P[State_{\II}(S\alpha)/z]
\end{align*}
\begin{proofs}\begin{proof}\checkt{alcc}
\begin{flalign*}
	&\exists z : State(S\alpha) \spot P \land (\bigwedge x : S\alpha \spot z.x = x)
	&&\ptext{Equality of records}\\
	&=\exists z : State(S\alpha) \spot P \land z = \{ x_0 \mapsto x_0, \ldots, x_n \mapsto x_n \}
	&&\ptext{Definition of $State_{\II}$}\\
	&=\exists z : State(S\alpha) \spot P \land State_{\II} (S\alpha) = z
	&&\ptext{One-point rule}\\
	&=P[State_{\II} (S\alpha)/z]
\end{flalign*}
\end{proof}\end{proofs}
\end{lemma}

\begin{lemma}\label{lemma:state-sub:P-z-S:S-z} Provided $z$ is not free in $P$,
\begin{align*}
	&P[\mathbf{z}/S\alpha][State_{\II}(S\alpha)/z] = P
\end{align*}
\begin{proofs}\begin{proof}\checkt{alcc}
\begin{flalign*}
	&P[\mathbf{z}/S\alpha][State_{\II}(S\alpha)/z]
	&&\ptext{Definition of state substitution}\\
	&=P[z.x_0,\ldots,z.x_n/x_0,\ldots,x_n][State_{\II}(S\alpha)/z]
	&&\ptext{Definition of $State_{\II}(S\alpha)$}\\
	&=P[z.x_0,\ldots,z.x_n/x_0,\ldots,x_n][\{ x_0 \mapsto x_0, \ldots, x_n \mapsto x_n \}/z]
	&&\ptext{Substitution: $z$ is not free in $P$}\\
	&=P[\{ x_0 \mapsto x_0, \ldots, x_n \mapsto x_n \}.x_0,\ldots,\{ x_0 \mapsto x_0, \ldots, x_n \mapsto x_n \}.x_n/x_0,\ldots,x_n]
	&&\ptext{Value of state component}\\
	&=P[x_0,\ldots,x_n/x_0,\ldots,x_n]
	&&\ptext{Property of substitution}\\
	&=P
\end{flalign*}
\end{proof}\end{proofs}
\end{lemma}

\begin{lemma}\label{lemma:state-sub:P-S-z:z-S} Provided none of the varibles in $S\alpha$ are free in $P$,
\begin{align*}
	&P[State_{\II}(S\alpha)/z][\mathbf{z}/S\alpha] = P
\end{align*}
\begin{proofs}\begin{proof}
\begin{flalign*}
	&P[State_{\II}(S\alpha)/z][\mathbf{z}/S\alpha]
	&&\ptext{Definition of $State_{\II}(S\alpha)$ and state substitution}\\
	&=P[\{ x_0 \mapsto x_0, \ldots, x_n \mapsto x_n\}/z][z.x_0,\ldots,z.x_n/x_0,\ldots,x_n]
	&&\ptext{Substitution: $x_i \notin fv(P)$}\\
	&=P[\{ x_0 \mapsto z.x_0, \ldots, x_n \mapsto z.x_n\}/z]
	&&\ptext{Equality of records}\\
	&=P[z/z]
	&&\ptext{Property of substitution}\\
	&=P
\end{flalign*}
\end{proof}\end{proofs}
\end{lemma}

\begin{lemma}\label{lemma:state-sub:S-z-e-xi:z-oplus-xi-e-S-z} Provided $x_i \in S\alpha$ and $x_i$ is not free in $P$ nor in $e$,
\begin{align*}
	&P[State_{\II}(S\alpha)/z][e/x_i] = P[z\oplus x_i\mapsto e\}/z][State_{\II}(S\alpha)/z]
\end{align*}
\begin{proofs}\begin{proof}\checkt{alcc}
\begin{flalign*}
	&P[State_{\II}(S\alpha)/z][e/x_i]
	&&\ptext{Definition of $State_{\II}(S\alpha)$}\\
	&=P[\{ x_0 \mapsto x_0, \ldots, x_i \mapsto x_i, \ldots, x_n \mapsto x_n \}/z][e/x_i]
	&&\ptext{Substitution: $x_i$ not free in $P$}\\
	&=P[\{ x_0 \mapsto x_0, \ldots, x_i \mapsto e, \ldots, x_n \mapsto x_n \}/z]
	&&\ptext{Property of sets}\\
	&=P[\{ x_0 \mapsto x_0, \ldots, x_i \mapsto x_i, \ldots, x_n \mapsto x_n \} \oplus \{ x_i \mapsto e \}/z]
	&&\ptext{Definition of $State_{\II}(S\alpha)$}\\
	&=P[State_{\II}(S\alpha) \oplus \{ x_i \mapsto e \}/z]
	&&\ptext{Substitution}\\
	&=P[z \oplus \{ x_i \mapsto e \}/z][State_{\II}(S\alpha)/z]
\end{flalign*}
\end{proof}\end{proofs}
\end{lemma}

\section{$dash$ and $undash$}\label{sec:appendix-state-substituion:dash-and-undash}

\begin{define}
\begin{align*} 
	dash(z) 	&\circdef \{ x : S\alpha, e | (x \mapsto e) \in z @ x' \mapsto e \} \\
	undash(z) 	&\circdef \{ x : S\alpha, e | (x' \mapsto e) \in z @ x \mapsto e \}
\end{align*}
\end{define}\noindent%
The function $dash$ considers every pair $(x, e)$ in $z$, where $x$ is a variable name and $e$ the corresponding expression or value associated with $x$, and dashes the name of $x$ into $x'$. Function $undash$ is similar except for the undash of $x'$ to $x$.

\begin{lemma}\label{lemma:dash(z).x':z.x}
\begin{statement}$dash(z).x' = z.x$\end{statement}
\begin{proofs}\begin{proof}
\begin{flalign*}
	&dash(z).x'
	&&\ptext{Definition of $dash$}\\
	&=\{ y : S\alpha, e | (y \mapsto e) \in z @ y' \mapsto e \}.x'
	&&\ptext{Value of record component $x'$}\\
	&=\{ y : S\alpha, e | (y \mapsto e) \in z \}.x'
	&&\ptext{Definition of record}\\
	&=z.x
\end{flalign*}
\end{proof}\end{proofs}
\end{lemma}

\begin{lemma}\label{lemma:undash(z).x:x.x'}
\begin{statement}$undash(z).x = z.x'$\end{statement}
\begin{proofs}\begin{proof}
\begin{flalign*}
	&undash(z).x
	&&\ptext{Definition of $undash$}\\
	&=\{ y : S\alpha, e | (y' \mapsto e) \in z @ y \mapsto e \}.x
	&&\ptext{Value of record component $x'$}\\
	&=\{ y : S\alpha, e | (y' \mapsto e) \in z \}.x'
	&&\ptext{Definition of record}\\
	&=z.x'
\end{flalign*}
\end{proof}\end{proofs}
\end{lemma}

\begin{lemma}\label{lemma:undash-o-dash(ss):ss}
\begin{statement}$undash\circ dash(z) = z$\end{statement}
\begin{proofs}\begin{proof}
\begin{flalign*}
	&undash\circ dash(z)
	&&\ptext{Definition of $undash$}\\
	&=\{ y_0 : S\alpha, e_0 | (y_0' \mapsto e_0) \in dash(z) @ y_0 \mapsto e_0 \}
	&&\ptext{Definition of $dash$}\\
	&=\left\{ y_0 : S\alpha, e_0 \left|\begin{array}{l}
		 (y_0' \mapsto e_0) \in \{ x : S\alpha, e | (x \mapsto e) \in z @ x' \mapsto e \} 
		 \\ @ y_0 \mapsto e_0
	\end{array}\right.\right\}
	&&\ptext{Property of sets}\\
	&=\left\{ y_0 : S\alpha, e_0 \left|\begin{array}{l}
		 \exists x, e @ (x \mapsto e) \in z \land (x' \mapsto e) = (y_0' \mapsto e_0) 
		 \\ @ y_0 \mapsto e_0
	\end{array}\right.\right\}
	&&\ptext{Undash variables}\\
	&=\left\{ y_0 : S\alpha, e_0 \left|\begin{array}{l}
		 \exists x, e @ (x \mapsto e) \in z \land (x \mapsto e) = (y_0 \mapsto e_0) 
		 \\ @ y_0 \mapsto e_0
	\end{array}\right.\right\}
	&&\ptext{One-point rule}\\
	&=\{ y_0 : S\alpha, e_0 | (y_0 \mapsto e_0) \in z @ y_0 \mapsto e_0\}
	&&\ptext{Property of sets}\\
	&=\{ y_0 : S\alpha, e_0 | (y_0 \mapsto e_0) \in z\}
	&&\ptext{Property of sets}\\
	&=z
\end{flalign*}
\end{proof}\end{proofs}
\end{lemma}

\begin{lemma}\label{lemma:dash-o-undash(ss):ss}
\begin{statement}$dash\circ undash(z) = z$\end{statement}
\begin{proofs}\begin{proof}
\begin{flalign*}
	&dash\circ undash(z)
	&&\ptext{Definition of $dash$}\\
	&=\{ y_0 : S\alpha, e_0 | (y_0 \mapsto e_0) \in undash(z) @ y_0' \mapsto e_0 \}
	&&\ptext{Definition of $undash$}\\
	&=\left\{ y_0 : S\alpha, e_0 \left|\begin{array}{l}
		 (y_0 \mapsto e_0) \in \{ x : S\alpha, e | (x' \mapsto e) \in z @ x \mapsto e \} 
		 \\ @ y_0' \mapsto e_0
	\end{array}\right.\right\}
	&&\ptext{Property of sets}\\
	&=\left\{ y_0 : S\alpha, e_0 \left|\begin{array}{l}
		 \exists x, e @ (x' \mapsto e) \in z \land (x \mapsto e) = (y_0 \mapsto e_0) 
		 \\ @ y_0' \mapsto e_0
	\end{array}\right.\right\}
	&&\ptext{Dash variables}\\
	&=\left\{ y_0 : S\alpha, e_0 \left|\begin{array}{l}
		 \exists x, e @ (x' \mapsto e) \in z \land (x' \mapsto e) = (y_0' \mapsto e_0) 
		 \\ @ y_0' \mapsto e_0
	\end{array}\right.\right\}
	&&\ptext{One-point rule}\\
	&=\{ y_0 : S\alpha, e_0 | (y_0' \mapsto e_0) \in z @ y_0' \mapsto e_0\}
	&&\ptext{Property of sets}\\
	&=\{ y_0 : S\alpha, e_0 | (y_0' \mapsto e_0) \in z\}
	&&\ptext{Property of sets}\\
	&=z
\end{flalign*}
\end{proof}\end{proofs}
\end{lemma}

\begin{lemma}\label{lemma:undash:exists-z-undash-in-ac'} Provided $y$ is fresh,
\begin{align*}
	&\exists z \spot P \land undash(z) \in ac' = \exists y \spot P[dash(y)/z] \land y \in ac'
\end{align*}
\begin{proofs}\begin{proof}
\begin{flalign*}
	&\exists z \spot P \land undash(z) \in ac'
	&&\ptext{Introduce fresh variable}\\
	&=\exists z,y \spot P \land y \in ac' \land y = undash(z)
	&&\ptext{Property of $dash$}\\
	&=\exists z,y \spot P \land y \in ac' \land dash(y) = dash \circ undash(z)
	&&\ptext{$dash \circ undash (z) = z$}\\
	&=\exists z,y \spot P \land y \in ac' \land dash(y) = z
	&&\ptext{One-point rule}\\
	&=\exists y \spot P[dash(y)/z] \land y \in ac'
\end{flalign*}
\end{proof}\end{proofs}
\end{lemma}

\chapter{$\mathbf{PBMH}$}\label{appendix:pbmh}

\section{Definition}

\theoremstatementref{def:PBMH}

\section{Properties}

\begin{lemma}\label{lemma:P-implies-PBMH(P)}\label{lemma:P-implies-PBMH}
$P \implies \mathbf{PBMH} (P)$
\begin{proofs}\begin{proof}\checkt{pfr}\checkt{alcc}
\begin{xflalign*}
	&P
	&&\ptext{Predicate calculus}\\
	&=\exists ac_0 \spot P[ac_0/ac'] \land ac_0 = ac'
	&&\ptext{Property of sets}\\
	&\implies \exists ac_0 \spot P[ac_0/ac'] \land ac_0 \subseteq ac'
	&&\ptext{Definition of $\mathbf{PBMH}$ (\cref{lemma:PBMH:alternative-1})}\\
	&=\mathbf{PBMH} (P)
\end{xflalign*}
\end{proof}\end{proofs}
\end{lemma}

\begin{theorem}\label{law:pbmh:idempotent}
$\mathbf{PBMH} \circ \mathbf{PBMH} (P) = \mathbf{PBMH} (P)$
\begin{proofs}\begin{proof}\checkt{alcc}\checkt{pfr}
\begin{xflalign*}
	&\mathbf{PBMH} \circ \mathbf{PBMH} (P)
	&&\ptext{Definition of $\mathbf{PBMH}$}\\
	&=\mathbf{PBMH} (P \circseq ac \subseteq ac' \land v'=v)
	&&\ptext{Definition of $\mathbf{PBMH}$}\\
	&=((P \circseq ac \subseteq ac' \land v'=v) \circseq ac \subseteq ac' \land v'=v)
	&&\ptext{Associativity of sequential composition}\\
	&=(P \circseq (ac \subseteq ac' \land v'=v \circseq ac \subseteq ac' \land v'=v))
	&&\ptext{Definition of sequential composition}\\
	&=(P \circseq (\exists ac_0 \spot ac \subseteq ac_0 \land ac_0 \subseteq ac'))
	&&\ptext{Transitivity of subset inclusion}\\
	&=(P \circseq ac \subseteq ac')
	&&\ptext{Definition of sequential composition}\\
	&=(P \circseq ac \subseteq ac' \land v'=v)
	&&\ptext{Definition of $\mathbf{PBMH}$}\\
	&=\mathbf{PBMH} (P)
\end{xflalign*}
\end{proof}\end{proofs}
\end{theorem}

\begin{theorem}\label{theorem:PBMH(P-lor-Q):PBMH(P)-lor-PBMH(Q)}\label{law:pbmh:distribute-disjunction}
$\mathbf{PBMH} (P \lor Q) = \mathbf{PBMH} (P) \lor \mathbf{PBMH} (Q)$
\begin{proofs}\begin{proof}\checkt{pfr}\checkt{alcc}
\begin{xflalign*}
	&\mathbf{PBMH} (P \lor Q)
	&&\ptext{Definition of $\mathbf{PBMH}$ (\cref{lemma:PBMH:alternative-1})}\\
	&=\exists ac_0 \spot (P \lor Q)[ac_0/ac'] \land ac_0 \subseteq ac'
	&&\ptext{Property of substition}\\
	&=\exists ac_0 \spot (P[ac_0/ac'] \lor Q[ac_0/ac']) \land ac_0 \subseteq ac'
	&&\ptext{Predicate calculus}\\
	&=\exists ac_0 \spot (P[ac_0/ac'] \land ac_0 \subseteq ac') \lor (Q[ac_0/ac'] \land ac_0 \subseteq ac')
	&&\ptext{Predicate calculus}\\
	&=\left(\begin{array}{l}
		(\exists ac_0 \spot P[ac_0/ac'] \land ac_0 \subseteq ac')
		\\ \lor \\
		(\exists ac_0 \spot Q[ac_0/ac'] \land ac_0 \subseteq ac')
	\end{array}\right)
	&&\ptext{Definition of $\mathbf{PBMH}$ (\cref{lemma:PBMH:alternative-1})}\\
	&=\mathbf{PBMH} (P) \lor \mathbf{PBMH} (Q)
\end{xflalign*}
\end{proof}\end{proofs}
\end{theorem} 

\begin{lemma}\label{law:pmbh:P-empty-ac'-or-P} Provided $P$ satisfies $\mathbf{PBMH}$,
$P[\emptyset/ac'] \lor P = P$
\begin{proofs}\begin{proof}
\begin{flalign*}
	&P[\emptyset/ac'] \lor P
	&&\ptext{Assumption: $P$ is $\mathbf{PBMH}$-healthy}\\
	&=(P \circseq ac \subseteq ac')[\emptyset/ac'] \lor (P \circseq ac \subseteq ac')
	&&\ptext{Substitution}\\
	&=(P \circseq ac \subseteq \emptyset) \lor (P \circseq ac \subseteq ac')
	&&\ptext{Distributivity of sequential composition w.r.t. disjunction}\\
	&=P \circseq (ac \subseteq \emptyset \lor ac \subseteq ac')
	&&\ptext{Property of subset inclusion}\\
	&=P \circseq (ac \subseteq ac')
	&&\ptext{Assumption: $P$ is $\mathbf{PBMH}$-healthy}\\
	&=P
\end{flalign*}
\end{proof}\end{proofs}
\end{lemma}

\section{Closure Properties}

\begin{lemma}\label{law:pbmh:distribute-conjunction}
Provided $P$ and $Q$ satisfy $\mathbf{PBMH}$,
\begin{align*}
	&\mathbf{PBMH} (P \land Q) = \mathbf{PBMH} (P) \land \mathbf{PBMH} (Q)
\end{align*}
\begin{proofs}\begin{proof}\checkt{alcc}\checkt{pfr}
\begin{xflalign*}
	&\mathbf{PBMH} (P \land Q)
	&&\ptext{Assumption: $P$ and $Q$ are $\mathbf{PBMH}$-healthy and~\cref{law:pbmh:conjunction-closure}}\\
	&=\mathbf{PBMH} (P) \land \mathbf{PBMH} (Q)
\end{xflalign*}
\end{proof}\end{proofs}
\end{lemma}

\begin{lemma}\label{lemma:PBMH(P-land-Q):implies:PBMH(P)-land-PBMH(Q)}
$\mathbf{PBMH} (P \land Q) \implies \mathbf{PBMH} (P) \land \mathbf{PBMH} (Q)$
\begin{proofs}\begin{proof}\checkt{pfr}
\begin{xflalign*}
	&\mathbf{PBMH} (P \land Q)
	&&\ptext{Definition of $\mathbf{PBMH}$ (\cref{lemma:PBMH:alternative-1})}\\
	&=\exists ac_0 \spot (P \land Q)[ac_0/ac'] \land ac_0 \subseteq ac'
	&&\ptext{Substitution}\\
	&=\exists ac_0 \spot P[ac_0/ac'] \land Q[ac_0/ac'] \land ac_0 \subseteq ac'
	&&\ptext{Predicate calculus}\\
	&\implies\left(\begin{array}{l}
		(\exists ac_0 \spot P[ac_0/ac'] \land ac_0 \subseteq ac')
		\\ \land \\
		(\exists ac_0 \spot Q[ac_0/ac'] \land ac_0 \subseteq ac')
	\end{array}\right)
	&&\ptext{Definition of $\mathbf{PBMH}$ (\cref{lemma:PBMH:alternative-1})}\\
	&=\mathbf{PBMH} (P) \land \mathbf{PBMH} (Q)
\end{xflalign*}
\end{proof}\end{proofs}
\end{lemma}

\begin{theorem}\label{law:pbmh:conjunction-closure} Provided $P$ and $Q$ are $\mathbf{PBMH}$-healthy,
\begin{align*}
	&\mathbf{PBMH} (P \land Q) = P \land Q
\end{align*}
\begin{proofs}\begin{proof}\checkt{alcc}\checkt{pfr}
\begin{xflalign*}
	&\mathbf{PBMH} (P \land Q)
	&&\ptext{Assumption: $P$ and $Q$ are $\mathbf{PBMH}$-healthy}\\
	&=\mathbf{PBMH} (\mathbf{PBMH} (P) \land \mathbf{PBMH} (Q))
	&&\ptext{Definition of $\mathbf{PBMH}$ (\cref{lemma:PBMH:alternative-1})}\\
	&=\exists ac_0 \spot (\mathbf{PBMH} (P) \land \mathbf{PBMH} (Q))[ac_0/ac'] \land ac_0 \subseteq ac'
	&&\ptext{Definition of $\mathbf{PBMH}$ (\cref{lemma:PBMH:alternative-1})}\\
	&=\exists ac_0 \spot \left(\begin{array}{l}
		(\exists ac_0 \spot P[ac_0/ac'] \land ac_0 \subseteq ac') 
		\\ \land \\ 
		(\exists ac_0 \spot Q[ac_0/ac'] \land ac_0 \subseteq ac')
	\end{array}\right)[ac_0/ac'] \land ac_0 \subseteq ac'
	&&\ptext{Variable renaming}\\
	&=\exists ac_0 \spot \left(\begin{array}{l}
		(\exists ac_1 \spot P[ac_1/ac'] \land ac_1 \subseteq ac') 
		\\ \land \\ 
		(\exists ac_2 \spot Q[ac_1/ac'] \land ac_2 \subseteq ac')
	\end{array}\right)[ac_0/ac'] \land ac_0 \subseteq ac'
	&&\ptext{Substitution}\\
	&=\exists ac_0 \spot \left(\begin{array}{l}
		(\exists ac_1 \spot P[ac_1/ac'] \land ac_1 \subseteq ac_0) 
		\\ \land \\ 
		(\exists ac_2 \spot Q[ac_2/ac'] \land ac_2 \subseteq ac_0)
	\end{array}\right) \land ac_0 \subseteq ac'
	&&\ptext{Predicate calculus}\\
	&=\left(\begin{array}{l}
		(\exists ac_1 \spot P[ac_1/ac'] \land ac_1 \subseteq ac') 
		\\ \land \\ 
		(\exists ac_2 \spot Q[ac_2/ac'] \land ac_2 \subseteq ac')
	\end{array}\right)
	&&\ptext{Definition of $\mathbf{PBMH}$ (\cref{lemma:PBMH:alternative-1})}\\
	&=\mathbf{PBMH} (P) \land \mathbf{PBMH} (Q)
	&&\ptext{Assumption: $P$ and $Q$ are $\mathbf{PBMH}$-healthy}\\
	&=P \land Q
\end{xflalign*}
\end{proof}\end{proofs}
\end{theorem}

\begin{theorem}\label{law:pbmh:disjunction-closure}
Provided $P$ and $Q$ satisfy $\mathbf{PBMH}$,
\begin{align*}
	&\mathbf{PBMH} (P \lor Q) = P \lor Q
\end{align*}
\begin{proofs}\begin{proof}\checkt{pfr}
\begin{flalign*}
	&\mathbf{PBMH} (P \lor Q)
	&&\ptext{\cref{law:pbmh:distribute-disjunction}}\\
	&=\mathbf{PBMH} (P) \lor \mathbf{PBMH} (Q)
	&&\ptext{Assumption: $P$ and $Q$ satisfy $\mathbf{PBMH}$}\\
	&=P \lor Q
\end{flalign*}
\end{proof}\end{proofs}
\end{theorem}

\section{Lemmas}

\begin{lemma}\label{lemma:PBMH:alternative-1}
\begin{statement}
$\mathbf{PBMH} (P) = \exists ac_0 \spot P[ac_0/ac'] \land ac_0 \subseteq ac'$
\end{statement}
\begin{proofs}
\begin{proof}\checkt{alcc}\checkt{pfr}
\begin{flalign*}
	&\mathbf{PBMH} (P)
	&&\ptext{Definition of $\mathbf{PBMH}$}\\
	&=P \circseq ac \subseteq ac' \land v'=v
	&&\ptext{Definition of sequential composition}\\
	&=\exists ac_0, v_0 \spot P[ac_0,v_0/ac',v'] \land ac_0 \subseteq ac' \land v'=v_0
	&&\ptext{One-point rule}\\
	&=\exists ac_0 \spot P[ac_0/ac'] \land ac_0 \subseteq ac'
\end{flalign*}
\end{proof}
\end{proofs}
\end{lemma}

\begin{lemma}\label{law:pbmh:true}
$\mathbf{PBMH} (true) = true$
\begin{proofs}\begin{proof}\checkt{pfr}\checkt{alcc}
\begin{xflalign*}
	&\mathbf{PBMH} (true)
	&&\ptext{Definition of $\mathbf{PBMH}$~(\cref{lemma:PBMH:alternative-1})}\\
	&=\exists ac_0 \spot true[ac_0/ac'] \land ac_0 \subseteq ac'
	&&\ptext{Property of substitution and predicate calculus}\\
	&=true
\end{xflalign*}
\end{proof}\end{proofs}
\end{lemma}

\begin{lemma}\label{law:pbmh:false}
$\mathbf{PBMH} (false) = false$
\begin{proofs}\begin{proof}\checkt{pfr}\checkt{alcc}
\begin{xflalign*}
	&\mathbf{PBMH} (false)
	&&\ptext{Definition of $\mathbf{PBMH}$ (\cref{lemma:PBMH:alternative-1})}\\
	&=\exists ac_0 \spot false[ac_0/ac'] \land ac_0 \subseteq ac'
	&&\ptext{Substitution and predicate calculus}\\
	&=false
\end{xflalign*}
\end{proof}\end{proofs}
\end{lemma}

\begin{lemma}\label{law:pbmh:s-in-ac'}
$\mathbf{PBMH} (s \in ac') = s \in ac'$
\begin{proofs}\begin{proof}\checkt{pfr}\checkt{alcc}
\begin{xflalign*}
	&\mathbf{PBMH} (s \in ac')
	&&\ptext{Definition of $\mathbf{PBMH}$~(\cref{lemma:PBMH:alternative-1})}\\
	&=\exists ac_0 \spot (s \in ac')[ac_0/ac'] \land ac_0 \subseteq ac'
	&&\ptext{Substitution}\\
	&=\exists ac_0 \spot s \in ac_0 \land ac_0 \subseteq ac'
	&&\ptext{Property of sets}\\
	&=s \in ac'
\end{xflalign*}
\end{proof}\end{proofs}
\end{lemma}

\begin{lemma}\label{law:pbmh:ac'-neq-emptyset}
$\mathbf{PBMH} (ac'\neq\emptyset) = ac'\neq\emptyset$
\begin{proofs}\begin{proof}\checkt{pfr}
\begin{xflalign*}
	&\mathbf{PBMH} (ac'\neq\emptyset)
	&&\ptext{Definition of $\mathbf{PBMH}$ (\cref{lemma:PBMH:alternative-1})}\\
	&=\exists ac_0 \spot ac_0\neq\emptyset \land ac_0 \subseteq ac'
	&&\ptext{Property of sets (\cref{law:set-theory:transitivity-non-empty})}&\\
	&=ac' \neq\emptyset
\end{xflalign*}
\end{proof}\end{proofs}
\end{lemma}

\begin{lemma}\label{law:pbmh:P:ac'-not-free}
Provided $ac'$ is not free in $P$, $\mathbf{PBMH} (P) = P$.
\begin{proofs}\begin{proof}\checkt{alcc}\checkt{pfr}
\begin{xflalign*}
	&\mathbf{PBMH} (P)
	&&\ptext{Definition of $\mathbf{PBMH}$ (\cref{lemma:PBMH:alternative-1})}\\
	&=\exists ac_0 \spot P[ac_0/ac'] \land ac_0 \subseteq ac'
	&&\ptext{Assumption: $ac'$ not free in $P$ and predicate calculus}\\
	&=P \land \exists ac_0 \spot ac_0 \subseteq ac'
	&&\ptext{Case-analysis on $ac_0$}\\
	&=P
\end{xflalign*}
\end{proof}\end{proofs}
\end{lemma}

\begin{lemma} Provided $c$ is a condition,
\label{lemma:PBMH(c)-condition:c}
$\mathbf{PBMH} (c) = c$.
\begin{proofs}\begin{proof}\checkt{alcc}\checkt{pfr}
\begin{flalign*}
	&\mathbf{PBMH} (c)
	&&\ptext{Definition of $\mathbf{PBMH}$ (\cref{lemma:PBMH:alternative-1})}\\
	&=\exists ac_0 \spot c[ac_0/ac'] \land ac_0 \subseteq ac'
	&&\ptext{Assumption: $c$ is a condition, hence $ac'$ is not free}\\
	&=\exists ac_0 \spot c \land ac_0 \subseteq ac'
	&&\ptext{Predicate calculus}\\
	&=c
\end{flalign*}
\end{proof}\end{proofs}
\end{lemma}

\begin{lemma}\label{lemma:PBMH(x-in-ac'):x-in-ac'}
$\mathbf{PBMH} (x \in ac') = x \in ac'$
\begin{proofs}\begin{proof}
\begin{flalign*}
	&\mathbf{PBMH} (x \in ac')
	&&\ptext{\cref{lemma:PBMH:alternative-1}}\\
	&=\exists ac_0 \spot x \in ac_0 \land ac_0 \subseteq ac'
	&&\ptext{Predicate calculus}\\
	&=x \in ac'
\end{flalign*}
\end{proof}\end{proofs}
\end{lemma}

\begin{lemma}\label{lemma:PBMH(c-land-P):c-land-PBMH(P)}
Provided $ac'$ is not free in $c$,
$\mathbf{PBMH} (c \land P) = c \land \mathbf{PBMH} (P)$
\begin{proofs}\begin{proof}\checkt{alcc}
\begin{flalign*}
	&\mathbf{PBMH} (c \land P)
	&&\ptext{\cref{lemma:PBMH:alternative-1}}\\
	&=\exists ac_0 \spot (c \land P)[ac_0/ac'] \land ac_0 \subseteq ac'
	&&\ptext{Assumption: $c$ is a condition, hence $ac'$ is not free}\\
	&=\exists ac_0 \spot c \land P[ac_0/ac'] \land ac_0 \subseteq ac'
	&&\ptext{Predicate calculus}\\
	&=c \land \exists ac_0 \spot P[ac_0/ac'] \land ac_0 \subseteq ac'
	&&\ptext{\cref{lemma:PBMH:alternative-1}}\\
	&=c \land \mathbf{PBMH} (P)
\end{flalign*}
\end{proof}\end{proofs}
\end{lemma}

\begin{lemma}\label{lemma:PBMH(conditional)} Provided $ac'$ is not free in $c$,
\begin{align*}
	&\mathbf{PBMH} (P \dres c \rres Q) = \mathbf{PBMH} (P) \dres c \rres \mathbf{PBMH} (Q)
\end{align*}
\begin{proofs}\begin{proof}\checkt{alcc}\checkt{pfr}
\begin{flalign*}
	&\mathbf{PBMH} (P \dres c \rres Q)
	&&\ptext{Definition of conditional}\\
	&=\mathbf{PBMH} ((c \land P) \lor (\lnot c \land Q))
	&&\ptext{Distributivity of $\mathbf{PBMH}$}\\
	&=\mathbf{PBMH} (c \land P) \lor \mathbf{PBMH} (\lnot c \land Q)
	&&\ptext{\cref{lemma:PBMH(c-land-P):c-land-PBMH(P)}}\\
	&=(c \land \mathbf{PBMH} (P)) \lor (\lnot c \land \mathbf{PBMH} (Q))
	&&\ptext{Definition of conditional}\\
	&=\mathbf{PBMH} (P) \dres c \rres \mathbf{PBMH} (Q)
\end{flalign*}
\end{proof}\end{proofs}
\end{lemma}

\begin{lemma}\label{lemma:PBMH(exists-y-in-ac'-land-e)} Provided $ac'$ is not free in $e$,
\begin{align*}
	&\mathbf{PBMH} (\exists y \spot y \in ac' \land e) = \exists y \spot y \in ac' \land e 
\end{align*}
\begin{proofs}\begin{proof}\checkt{alcc}\checkt{pfr}
\begin{flalign*}
	&\mathbf{PBMH} (\exists y \spot y \in ac' \land e)
	&&\ptext{Definition of $\mathbf{PBMH}$ (\cref{lemma:PBMH:alternative-1})}\\
	&=\exists ac_0 \spot (\exists y \spot y \in ac' \land e)[ac_0/ac'] \land ac_0 \subseteq ac'
	&&\ptext{Substitution: $ac'$ not free in $e$}\\
	&=\exists ac_0 \spot (\exists y \spot y \in ac_0 \land e) \land ac_0 \subseteq ac'
	&&\ptext{Property of sets}\\
	&=\exists y \spot y \in ac' \land e
\end{flalign*}
\end{proof}\end{proofs}
\end{lemma}

\begin{lemma}\label{law:P-ac'neq-emptyset:seqA:Q-ac'neq-emptyset}
\begin{align*}
	&(P \land ac'\neq\emptyset) \seqA (Q \land ac'\neq\emptyset)\\
	&=\\
	&(P \land ac'\neq\emptyset) \seqA (Q \land ac'\neq\emptyset)) \land ac'\neq\emptyset
\end{align*}
\begin{proofs}\begin{proof}
\begin{flalign*}
	&(P \land ac'\neq\emptyset) \seqA (Q \land ac'\neq\emptyset)\\
	&&\ptext{Definition of $\seqA$}\\
	&=(P \land ac'\neq\emptyset)[\{ z | Q \land ac'\neq\emptyset)[z/s] \}/ac']
	&&\ptext{Substitution}\\
	&=(P \land ac'\neq\emptyset)[\{ z | Q[z/s] \land ac'\neq\emptyset \}/ac']
	&&\ptext{Substitution}\\
	&=\left(\begin{array}{l}
		P[\{ z | Q[z/s] \land ac'\neq\emptyset \}/ac'] 
		\\ \land \\
		\{ z | Q[z/s] \land ac'\neq\emptyset \}\neq\emptyset
	\end{array}\right)
	&&\ptext{Propositional calculus}\\
	&=\left(\begin{array}{l}
		P[\{ z | Q[z/s] \land ac'\neq\emptyset \}/ac'] 
		\\ \land \\
		\exists z \spot z \in \{ z | Q[z/s] \land ac'\neq\emptyset \}
	\end{array}\right)
	&&\ptext{Property of sets}\\
	&=\left(\begin{array}{l}
		P[\{ z | Q[z/s] \land ac'\neq\emptyset \}/ac'] 
		\\ \land \\
		\exists z \spot Q[z/s] \land ac'\neq\emptyset
	\end{array}\right)
	&&\ptext{Predicate calculus: quantifier scope and duplicate term}\\
	&=\left(\begin{array}{l}
		P[\{ z | Q[z/s] \land ac'\neq\emptyset \}/ac'] 
		\\ \land \\
		(\exists z \spot Q[z/s] \land ac'\neq\emptyset)
	\end{array}\right) \land ac'\neq\emptyset
	&&\ptext{Property of sets}\\
	&=\left(\begin{array}{l}
		P[\{ z | Q[z/s] \land ac'\neq\emptyset \}/ac'] 
		\\ \land \\
		\{ z | Q[z/s] \land ac'\neq\emptyset\}\neq\emptyset
	\end{array}\right) \land ac'\neq\emptyset
	&&\ptext{Re-introduce $ac'$ and substitution}\\
	&=((P \land ac'\neq\emptyset)[\{ z | Q[z/s] \land ac'\neq\emptyset \}/ac']) \land ac'\neq\emptyset
	&&\ptext{Substitution}\\
	&=((P \land ac'\neq\emptyset)[\{ z | (Q \land ac'\neq\emptyset)[z/s] \}/ac']) \land ac'\neq\emptyset
	&&\ptext{Definition of $\seqA$}\\
	&=((P \land ac'\neq\emptyset) \seqA (Q \land ac'\neq\emptyset)) \land ac'\neq\emptyset
\end{flalign*}
\end{proof}\end{proofs}
\end{lemma}

\begin{lemma}
$\mathbf{PBMH} (P \circseq ac = \emptyset) = P \circseq ac = \emptyset$
\begin{proofs}\begin{proof}\checkt{pfr}
\begin{xflalign*}
	&\mathbf{PBMH} (P \circseq ac = \emptyset)
	&&\ptext{Definition of $\mathbf{PBMH}$}\\
	&=(P \circseq ac = \emptyset) \circseq ac \subseteq ac' \land v'=v
	&&\ptext{Associativity of sequential composition}\\
	&=P \circseq (ac = \emptyset \circseq ac \subseteq ac' \land v'=v)
	&&\ptext{Definition of sequential composition}\\
	&=P \circseq (\exists ac_0, v_0 \spot ac = \emptyset \land ac_0 \subseteq ac' \land v'=v_0)
	&&\ptext{One-point rule}\\
	&=P \circseq (\exists ac_0 \spot ac = \emptyset \land ac_0 \subseteq ac')
	&&\ptext{Propositional calculus}\\
	&=P \circseq (ac = \emptyset \land \exists ac_0 \spot ac_0 \subseteq ac')
	&&\ptext{Choose $ac_0 = \emptyset$}\\
	&=P \circseq (ac = \emptyset \land true)
	&&\ptext{Propositional calculus}\\
	&=P \circseq ac = \emptyset
\end{xflalign*}
\end{proof}\end{proofs}
\end{lemma}


\begin{lemma}\label{lemma:exists-ac1-forall-in-ac'} Provided $ac_1$ is not free in $F(x)$,
\begin{align*}
	&\exists ac_1 \spot (\forall x \spot x \in ac_0 \implies F(x) \in ac_1) \land ac_1 \subseteq ac' \\
	&\iff \\
	&\forall x \spot x \in ac_0 \implies F(x) \in ac'
\end{align*}
\begin{proofs}\begin{proof}(Implication)
\begin{flalign*}
	&\exists ac_1 \spot (\forall x \spot x \in ac_0 \implies F(x) \in ac_1) \land ac_1 \subseteq ac'
	&&\ptext{Predicate calculus}\\
	&\implies \forall x \spot \exists ac_1 \spot (x \in ac_0 \implies F(x) \in ac_1) \land ac_1 \subseteq ac'
	&&\ptext{Predicate calculus}\\
	&=\forall x \spot \exists ac_1 \spot (x \notin ac_0 \land ac_1 \subseteq ac') \lor (F(x) \in ac_1 \land ac_1 \subseteq ac')
	&&\ptext{Predicate calculus}\\
	&=\forall x \spot (\exists ac_1 \spot x \notin ac_0 \land ac_1 \subseteq ac') \lor (\exists ac_1 \spot F(x) \in ac_1 \land ac_1 \subseteq ac')
	&&\ptext{Predicate calculus}\\
	&=\forall x \spot (x \notin ac_0) \lor (\exists ac_1 \spot F(x) \in ac_1 \land ac_1 \subseteq ac')
	&&\ptext{Assumption: $ac_1$ not free in $F(x)$ and predicate calculus}\\
	&=\forall x \spot x \notin ac_0 \lor F(x) \in ac'
	&&\ptext{Predicate calculus}\\
	&=\forall x \spot x \in ac_0 \implies F(x) \in ac'
\end{flalign*}
\end{proof}
\begin{proof}(Reverse implication)
\begin{flalign*}
	&\forall x \spot x \in ac_0 \implies f(x) \in ac'
	&&\ptext{Introduce fresh variable}\\
	&=\exists ac_1 \spot (\forall x \spot x \in ac_0 \implies f(x) \in ac_1) \land ac_1=ac'
	&&\ptext{Predicate calculus}\\
	&\implies \exists ac_1 \spot (\forall x \spot x \in ac_0 \implies f(x) \in ac_1) \land ac_1 \subseteq ac'
\end{flalign*}
\end{proof}\end{proofs}
\end{lemma}

\begin{lemma}\label{lemma:P-refinedby-Q:iff:ac'-Q-subseteq-ac'-P}
$P \sqsubseteq Q \iff [\{ ac' | Q \} \subseteq \{ ac' | P \}]$
\begin{proofs}\begin{proof}
\begin{flalign*}
	&P \sqsubseteq Q
	&&\ptext{Definition of $\sqsubseteq$}\\
	&\iff [Q \implies P]
	&&\ptext{Universal quantification}\\
	&\iff \forall ac', ok', ok, s \spot Q \implies P
	&&\ptext{Property of sets}\\
	&\iff \forall ac', ok', ok, s \spot ac' \in \{ ac' | Q \} \implies ac' \in \{ ac' | P \}
	&&\ptext{Property of sets}\\
	&\iff \forall ac', ok', ok, s \spot \{ ac' | Q \} \subseteq \{ ac' | P \}
	&&\ptext{Universal quantification}\\
	&\iff [\{ ac' | Q \} \subseteq \{ ac' | P \}]
\end{flalign*}
\end{proof}\end{proofs}
\end{lemma}

\begin{lemma}\label{lemma:PBMH(P):implies:exists-ac'-P}
$\mathbf{PBMH} (P) \implies \exists ac' \spot P$
\begin{proofs}\begin{proof}\checkt{alcc}\checkt{pfr}
\begin{xflalign*}
	&\mathbf{PBMH} (P)
	&&\ptext{Definition of $\mathbf{PBMH}$ (\cref{lemma:PBMH:alternative-1})}\\
	&=\exists ac_0 \spot P[ac_0/ac'] \land ac_0 \subseteq ac'
	&&\ptext{Predicate calculus}\\
	&\implies (\exists ac_0 \spot P[ac_0/ac']) \land (\exists ac_0 \spot ac_0 \subseteq ac')
	&&\ptext{Property of sets}\\
	&=\exists ac_0 \spot P[ac_0/ac']
	&&\ptext{Predicate calculus}\\
	&=\exists ac' \spot P
\end{xflalign*}
\end{proof}\end{proofs}
\end{lemma}

\begin{lemma}\label{lemma:PBMH(P)-seqA-true:exists-ac'-P}
$\mathbf{PBMH} (P) \seqA true = \exists ac' \spot P$
\begin{proofs}\begin{proof}\checkt{alcc}\checkt{pfr}
\begin{xflalign*}
	&\mathbf{PBMH} (P) \seqA true
	&&\ptext{Definition of $\mathbf{PBMH}$ (\cref{lemma:PBMH:alternative-1})}\\
	&=(\exists ac_0 \spot P[ac_0/ac'] \land ac_0 \subseteq ac') \seqA true
	&&\ptext{Definition of $\seqA$ and substitution}\\
	&=\exists ac_0 \spot P[ac_0/ac'] \land ac_0 \subseteq \{ s | true \}
	&&\ptext{Property of sets}\\
	&=\exists ac_0 \spot P[ac_0/ac']
	&&\ptext{Predicate calculus}\\
	&=\exists ac' \spot P
\end{xflalign*}
\end{proof}\end{proofs}
\end{lemma}

\section{Substitution Lemmas}

\begin{lemma}\label{lemma:PBMH(P)-ow:PBMH(P-ow)}
$\mathbf{PBMH} (P)^o_w = \mathbf{PBMH} (P^o_w)$
\begin{proofs}\begin{proof}\checkt{alcc}
\begin{xflalign*}
	&\mathbf{PBMH} (P)^o_w
	&&\ptext{Definition of $\mathbf{PBMH}$ (\cref{lemma:PBMH:alternative-1})}\\
	&=(\exists ac_0 \spot P[ac_0/ac'] \land ac_0 \subseteq ac')^o_w
	&&\ptext{Substitution abbreviation}\\
	&=(\exists ac_0 \spot P[ac_0/ac'] \land ac_0 \subseteq ac')[o,s\oplus\{wait\mapsto w\}/ok',s]
	&&\ptext{Substitution}\\
	&=\exists ac_0 \spot P[ac_0/ac'][o,s\oplus\{wait\mapsto w\}/ok',s] \land ac_0 \subseteq ac'
	&&\ptext{Substitution}\\
	&=\exists ac_0 \spot P[o,s\oplus\{wait\mapsto w\}/ok',s][ac_0/ac'] \land ac_0 \subseteq ac'
	&&\ptext{Substitution abbreviation}\\
	&=\exists ac_0 \spot P^o_w[ac_0/ac'] \land ac_0 \subseteq ac'
	&&\ptext{Definition of $\mathbf{PBMH}$ (\cref{lemma:PBMH:alternative-1})}\\
	&=\mathbf{PBMH} (P^o_w)
\end{xflalign*}
\end{proof}\end{proofs}
\end{lemma}

\begin{lemma}\label{lemma:PBMH-substitution-s}
Provided $ac'$ is not free in $e$, 
\begin{align*}
	&\mathbf{PBMH} (P)[e/s] = \mathbf{PBMH} (P[e/s])
\end{align*}
\begin{proofs}\begin{proof}\checkt{alcc}
\begin{xflalign*}
	&\mathbf{PBMH} (P)[e/s]
	&&\ptext{Definition of $\mathbf{PBMH}$ (\cref{lemma:PBMH:alternative-1})}\\
	&=(\exists ac_0 \spot P[ac_0/ac'] \land ac_0 \subseteq ac')[e/s]
	&&\ptext{Property of substitution}\\
	&=(\exists ac_0 \spot P[ac_0/ac'][e/s] \land ac_0 \subseteq ac')
	&&\ptext{Property of substitution: $ac'$ not free in $e$ and $ac_0$ is fresh}\\
	&=(\exists ac_0 \spot P[e/s][ac_0/ac'] \land ac_0 \subseteq ac')
	&&\ptext{Definition of $\mathbf{PBMH}$ (\cref{lemma:PBMH:alternative-1})}\\
	&=\mathbf{PBMH} (P[e/s])
\end{xflalign*}
\end{proof}\end{proofs}
\end{lemma}

\begin{lemma}\label{lemma:PBMH(exsits-x):exists-x(PBMH(x))}
\begin{statement}
Provided $x$ is not $ac'$,
$\mathbf{PBMH} (\exists x @ P) = \exists x @ \mathbf{PBMH} (x)$
\end{statement}
\begin{proofs}
\begin{proof}\checkt{alcc}
\begin{xflalign*}
	&\mathbf{PBMH} (\exists x @ P)
	&&\ptext{Definition of $\mathbf{PBMH}$ (\cref{lemma:PBMH:alternative-1})}\\
	&=\exists ac_0 @ (\exists x @ P)[ac_0/ac'] \land ac_0\subseteq ac'
	&&\ptext{Assumption: $x$ is not $ac'$}\\
	&=\exists ac_0 @ (\exists x @ P[ac_0/ac']) \land ac_0\subseteq ac'
	&&\ptext{Predicate calculus}\\
	&=\exists ac_0 @ (\exists x @ P[ac_0/ac'] \land ac_0\subseteq ac')
	&&\ptext{Predicate calculus}\\
	&=\exists x @ (\exists ac_0 @ P[ac_0/ac'] \land ac_0\subseteq ac')
	&&\ptext{Definition of $\mathbf{PBMH}$ (\cref{lemma:PBMH:alternative-1})}\\
	&=\exists x @ \mathbf{PBMH} (P) 
\end{xflalign*}
\end{proof}
\end{proofs}
\end{lemma}

\begin{lemma}\label{lemma:PBMH(P-y-cap-ac'):PBMH(P)-y-cap-ac'}
\begin{statement}
Provided $P$ is $\mathbf{PBMH}$-healthy,
\begin{align*}
	&\mathbf{PBMH} (P[\{y\}\cap ac'/ac']) = \mathbf{PBMH} (P)[\{y\}\cap ac'/ac']
\end{align*}
\end{statement}
\begin{proofs}
\begin{proof}\checkt{alcc}
\begin{xflalign*}
	&\mathbf{PBMH} (P[\{y\}\cap ac'/ac'])
	&&\ptext{Definition of $\mathbf{PBMH}$ (\cref{lemma:PBMH:alternative-1})}\\
	&=\exists ac_0 @ P[\{y\}\cap ac'/ac'][ac_0/ac'] \land ac_0 \subseteq ac'
	&&\ptext{Substitution}\\
	&=\exists ac_0 @ P[\{y\}\cap ac_0/ac'] \land ac_0 \subseteq ac'
	&&\ptext{Assumption: $P$ is $\mathbf{PBMH}$-healthy}\\
	&=\exists ac_0 @ (\exists ac_1 @ P[ac_1/ac'] \land ac_1\subseteq ac')[\{y\}\cap ac_0/ac'] \land ac_0 \subseteq ac'
	&&\ptext{Substitution}\\
	&=\exists ac_0 @ (\exists ac_1 @ P[ac_1/ac'] \land ac_1\subseteq \{y\}\cap ac_0) \land ac_0 \subseteq ac'
	&&\ptext{Predicate calculus}\\
	&=\exists ac_0, ac_1 @ P[ac_1/ac'] \land ac_1\subseteq \{y\}\cap ac_0 \land ac_0 \subseteq ac'
	&&\ptext{Property of sets}\\
	&=\exists ac_0, ac_1 @ P[ac_1/ac'] \land ac_1\subseteq\{y\} \land ac_1\subseteq ac_0 \land ac_0 \subseteq ac'
	&&\ptext{Property of sets}\\
	&=\exists ac_1 @ P[ac_1/ac'] \land ac_1\subseteq\{y\} \land ac_1\subseteq ac'
	&&\ptext{Property of sets}\\
	&=\exists ac_1 @ P[ac_1/ac'] \land ac_1\subseteq\{y\}\cap ac'
	&&\ptext{Substitution}\\
	&=(\exists ac_1 @ P[ac_1/ac'] \land ac_1\subseteq ac')[\{y\}\cap ac'/ac']
	&&\ptext{Definition of $\mathbf{PBMH}$ (\cref{lemma:PBMH:alternative-1})}\\
	&=\mathbf{PBMH} (P)[\{y\}\cap ac'/ac']	
\end{xflalign*}
\end{proof}
\end{proofs}
\end{lemma}

\begin{lemma}\label{lemma:PBMH(P)-o-ok:PBMH(o-ok)}
\begin{statement}
\begin{align*}
	&\mathbf{PBMH} (P)[o/ok] = \mathbf{PBMH} (P[o/ok])
\end{align*}
\end{statement}
\begin{proofs}
\begin{proof}
\begin{xflalign*}
	&\mathbf{PBMH} (P)[o/ok]
	&&\ptext{Definition of $\mathbf{PBMH}$ (\cref{lemma:PBMH:alternative-1})}\\
	&=(\exists ac_0 @ P[ac_0/ac'] \land ac_0 \subseteq ac')[o/ok]
	&&\ptext{Substitution}\\
	&=\exists ac_0 @ P[o/ok][ac_0/ac'] \land ac_0 \subseteq ac'
	&&\ptext{Definition of $\mathbf{PBMH}$ (\cref{lemma:PBMH:alternative-1})}\\
	&=\mathbf{PBMH} (P[o/ok])
\end{xflalign*}
\end{proof}
\end{proofs}
\end{lemma}

\section{Properties with respect to Designs}

\begin{lemma}\label{lemma:PBMH(design):(lnot-PBMH(pre)|-PBMH(post))}
\begin{statement}
	$\mathbf{PBMH} (P \vdash Q) = (\lnot \mathbf{PBMH} (\lnot P) \vdash \mathbf{PBMH} (Q))$
\end{statement}
\begin{proofs}
\begin{proof}\checkt{alcc}
\begin{xflalign*}
	&\mathbf{PBMH} (P \vdash Q)
	&&\ptext{Definition of design}\\
	&=\mathbf{PBMH} ((ok \land P) \implies (Q \land ok'))
	&&\ptext{Predicate calculus}\\
	&=\mathbf{PBMH} (\lnot ok \lor \lnot P \lor (Q \land ok'))
	&&\ptext{\cref{law:pbmh:distribute-disjunction}}\\
	&=\mathbf{PBMH} (\lnot ok) \lor \mathbf{PBMH} (\lnot P) \lor \mathbf{PBMH} (Q \land ok')
	&&\ptext{\cref{lemma:PBMH(c)-condition:c}}\\
	&=\lnot ok \lor \mathbf{PBMH} (\lnot P) \lor \mathbf{PBMH} (Q \land ok')
	&&\ptext{\cref{lemma:PBMH(c-land-P):c-land-PBMH(P)}}\\
	&=\lnot ok \lor \mathbf{PBMH} (\lnot P) \lor (\mathbf{PBMH} (Q) \land ok')
	&&\ptext{Predicate calculus}\\
	&=(ok \land \lnot \mathbf{PBMH} (\lnot P)) \implies (\mathbf{PBMH} (Q) \land ok')
	&&\ptext{Definition of design}\\
	&=(\lnot \mathbf{PBMH} (\lnot P) \vdash \mathbf{PBMH} (Q))
\end{xflalign*}
\end{proof}
\end{proofs}
\end{lemma}

\begin{lemma}\label{lemma:PBMH-o-J:J-o-PBMH}
$J \circseq (ac \subseteq ac' \land ok'=ok) = (ac \subseteq ac' \land ok'=ok) \circseq J$
\begin{proofs}\begin{proof}\checkt{alcc}\checkt{pfr}
\begin{flalign*}
	&J \circseq (ac \subseteq ac' \land ok'=ok)
	&&\ptext{Definition of $J$}\\
	&=(ac'=ac \land ok \implies ok') \circseq (ac \subseteq ac' \land ok'=ok)
	&&\ptext{Definition of sequential composition}\\
	&=\exists ac_0, ok_0 \spot ac_0=ac \land (ok \implies ok_0) \land ac_0 \subseteq ac' \land ok'=ok_0
	&&\ptext{One-point rule}\\
	&=(ok \implies ok') \land ac \subseteq ac'
	&&\ptext{One-point rule}\\
	&=\exists ac_0, ok_0 \spot ac \subseteq ac_0 \land ok_0 = ok \land ac' = ac_0 \land ok_0 \implies ok'
	&&\ptext{Definition of sequential composition}\\
	&=(ac \subseteq ac' \land ok'=ok) \circseq (ac' = ac \land ok \implies ok')
	&&\ptext{Definition of $J$}\\
	&=(ac \subseteq ac' \land ok'=ok) \circseq J
\end{flalign*}
\end{proof}\end{proofs}
\end{lemma}

\begin{lemma}\label{lemma:PBMH(lnot-PBMH(lnot-P)|-Q):PBMH(P|-Q)}
\begin{statement}
	$\mathbf{PBMH} (\lnot \mathbf{PBMH} (\lnot P) \vdash Q) = \mathbf{PBMH} (P \vdash Q)$ 
\end{statement}
\begin{proofs}
\begin{proof}
\begin{xflalign*}
	&\mathbf{PBMH} (\lnot \mathbf{PBMH} (\lnot P) \vdash Q)
	&&\ptext{Definition of design}\\
	&=\mathbf{PBMH} ((ok \land \lnot \mathbf{PBMH} (\lnot P)) \implies (Q \land ok'))
	&&\ptext{Predicate calculus}\\
	&=\mathbf{PBMH} (\lnot ok \lor \mathbf{PBMH} (\lnot P) \lor (Q \land ok'))
	&&\ptext{\cref{law:pbmh:distribute-disjunction,law:pbmh:idempotent}}\\
	&=\mathbf{PBMH} (\lnot ok \lor \lnot P \lor (Q \land ok'))
	&&\ptext{Predicate calculus}\\
	&=\mathbf{PBMH} ((ok \land P) \implies (Q \land ok'))
	&&\ptext{Definition of design}\\
	&=\mathbf{PBMH} (P \vdash Q)
\end{xflalign*}
\end{proof}
\end{proofs}
\end{lemma}

\begin{theorem}\label{theorem:H2-o-PBMH:PBMH-o-H2}
$\mathbf{H2} \circ \mathbf{PBMH} (P) = \mathbf{PBMH} \circ \mathbf{H2} (P)$
\begin{proofs}\begin{proof}\checkt{alcc}\checkt{pfr}
\begin{flalign*}
	&\mathbf{H2} \circ \mathbf{PBMH} (P)
	&&\ptext{Definition of $\mathbf{H2}$ (J-split)}\\
	&=\mathbf{PBMH} (P) \circseq J
	&&\ptext{Definition of $\mathbf{PBMH}$}\\
	&=(P \circseq ac \subseteq ac' \land ok'=ok) \circseq J
	&&\ptext{Associativity of sequential composition}\\
	&=P \circseq ((ac \subseteq ac' \land ok'=ok) \circseq J)
	&&\ptext{\cref{lemma:PBMH-o-J:J-o-PBMH}}\\
	&=P \circseq (J \circseq (ac \subseteq ac' \land ok'=ok))
	&&\ptext{Associativity of sequential composition}\\
	&=(P \circseq J) \circseq (ac \subseteq ac' \land ok'=ok)
	&&\ptext{Definition of $\mathbf{PBMH}$}\\
	&=\mathbf{PBMH} (P \circseq J)
	&&\ptext{Definition of $\mathbf{H2}$ (J-split)}\\
	&=\mathbf{PBMH} \circ \mathbf{H2} (P)
\end{flalign*}
\end{proof}\end{proofs}
\end{theorem}

\begin{theorem}\label{theorem:H1-o-PBMH:PBMH-o-H1}
$\mathbf{H1} \circ \mathbf{PBMH} (P) = \mathbf{PBMH} \circ \mathbf{H1} (P)$
\begin{proofs}\begin{proof}\checkt{alcc}\checkt{pfr}
\begin{flalign*}
	&\mathbf{PBMH} \circ \mathbf{H1} (P)
	&&\ptext{Definition of $\mathbf{H1}$}\\
	&=\mathbf{PBMH} (ok \implies P)
	&&\ptext{Predicate calculus}\\
	&=\mathbf{PBMH} (\lnot ok \lor P)
	&&\ptext{Distributivity of $\mathbf{PBMH}$}\\
	&=\mathbf{PBMH} (\lnot ok) \lor \mathbf{PBMH} (P)
	&&\ptext{\cref{lemma:PBMH(c)-condition:c}}\\
	&=\lnot ok \lor \mathbf{PBMH} (P)
	&&\ptext{Predicate calculus}\\
	&=ok \implies \mathbf{PBMH} (P)
	&&\ptext{Definition of $\mathbf{H1}$}\\
	&=\mathbf{H1} \circ \mathbf{PBMH} (P)
\end{flalign*}
\end{proof}\end{proofs}
\end{theorem}

\section{Properties with respect to $\mathbf{A2}$}

\begin{lemma}\label{lemma:A2-o-PBMH(P)} Provided $P$ is $\mathbf{PBMH}$-healthy.
\begin{align*}
	&\mathbf{PBMH} (P \seqA \{ s | \{s\} = ac'\}) \\
	&= \\
	&\exists ac_1, ac_0 \spot P[ac_0/ac'] \land ac_0 \subseteq \{ s | \{s\} = ac_1 \} \land ac_1 \subseteq ac'
\end{align*}
\begin{proofs}\begin{proof}
\begin{xflalign*}
	&\mathbf{PBMH} (P \seqA \{ s | \{s\} = ac'\})
	&&\ptext{Assumption: $P$ is $\mathbf{PBMH}$-healthy}\\
	&=\mathbf{PBMH} (\mathbf{PBMH} (P) \seqA \{ s | \{s\} = ac'\})
	&&\ptext{Definition of $\mathbf{PBMH}$ (\cref{lemma:PBMH:alternative-1})}\\
	&=\mathbf{PBMH} ((\exists ac_0 \spot P[ac_0/ac'] \land ac_0 \subseteq ac') \seqA \{ s | \{s\} = ac'\})
	&&\ptext{Definition of $\seqA$ and substitution}\\ 
	&=\mathbf{PBMH} (\exists ac_0 \spot P[ac_0/ac'] \land ac_0 \subseteq \{ s | \{s\} = ac'\})
	&&\ptext{Definition of $\mathbf{PBMH}$ (\cref{lemma:PBMH:alternative-1})}\\
	&=(\exists ac_1 \spot (\exists ac_0 \spot P[ac_0/ac'] \land ac_0 \subseteq \{ s | \{s\} = ac'\})[ac_1/ac'] \land ac_1 \subseteq ac')
	&&\ptext{Substitution and predicate calculus}\\
	&=\exists ac_1, ac_0 \spot P[ac_0/ac'] \land ac_0 \subseteq \{ s | \{s\} = ac_1 \} \land ac_1 \subseteq ac'
\end{xflalign*}
\end{proof}\end{proofs}
\end{lemma}

\begin{theorem} Provided $P$ is $\mathbf{PBMH}$-healthy and $v$ is not free in $P$,
\begin{align*}
	&\exists v \spot (P \seqA Q) \implies P \seqA (\exists v \spot Q)
\end{align*}
\begin{proofs}\begin{proof}
\begin{xflalign*}
	&\exists v \spot (P \seqA Q)
	&&\ptext{Assumption: $P$ is $\mathbf{PBMH}$-healthy}\\
	&=\exists v \spot (\mathbf{PBMH} (P) \seqA Q)
	&&\ptext{Definition of $\mathbf{PBMH}$ (\cref{lemma:PBMH:alternative-1})}\\
	&=\exists v \spot ((\exists ac_0 \spot P[ac_0/ac'] \land ac_0 \subseteq ac') \seqA Q)
	&&\ptext{Definition of $\seqA$ and substitution}\\
	&=\exists v \spot (\exists ac_0 \spot P[ac_0/ac'] \land ac_0 \subseteq \{ s | Q \})
	&&\ptext{Predicate calculus: $v$ is not free in $P$}\\
	&=\exists ac_0 \spot P[ac_0/ac'] \land (\exists v \spot ac_0 \subseteq \{ s | Q \})
	&&\ptext{\cref{lemma:set-theory:exists-v-subset-1}}\\
	&\implies \exists ac_0 \spot P[ac_0/ac'] \land ac_0 \subseteq \{ s | \exists v \spot Q \}
	&&\ptext{Definition of $\seqA$ and substitution}\\
	&=(\exists ac_0 \spot P[ac_0/ac'] \land ac_0 \subseteq ac') \seqA (\exists v \spot Q)
	&&\ptext{Definition of $\mathbf{PBMH}$ (\cref{lemma:PBMH:alternative-1})}\\
	&=\mathbf{PBMH} (P) \seqA (\exists v \spot Q)
	&&\ptext{Assumption: $P$ is $\mathbf{PBMH}$-healthy}\\
	&=P \seqA (\exists v \spot Q)
\end{xflalign*}
\end{proof}\end{proofs}
\end{theorem}

\chapter{Sequential Composition ($\mathcal{A}$)}
\label{appendix:seqA}


\section{Properties}

\begin{lemma}\label{law:seqA-ac'-not-free}
Provided $ac'$ is not free in $P$,
$P \seqA Q = P$.
\begin{proofs}\begin{proof}\checkt{alcc}\checkt{pfr}
\begin{flalign*}
	&P \seqA Q
	&&\ptext{Definition of $\seqA$}\\
	&=P[\{ z : State | Q[z/s]\}/ac']
	&&\ptext{Assumption: $ac'$ not free in $P$}\\
	&=P
\end{flalign*}
\end{proof}\end{proofs}
\end{lemma}

\begin{lemma}\label{law:seqA-negation}
$\lnot (P \seqA Q) = (\lnot P \seqA Q)$
\begin{proofs}\begin{proof}
\begin{flalign*}
	&\lnot (P \seqA Q)
	&&\ptext{Definition of sequential composition}\\
	&=\lnot (P[\{ z | Q[z/s] \}/ac'])
	&&\ptext{Propositional calculus}\\
	&=(\lnot P[\{ z | Q[z/s] \}/ac'])
	&&\ptext{Definition of sequential composition}\\
	&=(\lnot P \seqA Q)
\end{flalign*}
\end{proof}\end{proofs}
\end{lemma}

\begin{lemma}\label{law:seqA-associativity}
Provided $P$ and $Q$ satisfy $\mathbf{PBMH}$,
\begin{align*}
	&P \seqA (Q \seqA R) = (P \seqA Q) \seqA R
\end{align*}
\begin{proofs}\begin{proof}
\begin{xflalign*}
	&P \seqA (Q \seqA R)
	&&\ptext{Definition of $\seqA$}\\
	&=P[\{ s | Q \seqA R\}/ac']
	&&\ptext{Definition of $\seqA$}\\
	&=P[\{ s | Q[\{ s | R\}/ac']\}/ac']
	&&\ptext{Property of substitution}\\
	&=P[\{ s | Q\}/ac'][\{ s | R\}/ac']
	&&\ptext{Definition of $\seqA$}\\
	&=P[\{ s | Q\}/ac'] \seqA R
	&&\ptext{Definition of $\seqA$}\\
	&=(P \seqA Q) \seqA R
\end{xflalign*}
\end{proof}\end{proofs}
\end{lemma}

\begin{lemma}\label{law:seqA-right-distributivity}
$(P \lor Q) \seqA R = (P \seqA R) \lor (Q \seqA R)$
\begin{proofs}\begin{proof}\checkt{alcc}\checkt{pfr}
\begin{flalign*}
	&(P \lor Q) \seqA R
	&&\ptext{Definition of $\seqA$}\\
	&=(P \lor Q)[\{ z | R[z/s]\}/ac']
	&&\ptext{Substitution}\\
	&=(P[\{ z | R[z/s]\}/ac'] \lor Q[\{ z | R[z/s]\}/ac'])
	&&\ptext{Definition of $\seqA$}\\
	&=(P \seqA R) \lor (Q \seqA R)
\end{flalign*}
\end{proof}\end{proofs}
\end{lemma}

\begin{lemma}\label{law:seqA-right-distributivity-conjunction}
$(P \land Q) \seqA R = (P \seqA R) \land (Q \seqA R)$
\begin{proofs}\begin{proof}\checkt{alcc}\checkt{pfr}
\begin{flalign*}
	&(P \land Q) \seqA R
	&&\ptext{Definition of $\seqA$}\\
	&=(P \land Q)[\{ z | R[z/s]\}/ac']
	&&\ptext{Property of substitution}\\
	&=(P[\{ z | R[z/s]\}/ac'] \land Q[\{ z | R[z/s]\}/ac'])
	&&\ptext{Definition of $\seqA$}\\
	&=(P \seqA R) \land (Q \seqA R)
\end{flalign*}
\end{proof}\end{proofs}
\end{lemma}

\begin{lemma}\label{lemma:seqA:P-seqA(Q-land-R):implies:(P-seqA-Q)-land-(P-seqA-R)} Provided $P$ is $\mathbf{PBMH}$-healthy,
\begin{align*}
	&P \seqA (Q \land R) \implies (P \seqA Q) \land (P \seqA R)
\end{align*}
\begin{proofs}\begin{proof}\checkt{alcc}
\begin{xflalign*}
	&P \seqA (Q \land R)
	&&\ptext{Assumption: $P$ is $\mathbf{PBMH}$-healthy}\\
	&=\mathbf{PBMH} (P) \seqA (Q \land R)
	&&\ptext{Definition of $\mathbf{PBMH}$ (\cref{lemma:PBMH:alternative-1})}\\
	&=(\exists ac_0 \spot P[ac_0/ac'] \land ac_0\subseteq ac') \seqA (Q \land R)
	&&\ptext{Definition of $\seqA$ and substitution}\\
	&=\exists ac_0 \spot P[ac_0/ac'] \land ac_0\subseteq \{ s | Q \land R \}
	&&\ptext{Property of sets}\\
	&=\exists ac_0 \spot P[ac_0/ac'] \land ac_0\subseteq \{ s | Q \} \land ac_0\subseteq \{ s | R \}
	&&\ptext{Predicate calculus}\\
	&\implies \left(\begin{array}{l}
		(\exists ac_0 \spot P[ac_0/ac'] \land ac_0\subseteq \{ s | Q \}) 
		\\ \land \\
		(\exists ac_0 \spot P[ac_0/ac'] \land ac_0\subseteq \{ s | R \})
	\end{array}\right)
	&&\ptext{Definition of $\seqA$ and substitution}\\
	&=\left(\begin{array}{l}
		((\exists ac_0 \spot P[ac_0/ac'] \land ac_0\subseteq ac') \seqA Q) 
		\\ \land \\
		((\exists ac_0 \spot P[ac_0/ac'] \land ac_0\subseteq ac') \seqA R)
	\end{array}\right)
	&&\ptext{Definition of $\mathbf{PBMH}$ (\cref{lemma:PBMH:alternative-1})}\\
	&=(\mathbf{PBMH} (P) \seqA Q) \land (\mathbf{PBMH} (P) \seqA R)
	&&\ptext{Assumption: $P$ is $\mathbf{PBMH}$-healthy}\\
	&=(P \seqA Q) \land (P \seqA R)
\end{xflalign*}
\end{proof}\end{proofs}
\end{lemma}

\section{Lemmas}

\begin{lemma}\label{lemma:(P-seqA-Q)-lor-(P-seqA-R):implies:(P-seqA-(Q-lor-R))} Provided $P$ is $\mathbf{PBMH}$-healthy,
\begin{align*}
	&(P \seqA Q) \lor (P \seqA R) \implies (P \seqA (Q \lor R))
\end{align*}
\begin{proofs}\begin{proof}\checkt{alcc}\checkt{pfr}
\begin{xflalign*}
	&(P \seqA Q) \lor (P \seqA R)
	&&\ptext{Assumption: $P$ is $\mathbf{PBMH}$-healthy (\cref{lemma:PBMH:alternative-1})}\\
	&=\left(\begin{array}{l}
		((\exists ac_0 \spot P[ac_0/ac'] \land ac_0 \subseteq ac') \seqA Q) 
		\\ \lor \\
		((\exists ac_0 \spot P[ac_0/ac'] \land ac_0 \subseteq ac') \seqA R)
	\end{array}\right)
	&&\ptext{Definition of $\seqA$ and substitution}\\
	&=\left(\begin{array}{l}
		(\exists ac_0 \spot P[ac_0/ac'] \land ac_0 \subseteq \{ s | Q \}) 
		\\ \lor \\
		(\exists ac_0 \spot P[ac_0/ac'] \land ac_0 \subseteq \{ s | R \})
	\end{array}\right)
	&&\ptext{Predicate calculus}\\
	&=\exists ac_0 \spot P[ac_0/ac'] \land (ac_0 \subseteq \{ s | Q \} \lor ac_0 \subseteq \{ s | R \})
	&&\ptext{Property of sets and predicate calculus}\\
	&\implies \exists ac_0 \spot P[ac_0/ac'] \land ac_0 \subseteq \{ s | Q \} \cup \{ s | R \}
	&&\ptext{Property of sets}\\
	&=\exists ac_0 \spot P[ac_0/ac'] \land ac_0 \subseteq \{ s | Q \lor R \}
	&&\ptext{Definition of $\seqA$ and substitution}\\
	&=(\exists ac_0 \spot P[ac_0/ac'] \land ac_0 \subseteq ac') \seqA (Q \lor R)
	&&\ptext{Assumption: $P$ is $\mathbf{PBMH}$-healthy (\cref{lemma:PBMH:alternative-1})}\\
	&=P \seqA (Q \lor R)
\end{xflalign*}
\end{proof}\end{proofs}
\end{lemma}

\begin{lemma}\label{theorem:(P-seqA-Q)-lor-(P-seqA-true):P-seqA-true} Provided $P$ is $\mathbf{PBMH}$-healthy,
\begin{align*}
	&(P \seqA Q) \lor (P \seqA true) = P \seqA true
\end{align*}
\begin{proofs}\begin{proof}\checkt{alcc}\checkt{pfr}
\begin{xflalign*}
	&(P \seqA Q) \lor (P \seqA true)
	&&\ptext{\cref{lemma:(P-seqA-Q)-lor-(P-seqA-R):implies:(P-seqA-(Q-lor-R))}}\\
	&=((P \seqA Q) \lor (P \seqA true)) \land (P \seqA (Q \lor true))
	&&\ptext{Predicate calculus}\\
	&=((P \seqA Q) \lor (P \seqA true)) \land (P \seqA true)
	&&\ptext{Predicate calculus: absorption law}\\
	&=(P \seqA true)
\end{xflalign*}
\end{proof}\end{proofs}
\end{lemma}

\begin{lemma}\label{theorem:(P-seqA-Q)-lor-(P-seqA-false):P-seqA-Q} Provided $P$ is $\mathbf{PBMH}$-healthy,
\begin{align*}
	&(P \seqA Q) \lor (P \seqA false) = P \seqA Q
\end{align*}
\begin{proofs}\begin{proof}\checkt{alcc}\checkt{pfr}
\begin{xflalign*}
	&(P \seqA Q) \lor (P \seqA false)
	&&\ptext{\cref{lemma:(P-seqA-Q)-lor-(P-seqA-R):implies:(P-seqA-(Q-lor-R))}}\\
	&=((P \seqA Q) \lor (P \seqA false)) \land (P \seqA (Q \lor false))
	&&\ptext{Predicate calculus}\\
	&=((P \seqA Q) \lor (P \seqA false)) \land (P \seqA Q)
	&&\ptext{Predicate calculus: absorption law}\\
	&=P \seqA Q
\end{xflalign*}
\end{proof}\end{proofs}
\end{lemma}

\begin{lemma}\label{theorem:P-seqA-(Q-implies-(R-land-ok'))} Provided $P$ is $\mathbf{PBMH}$-healthy,
\begin{align*}
	&P \seqA (Q \implies (R \land ok')) = (P \seqA \lnot Q) \lor ((P \seqA (Q \implies R)) \land ok')
\end{align*}
\begin{proofs}\begin{proof}\checkt{alcc}
\begin{xflalign*}
	&P \seqA (Q \implies (R \land ok'))
	&&\ptext{Assumption: $P$ is $\mathbf{PBMH}$-healthy (\cref{lemma:PBMH:alternative-1})}\\
	&=(\exists ac_0 \spot P[ac_0/ac'] \land ac_0 \subseteq ac') \seqA (Q \implies (R \land ok'))
	&&\ptext{Definition of $\seqA$ and substitution}\\
	&=\exists ac_0 \spot P[ac_0/ac'] \land ac_0 \subseteq \{ s | Q \implies (R \land ok') \}
	&&\ptext{Property of sets}\\
	&=\exists ac_0 \spot P[ac_0/ac'] \land \forall z \spot z \in ac_0 \implies (Q[z/s] \implies (R[z/s] \land ok'))
	&&\ptext{\cref{lemma:forall-x-P-implies(Q-implies(R-land-e))}}\\
	&=\exists ac_0 \spot P[ac_0/ac'] \land \left(\begin{array}{l}
		(\forall z \spot z \in ac_0 \implies \lnot Q[z/s])
		\\ \lor \\
		((\forall z \spot z \in ac_0 \implies (Q[z/s] \implies R[z/s])) \land ok')
	\end{array}\right)
	&&\ptext{Predicate calculus}\\
	&=\left(\begin{array}{l}
		(\exists ac_0 \spot P[ac_0/ac'] \land (\forall z \spot z \in ac_0 \implies \lnot Q[z/s]))
		\\ \lor \\
		(\exists ac_0 \spot P[ac_0/ac'] \land ((\forall z \spot z \in ac_0 \implies (Q[z/s] \implies R[z/s])) \land ok'))
	\end{array}\right)
	&&\ptext{Property of sets}\\
	&=\left(\begin{array}{l}
		(\exists ac_0 \spot P[ac_0/ac'] \land (\forall z \spot z \in ac_0 \implies z \in \{ s | \lnot Q\}))
		\\ \lor \\
		(\exists ac_0 \spot P[ac_0/ac'] \land ((\forall z \spot z \in ac_0 \implies z \in \{ s | Q \implies R \}) \land ok'))
	\end{array}\right)
	&&\ptext{Property of sets}\\
	&=\left(\begin{array}{l}
		(\exists ac_0 \spot P[ac_0/ac'] \land ac_0 \subseteq \{ s | \lnot Q\})
		\\ \lor \\
		(\exists ac_0 \spot P[ac_0/ac'] \land ac_0 \subseteq \{ s | Q \implies R \} \land ok')
	\end{array}\right)
	&&\ptext{Predicate calculus}\\
	&=\left(\begin{array}{l}
		(\exists ac_0 \spot P[ac_0/ac'] \land ac_0 \subseteq \{ s | \lnot Q\})
		\\ \lor \\
		((\exists ac_0 \spot P[ac_0/ac'] \land ac_0 \subseteq \{ s | Q \implies R \}) \land ok')
	\end{array}\right)
	&&\ptext{Definition of $\seqA$ and substitution}\\
	&=\left(\begin{array}{l}
		((\exists ac_0 \spot P[ac_0/ac'] \land ac_0 \subseteq ac') \seqA \lnot Q)
		\\ \lor \\
		(((\exists ac_0 \spot P[ac_0/ac'] \land ac_0 \subseteq ac') \seqA (Q \implies R)) \land ok')
	\end{array}\right)
	&&\ptext{Assumption: $P$ is $\mathbf{PBMH}$-healthy (\cref{lemma:PBMH:alternative-1})}\\
	&=(P \seqA \lnot Q) \lor ((P \seqA (Q \implies R)) \land ok')
\end{xflalign*}
\end{proof}\end{proofs}
\end{lemma}

\begin{lemma}\label{lemma:forall-x-P-implies(Q-implies(R-land-e))} Provided $x$ is not free in $e$,
\begin{align*}
	&\forall x \spot P \implies (Q \implies (R \land e))\\
	&=\\
	&(\forall x \spot P \implies \lnot Q) \lor ((\forall x \spot P \implies (Q \implies R)) \land e)
\end{align*}
\begin{proofs}\begin{proof}\checkt{alcc}\checkt{pfr}
\begin{xflalign*}
	&\forall x \spot P \implies (Q \implies (R \land e))
	&&\ptext{Predicate calculus}\\
	&=\forall x \spot (P \land Q) \implies (R \land e)
	&&\ptext{Predicate calculus}\\
	&=\forall x \spot ((P \land Q) \implies R) \land ((P \land Q) \implies e)
	&&\ptext{Predicate calculus}\\
	&=\forall x \spot ((P \land Q) \implies R) \land (\lnot (P \land Q) \lor e)
	&&\ptext{Predicate calculus}\\
	&=(\forall x \spot (P \land Q) \implies R) \land (\forall x \spot \lnot (P \land Q) \lor e)
	&&\ptext{Predicate calculus: $x$ is not free in $e$}\\
	&=(\forall x \spot (P \land Q) \implies R) \land ((\forall x \spot \lnot (P \land Q)) \lor e)
	&&\ptext{Predicate calculus}\\
	&=\left(\begin{array}{l}
		((\forall x \spot (P \land Q) \implies R) \land (\forall x \spot \lnot (P \land Q)))
		\\ \lor \\
		((\forall x \spot (P \land Q) \implies R) \land e)
	\end{array}\right)	
	&&\ptext{Predicate calculus}\\
	&=\left(\begin{array}{l}
		(\forall x \spot ((P \land Q) \implies R) \land \lnot (P \land Q))
		\\ \lor \\
		((\forall x \spot ((P \land Q) \implies R)) \land e)
	\end{array}\right)
	&&\ptext{Predicate calculus}\\
	&=\left(\begin{array}{l}
		(\forall x \spot \lnot (P \land Q))
		\\ \lor \\
		((\forall x \spot ((P \land Q) \implies R)) \land e)
	\end{array}\right)
		&&\ptext{Predicate calculus}\\
	&=(\forall x \spot P \implies \lnot Q) \lor ((\forall x \spot P \implies (Q \implies R)) \land e)
\end{xflalign*}
\end{proof}\end{proofs}
\end{lemma}

\begin{lemma}\label{law:seqA:PBMH-PQ-and-ok'}Provided $P$ is $\mathbf{PBMH}$-healthy,
\begin{align*}
	&P \seqA (Q \land ok') = (P \seqA false) \lor ((P \seqA Q) \land ok')&
\end{align*}
\begin{proofs}\begin{proof}\checkt{alcc}\checkt{pfr}
\begin{xflalign*}
	&P \seqA (Q \land ok')
	&&\ptext{Assumption: $P$ is $\mathbf{PBMH}$-healthy}\\
	&=\mathbf{PBMH} (P) \seqA (Q \land ok')
	&&\ptext{Definition of $\mathbf{PBMH}$ (\cref{lemma:PBMH:alternative-1})}\\
	&=(\exists ac_0 \spot P[ac_0/ac'] \land ac_0 \subseteq ac') \seqA (Q \land ok')
	&&\ptext{Definition of $\seqA$ and substitution}\\
	&=\exists ac_0 \spot P[ac_0/ac'] \land ac_0 \subseteq \{ z | (Q \land ok')[z/s] \}
	&&\ptext{Property of substitution}\\
	&=\exists ac_0 \spot P[ac_0/ac'] \land ac_0 \subseteq \{ z | Q[z/s] \land ok' \}
	&&\ptext{Property of sets}\\
	&=\exists ac_0 \spot P[ac_0/ac'] \land (\forall z \spot z \in ac_0 \implies (Q[z/s] \land ok'))
	&&\ptext{Propositional calculus}\\
	&=\exists ac_0 \spot P[ac_0/ac'] \land (\forall z \spot z \in ac_0 \implies Q[z/s]) \land (\forall z \spot z \in ac_0 \implies ok')
	&&\ptext{Propositional calculus}\\
	&=\exists ac_0 \spot P[ac_0/ac'] \land (\forall z \spot z \in ac_0 \implies Q[z/s]) \land (\forall z \spot z \notin ac_0 \lor ok')
	&&\ptext{Predicate calculus: $ok' \neq z$, move quantifier}\\
	&=\exists ac_0 \spot P[ac_0/ac'] \land (\forall z \spot z \in ac_0 \implies Q[z/s]) \land ((\forall z \spot z \notin ac_0) \lor ok')
	&&\ptext{Predicate calculus: distribution}\\
	&=\exists ac_0 \spot P[ac_0/ac'] \land \left(\begin{array}{l}
		((\forall z \spot z \in ac_0 \implies Q[z/s]) \land (\forall z \spot z \notin ac_0))\\
		\lor \\
		((\forall z \spot z \in ac_0 \implies Q[z/s]) \land ok')
	\end{array}\right)
	&&\ptext{Predicate calculus}\\
	&=\exists ac_0 \spot P[ac_0/ac'] \land \left(\begin{array}{l}
		(\forall z \spot (z \in ac_0 \implies Q[z/s]) \land z \notin ac_0)\\
		\lor \\
		((\forall z \spot z \in ac_0 \implies Q[z/s]) \land ok')
		\end{array}\right)
	&&\ptext{Propositional calculus}\\
	&=\exists ac_0 \spot P[ac_0/ac'] \land ((\forall z \spot z \notin ac_0) \lor ((\forall z \spot z \in ac_0 \implies Q[z/s]) \land ok'))
	&&\ptext{Propositional calculus}\\
	&=\left(\begin{array}{l}
		(\exists ac_0 \spot P[ac_0/ac'] \land \forall z \spot z \notin ac_0) \\
	  	\lor \\
		(\exists ac_0 \spot P[ac_0/ac'] \land (\forall z \spot z \in ac_0 \implies Q[z/s]) \land ok')
	  \end{array}\right)
	&&\ptext{Property of sets and introduce set comprehension}\\
	&=\left(\begin{array}{l}
		(\exists ac_0 \spot P[ac_0/ac'] \land ac_0=\emptyset) \\
	  	\lor \\
		((\exists ac_0 \spot P[ac_0/ac'] \land ac_0 \subseteq \{ z | Q[z/s]\}) \land ok')
	  \end{array}\right)
	&&\ptext{One-point rule and substitution}\\
	&=\left(\begin{array}{l}
		P[\emptyset/ac'] \\
	  	\lor \\
		((\exists ac_0 \spot P[ac_0/ac'] \land ac_0 \subseteq \{ z | Q[z/s]\}) \land ok')
	  \end{array}\right)
	&&\ptext{Re-introduce $ac'$}\\
	&=\left(\begin{array}{l}
		P[\emptyset/ac'] \\
	  	\lor \\
		((\exists ac_0 \spot P[ac_0/ac'] \land ac_0 \subseteq ac')[\{ z | Q[z/s]\}/ac'] \land ok')
	  \end{array}\right)
	&&\ptext{Definition of $\seqA$}\\
	&=\left(\begin{array}{l}
		P[\emptyset/ac'] \\
	  	\lor \\
		(((\exists ac_0 \spot P[ac_0/ac'] \land ac_0 \subseteq ac') \seqA Q) \land ok')
	  \end{array}\right)
	&&\ptext{Definition of $\mathbf{PBMH}$ (\cref{lemma:PBMH:alternative-1})}\\
	&=P[\emptyset/ac'] \lor ((\mathbf{PBMH} (P) \seqA Q) \land ok')
	&&\ptext{Assumption: $P$ is $\mathbf{PBMH}$-healthy}\\
	&=P[\emptyset/ac'] \lor ((P \seqA Q) \land ok')
	&&\ptext{\cref{law:seqA-P-sequence-false:1}}\\
	&=(P \seqA false) \lor ((P \seqA Q) \land ok')
\end{xflalign*}
\end{proof}\end{proofs}
\end{lemma}

\begin{lemma}\label{lemma:P-seqA-(Q-land-R)-land-R:(P-seqA-Q)-land-R} 
Provided $s$ is not free in $R$ and $P$ is $\mathbf{PBMH}$-healthy,
\begin{align*}
	&(P \seqA (Q \land R)) \land R = (P \seqA Q) \land R
\end{align*}
\begin{proofs}\begin{proof}\checkt{alcc}\checkt{pfr}
\begin{xflalign*}
	&(P \seqA (Q \land R)) \land R
	&&\ptext{Assumption: $P$ is $\mathbf{PBMH}$-healthy (\cref{lemma:PBMH:alternative-1})}\\
	&=((\exists ac_0 \spot P[ac_0/ac'] \land ac_0 \subseteq ac') \seqA (Q \land R)) \land R
	&&\ptext{Definition of $\seqA$ and substitution}\\
	&=(\exists ac_0 \spot P[ac_0/ac'] \land ac_0 \subseteq \{ s | Q \land R \}) \land R
	&&\ptext{Property of sets}\\
	&=\exists ac_0 \spot P[ac_0/ac'] \land ac_0 \subseteq \{ s | Q \} \land ac_0 \subseteq \{ s | R \} \land R
	&&\ptext{Property of sets}\\
	&=\exists ac_0 \spot P[ac_0/ac'] \land ac_0 \subseteq \{ s | Q \} \land (\forall s \spot s \in ac_0 \implies R) \land R
	&&\ptext{Assumption: $s$ is not free in $R$ and predicate calculus}\\
	&=\exists ac_0 \spot P[ac_0/ac'] \land ac_0 \subseteq \{ s | Q \} \land ((\forall s \spot s \notin ac_0) \lor R) \land R
	&&\ptext{Predicate calculus: absorption law}\\
	&=\exists ac_0 \spot P[ac_0/ac'] \land ac_0 \subseteq \{ s | Q \} \land R
	&&\ptext{Predicate calculus}\\
	&=(\exists ac_0 \spot P[ac_0/ac'] \land ac_0 \subseteq \{ s | Q \}) \land R
	&&\ptext{Definition of $\seqA$ and substitution}\\
	&=((\exists ac_0 \spot P[ac_0/ac'] \land ac_0 \subseteq ac') \seqA Q) \land R
	&&\ptext{Assumption: $P$ is $\mathbf{PBMH}$-healthy (\cref{lemma:PBMH:alternative-1})}\\
	&=(P \seqA Q) \land R
\end{xflalign*}
\end{proof}\end{proofs}
\end{lemma}

\begin{lemma}\label{lemma:(P-land-Q)-seqA-R:P-land-(Q-seqA-R)} Provided $ac'$ is not free in $P$,
\begin{align*}
	&(P \land Q) \seqA R = P \land (Q \seqA R)
\end{align*}
\begin{proofs}\begin{proof}\checkt{alcc}\checkt{pfr}
\begin{xflalign*}
	&(P \land Q) \seqA R
	&&\ptext{\cref{law:seqA-right-distributivity-conjunction}}\\
	&=(P \seqA R) \land (Q \seqA R)
	&&\ptext{Assumption: $ac'$ not free in $P$ and~\cref{law:seqA-ac'-not-free}}\\
	&=P \land (Q \seqA R)
\end{xflalign*}
\end{proof}\end{proofs}
\end{lemma}

\begin{lemma}\label{theorem:P-seqA-(Q-implies-(R-land-e))} Provided $P$ is $\mathbf{PBMH}$-healthy and $s$ is not free in $e$,
\begin{align*}
	&P \seqA (Q \implies (R \land e)) = (P \seqA \lnot Q) \lor ((P \seqA (Q \implies R)) \land e)
\end{align*}
\begin{proofs}\begin{proof}\checkt{alcc}\checkt{pfr}
\begin{xflalign*}
	&P \seqA (Q \implies (R \land e))
	&&\ptext{Assumption: $P$ is $\mathbf{PBMH}$-healthy (\cref{lemma:PBMH:alternative-1})}\\
	&=(\exists ac_0 \spot P[ac_0/ac'] \land ac_0 \subseteq ac') \seqA (Q \implies (R \land e))
	&&\ptext{Definition of $\seqA$ and substitution}\\
	&=\exists ac_0 \spot P[ac_0/ac'] \land ac_0 \subseteq \{ s | Q \implies (R \land e) \}
	&&\ptext{Property of sets and $s$ not free in $e$}\\
	&=\exists ac_0 \spot P[ac_0/ac'] \land \forall z \spot z \in ac_0 \implies (Q[z/s] \implies (R[z/s] \land e))
	&&\ptext{\cref{lemma:forall-x-P-implies(Q-implies(R-land-e))}}\\
	&=\exists ac_0 \spot P[ac_0/ac'] \land \left(\begin{array}{l}
		(\forall z \spot z \in ac_0 \implies \lnot Q[z/s])
		\\ \lor \\
		((\forall z \spot z \in ac_0 \implies (Q[z/s] \implies R[z/s])) \land e)
	\end{array}\right)
	&&\ptext{Predicate calculus}\\
	&=\left(\begin{array}{l}
		(\exists ac_0 \spot P[ac_0/ac'] \land (\forall z \spot z \in ac_0 \implies \lnot Q[z/s]))
		\\ \lor \\
		(\exists ac_0 \spot P[ac_0/ac'] \land ((\forall z \spot z \in ac_0 \implies (Q[z/s] \implies R[z/s])) \land e))
	\end{array}\right)
	&&\ptext{Property of sets}\\
	&=\left(\begin{array}{l}
		(\exists ac_0 \spot P[ac_0/ac'] \land (\forall z \spot z \in ac_0 \implies z \in \{ s | \lnot Q\}))
		\\ \lor \\
		(\exists ac_0 \spot P[ac_0/ac'] \land ((\forall z \spot z \in ac_0 \implies z \in \{ s | Q \implies R \}) \land e))
	\end{array}\right)
	&&\ptext{Property of sets}\\
	&=\left(\begin{array}{l}
		(\exists ac_0 \spot P[ac_0/ac'] \land ac_0 \subseteq \{ s | \lnot Q\})
		\\ \lor \\
		(\exists ac_0 \spot P[ac_0/ac'] \land ac_0 \subseteq \{ s | Q \implies R \} \land e)
	\end{array}\right)
	&&\ptext{Predicate calculus}\\
	&=\left(\begin{array}{l}
		(\exists ac_0 \spot P[ac_0/ac'] \land ac_0 \subseteq \{ s | \lnot Q\})
		\\ \lor \\
		((\exists ac_0 \spot P[ac_0/ac'] \land ac_0 \subseteq \{ s | Q \implies R \}) \land e)
	\end{array}\right)
	&&\ptext{Definition of $\seqA$ and substitution}\\
	&=\left(\begin{array}{l}
		((\exists ac_0 \spot P[ac_0/ac'] \land ac_0 \subseteq ac') \seqA \lnot Q)
		\\ \lor \\
		(((\exists ac_0 \spot P[ac_0/ac'] \land ac_0 \subseteq ac') \seqA (Q \implies R)) \land e)
	\end{array}\right)
	&&\ptext{Assumption: $P$ is $\mathbf{PBMH}$-healthy (\cref{lemma:PBMH:alternative-1})}\\
	&=(P \seqA \lnot Q) \lor ((P \seqA (Q \implies R)) \land e)
\end{xflalign*}
\end{proof}\end{proofs}
\end{lemma}

\section{Closure Properties}

\begin{theorem}\label{law:seqA-closure} Provided $P$ and $Q$ are $\mathbf{PBMH}$-healthy,
\begin{align*}
	&\mathbf{PBMH} (P \seqA Q) = P \seqA Q
\end{align*}
\begin{proofs}\begin{proof}\checkt{alcc}\checkt{pfr} (Implication)
\begin{xflalign*}
	&\mathbf{PBMH} (P \seqA Q)
	&&\ptext{Assumption: $P$ is $\mathbf{PBMH}$-healthy (\cref{lemma:PBMH:alternative-1})}\\
	&=\mathbf{PBMH} ((\exists ac_1 \spot P[ac_1/ac'] \land ac_1 \subseteq ac') \seqA Q)
	&&\ptext{Definition of $\seqA$ and substitution}\\
	&=\mathbf{PBMH} (\exists ac_1 \spot P[ac_1/ac'] \land ac_1 \subseteq \{ s | Q \})
	&&\ptext{Property of sets}\\
	&=\mathbf{PBMH} (\exists ac_1 \spot P[ac_1/ac'] \land (\forall z \spot z \in ac_1 \implies Q[z/s]))
	&&\ptext{Definition of $\mathbf{PBMH}$ (\cref{lemma:PBMH:alternative-1})}\\
	&=\exists ac_0 \spot (\exists ac_1 \spot P[ac_1/ac'] \land (\forall z \spot z \in ac_1 \implies Q[z/s]))[ac_0/ac'] \land ac_0 \subseteq ac'
	&&\ptext{Predicate calculus and substitution}\\
	&=\exists ac_0, ac_1 \spot P[ac_1/ac'] \land (\forall z \spot z \in ac_1 \implies Q[z/s][ac_0/ac']) \land ac_0 \subseteq ac'
	&&\ptext{Predicate calculus: quantifier scope}\\
	&=\exists ac_1 \spot P[ac_1/ac'] \land (\exists ac_0 \spot (\forall z \spot z \in ac_1 \implies Q[z/s][ac_0/ac']) \land ac_0 \subseteq ac')
	&&\ptext{Predicate calculus: quantifier scope}\\
	&=\exists ac_1 \spot P[ac_1/ac'] \land (\exists ac_0 \spot (\forall z \spot (z \in ac_1 \implies Q[z/s][ac_0/ac']) \land ac_0 \subseteq ac'))
	&&\ptext{Predicate calculus}\\
	&=\exists ac_1 \spot \left(\begin{array}{l}
		P[ac_1/ac'] 
		\\ \land \\ 
		\exists ac_0 \spot \left(\forall z \spot \left(\begin{array}{l}
			(z \notin ac_1 \land ac_0 \subseteq ac') 
			\\ \lor \\
			(Q[z/s][ac_0/ac'] \land ac_0 \subseteq ac')
		\end{array}\right)\right)
	\end{array}\right)
	&&\ptext{Predicate calculus}\\
	&\implies \exists ac_1 \spot \left(\begin{array}{l}
		P[ac_1/ac'] 
		\\ \land \\ 
		\left(\forall z \spot \exists ac_0 \spot \left(\begin{array}{l}
			(z \notin ac_1 \land ac_0 \subseteq ac') 
			\\ \lor \\
			(Q[z/s][ac_0/ac'] \land ac_0 \subseteq ac')
		\end{array}\right)\right)
	\end{array}\right)
	&&\ptext{Predicate calculus}\\
	&=\exists ac_1 \spot \left(\begin{array}{l}
		P[ac_1/ac'] 
		\\ \land \\ 
		\left(\forall z \spot \left(\begin{array}{l}
			(z \notin ac_1 \land \exists ac_0 \spot ac_0 \subseteq ac') 
			\\ \lor \\
			(\exists ac_0 \spot Q[z/s][ac_0/ac'] \land ac_0 \subseteq ac')
		\end{array}\right)\right)
	\end{array}\right)
	&&\ptext{Property of sets and predicate calculus}\\
	&=\exists ac_1 \spot \left(\begin{array}{l}
		P[ac_1/ac'] 
		\\ \land \\ 
		\left(\forall z \spot \left(\begin{array}{l}
			(z \in ac_1) 
			\\ \implies \\
			(\exists ac_0 \spot Q[z/s][ac_0/ac'] \land ac_0 \subseteq ac')
		\end{array}\right)\right)
	\end{array}\right)
	&&\ptext{Substitution}\\
	&=\exists ac_1 \spot \left(\begin{array}{l}
		P[ac_1/ac'] 
		\\ \land \\ 
		\left(\forall z \spot \left(\begin{array}{l}
			(z \in ac_1) 
			\\ \implies \\
			(\exists ac_0 \spot Q[ac_0/ac'] \land ac_0 \subseteq ac')[z/s]
		\end{array}\right)\right)
	\end{array}\right)
	&&\ptext{Definition of $\mathbf{PBMH}$ (\cref{lemma:PBMH:alternative-1})}\\
	&=\exists ac_1 \spot P[ac_1/ac'] \land (\forall z \spot z \in ac_1 \implies \mathbf{PBMH} (Q)[z/s])
	&&\ptext{Property of sets}\\
	&=\exists ac_1 \spot P[ac_1/ac'] \land ac_1 \subseteq \{ s | \mathbf{PBMH} (Q) \}
	&&\ptext{Definition of $\seqA$ and substitution}\\
	&=(\exists ac_1 \spot P[ac_1/ac'] \land ac_1 \subseteq ac') \seqA \mathbf{PBMH} (Q)
	&&\ptext{Definition of $\mathbf{PBMH}$ (\cref{lemma:PBMH:alternative-1})}\\
	&=\mathbf{PBMH} (P) \seqA \mathbf{PBMH} (Q)
	&&\ptext{Assumption: $P$ and $Q$ are $\mathbf{PBMH}$-healthy (\cref{lemma:PBMH:alternative-1})}\\
	&=P \seqA Q
\end{xflalign*}
\end{proof}
\begin{proof} (Reverse implication)
\begin{xflalign*}
	&P \seqA Q
	&&\ptext{\cref{lemma:P-implies-PBMH(P)}}\\
	&\implies \mathbf{PBMH} (P \seqA Q)
\end{xflalign*}
\end{proof}\end{proofs}
\end{theorem}

\section{Extreme Points}
\begin{lemma}\label{law:seqA-P-sequence-false:1}
Provided $P$ is $\mathbf{PBMH}$-healthy,
\begin{align*}
	&P \seqA \mathbf{false} = P[\emptyset/ac']&
\end{align*}
\begin{proofs}\begin{proof}\checkt{alcc}\checkt{pfr}
\begin{xflalign*}
	&P \seqA \mathbf{false}
	&&\ptext{Assumption: $P$ is $\mathbf{PBMH}$-healthy}\\
	&=\mathbf{PBMH} (P) \seqA false
	&&\ptext{Definition of $\mathbf{PBMH}$ (\cref{lemma:PBMH:alternative-1})}\\
	&=(\exists ac_0 \spot P[ac_0/ac'] \land ac_0 \subseteq ac') \seqA false
	&&\ptext{Definition of $\seqA$}\\
	&=\exists ac_0 \spot P[ac_0/ac'] \land ac_0 \subseteq \emptyset
	&&\ptext{Property of sets and one-point rule}\\
	&=P[\emptyset/ac']
\end{xflalign*}
\end{proof}\end{proofs}
\end{lemma}
\begin{lemma}\label{law:seqA-P-sequence-true}
Provided $P$ is $\mathbf{PBMH}$-healthy,
\begin{align*}
	&P \seqA \mathbf{true} = \exists ac' \spot P&
\end{align*}
\begin{proofs}\begin{proof}\checkt{pfr}
\begin{xflalign*}
	&P \seqA \mathbf{true}
	&&\ptext{Assumption: $P$ is $\mathbf{PBMH}$-healthy}\\
	&=\mathbf{PBMH} (P) \seqA true
	&&\ptext{\cref{lemma:PBMH:alternative-1}}\\
	&=(\exists ac_0 \spot P[ac_0/ac'] \land ac_0 \subseteq ac') \seqA true
	&&\ptext{Definition of $\seqA$}\\
	&=\exists ac_0 \spot P[ac_0/ac'] \land ac_0 \subseteq \{ z | true \}
	&&\ptext{Property of sets}\\
	&=\exists ac_0 \spot P[ac_0/ac'] \land (\forall z \spot z \in ac_0 \implies true)
	&&\ptext{Propositional calculus}\\
	&=\exists ac_0 \spot P[ac_0/ac']
	&&\ptext{One-point rule}\\
	&=\exists ac_0 \spot (\exists ac' \spot P \land ac' = ac_0)
	&&\ptext{One-point rule: $ac_0$ not free in $P$}\\
	&=\exists ac' \spot P
\end{xflalign*}
\end{proof}\end{proofs}
\end{lemma}
\section[Algebraic Properties and Sequential Composition]{Algebraic Properties \\ and Sequential Composition}
\begin{lemma}\label{law:seqA-left-associative}
Provided $ok$ and $ac$ are not free in $R$,
\begin{align*}
	& (P \circseq Q) \seqA R = P \circseq (Q \seqA R) &
\end{align*}
\begin{proofs}\begin{proof}
\begin{flalign*}
	& (P \circseq Q) \seqA R
	&&\ptext{Definition of sequential composition}\\
	&= (\exists ok_0, ac_0 \spot P[ok_0,ac_0/ok,ac'] \land Q[ok_0,ac_0/ok,ac]) \seqA R
	&&\ptext{Definition of $\seqA$}\\
	&= (\exists ok_0, ac_0 \spot P[ok_0,ac_0/ok,ac'] \land Q[ok_0,ac_0/ok,ac])[\{ z | R[z/s]\}/ac']
	&&\ptext{Substitution: $ac'$ not free in $ac_0$}\\
	&= (\exists ok_0, ac_0 \spot P[ok_0,ac_0/ok,ac'] \land Q[ok_0,ac_0/ok,ac][\{ z | R[z/s]\}/ac'])
	&&\ptext{Assumption: $\{ok,ac\}$ not free in $R$}\\
	&= (\exists ok_0, ac_0 \spot P[ok_0,ac_0/ok,ac'] \land Q[\{ z | R[z/s]\}/ac'][ok_0,ac_0/ok,ac])
	&&\ptext{Definition of sequential composition}\\
	&= P \circseq Q[\{ z | R[z/s]\}/ac']
	&&\ptext{Definition of $\seqA$}\\
	&= P \circseq (Q \seqA R)
\end{flalign*}
\end{proof}\end{proofs}
\end{lemma}

\section{Skip}

\begin{define}
$\IIA \circdef s \in ac'$
\end{define}
\begin{lemma} $\IIA$ is a fixed point of $\mathbf{PBMH}$,
$\mathbf{PBMH} (\IIA) = \IIA$.
\begin{proofs}\begin{proof}\checkt{pfr}
\begin{xflalign*}
	&\mathbf{PBMH} (\IIA)
	&&\ptext{Definition of $\IIA$ and $\mathbf{PBMH}$ (\cref{lemma:PBMH:alternative-1})}\\
	&=\exists ac_0 \spot s \in ac_0 \land ac_0 \subseteq ac'
	&&\ptext{\cref{law:aux:ac-sequence}}\\
	&=s \in ac'
\end{xflalign*}
\end{proof}\end{proofs}
\end{lemma}
\begin{lemma}\label{law:seqA:IIA:left-unit}
$\IIA \seqA P = P$
\begin{proofs}\begin{proof}\checkt{alcc}\checkt{pfr}
\begin{flalign*}
	&\IIA \seqA P
	&&\ptext{Definition of $\IIA$}\\
	&=s \in ac' \seqA P
	&&\ptext{Definition of $\seqA$ and substitution}\\
	&=s \in \{ z | P[z/s] \}
	&&\ptext{Property of sets}\\
	&=P[z/s][s/z]
	&&\ptext{Substitution}\\
	&=P
\end{flalign*}
\end{proof}\end{proofs}
\end{lemma}
\begin{lemma}\label{law:seqA:IIA:right-unit}
Provided $P$ is $\mathbf{PBMH}$-healthy,
$P \seqA \IIA$.
\begin{proofs}\begin{proof}\checkt{pfr}
\begin{xflalign*}
	&P \seqA \IIA
	&&\ptext{Definition of $\IIA$}\\
	&=P \seqA (s \in ac')
	&&\ptext{Assumption: $P$ is $\mathbf{PBMH}$-healthy}\\
	&=\mathbf{PBMH} (P) \seqA (s \in ac')
	&&\ptext{\cref{lemma:PBMH:alternative-1}}\\
	&=(\exists ac_0 \spot P[ac_0/ac'] \land ac_0 \subseteq ac') \seqA (s \in ac')
	&&\ptext{Definition of $\seqA$}\\
	&=\exists ac_0 \spot P[ac_0/ac'] \land ac_0 \subseteq \{ z | z \in ac' \}
	&&\ptext{Property of sets}\\
	&=\exists ac_0 \spot P[ac_0/ac'] \land ac_0 \subseteq ac'
	&&\ptext{\cref{lemma:PBMH:alternative-1}}\\
	&=\mathbf{PBMH} (P)
	&&\ptext{Assumption: $P$ satisfies $\mathbf{PBMH}$}\\
	&=P
\end{xflalign*}
\end{proof}\end{proofs}
\end{lemma}

\chapter{Reactive Angelic Designs ($\mathbf{RAD}$)}\label{appendix:RAD}

\section{$\mathbf{RA1}$}

\subsection{Definition}
\theoremstatementref{def:RA1}

\subsection{Properties}

\begin{theorem}\label{theorem:RA1-o-A0:RA1}
\begin{statement}$\mathbf{RA1} \circ \mathbf{A0} (P) = \mathbf{RA1} (P)$\end{statement}
\begin{proofs}
\begin{proof}\checkt{alcc}
\begin{flalign*}
	&\mathbf{RA1} \circ \mathbf{A0} (P)
	&&\ptext{\cref{lemma:RA1:alternative-1}}\\
	&=\mathbf{A0} (P)[\{z | z \in ac' \land s.tr \le z.tr\}/ac'] \land \exists z \spot s.tr \le z.tr \land z \in ac'
	&&\ptext{Definition of $\mathbf{A0}$}\\
	&=\left(\begin{array}{l}
		(P \land ((ok \land \lnot P^f) \implies (ok' \implies ac'\neq\emptyset)))[\{z | z \in ac' \land s.tr \le z.tr\}/ac'] 
		\\ \land \\
		\exists z \spot s.tr \le z.tr \land z \in ac'
	\end{array}\right)
	&&\ptext{Substitution}\\
	&=\left(\begin{array}{l}
		P[\{z | z \in ac' \land s.tr \le z.tr\}/ac']
		\\ \land \\
		\left(\begin{array}{l}
			(ok \land \lnot P^f[\{z | z \in ac' \land s.tr \le z.tr\}/ac'])
			\\ \implies \\
			(ok' \implies \{z | z \in ac' \land s.tr \le z.tr\}\neq\emptyset)
		\end{array}\right)
		\\ \land \\
		\exists z \spot s.tr \le z.tr \land z \in ac'
	\end{array}\right)
	&&\ptext{Property of sets}\\
	&=\left(\begin{array}{l}
		P[\{z | z \in ac' \land s.tr \le z.tr\}/ac']
		\\ \land \\
		\left(\begin{array}{l}
			(ok \land \lnot P^f[\{z | z \in ac' \land s.tr \le z.tr\}/ac'])
			\\ \implies \\
			(ok' \implies (\exists z \spot z \in ac' \land s.tr \le z.tr))
		\end{array}\right)
		\\ \land \\
		\exists z \spot s.tr \le z.tr \land z \in ac'
	\end{array}\right)	
	&&\ptext{Predicate calculus}\\
	&=\left(\begin{array}{l}
		P[\{z | z \in ac' \land s.tr \le z.tr\}/ac']
		\\ \land \\
		\left(\begin{array}{l}
			(\lnot ok \lor P^f[\{z | z \in ac' \land s.tr \le z.tr\}/ac'])
			\\ \lor \\
			(\lnot ok' \lor (\exists z \spot z \in ac' \land s.tr \le z.tr))
		\end{array}\right)
		\\ \land \\
		\exists z \spot s.tr \le z.tr \land z \in ac'
	\end{array}\right)
	&&\ptext{Predicate calculus: absorption law}\\
	&=\left(\begin{array}{l}
		P[\{z | z \in ac' \land s.tr \le z.tr\}/ac']
		\\ \land \\
		\exists z \spot s.tr \le z.tr \land z \in ac'
	\end{array}\right)	
	&&\ptext{\cref{lemma:RA1:alternative-1}}\\
	&=\mathbf{RA1} (P)
\end{flalign*}
\end{proof}
\end{proofs}
\end{theorem}

\begin{theorem}\label{lemma:RA1(P-land-Q):RA1(P)-land-RA1(Q)}
\begin{statement}$\mathbf{RA1} (P \land Q) = \mathbf{RA1} (P) \land \mathbf{RA1} (Q)$\end{statement}
\begin{proofs}
\begin{proof}\checkt{alcc}
\begin{xflalign*}
	&\mathbf{RA1} (P \land Q)
	&&\ptext{Definition of $\mathbf{RA1}$ (\cref{lemma:RA1:alternative-1})}\\
	&=(P \land Q)[\{z | z \in ac' \land s.tr \le z.tr\}/ac'] \land \exists z \spot s.tr \le z.tr \land z \in ac'
	&&\ptext{Substitution}\\
	&=\left(\begin{array}{l}
		P[\{z | z \in ac' \land s.tr \le z.tr\}/ac'] 
		\\ \land \\ 
		Q[\{z | z \in ac' \land s.tr \le z.tr\}/ac']
		\\ \land \\
		\exists z \spot s.tr \le z.tr \land z \in ac'
	\end{array}\right)
	&&\ptext{Predicate calculus}\\
	&=\left(\begin{array}{l}
		(P[\{z | z \in ac' \land s.tr \le z.tr\}/ac'] \land \exists z \spot s.tr \le z.tr \land z \in ac') 
		\\ \land \\ 
		(Q[\{z | z \in ac' \land s.tr \le z.tr\}/ac'] \land \exists z \spot s.tr \le z.tr \land z \in ac')		
	\end{array}\right)
	&&\ptext{Definition of $\mathbf{RA1}$ (\cref{lemma:RA1:alternative-1})}\\
	&=\mathbf{RA1} (P) \land \mathbf{RA1} (Q)
\end{xflalign*}
\end{proof}
\end{proofs}
\end{theorem}

\begin{theorem}\label{lemma:RA1(P-lor-Q):RA1(P)-lor-RA1(Q)}
\begin{statement}$\mathbf{RA1} (P \lor Q) = \mathbf{RA1} (P) \lor \mathbf{RA1} (Q)$\end{statement}
\begin{proofs}
\begin{proof}\checkt{alcc}
\begin{flalign*}
	&\mathbf{RA1} (P \lor Q)
	&&\ptext{Definition of $\mathbf{RA1}$ (\cref{lemma:RA1:alternative-1})}\\
	&=(P \lor Q)[\{z | z \in ac' \land s.tr \le z.tr\}/ac'] \land \exists z \spot s.tr \le z.tr \land z \in ac'
	&&\ptext{Substitution}\\
	&=\left(\begin{array}{l}
		(P[\{z | z \in ac' \land s.tr \le z.tr\}/ac'] \lor Q[\{z | z \in ac' \land s.tr \le z.tr\}/ac'])
	\\ \land \\
	\exists z \spot s.tr \le z.tr \land z \in ac'
	\end{array}\right)
	&&\ptext{Predicate calculus}\\
	&=\left(\begin{array}{l}
		(P[\{z | z \in ac' \land s.tr \le z.tr\}/ac'] \land \exists z \spot s.tr \le z.tr \land z \in ac')
		\\ \lor \\
		(Q[\{z | z \in ac' \land s.tr \le z.tr\}/ac'] \land \exists z \spot s.tr \le z.tr \land z \in ac')
	\end{array}\right) 	
	&&\ptext{Definition of $\mathbf{RA1}$ (\cref{lemma:RA1:alternative-1})}\\
	&=\mathbf{RA1} (P) \lor \mathbf{RA1} (Q)
\end{flalign*}
\end{proof}
\end{proofs}
\end{theorem}

\begin{theorem}\label{theorem:RA1(P-seqA-Q):closure}
\begin{statement}
Provided $P$ and $Q$ are $\mathbf{RA1}$-healthy and $Q$ is $\mathbf{PBMH}$-healthy,
\begin{align*}
	&\mathbf{RA1} (P \seqA Q) = P \seqA Q
\end{align*}
\end{statement}
\begin{proofs}
\begin{proof}\checkt{alcc}\checkt{pfr}
\begin{xflalign*}
	&P \seqA Q
	&&\ptext{Assumption: $P$ and $Q$ are $\mathbf{RA1}$-healthy}\\
	&=\mathbf{RA1} (P) \seqA \mathbf{RA1} (Q)
	&&\ptext{Definition of $\mathbf{RA1}$}\\
	&=(P[States_{tr\le tr'} (s)\cap ac'/ac'] \land \mathbf{RA1} (true)) \seqA \mathbf{RA1} (Q)
	&&\ptext{Right-distributivity of $\seqA$ (\cref{law:seqA-right-distributivity-conjunction})}\\
	&=\left(
\begin{array}{l}
		(P[States_{tr\le tr'} (s)\cap ac'/ac'] \seqA \mathbf{RA1} (Q))
		\\ \land \\
		(\mathbf{RA1} (true) \seqA \mathbf{RA1} (Q)) 
	\end{array}%
\right)
	&&\ptext{Property of sets}\\
	&=\left(%
\right)
	&&\ptext{Property of sets and definition of $States_{tr\le tr'}$}\\
	&=\left(%
\right)
	&&\ptext{Property of sets and substitution}\\
	&=\left(%
\right)
	&&\ptext{Property of sets and definition of $\mathbf{PBMH}$}\\
	&=\mathbf{RA1} \left(%
\right)[z/s]\right. \right\}\end{array}\right/ac'\right] 		
		\end{array}\right)
	&&\ptext{Assumption: $Q$ is $\mathbf{PBMH}$-healthy and~\cref{lemma:RA1(P)-P-is-PBMH}}\\
	&=\mathbf{RA1} ((P \land ac'\neq\emptyset)[States_{tr\le tr'} (s) \cap	\{ z | \mathbf{RA1} (Q)[z/s] \}/ac'])
	&&\ptext{Definition of $\seqA$ and substitution}\\
	&=\mathbf{RA1} ((P \land ac'\neq\emptyset)[States_{tr\le tr'} (s) \cap ac'/ac'] \seqA \mathbf{RA1} (Q))
	&&\ptext{Definition of $\mathbf{RA1}$}\\
	&=\mathbf{RA1} (\mathbf{RA1} (P) \seqA \mathbf{RA1} (Q))
	&&\ptext{Assumption: $P$ and $Q$ are $\mathbf{RA1}$-healthy}\\
	&=\mathbf{RA1} (P \seqA Q)	
\end{xflalign*}
\end{proof}
\end{proofs}
\end{theorem}

\begin{theorem}\label{theorem:PBMH-o-RA1(P):RA1(P)}
\begin{statement}$\mathbf{PBMH} \circ \mathbf{RA1} \circ \mathbf{PBMH} (P) = \mathbf{RA1} \circ \mathbf{PBMH} (P)$\end{statement}
\begin{proofs}
\begin{proof}\checkt{alcc}
\begin{xflalign*}
	&\mathbf{PBMH} \circ \mathbf{RA1} \circ \mathbf{PBMH} (P)
	&&\ptext{\cref{lemma:RA1:alternative-1}}\\
	&=\mathbf{PBMH}\left(
\right)
	&&\ptext{Property of sets}\\
	&=\mathbf{PBMH}\left(%
\right)
	&&\ptext{Property of sets}\\
	&=\left(%
\right)
	&&\ptext{\cref{lemma:PBMH:alternative-1}}\\
	&=\mathbf{PBMH} (P)[\{ z | z \in ac' \land s.tr \le z.tr\}/ac'] \land \exists z \spot s.tr \le z.tr \land z \in ac'
	&&\ptext{\cref{lemma:RA1:alternative-1}}\\
	&=\mathbf{RA1} \circ \mathbf{PBMH} (P)
\end{xflalign*}
\end{proof}
\end{proofs}
\end{theorem}

\begin{theorem}\label{theorem:RA1-idempotent}
\begin{statement}$\mathbf{RA1} \circ \mathbf{RA1} (P) = \mathbf{RA1} (P)$\end{statement}
\begin{proofs}
\begin{proof}\checkt{alcc}
\begin{flalign*}
	&\mathbf{RA1} \circ \mathbf{RA1} (P)
	&&\ptext{Definition of $\mathbf{RA1}$ (\cref{lemma:RA1:alternative-1})}\\
	&=\mathbf{RA1} (P)[\{z | z \in ac' \land s.tr \le z.tr \}/ac'] \land \exists z \spot z \in ac' \land s.tr \le z.tr
	&&\ptext{Definition of $\mathbf{RA1}$ (\cref{lemma:RA1:alternative-1})}\\
	&=\left(\begin{array}{l}
		\left(\begin{array}{l}
		P[\{z | z \in ac' \land s.tr \le z.tr \}/ac'] 
		\\ \land \\
		\exists z \spot z \in ac' \land s.tr \le z.tr
	\end{array}\right)[\{z | z \in ac' \land s.tr \le z.tr \}/ac'] 
	\\ \land \exists z \spot z \in ac' \land s.tr \le z.tr
	\end{array}\right)
	&&\ptext{Substitution}\\
	&=\left(\begin{array}{l}
		\left(\begin{array}{l}
		P[\{z | z \in \{z | z \in ac' \land s.tr \le z.tr \} \land s.tr \le z.tr \}/ac'] 
		\\ \land \\
		\exists z \spot z \in \{z | z \in ac' \land s.tr \le z.tr \} \land s.tr \le z.tr
	\end{array}\right)
	\\ \land \exists z \spot z \in ac' \land s.tr \le z.tr
	\end{array}\right)
	&&\ptext{Variable renaming}\\
	&=\left(\begin{array}{l}
		\left(\begin{array}{l}
		P[\{z | z \in \{y | y \in ac' \land s.tr \le y.tr \} \land s.tr \le z.tr \}/ac'] 
		\\ \land \\
		\exists z \spot z \in \{y | y \in ac' \land s.tr \le y.tr \} \land s.tr \le z.tr
	\end{array}\right)
	\\ \land \exists z \spot z \in ac' \land s.tr \le z.tr
	\end{array}\right)
	&&\ptext{Property of sets}\\
	&=\left(\begin{array}{l}
		P[\{z | z \in ac' \land s.tr \le z.tr \land s.tr \le z.tr \}/ac'] 
		\\ \land \\
		\exists z \spot z \in ac' \land s.tr \le z.tr \land s.tr \le z.tr
		\\ \land \\
		\exists z \spot z \in ac' \land s.tr \le z.tr
	\end{array}\right)
	&&\ptext{Predicate calculus}\\
	&=\left(\begin{array}{l}
		P[\{z | z \in ac' \land s.tr \le z.tr \}/ac'] 
		\\ \land \\
		\exists z \spot z \in ac' \land s.tr \le z.tr
	\end{array}\right)
	&&\ptext{Definition of $\mathbf{RA1}$ (\cref{lemma:RA1:alternative-1})}\\
	&=\mathbf{RA1} (P)
\end{flalign*}
\end{proof}
\end{proofs}
\end{theorem}

\begin{theorem}\label{theorem:RA1-monotonic}
\begin{statement}$P \sqsubseteq Q \implies \mathbf{RA1} (P) \sqsubseteq \mathbf{RA1} (Q)$\end{statement}
\begin{proofs}
\begin{proof}\checkt{alcc}
\begin{xflalign*}
	&\mathbf{RA1} (Q)
	&&\ptext{Assumption: $P \sqsubseteq Q = [Q \implies P]$}\\
	&=\mathbf{RA1} (Q \land P)
	&&\ptext{\cref{lemma:RA1(P-land-Q):RA1(P)-land-RA1(Q)}}\\
	&=\mathbf{RA1} (Q) \land \mathbf{RA1} (P)
	&&\ptext{Predicate calculus}\\
	&\implies \mathbf{RA1} (P)
\end{xflalign*}
\end{proof}
\end{proofs}
\end{theorem}

\subsection{Lemmas}

\begin{lemma}\label{lemma:RA1:alternative-1}
\checkt{alcc}
\begin{align*}
	&\mathbf{RA1} (P) \\
	&=\\
	&P[\{z | z \in ac' \land s.tr \le z.tr\}/ac'] \land \exists z \spot s.tr \le z.tr \land z \in ac'
\end{align*}
\begin{proofs}\begin{proof}
\begin{flalign*}
	&\mathbf{RA1} (P)
	&&\ptext{Definition of $\mathbf{RA1}$ (\cref{def:RA1})}\\
	&=(P \land ac'\neq\emptyset)[States_{tr\le tr'} (s) \cap ac'/ac']
	&&\ptext{Property of sets and definition of $States_{tr\le tr'}$}\\
	&=(P \land ac'\neq\emptyset)[\{z | z \in ac' \land s.tr \le z.tr \}/ac']
	&&\ptext{Substitution}\\
	&=P[\{z | z \in ac' \land s.tr \le z.tr \}/ac'] \land \{z | z \in ac' \land s.tr \le z.tr \}\neq\emptyset
	&&\ptext{Property of sets}\\
	&=P[\{z | z \in ac' \land s.tr \le z.tr \}/ac'] \land \exists z \spot z \in \{z | z \in ac' \land s.tr \le z.tr \}
	&&\ptext{Property of sets}\\
	&=P[\{z | z \in ac' \land s.tr \le z.tr \}/ac'] \land \exists z \spot z \in ac' \land s.tr \le z.tr
\end{flalign*}
\end{proof}\end{proofs}
\end{lemma}

\begin{lemma}\label{lemma:RA1:alternative-2}
\checkt{alcc}
\begin{align*}
	&\mathbf{RA1} (P) = (P \land ac'\neq\emptyset)[\{z | z \in ac' \land z \in \{ z | s.tr \le z.tr \}\}/ac']
\end{align*}
\begin{proofs}\begin{proof}
\begin{flalign*}
	&\mathbf{RA1} (P)
	&&\ptext{Definition of $\mathbf{RA1}$ (\cref{def:RA1})}\\
	&=(P \land ac'\neq\emptyset)[States_{tr\le tr'} (s)\cap ac'/ac']
	&&\ptext{Property of sets and definition of $States_{tr\le tr'}$}\\
	&=(P \land ac'\neq\emptyset)[\{z | z \in ac' \land z \in \{ z | s.tr \le z.tr \}\}/ac']
\end{flalign*}
\end{proof}\end{proofs}
\end{lemma}\noindent

\begin{lemma}\label{lemma:RA1(P)-ac'-emptyset:false}
\begin{statement}
$\mathbf{RA1} (P)[\emptyset/ac'] = false$
\end{statement}
\begin{proofs}
\begin{proof}\checkt{alcc}
\begin{xflalign*}
	&\mathbf{RA1} (P)[\emptyset/ac']
	&&\ptext{Definition of $\mathbf{RA1}$}\\
	&=(P \land ac'\neq\emptyset)[\{z | s.tr \le z.tr\} \cap ac'/ac'][\emptyset/ac']
	&&\ptext{Substitution}\\
	&=(P \land ac'\neq\emptyset)[\{z | s.tr \le z.tr\} \cap \emptyset/ac']
	&&\ptext{Property of sets}\\
	&=(P \land ac'\neq\emptyset)[\emptyset/ac']
	&&\ptext{Substitution}\\
	&=P[\emptyset/ac'] \land \emptyset\neq\emptyset
	&&\ptext{Predicate calculus}\\
	&=false
\end{xflalign*}
\end{proof}
\end{proofs}
\end{lemma}

\begin{lemma}\label{lemma:RA1(true)-y-for-ac':str-le-ytr}
\begin{statement}
$\mathbf{RA1} (true)[\{y\}/ac'] = s.tr \le y.tr$
\end{statement}
\begin{proofs}
\begin{proof}\checkt{alcc}
\begin{xflalign*}
	&\mathbf{RA1} (true)[\{y\}/ac']
	&&\ptext{\cref{lemma:RA1(true)}}\\
	&=(\exists z \spot s.tr \le z.tr \land z \in ac')[\{y\}/ac']
	&&\ptext{Substitution}\\
	&=\exists z \spot s.tr \le z.tr \land z \in \{y\}
	&&\ptext{Property of sets}\\
	&=\exists z \spot s.tr \le z.tr \land z = y
	&&\ptext{One-point rule}\\
	&=s.tr \le y.tr
\end{xflalign*}
\end{proof}
\end{proofs}
\end{lemma}

\begin{lemma}\label{lemma:RA1(exists-y-P-y-for-ac'-y-in-ac')}
\begin{statement}
Provided $y$ is not $s$ and not $ac'$,
\begin{align*}
	&\mathbf{RA1} (\exists y @ P[\{y\}/ac'] \land y \in ac')\\ 
	&=\\
	&\exists y @ P[\{y\}/ac'] \land s.tr \le y.tr \land y \in ac'
\end{align*}
\end{statement}
\begin{proofs}
\begin{proof}\checkt{alcc}
\begin{xflalign*}
	&\mathbf{RA1} (\exists y @ P[\{y\}/ac'] \land y \in ac')
	&&\ptext{\cref{lemma:RA1(exists-P):exists-RA1(P)}}\\
	&=\exists y @ \mathbf{RA1}(P[\{y\}/ac'] \land y \in ac')
	&&\ptext{\cref{lemma:RA1(P-land-Q):ac'-not-free}}\\
	&=\exists y @ P[\{y\}/ac'] \land \mathbf{RA1} (y \in ac')
	&&\ptext{\cref{lemma:RA1(x-in-ac'):x-in-ac'-land-prefix-tr}}\\
	&=\exists y @ P[\{y\}/ac'] \land s.tr \le y.tr \land y \in ac'
\end{xflalign*}
\end{proof}
\end{proofs}
\end{lemma}

\begin{lemma}\label{lemma:RA1-implies-ac'-neq-emptyset}
\checkt{alcc}
\checkt{pfr}
$\mathbf{RA1} (P) \implies ac'\neq\emptyset$
\begin{proofs}\begin{proof}
\begin{xflalign*}
	&\mathbf{RA1} (P)
	&&\ptext{Definition of $\mathbf{RA1}$ (\cref{lemma:RA1:alternative-1})}\\
	&=P[\{z | z \in ac' \land s.tr \le z.tr\}/ac'] \land \exists z \spot s.tr \le z.tr \land z \in ac'
	&&\ptext{Predicate calculus}\\
	&\implies \exists z \spot s.tr \le z.tr \land z \in ac'
	&&\ptext{Predicate calculus}\\
	&\implies \exists z \spot z \in ac'
	&&\ptext{Property of sets}\\
	&=ac'\neq\emptyset
\end{xflalign*}
\end{proof}\end{proofs}
\end{lemma}

\begin{lemma}\label{lemma:s-in-ac'-implies-RA1}
\checkt{alcc}
\checkt{pfr}
$s \in ac' \implies \exists z \spot s.tr \le z.tr \land z \in ac'$
\begin{proofs}\begin{proof}
\begin{flalign*}
	&s \in ac'
	&&\ptext{Property of sequences}\\
	&=s.tr \le s.tr \land s \in ac'
	&&\ptext{Predicate calculus}\\
	&\implies \exists z \spot s.tr \le z.tr \land s \in ac'
\end{flalign*}
\end{proof}\end{proofs}
\end{lemma}

\begin{lemma}\label{lemma:x-oplus-tr-in-ac}
\checkt{alcc}
\checkt{pfr}
\begin{align*}
	&\exists z \spot z \in ac' \land tr_0 \le z.tr \land x = z \oplus \{ tr \mapsto z.tr - tr_0 \} \\
	&\iff \\
	&x \oplus \{ tr \mapsto tr_0 \cat x.tr \} \in ac'
\end{align*}
\begin{proofs}\begin{proof}
\begin{flalign*}
	&\exists z \spot z \in ac' \land tr_0 \le z.tr \land x = z \oplus \{ tr \mapsto z.tr - tr_0 \}
	&&\ptext{Definition of $\oplus$}\\
	&\iff \exists z \spot z \in ac' \land tr_0 \le z.tr \land x = \{ tr \} \ndres z \cup \{ tr \mapsto z.tr - tr_0 \}
	&&\ptext{Property of relations}\\
	&\iff \left(\begin{array}{l}
		\exists z \spot z \in ac' \land tr_0 \le z.tr \land \{ tr \} \ndres x = \{ tr \} \ndres z 
		\\ \land x.tr = z.tr - tr_0 \land \dom x = \dom z \cup \{tr\}
	\end{array}\right)
	&&\ptext{Property of sequences}\\
	&\iff \left(\begin{array}{l}
		\exists z \spot z \in ac' \land tr_0 \le z.tr \land \{ tr \} \ndres x = \{ tr \} \ndres z 
		\\ \land tr_0 \cat x.tr = z.tr \land \dom x = \dom z \cup \{tr\}
	\end{array}\right)
	&&\ptext{Property of relations}\\
	&\iff \exists z \spot z \in ac' \land tr_0 \le z.tr \land z = \{ tr \} \ndres x \cup \{ tr \mapsto tr_0 \cat x.tr \}
	&&\ptext{Definition of $\oplus$}\\
	&\iff \exists z \spot z \in ac' \land tr_0 \le z.tr \land z = x \oplus \{ tr \mapsto tr_0 \cat x.tr \}
	&&\ptext{One-point rule}\\
	&\iff x \oplus \{ tr \mapsto tr_0 \cat x.tr \} \in ac' \land tr_0 \le (x \oplus \{ tr \mapsto tr_0 \cat x.tr \}).tr
	&&\ptext{Property of $\oplus$ and value of $tr$}\\
	&\iff x \oplus \{ tr \mapsto tr_0 \cat x.tr \} \in ac' \land tr_0 \le tr_0 \cat x.tr
	&&\ptext{Property of sequence}\\
	&\iff x \oplus \{ tr \mapsto tr_0 \cat x.tr \} \in ac'
\end{flalign*}
 \end{proof}\end{proofs}
\end{lemma}

\begin{lemma}\label{lemma:RA1(false)}
$\mathbf{RA1} (false) = false$
\begin{proofs}\begin{proof}\checkt{pfr}\checkt{alcc}
\begin{flalign*}
	&\mathbf{RA1} (false)
	&&\ptext{Definition of $\mathbf{RA1}$}\\
	&=(false \land ac'\neq\emptyset)[States_{tr\le tr'}(s)\cap ac'/ac']
	&&\ptext{Predicate calculus}\\
	&=false[States_{tr\le tr'}(s)\cap ac'/ac']
	&&\ptext{Substitution}\\
	&=false
\end{flalign*}
\end{proof}\end{proofs}
\end{lemma}

\begin{lemma}\label{lemma:RA1(true)}
\begin{align*}
	&\mathbf{RA1} (true) = \exists z \spot s.tr \le z.tr \land z \in ac'
\end{align*}
\begin{proofs}\begin{proof}\checkt{pfr}\checkt{alcc}
\begin{xflalign*}
	&\mathbf{RA1} (true)
	&&\ptext{Definition of $\mathbf{RA1}$ (\cref{lemma:RA1:alternative-1})}\\
	&=true[\{z | z \in ac' \land s.tr \le z.tr\}/ac'] \land \exists z \spot s.tr \le z.tr \land z \in ac'
	&&\ptext{Property of substitution}\\
	&=\exists z \spot s.tr \le z.tr \land z \in ac'
\end{xflalign*}
\end{proof}\end{proofs}
\end{lemma}

\begin{lemma}\label{lemma:RA1(true):alternative-1}
\begin{align*}
	&\mathbf{RA1} (true) = States_{tr\le tr'} (s) \cap ac' \neq\emptyset
\end{align*}
\begin{proofs}\begin{proof}
\begin{xflalign*}
	&\mathbf{RA1} (true)
	&&\ptext{Definition of $\mathbf{RA1}$}\\
	&=(true \land ac'\neq\emptyset)[States_{tr\le tr'} (s) \cap ac'/ac']
	&&\ptext{Predicate calculus}\\
	&=(ac'\neq\emptyset)[States_{tr\le tr'} (s) \cap ac'/ac']
	&&\ptext{Substitution}\\
	&=States_{tr\le tr'} (s) \cap ac'\neq\emptyset
\end{xflalign*}
\end{proof}\end{proofs}
\end{lemma}

\begin{lemma}\label{lemma:RA1(exists-P):exists-RA1(P)} Provided $x$ is not in the set $\{s, ac'\}$,
\begin{align*}
	&\mathbf{RA1} (\exists x \spot P) = \exists x \spot \mathbf{RA1} (P)
\end{align*}
\begin{proofs}\begin{proof}\checkt{alcc}\checkt{pfr}
\begin{xflalign*}
	&\mathbf{RA1} (\exists x \spot P)
	&&\ptext{Definition of $\mathbf{RA1}$}\\
	&=((\exists x \spot P) \land ac'\neq\emptyset)[\{z | z \in ac' \land s.tr \le z.tr\}/ac']
	&&\ptext{Assumption: $x$ is not $ac'$ and predicate calculus}\\
	&=(\exists x \spot P \land ac'\neq\emptyset)[\{z | z \in ac' \land s.tr \le z.tr\}/ac']
	&&\ptext{Assumption: $x$ is not $s$ and predicate calculus}\\
	&=\exists x \spot (P \land (ac'\neq\emptyset)[\{z | z \in ac' \land s.tr \le z.tr\}/ac'])
	&&\ptext{Definition of $\mathbf{RA1}$}\\
	&=\exists x \spot \mathbf{RA1} (P)
\end{xflalign*}
\end{proof}\end{proofs}
\end{lemma}

\begin{lemma}\label{lemma:RA1(x-in-ac'):x-in-ac'-land-prefix-tr}
$\mathbf{RA1} (x \in ac') = s.tr \le x.tr \land x \in ac'$
\begin{proofs}\begin{proof}\checkt{alcc}\checkt{pfr}
\begin{flalign*}
	&\mathbf{RA1} (x \in ac')
	&&\ptext{Definition of $\mathbf{RA1}$ (\cref{lemma:RA1:alternative-1})}\\
	&=(x \in ac')[\{z | z \in ac' \land s.tr \le z.tr\}/ac'] \land \exists z \spot s.tr \le z.tr \land z \in ac'
	&&\ptext{Substitution}\\
	&=x \in \{z | z \in ac' \land s.tr \le z.tr\} \land \exists z \spot s.tr \le z.tr \land z \in ac
	&&\ptext{Property of sets}\\
	&=x \in ac' \land s.tr \le x.tr \land \exists z \spot s.tr \le z.tr \land z \in ac'
	&&\ptext{Predicate calculus}\\
	&=x \in ac' \land s.tr \le x.tr
\end{flalign*}
\end{proof}\end{proofs}
\end{lemma}

\begin{lemma}\label{lemma:RA1(s-in-ac'):s-in-ac'}
\begin{align*}
	&\mathbf{RA1} (s \in ac') = s \in ac'
\end{align*}
\begin{proofs}\begin{proof}\checkt{alcc}\checkt{pfr}
\begin{flalign*}
	&\mathbf{RA1} (s \in ac')
	&&\ptext{\cref{lemma:RA1(x-in-ac'):x-in-ac'-land-prefix-tr}}\\
	&=s.tr \le s.tr \land s \in ac'
	&&\ptext{Property of sequences}\\
	&=s \in ac'
\end{flalign*}
\end{proof}\end{proofs}
\end{lemma}

\begin{lemma}\label{lemma:RA1(conditional)} Provided $c$ is a condition,
\checkt{alcc}
\checkt{pfr}
\begin{align*}
	&\mathbf{RA1} (P \dres c \rres Q) = \mathbf{RA1} (P) \dres c \rres \mathbf{RA1} (Q)
\end{align*}
\begin{proofs}\begin{proof}
\begin{flalign*}
	&\mathbf{RA1} (P \dres c \rres Q)
	&&\ptext{Definition of conditional}\\
	&=\mathbf{RA1} ((c \land P) \lor (\lnot c \land Q))
	&&\ptext{\cref{lemma:RA1(P-lor-Q):RA1(P)-lor-RA1(Q)}}\\
	&=\mathbf{RA1} (c \land P) \lor \mathbf{RA1} (\lnot c \land Q)
	&&\ptext{Assumption: $c$ is a condition and \cref{lemma:RA1(P-land-Q):ac'-not-free}}\\
	&=(c \land \mathbf{RA1} (P)) \lor (\lnot c \land \mathbf{RA1} (Q))
	&&\ptext{Definition of conditional}\\
	&=\mathbf{RA1} (P) \dres c \rres \mathbf{RA1} (Q)
\end{flalign*}
\end{proof}\end{proofs}
\end{lemma}

\begin{lemma}\label{lemma:RA1(P-land-Q):ac'-not-free} Provided $ac'$ is not free in $P$,
\begin{align*}
	&\mathbf{RA1} (P \land Q) = P \land \mathbf{RA1} (Q)
\end{align*}
\begin{proofs}\begin{proof}\checkt{pfr}\checkt{alcc}
\begin{flalign*}
	&\mathbf{RA1} (P \land Q)
	&&\ptext{Definition of $\mathbf{RA1}$ (\cref{lemma:RA1:alternative-1})}\\
	&=(P \land Q)[\{z | z \in ac' \land s.tr \le z.tr\}/ac'] \land \exists z \spot s.tr \le z.tr \land z \in ac'
	&&\ptext{Substitution: $ac'$ not free in $P$}\\
	&=P \land Q[\{z | z \in ac' \land s.tr \le z.tr\}/ac'] \land \exists z \spot s.tr \le z.tr \land z \in ac'
	&&\ptext{Definition of $\mathbf{RA1}$ (\cref{lemma:RA1:alternative-1})}\\
	&=P \land \mathbf{RA1} (Q)
\end{flalign*}
\end{proof}\end{proofs}
\end{lemma}

\begin{lemma}\label{lemma:RA1(lnot-ok):lnot-ok-land-RA1(true)}
$\mathbf{RA1} (\lnot ok) = \lnot ok \land \mathbf{RA1} (true)$
\begin{proofs}\begin{proof}
\begin{xflalign*}
	&\mathbf{RA1} (\lnot ok)
	&&\ptext{Definition of $\mathbf{RA1}$ (\cref{lemma:RA1:alternative-1})}\\
	&=(\lnot ok)[\{z | z \in ac' \land s.tr \le z.tr\}/ac'] \land \exists z \spot s.tr \le z.tr \land z \in ac'
	&&\ptext{Substitution}\\
	&=\lnot ok \land \exists z \spot s.tr \le z.tr \land z \in ac'
	&&\ptext{\cref{lemma:RA1(true)}}\\
	&=\lnot ok \land \mathbf{RA1} (true)
\end{xflalign*}
\end{proof}\end{proofs}
\end{lemma}

\begin{lemma}\label{lemma:RA1(Pff|-Ptf):RA1(Pff-ac'-neq-emptyset|-Ptf-ac'-neq-emptyset)}
\begin{align*}
	&\mathbf{RA1} (\lnot P^f_f \vdash P^t_f) = \mathbf{RA1} (\lnot (P^f_f \land ac'\neq\emptyset) \vdash P^t_f \land ac'\neq\emptyset)
\end{align*}
\begin{proofs}\begin{proof}\checkt{pfr}\checkt{alcc}
\begin{xflalign*}
	&\mathbf{RA1} (\lnot P^f_f \vdash P^t_f)
	&&\ptext{Definition of $\mathbf{RA1}$}\\
	&=((\lnot P^f_f \vdash P^t_f) \land ac'\neq\emptyset)[States_{tr\le tr'}(s)\cap ac'/ac']
	&&\ptext{Definition of design}\\
	&=(((ok \land \lnot P^f_f) \implies (P^t_f \land ok')) \land ac'\neq\emptyset)[States_{tr\le tr'}(s)\cap ac'/ac']
	&&\ptext{Predicate calculus}\\
	&=((\lnot ok \lor P^f_f \lor (P^t_f \land ok')) \land ac'\neq\emptyset)[States_{tr\le tr'}(s)\cap ac'/ac']
	&&\ptext{Predicate calculus}\\
	&=\left(\left(\begin{array}{l}
			\lnot ok \lor (P^f_f \land ac'\neq\emptyset) 
			\\ \lor \\
			(P^t_f \land ac'\neq\emptyset \land ok')			
		\end{array}\right) 
		\land ac'\neq\emptyset\right)[States_{tr\le tr'}(s)\cap ac'/ac']
	&&\ptext{Predicate calculus}\\
	&=\left(\left(\begin{array}{l}
		(ok \land \lnot (P^f_f \land ac'\neq\emptyset)) 
		\\ \implies \\
		(P^t_f \land ac'\neq\emptyset \land ok')
	\end{array}\right) \land ac'\neq\emptyset\right)[States_{tr\le tr'}(s)\cap ac'/ac']
	&&\ptext{Definition of design}\\
	&=((\lnot (P^f_f \land ac'\neq\emptyset) \vdash P^t_f \land ac'\neq\emptyset) \land ac'\neq\emptyset)[States_{tr\le tr'}(s)\cap ac'/ac']
	&&\ptext{Definition of $\mathbf{RA1}$}\\
	&=\mathbf{RA1} (\lnot (P^f_f \land ac'\neq\emptyset) \vdash P^t_f \land ac'\neq\emptyset)
\end{xflalign*}
\end{proof}\end{proofs}
\end{lemma}

\begin{lemma}\label{lemma:RA1(P)-ac'-not-free:P-land-RA1(true)} Provided $ac'$ is not free in $P$,
\begin{align*}
	&\mathbf{RA1} (P) = P \land \mathbf{RA1} (true)
\end{align*}
\begin{proofs}\begin{proof}\checkt{alcc}
\begin{xflalign*}
	&\mathbf{RA1} (P)
	&&\ptext{Predicate calculus}\\
	&=\mathbf{RA1} (P \land true)
	&&\ptext{Assumption: $ac'$ not free in $P$ and~\cref{lemma:RA1(P-land-Q):ac'-not-free}}\\
	&=P \land \mathbf{RA1} (true)
\end{xflalign*}
\end{proof}\end{proofs}
\end{lemma}

\begin{lemma}\label{lemma:RA1(P|-Q):RA1(P|-RA1(Q))}
$\mathbf{RA1} (P \vdash Q) = \mathbf{RA1} (P \vdash \mathbf{RA1} (Q))$
\begin{proofs}\begin{proof}\checkt{alcc}\checkt{pfr}
\begin{xflalign*}
	&\mathbf{RA1} (P \vdash Q)
	&&\ptext{Definition of design}\\
	&=\mathbf{RA1} ((ok \land P) \implies (Q \land ok'))
	&&\ptext{Predicate calculus}\\
	&=\mathbf{RA1} (\lnot ok \lor \lnot P \lor (Q \land ok'))
	&&\ptext{\cref{lemma:RA1(P-lor-Q):RA1(P)-lor-RA1(Q),theorem:RA1-idempotent}}\\
	&=\mathbf{RA1} (\lnot ok \lor \lnot P \lor \mathbf{RA1} (Q \land ok'))
	&&\ptext{\cref{lemma:RA1(P-land-Q):ac'-not-free}}\\
	&=\mathbf{RA1} (\lnot ok \lor \lnot P \lor (\mathbf{RA1} (Q) \land ok'))
	&&\ptext{Predicate calculus}\\
	&=\mathbf{RA1} ((ok \land P) \implies (\mathbf{RA1} (Q) \land ok'))
	&&\ptext{Definition of design}\\
	&=\mathbf{RA1} (P \vdash \mathbf{RA1} (Q))
\end{xflalign*}
\end{proof}\end{proofs}
\end{lemma}

\begin{lemma}\label{lemma:RA1(P):implies:P} Provided $P$ is $\mathbf{PBMH}$-healthy,
\begin{align*}
	&\mathbf{RA1} (P) \implies P
\end{align*}
\begin{proofs}\begin{proof}
\begin{xflalign*}
	&\mathbf{RA1} (P)
	&&\ptext{Definition of $\mathbf{RA1}$ and~\cref{lemma:RA1(true)}}\\
	&=P[States_{tr\le tr'} (s) \cap ac'/ac'] \land \mathbf{RA1} (true)
	&&\ptext{Predicate calculus}\\
	&\implies P[States_{tr\le tr'} (s) \cap ac'/ac']
	&&\ptext{Assumption: $P$ is $\mathbf{PBMH}$-healthy (\cref{lemma:PBMH:alternative-1})}\\
	&=(\exists ac_0 \spot P[ac_0/ac'] \land ac_0 \subseteq ac')[States_{tr\le tr'} (s) \cap ac'/ac']
	&&\ptext{Substitution}\\
	&=\exists ac_0 \spot P[ac_0/ac'] \land ac_0 \subseteq States_{tr\le tr'} (s) \cap ac'
	&&\ptext{Property of sets}\\
	&=\exists ac_0 \spot P[ac_0/ac'] \land ac_0 \subseteq States_{tr\le tr'} (s) \land ac_0 \subseteq ac'
	&&\ptext{Predicate calculus}\\
	&\implies \exists ac_0 \spot P[ac_0/ac'] \land ac_0 \subseteq ac'
	&&\ptext{\cref{lemma:PBMH:alternative-1}}\\
	&=\mathbf{PBMH} (P)
	&&\ptext{Assumption: $P$ is $\mathbf{PBMH}$-healthy}\\
	&=P
\end{xflalign*}
\end{proof}\end{proofs}
\end{lemma}

\begin{lemma}\label{lemma:RA1(ac'-neq-emptyset):RA1(true)}
$\mathbf{RA1} (ac'\neq\emptyset) = \mathbf{RA1} (true)$
\begin{proofs}\begin{proof}
\begin{xflalign*}
	&\mathbf{RA1} (ac'\neq\emptyset)
	&&\ptext{Definition of $\mathbf{RA1}$}\\
	&=(ac'\neq\emptyset \land ac'\neq\emptyset)[States_{tr\le tr'} (s) \cap ac'/ac']
	&&\ptext{Predicate calculus}\\
	&=(true \land ac'\neq\emptyset)[States_{tr\le tr'} (s) \cap ac'/ac']
	&&\ptext{Definition of $\mathbf{RA1}$}\\
	&=\mathbf{RA1} (true)
\end{xflalign*}
\end{proof}\end{proofs}
\end{lemma}

\begin{lemma}\label{lemma:RA1(P|-Q):RA1(lnot-RA1(lnot-P)|-Q)}
$\mathbf{RA1} (P \vdash Q) = \mathbf{RA1} (\lnot \mathbf{RA1} (\lnot P) \vdash Q)$
\begin{proofs}\begin{proof}\checkt{alcc}
\begin{xflalign*}
	&\mathbf{RA1} (P \vdash Q)
	&&\ptext{Definition of design}\\
	&=\mathbf{RA1} ((ok \land P) \implies (Q \land ok'))
	&&\ptext{Predicate calculus}\\
	&=\mathbf{RA1} (\lnot ok \lor \lnot P \lor (Q \land ok'))
	&&\ptext{\cref{lemma:RA1(P-lor-Q):RA1(P)-lor-RA1(Q)}}\\
	&=\mathbf{RA1} (\lnot ok) \lor \mathbf{RA1} (\lnot P) \lor \mathbf{RA1} (Q \land ok')
	&&\ptext{\cref{theorem:RA1-idempotent}}\\
	&=\mathbf{RA1} (\lnot ok) \lor \mathbf{RA1} \circ \mathbf{RA1} (\lnot P) \lor \mathbf{RA1} (Q \land ok')
	&&\ptext{\cref{lemma:RA1(P-lor-Q):RA1(P)-lor-RA1(Q)}}\\
	&=\mathbf{RA1} (\lnot ok \lor \mathbf{RA1} (\lnot P) \lor (Q \land ok'))
	&&\ptext{Predicate calculus}\\
	&=\mathbf{RA1} ((ok \land \lnot \mathbf{RA1} (\lnot P)) \implies (Q \land ok'))
	&&\ptext{Definition of design}\\
	&=\mathbf{RA1} (\lnot \mathbf{RA1} (\lnot P) \vdash Q)
\end{xflalign*}
\end{proof}\end{proofs}
\end{lemma}

\subsection{Substitution Properties}

\begin{lemma}\label{lemma:RA1(P)-o-w-subs:RA1(P-o-w-subs)}
$\mathbf{RA1} (P)^o_w = \mathbf{RA1} (P^o_w)$
\begin{proofs}\begin{proof}\checkt{pfr}\checkt{alcc}
\begin{xflalign*}
	&\mathbf{RA1} (P)^o_w
	&&\ptext{Definition of $\mathbf{RA1}$}\\
	&=((P \land ac'\neq\emptyset)[\{z | z \in ac' \land s.tr \le z.tr\}/ac'])^o_w
	&&\ptext{Substitution abbreviation}\\
	&=((P \land ac'\neq\emptyset)[\{z | z \in ac' \land s.tr \le z.tr\}/ac'])[o,s\oplus\{wait\mapsto w\}/ok',s]
	&&\ptext{Substitution}\\
	&=(P[o,s\oplus\{wait\mapsto w\}/ok',s] \land ac'\neq\emptyset)[\{z | z \in ac' \land (s\oplus\{wait\mapsto w\}).tr \le z.tr\}/ac']
	&&\ptext{Property of $\oplus$}\\
	&=(P[o,s\oplus\{wait\mapsto w\}/ok',s] \land ac'\neq\emptyset)[\{z | z \in ac' \land s.tr \le z.tr\}/ac']
	&&\ptext{Substitution abbreviation}\\
	&=(P^o_w \land ac'\neq\emptyset)[\{z | z \in ac' \land s.tr \le z.tr\}/ac']
	&&\ptext{Definition of $\mathbf{RA1}$}\\
	&=\mathbf{RA1} (P^o_w)
\end{xflalign*}
\end{proof}\end{proofs}
\end{lemma}

\subsection{Properties with respect to $\seqA$}

\begin{theorem}\label{theorem:RA1(true)-seqA-(P-lor-Q):(RA1(true)-seqA-P)-lor-(RA1(true)-seqA-Q)}
\begin{align*}
	&\mathbf{RA1} (true) \seqA (P \lor Q) = (\mathbf{RA1} (true) \seqA P) \lor (\mathbf{RA1} (true) \seqA Q)
\end{align*}
\begin{proofs}\begin{proof}
\begin{xflalign*}
	&\mathbf{RA1} (true) \seqA (P \lor Q)
	&&\ptext{Definition of $\mathbf{RA1}$ (\cref{lemma:RA1(true)})}\\
	&=(\exists z \spot s.tr \le z.tr \land z \in ac') \seqA (P \lor Q)
	&&\ptext{Definition of $\seqA$ and substitution}\\
	&=\exists z \spot s.tr \le z.tr \land z \in \{ s | P \lor Q \}
	&&\ptext{Property of sets}\\
	&=\exists z \spot s.tr \le z.tr \land (P[z/s] \lor Q[z/s])
	&&\ptext{Predicate calculus}\\
	&=(\exists z \spot s.tr \le z.tr \land P[z/s]) \lor (\exists z \spot s.tr \le z.tr \land Q[z/s])
	&&\ptext{Property of sets}\\
	&=(\exists z \spot s.tr \le z.tr \land z \in \{ s | P \}) \lor (\exists z \spot s.tr \le z.tr \land z \in \{ s | Q \})
	&&\ptext{Definition of $\seqA$ and substitution}\\
	&=((\exists z \spot s.tr \le z.tr \land z \in ac') \seqA P) \lor ((\exists z \spot s.tr \le z.tr \land z \in ac') \seqA Q)
	&&\ptext{Definition of $\mathbf{RA1}$ (\cref{lemma:RA1(true)})}\\
	&=(\mathbf{RA1} (true) \seqA P) \lor (\mathbf{RA1} (true) \seqA Q)
\end{xflalign*}
\end{proof}\end{proofs}
\end{theorem}

\begin{theorem}\label{theorem:RA1(P)-seqA-(Q-lor-R):(RA1(P)-seqA-Q)-lor-(RA1(P)-seqA-R)} Provided $ac'$ is not free in $P$,
\begin{align*}
	&\mathbf{RA1} (P) \seqA (Q \lor R) = (\mathbf{RA1} (P) \seqA Q) \lor (\mathbf{RA1} (P) \seqA R)
\end{align*}
\begin{proofs}\begin{proof}
\begin{xflalign*}
	&\mathbf{RA1} (P) \seqA (Q \lor R)
	&&\ptext{Predicate calculus}\\
	&=\mathbf{RA1} (P \land true) \seqA (Q \lor R)
	&&\ptext{Assumption: $ac'$ not free in $P$ and~\cref{lemma:RA1(P-land-Q):ac'-not-free}}\\
	&=(P \land \mathbf{RA1} (true)) \seqA (Q \lor R)
	&&\ptext{\cref{law:seqA-right-distributivity-conjunction}}\\
	&=(P \seqA (Q \lor R)) \land (\mathbf{RA1} (true) \seqA (Q \lor R))
	&&\ptext{\cref{law:seqA-ac'-not-free}}\\
	&=P \land (\mathbf{RA1} (true) \seqA (Q \lor R))
	&&\ptext{\cref{theorem:RA1(true)-seqA-(P-lor-Q):(RA1(true)-seqA-P)-lor-(RA1(true)-seqA-Q)}}\\
	&=P \land ((\mathbf{RA1} (true) \seqA Q) \lor (\mathbf{RA1} (true) \seqA R))
	&&\ptext{Predicate calculus}\\
	&=(P \land (\mathbf{RA1} (true) \seqA Q)) \lor (P \land (\mathbf{RA1} (true) \seqA R))
	&&\ptext{\cref{law:seqA-ac'-not-free}}\\
	&=((P \seqA Q) \land (\mathbf{RA1} (true) \seqA Q)) \lor ((P \seqA R) \land (\mathbf{RA1} (true) \seqA R))
	&&\ptext{\cref{law:seqA-right-distributivity-conjunction}}\\
	&=((P \land \mathbf{RA1} (true)) \seqA Q) \lor ((P \land \mathbf{RA1} (true)) \seqA R)
	&&\ptext{Assumption: $ac'$ not free in $P$ and~\cref{lemma:RA1(P-land-Q):ac'-not-free}}\\
	&=(\mathbf{RA1} (P \land true) \seqA Q) \lor (\mathbf{RA1} (P \land true) \seqA R)
	&&\ptext{Predicate calculus}\\
	&=(\mathbf{RA1} (P) \seqA Q) \lor (\mathbf{RA1} (P) \seqA R)
\end{xflalign*}
\end{proof}\end{proofs}
\end{theorem}

\begin{theorem}\label{theorem:(P-seqA-RA1(true))-lor-(P-seqA-RA1(Q)):P-seqA-RA1(true)} Provided $P$ is $\mathbf{PBMH}$-healthy,
\begin{align*}
	&(P \seqA \mathbf{RA1} (true)) \lor (P \seqA \mathbf{RA1} (Q))\\
	&=\\
	&(P \seqA \mathbf{RA1} (true))
\end{align*}
\begin{proofs}\begin{proof}\checkt{alcc}\checkt{pfr}
\begin{xflalign*}
	&(P \seqA \mathbf{RA1} (true)) \lor (P \seqA \mathbf{RA1} (Q))
	&&\ptext{Assumption: $P$ is $\mathbf{PBMH}$-healthy and \cref{lemma:(P-seqA-Q)-lor-(P-seqA-R):implies:(P-seqA-(Q-lor-R))}}\\
	&=((P \seqA \mathbf{RA1} (true)) \lor (P \seqA \mathbf{RA1} (Q))) \land (P \seqA (\mathbf{RA1} (true) \lor \mathbf{RA1} (Q))
	&&\ptext{\cref{lemma:RA1(P-lor-Q):RA1(P)-lor-RA1(Q)}}\\
	&=((P \seqA \mathbf{RA1} (true)) \lor (P \seqA \mathbf{RA1} (Q))) \land (P \seqA \mathbf{RA1} (true \lor Q))
	&&\ptext{Predicate calculus}\\
	&=((P \seqA \mathbf{RA1} (true)) \lor (P \seqA \mathbf{RA1} (Q))) \land (P \seqA \mathbf{RA1} (true))
	&&\ptext{Predicate calculus: absorption law}\\
	&=P \seqA \mathbf{RA1} (true)
\end{xflalign*}
\end{proof}\end{proofs}
\end{theorem}

\begin{lemma}\label{lemma:RA1(true)-seqA-true:true}
$\mathbf{RA1} (true) \seqA true$
\begin{proofs}\begin{proof}
\begin{xflalign*}
	&\mathbf{RA1} (true) \seqA true
	&&\ptext{\cref{lemma:RA1(true)}}\\
	&=(\exists z \spot s.tr \le z.tr \land z \in ac') \seqA true
	&&\ptext{Definition of $\seqA$ and substitution}\\
	&=\exists z \spot s.tr \le z.tr \land z \in \{ s | true \}
	&&\ptext{Property of sets}\\
	&=\exists z \spot s.tr \le z.tr \land true
	&&\ptext{Predicate calculus}\\
	&=\exists z \spot s.tr \le z.tr
	&&\ptext{Predicate calculus}\\
	&=true
\end{xflalign*}
\end{proof}\end{proofs}
\end{lemma}

\begin{lemma}\label{lemma:RA1(true)-seqA-s.wait:lnot-ok-land-RA1(true)}
\begin{align*}
	&\mathbf{RA1} (true) \seqA (s.wait \land \lnot ok \land \mathbf{RA1} (true)) = \lnot ok \land \mathbf{RA1} (true)
\end{align*}
\begin{proofs}\begin{proof}
\begin{xflalign*}
	&\mathbf{RA1} (true) \seqA (s.wait \land \lnot ok \land \mathbf{RA1} (true))
	&&\ptext{Definition of $\mathbf{RA1}$ (\cref{lemma:RA1(true)})}\\
	&=(\exists z \spot s.tr \le z.tr \land z \in ac') \seqA (s.wait \land \lnot ok \land \mathbf{RA1} (true))
	&&\ptext{Definition of $\seqA$ and substitution}\\
	&=\exists z \spot s.tr \le z.tr \land z \in \{ s | s.wait \land \lnot ok \land \mathbf{RA1} (true) \}
	&&\ptext{Property of sets and substitution}\\
	&=\exists z \spot s.tr \le z.tr \land z.wait \land \lnot ok \land \mathbf{RA1} (true)[z/s]
	&&\ptext{Predicate calculus: quantifier scope}\\
	&=\lnot ok \land \exists z \spot s.tr \le z.tr \land z.wait \land \mathbf{RA1} (true)[z/s]
	&&\ptext{Definition of $\mathbf{RA1}$ (\cref{lemma:RA1(true)})}\\
	&=\lnot ok \land \exists z \spot s.tr \le z.tr \land z.wait \land (\exists y \spot z.tr \le y.tr \land y \in ac')
	&&\ptext{Introduce fresh variables}\\
	&=\lnot ok \land \exists z, w, t, o \spot \left(\begin{array}{l}
		s.tr \le t \land w
		\\ \land \\
		t = z.tr \land w = z.wait 
		\\ \land \\
		(\exists y \spot t \le y.tr \land y \in ac')
	\end{array}\right)
	&&\ptext{Property of records}\\
	&=\lnot ok \land \exists z, w, t, o \spot \left(\begin{array}{l}
		s.tr \le t \land w
		\\ \land \\
		z = o \oplus \{ tr \mapsto t, wait \mapsto w \} 
		\\ \land \\
		(\exists y \spot t \le y.tr \land y \in ac')
	\end{array}\right)
	&&\ptext{One-point rule}\\
	&=\lnot ok \land \exists w, t \spot s.tr \le t \land w \land (\exists y \spot t \le y.tr \land y \in ac')
	&&\ptext{Predicate calculus}\\
	&=\lnot ok \land \exists w, t, y \spot s.tr \le t \land t \le y.tr \land y \in ac' \land w
	&&\ptext{One-point rule}\\
	&=\lnot ok \land \exists t, y \spot s.tr \le t \land t \le y.tr \land y \in ac'
	&&\ptext{Property of sets}\\
	&=\lnot ok \land \exists y \spot s.tr \le y.tr \land y \in ac'
	&&\ptext{Definition of $\mathbf{RA1}$ (\cref{lemma:RA1(true)})}\\
	&=\lnot ok \land \mathbf{RA1} (true)
\end{xflalign*}
\end{proof}\end{proofs}
\end{lemma}

\begin{lemma}\label{theorem:RA1(lnot-ok)-seqA-P:RA1(lnot-ok)}Provided $P$ is $\mathbf{RA3}$ and $\mathbf{RA1}$-healthy,
\begin{align*}
	&\mathbf{RA1} (\lnot ok) \seqA P = \mathbf{RA1} (\lnot ok)
\end{align*}
\begin{proofs}\begin{proof}
\begin{xflalign*}
	&\mathbf{RA1} (\lnot ok) \seqA P
	&&\ptext{\cref{lemma:RA1(P-land-Q):ac'-not-free}}\\
	&=(\lnot ok \land \mathbf{RA1} (true)) \seqA P
	&&\ptext{\cref{law:seqA-right-distributivity-conjunction}}\\
	&=(\lnot ok \seqA P) \land (\mathbf{RA1} (true) \seqA P)
	&&\ptext{\cref{law:seqA-ac'-not-free}}\\
	&=\lnot ok \land (\mathbf{RA1} (true) \seqA P)
	&&\ptext{Assumption: $P$ is $\mathbf{RA1}$-healthy and~\cref{lemma:RA1(true)-seqA-RA1(true):implies:RA1(true)}}\\
	&=\lnot ok \land \mathbf{RA1} (true) \land (\mathbf{RA1} (true) \seqA P)
	&&\ptext{Assumption: $P$ is $\mathbf{RA3}$-healthy}\\
	&=\lnot ok \land \mathbf{RA1} (true) \land (\mathbf{RA1} (true) \seqA (\IIRac \dres s.wait \rres P))
	&&\ptext{Definition of conditional and $\IIRac$}\\
	&=\lnot ok \land \mathbf{RA1} (true) \left(\mathbf{RA1} (true) \seqA \left(\begin{array}{l}
		(s.wait \land \mathbf{RA1} (\lnot ok)) 
		\\ \lor \\
		(s.wait \land ok' \land s \in ac') 
		\\ \lor \\
		(\lnot s.wait \land P)
	\end{array}\right)\right)
	&&\ptext{\cref{theorem:RA1(P)-seqA-(Q-lor-R):(RA1(P)-seqA-Q)-lor-(RA1(P)-seqA-R)}}\\
	&=\lnot ok \land \mathbf{RA1} (true) \land \left(\begin{array}{l} 
		(\mathbf{RA1} (true) \seqA (s.wait \land \mathbf{RA1} (\lnot ok))) 
		\\ \lor \\
		(\mathbf{RA1} (true) \seqA (s.wait \land ok' \land s \in ac'))
		\\ \lor \\
		(\mathbf{RA1} (true) \seqA (\lnot s.wait \land P))
	\end{array}\right)
	&&\ptext{Predicate calculus and~\cref{lemma:RA1(P-land-Q):ac'-not-free}}\\
	&=\lnot ok \land \mathbf{RA1} (true) \land \left(\begin{array}{l} 
		(\mathbf{RA1} (true) \seqA (s.wait \land \lnot ok \land \mathbf{RA1} (true))) 
		\\ \lor \\
		(\mathbf{RA1} (true) \seqA (s.wait \land ok' \land s \in ac'))
		\\ \lor \\
		(\mathbf{RA1} (true) \seqA (\lnot s.wait \land P))
	\end{array}\right)
	&&\ptext{\cref{lemma:RA1(true)-seqA-s.wait:lnot-ok-land-RA1(true)}}\\
	&=\lnot ok \land \mathbf{RA1} (true) \land \left(\begin{array}{l} 
		(\lnot ok \land \mathbf{RA1} (true)) 
		\\ \lor \\
		(\mathbf{RA1} (true) \seqA (s.wait \land ok' \land s \in ac'))
		\\ \lor \\
		(\mathbf{RA1} (true) \seqA (\lnot s.wait \land P))
	\end{array}\right)
	&&\ptext{Predicate calculus: absorption law}\\
	&=\lnot ok \land \mathbf{RA1} (true)
	&&\ptext{\cref{lemma:RA1(P-land-Q):ac'-not-free}}\\
	&=\mathbf{RA1} (\lnot ok \land true)
	&&\ptext{Predicate calculus}\\
	&=\mathbf{RA1} (\lnot ok)
\end{xflalign*}
\end{proof}\end{proofs}
\end{lemma}

\begin{lemma}\label{lemma:RA1(true)-seqA-RA1(true):RA1(true)}
$\mathbf{RA1} (true) \seqA \mathbf{RA1} (true) = \mathbf{RA1} (true)$
\begin{proofs}\begin{proof}\checkt{alcc}
\begin{xflalign*}
	&\mathbf{RA1} (true) \seqA \mathbf{RA1} (true)
	&&\ptext{\cref{lemma:RA1(true)}}\\
	&=(\exists z \spot s.tr \le z.tr \land z \in ac') \seqA (\exists z \spot s.tr \le z.tr \land z \in ac')
	&&\ptext{Definition of $\seqA$}\\
	&=(\exists z \spot s.tr \le z.tr \land z \in ac')[\{ s | \exists z \spot s.tr \le z.tr \land z \in ac' \}/ac']
	&&\ptext{Substitution}\\
	&=\exists z \spot s.tr \le z.tr \land z \in \{ s | \exists z \spot s.tr \le z.tr \land z \in ac' \}
	&&\ptext{Variable renaming}\\
	&=\exists z \spot s.tr \le z.tr \land z \in \{ s | \exists y \spot s.tr \le y.tr \land y \in ac' \}
	&&\ptext{Property of sets}\\
	&=\exists z \spot s.tr \le z.tr \land (\exists y \spot z.tr \le y.tr \land y \in ac')
	&&\ptext{Predicate calculus}\\
	&=\exists z, y \spot s.tr \le z.tr \land z.tr \le y.tr \land y \in ac'
	&&\ptext{Transitivity of sequence prefixing}\\
	&=\exists y \spot s.tr \le y.tr \land y \in ac'
	&&\ptext{\cref{lemma:RA1(true)}}\\
	&=\mathbf{RA1} (true)
\end{xflalign*}
\end{proof}\end{proofs}
\end{lemma}

\begin{lemma}\label{lemma:RA1(P)-seqA-RA1(true)-ac'-not-free:RA1(P)} Provided $ac'$ is not free in $P$,
\begin{align*}
	&\mathbf{RA1} (P) \seqA \mathbf{RA1} (true) = \mathbf{RA1} (P)
\end{align*}
\begin{proofs}\begin{proof}\checkt{alcc}\checkt{pfr}
\begin{xflalign*}
	&\mathbf{RA1} (P) \seqA \mathbf{RA1} (true)
	&&\ptext{Assumption: $ac'$ not free in $P$ and~\cref{lemma:RA1(P)-ac'-not-free:P-land-RA1(true)}}\\
	&=(P \land \mathbf{RA1} (true)) \seqA \mathbf{RA1} (true)
	&&\ptext{Distributivity of $\seqA$}\\
	&=(P \seqA \mathbf{RA1} (true)) \land (\mathbf{RA1} (true) \seqA \mathbf{RA1} (true))
	&&\ptext{Property of $\seqA$ when $ac'$ not free}\\
	&=P \land (\mathbf{RA1} (true) \seqA \mathbf{RA1} (true))
	&&\ptext{\cref{lemma:RA1(true)-seqA-RA1(true):RA1(true)}}\\
	&=P \land \mathbf{RA1} (true)
	&&\ptext{\cref{lemma:RA1(P-land-Q):ac'-not-free}}\\
	&=\mathbf{RA1} (P \land true)
	&&\ptext{Predicate calculus}\\
	&=\mathbf{RA1} (P)
\end{xflalign*}
\end{proof}\end{proofs}
\end{lemma}

\begin{lemma}\label{lemma:RA1(P)-seqA-RA1(true):implies:RA1(P)-seqA-true} Provided $P$ is $\mathbf{PBMH}$-healthy,
\begin{align*}
	&\mathbf{RA1} (P) \seqA \mathbf{RA1} (true) \implies \mathbf{RA1} (P) \seqA true
\end{align*}
\begin{proofs}\begin{proof}
\begin{xflalign*}
	&\mathbf{RA1} (P) \seqA \mathbf{RA1} (true)
	&&\ptext{Assumption: $P$ is $\mathbf{PBMH}$-healthy and~\cref{theorem:PBMH-o-RA1(P):RA1(P),lemma:seqA:P-seqA-Q:implies:P-seqA-true}}\\
	&\implies \mathbf{RA1} (P) \seqA true
\end{xflalign*}
\end{proof}\end{proofs}
\end{lemma}

\begin{lemma}\label{lemma:seqA:P-seqA-Q:implies:P-seqA-true} Provided $P$ is $\mathbf{PBMH}$-healthy,
\begin{align*}
	&P \seqA Q \implies P \seqA true
\end{align*}
\begin{proofs}\begin{proof}
\begin{xflalign*}
	&P \seqA Q
	&&\ptext{Predicate calculus}\\
	&=P \seqA (Q \land true)
	&&\ptext{Assumption: $P$ is $\mathbf{PBMH}$-healthy and~\cref{lemma:seqA:P-seqA(Q-land-R):implies:(P-seqA-Q)-land-(P-seqA-R)}}\\
	&\implies (P \seqA Q) \land (P \seqA true)
	&&\ptext{Predicate calculus}\\
	&\implies (P \seqA true)
\end{xflalign*}
\end{proof}\end{proofs}
\end{lemma}

\begin{lemma}\label{lemma:RA1(true)-seqA-RA1(true):implies:RA1(true)}
$\mathbf{RA1} (true) \seqA \mathbf{RA1} (P) \implies \mathbf{RA1} (true)$
\begin{proofs}\begin{proof}
\begin{xflalign*}
	&\mathbf{RA1} (true) \seqA \mathbf{RA1} (P)
	&&\ptext{Definition of $\mathbf{RA1}$}\\
	&=\mathbf{RA1} (true) \seqA (P[States_{tr\le tr'} (s)\cap ac'/ac'] \land \mathbf{RA1} (true))
	&&\ptext{\cref{lemma:seqA:P-seqA(Q-land-R):implies:(P-seqA-Q)-land-(P-seqA-R)}}\\
	&\implies \mathbf{RA1} (true) \seqA \mathbf{RA1} (true)
	&&\ptext{\cref{lemma:RA1(true)-seqA-RA1(true):RA1(true)}}\\
	&=\mathbf{RA1} (true)
\end{xflalign*}
\end{proof}\end{proofs}
\end{lemma}

\subsection{Properties with respect to $\mathbf{RA2}$}

\begin{lemma}\label{lemma:RA1-o-RA2(P):RA2(P)-land-RA1-subs}
\checkt{pfr}
\checkt{alcc}
\begin{align*}
	&\mathbf{RA1} \circ \mathbf{RA2} (P)\\
	&=\\
	&\mathbf{RA2} (P) \land \exists z \spot s.tr \le z.tr \land z \in ac'
\end{align*}
\begin{proofs}\begin{proof}
\begin{xflalign*}
	&\mathbf{RA1} \circ \mathbf{RA2} (P)
	&&\ptext{Definition of $\mathbf{RA1}$ (\cref{lemma:RA1:alternative-1})}\\
	&=\mathbf{RA2} (P)[\{ z | z \in ac' \land s.tr \le z.tr\}/ac'] \land \exists z \spot s.tr \le z.tr \land z \in ac'
	&&\ptext{\cref{lemma:RA2(P)-subs-RA1:RA2(P)}}\\
	&=\mathbf{RA2} (P) \land \exists z \spot s.tr \le z.tr \land z \in ac'
\end{xflalign*}
\end{proof}\end{proofs}
\end{lemma}

\begin{lemma}\label{lemma:RA2(P)-subs-RA1:RA2(P)}
\checkt{pfr}
\checkt{alcc}
\begin{align*}
	&\mathbf{RA2} (P)[\{ z | z \in ac' \land s.tr \le z.tr\}/ac'] = \mathbf{RA2} (P)
\end{align*}
\begin{proofs}\begin{proof}
\begin{flalign*}
	&\mathbf{RA2} (P)[\{ z | z \in ac' \land s.tr \le z.tr\}/ac']
	&&\ptext{Definition of $\mathbf{RA2}$}\\
	&=\left(P\left)\begin{aligned}
		\left[	s\oplus\{tr\mapsto\lseq\rseq\}
				,\{z|z \in ac' \land s.tr \le z.tr \spot z \oplus \{ tr \mapsto z.tr - s.tr\}\}
				/
				s,ac']\right. \\
		[\{ z | z \in ac' \land s.tr \le z.tr\}/ac']
	\end{aligned}\right.\right.
	&&\ptext{Substitution}\\
	&=\left(P\left)\begin{aligned}
		\left[	s\oplus\{tr\mapsto\lseq\rseq\}
				,\left\{\begin{array}{l}
					z \left|\begin{array}{l}
						z \in \{ z | z \in ac' \land s.tr \le z.tr\} \land s.tr \le z.tr 
						\\ \spot z \oplus \{ tr \mapsto z.tr - s.tr\}
					\end{array}\right.
				\end{array}\right\}
				/
				s,ac'\right] \\
	\end{aligned}\right.\right.
	&&\ptext{Property of sets}\\
	&=\left(P\left)\begin{aligned}
		\left[	s\oplus\{tr\mapsto\lseq\rseq\}
				,\left\{\begin{array}{l}
					z \left|\begin{array}{l}
						z \in ac' \land s.tr \le z.tr \land s.tr \le z.tr 
						\\ \spot z \oplus \{ tr \mapsto z.tr - s.tr\}
					\end{array}\right.
				\end{array}\right\}
				/
				s,ac'\right] \\
	\end{aligned}\right.\right.
	&&\ptext{Predicate calculus}\\
	&=P[s\oplus\{tr\mapsto\lseq\rseq\},\{ z | z \in ac' \land s.tr \le z.tr \spot z \oplus \{ tr \mapsto z.tr - s.tr\}\}/s,ac']
	&&\ptext{Definition of $\mathbf{RA2}$}\\
	&=\mathbf{RA2} (P)
\end{flalign*}
\end{proof}\end{proofs}
\end{lemma}

\begin{lemma}\label{lemma:RA1(P):implies:RA1(true)}
$\mathbf{RA1} (P) \implies \mathbf{RA1} (true)$
\begin{proofs}\begin{proof}
\begin{xflalign*}
	&\mathbf{RA1} (P)
	&&\ptext{Predicate calculus}\\
	&=\mathbf{RA1} (P \land true)
	&&\ptext{\cref{lemma:RA1(P-land-Q):RA1(P)-land-RA1(Q)}}\\
	&=\mathbf{RA1} (P) \land \mathbf{RA1} (true)
	&&\ptext{Predicate calculus}\\
	&\implies \mathbf{RA1} (true)
\end{xflalign*}
\end{proof}\end{proofs}
\end{lemma}

\begin{lemma}\label{lemma:RA1-o-RA2(P):implies:RA1(true)}
$\mathbf{RA1} \circ \mathbf{RA2} (P) \implies \mathbf{RA1} (true)$
\begin{proofs}\begin{proof}
\begin{xflalign*}
	&\mathbf{RA1} \circ \mathbf{RA2} (P)
	&&\ptext{\cref{lemma:RA1-o-RA2(P):RA2(P)-land-RA1-subs}}\\
	&=\mathbf{RA2} (P) \land \mathbf{RA1} (true)
	&&\ptext{Predicate calculus}\\
	&=\implies \mathbf{RA1} (true)
\end{xflalign*}
\end{proof}\end{proofs}
\end{lemma}

\subsection{Properties with respect to $\mathbf{PBMH}$}

\begin{theorem}\label{theorem:RA-o-A(P):RA-o-PBMH(P)}
$\mathbf{RA} \circ \mathbf{A} (P) = \mathbf{RA} \circ \mathbf{PBMH} (P)$
\begin{proofs}\begin{proof}\checkt{pfr}\checkt{alcc}
\begin{xflalign*}
	&\mathbf{RA} \circ \mathbf{A} (P)
	&&\ptext{Definition of $\mathbf{RA}$ and $\mathbf{A}$}\\
	&=\mathbf{RA3} \circ \mathbf{RA2} \circ \mathbf{RA1} \circ \mathbf{A0} \circ \mathbf{A1} (P)
	&&\ptext{$\mathbf{A1}$ is $\mathbf{PBMH}$}\\
	&=\mathbf{RA3} \circ \mathbf{RA2} \circ \mathbf{RA1} \circ \mathbf{A0} \circ \mathbf{PBMH} (P)
	&&\ptext{\cref{theorem:RA1-o-A0:RA1}}\\
	&=\mathbf{RA3} \circ \mathbf{RA2} \circ \mathbf{RA1} \circ \mathbf{PBMH} (P)
	&&\ptext{Definition of $\mathbf{RA}$}\\
	&=\mathbf{RA} \circ \mathbf{PBMH} (P)
\end{xflalign*}
\end{proof}\end{proofs}
\end{theorem}

\begin{lemma}\label{lemma:RA1(P)-P-is-PBMH} Provided $P$ is $\mathbf{PBMH}$-healthy,
\begin{align*}
	&\mathbf{RA1} (P) = \mathbf{PBMH} (P \land ac'\neq\emptyset \land ac'\subseteq States_{tr\le tr'}(s))
\end{align*}
\begin{proofs}\begin{proof}\checkt{alcc}\checkt{pfr}
\begin{xflalign*}
	&\mathbf{RA1} (P)
	&&\ptext{Definition of $\mathbf{RA1}$ (\cref{lemma:RA1:alternative-2})}\\
	&=(P \land ac'\neq\emptyset)[States_{tr\le tr'}(s)\cap ac'/ac']
	&&\ptext{Assumption: $P$ is $\mathbf{PBMH}$-healthy}\\
	&=(\mathbf{PBMH} (P) \land ac'\neq\emptyset)[States_{tr\le tr'}(s)\cap ac'/ac']
	&&\ptext{$ac'\neq\emptyset$ is $\mathbf{PBMH}$-healthy and closure (\cref{law:pbmh:conjunction-closure})}\\
	&=\mathbf{PBMH} (\mathbf{PBMH} (P) \land ac'\neq\emptyset)[States_{tr\le tr'}(s)\cap ac'/ac']
	&&\ptext{Assumption: $P$ is $\mathbf{PBMH}$-healthy}\\
	&=\mathbf{PBMH} (P \land ac'\neq\emptyset)[States_{tr\le tr'}(s)\cap ac'/ac']
	&&\ptext{Definition of $\mathbf{PBMH}$ (\cref{lemma:PBMH:alternative-1})}\\
	&=(\exists ac_0 \spot P[ac_0/ac'] \land ac_0\neq\emptyset \land ac_0 \subseteq ac')[States_{tr\le tr'}(s)\cap ac'/ac']
	&&\ptext{Substitution}\\
	&=\exists ac_0 \spot P[ac_0/ac'] \land ac_0\neq\emptyset \land ac_0 \subseteq (States_{tr\le tr'}(s)\cap ac')
	&&\ptext{Property of sets}\\
	&=\exists ac_0 \spot P[ac_0/ac'] \land ac_0\neq\emptyset \land ac_0 \subseteq States_{tr\le tr'}(s) \land ac_0 \subseteq ac'
	&&\ptext{Substitution}\\
	&=\exists ac_0 \spot (P \land ac'\neq\emptyset \land ac'\subseteq States_{tr\le tr'}(s))[ac_0/ac'] \land ac_0 \subseteq ac'
	&&\ptext{Definition of $\mathbf{PBMH}$ (\cref{lemma:PBMH:alternative-1})}\\
	&=\mathbf{PBMH} (P \land ac'\neq\emptyset \land ac'\subseteq States_{tr\le tr'}(s))
\end{xflalign*}
\end{proof}\end{proofs}
\end{lemma}

\begin{lemma}\label{lemma:PBMH(P-ac'-neq-emptyset-ac'-subset-states):implies}
\begin{align*}
	&\mathbf{PBMH} (P \land ac'\neq\emptyset \land ac'\subseteq States_{tr\le tr'}(s)) \implies ac'\cap States_{tr\le tr'}(s) \neq\emptyset
\end{align*}
\begin{proofs}\begin{proof}\checkt{pfr}
\begin{xflalign*}
	&\mathbf{PBMH} (P \land ac'\neq\emptyset \land ac'\subseteq States_{tr\le tr'}(s))
	&&\ptext{Definition of $\mathbf{PBMH}$ (\cref{lemma:PBMH:alternative-1})}\\
	&=\exists ac_0 \spot (P \land ac'\neq\emptyset \land ac'\subseteq States_{tr\le tr'}(s))[ac_0/ac'] \land ac_0 \subseteq ac'
	&&\ptext{Substitution}\\
	&=\exists ac_0 \spot P[ac_0/ac'] \land ac_0\neq\emptyset \land ac_0\subseteq States_{tr\le tr'}(s) \land ac_0 \subseteq ac'
	&&\ptext{Property of sets}\\
	&=\exists ac_0 \spot P[ac_0/ac'] \land ac_0\neq\emptyset \land ac_0\subseteq (States_{tr\le tr'}(s) \cap ac')
	&&\ptext{Property of sets}\\
	&=\exists ac_0 \spot P[ac_0/ac'] \land ac_0\neq\emptyset \land ac_0\subseteq (States_{tr\le tr'}(s) \cap ac') \land States_{tr\le tr'}(s) \cap ac' \neq\emptyset
	&&\ptext{Predicate calculus}\\
	&\implies States_{tr\le tr'}(s) \cap ac' \neq\emptyset 
\end{xflalign*}
\end{proof}\end{proofs}
\end{lemma}

\begin{lemma}\label{lemma:ac'-States-seqA-RA1-o-PBMH:implies}
\begin{align*}
	&ac'\cap States_{tr\le tr'}(s) \neq\emptyset \seqA \mathbf{PBMH} (P \land ac'\neq\emptyset \land ac'\subseteq States_{tr\le tr'}(s))\\
	&\implies \\
	&ac'\cap States_{tr\le tr'}(s) \neq\emptyset
\end{align*}
\begin{proofs}\begin{proof}\checkt{pfr}
\begin{xflalign*}
	&ac'\cap States_{tr\le tr'}(s) \neq\emptyset \seqA \mathbf{PBMH} (P \land ac'\neq\emptyset \land ac'\subseteq States_{tr\le tr'}(s))
	&&\ptext{Property of sets}\\
	&=(\exists z \spot z \in States_{tr\le tr'}(s) \land z \in ac') \seqA \mathbf{PBMH} (P \land ac'\neq\emptyset \land ac'\subseteq States_{tr\le tr'}(s))
	&&\ptext{Property of sets and definition of $States_{tr\le tr'}(s)$}\\
	&=(\exists z \spot s.tr \le z.tr \land z \in ac') \seqA \mathbf{PBMH} (P \land ac'\neq\emptyset \land ac'\subseteq States_{tr\le tr'}(s))
	&&\ptext{Definition of $\seqA$ and substitution}\\
	&=(\exists z \spot s.tr \le z.tr \land z \in \{ s | \mathbf{PBMH} (P \land ac'\neq\emptyset \land ac'\subseteq States_{tr\le tr'}(s)) \}
	&&\ptext{Variable renaming and property of sets}\\
	&=(\exists z \spot s.tr \le z.tr \land \mathbf{PBMH} (P \land ac'\neq\emptyset \land ac'\subseteq States_{tr\le tr'}(s))[z/s]
	&&\ptext{\cref{lemma:PBMH(P-ac'-neq-emptyset-ac'-subset-states):implies}}\\
	&=\exists z \spot \left(\begin{array}{l}
		s.tr \le z.tr \\ \land \\ \left(\begin{array}{l}
			\mathbf{PBMH} (P \land ac'\neq\emptyset \land ac'\subseteq States_{tr\le tr'}(s)) 
			\\ \land \\
			ac'\cap States_{tr\le tr'}(s) \neq\emptyset 
		\end{array}\right)[z/s]
	\end{array}\right)
	&&\ptext{Substitution}\\
	&=\exists z \spot \left(\begin{array}{l}
		s.tr \le z.tr \\ \land \\ \left(\begin{array}{l}
			\mathbf{PBMH} (P \land ac'\neq\emptyset \land ac'\subseteq States_{tr\le tr'}(s))[z/s]
			\\ \land \\
			ac'\cap States_{tr\le tr'}(z) \neq\emptyset 
		\end{array}\right)
	\end{array}\right)
	&&\ptext{Predicate calculus}\\
	&\implies \left(\begin{array}{l}
		\exists z \spot s.tr \le z.tr \land \mathbf{PBMH} (P \land ac'\neq\emptyset \land ac'\subseteq States_{tr\le tr'}(s))[z/s]
			\\ \land \\
		\exists z \spot s.tr \le z.tr \land	ac'\cap States_{tr\le tr'}(z) \neq\emptyset 
	\end{array}\right)
	&&\ptext{Predicate calculus}\\
	&\implies \exists z \spot s.tr \le z.tr \land ac'\cap States_{tr\le tr'}(z) \neq\emptyset
	&&\ptext{Property of sets and definition of $States_{tr\le tr'}$}\\
	&=\exists z \spot s.tr \le z.tr \land (\exists y \spot z.tr \le y.tr \land y \in ac')
	&&\ptext{Predicate calculus}\\
	&=\exists z,y \spot s.tr \le z.tr \land z.tr \le y.tr \land y \in ac'
	&&\ptext{Predicate calculus and transitivity of sequence prefixing}\\
	&=\exists y \spot s.tr \le y.tr \land y \in ac'
	&&\ptext{Property of sets and definition of $States_{tr\le tr'}$}\\
	&=ac'\cap States_{tr\le tr'}(s) \neq\emptyset
\end{xflalign*}
\end{proof}\end{proofs}
\end{lemma}


\subsection{Properties with respect to $\mathbf{A2}$}

\begin{lemma}\label{lemma:RA1-o-A2(P):RA1(true)-land-((P-emptyset-for-ac')-lor-exists-y-P)}
\begin{statement}
\begin{align*}
	&\mathbf{RA1} \circ \mathbf{A2} (P)\\
	&=\\
	&\mathbf{RA1} (true) \land \left(\begin{array}{l}
		(P[\emptyset/ac']) 
		\\ \lor \\
		(\exists y \spot P[\{y\}/ac'] \land s.tr \le y.tr \land y \in ac')
	\end{array}\right)
\end{align*}
\end{statement}
\begin{proofs}
\begin{proof}\checkt{alcc}
\begin{xflalign*}
	&\mathbf{RA1} \circ \mathbf{A2} (P)
	&&\ptext{\cref{lemma:A2:alternative-2:disjunction}}\\
	&=\mathbf{RA1} (P[\emptyset/ac'] \lor (\exists y \spot P[\{y\}/ac'] \land y \in ac'))
	&&\ptext{\cref{lemma:RA1(P-lor-Q):RA1(P)-lor-RA1(Q)}}\\
	&=\mathbf{RA1} (P[\emptyset/ac']) \lor \mathbf{RA1} (\exists y \spot P[\{y\}/ac'] \land s.tr \le y.tr \land y \in ac'))
	&&\ptext{\cref{lemma:RA1(exists-y-P-y-for-ac'-y-in-ac')}}\\
	&=\mathbf{RA1} (P[\emptyset/ac']) \lor (\exists y \spot P[\{y\}/ac'] \land s.tr \le y.tr \land y \in ac'))
	&&\ptext{\cref{lemma:RA1(P-land-Q):ac'-not-free}}\\
	&=(\mathbf{RA1} (true) \land P[\emptyset/ac']) \lor (\exists y \spot P[\{y\}/ac'] \land s.tr \le y.tr \land y \in ac'))
	&&\ptext{Predicate calculus}\\
	&=\left(
\right)
	&&\ptext{Predicate calculus and~\cref{lemma:RA1(true)}}\\
	&=\mathbf{RA1} (true) \land \left(%
\right)
	&&\ptext{\cref{lemma:RA1(P)-ac'-emptyset:false} and predicate calculus}\\
	&=\left(\exists z \spot \left(%
\right) \land z \in ac'\right)
	&&\ptext{Property of sets}\\
	&=\left(\exists z \spot \left(%
\right)
	&&\ptext{Predicate calculus and~\cref{lemma:RA1(true)}}\\
	&=\left(%
\right)
	&&\ptext{Predicate calculus and~\cref{lemma:RA1(true)}}\\
	&=\left(%
\right)
	&&\ptext{\cref{lemma:RA1-o-A2(P):RA1(true)-land-((P-emptyset-for-ac')-lor-exists-y-P)}}\\
	&=\mathbf{RA1}\circ\mathbf{A2} (P)
\end{xflalign*}
\end{proof}
\end{proofs}
\end{theorem}

\section{$\mathbf{RA2}$}

\subsection{Definition}
\theoremstatementref{def:RA2}

\subsection{Properties}

\begin{theorem}\label{theorem:RA2(P-land-Q):RA2(P)-land-RA2(Q)}
\begin{statement}$\mathbf{RA2} (P \land Q) = \mathbf{RA2} (P) \land \mathbf{RA2} (Q)$\end{statement}
\begin{proofs}
\begin{proof}\checkt{pfr}\checkt{alcc}
\begin{xflalign*}
	&\mathbf{RA2} (P \land Q)
	&&\ptext{Definition of $\mathbf{RA2}$}\\
	&=(P \land Q)\left[
\right]
	&&\ptext{Property of substitution}\\
	&=\left(%
\right]
	&&\ptext{Property of substitution}\\
	&=\left(%
\right]
	\end{array}\right)
	&&\ptext{Definition of $\mathbf{RA2}$}\\
	&=\mathbf{RA2} (P) \lor \mathbf{RA2} (Q)
\end{xflalign*}
\end{proof}
\end{proofs}
\end{theorem}

\begin{theorem}\label{theorem:RA2(P-seqA-Q)-closure}
\begin{statement}Provided $P$ and $Q$ are $\mathbf{RA2}$-healthy,
\begin{align*}
	&\mathbf{RA2} (P \seqA Q) = P \seqA Q 
\end{align*}
\end{statement}
\begin{proofs}
\begin{proof}
\begin{xflalign*}
	&\mathbf{RA2} (P \seqA Q)
	&&\ptext{Assumption: $P$ and $Q$ are $\mathbf{RA2}$-healthy}\\
	&=\mathbf{RA2} (\mathbf{RA2} (P) \seqA  \mathbf{RA2} (Q))
	&&\ptext{\cref{lemma:RA2-seqA-RA2}}\\
	&=\mathbf{RA2} \left(
	\left(P\left)\begin{aligned}
			&[s\oplus\{tr\mapsto\lseq\rseq\}/s] \\
			&\left[\left\{t \left|
\right\}\right/ac'\right] \\
		\end{aligned}\right.\right.
	&&\ptext{Variable renaming, property of $\oplus$ and value of record component $tr$}\\
	&=\left(P\left)\begin{aligned}
			&[s\oplus\{tr\mapsto\lseq\rseq\}/s] \\
			&\left[\left\{t \left|%
\right\}\right/ac'\right] \\
		\end{aligned}\right.\right.
	&&\ptext{Property of sequences}\\
	&=\left(P\left)\begin{aligned}
			&[s\oplus\{tr\mapsto\lseq\rseq\}/s] \\
			&\left[\left\{t \left|%
\right\}\right/ac'\right] \\
	\end{aligned}\right.\right.
	&&\ptext{Property of sets, $\oplus$ and value of record component $tr$}\\
	&=\left(P\left)\begin{aligned}
			&[s\oplus\{tr\mapsto\lseq\rseq\}/s] \\
			&\left[\left\{t \left|%
\right.\right\}\right/ac'\right]
				  			\end{aligned}\right.\right.
			\end{array}\right\}\right/ac'\right] \\
	\end{aligned}\right.\right.
	&&\ptext{\cref{lemma:RA2-seqA-RA2}}\\
	&=\mathbf{RA2} (P) \seqA \mathbf{RA2} (Q)
	&&\ptext{Assumption: $P$ and $Q$ are $\mathbf{RA2}$-healthy}\\
	&=P \seqA Q
\end{xflalign*}
\end{proof}
\end{proofs}
\end{theorem}

\begin{theorem}\label{lemma:RA2(ac'-neq-emptyset):RA1(true)}
\begin{statement}
$\mathbf{RA2} (ac'\neq\emptyset) = \mathbf{RA1} (true)$
\end{statement}
\begin{proofs}
\begin{proof}\checkt{alcc}
\begin{xflalign*}	
	&\mathbf{RA2} (ac'\neq\emptyset)
	&&\ptext{Definition of $\mathbf{RA2}$}\\
	&=(ac'\neq\emptyset)[s\oplus\{tr\mapsto\lseq\rseq,\{ z | z \in ac' \land s.tr \le z.tr \spot z \oplus \{ tr \mapsto z.tr - s.tr\}\}/s,ac']
	&&\ptext{Substitution}\\
	&=\{ z | z \in ac' \land s.tr \le z.tr \spot z \oplus \{ tr \mapsto z.tr - s.tr\}\}\neq\emptyset
	&&\ptext{Property of sets}\\
	&=\exists y \spot y \in \{ z | z \in ac' \land s.tr \le z.tr \spot z \oplus \{ tr \mapsto z.tr - s.tr\}\}
	&&\raisetag{18pt}\ptext{Property of sets}\\
	&=\exists y,z \spot z \in ac' \land s.tr \le z.tr \land y = z \oplus \{ tr \mapsto z.tr - s.tr\}
	&&\ptext{One-point rule}\\
	&=\exists z \spot z \in ac' \land s.tr \le z.tr
	&&\ptext{\cref{lemma:RA1(true)}}\\
	&=\mathbf{RA1} (true)
\end{xflalign*}
\end{proof}
\end{proofs}
\end{theorem}

\begin{theorem}\label{theorem:RA2-o-RA1:RA1-o-RA2}
\begin{statement}
$\mathbf{RA2} \circ \mathbf{RA1} (P) = \mathbf{RA1} \circ \mathbf{RA2} (P)$
\end{statement}
\begin{proofs}
\begin{proof}\checkt{pfr}\checkt{alcc}
\begin{flalign*}
	&\mathbf{RA2} \circ \mathbf{RA1} (P)
	&&\ptext{Definition of $\mathbf{RA2}$}\\
	&=\mathbf{RA1} (P)[s\oplus\{tr\mapsto\lseq\rseq\},\{z|z\in ac' \land s.tr \le z.tr \spot z\oplus\{tr\mapsto z.tr-s.tr\}\}/s,ac']
	&&\ptext{Definition of $\mathbf{RA1}$}\\
	&=\left(P \land ac'\neq\emptyset\left)\begin{aligned}
		&[\{z | z\in ac' \land s.tr \le z.tr\}/ac']\\
		&[s\oplus\{tr\mapsto\lseq\rseq\},\{z|z\in ac' \land s.tr \le z.tr \spot z\oplus\{tr\mapsto z.tr-s.tr\}\}/s,ac']
	\end{aligned}\right.\right.
	&&\ptext{Substitution of $s$}\\
	&=\left(
\right)\begin{aligned}
		&[\{z | z\in ac' \land (s\oplus\{tr\mapsto\lseq\rseq\}).tr \le z.tr\}/ac']\\
		&[\{z|z\in ac' \land s.tr \le z.tr \spot z\oplus\{tr\mapsto z.tr-s.tr\}\}/ac']
	\end{aligned}
	&&\ptext{Value of state component $tr$}\\
	&=\left(%
\right)\begin{aligned}
		&[\{z | z\in ac' \land \lseq\rseq \le z.tr\}/ac']\\
		&[\{z|z\in ac' \land s.tr \le z.tr \spot z\oplus\{tr\mapsto z.tr-s.tr\}\}/ac']
	\end{aligned}
	&&\ptext{Property of sequence prefixing}\\
	&=\left(%
\right)\begin{aligned}
		&[\{z | z\in ac'\}/ac']\\
		&[\{z|z\in ac' \land s.tr \le z.tr \spot z\oplus\{tr\mapsto z.tr-s.tr\}\}/ac']
	\end{aligned}
	&&\ptext{Property of sets}\\
	&=\left(%
\right)\begin{aligned}
		&[ac'/ac']\\
		&[\{z|z\in ac' \land s.tr \le z.tr \spot z\oplus\{tr\mapsto z.tr-s.tr\}\}/ac']
	\end{aligned}
	&&\ptext{Property of substitution}\\
	&=\left(%
\right)
	&&\ptext{Property of sets}\\
	&=\left(%
\right)
	&&\ptext{Property of sets}\\
	&=\left(%
\right)
	&&\ptext{Property of substitution}\\
	&=\left(%
\right)
	&&\ptext{\cref{lemma:RA1:alternative-1}}\\
	&=\mathbf{RA1} \circ \mathbf{RA2} (P)	
\end{flalign*}
\end{proof}
\end{proofs}
\end{theorem}

\begin{theorem}\label{theorem:PBMH-o-RA2(P):RA2(P)}
\begin{statement}
$\mathbf{PBMH} \circ \mathbf{RA2} \circ \mathbf{PBMH} (P) = \mathbf{RA2} \circ \mathbf{PBMH} (P)$
\end{statement}
\begin{proofs}
\begin{proof}\checkt{pfr}\checkt{alcc}
\begin{flalign*}
	&\mathbf{PBMH} \circ \mathbf{RA2} \circ \mathbf{PBMH} (P)
	&&\ptext{Definition of $\mathbf{PBMH}$ (\cref{lemma:PBMH:alternative-1})}\\
	&=\mathbf{PBMH} \circ \mathbf{RA2} (\exists ac_0 \spot P[ac_0/ac'] \land ac_0 \subseteq ac')
	&&\ptext{Definition of $\mathbf{RA2}$}\\
	&=\mathbf{PBMH} \left(
\right)	
	&&\ptext{Definition of subset inclusion}\\
	&=\left(%
\right)
	&&\ptext{Property of sets}\\
	&=\left(%
\right)
	&&\ptext{Property of sets}\\
	&=\left(%
\right]
	\end{array}\right)
	&&\ptext{Definition of $\mathbf{RA2}$}\\
	&=\mathbf{RA2} (\exists ac_0 \spot P[ac_0/ac'] \land ac_0 \subseteq ac')
	&&\ptext{Definition of $\mathbf{PBMH}$ (\cref{lemma:PBMH:alternative-1})}\\
	&=\mathbf{RA2} \circ \mathbf{PBMH} (P)
\end{flalign*}
\end{proof}
\end{proofs}
\end{theorem}

\begin{theorem}\label{theorem:RA2:idempotent}
\begin{statement}$\mathbf{RA2} \circ \mathbf{RA2} (P) = \mathbf{RA2} (P)$\end{statement}
\begin{proofs}
\begin{proof}\checkt{pfr}\checkt{alcc}
\begin{flalign*}
	&\mathbf{RA2} \circ \mathbf{RA2} (P)
	&&\ptext{Definition of $\mathbf{RA2}$ twice}\\
	&=P\left[\begin{array}{l}
				(s \oplus \{ tr \mapsto \lseq\rseq\}) \oplus \{ tr \mapsto \lseq\rseq\} \\
				\left\{\begin{array}{l}
					z \left|\begin{array}{l}
					z \in \{ z | z \in ac' \land s.tr \le z.tr \spot z \oplus \{ tr \mapsto z.tr - s.tr \} \} 
					\\ \land (s \oplus \{ tr \mapsto \lseq\rseq\}).tr \le z.tr
					\\ \spot z \oplus \{ tr \mapsto z.tr - (s \oplus \{ tr \mapsto \lseq\rseq\}).tr \}
					\end{array}\right.
				\end{array}\right\}
			\end{array}
			\right/\left.
			\begin{array}{r}
				s \\ \\ \\ ac'
			\end{array}\right] 
	&&\ptext{Property of $\oplus$ and value of $tr$ component}\\
	&=P\left[\begin{array}{l}
				s \oplus \{ tr \mapsto \lseq\rseq\} \\
				\left\{\begin{array}{l}
					z \left|\begin{array}{l}
						z \in \{ z | z \in ac' \land s.tr \le z.tr \spot z \oplus \{ tr \mapsto z.tr - s.tr \} \} 
						\\ \spot z \oplus \{ tr \mapsto z.tr - \lseq\rseq \}
						\end{array}\right.
				\end{array}\right\}
			\end{array}
			\right/\left.
			\begin{array}{r}
				s \\ \\ ac'
			\end{array}\right]
	&&\ptext{Property of sequence difference}\\
	&=P\left[\begin{array}{l}
				s \oplus \{ tr \mapsto \lseq\rseq\} \\
				\left\{\begin{array}{l}
					z \left|\begin{array}{l}
					z \in \{ z | z \in ac' \land s.tr \le z.tr \spot z \oplus \{ tr \mapsto z.tr - s.tr \} \} 
					\\ \spot z \oplus \{ tr \mapsto z.tr \}
					\end{array}\right.
				\end{array}\right\}
			\end{array}
			\right/\left.
			\begin{array}{r}
				s \\ \\ ac'
			\end{array}\right]
	&&\ptext{Property of $\oplus$}\\
	&=P\left[\begin{array}{l}
			s \oplus \{ tr \mapsto \lseq\rseq\} \\
			\{ z | z \in \{ z | z \in ac' \land s.tr \le z.tr \spot z \oplus \{ tr \mapsto z.tr - s.tr \} \} \}
			\end{array}
			\right/\left.
			\begin{array}{r}
				s \\ ac'
			\end{array}\right]
	&&\ptext{Property of sets}\\
	&=P[s \oplus \{ tr \mapsto \lseq\rseq\}, 
		\{ z | z \in ac' \land s.tr \le z.tr \spot z \oplus \{ tr \mapsto z.tr - s.tr \} \}/s,ac']
	&&\ptext{Definition of $\mathbf{RA2}$}\\
	&=\mathbf{RA2} (P)
\end{flalign*}
\end{proof}
\end{proofs}
\end{theorem}

\begin{theorem}\label{theorem:RA2:monotonic}
\begin{statement}$P \sqsubseteq Q \implies \mathbf{RA2} (P) \sqsubseteq \mathbf{RA2} (Q)$\end{statement}
\begin{proofs}
\begin{proof}\checkt{pfr}\checkt{alcc}
\begin{flalign*}
	&\mathbf{RA2} (Q)
	&&\ptext{Assumption: $P \sqsubseteq Q = [Q \implies P]$}\\
	&=\mathbf{RA2} (Q \land P)
	&&\ptext{Definition of $\mathbf{RA2}$ and property of substitution}\\
	&=\mathbf{RA2} (Q) \land \mathbf{RA2} (P)
	&&\ptext{Predicate calculus}\\
	&\implies \mathbf{RA2} (P)
\end{flalign*}
\end{proof}
\end{proofs}
\end{theorem}

\subsection{Lemmas}

\begin{lemma}\label{lemma:RA2:alternative-1}
\begin{align*}
	&\mathbf{RA2} (P) = P[s\oplus\{tr\mapsto\lseq\rseq\},\{y | y \oplus \{tr\mapsto s.tr\cat y.tr\} \in ac'\}/s,ac']
\end{align*}
\begin{proofs}\begin{proof}\checkt{alcc}\checkt{pfr}
\begin{xflalign*}
	&\mathbf{RA2} (P)
	&&\ptext{Definition of $\mathbf{RA2}$}\\
	&=P[s\oplus\{tr\mapsto\lseq\rseq\},\{z|z\in ac' \land s.tr\le z.tr \spot z\oplus \{tr\mapsto z.tr-s.tr\}\}/s,ac']
	&&\ptext{Property of sets}\\
	&=P\left[\left.s\oplus\{tr\mapsto\lseq\rseq\},\left\{ y \left|
		y \in \left\{z\left|\begin{array}{l}
			z\in ac' \land s.tr\le z.tr \\
			\spot z\oplus \{tr\mapsto z.tr-s.tr\}
		\end{array}\right.\right\}\right.\right\}\right/s,ac'\right]
	&&\ptext{Property of sets}\\
	&=P\left[\left.s\oplus\{tr\mapsto\lseq\rseq\},\left\{ y \left|\begin{array}{l}
		\exists z \spot z\in ac' \land s.tr\le z.tr \\
		\land y = z \oplus \{tr\mapsto z.tr-s.tr\}
		\end{array}\right.\right\}\right/s,ac'\right]
	&&\ptext{\cref{lemma:x-oplus-tr-in-ac}}\\
	&=P[s\oplus\{tr\mapsto\lseq\rseq\},\{y | y \oplus \{tr\mapsto s.tr\cat y.tr\} \in ac'\}/s,ac']	
\end{xflalign*}
\end{proof}\end{proofs}
\end{lemma}

\begin{lemma}\label{lemma:RA2(true):true}
$\mathbf{RA2} (true) = true$
\begin{proofs}\begin{proof}\checkt{pfr}\checkt{alcc}
\begin{xflalign*}
	&\mathbf{RA2} (true)
	&&\ptext{Definition of $\mathbf{RA2}$}\\
	&=true[s \oplus \{ tr \mapsto \lseq\rseq\},\{ z | z \in ac' \land s.tr \le z.tr \spot z \oplus \{ tr \mapsto z.tr - s.tr \} \}/s,ac']
	&&\ptext{Substitution}\\
	&=true
\end{xflalign*}
\end{proof}\end{proofs}
\end{lemma}

\begin{lemma}\label{lemma:RA2(s-in-ac'):s-in-ac'}
$\mathbf{RA2} (s \in ac') = s \in ac'$
\begin{proofs}\begin{proof}\checkt{pfr}\checkt{alcc}
\begin{flalign*}
	&\mathbf{RA2} (s \in ac')
	&&\ptext{Definition of $\mathbf{RA2}$}\\
	&=(s \in ac')[s \oplus \{ tr \mapsto \lseq\rseq\},\{ z | z \in ac' \land s.tr \le z.tr \spot z \oplus \{ tr \mapsto z.tr - s.tr \} \}/s,ac']
	&&\ptext{Substitution}\\
	&=s \oplus \{ tr \mapsto \lseq\rseq\} \in \{ z | z \in ac' \land s.tr \le z.tr \spot z \oplus \{ tr \mapsto z.tr - s.tr \} \}
	&&\ptext{Property of sets}\\
	&=\exists z \spot z \in ac' \land s.tr \le z.tr \land s \oplus \{ tr \mapsto \lseq\rseq\} = z \oplus \{ tr \mapsto z.tr - s.tr \} \}
	&&\ptext{Property of $\oplus$}\\
	&=\left(\begin{array}{l}
		\exists z \spot z \in ac' \land s.tr \le z.tr \\
		\land \{tr \} \ndres s \cup \{ tr \mapsto \lseq\rseq\} = \{tr \} \ndres z \cup \{ tr \mapsto z.tr - s.tr \}
	\end{array}\right)
	&&\ptext{Property of relations}\\
	&=\left(\begin{array}{l}
		\exists z \spot z \in ac' \land s.tr \le z.tr \\
		\land \{tr \} \ndres s = \{tr\} \ndres z 
		\land \{ tr \mapsto \lseq\rseq\} = \{ tr \mapsto z.tr - s.tr \}
	\end{array}\right)
	&&\ptext{Property of relations}\\
	&=\left(\begin{array}{l}
		\exists z \spot z \in ac' \land s.tr \le z.tr \\
		\land \{tr \} \ndres s = \{tr\} \ndres z 
		\land \lseq\rseq = z.tr - s.tr
	\end{array}\right)
	&&\ptext{Property of sequences}\\
	&=\left(\begin{array}{l}
		\exists z \spot z \in ac' \land s.tr \le z.tr \\
		\land \{tr \} \ndres s = \{tr\} \ndres z 
		\land z.tr = s.tr
	\end{array}\right)
	&&\ptext{Property of relations}\\
	&=\exists z \spot z \in ac' \land s.tr \le z.tr \land s = z
	&&\ptext{One-point rule}\\
	&=s \in ac' \land s.tr \le s.tr
	&&\ptext{Property of sequences}\\
	&=s \in ac'
\end{flalign*}
\end{proof}\end{proofs}
\end{lemma}

\begin{lemma}\label{lemma:RA2(P):P:s-ac'-not-free} Provided $s$ and $ac'$ are not free in $P$,
$\mathbf{RA2} (P) = P$.
\begin{proofs}\begin{proof}\checkt{alcc}\checkt{pfr}
\begin{flalign*}
	&\mathbf{RA2} (P)
	&&\ptext{Definition of $\mathbf{RA2}$}\\
	&=P[s \oplus \{ tr \mapsto \lseq\rseq\},\{ z | z \in ac' \land s.tr \le z.tr \spot z \oplus \{ tr \mapsto z.tr - s.tr \} \}/s,ac']
	&&\ptext{Assumption and substitution}\\
	&=P
\end{flalign*}
\end{proof}\end{proofs}
\end{lemma}

\begin{lemma}\label{lemma:RA2(P-conditional-Q):RA2(P)-RA2(conditional)-RA2(Q)}
\begin{align*}
	&\mathbf{RA2} (P \dres c \rres Q) = \mathbf{RA2} (P) \dres \mathbf{RA2} (c) \rres \mathbf{RA2} (Q)
\end{align*}
\begin{proofs}\begin{proof}\checkt{pfr}
\begin{xflalign*}
	&\mathbf{RA2} (P \dres c \rres Q)
	&&\ptext{Definition of conditional}\\
	&=\mathbf{RA2} ((c \land P) \lor (\lnot c \land Q))
	&&\ptext{\cref{theorem:RA2(P-lor-Q):RA2(P)-lor-RA2(Q)}}\\
	&=\mathbf{RA2} (c \land P) \lor \mathbf{RA2} (\lnot c \land Q)
	&&\ptext{\cref{theorem:RA2(P-land-Q):RA2(P)-land-RA2(Q)}}\\
	&=(\mathbf{RA2} (c) \land \mathbf{RA2} (P)) \lor (\mathbf{RA2} (\lnot c) \land \mathbf{RA2} (Q))
	&&\ptext{\cref{lemma:RA2(lnot-P):lnot-RA2(P)}}\\
	&=(\mathbf{RA2} (c) \land \mathbf{RA2} (P)) \lor (\lnot \mathbf{RA2} (c) \land \mathbf{RA2} (Q))
	&&\ptext{Definition of conditional}\\
	&=\mathbf{RA2} (P) \dres \mathbf{RA2} (c) \rres \mathbf{RA2} (Q)
\end{xflalign*}
\end{proof}\end{proofs}
\end{lemma}

\begin{lemma}\label{lemma:RA2:conditional-no-s.tr} Provided $c$ is $\mathbf{RA2}$-healthy,
\checkt{alcc}
\checkt{pfr}
\begin{align*}
	&\mathbf{RA2} (P \dres c \rres Q) = \mathbf{RA2} (P) \dres c \rres \mathbf{RA2} (Q)
\end{align*}
\begin{proofs}\begin{proof}
\begin{flalign*}
	&\mathbf{RA2} (P \dres c \rres Q)
	&&\ptext{\cref{lemma:RA2(P-conditional-Q):RA2(P)-RA2(conditional)-RA2(Q)}}\\
	&=\mathbf{RA2} (P) \dres \mathbf{RA2} (c) \rres \mathbf{RA2} (Q)
	&&\ptext{Assumption: $c$ is $\mathbf{RA2}$-healthy}\\
	&=\mathbf{RA2} (P) \dres c \rres \mathbf{RA2}(Q)
\end{flalign*}
\end{proof}\end{proofs}
\end{lemma}

\begin{lemma}\label{lemma:RA2(lnot-P):lnot-RA2(P)}
$\mathbf{RA2} (\lnot P) = \lnot \mathbf{RA2} (P)$
\begin{proofs}\begin{proof}\checkt{alcc}
\begin{flalign*}
	&\mathbf{RA2} (\lnot P)
	&&\ptext{Definition of $\mathbf{RA2}$}\\
	&=(\lnot P)[s \oplus \{ tr \mapsto \lseq\rseq\} ,\{ z | z \in ac' \land s.tr \le z.tr \spot z \oplus \{ tr \mapsto z.tr - s.tr \} \}/s,ac']
	&&\ptext{Property of substitution}\\
	&=\lnot P[s \oplus \{ tr \mapsto \lseq\rseq\} ,\{ z | z \in ac' \land s.tr \le z.tr \spot z \oplus \{ tr \mapsto z.tr - s.tr \} \}/s,ac']
	&&\ptext{Definition of $\mathbf{RA2}$}\\
	&=\lnot \mathbf{RA2} (P)
\end{flalign*}
\end{proof}\end{proofs}
\end{lemma}

\begin{lemma}\label{lemma:RA2(s.c)-s-not-tr:s.c} Where $c$ is not $tr$,
$\mathbf{RA2} (s.c) = s.c$
\begin{proofs}\begin{proof}\checkt{alcc}
\begin{flalign*}
	&\mathbf{RA2} (s.c)
	&&\ptext{Definition of $\mathbf{RA2}$}\\
	&=s.c[s \oplus \{ tr \mapsto \lseq\rseq\} ,\{ z | z \in ac' \land s.tr \le z.tr \spot z \oplus \{ tr \mapsto z.tr - s.tr \} \}/s,ac']
	&&\ptext{Substitution}\\
	&=(s\oplus\{tr\mapsto\lseq\rseq\}).c
	&&\ptext{Property of $\oplus$}\\
	&=s.c
\end{flalign*}
\end{proof}\end{proofs}
\end{lemma}

\begin{lemma}\label{lemma:RA2(P-land-ac'-neq-emptyset):RA2-o-RA1(P)}
$\mathbf{RA2} (P \land ac'\neq\emptyset) = \mathbf{RA2} \circ \mathbf{RA1} (P)$
\begin{proofs}\begin{proof}\checkt{alcc}\checkt{pfr}
\begin{xflalign*}
	&\mathbf{RA2} (P \land ac'\neq\emptyset)
	&&\ptext{\cref{theorem:RA2(P-land-Q):RA2(P)-land-RA2(Q)}}\\
	&=\mathbf{RA2} (P) \land \mathbf{RA2} (ac'\neq\emptyset)
	&&\ptext{\cref{lemma:RA2(ac'-neq-emptyset):RA1(true)}}\\
	&=\mathbf{RA2} (P) \land \mathbf{RA1} (true)
	&&\ptext{\cref{lemma:RA1(true),lemma:RA1-o-RA2(P):RA2(P)-land-RA1-subs}}\\
	&=\mathbf{RA1} \circ \mathbf{RA2} (P)
\end{xflalign*}
\end{proof}\end{proofs}
\end{lemma}

\begin{lemma}\label{lemma:RA2(P)-y-ac'-land-str-ytr}
\begin{statement}
\begin{align*}
	&\mathbf{RA2} (P)[\{y\}/ac'] \land s.tr \le y.tr\\
	&=\\
	&P[s\oplus\{tr \mapsto \lseq\rseq\},\{y\oplus\{tr\mapsto y.tr-s.tr\}\}/s,ac'] \land s.tr \le y.tr
\end{align*}
\end{statement}
\begin{proofs}
\begin{proof}
\begin{xflalign*}
	&\mathbf{RA2} (P)[\{y\}/ac'] \land s.tr \le y.tr
	&&\ptext{Definition of $\mathbf{RA2}$}\\
	&=\left(P\left[s\oplus\{tr \mapsto \lseq\rseq\},\left.\left\{z\left|\begin{array}{l}
		z\in ac'\land s.tr\le z.tr\\
		\spot z\oplus\{tr\mapsto z.tr-s.tr\}
	\end{array}\right.\right\}\right/s,ac'\right]\right)[\{y\}/ac'] \land s.tr \le y.tr
	&&\ptext{Substitution}\\
	&=P\left[s\oplus\{tr \mapsto \lseq\rseq\},\left.\left\{z\left|\begin{array}{l}
		z\in \{y\}\land s.tr\le z.tr\\
		\spot z\oplus\{tr\mapsto z.tr-s.tr\}
	\end{array}\right.\right\}\right/s,ac'\right] \land s.tr \le y.tr
	&&\ptext{Property of sets}\\
	&=P\left[s\oplus\{tr \mapsto \lseq\rseq\},\left.\left\{z\left|\begin{array}{l}
		z=y\land s.tr\le z.tr\\
		\spot z\oplus\{tr\mapsto z.tr-s.tr\}
	\end{array}\right.\right\}\right/s,ac'\right] \land s.tr \le y.tr
	&&\ptext{\cref{lemma:set-theory:P-land-z-eq-y-Q:Q-y-z}}\\
	&=P[s\oplus\{tr \mapsto \lseq\rseq\},\{y\oplus\{tr\mapsto y.tr-s.tr\}\}/s,ac'] \land s.tr \le y.tr
\end{xflalign*}
\end{proof}
\end{proofs}
\end{lemma}

\begin{lemma}\label{lemma:circledIn(RA1-o-RA2(P)-land-Q)}
\begin{statement}
Provided $ac'$ is not free in $Q$ and $P$ is $\mathbf{PBMH}$-healthy,
\begin{align*}
	&\circledIn{y}{ac'} (\mathbf{RA1}\circ\mathbf{RA2}(P) \land Q)\\
	&=\\
	&\exists y @ \left(\begin{array}{l}
		P[s\oplus\{tr \mapsto \lseq\rseq\},\{y\oplus\{tr\mapsto y.tr-s.tr\}\}/s,ac']
		\\ \land s.tr \le y.tr \land Q \land y \in ac'
	\end{array}\right)
\end{align*}
\end{statement}
\begin{proofs}
\begin{proof}
\begin{xflalign*}
	&\circledIn{y}{ac'} (\mathbf{RA1}\circ\mathbf{RA2}(P) \land Q)
	&&\ptext{Definition of $\circledIn{y}{ac'}$ (\cref{lemma:circledIn(P)-PBMH:exists-y-P})}\\
	&=\exists y @ (\mathbf{RA1}\circ\mathbf{RA2} (P) \land Q)[\{y\}/ac'] \land y \in ac'
	&&\ptext{Assumption: $ac'$ is not free in $Q$ and substitution}\\
	&=\exists y @ \mathbf{RA1}\circ\mathbf{RA2} (P)[\{y\}/ac'] \land Q \land y \in ac'
	&&\ptext{\cref{lemma:RA1-o-RA2(P):RA2(P)-land-RA1-subs,lemma:RA1(true)}}\\
	&=\exists y @ (\mathbf{RA2} (P) \land \mathbf{RA1} (true))[\{y\}/ac'] \land Q \land y \in ac'
	&&\ptext{Substitution}\\
	&=\exists y @ (\mathbf{RA2} (P)[\{y\}/ac'] \land \mathbf{RA1} (true)[\{y\}/ac']) \land Q \land y \in ac'
	&&\ptext{\cref{lemma:RA1(true)-y-for-ac':str-le-ytr}}\\
	&=\exists y @ (\mathbf{RA2} (P)[\{y\}/ac'] \land s.tr \le y.tr) \land Q \land y \in ac'
	&&\ptext{\cref{lemma:RA2(P)-y-ac'-land-str-ytr}}\\
	&=\exists y @ \left(\begin{array}{l}
		P[s\oplus\{tr \mapsto \lseq\rseq\},\{y\oplus\{tr\mapsto y.tr-s.tr\}\}/s,ac']
		\\ \land s.tr \le y.tr \land Q \land y \in ac'
	\end{array}\right)
\end{xflalign*}
\end{proof}
\end{proofs}
\end{lemma}

\begin{lemma}\label{lemma:RA2(x-in-ac'):exists-z-land-x-eq-z-oplus}
\begin{statement}
\begin{align*}
	&\mathbf{RA2} (x \in ac')\\
	&=\\
	&\exists z @ z\in ac'\land s.tr\le z.tr \land x = z\oplus\{tr\mapsto z.tr-s.tr\}
\end{align*}
\end{statement}
\begin{proofs}
\begin{proof}
\begin{xflalign*}
	&\mathbf{RA2} (x \in ac')
	&&\ptext{Definition of $\mathbf{RA2}$}\\
	&=(x \in ac')\left[s\oplus\{tr \mapsto \lseq\rseq\},\left.\left\{z\left|\begin{array}{l}
		z\in ac'\land s.tr\le z.tr\\
		\spot z\oplus\{tr\mapsto z.tr-s.tr\}
	\end{array}\right.\right\}\right/s,ac'\right]
	&&\ptext{Substitution}\\
	&=x \in \{z | z\in ac'\land s.tr\le z.tr \spot z\oplus\{tr\mapsto z.tr-s.tr\} \}
	&&\ptext{Property of sets}\\
	&=\exists z @ z\in ac'\land s.tr\le z.tr \land x = z\oplus\{tr\mapsto z.tr-s.tr\}
\end{xflalign*}
\end{proof} 
\end{proofs}
\end{lemma}

\begin{lemma}\label{lemma:RA2(circledIn(P-land-Q))-Q-ac'-not-free}
\begin{statement}
Provided $ac'$ is not free in $Q$ and $P$ is $\mathbf{PBMH}$-healthy,
\begin{align*}
	&\mathbf{RA2} (\circledIn{y}{ac'} (P \land Q))\\
	&=\\
	&\exists y @ \left(\begin{array}{l}
		P[s\oplus\{tr\mapsto\lseq\rseq\}/s][\{y\oplus\{tr\mapsto y.tr-s.tr\}\}/ac']
		\\ \land \\
		Q[s\oplus\{tr\mapsto\lseq\rseq\}/s][y\oplus\{tr\mapsto y.tr-s.tr\}/y]
		\\ \land \\
		y\in ac'\land s.tr\le y.tr 
	\end{array}\right)
\end{align*}
\end{statement}
\begin{proofs}
\begin{proof}
\begin{xflalign*}
	&\mathbf{RA2} (\circledIn{y}{ac'} (P \land Q))
	&&\ptext{Assumption: $P$ is $\mathbf{PBMH}$-healthy and $ac'$ is not free in $Q$}\\
	&&\ptext{\cref{law:pbmh:P:ac'-not-free,law:pbmh:conjunction-closure,lemma:circledIn(P)-PBMH:exists-y-P}}\\
	&=\mathbf{RA2} (\exists y @ (P \land Q)[\{y\}/ac'] \land y \in ac')
	&&\ptext{Assumption: $ac'$ not free in $Q$ and substitution}\\
	&=\mathbf{RA2} (\exists y @ P[\{y\}/ac'] \land Q \land y \in ac')
	&&\ptext{\cref{lemma:RA2(exists-P):exists-RA2(P)}}\\
	&=\exists y @ \mathbf{RA2} (P[\{y\}/ac'] \land Q \land y \in ac')
	&&\ptext{\cref{theorem:RA2(P-land-Q):RA2(P)-land-RA2(Q)}}\\
	&=\exists y @ \mathbf{RA2} (P[\{y\}/ac']) \land \mathbf{RA2} (Q) \land \mathbf{RA2} (y \in ac')
	&&\ptext{\cref{lemma:RA2(P):P:ac'-not-free}}\\
	&=\exists y @ P[s\oplus\{tr\mapsto\lseq\rseq\}/s][\{y\}/ac'] \land \mathbf{RA2} (Q) \land \mathbf{RA2} (y \in ac')
	&&\ptext{\cref{lemma:RA2(x-in-ac'):x-oplus-in-ac'}}\\
	&=\exists y @ P[s\oplus\{tr\mapsto\lseq\rseq\}/s][\{y\}/ac'] \land \mathbf{RA2} (Q) \land \mathbf{RA2} (y \in ac')
	&&\ptext{\cref{lemma:RA2(x-in-ac'):exists-z-land-x-eq-z-oplus}}\\
	&=\exists y @ \left(\begin{array}{l}
		P[s\oplus\{tr\mapsto\lseq\rseq\}/s][\{y\}/ac'] \land \mathbf{RA2} (Q) 
		\\ \land \\
		\exists z @ z\in ac'\land s.tr\le z.tr \land y = z\oplus\{tr\mapsto z.tr-s.tr\}
	\end{array}\right)
	&&\ptext{Predicate calculus}\\
	&=\exists y,z @ \left(\begin{array}{l}
		P[s\oplus\{tr\mapsto\lseq\rseq\}/s][\{y\}/ac'] \land \mathbf{RA2} (Q) 
		\\ \land \\
		z\in ac'\land s.tr\le z.tr \land y = z\oplus\{tr\mapsto z.tr-s.tr\}
	\end{array}\right)
	&&\ptext{One-point rule}\\
	&=\exists z @ \left(\begin{array}{l}
		P[s\oplus\{tr\mapsto\lseq\rseq\}/s][\{z\oplus\{tr\mapsto z.tr-s.tr\}\}/ac']
		\\ \land \\
		\mathbf{RA2} (Q)[z\oplus\{tr\mapsto z.tr-s.tr\}/y]
		\\ \land \\
		z\in ac'\land s.tr\le z.tr 
	\end{array}\right)
	&&\ptext{Assumption: $ac'$ is not free in $Q$ and~\cref{lemma:RA2(P):P:ac'-not-free}}\\
	&=\exists z @ \left(\begin{array}{l}
		P[s\oplus\{tr\mapsto\lseq\rseq\}/s][\{z\oplus\{tr\mapsto z.tr-s.tr\}\}/ac']
		\\ \land \\
		Q[s\oplus\{tr\mapsto\lseq\rseq\}/s][z\oplus\{tr\mapsto z.tr-s.tr\}/y]
		\\ \land \\
		z\in ac'\land s.tr\le z.tr 
	\end{array}\right)
	&&\ptext{Variable renaming $z$ to $y$}\\
	&=\exists y @ \left(\begin{array}{l}
		P[s\oplus\{tr\mapsto\lseq\rseq\}/s][\{y\oplus\{tr\mapsto y.tr-s.tr\}\}/ac']
		\\ \land \\
		Q[s\oplus\{tr\mapsto\lseq\rseq\}/s][y\oplus\{tr\mapsto y.tr-s.tr\}/y]
		\\ \land \\
		y\in ac'\land s.tr\le y.tr 
	\end{array}\right)
\end{xflalign*}
\end{proof}
\end{proofs}
\end{lemma}

\begin{theorem}\label{theorem:RA2(circledIn(P-land-Q)):circledIn(RA1-o-RA2(P)-land-Q)}
\begin{statement}
Provided $ac'$ is not free in $Q$, $P$ is $\mathbf{PBMH}$-healthy, and $Q = [s\oplus\{tr\mapsto\lseq\rseq\}/s][y\oplus\{tr\mapsto y.tr-s.tr\}/y]$,
\begin{align*}
	&\mathbf{RA2} (\circledIn{y}{ac'} (P \land Q))\\
	&=\\
	&\circledIn{y}{ac'} (\mathbf{RA1} \circ \mathbf{RA2} (P) \land Q)
\end{align*}
\end{statement}
\begin{proofs}
\begin{proof}
\begin{xflalign*}
	&\mathbf{RA2} (\circledIn{y}{ac'} (P \land Q))
	&&\ptext{Assumption and~\cref{lemma:RA2(circledIn(P-land-Q))-Q-ac'-not-free}}\\
	&=\exists y @ \left(\begin{array}{l}
		P[s\oplus\{tr\mapsto\lseq\rseq\}/s][\{y\oplus\{tr\mapsto y.tr-s.tr\}\}/ac']
		\\ \land \\
		Q[s\oplus\{tr\mapsto\lseq\rseq\}/s][y\oplus\{tr\mapsto y.tr-s.tr\}/y]
		\\ \land \\
		y\in ac'\land s.tr\le y.tr 
	\end{array}\right)
	&&\ptext{Assumption on $Q$}\\
	&=\exists y @ \left(\begin{array}{l}
		P[s\oplus\{tr\mapsto\lseq\rseq\}/s][\{y\oplus\{tr\mapsto y.tr-s.tr\}\}/ac']
		\\ \land \\
		Q \land y\in ac'\land s.tr\le y.tr 
	\end{array}\right)
	&&\ptext{\cref{lemma:circledIn(RA1-o-RA2(P)-land-Q)}}\\
	&=\circledIn{y}{ac'} (\mathbf{RA1} \circ \mathbf{RA2} (P) \land Q)
\end{xflalign*}
\end{proof}
\end{proofs}
\end{theorem}

\subsection{Substitution Properties}

\begin{lemma}\label{lemma:RA2(P)-o-w-subs:RA2(P-o-w-subs)}
$\mathbf{RA2} (P)^o_w = \mathbf{RA2} (P^o_w)$
\begin{proofs}\begin{proof}\checkt{pfr}\checkt{alcc}
\begin{flalign*}
	&\mathbf{RA2} (P)^o_w
	&&\ptext{Definition of $\mathbf{RA2}$}\\
	&=P[s \oplus \{ tr \mapsto \lseq\rseq\},\{ z | z \in ac' \land s.tr \le z.tr \spot z \oplus \{ tr \mapsto z.tr - s.tr \} \}/s,ac']^o_w
	&&\ptext{Substitution abbreviation}\\
	&=\left(P\left)\begin{aligned}
	&[s \oplus \{ tr \mapsto \lseq\rseq\},\{ z | z \in ac' \land s.tr \le z.tr \spot z \oplus \{ tr \mapsto z.tr - s.tr \} \}/s,ac']
	\\
	&[o,s\oplus\{wait\mapsto w\}/ok',s]
	\end{aligned}\right.\right.
	&&\ptext{Substitution}\\
	&=\left(P\left)\begin{aligned}
	&[o/ok'] \\
	&[s \oplus \{ tr \mapsto \lseq\rseq\},\{ z | z \in ac' \land s.tr \le z.tr \spot z \oplus \{ tr \mapsto z.tr - s.tr \} \}/s,ac']
	\\
	&[s\oplus\{wait\mapsto w\}/s]
	\end{aligned}\right.\right.
	&&\ptext{Substitution}\\
	&=\left(P\left)\begin{aligned}
		&\left[o/ok'\right] \\
		&\left[s\oplus\{wait\mapsto w\} \oplus \{ tr \mapsto \lseq\rseq\}/s\right] \\
		&\left[\left\{ z \left|\begin{array}{l}
				 z \in ac' \land s\oplus\{wait\mapsto w\}.tr \le z.tr \\
					\spot z \oplus \{ tr \mapsto z.tr - s\oplus\{wait\mapsto w\}.tr \}
			\end{array}\right.\right\}/ac'\right]
	\end{aligned}\right.\right.
	&&\ptext{Property of $\oplus$}\\
	&=\left(P\left)\begin{aligned}
		&\left[o/ok'\right] \\
		&\left[s\oplus\{wait\mapsto w\} \oplus \{ tr \mapsto \lseq\rseq\}/s\right]\\
		&\left[\{ z | z \in ac' \land s.tr \le z.tr \spot z \oplus \{ tr \mapsto z.tr - s.tr \}\}/ac'\right]
	\end{aligned}\right.\right.
	&&\ptext{Property of $\oplus$: distinct record components}\\
	&=\left(P\left)\begin{aligned}
		&\left[o/ok'\right] \\
		&\left[s\oplus \{ tr \mapsto \lseq\rseq\}\oplus\{wait\mapsto w\}/s\right]\\
		&\left[\{ z | z \in ac' \land s.tr \le z.tr \spot z \oplus \{ tr \mapsto z.tr - s.tr \}\}/ac'\right]
	\end{aligned}\right.\right.
	&&\ptext{Substitution}\\
	&=\left(P\left)\begin{aligned}
		&\left[o/ok'\right] \\
		&\left[s\oplus\{wait\mapsto w\}/s\right] \\
		&\left[s\oplus \{ tr \mapsto \lseq\rseq\}/s\right] \\
		&\left[\{ z | z \in ac' \land s.tr \le z.tr \spot z \oplus \{ tr \mapsto z.tr - s.tr \}\}/ac'\right]
	\end{aligned}\right.\right.
	&&\ptext{Substitution}\\
	&=\left(P\left)\begin{aligned}
		&\left[o,s\oplus\{wait\mapsto w\}/ok',s\right] \\
		&\left[s\oplus \{ tr \mapsto \lseq\rseq\},\{ z | z \in ac' \land s.tr \le z.tr \spot z \oplus \{ tr \mapsto z.tr - s.tr \}\}/s,ac'\right]
	\end{aligned}\right.\right.
	&&\ptext{Substitution abbreviation}\\
	&=\left(P^o_w\left)\begin{aligned}
		&\left[s\oplus \{ tr \mapsto \lseq\rseq\},\{ z | z \in ac' \land s.tr \le z.tr \spot z \oplus \{ tr \mapsto z.tr - s.tr \}\}/s,ac'\right]
	\end{aligned}\right.\right.
	&&\ptext{Definition of $\mathbf{RA2}$}\\
	&=\mathbf{RA2} (P^o_w)
\end{flalign*}
\end{proof}\end{proofs}
\end{lemma}

\subsection{Properties with respect to Designs}

\begin{lemma}\label{lemma:RA2(P|-Q):(lnot-RA2(lnot-P)|-RA2(Q))}
\checkt{pfr}
$\mathbf{RA2} (P \vdash Q) = (\lnot \mathbf{RA2} (\lnot P) \vdash \mathbf{RA2} (Q))$
\begin{proofs}\begin{proof}\checkt{alcc}
\begin{xflalign*}
	&\mathbf{RA2} (P \vdash Q)
	&&\ptext{Definition of design}\\
	&=\mathbf{RA2} ((ok \land P) \implies (Q \land ok'))
	&&\ptext{Predicate calculus}\\
	&=\mathbf{RA2} (\lnot ok \lor \lnot P \lor (Q \land ok'))
	&&\ptext{\cref{theorem:RA2(P-lor-Q):RA2(P)-lor-RA2(Q)}}\\
	&=\mathbf{RA2} (\lnot ok) \lor \mathbf{RA2} (\lnot P) \lor \mathbf{RA2} (Q \land ok')
	&&\ptext{\cref{theorem:RA2(P-land-Q):RA2(P)-land-RA2(Q)}}\\
	&=\mathbf{RA2} (\lnot ok) \lor \mathbf{RA2} (\lnot P) \lor (\mathbf{RA2} (Q) \land \mathbf{RA2} (ok'))
	&&\ptext{\cref{lemma:RA2(P):P:s-ac'-not-free}}\\
	&=\lnot ok \lor \mathbf{RA2} (\lnot P) \lor (\mathbf{RA2} (Q) \land ok')
	&&\ptext{Predicate calculus}\\
	&=(ok \land \lnot \mathbf{RA2} (\lnot P)) \implies (\mathbf{RA2} (Q) \land ok')
	&&\ptext{Definition of design}\\
	&=(\lnot \mathbf{RA2} (\lnot P) \vdash \mathbf{RA2} (Q))	
\end{xflalign*}
\end{proof}\end{proofs}
\end{lemma}

\begin{lemma}\label{lemma:RA2(P|-Q):RA2(P|-RA2(Q))}
$\mathbf{RA2} (P \vdash Q) = \mathbf{RA2} (P \vdash \mathbf{RA2} (Q))$
\begin{proofs}\begin{proof}\checkt{alcc}
\begin{xflalign*}
	&\mathbf{RA2} (P \vdash Q)
	&&\ptext{Definition of design}\\
	&=\mathbf{RA2} ((ok \land P) \implies (Q \land ok'))
	&&\ptext{Predicate calculus}\\
	&=\mathbf{RA2} (\lnot ok \lor \lnot P \lor (Q \land ok'))
	&&\ptext{\cref{theorem:RA2(P-lor-Q):RA2(P)-lor-RA2(Q),theorem:RA2:idempotent}}\\
	&=\mathbf{RA2} (\lnot ok \lor \lnot P \lor \mathbf{RA2} (Q \land ok'))
	&&\ptext{\cref{theorem:RA2(P-land-Q):RA2(P)-land-RA2(Q),lemma:RA2(P):P:s-ac'-not-free}}\\
	&=\mathbf{RA2} (\lnot ok \lor \lnot P \lor (\mathbf{RA2} (Q) \land ok'))
	&&\ptext{Predicate calculus}\\
	&=\mathbf{RA2} ((ok \land P) \implies (\mathbf{RA2} (Q) \land ok'))
	&&\ptext{Definition of design}\\
	&=\mathbf{RA2} (P \vdash \mathbf{RA2} (Q))
\end{xflalign*}
\end{proof}\end{proofs}
\end{lemma}

\subsection{Properties with respect to $\seqA$}

\begin{theorem}\label{theorem:RA2(P-seqA-RA2(Q)):RA2(P)-seqA-RA2(Q)}
\begin{statement}
$\mathbf{RA2} (P \seqA \mathbf{RA2} (Q)) = \mathbf{RA2} (P) \seqA \mathbf{RA2} (Q)$
\end{statement}
\begin{proofs}
\begin{proof}\checkt{alcc}
\begin{xflalign*}
	&\mathbf{RA2} (P \seqA \mathbf{RA2} (Q))
	&&\ptext{Definition of $\mathbf{RA2}$ (\cref{lemma:RA2:alternative-1})}\\
	&=\mathbf{RA2} (P \seqA Q[s\oplus\{tr\mapsto\lseq\rseq\},\{y | y \oplus \{tr\mapsto s.tr\cat y.tr\} \in ac'\}/s,ac'])
	&&\ptext{Definition of $\seqA$ and substitution}\\
	&=\mathbf{RA2} \left(P\begin{aligned}
			&\left[\left\{s \left|
\right\}\right/ac'\right] \\
	\end{aligned}\right.\right.
	&&\ptext{Property of sets, $\oplus$ and value of record component $tr$}\\
	&=\left(P\left)\begin{aligned}
			&\left[s\oplus\{tr\mapsto\lseq\rseq\}/s\right]\\
			&\left[\left\{z \left|%
	&&\ptext{Property of sets}\\
	&=%
	&&\ptext{Property of $\oplus$ and record component}\\
	&=%
\right\}\right/ac'\right] 
			\\
		\end{aligned}\right.\right.
	\end{array}	
\end{xflalign*}
\end{proof}
\end{proofs}
\end{lemma}

\begin{lemma}\label{lemma:RA2(P)-seqA-true:P-subs-seqA-true}
\begin{statement}
$\mathbf{RA2} (P) \seqA true = P[s\oplus\{tr\mapsto\lseq\rseq\}/s] \seqA true$
\end{statement}
\begin{proofs}
\begin{proof}\checkt{alcc}
\begin{xflalign*}
	&\mathbf{RA2} (P) \seqA true
	&&\ptext{\cref{lemma:RA2:alternative-1}}\\
	&=P[s\oplus\{tr\mapsto\lseq\rseq\},\{y | y \oplus \{tr\mapsto s.tr\cat y.tr\} \in ac'\}/s,ac'] \seqA true
	&&\ptext{Definition of $\seqA$}\\
	&=P[s\oplus\{tr\mapsto\lseq\rseq\},\{y | y \oplus \{tr\mapsto s.tr\cat y.tr\} \in ac'\}/s,ac'][\{s|true\}/ac']
	&&\ptext{Substitution}\\
	&=P[s\oplus\{tr\mapsto\lseq\rseq\},\{y | y \oplus \{tr\mapsto s.tr\cat y.tr\} \in \{s|true\}\}/s,ac']
	&&\ptext{Property of sets}\\
	&=P[s\oplus\{tr\mapsto\lseq\rseq\},\{y | true\}/s,ac']
	&&\ptext{Property of substitution: $ac'$ not free in $s$}\\
	&=P[s\oplus\{tr\mapsto\lseq\rseq\}/s][\{y | true\}/ac']
	&&\ptext{Definition of $\seqA$}\\
	&=P[s\oplus\{tr\mapsto\lseq\rseq\}/s] \seqA true
\end{xflalign*}
\end{proof}
\end{proofs}
\end{lemma}

\subsection{Properties with respect to $\mathbf{A2}$}

\begin{theorem}\label{theorem:A2-o-RA2-o-A2(P):RA2-o-A2(P)}
\begin{statement}
$\mathbf{A2} \circ \mathbf{RA2} \circ \mathbf{A2} (P) = \mathbf{RA2} \circ \mathbf{A2} (P)$
\end{statement}
\begin{proofs}
\begin{proof}\checkt{alcc}
\begin{xflalign*}
	&\mathbf{A2} \circ \mathbf{RA2} \circ \mathbf{A2} (P)
	&&\ptext{\cref{lemma:RA2-o-A2(P)}}\\
	&=\mathbf{A2} \left(
\right)
	&&\ptext{Substitution and predicate calculus}\\
	&=\left(%
\right)
	&&\ptext{Property of sets}\\
	&=\left(%
\right)
	&&\ptext{Predicate calculus and property of sets}\\
	&=\left(%
\right)
	&&\ptext{Predicate calculus: absorption law}\\
	&=\left(%
\right)
\end{align*}
\end{statement}
\begin{proofs}
\begin{proof}\checkt{alcc}
\begin{xflalign*}
	&\mathbf{RA2} \circ \mathbf{A2} (P)
	&&\ptext{Definition of $\mathbf{A2}$ (\cref{lemma:A2:alternative-2:disjunction})}\\
	&=\mathbf{RA2} (P[\emptyset/ac'] \lor (\exists y \spot P[\{y\}/ac'] \land y \in ac'))
	&&\ptext{\cref{theorem:RA2(P-lor-Q):RA2(P)-lor-RA2(Q)}}\\
	&=\mathbf{RA2} (P[\emptyset/ac']) \lor \mathbf{RA2} (\exists y \spot P[\{y\}/ac'] \land y \in ac')
	&&\ptext{\cref{lemma:RA2(exists-P):exists-RA2(P)}}\\
	&=\mathbf{RA2} (P[\emptyset/ac']) \lor (\exists y \spot \mathbf{RA2} (P[\{y\}/ac'] \land y \in ac'))
	&&\ptext{\cref{theorem:RA2(P-land-Q):RA2(P)-land-RA2(Q)}}\\
	&=\mathbf{RA2} (P[\emptyset/ac']) \lor (\exists y \spot \mathbf{RA2} (P[\{y\}/ac']) \land \mathbf{RA2} (y \in ac'))
	&&\ptext{\cref{lemma:RA2(P):P:ac'-not-free}}\\
	&=P[\emptyset/ac'][s\oplus\{tr\mapsto\lseq\rseq\}/s] \lor (\exists y \spot P[\{y\}/ac'][s\oplus\{tr\mapsto\lseq\rseq\}/s] \land \mathbf{RA2} (y \in ac'))
	&&\ptext{\cref{lemma:RA2(x-in-ac'):x-oplus-in-ac'}}\\
	&=\left(\begin{array}{l}
		P[\emptyset/ac'][s\oplus\{tr\mapsto\lseq\rseq\}/s]
		\\ \lor \\
		(\exists y \spot P[\{y\}/ac'][s\oplus\{tr\mapsto\lseq\rseq\}/s] \land y \oplus \{tr\mapsto s.tr\cat y.tr\} \in ac')
	\end{array}\right)
\end{xflalign*}
\end{proof}
\end{proofs}
\end{lemma}

\section{$\mathbf{RA3}$}

\subsection{Definition}
\theoremstatementref{def:RA3}

\subsection{Properties}

\begin{theorem}\label{theorem:RA3(P-land-Q):RA3(P)-land-RA3(Q)}
\begin{statement}
$\mathbf{RA3} (P \land Q) = \mathbf{RA3} (P) \land \mathbf{RA3} (Q)$
\end{statement}
\begin{proofs}
\begin{proof}\checkt{alcc}\checkt{pfr}
\begin{xflalign*}
	&\mathbf{RA3} (P \land Q)
	&&\ptext{Definition of $\mathbf{RA3}$}\\
	&=\IIRac \dres s.wait \rres (P \land Q)
	&&\ptext{Definition of conditional}\\
	&=(s.wait \land \IIRac) \lor (\lnot s.wait \land P \land Q)
	&&\ptext{Predicate calculus}\\
	&=(s.wait \land \IIRac) \lor ((\lnot s.wait \land P) \land (\lnot s.wait \land Q))
	&&\ptext{Predicate calculus}\\
	&=((s.wait \land \IIRac) \lor (\lnot s.wait \land P)) \land ((s.wait \land \IIRac) \lor (\lnot s.wait \land Q))
	&&\ptext{Definition of conditional}\\
	&=(\IIRac \dres s.wait \rres P) \land (\IIRac \dres s.wait \rres Q)
	&&\ptext{Definition of $\mathbf{RA3}$}\\
	&=\mathbf{RA3} (P) \land \mathbf{RA3} (Q)	
\end{xflalign*}
\end{proof}
\end{proofs}
\end{theorem}

\begin{theorem}\label{theorem:RA3(P-lor-Q):RA3(P)-lor-RA3(Q)}
\begin{statement}
$\mathbf{RA3} (P \lor Q) = \mathbf{RA3} (P) \lor \mathbf{RA3} (Q)$
\end{statement}
\begin{proofs}
\begin{proof}\checkt{pfr}\checkt{alcc}
\begin{xflalign*}
	&\mathbf{RA3} (P \lor Q)
	&&\ptext{Definition of $\mathbf{RA3}$}\\
	&=\IIRac \dres s.wait \rres (P \lor Q)
	&&\ptext{Definition of conditional}\\
	&=(s.wait \land \IIRac) \lor (\lnot s.wait \land (P \lor Q))
	&&\ptext{Predicate calculus}\\
	&=(s.wait \land \IIRac) \lor (\lnot s.wait \land P) \lor (\lnot s.wait \land Q)
	&&\ptext{Predicate calculus}\\
	&=(s.wait \land \IIRac) \lor (\lnot s.wait \land P) \lor (s.wait \land \IIRac) \lor (\lnot s.wait \land Q)
	&&\ptext{Definition of conditional}\\
	&=(\IIRac \dres s.wait \rres P) \lor (\IIRac \dres s.wait \rres Q)
	&&\ptext{Definition of $\mathbf{RA3}$}\\
	&=\mathbf{RA3} (P) \lor \mathbf{RA3} (Q)	
\end{xflalign*}
\end{proof}
\end{proofs}
\end{theorem}

\begin{theorem}\label{theorem:RA3(P-seqA-Q)-closure} 
\begin{statement}
Provided $P$ and $Q$ are $\mathbf{RA3}$-healthy and $Q$ is $\mathbf{RA1}$-healthy,
\begin{align*}
	&\mathbf{RA3} (P \seqA Q) = P \seqA Q
\end{align*}
\end{statement}
\begin{proofs}
\begin{proof}
\begin{xflalign*}
	&P \seqA Q
	&&\ptext{Assumption: $P$ is $\mathbf{RA3}$-healthy}\\
	&=\mathbf{RA3} (P) \seqA Q
	&&\ptext{Definition of $\mathbf{RA3}$}\\
	&=(\IIRac \dres s.wait \rres P) \seqA Q
	&&\ptext{\cref{lemma:seqA:conditional-right}}\\
	&=(\IIRac \seqA Q) \dres s.wait \rres (P \seqA Q)
	&&\ptext{Definition of $\IIRac$}\\
	&=((\mathbf{RA1} (\lnot ok) \lor (ok' \land s \in ac')) \seqA Q) \dres s.wait \rres (P \seqA Q)
	&&\ptext{\cref{law:seqA-right-distributivity}}\\
	&=((\mathbf{RA1} (\lnot ok) \seqA Q) \lor ((ok' \land s \in ac') \seqA Q)) \dres s.wait \rres (P \seqA Q)
	&&\ptext{Assumption: $Q$ is $\mathbf{RA1}$ and $\mathbf{RA3}$-healthy and~\cref{theorem:RA1(lnot-ok)-seqA-P:RA1(lnot-ok)}}\\
	&=(\mathbf{RA1} (\lnot ok) \lor ((ok' \land s \in ac') \seqA Q)) \dres s.wait \rres (P \seqA Q)
	&&\ptext{\cref{law:seqA-right-distributivity-conjunction}}\\
	&=(\mathbf{RA1} (\lnot ok) \lor ((ok' \seqA Q) \land (s \in ac' \seqA Q))) \dres s.wait \rres (P \seqA Q)
	&&\ptext{\cref{law:seqA-ac'-not-free}}\\
	&=(\mathbf{RA1} (\lnot ok) \lor (ok' \land (s \in ac' \seqA Q))) \dres s.wait \rres (P \seqA Q)
	&&\ptext{\cref{law:seqA:IIA:left-unit}}\\
	&=(\mathbf{RA1} (\lnot ok) \lor (ok' \land Q)) \dres s.wait \rres (P \seqA Q)
	&&\ptext{Assumption: $Q$ is $\mathbf{RA3}$-healthy}\\
	&=(\mathbf{RA1} (\lnot ok) \lor (ok' \land (\IIRac \dres s.wait \rres Q))) \dres s.wait \rres (P \seqA Q)
	&&\ptext{Property of conditional}\\
	&=(\mathbf{RA1} (\lnot ok) \lor (ok' \land \IIRac)) \dres s.wait \rres (P \seqA Q)
	&&\ptext{Definition of $\IIRac$}\\
	&=\left(\begin{array}{l}
		\mathbf{RA1} (\lnot ok) 
		\\ \lor \\
		(ok' \land (\mathbf{RA1} (\lnot ok) \lor (ok' \land s \in ac')))
	\end{array}\right) \dres s.wait \rres (P \seqA Q)
	&&\ptext{Predicate calculus}\\
	&=\left(\begin{array}{l}
		\mathbf{RA1} (\lnot ok) 
		\\ \lor \\
		(ok' \land \mathbf{RA1} (\lnot ok))
		\\ \lor \\
		(ok' \land s \in ac')
	\end{array}\right) \dres s.wait \rres (P \seqA Q)
	&&\ptext{Predicate calculus: absorption law}\\
	&=\left(\begin{array}{l}
		\mathbf{RA1} (\lnot ok) 
		\\ \lor \\
		(ok' \land s \in ac')
	\end{array}\right) \dres s.wait \rres (P \seqA Q)
	&&\ptext{Definition of $\IIRac$}\\
	&=\IIRac \dres s.wait \rres (P \seqA Q)
	&&\ptext{Definition of $\mathbf{RA3}$}\\
	&=\mathbf{RA3} (P \seqA Q)
\end{xflalign*}
\end{proof}
\end{proofs}
\end{theorem}

\begin{theorem}\label{theorem:PBMH-o-RA3(P):RA3(P)}
\begin{statement}$\mathbf{PBMH} \circ \mathbf{RA3} \circ \mathbf{PBMH} (P) = \mathbf{RA3} \circ \mathbf{PBMH} (P)$\end{statement}
\begin{proofs}
\begin{proof}\checkt{alcc}\checkt{pfr}
\begin{xflalign*}
	&\mathbf{PBMH} \circ \mathbf{RA3} \circ \mathbf{PBMH} (P)
	&&\ptext{Definition of $\mathbf{RA3}$}\\
	&=\mathbf{PBMH} (\IIRac \dres s.wait \rres \mathbf{PBMH} (P))
	&&\ptext{\cref{lemma:PBMH(conditional)}}\\
	&=\mathbf{PBMH} (\IIRac) \dres s.wait \rres \mathbf{PBMH} \circ \mathbf{PBMH} (P)
	&&\ptext{\cref{theorem:PBMH(IIRac):IIRac}}\\
	&=\IIRac \dres s.wait \rres \mathbf{PBMH} \circ \mathbf{PBMH} (P)
	&&\ptext{\cref{law:pbmh:idempotent}}\\
	&=\IIRac \dres s.wait \rres \mathbf{PBMH} (P)
	&&\ptext{Definition of $\mathbf{RA3}$}\\
	&=\mathbf{RA3} \circ \mathbf{PBMH} (P)
\end{xflalign*}
\end{proof}
\end{proofs}
\end{theorem}

\begin{theorem}\label{theorem:RA3-o-RA1:RA1-o-RA3}
\begin{statement}
$\mathbf{RA3} \circ \mathbf{RA1} (P) = \mathbf{RA3} \circ \mathbf{RA1} (P)$
\end{statement}
\begin{proofs}
\begin{proof}\checkt{alcc}\checkt{pfr}
\begin{xflalign*}
	&\mathbf{RA1} \circ \mathbf{RA3} (P)
	&&\ptext{Definition of $\mathbf{RA3}$}\\
	&=\mathbf{RA1} (\IIRac \dres s.wait \rres P)
	&&\ptext{\cref{lemma:RA1(conditional)}}\\
	&=\mathbf{RA1} (\IIRac) \dres s.wait \rres \mathbf{RA1} (P)
	&&\ptext{\cref{theorem:RA1(IIRac):IIRac}}\\
	&=\IIRac \dres s.wait \rres \mathbf{RA1} (P)
	&&\ptext{Definition of $\mathbf{RA3}$}\\
	&=\mathbf{RA3} \circ \mathbf{RA1} (P)
\end{xflalign*}
\end{proof}
\end{proofs}
\end{theorem}

\begin{theorem}\label{theorem:RA3-o-RA2:RA2-o-RA3}
\begin{statement}
$\mathbf{RA2} \circ \mathbf{RA3} (P) = \mathbf{RA3} \circ \mathbf{RA2} (P)$
\end{statement}
\begin{proofs}
\begin{proof}\checkt{alcc}\checkt{pfr}
\begin{flalign*}
	&\mathbf{RA2} \circ \mathbf{RA3} (P)
	&&\ptext{Definition of $\mathbf{RA3}$}\\
	&=\mathbf{RA2} (\IIRac \dres s.wait \rres P)
	&&\ptext{\cref{lemma:RA2:conditional-no-s.tr} and $s.wait$ is $\mathbf{RA2}$-healthy}\\
	&=\mathbf{RA2} (\IIRac) \dres s.wait \rres \mathbf{RA2} (P)
	&&\ptext{\cref{theorem:RA2(IIRac):IIRac}}\\
	&=\IIRac \dres s.wait \rres \mathbf{RA2} (P)
	&&\ptext{Definition of $\mathbf{RA3}$}\\
	&=\mathbf{RA3} \circ \mathbf{RA2} (P)
\end{flalign*}
\end{proof}
\end{proofs}
\end{theorem}

\begin{theorem}\label{theorem:RA1(IIRac):IIRac}
$\mathbf{RA1} (\IIRac) = \IIRac$
\begin{proofs}\begin{proof}\checkt{alcc}\checkt{pfr}
\begin{xflalign*}
	&\mathbf{RA1} (\IIRac)
	&&\ptext{Definition of $\IIRac$}\\
	&=\mathbf{RA1} (\mathbf{RA1} (\lnot ok) \lor (ok' \land s \in ac'))
	&&\ptext{Distributivity of $\mathbf{RA1}$ (\cref{lemma:RA1(P-lor-Q):RA1(P)-lor-RA1(Q)})}\\
	&=\mathbf{RA1} \circ \mathbf{RA1} (\lnot ok) \lor \mathbf{RA1} (ok' \land s \in ac')
	&&\ptext{\cref{lemma:RA1(P-land-Q):ac'-not-free}}\\
	&=\mathbf{RA1} \circ \mathbf{RA1} (\lnot ok) \lor (ok' \land \mathbf{RA1} (s \in ac'))
	&&\ptext{\cref{lemma:RA1(s-in-ac'):s-in-ac'}}\\
	&=\mathbf{RA1} \circ \mathbf{RA1} (\lnot ok) \lor (ok' \land s \in ac')
	&&\ptext{$\mathbf{RA1}$-idempotent (\cref{theorem:RA1-idempotent})}\\
	&=\mathbf{RA1} (\lnot ok) \lor (ok' \land s \in ac')
	&&\ptext{Definition of $\IIRac$}\\
	&=\IIRac
\end{xflalign*}
\end{proof}\end{proofs}
\end{theorem}

\begin{theorem}\label{theorem:RA2(IIRac):IIRac}
$\mathbf{RA2} (\IIRac) = \IIRac$
\begin{proofs}\begin{proof}\checkt{alcc}\checkt{pfr}
\begin{xflalign*}
	&\mathbf{RA2} (\IIRac)
	&&\ptext{Definition of $\IIRac$}\\
	&=\mathbf{RA2} ((\lnot ok \land \mathbf{RA1} (true)) \lor (ok' \land s \in ac'))
	&&\ptext{Distributivity of $\mathbf{RA2}$ (\cref{theorem:RA2(P-lor-Q):RA2(P)-lor-RA2(Q)})}\\
	&=\mathbf{RA2} (\lnot ok \land \mathbf{RA1} (true)) \lor \mathbf{RA2} (ok' \land s \in ac')
	&&\ptext{Distributivity of $\mathbf{RA2}$ (\cref{theorem:RA2(P-land-Q):RA2(P)-land-RA2(Q)})}\\
	&=(\mathbf{RA2} (\lnot ok) \land \mathbf{RA2} \circ \mathbf{RA1} (true)) \lor (\mathbf{RA2} (ok') \land \mathbf{RA2} (s \in ac'))
	&&\ptext{\cref{lemma:RA2(P):P:s-ac'-not-free}}\\
	&=(\lnot ok \land \mathbf{RA2} \circ \mathbf{RA1} (true)) \lor (ok' \land \mathbf{RA2} (s \in ac'))
	&&\ptext{\cref{lemma:RA2(s-in-ac'):s-in-ac'}}\\
	&=(\lnot ok \land \mathbf{RA2} \circ \mathbf{RA1} (true)) \lor (ok' \land s \in ac')
	&&\ptext{\cref{theorem:RA2-o-RA1:RA1-o-RA2}}\\
	&=(\lnot ok \land \mathbf{RA1} \circ \mathbf{RA2} (true)) \lor (ok' \land s \in ac')
	&&\ptext{\cref{lemma:RA2(true):true}}\\
	&=(\lnot ok \land \mathbf{RA1} (true)) \lor (ok' \land s \in ac')	
	&&\ptext{Definition of $\IIRac$}\\
	&=\IIRac
\end{xflalign*}
\end{proof}\end{proofs}
\end{theorem}

\begin{theorem}\label{theorem:RA3(IIRac):IIRac}
$\mathbf{RA3} (\IIRac) = \IIRac$
\begin{proofs}\begin{proof}\checkt{alcc}\checkt{pfr}
\begin{xflalign*}
	&\mathbf{RA3} (\IIRac)
	&&\ptext{Definition of $\mathbf{RA3}$}\\
	&=\IIRac \dres s.wait \rres \IIRac
	&&\ptext{Property of conditonal}\\
	&=\IIRac
\end{xflalign*}
\end{proof}\end{proofs}
\end{theorem}

\begin{theorem}\label{theorem:PBMH(IIRac):IIRac}
$\mathbf{PBMH} (\IIRac) = \IIRac$
\begin{proofs}\begin{proof}\checkt{alcc}\checkt{pfr}
\begin{flalign*}
	&\mathbf{PBMH} (\IIRac)
	&&\ptext{Definition of $\IIRac$}\\
	&=\mathbf{PBMH} ((\lnot ok \land \exists z \spot s.tr \le z.tr \land z \in ac') \lor (ok' \land s \in ac'))
	&&\ptext{Distributivity of $\mathbf{PBMH}$}\\
	&=\left(\begin{array}{l}
		\mathbf{PBMH} (\lnot ok \land \exists z \spot s.tr \le z.tr \land z \in ac')
		\\ \lor \\
		\mathbf{PBMH} (ok' \land s \in ac')
	\end{array}\right)
	&&\ptext{\cref{lemma:PBMH(c-land-P):c-land-PBMH(P)}}\\
	&=\left(\begin{array}{l}
		(\lnot ok \land \mathbf{PBMH} (\exists z \spot s.tr \le z.tr \land z \in ac'))
		\\ \lor \\
		(ok' \land \mathbf{PBMH} (s \in ac'))
	\end{array}\right)
	&&\ptext{\cref{lemma:PBMH(x-in-ac'):x-in-ac'}}\\
	&=\left(\begin{array}{l}
		(\lnot ok \land \exists z \spot s.tr \le z.tr \land z \in ac')
		\\ \lor \\
		(ok' \land s \in ac')
	\end{array}\right)
	&&\ptext{Definition of $\IIRac$}\\
	&=\IIRac
\end{flalign*}
\end{proof}\end{proofs}
\end{theorem}

\begin{theorem}\label{theorem:RA3-idempotent}
\begin{statement}
$\mathbf{RA3} \circ \mathbf{RA3} (P) = \mathbf{RA3} (P)$
\end{statement}
\begin{proofs}
\begin{proof}\checkt{pfr}\checkt{alcc}
\begin{flalign*}
	&\mathbf{RA3} \circ \mathbf{RA3} (P)
	&&\ptext{Definition of $\mathbf{RA3}$}\\
	&=\IIRac \dres s.wait \rres \mathbf{RA3} (P)
	&&\ptext{Definition of $\mathbf{RA3}$}\\
	&=\IIRac \dres s.wait \rres (\IIRac \dres s.wait \rres P)
	&&\ptext{Defiition of conditional}\\
	&=(s.wait \land \IIRac) \lor (\lnot s.wait \land (\IIRac \dres s.wait \rres P))
	&&\ptext{Property of conditional}\\
	&=(s.wait \land \IIRac) \lor (\lnot s.wait \land P)
	&&\ptext{Defiition of conditional}\\
	&=\IIRac \dres s.wait \rres P
	&&\ptext{Definition of $\mathbf{RA3}$}\\
	&=\mathbf{RA3} (P)
\end{flalign*}
\end{proof}
\end{proofs}
\end{theorem}

\begin{theorem}\label{theorem:RA3-monotonic}
\begin{statement}
$P \sqsubseteq Q \implies \mathbf{RA3} (P) \sqsubseteq \mathbf{RA3} (Q)$
\end{statement}
\begin{proofs}
\begin{proof}\checkt{pfr}\checkt{alcc}
\begin{xflalign*}
	&\mathbf{RA3} (Q)
	&&\ptext{Assumption: $P \sqsubseteq Q = [Q \implies P]$}\\
	&=\mathbf{RA3} (Q \land P)
	&&\ptext{\cref{theorem:RA3(P-land-Q):RA3(P)-land-RA3(Q)}}\\
	&=\mathbf{RA3} (Q) \land \mathbf{RA3} (P)
	&&\ptext{Predicate calculus}\\
	&\sqsupseteq\mathbf{RA3} (P)
\end{xflalign*}
\end{proof}
\end{proofs}
\end{theorem}

\subsubsection{Properties with respect to $\mathbf{PBMH}$}

\begin{theorem}\label{theorem:PBMH-o-RA3(P):RA3-o-PBMH(P)}
$\mathbf{PBMH} \circ \mathbf{RA3} (P) = \mathbf{RA3} \circ \mathbf{PBMH} (P)$
\begin{proofs}\begin{proof}\checkt{pfr}
\begin{xflalign*}
	&\mathbf{PBMH} \circ \mathbf{RA3} (P)
	&&\ptext{Definition of $\mathbf{RA3}$}\\
	&=\mathbf{PBMH} (\IIRac \dres s.wait \rres P)
	&&\ptext{\cref{lemma:PBMH(conditional)}}\\
	&=\mathbf{PBMH} (\IIRac) \dres s.wait \rres \mathbf{PBMH} (P)
	&&\ptext{\cref{theorem:PBMH(IIRac):IIRac}}\\
	&=\IIRac \dres s.wait \rres \mathbf{PBMH} (P)
	&&\ptext{Definition of $\mathbf{RA3}$}\\
	&=\mathbf{RA3} \circ \mathbf{PBMH} (P)
\end{xflalign*}
\end{proof}\end{proofs}
\end{theorem}

\subsubsection{Properties with respect to $\mathbf{A2}$}

\begin{theorem}\label{theorem:A2-o-RA3(P):RA3-o-A2(P)}
\begin{statement}
$\mathbf{A2} \circ \mathbf{RA3} (P) = \mathbf{RA3} \circ \mathbf{A2} (P)$
\end{statement}
\begin{proofs}
\begin{proof}\checkt{alcc}
\begin{xflalign*}
	&\mathbf{A2} \circ \mathbf{RA3} (P)
	&&\ptext{Definition of $\mathbf{RA3}$}\\
	&=\mathbf{A2} (\IIRac \dres s.wait \rres P)
	&&\ptext{\cref{lemma:A2(P-cond-Q):A2(P)-cond-A2(Q)}}\\
	&=\mathbf{A2} (\IIRac) \dres s.wait \rres \mathbf{A2} (P)
	&&\ptext{\cref{lemma:A2(IIRac):IIRac}}\\
	&=\IIRac \dres s.wait \rres \mathbf{A2} (P)
	&&\ptext{Definition of $\mathbf{RA3}$}\\
	&=\mathbf{RA3}\circ\mathbf{A2} (P)
\end{xflalign*}
\end{proof}
\end{proofs}
\end{theorem}

\begin{theorem}\label{theorem:A2-o-RA3-o-A2(P):RA3-o-A2(P)}
\begin{statement}
$\mathbf{A2} \circ \mathbf{RA3} \circ \mathbf{A2} (P) = \mathbf{RA3} \circ \mathbf{A2} (P)$
\end{statement}
\begin{proofs}
\begin{proof}\checkt{alcc}
\begin{xflalign*}
	&\mathbf{RA3} \circ \mathbf{A2} (P)
	&&\ptext{\cref{theorem:A2-o-A2(P):A2(P)}}\\
	&=\mathbf{RA3} \circ \mathbf{A2} \circ \mathbf{A2} (P)
	&&\ptext{\cref{theorem:A2-o-RA3(P):RA3-o-A2(P)}}\\
	&=\mathbf{A2} \circ \mathbf{RA3} \circ \mathbf{A2} (P)
\end{xflalign*}
\end{proof}
\end{proofs}
\end{theorem}

\begin{lemma}\label{lemma:A2(IIRac):IIRac}
\begin{statement}
$\mathbf{A2} (\IIRac) = \IIRac$
\end{statement}
\begin{proofs}
\begin{proof}\checkt{alcc}
\begin{xflalign*}
	&\mathbf{A2} (\IIRac)
	&&\ptext{Definition of $\IIRac$}\\
	&=\mathbf{A2} (\mathbf{RA1} (\lnot ok) \lor (ok' \land s \in ac'))
	&&\ptext{\cref{theorem:A2(P-lor-Q):A2(P)-lor-A2(Q)}}\\
	&=\mathbf{A2} \circ \mathbf{RA1} (\lnot ok) \lor \mathbf{A2} (ok' \land s \in ac')
	&&\ptext{\cref{lemma:A2(P)-ac'-not-free:P,theorem:A2-o-RA1-o-A2(P):RA1-o-A2(P)}}\\
	&=\mathbf{RA1} (\lnot ok) \lor \mathbf{A2} (ok' \land s \in ac')
	&&\ptext{\cref{lemma:A2(P-land-Q)-ac'-not-free:P-land-A2(Q)}}\\
	&=\mathbf{RA1} (\lnot ok) \lor (ok' \land \mathbf{A2} (s \in ac'))
	&&\ptext{\cref{lemma:A2(x-in-ac'):x-in-ac'}}\\
	&=\mathbf{RA1} (\lnot ok) \lor (ok' \land s \in ac'))
	&&\ptext{Definition of $\IIRac$}\\
	&=\IIRac
\end{xflalign*}
\end{proof}
\end{proofs}
\end{lemma}

\subsection{Substitution Lemmas}

\begin{lemma}\label{lemma:RA3:s-oplus-wait-false}
\begin{statement}$\mathbf{RA3} (P) = \mathbf{RA3} (P_f)$\end{statement}
\begin{proofs}
\begin{proof}\checkt{alcc}\checkt{pfr}
\begin{xflalign*}
	&\mathbf{RA3} (P)
	&&\ptext{Definition of $\mathbf{RA3}$}\\
	&=(\IIRac \dres s.wait \rres P)
	&&\ptext{Definition of conditional and predicate calculus}\\
	&=(\IIRac \dres s.wait \rres (\lnot s.wait \land P))
	&&\ptext{Predicate calculus}\\
	&=(\IIRac \dres s.wait \rres (s.wait = false \land P))
	&&\ptext{\cref{lemma:state-s-Leibniz-P}}\\
	&=(\IIRac \dres s.wait \rres (s.wait = false \land P[s \oplus \{wait \mapsto false\}/s]))
	&&\ptext{Definition of conditional and predicate calculus}\\
	&=(\IIRac \dres s.wait \rres P[s \oplus \{wait \mapsto false\}/s])
	&&\ptext{Definition of $\mathbf{RA3}$}\\
	&=\mathbf{RA3} (P[s \oplus \{wait \mapsto false\}/s])
	&&\ptext{Substitution abbreviation}\\
	&=\mathbf{RA3} (P_f)
\end{xflalign*}
\end{proof}
\end{proofs}
\end{lemma}

\begin{lemma}\label{lemma:RA3-o-f-subs:P-o-f}
$\mathbf{RA3} (P)^o_f = P^o_f$
\begin{proofs}\begin{proof}\checkt{alcc}
\begin{flalign*}
	&\mathbf{RA3} (P)^o_f
	&&\ptext{\cref{lemma:RA3-o-w-subs}}\\
	&=(\IIRac)^o_f \dres false \rres P^o_f
	&&\ptext{Property of conditional}\\
	&=P^o_f
\end{flalign*}
\end{proof}\end{proofs}
\end{lemma}

\begin{lemma}\label{lemma:RA3-o-w-subs}
$\mathbf{RA3} (P)^o_w = (\IIRac)^o_w \dres w \rres P^o_w$
\begin{proofs}\begin{proof}
\begin{flalign*}
	&\mathbf{RA3} (P)^o_w
	&&\ptext{Definition of $\mathbf{RA3}$}\\
	&=(\IIRac \dres s.wait \rres P)^o_w
	&&\ptext{Substitution abbreviation}\\
	&=(\IIRac \dres s.wait \rres P)[o,s\oplus\{wait\mapsto w\}/ok',s]
	&&\ptext{Substitution}\\
	&=(\IIRac[o,s\oplus\{wait\mapsto w\}/ok',s] \dres (s\oplus\{wait\mapsto w\}).wait \rres P[o,s\oplus\{wait\mapsto w\}/ok',s])
	&&\ptext{Value of record component}\\
	&=(\IIRac[o,s\oplus\{wait\mapsto w\}/ok',s] \dres w \rres P[o,s\oplus\{wait\mapsto w\}/ok',s])
	&&\ptext{Substitution abbreviation}\\
	&=(\IIRac)^o_w \dres w \rres P^o_w
\end{flalign*}
\end{proof}\end{proofs}
\end{lemma}

\section{$\mathbf{RA}$}

\subsection{Definition}
\theoremstatementref{def:RA}

\begin{theorem}\label{theorem:RA-o-A(design):RA-CSPA-PBMH}
\begin{statement}
$\mathbf{RAD} (P) = \mathbf{RA} \circ \mathbf{A} (\lnot P^f_f \vdash P^t_f)$
\end{statement}
\begin{proofs}
\begin{proof}
\begin{xflalign*}
	&\mathbf{RAD} (P)
	&&\ptext{Definition of $\mathbf{RAD}$}\\
	&=\mathbf{RA3} \circ \mathbf{RA2} \circ \mathbf{RA1} \circ \mathbf{CSPA1} \circ \mathbf{CSPA2} \circ \mathbf {PBMH} (P)
	&&\ptext{\cref{theorem:RA1-o-CSPA1:RA1-o-H1}}\\
	&=\mathbf{RA3} \circ \mathbf{RA2} \circ \mathbf{RA1} \circ \mathbf{H1} \circ \mathbf{CSPA2} \circ \mathbf {PBMH} (P)
	&&\ptext{$\mathbf{CSPA2}$ is $\mathbf{H2}$}\\
	&=\mathbf{RA3} \circ \mathbf{RA2} \circ \mathbf{RA1} \circ \mathbf{H1} \circ \mathbf{H2} \circ \mathbf {PBMH} (P)
	&&\ptext{\cref{theorem:RA1-o-A0:RA1}}\\
	&=\mathbf{RA3} \circ \mathbf{RA2} \circ \mathbf{RA1} \circ \mathbf{A0} \circ \mathbf{H1} \circ \mathbf{H2} \circ \mathbf {PBMH} (P)
	&&\ptext{\cref{theorem:H2-o-PBMH:PBMH-o-H2,theorem:H1-o-PBMH:PBMH-o-H1}}\\
	&=\mathbf{RA3} \circ \mathbf{RA2} \circ \mathbf{RA1} \circ \mathbf{A0} \circ \mathbf {PBMH} \circ \mathbf{H1} \circ \mathbf{H2} (P)
	&&\raisetag{12pt}\ptext{Definition of design}\\
	&=\mathbf{RA3} \circ \mathbf{RA2} \circ \mathbf{RA1} \circ \mathbf{A0} \circ \mathbf {PBMH} (\lnot P^f \vdash P^t)
	&&\ptext{Definition of $\mathbf{A}$}\\
	&=\mathbf{RA3} \circ \mathbf{RA2} \circ \mathbf{RA1} \circ \mathbf{A} (\lnot P^f \vdash P^t)
	&&\ptext{\cref{theorem:RA3-o-RA2:RA2-o-RA3,theorem:RA3-o-RA1:RA1-o-RA3,theorem:RA2-o-RA1:RA1-o-RA2}}\\
	&=\mathbf{RA1} \circ \mathbf{RA2} \circ \mathbf{RA3} \circ \mathbf{A} (\lnot P^f \vdash P^t)
	&&\ptext{\cref{lemma:RA3:s-oplus-wait-false,lemma:A-substitution-s}}\\
	&=\mathbf{RA1} \circ \mathbf{RA2} \circ \mathbf{RA3} \circ \mathbf{A} ((\lnot P^f \vdash P^t)_f)
	&&\ptext{Substitution}\\
	&=\mathbf{RA1} \circ \mathbf{RA2} \circ \mathbf{RA3} \circ \mathbf{A} (\lnot P^f_f \vdash P^t_f)
	&&\ptext{Definition of $\mathbf{RA}$}\\
	&=\mathbf{RA} \circ \mathbf{A} (\lnot P^f_f \vdash P^t_f)
\end{xflalign*}
\end{proof}
\end{proofs}
\end{theorem}

\begin{theorem}\label{theorem:PBMH(P)-RAP:P}
\begin{statement} Provided $P$ is $\mathbf{RAD}$-healthy,
$\mathbf{PBMH} (P) = P$.
\end{statement}
\begin{proofs}
\begin{proof}\checkt{alcc}
\begin{xflalign*}
	&\mathbf{PBMH} (P)
	&&\ptext{Assumption: $P$ is $\mathbf{RAP}$-healthy}\\
	&=\mathbf{PBMH} \circ \mathbf{RAP} (P)
	&&\ptext{Definition of $\mathbf{RAP}$}\\
	&=\mathbf{PBMH} \circ \mathbf{RA} \circ \mathbf{A} (\lnot P^f_f \vdash P^t_f)
	&&\ptext{\cref{theorem:RA-o-A(P):RA-o-PBMH(P)}}\\
	&=\mathbf{PBMH} \circ \mathbf{RA} \circ \mathbf{PBMH} (\lnot P^f_f \vdash P^t_f)
	&&\ptext{\cref{theorem:PBMH-o-RA(P):RA(P)}}\\
	&=\mathbf{RA} \circ \mathbf{PBMH} (\lnot P^f_f \vdash P^t_f)
	&&\ptext{\cref{theorem:RA-o-A(P):RA-o-PBMH(P)}}\\
	&=\mathbf{RA} \circ \mathbf{A} (\lnot P^f_f \vdash P^t_f)
	&&\ptext{Definition of $\mathbf{RAP}$}\\
	&=\mathbf{RAP} (P)
	&&\ptext{Assumption: $P$ is $\mathbf{RAP}$-healthy}\\
	&=P
\end{xflalign*}
\end{proof}
\end{proofs}
\end{theorem}

\begin{lemma}\label{lemma:RAD(P):RA(lnot-PBMH(P)ff|-PBMH(P)tf)}
\begin{statement}
$\mathbf{RAD} (P) = \mathbf{RA} (\lnot \mathbf{PBMH} (P)^f_f \vdash \mathbf{PBMH} (P)^t_f)$
\end{statement}
\begin{proofs}
\begin{proof}\checkt{alcc}
\begin{xflalign*}
	&\mathbf{RAD} (P)
	&&\ptext{\cref{theorem:RA-o-A(design):RA-CSPA-PBMH}}\\
	&=\mathbf{RA} \circ \mathbf{A} (\lnot P^f_f \vdash P^t_f)
	&&\ptext{\cref{theorem:RA-o-A(P):RA-o-PBMH(P)}}\\
	&=\mathbf{RA} \circ \mathbf{PBMH} (\lnot P^f_f \vdash P^t_f)
	&&\ptext{$\mathbf{PBMH}$ is $\mathbf{A1}$}\\
	&=\mathbf{RA} (\lnot \mathbf{PBMH} (P^f_f) \vdash \mathbf{PBMH} (P^t_f))
	&&\ptext{\cref{lemma:PBMH(P)-ow:PBMH(P-ow)}}\\
	&=\mathbf{RA} (\lnot \mathbf{PBMH} (P)^f_f \vdash \mathbf{PBMH} (P)^t_f)
\end{xflalign*}
\end{proof}
\end{proofs}
\end{lemma}

\begin{theorem}\label{theorem:RA(P-land-Q):RA(P)-land-RA(Q)}
$\mathbf{RA} (P \land Q) = \mathbf{RA} (P) \land \mathbf{RA} (Q)$
\begin{proofs}\begin{proof}\checkt{pfr}\checkt{alcc}
\begin{xflalign*}
	&\mathbf{RA} (P \land Q)
	&&\ptext{Definition of $\mathbf{RA}$}\\
	&=\mathbf{RA1} \circ \mathbf{RA2} \circ \mathbf{RA3} (P \land Q)
	&&\ptext{\cref{theorem:RA3(P-land-Q):RA3(P)-land-RA3(Q)}}\\
	&=\mathbf{RA1} \circ \mathbf{RA2} (\mathbf{RA3} (P) \land \mathbf{RA3} (Q))
	&&\ptext{\cref{theorem:RA2(P-land-Q):RA2(P)-land-RA2(Q)}}\\
	&=\mathbf{RA1} (\mathbf{RA2} \circ \mathbf{RA3} (P) \land \mathbf{RA2} \circ \mathbf{RA3} (Q))
	&&\ptext{\cref{lemma:RA1(P-land-Q):RA1(P)-land-RA1(Q)}}\\
	&=\mathbf{RA1} \circ \mathbf{RA2} \circ \mathbf{RA3} (P) \land \mathbf{RA1} \circ \mathbf{RA2} \circ \mathbf{RA3} (Q)
	&&\ptext{Definition of $\mathbf{RA}$}\\
	&=\mathbf{RA} (P) \land \mathbf{RA} (Q)	
\end{xflalign*}
\end{proof}\end{proofs}
\end{theorem}

\begin{theorem}\label{theorem:RA(P-lor-Q):RA(P)-lor-RA(Q)}
$\mathbf{RA} (P \lor Q) = \mathbf{RA} (P) \lor \mathbf{RA} (Q)$
\begin{proofs}\begin{proof}\checkt{pfr}\checkt{alcc}
\begin{xflalign*}
	&\mathbf{RA} (P \lor Q)
	&&\ptext{Definition of $\mathbf{RA}$}\\
	&=\mathbf{RA1} \circ \mathbf{RA2} \circ \mathbf{RA3} (P \lor Q)
	&&\ptext{\cref{theorem:RA3(P-lor-Q):RA3(P)-lor-RA3(Q)}}\\
	&=\mathbf{RA1} \circ \mathbf{RA2} (\mathbf{RA3} (P) \lor \mathbf{RA3} (Q))
	&&\ptext{\cref{theorem:RA2(P-lor-Q):RA2(P)-lor-RA2(Q)}}\\
	&=\mathbf{RA1} (\mathbf{RA2} \circ \mathbf{RA3} (P) \lor \mathbf{RA2} \circ \mathbf{RA3} (Q))
	&&\ptext{\cref{lemma:RA1(P-lor-Q):RA1(P)-lor-RA1(Q)}}\\
	&=\mathbf{RA1} \circ \mathbf{RA2} \circ \mathbf{RA3} (P) \lor \mathbf{RA1} \circ \mathbf{RA2} \circ \mathbf{RA3} (Q)
	&&\ptext{Definition of $\mathbf{RA}$}\\
	&=\mathbf{RA} (P) \lor \mathbf{RA} (Q)	
\end{xflalign*}
\end{proof}\end{proofs}
\end{theorem}

\begin{theorem}\label{theorem:RA-idempotent}
$\mathbf{RA} \circ \mathbf{RA} (P) = \mathbf{RA} (P)$
\begin{proofs}\begin{proof}\checkt{pfr}\checkt{alcc}
\begin{xflalign*}
	&\mathbf{RA} \circ \mathbf{RA} (P)
	&&\ptext{Definition of $\mathbf{RA}$}\\
	&=\mathbf{RA3} \circ \mathbf{RA2} \circ \mathbf{RA1} \circ \mathbf{RA3} \circ \mathbf{RA2} \circ \mathbf{RA1} (P)
	&&\ptext{\cref{theorem:RA2-o-RA1:RA1-o-RA2}}\\
	&=\mathbf{RA3} \circ \mathbf{RA2} \circ \mathbf{RA1} \circ \mathbf{RA3} \circ \mathbf{RA1} \circ \mathbf{RA2} (P)
	&&\ptext{\cref{theorem:RA3-o-RA1:RA1-o-RA3}}\\
	&=\mathbf{RA3} \circ \mathbf{RA2} \circ \mathbf{RA1} \circ \mathbf{RA1} \circ \mathbf{RA3} \circ \mathbf{RA2} (P)
	&&\ptext{\cref{theorem:RA1-idempotent}}\\
	&=\mathbf{RA3} \circ \mathbf{RA2} \circ \mathbf{RA1} \circ \mathbf{RA3} \circ \mathbf{RA2} (P)
	&&\ptext{\cref{theorem:RA3-o-RA2:RA2-o-RA3}}\\
	&=\mathbf{RA2} \circ \mathbf{RA3} \circ \mathbf{RA1} \circ \mathbf{RA3} \circ \mathbf{RA2} (P)
	&&\ptext{\cref{theorem:RA3-o-RA1:RA1-o-RA3}}\\
	&=\mathbf{RA2} \circ \mathbf{RA1} \circ \mathbf{RA3} \circ \mathbf{RA3} \circ \mathbf{RA2} (P)
	&&\ptext{\cref{theorem:RA3-idempotent}}\\
	&=\mathbf{RA2} \circ \mathbf{RA1} \circ \mathbf{RA3} \circ \mathbf{RA2} (P)
	&&\ptext{\cref{theorem:RA3-o-RA1:RA1-o-RA3}}\\
	&=\mathbf{RA2} \circ \mathbf{RA3} \circ \mathbf{RA1} \circ \mathbf{RA2} (P)
	&&\ptext{\cref{theorem:RA3-o-RA2:RA2-o-RA3}}\\
	&=\mathbf{RA3} \circ \mathbf{RA2} \circ \mathbf{RA1} \circ \mathbf{RA2} (P)
	&&\ptext{\cref{theorem:RA2-o-RA1:RA1-o-RA2}}\\
	&=\mathbf{RA3} \circ \mathbf{RA1} \circ \mathbf{RA2} \circ \mathbf{RA2} (P)
	&&\ptext{\cref{theorem:RA2:idempotent}}\\
	&=\mathbf{RA3} \circ \mathbf{RA1} \circ \mathbf{RA2} (P)
	&&\ptext{\cref{theorem:RA2-o-RA1:RA1-o-RA2}}\\
	&=\mathbf{RA3} \circ \mathbf{RA2} \circ \mathbf{RA1} (P)
	&&\ptext{Definition of $\mathbf{RA}$}\\
	&=\mathbf{RA} (P)
\end{xflalign*}
\end{proof}\end{proofs}
\end{theorem}

\begin{theorem}\label{theorem:PBMH-o-RA(P):RA(P)} Provided $P$ is $\mathbf{PBMH}$-healthy,
\begin{align*}
	&\mathbf{PBMH} \circ \mathbf{RA} (P) = \mathbf{RA} (P)
\end{align*}
\begin{proofs}\begin{proof}\checkt{pfr}\checkt{alcc}
\begin{xflalign*}
	&\mathbf{RA} (P)
	&&\ptext{Definition of $\mathbf{RA}$}\\
	&=\mathbf{RA3} \circ \mathbf{RA2} \circ \mathbf{RA1} (P)
	&&\ptext{Assumption: $P$ is $\mathbf{PBMH}$-healthy and \cref{theorem:PBMH-o-RA1(P):RA1(P)}}\\
	&=\mathbf{RA3} \circ \mathbf{RA2} \circ \mathbf{PBMH} \circ \mathbf{RA1} (P)
	&&\ptext{\cref{theorem:PBMH-o-RA2(P):RA2(P)}}\\
	&=\mathbf{RA3} \circ \mathbf{PBMH} \circ \mathbf{RA2} \circ \mathbf{PBMH} \circ \mathbf{RA1} (P)
	&&\ptext{\cref{theorem:PBMH-o-RA3(P):RA3(P)}}\\
	&=\mathbf{PBMH} \circ \mathbf{RA3} \circ \mathbf{PBMH} \circ \mathbf{RA2} \circ \mathbf{PBMH} \circ \mathbf{RA1} (P)
	&&\ptext{\cref{theorem:PBMH-o-RA2(P):RA2(P)}}\\
	&=\mathbf{PBMH} \circ \mathbf{RA3} \circ \mathbf{RA2} \circ \mathbf{PBMH} \circ \mathbf{RA1} (P)
	&&\ptext{Assumption: $P$ is $\mathbf{PBMH}$-healthy and \cref{theorem:PBMH-o-RA1(P):RA1(P)}}\\
	&=\mathbf{PBMH} \circ \mathbf{RA3} \circ \mathbf{RA2} \circ \mathbf{RA1} (P)
	&&\ptext{Definition of $\mathbf{RA}$}\\
	&=\mathbf{PBMH} \circ \mathbf{RA} (P)
\end{xflalign*}
\end{proof}\end{proofs}
\end{theorem}\noindent

\begin{theorem}
\begin{align*}
	&\mathbf{RA} \circ \mathbf{A} (\lnot (\mathbf{RA} \circ \mathbf{A} (\lnot P^f_f \vdash P^t_f))^f_f \vdash (\mathbf{RA} \circ \mathbf{A} (\lnot P^f_f \vdash P^t_f))^t_f)\\
	&=\\
	&\mathbf{RA} \circ \mathbf{A} (\lnot P^f_f \vdash P^t_f)
\end{align*}
\begin{proofs}\begin{proof}
\begin{xflalign*}
	&\mathbf{RA} \circ \mathbf{A} (\lnot (\mathbf{RA} \circ \mathbf{A} (\lnot P^f_f \vdash P^t_f))^f_f \vdash (\mathbf{RA} \circ \mathbf{A} (\lnot P^f_f \vdash P^t_f))^t_f)
	&&\ptext{\cref{lemma:RA-o-A(design)-ow-ok'-false-wait-false,lemma:RA-o-A(design)-ow-ok'-true-wait-false}}\\
	&=\mathbf{RA} \circ \mathbf{A} \left(
\right)
	&&\ptext{Definition of design}\\
	&=\mathbf{RA} \circ \mathbf{A} \left(%
\right)
	&&\ptext{Predicate calculus: absorption law}\\
	&=\mathbf{RA} \circ \mathbf{PBMH} (\lnot ok \lor P^f_f \lor (P^t_f \land ok'))
	&&\ptext{Predicate calculus}\\
	&=\mathbf{RA} \circ \mathbf{PBMH} ((ok \land \lnot P^f_f) \implies (P^t_f \land ok'))
	&&\ptext{Definition of design}\\
	&=\mathbf{RA} \circ \mathbf{PBMH} (\lnot P^f_f \vdash P^t_f)
	&&\ptext{\cref{theorem:RA-o-A(P):RA-o-PBMH(P)}}\\
	&=\mathbf{RA} \circ \mathbf{A} (\lnot P^f_f \vdash P^t_f)
\end{xflalign*}
\end{proof}\end{proofs}
\end{theorem}

\begin{lemma}\label{lemma:RA1-o-RA3(design):RA1(design)}
\begin{align*}
	&\mathbf{RA1} \circ \mathbf{RA3} (P \vdash Q)\\
	&=\\
	&\mathbf{RA1} ((true \dres s.wait \rres P) \vdash (s \in ac' \dres s.wait \rres Q))
\end{align*}
\begin{proofs}\begin{proof}\checkt{alcc}\checkt{pfr}
\begin{xflalign*}
	&\mathbf{RA1} \circ \mathbf{RA3} (P \vdash Q)
	&&\ptext{Definition of design}\\
	&=\mathbf{RA1} \circ \mathbf{RA3} (((ok \land P) \implies (Q \land ok')))
	&&\ptext{Predicate calculus}\\
	&=\mathbf{RA1} \circ \mathbf{RA3} ((\lnot ok \lor \lnot P \lor (Q \land ok')))
	&&\ptext{\cref{theorem:RA3(P-lor-Q):RA3(P)-lor-RA3(Q)}}\\
	&=\mathbf{RA1} (\mathbf{RA3} (\lnot ok) \lor \mathbf{RA3} (\lnot P) \lor \mathbf{RA3} (Q \land ok'))
	&&\ptext{\cref{lemma:RA1(P-lor-Q):RA1(P)-lor-RA1(Q)}}\\
	&=\mathbf{RA1} \circ \mathbf{RA3} (\lnot ok) \lor \mathbf{RA1} \circ \mathbf{RA3} (\lnot P) \lor \mathbf{RA1} \circ \mathbf{RA3} (Q \land ok'))
	&&\ptext{\cref{lemma:RA1-o-RA3(lnot-ok):RA1(lnot-ok)-lor-(s.wait-land-IIRac)}}\\
	&=\left(
\right)
	&&\ptext{Definition of conditional and predicate calculus}\\	
	&=\left(%
\right)
	&&\ptext{Definition of $\IIRac$ and predicate calculus}\\
	&=\left(%
\right)
	&&\ptext{Predicate calculus: absorption law}\\
	&=\left(%
\right)
	&&\ptext{Definition of conditional}\\
	&=\left(%
\right)
	&&\ptext{\cref{lemma:RA1(P-lor-Q):RA1(P)-lor-RA1(Q)}}\\
	&=\mathbf{RA1} (\lnot ok \lor (false \dres s.wait \rres \lnot P) \lor ((s \in ac' \dres s.wait \rres Q) \land ok'))
	&&\ptext{Predicate calculus}\\
	&=\mathbf{RA1} ((ok \land \lnot (false \dres s.wait \rres \lnot P)) \implies ((s \in ac' \dres s.wait \rres Q) \land ok'))
	&&\ptext{\cref{lemma:lnot-(false-cond-Q):(true-cond-lnot-Q)}}\\
	&=\mathbf{RA1} ((ok \land (true \dres s.wait \rres P)) \implies ((s \in ac' \dres s.wait \rres Q) \land ok'))
	&&\ptext{Definition of design}\\
	&=\mathbf{RA1} ((true \dres s.wait \rres P) \vdash (s \in ac' \dres s.wait \rres Q))
\end{xflalign*}
\end{proof}\end{proofs}
\end{lemma}

\begin{lemma}\label{lemma:RA1-o-RA3(lnot-ok):RA1(lnot-ok)-lor-(s.wait-land-IIRac)}
\begin{align*}
	&\mathbf{RA1} \circ \mathbf{RA3} (\lnot ok) = \mathbf{RA1} (\lnot ok) \lor (s.wait \land \IIRac)
\end{align*}
\begin{proofs}\begin{proof}\checkt{alcc}\checkt{pfr}
\begin{xflalign*}
	&\mathbf{RA1} \circ \mathbf{RA3} (\lnot ok)
	&&\ptext{Definition of $\mathbf{RA3}$}\\
	&=\mathbf{RA1} (\IIRac \dres s.wait \rres (\lnot ok))
	&&\ptext{\cref{lemma:RA1(conditional)}}\\
	&=\mathbf{RA1} (\IIRac) \dres s.wait \rres \mathbf{RA1} (\lnot ok)
	&&\ptext{\cref{theorem:RA1(IIRac):IIRac}}\\
	&=\IIRac \dres s.wait \rres \mathbf{RA1} (\lnot ok)
	&&\ptext{\cref{lemma:IIRac:alternative-2}}\\
	&=(\mathbf{RA1} (\lnot ok) \lor (ok' \land s \in ac')) \dres s.wait \rres \mathbf{RA1} (\lnot ok)
	&&\ptext{Definition of conditional}\\
	&=(s.wait \land \mathbf{RA1} (\lnot ok)) \lor (s.wait \land ok' \land s \in ac') \lor (\lnot s.wait \land \mathbf{RA1} (\lnot ok))
	&&\ptext{Predicate calculus}\\
	&=\mathbf{RA1} (\lnot ok) \lor (s.wait \land ok' \land s \in ac')
	&&\ptext{Predicate calculus: absorption law}\\
	&=\mathbf{RA1} (\lnot ok) \lor (s.wait \land \mathbf{RA1} (\lnot ok)) \lor (s.wait \land ok' \land s \in ac')
	&&\ptext{Predicate calculus}\\
	&=\mathbf{RA1} (\lnot ok) \lor (s.wait \land (\mathbf{RA1} (\lnot ok) \lor (ok' \land s \in ac')))
	&&\ptext{\cref{lemma:IIRac:alternative-2}}\\
	&=\mathbf{RA1} (\lnot ok) \lor (s.wait \land \IIRac)
\end{xflalign*}
\end{proof}\end{proofs}
\end{lemma}

\begin{lemma}\label{lemma:IIRac:alternative-2}
$\IIRac = \mathbf{RA1} (\lnot ok) \lor (ok' \land s \in ac')$
\begin{proofs}\begin{proof}
\begin{xflalign*}
	&\IIRac
	&&\ptext{Definition of $\IIRac$}\\
	&=(\lnot ok \land \mathbf{RA1} (true)) \lor (ok' \land s \in ac')	
	&&\ptext{\cref{lemma:RA1(lnot-ok):lnot-ok-land-RA1(true)}}\\
	&=\mathbf{RA1} (\lnot ok) \lor (ok' \land s \in ac')
\end{xflalign*}
\end{proof}\end{proofs}
\end{lemma}

\begin{lemma}\label{lemma:RA1-o-RA3(P):(s.wait-land-IIRac)-lor-RA1-o-RA3(P)}
\begin{align*}
	&\mathbf{RA1} \circ \mathbf{RA3} (P) = (s.wait \land \IIRac) \lor \mathbf{RA1} \circ \mathbf{RA3} (P)
\end{align*}
\begin{proofs}\begin{proof}
\begin{xflalign*}
	&\mathbf{RA1} \circ \mathbf{RA3} (P)
	&&\ptext{Definition of $\mathbf{RA3}$}\\
	&=\mathbf{RA1} (\IIRac \dres s.wait \rres P)
	&&\ptext{Definition of conditional and predicate calculus}\\
	&=\mathbf{RA1} ((s.wait \land \IIRac) \lor (\IIRac \dres s.wait \rres P))
	&&\ptext{\cref{lemma:RA1(P-lor-Q):RA1(P)-lor-RA1(Q)}}\\
	&=\mathbf{RA1} (s.wait \land \IIRac) \lor \mathbf{RA1} (\IIRac \dres s.wait \rres P)
	&&\ptext{\cref{lemma:RA1(P-land-Q):ac'-not-free}}\\
	&=(s.wait \land \mathbf{RA1} (\IIRac)) \lor \mathbf{RA1} (\IIRac \dres s.wait \rres P)
	&&\ptext{\cref{theorem:RA1(IIRac):IIRac}}\\
	&=(s.wait \land \IIRac) \lor \mathbf{RA1} (\IIRac \dres s.wait \rres P)
	&&\ptext{Definition of $\mathbf{RA3}$}\\
	&=(s.wait \land \IIRac) \lor \mathbf{RA1} \circ \mathbf{RA3} (P)
\end{xflalign*}
\end{proof}\end{proofs}
\end{lemma}

\begin{lemma}\label{lemma:RA1-o-RA3(P):IIrac-RA1(P)}
$\mathbf{RA1} \circ \mathbf{RA3} (P) = \IIRac \dres s.wait \rres \mathbf{RA1} (P)$
\begin{proofs}\begin{proof}\checkt{alcc}\checkt{pfr}
\begin{xflalign*}
	&\mathbf{RA1} \circ \mathbf{RA3} (P)
	&&\ptext{Definition of $\mathbf{RA3}$}\\
	&=\mathbf{RA1} (\IIRac \dres s.wait \rres P)
	&&\ptext{\cref{lemma:RA1(conditional)}}\\
	&=\mathbf{RA1} (\IIRac) \dres s.wait \rres \mathbf{RA1} (P)
	&&\ptext{\cref{theorem:RA1(IIRac):IIRac}}\\
	&=\IIRac \dres s.wait \rres \mathbf{RA1} (P)
\end{xflalign*}
\end{proof}\end{proofs}
\end{lemma}

\begin{lemma}\label{lemma:RA(P)-ow-subs:RA2-o-RA1(P-ow-subs)}
$\mathbf{RA} (P)^o_f = \mathbf{RA2} \circ \mathbf{RA1} (P^o_f)$
\begin{proofs}\begin{proof}\checkt{alcc}\checkt{pfr}
\begin{xflalign*}
	&\mathbf{RA} (P)^o_f
	&&\ptext{Definition of $\mathbf{RA}$}\\
	&=(\mathbf{RA3} \circ \mathbf{RA2} \circ \mathbf{RA1} (P))^o_f
	&&\ptext{\cref{lemma:RA3-o-f-subs:P-o-f}}\\
	&=(\mathbf{RA2} \circ \mathbf{RA1} (P))^o_f
	&&\ptext{\cref{lemma:RA2(P)-o-w-subs:RA2(P-o-w-subs)}}\\
	&=\mathbf{RA2} \circ (\mathbf{RA1} (P))^o_f
	&&\ptext{\cref{lemma:RA1(P)-o-w-subs:RA1(P-o-w-subs)}}\\
	&=\mathbf{RA2} \circ \mathbf{RA1} (P^o_f)
\end{xflalign*}
\end{proof}\end{proofs}
\end{lemma}

\begin{lemma}\label{lemma:RA-o-A(design)-ow}
\begin{align*}
	&(\mathbf{RA} \circ \mathbf{A} (\lnot P^f_f \vdash P^t_f))^o_w\\
	&=\\
	&\mathbf{RA2} \circ \mathbf{RA1} \circ \mathbf{PBMH} (\lnot ok \lor P^f_f \lor (P^t_f \land o)) 
\end{align*}
\begin{proofs}\begin{proof}\checkt{alcc}
\begin{xflalign*}
	&(\mathbf{RA} \circ \mathbf{A} (\lnot P^f_f \vdash P^t_f))^o_w
	&&\ptext{\cref{theorem:RA-o-A(P):RA-o-PBMH(P)}}\\
	&=(\mathbf{RA} \circ \mathbf{PBMH} (\lnot P^f_f \vdash P^t_f))^o_w
	&&\ptext{\cref{lemma:RA(P)-ow-subs:RA2-o-RA1(P-ow-subs)}}\\
	&=\mathbf{RA2} \circ \mathbf{RA1} \circ (\mathbf{PBMH} (\lnot P^f_f \vdash P^t_f))^o_w
	&&\ptext{\cref{lemma:PBMH(P)-ow:PBMH(P-ow)}}\\
	&=\mathbf{RA2} \circ \mathbf{RA1} \circ \mathbf{PBMH} (\lnot P^f_f \vdash P^t_f)^o_w
	&&\ptext{Definition of design}\\
	&=\mathbf{RA2} \circ \mathbf{RA1} \circ \mathbf{PBMH} ((ok \land \lnot P^f_f) \implies (P^t_f \land ok'))^o_w
	&&\ptext{Substitution}\\
	&=\mathbf{RA2} \circ \mathbf{RA1} \circ \mathbf{PBMH} ((ok \land \lnot (P^f_f)^o_w) \implies ((P^t_f)^o_w \land o))
	&&\ptext{Substitution: $ok'$ not free and property of $\oplus$}\\
	&=\mathbf{RA2} \circ \mathbf{RA1} \circ \mathbf{PBMH} ((ok \land \lnot P^f_f) \implies (P^t_f \land o))
	&&\ptext{Predicate calculus}\\
	&=\mathbf{RA2} \circ \mathbf{RA1} \circ \mathbf{PBMH} (\lnot ok \lor P^f_f \lor (P^t_f \land o))
\end{xflalign*}
\end{proof}\end{proofs}
\end{lemma}

\begin{lemma}\label{lemma:RA-o-A(design)-ow-ok'-false-wait-false}
\begin{align*}
	&(\mathbf{RA} \circ \mathbf{A} (\lnot P^f_f \vdash P^t_f))^f_f\\
	&=\\
	&\mathbf{RA2} \circ \mathbf{RA1} \circ \mathbf{PBMH} (\lnot ok \lor P^f_f)
\end{align*}
\begin{proofs}\begin{proof}\checkt{alcc}
\begin{flalign*}
	&(\mathbf{RA} \circ \mathbf{A} (\lnot P^f_f \vdash P^t_f))^f_f
	&&\ptext{\cref{lemma:RA-o-A(design)-ow}}\\
	&=\mathbf{RA2} \circ \mathbf{RA1} \circ \mathbf{PBMH} (\lnot ok \lor P^f_f \lor (P^t_f \land false))
	&&\ptext{Predicate calculus}\\
	&=\mathbf{RA2} \circ \mathbf{RA1} \circ \mathbf{PBMH} (\lnot ok \lor P^f_f)
\end{flalign*}
\end{proof}\end{proofs}
\end{lemma}

\begin{lemma}\label{lemma:RA-o-A(design)-ow-ok'-true-wait-false}
\begin{align*}
	&(\mathbf{RA} \circ \mathbf{A} (\lnot P^f_f \vdash P^t_f))^t_f\\
	&=\\
	&\mathbf{RA2} \circ \mathbf{RA1} \circ \mathbf{PBMH} (\lnot ok \lor P^f_f \lor P^t_f)
\end{align*}
\begin{proofs}\begin{proof}\checkt{alcc}
\begin{flalign*}
	&(\mathbf{RA} \circ \mathbf{A} (\lnot P^f_f \vdash P^t_f))^t_f
	&&\ptext{\cref{lemma:RA-o-A(design)-ow}}\\
	&=\mathbf{RA2} \circ \mathbf{RA1} \circ \mathbf{PBMH} (\lnot ok \lor P^f_f \lor (P^t_f \land true))
	&&\ptext{Predicate calculus}\\
	&=\mathbf{RA2} \circ \mathbf{RA1} \circ \mathbf{PBMH} (\lnot ok \lor P^f_f \lor P^t_f)
\end{flalign*}
\end{proof}\end{proofs}
\end{lemma}

\begin{lemma}\label{lemma:exists-ac'-RA1-o-RA2-o-PBMH:exists-ac'-RA2-o-PBMH}
\begin{align*}
	&\exists ac' \spot \mathbf{RA1} \circ \mathbf{RA2} \circ \mathbf{PBMH} (P) = \exists ac' \spot \mathbf{RA2} \circ \mathbf{PBMH} (P)
\end{align*}
\begin{proofs}\begin{proof}\checkt{alcc}
\begin{xflalign*}
	&\exists ac' \spot \mathbf{RA1} \circ \mathbf{RA2} \circ \mathbf{PBMH} (P)
	&&\ptext{\cref{theorem:PBMH-o-RA2(P):RA2(P),theorem:PBMH-o-RA1(P):RA1(P)}}\\
	&=\exists ac' \spot \mathbf{PBMH} \circ \mathbf{RA1} \circ \mathbf{RA2} \circ \mathbf{PBMH} (P)
	&&\ptext{\cref{lemma:PBMH(P)-seqA-true:exists-ac'-P}}\\
	&=\mathbf{PBMH} \circ \mathbf{RA1} \circ \mathbf{RA2} \circ \mathbf{PBMH} (P) \seqA true
	&&\ptext{\cref{theorem:PBMH-o-RA2(P):RA2(P),theorem:PBMH-o-RA1(P):RA1(P)}}\\
	&=\mathbf{RA1} \circ \mathbf{RA2} \circ \mathbf{PBMH} (P) \seqA true
	&&\ptext{\cref{lemma:RA1-o-RA2(P):RA2(P)-land-RA1-subs}}\\
	&=(\mathbf{RA2} \circ \mathbf{PBMH} (P) \land \mathbf{RA1} (true)) \seqA true
	&&\ptext{Distributivity of $\seqA$ (\cref{law:seqA-right-distributivity-conjunction})}\\
	&=(\mathbf{RA2} \circ \mathbf{PBMH} (P) \seqA true) \land (\mathbf{RA1} (true) \seqA true)
	&&\ptext{\cref{lemma:RA1(true)-seqA-true:true} and predicate calculus}\\
	&=(\mathbf{RA2} \circ \mathbf{PBMH} (P) \seqA true)
	&&\ptext{\cref{theorem:PBMH-o-RA2(P):RA2(P)}}\\
	&=\mathbf{PBMH} \circ \mathbf{RA2} \circ \mathbf{PBMH} (P) \seqA true
	&&\ptext{\cref{lemma:PBMH(P)-seqA-true:exists-ac'-P}}\\
	&=\exists ac' \spot \mathbf{PBMH} \circ \mathbf{RA2} \circ \mathbf{PBMH} (P)
	&&\ptext{\cref{theorem:PBMH-o-RA2(P):RA2(P)}}\\
	&=\exists ac' \spot \mathbf{RA2} \circ \mathbf{PBMH} (P)
\end{xflalign*}
\end{proof}\end{proofs}
\end{lemma}

\begin{lemma}\label{lemma:RA-o-A(lnot-RA2-PBMH(P)|-RA2-o-PBMH(Q)):RA-o-A(lnot-P|-Q)}
\begin{statement}
\begin{align*}
	&\mathbf{RA}\circ\mathbf{A} (\lnot \mathbf{RA2} \circ \mathbf{PBMH} (P) \vdash \mathbf{RA2} \circ \mathbf{PBMH} (Q))\\
	&=\\
	&\mathbf{RA}\circ\mathbf{A} (\lnot P \vdash Q)
\end{align*}
\end{statement}
\begin{proofs}
\begin{proof}\checkt{alcc}
\begin{xflalign*}
	&\mathbf{RA}\circ\mathbf{A} (\lnot \mathbf{RA2} \circ \mathbf{PBMH} (P) \vdash \mathbf{RA2} \circ \mathbf{PBMH} (Q))
	&&\ptext{\cref{theorem:RA-o-A(P):RA-o-PBMH(P)}}\\
	&=\mathbf{RA}\circ\mathbf{PBMH} (\lnot \mathbf{RA2} \circ \mathbf{PBMH} (P) \vdash \mathbf{RA2} \circ \mathbf{PBMH} (Q))
	&&\ptext{\cref{lemma:PBMH(design):(lnot-PBMH(pre)|-PBMH(post))}}\\
	&=\mathbf{RA} (\lnot \mathbf{PBMH}\circ\mathbf{RA2} \circ \mathbf{PBMH} (P) \vdash \mathbf{PBMH}\mathbf{RA2} \circ \mathbf{PBMH} (Q))
	&&\ptext{\cref{theorem:PBMH-o-RA2(P):RA2(P)}}\\
	&=\mathbf{RA} (\lnot \mathbf{RA2} \circ \mathbf{PBMH} (P) \vdash \mathbf{RA2} \circ \mathbf{PBMH} (Q))
	&&\ptext{Definition of $\mathbf{RA}$ and~\cref{lemma:RA2(P|-Q):(lnot-RA2(lnot-P)|-RA2(Q))}}\\
	&=\mathbf{RA} (\lnot \mathbf{PBMH} (P) \vdash \mathbf{PBMH} (Q))
	&&\ptext{\cref{lemma:PBMH(design):(lnot-PBMH(pre)|-PBMH(post))}}\\
	&=\mathbf{RA}\circ\mathbf{PBMH} (\lnot P \vdash Q)
	&&\ptext{\cref{theorem:RA-o-A(P):RA-o-PBMH(P)}}\\
	&=\mathbf{RA}\circ\mathbf{A} (\lnot P \vdash Q)
\end{xflalign*}
\end{proof}
\end{proofs}
\end{lemma}

\begin{lemma}\label{lemma:RA-o-A(lnot-RA2-o-PBMH(P)|-Q):RA-o-A(lnot-P|-Q)}
\begin{statement}
\begin{align*}
	&\mathbf{RA}\circ\mathbf{A} (\lnot \mathbf{RA2} \circ \mathbf{PBMH} (P) \vdash Q)\\
	&=\\
	&\mathbf{RA}\circ\mathbf{A} (\lnot P \vdash Q)
\end{align*}
\end{statement}
\begin{proofs}
\begin{proof}\checkt{alcc}
\begin{xflalign*}
	&\mathbf{RA}\circ\mathbf{A} (\lnot \mathbf{RA2} \circ \mathbf{PBMH} (P) \vdash Q)
	&&\ptext{\cref{theorem:RA-o-A(P):RA-o-PBMH(P)}}\\
	&=\mathbf{RA}\circ\mathbf{PBMH} (\lnot \mathbf{RA2} \circ \mathbf{PBMH} (P) \vdash Q)
	&&\ptext{\cref{lemma:PBMH(design):(lnot-PBMH(pre)|-PBMH(post))}}\\
	&=\mathbf{RA} (\lnot \mathbf{PBMH}\circ\mathbf{RA2} \circ \mathbf{PBMH} (P) \vdash \mathbf{PBMH} (Q))
	&&\ptext{\cref{theorem:PBMH-o-RA2(P):RA2(P)}}\\
	&=\mathbf{RA} (\lnot \mathbf{RA2} \circ \mathbf{PBMH} (P) \vdash \mathbf{PBMH} (Q))
	&&\ptext{Definition of~$\mathbf{RA}$ and~\cref{lemma:RA2(P|-Q):(lnot-RA2(lnot-P)|-RA2(Q))}}\\
	&=\mathbf{RA} (\lnot \mathbf{PBMH} (P) \vdash \mathbf{PBMH} (Q))
	&&\ptext{\cref{lemma:PBMH(design):(lnot-PBMH(pre)|-PBMH(post))}}\\
	&=\mathbf{RA}\circ\mathbf{PBMH} (\lnot P \vdash Q)
	&&\ptext{\cref{theorem:RA-o-A(P):RA-o-PBMH(P)}}\\
	&=\mathbf{RA}\circ\mathbf{A} (\lnot P \vdash Q)
\end{xflalign*}
\end{proof}
\end{proofs}
\end{lemma}

\begin{lemma}\label{lemma:RA-o-A(P|-RA2-o-PBMH(Q)):RA-o-A(P|-Q)}
\begin{statement}
\begin{align*}
	&\mathbf{RA}\circ\mathbf{A} (P \vdash \mathbf{RA2} \circ \mathbf{PBMH} (Q))\\
	&=\\
	&\mathbf{RA}\circ\mathbf{A} (P \vdash Q)
\end{align*}
\end{statement}
\begin{proofs}
\begin{proof}\checkt{alcc}
\begin{xflalign*}
	&\mathbf{RA}\circ\mathbf{A} (P \vdash \mathbf{RA2} \circ \mathbf{PBMH} (Q))
	&&\ptext{\cref{theorem:RA-o-A(P):RA-o-PBMH(P)}}\\
	&=\mathbf{RA}\circ\mathbf{PBMH} (P \vdash \mathbf{RA2} \circ \mathbf{PBMH} (Q))
	&&\ptext{\cref{lemma:PBMH(design):(lnot-PBMH(pre)|-PBMH(post))}}\\
	&=\mathbf{RA} (\lnot \mathbf{PBMH} (\lnot P) \vdash \mathbf{PBMH}\circ\mathbf{RA2}\circ\mathbf{PBMH} (Q))
	&&\ptext{\cref{theorem:PBMH-o-RA2(P):RA2(P)}}\\
	&=\mathbf{RA} (\lnot \mathbf{PBMH} (\lnot P) \vdash \mathbf{RA2}\circ\mathbf{PBMH} (Q))
	&&\ptext{Definition of $\mathbf{RA}$ and~\cref{lemma:RA2(P|-Q):(lnot-RA2(lnot-P)|-RA2(Q))}}\\
	&=\mathbf{RA} (\lnot \mathbf{PBMH} (\lnot P) \vdash \mathbf{PBMH} (Q))
	&&\ptext{\cref{lemma:PBMH(design):(lnot-PBMH(pre)|-PBMH(post))}}\\
	&=\mathbf{RA}\circ\mathbf{PBMH} (P \vdash Q)
\end{xflalign*}
\end{proof}
\end{proofs}
\end{lemma}

\begin{lemma}\label{lemma:RA-o-A(P|-RA1-o-PBMH(Q)):RA-o-A(P|-Q)}
\begin{statement}
\begin{align*}
	&\mathbf{RA}\circ\mathbf{A} (P \vdash \mathbf{RA1} \circ \mathbf{PBMH} (Q))\\
	&=\\
	&\mathbf{RA}\circ\mathbf{A} (P \vdash Q)
\end{align*}
\end{statement}
\begin{proofs}
\begin{proof}
\begin{xflalign*}
	&\mathbf{RA}\circ\mathbf{A} (P \vdash \mathbf{RA1} \circ \mathbf{PBMH} (Q))
	&&\ptext{\cref{theorem:RA-o-A(P):RA-o-PBMH(P)}}\\
	&=\mathbf{RA}\circ\mathbf{PBMH} (P \vdash \mathbf{RA1} \circ \mathbf{PBMH} (Q))
	&&\ptext{\cref{lemma:PBMH(design):(lnot-PBMH(pre)|-PBMH(post))}}\\
	&=\mathbf{RA} (\lnot \mathbf{PBMH} (\lnot P) \vdash \mathbf{PBMH}\circ\mathbf{RA1}\circ\mathbf{PBMH} (Q))
	&&\ptext{\cref{theorem:PBMH-o-RA1(P):RA1(P)}}\\
	&=\mathbf{RA} (\lnot \mathbf{PBMH} (\lnot P) \vdash \mathbf{RA1}\circ\mathbf{PBMH} (Q))
	&&\ptext{Definition of $\mathbf{RA}$ and~\cref{lemma:RA1(P|-Q):RA1(P|-RA1(Q))}}\\
	&=\mathbf{RA} (\lnot \mathbf{PBMH} (\lnot P) \vdash \mathbf{PBMH} (Q))
	&&\ptext{\cref{lemma:PBMH(design):(lnot-PBMH(pre)|-PBMH(post))}}\\
	&=\mathbf{RA}\circ\mathbf{PBMH} (P \vdash Q)
	&&\ptext{\cref{theorem:RA-o-A(P):RA-o-PBMH(P)}}\\
	&=\mathbf{RA}\circ\mathbf{A} (P \vdash Q)
\end{xflalign*}
\end{proof}
\end{proofs}
\end{lemma}

\begin{lemma}\label{lemma:RA-o-A(lnot-RA1-o-PBMH(P)|-Q):RA-o-A(lnot-P|-Q)}
\begin{statement}
\begin{align*}
	&\mathbf{RA}\circ\mathbf{A} (\lnot \mathbf{RA1} \circ \mathbf{PBMH} (P) \vdash Q)\\
	&=\\
	&\mathbf{RA}\circ\mathbf{A} (\lnot P \vdash Q)
\end{align*}
\end{statement}
\begin{proofs}
\begin{proof}\checkt{alcc}
\begin{xflalign*}
	&\mathbf{RA}\circ\mathbf{A} (\lnot \mathbf{RA1} \circ \mathbf{PBMH} (P) \vdash Q)
	&&\ptext{\cref{theorem:RA-o-A(P):RA-o-PBMH(P)}}\\
	&=\mathbf{RA}\circ\mathbf{PBMH} (\lnot \mathbf{RA1} \circ \mathbf{PBMH} (P) \vdash Q)
	&&\ptext{\cref{lemma:PBMH(design):(lnot-PBMH(pre)|-PBMH(post))}}\\
	&=\mathbf{RA} (\lnot \mathbf{PBMH}\circ\mathbf{RA1} \circ \mathbf{PBMH} (P) \vdash \mathbf{PBMH} (Q))
	&&\ptext{\cref{theorem:PBMH-o-RA1(P):RA1(P)}}\\
	&=\mathbf{RA} (\lnot \mathbf{RA1} \circ \mathbf{PBMH} (P) \vdash \mathbf{PBMH} (Q))
	&&\ptext{Definition of $\mathbf{RA}$ and~\cref{lemma:RA1(P|-Q):RA1(lnot-RA1(lnot-P)|-Q)}}\\
	&=\mathbf{RA} (\lnot \mathbf{PBMH} (P) \vdash \mathbf{PBMH} (Q))
	&&\ptext{\cref{lemma:PBMH(design):(lnot-PBMH(pre)|-PBMH(post))}}\\
	&=\mathbf{RA}\circ\mathbf{PBMH} (\lnot P \vdash Q)
	&&\ptext{\cref{theorem:RA-o-A(P):RA-o-PBMH(P)}}\\
	&=\mathbf{RA}\circ\mathbf{A} (\lnot P \vdash Q)
\end{xflalign*}
\end{proof}
\end{proofs}
\end{lemma}

\subsection{Properties with respect to $\mathbf{A2}$}

\begin{theorem}\label{theorem:A2-o-RA-o-A-o-A2(P):RA-o-A-o-A2(P)}
\begin{statement}
$\mathbf{RA}\circ\mathbf{A}\circ\mathbf{A2} (P) = \mathbf{A2}\circ\mathbf{RA}\circ\mathbf{A}\circ\mathbf{A2} (P)$
\end{statement}
\begin{proofs}
\begin{proof}\checkt{alcc}
\begin{xflalign*}
	&\mathbf{RA}\circ\mathbf{A}\circ\mathbf{A2} (P)
	&&\ptext{\cref{theorem:RA-o-A(P):RA-o-PBMH(P)}}\\
	&=\mathbf{RA}\circ\mathbf{PBMH}\circ\mathbf{A2} (P)
	&&\ptext{\cref{lemma:PBMH-o-A2(P):A2(P)}}\\
	&=\mathbf{RA}\circ\mathbf{A2} (P)
	&&\ptext{Definition of $\mathbf{RA}$}\\
	&=\mathbf{RA3}\circ\mathbf{RA2}\circ\mathbf{RA1}\circ\mathbf{A2} (P)
	&&\ptext{\cref{theorem:A2-o-RA1-o-A2(P):RA1-o-A2(P)}}\\
	&=\mathbf{RA3}\circ\mathbf{RA2}\circ\mathbf{A2}\circ\mathbf{RA1}\circ\mathbf{A2} (P)
	&&\ptext{\cref{theorem:A2-o-RA2-o-A2(P):RA2-o-A2(P)}}\\
	&=\mathbf{RA3}\circ\mathbf{A2}\circ\mathbf{RA2}\circ\mathbf{A2}\circ\mathbf{RA1}\circ\mathbf{A2} (P)
	&&\ptext{\cref{theorem:A2-o-RA3-o-A2(P):RA3-o-A2(P)}}\\
	&=\mathbf{A2}\circ\mathbf{RA3}\circ\mathbf{A2}\circ\mathbf{RA2}\circ\mathbf{A2}\circ\mathbf{RA1}\circ\mathbf{A2} (P)
	&&\ptext{\cref{theorem:A2-o-RA2-o-A2(P):RA2-o-A2(P)}}\\
	&=\mathbf{A2}\circ\mathbf{RA3}\circ\mathbf{RA2}\circ\mathbf{A2}\circ\mathbf{RA1}\circ\mathbf{A2} (P)
	&&\ptext{\cref{theorem:A2-o-RA1-o-A2(P):RA1-o-A2(P)}}\\
	&=\mathbf{A2}\circ\mathbf{RA3}\circ\mathbf{RA2}\circ\mathbf{RA1}\circ\mathbf{A2} (P)
	&&\ptext{Definition of $\mathbf{RA}$}\\
	&=\mathbf{A2}\circ\mathbf{RA}\circ\mathbf{A2} (P)
	&&\ptext{\cref{lemma:PBMH-o-A2(P):A2(P)}}\\
	&=\mathbf{A2}\circ\mathbf{RA}\circ\mathbf{PBMH}\circ\mathbf{A2} (P)
	&&\ptext{\cref{theorem:RA-o-A(P):RA-o-PBMH(P)}}\\
	&=\mathbf{A2}\circ\mathbf{RA}\circ\mathbf{A}\circ\mathbf{A2} (P)
\end{xflalign*}
\end{proof}
\end{proofs}
\end{theorem}

\begin{theorem}\label{theorem:A2-o-RA-o-A(lnot-Pff|-Ptf):RA-o-A(lnot-Pff|-Ptf)}
\begin{statement}
Provided $P$ is $\mathbf{A2}$-healthy,
\begin{align*}
 	&\mathbf{RA}\circ\mathbf{A} (\lnot P^f_f \vdash P^t_f) = \mathbf{A2}\circ\mathbf{RA}\circ\mathbf{A} (\lnot P^f_f \vdash P^t_f)
\end{align*}
\end{statement}
\begin{proofs}
\begin{proof}\checkt{alcc}
\begin{xflalign*}
	&\mathbf{RA}\circ\mathbf{A} (\lnot P^f_f \vdash P^t_f)
	&&\ptext{Assumption: $P$ is $\mathbf{A2}$-healthy}\\
	&=\mathbf{RA}\circ\mathbf{A} (\lnot \mathbf{A2} (P)^f_f \vdash \mathbf{A2} (P)^t_f)
	&&\ptext{\cref{lemma:A2(P)-o-w:A2(P-o-w)}}\\
	&=\mathbf{RA}\circ\mathbf{A} (\lnot \mathbf{A2} (P^f_f) \vdash \mathbf{A2} (P^t_f))
	&&\ptext{\cref{lemma:A2-o-H1-o-H2}}\\
	&=\mathbf{RA}\circ\mathbf{A}\circ\mathbf{A2}(\lnot P^f_f \vdash P^t_f)
	&&\ptext{\cref{theorem:A2-o-RA-o-A-o-A2(P):RA-o-A-o-A2(P)}}\\
	&=\mathbf{A2}\circ\mathbf{RA}\circ\mathbf{A}\circ\mathbf{A2}(\lnot P^f_f \vdash P^t_f)
	&&\ptext{\cref{lemma:A2-o-H1-o-H2}}\\
	&=\mathbf{A2}\circ\mathbf{RA}\circ\mathbf{A} (\lnot \mathbf{A2} (P^f_f) \vdash \mathbf{A2} (P^t_f))
	&&\ptext{\cref{lemma:A2(P)-o-w:A2(P-o-w)}}\\
	&=\mathbf{A2}\circ\mathbf{RA}\circ\mathbf{A} (\lnot \mathbf{A2} (P)^f_f \vdash \mathbf{A2} (P)^t_f)
	&&\ptext{Assumption: $P$ is $\mathbf{A2}$-healthy}\\
	&=\mathbf{A2}\circ\mathbf{RA}\circ\mathbf{A} (\lnot P^f_f \vdash P^t_f)
\end{xflalign*}
\end{proof}
\end{proofs}
\end{theorem}

\begin{lemma}\label{lemma:A2-o-RA-o-A(lnot-A2(P)|-A2(Q)):RA-o-A(lnot-A2(P)|-A2(Q))}
\begin{statement}
\begin{align*}
	&\mathbf{RA}\circ\mathbf{A} (\lnot \mathbf{A2} (P) \vdash \mathbf{A2} (Q))\\
	&=\\
	&\mathbf{A2}\circ\mathbf{RA}\circ\mathbf{A} (\lnot \mathbf{A2} (P) \vdash \mathbf{A2} (Q))
\end{align*}
\end{statement}
\begin{proofs}
\begin{proof}\checkt{alcc}
\begin{xflalign*}
	&\mathbf{RA}\circ\mathbf{A} (\lnot \mathbf{A2} (P) \vdash \mathbf{A2} (Q))
	&&\ptext{\cref{lemma:A2-o-H1-o-H2,theorem:A2-o-RA-o-A-o-A2(P):RA-o-A-o-A2(P)}}\\
	&=\mathbf{A2}\circ\mathbf{RA}\circ\mathbf{A} (\lnot \mathbf{A2} (P) \vdash \mathbf{A2} (Q))
\end{xflalign*}
\end{proof}
\end{proofs}
\end{lemma}

\section{$\mathbf{CSPA1}$}

\begin{lemma}\label{lemma:CSPA1:alternative-1}
\begin{statement}
$\mathbf{CSPA1} (P) = P \lor (\lnot ok \land \exists z \spot s.tr \le z.tr \land z \in ac')$
\end{statement}
\begin{proofs}
\begin{proof}\checkt{pfr}\checkt{alcc}
\begin{xflalign*}
	&\mathbf{CSPA1} (P)
	&&\ptext{Definition of $\mathbf{CSPA1}$}\\
	&=P \lor \mathbf{RA1} (\lnot ok)
	&&\ptext{\cref{lemma:RA1(lnot-ok):lnot-ok-land-RA1(true)}}\\
	&=P \lor (\lnot ok \land \mathbf{RA1} (true))
	&&\ptext{\cref{lemma:RA1(true)}}\\
	&=P \lor (\lnot ok \land \exists z \spot s.tr \le z.tr \land z \in ac')
\end{xflalign*}
\end{proof}
\end{proofs}
\end{lemma}

\subsection{Properties}

\begin{theorem}\label{theorem:CSPA1-o-RA1:RA1-o-H1}
\begin{statement}
$\mathbf{CSPA1} \circ \mathbf{RA1} (P) = \mathbf{RA1} \circ \mathbf{H1} (P)$
\end{statement}
\begin{proofs}
\begin{proof}
\begin{xflalign*}
	&\mathbf{CSPA1} \circ \mathbf{RA1} (P)
	&&\ptext{Definition of $\mathbf{CSPA1}$}\\
	&=\mathbf{RA1} (P) \lor \mathbf{RA1} (\lnot ok)
	&&\ptext{\cref{lemma:RA1(P-lor-Q):RA1(P)-lor-RA1(Q)}}\\
	&=\mathbf{RA1} (P \lor \lnot ok)
	&&\ptext{Predicate calculus}\\
	&=\mathbf{RA1} (ok \implies P)
	&&\ptext{Definition of $\mathbf{H1}$}\\
	&=\mathbf{RA1}\circ\mathbf{H1} (P)
\end{xflalign*}
\end{proof}
\end{proofs}
\end{theorem}

\begin{theorem}\label{theorem:PBMH-o-CSPA1:CSPA1}
\begin{statement}Provided $P$ is $\mathbf{PBMH}$-healthy,
\begin{align*}
	&\mathbf{PBMH} \circ \mathbf{CSPA1} (P) = \mathbf{CSPA1} (P)
\end{align*}
\end{statement}
\begin{proofs}
\begin{proof}
\begin{xflalign*}
	&\mathbf{PBMH} \circ \mathbf{CSPA1} (P)
	&&\ptext{Definition of $\mathbf{CSPA1}$}\\
	&=\mathbf{PBMH} (P \lor (\mathbf{RA1} (\lnot ok)))
	&&\ptext{Distributivity of $\mathbf{PBMH}$}\\
	&=\mathbf{PBMH} (P) \lor \mathbf{PBMH} \circ \mathbf{RA1} (\lnot ok)
	&&\ptext{\cref{lemma:PBMH(c)-condition:c}}\\
	&=\mathbf{PBMH} (P) \lor \mathbf{PBMH} \circ \mathbf{RA1} \circ \mathbf{PBMH} (\lnot ok)
	&&\ptext{$\lnot ok$ is $\mathbf{PBMH}$-healthy and~\cref{theorem:PBMH-o-RA1(P):RA1(P)}}\\
	&=\mathbf{PBMH} (P) \lor \mathbf{RA1} \circ \mathbf{PBMH} (\lnot ok)
	&&\ptext{\cref{lemma:PBMH(c)-condition:c}}\\
	&=\mathbf{PBMH} (P) \lor \mathbf{RA1} (\lnot ok)
	&&\ptext{Assumption: $P$ is $\mathbf{PBMH}$-healthy}\\
	&=P \lor \mathbf{RA1} (\lnot ok)
	&&\ptext{Definition of $\mathbf{CSPA1}$}\\
	&=\mathbf{CSPA1} (P)	
\end{xflalign*}
\end{proof}
\end{proofs}
\end{theorem}

\begin{theorem}\label{theorem:CSPA1:idempotent}
\begin{statement}
$\mathbf{CSPA1} \circ \mathbf{CSPA1} (P) = \mathbf{CSPA1} (P)$
\end{statement}
\begin{proofs}
\begin{proof}\checkt{pfr}\checkt{alcc}
\begin{flalign*}
	&\mathbf{CSPA1} \circ \mathbf{CSPA1} (P)
	&&\ptext{Definition of $\mathbf{CSPA1}$}\\
	&=\mathbf{CSPA1} (P \lor (\lnot ok \land \mathbf{RA1} (true)))
	&&\ptext{Definition of $\mathbf{CSPA1}$}\\
	&=(P \lor (\lnot ok \land \mathbf{RA1} (true))) \lor (\lnot ok \land \mathbf{RA1} (true))
	&&\ptext{Predicate calculus}\\
	&=P \lor (\lnot ok \land \mathbf{RA1} (true))
	&&\ptext{Definition of $\mathbf{CSPA1}$}\\
	&=\mathbf{CSPA1} (P)
\end{flalign*}
\end{proof}
\end{proofs}
\end{theorem}

\begin{theorem}\label{theorem:CSPA1:monotonic}
\begin{statement}
$P \sqsubseteq Q \implies \mathbf{CSPA1} (P) \sqsubseteq \mathbf{CSPA1} (Q)$
\end{statement}
\begin{proofs}
\begin{proof}\checkt{pfr}\checkt{alcc}
\begin{flalign*}
	&\mathbf{CSPA1} (Q)
	&&\ptext{Definition of $\mathbf{CSPA1}$}\\
	&=Q \lor (\lnot ok \land \mathbf{RA1} (true))
	&&\ptext{Assumption: $P \sqsubseteq Q = [Q \implies P]$}\\
	&=(Q \land P) \lor (\lnot ok \land \mathbf{RA1} (true))
	&&\ptext{Predicate calculus}\\
	&=(Q \lor (\lnot ok \land \mathbf{RA1} (true))) \land (P \lor (\lnot ok \land \mathbf{RA1} (true)))
	&&\ptext{Predicate calculus}\\
	&\implies (P \lor (\lnot ok \land \mathbf{RA1} (true)))
	&&\ptext{Definition of $\mathbf{CSPA1}$}\\
	&=\mathbf{CSPA1} (P)
\end{flalign*}
\end{proof}
\end{proofs}
\end{theorem}

\subsubsection{Properties with respect to $\mathbf{RA1}$ and $\mathbf{H1}$}

\begin{theorem}\label{theorem:RA1-o-CSPA1:RA1-o-H1}
\begin{statement}$\mathbf{RA1} \circ \mathbf{CSPA1} (P) = \mathbf{RA1} \circ \mathbf{H1} (P)$\end{statement}
\begin{proofs}
\begin{proof}
\begin{xflalign*}
	&\mathbf{RA1} \circ \mathbf{H1} (P)
	&&\ptext{Definition of $\mathbf{H1}$}\\
	&=\mathbf{RA1} (ok \implies P)
	&&\ptext{Predicate calculus}\\
	&=\mathbf{RA1} (\lnot ok \lor P)
	&&\ptext{\cref{lemma:RA1(P-lor-Q):RA1(P)-lor-RA1(Q)}}\\
	&=\mathbf{RA1} (\lnot ok) \lor \mathbf{RA1} (P)
	&&\ptext{$\mathbf{RA1}$-idempotent (\cref{theorem:RA1-idempotent})}\\
	&=\mathbf{RA1} \circ \mathbf{RA1} (\lnot ok) \lor \mathbf{RA1} (P)
	&&\ptext{\cref{lemma:RA1(P-lor-Q):RA1(P)-lor-RA1(Q)}}\\
	&=\mathbf{RA1} (\mathbf{RA1} (\lnot ok) \lor P)
	&&\ptext{Definition of $\mathbf{CSPA1}$}\\	
	&=\mathbf{RA1} \circ \mathbf{CSPA1} (P)
\end{xflalign*}
\end{proof}
\end{proofs}
\end{theorem}

\begin{theorem}\label{theorem:RA1-o-CSPA1(P):CSPA1-o-RA1(P)}
\begin{statement}
$\mathbf{RA1}\circ\mathbf{CSPA} (P) = \mathbf{CSPA1} \circ \mathbf{RA1} (P)$
\end{statement}
\begin{proofs}
\begin{proof}
\begin{xflalign*}
	&\mathbf{RA1}\circ\mathbf{CSPA1} (P)
	&&\ptext{Definition of $\mathbf{CSPA1}$}\\
	&=\mathbf{RA1} (P \lor \mathbf{RA1} (\lnot ok))
	&&\ptext{\cref{lemma:RA1(P-lor-Q):RA1(P)-lor-RA1(Q)}}\\
	&=\mathbf{RA1} (P) \lor \mathbf{RA1}\circ\mathbf{RA1} (\lnot ok)
	&&\ptext{\cref{theorem:RA1-idempotent}}\\
	&=\mathbf{RA1} (P) \lor \mathbf{RA1} (\lnot ok)
	&&\ptext{Definition of $\mathbf{CSPA1}$}\\
	&=\mathbf{CSPA1} \circ \mathbf{RA1} (P)
\end{xflalign*}
\end{proof}
\end{proofs}
\end{theorem}

\section{$\mathbf{ND_{RAD}}$}

\begin{theorem}\label{theorem:RAP:P-sqcup-Choice:(true|Pt)}
\begin{statement}
Provided $P$ is $\mathbf{RAD}$-healthy,
\begin{align*}
	&\mathbf{ND_{RAD}} (P) = \mathbf{RA} \circ \mathbf{A} (true \vdash P^t_f)
\end{align*}
\end{statement}
\begin{proofs}
\begin{proof}\checkt{alcc}\checkt{pfr}
\begin{xflalign*}
	&\mathbf{ND_{RAP}} (P)
	&&\ptext{Assumption: $P$ is $\mathbf{RAP}$-healthy}\\
	&=\mathbf{ND_{RAP}} \circ \mathbf{RA} \circ \mathbf{A} (\lnot P^f_f \vdash P^t_f)
	&&\ptext{Definition of $\mathbf{ND_{RAP}}$ and $Choice_{\mathbf{RAP}}$}\\
	&=\mathbf{RA} \circ \mathbf{A} (\lnot P^f_f \vdash P^t_f) \sqcup \mathbf{RA} \circ \mathbf{A} (true \vdash true)
	&&\ptext{\cref{theorem:RAP:P-sqcup-Q}}\\
	&=\mathbf{RA} \circ \mathbf{A} (true \vdash \lnot P^f_f \implies P^t_f)
	&&\ptext{Definition of design and predicate calculus}\\
	&=\mathbf{RA} \circ \mathbf{A} (true \vdash (ok \land \lnot P^f_f) \implies P^t_f)
	&&\ptext{\cref{theorem:RA-o-A(P):RA-o-PBMH(P)}}\\
	&=\mathbf{RA} \circ \mathbf{PBMH} (true \vdash (ok \land \lnot P^f_f) \implies P^t_f)
	&&\ptext{\cref{lemma:PBMH(design):(lnot-PBMH(pre)|-PBMH(post))}}\\
	&=\mathbf{RA} (\lnot \mathbf{PBMH} (false) \vdash \mathbf{PBMH} ((ok \land \lnot P^f_f) \implies P^t_f))
	&&\ptext{Definition of $\mathbf{RA}$ and~\cref{lemma:RA2(P|-Q):RA2(P|-RA2(Q)),lemma:RA1(P|-Q):RA1(P|-RA1(Q))}}\\
	&=\mathbf{RA} \left(\begin{array}{l}
			\lnot \mathbf{PBMH} (false) 
			\\ \vdash \\
			\mathbf{RA2}\circ\mathbf{RA1}\circ\mathbf{PBMH} ((ok \land \lnot P^f_f) \implies P^t_f)
		\end{array}\right)
	&&\ptext{\cref{theorem:PBMH-o-RA1(P):RA1(P),theorem:PBMH-o-RA2(P):RA2(P)}}\\
	&=\mathbf{RA} \left(\begin{array}{l}
		\lnot \mathbf{PBMH} (false) 
		\\ \vdash \\
		\mathbf{PBMH}\circ\mathbf{RA2}\circ\mathbf{RA1}\circ\mathbf{PBMH} ((ok \land \lnot P^f_f) \implies P^t_f)
	\end{array}\right)
	&&\ptext{\cref{theorem:RA2-o-RA1:RA1-o-RA2,lemma:RA-o-A(design)-ow-ok'-true-wait-false}}\\
	&=\mathbf{RA} (\lnot \mathbf{PBMH} (false) \vdash \mathbf{PBMH} ((\mathbf{RA}\circ\mathbf{A} (\lnot P^f_f \vdash P^t_f))^t_f))
	&&\ptext{Assumption: $P$ is $\mathbf{RAP}$-healthy}\\
	&=\mathbf{RA} (\lnot \mathbf{PBMH} (false) \vdash \mathbf{PBMH} (P^t_f))
	&&\ptext{\cref{lemma:PBMH(design):(lnot-PBMH(pre)|-PBMH(post))}}\\
	&=\mathbf{RA} \circ \mathbf{PBMH} (true \vdash P^t_f)
	&&\ptext{\cref{theorem:RA-o-A(P):RA-o-PBMH(P)}}\\
	&=\mathbf{RA} \circ \mathbf{A} (true \vdash P^t_f)
\end{xflalign*}
\end{proof}
\end{proofs}
\end{theorem}


\begin{theorem}\label{theorem:RAP:P-sqcup-Choice:iff:forall-s-ac'-lnot-Pff}
\begin{statement}
Provided $P$ is $\mathbf{RAD}$-healthy,
\begin{align*}
	&\mathbf{ND_{RAD}} (P) = P \iff \forall s, ac' \spot \lnot P^f_f
\end{align*}
\end{statement}
\begin{proofs}
\begin{proof}\checkt{pfr}\checkt{alcc}
\begin{xflalign*}
	&P = P \sqcup Choice
	&&\ptext{Definition of $Choice$}\\
	&\iff P = P \sqcup \mathbf{RA}\circ\mathbf{A} (true \vdash true)
	&&\ptext{Assumption: $P$ is $\mathbf{RAP}$-healthy}\\
	&\iff \mathbf{RA}\circ\mathbf{A} (\lnot P^f_f \vdash P^t_f) = \mathbf{RA}\circ\mathbf{A} (\lnot P^f_f \vdash P^t_f) \sqcup \mathbf{RA}\circ\mathbf{A} (true \vdash true)
	&&\ptext{\cref{theorem:RAP:P-sqcup-Q}}\\
	&\iff \mathbf{RA}\circ\mathbf{A} (\lnot P^f_f \vdash P^t_f) = \mathbf{RA}\circ\mathbf{A} (\lnot P^f_f \lor true \vdash (\lnot P^f_f \implies P^t_f) \land true\implies true)
	&&\ptext{Predicate calculus}\\
	&\iff \mathbf{RA}\circ\mathbf{A} (\lnot P^f_f \vdash P^t_f) = \mathbf{RA}\circ\mathbf{A} (true \vdash \lnot P^f_f \implies P^t_f)
	&&\ptext{\cref{theorem:RA-o-A(P):RA-o-PBMH(P)}}\\
	&\iff \mathbf{RA}\circ\mathbf{PBMH} (\lnot P^f_f \vdash P^t_f) = \mathbf{RA}\circ\mathbf{PBMH} (true \vdash \lnot P^f_f \implies P^t_f)
	&&\ptext{Definition of $\mathbf{RA}$}\\
	&\iff \left(
\right)
	&&\ptext{\cref{law:pbmh:false} and predicate calculus}\\
	&\iff \left(%
\right)
	&&\ptext{Predicate calculus and~\cref{theorem:PBMH(P-lor-Q):PBMH(P)-lor-PBMH(Q)}}\\
	&\iff \left(%
\right)
	&&\ptext{\cref{lemma:RA2(P):P:s-ac'-not-free} and predicate calculus}\\
	&\iff \left(%
\right)
	&&\ptext{Definition of design and predicate calculus}\\
	&\iff \left(%
\right)
	&&\ptext{Property of conditional}\\
	&\iff \left(%
\right)
	&&\ptext{Equality of relations}\\
	&\iff \left[%
\right]
	&&\ptext{Predicate calculus and~\cref{theorem:RA2(P-lor-Q):RA2(P)-lor-RA2(Q),lemma:RA1(P-lor-Q):RA1(P)-lor-RA1(Q)}}\\
	&\iff \left[%
\right]
	&&\ptext{Property of conditional}\\
	&\iff \left[%
\right]
	&&\ptext{Property of conditional and predicate calculus}\\
	&\iff \left[%
\right]
	&&\ptext{Definition of universal quantification}\\
	&\iff \forall ok',ok,s,ac' \spot \left(%
\right)
	&&\ptext{Case-analysis on $ok'$ and predicate calculus}\\
	&\iff \forall ok,s,ac' \spot \left(%
\right)
	&&\ptext{Predicate calculus}\\
	&\iff \forall ok,s,ac' \spot (
		\mathbf{RA1} (\lnot ok) \lor (s.wait \lor \lnot \mathbf{RA1}\circ\mathbf{RA2} (P^f_f)) 
	)
	&&\ptext{Case-analysis on $ok$ and predicate calculus}\\
	&\iff \forall s,ac' \spot \left(%
\right)
	&&\ptext{\cref{lemma:RA1(false)} and predicate calculus}\\
	&\iff \forall s,ac' \spot \left(%
\right)
	&&\ptext{Predicate calculus: absorption law}\\
	&\iff \forall s,ac' \spot (s.wait \lor \lnot \mathbf{RA1}\circ\mathbf{RA2} (P^f_f))
	&&\ptext{Predicate calculus}\\
	&\iff \forall s \spot (\lnot s.wait) \implies (\forall ac' \spot \lnot \mathbf{RA1}\circ\mathbf{RA2} (P^f_f))
	&&\ptext{Predicate calculus}\\
	&\iff \forall s \spot (\exists z \spot s = z\oplus \{wait\mapsto false\}) \implies (\forall ac' \spot \lnot \mathbf{RA1}\circ\mathbf{RA2} (P^f_f))
	&&\ptext{Predicate calculus}\\
	&\iff \forall s,z \spot (s = z\oplus \{wait\mapsto false\}) \implies (\forall ac' \spot \lnot \mathbf{RA1}\circ\mathbf{RA2} (P^f_f))
	&&\ptext{Predicate calculus}\\
	&\iff \forall z \spot (\forall ac' \spot \lnot \mathbf{RA1}\circ\mathbf{RA2} (P^f_f))[z\oplus \{wait\mapsto false\}/s]
	&&\ptext{Predicate calculus}\\
	&\iff \forall s \spot (\forall ac' \spot \lnot \mathbf{RA1}\circ\mathbf{RA2} (P^f_f))[z\oplus \{wait\mapsto false\}/s][s/z]
	&&\ptext{Property of substitution}\\
	&\iff \forall s \spot (\forall ac' \spot \lnot \mathbf{RA1}\circ\mathbf{RA2} (P^f_f))[s\oplus \{wait\mapsto false\}/s]
	&&\ptext{\cref{lemma:RA2(P)-o-w-subs:RA2(P-o-w-subs),lemma:RA1(P)-o-w-subs:RA1(P-o-w-subs)} and property of substitution}\\
	&\iff \forall s \spot (\forall ac' \spot (\lnot \mathbf{RA1}\circ\mathbf{RA2} (P))^f_f)[s\oplus \{wait\mapsto false\}/s]
	&&\ptext{Substitution abbreviation}\\
	&\iff \forall s \spot (\forall ac' \spot (\lnot \mathbf{RA1}\circ\mathbf{RA2} (P))^f[s\oplus\{wait\mapsto false\}/s])[s\oplus \{wait\mapsto false\}/s]
	&&\ptext{Property of substitution and $\oplus$}\\
	&\iff \forall s \spot (\forall ac' \spot (\lnot \mathbf{RA1}\circ\mathbf{RA2} (P))^f[s\oplus\{wait\mapsto false\}/s])
	&&\ptext{Substitution abbreviation}\\
	&\iff \forall s \spot (\forall ac' \spot (\lnot \mathbf{RA1}\circ\mathbf{RA2} (P))^f_f)
	&&\ptext{Predicate calculus}\\
	&\iff \forall s, ac' \spot (\lnot \mathbf{RA1}\circ\mathbf{RA2} (P))^f_f
	&&\ptext{Assumption: $P$ is $\mathbf{RAP}$-healthy, hence $\mathbf{RA1}$ and $\mathbf{RA2}$-healthy}\\
	&\iff \forall s, ac' \spot (\lnot P)^f_f
	&&\ptext{Property of substitution}\\
	&\iff \forall s, ac' \spot \lnot P^f_f
\end{xflalign*}
\end{proof}
\end{proofs}
\end{theorem}

\begin{theorem}\label{theorem:NDRA-idempotent}
\begin{statement}
$\mathbf{ND_{RAD}} \circ \mathbf{ND_{RAD}} (P) = \mathbf{ND_{RAD}} (P)$
\end{statement}
\begin{proofs}
\begin{proof}\checkt{pfr}
\begin{xflalign*}
	&\mathbf{ND_{RAP}} \circ \mathbf{ND_{RAP}} (P)
	&&\ptext{Definition of $\mathbf{ND_{RAP}}$}\\
	&=\mathbf{ND_{RAP}} (P) \sqcup Choice_{\mathbf{RAP}}
	&&\ptext{Definition of $\mathbf{ND_{RAP}}$}\\
	&=P \sqcup Choice_{\mathbf{RAP}} \sqcup Choice_{\mathbf{RAP}}
	&&\ptext{Predicate calculus}\\
	&=P \sqcup Choice_{\mathbf{RAP}}
	&&\ptext{Definition of $\mathbf{ND_{RAP}}$}\\
	&=\mathbf{ND_{RAP}} (P)
\end{xflalign*}
\end{proof}
\end{proofs}
\end{theorem}

\begin{theorem}\label{theorem:RAP:P-seqDac-Q-ND}
\begin{statement}
Provided $P$ and $Q$ are reactive angelic designs and $\mathbf{ND_{RAD}}$-healthy,
\begin{align*}
	&P \seqDac Q\\
	&=\\
	&\mathbf{RA} \circ \mathbf{A} \left(
\right)
	&&\ptext{\cref{lemma:RA1(false),lemma:RA2(P):P:s-ac'-not-free} and predicate calculus}\\
	&=\mathbf{RA} \circ \mathbf{A} \left(%
\right)
	&&\ptext{\cref{law:seqA-ac'-not-free} and predicate calculus}\\
	&=\mathbf{RA} \circ \mathbf{A} \left(%
\right)
	&&\ptext{Property of sets and predicate calculus}\\
	&=\mathbf{RA} \circ \mathbf{A} \left(%
\right)
\end{xflalign*}
\end{proof}
\end{proofs}
\end{theorem}

\begin{lemma}\label{lemma:RAD:NDRAD(Chaos):Choice}
\begin{statement}
$\mathbf{ND_{RAD}} (Chaos_{\mathbf{RAD}}) = Choice_{\mathbf{RAD}}$
\end{statement}
\begin{proofs}
\begin{proof}
\begin{xflalign*}
	&\mathbf{ND_{RAD}} (Chaos_{\mathbf{RAD}})
	&&\ptext{Definition of $\mathbf{ND_{RAD}}$}\\
	&=Chaos_{\mathbf{RAD}} \sqcup_{\mathbf{RAD}} Choice_{\mathbf{RAD}}
	&&\ptext{\cref{theorem:Choice-RAD-sqcup-P}}\\
	&=\mathbf{RA} \circ \mathbf{A} (true \vdash ac'\neq\emptyset)
	&&\ptext{Definition of $Choice_{\mathbf{RAD}}$}\\
	&=Choice_{\mathbf{RAD}}
\end{xflalign*}
\end{proof}
\end{proofs}
\end{lemma}

\begin{lemma}\label{lemma:RAD:NDRAD(a-then-Skip):a-then-Skip}
\begin{statement}
$\mathbf{ND_{RAD}} (a\circthen_{\mathbf{RAD}} Skip_{\mathbf{RAD}}) = a\circthen_{\mathbf{RAD}} Skip_{\mathbf{RAD}}$
\end{statement}
\begin{proofs}
\begin{proof}
\begin{xflalign*}
	&\mathbf{ND_{RAD}} (a\circthen_{\mathbf{RAD}} Skip_{\mathbf{RAD}})
	&&\ptext{Definition of $\mathbf{ND_{RAD}}$}\\
	&=a\circthen_{\mathbf{RAD}} Skip_{\mathbf{RAD}} \sqcup_{\mathbf{RAD}} Choice_{\mathbf{RAD}}
	&&\ptext{\cref{theorem:Choice-RAD-sqcup-P} and definition of $a\circthen_{\mathbf{RAD}} Skip_{\mathbf{RAD}}$}\\
	&=\mathbf{RA} \circ \mathbf{A} 
		\left(true 
				\vdash 
					\circledIn{y}{ac'} \left(\begin{array}{l}(y.tr=s.tr \land a \notin y.ref)
						\\ \dres y.wait \rres \\
					(y.tr = s.tr \cat \lseq a \rseq)
					\end{array}\right)
		\right)
	&&\ptext{Definition of $a\circthen_{\mathbf{RAD}} Skip_{\mathbf{RAD}}$}\\
	&=a\circthen_{\mathbf{RAD}} Skip_{\mathbf{RAD}}
\end{xflalign*}
\end{proof}
\end{proofs}
\end{lemma}

\section{Relationship with CSP}

\subsection{Results with respect to $\mathbf{R}$}

\begin{theorem}\label{theorem:ac2p-o-RA(P):R-o-ac2p(P)}
\begin{statement}
Provided $P$ is $\mathbf{PBMH}$-healthy,
$ac2p \circ \mathbf{RA} (P) = \mathbf{R} \circ ac2p(P)$
\end{statement}
\begin{proofs}
\begin{proof}
\begin{xflalign*}
	&ac2p \circ \mathbf{RA} (P)
	&&\ptext{Definition of $\mathbf{RA}$}\\
	&=ac2p \circ \mathbf{RA3} \circ \mathbf{RA2} \circ \mathbf{RA1} (P)
	&&\ptext{\cref{theorem:ac2p-o-RA3(P):R3-o-ac2p(P)}}\\
	&=\mathbf{R3} \circ ac2p \circ \mathbf{RA2} \circ \mathbf{RA1} (P)
	&&\ptext{\cref{theorem:RA2-o-RA1:RA1-o-RA2}}\\
	&=\mathbf{R3} \circ ac2p \circ \mathbf{RA1} \circ \mathbf{RA2} (P)
	&&\ptext{\cref{theorem:ac2p-o-RA1-o-RA2(P):R1-o-R2-o-ac2p(P)}}\\
	&=\mathbf{R3} \circ \mathbf{R1} \circ \mathbf{R2} \circ ac2p (P)
	&&\ptext{Definition of $\mathbf{R}$}\\
	&=\mathbf{R} \circ ac2p (P)
\end{xflalign*}
\end{proof}
\end{proofs}
\end{theorem}

\begin{theorem}\label{theorem:ac2p-o-RA-o-A(design):R(lnot-ac2p(pre)|-ac2p(post))}
\begin{statement}
	$ac2p \circ \mathbf{RA} \circ \mathbf{A} (\lnot P^f_f \vdash P^t_f) = \mathbf{R} (\lnot ac2p(P^f_f) \vdash ac2p(P^t_f))$
\end{statement}
\begin{proofs}
\begin{proof}\checkt{alcc}
\begin{xflalign*}
	&ac2p \circ \mathbf{RA} \circ \mathbf{A} (\lnot P^f_f \vdash P^t_f)
	&&\ptext{\cref{theorem:RA-o-A(P):RA-o-PBMH(P)}}\\
	&=ac2p \circ \mathbf{RA} \circ \mathbf{PBMH} (\lnot P^f_f \vdash P^t_f)
	&&\ptext{\cref{theorem:ac2p-o-RA(P):R-o-ac2p(P)}}\\
	&=\mathbf{R} \circ ac2p \circ \mathbf{PBMH} (\lnot P^f_f \vdash P^t_f)
	&&\ptext{\cref{lemma:ac2p-o-PBMH(P):ac2p(P)}}\\
	&=\mathbf{R} \circ ac2p(\lnot P^f_f \vdash P^t_f)
	&&\ptext{\cref{lemma:ac2p(P-design):(ac2p(lnot-Pf)--ac2p(Pt))}}\\
	&=\mathbf{R} (\lnot ac2p(P^f_f) \vdash ac2p(P^t_f))
\end{xflalign*}
\end{proof}
\end{proofs}
\end{theorem}

\begin{theorem}\label{theorem:ac2p-o-RA1(P):RA1-o-ac2p(P)} Provided $P$ is $\mathbf{PBMH}$-healthy,
\begin{align*}
	&ac2p \circ \mathbf{RA1} (P) = \mathbf{R1} \circ ac2p(P)
\end{align*}
\begin{proofs}\begin{proof}\checkt{pfr}\checkt{alcc}
\begin{xflalign*}
	&ac2p \circ \mathbf{RA1} (P)
	&&\ptext{Definition of $ac2p$}\\
	&=\mathbf{PBMH} (\mathbf{RA1} (P))[State_{\II}(in\alpha_{-ok})/s] \seqA \bigwedge x : out\alpha_{-ok'} \spot dash(s).x = x
	&&\ptext{Assumption: $P$ is $\mathbf{PBMH}$-healthy and~\cref{theorem:PBMH-o-RA1(P):RA1(P)}}\\
	&=\mathbf{RA1} (P)[State_{\II}(in\alpha_{-ok})/s] \seqA \bigwedge x : out\alpha_{-ok'} \spot dash(s).x = x
	&&\ptext{Definition of $\mathbf{RA1}$ (\cref{lemma:RA1:alternative-1})}\\
	&=\left(
\right)
	&&\ptext{Property of sets}\\
	&=\left(%
\right)
	&&\ptext{Transitivity of equality on $dash(z).tr'=tr'$}\\
	&=\left(%
\right)
	&&\ptext{Definition of $ac2p$}\\
	&=ac2p(P) \land tr\le tr'
	&&\ptext{Definition of $\mathbf{R1}$}\\
	&=\mathbf{R1} \circ ac2p(P)
\end{xflalign*}
\end{proof}\end{proofs}
\end{theorem}

\begin{theorem}\label{theorem:ac2p-o-RA1-o-RA2(P):R1-o-R2-o-ac2p(P)} Provided $P$ is $\mathbf{PBMH}$-healthy,
\begin{align*}
	&ac2p \circ \mathbf{RA1} \circ \mathbf{RA2} (P) = \mathbf{R1} \circ \mathbf{R2} \circ ac2p(P) 
\end{align*}
\begin{proofs}\begin{proof}\checkt{alcc}
\begin{xflalign*}
	&ac2p \circ \mathbf{RA1} \circ \mathbf{RA2} (P)
	&&\ptext{Definition of $ac2p$}\\
	&=\mathbf{PBMH} (\mathbf{RA1} \circ \mathbf{RA2} (P))[State_{\II}(in\alpha_{-ok})/s] \seqA \bigwedge x:out\alpha_{-ok'} \spot dash(s).x = x
	&&\ptext{Assumption: $P$ is $\mathbf{PBMH}$-healthy and~\cref{theorem:PBMH-o-RA1(P):RA1(P),theorem:PBMH-o-RA2(P):RA2(P)}}\\
	&=(\mathbf{RA1} \circ \mathbf{RA2} (P))[State_{\II}(in\alpha_{-ok})/s] \seqA \bigwedge x:out\alpha_{-ok'} \spot dash(s).x = x
	&&\ptext{\cref{lemma:RA1-o-RA2(P):RA2(P)-land-RA1-subs} and definition of $\mathbf{RA2}$}\\
	&=\left(
\right)
	&&\ptext{Property of sets}\\
	&=\left(%
\right)
	&&\ptext{Property of sets}\\
	&=\left(%
\right)
	&&\ptext{Property of sets}\\
	&=\left(%
\right)
	&&\ptext{Value of record component $tr'$}\\
	&=\left(%
\right)
	&&\ptext{Property of sequences}\\
	&=\left(%
\right)
	&&\ptext{Transitivity of equality on $tr'\in out\alpha_{-ok'}$}\\
	&=\left(%
\right)
	&&\ptext{Property of substitution: $tr$ not free in set comprehension}\\
	&=\left(%
\right)
	&&\ptext{Assumption: $P$ is $\mathbf{PBMH}$ and definition of $ac2p$}\\
	&=ac2p(P)[\lseq\rseq,tr'-tr/tr,tr'] \land tr \le tr'
	&&\ptext{Definition of $\mathbf{R2}$ and $\mathbf{R1}$}\\
	&=\mathbf{R1} \circ \mathbf{R2} \circ ac2p(P)
\end{xflalign*}
\end{proof}\end{proofs}
\end{theorem}

\begin{theorem}\label{theorem:ac2p-o-RA3(P):R3-o-ac2p(P)}
$ac2p \circ \mathbf{RA3} (P) = \mathbf{R3} \circ ac2p(P)$
\begin{proofs}\begin{proof}\checkt{pfr}\checkt{alcc}
\begin{xflalign*}
	&ac2p \circ \mathbf{RA3} (P)
	&&\ptext{Definition of $\mathbf{RA3}$}\\
	&=ac2p(\IIRac \dres s.wait \rres P)
	&&\ptext{\cref{lemma:ac2p(conditional)}}\\
	&=ac2p(\IIRac) \dres s.wait[State_{\II}(in\alpha_{-ok})/s] \rres ac2p(P)
	&&\ptext{Definition of $State_{\II}$ and substitution}\\
	&=ac2p(\IIRac) \dres wait \rres ac2p(P)
	&&\ptext{\cref{theorem:ac2p(IIRac):IIrea}}\\
	&=\IIrea \dres wait \rres ac2p(P)
	&&\ptext{Definition of $\mathbf{R3}$}\\
	&=\mathbf{R3} \circ ac2p(P)
\end{xflalign*}
\end{proof}\end{proofs}
\end{theorem}

\begin{theorem}\label{theorem:ac2p(IIRac):IIrea} Provided $out\alpha = \{ tr', ref', wait' \}$,
\begin{align*}
	&ac2p (\IIRac) = \IIrea
\end{align*}
\begin{proofs}\begin{proof}\checkt{pfr}\checkt{alcc}
\begin{xflalign*}
	&ac2p(\IIRac)
	&&\ptext{Definition of $\IIRac$}\\
	&=ac2p(\mathbf{RA1} (\lnot ok) \lor (ok' \land s \in ac'))
	&&\ptext{Distributivity of $ac2p$ (\cref{theorem:ac2p(P-lor-Q):ac2p(P)-lor-ac2p(Q)})}\\
	&=ac2p \circ \mathbf{RA1} (\lnot ok) \lor ac2p(ok' \land s \in ac')
	&&\ptext{$\lnot ok$ is $\mathbf{PBMH}$-healthy and~\cref{theorem:ac2p-o-RA1(P):RA1-o-ac2p(P)}}\\
	&=\mathbf{R1} \circ ac2p(\lnot ok) \lor ac2p(ok' \land s \in ac')
	&&\ptext{\cref{lemma:ac2p(P)-s-ac'-not-free:P}}\\
	&=\mathbf{R1} (\lnot ok) \lor ac2p(ok' \land s \in ac')
	&&\ptext{\cref{lemma:ac2p(P-land-Q)-s-ac'-not-free:P-land-ac2p(Q)}}\\
	&=\mathbf{R1} (\lnot ok) \lor (ok' \land ac2p(s \in ac'))
	&&\ptext{Assumption: $in\alpha_{-ok} = \{ tr, ref, wait\}$ and~\cref{lemma:ac2p(s-in-ac'):x0-xi}}\\
	&=\mathbf{R1} (\lnot ok) \lor (ok' \land tr'=tr \land ref'=ref \land wait'=wait)
	&&\ptext{Definition of $\IIrea$}\\
	&=\IIrea
\end{xflalign*}
\end{proof}\end{proofs}
\end{theorem}

\begin{theorem}\label{theorem:p2ac-o-R:RA-o-p2ac}
\begin{statement}
$p2ac \circ \mathbf{R} (P) = \mathbf{RA} \circ p2ac(P)$
\end{statement}
\begin{proofs}
\begin{proof}\checkt{alcc}\checkt{pfr}
\begin{xflalign*}
	&p2ac \circ \mathbf{R} (P)
	&&\ptext{Definition of $\mathbf{R}$}\\
	&=p2ac \circ \mathbf{R3} \circ \mathbf{R1} \circ \mathbf{R2} (P)
	&&\ptext{\cref{theorem:p2ac-o-R3(P):RA3-o-p2ac(P)}}\\
	&=\mathbf{RA3} \circ p2ac \circ \mathbf{R1} \circ \mathbf{R2} (P)
	&&\ptext{$\mathbf{R1}$-idempotent}\\
	&=\mathbf{RA3} \circ p2ac \circ \mathbf{R1} \circ \mathbf{R1} \circ \mathbf{R2} (P)
	&&\ptext{\cref{theorem:p2ac-o-R1(P):RA1-o-p2ac(P)}}\\
	&=\mathbf{RA3} \circ \mathbf{RA1} \circ p2ac \circ \mathbf{R1} \circ \mathbf{R2} (P)
	&&\ptext{\cref{theorem:p2ac-o-R1-o-R2(P):RA2-o-p2ac(P)}}\\
	&=\mathbf{RA3} \circ \mathbf{RA1} \circ \mathbf{RA2} \circ p2ac (P)
	&&\ptext{\cref{theorem:RA2-o-RA1:RA1-o-RA2}}\\
	&=\mathbf{RA3} \circ \mathbf{RA2} \circ \mathbf{RA1} \circ p2ac (P)
	&&\ptext{Definition of $\mathbf{RA}$}\\
	&=\mathbf{RA} \circ p2ac(P)
\end{xflalign*}
\end{proof}
\end{proofs}
\end{theorem}

\begin{theorem}\label{theorem:p2ac-o-R(design):RA-o-A(lnot-p2ac(pre)|-p2ac(post))}
\begin{statement}
$p2ac \circ \mathbf{R} (\lnot P^f_f \vdash P^t_f) = \mathbf{RA} \circ \mathbf{A} (\lnot p2ac(P^f_f) \vdash p2ac(P^t_f))$
\end{statement}
\begin{proofs}
\begin{proof}\checkt{alcc}
\begin{xflalign*}
	&p2ac \circ \mathbf{R} (\lnot P^f_f \vdash P^t_f)
	&&\ptext{\cref{theorem:p2ac-o-R:RA-o-p2ac} and definition of~$\mathbf{RA}$}\\
	&=\mathbf{RA3} \circ \mathbf{RA2} \circ \mathbf{RA1} \circ p2ac(\lnot P^f_f \vdash P^t_f)
	&&\ptext{Definition of $\mathbf{RA1}$}\\
	&=\mathbf{RA3} \circ \mathbf{RA2} \circ \mathbf{RA1} (p2ac(\lnot P^f_f \vdash P^t_f) \land ac'\neq\emptyset)
	&&\ptext{\cref{theorem:p2ac(design)-land-ac'-neq-emptyset}}\\
	&=\mathbf{RA3} \circ \mathbf{RA2} \circ \mathbf{RA1} ((\lnot p2ac(P^f_f) \vdash p2ac(P^t_f)) \land ac'\neq\emptyset)
	&&\raisetag{14pt}\ptext{$\mathbf{RA1}$ and $\mathbf{RA}$}\\
	&=\mathbf{RA} (\lnot p2ac(P^f_f) \vdash p2ac(P^t_f))
	&&\ptext{\cref{lemma:PBMH-o-p2ac(P):p2ac(P)}}\\
	&=\mathbf{RA} (\lnot \mathbf{PBMH} \circ p2ac(P^f_f) \vdash \mathbf{PBMH} \circ p2ac(P^t_f))
	&&\ptext{Definition of $\mathbf{A1}$}\\
	&=\mathbf{RA} \circ \mathbf{A1} (\lnot p2ac(P^f_f) \vdash p2ac(P^t))
	&&\ptext{Definition of $\mathbf{RA}$ and~\cref{theorem:RA1-o-A0:RA1}}\\
	&=\mathbf{RA} \circ \mathbf{A0} \circ \mathbf{A1} (\lnot p2ac(P^f_f) \vdash p2ac(P^t))
	&&\ptext{Definition of $\mathbf{A}$}\\
	&=\mathbf{RA} \circ \mathbf{A} (\lnot p2ac(P^f_f) \vdash p2ac(P^t))
\end{xflalign*}
\end{proof}
\end{proofs}
\end{theorem}

\begin{theorem}\label{theorem:p2ac-o-R(design):RA(p2ac-design)}
$p2ac \circ \mathbf{R} (\lnot P^f \vdash P^t) = \mathbf{RA} (\lnot p2ac(P^f) \vdash p2ac(P^t))$
\begin{proofs}\begin{proof}\checkt{pfr}\checkt{alcc}
\begin{xflalign*}
	&p2ac \circ \mathbf{R} (\lnot P^f \vdash P^t)
	&&\ptext{\cref{theorem:p2ac-o-R:RA-o-p2ac}}\\
	&=\mathbf{RA} \circ p2ac(\lnot P^f \vdash P^t)
	&&\ptext{Definition of $\mathbf{RA}$}\\
	&=\mathbf{RA3} \circ \mathbf{RA2} \circ \mathbf{RA1} \circ p2ac(\lnot P^f \vdash P^t)
	&&\ptext{Definition of $\mathbf{RA1}$}\\
	&=\mathbf{RA3} \circ \mathbf{RA2} ((p2ac(\lnot P^f \vdash P^t) \land ac'\neq\emptyset)[States_{tr\le tr'}(s) \cap ac'/ac'])
	&&\ptext{\cref{theorem:p2ac(design)-land-ac'-neq-emptyset}}\\
	&=\mathbf{RA3} \circ \mathbf{RA2} ((\lnot p2ac(P^f) \vdash p2ac(P^t)) \land ac'\neq\emptyset)[States_{tr\le tr'}(s) \cap ac'/ac'])
	&&\ptext{Definition of $\mathbf{RA1}$}\\
	&=\mathbf{RA3} \circ \mathbf{RA2} \circ \mathbf{RA1} (\lnot p2ac(P^f) \vdash p2ac(P^t))
	&&\ptext{Definition of $\mathbf{RA}$}\\
	&=\mathbf{RA} (\lnot p2ac(P^f) \vdash p2ac(P^t))
\end{xflalign*}
\end{proof}\end{proofs}
\end{theorem}

\begin{theorem}\label{theorem:p2ac-o-R(design):RA-o-A(d2ac)}
\begin{align*}
	&p2ac \circ \mathbf{R} (\lnot P^f \vdash P^t)\\
	&=\\
	&\mathbf{RA} \circ \mathbf{A} (\lnot p2ac(P^f) \land (\lnot P^f[\mathbf{s}/in\alpha] \circseq true) \vdash p2ac(P^t))
\end{align*}
\begin{proofs}\begin{proof}\checkt{alcc}
\begin{xflalign*}
	&p2ac \circ \mathbf{R} (\lnot P^f \vdash P^t)
	&&\ptext{\cref{theorem:p2ac-o-R(design):RA-d2ac(design)}}\\
	&=\mathbf{RA} (\lnot p2ac(P^f) \land (\lnot P^f[\mathbf{s}/in\alpha] \circseq true) \vdash p2ac(P^t))
	&&\ptext{Predicate calculus}\\
	&=\mathbf{RA} (\lnot (p2ac(P^f) \lor \lnot (\lnot P^f[\mathbf{s}/in\alpha] \circseq true)) \vdash p2ac(P^t))
	&&\ptext{\cref{lemma:PBMH-o-p2ac(P):p2ac(P),law:pbmh:P:ac'-not-free}}\\
	&=\mathbf{RA} \left(\begin{array}{l}
		\lnot (\mathbf{PBMH} \circ p2ac(P^f) \lor \mathbf{PBMH} \circ (\lnot (\lnot P^f[\mathbf{s}/in\alpha] \circseq true))) 		\\ \vdash \\
		\mathbf{PBMH} \circ p2ac(P^t)
	\end{array}\right)
	&&\ptext{Distributivity of $\mathbf{PBMH}$ (\cref{law:pbmh:distribute-disjunction})}\\
	&=\mathbf{RA} \left(\begin{array}{l}
		\lnot \mathbf{PBMH} (p2ac(P^f) \lor (\lnot (\lnot P^f[\mathbf{s}/in\alpha] \circseq true)))
		\\ \vdash \\
		\mathbf{PBMH} \circ p2ac(P^t)
	\end{array}\right)
	&&\ptext{\cref{lemma:PBMH(design):(lnot-PBMH(pre)|-PBMH(post))}}\\
	&=\mathbf{RA} \circ \mathbf{PBMH} (\lnot p2ac(P^f) \land (\lnot P^f[\mathbf{s}/in\alpha] \circseq true) \vdash p2ac(P^t))
	&&\ptext{\cref{theorem:RA-o-A(P):RA-o-PBMH(P)}}\\
	&=\mathbf{RA} \circ \mathbf{A} (\lnot p2ac(P^f) \land (\lnot P^f[\mathbf{s}/in\alpha] \circseq true) \vdash p2ac(P^t))
\end{xflalign*}
\end{proof}\end{proofs}
\end{theorem}

\begin{theorem}\label{theorem:p2ac-o-R(design):RA-d2ac(design)}
\begin{align*}
	&p2ac \circ \mathbf{R} (\lnot P^f \vdash P^t)\\
	&=\\
	&\mathbf{RA} (\lnot p2ac(P^f) \land (\lnot P^f[\mathbf{s}/in\alpha] \circseq true) \vdash p2ac(P^t))
\end{align*}
\begin{proofs}\begin{proof}\checkt{alcc}
\begin{xflalign*}
	&p2ac \circ \mathbf{R} (\lnot P^f \vdash P^t)
	&&\ptext{\cref{theorem:p2ac-o-R:RA-o-p2ac}}\\
	&=\mathbf{RA} \circ p2ac(\lnot P^f \vdash P^t)
	&&\ptext{Definition of $\mathbf{RA}$}\\
	&=\mathbf{RA3} \circ \mathbf{RA2} \circ \mathbf{RA1} \circ p2ac(\lnot P^f \vdash P^t)
	&&\ptext{Definition of $\mathbf{RA1}$}\\
	&=\mathbf{RA3} \circ \mathbf{RA2} ((p2ac(\lnot P^f \vdash P^t) \land ac'\neq\emptyset)[\{ z | z \in ac' \land s.tr \le z.tr\}/ac'])
	&&\ptext{\cref{theorem:p2ac(P)-land-ac'-neq-emptyset:d2ac(P)-land-ac'-neq-emptyset}}\\
	&=\mathbf{RA3} \circ \mathbf{RA2} ((d2ac(\lnot P^f \vdash P^t) \land ac'\neq\emptyset)[\{ z | z \in ac' \land s.tr \le z.tr\}/ac'])
	&&\ptext{Definition of $\mathbf{RA1}$}\\
	&=\mathbf{RA3} \circ \mathbf{RA2} \circ \mathbf{RA1} \circ d2ac(\lnot P^f \vdash P^t)
	&&\ptext{Definition of $\mathbf{RA}$ and $d2ac$}\\
	&=\mathbf{RA} (\lnot p2ac(P^f) \land (\lnot P^f[\mathbf{s}/in\alpha] \circseq true) \vdash p2ac(P^t))
\end{xflalign*}
\end{proof}\end{proofs}
\end{theorem}

\begin{theorem}\label{theorem:p2ac-o-R1(P):RA1-o-p2ac(P)}
$\mathbf{RA1} \circ p2ac(P) = p2ac \circ \mathbf{R1} (P)$
\begin{proofs}\begin{proof}\checkt{pfr}\checkt{alcc}
\begin{xflalign*}
	&\mathbf{RA1} \circ p2ac(P)
	&&\ptext{Definition of $p2ac$}\\
	&=\mathbf{RA1} (\exists z \spot P[\mathbf{s},\mathbf{z}/in\alpha_{-ok},out\alpha_{-ok'}] \land undash(z) \in ac')
	&&\ptext{Definition of $\mathbf{RA1}$ (\cref{lemma:RA1:alternative-1})}\\
	&=\left(\begin{array}{l}
		\left(\begin{array}{l}
			\exists z \spot P[\mathbf{s},\mathbf{z}/in\alpha_{-ok},out\alpha_{-ok'}]
			\\ \land \\
			undash(z) \in ac'
		\end{array}\right)[\{z | z \in ac' \land s.tr \le z.tr\}/ac'] 
		\\ \land \exists z \spot s.tr \le z.tr \land z \in ac'
	\end{array}\right)
	&&\ptext{Substitution: $ac'$ not free in $P$}\\
	&=\left(\begin{array}{l}
		\left(\begin{array}{l}
			\exists z \spot P[\mathbf{s},\mathbf{z}/in\alpha_{-ok},out\alpha_{-ok'}]
			\\ \land \\
			undash(z) \in \{z | z \in ac' \land s.tr \le z.tr\}
		\end{array}\right)
		\\ \land \exists z \spot s.tr \le z.tr \land z \in ac'
	\end{array}\right)
	&&\ptext{Property of sets}\\
	&=\left(\begin{array}{l}
		\left(\begin{array}{l}
			\exists z \spot P[\mathbf{s},\mathbf{z}/in\alpha_{-ok},out\alpha_{-ok'}]
			\\ \land \\
			undash(z) \in ac' \land s.tr \le undash(z).tr
		\end{array}\right)
		\\ \land \exists z \spot s.tr \le z.tr \land z \in ac'
	\end{array}\right)
	&&\ptext{Predicate calculus: implication}\\
	&=\left(\begin{array}{l}
			\exists z \spot P[\mathbf{s},\mathbf{z}/in\alpha_{-ok},out\alpha_{-ok'}]
			\\ \land \\
			undash(z) \in ac' \land s.tr \le undash(z).tr
	\end{array}\right)
	&&\ptext{Property of $undash$}\\
	&=\exists z \spot P[\mathbf{s},\mathbf{z}/in\alpha_{-ok},out\alpha_{-ok'}] \land s.tr \le z.tr' \land undash(z) \in ac'
	&&\ptext{Substitution}\\
	&=\exists z \spot (P \land tr \le tr')[\mathbf{s},\mathbf{z}/in\alpha_{-ok},out\alpha_{-ok'}] \land undash(z) \in ac'
	&&\ptext{Definition of $p2ac$}\\
	&=p2ac(P \land tr \le tr')
	&&\ptext{Definition of $\mathbf{R1}$}\\
	&=p2ac \circ \mathbf{R1} (P)
\end{xflalign*}
\end{proof}\end{proofs}
\end{theorem}

\begin{theorem}\label{theorem:p2ac-o-R1-o-R2(P):RA2-o-p2ac(P)}
$p2ac \circ \mathbf{R1} \circ \mathbf{R2} (P) = \mathbf{RA2} \circ p2ac(P)$
\begin{proofs}\begin{proof}\checkt{pfr}\checkt{alcc}
\begin{flalign*}
	&\mathbf{RA2} \circ p2ac(P)
	&&\ptext{Definition of $p2ac$}\\
	&=\mathbf{RA2} (\exists z \spot P[\mathbf{s},\mathbf{z}/in\alpha_{-ok},out\alpha_{-ok'}] \land undash(z) \in ac')
	&&\ptext{Definition of $\mathbf{RA2}$}\\
	&=\left(
\right) \left[\begin{aligned}
	 &s\oplus\{tr\mapsto\lseq\rseq\} / s,\\
	 &\{z | z \in ac' \land s.tr \le z.tr \spot z \oplus \{ tr \mapsto z.tr - s.tr\} / ac' \\
	 \end{aligned}\right]
	&&\ptext{Substitution: $ac'$ not free in $P$}\\
	&=\left(%
\right)
	&&\ptext{Property of sets and variable renaming}\\
	&=\left(%
\right)
	&&\ptext{Introduce fresh variable}\\
	&=\left(%
\right)
	&&\ptext{\cref{lemma:state-substitution:e-si:z-Salpha} and substitution}\\
	&=\exists z \spot (P[\lseq\rseq,tr'-tr/tr,tr'][\mathbf{s},\mathbf{z}/in\alpha_{-ok},out\alpha_{-ok'}] \land s.tr \le z.tr') \land undash(z) \in ac'
	&&\ptext{Substitution}\\
	&=\exists z \spot (P[\lseq\rseq,tr'-tr/tr,tr'] \land tr \le tr')[\mathbf{s},\mathbf{z}/in\alpha_{-ok},out\alpha_{-ok'}] \land undash(z) \in ac'
	&&\ptext{Definition of $\mathbf{R2}$ and $\mathbf{R1}$}\\
	&=\exists z \spot (\mathbf{R1} \circ \mathbf{R2} (P))[\mathbf{s},\mathbf{z}/in\alpha_{-ok},out\alpha_{-ok'}] \land undash(z) \in ac'
	&&\ptext{Definition of $p2ac$}\\
	&=p2ac \circ \mathbf{R1} \circ \mathbf{R2} (P)	
\end{flalign*}
\end{proof}\end{proofs}
\end{theorem}

\begin{theorem}\label{theorem:p2ac-o-R3(P):RA3-o-p2ac(P)}
$p2ac \circ \mathbf{R3} (P) = \mathbf{RA3} \circ p2ac(P)$
\begin{proofs}\begin{proof}\checkt{pfr}\checkt{alcc}
\begin{xflalign*}
	&p2ac \circ \mathbf{R3} (P)
	&&\ptext{Definition of $\mathbf{R3}$}\\
	&=p2ac(\IIrea \dres wait \rres P)
	&&\ptext{\cref{lemma:p2ac(conditional)}}\\
	&=p2ac(\IIrea) \dres s.wait \rres p2ac(P)
	&&\ptext{\cref{lemma:p2ac(IIRea):IIRac}}\\
	&=\IIRac \dres s.wait \rres p2ac(P)
	&&\ptext{Definition of $\mathbf{RA3}$}\\
	&=\mathbf{RA3} \circ p2ac(P)
\end{xflalign*}
\end{proof}\end{proofs}
\end{theorem}

\begin{lemma}\label{lemma:IIRac:alternative-1}
\begin{align*}
	&\IIRac = (\lnot ok \land \exists z \spot s.tr \le z.tr \land z \in ac') \lor (ok' \land s \in ac')
\end{align*}
\begin{proofs}\begin{proof}\checkt{pfr}\checkt{alcc}
\begin{xflalign*}
	&\IIRac
	&&\ptext{Definition of $\IIRac$}\\
	&=\mathbf{RA1} (\lnot ok) \lor (ok' \land s \in ac')
	&&\ptext{Definition of $\mathbf{RA1}$ (\cref{lemma:RA1:alternative-1})}\\
	&=\left(
\right)
	\\ \land y \in ac'
	\end{array}\right)
	&&\ptext{Equality of records}\\
	&=\exists y \spot ((\lnot ok \land s.tr \le y.tr) \lor (ok' \land y = s)) \land y \in ac'
	&&\ptext{Predicate calculus}\\
	&=(\lnot ok \land \exists y \spot s.tr \le y.tr \land y \in ac') \lor (\exists y \spot ok' \land y = s \land y \in ac')
	&&\ptext{One-point rule}\\
	&=(\lnot ok \land \exists y \spot s.tr \le y.tr \land y \in ac') \lor (ok' \land s \in ac')
	&&\ptext{Definition of $\IIRac$ (\cref{lemma:IIRac:alternative-1})}\\
	&=\IIRac
\end{xflalign*}
\end{proof}\end{proofs}
\end{lemma}

\subsection{$ac2p$}

\begin{lemma}\label{lemma:ac2p(circledIn(P))-ac'-not-free:P-s-y-sub}
\begin{statement}
Provided $ac'$ is not free in $P$,
\begin{align*}
	&ac2p(\circledIn{y}{ac'} (P)) = P[State_{II}(in\alpha)/s][undash(State_{\II}(out\alpha_{-ok'}))/y]
\end{align*}
\end{statement}
\begin{proofs}
\begin{proof}
\begin{xflalign*}
	&ac2p(\circledIn{y}{ac'} (P))
	&&\ptext{\cref{lemma:ac2p(circledIn(P)):ac2p(P)-y-subs:new}}\\
	&=ac2p(P[\{y\}\cap ac'/ac'])[undash(State_{\II}(out\alpha_{-ok'}))/y]
	&&\ptext{Assumption: $ac'$ is not free in $P$}\\
	&=ac2p(P)[undash(State_{\II}(out\alpha_{-ok'}))/y]
	&&\ptext{\cref{lemma:ac2p(P):ac'-not-free}}\\
	&=P[State_{II}(in\alpha)/s][undash(State_{\II}(out\alpha_{-ok'}))/y]
\end{xflalign*}
\end{proof}
\end{proofs}
\end{lemma}

\begin{lemma}\label{theorem:ac2p(P)-seq-ac2p(Q)}
\begin{statement}
\begin{align*}
	&ac2p(P) \circseq ac2p(Q)\\
	&=\\
	&\exists ok_0, y @ \left(
\right)
	&&\ptext{Expand conjunction, where $x$ ranges over $x_0'$ to $x_n'$}\\
	&=\left(%
\right)
	&&\ptext{Definition of sequential composition, where the vector $\hatop{x}$ ranges over $x_0$ to $x_n$}\\
	&=\exists ok_0, \hatop{x} @ \left(%
\right)
	&&\ptext{Introduce fresh variable $y$}\\
	&=\exists ok_0, \hatop{x} @ \left(%
\right)
	&&\ptext{Equality of records}\\
	&=\exists ok_0, y @ \left(%
\right)
	&&\ptext{Property of sets}\\
	&=\exists ok_0, y @ \left(%
\right)
	&&\ptext{Case analysis on $ac'$ and substitution}\\
	&=\exists ok_0, y @ \left(%
\right)
	&&\ptext{Introduce fresh variable $s'$}\\
	&=\exists ok_0, y @ \left(%
\right)
	&&\ptext{Definition of sequential composition}\\
	&=\left(%
\right)
	&&\ptext{Predicate calculus and property of substitution}\\
	&=\left(%
\right)
	&&\ptext{Introduce fresh variable $s'$}\\
	&=\left(%
\right)
	&&\ptext{Definition of sequential composition}\\
	&=\left(%
\right)
	\end{array}\right)
\end{xflalign*}
\end{proof}
\end{proofs}
\end{lemma}

\subsection{$p2ac$}

\begin{theorem}\label{theorem:p2ac(P-seq-Q):p2ac(P)-seqDac-p2ac(Q)}
\begin{statement}
$p2ac(P \circseq Q) = p2ac(P) \seqDac p2ac(Q)$
\end{statement}
\begin{proofs}
\begin{proof}\checkt{alcc}
\begin{xflalign*}
	&p2ac(P) \seqDac p2ac(Q)
	&&\ptext{Definition of $\seqDac$}\\
	&=\exists ok_0 @ p2ac(P)[ok_0/ok'] \seqA p2ac(Q)[ok_0/ok]
	&&\ptext{\cref{lemma:p2ac(P)-o-ok:p2ac(P-o-ok),lemma:p2ac(P)-sub-ok'-wait:p2ac(sub-ok'-wait)}}\\
	&=\exists ok_0 @ p2ac(P[ok_0/ok']) \seqA p2ac(Q[ok_0/ok])
	&&\ptext{\cref{theorem:p2ac(P)-seqA-p2ac(Q):p2ac(P-seq-Q)}}\\
	&=\exists ok_0 @ p2ac(P[ok_0/ok'] \circseq Q[ok_0/ok])
	&&\ptext{\cref{lemma:exists-x-p2ac(P):p2ac(exists-x-P)}}\\
	&=p2ac(\exists ok_0 @ P[ok_0/ok'] \circseq Q[ok_0/ok])
	&&\ptext{Definition of sequential composition}\\
	&=p2ac(P \circseq Q)
\end{xflalign*}
\end{proof}
\end{proofs}
\end{theorem}

\begin{theorem}\label{theorem:p2ac(P)-seqA-p2ac(Q):p2ac(P-seq-Q)}
\begin{statement}
Provided $ok'$ is not free in $P$ and $ok$ is not free in $Q$,
\begin{align*}
	&p2ac(P) \seqA p2ac(Q) = p2ac(P \circseq Q)
\end{align*}
\end{statement}
\begin{proofs}\begin{proof}\checkt{alcc}
\begin{xflalign*}
	&p2ac(P) \seqA p2ac(Q)
	&&\ptext{Definition of $p2ac$}\\
	&=(\exists z @ P[\mathbf{s},\mathbf{z}/in\alpha_{-ok},out\alpha_{-ok'}] \land undash(z) \in ac') \seqA p2ac(Q)
	&&\ptext{Definition of $\seqA$ and substitution}\\
	&=\exists z @ P[\mathbf{s},\mathbf{z}/in\alpha_{-ok},out\alpha_{-ok'}] \land undash(z) \in \{ s | p2ac(Q)\}
	&&\ptext{Property of sets}\\
	&=\exists z @ P[\mathbf{s},\mathbf{z}/in\alpha_{-ok},out\alpha_{-ok'}] \land p2ac(Q)[undash(z)/s]
	&&\ptext{Definition of $p2ac$}\\
	&=\exists z @ \left(
\right)
	&&\ptext{Equality of records}\\
	&=\exists z @ \left(%
\right)
	&&\ptext{Introduce fresh vector of variables $\hatop{x}$}\\
	&=\exists z, \hatop{x} @ \left(%
\right)
	&&\ptext{Equality of records}\\
	&=\exists z, \hatop{x} @ \left(%
\right)
	&&\ptext{Substitution and value of record components}\\
	&=\exists \hatop{x} @ \left(%
\right)
	&&\ptext{Vector of variables $\hatop{x}$}\\
	&=\exists \hatop{x} @ \left(%
\right)
	&&\ptext{Property of substitution}\\
	&=\exists \hatop{x} @ \left(%
\right)  \land undash(t) \in ac'
	&&\ptext{Property of substitution}\\
	&=\exists t @ (\exists \hatop{x} @ P[\hatop{x}/x'] \land Q[\hatop{x}/x])[\mathbf{s},\mathbf{t}/in\alpha_{-ok},out\alpha_{-ok'}]  \land undash(t) \in ac'
	&&\ptext{Definition of sequential composition assuming $ok' \notin fv(P)$ and $ok \notin fv(Q)$}\\
	&=\exists t @ (P \circseq Q)[\mathbf{s},\mathbf{t}/in\alpha_{-ok},out\alpha_{-ok'}]  \land undash(t) \in ac'
	&&\ptext{Definition of $p2ac$}\\
	&=p2ac(P \circseq Q)
\end{xflalign*}
\end{proof}\end{proofs}
\end{theorem}

\begin{lemma}\label{lemma:p2ac(P)-z-ac'-land-z-in-ac'}
$p2ac(P)[\{z\}/ac'] \land z \in ac' = p2ac(P)[\{z\}\cap ac'/ac'] \land z \in ac'$
\begin{proofs}\begin{proof}
\begin{xflalign*}
	&p2ac(P)[\{z\}\cap ac'/ac'] \land z \in ac'
	&&\ptext{\cref{lemma:PBMH-o-p2ac(P):p2ac(P)}}\\
	&=(\mathbf{PBMH} \circ p2ac(P))[\{z\}\cap ac'/ac'] \land z \in ac'
	&&\ptext{Definition of $\mathbf{PBMH}$ (\cref{lemma:PBMH:alternative-1})}\\
	&=(\exists ac_0 @ p2ac(P)[ac_0/ac'] \land ac_0\subseteq ac')[\{z\}\cap ac'/ac'] \land z \in ac'
	&&\ptext{Substitution}\\
	&=\exists ac_0 @ p2ac(P)[ac_0/ac'] \land ac_0\subseteq (\{z\}\cap ac') \land z \in ac'
	&&\ptext{Property of sets}\\
	&=\exists ac_0 @ p2ac(P)[ac_0/ac'] \land ac_0\subseteq \{z\} \land ac_0\subseteq ac' \land z \in ac'
	&&\ptext{Property of sets}\\
	&=\exists ac_0 @ p2ac(P)[ac_0/ac'] \land ac_0\subseteq \{z\} \land z \in ac'
	&&\ptext{Substitution}\\
	&=(\exists ac_0 @ p2ac(P)[ac_0/ac'] \land ac_0\subseteq ac')[\{z\}/ac'] \land z \in ac'
	&&\ptext{Definition of $\mathbf{PBMH}$ (\cref{lemma:PBMH:alternative-1})}\\
	&=(\mathbf{PBMH} \circ p2ac(P))[\{z\}/ac'] \land z \in ac'
	&&\ptext{\cref{lemma:PBMH-o-p2ac(P):p2ac(P)}}\\
	&=p2ac(P)[\{z\}/ac'] \land z \in ac'
\end{xflalign*}
\end{proof}\end{proofs}
\end{lemma}

\subsection{$p2ac$ and $ac2p$}

\begin{theorem}\label{theorem:ac2p-o-p2ac(P):P}
\begin{statement}
$ac2p \circ p2ac (P) = P$
\end{statement}
\begin{proofs}
\begin{proof}\checkt{pfr}\checkt{alcc}
\begin{xflalign*}
	&ac2p \circ p2ac(P)
	&&\ptext{Definition of $ac2p$}\\
	&=(\mathbf{PBMH} \circ p2ac(P))[State_{\II}(in\alpha_{-ok})/s] \seqA \bigwedge x : out\alpha_{-ok'} \spot dash(s).x = x
	&&\ptext{\cref{lemma:PBMH-o-p2ac(P):p2ac(P)}}\\
	&=p2ac(P)[State_{\II}(in\alpha_{-ok})/s] \seqA \bigwedge x : out\alpha_{-ok'} \spot dash(s).x = x
	&&\ptext{Definition of $p2ac$}\\
	&=\left(\begin{array}{l}
		(\exists z @ P[\mathbf{s},\mathbf{z}/in\alpha_{-ok},out\alpha_{-ok'}] \land undash(z) \in ac')[State_{\II}(in\alpha_{-ok})/s] 
		\\ \seqA \\
		\bigwedge x : out\alpha_{-ok'} \spot dash(s).x = x
	\end{array}\right)
	&&\ptext{Substitution}\\
	&=\left(\begin{array}{l}
		(\exists z @ P[\mathbf{s},\mathbf{z}/in\alpha_{-ok},out\alpha_{-ok'}][State_{\II}(in\alpha_{-ok})/s]  \land undash(z) \in ac')
		\\ \seqA \\
		\bigwedge x : out\alpha_{-ok'} \spot dash(s).x = x
	\end{array}\right)
	&&\ptext{\cref{lemma:state-sub:P-z-S:S-z}}\\
	&=(\exists z @ P[\mathbf{z}/out\alpha_{-ok'}] \land undash(z) \in ac') \seqA \bigwedge x : out\alpha_{-ok'} \spot dash(s).x = x
	&&\ptext{Definition of $\seqA$ and substitution}\\
	&=\exists z @ P[\mathbf{z}/out\alpha_{-ok'}] \land undash(z) \in \{ s | \bigwedge x : out\alpha_{-ok'} \spot dash(s).x = x \}
	&&\ptext{Property of sets}\\
	&=\exists z @ P[\mathbf{z}/out\alpha_{-ok'}] \land \bigwedge x : out\alpha_{-ok'} \spot dash(undash(z)).x = x
	&&\ptext{Property of $dash$ and $undash$}\\
	&=\exists z @ P[\mathbf{z}/out\alpha_{-ok'}] \land \bigwedge x : out\alpha_{-ok'} \spot z.x = x
	&&\ptext{\cref{lemma:state-sub:exists-z-State-P}}\\
	&=P[\mathbf{z}/out\alpha_{-ok'}][State_{\II} (out\alpha_{-ok'})/z]
	&&\ptext{\cref{lemma:state-sub:P-z-S:S-z}}\\
 	&=P
\end{xflalign*}
\end{proof}
\end{proofs}
\end{theorem}

\begin{theorem}\label{theorem:p2ac-o-ac2p:implies:P}
\begin{statement}
Provided $P$ is $\mathbf{PBMH}$-healthy, $p2ac \circ ac2p (P) \sqsupseteq P$.
\end{statement}
\begin{proofs}
\begin{proof}\checkt{pfr}\checkt{alcc}
\begin{xflalign*}
	&p2ac \circ ac2p (P)
	&&\ptext{\cref{lemma:p2ac-o-ac2p(P)}}\\
	&=\exists ac_0, y \spot P[ac_0/ac'] \land ac_0 \subseteq \{ y \} \land y \in ac'
	&&\ptext{Property of sets}\\
	&=\exists ac_0, y \spot P[ac_0/ac'] \land ac_0 \subseteq \{ y \} \land \{y\} \subseteq ac'
	&&\ptext{Predicate calculus}\\
	&\implies \exists ac_0 \spot P[ac_0/ac'] \land ac_0 \subseteq ac'
	&&\ptext{Definition of $\mathbf{PBMH}$ (\cref{lemma:PBMH:alternative-1})}\\
	&=\mathbf{PBMH} (P)
	&&\ptext{Assumption: $P$ is $\mathbf{PBMH}$-healthy}\\
	&=P
\end{xflalign*}
\end{proof}
\end{proofs}
\end{theorem}

\begin{theorem}\label{theorem:p2ac-o-ac2p:implies:PBMH(P)}
$p2ac \circ ac2p (P) \sqsupseteq \mathbf{PBMH} (P)$
\begin{proofs}\begin{proof}\checkt{pfr}\checkt{alcc}
\begin{xflalign*}
	&p2ac \circ ac2p (P)
	&&\ptext{\cref{lemma:p2ac-o-ac2p(P)}}\\
	&=\exists ac_0, y \spot P[ac_0/ac'] \land ac_0 \subseteq \{ y \} \land y \in ac'
	&&\ptext{Property of sets}\\
	&=\exists ac_0, y \spot P[ac_0/ac'] \land ac_0 \subseteq \{ y \} \land \{y\} \subseteq ac'
	&&\ptext{Predicate calculus}\\
	&\implies \exists ac_0 \spot P[ac_0/ac'] \land ac_0 \subseteq ac'
	&&\ptext{Definition of $\mathbf{PBMH}$ (\cref{lemma:PBMH:alternative-1})}\\
	&=\mathbf{PBMH} (P)
\end{xflalign*}
\end{proof}\end{proofs}
\end{theorem}

\begin{theorem}\label{theorem:p2ac(ac2p(P)-seq-ac2p(Q)):(exists-ac'-P)-seq-p2ac-o-ac2p(Q)}
\begin{statement}
\begin{align*}
	&p2ac(ac2p(P) \circseq ac2p(Q)) = (\exists ac' @ P \land ac' \subseteq \{ s'\}) \circseq p2ac\circ ac2p(Q)
\end{align*}
\end{statement}
\begin{proofs}
\begin{proof}
\begin{xflalign*}
	&p2ac(ac2p(P) \circseq ac2p(Q))
	&&\ptext{\cref{lemma:ac2p(P)-seq-ac2p(Q):seq-alternative}}\\
	&=ac2p\left(
\right)
	&&\ptext{Change variable name}\\
	&=\left(%
\right)
	&&\ptext{Introduce fresh variable $t$}\\
	&=\left(%
\right)
	&&\ptext{Property of $dash$ and equality of records}\\
	&=\left(%
\right)
	&&\ptext{Property of records}\\
	&=\left(%
\right)
	&&\ptext{Introduce fresh variable $ac_0$}\\
	&=\left(%
\right)
	&&\ptext{\cref{lemma:p2ac-o-ac2p(P)}}\\
	&=(\exists ac' @ P \land ac' \subseteq \{ s'\}) \circseq p2ac\circ ac2p(Q)
\end{xflalign*}
\end{proof}
\end{proofs}
\end{theorem}

\begin{theorem}\label{theorem:p2ac(ac2p(P)-seq-ac2p(Q)):implies:PBMH(P)-seqDac-Q}
\begin{statement}
Provided $Q$ is $\mathbf{PBMH}$-healthy and $s'$ is not free in $P$,
\begin{align*}
	&p2ac(ac2p(P) \circseq ac2p(Q)) \implies \mathbf{PBMH} (P) \seqDac Q
\end{align*}
\end{statement}
\begin{proofs}
\begin{proof}
\begin{xflalign*}
	&p2ac(ac2p(P) \circseq ac2p(Q))
	&&\ptext{\cref{theorem:p2ac(ac2p(P)-seq-ac2p(Q)):(exists-ac'-P)-seq-p2ac-o-ac2p(Q)}}\\
	&=(\exists ac' @ P \land ac' \subseteq \{ s'\}) \circseq p2ac\circ ac2p(Q)
	&&\ptext{Assumption: $Q$ is $\mathbf{PBMH}$-healthy and~\cref{theorem:p2ac-o-ac2p:implies:P}}\\
	&\implies (\exists ac' @ P \land ac' \subseteq \{ s'\}) \circseq Q
	&&\ptext{Definition of sequential composition}\\
	&=\exists s_0, ok_0 @ (\exists ac' @ P \land ac' \subseteq \{ s'\})[ok_0,s_0/ok',s'] \land Q[ok_0,s_0/ok,s]
	&&\ptext{Substitution}\\
	&=\exists s_0, ok_0 @ (\exists ac' @ P[ok_0/ok'] \land ac' \subseteq \{ s_0\}) \land Q[ok_0/ok][s_0/s]
	&&\ptext{Introduce fresh variable $ac_0$}\\
	&=\exists s_0, ok_0, ac_0 @ P[ok_0/ok'][ac_0/ac'] \land ac_0 \subseteq \{ s_0\} \land Q[ok_0/ok][s_0/s]
	&&\ptext{Property of sets}\\
	&=\exists s_0, ok_0, ac_0 @ P[ok_0/ok'][ac_0/ac'] \land ac_0 \subseteq \{ s_0\} \land s_0 \in \{ s | Q[ok_0/ok] \}
	&&\ptext{Property of sets}\\
	&=\exists s_0, ok_0, ac_0 @ P[ok_0/ok'][ac_0/ac'] \land ac_0 \subseteq \{ s_0\} \land \{ s_0 \} \subseteq \{ s | Q[ok_0/ok] \}
	&&\ptext{Property of sets and predicate calculus}\\
	&\implies \exists ok_0, ac_0 @ P[ok_0/ok'][ac_0/ac'] \land ac_0 \subseteq \{ s | Q[ok_0/ok] \}
	&&\ptext{Substitution}\\
	&=\exists ok_0, ac_0 @ (P[ok_0/ok'][ac_0/ac'] \land ac_0 \subseteq ac')[\{ s | Q[ok_0/ok] \}/ac']
	&&\ptext{Predicate calculus}\\
	&=\exists ok_0 @ (\exists ac_0 @ P[ok_0/ok'][ac_0/ac'] \land ac_0 \subseteq ac')[\{ s | Q[ok_0/ok] \}/ac']
	&&\ptext{\cref{lemma:PBMH:alternative-1}}\\
	&=\exists ok_0 @ (\mathbf{PBMH} (P[ok_0/ok']))[\{ s | Q[ok_0/ok] \}/ac']
	&&\ptext{Definition of $\seqA$}\\
	&=\exists ok_0 @ (\mathbf{PBMH} (P[ok_0/ok'])) \seqA Q[ok_0/ok]
	&&\ptext{\cref{lemma:PBMH(P)-ow:PBMH(P-ow)}}\\
	&=\exists ok_0 @ (\mathbf{PBMH} (P)[ok_0/ok']) \seqA Q[ok_0/ok]
	&&\ptext{Definition of $\seqDac$}\\
	&=\mathbf{PBMH} (P) \seqDac Q
\end{xflalign*}
\end{proof}
\end{proofs}
\end{theorem}

\begin{lemma}\label{lemma:p2ac(ac2p(P)-seq-ac2p(Q)):implies:P-seqA-Q}
\begin{statement}
Provided $P$ and $Q$ are $\mathbf{PBMH}$-healthy, $s'$ is not free in $P$, $ok'$ is not free in $P$ and $ok$ is not free in $Q$,
\begin{align*}
	&p2ac(ac2p(P) \circseq ac2p(Q)) \implies P \seqA Q
\end{align*}
\end{statement}
\begin{proofs}
\begin{proof}
\begin{xflalign*}
	&p2ac(ac2p(P) \circseq ac2p(Q))
	&&\ptext{\cref{theorem:p2ac(ac2p(P)-seq-ac2p(Q)):implies:PBMH(P)-seqDac-Q}}\\
	&\implies \mathbf{PBMH} (P) \seqDac Q
	&&\ptext{Assumption: $P$ is $\mathbf{PBMH}$-healthy}\\
	&=P \seqDac Q
	&&\ptext{Assumption: $ok' \notin fv(P)$, $ok \notin fv(Q)$ and~\cref{lemma:P-seqDac-Q:ok-ok'-not-free:P-seqA-Q}}\\
	&=P \seqA Q
\end{xflalign*}
\end{proof}
\end{proofs}
\end{lemma}

\begin{lemma}\label{lemma:p2ac(ac2p(P)-seq-ac2p(Q)):ok-free-in-Q:implies:P-seqA-Q}
\begin{statement}
Provided $P$ and $Q$ are $\mathbf{PBMH}$-healthy, $s'$ is not free in $P$, $ok'$ is not free in $P$.
\begin{align*}
	&p2ac(ac2p(P) \circseq ac2p(Q)) \implies P \seqA (\exists ok @ Q)
\end{align*}
\end{statement}
\begin{proofs}
\begin{proof}
\begin{xflalign*}
	&p2ac(ac2p(P) \circseq ac2p(Q))
	&&\ptext{\cref{theorem:p2ac(ac2p(P)-seq-ac2p(Q)):implies:PBMH(P)-seqDac-Q}}\\
	&\implies \mathbf{PBMH} (P) \seqDac Q
	&&\ptext{\cref{lemma:P-seqDac-Q:implies:P-seqA-(exists-ok-Q)}}\\
	&\implies \mathbf{PBMH} (P) \seqA (\exists ok @ Q)
	&&\ptext{Assumption: $P$ is $\mathbf{PBMH}$-healthy}\\
	&=P \seqA (\exists ok @ Q)
\end{xflalign*}
\end{proof}
\end{proofs}
\end{lemma}

\subsubsection{Results with respect to $\mathbf{A2}$}

\begin{theorem}\label{theorem:p2ac-o-ac2p-RA-o-A:RA-o-A}
\begin{statement}Provided $P^f_f$ and $P^t_f$ are $\mathbf{A2}$-healthy,
\begin{align*}
	&p2ac \circ ac2p \circ \mathbf{RA} \circ \mathbf{A} (\lnot P^f_f \vdash P^t_f) = \mathbf{RA} \circ \mathbf{A} (\lnot P^f_f \vdash P^t_f)
\end{align*}
\end{statement}
\begin{proofs}
\begin{proof}
\begin{xflalign*}
	&p2ac \circ ac2p \circ \mathbf{RA} \circ \mathbf{A} (\lnot P^f_f \vdash P^t_f)
	&&\ptext{\cref{theorem:ac2p-o-RA-o-A(design):R(lnot-ac2p(pre)|-ac2p(post))}}\\
	&=p2ac \circ \mathbf{R} (\lnot ac2p(P^f_f) \vdash ac2p(P^t_f))
	&&\ptext{\cref{theorem:p2ac-o-R(design):RA(p2ac-design)}}\\
	&=\mathbf{RA} (\lnot p2ac \circ ac2p(P^f_f) \vdash p2ac \circ ac2p(P^t_f))
	&&\ptext{Definition of $\mathbf{RA}$}\\
	&=\mathbf{RA3} \circ \mathbf{RA2} \circ \mathbf{RA1} (\lnot p2ac \circ ac2p(P^f_f) \vdash p2ac \circ ac2p(P^t_f))
	&&\ptext{\cref{lemma:RA1(Pff|-Ptf):RA1(Pff-ac'-neq-emptyset|-Ptf-ac'-neq-emptyset)}}\\
	&=\mathbf{RA3} \circ \mathbf{RA2} \circ \mathbf{RA1} 
		\left(\begin{array}{l}
			\lnot (p2ac \circ ac2p(P^f_f) \land ac'\neq\emptyset)
			\\ \vdash \\
			p2ac \circ ac2p(P^t_f) \land ac'\neq\emptyset
		\end{array}\right)
	&&\ptext{Assumption: $P^t_f$ and $P^f_f$ are $\mathbf{A2}$-healthy}\\
	&=\mathbf{RA3} \circ \mathbf{RA2} \circ \mathbf{RA1} 
		\left(\begin{array}{l}
			\lnot (p2ac \circ ac2p \circ \mathbf{A2} (P^f_f) \land ac'\neq\emptyset)
			\\ \vdash \\
			p2ac \circ ac2p \circ \mathbf{A2} (P^t_f) \land ac'\neq\emptyset
		\end{array}\right)	
	&&\ptext{\cref{lemma:p2ac-o-ac2p-o-A2(P):A2(P)-land-ac'-neq-emptyset}}\\
	&=\mathbf{RA3} \circ \mathbf{RA2} \circ \mathbf{RA1} 
		\left(\begin{array}{l}
			\lnot (\mathbf{A2} (P^f_f) \land ac'\neq\emptyset)
			\\ \vdash \\
			\mathbf{A2} (P^t_f) \land ac'\neq\emptyset
		\end{array}\right)
	&&\ptext{\cref{lemma:PBMH-o-A2(P):A2(P)}}\\
	&=\mathbf{RA3} \circ \mathbf{RA2} \circ \mathbf{RA1} 
		\left(\begin{array}{l}
			\lnot (\mathbf{PBMH} \circ \mathbf{A2} (P^f_f) \land ac'\neq\emptyset)
			\\ \vdash \\
			\mathbf{PBMH} \circ \mathbf{A2} (P^t_f) \land ac'\neq\emptyset
		\end{array}\right)
	&&\ptext{Assumption: $P^t_f$ and $P^f_f$ are $\mathbf{A2}$-healthy}\\
	&=\mathbf{RA3} \circ \mathbf{RA2} \circ \mathbf{RA1} 
		\left(\begin{array}{l}
			\lnot (\mathbf{PBMH} (P^f_f) \land ac'\neq\emptyset)
			\\ \vdash \\
			\mathbf{PBMH} (P^t_f) \land ac'\neq\emptyset
		\end{array}\right)
	&&\ptext{\cref{lemma:RA1(Pff|-Ptf):RA1(Pff-ac'-neq-emptyset|-Ptf-ac'-neq-emptyset)}}\\
	&=\mathbf{RA3} \circ \mathbf{RA2} \circ \mathbf{RA1} (\lnot \mathbf{PBMH} (P^f_f) \vdash \mathbf{PBMH} (P^t_f))
	&&\ptext{Definition of $\mathbf{A1}$}\\
	&=\mathbf{RA3} \circ \mathbf{RA2} \circ \mathbf{RA1} \circ \mathbf{A1} (\lnot P^f_f \vdash P^t_f)
	&&\ptext{\cref{theorem:RA1-o-A0:RA1}}\\
	&=\mathbf{RA3} \circ \mathbf{RA2} \circ \mathbf{RA1} \circ \mathbf{A0} \circ \mathbf{A1} (\lnot P^f_f \vdash P^t_f)
	&&\ptext{Definition of $\mathbf{RA}$}\\
	&=\mathbf{RA} \circ \mathbf{A0} \circ \mathbf{A1} (\lnot P^f_f \vdash P^t_f)
	&&\ptext{Definition of $\mathbf{A}$}\\
	&=\mathbf{RA} \circ \mathbf{A} (\lnot P^f_f \vdash P^t_f)
\end{xflalign*}
\end{proof}
\end{proofs}
\end{theorem}

\begin{lemma}\label{lemma:p2ac-o-ac2p(P)-A2:P-land-ac'-neq-emptyset}
\begin{statement}
Provided $P$ is $\mathbf{A2}$-healthy,
$p2ac\circ ac2p(P) = P \land ac'\neq\emptyset$
\end{statement}
\begin{proofs}
\begin{proof}\checkt{alcc}
\begin{xflalign*}
	&p2ac\circ ac2p(P)
	&&\ptext{Assumption: $P$ is $\mathbf{A2}$-healthy}\\
	&=p2ac\circ ac2p\circ \mathbf{A2} (P)
	&&\ptext{\cref{lemma:p2ac-o-ac2p-o-A2(P):A2(P)-land-ac'-neq-emptyset}}\\
	&=\mathbf{A2} (P) \land ac'\neq\emptyset
	&&\ptext{Assumption: $P$ is $\mathbf{A2}$-healthy}\\
	&=P \land ac'\neq\emptyset
\end{xflalign*}
\end{proof}
\end{proofs}
\end{lemma}

\begin{lemma}\label{lemma:p2ac-o-ac2p(P)-x-for-ac'-A2:P-x-for-ac'}
\begin{statement}
Provided $P$ is $\mathbf{A2}$-healthy,
\begin{align*}
	&p2ac\circ ac2p(P)[\{x\}/ac'] = P[\{x\}/ac']
\end{align*}
\end{statement}
\begin{proofs}
\begin{proof}\checkt{alcc}
\begin{xflalign*}
	&p2ac\circ ac2p(P)[\{x\}/ac']
	&&\ptext{Assumption: $P$ is $\mathbf{A2}$-healthy and~\cref{lemma:p2ac-o-ac2p(P)-A2:P-land-ac'-neq-emptyset}}\\
	&=(P \land ac'\neq\emptyset)[\{x\}/ac']
	&&\ptext{Substitution}\\
	&=P[\{x\}/ac'] \land \{x\}\neq\emptyset
	&&\ptext{Property of sets}\\
	&=P[\{x\}/ac']
\end{xflalign*}
\end{proof}
\end{proofs}
\end{lemma}

\subsection{Lifting}

\theoremstatementref{def:circledIn}

\begin{lemma}\label{lemma:PBMH(circledIn):circledIn} Provided $ac'$ is not free in $P$,
\begin{align*}
	&\mathbf{PBMH} (\circledIn{y}{ac'} (P)) = \circledIn{y}{ac'} (P) 
\end{align*}
\begin{proofs}\begin{proof}
\begin{xflalign*}
	&\mathbf{PBMH} (\circledIn{y}{ac'} (P))
	&&\ptext{Assumption and~\cref{lemma:circledIn:ac'-not-free}}\\
	&=\mathbf{PBMH} (\exists y \spot P \land y \in ac')
	&&\ptext{Definition of $\mathbf{PBMH}$ (\cref{lemma:PBMH:alternative-1})}\\
	&=\exists ac_0 \spot (\exists y \spot P \land y \in ac')[ac_0/ac'] \land ac_0 \subseteq ac'
	&&\ptext{Assumption: $ac'$ not free in $P$ and substitution}\\
	&=\exists ac_0 \spot (\exists y \spot P \land y \in ac_0) \land ac_0 \subseteq ac'
	&&\ptext{Predicate calculus}\\
	&=\exists ac_0, y \spot P \land y \in ac_0 \land ac_0 \subseteq ac'
	&&\ptext{Predicate calculus and property of sets}\\
	&=\exists y \spot P \land y \in ac'
	&&\ptext{Definition of $\circledIn{y}{ac'}$}\\
	&=\circledIn{y}{ac'} (P)
\end{xflalign*}
\end{proof}\end{proofs}
\end{lemma}

\begin{lemma}\label{lemma:RA1(circledIn):circledIn(P-land-s.tr-le-y.tr)}
\begin{align*}
	&\mathbf{RA1} (\circledIn{y}{ac'} (P)) = \circledIn{y}{ac'} (\mathbf{RA1} (P[\{y\}\cap ac'/ac']) \land s.tr \le y.tr)
\end{align*}
\begin{proofs}\begin{proof}\checkt{alcc}\checkt{pfr}
\begin{xflalign*}
	&\mathbf{RA1} (\circledIn{y}{ac'} (P))
	&&\ptext{Definition of $\circledIn{y}{ac'}$}\\
	&=\mathbf{RA1} (\exists y \spot P[\{y\}\cap ac'/ac'] \land y \in ac')
	&&\ptext{\cref{lemma:RA1(exists-P):exists-RA1(P)}}\\
	&=\exists y \spot \mathbf{RA1} (P[\{y\}\cap ac'/ac'] \land y \in ac')
	&&\ptext{\cref{lemma:RA1(P-land-Q):RA1(P)-land-RA1(Q)}}\\
	&=\exists y \spot \mathbf{RA1} (P[\{y\}\cap ac'/ac']) \land \mathbf{RA1} (y \in ac')
	&&\ptext{\cref{lemma:RA1(x-in-ac'):x-in-ac'-land-prefix-tr}}\\
	&=\exists y \spot \mathbf{RA1} (P[\{y\}\cap ac'/ac']) \land s.tr \le y.tr \land y \in ac'
	&&\ptext{Definition of $\circledIn{y}{ac'}$}\\
	&=\circledIn{y}{ac'} (\mathbf{RA1} (P[\{y\}\cap ac'/ac']) \land s.tr \le y.tr)
\end{xflalign*}
\end{proof}\end{proofs}
\end{lemma}

\begin{lemma}\label{lemma:RA1(circledIn)-ac'-not-free:circledIn(P-land-s.tr-le-y.tr)}
Provided $ac'$ is not free in $P$,
\begin{align*}
	&\mathbf{RA1} (\circledIn{y}{ac'} (P)) = \circledIn{y}{ac'} (P \land s.tr \le y.tr)
\end{align*}
\begin{proofs}\begin{proof}\checkt{alcc}\checkt{pfr}
\begin{xflalign*}
	&\mathbf{RA1} (\circledIn{y}{ac'} (P))
	&&\ptext{\cref{lemma:RA1(circledIn):circledIn(P-land-s.tr-le-y.tr)}}\\
	&=\circledIn{y}{ac'} (\mathbf{RA1} (P[\{y\}\cap ac'/ac']) \land s.tr \le y.tr)
	&&\ptext{Substitution: $ac'$ not free in $P$}\\
	&=\circledIn{y}{ac'} (\mathbf{RA1} (P) \land s.tr \le y.tr)
	&&\ptext{Assumption: $ac'$ is not free in $P$ and~\cref{lemma:RA1(P)-ac'-not-free:P-land-RA1(true)}}\\
	&=\circledIn{y}{ac'} (P \land \mathbf{RA1} (true) \land s.tr \le y.tr)
	&&\ptext{\cref{lemma:RA1(true)}}\\
	&=\circledIn{y}{ac'} (P \land (\exists z \spot s.tr \le z.tr \land z \in ac') \land s.tr \le y.tr)
	&&\ptext{Definition of $\circledIn{y}{ac'}$}\\
	&=\exists y \spot P \land (\exists z \spot s.tr \le z.tr \land z \in ac') \land s.tr \le y.tr \land y \in ac'
	&&\ptext{Predicate calculus}\\
	&=(\exists y \spot P \land s.tr \le y.tr \land y \in ac') \land (\exists z \spot s.tr \le z.tr \land z \in ac')
	&&\ptext{Predicate calculus}\\
	&=\exists y \spot P \land s.tr \le y.tr \land y \in ac'
	&&\ptext{Definition of $\circledIn{y}{ac'}$}\\
	&=\circledIn{y}{ac'} (P \land s.tr \le y.tr)
\end{xflalign*}
\end{proof}\end{proofs}
\end{lemma}

\begin{lemma}\label{lemma:RA2(circledIn)}
Provided $ac'$ is not free in $P$,
\begin{align*}
	&\mathbf{RA2} (\circledIn{y}{ac'} (P))\\
	&=\\
	&\exists y \spot \mathbf{RA2} (P) \land \circledIn{z}{ac'} (s.tr \le z.tr \land y = z\oplus\{tr\mapsto z.tr-s.tr\})
\end{align*}
\begin{proofs}\begin{proof}\checkt{alcc}
\begin{xflalign*}
	&\mathbf{RA2} (\circledIn{y}{ac'} (P))
	&&\ptext{Assumption: $ac'$ is not free in $P$ and~\cref{lemma:circledIn:ac'-not-free}}\\
	&=\mathbf{RA2} (\exists y \spot P \land y \in ac')
	&&\ptext{\cref{lemma:RA2(exists-P):exists-RA2(P)}}\\
	&=\exists y \spot \mathbf{RA2} (P \land y \in ac')
	&&\ptext{\cref{theorem:RA2(P-land-Q):RA2(P)-land-RA2(Q)}}\\
	&=\exists y \spot \mathbf{RA2} (P) \land \mathbf{RA2} (y \in ac')
	&&\ptext{\cref{lemma:RA2(x-in-ac'):circledIn}}\\
	&=\exists y \spot \mathbf{RA2} (P) \land \circledIn{z}{ac'} (s.tr \le z.tr \land y = z\oplus\{tr\mapsto z.tr-s.tr\})
\end{xflalign*}
\end{proof}\end{proofs}
\end{lemma}

\begin{lemma}\label{lemma:RA2(x-in-ac'):circledIn}
Provided $x$ is not $s$,
\begin{align*}
	&\mathbf{RA2} (x \in ac') = \circledIn{z}{ac'} (s.tr \le z.tr \land x = z\oplus\{tr\mapsto z.tr-s.tr\})
\end{align*}
\begin{proofs}\begin{proof}\checkt{alcc}
\begin{xflalign*}
	&\mathbf{RA2} (x \in ac')
	&&\ptext{Definition of $\mathbf{RA2}$}\\
	&=(x \in ac')[s\oplus\{tr \mapsto \lseq\rseq\},\{z|z\in ac'\land s.tr\le z.tr\spot z\oplus\{tr\mapsto z.tr-s.tr\}\}/s,ac']
	&&\ptext{Substitution: $x$ is not $s$}\\
	&=x \in \{z|z\in ac'\land s.tr\le z.tr\spot z\oplus\{tr\mapsto z.tr-s.tr\}\}
	&&\ptext{Property of sets}\\
	&=\exists z \spot z \in ac' \land s.tr \le z.tr \land x = z\oplus\{tr\mapsto z.tr-s.tr\}
	&&\ptext{Definition of $\circledIn{z}{ac'}$}\\
	&=\circledIn{z}{ac'} (s.tr \le z.tr \land x = z\oplus\{tr\mapsto z.tr-s.tr\})
\end{xflalign*}
\end{proof}\end{proofs}
\end{lemma}

\begin{lemma}\label{lemma:RA2(x-in-ac'):x-oplus-in-ac'}
$\mathbf{RA2} (x \in ac') = x \oplus \{tr\mapsto s.tr\cat x.tr\} \in ac'$
\begin{proofs}\begin{proof}\checkt{alcc}
\begin{xflalign*}
	&\mathbf{RA2} (x \in ac')
	&&\ptext{Definition of $\mathbf{RA2}$ (\cref{lemma:RA2:alternative-1})}\\
	&=(x \in ac')[s\oplus\{tr\mapsto\lseq\rseq\},\{y | y \oplus \{tr\mapsto s.tr\cat y.tr\} \in ac'\}/s,ac']
	&&\ptext{Substitution}\\
	&=x \in \{y | y \oplus \{tr\mapsto s.tr\cat y.tr\} \in ac'\}
	&&\ptext{Property of sets}\\
	&=x \oplus \{tr\mapsto s.tr\cat x.tr\} \in ac'
\end{xflalign*}
\end{proof}\end{proofs}
\end{lemma}

\begin{lemma}\label{lemma:RA2(exists-P):exists-RA2(P)} Provided $x$ is not in the set $\{s,ac'\}$,
\begin{align*}
	&\mathbf{RA2} (\exists x \spot P) = \exists x \spot \mathbf{RA2} (P)
\end{align*}
\begin{proofs}\begin{proof}\checkt{alcc}
\begin{xflalign*}
	&\mathbf{RA2} (\exists x \spot P)
	&&\ptext{Definition of $\mathbf{RA2}$}\\
	&=(\exists x \spot P)[s\oplus\{tr \mapsto \lseq\rseq\},\{z|z\in ac'\land s.tr\le z.tr\spot z\oplus\{tr\mapsto z.tr-s.tr\}\}/s,ac']
	&&\ptext{Assumption: $x$ is not $ac'$ nor $s$ and predicate calculus}\\
	&=\exists x \spot P[s\oplus\{tr \mapsto \lseq\rseq\},\{z|z\in ac'\land s.tr\le z.tr\spot z\oplus\{tr\mapsto z.tr-s.tr\}\}/s,ac']
	&&\ptext{Definition of $\mathbf{RA2}$}\\
	&=\exists x \spot \mathbf{RA2} (P)
\end{xflalign*}
\end{proof}\end{proofs}
\end{lemma}

\begin{lemma}\label{lemma:RA1-o-RA2-o-PBMH(circledIn)-ac'-not-free} Provided $ac'$ is not free in $P$,
\begin{align*}
	&\mathbf{RA1}\circ\mathbf{RA2}\circ\mathbf{PBMH} (\circledIn{y}{ac'} (P)) \\
	&=\\
	&\circledIn{z}{ac'} (P[s\oplus\{tr\mapsto\lseq\rseq\}/s][z\oplus\{tr\mapsto z.tr-s.tr\}/y] \land s.tr \le z.tr)
\end{align*}
\begin{proofs}\begin{proof}\checkt{alcc}
\begin{xflalign*}
	&\mathbf{RA1}\circ\mathbf{RA2}\circ\mathbf{PBMH} (\circledIn{y}{ac'} (P))
	&&\ptext{Assumption: $ac'$ is not free in $P$ and~\cref{lemma:PBMH(circledIn):circledIn}}\\
	&=\mathbf{RA1}\circ\mathbf{RA2} (\circledIn{y}{ac'} (P))
	&&\ptext{\cref{theorem:RA2-o-RA1:RA1-o-RA2}}\\
	&=\mathbf{RA2}\circ\mathbf{RA1} (\circledIn{y}{ac'} (P))
	&&\ptext{Assumption: $ac'$ is not free in $P$ and~\cref{lemma:RA1(circledIn)-ac'-not-free:circledIn(P-land-s.tr-le-y.tr)}}\\
	&=\mathbf{RA2} (\circledIn{y}{ac'} (P \land s.tr \le y.tr))
	&&\ptext{\cref{lemma:RA2(circledIn)}}\\
	&=\exists y \spot \mathbf{RA2} (P \land s.tr \le y.tr) \land \circledIn{z}{ac'} (s.tr \le z.tr \land y = z\oplus\{tr\mapsto z.tr-s.tr\})
	&&\ptext{\cref{theorem:RA2(P-land-Q):RA2(P)-land-RA2(Q)}}\\
	&=\exists y \spot \mathbf{RA2} (P) \land \mathbf{RA2} (s.tr \le y.tr) \land \circledIn{z}{ac'} (s.tr \le z.tr \land y = z\oplus\{tr\mapsto z.tr-s.tr\})
	&&\ptext{Definition of $\mathbf{RA2}$ and substitution}\\
	&=\exists y \spot \mathbf{RA2} (P) \land \lseq\rseq \le y.tr \land \circledIn{z}{ac'} (s.tr \le z.tr \land y = z\oplus\{tr\mapsto z.tr-s.tr\})
	&&\ptext{Property of sequences}\\
	&=\exists y \spot \mathbf{RA2} (P) \land \circledIn{z}{ac'} (s.tr \le z.tr \land y = z\oplus\{tr\mapsto z.tr-s.tr\})
	&&\ptext{Assumption: $ac'$ is not free in $P$ and~\cref{lemma:RA2(P):P:ac'-not-free}}\\
	&=\exists y \spot P[s\oplus\{tr\mapsto\lseq\rseq\}/s] \land \circledIn{z}{ac'} (s.tr \le z.tr \land y = z\oplus\{tr\mapsto z.tr-s.tr\})
	&&\ptext{Definition of $\circledIn{z}{ac'}$}\\
	&=\exists y \spot P[s\oplus\{tr\mapsto\lseq\rseq\}/s] \land (\exists z \spot s.tr \le z.tr \land y = z\oplus\{tr\mapsto z.tr-s.tr\} \land z \in ac')
	&&\ptext{Predicate calculus}\\
	&=\exists y, z \spot P[s\oplus\{tr\mapsto\lseq\rseq\}/s] \land s.tr \le z.tr \land y = z\oplus\{tr\mapsto z.tr-s.tr\} \land z \in ac'
	&&\ptext{One-point rule}\\
	&=\exists z \spot P[s\oplus\{tr\mapsto\lseq\rseq\}/s][z\oplus\{tr\mapsto z.tr-s.tr\}/y] \land s.tr \le z.tr \land z \in ac'
	&&\ptext{Definition of $\circledIn{z}{ac'}$}\\
	&=\circledIn{z}{ac'} (P[s\oplus\{tr\mapsto\lseq\rseq\}/s][z\oplus\{tr\mapsto z.tr-s.tr\}/y] \land s.tr \le z.tr)
\end{xflalign*}
\end{proof}\end{proofs}
\end{lemma}

\begin{lemma}\label{lemma:RA2(P):P:ac'-not-free} Provided $ac'$ is not free in $P$,
\begin{align*}
	&\mathbf{RA2} (P) = P[s\oplus\{tr\mapsto\lseq\rseq\}/s]
\end{align*}
\begin{proofs}
\begin{proof}
\begin{xflalign*}
	&\mathbf{RA2} (P)
	&&\ptext{Definition of $\mathbf{RA2}$}\\
	&=P\left[s\oplus\{tr \mapsto \lseq\rseq\},\left.\left\{z\left|\begin{array}{l}
		z\in ac'\land s.tr\le z.tr\\
		\spot z\oplus\{tr\mapsto z.tr-s.tr\}
	\end{array}\right.\right\}\right/s,ac'\right]
	&&\ptext{Assumption: $ac'$ is not free in $P$}\\
	&=P[s\oplus\{tr\mapsto\lseq\rseq\}/s]
\end{xflalign*}
\end{proof}
\end{proofs}
\end{lemma}

\begin{lemma}\label{lemma:RA1-o-RA2-o-PBMH(circledIn(y.tr=s.tr))}
\begin{align*}
	&\mathbf{RA1}\circ\mathbf{RA2}\circ\mathbf{PBMH} (\circledIn{y}{ac'} (y.tr=s.tr \land a \notin y.ref \land y.wait))\\
	&=\\
	&\circledIn{y}{ac'} (y.tr=s.tr \land a \notin y.ref \land y.wait)
\end{align*}
\begin{proofs}\begin{proof}\checkt{alcc}
\begin{xflalign*}
	&\mathbf{RA1}\circ\mathbf{RA2}\circ\mathbf{PBMH} (\circledIn{y}{ac'} (y.tr=s.tr \land a \notin y.ref \land y.wait))
	&&\ptext{\cref{lemma:RA1-o-RA2-o-PBMH(circledIn)-ac'-not-free}}\\
	&=\circledIn{z}{ac'} \left(\begin{array}{l}
		\left(\begin{array}{l}
			y.tr=s.tr 
			\\ \land \\ 
			a \notin y.ref 
			\\ \land \\
			y.wait
		\end{array}\right)[s\oplus\{tr\mapsto\lseq\rseq\}/s][z\oplus\{tr\mapsto z.tr-s.tr\}/y] 
		\\ \land \\
		s.tr \le z.tr
	\end{array}\right)
	&&\ptext{Substitution and value of record component $tr$}\\
	&=\circledIn{z}{ac'} \left(\begin{array}{l}
		\left(\begin{array}{l}
			y.tr=\lseq\rseq\
			\\ \land \\ 
			a \notin y.ref 
			\\ \land \\
			y.wait
		\end{array}\right)[z\oplus\{tr\mapsto z.tr-s.tr\}/y] 
		\\ \land \\
		s.tr \le z.tr
	\end{array}\right)
	&&\ptext{Substitution and value of record component $tr$}\\
	&=\circledIn{z}{ac'} (z.tr-s.tr=\lseq\rseq \land a \notin z.ref \land z.wait \land s.tr \le z.tr)
	&&\ptext{Property of sequences}\\
	&=\circledIn{z}{ac'} (z.tr=s.tr \land a \notin z.ref \land z.wait \land s.tr \le z.tr)
	&&\ptext{Predicate calculus}\\
	&=\circledIn{z}{ac'} (z.tr=s.tr \land a \notin z.ref \land z.wait)
	&&\ptext{Variable renaming}\\
	&=\circledIn{y}{ac'} (y.tr=s.tr \land a \notin y.ref \land y.wait)
\end{xflalign*}
\end{proof}\end{proofs}
\end{lemma}

\begin{lemma}\label{lemma:RA1-o-RA2-o-PBMH(circledIn(y.tr=s.tr-cat-a))}
\begin{align*}
	&\mathbf{RA1}\circ\mathbf{RA2}\circ\mathbf{PBMH} (\circledIn{y}{ac'} (y.tr=s.tr\cat\lseq a \rseq \land \lnot y.wait))\\
	&=\\
	&\circledIn{y}{ac'} (y.tr=s.tr\cat\lseq a \rseq \land \lnot y.wait)
\end{align*}
\begin{proofs}\begin{proof}\checkt{alcc}
\begin{xflalign*}
	&\mathbf{RA1}\circ\mathbf{RA2}\circ\mathbf{PBMH} (\circledIn{y}{ac'} (y.tr=s.tr\cat\lseq a \rseq \land \lnot y.wait))
	&&\ptext{\cref{lemma:RA1-o-RA2-o-PBMH(circledIn)-ac'-not-free}}\\
	&=\circledIn{z}{ac'} \left(\begin{array}{l}
		\left(\begin{array}{l}
			y.tr=s.tr \cat\lseq a \rseq
			\\ \land \\
			\lnot y.wait
		\end{array}\right)[s\oplus\{tr\mapsto\lseq\rseq\}/s][z\oplus\{tr\mapsto z.tr-s.tr\}/y] 
		\\ \land \\
		s.tr \le z.tr
	\end{array}\right)
	&&\ptext{Substitution and value of record component $tr$}\\
	&=\circledIn{z}{ac'} \left(\begin{array}{l}
		(y.tr=\lseq\rseq \cat\lseq a \rseq \land \lnot y.wait)[z\oplus\{tr\mapsto z.tr-s.tr\}/y] 
		\\ \land \\
		s.tr \le z.tr
	\end{array}\right)
	&&\ptext{Property of sequences}\\
	&=\circledIn{z}{ac'} \left(\begin{array}{l}
		(y.tr=\lseq a \rseq \land \lnot y.wait)[z\oplus\{tr\mapsto z.tr-s.tr\}/y] 
		\\ \land \\
		s.tr \le z.tr
	\end{array}\right)
	&&\ptext{Substitution and value of record component $tr$}\\
	&=\circledIn{z}{ac'} (z.tr-s.tr=\lseq a \rseq \land \lnot z.wait \land s.tr \le z.tr)
	&&\ptext{Property of sequences}\\
	&=\circledIn{z}{ac'} (z.tr=s.tr\cat\lseq a \rseq \land \lnot z.wait \land s.tr \le z.tr)
	&&\ptext{Property of sequences and predicate calculus}\\
	&=\circledIn{z}{ac'} (z.tr=s.tr\cat\lseq a \rseq \land \lnot z.wait)
	&&\ptext{Variable renaming}\\
	&=\circledIn{y}{ac'} (y.tr=s.tr\cat\lseq a \rseq \land \lnot y.wait)
\end{xflalign*}
\end{proof}\end{proofs}
\end{lemma}

\begin{lemma}\label{lemma:RA1-o-RA2-o-PBMH(circledIn(s.tr-cat-a-le-y.tr))}
\begin{align*}
	&\mathbf{RA1}\circ\mathbf{RA2}\circ\mathbf{PBMH} (\circledIn{y}{ac'} (s.tr\cat\lseq a \rseq \le y.tr))\\
	&=\\
	&\circledIn{y}{ac'} (s.tr \cat\lseq a \rseq \le y.tr)
\end{align*}
\begin{proofs}\begin{proof}\checkt{alcc}
\begin{xflalign*}
	&\mathbf{RA1}\circ\mathbf{RA2}\circ\mathbf{PBMH} (\circledIn{y}{ac'} (s.tr\cat\lseq a \rseq \le y.tr))
	&&\ptext{\cref{lemma:RA1-o-RA2-o-PBMH(circledIn)-ac'-not-free}}\\
	&=\circledIn{z}{ac'} \left(\begin{array}{l}
		(s.tr\cat\lseq a \rseq \le y.tr)[s\oplus\{tr\mapsto\lseq\rseq\}/s][z\oplus\{tr\mapsto z.tr-s.tr\}/y] 
		\\ \land \\
		s.tr \le z.tr
	\end{array}\right)
	&&\ptext{Substitution and value of record component $tr$}\\
	&=\circledIn{z}{ac'} \left(\begin{array}{l}
		(\lseq\rseq\cat\lseq a \rseq \le y.tr)[z\oplus\{tr\mapsto z.tr-s.tr\}/y] 
		\\ \land \\
		s.tr \le z.tr
	\end{array}\right)
	&&\ptext{Substitution and value of record component $tr$}\\
	&=\circledIn{z}{ac'} (\lseq\rseq\cat\lseq a \rseq \le z.tr-s.tr \land s.tr \le z.tr)
	&&\ptext{Property of sequences}\\
	&=\circledIn{z}{ac'} (\lseq a \rseq \le z.tr-s.tr \land s.tr \le z.tr)
	&&\ptext{Property of sequences}\\
	&=\circledIn{z}{ac'} (s.tr \cat\lseq a \rseq \le z.tr \land s.tr \le z.tr)
	&&\ptext{Property of sequences and predicate calculus}\\
	&=\circledIn{z}{ac'} (s.tr \cat\lseq a \rseq \le z.tr)
	&&\ptext{Variable renaming}\\
	&=\circledIn{y}{ac'} (s.tr \cat\lseq a \rseq \le y.tr)
\end{xflalign*}
\end{proof}\end{proofs}
\end{lemma}

\begin{lemma}\label{lemma:circledIn(P-lor-Q):circledIn(P)-lor-circledIn(Q)}
$\circledIn{y}{ac'} (P \lor Q) = \circledIn{y}{ac'} (P) \lor \circledIn{y}{ac'} (Q)$
\begin{proofs}\begin{proof}\checkt{alcc}
\begin{xflalign*}
	&\circledIn{y}{ac'} (P \lor Q)
	&&\ptext{Definition of $\circledIn{y}{ac'}$}\\
	&=\exists y \spot (P \lor Q)[\{y\}\cap ac'/ac'] \land y \in ac'
	&&\ptext{Substitution}\\
	&=\exists y \spot (P[\{y\}\cap ac'/ac'] \lor Q[\{y\}\cap ac'/ac']) \land y \in ac'
	&&\ptext{Predicate calculus}\\
	&=(\exists y \spot P[\{y\}\cap ac'/ac'] \land y \in ac') \lor (\exists y \spot Q[\{y\}\cap ac'/ac'] \land y \in ac')
	&&\ptext{Definition of $\circledIn{y}{ac'}$}\\
	&=\circledIn{y}{ac'} (P) \lor \circledIn{y}{ac'} (Q)
\end{xflalign*}
\end{proof}\end{proofs}
\end{lemma}

\begin{lemma}\label{lemma:RA1-o-RA2-o-PBMH(circledIn(a-then-Skip-post))}
\begin{align*}
	&\mathbf{RA1}\circ\mathbf{RA2}\circ\mathbf{PBMH} \left(\circledIn{y}{ac'} \left(
\right)\right)
	&&\ptext{Definition of conditional}\\
	&=\mathbf{RA1}\circ\mathbf{RA2}\circ\mathbf{PBMH} \left(\circledIn{y}{ac'} \left(%
\right)
	&&\ptext{Definition of conditional}\\
	&=\circledIn{y}{ac'} \left(%
\right)
\end{xflalign*}
\end{proof}\end{proofs}
\end{lemma}

\begin{lemma}\label{lemma:circledIn(P)-seqA-Q:exists-y-P-land-Q} Provided $ac'$ is not free in $P$,
\begin{align*}
	&\circledIn{y}{ac'} (P) \seqA Q = \exists y \spot P \land Q[y/s]
\end{align*}
\begin{proofs}\begin{proof}\checkt{alcc}
\begin{xflalign*}
	&\circledIn{y}{ac'} (P) \seqA Q
	&&\ptext{\cref{lemma:circledIn:ac'-not-free}}\\
	&=(\exists y \spot P \land y \in ac') \seqA Q
	&&\ptext{Definition of $\seqA$}\\
	&=(\exists y \spot P \land y \in ac')[\{ s | Q \}/ac']
	&&\ptext{Assumption: $ac'$ is not free in $P$ and substitution}\\
	&=\exists y \spot P \land y \in \{ s | Q \}
	&&\ptext{Property of sets}\\
	&=\exists y \spot P \land Q[y/s]
\end{xflalign*}
\end{proof}\end{proofs}
\end{lemma}

\begin{lemma}\label{lemma:circledIn(P)-PBMH:exists-y-P}
\begin{statement}
Provided $P$ is $\mathbf{PBMH}$-healthy,
\begin{align*}
	&\circledIn{y}{ac'} (P) = \exists y @ P[\{y\}/ac'] \land y \in ac'
\end{align*}
\end{statement}
\begin{proofs}
\begin{proof}
\begin{xflalign*}
	&\circledIn{y}{ac'} (P)
	&&\ptext{Definition of $\circledIn{y}{ac'}$}\\
	&=\exists y @ P[\{y\}\cap ac'/ac'] \land y \in ac'
	&&\ptext{Assumption: $P$ is $\mathbf{PBMH}$-healthy}\\
	&=\exists y @ (\mathbf{PBMH} (P))[\{y\}\cap ac'/ac'] \land y \in ac'
	&&\ptext{Definition of $\mathbf{PBMH}$ (\cref{lemma:PBMH:alternative-1})}\\
	&=\exists y @ (\exists ac_0 @ P[ac_0/ac'] \land ac_0 \subseteq ac')[\{y\}\cap ac'/ac'] \land y \in ac'
	&&\ptext{Substitution}\\
	&=\exists y @ (\exists ac_0 @ P[ac_0/ac'] \land ac_0 \subseteq \{y\}\cap ac') \land y \in ac'
	&&\ptext{Property of sets}\\
	&=\exists y @ (\exists ac_0 @ P[ac_0/ac'] \land ac_0 \subseteq ac' \land ac_0 \subseteq \{y\}) \land y \in ac'
	&&\ptext{Predicate calculus}\\
	&=\exists y, ac_0 @ P[ac_0/ac'] \land ac_0 \subseteq ac' \land ac_0 \subseteq \{y\} \land y \in ac'
	&&\ptext{Property of sets and predicate calculus}\\
	&=\exists y, ac_0 @ P[ac_0/ac'] \land ac_0 \subseteq \{y\} \land y \in ac'
	&&\ptext{Predicate calculus}\\
	&=\exists y @ (\exists ac_0 @ P[ac_0/ac'] \land ac_0 \subseteq \{y\}) \land y \in ac'
	&&\ptext{Substitution}\\
	&=\exists y @ (\exists ac_0 @ P[ac_0/ac'] \land ac_0 \subseteq ac')[\{y\}/ac'] \land y \in ac'
	&&\ptext{Definition of $\mathbf{PBMH}$ (\cref{lemma:PBMH:alternative-1})}\\
	&=\exists y @ \mathbf{PBMH} (P)[\{y\}/ac'] \land y \in ac'
	&&\ptext{Assumption: $P$ is $\mathbf{PBMH}$-healthy}\\
	&=\exists y @ P[\{y\}/ac'] \land y \in ac'
\end{xflalign*}
\end{proof}
\end{proofs}
\end{lemma}

\begin{lemma}\label{lemma:circledIn:ac'-not-free}
\begin{statement}
Provided $ac'$ is not free in $P$,
$\circledIn{y}{ac'} (P) = \exists y @ P \land y \in ac'$	
\end{statement}
\begin{proofs}
\begin{proof}
\begin{xflalign*}
	&\circledIn{y}{ac'} (P)
	&&\ptext{Definition of $\circledIn{y}{ac'}$}\\
	&=\exists y @ P[\{y\}\cap ac'/ac'] \land y \in ac'
	&&\ptext{Assumption: $ac'$ is not free in $P$}\\
	&=\exists y @ P \land y \in ac'
\end{xflalign*}
\end{proof}
\end{proofs}
\end{lemma}

\begin{lemma}\label{lemma:circledIn(P-lor-Q):circledIn(P)-lor-circledIn(Q):new}
\begin{statement}
$\circledIn{y}{ac'} (P \lor Q) = \circledIn{y}{ac'} (P) \lor \circledIn{y}{ac'} (Q)$
\end{statement}
\begin{proofs}
\begin{proof}\checkt{alcc}
\begin{xflalign*}
	&\circledIn{y}{ac'} (P \lor Q)
	&&\ptext{Definition of $\circledIn{y}{ac'}$}\\
	&=(\exists y @ (P \lor Q)[\{y\}\cap ac'/ac'] \land y \in ac')
	&&\ptext{Substitution}\\
	&=(\exists y @ (P[\{y\}\cap ac'/ac'] \lor Q[\{y\}\cap ac'/ac']) \land y \in ac')
	&&\ptext{Predicate calculus}\\
	&=\exists y @ (P[\{y\}\cap ac'/ac'] \land y \in ac') \lor (Q[\{y\}\cap ac'/ac'] \land y \in ac')
	&&\ptext{Predicate calculus}\\
	&=\exists y @ (P[\{y\}\cap ac'/ac'] \land y \in ac') \lor \exists y @ (Q[\{y\}\cap ac'/ac'] \land y \in ac')
	&&\ptext{Definition of $\circledIn{y}{ac'}$}\\
	&=\circledIn{y}{ac'} (P) \lor \circledIn{y}{ac'} (Q)
\end{xflalign*}
\end{proof}
\end{proofs}
\end{lemma}

\begin{lemma}\label{lemma:circledIn(conditional-and)}
\begin{statement}
\begin{align*}
	&\circledIn{y}{ac'} (P \dres c_0 \land \ldots \land c_n \rres Q)\\
	&=\\
	&\circledIn{y}{ac'} (c_0 \land \ldots \land c_n \land P) 
		\lor 
		\circledIn{y}{ac'} (\lnot c_0 \land Q)
		\lor
		\ldots 
		\lor
		\circledIn{y}{ac'} (\lnot c_n \land Q)
\end{align*}
\end{statement}
\begin{proofs}
\begin{proof}\checkt{alcc}
\begin{xflalign*}
	&\circledIn{y}{ac'} (P \dres c_0 \land \ldots \land c_n \rres Q)
	&&\ptext{Definition of conditional}\\
	&=\circledIn{y}{ac'} \left(\begin{array}{l}
		(c_0 \land \ldots \land c_n \land P) 
		\\ \lor \\
		(\lnot (c_0 \land \ldots \land c_n) \land Q)
	\end{array}\right)
	&&\ptext{Predicate calculus}\\
	&=\circledIn{y}{ac'} \left(\begin{array}{l}
		(c_0 \land \ldots \land c_n \land P) 
		\\ \lor \\
		((\lnot c_0 \lor \ldots \lor \lnot c_n) \land Q)
	\end{array}\right)
	&&\ptext{Predicate calculus}\\
	&=\circledIn{y}{ac'} \left(\begin{array}{l}
		(c_0 \land \ldots \land c_n \land P) 
		\lor
		(\lnot c_0 \land Q)
		\lor
		\ldots 
		\lor
		(\lnot c_n \land Q)
	\end{array}\right)
	&&\ptext{\cref{lemma:circledIn(P-lor-Q):circledIn(P)-lor-circledIn(Q):new}}\\
	&=\circledIn{y}{ac'} (c_0 \land \ldots \land c_n \land P) 
		\lor
		\circledIn{y}{ac'} (\lnot c_0 \land Q)
		\lor
		\ldots 
		\lor
		\circledIn{y}{ac'} (\lnot c_n \land Q)
\end{xflalign*}
\end{proof}
\end{proofs}
\end{lemma}

\begin{lemma}\label{lemma:(str-ytr-land-ywait)-s-y-oplus}
\begin{statement}
Provided $s.tr \le z.tr$,
\begin{align*}
	&(s.tr=y.tr \land y.wait)[s\oplus\{tr\mapsto\lseq\rseq\}/s][y\oplus\{tr\mapsto y.tr-s.tr\}/y]\\
	&=\\
	&(s.tr=y.tr \land y.wait)
\end{align*}
\end{statement}
\begin{proofs}
\begin{proof}
\begin{xflalign*}
	&(s.tr=y.tr \land y.wait)[s\oplus\{tr\mapsto\lseq\rseq\}/s][y\oplus\{tr\mapsto y.tr-s.tr\}/y]
	&&\ptext{Substitution}\\
	&=s\oplus\{tr\mapsto\lseq\rseq\}.tr=y\oplus\{tr\mapsto y.tr-s.tr\}.tr \land y\oplus\{tr\mapsto y.tr-s.tr\}.wait
	&&\ptext{Property of $\oplus$ and value of component $tr$}\\
	&=\lseq\rseq=y.tr-s.tr \land y.wait
	&&\ptext{Assumption and property of sequences}\\
	&=s.tr=y.tr \land y.wait
\end{xflalign*}
\end{proof}
\end{proofs}
\end{lemma}

\begin{lemma}\label{lemma:(str-neq-ytr)-s-y-oplus}
\begin{statement}
Provided $s.tr \le y.tr$,
\begin{align*}
	&(s.tr\neq y.tr)[s\oplus\{tr\mapsto\lseq\rseq\}/s][y\oplus\{tr\mapsto y.tr-s.tr\}/y]\\
	&=\\
	&(s.tr\neq y.tr)
\end{align*}
\end{statement}
\begin{proofs}
\begin{proof}
\begin{xflalign*}
	&(s.tr\neq y.tr)[s\oplus\{tr\mapsto\lseq\rseq\}/s][y\oplus\{tr\mapsto y.tr-s.tr\}/y]
	&&\ptext{Substitution}\\
	&=(s\oplus\{tr\mapsto\lseq\rseq\}).tr\neq (y\oplus\{tr\mapsto y.tr-s.tr\}).tr
	&&\ptext{Property of $\oplus$ and value of component $tr$}\\
	&=\lseq\rseq\neq y.tr-s.tr
	&&\ptext{Assumption and property of sequences}\\
	&=s.tr\neq y.tr
\end{xflalign*}
\end{proof}
\end{proofs}
\end{lemma}

\begin{lemma}\label{lemma:(zx)-s-y-oplus}
\begin{statement}
Provided $x$ is not $tr$,
\begin{align*}
	&(y.x)[s\oplus\{tr\mapsto\lseq\rseq\}/s][y\oplus\{tr\mapsto y.tr-s.tr\}/y]\\
	&=\\
	&(y.x)
\end{align*}
\end{statement}
\begin{proofs}
\begin{proof}
\begin{xflalign*}
	&(y.x)[s\oplus\{tr\mapsto\lseq\rseq\}/s][y\oplus\{tr\mapsto y.tr-s.tr\}/y]
	&&\ptext{Substitution}\\
	&=(y\oplus\{tr\mapsto y.tr-s.tr\}).x
	&&\ptext{Assumption, property of $\oplus$ and value of component $x$}\\
	&=y.x
\end{xflalign*}
\end{proof}
\end{proofs}
\end{lemma}

\begin{lemma}\label{lemma:circledIn(RA2-o-RA1(P)-cond-RA2-o-RA1(Q)):RA2(circledIn(P-cond-Q))}
\begin{statement}
Provided:
\begin{itemize}
	\item $P$ and $Q$ are $\mathbf{PBMH}$-healthy
	\item For $0 \leq i \leq n$: $ac'$ is not free in $c_i$
	\item $(c_0 \land \ldots \land c_n)[s\oplus\{tr\mapsto\lseq\rseq\}/s][y\oplus\{tr\mapsto y.tr-s.tr\}/y] = (c_0 \land \ldots \land c_n)$, assuming $s.tr \le y.tr$
	\item For $0 \leq i \leq n$: $(\lnot c_i)[s\oplus\{tr\mapsto\lseq\rseq\}/s][y\oplus\{tr\mapsto y.tr-s.tr\}/y] = \lnot c_i$, assuming $s.tr \le y.tr$
\end{itemize}
\begin{align*}
	&\circledIn{y}{ac'} (\mathbf{RA2}\circ\mathbf{RA1} (P) \dres c_0 \land \ldots \land c_n \rres \mathbf{RA2}\circ\mathbf{RA1} (Q))\\
	&=\\
	&\mathbf{RA2} (\circledIn{y}{ac'} (P \dres (c_0 \land \ldots \land c_n) \rres Q))
\end{align*}
\end{statement}
\begin{proofs}
\begin{proof}
\begin{xflalign*}
	&\circledIn{y}{ac'} (\mathbf{RA2}\circ\mathbf{RA1} (P) \dres c_0 \land \ldots \land c_n \rres \mathbf{RA2}\circ\mathbf{RA1} (Q))
	&&\ptext{Definition of conditional}\\
	&=\circledIn{y}{ac'} \left(\begin{array}{l}
		(\mathbf{RA2}\circ\mathbf{RA1} (P) \land (c_0 \land \ldots \land c_n))
		\\ \lor \\
		(\mathbf{RA2}\circ\mathbf{RA1} (Q) \land \lnot c_0)
		\\ \ldots \\
		\\ \lor \\
		(\mathbf{RA2}\circ\mathbf{RA1} (Q) \land \lnot c_n)
	\end{array}\right)
	&&\ptext{\cref{lemma:circledIn(P-lor-Q):circledIn(P)-lor-circledIn(Q):new}}\\
	&=\left(\begin{array}{l}
		\circledIn{y}{ac'} (\mathbf{RA2}\circ\mathbf{RA1} (P) \land (c_0 \land \ldots \land c_n))
		\\ \lor \\
		\circledIn{y}{ac'} (\mathbf{RA2}\circ\mathbf{RA1} (Q) \land \lnot c_0)
		\\ \ldots \\
		\\ \lor \\
		\circledIn{y}{ac'} (\mathbf{RA2}\circ\mathbf{RA1} (Q) \land \lnot c_n)
	\end{array}\right)
	&&\ptext{Assumption and~\cref{lemma:RA2(circledIn(P-land-Q))-Q-ac'-not-free}}\\
	&=\left(\begin{array}{l}
		\mathbf{RA2} (\circledIn{y}{ac'} (P \land (c_0 \land \ldots \land c_n)))
		\\ \lor \\
		\mathbf{RA2} (\circledIn{y}{ac'} (Q \land \lnot c_0))
		\\ \ldots \\
		\\ \lor \\
		\mathbf{RA2} (\circledIn{y}{ac'} (Q \land \lnot c_n))
	\end{array}\right)
	&&\ptext{\cref{theorem:RA2(P-lor-Q):RA2(P)-lor-RA2(Q)}}\\
	&=\mathbf{RA2} \left(\begin{array}{l}
		\circledIn{y}{ac'} (P \land (c_0 \land \ldots \land c_n))
		\\ \lor \\
		\circledIn{y}{ac'} (Q \land \lnot c_0)
		\\ \ldots \\
		\\ \lor \\
		\circledIn{y}{ac'} (Q \land \lnot c_n)
	\end{array}\right)
	&&\ptext{\cref{lemma:circledIn(P-lor-Q):circledIn(P)-lor-circledIn(Q):new}}\\
	&=\mathbf{RA2} \left(\circledIn{y}{ac'}\left(\begin{array}{l}
		(P \land (c_0 \land \ldots \land c_n))
		\\ \lor \\
		(Q \land \lnot c_0)
		\\ \ldots \\
		\\ \lor \\
		(Q \land \lnot c_n)
	\end{array}\right)\right)
	&&\ptext{Definition of conditional}\\
	&=\mathbf{RA2} (\circledIn{y}{ac'} (P \dres (c_0 \land \ldots \land c_n) \rres Q))
\end{xflalign*}
\end{proof}
\end{proofs}
\end{lemma}

\begin{lemma}\label{lemma:circledIn(post-extchoice):RA2(circledIn(post-extchoice))}
\begin{statement}
Provided that $P$ and $Q$ are $\mathbf{PBMH}$-healthy,
\begin{align*}
	&\circledIn{y}{ac'} (\mathbf{RA2}\circ\mathbf{RA1} (P) \dres ytr=s.tr \land y.wait \rres \mathbf{RA2}\circ\mathbf{RA1} (Q))\\
	&=\\
	&\mathbf{RA2} (\circledIn{y}{ac'} (P \dres ytr=s.tr \land y.wait \rres Q))
\end{align*}
\end{statement}
\begin{proofs}
\begin{proof}
\begin{xflalign*}
	&\circledIn{y}{ac'} (\mathbf{RA2}\circ\mathbf{RA1} (P) \dres ytr=s.tr \land y.wait \rres \mathbf{RA2}\circ\mathbf{RA1} (Q))
	&&\ptext{Assumption: $P$ and $Q$ are $\mathbf{PBMH}$-healthy}\\
	&&\ptext{\cref{lemma:circledIn(RA2-o-RA1(P)-cond-RA2-o-RA1(Q)):RA2(circledIn(P-cond-Q)),lemma:(str-ytr-land-ywait)-s-y-oplus,lemma:(str-neq-ytr)-s-y-oplus,lemma:(zx)-s-y-oplus}}\\
	&=\mathbf{RA2} \circledIn{y}{ac'} (P \dres ytr=s.tr \land y.wait \rres Q)
\end{xflalign*}
\end{proof}
\end{proofs}
\end{lemma}

\begin{lemma}\label{lemma:circledIn(P-land-circledIn(Q)):circledIn(P-land-Q-sub-y-for-z)}
$\circledIn{y}{ac'} (P \land \circledIn{z}{ac'} (Q)) = \circledIn{y}{ac'} (P \land Q[y/z])$
\begin{proofs}\begin{proof}
\begin{xflalign*}
	&\circledIn{y}{ac'} (P \land \circledIn{z}{ac'} (Q))
	&&\ptext{Definition of $\circledIn{y}{ac'}$}\\
	&=\exists y @ (P \land \circledIn{z}{ac'} (Q))[\{y\}\cap ac'/ac'] \land y \in ac'
	&&\ptext{Substitution}\\
	&=\exists y @ P[\{y\}\cap ac'/ac'] \land \circledIn{z}{ac'} (Q)[\{y\}\cap ac'/ac'] \land y \in ac'
	&&\ptext{\cref{lemma:circledIn:Q-subs-y-ac'}}\\
	&=\exists y @ P[\{y\}\cap ac'/ac'] \land Q[y/z][\{y\}\cap ac'/ac'] \land y \in ac' \land y \in ac'
	&&\ptext{Predicate calculus}\\
	&=\exists y @ P[\{y\}\cap ac'/ac'] \land Q[y/z][\{y\}\cap ac'/ac'] \land y \in ac'
	&&\ptext{Substitution}\\
	&=\exists y @ (P \land Q[y/z])[\{y\}\cap ac'/ac'] \land y \in ac'
	&&\ptext{Definition of $\circledIn{y}{ac'}$}\\
	&=\circledIn{y}{ac'} (P \land Q[y/z])
\end{xflalign*}
\end{proof}\end{proofs}
\end{lemma}

\begin{lemma}\label{lemma:circledIn:Q-subs-y-ac'}
$\circledIn{z}{ac'} (Q)[\{y\}\cap ac'/ac'] = Q[y/z][\{y\}\cap ac'/ac'] \land y \in ac'$
\begin{proofs}\begin{proof}
\begin{xflalign*}
	&\circledIn{z}{ac'} (Q)[\{y\}\cap ac'/ac']
	&&\ptext{Definition of $\circledIn{y}{ac'}$}\\
	&=(\exists z @ Q[\{z\}\cap ac'/ac'] \land z \in ac')[\{y\}\cap ac'/ac']
	&&\ptext{Substitution}\\
	&=\exists z @ Q[\{z\}\cap \{y\}\cap ac'/ac'] \land z \in \{y\}\cap ac'
	&&\ptext{Property of sets}\\
	&=\exists z @ Q[\{z\}\cap \{y\}\cap ac'/ac'] \land z \in \{y\} \land z \in ac'
	&&\ptext{Property of sets}\\
	&=\exists z @ Q[\{z\}\cap \{y\}\cap ac'/ac'] \land z=y \land z \in ac'
	&&\ptext{One-point rule}\\
	&=Q[\{z\}\cap \{y\}\cap ac'/ac'][y/z] \land y \in ac'
	&&\ptext{Substitution}\\
	&=Q[y/z][\{y\}\cap \{y\}\cap ac'/ac'] \land y \in ac'
	&&\ptext{Property of sets}\\
	&=Q[y/z][\{y\}\cap ac'/ac'] \land y \in ac'
\end{xflalign*}
\end{proof}\end{proofs}
\end{lemma}

\subsubsection{Properties with respect to $\mathbf{PBMH}$}

\begin{lemma}\label{lemma:circledIn(PBMH(P)-land-Q):implies:PBMH(P)}
\begin{statement}
$\circledIn{y}{ac'} (\mathbf{PBMH} (P) \land Q) \implies \mathbf{PBMH} (P)$
\end{statement}
\begin{proofs}
\begin{proof}
\begin{xflalign*}
	&\circledIn{y}{ac'} (\mathbf{PBMH} (P) \land Q)
	&&\ptext{Definition of $\circledIn{y}{ac'}$}\\
	&=\exists y @ (\mathbf{PBMH} (P) \land Q)[\{y\}\cap ac'/ac'] \land y \in ac'
	&&\ptext{Substitution}\\
	&=\exists y @ \mathbf{PBMH} (P)[\{y\}\cap ac'/ac'] \land Q[\{y\}\cap ac'/ac'] \land y \in ac'
	&&\ptext{Predicate calculus}\\
	&\implies \exists y @ \mathbf{PBMH} (P)[\{y\}\cap ac'/ac'] \land y \in ac'
	&&\ptext{Definition of $\circledIn{y}{ac'}$ and~\cref{lemma:circledIn(P)-PBMH:exists-y-P}}\\
	&=\exists y @ \mathbf{PBMH} (P)[\{y\}/ac'] \land y \in ac'
	&&\ptext{Introduce fresh variable $z$}\\
	&=\exists y, z @ \mathbf{PBMH} (P)[z/ac'] \land z = \{y\} \land y \in ac'
	&&\ptext{Property of sets}\\
	&=\exists y, z @ \mathbf{PBMH} (P)[z/ac'] \land z = \{y\} \land y \in ac' \land z\in ac'
	&&\ptext{Predicate calculus}\\
	&\implies \exists z @ \mathbf{PBMH} (P)[z/ac'] \land z\in ac'
	&&\ptext{Definition of $\mathbf{PBMH}$ (\cref{lemma:PBMH:alternative-1})}\\
	&=\mathbf{PBMH} \circ \mathbf{PBMH} (P)
	&&\ptext{\cref{law:pbmh:idempotent}}\\
	&=\mathbf{PBMH} (P)
\end{xflalign*}
\end{proof}
\end{proofs}
\end{lemma}

\begin{lemma}\label{lemma:lnot-PBMH(P)-land-circledIn(PBMH(P)-cond-R-T)}
\begin{statement}
\begin{align*}
	&\lnot \mathbf{PBMH} (P) \land \circledIn{y}{ac'} (((\mathbf{PBMH} (P) \land Q) \lor R) \dres c \rres T)\\
	&=\\
	&\lnot \mathbf{PBMH} (P) \land \circledIn{y}{ac'} (R \dres c \rres T)\\
\end{align*}
\end{statement}
\begin{proofs}
\begin{proof}\checkt{alcc}
\begin{xflalign*}
	&\lnot \mathbf{PBMH} (P) \land \circledIn{y}{ac'} (((\mathbf{PBMH} (P) \land Q) \lor R) \dres c \rres T)
	&&\ptext{Definition of conditional}\\
	&=\lnot \mathbf{PBMH} (P) \land \circledIn{y}{ac'} \left(\begin{array}{l}
		(c \land ((\mathbf{PBMH} (P) \land Q) \lor R))
		\\ \lor \\
		(\lnot c \land T)
	\end{array}\right)
	&&\ptext{Predicate calculus}\\
	&=\lnot \mathbf{PBMH} (P) \land \circledIn{y}{ac'} \left(\begin{array}{l}
		(c \land \mathbf{PBMH} (P) \land Q)
		\\ \lor \\
		(c \land R)
		\\ \lor \\
		(\lnot c \land T)
	\end{array}\right)
	&&\ptext{\cref{lemma:circledIn(P-lor-Q):circledIn(P)-lor-circledIn(Q):new}}\\
	&=\lnot \mathbf{PBMH} (P) \land \left(\begin{array}{l}
		\circledIn{y}{ac'} (c \land \mathbf{PBMH} (P) \land Q)
		\\ \lor \\
		\circledIn{y}{ac'} (c \land R)
		\\ \lor \\
		\circledIn{y}{ac'} (\lnot c \land T)
	\end{array}\right)
	&&\ptext{\cref{lemma:circledIn(PBMH(P)-land-Q):implies:PBMH(P)} and predicate calculus}\\
	&=\lnot \mathbf{PBMH} (P) \land \left(\begin{array}{l}
		(\circledIn{y}{ac'} (c \land \mathbf{PBMH} (P) \land Q) \land \mathbf{PBMH} (P))
		\\ \lor \\
		\circledIn{y}{ac'} (c \land R)
		\\ \lor \\
		\circledIn{y}{ac'} (\lnot c \land T)
	\end{array}\right)
	&&\ptext{Predicate calculus}\\
	&=\lnot \mathbf{PBMH} (P) \land \left(\begin{array}{l}
		\circledIn{y}{ac'} (c \land R)
		\\ \lor \\
		\circledIn{y}{ac'} (\lnot c \land T)
	\end{array}\right)
	&&\ptext{\cref{lemma:circledIn(P-lor-Q):circledIn(P)-lor-circledIn(Q):new}}\\
	&=\lnot \mathbf{PBMH} (P) \land \circledIn{y}{ac'} ((c \land R) \lor (\lnot c \land T))
	&&\ptext{Definition of conditional}\\
	&=\lnot \mathbf{PBMH} (P) \land \circledIn{y}{ac'} (R \dres c \rres T)
\end{xflalign*}
\end{proof}
\end{proofs}
\end{lemma}

\begin{lemma}\label{lemma:lnot-PBMH(P)-land-circledIn(Q-cond-(PBMH(P)-lor-R))}
\begin{statement}
\begin{align*}
	&\lnot \mathbf{PBMH} (P) \land \circledIn{y}{ac'} (Q \dres c \rres (\mathbf{PBMH} (P) \lor R))\\
	&=\\
	&\lnot \mathbf{PBMH} (P) \land \circledIn{y}{ac'} (Q \dres c \rres R)\\
\end{align*}
\end{statement}
\begin{proofs}
\begin{proof}\checkt{alcc}
\begin{xflalign*}
	&\lnot \mathbf{PBMH} (P) \land \circledIn{y}{ac'} (Q \dres c \rres (\mathbf{PBMH} (P) \lor R))
	&&\ptext{Definition of conditional}\\
	&=\lnot \mathbf{PBMH} (P) \land \circledIn{y}{ac'} \left(\begin{array}{l}
		(c \land Q)
		\\ \lor \\
		(\lnot c \land (\mathbf{PBMH} (P) \lor R))
	\end{array}\right)
	&&\ptext{Predicate calculus}\\
	&=\lnot \mathbf{PBMH} (P) \land \circledIn{y}{ac'} \left(\begin{array}{l}
		(c \land Q)
		\\ \lor \\
		(\lnot c \land \mathbf{PBMH} (P))
		\\ \lor \\
		(\lnot c \land R)
	\end{array}\right)
	&&\ptext{\cref{lemma:circledIn(P-lor-Q):circledIn(P)-lor-circledIn(Q):new}}\\
	&=\lnot \mathbf{PBMH} (P) \land \left(\begin{array}{l}
		\circledIn{y}{ac'} (c \land Q)
		\\ \lor \\
		\circledIn{y}{ac'} (\lnot c \land \mathbf{PBMH} (P))
		\\ \lor \\
		\circledIn{y}{ac'} (\lnot c \land R)
	\end{array}\right)
	&&\ptext{\cref{lemma:circledIn(PBMH(P)-land-Q):implies:PBMH(P)} and predicate calculus}\\
	&=\lnot \mathbf{PBMH} (P) \land \left(\begin{array}{l}
		\circledIn{y}{ac'} (c \land Q)
		\\ \lor \\
		(\circledIn{y}{ac'} (\lnot c \land \mathbf{PBMH} (P)) \land \mathbf{PBMH} (P))
		\\ \lor \\
		\circledIn{y}{ac'} (\lnot c \land R)
	\end{array}\right)
	&&\ptext{Predicate calculus}\\
	&=\lnot \mathbf{PBMH} (P) \land \left(\begin{array}{l}
		\circledIn{y}{ac'} (c \land Q)
		\\ \lor \\
		\circledIn{y}{ac'} (\lnot c \land R)
	\end{array}\right)
	&&\ptext{\cref{lemma:circledIn(P-lor-Q):circledIn(P)-lor-circledIn(Q):new}}\\
	&=\lnot \mathbf{PBMH} (P) \land \circledIn{y}{ac'} ((c \land Q) \lor (\lnot c \land R))
	&&\ptext{Definition of conditional}\\
	&=\lnot \mathbf{PBMH} (P) \land \circledIn{y}{ac'} (Q \dres c \rres R)
\end{xflalign*}
\end{proof}
\end{proofs}
\end{lemma}

\subsubsection{Properties with respect to $ac2p$}

\begin{theorem}\label{theorem:ac2p(circledIn(p2ac(P)-land-p2ac(Q)-land-R)):P-land-Q-land-R}
\begin{statement}
Provided $ac'$ is not free in $P$, $Q$ and $R$, and $y$ is not free in $P$ nor $Q$,
\begin{align*}
	&ac2p(\circledIn{y}{ac'} (p2ac(P) \land p2ac(Q) \land R))\\
	&=\\
	&P \land Q \land R[undash(State_{\II}(out\alpha_{-ok'}))/y][State_{\II}(in\alpha_{-ok})/s]
\end{align*}
\end{statement}
\begin{proofs}
\begin{proof}\checkt{alcc}
\begin{xflalign*}
	&ac2p(\circledIn{y}{ac'} (p2ac(P) \land p2ac(Q) \land R))
	&&\ptext{\cref{lemma:PBMH-o-p2ac(P):p2ac(P),lemma:ac2p(circledIn(P-land-Q)):ac'-not-free-in-Q-y-not-free-in-P}}\\
	&=\left(\begin{array}{l}
		ac2p((p2ac(P) \land p2ac(Q))[\{undash(State_{\II}(out\alpha_{-ok'}))\}\cap ac'/ac']) 
		\\ \land \\
		R[undash(State_{\II}(out\alpha_{-ok'}))/y][State_{\II}(in\alpha_{-ok})/s]
	\end{array}\right)
	&&\ptext{Assumption: $ac'$ is not free in $P$ nor $Q$ and~\cref{lemma:p2ac(P-land-Q)-y-cap-ac':(p2ac(P)-land-p2ac(Q))-y-cap-ac'}}\\
	&=\left(\begin{array}{l}
		ac2p(p2ac(P \land Q)[\{undash(State_{\II}(out\alpha_{-ok'}))\}\cap ac'/ac']) 
		\\ \land \\
		R[undash(State_{\II}(out\alpha_{-ok'}))/y][State_{\II}(in\alpha_{-ok})/s]
	\end{array}\right)
	&&\ptext{\cref{lemma:p2ac(P)-undash-out-alpha-cap-ac'}}\\
	&=\left(\begin{array}{l}
		ac2p((P \land Q)[State_{\II}(in\alpha_{-ok})/s] \land undash(State_{\II}(out\alpha_{-ok'})) \in ac') 
		\\ \land \\
		R[undash(State_{\II}(out\alpha_{-ok'}))/y][State_{\II}(in\alpha_{-ok})/s]
	\end{array}\right)
	&&\ptext{Assumption: $ac'$ is not free in $P$ nor $Q$ and~\cref{lemma:ac2p(P-in-alpha-s-land-undash-state-II-out-alpha-in-ac'):P}}\\
	&=P \land Q \land R[undash(State_{\II}(out\alpha_{-ok'}))/y][State_{\II}(in\alpha_{-ok})/s]
\end{xflalign*}
\end{proof}
\end{proofs}
\end{theorem}

\begin{lemma}\label{lemma:ac2p(circledIn(p2ac(P)-land-R)):P-land-R}
\begin{statement}
Provided $ac'$ is not free in $P$, $Q$ and $R$, and $y$ is not free in $P$ nor $Q$,
\begin{align*}
	&ac2p(\circledIn{y}{ac'} (p2ac(P) \land R))\\
	&=\\
	&P \land R[undash(State_{\II}(out\alpha_{-ok'}))/y][State_{\II}(in\alpha_{-ok})/s]
\end{align*}
\end{statement}
\begin{proofs}
\begin{proof}\checkt{alcc}
\begin{xflalign*}
	&ac2p(\circledIn{y}{ac'} (p2ac(P) \land R))
	&&\ptext{\cref{lemma:p2ac(P):implies:ac'-neq-emptyset}}\\
	&=ac2p(\circledIn{y}{ac'} (p2ac(P) \land ac'\neq\emptyset \land R))
	&&\ptext{\cref{lemma:p2ac(true)}}\\
	&=ac2p(\circledIn{y}{ac'} (p2ac(P) \land p2ac(true) \land R))
	&&\ptext{Assumption and~\cref{theorem:ac2p(circledIn(p2ac(P)-land-p2ac(Q)-land-R)):P-land-Q-land-R}}\\
	&=P \land true \land R[undash(State_{\II}(out\alpha_{-ok'}))/y][State_{\II}(in\alpha_{-ok})/s]
	&&\ptext{Predicate calculus}\\
	&=P \land R[undash(State_{\II}(out\alpha_{-ok'}))/y][State_{\II}(in\alpha_{-ok})/s]
\end{xflalign*}
\end{proof}
\end{proofs}
\end{lemma}

\begin{lemma}\label{lemma:PBMH(circledIn-P-PBMH):circledIn}
\begin{statement} Provided $P$ is $\mathbf{PBMH}$-healthy,
$\mathbf{PBMH} (\circledIn{y}{ac'} (P)) = \circledIn{y}{ac'} (P)$
\end{statement}
\begin{proofs}
\begin{proof}
\begin{xflalign*}
	&\mathbf{PBMH} (\circledIn{y}{ac'} (P))
	&&\ptext{Assumption: $P$ is $\mathbf{PBMH}$-healthy and~\cref{lemma:circledIn(P)-PBMH:exists-y-P}}\\
	&=\mathbf{PBMH} (\exists y @ P[\{y\}/ac'] \land y \in ac')
	&&\ptext{Definition of $\mathbf{PBMH}$ (\cref{lemma:PBMH:alternative-1})}\\
	&=\exists ac_0 @ (\exists y @ P[\{y\}/ac'] \land y \in ac')[ac_0/ac'] \land ac_0\subseteq ac'
	&&\ptext{Substitution}\\
	&=\exists ac_0 @ (\exists y @ P[\{y\}/ac'] \land y \in ac_0 \land ac_0\subseteq ac')
	&&\ptext{Property of sets}\\
	&=\exists y @ P[\{y\}/ac'] \land y \in ac'
	&&\ptext{Assumption: $P$ is $\mathbf{PBMH}$-healthy and~\cref{lemma:circledIn(P)-PBMH:exists-y-P}}\\
	&=\circledIn{y}{ac'} (P)
\end{xflalign*}
\end{proof}
\end{proofs}
\end{lemma}

\begin{lemma}\label{lemma:RA2(circledIn(ywait-land-ytr-eq-str))}
\begin{statement}
\begin{align*}
	&\mathbf{RA2} (\circledIn{y}{ac'} (y.wait \land y.tr=s.tr))\\
	&=\\
	&\circledIn{y}{ac'} (y.wait \land y.tr=s.tr)
\end{align*}
\end{statement}
\begin{proofs}
\begin{proof}
\begin{xflalign*}
	&\mathbf{RA2} (\circledIn{y}{ac'} (y.wait \land y.tr=s.tr))
	&&\ptext{\cref{lemma:RA2(circledIn)}}\\
	&=\exists y \spot \mathbf{RA2} (y.wait \land y.tr=s.tr) \land \circledIn{z}{ac'} (s.tr \le z.tr \land y = z\oplus\{tr\mapsto z.tr-s.tr\})
	&&\ptext{\cref{lemma:RA2(P):P:ac'-not-free}}\\
	&=\left(\begin{array}{l}
			\exists y \spot (y.wait \land y.tr=s.tr)[s\oplus\{tr\mapsto\lseq\rseq\}/s] 
			\\ \land \\
			\circledIn{z}{ac'} (s.tr \le z.tr \land y = z\oplus\{tr\mapsto z.tr-s.tr\})
	\end{array}\right)
	&&\ptext{Substitution and value of record component $tr$}\\
	&=\exists y \spot y.wait \land y.tr=\lseq\rseq \land \circledIn{z}{ac'} (s.tr \le z.tr \land y = z\oplus\{tr\mapsto z.tr-s.tr\})
	&&\ptext{Definition of $\circledIn{y}{ac'}$ and predicate calculus}\\
	&=\exists y, z \spot y.wait \land y.tr=\lseq\rseq \land s.tr \le z.tr \land y = z\oplus\{tr\mapsto z.tr-s.tr\} \land z \in ac'
	&&\ptext{One-point rule and substitution}\\
	&=\left(\begin{array}{l}
		\exists z \spot (z\oplus\{tr\mapsto z.tr-s.tr\}).wait
		\\ \land \\
		(z\oplus\{tr\mapsto z.tr-s.tr\}).tr=\lseq\rseq \land s.tr \le z.tr \land z \in ac'
	\end{array}\right)
	&&\ptext{Value of record component $tr$ and $wait$}\\
	&=\exists z \spot z.wait \land z.tr-s.tr=\lseq\rseq \land s.tr \le z.tr \land z \in ac'
	&&\ptext{Property of sequences}\\
	&=\exists z \spot z.wait \land z.tr=s.tr \land z \in ac'
	&&\ptext{Variable renaming and definition of $\circledIn{y}{ac'}$}\\
	&=\circledIn{y}{ac'} (y.wait \land y.tr=s.tr)
\end{xflalign*}
\end{proof}
\end{proofs}
\end{lemma}

\subsubsection{Properties with respect to $\mathbf{A2}$}

\begin{lemma}\label{lemma:A2(circledIn(P)):exists-y-P}
\begin{statement}
$\mathbf{A2} (\circledIn{y}{ac'} (P)) = \exists y @ P[\{y\}/ac'] \land y \in ac'$
\end{statement}
\begin{proofs}
\begin{proof}\checkt{alcc}
\begin{xflalign*}
	&\mathbf{A2} (\circledIn{y}{ac'} (P))
	&&\ptext{Definition of $\mathbf{A2}$}\\
	&=\mathbf{PBMH} (\circledIn{y}{ac'} (P) \seqA \{s\} = ac')
	&&\ptext{Definition of $\circledIn{y}{ac'}$}\\
	&=\mathbf{PBMH} ((\exists y @ P[\{y\}\cap ac'/ac'] \land y \in ac') \seqA \{s\} = ac')
	&&\ptext{Definition of $\seqA$}\\
	&=\mathbf{PBMH} ((\exists y @ P[\{y\}\cap ac'/ac'] \land y \in ac')[\{ s | \{s\} = ac'\}/ac'])
	&&\ptext{Substitution}\\
	&=\mathbf{PBMH} (\exists y @ P[\{y\}\cap \{ s | \{s\} = ac'\}/ac'] \land y \in \{ s | \{s\} = ac'\})
	&&\ptext{Property of sets}\\
	&=\mathbf{PBMH} (\exists y @ P[\{ s | s = y \land \{s\} = ac'\}/ac'] \land \{ y\} = ac')
	&&\ptext{Transitivity of equality}\\
	&=\mathbf{PBMH} (\exists y @ P[\{ s | s = y \land \{s\} = \{y\}\}/ac'] \land \{ y\} = ac')
	&&\ptext{Property of sets}\\
	&=\mathbf{PBMH} (\exists y @ P[\{y\}/ac'] \land \{ y\} = ac')
	&&\ptext{Definition of $\mathbf{PBMH}$ (\cref{lemma:PBMH:alternative-1})}\\
	&=\exists ac_0 @ (\exists y @ P[\{y\}/ac'] \land \{ y\} = ac')[ac_0/ac'] \land ac_0 \subseteq ac'
	&&\ptext{Substitution}\\
	&=\exists ac_0 @ \exists y @ P[\{y\}/ac'] \land \{ y\} = ac_0 \land ac_0 \subseteq ac'
	&&\ptext{One-point rule}\\
	&=\exists y @ P[\{y\}/ac'] \land \{ y\} \subseteq ac'
	&&\ptext{Property of sets}\\
	&=\exists y @ P[\{y\}/ac'] \land y \in ac'
\end{xflalign*}
\end{proof}
\end{proofs}
\end{lemma}

\begin{theorem}\label{theorem:A2(circledIn(P)):circledIn(P)}
\begin{statement}
Provided $P$ is $\mathbf{PBMH}$-healthy,
$\mathbf{A2} (\circledIn{y}{ac'} (P)) = \circledIn{y}{ac'} (P)$.
\end{statement}
\begin{proofs}
\begin{proof}\checkt{alcc}
\begin{xflalign*}
	&\mathbf{A2} (\circledIn{y}{ac'} (P))
	&&\ptext{\cref{lemma:A2(circledIn(P)):exists-y-P}}\\
	&=\exists y @ P[\{y\}/ac'] \land y \in ac'
	&&\ptext{Assumption: $P$ is $\mathbf{PBMH}$-healthy and~\cref{lemma:circledIn(P)-PBMH:exists-y-P}}\\
	&=\circledIn{y}{ac'} (P)
\end{xflalign*}
\end{proof}
\end{proofs}
\end{theorem}


\section{Operators}

\subsection{Angelic Choice}

\begin{theorem}\label{theorem:RAP:P-sqcup-Q}
\begin{statement}Provided $P$ and $Q$ are reactive angelic designs, 
\begin{align*}
	&P \sqcup Q = \mathbf{RA} \circ \mathbf{A} (
			\lnot P^f_f \lor \lnot Q^f_f
			\vdash
			(\lnot P^f_f \implies P^t_f) \land (\lnot Q^f_f \implies Q^t_f))
\end{align*}
\end{statement}
\begin{proofs}
\begin{proof}\checkt{pfr}\checkt{alcc}
\begin{xflalign*}
	&P \sqcup Q
	&&\ptext{Assumption: $P$ and $Q$ are $\mathbf{RAP}$-healthy}\\
	&=\mathbf{RA} \circ \mathbf{A} (\lnot P^f_f \vdash P^t_f) \sqcup \mathbf{RA} \circ \mathbf{A} (\lnot Q^f_f \vdash Q^t_f)
	&&\ptext{\cref{theorem:RAP(P|-Q)-sqcup-RAP(R|-S)}}\\
	&=\mathbf{RA} \circ \mathbf{A} \left(\begin{array}{l}
		(\lnot \mathbf{PBMH} (P^f_f) \lor \lnot \mathbf{PBMH} (Q^f_f))
		\\ \vdash \\
		\left(\begin{array}{l}
			(\lnot \mathbf{PBMH} (P^f_f) \implies \mathbf{PBMH} (P^t_f)) 
			\\ \land \\ 
			(\lnot \mathbf{PBMH} (Q^f_f) \implies \mathbf{PBMH} (Q^t_f))
		\end{array}\right)
	\end{array}\right)
	&&\ptext{\cref{lemma:PBMH(P)-ow:PBMH(P-ow)}}\\
	&=\mathbf{RA} \circ \mathbf{A} \left(\begin{array}{l}
		(\lnot \mathbf{PBMH} (P)^f_f \lor \lnot \mathbf{PBMH} (Q)^f_f)
		\\ \vdash \\
		\left(\begin{array}{l}
			(\lnot \mathbf{PBMH} (P)^f_f \implies \mathbf{PBMH} (P)^t_f) 
			\\ \land \\ 
			(\lnot \mathbf{PBMH} (Q)^f_f \implies \mathbf{PBMH} (Q)^t_f)
		\end{array}\right)
	\end{array}\right)
	&&\ptext{Assumption: $P$ and $Q$ are $\mathbf{RAP}$-healthy and~\cref{theorem:PBMH(P)-RAP:P}}\\
	&=\mathbf{RA} \circ \mathbf{A} (
		(\lnot P^f_f \lor \lnot Q^f_f)
		\vdash
		(\lnot P^f_f \implies P^t_f) \land (\lnot Q^f_f \implies Q^t_f))
\end{xflalign*}
\end{proof}
\end{proofs}
\end{theorem}

\begin{theorem}\label{theorem:ac2p(p2ac(P)-sqcup-RAD-p2ac(Q)):P-sqcup-R-Q}
\begin{statement}
$ac2p(p2ac(P) \sqcup_{\mathbf{RAD}} p2ac(Q)) = P \sqcup_{\mathbf{R}} Q$
\end{statement}
\begin{proofs}
\begin{proof}\checkt{alcc}
\begin{xflalign*}
	&ac2p(p2ac(P) \sqcup_{\mathbf{RAD}} p2ac(Q))
	&&\ptext{Definition of $\sqcup_{\mathbf{RAD}}$}\\
	&=ac2p(p2ac(P) \land p2ac(Q))
	&&\ptext{\cref{theorem:ac2p(P-land-Q):ac2p(P)-land-ac2p(Q)}}\\
	&=ac2p\circ p2ac(P) \land ac2p\circ p2ac(Q)
	&&\ptext{\cref{theorem:ac2p-o-p2ac(P):P}}\\
	&=P \land Q
	&&\ptext{Definition of $\sqcup_{\mathbf{R}}$}\\
	&=P \sqcup_{\mathbf{R}} Q
\end{xflalign*}
\end{proof}
\end{proofs}
\end{theorem}

\begin{theorem}\label{theorem:p2ac(ac2p(P)-sqcup-R-ac2p(Q)):P-sqcup-RAD-Q}
\begin{statement}
Provided that $P$ and $Q$ are reactive angelic designs,
\begin{align*}
	&p2ac(ac2p(P) \sqcup_{\mathbf{R}} ac2p(Q)) \sqsupseteq P \sqcup_{\mathbf{RAD}} Q
\end{align*}
\end{statement}
\begin{proofs}
\begin{proof}\checkt{alcc}
\begin{xflalign*}
	&p2ac(ac2p(P) \sqcup_{\mathbf{R}} ac2p(Q))
	&&\ptext{Definition of $\sqcup_{\mathbf{R}}$}\\
	&=p2ac(ac2p(P) \land ac2p(Q))
	&&\ptext{\cref{theorem:p2ac(P-land-Q):implies:p2ac(P)-land-p2ac(Q)}}\\
	&\sqsupseteq p2ac\circ ac2p(P) \land p2ac\circ ac2p(Q)
	&&\ptext{\cref{theorem:p2ac-o-ac2p:implies:PBMH(P)}}\\
	&\sqsupseteq \mathbf{PBMH} (P) \land \mathbf{PBMH} (Q)
	&&\raisetag{18pt}\ptext{$P$ and $Q$ are $\mathbf{RAD}$-healthy and~\cref{theorem:PBMH(P)-RAP:P}}\\
	&=P \land Q
	&&\ptext{Definition of $\sqcup_{\mathbf{RAD}}$}\\
	&=P \sqcup_{\mathbf{RAD}} Q 
\end{xflalign*}
\end{proof}
\end{proofs}
\end{theorem}

\begin{theorem}\label{theorem:RAP(P|-Q)-sqcup-RAP(R|-S)}
\begin{align*}
	&\mathbf{RA} \circ \mathbf{A} (P \vdash Q) \sqcup \mathbf{RA} \circ \mathbf{A} (R \vdash S)\\
	&=\\
	&\mathbf{RA} \circ \mathbf{A} \left(\begin{array}{l}
		(\lnot \mathbf{PBMH} (\lnot P) \lor \lnot \mathbf{PBMH} (\lnot R))
		\\ \vdash \\
		\left(\begin{array}{l}
			(\lnot \mathbf{PBMH} (\lnot P) \implies \mathbf{PBMH} (Q)) 
			\\ \land \\ 
			(\lnot \mathbf{PBMH} (\lnot R) \implies \mathbf{PBMH} (S)
		\end{array}\right)
	\end{array}\right)
\end{align*}
\begin{proofs}\begin{proof}\checkt{pfr}\checkt{alcc}
\begin{xflalign*}
	&\mathbf{RA} \circ \mathbf{A} (P \vdash Q) \sqcup \mathbf{RA} \circ \mathbf{A} (R \vdash S)
	&&\ptext{Definition of $\sqcup$}\\
	&=\mathbf{RA} \circ \mathbf{A} (P \vdash Q) \land \mathbf{RA} \circ \mathbf{A} (R \vdash S)
	&&\ptext{\cref{theorem:RA-o-A(P):RA-o-PBMH(P)}}\\
	&=\mathbf{RA} \circ \mathbf{PBMH} (P \vdash Q) \land \mathbf{RA} \circ \mathbf{PBMH} (R \vdash S)
	&&\ptext{\cref{theorem:RA(P-land-Q):RA(P)-land-RA(Q)}}\\
	&=\mathbf{RA} (\mathbf{PBMH} (P \vdash Q) \land \mathbf{PBMH} (R \vdash S))
	&&\ptext{\cref{law:pbmh:conjunction-closure}}\\
	&=\mathbf{RA} \circ \mathbf{PBMH} (\mathbf{PBMH} (P \vdash Q) \land \mathbf{PBMH} (R \vdash S))
	&&\ptext{\cref{lemma:PBMH(design):(lnot-PBMH(pre)|-PBMH(post))}}\\
	&=\mathbf{RA} \circ \mathbf{PBMH} \left(\begin{array}{l}
		(\lnot \mathbf{PBMH} (\lnot P) \vdash \mathbf{PBMH} (Q)) 
		\\ \land \\
		(\lnot \mathbf{PBMH} (\lnot R) \vdash \mathbf{PBMH}(S))
	\end{array}\right)
	&&\ptext{\cref{theorem:RA-o-A(P):RA-o-PBMH(P)}}\\
	&=\mathbf{RA} \circ \mathbf{A} \left(\begin{array}{l}
		(\lnot \mathbf{PBMH} (\lnot P) \vdash \mathbf{PBMH} (Q)) 
		\\ \land \\
		(\lnot \mathbf{PBMH} (\lnot R) \vdash \mathbf{PBMH}(S))
	\end{array}\right)
	&&\ptext{Conjunction of designs}\\
	&=\mathbf{RA} \circ \mathbf{A} \left(\begin{array}{l}
		(\lnot \mathbf{PBMH} (\lnot P) \lor \lnot \mathbf{PBMH} (R))
		\\ \vdash \\
		\left(\begin{array}{l}
			(\lnot \mathbf{PBMH} (\lnot P) \implies \mathbf{PBMH} (Q)) 
			\\ \land \\ 
			(\lnot \mathbf{PBMH} (\lnot R) \implies \mathbf{PBMH} (S))
		\end{array}\right)
	\end{array}\right)
\end{xflalign*}
\end{proof}\end{proofs}
\end{theorem}

\begin{theorem}\label{theorem:RAP(P|-Q)-sqcup-RAP(R|-S)-healthy-PBMH} Provided $\lnot P$, $\lnot Q$, $R$ and $S$ are $\mathbf{PBMH}$-healthy.
\begin{align*}
	&\mathbf{RA} \circ \mathbf{A} (P \vdash Q) \sqcup \mathbf{RA} \circ \mathbf{A} (R \vdash S)\\
	&=\\
	&\mathbf{RA} \circ \mathbf{A} (P \lor R \vdash (P \implies Q) \land (R \implies S))
\end{align*}
\begin{proofs}\begin{proof}\checkt{alcc}
\begin{xflalign*}
	&\mathbf{RA} \circ \mathbf{A} (P \vdash Q) \sqcup \mathbf{RA} \circ \mathbf{A} (R \vdash S)
	&&\ptext{\cref{theorem:RAP(P|-Q)-sqcup-RAP(R|-S)}}\\
	&=\mathbf{RA} \circ \mathbf{A} \left(\begin{array}{l}
		(\lnot \mathbf{PBMH} (\lnot P) \lor \lnot \mathbf{PBMH} (\lnot R))
		\\ \vdash \\
		\left(\begin{array}{l}
			(\lnot \mathbf{PBMH} (\lnot P) \implies \mathbf{PBMH} (Q)) 
			\\ \land \\ 
			(\lnot \mathbf{PBMH} (\lnot R) \implies \mathbf{PBMH} (S)
		\end{array}\right)
	\end{array}\right)
	&&\ptext{Assumption: $\lnot P$, $\lnot R$, $Q$ and $S$ are $\mathbf{PBMH}$-healthy}\\
	&=\mathbf{RA} \circ \mathbf{A} \left(\begin{array}{l}
		(\lnot \lnot P \lor \lnot \lnot R)
		\\ \vdash \\
		(\lnot \lnot P \implies Q) \land (\lnot \lnot R \implies S)
	\end{array}\right)
	&&\ptext{Predicate calculus}\\
	&=\mathbf{RA} \circ \mathbf{A} (P \lor R \vdash (P \implies Q) \land (R \implies S))
\end{xflalign*}
\end{proof}\end{proofs}
\end{theorem}

\begin{theorem}\label{theorem:RAP:Chaos-sqcup-P}
\begin{statement}Provided $P$ is a reactive angelic design,
$Chaos_{\mathbf{RAD}} \sqcup_{\mathbf{RAD}} \mathbf{RA} \circ \mathbf{A} (\lnot P^f_f \vdash P^t_f) = \mathbf{RA} \circ \mathbf{A} (\lnot P^f_f \vdash P^t_f)$
\end{statement}
\begin{proofs}
\begin{proof}
\begin{flalign*}
	&Chaos \sqcup \mathbf{RA} \circ \mathbf{A} (\lnot P^f_f \vdash P^t_f)
	&&\ptext{Definition of $Chaos$}\\
	&=\mathbf{RA} \circ \mathbf{A} (false \vdash ac'\neq\emptyset) \sqcup \mathbf{RA} \circ \mathbf{A} (\lnot P^f_f \vdash P^t_f)
	&&\ptext{\cref{theorem:RAP:P-sqcup-Q}}\\
	&=\mathbf{RA} \circ \mathbf{A} (
			false \lor \lnot P^f_f
			\vdash (false \implies ac'\neq\emptyset)\land (\lnot P^f_f \implies P^t_f))
	&&\ptext{Predicate calculus}\\
	&=\mathbf{RA} \circ \mathbf{A} (\lnot P^f_f \vdash (\lnot P^f_f \implies P^t_f))
\end{flalign*}
\end{proof}
\end{proofs}
\end{theorem}

\subsection{Demonic Choice}

\begin{theorem}\label{theorem:RAP:P-sqcap-Q}
\begin{statement}
Provided $P$ and $Q$ are reactive angelic processes,
\begin{align*}
	&P \sqcap_{\mathbf{RAD}} Q = \mathbf{RA} \circ \mathbf{A} (\lnot P^f_f \land \lnot Q^f_f \vdash P^t_f \lor Q^t_f)
\end{align*}
\end{statement}
\begin{proofs}
\begin{proof}\checkt{pfr}\checkt{alcc}
\begin{xflalign*}
	&P \sqcap_{\mathbf{RAD}} Q
	&&\ptext{Assumption: $P$ and $Q$ are $\mathbf{RAD}$-healthy}\\
	&=\mathbf{RA} \circ \mathbf{A} (\lnot P^f_f \vdash P^t_f) \sqcap_{\mathbf{RAD}} \mathbf{RA} \circ \mathbf{A} (\lnot Q^f_f \vdash Q^t_f)
	&&\ptext{\cref{theorem:RAP(P|-Q)-sqcap-RAP(S|-R)}}\\
	&=\mathbf{RA} \circ \mathbf{A} (\lnot P^f_f \land \lnot Q^f_f \vdash P^t_f \lor Q^t_f)
\end{xflalign*}
\end{proof}
\end{proofs}
\end{theorem}

\begin{theorem}\label{theorem:p2ac(ac2p(P)-sqcap-ac2p(Q)):p2ac-o-ac2p(P)-sqcap-p2ac-o-ac2p(Q)}
\begin{statement}
\begin{align*}
	&p2ac(ac2p(P) \sqcap_{\mathbf{R}} ac2p(Q)) = p2ac \circ ac2p(P) \sqcap_{\mathbf{RAD}} p2ac \circ ac2p(Q)
\end{align*}
\end{statement}
\begin{proofs}
\begin{proof}
\begin{xflalign*}
	&p2ac(ac2p(P) \sqcap_{\mathbf{R}} ac2p(Q))
	&&\ptext{Definition of $\sqcap$}\\
	&=p2ac(ac2p(P) \lor ac2p(Q))
	&&\ptext{\cref{theorem:p2ac(P-lor-Q):p2ac(P)-lor-p2ac(Q)}}\\
	&=p2ac\circ ac2p(P) \lor p2ac \circ ac2p(Q)
	&&\ptext{Definition of $\sqcap$}\\
	&=p2ac\circ ac2p(P) \sqcap_{\mathbf{RAD}} p2ac \circ ac2p(Q)
\end{xflalign*}
\end{proof}
\end{proofs}
\end{theorem}

\begin{theorem}\label{theorem:ac2p(p2ac(P)-sqcap-RAD-p2ac(Q)):P-sqcap-R-Q}
\begin{statement}
$ac2p(p2ac(P) \sqcap_{\mathbf{RAD}} p2ac(Q)) = P \sqcap_{\mathbf{R}} Q$
\end{statement}
\begin{proofs}\begin{proof}
\begin{xflalign*}
	&ac2p(p2ac(P) \sqcap_{\mathbf{RAD}} p2ac(Q))
	&&\ptext{Definition of $\sqcap_{\mathbf{RAD}}$ and~ \cref{theorem:ac2p(P-lor-Q):ac2p(P)-lor-ac2p(Q)}}\\
	&=ac2p \circ p2ac(P) \sqcap_{\mathbf{RAD}} ac2p \circ p2ac(Q)
	&&\ptext{Definition of $\sqcap_{\mathbf{R}}$ and~\cref{theorem:ac2p-o-p2ac(P):P}}
	&=P \sqcap_{\mathbf{R}} Q
\end{xflalign*}
\end{proof}\end{proofs}
\end{theorem}

\begin{theorem}\label{theorem:Chaos-RAD-sqcap-P:Chaor-RAD}
\begin{statement}
Provided $P$ is a reactive angelic design,
\begin{align*}
	&Chaos_{\mathbf{RAD}} \sqcap_{\mathbf{RAD}} P = Chaos_{\mathbf{RAD}}
\end{align*}
\end{statement}
\begin{proofs} 
\begin{proof}\checkt{alcc}
\begin{xflalign*}
	&Chaos_{\mathbf{RAD}} \sqcap P
	&&\ptext{Definition of $Chaos_{\mathbf{RAD}}$}\\
	&=\mathbf{RA} \circ \mathbf{A} (false \vdash ac'\neq\emptyset) \sqcap P
	&&\ptext{Assumption: $P$ is $\mathbf{RAD}$-healthy}\\
	&=\mathbf{RA} \circ \mathbf{A} (false \vdash ac'\neq\emptyset) \sqcap \mathbf{RA} \circ \mathbf{A} (\lnot P^f_f \vdash P^t_f)
	&&\ptext{\cref{theorem:RAP(P|-Q)-sqcap-RAP(S|-R)}}\\
	&=\mathbf{RA} \circ \mathbf{A} (false \land \lnot P^f_f \vdash ac'\neq\emptyset \lor P^t_f)
	&&\ptext{Predicate calculus}\\
	&=\mathbf{RA} \circ \mathbf{A} (false \vdash ac'\neq\emptyset \lor P^t_f)
	&&\ptext{Predicate calculus}\\
	&=\mathbf{RA} \circ \mathbf{A} (false \vdash ac'\neq\emptyset)
	&&\ptext{Definition of $Chaos_{\mathbf{RAP}}$}\\
	&=Chaos_{\mathbf{RAD}}
\end{xflalign*}
\end{proof}
\end{proofs}
\end{theorem}

\begin{theorem}\label{theorem:RAP(P|-Q)-sqcap-RAP(S|-R)}
\begin{statement}
\begin{align*}
	\mathbf{RA} \circ \mathbf{A} (P \vdash Q) \sqcap \mathbf{RA} \circ \mathbf{A} (R \vdash S)
	=
	\mathbf{RA} \circ \mathbf{A} (P \land R \vdash Q \lor S)
\end{align*}
\end{statement}
\begin{proofs}
\begin{proof}\checkt{pfr}\checkt{alcc}
\begin{xflalign*}
	&\mathbf{RA} \circ \mathbf{A} (P \vdash Q) \sqcap \mathbf{RA} \circ \mathbf{A} (R \vdash S)
	&&\ptext{\cref{theorem:RA-o-A(P):RA-o-PBMH(P)}}\\
	&=\mathbf{RA} \circ \mathbf{PBMH} (P \vdash Q) \sqcap \mathbf{RA} \circ \mathbf{PBMH} (R \vdash S)
	&&\ptext{Definition of $\sqcup$}\\
	&=\mathbf{RA} \circ \mathbf{PBMH} (P \vdash Q) \lor \mathbf{RA} \circ \mathbf{PBMH} (R \vdash S)
	&&\ptext{\cref{theorem:RA(P-lor-Q):RA(P)-lor-RA(Q)}}\\
	&=\mathbf{RA} (\mathbf{PBMH} (P \vdash Q) \lor \mathbf{PBMH} (R \vdash S))
	&&\ptext{\cref{law:pbmh:distribute-disjunction}}\\
	&=\mathbf{RA} \circ \mathbf{PBMH} ((P \vdash Q) \lor (R \vdash S))
	&&\ptext{Disjunction of designs}\\
	&=\mathbf{RA} \circ \mathbf{PBMH} (P \land R \vdash Q \lor S)
\end{xflalign*}
\end{proof}
\end{proofs}
\end{theorem}

\begin{lemma}\label{lemma:p2ac(ac2p(P)-sqcap-ac2p(Q)):A2-healthy}
\begin{statement}
Provided $P$ and $Q$ are reactive angelic designs and $\mathbf{A2}$-healthy,
\begin{align*}
	&p2ac(ac2p(P) \sqcap_{\mathbf{R}} ac2p(Q)) = P \sqcap_{\mathbf{RAD}} Q
\end{align*}
\end{statement}
\begin{proofs}
\begin{proof}
\begin{xflalign*}
	&p2ac(ac2p(P) \sqcap_{\mathbf{R}} ac2p(Q))
	&&\ptext{\cref{theorem:p2ac(ac2p(P)-sqcap-ac2p(Q)):p2ac-o-ac2p(P)-sqcap-p2ac-o-ac2p(Q)}}\\
	&=p2ac \circ ac2p(P) \sqcap_{\mathbf{RAD}} p2ac \circ ac2p(Q)
	&&\ptext{Assumption: $P$ and $Q$ are $\mathbf{RAD}$ and $\mathbf{A2}$-healthy and~\cref{theorem:p2ac-o-ac2p-RA-o-A:RA-o-A}}\\
	&=P \sqcap_{\mathbf{RAD}} Q
\end{xflalign*}
\end{proof}
\end{proofs}
\end{lemma}

\subsection{Chaos}

\begin{theorem}\label{theorem:RAP:Chaos-sqcup-P-RAD}
\begin{statement}Provided $P$ is a reactive angelic design,
\begin{align*}
 	&Chaos_{\mathbf{RAD}} \sqcup_{\mathbf{RAD}} P = P
\end{align*}
\end{statement}
\begin{proofs}
\begin{proof}
\begin{xflalign*}
	&Chaos_{\mathbf{RAD}} \sqcup_{\mathbf{RAD}} P
	&&\ptext{Assumption: $P$ is $\mathbf{RAD}$-healthy}\\
	&Chaos \sqcup \mathbf{RA} \circ \mathbf{A} (\lnot P^f_f \vdash P^t_f)
	&&\ptext{Definition of $Chaos$}\\
	&=\mathbf{RA} \circ \mathbf{A} (false \vdash ac'\neq\emptyset) \sqcup \mathbf{RA} \circ \mathbf{A} (\lnot P^f_f \vdash P^t_f)
	&&\ptext{\cref{theorem:RAP:P-sqcup-Q}}\\
	&=\mathbf{RA} \circ \mathbf{A} (
			false \lor \lnot P^f_f
			\vdash (false \implies ac'\neq\emptyset)\land (\lnot P^f_f \implies P^t_f))
	&&\raisetag{18pt}\ptext{Predicate calculus}\\
	&=\mathbf{RA} \circ \mathbf{A} (\lnot P^f_f \vdash (\lnot P^f_f \implies P^t_f))
	&&\ptext{Definition of design and predicate calculus}\\
	&=\mathbf{RA} \circ \mathbf{A} (\lnot P^f_f \vdash P^t_f)
	&&\ptext{Assumption: $P$ is $\mathbf{RAD}$-healthy}\\
	&=P
\end{xflalign*}
\end{proof}
\end{proofs}
\end{theorem}

\begin{theorem}\label{theorem:ac2p(Chaos-RAD):Chaos-R}
\begin{statement}
$ac2p(Chaos_{\mathbf{RAD}}) = Chaos_{\mathbf{R}}$
\end{statement}
\begin{proofs}
\begin{proof}
\begin{xflalign*}
	&ac2p(Chaos_{\mathbf{RAD}})
	&&\ptext{Definition of $Chaos_\mathbf{RAD}$}\\
	&=ac2p(\mathbf{RA} \circ \mathbf{A} (false \vdash ac'\neq\emptyset))
	&&\ptext{\cref{theorem:ac2p-o-RA-o-A(design):R(lnot-ac2p(pre)|-ac2p(post))}}\\
	&=\mathbf{R} (\lnot ac2p(true) \vdash ac2p(ac'\neq\emptyset))
	&&\ptext{\cref{lemma:ac2p(P)-s-ac'-not-free:P}}\\
	&=\mathbf{R} (\lnot true \vdash ac2p(ac'\neq\emptyset))
	&&\ptext{Predicate calculus and definition of design}\\
	&=\mathbf{R} (false \vdash true)
	&&\ptext{Definition of $Chaos_{\mathbf{R}}$}\\
	&=Chaos_{\mathbf{R}}
\end{xflalign*}
\end{proof}
\end{proofs}
\end{theorem}

\begin{theorem}\label{theorem:p2ac(Chaos-R):Chaos-RAD}
\begin{statement}$p2ac(Chaos_{\mathbf{R}}) = Chaos_{\mathbf{RAD}}$\end{statement}
\begin{proofs}
\begin{proof}\checkt{pfr}\checkt{alcc}
\begin{xflalign*}
	&p2ac(Chaos_{\mathbf{R}})
	&&\ptext{Definition of $Chaos_{\mathbf{R}}$}\\
	&=p2ac \circ \mathbf{R} (false \vdash true)
	&&\ptext{\cref{theorem:p2ac-o-R(design):RA-o-A(lnot-p2ac(pre)|-p2ac(post))}}\\
	&=\mathbf{RA} \circ \mathbf{A} (\lnot p2ac(true) \vdash p2ac(true))
	&&\ptext{\cref{lemma:p2ac(true)}}\\
	&=\mathbf{RA} \circ \mathbf{A} (\lnot ac'\neq\emptyset \vdash ac'\neq\emptyset)
	&&\ptext{Definition of $\mathbf{A}$ and $\mathbf{PBMH}$-idempotent (\cref{law:pbmh:idempotent})}\\
	&=\mathbf{RA} \circ \mathbf{A} \circ \mathbf{PBMH} (\lnot ac'\neq\emptyset \vdash ac'\neq\emptyset)
	&&\ptext{\cref{lemma:PBMH(design):(lnot-PBMH(pre)|-PBMH(post))}}\\
	&=\mathbf{RA} \circ \mathbf{A} (\lnot \mathbf{PBMH} (ac'=\emptyset) \vdash \mathbf{PBMH} (ac'\neq\emptyset))
	&&\ptext{$\mathbf{PBMH} (ac'=\emptyset) = true$}\\
	&=\mathbf{RA} \circ \mathbf{A} (\lnot true \vdash ac'\neq\emptyset)
	&&\ptext{Predicate calculus}\\
	&=\mathbf{RA} \circ \mathbf{A} (false \vdash ac'\neq\emptyset)
	&&\ptext{Definition of $Chaos_{\mathbf{RAD}}$}\\
	&=Chaos_{\mathbf{RAD}}
\end{xflalign*}
\end{proof}
\end{proofs}
\end{theorem}

\subsection{Choice}

\begin{theorem}\label{theorem:p2ac(Choice-R):Choice-RAD}
\begin{statement}
$p2ac(Choice_{\mathbf{R}}) = Choice_{\mathbf{RAD}}$
\end{statement}
\begin{proofs}
\begin{proof}\checkt{pfr}
\begin{xflalign*}
	&p2ac(Choice_{\mathbf{R}})
	&&\ptext{Definition of $Choice_{\mathbf{R}}$}\\
	&=p2ac \circ \mathbf{R} (true \vdash true)
	&&\ptext{\cref{theorem:p2ac-o-R(design):RA-o-A(lnot-p2ac(pre)|-p2ac(post))}}\\
	&=\mathbf{RA} \circ \mathbf{A} (\lnot p2ac(false) \vdash p2ac(true))\\
	&&\raisetag{18pt}\ptext{\cref{lemma:p2ac(false),lemma:p2ac(true)}}\\
	&=\mathbf{RA} \circ \mathbf{A} (\lnot false \vdash ac'\neq\emptyset)
	&&\ptext{Predicate calculus}\\
	&=\mathbf{RA} \circ \mathbf{A} (true \vdash ac'\neq\emptyset)
	&&\ptext{Definition of $Choice_{\mathbf{RAD}}$}\\
	&=Choice_{\mathbf{RAD}}
\end{xflalign*}
\end{proof}
\end{proofs}
\end{theorem}

\begin{theorem}\label{theorem:ac2p(Choice-RAD):Choice-R}
\begin{statement}
$ac2p(Choice_{\mathbf{RAD}}) = Choice_{\mathbf{R}}$
\end{statement}
\begin{proofs}\begin{proof}\checkt{alcc}
\begin{xflalign*}
	&ac2p(Choice_{\mathbf{RAD}})
	&&\ptext{Definition of $Choice_{\mathbf{RAD}}$}\\
	&=ac2p\mathbf{RA}\circ\mathbf{A} (true \vdash ac'\neq\emptyset)
	&&\ptext{\cref{theorem:ac2p-o-RA-o-A(design):R(lnot-ac2p(pre)|-ac2p(post))}}\\
	&=\mathbf{R} (\lnot ac2p(false) \vdash ac2p(\circledIn{y}{ac'} (ac'\neq\emptyset)))
	&&\ptext{Predicate calculus}\\
	&=\mathbf{R} (\lnot ac2p(false) \vdash ac2p(true \land ac'\neq\emptyset))
	&&\ptext{\cref{lemma:ac2p(P)-s-ac'-not-free:P,lemma:ac2p(P-land-ac'-neq-emptyset):ac2p(P)}}\\
	&=\mathbf{R} (true \vdash true)
	&&\ptext{Definition of $Choice_{\mathbf{R}}$}\\
	&=Choice_{\mathbf{R}}
\end{xflalign*}
\end{proof}\end{proofs}
\end{theorem}

\begin{theorem}\label{theorem:Choice-RAD-sqcup-P}
\begin{statement}
Provided $P$ is $\mathbf{RAD}$-healthy,
\begin{align*}
	&Choice_{\mathbf{RAD}} \sqcup_{\mathbf{RAD}} P = \mathbf{RA} \circ \mathbf{A} (true \vdash P^t_f)
\end{align*}
\end{statement}
\begin{proofs}
\begin{proof}
\begin{xflalign*}
	&Choice_{\mathbf{RAD}} \sqcup_{\mathbf{RAD}} P
	&&\ptext{Definition of $\mathbf{ND_{RAD}}$ and~\cref{theorem:RAP:P-sqcup-Choice:(true|Pt)}}\\
	&=\mathbf{RA} \circ \mathbf{A} (true \vdash P^t_f)
\end{xflalign*}
\end{proof}
\end{proofs}
\end{theorem}

\begin{theorem}\label{theorem:Choice-RAD-sqcap-P}
\begin{statement}
Provided $P$ is $\mathbf{RAD}$-healthy,
\begin{align*}
		&Choice_{\mathbf{RAD}} \sqcap_{\mathbf{RAD}} P = \mathbf{RA}\circ\mathbf{A} (\lnot P^f_f \vdash ac'\neq\emptyset)
\end{align*}
\end{statement}
\begin{proofs}
\begin{proof}
\begin{xflalign*}
	&Choice_{\mathbf{RAD}} \sqcap_{\mathbf{RAD}} P
	&&\ptext{Definition of $Choice_{\mathbf{RAD}}$}\\
	&=\mathbf{RA}\circ\mathbf{A} (true \vdash ac'\neq\emptyset) \sqcap_{\mathbf{RAD}} P
	&&\ptext{Assumption: $P$ is $\mathbf{RAD}$-healthy}\\
	&=\mathbf{RA}\circ\mathbf{A} (true \vdash ac'\neq\emptyset) \sqcap_{\mathbf{RAD}} \mathbf{RA}\circ\mathbf{A} (\lnot P^f_f \vdash P^t_f)
	&&\ptext{\cref{theorem:RAP:P-sqcap-Q}}\\
	&=\mathbf{RA}\circ\mathbf{A} (true \land \lnot P^f_f \vdash ac'\neq\emptyset \lor P^t_f)
	&&\ptext{Predicate calculus}\\
	&=\mathbf{RA}\circ\mathbf{A} (\lnot P^f_f \vdash ac'\neq\emptyset \lor P^t_f)
	&&\ptext{Definition of $\mathbf{A}$, $\mathbf{A0}$ and predicate calculus}\\
	&=\mathbf{RA}\circ\mathbf{A} (\lnot P^f_f \vdash ac'\neq\emptyset)
\end{xflalign*}
\end{proof}
\end{proofs}
\end{theorem}

\subsection{Stop}

\begin{theorem}\label{theorem:Stop-RAD-sqcup-P}
\begin{statement}
Provided $P$ is $\mathbf{RAD}$-healthy,
\begin{align*}
	&Stop_{\mathbf{RAD}} \sqcup_{\mathbf{RAD}} P\\
	&=\\
	&\mathbf{RA}\circ\mathbf{A} (true \vdash (\lnot P^f_f \implies P^t_f) \land \circledIn{y}{ac'} (y.tr=s.tr \land y.wait))
\end{align*}
\end{statement}
\begin{proofs}
\begin{proof}
\begin{xflalign*}
	&Stop_{\mathbf{RAD}} \sqcup_{\mathbf{RAD}} P
	&&\ptext{Definition of $Stop_{\mathbf{RAD}}$}\\
	&=\mathbf{RA}\circ \mathbf{A} (true \vdash \circledIn{y}{ac'} (y.tr=s.tr \land y.wait)) \sqcup_{\mathbf{RAD}} P
	&&\ptext{Assumption: $P$ is $\mathbf{RAD}$-healthy}\\
	&=\left(\begin{array}{l}
		\mathbf{RA}\circ \mathbf{A} (true \vdash \circledIn{y}{ac'} (y.tr=s.tr \land y.wait)) 
		\\ \sqcup_{\mathbf{RAD}} \\
		\mathbf{RA}\circ\mathbf{A} (\lnot P^f_f \vdash P^t_f)
	\end{array}\right)
	&&\ptext{\cref{theorem:RAP:P-sqcup-Q}}\\
	&=\mathbf{RA}\circ\mathbf{A} (true \lor \lnot P^f_f \vdash (\lnot P^f_f \implies P^t_f) \land \circledIn{y}{ac'} (y.tr=s.tr \land y.wait))
	&&\ptext{Predicate calculus}\\
	&=\mathbf{RA}\circ\mathbf{A} (true \vdash (\lnot P^f_f \implies P^t_f) \land \circledIn{y}{ac'} (y.tr=s.tr \land y.wait))
\end{xflalign*}
\end{proof}
\end{proofs}
\end{theorem}

\begin{theorem}\label{theorem:p2ac(Stop-R):Stop-RAD}
\begin{statement}$p2ac(Stop_{\mathbf{R}}) = Stop_{\mathbf{RAD}}$\end{statement}
\begin{proofs}
\begin{proof}\checkt{pfr}\checkt{alcc}
\begin{xflalign*}
	&p2ac(Stop_{\mathbf{R}})
	&&\ptext{Definition of $Stop_{\mathbf{R}}$}\\
	&=p2ac \circ \mathbf{R} (true \vdash tr'=tr \land wait')
	&&\ptext{\cref{theorem:p2ac-o-R(design):RA-o-A(lnot-p2ac(pre)|-p2ac(post))}}\\
	&=\mathbf{RA} \circ \mathbf{A} (\lnot p2ac(false) \vdash p2ac(tr'=tr \land wait'))
	&&\ptext{\cref{lemma:p2ac(false)}}\\
	&=\mathbf{RA} \circ \mathbf{A} (\lnot false \vdash p2ac(tr'=tr \land wait'))
	&&\ptext{Predicate calculus}\\
	&=\mathbf{RA} \circ \mathbf{A} (true \vdash p2ac(tr'=tr \land wait'))
	&&\ptext{Definition of $p2ac$}\\
	&=\mathbf{RA} \circ \mathbf{A} (true \vdash \exists z \spot z.tr'=s.tr \land z.wait' \land undash(z) \in ac')
	&&\ptext{Introduce fresh variable}\\
	&=\mathbf{RA} \circ \mathbf{A} (true \vdash \exists z, y \spot z.tr'=s.tr \land z.wait' \land y = undash(z) \land y \in ac')
	&&\ptext{Property of $dash$}\\
	&=\mathbf{RA} \circ \mathbf{A} (true \vdash \exists z, y \spot z.tr'=s.tr \land z.wait' \land dash(y) = z \land y \in ac')
	&&\ptext{One-point rule}\\
	&=\mathbf{RA} \circ \mathbf{A} (true \vdash \exists y \spot dash(y).tr'=s.tr \land dash(y).wait' \land y \in ac')
	&&\ptext{Property of $dash$}\\
	&=\mathbf{RA} \circ \mathbf{A} (true \vdash \exists y \spot y.tr=s.tr \land y.wait \land y \in ac')
	&&\ptext{Definition of $\circledIn{y}{ac'}$}\\
	&=\mathbf{RA} \circ \mathbf{A} (true \vdash \circledIn{y}{ac'} (y.tr=s.tr \land y.wait))
	&&\ptext{Definition of $Stop_{\mathbf{RAD}}$}\\
	&=Stop_{\mathbf{RAD}}
\end{xflalign*}
\end{proof}
\end{proofs}
\end{theorem}

\begin{theorem}\label{theorem:ac2p(Stop-RAD):Stop-R}
\begin{statement}
$ac2p(Stop_{\mathbf{RAD}}) = Stop_{\mathbf{R}}$
\end{statement}
\begin{proofs}
\begin{proof}\checkt{alcc}
\begin{xflalign*}
	&ac2p(Stop_{\mathbf{RAD}})
	&&\ptext{Definition of $Stop_{\mathbf{RAD}}$}\\
	&=ac2p\mathbf{RA}\circ\mathbf{A} (true \vdash \circledIn{y}{ac'} (y.wait \land y.tr=s.tr))
	&&\ptext{\cref{theorem:ac2p-o-RA-o-A(design):R(lnot-ac2p(pre)|-ac2p(post))}}\\
	&=\mathbf{R} (\lnot ac2p(false) \vdash ac2p(\circledIn{y}{ac'} (y.wait \land y.tr=s.tr)))
	&&\ptext{\cref{lemma:ac2p(P)-s-ac'-not-free:P} and predicate calculus}\\
	&=\mathbf{R} (true \vdash ac2p(\circledIn{y}{ac'} (y.wait \land y.tr=s.tr)))
	&&\ptext{Definition of $\circledIn{y}{ac'}$ and~\cref{lemma:ac2p(exists-y-in-ac'-e)}}\\
	&=\mathbf{R} (true \vdash wait' \land tr'=tr)
	&&\ptext{Definition of $Stop_{\mathbf{R}}$}\\
	&=Stop_{\mathbf{R}}
\end{xflalign*}
\end{proof}
\end{proofs}
\end{theorem}

\subsection{Skip}

\begin{theorem}\label{theorem:Skip-RAD-sqcup-P}
\begin{statement}
Provided $P$ is $\mathbf{RAD}$-healthy,
\begin{align*}
	&Skip_{\mathbf{RAD}} \sqcup_{\mathbf{RAD}} P\\
	&=\\
	&\mathbf{RA}\circ\mathbf{A}(true 
		\vdash 
		\circledIn{y}{ac'} (\lnot y.wait \land y.tr=s.tr)) \land (\lnot P^f_f \implies P^t_t))
\end{align*}
\end{statement}
\begin{proofs}
\begin{proof}
\begin{xflalign*}
	&Skip_{\mathbf{RAD}} \sqcup_{\mathbf{RAD}} P
	&&\ptext{Definition of $Skip_{\mathbf{RAD}}$}\\
	&=\mathbf{RA}\circ\mathbf{A}(true \vdash \circledIn{y}{ac'} (\lnot y.wait \land y.tr=s.tr)) \sqcup_{\mathbf{RAD}} P
	&&\ptext{Assumption: $P$ is $\mathbf{RAD}$-healthy}\\
	&=\mathbf{RA}\circ\mathbf{A}(true \vdash \circledIn{y}{ac'} (\lnot y.wait \land y.tr=s.tr)) \sqcup_{\mathbf{RAD}} \mathbf{RA}\circ\mathbf{A} (\lnot P^f_f \vdash P^t_t)
	&&\ptext{\cref{theorem:RAP:P-sqcup-Q}}\\
	&=\mathbf{RA}\circ\mathbf{A}\left(\begin{array}{l}
		true \lor \lnot P^f_f 
		\\ \vdash \\
		true \implies \circledIn{y}{ac'} (\lnot y.wait \land y.tr=s.tr)) \land (\lnot P^f_f \implies P^t_t)
	\end{array}\right)
	&&\ptext{Predicate calculus}\\
	&=\mathbf{RA}\circ\mathbf{A}\left(\begin{array}{l}
		true 
		\\ \vdash \\
		\circledIn{y}{ac'} (\lnot y.wait \land y.tr=s.tr)) \land (\lnot P^f_f \implies P^t_t)
	\end{array}\right)
\end{xflalign*}
\end{proof}
\end{proofs}
\end{theorem}

\begin{theorem}\label{theorem:p2ac(Skip-R):Skip-RAD}
\begin{statement}
$p2ac(Skip_{\mathbf{R}}) = Skip_{\mathbf{RAD}}$
\end{statement}
\begin{proofs}
\begin{proof}
\begin{xflalign*}
	&p2ac(Skip_{\mathbf{R}})
	&&\ptext{Definition of $Skip_{\mathbf{R}}$}\\
	&=p2ac\circ \mathbf{R} (true \vdash tr'=tr \land \lnot wait')
	&&\ptext{\cref{theorem:p2ac-o-R(design):RA-o-A(lnot-p2ac(pre)|-p2ac(post))}}\\
	&=\mathbf{RA}\circ\mathbf{A} (\lnot p2ac(false) \vdash p2ac(tr'=tr \land \lnot wait'))
	&&\ptext{\cref{lemma:p2ac(false)} and predicate calculus}\\
	&=\mathbf{RA}\circ\mathbf{A} (true \vdash p2ac(tr'=tr \land \lnot wait'))
	&&\ptext{Definition of $p2ac$ and sustitution}\\
	&=\mathbf{RA}\circ\mathbf{A} (true \vdash \exists z @ z.tr'=s.tr \land \lnot z.wait' \land undash(z) \in ac')
	&&\ptext{Introduce fresh variable $y$ and property of $dash$ and $undash$}\\
	&=\mathbf{RA}\circ\mathbf{A} (true \vdash \exists y @ y.tr=s.tr \land \lnot y.wait \land y \in ac')
	&&\ptext{Definition of $\circledIn{y}{ac'}$}\\
	&=\mathbf{RA}\circ\mathbf{A} (true \vdash \circledIn{y}{ac'} (y.tr=s.tr \land \lnot y.wait))
	&&\ptext{Definition of $Skip_{\mathbf{RAD}}$}\\
	&=Skip_{\mathbf{RAD}}	
\end{xflalign*}
\end{proof}
\end{proofs}
\end{theorem}

\begin{theorem}\label{theorem:ac2p(Skip-RAD):Skip-R}
\begin{statement}
$ac2p(Skip_{\mathbf{RAD}}) = Skip_{\mathbf{R}}$
\end{statement}
\begin{proofs}
\begin{proof}
\begin{xflalign*}
	&ac2p(Skip_{\mathbf{RAD}})
	&&\ptext{Definition of $Skip_{\mathbf{RAD}}$}\\
	&=ac2p\mathbf{RA}\circ\mathbf{A} (true \vdash \circledIn{y}{ac'} (\lnot y.wait \land y.tr=s.tr))
	&&\ptext{\cref{theorem:ac2p-o-RA-o-A(design):R(lnot-ac2p(pre)|-ac2p(post))}}\\
	&=\mathbf{R} (\lnot ac2p(false) \vdash ac2p(\circledIn{y}{ac'} (\lnot y.wait \land y.tr=s.tr)))
	&&\ptext{\cref{lemma:ac2p(P)-s-ac'-not-free:P} and predicate calculus}\\
	&=\mathbf{R} (true \vdash ac2p(\circledIn{y}{ac'} (\lnot y.wait \land y.tr=s.tr)))
	&&\ptext{Definition of $\circledIn{y}{ac'}$ and~\cref{lemma:ac2p(exists-y-in-ac'-e)}}\\
	&=\mathbf{R} (true \vdash \lnot wait' \land tr'=tr)
	&&\ptext{Definition of $Skip_{\mathbf{R}}$}\\
	&=Skip_{\mathbf{R}}
\end{xflalign*}
\end{proof}
\end{proofs}
\end{theorem}

\begin{lemma}\label{lemma:ac2p(Stop-RAD-sqcup-Skip-RAD):top-R}
\begin{statement}
$ac2p(Stop_{\mathbf{RAD}} \sqcup_{\mathbf{RAD}} Skip_{\mathbf{RAD}}) = \top_{\mathbf{R}}$
\end{statement}
\begin{proofs}
\begin{proof}
\begin{xflalign*}
	&ac2p(Stop_{\mathbf{RAD}} \sqcup_{\mathbf{RAD}} Skip_{\mathbf{RAD}})
	&&\ptext{Definition of $Stop_{\mathbf{RAD}}$ and $Skip_{\mathbf{RAD}}$}\\
	&=ac2p\left(
		\begin{array}{l}
			\mathbf{RA} \circ \mathbf{A} (true \vdash \circledIn{y}{ac'} (y.tr=s.tr \land y.wait))
			\\ \sqcup_{\mathbf{RAD}} \\
			\mathbf{RA}\circ\mathbf{A} (true \vdash \circledIn{y}{ac'} (\lnot y.wait \land y.tr=s.tr))
		\end{array}
		\right)
	&&\ptext{\cref{theorem:RAP:P-sqcup-Q}}\\
	&=ac2p\circ\mathbf{RA} \circ \mathbf{A}\left(\begin{array}{l}
		true \lor true
		\\ \vdash \\
		\left(\begin{array}{l}
			(true \implies \circledIn{y}{ac'} (y.tr=s.tr \land y.wait))
			\\ \land \\
			(true \implies \circledIn{y}{ac'} (\lnot y.wait \land y.tr=s.tr))
		\end{array}\right)
	\end{array}\right)
	&&\ptext{Predicate calculus}\\
	&=ac2p\circ\mathbf{RA} \circ \mathbf{A}\left(\begin{array}{l}
		true
		\\ \vdash \\
		\left(\begin{array}{l}
			\circledIn{y}{ac'} (y.tr=s.tr \land y.wait)
			\\ \land \\
			\circledIn{y}{ac'} (\lnot y.wait \land y.tr=s.tr)
		\end{array}\right)
	\end{array}\right)
	&&\ptext{\cref{theorem:ac2p-o-RA-o-A(design):R(lnot-ac2p(pre)|-ac2p(post))}}\\
	&=\mathbf{R}\left(\begin{array}{l}
		\lnot ac2p(false)
		\\ \vdash \\
		ac2p\left(\begin{array}{l}
			\circledIn{y}{ac'} (y.tr=s.tr \land y.wait)
			\\ \land \\
			\circledIn{y}{ac'} (\lnot y.wait \land y.tr=s.tr)
		\end{array}\right)
	\end{array}\right)
	&&\ptext{\cref{lemma:ac2p(P)-s-ac'-not-free:P,theorem:ac2p(P-land-Q):ac2p(P)-land-ac2p(Q)}}\\
	&=\mathbf{R}\left(\begin{array}{l}
		\lnot false
		\\ \vdash \\
		\left(\begin{array}{l}
			ac2p(\circledIn{y}{ac'} (y.tr=s.tr \land y.wait))
			\\ \land \\
			ac2p(\circledIn{y}{ac'} (\lnot y.wait \land y.tr=s.tr))
		\end{array}\right)
	\end{array}\right)
	&&\ptext{Predicate calculus and~\cref{lemma:ac2p(exists-y-in-ac'-e)}}\\
	&=\mathbf{R}(true \vdash tr'=tr \land wait' \land \lnot wait' \land tr'=tr)
	&&\ptext{Predicate calculus}\\
	&=\mathbf{R}(true \vdash false)
	&&\ptext{Definition of $\top_{\mathbf{R}}$}
\end{xflalign*}
\end{proof}
\end{proofs}
\end{lemma}

\subsection{Sequential Composition}

\begin{theorem}\label{theorem:RAP:seqDac}
\begin{statement}
Provided $P$ and $Q$ are reactive angelic designs,
\begin{align*}
	&P \seqDac Q\\
	&=\\
	&\mathbf{RA} \circ \mathbf{A} \left(\begin{array}{l}
			\left(\begin{array}{l} 
				\lnot (\mathbf{RA1} (P^f_f) \seqA \mathbf{RA1} (true))
				\\ \land \\
				\lnot (\mathbf{RA1} (P^t_f) \seqA (\lnot s.wait \land \mathbf{RA2} \circ \mathbf{RA1} (Q^f_f))) 
			\end{array}\right)
		\\ \vdash \\
		\mathbf{RA1} (P^t_f) \seqA (s \in ac' \dres s.wait \rres (\mathbf{RA2} \circ \mathbf{RA1} (\lnot Q^f_f \implies Q^t_f)))
	\end{array}\right)
\end{align*}
\end{statement}
\begin{proofsbig}
\begin{proof}\checkt{alcc}\checkt{pfr}
\begin{xflalign*}
	&=P \seqDac Q
	&&\ptext{Assumption: $P$ and $Q$ are $\mathbf{RAP}$-healthy}\\
	&=\mathbf{RA} \circ \mathbf{A} (\lnot P^f_f \vdash P^t_f) \seqDac \mathbf{RA} \circ \mathbf{A} (\lnot Q^f_f \vdash Q^t_f)
	&&\ptext{\cref{theorem:RA-o-A(P):RA-o-PBMH(P)}}\\
	&=\mathbf{RA} \circ \mathbf{PBMH} (\lnot P^f_f \vdash P^t_f) \seqDac \mathbf{RA} \circ \mathbf{PBMH} (\lnot Q^f_f \vdash Q^t_f)
	&&\ptext{\cref{lemma:PBMH(design):(lnot-PBMH(pre)|-PBMH(post))}}\\
	&=\left(
\right)
	&&\ptext{Predicate calculus and~\cref{law:pbmh:distribute-disjunction}}\\
	&=\mathbf{RA} \left(%
\right)
	&&\ptext{\cref{lemma:PBMH(design):(lnot-PBMH(pre)|-PBMH(post))} and predicate calculus}\\
	&=\mathbf{RA} \circ \mathbf{PBMH} \left(%
\right)
	&&\ptext{Predicate calculus and~\cref{law:pbmh:distribute-disjunction}}\\
	&=\mathbf{RA} \circ \mathbf{A} \left(%
\right)
		\\ \vdash \\
		\mathbf{RA1} (P^t_f) \seqA (s \in ac' \dres s.wait \rres (\mathbf{RA2} \circ \mathbf{RA1} (\lnot Q^f_f \implies Q^t_f)))
	\end{array}\right)
\end{xflalign*}
\end{proof}
\end{proofsbig}
\end{theorem}

\begin{theorem}\label{theorem:p2ac(ac2p(P)-seq-ac2p(Q)):sqsupseteq:P-seqDac-Q}
\begin{statement}
Provided $P$ and $Q$ are reactive angelic designs,
\begin{align*}
	&p2ac(ac2p(P) \circseq ac2p(Q)) \sqsupseteq P \seqDac Q
\end{align*}	
\end{statement}
\begin{proofs}
\begin{proof}\checkt{alcc}
\begin{xflalign*}
	&p2ac(ac2p(P) \circseq ac2p(Q))
	&&\ptext{\cref{theorem:p2ac(P-seq-Q):p2ac(P)-seqDac-p2ac(Q)}}\\
	&=p2ac \circ ac2p(P) \seqDac p2ac \circ ac2p(Q)
	&&\ptext{\cref{theorem:p2ac-o-ac2p:implies:PBMH(P),lemma:P-seqDac-Q:implies:P-seqDac-R,lemma:P-seqDac-Q:implies:R-seqDac-Q}}\\
	&\sqsupseteq \mathbf{PBMH} (P) \seqDac \mathbf{PBMH}(Q)
	&&\ptext{Assumption: $P$ and $Q$ are $\mathbf{RAD}$-healthy and~\cref{theorem:PBMH(P)-RAP:P}}\\
	&=P \seqDac Q
\end{xflalign*}
\end{proof}
\end{proofs}
\end{theorem}

\begin{theorem}\label{theorem:p2ac(ac2p(P)-seq-ac2p(Q))-A2:P-seqDac-Q}
\begin{statement}
Provided $P$ and $Q$ are $\mathbf{RAD}$-healthy and $\mathbf{A2}$-healthy,
\begin{align*}
	&p2ac(ac2p(P) \circseq ac2p(Q)) = P \seqDac Q
\end{align*}
\end{statement}
\begin{proofs}
\begin{proof}
\begin{xflalign*}
	&p2ac(ac2p(P) \circseq ac2p(Q))
	&&\ptext{\cref{theorem:p2ac(P-seq-Q):p2ac(P)-seqDac-p2ac(Q)}}\\
	&=p2ac\circ ac2p(P) \seqDac p2ac\circ ac2p(Q)
	&&\ptext{Assumption: $P$ and $Q$ are $\mathbf{A2}$-healthy and~\cref{lemma:p2ac-o-ac2p(P)-A2:P-land-ac'-neq-emptyset}}\\
	&=(P \land ac'\neq\emptyset) \seqDac (Q \land ac'\neq\emptyset)
	&&\ptext{Assumption: $P$ and $Q$ are $\mathbf{RAD}$-healthy and~\cref{lemma:RA1-implies-ac'-neq-emptyset}}\\
	&=P \seqDac Q
\end{xflalign*}
\end{proof}
\end{proofs}
\end{theorem}

\begin{theorem}\label{theorem:ac2p(p2ac(P)-seqDac-p2ac(Q)):P-seq-Q}
\begin{statement}
$ac2p(p2ac(P) \seqDac p2ac(Q)) = P \circseq Q$
\end{statement}
\begin{proofs}
\begin{proof}\checkt{alcc}
\begin{xflalign*}
	&ac2p(p2ac(P) \seqDac p2ac(Q))
	&&\ptext{\cref{theorem:p2ac(P-seq-Q):p2ac(P)-seqDac-p2ac(Q)}}\\
	&=ac2p\circ p2ac(P \circseq Q)
	&&\ptext{\cref{theorem:ac2p-o-p2ac(P):P}}\\
	&=P
\end{xflalign*}
\end{proof}
\end{proofs}
\end{theorem}

\begin{theorem}\label{theorem:RAD:A2(P-seqDac-Q):P-seqDac-Q}
\begin{statement}
Provided $P$ and $Q$ are reactive angelic designs and $\mathbf{A2}$-healthy,
$\mathbf{A2} (P \seqDac Q) = P \seqDac Q$
\end{statement}
\begin{proofs}
\begin{proof}\checkt{alcc}
\begin{xflalign*}
	&P \seqDac Q
	&&\ptext{Assumption: $P$ and $Q$ are $\mathbf{RAD}$-healthy and~\cref{theorem:RA-o-A(design):RA-CSPA-PBMH}}\\
	&=\mathbf{RA}\circ\mathbf{A} (\lnot P^f_f \vdash P^t_f) \seqDac \mathbf{RA}\circ\mathbf{A} (\lnot Q^f_f \vdash Q^t_f)
	&&\ptext{Assumption: $P$ and $Q$ are $\mathbf{A2}$-healthy and~\cref{theorem:A2-o-RA-o-A(lnot-Pff|-Ptf):RA-o-A(lnot-Pff|-Ptf)}}\\
	&=\mathbf{A2}\circ\mathbf{RA}\circ\mathbf{A} (\lnot P^f_f \vdash P^t_f) \seqDac \mathbf{A2}\circ\mathbf{RA}\circ\mathbf{A} (\lnot Q^f_f \vdash Q^t_f)
	&&\ptext{\cref{theorem:A2(P-seqDac-Q):P-seqDac-Q}}\\
	&=\mathbf{A2} (\mathbf{A2}\circ\mathbf{RA}\circ\mathbf{A} (\lnot P^f_f \vdash P^t_f) \seqDac \mathbf{A2}\circ\mathbf{RA}\circ\mathbf{A} (\lnot Q^f_f \vdash Q^t_f))
	&&\ptext{Assumption: $P$ and $Q$ are $\mathbf{A2}$-healthy and~\cref{theorem:A2-o-RA-o-A(lnot-Pff|-Ptf):RA-o-A(lnot-Pff|-Ptf)}}\\
	&=\mathbf{A2} (\mathbf{RA}\circ\mathbf{A} (\lnot P^f_f \vdash P^t_f) \seqDac \mathbf{RA}\circ\mathbf{A} (\lnot Q^f_f \vdash Q^t_f))
	&&\ptext{Assumption: $P$ and $Q$ are $\mathbf{RAD}$-healthy and~\cref{theorem:RA-o-A(design):RA-CSPA-PBMH}}\\
	&=\mathbf{A2} (P \seqDac Q)
\end{xflalign*}
\end{proof}
\end{proofs}
\end{theorem}

\begin{theorem}\label{theorem:RA(design)-seqDac-RA(design)} Provided $\lnot P$, $\lnot R$, $Q$ and $S$ are $\mathbf{PBMH}$-healthy and $ok, ok'$ are not free in $P$, $Q$, $R$ and $S$,
\begin{align*}
	&\mathbf{RA} (P \vdash Q) \seqDac \mathbf{RA} (R \vdash S)\\
	&=\\
	&\mathbf{RA} \left(\begin{array}{l} 
		\left(\begin{array}{l}
			\lnot (\mathbf{RA1} (\lnot P) \seqA \mathbf{RA1} (true))
			\\ \land \\
			\lnot (\mathbf{RA1} (Q) \seqA (\lnot s.wait \land \mathbf{RA2} \circ \mathbf{RA1} (\lnot R)))
		\end{array}\right)
		\\ \vdash \\
		\mathbf{RA1} (Q) \seqA (s \in ac' \dres s.wait \rres \mathbf{RA2} \circ \mathbf{RA1} (R \implies S))
	\end{array}\right)
\end{align*}
\begin{proofsbig}\begin{proof}\checkt{alcc}\checkt{pfr}
\begin{xflalign*}
	&\mathbf{RA} (P \vdash Q) \seqDac \mathbf{RA} (R \vdash S)
	&&\ptext{Definition of $\mathbf{RA}$}\\
	&=\mathbf{RA3} \circ \mathbf{RA2} \circ \mathbf{RA1} (P \vdash Q) \seqDac \mathbf{RA3} \circ \mathbf{RA2} \circ \mathbf{RA1} (R \vdash S)
	&&\ptext{Commutativity of $\mathbf{RA1}$-$\mathbf{RA2}$ (\cref{theorem:RA2-o-RA1:RA1-o-RA2})}\\
	&=\mathbf{RA3} \circ \mathbf{RA1} \circ \mathbf{RA2} (P \vdash Q) \seqDac \mathbf{RA3} \circ \mathbf{RA1} \circ \mathbf{RA2} (R \vdash S)
	&&\ptext{Commutativity of $\mathbf{RA1}$-$\mathbf{RA3}$ (\cref{theorem:RA3-o-RA1:RA1-o-RA3})}\\
	&=\mathbf{RA1} \circ \mathbf{RA3} \circ \mathbf{RA2} (P \vdash Q) \seqDac \mathbf{RA1} \circ \mathbf{RA3} \circ \mathbf{RA2} (R \vdash S)
	&&\ptext{\cref{lemma:RA2(P|-Q):(lnot-RA2(lnot-P)|-RA2(Q))}}\\
	&=\left(
\right)
	&&\ptext{\cref{lemma:lnot(conditional)} and predicate calculus}\\
	&=\mathbf{RA1} \left(%
\right)
	&&\ptext{Property of conditional and predicate calculus}\\
	&=\mathbf{RA1} \left(%
\right)
	&&\ptext{\cref{theorem:RA2(P-lor-Q):RA2(P)-lor-RA2(Q)} and predicate calculus}\\
	&=\mathbf{RA1} \left(%
\right)
	&&\ptext{Property of conditional}\\
	&=\mathbf{RA1} \left(%
\right)
	&&\ptext{Property of conditional}\\
	&=\mathbf{RA1} \left(%
\right)
	&&\ptext{Predicate calculus and property of conditional}\\
	&=\mathbf{RA1} \left(%
\right)
\end{xflalign*}
\end{proof}\end{proofsbig}
\end{theorem}

\begin{theorem}\label{theorem:RA1(design)-seqDac-RA1(design)} Provided $\lnot P$,$Q$,$\lnot R$ and $S$ are $\mathbf{PBMH}$-healthy, and $ok$ and $ok'$ are not free in $P$,$Q$,$R$ and $S$,
\begin{align*}
	&\mathbf{RA1} (P \vdash Q) \seqDac \mathbf{RA1} (R \vdash S)\\
	&=\\
	&\mathbf{RA1} \left(%
\right)
\end{align*}
\begin{proofsbig}\begin{proof}\checkt{alcc}\checkt{pfr}
\begin{xflalign*}
	&\mathbf{RA1} (P \vdash Q) \seqDac \mathbf{RA1} (R \vdash S)
	&&\ptext{Definition of $\seqDac$}\\
	&=\exists ok_0 \spot \mathbf{RA1} (P \vdash Q)[ok_0/ok'] \seqA \mathbf{RA1} (R \vdash S)[ok_0/ok]
	&&\ptext{Definition of design}\\
	&=\exists ok_0 \spot \left(%
\right)
	&&\ptext{Definition of design}\\
	&=\mathbf{RA1} \left(%
\right)
\end{xflalign*}
\end{proof}\end{proofsbig}
\end{theorem}

\begin{lemma}\label{lemma:(Skip-RAD-sqcup-Stop-RAD)-seqDac-Chaos-RAD}
\begin{statement}
$(Stop_{\mathbf{RAD}} \sqcup_{\mathbf{RAD}} Skip_{\mathbf{RAD}}) \seqDac Chaos_{\mathbf{RAD}} = Stop_{\mathbf{RAD}}$
\end{statement}
\begin{proofs}
\begin{proof}
\begin{xflalign*}
	&(Stop_{\mathbf{RAD}} \sqcup_{\mathbf{RAD}} Skip_{\mathbf{RAD}}) \seqDac Chaos_{\mathbf{RAD}}
	&&\ptext{Result of~\cref{example:Stop-RAD-sqcup-Skip-RAD} and definition of $Chaos_{\mathbf{RAD}}$}\\
	&=\left(
\right)
	&&\ptext{\cref{law:seqA-ac'-not-free} and predicate calculus}\\
	&=\mathbf{RA} \circ \mathbf{A} \left(%
\right)
	&&\ptext{Definition of $\seqA$, substitution and property of sets}\\
	&=\mathbf{RA} \circ \mathbf{A} \left(%
\right)
	&&\ptext{Predicate calculus and property of conditional}\\
	&=\mathbf{RA} \circ \mathbf{A} \left(%
\right)
	&&\ptext{Transitivity of equality}\\
	&=\mathbf{RA} \circ \mathbf{A} \left(%
\right)
	\end{array}\right)
	&&\ptext{Definition of $\circledIn{y}{ac'}$}\\
	&=\mathbf{RA} \circ \mathbf{A} (
		true 
		\vdash 
		\circledIn{y}{ac'} (y.tr=s.tr \land y.wait) \land \mathbf{RA1} (true))
	&&\ptext{\cref{lemma:RA1(circledIn)-ac'-not-free:circledIn(P-land-s.tr-le-y.tr),lemma:RA1(P-land-Q):RA1(P)-land-RA1(Q)}}\\
	&=\mathbf{RA} \circ \mathbf{A} (
		true 
		\vdash 
		\circledIn{y}{ac'} (y.tr=s.tr \land y.wait)
	&&\ptext{Definition of $Stop_{\mathbf{RAD}}$}\\
	&=Stop_{\mathbf{RAD}}
\end{xflalign*}
\end{proof}
\end{proofs}
\end{lemma}

\subsection{Event Prefixing}

\begin{theorem}\label{theorem:a-then-Skip-RAD-sqcup-P}
\begin{statement}
Provided $P$ is a reactive angelic design,
\begin{align*}
	&a \circthen_{\mathbf{RAD}} Skip_{\mathbf{RAD}} \sqcup_{\mathbf{RAD}} P\\
	&=\\
	&\mathbf{RA} \circ \mathbf{A} \left(
			true \vdash 
			\circledIn{y}{ac'} \left(\begin{array}{l}(y.tr=s.tr \land a \notin y.ref)
						\\ \dres y.wait \rres \\
					(y.tr = s.tr \cat \lseq a \rseq)
					\end{array}\right)
				\land
			(\lnot P^f_f \implies P^t_f)\right)
\end{align*}
\end{statement}
\begin{proofs}
\begin{proof}
\begin{xflalign*}
	&a \circthen_{\mathbf{RAD}} Skip_{\mathbf{RAD}} \sqcup_{\mathbf{RAD}} P
	&&\ptext{Definition of prefixing}\\
	&=\mathbf{RA} \circ \mathbf{A} 
		\left(true 
				\vdash 
					\circledIn{y}{ac'} \left(\begin{array}{l}(y.tr=s.tr \land a \notin y.ref)
						\\ \dres y.wait \rres \\
					(y.tr = s.tr \cat \lseq a \rseq)
					\end{array}\right)
		\right) \sqcup_{\mathbf{RAD}} P
	&&\ptext{Assumption: $P$ is $\mathbf{RAD}$-healthy}\\
	&=\mathbf{RA} \circ \mathbf{A} 
		\left(true 
				\vdash 
					\circledIn{y}{ac'} \left(\begin{array}{l}(y.tr=s.tr \land a \notin y.ref)
						\\ \dres y.wait \rres \\
					(y.tr = s.tr \cat \lseq a \rseq)
					\end{array}\right)
		\right) \sqcup_{\mathbf{RAD}} \mathbf{RA}\circ\mathbf{A} (\lnot P^f_f \vdash P^t_t)
	&&\ptext{\cref{theorem:RAP:P-sqcup-Q}}\\
	&=\mathbf{RA} \circ \mathbf{A} 
		\left(\begin{array}{l}
			true \lor \lnot P^f_f 
			\\ \vdash \\ 
			\left(true \implies \circledIn{y}{ac'} \left(\begin{array}{l}(y.tr=s.tr \land a \notin y.ref)
						\\ \dres y.wait \rres \\
					(y.tr = s.tr \cat \lseq a \rseq)
					\end{array}\right)\right)
				\land
			(\lnot P^f_f \implies P^t_f)
	\end{array}\right)
	&&\ptext{Predicate calculus}\\
	&=\mathbf{RA} \circ \mathbf{A} \left(
			true \vdash 
			\circledIn{y}{ac'} \left(\begin{array}{l}(y.tr=s.tr \land a \notin y.ref)
						\\ \dres y.wait \rres \\
					(y.tr = s.tr \cat \lseq a \rseq)
					\end{array}\right)
				\land
			(\lnot P^f_f \implies P^t_f)\right)
\end{xflalign*}
\end{proof}
\end{proofs}
\end{theorem}

\subsubsection{Relationship with CSP}

\begin{theorem}\label{theorem:ac2p(a-then-Skip-RAD):a-then-Skip-R}
\begin{statement}
$ac2p(a \circthen_\mathbf{RAD} Skip_\mathbf{RAD}) = a \circthen_\mathbf{R} Skip_\mathbf{R}$
\end{statement}
\begin{proofs}
\begin{proof}\checkt{pfr}
\begin{xflalign*}
	&ac2p(a \circthen_\mathbf{RAD} Skip_\mathbf{RAD})
	&&\ptext{Definition of $a \circthen_{\mathbf{RAD}} Skip_{\mathbf{RAD}}$}\\
	&=ac2p\circ \mathbf{RA}\circ\mathbf{A} \left(true 
				\vdash 
					\circledIn{y}{ac'} \left(
\right)\right)
		\right)
	&&\ptext{\cref{lemma:ac2p(P)-s-ac'-not-free:P} and predicate calculus}\\
	&=\mathbf{R} \left(true 
				\vdash 
					ac2p\left(\circledIn{y}{ac'} \left(%
\right)
	\right)
	&&\ptext{Definition of conditional}\\
	&=\mathbf{R} \left(true 
			\vdash 
				\left(%
\right)
	\right)
	&&\ptext{Definition of $Skip_{\mathbf{R}}$}\\
	&=a \circthen_{\mathbf{R}} Skip_{\mathbf{R}}
\end{xflalign*}
\end{proof}
\end{proofs}
\end{theorem}

\begin{theorem}\label{theorem:p2ac(a-then-Skip-R):a-then-Skip-RAD}
\begin{statement}
$p2ac(a \circthen_\mathbf{R} Skip_\mathbf{R}) = a\circthen_\mathbf{RAD} Skip_\mathbf{RAD}$
\end{statement}
\begin{proofs}
\begin{proof}
\begin{xflalign*}
	&p2ac(a \circthen_{\mathbf{R}} Skip_{\mathbf{R}})
	&&\ptext{Definition of $\circthen_{\mathbf{R}} Skip_{\mathbf{R}}$}\\
	&=p2ac\circ\mathbf{R}\left(true 
			\vdash 
				\left(
\right)
	\right)
	&&\ptext{\cref{lemma:p2ac(false)} and predicate calculus}\\
	&=\mathbf{RA} \circ \mathbf{A} \left(true 
			\vdash 
				p2ac\left(%
\right)
	\right)
	&&\ptext{Definition of conditional}\\
	&=\mathbf{RA} \circ \mathbf{A} \left(true 
			\vdash 
				p2ac\left(%
\right)
	\right)	
	&&\ptext{Introduce auxiliary variable}\\
	&=\mathbf{RA} \circ \mathbf{A} \left(true 
			\vdash 
				\left(%
\right)
	\right)
	&&\ptext{Predicate calculus and definition of $\circledIn{y}{ac'}$}\\
	&=\mathbf{RA} \circ \mathbf{A} \left(true 
			\vdash 
				\circledIn{y}{ac'} \left(%
\right)
	\right)
	&&\ptext{Definition of conditional}\\
	&=\mathbf{RA} \circ \mathbf{A} \left(true 
			\vdash 
				\circledIn{y}{ac'} \left(%
\right)
	\end{array}\right)
\end{align*}
\end{statement}
\begin{proofsbig}
\begin{proof}
\begin{xflalign*}
	&a \circthen_{\mathbf{RAD}} P
	&&\ptext{Definition of $a \circthen_{\mathbf{RAD}} P$ event prefixing}\\
	&=a \circthen_{\mathbf{RAD}} Skip_{\mathbf{RAD}} \seqDac P
	&&\ptext{Assumption: $P$ is $\mathbf{RAD}$-healthy (\cref{theorem:RA-o-A(design):RA-CSPA-PBMH})}\\
	&=a \circthen_{\mathbf{RAD}} Skip_{\mathbf{RAD}} \seqDac \mathbf{RA}\circ\mathbf{A} (\lnot P^f_f \vdash P^t_f)
	&&\ptext{Definition of $a \circthen_{\mathbf{RAD}} Skip_{\mathbf{RAD}}$}\\
	&=\left(
\right)
	&&\ptext{\cref{law:seqA-ac'-not-free} and predicate calculus}\\
	&=\mathbf{RA} \circ \mathbf{A} \left(%
\right)
	&&\ptext{Property of conditional and predicate calculus}\\
	&=\mathbf{RA} \circ \mathbf{A} \left(%
\right)
	&&\ptext{Definition of conditional and predicate calculus}\\
	&=\mathbf{RA} \circ \mathbf{A} \left(%
\right)
	&&\ptext{Property of substutiton and predicate calculus}\\
	&=\mathbf{RA} \circ \mathbf{A} \left(%
\right)
	&&\ptext{Definition of design and predicate calculus}\\
	&=\mathbf{RA} \circ \mathbf{A} \left(%
\right)
	&&\ptext{Predicate calculus and definition of conditional}\\
	&=\mathbf{RA} \circ \mathbf{A} \left(%
\right)
	\end{array}\right)
\end{xflalign*}
\end{proof}
\end{proofsbig}
\end{theorem}

\begin{lemma}\label{lemma:ac2p(a-circthen-Chaos-sqcuo-b-circthen-Chaos)}
\begin{statement}
\begin{align*}
	&ac2p(a \circthen_{\mathbf{RAD}} Chaos_{\mathbf{RAD}} \sqcup_{\mathbf{RAD}} b \circthen_{\mathbf{RAD}} Chaos_{\mathbf{RAD}})\\
	&=\\
	&\mathbf{R} (true \vdash tr'=tr \land wait' \land a \notin ref' \land b \notin ref')
\end{align*}
\end{statement}
\begin{proofs}
\begin{proof}
\begin{xflalign*}
	&ac2p\left(a \circthen_{\mathbf{RAD}} Chaos_{\mathbf{RAD}} \sqcup_{\mathbf{RAD}} b \circthen_{\mathbf{RAD}} Chaos_{\mathbf{RAD}})\right)
	&&\ptext{\cref{lemma:RAD:a-circthen-Chaos-sqcup-b-circthen-Chaos}}\\
	&=ac2p\circ\mathbf{RA}\circ\mathbf{A} \left(
\right)
	&&\ptext{Property of conditional}\\
	&=\mathbf{R} \left(%
\right)
	&&\ptext{Property of sequences}\\
	&=\mathbf{R} \left(%
\right)
	&&\ptext{Predicate calculus and property of conditional}\\
	&=\mathbf{R} (true \vdash tr'=tr \land wait' \land a \notin ref' \land b \notin ref')
\end{xflalign*}
\end{proof}
\end{proofs}
\end{lemma}

\begin{lemma}\label{lemma:ac2p(RAD:a-then-Stop-sqcup-b-then-Stop):R:a-then-Stop-sqcup-b-then-Stop}
\begin{statement}
\begin{align*}
	&ac2p(a \circthen_{\mathbf{RAD}} Stop_{\mathbf{RAD}} \sqcup_{\mathbf{RAD}} b \circthen_{\mathbf{RAD}} Stop_{\mathbf{RAD}})\\
	&=\\
	&a \circthen_{\mathbf{R}} Stop_{\mathbf{R}} \sqcup_{\mathbf{R}} b \circthen_{\mathbf{R}} Stop_{\mathbf{R}}
\end{align*}
\end{statement}
\begin{proofs}
\begin{proof}
\begin{xflalign*}
	&ac2p(a \circthen_{\mathbf{RAD}} Stop_{\mathbf{RAD}} \sqcup_{\mathbf{RAD}} b \circthen_{\mathbf{RAD}} Stop_{\mathbf{RAD}})\\
	&&\ptext{Definition of $\sqcup_{\mathbf{RAD}}$ and~\cref{theorem:ac2p(P-land-Q):ac2p(P)-land-ac2p(Q)}}\\
	&=\left(
\right)
	&&\ptext{Definition of prefixing and~\cref{lemma:exp:a-circthen-Stop}}\\
	&=p2ac\left(%
\right)
	&&\ptext{Conjunction of designs}\\
	&=p2ac\circ\mathbf{R}\left(%
\right)
	&&\ptext{Predicate calculus and property of sequences}\\
	&=p2ac\circ\mathbf{R}\left(%
\right)
	&&\ptext{\cref{lemma:p2ac(false)} and predicate calculus}\\
	&=\mathbf{RA}\circ\mathbf{A}\left(%
\right)
	&&\ptext{Introduce fresh variable $y$}\\
	&=\mathbf{RA}\circ\mathbf{A}\left(%
\right)
	&&\ptext{One-point rule and substitution}\\
	&=\mathbf{RA}\circ\mathbf{A}\left(%
\right)
	&&\ptext{Predicate calculus and~\cref{lemma:ac2p(P)-s-ac'-not-free:P}}\\
	&=\mathbf{R} \left(%
\right)
	&&\ptext{Substitution and value of record components}\\
	&=\mathbf{R} \left(%
\right)
	&&\ptext{\cref{lemma:exp:a-circthen-Stop}}\\
	&=a \circthen_{\mathbf{R}} Stop_{\mathbf{R}}
\end{xflalign*}
\end{proof}
\end{proofs}
\end{lemma}

\subsubsection{Properties and Examples}

\begin{theorem}\label{theorem:P-seqRac-ChaosRA} Provided $P$ is $\mathbf{RAD}$-healthy,
\begin{align*}
	&P \seqRac Chaos_{\mathbf{RAD}}\\
	&=\\
	&\mathbf{RA}\circ\mathbf{A} \left(%
\right)
		\\ \vdash \\
		\mathbf{RA1} (P^t_f) \seqA (s \in ac' \dres s.wait \rres \mathbf{RA2}\circ\mathbf{RA1} (true))
	\end{array}\right)
\end{align*}
\begin{proofs}\begin{proof}
\begin{xflalign*}
	&P \seqRac Chaos_{\mathbf{RAD}}
	&&\ptext{Definition of $Chaos_{\mathbf{RAD}}$}\\
	&=P \seqRac \mathbf{RA}\circ\mathbf{A} (false \vdash true)
	&&\ptext{Assumption: $P$ is $\mathbf{RAD}$-healthy and~\cref{theorem:RA-o-A(design):RA-CSPA-PBMH}}\\
	&=\mathbf{RA}\circ\mathbf{A} (\lnot P^f_f \vdash P^t_f) \seqRac \mathbf{RA}\circ\mathbf{A} (false \vdash true)
	&&\ptext{\cref{theorem:RAP:seqDac}}\\
	&=\mathbf{RA}\circ\mathbf{A} \left(
\right)
	&&\ptext{Definition of design and predicate calculus}\\
	&=\mathbf{RA} \circ \mathbf{A} \left(%
\right)
\end{align*}
\end{statement}
\begin{proofs}
\begin{proof}
\begin{xflalign*}
	&a \circthen_{\mathbf{RAD}} Chaos_{\mathbf{RAD}}
	&&\ptext{Definition of prefixing}\\
	&=a \circthen_{\mathbf{RAD}} Skip_{\mathbf{RAD}} \seqDac Chaos_{\mathbf{RAD}}
	&&\ptext{Definition of prefixing and $Chaos_{\mathbf{RAD}}$}\\
	&=\left(%
\right)
	&&\ptext{Property of sequences and conditional}\\
	&=\mathbf{RA}\circ\mathbf{A} \left(%
\right)
	&&\ptext{Predicate calculus and~\cref{theorem:RA2-o-RA1:RA1-o-RA2,lemma:RA2(P):P:ac'-not-free}}\\
	&=\mathbf{RA}\circ\mathbf{A} \left(%
\right)
	&&\ptext{\cref{lemma:RA1(false),law:seqA-ac'-not-free} and predicate calculus}\\
	&=\mathbf{RA}\circ\mathbf{A} \left(%
\right)
	&&\ptext{Definition of design and predicate calculus}\\
	&=\mathbf{RA}\circ\mathbf{A} \left(%
\right)
	&&\ptext{Introduce fresh variable $t$}\\
	&=\mathbf{RA}\circ\mathbf{A} \left(%
\right)
	&&\ptext{One-point rule, substitution and value of record component $tr$}\\
	&=\mathbf{RA}\circ\mathbf{A} \left(%
\right)
	&&\ptext{Rename variable, definition of $\circledIn{y}{ac'}$ and~\cref{lemma:circledIn:ac'-not-free}}\\
	&=\mathbf{RA}\circ\mathbf{A} \left(%
\right)
	&&\ptext{\cref{lemma:circledIn(P-lor-Q):circledIn(P)-lor-circledIn(Q):new} and definition of conditional}\\
	&=\mathbf{RA}\circ\mathbf{A} \left(%
\right)
	&&\ptext{Substituion, property of sets and one-point rule}\\
	&=\mathbf{RA}\circ\mathbf{A} \left(%
\right)
	&&\ptext{Property of conditional}\\
	&=\mathbf{RA}\circ\mathbf{A} \left(%
\right)
	&&\ptext{Property of sequences}\\
	&=\mathbf{RA}\circ\mathbf{A} \left(%
\right)
	&&\ptext{Property of conditional}\\
	&=\mathbf{RA}\circ\mathbf{A} \left(%
\right)
	&&\ptext{Variable renaming, definition of $\circledIn{y}{ac'}$ and  \cref{lemma:circledIn:ac'-not-free}}\\
	&=\mathbf{RA}\circ\mathbf{A} \left(%
\right)
	\end{array}\right)
\end{xflalign*}
\end{proof}
\end{proofs}
\end{lemma}

\begin{lemma}\label{lemma:RAD:(a-circthen-Skip)-sqcup-(b-circthen-Chaos)}
\begin{statement}
\begin{align*}
	&(a \circthen_{\mathbf{RAD}} Skip_{\mathbf{RAD}}) \sqcup_{\mathbf{RAD}} (b \circthen_{\mathbf{RAD}} Chaos_{\mathbf{RAD}})\\
	&=\\
	&(a \circthen_{\mathbf{RAD}} Skip_{\mathbf{RAD}}) \sqcup_{\mathbf{RAD}} (b \circthen_{\mathbf{RAD}} Choice_{\mathbf{RAD}})
\end{align*}
\end{statement}
\begin{proofs}
\begin{proof}
\begin{xflalign*}
	&(a \circthen_{\mathbf{RAD}} Skip_{\mathbf{RAD}}) \sqcup_{\mathbf{RAD}} (b \circthen_{\mathbf{RAD}} Chaos_{\mathbf{RAD}})
	&&\ptext{Definition of $a \circthen Skip$ and~\cref{lemma:RAD:a-circthen-Choice}}\\
	&=\left(
\right)
	&&\ptext{Predicate calculus and~\cref{lemma:circledIn(P-lor-Q):circledIn(P)-lor-circledIn(Q):new}}\\
	&=\mathbf{RA} \circ \mathbf{A} \left(%
\right)
	&&\ptext{Definition of conditional and predicate calculus}\\
	&=\mathbf{RA} \circ \mathbf{A} \left(%
\right)
	&&\ptext{Property of records}\\
	&=\mathbf{RA} \circ \mathbf{A} \left(%
\right)
	&&\ptext{One-point rule, substitution and value of record component $tr$}\\
	&=\mathbf{RA} \circ \mathbf{A} \left(%
\right)
	&&\ptext{Predicate calculus and definition of $\circledIn{y}{ac'}$}\\
	&=\mathbf{RA} \circ \mathbf{A} \left(%
\right)
	&&\ptext{\cref{lemma:RA2(P):P:s-ac'-not-free} and predicate calculus}\\
	&=\mathbf{RA} \circ \mathbf{A} \left(%
\right)
	&&\ptext{Definition of $\circledIn{y}{ac'}$, substitution and value of component $tr$}\\
	&=\mathbf{RA} \circ \mathbf{A} \left(%
\right)
	&&\ptext{Definition of $\circledIn{y}{ac'}$ and predicate calculus}\\
	&=\mathbf{RA} \circ \mathbf{A} \left(%
\right)
	&&\ptext{Definition of $\circledIn{y}{ac'}$, \cref{lemma:circledIn(P-lor-Q):circledIn(P)-lor-circledIn(Q):new} and predicate calculus}\\
	&=\mathbf{RA} \circ \mathbf{A} \left(%
\right)	
	\end{array}\right)
\end{xflalign*}
\end{proof}
\end{proofs}
\end{lemma}

\begin{lemma}\label{lemma:RAD:((a-circthen-Stop)-sqcup-Skip)-seqDac-Chaos:a-circthen-Stop}
\begin{statement}
\begin{align*}
	&((a \circthen_{\mathbf{RAD}} Stop_{\mathbf{RAD}}) \sqcup_{\mathbf{RAD}} Skip_{\mathbf{RAD}}) \seqDac Chaos_{\mathbf{RAD}}\\
	&=\\
	&a \circthen_{\mathbf{RAD}} Stop_{\mathbf{RAD}}
\end{align*}
\end{statement}
\begin{proofsbig}
\begin{proof}
\begin{xflalign*}
	&((a \circthen_{\mathbf{RAD}} Stop_{\mathbf{RAD}}) \sqcup_{\mathbf{RAD}} Skip_{\mathbf{RAD}}) \seqDac Chaos_{\mathbf{RAD}}
	&&\ptext{\cref{lemma:RAD:(a-circthen-Stop)-sqcup-Skip,theorem:P-seqRac-ChaosRA}}\\
	&=\mathbf{RA}\circ\mathbf{A} \left(
\right)
	&&\ptext{Predicate calculus and~\cref{lemma:circledIn(P-lor-Q):circledIn(P)-lor-circledIn(Q):new}}\\
	&=\mathbf{RA}\circ\mathbf{A} \left(%
\right)
	&&\ptext{Property of sequences and predicate calculus}\\
	&=\mathbf{RA}\circ\mathbf{A} \left(%
\right)
	&&\ptext{Definition of $\circledIn{y}{ac'}$, $\seqA$ and property of sets}\\
	&=\mathbf{RA}\circ\mathbf{A} \left(%
\right)
	&&\ptext{Predicate calculus and property of conditional}\\
	&=\mathbf{RA}\circ\mathbf{A} \left(%
\right)
	&&\ptext{Definition of $\circledIn{y}{ac'}$ and predicate calculus}\\
	&=\mathbf{RA}\circ\mathbf{A} \left(%
\right)
	&&\ptext{\cref{lemma:RA1(false),law:seqA-ac'-not-free} and predicate calculus}\\
	&=\mathbf{RA}\circ\mathbf{A} \left(%
\right)
	&&\ptext{\cref{lemma:RA1(circledIn)-ac'-not-free:circledIn(P-land-s.tr-le-y.tr),lemma:RA1(P):implies:RA1(true)} and predicate calculus}\\
	&=\mathbf{RA}\circ\mathbf{A} \left(%
\right)
	&&\ptext{\cref{lemma:circledIn(P-lor-Q):circledIn(P)-lor-circledIn(Q)} and predicate calculus}\\
	&=\mathbf{RA}\circ\mathbf{A} \left(%
\right)
	&&\ptext{\cref{lemma:RA2(P):P:s-ac'-not-free} and predicate calculus}\\
	&=\mathbf{RA} \circ \mathbf{A} \left(%
\right)
	&&\ptext{Substitution and value of record component $tr$}\\
	&=\mathbf{RA} \circ \mathbf{A} \left(%
\right)
	&&\ptext{Predicate calculus and definition of $\circledIn{y}{ac'}$}\\
	&=\mathbf{RA} \circ \mathbf{A} \left(%
\right)
	&&\ptext{\cref{lemma:circledIn(P-lor-Q):circledIn(P)-lor-circledIn(Q)} and predicate calculus}\\
	&=\mathbf{RA} \circ \mathbf{A} \left(%
\right)
\end{xflalign*}
\end{proof}
\end{proofs}
\end{lemma}

\subsection{External Choice}

\begin{theorem}\label{theorem:RAD:P-extchoice-Stop}
\begin{statement}
Provided $P$ is a reactive angelic design,
\begin{align*}
	&P \extchoice_{\mathbf{RAD}} Stop_{\mathbf{RAD}} =
	\mathbf{RA}\circ\mathbf{A}(
		\lnot P^f_f
		\vdash 
		\exists y @ (P^t_f)[\{y\}/ac'] \land y \in ac')
\end{align*}
\end{statement}
\begin{proofs}
\begin{proof}\checkt{alcc}
\begin{xflalign*}
	&P \extchoice_{\mathbf{RAD}} Stop_{\mathbf{RAD}}
	&&\ptext{Assumption: $P$ is $\mathbf{RAD}$-healthy and definition of $Stop_{\mathbf{RAD}}$}\\
	&=\mathbf{RA}\circ\mathbf{A} (\lnot P^f_f \vdash P^t_f) \extchoice_\mathbf{RAD} \mathbf{RA}\circ\mathbf{A} (true \vdash \circledIn{y}{ac'} (y.tr=s.tr \land y.wait))
	&&\ptext{\cref{theorem:RAD:P-extchoice-Q}}\\
	&=\mathbf{RA}\circ\mathbf{A}\left(
\right)
	&&\ptext{Predicate calculus and~\cref{lemma:circledIn(P-lor-Q):circledIn(P)-lor-circledIn(Q):new}}\\
	&=\mathbf{RA}\circ\mathbf{A}\left(%
\right)
	&&\ptext{Property of conditional}\\
	&=\mathbf{RA}\circ\mathbf{A}\left(%
\right)
\end{xflalign*}
\end{proof}
\end{proofs}
\end{theorem}

\begin{theorem}\label{theorem:RAD:P-A2-extchoice-Stop:P-A2}
\begin{statement}
Provided $P$ is a reactive angelic design and $\mathbf{A2}$-healthy,
\begin{align*}
	&P \extchoice_{\mathbf{RAD}} Stop_{\mathbf{RAD}} = P
\end{align*}
\end{statement}
\begin{proofs}
\begin{proof}
\begin{xflalign*}
	&P \extchoice_{\mathbf{RAD}} Stop_{\mathbf{RAD}}
	&&\ptext{\cref{theorem:RAD:P-extchoice-Stop}}\\
	&=\mathbf{RA}\circ\mathbf{A}(
		\lnot P^f_f
		\vdash 
		\exists y @ (P^t_f)[\{y\}/ac'] \land y \in ac')
	&&\ptext{Assumption: $P$ is $\mathbf{A2}$-healthy}\\
	&=\mathbf{RA}\circ\mathbf{A}(
		\lnot P^f_f
		\vdash 
		\exists y @ (\mathbf{A2} (P)^t_f)[\{y\}/ac'] \land y \in ac')
	&&\ptext{\cref{lemma:A2(P)-o-w:A2(P-o-w)}}\\
	&=\mathbf{RA}\circ\mathbf{A}(
		\lnot P^f_f
		\vdash 
		\exists y @ (\mathbf{A2} (P^t_f))[\{y\}/ac'] \land y \in ac')
	&&\ptext{\cref{lemma:A2:alternative-2:disjunction}}\\
	&=\mathbf{RA}\circ\mathbf{A}\left(
		\lnot P^f_f
		\vdash 
		\left(\exists y @ \left(\begin{array}{l}
			P^t_f[\emptyset/ac'] 
			\\ \lor \\
			(\exists z @ P^t_f[\{z\}/ac'] \land z \in ac')
		\end{array}\right)\right)[\{y\}/ac'] \land y \in ac'\right)
	&&\ptext{Substitution}\\
	&=\mathbf{RA}\circ\mathbf{A}\left(
		\lnot P^f_f
		\vdash 
		\left(\exists y @ \left(\begin{array}{l}
			P^t_f[\emptyset/ac'] 
			\\ \lor \\
			(\exists z @ P^t_f[\{z\}/ac'] \land z \in \{y\})
		\end{array}\right)\right)\land y \in ac'\right)
	&&\ptext{Predicate calculus}\\
	&=\mathbf{RA}\circ\mathbf{A}\left(
		\lnot P^f_f
		\vdash 
		\left(\begin{array}{l}
			(\exists y @ P^t_f[\emptyset/ac'] \land y \in ac')
			\\ \lor \\
			(\exists y @ \exists z @ P^t_f[\{z\}/ac'] \land z \in \{y\} \land y \in ac')
		\end{array}\right)\right)
	&&\ptext{Predicate calculus}\\
	&=\mathbf{RA}\circ\mathbf{A}\left(
		\lnot P^f_f
		\vdash 
		\left(\begin{array}{l}
			(P^t_f[\emptyset/ac'] \land \exists y @ y \in ac')
			\\ \lor \\
			(\exists y @ \exists z @ P^t_f[\{z\}/ac'] \land z \in \{y\} \land y \in ac')
		\end{array}\right)\right)
	&&\ptext{Property of sets and one-point rule}\\
	&=\mathbf{RA}\circ\mathbf{A}\left(
		\lnot P^f_f
		\vdash 
		\left(\begin{array}{l}
			(P^t_f[\emptyset/ac'] \land ac'\neq\emptyset)
			\\ \lor \\
			(\exists y @ P^t_f[\{y\}/ac'] \land y \in ac')
		\end{array}\right)\right)
	&&\ptext{Definition of $\mathbf{A1}$ and predicate calculus}\\
	&=\mathbf{RA}\circ\mathbf{A}(
		\lnot P^f_f
		\vdash 
		(P^t_f[\emptyset/ac'] \lor (\exists y @ P^t_f[\{y\}/ac'] \land y \in ac')))
	&&\ptext{\cref{lemma:A2:alternative-2:disjunction}}\\
	&=\mathbf{RA}\circ\mathbf{A}(
		\lnot P^f_f
		\vdash 
		\mathbf{A2} (P^t_f))
	&&\ptext{\cref{lemma:A2(P)-o-w:A2(P-o-w)}}\\
	&=\mathbf{RA}\circ\mathbf{A}(
		\lnot P^f_f
		\vdash 
		\mathbf{A2} (P)^t_f)
	&&\ptext{Assumption: $P$ is $\mathbf{A2}$-healthy}\\
	&=\mathbf{RA}\circ\mathbf{A}(
		\lnot P^f_f
		\vdash 
		P^t_f)
	&&\ptext{Assumption: $P$ is $\mathbf{RAD}$-healthy}\\
	&=P
\end{xflalign*}
\end{proof}
\end{proofs}
\end{theorem}

\begin{theorem}\label{theorem:RA-o-A(lnot-Pff|-Ptf)-extchoice-RA-o-A(lnot-Qff|-Qtf)}
\begin{statement}
\begin{align*}
	&\mathbf{RA}\circ\mathbf{A} (\lnot P^f_f \vdash P^t_f) \extchoice_{\mathbf{RAD}} \mathbf{RA}\circ\mathbf{A} (\lnot Q^f_f \vdash Q^t_f)\\
	&=\\
	&\mathbf{RA}\circ\mathbf{A}\left(
\right)
	&&\ptext{\cref{law:pbmh:P:ac'-not-free} and predicate calculus}\\
	&=\mathbf{RA}\circ\mathbf{A}\left(%
\right)
	&&\ptext{Definition of design and predicate calculus}\\
	&=\mathbf{RA}\circ\mathbf{A}\left(%
\right)
	&&\ptext{Definition of design, predicate calculus and~\cref{lemma:lnot-PBMH(P)-land-circledIn(PBMH(P)-cond-R-T)}}\\
	&=\mathbf{RA}\circ\mathbf{A}\left(%
\right)
	&&\ptext{Definition of design, predicate calculus and~\cref{lemma:lnot-PBMH(P)-land-circledIn(Q-cond-(PBMH(P)-lor-R))}}\\
	&=\mathbf{RA}\circ\mathbf{A}\left(%
\right)
	&&\ptext{Predicate calculus and~\cref{law:pbmh:distribute-disjunction}}\\
	&=\mathbf{RA}\circ\mathbf{A}\left(%
\right)
\end{xflalign*}
\end{proof}
\end{proofsbig}
\end{theorem}

\begin{theorem}\label{theorem:RAD:P-extchoice-Q}
\begin{statement}
Provided $P$ and $Q$ are reactive angelic designs,
\begin{align*}
	&P \extchoice_{\mathbf{RAD}} Q\\
	&=\\
	&\mathbf{RA}\circ\mathbf{A}\left(%
\right)
	\end{array}\right)
\end{xflalign*}
\end{proof}
\end{proofs}
\end{theorem}

\subsubsection{Relationship with CSP}

\begin{theorem}
\label{theorem:ac2p(p2ac(P)-extchoice-RAD-p2ac(Q)):P-extchoice-R-Q}
\label{theorem:ac2p(p2ac(P)-extchoiceONE-p2ac(Q)):P-extchoice-R-Q} 
\begin{statement}
Provided that $P$ and $Q$ are~\ac{CSP} processes,
\begin{align*}
	&ac2p(p2ac(P) \extchoice_{\mathbf{RAD}} p2ac(Q)) = P \extchoice_{\mathbf{R}} Q
\end{align*}
\end{statement}
\begin{proofs}
\begin{proof}\checkt{alcc}
\begin{xflalign*}
	&ac2p(p2ac(P) \extchoiceONE_{\mathbf{RAD}} p2ac(Q))
	&&\ptext{Definition of $\extchoiceONE_{\mathbf{RAD}}$}\\
	&=ac2p\circ\mathbf{RA}\circ\mathbf{A}\left(
\right)
	&&\ptext{Predicate calculus and property of conditional}\\
	&=\mathbf{R} \left(%
\right)
	&&\ptext{Assumption: $P$ and $Q$ are~\ac{CSP} processes and definition of $\extchoice_{\mathbf{R}}$}\\
	&=P \extchoice_{\mathbf{R}} Q
\end{xflalign*}
\end{proof}
\end{proofs}
\end{theorem}

\begin{theorem}\label{theorem:p2ac(ac2p(P)-extchoice-acp(Q))}
\begin{statement}Provided $P$ and $Q$ are reactive angelic designs,
\begin{align*}
	&p2ac(ac2p(P) \extchoice_{\mathbf{R}} ac2p(Q)) \sqsupseteq P \extchoice_{\mathbf{RAD}} Q
\end{align*}
\end{statement}
\begin{proofs}
\begin{proof}\checkt{pfr}\checkt{alcc}
\begin{xflalign*}
	&p2ac(ac2p(P) \extchoice_{\mathbf{R}} ac2p(Q))
	&&\ptext{Definition of $\extchoice_{\mathbf{R}}$}\\
	&=p2ac \circ \mathbf{R}\left(
\right)
	&&\ptext{Definition of conditional and predicate calculus}\\
	&=\mathbf{RA} \circ \mathbf{A} \left(%
\right)
	&&\ptext{Refinement of designs}\\
	&\sqsupseteq \mathbf{RA} \circ \mathbf{A} \left(%
\right)
	&&\ptext{\cref{theorem:p2ac(P-land-Q):implies:p2ac(P)-land-p2ac(Q)} and weaken postcondition}\\
	&\sqsupseteq \mathbf{RA} \circ \mathbf{A} \left(%
\right)
	&&\ptext{\cref{theorem:p2ac-o-ac2p:implies:PBMH(P)} and weaken postcondition}\\
	&\sqsupseteq \mathbf{RA} \circ \mathbf{A} \left(%
\right)
	&&\ptext{Property of dashed state variable}\\
	&=\mathbf{RA} \circ \mathbf{A} \left(%
\right)
	&&\ptext{Predicate calculus and distributivity of $\mathbf{PBMH}$}\\
	&=\mathbf{RA} \circ \mathbf{A} \left(%
\right)
	&&\ptext{Definition of conditional}\\
	&=\mathbf{RA} \circ \mathbf{A} \left(%
\right)
	&&\ptext{Definition of $\extchoice_{\mathbf{RAD}}$}\\
	&=P \extchoice_{\mathbf{RAD}} Q
\end{xflalign*}
\end{proof}
\end{proofs}
\end{theorem}\noindent

\begin{theorem}\label{theorem:p2ac(ac2p(P)-exthoice-R-ac2p(Q)):P-extchoiceONE-Q}
\begin{statement}
Provided $P$ and $Q$ are~$\mathbf{RAD}$-healthy and~$\mathbf{A2}$-healthy,
\begin{align*}
	&p2ac(ac2p(P) \extchoice_{\mathbf{R}} ac2p(Q)) = P \extchoiceONE_{\mathbf{RAD}} Q
\end{align*}
\end{statement}
\begin{proofsbig}
\begin{proof}\checkt{alcc}
\begin{xflalign*}
	&p2ac(ac2p(P) \extchoice_{\mathbf{R}} ac2p(Q))
	&&\ptext{Definition of $\extchoice_{\mathbf{R}}$}\\
	&=p2ac\circ \mathbf{R} \left(
\right)
	&&\ptext{\cref{theorem:p2ac-o-R(design):RA-o-A(lnot-p2ac(pre)|-p2ac(post))} and predicate calculus}\\
	&=\mathbf{RA}\circ\mathbf{A} \left(%
\right)
	&&\ptext{Definition of conditional and predicate calculus}\\
	&=\mathbf{RA}\circ\mathbf{A} \left(%
\right)
	&&\ptext{Introduce fresh variable $y$}\\
	&=\mathbf{RA}\circ\mathbf{A} \left(%
\right)
	&&\ptext{Property of substitution}\\
	&=\mathbf{RA}\circ\mathbf{A} \left(%
\right)
	&&\ptext{Assumption: $P$ and $Q$ are $\mathbf{RAD}$-healthy and so $\mathbf{PBMH}$-healthy (\cref{theorem:PBMH(P)-RAP:P})}\\
	&&\ptext{\cref{law:pbmh:P:ac'-not-free,law:pbmh:disjunction-closure,law:pbmh:conjunction-closure,lemma:circledIn(P)-PBMH:exists-y-P}}\\
	&=\mathbf{RA}\circ\mathbf{A} \left(%
\right)
	&&\ptext{Assumption: $P$ and $Q$ are $\mathbf{RAD}$-healthy and so $\mathbf{PBMH}$-healthy (\cref{theorem:PBMH(P)-RAP:P})}\\
	&&\ptext{\cref{lemma:PBMH(P)-ow:PBMH(P-ow),law:pbmh:ac'-neq-emptyset,law:pbmh:disjunction-closure,law:pbmh:conjunction-closure}}\\
	&=\mathbf{RA} \left(%
\right)
	&&\ptext{Assumption: $P$ and $Q$ are $\mathbf{RAD}$-healthy and so $\mathbf{PBMH}$-healthy (\cref{theorem:PBMH(P)-RAP:P})}\\
	&&\ptext{\cref{lemma:PBMH(P)-ow:PBMH(P-ow),law:pbmh:disjunction-closure}}\\
	&=\mathbf{RA} \left(%
\right)
	&&\ptext{Definition of $\extchoiceONE_{\mathbf{RAD}}$}\\
	&=P \extchoiceONE_{\mathbf{RAD}} Q
\end{xflalign*}
\end{proof}
\end{proofsbig}
\end{theorem}

\subsubsection{Closure}

\begin{theorem}\label{theorem:A2(P-extchoiceOne-Q):P-extchoiceOne-Q}
\begin{statement}
Provided $P$ and $Q$ are reactive angelic designs and~$\mathbf{A2}$-healthy,
\begin{align*}
	&\mathbf{A2} (P \extchoiceONE_{\mathbf{RAD}} Q) = P \extchoiceONE_{\mathbf{RAD}} Q
\end{align*}
\end{statement}
\begin{proofs}
\begin{proof}\checkt{alcc}
\begin{xflalign*}
	&P \extchoiceONE_{\mathbf{RAD}} Q
	&&\ptext{Definition of $\extchoiceONE_{\mathbf{RAD}}$}\\
	&=\mathbf{RA}\circ\mathbf{A}\left(
\right)
	&&\ptext{Predicate calculus and~\cref{theorem:A2(P-lor-Q):A2(P)-lor-A2(Q)}}\\
	&=\mathbf{RA}\circ\mathbf{A}\left(%
\right)
	&&\ptext{Predicate calculus and~\cref{theorem:A2(P-lor-Q):A2(P)-lor-A2(Q)}}\\
	&=\mathbf{A2}\circ\mathbf{RA}\circ\mathbf{A}\left(%
\right)
	&&\ptext{Definition of $\extchoiceONE_{\mathbf{RAD}}$}\\
	&=\mathbf{A2} (P \extchoiceONE_{\mathbf{RAD}} Q)
\end{xflalign*}
\end{proof}
\end{proofs}
\end{theorem}

\subsubsection{Properties and Examples}

\begin{lemma}\label{lemma:RAD:(Skip-sqcup-Stop)-extchoice-Stop:Top}
\begin{statement}
	$(Skip_{\mathbf{RAD}} \sqcup_{\mathbf{RAD}} Stop_{\mathbf{RAD}}) \extchoice_{\mathbf{RAD}} Stop_{\mathbf{RAD}} = \top_{\mathbf{RAD}}$
\end{statement}
\begin{proofs}
\begin{proof}\checkt{alcc}
\begin{xflalign*}
	&(Skip_{\mathbf{RAD}} \sqcup_{\mathbf{RAD}} Stop_{\mathbf{RAD}}) \extchoice_{\mathbf{RAD}} Stop_{\mathbf{RAD}}
	&&\ptext{Result of~\cref{example:Stop-RAD-sqcup-Skip-RAD}}\\
	&=\left(\begin{array}{l}
		\mathbf{RA} \circ \mathbf{A} (true	\vdash
			\circledIn{y}{ac'} (y.tr=s.tr \land y.wait)	
			\land 
			\circledIn{y}{ac'} (\lnot y.wait \land y.tr=s.tr))
		\\ \extchoice_{\mathbf{RAD}} \\
		Stop_{\mathbf{RAD}}
	\end{array}\right)
	&&\ptext{\cref{theorem:RAD:P-extchoice-Stop}}\\
	&=\mathbf{RA} \circ \mathbf{A} \left(\begin{array}{l}
		true
		\\ \vdash \\
		\exists z @ \left(\begin{array}{l}
			\circledIn{y}{ac'} (y.tr=s.tr \land y.wait)	
			\\ \land \\
			\circledIn{y}{ac'} (\lnot y.wait \land y.tr=s.tr))
		\end{array}\right)[\{z\}/ac'] \land z \in ac'
	\end{array}\right)
	&&\ptext{Substitution and property of sets}\\
	&=\mathbf{RA} \circ \mathbf{A} \left(\begin{array}{l}
		true
		\\ \vdash \\
		\exists z @ z.tr=s.tr \land z.wait \land \lnot z.wait \land z.tr=s.tr \land z \in ac'
	\end{array}\right)
	&&\ptext{Predicate calculus}\\
	&=\mathbf{RA} \circ \mathbf{A} (true \vdash false)
	&&\ptext{Definition of $\top_{\mathbf{RAD}}$}\\
	&=\top_{\mathbf{RAD}}
\end{xflalign*}
\end{proof}
\end{proofs}
\end{lemma}

\begin{lemma}
\begin{statement}
	$(Skip_{\mathbf{RAD}} \sqcup_{\mathbf{RAD}} Stop_{\mathbf{RAD}}) \extchoice_{\mathbf{RAD}} Skip_{\mathbf{RAD}} = Skip_{\mathbf{RAD}}$
\end{statement}
\begin{proofs}
\begin{proof}\checkt{alcc}
\begin{xflalign*}
	&(Skip_{\mathbf{RAD}} \sqcup_{\mathbf{RAD}} Stop_{\mathbf{RAD}}) \extchoice_{\mathbf{RAD}} Skip_{\mathbf{RAD}}
	&&\ptext{Result of~\cref{example:Stop-RAD-sqcup-Skip-RAD} and definition of $Skip_{\mathbf{RAD}}$}\\
	&=\left(
\right)
	&&\ptext{Predicate calculus and absorption law}\\
	&=\left(%
\right)
	&&\ptext{Predicate calculus and definition of $\circledIn{y}{ac'}$}\\
	&=\left(%
\right)
	&&\ptext{Definition of $Skip_{\mathbf{RAD}}$}\\
	&=Skip_{\mathbf{RAD}}
\end{xflalign*}
\end{proof}
\end{proofs}
\end{lemma}

\chapter{Angelic Processes}\label{appendix:AP}

\section{Healthiness Conditions}

\subsection{$\IIRnew$}

\begin{lemma}\label{lemma:RA2(IIRnew):IIRnew}
\begin{statement}
$\mathbf{RA2} (\IIRnew) = \IIRnew$
\end{statement}
\begin{proofs}
\begin{proof}\checkt{alcc}\checkt{pfr}
\begin{xflalign*}
	&\mathbf{RA2} (\IIRnew)
	&&\ptext{Definition of $\IIRnew$}\\
	&=\mathbf{R2} \circ \mathbf{H1} (ok' \land s \in ac')
	&&\ptext{Definition of $\mathbf{H1}$}\\
	&=\mathbf{R2} (\lnot ok \lor (ok' \land s \in ac'))
	&&\ptext{\cref{theorem:RA2(P-lor-Q):RA2(P)-lor-RA2(Q)}}\\
	&=\mathbf{R2} (\lnot ok) \lor \mathbf{RA2} (ok' \land s \in ac')
	&&\ptext{\cref{theorem:RA2(P-land-Q):RA2(P)-land-RA2(Q)}}\\
	&=\mathbf{R2} (\lnot ok) \lor (\mathbf{RA2} (ok') \land \mathbf{RA2} (s \in ac'))
	&&\ptext{\cref{lemma:RA2(P):P:s-ac'-not-free}}\\
	&=\lnot ok \lor (ok' \land \mathbf{RA2} (s \in ac'))
	&&\ptext{\cref{lemma:RA2(s-in-ac'):s-in-ac'}}\\
	&=\lnot ok \lor (ok' \land s \in ac')
	&&\ptext{Definition of $\mathbf{H1}$ and $\IIRnew$}\\
	&=\IIRnew
\end{xflalign*}
\end{proof}
\end{proofs}
\end{lemma}

\begin{lemma}\label{lemma:RA1(IIRnew):IIRac}
$\mathbf{RA1} (\IIRnew) = \IIRac$
\begin{proofs}\begin{proof}\checkt{alcc}
\begin{xflalign*}
	&\mathbf{RA1} (\IIRnew)
	&&\ptext{Definition of $\IIRnew$}\\
	&=\mathbf{RA1} \circ \mathbf{H1} (ok' \land s \in ac')
	&&\ptext{Definition of $\mathbf{H1}$ and predicate calculus}\\
	&=\mathbf{RA1} (\lnot ok \lor (ok' \land s \in ac'))
	&&\ptext{\cref{lemma:RA1(P-lor-Q):RA1(P)-lor-RA1(Q)}}\\
	&=\mathbf{RA1} (\lnot ok) \lor \mathbf{RA1} (ok' \land s \in ac')
	&&\ptext{\cref{lemma:RA1(P-land-Q):ac'-not-free}}\\
	&=\mathbf{RA1} (\lnot ok) \lor (ok' \land \mathbf{RA1} (s \in ac'))
	&&\ptext{\cref{lemma:RA1(s-in-ac'):s-in-ac'}}\\
	&=\mathbf{RA1} (\lnot ok) \lor (ok' \land s \in ac')
	&&\ptext{Definition of $\IIRac$}\\
	&=\IIRac
\end{xflalign*}
\end{proof}\end{proofs}
\end{lemma}

\subsection{$\mathbf{RA3_{AP}}$}

\begin{theorem}\label{theorem:RA3N:idempotent}
\begin{statement}
$\mathbf{RA3_{AP}} \circ \mathbf{RA3_{AP}} (P) = \mathbf{RA3_{AP}} (P)$
\end{statement}
\begin{proofs}
\begin{proof}\checkt{alcc}\checkt{pfr}
\begin{xflalign*}
	&\mathbf{RA3_{AP}} \circ \mathbf{RA3_{AP}} (P)
	&&\ptext{Definition of $\mathbf{RA3_{AP}}$}\\
	&=\mathbf{RA3_{AP}} (\mathbf{H1} (ok' \land s\in ac') \dres s.wait \rres P)
	&&\ptext{Definition of $\mathbf{RA3_{AP}}$}\\
	&=\mathbf{H1} (ok' \land s\in ac') \dres s.wait \rres (\mathbf{H1} (ok' \land s\in ac') \dres s.wait \rres P)
	&&\ptext{Property of conditional}\\
	&=\mathbf{H1} (ok' \land s\in ac') \dres s.wait \rres P
	&&\ptext{Definition of $\mathbf{RA3_{AP}}$}\\
	&=\mathbf{RA3_{AP}} (P)
\end{xflalign*}
\end{proof}
\end{proofs}
\end{theorem}

\begin{theorem}\label{theorem:RA3N:monotonic}
\begin{statement}
$P \sqsubseteq Q \implies \mathbf{RA3_{AP}} (P) \sqsubseteq \mathbf{RA3_{AP}} (Q)$
\end{statement}
\begin{proofs}
\begin{proof}
\begin{xflalign*}
	&\mathbf{RA3_{AP}} (Q)
	&&\ptext{Assumption: $P \sqsubseteq Q = [Q \implies P]$}\\
	&=\mathbf{RA3_{AP}} (Q \land P)
	&&\ptext{\cref{theorem:RA3AP(P-land-Q):RA3AP(P)-land-RA3AP(Q)}}\\
	&=\mathbf{RA3_{AP}} (Q) \land \mathbf{RA3_{AP}} (P)
	&&\ptext{Predicate calculus}\\
	&\sqsupseteq \mathbf{RA3_{AP}} (P)
\end{xflalign*}
\end{proof}
\end{proofs}
\end{theorem}

\begin{theorem}\label{theorem:RA3AP(P-land-Q):RA3AP(P)-land-RA3AP(Q)}
\begin{statement}
$\mathbf{RA3_{AP}} (P \land Q) = \mathbf{RA3_{AP}} (P) \land \mathbf{RA3_{AP}} (Q)$
\end{statement}
\begin{proofs}
\begin{proof}
\begin{xflalign*}
	&\mathbf{RA3_{AP}} (P \land Q)
	&&\ptext{Definition of $\mathbf{RA3_{AP}}$}\\
	&=\IIRnew \dres s.wait \rres (P \land Q)
	&&\ptext{Predicate calculus}\\
	&=(\IIRnew \land \IIRnew) \dres s.wait \rres (P \land Q)
	&&\ptext{Property of conditional}\\
	&=(\IIRnew \dres s.wait \rres P) \land  (\IIRnew \dres s.wait \rres Q)
	&&\ptext{Definition of $\mathbf{RA3_{AP}}$}\\
	&=\mathbf{RA3_{AP}} (P) \land \mathbf{RA3_{AP}} (Q)
\end{xflalign*}
\end{proof}
\end{proofs}
\end{theorem}

\begin{theorem}\label{theorem:RA3AP(P-lor-Q):RA3AP(P)-lor-RA3AP(Q)}
\begin{statement}
$\mathbf{RA3_{AP}} (P \lor Q) = \mathbf{RA3_{AP}} (P) \lor \mathbf{RA3_{AP}} (Q)$
\end{statement}
\begin{proofs}
\begin{proof}
\begin{xflalign*}
	&\mathbf{RA3_{AP}} (P \lor Q)
	&&\ptext{Definition of $\mathbf{RA3_{AP}}$}\\
	&=\IIRnew \dres s.wait \rres (P \lor Q)
	&&\ptext{Predicate calculus}\\
	&=(\IIRnew \lor \IIRnew) \dres s.wait \rres (P \lor Q)
	&&\ptext{Property of conditional}\\
	&=(\IIRnew \dres s.wait \rres P) \lor  (\IIRnew \dres s.wait \rres Q)
	&&\ptext{Definition of $\mathbf{RA3_{AP}}$}\\
	&=\mathbf{RA3_{AP}} (P) \lor \mathbf{RA3_{AP}} (Q)
\end{xflalign*}
\end{proof}
\end{proofs}
\end{theorem}

\begin{theorem}\label{theorem:RA3AP(P-seqA-Q):P-seqA-Q}
\begin{statement}
Provided $P$ and $Q$ are $\mathbf{RA3_{AP}}$-healthy,
\begin{align*}
	&\mathbf{RA3_{AP}} (P \seqA Q) = P \seqA Q
\end{align*}
\end{statement}
\begin{proofs}
\begin{proof}
\begin{xflalign*}
	&P \seqA Q
	&&\ptext{Assumption: $P$ is $\mathbf{RA3_{AP}}$-healthy}\\
	&=\mathbf{RA3_{AP}} (P) \seqA Q
	&&\ptext{Definition of $\mathbf{RA3_{AP}}$}\\
	&=(\IIRnew \dres s.wait \rres P) \seqA Q
	&&\ptext{\cref{law:seqA-right-distributivity,law:seqA-right-distributivity-conjunction,law:seqA-ac'-not-free}}\\
	&=(\IIRnew \seqA Q) \dres s.wait \rres (P \seqA Q)
	&&\ptext{Definition of $\IIRnew$}\\
	&=(\mathbf{H1} (ok' \land s \in ac') \seqA Q) \dres s.wait \rres (P \seqA Q)
	&&\ptext{Definition of $\mathbf{H1}$}\\
	&=((ok \implies (ok' \land s \in ac')) \seqA Q) \dres s.wait \rres (P \seqA Q)
	&&\ptext{Predicate calculus and~\cref{law:seqA-right-distributivity,law:seqA-ac'-not-free}}\\
	&=(ok \implies ((ok' \land s \in ac') \seqA Q)) \dres s.wait \rres (P \seqA Q)
	&&\ptext{\cref{law:seqA-right-distributivity-conjunction,law:seqA-ac'-not-free}}\\
	&=(ok \implies (ok' \land (s \in ac' \seqA Q))) \dres s.wait \rres (P \seqA Q)
	&&\ptext{\cref{law:seqA:IIA:left-unit}}\\
	&=(ok \implies (ok' \land Q)) \dres s.wait \rres (P \seqA Q)
	&&\ptext{Assumption: $Q$ is $\mathbf{RA3_AP}$-healthy}\\
	&=(ok \implies (ok' \land \mathbf{RA3_{AP}} (Q))) \dres s.wait \rres (P \seqA Q)
	&&\ptext{Definition of $\mathbf{RA3_{AP}}$}\\
	&=(ok \implies (ok' \land (\IIRnew \dres s.wait \rres Q))) \dres s.wait \rres (P \seqA Q)
	&&\ptext{Property of conditional}\\
	&=(ok \implies (ok' \land \IIRnew)) \dres s.wait \rres (P \seqA Q)
	&&\ptext{Definition of $\IIRnew$}\\
	&=(ok \implies (ok' \land \mathbf{H1} (ok' \land s \in ac'))) \dres s.wait \rres (P \seqA Q)
	&&\ptext{Definition of $\mathbf{H1}$}\\
	&=(ok \implies (ok' \land ok \implies (ok' \land s \in ac'))) \dres s.wait \rres (P \seqA Q)
	&&\ptext{Predicate calculus}\\
	&=(ok \implies (ok' \land s \in ac')) \dres s.wait \rres (P \seqA Q)
	&&\ptext{Definition of $\mathbf{H1}$ and $\IIRnew$}\\
	&=\IIRnew \dres s.wait \rres (P \seqA Q)
	&&\ptext{Definition of $\mathbf{RA3_{AP}}$}\\
	&=\mathbf{RA3_{AP}} (P \seqA Q)
\end{xflalign*}
\end{proof}
\end{proofs}
\end{theorem}

\begin{theorem}\label{theorem:RA3N-o-PBMH(P):PBMH-o-RA3N(P)}
\begin{statement}
$\mathbf{RA3_{AP}} \circ \mathbf{PBMH} (P) = \mathbf{PBMH} \circ \mathbf{RA3_{AP}} (P)$
\end{statement}
\begin{proofs}
\begin{proof}
\begin{xflalign*}
	&\mathbf{RA3_{AP}} \circ \mathbf{PBMH} (P)
	&&\ptext{Definition of $\mathbf{RA3_{AP}}$}\\
	&=\mathbf{H1} (ok' \land s \in ac') \dres s.wait \rres \mathbf{PBMH} (P)
	&&\ptext{\cref{law:pbmh:s-in-ac'}}\\
	&=\mathbf{H1} (ok' \land \mathbf{PBMH} (s \in ac')) \dres s.wait \rres \mathbf{PBMH} (P)
	&&\ptext{\cref{lemma:PBMH(c-land-P):c-land-PBMH(P)}}\\
	&=\mathbf{H1} \circ \mathbf{PBMH} (ok' \land s \in ac') \dres s.wait \rres \mathbf{PBMH} (P)
	&&\ptext{\cref{theorem:H1-o-PBMH:PBMH-o-H1}}\\
	&=\mathbf{PBMH} \circ \mathbf{H1} (ok' \land s \in ac') \dres s.wait \rres \mathbf{PBMH} (P)
	&&\ptext{\cref{lemma:PBMH(conditional)}}\\
	&=\mathbf{PBMH} (\mathbf{H1} (ok' \land s \in ac') \dres s.wait \rres P)
	&&\ptext{Definition of $\mathbf{RA3_{AP}}$}\\
	&=\mathbf{PBMH} \circ \mathbf{RA3_{AP}} (P)
\end{xflalign*}
\end{proof}
\end{proofs}
\end{theorem}

\begin{theorem}\label{theorem:RA2-o-RA3N(P):RA3N-o-RA2(P)}
\begin{statement}
$\mathbf{RA2} \circ \mathbf{RA3_{AP}} (P) = \mathbf{RA3_{AP}} \circ \mathbf{RA2} (P)$
\end{statement} 
\begin{proofs}
\begin{proof}\checkt{alcc}
\begin{xflalign*}
	&\mathbf{RA2} \circ \mathbf{RA3_{AP}} (P)
	&&\ptext{Definition of $\mathbf{RA3_{AP}}$}\\
	&=\mathbf{RA2} (\IIRnew \dres s.wait \rres P)
	&&\ptext{\cref{lemma:RA2:conditional-no-s.tr} and $s.wait$ is $\mathbf{RA2}$-healthy}\\
	&=\mathbf{RA2} (\IIRnew) \dres s.wait \rres \mathbf{RA2} (P)
	&&\ptext{\cref{lemma:RA2(IIRnew):IIRnew}}\\
	&=\IIRnew \dres s.wait \rres \mathbf{RA2} (P)
	&&\ptext{Definition of $\mathbf{RA3_{AP}}$}\\
	&=\mathbf{RA3_{AP}} \circ \mathbf{RA2} (P)
\end{xflalign*}
\end{proof}
\end{proofs}
\end{theorem}

\begin{lemma}\label{lemma:PBMH-o-RA3N-o-PBMH(P):RA3N-o-PBMH(P)}
\begin{statement}
$\mathbf{PBMH} \circ \mathbf{RA3_{AP}} \circ \mathbf{PBMH} (P) = \mathbf{RA3_{AP}} \circ \mathbf{PBMH} (P)$
\end{statement}
\begin{proofs}
\begin{proof}
\begin{xflalign*}
	&\mathbf{PBMH} \circ \mathbf{RA3_{AP}} \circ \mathbf{PBMH} (P)
	&&\ptext{\cref{theorem:RA3N-o-PBMH(P):PBMH-o-RA3N(P)}}\\
	&=\mathbf{PBMH} \circ \mathbf{PBMH} \circ \mathbf{RA3_{AP}} (P)
	&&\ptext{\cref{law:pbmh:idempotent}}\\
	&=\mathbf{PBMH} \circ \mathbf{RA3_{AP}} (P)
	&&\ptext{\cref{theorem:RA3N-o-PBMH(P):PBMH-o-RA3N(P)}}\\
	&=\mathbf{RA3_{AP}} \circ \mathbf{PBMH} (P)
\end{xflalign*}
\end{proof}
\end{proofs}
\end{lemma}

\begin{theorem}\label{theorem:RA1-o-RA3N(P):RA3-o-RA1(P)}
\begin{statement}
$\mathbf{RA1} \circ \mathbf{RA3_{AP}} (P) = \mathbf{RA3} \circ \mathbf{RA1} (P)$
\end{statement}
\begin{proofs}
\begin{proof}\checkt{alcc}
\begin{xflalign*}
	&\mathbf{RA1} \circ \mathbf{RA3_{AP}} (P)
	&&\ptext{Definition of $\mathbf{RA3_{AP}}$}\\
	&=\mathbf{RA1} (\IIRnew \dres s.wait \rres P)
	&&\ptext{\cref{lemma:RA1(conditional)}}\\
	&=\mathbf{RA1} (\IIRnew) \dres s.wait \rres \mathbf{RA1} (P)
	&&\ptext{\cref{lemma:RA1(IIRnew):IIRac}}\\
	&=\IIRac \dres s.wait \rres \mathbf{RA1} (P)
	&&\ptext{Definition of $\mathbf{RA3}$}\\
	&=\mathbf{RA3} \circ \mathbf{RA1} (P)
\end{xflalign*}
\end{proof}
\end{proofs}
\end{theorem}

\subsubsection{Properties}

\begin{lemma}\label{theorem:RA3N-o-H1(P):H1}
\begin{align*}
	&\mathbf{RA3_{AP}} \circ \mathbf{H1} (P) = \mathbf{H1} ((ok' \land s \in ac') \dres s.wait \rres P)
\end{align*}
\begin{proofs}\begin{proof}\checkt{alcc}
\begin{xflalign*}
	&\mathbf{RA3_{AP}} \circ \mathbf{H1} (P)
	&&\ptext{Definition of $\mathbf{RA3_{AP}}$}\\
	&=\IIRnew \dres s.wait \rres \mathbf{H1} (P)
	&&\ptext{Definition of $\IIRnew$}\\
	&=\mathbf{H1} (ok' \land s \in ac') \dres s.wait \rres \mathbf{H1} (P)
	&&\ptext{\cref{lemma:H1(P-cond-Q):H1(P)-cond-H1(Q)}}\\
	&=\mathbf{H1} ((ok' \land s \in ac') \dres s.wait \rres P)
\end{xflalign*}
\end{proof}\end{proofs}
\end{lemma}

\begin{lemma}\label{theorem:RA3N(design)}
\begin{align*}
	&\mathbf{RA3_{AP}} (P \vdash Q) = (true \dres s.wait \rres P \vdash s \in ac' \dres s.wait \rres Q)
\end{align*}
\begin{proofs}\begin{proof}\checkt{alcc}
\begin{xflalign*}
	&\mathbf{RA3_{AP}} (P \vdash Q)
	&&\ptext{Definition of design}\\
	&=\mathbf{RA3_{AP}} ((ok \land P) \implies (Q \land ok'))
	&&\ptext{Predicate calculus}\\
	&=\mathbf{RA3_{AP}} (\lnot ok \lor \lnot P \lor (Q \land ok'))
	&&\ptext{Predicate calculus}\\
	&=\mathbf{RA3_{AP}} (ok \implies (P \implies (Q \land ok')))
	&&\ptext{Definition of $\mathbf{H1}$}\\
	&=\mathbf{RA3_{AP}} \circ \mathbf{H1} (P \implies (Q \land ok'))
	&&\ptext{\cref{theorem:RA3N-o-H1(P):H1}}\\
	&=\mathbf{H1} ((ok' \land s \in ac') \dres s.wait \rres (P \implies (Q \land ok')))
	&&\ptext{Definition of conditional}\\
	&=\mathbf{H1} ((s.wait \land ok' \land s \in ac') \lor (\lnot s.wait \land (P \implies (Q \land ok'))))
	&&\ptext{Predicate calculus}\\
	&=\mathbf{H1} ((s.wait \land ok' \land s \in ac') \lor (\lnot s.wait \land \lnot P) \lor (\lnot s.wait \land Q \land ok'))
	&&\ptext{Predicate calculus}\\
	&=\mathbf{H1} ((ok' \land ((s.wait \land s \in ac') \lor (\lnot s.wait \land Q))) \lor (\lnot s.wait \land \lnot P))
	&&\ptext{Property of conditional}\\
	&=\mathbf{H1} ((ok' \land (s \in ac' \dres s.wait \rres Q)) \lor (false \dres s.wait \rres \lnot P))
	&&\ptext{Predicate calculus and definition of $\mathbf{H1}$}\\
	&=(ok \land \lnot (false \dres s.wait \rres \lnot P)) \implies (ok' \land (s \in ac' \dres s.wait \rres Q))
	&&\ptext{Definition of design}\\
	&=(\lnot (false \dres s.wait \rres \lnot P) \vdash (s \in ac' \dres s.wait \rres Q))
	&&\ptext{\cref{lemma:lnot(conditional)}}\\
	&=(true \dres s.wait \rres P \vdash s \in ac' \dres s.wait \rres Q)
\end{xflalign*}
\end{proof}\end{proofs}
\end{lemma}

\subsection{$\mathbf{AP}$}

\subsubsection{Main Results}

\begin{theorem}\label{theorem:RAPN(P):RAPN(Pff-vdash-Ptf)}
\begin{statement}
$\mathbf{AP} (P) = \mathbf{RA3_{AP}}\circ\mathbf{RA2}\circ\mathbf{A} (\lnot P^f_f \vdash P^t_f)$
\end{statement}
\begin{proofs}
\begin{proof}\checkt{alcc}
\begin{xflalign*}
	&\mathbf{AP} (P)
	&&\ptext{Definition of $\mathbf{AP}$}\\
	&=\mathbf{RA3_{AP}} \circ \mathbf{RA2} \circ \mathbf{A} \circ \mathbf{H1} \circ \mathbf{CSPA2} (P)
	&&\ptext{Definition of $\mathbf{CSPA2}$}\\
	&=\mathbf{RA3_{AP}} \circ \mathbf{RA2} \circ \mathbf{A} \circ \mathbf{H1} \circ \mathbf{H2} (P)
	&&\ptext{Property of designs}\\
	&=\mathbf{RA3_{AP}} \circ \mathbf{RA2} \circ \mathbf{A} (\lnot P^f \vdash P^t)
	&&\ptext{\cref{theorem:RA2-o-RA3N(P):RA3N-o-RA2(P)}}\\
	&=\mathbf{RA2} \circ \mathbf{RA3_{AP}} \circ \mathbf{A} (\lnot P^f \vdash P^t)
	&&\ptext{\cref{lemma:RA3:s-oplus-wait-false}}\\
	&=\mathbf{RA2} \circ \mathbf{RA3_{AP}} \circ \mathbf{A} (\lnot P^f \vdash P^t)_f
	&&\ptext{\cref{lemma:A-substitution-s}}\\
	&=\mathbf{RA2} \circ \mathbf{RA3_{AP}} \circ \mathbf{A} ((\lnot P^f \vdash P^t)_f)
	&&\ptext{Substitution}\\
	&=\mathbf{RA2} \circ \mathbf{RA3_{AP}} \circ \mathbf{A} (\lnot P^f_f \vdash P^t_f)
	&&\ptext{\cref{theorem:RA2-o-RA3N(P):RA3N-o-RA2(P)}}\\
	&=\mathbf{RA3_{AP}} \circ \mathbf{RA2} \circ \mathbf{A} (\lnot P^f_f \vdash P^t_f)
\end{xflalign*}
\end{proof}
\end{proofs}
\end{theorem}

\begin{theorem}\label{theorem:RAPN(P)}
\begin{statement}
\begin{align*}
	&\mathbf{AP} (P) = \left(\begin{array}{l}
		true \dres s.wait \rres \lnot \mathbf{RA2}\circ\mathbf{PBMH} (P^f_f) 
		\\ \vdash \\
		s \in ac' \dres s.wait \rres \mathbf{RA2} \circ \mathbf{RA1} \circ \mathbf{PBMH} (P^t_f)
	\end{array}\right)
\end{align*}
\end{statement}
\begin{proofs}
\begin{proof}\checkt{alcc}
\begin{xflalign*}
	&\mathbf{AP} (P)
	&&\ptext{\cref{theorem:RAPN(P):RAPN(Pff-vdash-Ptf)}}\\
	&=\mathbf{RA3_{AP}}\circ\mathbf{RA2}\circ\mathbf{A} (\lnot P^f_f \vdash P^t_f)
	&&\ptext{Definition of $\mathbf{A}$}\\
	&=\mathbf{RA3_{AP}}\circ\mathbf{RA2} (\lnot \mathbf{PBMH} (P^f_f) \vdash \mathbf{PBMH} (P^t_f) \land ac'\neq\emptyset)
	&&\ptext{\cref{lemma:RA2(P|-Q):(lnot-RA2(lnot-P)|-RA2(Q))}}\\
	&=\mathbf{RA3_{AP}} (\lnot \mathbf{RA2}\circ\mathbf{PBMH} (P^f_f) \vdash \mathbf{RA2} (\mathbf{PBMH} (P^t_f) \land ac'\neq\emptyset))
	&&\ptext{\cref{lemma:RA2(P-land-ac'-neq-emptyset):RA2-o-RA1(P)}}\\
	&=\mathbf{RA3_{AP}} (\lnot \mathbf{RA2}\circ\mathbf{PBMH} (P^f_f) \vdash \mathbf{RA2} \circ \mathbf{RA1} \circ \mathbf{PBMH} (P^t_f))
	&&\ptext{\cref{theorem:RA3N(design)}}\\
	&=\left(\begin{array}{l}
		true \dres s.wait \rres \lnot \mathbf{RA2}\circ\mathbf{PBMH} (P^f_f) 
		\\ \vdash \\
		s \in ac' \dres s.wait \rres \mathbf{RA2} \circ \mathbf{RA1} \circ \mathbf{PBMH} (P^t_f)
	\end{array}\right)
\end{xflalign*}
\end{proof}
\end{proofs}
\end{theorem}

\begin{theorem}\label{theorem:RAPN:idempotent}
\begin{statement}
$\mathbf{AP} \circ \mathbf{AP} (P) = P$
\end{statement}
\begin{proofs}
\begin{proof}
\begin{xflalign*}
	&\mathbf{AP}\circ\mathbf{AP} (P)
	&&\ptext{Definition of $\mathbf{AP}$ (\cref{theorem:RAPN(P):RAPN(Pff-vdash-Ptf)})}\\
	&=\mathbf{RA3_{AP}}\circ\mathbf{RA2}\circ\mathbf{A} (\lnot \mathbf{AP} (P)^f_f \vdash \mathbf{AP} (P)^t_f)
	&&\ptext{\cref{lemma:RAPN(P)-subs-ff,lemma:RAPN(P)-subs-tf}}\\
	&=\mathbf{RA3_{AP}}\circ\mathbf{RA2}\circ\mathbf{A} \left(
\right)
	&&\ptext{Definition of design and predicate calculus}\\
	&=\mathbf{RA3_{AP}}\circ\mathbf{RA2}\circ\mathbf{A} \left(%
\right)
	&&\ptext{Definition of design and predicate calculus}\\
	&=\mathbf{RA3_{AP}}\circ\mathbf{RA2}\circ\mathbf{A} \left(%
\right)
	&&\ptext{\cref{law:A0:design}}\\
	&=\mathbf{RA3_{AP}}\circ\mathbf{RA2}\circ\mathbf{A0} (\lnot \mathbf{PBMH} (P^f_f) \vdash \mathbf{PBMH} (P^t_f))
	&&\ptext{\cref{lemma:PBMH(design):(lnot-PBMH(pre)|-PBMH(post))}}\\
	&=\mathbf{RA3_{AP}}\circ\mathbf{RA2}\circ\mathbf{A0}\circ\mathbf{PBMH} (\lnot P^f_f \vdash P^t_f)
	&&\ptext{Definition of $\mathbf{A}$}\\
	&=\mathbf{RA3_{AP}}\circ\mathbf{RA2}\circ\mathbf{A} (\lnot P^f_f \vdash P^t_f)
	&&\ptext{Definition of $\mathbf{AP}$ (\cref{theorem:RAPN(P):RAPN(Pff-vdash-Ptf)})}\\
	&=\mathbf{AP} (P)
\end{xflalign*}
\end{proof}
\end{proofs}
\end{theorem}

\begin{theorem}\label{theorem:PBMH-o-AP(P):AP(P)}
\begin{statement}
$\mathbf{PBMH} \circ \mathbf{AP} (P) = \mathbf{AP} (P)$
\end{statement}
\begin{proofs}
\begin{proof}
\begin{xflalign*}
	&\mathbf{PBMH} \circ \mathbf{AP} (P)
	&&\ptext{Definition of $\mathbf{AP}$}\\
	&=\mathbf{PBMH} \circ \mathbf{RA3_{AP}}\circ\mathbf{RA2}\circ\mathbf{A}\circ\mathbf{H1}\circ\mathbf{CSPA2} (P)
	&&\ptext{Definition of $\mathbf{A}$}\\
	&=\mathbf{PBMH} \circ \mathbf{RA3_{AP}}\circ\mathbf{RA2}\circ\mathbf{A0}\circ\mathbf{A1}\circ\mathbf{H1}\circ\mathbf{CSPA2} (P)
	&&\ptext{\cref{law:A1:idempotent}}\\
	&=\mathbf{PBMH} \circ \mathbf{RA3_{AP}}\circ\mathbf{RA2}\circ\mathbf{A0}\circ\mathbf{A1}\circ\mathbf{A1}\circ\mathbf{H1}\circ\mathbf{CSPA2} (P)
	&&\ptext{\cref{law:A0:commute-A0-healthy}}\\
	&=\mathbf{PBMH} \circ \mathbf{RA3_{AP}}\circ\mathbf{RA2}\circ\mathbf{A1}\circ\mathbf{A0}\circ\mathbf{A1}\circ\mathbf{H1}\circ\mathbf{CSPA2} (P)
	&&\ptext{$\mathbf{A1}$ is $\mathbf{PBMH}$}\\
	&=\mathbf{PBMH} \circ \mathbf{RA3_{AP}}\circ\mathbf{RA2}\circ\mathbf{PBMH}\circ\mathbf{A0}\circ\mathbf{A1}\circ\mathbf{H1}\circ\mathbf{CSPA2} (P)
	&&\ptext{\cref{theorem:PBMH-o-RA2(P):RA2(P)}}\\
	&=\mathbf{PBMH} \circ \mathbf{RA3_{AP}}\circ\mathbf{PBMH}\circ\mathbf{RA2}\circ\mathbf{PBMH}\circ\mathbf{A0}\circ\mathbf{A1}\circ\mathbf{H1}\circ\mathbf{CSPA2} (P)
	&&\ptext{\cref{theorem:PBMH-o-RA3(P):RA3-o-PBMH(P),law:pbmh:idempotent}}\\
	&=\mathbf{RA3_{AP}}\circ\mathbf{PBMH}\circ\mathbf{RA2}\circ\mathbf{PBMH}\circ\mathbf{A0}\circ\mathbf{A1}\circ\mathbf{H1}\circ\mathbf{CSPA2} (P)
	&&\ptext{\cref{theorem:PBMH-o-RA2(P):RA2(P)}}\\
	&=\mathbf{RA3_{AP}}\circ\mathbf{RA2}\circ\mathbf{PBMH}\circ\mathbf{A0}\circ\mathbf{A1}\circ\mathbf{H1}\circ\mathbf{CSPA2} (P)
	&&\ptext{$\mathbf{A1}$ is $\mathbf{PBMH}$}\\
	&=\mathbf{RA3_{AP}}\circ\mathbf{RA2}\circ\mathbf{A1}\circ\mathbf{A0}\circ\mathbf{A1}\circ\mathbf{H1}\circ\mathbf{CSPA2} (P)
	&&\ptext{\cref{law:A0:commute-A0-healthy,law:A1:idempotent}}\\
	&=\mathbf{RA3_{AP}}\circ\mathbf{RA2}\circ\mathbf{A0}\circ\mathbf{A1}\circ\mathbf{H1}\circ\mathbf{CSPA2} (P)
	&&\ptext{Definition of $\mathbf{A}$}\\
	&=\mathbf{RA3_{AP}}\circ\mathbf{RA2}\circ\mathbf{A}\circ\mathbf{H1}\circ\mathbf{CSPA2} (P)
	&&\ptext{Definition of $\mathbf{AP}$}\\
	&=\mathbf{AP} (P)
\end{xflalign*}
\end{proof}
\end{proofs}
\end{theorem}

\begin{theorem}\label{theorem:RA3N-o-RA2-o-A(P|-Q)}
\begin{align*}
	&\mathbf{RA3_{AP}} \circ \mathbf{RA2} \circ \mathbf{A} (P \vdash Q)\\
	&=\\
	&\left(\begin{array}{l} 
		true \dres s.wait \rres \lnot \mathbf{RA2} \circ \mathbf{PBMH} (\lnot P)
		\\ \vdash \\
		s \in ac' \dres s.wait \rres \mathbf{RA2} \circ \mathbf{RA1} \circ \mathbf{PBMH} (Q)
	\end{array}\right)
\end{align*}
\begin{proofs}\begin{proof}\checkt{alcc}
\begin{xflalign*}
	&\mathbf{RA3_{AP}}\circ\mathbf{RA2}\circ\mathbf{A} (P \vdash Q)
	&&\ptext{Definition of $\mathbf{A}$}\\
	&=\mathbf{RA3_{AP}}\circ\mathbf{RA2} (\lnot \mathbf{PBMH} (P) \vdash \mathbf{PBMH} (Q) \land ac'\neq\emptyset)
	&&\ptext{\cref{lemma:RA2(P|-Q):(lnot-RA2(lnot-P)|-RA2(Q))}}\\
	&=\mathbf{RA3_{AP}} (\lnot \mathbf{RA2}\circ\mathbf{PBMH} (P) \vdash \mathbf{RA2} (\mathbf{PBMH} (Q) \land ac'\neq\emptyset))
	&&\ptext{\cref{lemma:RA2(P-land-ac'-neq-emptyset):RA2-o-RA1(P)}}\\
	&=\mathbf{RA3_{AP}} (\lnot \mathbf{RA2}\circ\mathbf{PBMH} (P) \vdash \mathbf{RA2} \circ \mathbf{RA1} \circ \mathbf{PBMH} (Q))
	&&\ptext{\cref{theorem:RA3N(design)}}\\
	&=\left(
\right)^o_f
	&&\ptext{Definition of design}\\
	&=\left(%
\right)
	&&\ptext{Value of record component $wait$}\\
	&=\left(%
\right)
	&&\ptext{Property of conditional}\\
	&=\left(%
\right)
	&&\ptext{Property of substitution}\\
	&=\left(%
\right)
	&&\ptext{Predicate calculus}\\
	&=\lnot ok \lor \mathbf{RA2} \circ \mathbf{PBMH} (P^f_f)
	&&\ptext{Predicate calculus}\\
	&=ok \implies \mathbf{RA2} \circ \mathbf{PBMH} (P^f_f)
\end{xflalign*}
\end{proof}\end{proofs}
\end{lemma}

\begin{lemma}\label{lemma:RAPN(P)-subs-tf}
\begin{align*}
	&\mathbf{AP} (P)^t_f \\
	&=\\
	&(ok \land \lnot \mathbf{RA2} \circ \mathbf{PBMH} (P^f_f)) \implies \mathbf{RA2} \circ \mathbf{RA1} \circ \mathbf{PBMH} (P^t_f)
\end{align*}
\begin{proofs}\begin{proof}\checkt{alcc}\checkt{pfr}
\begin{xflalign*}
	&\mathbf{AP} (P)^t_f
	&&\ptext{\cref{lemma:RAPN(P)-o-f-subs}}\\
	&=\left(\begin{array}{l}
			(ok \land \lnot \mathbf{RA2} \circ \mathbf{PBMH} (P^f_f)) 
			\\ \implies \\
			(\mathbf{RA2} \circ \mathbf{RA1} \circ \mathbf{PBMH} (P^t_f) \land true)
	\end{array}\right)
	&&\ptext{Predicate calculus}\\
	&=(ok \land \lnot \mathbf{RA2} \circ \mathbf{PBMH} (P^f_f)) \implies \mathbf{RA2} \circ \mathbf{RA1} \circ \mathbf{PBMH} (P^t_f)
\end{xflalign*}
\end{proof}\end{proofs}
\end{lemma}

\begin{lemma}\label{lemma:RA2-o-PBMH(RAPN(P)-subs-tf):RAPN(P)-subs-tf}
\begin{align*}
	&\mathbf{RA2} \circ \mathbf{PBMH} (\mathbf{AP} (P)^t_f) = \mathbf{AP} (P)^t_f 
\end{align*}
\begin{proofs}\begin{proof}\checkt{alcc}\checkt{pfr}
\begin{xflalign*}
	&\mathbf{AP} (P)^t_f
	&&\ptext{\cref{lemma:RAPN(P)-subs-tf}}\\
	&=(ok \land \lnot \mathbf{RA2} \circ \mathbf{PBMH} (P^f_f)) \implies \mathbf{RA2} \circ \mathbf{RA1} \circ \mathbf{PBMH} (P^t_f)
	&&\ptext{Predicate calculus}\\
	&=\lnot ok \lor \mathbf{RA2} \circ \mathbf{PBMH} (P^f_f) \lor \mathbf{RA2} \circ \mathbf{RA1} \circ \mathbf{PBMH} (P^t_f)
	&&\ptext{\cref{lemma:RA2(P):P:s-ac'-not-free}}\\
	&=\mathbf{RA2} (\lnot ok) \lor \mathbf{RA2} \circ \mathbf{PBMH} (P^f_f) \lor \mathbf{RA2} \circ \mathbf{RA1} \circ \mathbf{PBMH} (P^t_f)
	&&\ptext{\cref{theorem:RA2:idempotent,theorem:RA2(P-lor-Q):RA2(P)-lor-RA2(Q)}}\\
	&=\mathbf{RA2} (\lnot ok \lor \mathbf{RA2} \circ \mathbf{PBMH} (P^f_f) \lor \mathbf{RA2} \circ \mathbf{RA1} \circ \mathbf{PBMH} (P^t_f))
	&&\ptext{\cref{law:pbmh:P:ac'-not-free}}\\
	&=\mathbf{RA2} (\mathbf{PBMH} (\lnot ok) \lor \mathbf{RA2} \circ \mathbf{PBMH} (P^f_f) \lor \mathbf{RA2} \circ \mathbf{RA1} \circ \mathbf{PBMH} (P^t_f))
	&&\ptext{\cref{theorem:PBMH-o-RA1(P):RA1(P),theorem:PBMH-o-RA2(P):RA2(P)}}\\
	&=\mathbf{RA2} \left(
\right)
	&&\ptext{Predicate calculus and property of conditional}\\
	&=(true \vdash s \in ac' \dres s.wait \rres \mathbf{RA2} \circ \mathbf{RA1} \circ \mathbf{PBMH} (P^t_f))
\end{xflalign*}
\end{proof}\end{proofs}
\end{lemma}

\begin{lemma}\label{lemma:RAPN(lnot-Pff|-Ptf):design}
\begin{align*}
	&\mathbf{AP} (\lnot P^f_f \vdash P^t_f) \\
	&=\\
	&\left(%
\right)
\end{xflalign*}
\end{proof}\end{proofs}
\end{lemma}

\begin{lemma}\label{lemma:RAPN(lnot-Pff|-Ptf):RA3N-o-RA2-o-A(lnot-Pff|-Ptf)}
$\mathbf{AP} (\lnot P^f_f \vdash P^t_f) = \mathbf{RA3_{AP}} \circ \mathbf{RA2} \circ \mathbf{A} (\lnot P^f_f \vdash P^t_f)$
\begin{proofs}\begin{proof}\checkt{alcc}
\begin{xflalign*}
	&\mathbf{AP} (\lnot P^f_f \vdash P^t_f)
	&&\ptext{\cref{theorem:RAPN(P):RAPN(Pff-vdash-Ptf)}}\\
	&=\mathbf{RA3_{AP}} \circ \mathbf{RA2} \circ \mathbf{A} (\lnot (\lnot P^f_f \vdash P^t_f)^f_f \vdash (\lnot P^f_f \vdash P^t_f)^t_f)
	&&\ptext{\cref{lemma:design:(P|-Q)f-|-(P|-Q)t}}\\
	&=\mathbf{RA3_{AP}} \circ \mathbf{RA2} \circ \mathbf{A} (\lnot P^f_f \vdash P^t_f)
\end{xflalign*}
\end{proof}\end{proofs}
\end{lemma}

\subsection{$\mathbf{ND_{AP_{N}}}$}

\begin{theorem}\label{theorem:RAPN:ChoiceN-sqcup-P}
\begin{statement} Provided $P$ is $\mathbf{AP}$-healthy.
\begin{align*}
	&Choice_{\mathbf{AP}} \sqcup P
	=
	(	true
		 \vdash 
		s \in ac' \dres s.wait \rres \mathbf{RA2}\circ\mathbf{RA1}\circ\mathbf{PBMH} (P^t_f))
\end{align*}
\end{statement}
\begin{proofs}
\begin{proof}\checkt{pfr}\checkt{alcc}
\begin{xflalign*}
	&Choice_{\mathbf{AP}} \sqcup P
	&&\ptext{Assumption: $P$ is $\mathbf{AP}$-healthy}\\
	&=Choice_{\mathbf{AP}} \sqcup \mathbf{AP} (P)
	&&\ptext{Definition of $Choice_{\mathbf{AP}}$ (\cref{lemma:RAPN(true|-ac'neq-emptyset)})}\\
	&=\left(
\right) \sqcup \mathbf{AP} (P)
	&&\ptext{Predicate calculus and property of conditional}\\
	&=\left(%
\right)
	&&\ptext{Property of conditional}\\
	&=\left(%
\right)
	&&\ptext{Predicate calculus and property of conditional}\\
	&=\left(%
\right)
	&&\ptext{Assumption: $P$ is $\mathbf{AP}$-healthy}\\
	&=(true
		 \vdash 
		s \in ac' \dres s.wait \rres \mathbf{RA2}\circ\mathbf{RA1}\circ\mathbf{PBMH} (P^t_f))
\end{xflalign*}
\end{proof}
\end{proofs}
\end{theorem}

\section{Relationship with Reactive Angelic Designs}

\subsection{From $\mathbf{RAD}$ to $\mathbf{AP}$}

\begin{theorem}\label{theorem:H1-o-RAP(P)}
\begin{statement}
\begin{align*}
	&\mathbf{H1} \circ \mathbf{RAD} (P)
	=
	\left(%
\right)
\end{align*}
\end{statement}
\begin{proofs}
\begin{proof}\checkt{alcc}
\begin{xflalign*}
	&\mathbf{H1} \circ \mathbf{RAD} (P)
	&&\ptext{Definition of $\mathbf{RAD}$}\\
	&=\mathbf{H1} \circ \mathbf{RA} \circ \mathbf{A} (\lnot P^f_f \vdash P^t_f)
	&&\ptext{\cref{theorem:RA-o-A(P):RA-o-PBMH(P)}}\\
	&=\mathbf{H1} \circ \mathbf{RA} \circ \mathbf{PBMH} (\lnot P^f_f \vdash P^t_f)
	&&\ptext{\cref{lemma:PBMH(design):(lnot-PBMH(pre)|-PBMH(post))}}\\
	&=\mathbf{H1} \circ \mathbf{RA} (\lnot \mathbf{PBMH} (P^f_f) \vdash \mathbf{PBMH} (P^t_f))
	&&\ptext{Definition of $\mathbf{RA}$}\\
	&=\mathbf{H1} \circ \mathbf{RA1} \circ \mathbf{RA2} \circ \mathbf{RA3} (\lnot \mathbf{PBMH} (P^f_f) \vdash \mathbf{PBMH} (P^t_f))
	&&\ptext{\cref{theorem:RA3-o-RA2:RA2-o-RA3}}\\
	&=\mathbf{H1} \circ \mathbf{RA1} \circ \mathbf{RA3} \circ \mathbf{RA2} (\lnot \mathbf{PBMH} (P^f_f) \vdash \mathbf{PBMH} (P^t_f))
	&&\ptext{\cref{lemma:RA2(P|-Q):(lnot-RA2(lnot-P)|-RA2(Q))}}\\
	&=\mathbf{H1} \circ \mathbf{RA1} \circ \mathbf{RA3} (\lnot \mathbf{RA2}\circ\mathbf{PBMH} (P^f_f) \vdash \mathbf{RA2}\circ\mathbf{PBMH} (P^t_f))
	&&\ptext{\cref{lemma:RA1-o-RA3(design):RA1(design)}}\\
	&=\mathbf{H1} \circ \mathbf{RA1} \left(
\right)
	&&\ptext{Definition of design}\\
	&=\mathbf{H1} \circ \mathbf{RA1} \left(%
\right)
	&&\ptext{Predicate calculus and~\cref{lemma:lnot(conditional)}}\\
	&=\mathbf{H1} \circ \mathbf{RA1} \left(%
\right)
	&&\ptext{Predicate calculus: absorption law}\\
	&=\left(%
\right)
	&&\ptext{Predicate calculus and~\cref{lemma:lnot(conditional)}}\\
	&=\left(%
\right)
	&&\ptext{Definition of design}\\
	&=\left(%
\right)
	&&\ptext{\cref{lemma:RA1-implies-ac'-neq-emptyset,lemma:s-in-ac'-implies-RA1} and predicate calculus}\\
	&=\left(%
\right)
	&&\ptext{Predicate calculus and property of conditional}\\
	&=\left(%
\right)
	&&\ptext{Property of conditional}\\
	&=\mathbf{A}\left(%
\right)
	&&\ptext{\cref{theorem:H1-o-RAP(P)}}\\
	&=\mathbf{A}\circ\mathbf{H1}\circ\mathbf{RAD} (P)
\end{xflalign*}
\end{proof}\end{proofs}
\end{theorem}

\begin{lemma}\label{lemma:H1-o-RA-o-A(true|-Ptf):(true|-Ptf)}
\begin{statement}
\begin{align*}
	&\mathbf{H1} \circ \mathbf{RA} \circ \mathbf{A} (true \vdash P^t_f)\\
 	&=\\
 	&(true \vdash s \in ac' \dres s.wait \rres \mathbf{RA2} \circ \mathbf{RA1} \circ \mathbf{PBMH} (P^t_f))
\end{align*}
\end{statement}
\begin{proofs}
\begin{proof}
\begin{xflalign*}
	&\mathbf{H1} \circ \mathbf{RA} \circ \mathbf{A} (true \vdash P^t_f)
	&&\ptext{\cref{theorem:RA-o-A(design):RA-CSPA-PBMH,lemma:H1-o-RAP(P):RAPN(lnot-RA1-o-PBMH(Pff)|-Ptf)}}\\
	&=\mathbf{AP} (\lnot \mathbf{RA1} \circ \mathbf{PBMH} (false) \vdash P^t_f)
	&&\ptext{\cref{law:pbmh:false,lemma:RA1(false)}}\\
	&=\mathbf{AP} (\lnot false \vdash P^t_f)
	&&\ptext{Predicate calculus}\\
	&=\mathbf{AP} (true \vdash P^t_f)
	&&\ptext{\cref{theorem:RAPN(P)}}\\
	&=\left(
\right)
	&&\ptext{\cref{lemma:RA2(P):P:s-ac'-not-free} and predicate calculus}\\
	&=\left(%
\right)
	&&\ptext{Property of conditional}\\
	&=(true \vdash s \in ac' \dres s.wait \rres \mathbf{RA2} \circ \mathbf{RA1} \circ \mathbf{PBMH} (P^t_f))
\end{xflalign*}
\end{proof}
\end{proofs}
\end{lemma}

\begin{lemma}\label{theorem:H3-o-H1-o-RAP(P)}
\begin{statement}
\begin{align*}
	&\mathbf{H3} \circ \mathbf{H1} \circ \mathbf{RAD} (P)\\
	&=\\
	&\left(%
\right)
	&&\ptext{Predicate calculus and property of conditional}\\
	&=\left(%
\right)
	&&\ptext{Predicate calculus and property of conditional}\\
	&=\left(%
\right)
\end{xflalign*}
\end{proof}
\end{proofs}
\end{lemma}

\begin{lemma}\label{lemma:H1-o-RA-o-A(true|-Ptf):RAPN(true|-Ptf)}
\begin{statement}
$\mathbf{H1} \circ \mathbf{RA} \circ \mathbf{A} (true \vdash P^t_f) = \mathbf{AP} (true \vdash P^t_f)$
\end{statement}
\begin{proofs}
\begin{proof}\checkt{alcc}
\begin{xflalign*}
	&\mathbf{H1} \circ \mathbf{RA} \circ \mathbf{A} (true \vdash P^t_f)
	&&\ptext{\cref{theorem:RA-o-A(design):RA-CSPA-PBMH,lemma:H1-o-RAP(P):RAPN(lnot-RA1-o-PBMH(Pff)|-Ptf)}}\\
	&=\mathbf{AP} (\lnot \mathbf{RA1} \circ \mathbf{PBMH} (false) \vdash P^t_f)
	&&\ptext{\cref{law:pbmh:false,lemma:RA1(false)}}\\
	&=\mathbf{AP} (\lnot false \vdash P^t_f)
	&&\ptext{Predicate calculus}\\
	&=\mathbf{AP} (true \vdash P^t_f)
\end{xflalign*}
\end{proof}
\end{proofs}
\end{lemma}

\begin{lemma}\label{theorem:H3-o-H1-o-RA-o-A(true|-Ptf)}
\begin{statement}
\begin{align*}
	&\mathbf{H3} \circ \mathbf{H1} \circ \mathbf{RA} \circ \mathbf{A} (true \vdash P^t_f) \\
	&=\\
	&(true \vdash s \in ac' \dres s.wait \rres \mathbf{RA1}\circ\mathbf{RA2}\circ\mathbf{PBMH} (P^t_f))
\end{align*}
\end{statement}
\begin{proofs}
\begin{proof}\checkt{pfr}\checkt{alcc}
\begin{xflalign*}
	&\mathbf{H3} \circ \mathbf{H1} \circ \mathbf{RA} \circ \mathbf{A} (true \vdash P^t_f)
	&&\ptext{\cref{theorem:RA-o-A(design):RA-CSPA-PBMH,theorem:H3-o-H1-o-RAP(P)}}\\
	&=\left(
\right)
	&&\ptext{Definition of $\mathbf{A}$}\\
	&=\mathbf{RA3_{AP}} \circ \mathbf{RA2} \circ \mathbf{A} (
		\lnot \mathbf{RA1}\circ\mathbf{PBMH} (P^f_f)
		\vdash
		P^t_f)
	&&\ptext{\cref{lemma:RAPN(lnot-Pff|-Ptf):RA3N-o-RA2-o-A(lnot-Pff|-Ptf)}}\\
	&=\mathbf{AP} (\lnot \mathbf{RA1}\circ\mathbf{PBMH} (P^f_f) \vdash P^t_f)
\end{xflalign*}
\end{proof}\end{proofs}
\end{lemma}

\begin{lemma}\label{lemma:H1(P)-RAP:RAPN(lnot-RA1(Pff)|-Ptf)} Provided $P$ is a reactive angelic process,
\begin{align*}
	&\mathbf{H1} (P) = \mathbf{AP} (\lnot \mathbf{RA1} (P^f_f) \vdash P^t_f)
\end{align*}
\begin{proofs}\begin{proof}\checkt{alcc}
\begin{xflalign*}
	&\mathbf{H1} (P)
	&&\ptext{Assumption: $P$ is $\mathbf{RAD}$-healthy}\\
	&=\mathbf{H1} \circ \mathbf{RAD} (P)
	&&\ptext{\cref{lemma:H1-o-RAP(P):RAPN(lnot-RA1-o-PBMH(Pff)|-Ptf)}}\\
	&=\mathbf{AP} (\lnot \mathbf{RA1} \circ \mathbf{PBMH} (P^f_f) \vdash P^t_f)
	&&\ptext{\cref{lemma:PBMH(P)-ow:PBMH(P-ow)}}\\
	&=\mathbf{AP} (\lnot \mathbf{RA1} (\mathbf{PBMH} (P)^f_f) \vdash P^t_f)
	&&\ptext{Assumption: $P$ is $\mathbf{RAD}$-healthy and~\cref{theorem:PBMH(P)-RAP:P}}\\
	&=\mathbf{AP} (\lnot \mathbf{RA1} (P^f_f) \vdash P^t_f)
\end{xflalign*}
\end{proof}\end{proofs}
\end{lemma}

\subsection{From $\mathbf{AP}$ to $\mathbf{RAD}$}

\begin{theorem}\label{theorem:RA1-o-RAPN(P):RA-o-A(Pff|-Ptf)}
\begin{statement}
$\mathbf{RA1} \circ \mathbf{AP} (P) = \mathbf{RA} \circ \mathbf{A} (\lnot P^f_f \vdash P^t_f)$
\end{statement} 
\begin{proofs}
\begin{proof}\checkt{alcc}
\begin{xflalign*}
	&\mathbf{RA1} \circ \mathbf{AP} (P)
	&&\ptext{\cref{theorem:RAPN(P)}}\\
	&=\mathbf{RA1} \left(\begin{array}{l}
		true \dres s.wait \rres \lnot \mathbf{RA2}\circ\mathbf{PBMH} (P^f_f) 
		\\ \vdash \\
		s \in ac' \dres s.wait \rres \mathbf{RA2} \circ \mathbf{RA1} \circ \mathbf{PBMH} (P^t_f)
	\end{array}\right)
	&&\ptext{\cref{lemma:RA1-o-RA3(design):RA1(design)}}\\
	&=\mathbf{RA1} \circ \mathbf{RA3} \left(\begin{array}{l}
		\lnot \mathbf{RA2}\circ\mathbf{PBMH} (P^f_f) 
		\\ \vdash \\
		\mathbf{RA2} \circ \mathbf{RA1} \circ \mathbf{PBMH} (P^t_f)
	\end{array}\right)
	&&\ptext{\cref{lemma:RA2(P|-Q):(lnot-RA2(lnot-P)|-RA2(Q))}}\\
	&=\mathbf{RA1} \circ \mathbf{RA3} \circ \mathbf{RA2} \left(\begin{array}{l}
		\lnot \mathbf{PBMH} (P^f_f) 
		\\ \vdash \\
		\mathbf{RA1} \circ \mathbf{PBMH} (P^t_f)
	\end{array}\right)
	&&\ptext{\cref{theorem:RA3-o-RA1:RA1-o-RA3,theorem:RA2-o-RA1:RA1-o-RA2}}\\
	&=\mathbf{RA3} \circ \mathbf{RA2} \circ \mathbf{RA1} \left(\begin{array}{l}
		\lnot \mathbf{PBMH} (P^f_f) 
		\\ \vdash \\
		\mathbf{RA1} \circ \mathbf{PBMH} (P^t_f)
	\end{array}\right)
	&&\ptext{\cref{lemma:RA1(P|-Q):RA1(P|-RA1(Q))}}\\
	&=\mathbf{RA3} \circ \mathbf{RA2} \circ \mathbf{RA1} (\lnot \mathbf{PBMH} (P^f_f) \vdash \mathbf{PBMH} (P^t_f))
	&&\ptext{\cref{lemma:PBMH(design):(lnot-PBMH(pre)|-PBMH(post))}}\\
	&=\mathbf{RA3} \circ \mathbf{RA2} \circ \mathbf{RA1} \circ \mathbf{PBMH} (\lnot P^f_f \vdash P^t_f)
	&&\ptext{Definition of $\mathbf{RA}$}\\
	&=\mathbf{RA} \circ \mathbf{PBMH} (\lnot P^f_f \vdash P^t_f)
	&&\ptext{\cref{theorem:RA-o-A(P):RA-o-PBMH(P)}}\\
	&=\mathbf{RA} \circ \mathbf{A} (\lnot P^f_f \vdash P^t_f)
\end{xflalign*}
\end{proof}
\end{proofs}
\end{theorem}

\subsection{Galois Connection and Isomorphism}

\begin{theorem}\label{theorem:RA1-o-H1-o-RAP(P):RAP(P)}
\begin{statement}
$\mathbf{RA1} \circ \mathbf{H1} \circ \mathbf{RAD} (P) = \mathbf{RAD} (P)$
\end{statement}
\begin{proofs}
\begin{proof}\checkt{alcc}
\begin{xflalign*}
	&\mathbf{RA1} \circ \mathbf{H1} \circ \mathbf{RAD} (P)
	&&\ptext{\cref{lemma:H1-o-RAP(P):RAPN(lnot-RA1-o-PBMH(Pff)|-Ptf)}}\\
	&=\mathbf{RA1} \circ \mathbf{AP} (\lnot \mathbf{RA1} \circ \mathbf{PBMH} (P^f_f) \vdash P^t_f)
	&&\ptext{\cref{theorem:RA1-o-RAPN(P):RA-o-A(Pff|-Ptf),law:design:true-ok',law:design:false-ok'}}\\
	&=\mathbf{RA} \circ \mathbf{A} (\lnot \mathbf{RA1} \circ \mathbf{PBMH} (P^f_f) \vdash P^t_f)
	&&\ptext{\cref{theorem:RA-o-A(P):RA-o-PBMH(P)}}\\
	&=\mathbf{RA} \circ \mathbf{PBMH} (\lnot \mathbf{RA1} \circ \mathbf{PBMH} (P^f_f) \vdash P^t_f)
	&&\ptext{\cref{lemma:PBMH(design):(lnot-PBMH(pre)|-PBMH(post))}}\\
	&=\mathbf{RA} (\lnot \mathbf{PBMH} \circ \mathbf{RA1} \circ \mathbf{PBMH} (P^f_f) \vdash \mathbf{PBMH} (P^t_f))
	&&\ptext{\cref{theorem:PBMH-o-RA1(P):RA1(P)}}\\
	&=\mathbf{RA} (\lnot \mathbf{RA1} \circ \mathbf{PBMH} (P^f_f) \vdash \mathbf{PBMH} (P^t_f))
	&&\ptext{Definition of $\mathbf{RA}$}\\
	&=\mathbf{RA3} \circ \mathbf{RA2} \circ \mathbf{RA1} (\lnot \mathbf{RA1} \circ \mathbf{PBMH} (P^f_f) \vdash \mathbf{PBMH} (P^t_f))
	&&\raisetag{18pt}\ptext{\cref{lemma:RA1(P|-Q):RA1(lnot-RA1(lnot-P)|-Q)}}\\
	&=\mathbf{RA3} \circ \mathbf{RA2} \circ \mathbf{RA1} (\lnot \mathbf{PBMH} (P^f_f) \vdash \mathbf{PBMH} (P^t_f))
	&&\ptext{Definition of $\mathbf{RA}$}\\
	&=\mathbf{RA} (\lnot \mathbf{PBMH} (P^f_f) \vdash \mathbf{PBMH} (P^t_f))
	&&\ptext{\cref{lemma:PBMH(design):(lnot-PBMH(pre)|-PBMH(post))}}\\
	&=\mathbf{RA} \circ \mathbf{PBMH} (\lnot P^f_f \vdash P^t_f)
	&&\ptext{\cref{theorem:RA-o-A(P):RA-o-PBMH(P)}}\\
	&=\mathbf{RA} \circ \mathbf{A} (\lnot P^f_f \vdash P^t_f)
	&&\ptext{\cref{theorem:RA-o-A(design):RA-CSPA-PBMH}}\\
	&=\mathbf{RAD} (P)
\end{xflalign*}
\end{proof}
\end{proofs}
\end{theorem}

\begin{theorem}\label{theorem:H1-o-RA1-o-RAPN(P):sqsupseteq:RAPN(P)}
\begin{statement}
$\mathbf{H1} \circ \mathbf{RA1} \circ \mathbf{AP} (P) \sqsupseteq \mathbf{AP} (P)$
\end{statement}
\begin{proofs}
\begin{proof}\checkt{alcc}
\begin{xflalign*}
	&\mathbf{H1} \circ \mathbf{RA1} \circ \mathbf{AP} (P)
	&&\ptext{\cref{theorem:RA1-o-RAPN(P):RA-o-A(Pff|-Ptf)}}\\
	&=\mathbf{H1} \circ \mathbf{RA} \circ \mathbf{A} (\lnot P^f_f \vdash P^t_f)
	&&\ptext{\cref{theorem:RA-o-A(design):RA-CSPA-PBMH,lemma:H1-o-RAP(P):RAPN(lnot-RA1-o-PBMH(Pff)|-Ptf)}}\\
	&=\mathbf{AP} (\lnot \mathbf{RA1} \circ \mathbf{PBMH} (P^f_f) \vdash P^t_f)
	&&\ptext{\cref{lemma:RA1(P):implies:P} and strengthen precondition}\\
	&\sqsupseteq \mathbf{AP} (\lnot \mathbf{PBMH} (P^f_f) \vdash P^t_f)
	&&\ptext{\cref{lemma:RAPN(lnot-Pff|-Ptf):RA3N-o-RA2-o-A(lnot-Pff|-Ptf)}}\\
	&=\mathbf{RA3_{AP}} \circ \mathbf{RA2} \circ \mathbf{A} (\lnot \mathbf{PBMH} (P^f_f) \vdash P^t_f)
	&&\ptext{Definition of $\mathbf{A}$ and~\cref{lemma:PBMH(design):(lnot-PBMH(pre)|-PBMH(post)),law:pbmh:idempotent}}\\
	&=\mathbf{RA3_{AP}} \circ \mathbf{RA2} \circ \mathbf{A} (\lnot P^f_f \vdash P^t_f)
	&&\ptext{\cref{lemma:RAPN(lnot-Pff|-Ptf):RA3N-o-RA2-o-A(lnot-Pff|-Ptf)}}\\
	&=\mathbf{AP} (P) 
\end{xflalign*}
\end{proof}
\end{proofs}
\end{theorem}

\begin{theorem}\label{theorem:H1-o-RA1-o-NDAP-o-AP(P):ND-o-AP(P)}
\begin{statement}
$\mathbf{H1}\circ\mathbf{RA1}\circ\mathbf{ND_{AP}}\circ\mathbf{AP} (P) = \mathbf{ND_{AP}}\circ\mathbf{AP} (P)$
\end{statement}
\begin{proofs}
\begin{proof}
\begin{xflalign*}
	&\mathbf{H1}\circ\mathbf{RA1}\circ\mathbf{ND_{AP}}\circ\mathbf{AP} (P)
	&&\ptext{Definition of $\mathbf{ND_{AP}}$ and~\cref{theorem:RAPN:ChoiceN-sqcup-P}}\\
	&=\mathbf{H1}\circ\mathbf{RA1}(true
		 \vdash 
		s \in ac' \dres s.wait \rres \mathbf{RA2}\circ\mathbf{RA1}\circ\mathbf{PBMH} (P^t_f))
	&&\ptext{\cref{lemma:H1-o-RA1(P|-Q):(lnot-RA1(lnot-P)|-RA1(Q))}}\\
	&=(\lnot \mathbf{RA1} (\lnot true)
		 \vdash 
		\mathbf{RA1} (s \in ac' \dres s.wait \rres \mathbf{RA2}\circ\mathbf{RA1}\circ\mathbf{PBMH} (P^t_f)))
	&&\ptext{Predicate calculus and~\cref{lemma:RA1(false)}}\\
	&=(true
		 \vdash 
		\mathbf{RA1} (s \in ac' \dres s.wait \rres \mathbf{RA2}\circ\mathbf{RA1}\circ\mathbf{PBMH} (P^t_f)))
	&&\ptext{\cref{lemma:RA1(conditional)}}\\
	&=(true
		 \vdash 
		\mathbf{RA1} (s \in ac') \dres s.wait \rres \mathbf{RA1}\circ\mathbf{RA2}\circ\mathbf{RA1}\circ\mathbf{PBMH} (P^t_f)))
	&&\ptext{\cref{theorem:RA2-o-RA1:RA1-o-RA2,theorem:RA1-idempotent}}\\
	&=(true
		 \vdash 
		\mathbf{RA1} (s \in ac') \dres s.wait \rres \mathbf{RA2}\circ\mathbf{RA1}\circ\mathbf{PBMH} (P^t_f)))
	&&\ptext{Definition of $\mathbf{ND_{AP}}$ and~\cref{theorem:RAPN:ChoiceN-sqcup-P}}\\
	&=\mathbf{ND_{AP}}\circ\mathbf{AP} (P)
\end{xflalign*}
\end{proof}
\end{proofs}
\end{theorem}

\begin{theorem}\label{theorem:RA1-o-H3-o-H1-o-RAD(P):sqsubseteq:RAD(P)}
\begin{statement}
$\mathbf{RA1} \circ \mathbf{H3} \circ \mathbf{H1} \circ \mathbf{RAD} (P) \sqsubseteq \mathbf{RAD} (P)$
\end{statement}
\begin{proofs}
\begin{proof}\checkt{alcc}
\begin{xflalign*}
	&\mathbf{RA1} \circ \mathbf{H3} \circ \mathbf{H1} \circ \mathbf{RAD} (P)
	&&\ptext{\cref{theorem:H3-o-H1-o-RAP(P)}}\\
	&=\mathbf{RA1} \left(
\right)
	&&\ptext{Predicate calculus and definition of conditional}\\
	&=\mathbf{RA1} \left(%
\right)
	&&\ptext{$\mathbf{RA1}$ and $\mathbf{RA3}$ are monotonic, weaken precondition as $(\lnot \exists ac' \spot P) \implies \lnot P$}\\
	&\sqsubseteq \mathbf{RA1} \circ \mathbf{RA3} \left(%
\right)
	&&\ptext{\cref{lemma:RA1(P|-Q):RA1(P|-RA1(Q)),lemma:RA1(P|-Q):RA1(lnot-RA1(lnot-P)|-Q)}}\\
	&=\mathbf{RA3} \circ \mathbf{RA1} (
		\lnot \mathbf{RA2}\circ\mathbf{PBMH} (P^f_f)
		\vdash 
		\mathbf{RA2}\circ\mathbf{PBMH} (P^t_f))
	&&\ptext{\cref{lemma:RA2(P|-Q):(lnot-RA2(lnot-P)|-RA2(Q))}}\\
	&=\mathbf{RA3} \circ \mathbf{RA1} \circ \mathbf{RA2} (
		\lnot \mathbf{PBMH} (P^f_f)
		\vdash
		\mathbf{PBMH} (P^t_f))
	&&\ptext{\cref{lemma:PBMH(design):(lnot-PBMH(pre)|-PBMH(post))}}\\
	&=\mathbf{RA3} \circ \mathbf{RA1} \circ \mathbf{RA2} \circ \mathbf{PBMH} (\lnot P^f_f \vdash P^t_f)
	&&\ptext{Definition of $\mathbf{RA}$}\\
	&=\mathbf{RA} \circ \mathbf{PBMH} (\lnot P^f_f \vdash P^t_f)
	&&\ptext{\cref{theorem:RA-o-A(P):RA-o-PBMH(P)}}\\
	&=\mathbf{RA} \circ \mathbf{A} (\lnot P^f_f \vdash P^t_f)
	&&\ptext{Definition of $\mathbf{RAD}$ (\cref{theorem:RA-o-A(design):RA-CSPA-PBMH})}\\
	&=\mathbf{RAD} (P)
\end{xflalign*}
\end{proof}
\end{proofs}
\end{theorem}

\begin{theorem}\label{theorem:H3-o-H1-o-RA1-o-AP(P):seqsubseteq:AP(P)}
\begin{statement}
$\mathbf{H3} \circ \mathbf{H1} \circ \mathbf{RA1} \circ \mathbf{AP} (P) \sqsubseteq \mathbf{AP} (P)$
\end{statement}
\begin{proofs}
\begin{proof}\checkt{alcc}
\begin{xflalign*}
	&\mathbf{H3} \circ \mathbf{H1} \circ \mathbf{RA1} \circ \mathbf{AP} (P)
	&&\ptext{\cref{theorem:RA1-o-RAPN(P):RA-o-A(Pff|-Ptf)}}\\
	&=\mathbf{H3} \circ \mathbf{H1} \circ \mathbf{RA} \circ \mathbf{A} (\lnot P^f_f \vdash P^t_f)
	&&\ptext{\cref{theorem:RA-o-A(design):RA-CSPA-PBMH}}\\
	&=\mathbf{H3} \circ \mathbf{H1} \circ \mathbf{RAD} (P)
	&&\ptext{\cref{theorem:H3-o-H1-o-RAP(P)}}\\
	&=\left(\begin{array}{l}
		\lnot s.wait \implies \lnot \exists ac' \spot \mathbf{RA1}\circ\mathbf{RA2}\circ\mathbf{PBMH} (P^f_f)
		\\ \vdash \\
		s \in ac' \dres s.wait \rres \mathbf{RA1}\circ\mathbf{RA2}\circ\mathbf{PBMH} (P^t_f)
	\end{array}\right)
	&&\ptext{\cref{lemma:exists-ac'-RA1-o-RA2-o-PBMH:exists-ac'-RA2-o-PBMH}}\\
	&=\left(\begin{array}{l}
		\lnot s.wait \implies \lnot \exists ac' \spot \mathbf{RA2}\circ\mathbf{PBMH} (P^f_f)
		\\ \vdash \\
		s \in ac' \dres s.wait \rres \mathbf{RA1}\circ\mathbf{RA2}\circ\mathbf{PBMH} (P^t_f)
	\end{array}\right)
	&&\ptext{Weaken precondition as $(\lnot \exists ac' \spot P) \implies \lnot P$}\\
	&\sqsubseteq \left(\begin{array}{l}
		\lnot s.wait \implies \lnot \mathbf{RA2}\circ\mathbf{PBMH} (P^f_f)
		\\ \vdash \\
		s \in ac' \dres s.wait \rres \mathbf{RA1}\circ\mathbf{RA2}\circ\mathbf{PBMH} (P^t_f)
	\end{array}\right)
	&&\ptext{Predicate calculus}\\
	&=\left(\begin{array}{l}
		\lnot s.wait \implies (\lnot s.wait \land \lnot \mathbf{RA2}\circ\mathbf{PBMH} (P^f_f))
		\\ \vdash \\
		s \in ac' \dres s.wait \rres \mathbf{RA1}\circ\mathbf{RA2}\circ\mathbf{PBMH} (P^t_f)
	\end{array}\right)
	&&\ptext{Predicate calculus and definition of conditional}\\
	&=\left(\begin{array}{l}
		true \dres s.wait \rres \lnot \mathbf{RA2}\circ\mathbf{PBMH} (P^f_f)
		\\ \vdash \\
		s \in ac' \dres s.wait \rres \mathbf{RA1}\circ\mathbf{RA2}\circ\mathbf{PBMH} (P^t_f)
	\end{array}\right)
	&&\ptext{\cref{theorem:RAPN(P)}}\\
	&=\mathbf{AP} (P)
\end{xflalign*}
\end{proof}
\end{proofs}
\end{theorem}

\begin{theorem}\label{theorem:RA1-o-H3-o-H1-o-RA-o-A(ND):RA-o-A(ND)}
\begin{statement}
\begin{align*}
	&\mathbf{RA1} \circ \mathbf{H3} \circ \mathbf{H1} \circ \mathbf{RA} \circ \mathbf{A} (true \vdash P^t_f)
	=
	\mathbf{RA}\circ\mathbf{A} (true \vdash P^t_f)
\end{align*}
\end{statement}
\begin{proofs}
\begin{proof}\checkt{alcc}
\begin{xflalign*}
	&=\mathbf{RA1} \circ \mathbf{H3} \circ \mathbf{H1} \circ \mathbf{RA} \circ \mathbf{A} (true \vdash P^t_f)
	&&\ptext{\cref{theorem:H3-o-H1-o-RA-o-A(true|-Ptf)}}\\
	&=\mathbf{RA1} \left(\begin{array}{l}
		true
		\\ \vdash \\
		s \in ac' \dres s.wait \rres \mathbf{RA1}\circ\mathbf{RA2}\circ\mathbf{PBMH} (P^t_f)
	\end{array}\right)
	&&\ptext{Predicate calculus and property of conditional}\\
	&=\mathbf{RA1} \left(\begin{array}{l}
		true \dres s.wait \rres true
		\\ \vdash \\
		s \in ac' \dres s.wait \rres \mathbf{RA1}\circ\mathbf{RA2}\circ\mathbf{PBMH} (P^t_f)
	\end{array}\right)
	&&\ptext{\cref{lemma:RA1-o-RA3(design):RA1(design)}}\\
	&=\mathbf{RA1}\circ\mathbf{RA3} (true \vdash \mathbf{RA1}\circ\mathbf{RA2}\circ\mathbf{PBMH} (P^t_f))
	&&\ptext{\cref{theorem:RA3-o-RA1:RA1-o-RA3}}\\
	&=\mathbf{RA3}\circ\mathbf{RA1} (true \vdash \mathbf{RA1}\circ\mathbf{RA2}\circ\mathbf{PBMH} (P^t_f))
	&&\ptext{\cref{lemma:RA1(P|-Q):RA1(P|-RA1(Q))}}\\
	&=\mathbf{RA3}\circ\mathbf{RA1} (true \vdash \mathbf{RA2}\circ\mathbf{PBMH} (P^t_f))
	&&\ptext{Predicate calculus and~\cref{lemma:RA2(P):P:s-ac'-not-free}}\\
	&=\mathbf{RA3}\circ\mathbf{RA1} (\lnot \mathbf{RA2} (false) \vdash \mathbf{RA2}\circ\mathbf{PBMH} (P^t_f))
	&&\ptext{\cref{lemma:RA2(P|-Q):(lnot-RA2(lnot-P)|-RA2(Q))}}\\
	&=\mathbf{RA3}\circ\mathbf{RA1}\circ\mathbf{RA2} (true \vdash \mathbf{PBMH} (P^t_f)))
	&&\ptext{Predicate calculus and~\cref{law:pbmh:false}}\\
	&=\mathbf{RA3}\circ\mathbf{RA1}\circ\mathbf{RA2} (\lnot \mathbf{PBMH} (false) \vdash \mathbf{PBMH} (P^t_f))
	&&\ptext{\cref{lemma:PBMH(design):(lnot-PBMH(pre)|-PBMH(post))}}\\
	&=\mathbf{RA3}\circ\mathbf{RA1}\circ\mathbf{RA2}\circ\mathbf{PBMH} (true \vdash P^t_f)
	&&\ptext{Definition of $\mathbf{RA}$ and~\cref{theorem:RA-o-A(P):RA-o-PBMH(P)}}\\
	&=\mathbf{RA}\circ\mathbf{A} (true \vdash P^t_f)
\end{xflalign*}
\end{proof}
\end{proofs}
\end{theorem}

\begin{theorem}\label{theorem:H3-o-H1-o-RA1-o-NDAP(P):ND(AP)} Provided $P$ is $\mathbf{AP}$-healthy,
\begin{statement}
\begin{align*}
	&\mathbf{H3}\circ\mathbf{H1}\circ\mathbf{RA1}\circ\mathbf{ND_{AP}}(P)
	=
	\mathbf{ND_{AP}}(P)
\end{align*}
\end{statement}
\begin{proofs}
\begin{proof}\checkt{alcc}
\begin{xflalign*}
	&\mathbf{H3}\circ\mathbf{H1}\circ\mathbf{RA1}\circ\mathbf{ND_{AP}}(P)
	&&\ptext{Definition of $\mathbf{ND_{AP}}$}\\
	&=\mathbf{H3}\circ\mathbf{H1}\circ\mathbf{RA1} (P \sqcup Choice_N)
	&&\ptext{Assumption: $P$ is $\mathbf{AP}$-healthy and~\cref{theorem:RAPN:ChoiceN-sqcup-P}}\\
	&=\mathbf{H3}\circ\mathbf{H1}\circ\mathbf{RA1} (
		true
		 \vdash 
		s \in ac' \dres s.wait \rres \mathbf{RA2}\circ\mathbf{RA1}\circ\mathbf{PBMH} (P^t_f))
	&&\ptext{\cref{lemma:RAPN(true|-Ptf)}}\\
	&=\mathbf{H3}\circ\mathbf{H1}\circ\mathbf{RA1}\circ\mathbf{AP} (true \vdash P^t_f)
	&&\ptext{\cref{theorem:RA1-o-RAPN(P):RA-o-A(Pff|-Ptf)}}\\
	&=\mathbf{H3}\circ\mathbf{H1}\circ\mathbf{RA}\circ\mathbf{A} (true \vdash P^t_f)
	&&\ptext{\cref{theorem:H3-o-H1-o-RA-o-A(true|-Ptf)}}\\
	&=(true \vdash s\in ac' \dres s.wait \rres \mathbf{RA1}\circ\mathbf{RA2}\circ\mathbf{PBMH} (P^t_f))
	&&\ptext{Assumption: $P$ is $\mathbf{AP}$-healthy and~\cref{theorem:RAPN:ChoiceN-sqcup-P}}\\
	&=Choice_{N} \sqcup P
	&&\ptext{Definition of $\mathbf{ND_{AP}}$}\\
	&=\mathbf{ND_{AP}}(P)
\end{xflalign*}
\end{proof}
\end{proofs}
\end{theorem}

\begin{lemma}\label{lemma:H1-o-RA1(P|-Q):(lnot-RA1(lnot-P)|-RA1(Q))}
\begin{statement}
$\mathbf{H1}\circ\mathbf{RA1} (P \vdash Q) = (\lnot \mathbf{RA1} (\lnot P) \vdash \mathbf{RA1} (Q))$
\end{statement}
\begin{proofs}
\begin{proof}
\begin{xflalign*}
	&\mathbf{H1}\circ\mathbf{RA1} (P \vdash Q)
	&&\ptext{Definition of design}\\
	&=\mathbf{H1}\circ\mathbf{RA1} ((ok \land P) \implies (Q \land ok'))
	&&\ptext{Predicate calculus}\\
	&=\mathbf{H1}\circ\mathbf{RA1} (\lnot ok \lor \lnot P \lor (Q \land ok'))
	&&\ptext{\cref{lemma:RA1(P-lor-Q):RA1(P)-lor-RA1(Q)}}\\
	&=\mathbf{H1} (\mathbf{RA1} (\lnot ok) \lor \mathbf{RA1} (\lnot P) \lor \mathbf{RA1} (Q \land ok'))
	&&\ptext{\cref{lemma:RA1(P-land-Q):ac'-not-free}}\\
	&=\mathbf{H1} (\mathbf{RA1} (\lnot ok) \lor \mathbf{RA1} (\lnot P) \lor (\mathbf{RA1} (Q) \land ok'))
	&&\ptext{\cref{lemma:RA1(P)-ac'-not-free:P-land-RA1(true)}}\\
	&=\mathbf{H1} ((\lnot ok \land \mathbf{RA1} (true)) \lor \mathbf{RA1} (\lnot P) \lor (\mathbf{RA1} (Q) \land ok'))
	&&\ptext{Definition of $\mathbf{H1}$}\\
	&=ok \implies ((\lnot ok \land \mathbf{RA1} (true)) \lor \mathbf{RA1} (\lnot P) \lor (\mathbf{RA1} (Q) \land ok'))
	&&\ptext{Predicate calculus}\\
	&=ok \implies (\mathbf{RA1} (\lnot P) \lor (\mathbf{RA1} (Q) \land ok'))
	&&\ptext{Predicate calculus}\\
	&=(ok \land \lnot \mathbf{RA1} (\lnot P)) \implies (\mathbf{RA1} (Q) \land ok')
	&&\ptext{Definition of design}\\
	&=(\lnot \mathbf{RA1} (\lnot P) \vdash \mathbf{RA1} (Q))
\end{xflalign*}
\end{proof}
\end{proofs}
\end{lemma}

\begin{lemma}\label{lemma:RA-o-A(true|-s-in-ac'-s.wait-Q):RA-o-A(true|-Q)}
\begin{align*}
	&\mathbf{RA}\circ\mathbf{A} (true \vdash s \in ac' \dres s.wait \rres \mathbf{RA2}\circ\mathbf{RA1}\circ\mathbf{PBMH} (Q))\\
	&=\\
	&\mathbf{RA} \circ \mathbf{A} (true \vdash Q)
\end{align*}
\begin{proofs}\begin{proof}\checkt{alcc}\checkt{pfr}
\begin{xflalign*}
	&\mathbf{RA}\circ\mathbf{A} (true \vdash s \in ac' \dres s.wait \rres \mathbf{RA2}\circ\mathbf{RA1}\circ\mathbf{PBMH} (Q))
	&&\ptext{\cref{theorem:RA-o-A(P):RA-o-PBMH(P)}}\\
	&=\mathbf{RA}\circ\mathbf{PBMH} (P \vdash s \in ac' \dres s.wait \rres \mathbf{RA2}\circ\mathbf{RA1}\circ\mathbf{PBMH} (Q))
	&&\ptext{\cref{lemma:PBMH(design):(lnot-PBMH(pre)|-PBMH(post))}}\\
	&=\mathbf{RA} \left(
\right)
	&&\ptext{Predicate calculus and~\cref{law:pbmh:false}}\\
	&=\mathbf{RA} \left(%
\right)
	&&\ptext{Property of conditional}\\
	&=\mathbf{RA2}\circ\mathbf{RA3}\circ\mathbf{RA1} \left(%
\right)
	&&\ptext{\cref{lemma:RA1-o-RA3(design):RA1(design)}}\\
	&=\mathbf{RA2}\circ\mathbf{RA3}\circ\mathbf{RA1}\circ\mathbf{RA3} (true \vdash \mathbf{PBMH} (Q))
	&&\ptext{\cref{theorem:RA3-o-RA1:RA1-o-RA3}}\\
	&=\mathbf{RA2}\circ\mathbf{RA3}\circ\mathbf{RA3}\circ\mathbf{RA1} (true \vdash \mathbf{PBMH} (Q))
	&&\ptext{\cref{theorem:RA3-idempotent}}\\
	&=\mathbf{RA2}\circ\mathbf{RA3}\circ\mathbf{RA1} (true \vdash \mathbf{PBMH} (Q))
	&&\ptext{Definition of $\mathbf{RA}$}\\
	&=\mathbf{RA} (true \vdash \mathbf{PBMH} (Q))
	&&\ptext{Predicate calculus and~\cref{law:pbmh:false}}\\
	&=\mathbf{RA} (\lnot \mathbf{PBMH} (false) \vdash \mathbf{PBMH} (Q))
	&&\ptext{Predicate calculus and~\cref{lemma:PBMH(design):(lnot-PBMH(pre)|-PBMH(post))}}\\
	&=\mathbf{RA} \circ \mathbf{PBMH} (true \vdash Q)
	&&\ptext{\cref{theorem:RA-o-A(P):RA-o-PBMH(P)}}\\
	&=\mathbf{RA} \circ \mathbf{A} (true \vdash Q)
\end{xflalign*}
\end{proof}\end{proofs}
\end{lemma}

\section{Operators}

\subsection{Angelic Choice}

\subsubsection{Closure}

\begin{theorem}\label{theorem:AP(P-sqcup-Q):P-sqcup-Q}
\begin{statement}
Provided $P$ and $Q$ are $\mathbf{AP}$-healthy,
\begin{align*}
	&\mathbf{AP} (P \sqcup_\mathbf{AP} Q) = P \sqcup_\mathbf{AP} Q
\end{align*}
\end{statement}
\begin{proofs}
\begin{proof}\checkt{alcc}
\begin{xflalign*}
	&P \sqcup Q
	&&\ptext{Assumption: $P$ and $Q$ are $\mathbf{AP}$-healthy}\\
	&=\mathbf{AP} (P) \sqcup \mathbf{AP} (Q)
	&&\ptext{\cref{lemma:RAPN(P)-sqcup-RAPN(Q)}}\\
	&=\left(
\right)
		\end{array}\right)	
	\end{array}\right)
	&&\ptext{\cref{lemma:RAPN(P)-sqcup-RAPN(Q)}}\\
	&=\mathbf{AP} (\mathbf{AP} (P) \sqcup \mathbf{AP} (Q))
	&&\ptext{Assumption: $P$ and $Q$ are $\mathbf{AP}$-healthy}\\
	&=\mathbf{AP} (P \sqcup Q)
\end{xflalign*}
\end{proof}
\end{proofs}
\end{theorem}

\begin{theorem}\label{theorem:NDAP(P-sqcup-Q):P-sqcup-Q}
\begin{statement}
Provided $P$ and $Q$ are $\mathbf{ND_{AP}}$-healthy,
\begin{align*}
	&\mathbf{ND_{AP}} (P \sqcup_{\mathbf{AP}} Q) = P \sqcup_{\mathbf{AP}} Q
\end{align*}
\end{statement}
\begin{proofs}
\begin{proof}
\begin{xflalign*}
	&\mathbf{ND_{AP}} (P \sqcup_{\mathbf{AP}} Q)
	&&\ptext{Definition of $\mathbf{ND_{AP}}$}\\
	&=Choice_{\mathbf{AP}} \sqcup_{\mathbf{AP}} (P \sqcup_{\mathbf{AP}} Q)
	&&\ptext{Definition of $\sqcup_{\mathbf{AP}}$}\\
	&=Choice_{\mathbf{AP}} \land (P \land Q)
	&&\ptext{Associativity of conjunction}\\
	&=Choice_{\mathbf{AP}} \land P \land Q
	&&\ptext{Predicate calculus}\\
	&=(Choice_{\mathbf{AP}} \land P) \land (Choice_{\mathbf{AP}} \land Q)
	&&\ptext{Definition of $\sqcup_{\mathbf{AP}}$}\\
	&=(Choice_{\mathbf{AP}} \sqcup_{\mathbf{AP}} P) \sqcup_{\mathbf{AP}} (Choice_{\mathbf{AP}} \sqcup_{\mathbf{AP}} Q)
	&&\ptext{Definition of $\mathbf{ND_{AP}}$}\\
	&=\mathbf{ND_{AP}} (P) \sqcup_{\mathbf{AP}} \mathbf{ND_{AP}} (Q)
	&&\ptext{Assumption: $P$ and $Q$ are $\mathbf{ND_{AP}}$-healthy}\\
	&=P \sqcup_{\mathbf{AP}} Q
\end{xflalign*}
\end{proof}
\end{proofs}
\end{theorem}

\begin{lemma}\label{lemma:RAPN(P)-sqcup-RAPN(Q)}
\begin{align*}
	&\mathbf{AP} (P) \sqcup \mathbf{AP} (Q) \\
	&=\\
	&\left(
\right)
	&&\ptext{Conjunction of designs (\cref{law:designs:conjunction-alternative})}\\
	&=\left(%
\right)
	&&\ptext{Property of conditional (\cref{lemma:lnot(conditional)}) and predicate calculus}\\
	&=\left(%
\right)
	&&\ptext{Property of conditional and predicate calculus}\\
	&=\left(%
\right)
	&&\ptext{Property of conditional and predicate calculus}\\
	&=\left(%
\right)
	&&\ptext{\cref{theorem:RA2-o-RA1:RA1-o-RA2}}\\
	&&\ptext{\cref{lemma:RA1-o-RA2(P):implies:RA1(true),lemma:RA1(s-in-ac'):s-in-ac',lemma:RA1(P):implies:RA1(true)} and predicate calculus}\\
	&=\left(%
\right)
		\end{array}\right)	
		\end{array}\right)
	\end{array}\right)
\end{xflalign*}
\end{proof}\end{proofsbig}
\end{lemma}

\subsubsection{Linking}

\begin{theorem}\label{theorem:RA1(H1(P)-sqcup-H1(Q)):P-sqcup-Q}
\begin{statement}
Provided $P$ and $Q$ are $\mathbf{RAD}$-healthy,
\begin{align*}
	&\mathbf{RA1} (\mathbf{H1} (P) \sqcup_\mathbf{AP} \mathbf{H1} (Q)) = P \sqcup_\mathbf{RAD} Q
\end{align*}
\end{statement}
\begin{proofs}
\begin{proof}\checkt{alcc}
\begin{xflalign*}
	&\mathbf{RA1} (\mathbf{H1} (P) \sqcup \mathbf{H1} (Q))
	&&\ptext{Definition of $\sqcup$}\\
	&=\mathbf{RA1} (\mathbf{H1} (P) \land \mathbf{H1} (Q))
	&&\ptext{\cref{lemma:RA1(P-land-Q):RA1(P)-land-RA1(Q)}}\\
	&=\mathbf{RA1} \circ \mathbf{H1} (P) \land \mathbf{RA1} \circ \mathbf{H1} (Q)
	&&\ptext{Assumption: $P$ and $Q$ are $\mathbf{RAD}$-healthy}\\
	&=\mathbf{RA1} \circ \mathbf{H1} \circ \mathbf{RAD} (P) \land \mathbf{RA1} \circ \mathbf{H1} \circ \mathbf{RAD} (Q)
	&&\ptext{\cref{theorem:RA1-o-H1-o-RAP(P):RAP(P)}}\\
	&=\mathbf{RAD} (P) \land \mathbf{RAD} (Q)
	&&\ptext{Assumption: $P$ and $Q$ are $\mathbf{RAD}$-healthy}\\
	&=P \land Q
	&&\ptext{Definition of $\sqcup$}\\
	&=P \sqcup Q
\end{xflalign*}
\end{proof}
\end{proofs}
\end{theorem}

\begin{theorem}\label{theorem:H1(RA1(P)-sqcup-RA1(Q)):sqsupseteq:P-sqcup-Q} 
\begin{statement}
Provided $P$ and $Q$ are $\mathbf{AP}$-healthy,
\begin{align*}
	&\mathbf{H1} (\mathbf{RA1} (P) \sqcup_\mathbf{RAD} \mathbf{RA1} (Q)) \sqsupseteq P \sqcup_\mathbf{AP} Q
\end{align*}
\end{statement}
\begin{proofs}
\begin{proof}\checkt{alcc}
\begin{xflalign*}
	&\mathbf{H1} (\mathbf{RA1} (P) \sqcup \mathbf{RA1} (Q))
	&&\ptext{Definition of $\sqcup$}\\
	&=\mathbf{H1} (\mathbf{RA1} (P) \land \mathbf{RA1} (Q))
	&&\ptext{\cref{lemma:H1(P-land-Q):H1(P)-land-H1(Q)}}\\
	&=\mathbf{H1} \circ \mathbf{RA1} (P) \land \mathbf{H1} \circ \mathbf{RA1} (Q)
	&&\ptext{Assumption: $P$ and $Q$ are $\mathbf{AP}$-healthy}\\
	&=\mathbf{H1} \circ \mathbf{RA1} \circ \mathbf{AP} (P) \land \mathbf{H1} \circ \mathbf{RA1} \circ \mathbf{AP} (Q)
	&&\ptext{\cref{theorem:H1-o-RA1-o-RAPN(P):sqsupseteq:RAPN(P)}}\\
	&\sqsupseteq \mathbf{AP} (P) \land \mathbf{AP} (Q)
	&&\ptext{Assumption: $P$ and $Q$ are $\mathbf{AP}$-healthy}\\
	&=P \land Q
	&&\ptext{Definition of $\sqcup$}\\
	&=P \sqcup Q
\end{xflalign*}
\end{proof}
\end{proofs}
\end{theorem}

\subsection{Demonic Choice}

\subsubsection{Closure}

\begin{theorem}\label{theorem:RAPN(P-sqcap-Q):closure}
\begin{statement}
Provided $P$ and $Q$ are $\mathbf{AP}$-healthy,
$\mathbf{AP} (P \sqcap Q) = P \sqcap Q$.
\end{statement}
\begin{proofs}
\begin{proof}\checkt{alcc}
\begin{xflalign*}
	&\mathbf{AP} (P \sqcap Q)
	&&\ptext{Assumption: $P$ and $Q$ are $\mathbf{AP}$-healthy}\\
	&=\mathbf{AP} (\mathbf{AP} (P) \sqcap \mathbf{AP} (Q))
	&&\ptext{\cref{lemma:RAPN(P)-sqcap-RAPN(Q)}}\\
	&=\mathbf{AP} \circ \mathbf{AP} (\lnot P^f_f \land \lnot Q^f_f \vdash P^t_f \lor Q^t_f)
	&&\ptext{\cref{theorem:RAPN:idempotent}}\\
	&=\mathbf{AP} (\lnot P^f_f \land \lnot Q^f_f \vdash P^t_f \lor Q^t_f)
	&&\ptext{\cref{lemma:RAPN(P)-sqcap-RAPN(Q)}}\\
	&=\mathbf{AP} (P) \sqcap \mathbf{AP} (Q)
	&&\ptext{Assumption: $P$ and $Q$ are $\mathbf{AP}$-healthy}\\
	&=P \sqcap Q
\end{xflalign*}
\end{proof}
\end{proofs}
\end{theorem}

\begin{theorem}\label{theorem:NDAP(P-sqcap-Q):P-sqcap-Q}
\begin{statement}
Provided $P$ and $Q$ are $\mathbf{ND_{AP}}$-healthy,
\begin{align*}
	&\mathbf{ND_{AP}} (P \sqcap_{\mathbf{AP}} Q) = P \sqcap_{\mathbf{AP}} Q
\end{align*}
\end{statement}
\begin{proofs}
\begin{proof}
\begin{xflalign*}
	&\mathbf{ND_{AP}} (P \sqcap_{\mathbf{AP}} Q)
	&&\ptext{Definition of $\mathbf{ND_{AP}}$}\\
	&=Choice_{\mathbf{AP}} \sqcup_{\mathbf{AP}} (P \sqcap_{\mathbf{AP}} Q)
	&&\ptext{Definition of $\sqcup_{\mathbf{AP}}$ and $\sqcap_{\mathbf{AP}}$}\\
	&=Choice_{\mathbf{AP}} \land (P \lor Q)
	&&\ptext{Predicate calculus}\\
	&=(Choice_{\mathbf{AP}} \land P) \lor (Choice_{\mathbf{AP}} \land Q)
	&&\ptext{Definition of $\sqcup_{\mathbf{AP}}$ and $\sqcap_{\mathbf{AP}}$}\\
	&=(Choice_{\mathbf{AP}} \sqcup_{\mathbf{AP}} P) \sqcap_{\mathbf{AP}} (Choice_{\mathbf{AP}} \sqcup_{\mathbf{AP}} Q)
	&&\ptext{Definition of $\mathbf{ND_{AP}}$}\\
	&=\mathbf{ND_{AP}} (P) \sqcap_{\mathbf{AP}} \mathbf{ND_{AP}} (Q)
	&&\ptext{Assumption: $P$ and $Q$ are $\mathbf{ND_{AP}}$-healthy}\\
	&=P \sqcap_{\mathbf{AP}} Q
\end{xflalign*}
\end{proof}
\end{proofs}
\end{theorem}

\begin{lemma}\label{lemma:RAPN(P)-sqcap-RAPN(Q)}
\begin{align*}
	&\mathbf{AP} (P) \sqcap \mathbf{AP} (Q) = \mathbf{AP} (\lnot P^f_f \land \lnot Q^f_f \vdash P^t_f \lor Q^t_f)
\end{align*}
\begin{proofs}\begin{proof}\checkt{alcc}
\begin{xflalign*}
	&\mathbf{AP} (P) \sqcap \mathbf{AP} (Q)
	&&\ptext{\cref{lemma:RAPN(lnot-Pff|-Ptf):design}}\\
	&=\left(
\right)
	&&\ptext{Disjunction of designs}\\
	&=\left(%
\right)
	&&\ptext{Property of conditional and predicate calculus}\\
	&=\left(%
\right)
	&&\ptext{\cref{lemma:RAPN(lnot-Pff|-Ptf):design}}\\
	&=\mathbf{AP} (\lnot (P^f_f \lor Q^f_f) \vdash P^t_f \lor Q^t_f)
	&&\ptext{Predicate calculus}\\
	&=\mathbf{AP} (\lnot P^f_f \land \lnot Q^f_f \vdash P^t_f \lor Q^t_f)
\end{xflalign*}
\end{proof}\end{proofs}
\end{lemma}

\subsubsection{Linking}

\begin{theorem}\label{theorem:RA1(H1(P)-sqcap-H1(Q)):P-sqcap-Q}
\begin{statement}
Provided $P$ and $Q$ $\mathbf{RAD}$-healthy,
\begin{align*}
	&\mathbf{RA1} (\mathbf{H1} (P) \sqcap_{\mathbf{AP}} \mathbf{H1} (Q)) = P \sqcap_{\mathbf{RAD}} Q
\end{align*}
\end{statement}
\begin{proofs}
\begin{proof}\checkt{alcc}
\begin{xflalign*}
	&\mathbf{RA1} (\mathbf{H1} (P) \sqcap_{\mathbf{AP}} \mathbf{H1} (Q))
	&&\ptext{Definition of $\sqcap_{\mathbf{AP}}$}\\
	&=\mathbf{RA1} (\mathbf{H1} (P) \lor \mathbf{H1} (Q))
	&&\ptext{\cref{lemma:RA1(P-lor-Q):RA1(P)-lor-RA1(Q)}}\\
	&=\mathbf{RA1} \circ \mathbf{H1} (P) \lor \mathbf{RA1}\circ\mathbf{H1} (Q)
	&&\ptext{Assumption: $P$ and $Q$ are $\mathbf{RAD}$-healthy}\\
	&=\mathbf{RA1} \circ \mathbf{H1}\circ\mathbf{RAD} (P) \lor \mathbf{RA1}\circ\mathbf{H1}\circ\mathbf{RAD} (Q)
	&&\ptext{\cref{theorem:RA1-o-H1-o-RAP(P):RAP(P)}}\\
	&=\mathbf{RAD} (P) \lor \mathbf{RAD} (Q)
	&&\ptext{Assumption: $P$ and $Q$ are $\mathbf{RAD}$-healthy}\\
	&=P \lor Q
	&&\ptext{Definition of $\sqcap_{\mathbf{RAD}}$}\\
	&=P \sqcap_{\mathbf{RAD}} Q
\end{xflalign*}
\end{proof}
\end{proofs}
\end{theorem}

\begin{theorem}\label{theorem:H1(RA1(P)-sqcap-RA1(Q)):sqsupseteq:P-sqcap-Q}
\begin{statement}
Provided $P$ and $Q$ are $\mathbf{AP}$-healthy,
\begin{align*}
	&\mathbf{H1} (\mathbf{RA1} (P) \sqcap_{\mathbf{RAD}} \mathbf{RA1} (Q)) \sqsupseteq P \sqcap_{\mathbf{AP}} Q
\end{align*}
\end{statement}
\begin{proofs}
\begin{proof}\checkt{alcc}
\begin{xflalign*}
	&\mathbf{H1} (\mathbf{RA1} (P) \sqcap_{\mathbf{RAD}} \mathbf{RA1} (Q))
	&&\ptext{Definition of $\sqcap_{\mathbf{RAD}}$}\\
	&=\mathbf{H1} (\mathbf{RA1} (P) \lor \mathbf{RA1} (Q))
	&&\ptext{\cref{lemma:H1(P-lor-Q):H1(P)-lor-H1(Q)}}\\
	&=\mathbf{H1}\circ\mathbf{RA1} (P) \lor \mathbf{H1}\circ\mathbf{RA1} (Q)
	&&\ptext{Assumption: $P$ and $Q$ are $\mathbf{AP}$-healthy}\\
	&=\mathbf{H1}\circ\mathbf{RA1}\circ\mathbf{AP} (P) \lor \mathbf{H1}\circ\mathbf{RA1}\circ\mathbf{AP} (Q) 
	&&\ptext{\cref{theorem:H1-o-RA1-o-RAPN(P):sqsupseteq:RAPN(P)}}\\
	&\sqsupseteq \mathbf{AP} (P) \lor \mathbf{AP} (Q)
	&&\ptext{Assumption: $P$ and $Q$ are $\mathbf{AP}$-healthy}\\
	&=P \lor Q
	&&\ptext{Definition of $\sqcap_{\mathbf{AP}}$}\\
	&=P \sqcap_{\mathbf{AP}} Q
\end{xflalign*}
\end{proof}
\end{proofs}
\end{theorem}

\subsection{Divergence: Chaos and Chaos of CSP}

\begin{theorem}\label{theorem:AP:P-sqcup-ChaosAP:P}
\begin{statement}
Provided $P$ is $\mathbf{AP}$-healthy,
$P \sqcup_{\mathbf{AP}} Chaos_{\mathbf{AP}} = P$
\end{statement}
\begin{proofs}
\begin{proof}
\begin{xflalign*}
	&P \sqcup_{\mathbf{AP}} Chaos_{\mathbf{AP}}
	&&\ptext{Assumption: $P$ is $\mathbf{AP}$-healthy and~\cref{theorem:RAPN(P)}}\\
	&=\left(
\right)
	&&\ptext{Definition of $\sqcup_{\mathbf{AP}}$ and conjunction of designs}\\
	&=\left(%
\right)
	&&\ptext{Definition of conditional and predicate calculus}\\
	&=\left(%
\right)
	&&\ptext{Definition of design and predicate calculus}\\
	&=\left(%
\right)
	&\ptext{Assumption: $P$ is $\mathbf{AP}$-healthy and~\cref{theorem:RAPN(P)}}\\
	&=P
\end{xflalign*}
\end{proof}
\end{proofs}
\end{theorem}

\begin{theorem}\label{theorem:H1(ChaosRA)}
\begin{statement}
$\mathbf{H1} (Chaos_{\mathbf{RAD}}) = ChaosCSP_{\mathbf{AP}}$
\end{statement}
\begin{proofs}
\begin{proof}\checkt{alcc}
\begin{xflalign*}
	&\mathbf{H1} (Chaos_{\mathbf{RAD}})
	&&\ptext{Definition of $Chaos_{\mathbf{RAD}}$}\\
	&=\mathbf{H1} \circ \mathbf{RA} \circ \mathbf{A} (false \vdash true)
	&&\ptext{\cref{theorem:H1-o-RAP(P)}}\\
	&=\left(\begin{array}{l}
		true \dres s.wait \rres \lnot \mathbf{RA1}\circ\mathbf{RA2}\circ\mathbf{PBMH} (true))
		\\ \vdash \\
		s \in ac' \dres s.wait \rres \mathbf{RA1}\circ\mathbf{RA2}\circ\mathbf{PBMH} (true))
	\end{array}\right)
	&&\ptext{\cref{law:pbmh:true}}\\
	&=\left(\begin{array}{l}
		true \dres s.wait \rres \lnot \mathbf{RA1}\circ\mathbf{RA2}(true))
		\\ \vdash \\
		s \in ac' \dres s.wait \rres \mathbf{RA1}\circ\mathbf{RA2}(true))
	\end{array}\right)
	&&\ptext{\cref{lemma:RA2(true):true}}\\
	&=\left(\begin{array}{l}
		true \dres s.wait \rres \lnot \mathbf{RA1}(true))
		\\ \vdash \\
		s \in ac' \dres s.wait \rres \mathbf{RA1}(true))
	\end{array}\right)
	&&\ptext{Predicate calculus}\\
	&=\left(\begin{array}{l}
		true \dres s.wait \rres \lnot \mathbf{RA1}(true))
		\\ \vdash \\
		(true \dres s.wait \rres \lnot \mathbf{RA1}(true)) \land (s \in ac' \dres s.wait \rres \mathbf{RA1}(true))
	\end{array}\right)
	&&\ptext{Property of conditional}\\
	&=\left(\begin{array}{l}
		true \dres s.wait \rres \lnot \mathbf{RA1}(true))
		\\ \vdash \\
		(true \land s\in ac') \dres s.wait \rres (\lnot \mathbf{RA1}(true) \land \mathbf{RA1}(true))
	\end{array}\right)
	&&\ptext{Predicate calculus}\\
	&=\left(\begin{array}{l}
		true \dres s.wait \rres \lnot \mathbf{RA1}(true))
		\\ \vdash \\
		(s\in ac' \dres s.wait \rres false)
	\end{array}\right)
	&&\ptext{Predicate calculus}\\
	&=\left(\begin{array}{l}
		s.wait \lor (\lnot s.wait \land \lnot \mathbf{RA1}(true))
		\\ \vdash \\
		(s.wait \land s\in ac')
	\end{array}\right)
	&&\ptext{Predicate calculus}\\
	&=(s.wait \lor \lnot \mathbf{RA1}(true) \vdash s.wait \land s\in ac')
	&&\ptext{Definition of $ChaosCSP_{\mathbf{AP}}$ and~\cref{lemma:AP:ChaosCSP}}\\
	&=ChaosCSP_{\mathbf{AP}}
\end{xflalign*}
\end{proof}
\end{proofs}
\end{theorem}

\begin{theorem}\label{theorem:RA1(ChaosCSPAP):ChaosRAD}
\begin{statement}
$\mathbf{RA1} (ChaosCSP_{\mathbf{AP}}) = Chaos_{\mathbf{RAD}}$
\end{statement}
\begin{proofs}
\begin{proof}
\begin{xflalign*}
	&\mathbf{RA1} (ChaosCSP_{\mathbf{AP}})
	&&\ptext{Definition of $Chaos_{\mathbf{AP}}$}\\
	&=\mathbf{RA1}\circ\mathbf{AP} (\lnot \mathbf{RA1} (true) \vdash true)
	&&\ptext{\cref{theorem:RA1-o-RAPN(P):RA-o-A(Pff|-Ptf)}}\\
	&=\mathbf{RA}\circ\mathbf{A} (\lnot (\lnot \mathbf{RA1} (true) \vdash true)^f_f \vdash (\lnot \mathbf{RA1} (true) \vdash true)^t_f)
	&&\ptext{\cref{lemma:design:(lnot-(P|-Q)ff|-(P|-Q)tf)}}\\
	&=\mathbf{RA}\circ\mathbf{A} (\lnot \mathbf{RA1} (true)_f \vdash true_f)
	&&\ptext{\cref{lemma:RA1(P)-o-w-subs:RA1(P-o-w-subs)}}\\
	&=\mathbf{RA}\circ\mathbf{A} (\lnot \mathbf{RA1} (true_f) \vdash true_f)
	&&\ptext{Substitution}\\
	&=\mathbf{RA}\circ\mathbf{A} (\lnot \mathbf{RA1} (true) \vdash true)
	&&\ptext{\cref{law:pbmh:P:ac'-not-free}}\\
	&=\mathbf{RA}\circ\mathbf{A} (\lnot \mathbf{RA1}\circ\mathbf{PBMH} (true) \vdash true)
	&&\ptext{\cref{lemma:RA-o-A(lnot-RA1-o-PBMH(P)|-Q):RA-o-A(lnot-P|-Q)}}\\
	&=\mathbf{RA}\circ\mathbf{A} (\lnot true \vdash true)
	&&\ptext{Predicate calculus}\\
	&=\mathbf{RA}\circ\mathbf{A} (false \vdash true)
	&&\ptext{Definition of $Chaos_{\mathbf{RAD}}$}\\
	&=Chaos_{\mathbf{RAD}}
\end{xflalign*}
\end{proof}
\end{proofs}
\end{theorem}

\begin{theorem}\label{theorem:H3-o-H1(ChaosRAD):ChaosAP}
\begin{statement}
$\mathbf{H3}\circ\mathbf{H1} (Chaos_{\mathbf{RAD}}) = Chaos_{\mathbf{AP}}$
\end{statement}
\begin{proofs}
\begin{proof}
\begin{xflalign*}
	&\mathbf{H3}\circ\mathbf{H1} (Chaos_{\mathbf{RAD}})
	&&\ptext{Definition of $Chaos_{\mathbf{RAD}}$}\\
	&=\mathbf{H3}\circ\mathbf{H1}\circ\mathbf{RA}\circ\mathbf{A} (false \vdash true)
	&&\ptext{\cref{theorem:RA-o-A(design):RA-CSPA-PBMH,theorem:H3-o-H1-o-RAP(P)}}\\
	&=\left(\begin{array}{l}
		true \dres s.wait \rres \lnot \exists ac' \spot \mathbf{RA1}\circ\mathbf{RA2}\circ\mathbf{PBMH} (true)
		\\ \vdash \\
		s \in ac' \dres s.wait \rres \mathbf{RA1}\circ\mathbf{RA2}\circ\mathbf{PBMH} (true)
	\end{array}\right)
	&&\ptext{\cref{law:pbmh:P:ac'-not-free}}\\
	&=\left(\begin{array}{l}
		true \dres s.wait \rres \lnot \exists ac' \spot \mathbf{RA1}\circ\mathbf{RA2} (true)
		\\ \vdash \\
		s \in ac' \dres s.wait \rres \mathbf{RA1}\circ\mathbf{RA2} (true)
	\end{array}\right)
	&&\ptext{\cref{lemma:RA2(P):P:s-ac'-not-free}}\\
	&=\left(\begin{array}{l}
		true \dres s.wait \rres \lnot \exists ac' \spot \mathbf{RA1} (true)
		\\ \vdash \\
		s \in ac' \dres s.wait \rres \mathbf{RA1} (true)
	\end{array}\right)
	&&\ptext{\cref{lemma:RA1(true)}}\\
	&=\left(\begin{array}{l}
		true \dres s.wait \rres \lnot \exists ac' \spot \exists z @ s.tr \le z.tr \land z \in ac'
		\\ \vdash \\
		s \in ac' \dres s.wait \rres \mathbf{RA1} (true)
	\end{array}\right)
	&&\ptext{Predicate calculus}\\
	&=\left(\begin{array}{l}
		true \dres s.wait \rres false
		\\ \vdash \\
		s \in ac' \dres s.wait \rres \mathbf{RA1} (true)
	\end{array}\right)
	&&\ptext{Definition of conditional and predicate calculus}\\
	&=(s.wait \vdash s \in ac' \dres s.wait \rres \mathbf{RA1} (true))
	&&\ptext{Definition of design and predicate calculus}\\
	&=(s.wait \vdash s \in ac')
	&&\ptext{Definition of $Chaos_{\mathbf{AP}}$}\\
	&=Chaos_{\mathbf{AP}}
\end{xflalign*}
\end{proof}
\end{proofs}
\end{theorem}

\begin{lemma}\label{lemma:ChaosRAPN:design}
\begin{statement}
$Chaos_{\mathbf{AP}} = (s.wait \vdash s \in ac')$
\end{statement}
\begin{proofs}
\begin{proof}\checkt{alcc}
\begin{xflalign*}
	&Chaos_{\mathbf{AP}}
	&&\ptext{Definition of $Chaos_{\mathbf{AP}}$}\\
	&=\mathbf{AP} (false \vdash true)
	&&\ptext{\cref{lemma:RAPN(lnot-Pff|-Ptf):design}}\\
	&=\left(\begin{array}{l} 
		true \dres s.wait \rres \lnot \mathbf{RA2} \circ \mathbf{PBMH} (true)
		\\ \vdash \\
		s \in ac' \dres s.wait \rres \mathbf{RA2} \circ \mathbf{RA1} \circ \mathbf{PBMH} (true)
	\end{array}\right)
	&&\ptext{\cref{law:pbmh:true}}\\
	&=\left(\begin{array}{l} 
		true \dres s.wait \rres \lnot \mathbf{RA2} (true)
		\\ \vdash \\
		s \in ac' \dres s.wait \rres \mathbf{RA2} \circ \mathbf{RA1} \circ \mathbf{PBMH} (true)
	\end{array}\right)
	&&\ptext{\cref{lemma:RA2(P):P:s-ac'-not-free} and predicate calculus}\\
	&=\left(\begin{array}{l} 
		true \dres s.wait \rres false
		\\ \vdash \\
		s \in ac' \dres s.wait \rres \mathbf{RA2} \circ \mathbf{RA1} \circ \mathbf{PBMH} (true)
	\end{array}\right)
	&&\ptext{Definition of conditional and predicate calculus}\\
	&=(s.wait \vdash s \in ac' \dres s.wait \rres \mathbf{RA2} \circ \mathbf{RA1} \circ \mathbf{PBMH} (true))
	&&\ptext{Definition of design, predicate calculus and definition of conditional}\\
	&=(s.wait \vdash s \in ac')
\end{xflalign*}
\end{proof}
\end{proofs}
\end{lemma}

\begin{lemma}\label{lemma:AP:ChaosCSP}
\begin{statement}
$ChaosCSP_{\mathbf{AP}} = (s.wait \lor \lnot \mathbf{RA1}(true) \vdash s.wait \land s\in ac')$
\end{statement}
\begin{proofs}
\begin{proof}\checkt{alcc}
\begin{xflalign*}
	&ChaosCSP_{\mathbf{AP}}
	&&\ptext{Definition of $ChaosCSP_{\mathbf{AP}}$}\\
	&=\mathbf{AP} (\lnot \mathbf{RA1} (true) \vdash true)
	&&\ptext{\cref{lemma:RAPN(lnot-Pff|-Ptf):design}}\\
	&=\left(\begin{array}{l} 
		true \dres s.wait \rres \lnot \mathbf{RA2} \circ \mathbf{PBMH} \circ \mathbf{RA1} (true)
		\\ \vdash \\
		s \in ac' \dres s.wait \rres \mathbf{RA2} \circ \mathbf{RA1} \circ \mathbf{PBMH} (true)
	\end{array}\right)
	&&\ptext{\cref{law:pbmh:true,theorem:PBMH-o-RA1(P):RA1(P)}}\\
	&=\left(\begin{array}{l} 
		true \dres s.wait \rres \lnot \mathbf{RA2} \circ \mathbf{RA1} (true)
		\\ \vdash \\
		s \in ac' \dres s.wait \rres \mathbf{RA2} \circ \mathbf{RA1} (true)
	\end{array}\right)
	&&\ptext{\cref{theorem:RA2-o-RA1:RA1-o-RA2}}\\
	&=\left(\begin{array}{l} 
		true \dres s.wait \rres \lnot \mathbf{RA1} \circ \mathbf{RA2} (true)
		\\ \vdash \\
		s \in ac' \dres s.wait \rres \mathbf{RA1} \circ \mathbf{RA2} (true)
	\end{array}\right)
	&&\ptext{\cref{lemma:RA2(P):P:s-ac'-not-free}}\\
	&=\left(\begin{array}{l} 
		true \dres s.wait \rres \lnot \mathbf{RA1} (true)
		\\ \vdash \\
		s \in ac' \dres s.wait \rres \mathbf{RA1} (true)
	\end{array}\right)
	&&\ptext{Property of designs}\\
	&=\left(\begin{array}{l}
		true \dres s.wait \rres \lnot \mathbf{RA1}(true))
		\\ \vdash \\
		(true \dres s.wait \rres \lnot \mathbf{RA1}(true)) \land (s \in ac' \dres s.wait \rres \mathbf{RA1}(true))
	\end{array}\right)
	&&\ptext{Property of conditional}\\
	&=\left(\begin{array}{l}
		true \dres s.wait \rres \lnot \mathbf{RA1}(true))
		\\ \vdash \\
		(true \land s\in ac') \dres s.wait \rres (\lnot \mathbf{RA1}(true) \land \mathbf{RA1}(true))
	\end{array}\right)
	&&\ptext{Predicate calculus}\\
	&=\left(\begin{array}{l}
		true \dres s.wait \rres \lnot \mathbf{RA1}(true))
		\\ \vdash \\
		(s\in ac' \dres s.wait \rres false)
	\end{array}\right)
	&&\ptext{Predicate calculus}\\
	&=\left(\begin{array}{l}
		s.wait \lor (\lnot s.wait \land \lnot \mathbf{RA1}(true))
		\\ \vdash \\
		(s.wait \land s\in ac')
	\end{array}\right)
	&&\ptext{Predicate calculus}\\
	&=(s.wait \lor \lnot \mathbf{RA1}(true) \vdash s.wait \land s\in ac')	
\end{xflalign*}
\end{proof}
\end{proofs}
\end{lemma}

\subsection{Choice}

\subsubsection{Properties}

\begin{lemma}\label{lemma:AP(true|-ac'neq-emptyset)}
\begin{statement}
$\mathbf{AP} (true \vdash ac'\neq\emptyset) = (true\vdash s \in ac' \dres s.wait \rres \mathbf{RA1} (true))$
\end{statement}
\begin{proofs}
\begin{proof}
\begin{xflalign*}
	&\mathbf{AP} (true \vdash ac'\neq\emptyset)
	&&\ptext{\cref{lemma:RAPN(true|-ac'neq-emptyset)}}\\
	&=\left(
\right)
	&&\ptext{Property of conditional}\\
	&=(true\vdash s \in ac' \dres s.wait \rres \mathbf{RA1} (true))
\end{xflalign*}
\end{proof}
\end{proofs}
\end{lemma}

\begin{lemma}\label{lemma:RAPN(true|-ac'neq-emptyset)}
\begin{statement}
\begin{align*}
	&\mathbf{AP} (true \vdash ac'\neq\emptyset) \\
	&=\\
	&\left(%
\right)
\end{xflalign*}
\end{proof}
\end{proofs}
\end{lemma}

\subsubsection{Linking}

\begin{theorem}\label{theorem:H1(ChoiceRAP):ChoiceAP}
\begin{statement}
$\mathbf{H1} (Choice_{\mathbf{RAD}}) = Choice_{\mathbf{AP}}$
\end{statement}
\begin{proofs}
\begin{proof}\checkt{alcc}
\begin{xflalign*}
	&\mathbf{H1} (Choice_{\mathbf{RAD}})
	&&\ptext{Definition of $Choice_{\mathbf{RAD}}$}\\
	&=\mathbf{H1} \circ \mathbf{RA} \circ \mathbf{A} (true \vdash ac'\neq\emptyset)
	&&\ptext{\cref{lemma:H1-o-RA-o-A(true|-Ptf):RAPN(true|-Ptf)}}\\
	&=\mathbf{AP} (true \vdash ac'\neq\emptyset)
	&&\ptext{Definition of $Choice_{\mathbf{AP}}$}\\
	&=Choice_{\mathbf{AP}}
\end{xflalign*}
\end{proof}
\end{proofs}
\end{theorem}

\begin{theorem}\label{theorem:RA1(ChoiceAP):ChoiceRAD}
\begin{statement}
$\mathbf{RA1} (Choice_{\mathbf{AP}}) = Choice_{\mathbf{RAD}}$
\end{statement}
\begin{proofs}
\begin{proof}
\begin{xflalign*}
	&\mathbf{RA1} (Choice_{\mathbf{AP}})
	&&\ptext{Definition of $Choice_{\mathbf{AP}}$}\\
	&=\mathbf{RA1}\circ\mathbf{AP} (true \vdash ac'\neq\emptyset)
	&&\ptext{\cref{theorem:RA1-o-RAPN(P):RA-o-A(Pff|-Ptf)}}\\
	&=\mathbf{RA}\circ\mathbf{A} (\lnot (true \vdash ac'\neq\emptyset)^f_f \vdash (true \vdash ac'\neq\emptyset)^t_f)
	&&\ptext{\cref{lemma:design:(lnot-(P|-Q)ff|-(P|-Q)tf)}}\\
	&=\mathbf{RA}\circ\mathbf{A} (true_f \vdash (ac'\neq\emptyset)_f)
	&&\ptext{Substitution}\\
	&=\mathbf{RA}\circ\mathbf{A} (true \vdash ac'\neq\emptyset)
	&&\ptext{Definition of $Choice_{\mathbf{RAD}}$}\\
	&=Choice_{\mathbf{RAD}}
\end{xflalign*}
\end{proof}
\end{proofs}
\end{theorem}

\subsection{Stop}

\begin{theorem}\label{theorem:H1(StopRA)}
\begin{statement}
$\mathbf{H1} (Stop_{\mathbf{RAD}}) = Stop_{\mathbf{AP}}$
\end{statement}
\begin{proofs}
\begin{proof}\checkt{alcc}
\begin{xflalign*}
	&\mathbf{H1} (Stop_{\mathbf{RAD}})
	&&\ptext{Definition of $Stop_{\mathbf{RAD}}$}\\
	&=\mathbf{H1} \circ \mathbf{RA} \circ \mathbf{A} (true \vdash \circledIn{y}{ac'} (y.tr=s.tr \land y.wait))
	&&\ptext{\cref{lemma:H1-o-RA-o-A(true|-Ptf):RAPN(true|-Ptf)}}\\
	&=\mathbf{AP} (true \vdash \circledIn{y}{ac'} (y.tr=s.tr \land y.wait))
	&&\ptext{Definition of $Stop_{\mathbf{AP}}$}\\
	&=Stop_{\mathbf{AP}}
\end{xflalign*}
\end{proof}
\end{proofs}
\end{theorem}

\begin{theorem}\label{theorem:RA1(StopAP)}
\begin{statement}
$\mathbf{RA1} (Stop_{\mathbf{AP}}) = Stop_{\mathbf{RAD}}$
\end{statement}
\begin{proofs}
\begin{proof}
\begin{xflalign*}
	&\mathbf{RA1} (Stop_{\mathbf{AP}})
	&&\ptext{Definition of $Stop_{\mathbf{AP}}$}\\
	&=\mathbf{RA1} \circ \mathbf{AP} (true \vdash \circledIn{y}{ac'} (y.tr=s.tr \land y.wait))
	&&\ptext{\cref{theorem:RA1-o-RAPN(P):RA-o-A(Pff|-Ptf)}}\\
	&=\mathbf{RA} \circ \mathbf{A} \left(\begin{array}{l}
		\lnot (true \vdash \circledIn{y}{ac'} (y.tr=s.tr \land y.wait))^f_f
		\\ \vdash\\
		(true \vdash \circledIn{y}{ac'} (y.tr=s.tr \land y.wait))^t_f
	\end{array}\right)
	&&\ptext{\cref{lemma:design:(lnot-(P|-Q)ff|-(P|-Q)tf)}}\\
	&=\mathbf{RA} \circ \mathbf{A} (true_f \vdash (\circledIn{y}{ac'} (y.tr=s.tr \land y.wait))_f)
	&&\ptext{Substitution}\\
	&=\mathbf{RA} \circ \mathbf{A} (true \vdash \circledIn{y}{ac'} (y.tr=s.tr \land y.wait))
	&&\ptext{Definition of $Stop_{\mathbf{RAD}}$}\\
	&=Stop_{\mathbf{RAD}}
\end{xflalign*}
\end{proof}
\end{proofs}
\end{theorem}

\subsection{Skip}

\begin{theorem}\label{theorem:H1(SkipRA)}
\begin{statement}
$\mathbf{H1} (Skip_{\mathbf{RAD}}) = Skip_{\mathbf{AP}}$
\end{statement}
\begin{proofs}
\begin{proof}\checkt{alcc}
\begin{xflalign*}
	&\mathbf{H1} (Skip_{\mathbf{RAD}})
	&&\ptext{Definition of $Skip_{\mathbf{RAD}}$}\\
	&=\mathbf{H1} \circ \mathbf{RA} \circ \mathbf{A} (true \vdash \circledIn{y}{ac'} (\lnot y.wait \land y.tr=s.tr))
	&&\ptext{\cref{lemma:H1-o-RA-o-A(true|-Ptf):RAPN(true|-Ptf)}}\\
	&=\mathbf{AP} (true \vdash \circledIn{y}{ac'} (\lnot y.wait \land y.tr=s.tr))
	&&\ptext{Definition of $Skip_{\mathbf{AP}}$}\\
	&=Skip_{\mathbf{AP}}
\end{xflalign*}
\end{proof}
\end{proofs}
\end{theorem}

\begin{theorem}\label{theorem:RA1(SkipAP)}
\begin{statement}
$\mathbf{RA1} (Skip_{\mathbf{AP}}) = Skip_{\mathbf{RAD}}$
\end{statement}
\begin{proofs}
\begin{proof}
\begin{xflalign*}
	&\mathbf{RA1} (Skip_{\mathbf{AP}})
	&&\ptext{Definition of $Skip_{\mathbf{AP}}$}\\
	&=\mathbf{RA1} \circ \mathbf{AP} (true \vdash \circledIn{y}{ac'} (y.tr=s.tr \land \lnot y.wait))
	&&\ptext{\cref{theorem:RA1-o-RAPN(P):RA-o-A(Pff|-Ptf)}}\\
	&=\mathbf{RA} \circ \mathbf{A} \left(\begin{array}{l}
		\lnot (true \vdash \circledIn{y}{ac'} (y.tr=s.tr \land \lnot y.wait))^f_f
		\\ \vdash\\
		(true \vdash \circledIn{y}{ac'} (y.tr=s.tr \land \lnot y.wait))^t_f
	\end{array}\right)
	&&\ptext{\cref{lemma:design:(lnot-(P|-Q)ff|-(P|-Q)tf)}}\\
	&=\mathbf{RA} \circ \mathbf{A} (true_f \vdash (\circledIn{y}{ac'} (y.tr=s.tr \land \lnot y.wait))_f)
	&&\ptext{Substitution}\\
	&=\mathbf{RA} \circ \mathbf{A} (true \vdash \circledIn{y}{ac'} (y.tr=s.tr \land \lnot y.wait))
	&&\ptext{Definition of $Skip_{\mathbf{RAD}}$}\\
	&=Skip_{\mathbf{RAD}}
\end{xflalign*}
\end{proof}
\end{proofs}
\end{theorem}

\subsection{Sequential Composition}

\begin{theorem}\label{theorem:AP(P)-seqDac-AP(Q)}
\begin{statement}
Provided $P$ and $Q$ are $\mathbf{AP}$-healthy,
\begin{align*}
	&P \seqDac Q \\
	&=\\
	&\mathbf{AP} \left(
\right)
	&&\ptext{Sequential composition of designs}\\
	&=\left(%
\right)
	&&\ptext{Property of conditional and predicate calculus}\\
	&=\left(%
\right)
	&&\ptext{Property of conditional and predicate calculus}\\
	&=\left(%
\right)
	&&\ptext{Property of conditional}\\
	&=\left(%
\right)
	&&\ptext{Property of conditional}\\
	&=\left(%
\right)
	&&\ptext{Predicate calculus and~\cref{theorem:RA2(P-lor-Q):RA2(P)-lor-RA2(Q)}}\\
	&=\mathbf{RA3_{AP}} \left(%
\right)
	&&\ptext{Predicate calculus and~\cref{theorem:RA2(P-lor-Q):RA2(P)-lor-RA2(Q)}}\\
	&=\mathbf{RA3_{AP}} \circ \mathbf{RA2} \left(%
\right)
	&&\ptext{\cref{lemma:RA1-implies-ac'-neq-emptyset} and predicate calculus}\\
	&=\mathbf{RA3_{AP}} \circ \mathbf{RA2} \left(%
\right)
	&&\ptext{Property of conditional and predicate calculus}\\
	&=\mathbf{RA3_{AP}} \circ \mathbf{RA2} \left(%
\right)
	&&\ptext{Property of designs and predicate calculus}\\
	&=\mathbf{RA3_{AP}} \circ \mathbf{RA2} \left(%
\right)
	&&\ptext{Predicate calculus and property of conditional}\\
	&=\mathbf{RA3_{AP}} \circ \mathbf{RA2} \circ \mathbf{A0} \left(%
\right)
	&&\ptext{Predicate calculus and~\cref{theorem:RA2-o-RA1:RA1-o-RA2,theorem:RA2(P-lor-Q):RA2(P)-lor-RA2(Q)}}\\
	&=\mathbf{RA_{3N,2}} \circ \mathbf{A} \left(%
\right)
	&&\ptext{Predicate calculus and~\cref{theorem:RA2(P-lor-Q):RA2(P)-lor-RA2(Q)}}\\
	&=\mathbf{AP} \left(%
\right)
	&&\ptext{Predicate calculus: absorption law}\\
	&=\mathbf{AP} \left(%
\right)
	&&\ptext{Property of conditional}\\
	&=\mathbf{AP} \left(%
\right)
	&&\ptext{Definition of $\sqcup_{\mathbf{AP}}$ and conjunction of designs}\\
	&=\left(%
\right)
	&&\ptext{Property of conditional}\\
	&=\left(%
\right)
	&&\ptext{\cref{law:seqA-ac'-not-free} and predicate calculus}\\
	&=\mathbf{AP} \left(%
\right)
	&&\ptext{Property of conditional and predicate calculus}\\
	&=\mathbf{AP} \left(%
\right)
	&&\ptext{Predicate calculus: absorption law}\\
	&=\mathbf{AP} \left(%
\right)
	&&\ptext{\cref{lemma:RAPN(true|-Ptf)}}\\
	&=\mathbf{AP} (true \vdash \circledIn{y}{ac'} (y.tr=s.tr \land y.wait))
	&&\ptext{Definition of $Stop_{\mathbf{AP}}$}\\
	&=Stop_{\mathbf{AP}}
\end{xflalign*}
\end{proof}
\end{proofsbig}
\end{lemma}

\subsubsection{Linking}

\begin{theorem}\label{theorem:RA1(H1(P)-seqDac-H1(Q)):sqsubseteq:P-seqDac-Q}
\begin{statement}
Provided $P$ and $Q$ are reactive angelic designs,
\begin{align*}
	&\mathbf{RA1} (\mathbf{H1} (P) \seqDac \mathbf{H1} (Q)) \sqsubseteq P \seqDac Q
\end{align*}
\end{statement}
\begin{proofsbig}
\begin{proof}\checkt{alcc}
\begin{xflalign*}
	&\mathbf{RA1} (\mathbf{H1} (P) \seqDac \mathbf{H1} (Q))
	&&\ptext{Assumption: $P$ and $Q$ are $\mathbf{RAD}$-healthy and~\cref{lemma:H1(P)-RAP:RAPN(lnot-RA1(Pff)|-Ptf)}}\\
	&=\mathbf{RA1} (\mathbf{AP} (\lnot \mathbf{RA1} (P^f_f) \vdash P^t_f)
		\seqDac 
		\mathbf{AP} (\lnot \mathbf{RA1} (Q^f_f) \vdash Q^t_f))
	&&\ptext{\cref{theorem:RAPN(P)-seqRac-RAPN(Q),lemma:RAPN(lnot-Pff|-Ptf):RA3N-o-RA2-o-A(lnot-Pff|-Ptf)}}\\
	&=\mathbf{RA1} \circ \mathbf{AP} \left(
\right)
	&&\ptext{Predicate calculus and~\cref{lemma:RA1(P-lor-Q):RA1(P)-lor-RA1(Q)}}\\
	&=\mathbf{RA}\circ\mathbf{A} \left(%
\right)
	&&\ptext{$P$ is $\mathbf{RAD}$-healthy, ~\cref{theorem:PBMH(P)-RAP:P,lemma:RA1(P)-seqA-RA1(true):implies:RA1(P)-seqA-true} and weaken precondition}\\
	&\sqsubseteq \mathbf{RA}\circ\mathbf{A} \left(%
\right)
		\\ \vdash \\
		\mathbf{RA1} (P^t_f) \seqA s \in ac' \dres s.wait \rres \mathbf{RA2} \circ \mathbf{RA1} (\lnot Q^f_f \implies Q^t_f)
	\end{array}\right)
	&&\ptext{Assumption: $P$ and $Q$ are $\mathbf{RAD}$-healthy and~\cref{theorem:RAP:seqDac}}\\
	&=P \seqDac Q
\end{xflalign*}
\end{proof}
\end{proofsbig}
\end{theorem}

\begin{theorem}\label{theorem:H1(RA1(P)-seqDac-RA1(Q)):sqsupseteq:P-seqDac-Q}
\begin{statement}
Provided $P$ and $Q$ are $\mathbf{AP}$-healthy,
\begin{align*}
	&\mathbf{H1} (\mathbf{RA1} (P) \seqDac \mathbf{RA1} (Q)) \sqsupseteq P \seqDac Q
\end{align*}
\end{statement}
\begin{proofsbig}
\begin{proof}\checkt{alcc}
\begin{xflalign*}
	&\mathbf{H1} (\mathbf{RA1} (P) \seqDac \mathbf{RA1} (Q))
	&&\ptext{Assumption: $P$ and $Q$ are $\mathbf{AP}$-healthy}\\
	&=\mathbf{H1} (\mathbf{RA1} \circ \mathbf{AP} (P) \seqDac \mathbf{RA1} \circ \mathbf{AP} (Q))
	&&\ptext{\cref{theorem:RA1-o-RAPN(P):RA-o-A(Pff|-Ptf)}}\\
	&=\mathbf{H1} (\mathbf{RA} \circ \mathbf{A} (\lnot P^f_f \vdash P^t_f) \seqDac \mathbf{RA} \circ \mathbf{A} (\lnot Q^f_f \vdash Q^t_f))
	&&\ptext{\cref{theorem:RAP:seqDac}}\\
	&=\mathbf{H1} \circ \mathbf{RA} \circ \mathbf{A} \left(
\right)
	&&\ptext{$P$ is $\mathbf{AP}$-healthy and~\cref{theorem:PBMH-o-AP(P):AP(P),lemma:PBMH(P)-ow:PBMH(P-ow)}}\\
	&&\ptext{\cref{lemma:RA1(P)-seqA-RA1(true):implies:RA1(P)-seqA-true} and strengthen precondition}\\
	&\sqsupseteq \mathbf{AP} \left(%
\right)
	&&\ptext{$P$ is $\mathbf{AP}$-healthy and~\cref{theorem:PBMH-o-AP(P):AP(P),lemma:PBMH(P)-ow:PBMH(P-ow)}}\\
	&&\ptext{\cref{lemma:seqA:P-seqA(Q-land-R):implies:(P-seqA-Q)-land-(P-seqA-R)} and stregthen precondition}\\
	&\sqsupseteq \mathbf{AP} \left(%
\right)
	&&\ptext{Predicate calculus and~\cref{lemma:RA1(P-lor-Q):RA1(P)-lor-RA1(Q),theorem:RA2(P-lor-Q):RA2(P)-lor-RA2(Q)}}\\
	&=\mathbf{AP} \left(%
\right)
	&&\ptext{Property of conditional}\\
	&=\mathbf{AP} \left(%
\right)
	&&\ptext{\cref{lemma:seqA:P-seqA(Q-land-R):implies:(P-seqA-Q)-land-(P-seqA-R)} and weaken postcondition}\\
	&=\mathbf{AP} \left(%
\right)
	&&\ptext{\cref{theorem:RA2(P-lor-Q):RA2(P)-lor-RA2(Q)} and predicate calculus}\\
	&=\mathbf{AP} \left(%
\right)
		\\ \vdash \\
		\mathbf{RA1} (P^t_f) \seqA (s \in ac' \dres s.wait \rres \mathbf{RA2} (\lnot Q^f_f \implies \mathbf{RA1} (Q^t_f)))
	\end{array}\right)
	&&\ptext{Assumption: $P$ and $Q$ are $\mathbf{AP}$-healthy and~\cref{theorem:RAPN(P)-seqRac-RAPN(Q)}}\\
	&=P \seqDac Q
\end{xflalign*}
\end{proof}
\end{proofsbig}
\end{theorem}

\begin{theorem}\label{theorem:RA1(H1(P)-seqDac-H1(Q)):P-seqDac-Q}
\begin{statement}
Provided $P$ and $Q$ are reactive angelic designs and $\mathbf{ND_{RAD}}$-healthy,
\begin{align*}
	&\mathbf{RA1} (\mathbf{H1} (P) \seqDac \mathbf{H1} (Q)) = P \seqDac Q
\end{align*}
\end{statement}
\begin{proofs}
\begin{proof}
\begin{xflalign*}
	&\mathbf{RA1} (\mathbf{H1} (P) \seqDac \mathbf{H1} (Q))
	&&\ptext{Assumption: $P$ and $Q$ are $\mathbf{RAD}$-healthy}\\
	&=\mathbf{RA1} (\mathbf{H1} \circ \mathbf{RAD} (P) \seqDac \mathbf{H1} \circ \mathbf{RAD} (Q))
	&&\ptext{\cref{theorem:RAP:P-sqcup-Choice:(true|Pt)}}\\
	&=\mathbf{RA1} (\mathbf{H1} \circ \mathbf{RA} \circ \mathbf{A} (true \vdash P^t_f) \seqDac \mathbf{H1} \circ \mathbf{RA}\circ\mathbf{A} (true \vdash Q^t_f))
	&&\ptext{\cref{lemma:H1-o-RA-o-A(true|-Ptf):RAPN(true|-Ptf)}}\\
	&=\mathbf{RA1} (\mathbf{AP} (true \vdash P^t_f) \seqDac \mathbf{AP} (true \vdash Q^t_f))
	&&\ptext{\cref{theorem:RAPN(P)-seqRac-RAPN(Q)}}\\
	&=\mathbf{RA1} \circ \mathbf{AP} \left(
\right)
	&&\ptext{\cref{law:seqA-ac'-not-free} and predicate calculus}\\
	&=\mathbf{RA1} \circ \mathbf{AP} \left(%
\right)
	&&\ptext{Assumption: $P$ and $Q$ are $\mathbf{RAD}$-healthy and $\mathbf{ND_{RAD}}$ and~\cref{theorem:PBMH(P)-RAP:P,theorem:RAP:P-seqDac-Q-ND}}\\
	&=P \seqDac Q
\end{xflalign*}
\end{proof}
\end{proofs}
\end{theorem}

\subsubsection{Closure}

\begin{theorem}\label{theorem:NDAP(P-seqDac-Q)}
\begin{statement}
Provided $P$ and $Q$ are angelic processes and $\mathbf{ND_{AP}}$-healthy,
\begin{align*}
	&\mathbf{ND_{AP}} (P \seqDac Q) = P \seqDac Q
\end{align*}
\end{statement}
\begin{proofs}
\begin{proof}\checkt{alcc}
\begin{xflalign*}
	&P \seqDac Q
	&&\ptext{Assumption and~\cref{theorem:NDAP(P)-seqDac-NDAP(Q)}}\\
	&=\left(
\right)
	&&\ptext{Property of conditional}\\
	&=\left(%
\right)
	&&\ptext{Conjunction of designs}\\
	&=\left(%
\right)
	&&\ptext{Definition of $Choice_{\mathbf{AP}}$}\\
	&=Choice_{\mathbf{AP}} \land (P \seqDac Q)
	&&\ptext{Definition of $\sqcup_{\mathbf{AP}}$}\\
	&=Choice_{\mathbf{AP}} \sqcup_{\mathbf{AP}} (P \seqDac Q)
	&&\ptext{Definition of $\mathbf{ND_{AP}}$}\\
	&=\mathbf{ND_{AP}} (P \seqDac Q)
\end{xflalign*}
\end{proof}
\end{proofs}
\end{theorem}

\begin{theorem}\label{theorem:NDAP(P)-seqDac-NDAP(Q)}
\begin{statement}
Provided $P$ and $Q$ are angelic processes,
\begin{align*}
	&\mathbf{ND_{AP}} (P) \seqDac \mathbf{ND_{AP}} (Q)\\
	&=\\
	&\left(%
\right)
	&&\ptext{Property of conditional}\\
	&=\left(%
\right)
	&&\ptext{\cref{law:seqA-ac'-not-free} and predicate calculus}\\
	&=\left(%
\right)
	&&\ptext{Definition of $\seqA$, substitution and predicate calculus}\\
	&=\left(%
\right)
	&&\ptext{Predicate calculus and property of conditional}\\
	&=\left(%
\right) \land \mathbf{RA1} (true)
	&&\ptext{Predicate calculus}\\
	&\implies \mathbf{RA1} (true)
\end{xflalign*}
\end{proof}
\end{proofs}
\end{lemma}

\subsection{Prefixing}

\begin{theorem}\label{theorem:AP:a-then-P}
\begin{statement}
Provided $P$ is $\mathbf{AP}$-healthy,
\begin{align*}
	&a \circthen P\\
	&=\\
	&\mathbf{AP} \left(%
\right)
\end{align*}
\end{statement}
\begin{proofsbig}
\begin{proof}
\begin{xflalign*}
	&a \circthen_\mathbf{AP} P
	&&\ptext{Definition of prefixing}\\
	&=a \circthen_\mathbf{AP} Skip_\mathbf{AP} \seqDac P
	&&\ptext{Assumption: $P$ is $\mathbf{AP}$-healthy}\\
	&=a \circthen_\mathbf{AP} Skip_\mathbf{AP} \seqDac \mathbf{AP} (P)
	&&\ptext{Definition of $a \circthen_\mathbf{AP} Skip_\mathbf{AP}$}\\
	&=\mathbf{AP} \left(true 
				\vdash 
					\circledIn{y}{ac'} \left(%
\right)
	&&\ptext{\cref{law:pbmh:false,law:seqA-ac'-not-free} and predicate calculus}\\
	&=\mathbf{RA_{3N,2}} \circ \mathbf{A} \left(%
\right)
	&&\ptext{\cref{lemma:PBMH(circledIn):circledIn,lemma:RA1(circledIn)-ac'-not-free:circledIn(P-land-s.tr-le-y.tr)} and predicate calculus}\\
	&=\mathbf{RA_{3N,2}} \circ \mathbf{A} \left(%
\right)
	&&\ptext{Predicate calculus and property of conditional}\\
	&=\mathbf{RA_{3N,2}} \circ \mathbf{A} \left(%
\right)
	&&\ptext{Definition of design and predicate calculus}\\
	&=\mathbf{RA_{3N,2}} \circ \mathbf{A} \left(%
\right)
	\end{array}\right)
\end{xflalign*}
\end{proof}
\end{proofsbig}
\end{theorem}

\begin{lemma}\label{lemma:H1(a-then-Skip-RAD):a-then-Skip-AP}
\begin{statement}
$\mathbf{H1} (a \circthen_{\mathbf{RAD}} Skip_{\mathbf{RAD}}) = a \circthen_{\mathbf{AP}} Skip_{\mathbf{AP}}$
\end{statement}
\begin{proofs}
\begin{proof}\checkt{alcc}
\begin{xflalign*}
	&\mathbf{H1} (a \circthen_{\mathbf{RAD}} Skip_{\mathbf{RAD}})
	&&\ptext{Definition of $a \circthen_{\mathbf{RAD}} Skip_{\mathbf{RAD}}$}\\
	&=\mathbf{H1} \circ \mathbf{RA} \circ \mathbf{A} \left(true 
				\vdash 
					\circledIn{y}{ac'} \left(\begin{array}{l}(y.tr=s.tr \land a \notin y.ref)
						\\ \dres y.wait \rres \\
					(y.tr = s.tr \cat \lseq a \rseq)
					\end{array}\right)
		\right)
	&&\ptext{\cref{lemma:H1-o-RA-o-A(true|-Ptf):RAPN(true|-Ptf)}}\\
	&=\mathbf{AP} \left(true 
				\vdash 
					\circledIn{y}{ac'} \left(\begin{array}{l}(y.tr=s.tr \land a \notin y.ref)
						\\ \dres y.wait \rres \\
					(y.tr = s.tr \cat \lseq a \rseq)
					\end{array}\right)
		\right)
	&&\ptext{Definition of $a \circthen_{\mathbf{AP}} Skip_{\mathbf{AP}}$}\\
	&=a \circthen_{\mathbf{AP}} Skip_{\mathbf{AP}}
\end{xflalign*}
\end{proof}
\end{proofs}
\end{lemma}

\begin{lemma}\label{lemma:RA1(a-then-Skip-AP):a-then-Skip-RAD}
\begin{statement}
$\mathbf{RA1} (a \circthen_{\mathbf{AP}} Skip_{\mathbf{AP}}) = a \circthen_{\mathbf{RAD}} Skip_{\mathbf{RAD}}$
\end{statement}
\begin{proofs}
\begin{proof}
\begin{xflalign*}
	&\mathbf{RA1} (a \circthen_{\mathbf{AP}} Skip_{\mathbf{AP}})
	&&\ptext{Definition of $a \circthen_{\mathbf{AP}} Skip_{\mathbf{AP}}$}\\
	&=\mathbf{RA1} \circ \mathbf{AP} \left(true 
				\vdash 
					\circledIn{y}{ac'} \left(\begin{array}{l}(y.tr=s.tr \land a \notin y.ref)
						\\ \dres y.wait \rres \\
					(y.tr = s.tr \cat \lseq a \rseq)
					\end{array}\right)
		\right)
	&&\ptext{\cref{theorem:RA1-o-RAPN(P):RA-o-A(Pff|-Ptf)}}\\
	&=\mathbf{RA}\circ\mathbf{A} \left(\begin{array}{l}
		\left(true 
				\vdash 
					\circledIn{y}{ac'} \left(\begin{array}{l}(y.tr=s.tr \land a \notin y.ref)
						\\ \dres y.wait \rres \\
					(y.tr = s.tr \cat \lseq a \rseq)
					\end{array}\right)
		\right)^f_f
		\\ \vdash \\
		\left(true 
				\vdash 
					\circledIn{y}{ac'} \left(\begin{array}{l}(y.tr=s.tr \land a \notin y.ref)
						\\ \dres y.wait \rres \\
					(y.tr = s.tr \cat \lseq a \rseq)
					\end{array}\right)
		\right)^t_f
	\end{array}\right)
	&&\ptext{\cref{lemma:design:(lnot-(P|-Q)ff|-(P|-Q)tf)}}\\
	&=\mathbf{RA}\circ\mathbf{A} \left(\begin{array}{l}
		true_f
		\\ \vdash \\ 
		\left(\circledIn{y}{ac'} \left(\begin{array}{l}(y.tr=s.tr \land a \notin y.ref)
					\\ \dres y.wait \rres \\
				(y.tr = s.tr \cat \lseq a \rseq)
				\end{array}\right)
		\right)_f
	\end{array}\right)
	&&\ptext{Substitution}\\
	&=\mathbf{RA}\circ\mathbf{A} \left(\begin{array}{l}
		true_f
		\\ \vdash \\ 
		\circledIn{y}{ac'} \left(\begin{array}{l}(y.tr=s.tr \land a \notin y.ref)
					\\ \dres y.wait \rres \\
				(y.tr = s.tr \cat \lseq a \rseq)
				\end{array}\right)
	\end{array}\right)
	&&\ptext{Definition of $a \circthen_{\mathbf{RAD}} Skip_{\mathbf{RAD}}$}\\
	&=a \circthen_{\mathbf{RAD}} Skip_{\mathbf{RAD}}
\end{xflalign*}
\end{proof}
\end{proofs}
\end{lemma}

\begin{lemma}\label{lemma:AP:a-then-ChaosCSPAP}
\begin{statement}
\begin{align*}
	&a \circthen_{\mathbf{AP}} ChaosCSP_{\mathbf{AP}}\\
	&=\\
	&\mathbf{AP} (\lnot \circledIn{y}{ac'} (s.tr \cat \lseq a \rseq \le y.tr)
		\vdash 
		\circledIn{y}{ac'} (y.wait \land y.tr=s.tr \land a \notin y.ref))
\end{align*}
\end{statement}
\begin{proofsbig}
\begin{proof}
\begin{xflalign*}
	&a \circthen_{\mathbf{AP}} ChaosCSP_{\mathbf{AP}}
	&&\ptext{\cref{theorem:AP:a-then-P} and definition of $ChaosCSP_{\mathbf{AP}}$}\\
	&=\mathbf{AP} \left(
\right)
	&&\ptext{Transitivity of equality}\\
	&=\mathbf{AP} \left(%
\right)
	&&\ptext{Definition of design and predicate calculus}\\
	&=\mathbf{AP} \left(%
\right)
	&&\ptext{Substitution and predicate calculus}\\
	&=\mathbf{AP} \left(%
\right)
	\end{array}\right)
	&&\ptext{Definition of design and predicate calculus}\\
	&=\mathbf{AP} (false \vdash true)
	&&\ptext{Definition of $Chaos_{\mathbf{AP}}$}\\
	&=Chaos_{\mathbf{AP}}
\end{xflalign*}
\end{proof}
\end{proofs}
\end{lemma}

\begin{lemma}\label{lemma:PBMH-o-RA1(true):RA1(true)}
\begin{statement}
$ \mathbf{PBMH}\circ\mathbf{RA1} (true) = \mathbf{RA1} (true)$
\end{statement}
\begin{proofs}
\begin{proof}
\begin{xflalign*}
	&\mathbf{PBMH}\circ\mathbf{RA1} (true)
	&&\ptext{\cref{lemma:RA1(true)}}\\
	&=\mathbf{PBMH} (\exists z @ s.tr \le z.tr \land z \in ac')
	&&\ptext{Definition of $\mathbf{PBMH}$ (\cref{lemma:PBMH:alternative-1})}\\
	&=\exists ac_0 @ (\exists z @ s.tr \le z.tr \land z \in ac')[ac_0/ac'] \land ac_0\subseteq ac'
	&&\ptext{Substitution}\\
	&=\exists ac_0 @ \exists z @ s.tr \le z.tr \land z \in ac_0 \land ac_0\subseteq ac'
	&&\ptext{Predicate calculus}\\
	&=\exists z @ s.tr \le z.tr \land \exists ac_0 @ z \in ac_0 \land ac_0\subseteq ac'
	&&\ptext{Property of sets}\\
	&=\exists z @ s.tr \le z.tr \land z \in ac'
	&&\ptext{\cref{lemma:RA1(true)}}\\
	&=\mathbf{RA1} (true)
\end{xflalign*}
\end{proof}
\end{proofs}
\end{lemma}

\subsubsection{Linking}

\begin{theorem}\label{theorem:H1(a-thenRA-SkipRA)}
\begin{align*}
	&\mathbf{H1} (a \circthen_{\mathbf{RAD}} Skip_{\mathbf{RAD}})\\
	&=\\
	&\mathbf{AP} \left(true 
				\vdash 
					\circledIn{y}{ac'} \left(
\right)\right)
\end{xflalign*}
\end{proof}\end{proofs}
\end{theorem}

\chapter{Set Theory}

\section*{Lemmas}

\begin{lemma}\label{lemma:set-theory:ac0-subseteq-A2:ac0-subseteq-ac1-land-ac0-subseteq-A2}
\begin{align*}
	&ac_0 \subseteq \{ s | \{s\} = ac_1 \} = ac_0 \subseteq ac_1 \land ac_0 \subseteq \{ s | ac_1 \subseteq \{ s \} \}
\end{align*}
\begin{proof}
\begin{xflalign*}
	&ac_0 \subseteq \{ s | \{s\} = ac_1 \}
	&&\ptext{Definition of subset inclusion}\\
	&=\forall x \spot x \in ac_0 \implies x \in \{ s | \{s\} = ac_1 \}
	&&\ptext{Property of sets}\\
	&=\forall x \spot x \in ac_0 \implies \{x\} = ac_1
	&&\ptext{Property of sets}\\
	&=\forall x \spot x \in ac_0 \implies (\{x\} \subseteq ac_1 \land ac_1 \subseteq \{x\})
	&&\ptext{Property of sets}\\
	&=\forall x \spot x \in ac_0 \implies (x \in ac_1 \land ac_1 \subseteq \{x\})
	&&\ptext{Predicate calculus}\\
	&=(\forall x \spot x \in ac_0 \implies x \in ac_1)
		 \land 
	  (\forall x \spot x \in ac_0 \implies ac_1 \subseteq \{x\})
	&&\ptext{Property of sets}\\
	&=(\forall x \spot x \in ac_0 \implies x \in ac_1)
		 \land 
	  (\forall x \spot x \in ac_0 \implies x \in \{ s | ac_1 \subseteq \{s\} \}
	&&\ptext{Definition of subset inclusion}\\
	&=ac_0 \subseteq ac_1 \land ac_0 \subseteq \{ s | ac_1 \subseteq \{s\} \}
\end{xflalign*}
\end{proof}
\end{lemma}

\begin{lemma}\label{lemma:set-theory:ac0-subseteq-A2:ac1-subseteq-A2}
\begin{align*}
	&ac_0 \subseteq \{ s | ac_1 \subseteq \{s\} \} = ac_1 \subseteq \{ s | ac_0 \subseteq \{s\} \}
\end{align*}
\begin{proof}
\begin{xflalign*}
	&ac_0 \subseteq \{ s | ac_1 \subseteq \{s\} \}
	&&\ptext{Definition of subset inclusion}\\
	&=\forall x \spot x \in ac_0 \implies x \in \{ s | ac_1 \subseteq \{s\} \}
	&&\ptext{Property of sets}\\
	&=\forall x \spot x \in ac_0 \implies ac_1 \subseteq \{x\}
	&&\ptext{Definition of subset inclusion}\\
	&=\forall x \spot x \in ac_0 \implies (\forall y \spot y \in ac_1 \implies y \in \{x\})
	&&\ptext{Property of sets}\\
	&=\forall x \spot x \in ac_0 \implies (\forall y \spot y \in ac_1 \implies y = x)
	&&\ptext{Predicate calculus}\\
	&=\forall x, y \spot x \in ac_0 \implies (y \in ac_1 \implies y \in y = x)
	&&\ptext{Predicate calculus}\\
	&=\forall x, y \spot x \in ac_0 \land y \in ac_1 \implies y = x
	&&\ptext{Predicate calculus}\\
	&=\forall x, y \spot y \in ac_1 \implies (x \in ac_0 \implies y = x)
	&&\ptext{Predicate calculus}\\
	&=\forall y \spot y \in ac_1 \implies (\forall x \spot x \in ac_0 \implies y = x)
	&&\ptext{Property of sets}\\
	&=\forall y \spot y \in ac_1 \implies (\forall x \spot x \in ac_0 \implies x \in \{y\})
	&&\ptext{Definition of subset inclusion}\\
	&=\forall y \spot y \in ac_1 \implies ac_0 \subseteq \{y\}
	&&\ptext{Property of sets}\\
	&=\forall y \spot y \in ac_1 \implies y \in \{ s | ac_0 \subseteq \{s\} \}
	&&\ptext{Definition of subset inclusion}\\
	&=ac_1 \subseteq \{ s | ac_0 \subseteq \{s\} \} 
\end{xflalign*}
\end{proof}
\end{lemma}

\begin{lemma}\label{lemma:set-theory:ac0-subseteq-A2:ac0-emptyset-lor-ac0-subseteq-ac'}
\begin{align*}
	&ac_0 \subseteq \{ s | ac_0 \subseteq ac' \} = ac_0 = \emptyset \lor ac_0 \subseteq ac'
\end{align*}
\begin{proof}
\begin{xflalign*}
	&ac_0 \subseteq \{ s | ac_0 \subseteq ac' \}
	&&\ptext{Definition of subset inclusion}\\
	&=\forall x \spot x \in ac_0 \implies x \in \{ s | ac_0 \subseteq ac' \}
	&&\ptext{Property of sets}\\
	&=\forall x \spot x \in ac_0 \implies ac_0 \subseteq ac'
	&&\ptext{Predicate calculus}\\
	&=\forall x \spot (x \notin ac_0 \lor ac_0 \subseteq ac')
	&&\ptext{Predicate calculus}\\
	&=(\forall x \spot x \notin ac_0) \lor ac_0 \subseteq ac'
	&&\ptext{Property of sets}\\
	&=ac_0 = \emptyset \lor ac_0 \subseteq ac'
\end{xflalign*}
\end{proof}
\end{lemma}

\begin{lemma}\label{lemma:set-theory:exists-v-subset-1} Provided $v$ is not $s$,
\begin{align*}
	&\exists v \spot t \subseteq \{ s | Q \} \implies t \subseteq \{ s | \exists v \spot Q \}
\end{align*}
\begin{proof}
\begin{flalign*}
	&\exists v \spot t \subseteq \{ s | Q \}
	&&\ptext{Property of sets, $x$ is fresh}\\
	&=\exists v \spot (\forall x \spot x \in t \implies (\exists s \spot Q \land x=s))
	&&\ptext{Predicate calculus}\\
	&\implies \forall x \spot (\exists v \spot x \in t \implies (\exists s \spot Q \land x=s))
	&&\ptext{Predicate calculus}\\
	&=\forall x \spot x \in t \implies (\exists v \spot (\exists s \spot Q \land x=s))
	&&\ptext{Predicate calculus: $v$ is not $s$}\\
	&=\forall x \spot x \in t \implies (\exists s \spot (\exists v \spot Q) \land x=s))
	&&\ptext{Property of sets}\\
	&=\forall x \spot x \in t \implies x \in \{ s | \exists v \spot Q \}
	&&\ptext{Property of sets}\\
	&=t \subseteq \{ s | \exists v \spot Q \}
\end{flalign*}
\end{proof}
\end{lemma}

\begin{lemma}\label{lemma:set-theory:x-preceq-y-land-set} Provided $\preceq$ is transitive,
\begin{align*}
	&x \preceq y \land A \subseteq \{ z | y \preceq z \land x \preceq z \land e \} = x \preceq y \land A \subseteq \{ z | y \preceq z \land e \} 
\end{align*}
\begin{proof}
\begin{xflalign*}
	&x \preceq y \land A \subseteq \{ z | x \preceq z \land e \}
	&&\ptext{Property of sets}\\
	&=x \preceq y \land \forall z \spot z \in A \implies (y \preceq z \land x \preceq z \land e)
	&&\ptext{Predicate calculus}\\
	&=\forall z \spot x \preceq y \land (z \in A \implies (y \preceq z \land x \preceq z \land e))
	&&\ptext{Predicate calculus: $\preceq$ is transitivite}\\
	&=\forall z \spot x \preceq y \land (z \in A \implies (y \preceq z \land e))
	&&\ptext{Predicate calculus}\\
	&=x \preceq y \land \forall z \spot z \in A \implies (y \preceq z \land e)
	&&\ptext{Property of sets}\\
	&=x \preceq y \land A \subseteq \{ z | y \preceq z \land e \}
\end{xflalign*}
\end{proof}
\end{lemma}

\begin{lemma}\label{law:set-theory:transitivity-non-empty}
\begin{flalign*}
	&\exists B \spot B \neq \emptyset \land B \subseteq C \iff C \neq \emptyset&
\end{flalign*}
\begin{proof} (Implication) By contradiction: Suppose the consequent is false yet the antecedent is true. Then $C = \emptyset$.
\begin{flalign*}
	&\exists B \spot B \neq \emptyset \land B \subseteq C
	&&\ptext{Assumption: $C = \emptyset$}\\
	&=\exists B \spot B \neq \emptyset \land B \subseteq \emptyset
	&&\ptext{Property of subset inclusion}\\
	&=\exists B \spot B \neq \emptyset \land B = \emptyset
	&&\ptext{Propositional calculus}\\
	&=false
\end{flalign*}
\end{proof}
\begin{proof} (Reverse implication)
\begin{flalign*}
	&C \neq \emptyset \implies \exists B \spot B \neq \emptyset \land B \subseteq C
	&&\ptext{Choose $B = C$}\\
	&=C \neq \emptyset \implies C \neq \emptyset \land C \subset C
	&&\ptext{Reflexivity of subset inclusion}\\
	&=C \neq \emptyset \implies C \neq \emptyset
	&&\ptext{Propositional calculus}\\
	&=true
\end{flalign*}
\end{proof}
\end{lemma}
\begin{lemma}\label{law:aux:ac-sequence}
\begin{flalign*}
	& \exists ac_0 \spot s \in ac_0 \land ac_0 \subseteq ac' \iff s \in ac' &
\end{flalign*}
\begin{proof}(Implication)
\begin{flalign*}
	&\exists ac_0 \spot s \in ac_0 \land ac_0 \subseteq ac'
	&&\ptext{Definition of subset inclusion}\\
	&= \exists ac_0 \spot s \in ac_0 \land (\forall z \spot z \in ac_0 \implies z \in ac')
	&&\ptext{Assume $s \in ac_0$ then there is a case when $z = s$}\\
	&= \exists ac_0 \spot s \in ac_0 \land (\forall z \spot z \in ac_0 \implies z \in ac') \land (s \in ac_0 \implies s \in ac')
	&&\ptext{Assume $s \in ac_0$ and propositional calculus}\\
	&\implies s \in ac'
\end{flalign*}
\end{proof}
\begin{proof}(Reverse implication)
\begin{flalign*}
	&s \in ac' \implies (\exists ac_0 \spot s \in ac_0 \land ac_0 \subseteq ac')
	&\ptext{Choose $ac_0 = ac'$}\\
	&= (s \in ac') \implies (s \in ac' \land ac' \subseteq ac')
	&\ptext{Reflexivity of subset inclusion and propositional calculus}\\
	&= true
\end{flalign*}
\end{proof}
\end{lemma}

\begin{lemma}\label{lemma:set-theory:P-land-z-eq-y-Q:Q-y-z}
\begin{statement}
Provided that $P[y/z]$ holds,
\begin{align*}
	&\{ z | P \land z = y @ Q \} = \{ Q[y/z] \} 
\end{align*}
\end{statement}
\begin{proofs}
\begin{proof}
\begin{xflalign*}
	&\{ z | P \land z = y @ Q \}
	&&\ptext{Property of sets}\\
	&=\{ x | x \in \{ z | P \land z = y @ Q \} \}
	&&\ptext{Property of sets}\\
	&=\{ x | \exists z @ P \land z = y \land x = Q \}
	&&\ptext{One-point rule}\\
	&=\{ x | P[y/z] \land x = Q[y/z] \}
	&&\ptext{Assumption: $P[y/z]$ holds}\\
	&=\{ x | x = Q[y/z] \}
	&&\ptext{Property of sets}\\
	&=\{ Q[y/z] \}
\end{xflalign*}
\end{proof}
\end{proofs}
\end{lemma}

\chapter{Definitions: Alphabets and Healthiness Conditions}

\section{Binary Multirelations}
\theoremstatementref{def:BM}
\subsection{Healthiness Conditions}
\theoremstatementref{def:bm:bmh}

\section{Designs}
\subsection{Alphabet}
$ok, ok' : \{ true, false \}$
\subsection{Healthiness Conditions}
\theoremstatementwithnameref[$\mathbf{H1}$]{def:H1}
\theoremstatementwithnameref[$\mathbf{H2}$]{def:H2}
\theoremstatementwithnameref[$\mathbf{H3}$]{def:H3}
\theoremstatementwithnameref{def:H4}

\section{Reactive Processes and CSP}
\subsection{Alphabet}
\begin{align*}
	&ok, ok', wait, wait' : \{ true, false \}\\
	&tr, tr' : \seq Event\\
	&ref, ref' : \power Event
\end{align*}
\subsection{Healthiness Conditions}
\theoremstatementref{def:R1-R2-R3}
\theoremstatementref{def:CSP1-CSP2}

\section{Extended Binary Multirelations}
\theoremstatementref{def:BMbot}
\subsection{Healthiness Conditions}
\theoremstatementwithnameref{def:BMH0}
\theoremstatementwithnameref{def:BMH1}
\theoremstatementwithnameref{def:BMH2}
\theoremstatementwithnameref{def:BMH3}

\section{Angelic Designs}
\subsection{Alphabet}
\theoremstatementref{def:angelic-designs:alphabet}
\subsection{Healthiness Conditions}
\theoremstatementref{def:A0}
\theoremstatementref{def:PBMH}
\theoremstatementref{def:A1}
\theoremstatementref{def:A}
\theoremstatementref{def:A2}

\section{Reactive Angelic Designs}
\subsection{Alphabet}
\theoremstatementref{def:RAD:alphabet}
\subsection{Healthiness Conditions}
\theoremstatementref{def:RA1}
\theoremstatementref{def:RA2}
\theoremstatementref{def:RA3}
\theoremstatementref{def:RA}
\theoremstatementref{def:CSPA1}
\theoremstatementref{def:CSPA2}
\theoremstatementref{def:RAD}
\theoremstatementref{def:NDRAD}

\section{Angelic Processes}
\subsection{Healthiness Conditions}
\theoremstatementref{def:AP}
\theoremstatementref{def:RA3AP}
\theoremstatementref{def:NDAP}

\cleardoublepage
\phantomsection
\addcontentsline{toc}{chapter}{Glossary}
\chapter*{Glossary}
\markboth{GLOSSARY}{}
\begin{acronym}

\acro{ACP}{Algebra of Communicating Processes}
\acro{ASM}{Abstract State Machine}
\acro{BNF}{Backus-Naur Normal Form}
\acro{CCS}{Calculus of Concurrent Systems}
\acro{CSP}{Communicating Sequential Processes}
\acro{FCD}{Free Completely Distributive}
\acro{FDR}{Failures-Divergence Refinement}	
\acro{FSM}{Finite State Machines}
\acro{JCSP}{Java Communicating Sequential Processes}
\acro{LTS}{Labelled Transition System}
\acro{SOS}{Structured Operational Semantics}
\acro{UTP}{Unifying Theories of Programming}
\acro{VDM}{Vienna Development Method}
\acro{ZRC}{Z Refinement Calculus}

\end{acronym}

\cleardoublepage
\phantomsection
\addcontentsline{toc}{chapter}{Bibliography}
\markboth{BIBLIOGRAPHY}{}
\bibliography{reading}


\end{document}